This work focuses on improving Airborne Gamma-Ray Spectrometry (AGRS), a critical tool in responding to radiological emergencies such as severe nuclear accidents or nuclear weapon detonations. AGRS systems use large-scale gamma-ray spectrometers mounted on aircraft to rapidly identify and quantify radiation hazards over extensive areas, providing essential data to guide emergency response efforts.

Current AGRS calibration and data evaluation methods struggle to accurately quantify many radioactive materials expected in radiological emergencies, limiting the risk assessment and hence the effectiveness of emergency response actions. The goal of this work is to address these challenges by developing and validating a novel Monte Carlo based full spectrum modeling approach for the Swiss AGRS system.

The methodology features three key innovations: high-fidelity Monte Carlo simulations that combine an advanced scintillation physics model with detailed geometric representations of the aircraft and detector system; a surrogate model that replicates the Monte Carlo simulations with significantly reduced computation time; and a new data evaluation method that leverages the surrogate model within a Bayesian inversion framework, enabling the quantification of any gamma-ray emitting radionuclide involved in radiological emergencies. These models were rigorously validated under various laboratory and field conditions, demonstrating strong agreement between model predictions and radiation measurements.

In conclusion, this work represents a significant advancement in AGRS calibration and data evaluation. The developed methodology not only enhances detection accuracy and sensitivity, but also significantly expands the operational capabilities of AGRS systems during radiological incidents. These innovations lay the groundwork for improving AGRS practices in Switzerland and may contribute to establishing a new global standard for AGRS systems, ultimately supporting better-informed protective actions and reducing health risks during radiological emergencies.

TOWARDS MONTE CARLO BASED FULL SPECTRUM MODELING OF AIRBORNE GAMMA-RAY SPECTROMETRY SYSTEMS

David Breitenmoser

David Breitenmoser

# Towards Monte Carlo based Full Spectrum Modeling of Airborne Gamma-Ray Spectrometry Systems

1st Edition

# Towards Monte Carlo based Full Spectrum Modeling of Airborne Gamma-Ray Spectrometry Systems

1st Edition

David Breitenmoser

*Typesetting* This work was typeset with LaTeX by adopting the `MiKTeX` distribution in combination with the `memoir` class. Writing and compilation was performed with the `Visual Studio Code` (VSC) using the following extensions: `LaTeX Workshop`, `GitHub Copilot` and `Grammarly`. Figures were created using `MATLAB` together with `Adobe Illustrator`.

*Cover information* The cover displays the angular dispersion of the Swiss AGRS system's detector response in the full spectrum band at a photon energy of 1618 keV using the Cassini projection (linear color encoding with a maximum response of 1945 $cm^2$).

*Funding* The work presented herein was co-financed by the Swiss Federal Nuclear Safety Inspectorate (ENSI) (grant no. CTR00836 & CTR00491) and the Paul Scherrer Institute (PSI).

*To my family for their never-ending support*
*and to you and your curiosity that brought you here.*

# Preface

This monograph, *Towards Monte Carlo based Full Spectrum Modeling of Airborne Gamma-Ray Spectrometry Systems*, builds on my doctoral dissertation completed at the Paul Scherrer Institute and ETH Zurich [1]. Originally developed as part of my PhD research, it addresses critical gaps in airborne gamma-ray spectrometry (AGRS), focusing on the calibration and data evaluation methods that are essential to its effective operation. Leveraging recent advancements in high-performance computing and numerical statistics, this work introduces a novel Monte Carlo based full spectrum modeling approach, which enables more accurate quantification of terrestrial radionuclides and significantly expands the scope of detectable gamma-ray sources. I truly believe that, considering the current limitations in emergency response, there is a clear need for a new global standard in AGRS calibration and data evaluation. It is my hope that this monograph will contribute to that effort, ultimately supporting better-informed protective actions and reducing health risks during radiological emergencies worldwide.

The material is intended for a readership with a general background in physics and mathematics at the undergraduate level. Specifically, readers are expected to have a solid understanding of basic concepts in calculus, linear algebra, probability and statistics. Familiarity with nuclear and particle physics can be helpful but is not strictly necessary.

Given the lack of a comprehensive up-to-date monograph in AGRS, it was my aim to not only introduce the technical details of the new methodology but also to provide an in-depth review of the fundamentals and current status of AGRS systems. The writing is structured to guide the readers through the theoretical foundations, methodological developments and practical applications, making it accessible to both experts seeking to deepen their knowledge and novices new to the field. To complement the material, I have included extensive literature references throughout the text. These references serve as valuable resources, directing interested readers to more specific literature on particular topics and providing a broader context for the discussions presented.

A key motivation to adapt the dissertation into a monograph was driven by a desire to make this material more accessible and adaptable for ongoing use. Unlike the static nature of a dissertation, this format offers the flexibility to integrate new findings and improvements as the field evolves. Readers can expect updated methodologies, fresh research findings and added publications that that reflect ongoing advances in AGRS. This is also where your engagement, as the reader, becomes invaluable. Should you identify any errors, omissions or areas that would benefit from further coverage, your insights would be greatly valued. You can find my current contact information on my ORCID profile https://orcid.org/0000-0003-0339-6592, where I am happy to receive any suggestions to improve this work. In this way, I hope this monograph serves not only as a technical resource but also as a collaborative, evolving document that grows with the discipline.

Villigen, Fall 2024 David Breitenmoser

# Abstract

AIRBORNE Gamma-Ray Spectrometry (AGRS) systems are essential tools in response to radiological emergencies, including severe nuclear accidents, lost-source scenarios, or dirty bomb and nuclear weapon explosions. By mounting large-scale gamma-ray spectrometers on aircraft, these systems enable rapid identification and quantification of radiation hazards over extensive areas, providing crucial data for guiding emergency response efforts.

The accuracy and reliability of AGRS systems depend heavily on their calibration and data evaluation methodologies. Traditional methods, based on empirical calibration and simplified physics models, are currently only able to quantify a limited range of radionuclides, excluding most sources expected to be released in severe nuclear accidents or nuclear weapon explosions. These limitations can severely hinder effective response and accurate risk assessment during radiological emergencies. Improving the current calibration and data evaluation protocols in AGRS is therefore of immediate practical relevance and importance for public safety.

The main objective of this work is to overcome the limitations of the current calibration and data evaluation methods by developing and validating a novel Monte Carlo based full spectrum modeling approach for the Swiss AGRS system. The proposed methodology combines high-fidelity Monte Carlo simulations with detailed geometric models of the aircraft and gamma-ray spectrometer. To account for deficiencies obtained with traditional Monte Carlo simulations, a non-proportional scintillation physics model (NPSM) was calibrated and subsequently integrated into the Monte Carlo code, using a novel NPSM inference method, Compton edge probing. Another key innovation in this work is the development of a surrogate model, capable of accurately emulating the brute-force Monte Carlo simulations with a factor $\mathcal{O}(10^6)$ lower computation time.

The developed models were rigorously validated under laboratory and field conditions across an extensive set of gamma-ray sources and source-detector configurations. The results showed good agreement between model predictions and measurements, with a median relative deviation <10 % over the detector's entire spectral range, even for worst-case source-detector scenarios involving significant attenuation from the aircraft fuselage and fuel tanks.

In addition to advancing simulation techniques, this work introduces a novel spectral inversion methodology that leverages Bayesian inference in conjunction with the developed surrogate model. Validation with the Swiss AGRS system using a series of field measurements demonstrated excellent accuracy and precision in quantifying the activities of deployed radionuclide sources with relative deviations <2 % for a measurement time of 1 s. Furthermore, the methodology was effectively extended to probe the cosmic-ray flux and radon progeny concentrations in the lower atmosphere, showcasing its versatility and broader applicability.

In conclusion, the Monte Carlo based full spectrum modeling approach developed and validated in this work significantly advances AGRS calibration and data evaluation by enhancing detection accuracy and sensitivity, while at the same time expanding the operational range of AGRS systems for radiological incidents like severe nuclear accidents or nuclear weapon explosions. This work not only lays the groundwork for incorporating these methods into routine practices for the Swiss AGRS system, but also sets the foundation for establishing a new global standard for AGRS systems, ultimately supporting better-informed protective actions and reducing health risks during radiological emergencies.

# Kurzfassung

AERO-Gammaspektrometriesysteme (AGS) sind wichtige Hilfsmittel zur Bewältigung radiologischer Notfälle, einschliesslich schwerer Nuklearunfälle, verlorener radioaktiver Quellen sowie dem Einsatz von schmutzigen Bomben oder Kernwaffen. Durch den Einsatz von Gammaspektrometern an Bord von Luftfahrzeugen ermöglichen diese Systeme die schnelle Identifizierung und Quantifizierung von Strahlungsgefahren über grosse Gebiete und liefern damit essentielle Daten zur Planung und Koordination von Notfallmassnahmen.

Die Genauigkeit und Zuverlässigkeit von AGS-Systemen hängt stark von den eingesetzten Kalibrier- und Datenauswertungsmethoden ab. Herkömmliche Methoden, die auf empirischer Kalibrierung und vereinfachten physikalischen Modellen basieren, sind derzeit nur in der Lage, eine begrenzte Anzahl von Radionukliden zu quantifizieren. Viele Radionuklide, welche bei schweren Nuklearunfällen oder Kernwaffenexplosionen freigesetzt werden, können mit den herkömmlichen Methoden nicht ausgewertet werden. Diese Einschränkungen können eine wirksame Reaktion und eine genaue Risikobewertung bei radiologischen Notfällen erheblich beeinträchtigen. Die Verbesserung der aktuellen Kalibrier- und Datenauswertungsmethoden in AGS ist daher von unmittelbarer praktischer Relevanz und Bedeutung für den Bevölkerungsschutz.

Das Hauptziel dieser Arbeit besteht darin, die Einschränkungen der aktuellen Kalibrier- und Datenauswertungsmethoden zu überwinden, indem ein neuartiger, auf Monte Carlo Simulationen basierender Vollspektrum-Modellansatz für das Schweizer AGS-System entwickelt und validiert wird. Die vorgeschlagene Methodik kombiniert präzise Monte Carlo Simulationen mit detaillierten geometrischen Modellen des Luftfahrzeugs und des Gam-maspektrometers. Um die bei herkömmlichen Monte Carlo Simulationen auftretenden Simulationsfehler zu beheben, wurde ein nichtproportionales Szintillationsmodell (NPSM) kalibriert und anschliessend in den Monte Carlo Code integriert, wobei eine neuartige NPSM-Inferenzmethode, die Compton-Kantensondierung, verwendet wurde. Eine weitere wichtige Innovation in dieser Arbeit ist die Entwicklung eines Surrogatmodells, das in der Lage ist, die Monte Carlo Simulationen mit einer um den Faktor $\mathcal{O}(10^6)$ geringeren Rechenzeit mit grosser Genauigkeit zu emulieren.

Die entwickelten Modelle wurden umfassend mit einer Vielzahl von Gammastrahlenquellen und Quelle-Detektor-Konfigurationen unter Labor- und Feldbedingungen validiert. Die Ergebnisse dieser Validierung zeigten eine gute Übereinstimmung zwischen den Modellvorhersagen und den Messungen mit einer mittleren relativen Abweichung <10 % über den gesamten Spektralbereich des Detektors. Dies gilt selbst für Worst-Case-Szenarien, bei welcher die Gammastrahlung durch den Flugzeugrumpf und die Treibstofftanks erheblich abgeschwächt wurde.

Neben der Weiterentwicklung von Simulationstechniken wird in dieser Arbeit auch eine neuartige Spektralinversionsmethode vorgestellt, welche auf Bayes'scher Inferenz in Verbindung mit dem entwickelten Surrogatmodell basiert. Die Validierung mit dem Schweizer AGS-System zeigte bei Feldmessungen eine ausgezeichnete Genauigkeit und Präzision bei der Quantifizierung der Aktivitäten der eingesetzten Radionuklidquellen mit relativen Abweichungen <2 % bei einer Messzeit von 1 s. Zudem wurde die Methode erfolgreich angepasst, um den kosmischen Strahlungsfluss und die Aktivitätskonzentration der Radonzerfallsprodukte in der unteren Erdatmosphäre zu untersuchen, was ihre Vielseitigkeit und breite Anwendbarkeit verdeutlicht.

Zusammenfassend hat der in dieser Arbeit entwickelte und validierte Monte Carlo basierte Vollspektrum-Modellansatz die Kalibrierung und Datenauswertung von AGS erheblich verbessert, indem er die Genauigkeit und Sensitivität erhöht und gleichzeitig den Anwendungsbereich der AGS-Systeme für radiologische Vorfälle, wie beispielsweise schwere Nuklearunfälle oder Kernwaffeneinsätze, erweitert hat. Diese Arbeit schafft nicht nur die Grundlage für die Integration dieser Methoden in die routinemässige Kalibrier- und Auswertepraxis des Schweizer AGS-Systems, sondern legt auch den Grundstein für die Einführung eines neuen Standards für AGS-Systeme weltweit, welcher letztlich zu besser abgestimmten Schutzmassnahmen und damit zur Verbesserung des Bevölkerungschutzes bei radiologischen Notfällen beiträgt.

# Contents

## Back Matter

> *"Begin at the beginning," the King said, gravely, "and go on till you come to the end; then stop."*
>
> — *Lewis Carroll, Alice in Wonderland*

# Chapter 1
# Introduction

### Contents

In this introductory chapter, I will outline the motivation, scope, and contributions of this work. The chapter begins with a discussion of the research motivation, followed by a description of the work's scope and unique contributions. Finally, the chapter provides an overview of the work's structure, offering a guide to assist readers in navigating the document.

## 1.1 Motivation

On December 2, 1942, deep below the stands of a football stadium at the University of Chicago, a team of physicists led by Enrico Fermi achieved a groundbreaking feat: the first controlled, self-sustaining nuclear chain reaction. This historic event marked the beginning of the nuclear age and laid the foundation for the development of nuclear technologies that have since become an integral part of our modern world. Today, nuclear technologies are used for a wide range of applications, ranging from fundamental research, energy production, medical diagnostics and cancer therapy to industrial quality control and product sterilization.

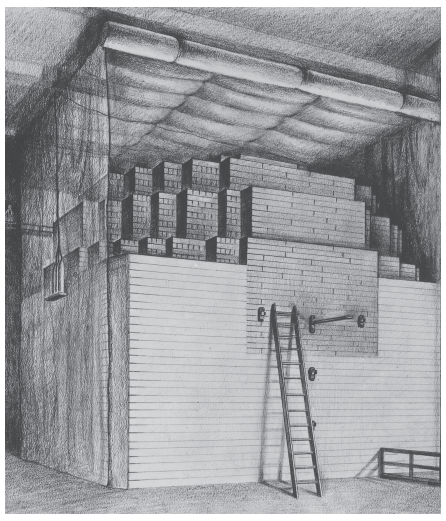

Chicago Pile-I
the world's first artificial nuclear reactor
Ⓢ Melvin A. Miller

However, with every new technology, there are also new risks and challenges. The nuclear weapon explosions in Hiroshima and Nagasaki or the severe nuclear accidents in Chernobyl and Fukushima have shown us the dark side of nuclear technologies by releasing and subsequently contaminating the environment with massive amounts of man-made radioactive material [2–4]. These materials can cause severe damage to living organisms, including humans. As a result, decision makers have to select appropriate measures to protect the afflicted population. These measures, like for example evacuation of affected regions, are accompanied by massive consequences for the people involved [5]. Therefore, the protective actions have to be precisely aligned with the radiological risk to effectively minimize harm to the population. To reach that goal, decision makers need a reliable overview of the radiological situation and they need it fast.

The method of choice for the rapid assessment of the radiological situation over a large area is airborne gamma-ray spectrometry (AGRS). This technique allows measuring the energy and intensity of high-energy photons using a gamma-ray spectrometer[1] mounted in a manned[2] aircraft. The raw data is collected in a series of spectra, data arrays that contain the number of registered photons as a function of their energy.

By analyzing these spectra, there are three fundamental tasks that AGRS systems aim to achieve:

1. **Localization** Find the location of elevated radiation levels.
2. **Identification** Determine the type of radioactive material causing the elevated radiation level.
3. **Quantification** Determine the amount of radioactive material causing the elevated radiation level.

1 *Combination from the Latin root "spectrum", which means "appearance" or "image", and the Ancient Greek root μέτρο (métron), which translates to "to measure".*

2 *AGRS performed by unmanned aerial vehicle (UAV) platforms is not the focus of the present work. That said, a brief introduction into the field of UAV-based AGRS is provided later in Section 5.1.3.*

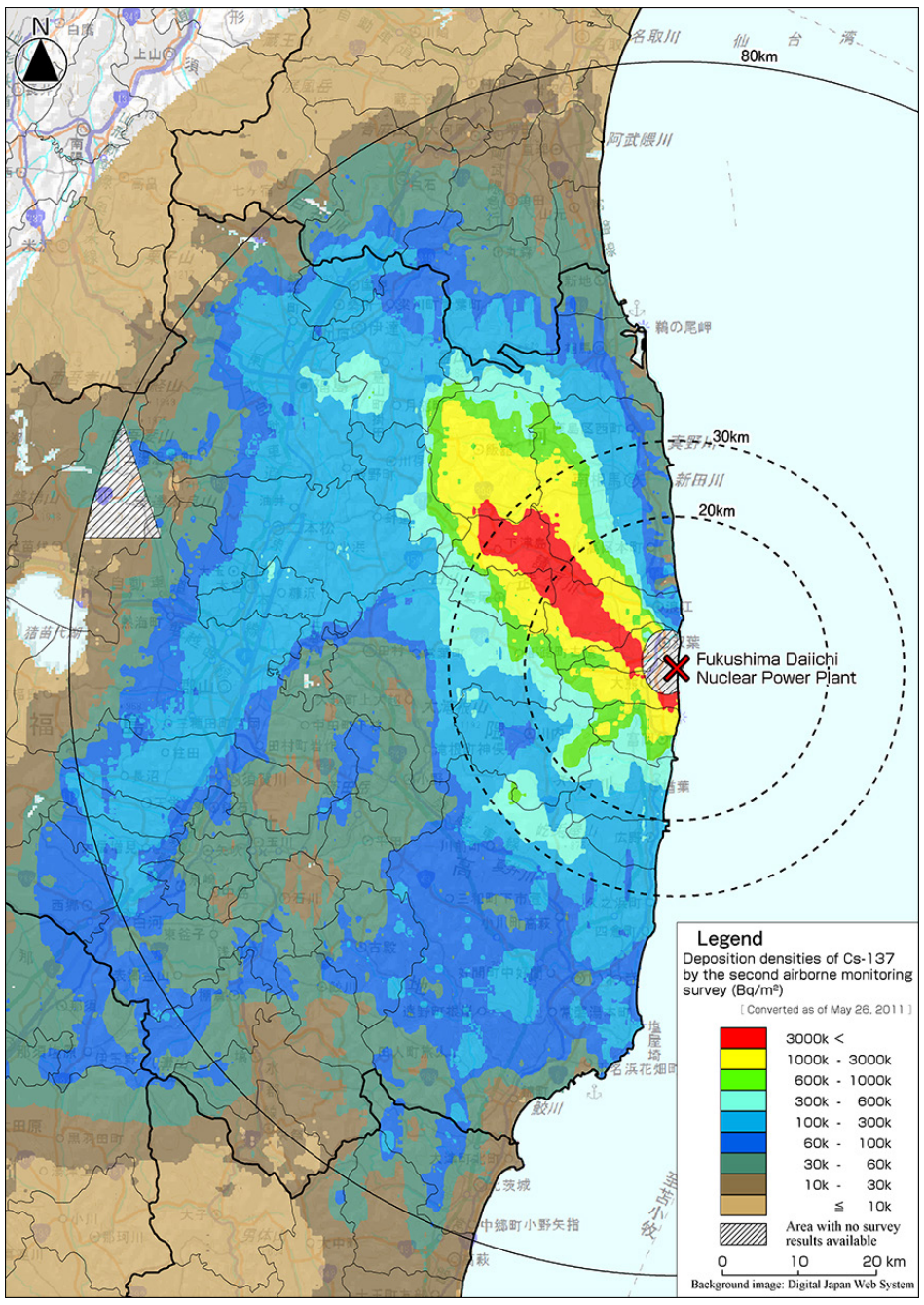

**Figure 1.1** This radiological map was generated in the aftermath of the Fukushima nuclear accident in 2011 by AGRS using a total of 13 survey flights performed by the DOE/NNSA in collaboration with the MEXT of Japan [6, 7]. It shows the deposition of the radionuclide $^{137}_{55}$Cs on the ground in units of $\mathrm{Bq\,m^{-2}}$ in a large area around the Fukushima Daiichi nuclear power plant in Japan. The data was decay corrected to the last day of the survey on 2011-03-26.

The product of these tasks is integrated into radiological maps, which provide decision makers with a comprehensive overview of the radiological situation and form the basis for the selection of appropriate protective actions. Fig. 1.1 shows an example of such a radiological map which was generated in the aftermath of the Fukushima nuclear accident in 2011 by the U.S. Department of Energy/National Nuclear Security Administration (DOE/NNSA) in collaboration with the Ministry of Education, Culture, Sports, Science and Technology (MEXT) of Japan [6, 7].

Given their importance for the protection of the population in case of radiological emergencies, AGRS systems are part of the emergency response organizations of many countries, including Switzerland. To fulfill the three main tasks of AGRS discussed above, these systems require dedicated calibration and data evaluation procedures to ensure that the radiological data they provide are accurate and reliable. Current calibration and data evaluation methods are standardized by the International Atomic Energy Agency (IAEA) in two technical reports [8, 9] and are based on simplified analytical physics models combined with empirical calibration procedures. In Switzerland, these methods were adapted and extended to the specific needs of the Swiss emergency response organization in two previous PhD studies [10, 11].

Despite their successful application in past radiological incidents, it is known that these methods have some significant drawbacks which currently limit the accuracy and reliability of the radiological data provided by AGRS systems. Specifically, there are the following two fundamental limitations in the current methodology proposed by the IAEA:

I. **Calibration** Given the empirical character, only a very limited number of sources can be calibrated using the current methods.[3] Furthermore, some of the calibration methods require a considerable amount of calibration time of up to 3 d to 5 d to cover all relevant source-detector configurations [12, 13], while others suffer from high costs in the preparation and/or decommissioning of the calibration sources [12, 14, 15]. Last but not least, all these methods are subject to significant systematic uncertainties, primarily originating from varying radiation backgrounds and required analytical corrections.

II. **Physics Modeling** The currently adopted analytical physics models deliver reliable results only for photon energies $\gtrsim$400 keV [8–10]. Furthermore, current data evaluation protocols focus on a small part of the spectrum provided by the gamma-ray spectrometer, which in turn significantly reduces the sensitivity of the AGRS systems [16, 17].

3 *In Switzerland, three natural and two anthropogenic sources are currently calibrated.*

In the context of severe nuclear accidents and nuclear weapon explosions, a large variety of different radionuclides is typically released, with a significant amount of them emitting photons with energies <400 keV [2–4]. It is therefore evident that the current calibration and data evaluation methods employed in Switzerland and abroad are not sufficient for such scenarios. Furthermore, considering the threat of nuclear terrorism and the recent conflicts in Ukraine and Israel — which involve parties possessing nuclear weapons and/or vulnerable nuclear infrastructures — the risk of nuclear incidents has regrettably escalated in recent years. Improving the current calibration and data evaluation protocols in AGRS to overcome the discussed limitations in the capabilities of AGRS systems is therefore of immediate practical relevance and importance for public safety.

To overcome the limitations in calibration and modeling discussed above, previous studies have proposed to exchange the empirical calibration methods and analytical physics models with a numerical approach, specifically high-fidelity Monte Carlo simulations [18, 19]. This approach allows for the comprehensive simulation and thus calibration of gamma-ray spectrometers across all source types and geometries [20–23]. However, practical implementations have struggled with significant challenges, including model complexity and high computational costs [18, 19, 24–28]. Simplifications in previous studies have led to substantial errors, especially at low energies [18, 27].[4] As a result, Monte Carlo-based calibration has not yet achieved the accuracy and applicability needed to replace traditional empirical-based methods for AGRS systems. Addressing these issues is essential for realizing the full potential of this approach.

4 *A more detailed literature review is provided in Section 5.5.2.*

## 1.2 Scope & Contribution

In response to the apparent need for improvement in the current calibration and data evaluation methods for AGRS systems, the Expert Group for Aeroradiometrics (FAR), in collaboration with the Swiss Federal Nuclear Safety Inspectorate (ENSI) and the Paul Scherrer Institute (PSI) in Switzerland, has initiated a project aimed at refining the existing methods specifically for the AGRS system currently in use in Switzerland. As this task was deemed too large for a single PhD project, it was divided into two steps: (1) the development of a numerical calibration model using state-of-the-art Monte-Carlo radiation transport codes and (2) integrating this model into the standard calibration and data evaluation protocols of the Swiss AGRS system. This work reports on the findings and advancements achieved in the first phase of the project.

To address the limitations of previous Monte Carlo-based calibration approaches for AGRS systems, this work was organized around two clearly defined objectives:

1. **AGRS Monte Carlo** Development and validation of a high-fidelity Monte Carlo model of the Swiss AGRS system including the aircraft and advanced physics models to accurately simulate the full spectrum response of the gamma-ray spectrometer for arbitrary source-detector configurations.
2. **Generalization** Derivation and validation of a surrogate model to emulate the developed Monte Carlo model with increased computational efficiency.

The first goal aims to reduce the systematic errors in Monte Carlo simulations that previous studies have identified. This is achieved by integrating a comprehensive model of the entire aircraft system and employing advanced physics models specifically tailored to the AGRS gamma-ray spectrometer's unique characteristics. These improvements will allow for the accurate simulation of the full spectrum response of the gamma-ray spectrometer for arbitrary source-detector configurations. To ensure the accuracy and precision of the developed Monte Carlo model, thorough validation is required. For this purpose, an extensive set of validation measurements was performed under laboratory and field conditions.

The second objective aims to overcome the main limitation of the Monte Carlo based calibration approach, the long computation times. For that purpose, I adopted a surrogate modeling approach that was originally developed for astrophysics and planetary science applications. The developed surrogate model allows for the emulation of the complex Monte Carlo model with model evaluation times of $\mathcal{O}(1)$ s, while at the same time maintaining a high degree of accuracy and precision. This is demonstrated by a series of verification studies using the validated Monte Carlo model from the first objective.

Given that, currently, no data evaluation framework exists to incorporate the newly developed surrogate models for AGRS applications, it became clear that at least a proof-of-concept study was required to demonstrate the feasibility and potential of the Monte Carlo based full spectrum modeling approach specifically for data evaluation purposes. This proof-of-concept study was limited to the quantification task which constitutes the third objective of this work:

3. **Quantification** Development and validation of a data evaluation algorithm to accurately quantify arbitrarily complex radioactive sources using a Monte Carlo based full spectrum modeling approach.

By completing the proposed goals, the present work will provide for the first time a comprehensive full spectrum numerical modeling framework to calibrate an AGRS system for an arbitrary number of sources and source-detector configurations, thereby overcoming the main limitations of the current empirical calibration methods proposed by the IAEA [8, 9].

Given that the AGRS systems currently in use feature all the same basic components and operate with similar flight characteristics, the methods presented in this work will be adaptable to other AGRS systems in use abroad. Consequently, the developed models will not only improve the capabilities of the Swiss AGRS system but will also provide a solid foundation for the establishment of a Monte Carlo based full spectrum calibration and data evaluation approach for AGRS systems worldwide that may replace the current methodology standardized by the IAEA in the future. I will discuss the potential and limitations of the developed models on the field of AGRS in Chapter 11.

**▶ How Contributions are Marked**

Active voice ("I") is consistently used throughout the document to clearly indicate the author's contributions, with the reader included in the discussion through the use of the pronoun "we" where appropriate. Contributions from other authors are explicitly acknowledged in the text. Supervisory and supporting contributions are recognized in the Acknowledgements.

Part of the work has already been published elsewhere. A concise List of Publications, which includes also scripts and codes developed in the context of this work, is provided at the end of this document. Specific publications are clearly referenced in the respective chapters (Chapters 6, 7 and 9).

## 1.3 Outline

Given the novelty of the proposed Monte Carlo based full spectrum modeling approach for AGRS systems, no comprehensive work exists currently in the literature that covers all the necessary background information in the context of AGRS. As a result, to ensure the accessibility of the present work to a broad audience and to provide a solid foundation for the continuation of the work presented in this monograph in the second phase of the project, the work is written in a textbook style with a high degree of detail.

An overview of the document and its general structure is now given. The document is divided into four main parts:

| | | |
|---|---|---|
| Part I | Chapter 2 | Gamma-Ray Sources |
| | Chapter 3 | Interaction with Matter |
| | Chapter 4 | Gamma-Ray Spectrometry |
| | Chapter 5 | Airborne Gamma-Ray Spectrometry |
| Part II | Chapter 6 | Proportional Scintillation Monte Carlo |
| | Chapter 7 | Non-Proportional Scintillation Monte Carlo |
| Part III | Chapter 8 | AGRS Monte Carlo |
| | Chapter 9 | Detector Response Model |
| Part IV | Chapter 10 | Full Spectrum Bayesian Inversion |

In Part I, I will review the physics background and fundamental concepts needed to fulfill the goals of the present work. The first three chapters of this part will provide a comprehensive introduction to the fundamental physics concepts of AGRS starting with the emission characteristics of the high-energy photon sources relevant in AGRS in Chapter 2, continuing with the interaction of the emitted photons with the environment and the aircraft structure in Chapter 3 and ending with the complex detection processes in the adopted gamma-ray spectrometer in AGRS in Chapter 4. Following these foundational chapters, I present a detailed review of the current state of AGRS worldwide in Chapter 5 with a special focus on the calibration and data evaluation methods currently in use. This includes also a detailed overview of the Swiss AGRS system as well as a thorough literature review on the past work in the field of Monte Carlo simulations in AGRS. The part will conclude with a refined problem statement and research questions that will be addressed in the subsequent parts of the document using the concepts and terminology introduced in the first part.

The three following parts contain the main contributions of the present work. To fulfill the goals stated in Section 1.2, I followed a bottom-up modeling approach. In Part II, I started by developing a high-fidelity Monte Carlo model of the Swiss AGRS system excluding the aircraft. This allowed me to validate the adopted physics models under controlled laboratory conditions with high-counting statistics. The validation measurements in Chapter 6 revealed that the current physics models commonly adopted to simulate the full spectrum response of gamma-ray spectrometers are not sufficient to accurately reproduce the detector signals of the Swiss AGRS system. As a result, a new set of advanced physics models tailored to the specific characteristics of the Swiss AGRS system was developed and validated in Chapter 7.

In Part III, following the bottom-up modeling approach, the validated Monte Carlo model from Part II is extended to the entire Swiss AGRS system including the aircraft and the environment. In Chapter 8, I will validate the developed Monte Carlo model using a series of field measurements performed with the Swiss AGRS system. In Chapter 9, using the validated Monte Carlo model from the previous chapter, a surrogate model is derived and validated to emulate the developed Monte Carlo model for arbitrary source-detector configurations, which will allow for the implementation of the developed models in a practical calibration procedure.

In Part IV, I will address the third goal of the present work by developing and validating a novel data evaluation approach using the surrogate model derived in Chapter 9.

The document will end with a conclusion in Chapter 11, where I will discuss the potential and limitations of the developed models on the field of AGRS and provide an outlook on future research directions, particularly the implementation of the developed methods for the routine calibration and data evaluation of the Swiss AGRS system.

For the reader's convenience, extensive supporting material is attached in the Appendices A–C, which provides additional information on the methods and techniques developed in this work. Last but not least, the Back Matter contains supporting material to facilitate the accessibility of the present work to the reader, including an extensive Bibliography, List of Figures, List of Tables, an Index as well as Glossaries containing List of Symbols, List of Abbreviations and which details also the formal rules and terminological conventions used in this work. A last note for readers who use the electronic version of this document: all references and cross-references are hyperlinked to facilitate the navigation through the document. If you should find yourself lost, remember that as an Adobe Acrobat user, you can always move back to a previously opened location in the document by the keyboard shortcut `<Alt> + <Left Arrow>` on Windows/UNIX systems and `<Command> + <Left Arrow>` on macOS systems.

# Part I

# Fundamentals

*"Nothing in life is to be feared, it is only to be understood. Now is the time to understand more so that we may fear less."*

*— Marie Curie*

# Chapter 2
# Gamma-Ray Sources

**Contents**

In this chapter, I provide an overview of gamma-ray sources in the Earth's environment, with a focus on radionuclides and cosmic rays. The purpose of this chapter is not to provide a detailed discussion of all possibilities how gamma rays might get created but rather to give a general idea of the diversity of gamma-ray sources and their associated energies with a special focus on those sources relevant for airborne gamma-ray spectrometry (AGRS). The concepts and data introduced here will be used later to interpret measurement and simulation results and to provide the basis for data evaluation.

## 2.1 Radionuclides

### 2.1.1 Radioactivity and High-Energy Photons

Sometimes, scientific results are the product of an unplanned discovery. This was also the case in early 1896, when the French physicist Henri Becquerel[1] was investigating the phosphorescence of uranium salts by exposing them to sunlight. However, on February 26th, because of bad weather, he had to postpone one of his experiments and stored the prepared uranium salts together with the photographic plates in a drawer. When he later developed the plates, he discovered that they were exposed, although the samples were never placed in direct sunlight. After conducting further experiments, he concluded that the uranium salts emitted radiation themselves. This was the discovery of radioactivity, i.e. the decay of unstable nuclides and subsequent emission of ionizing radiation[2] [29]. We call these unstable nuclides also "radionuclides". In AGRS, we are mainly interested in the high-energy electromagnetic radiation emitted by these radionuclides. But before we come to this point, let us have first a brief look at the decay of these radionuclides.

Today, we know several modes how radionuclides decay. The three most common decay modes are the alpha decay ($\alpha$), the beta decay ($\beta^-, \beta^+$) and electron capture (EC). These decays are commonly written as $\mathrm{X} \xrightarrow{m} \mathrm{Y} + \sum_i \mathrm{p}_i$ with X being the parent radionuclide, Y the product nuclide[3], $m$ the decay mode and $\mathrm{p}_i$ the secondary particles emitted during the decay [30, 31]. Radionuclides are defined by their

1 *Antoine Henri Becquerel (*1852, †1908), a French physicist, who discovered radioactivity in 1896. For his discovery, he was awarded the Nobel Prize in Physics in 1903 together with Pierre and Marie Curie, who showed that not only uranium salts but also other elements such as thorium are radioactive.*

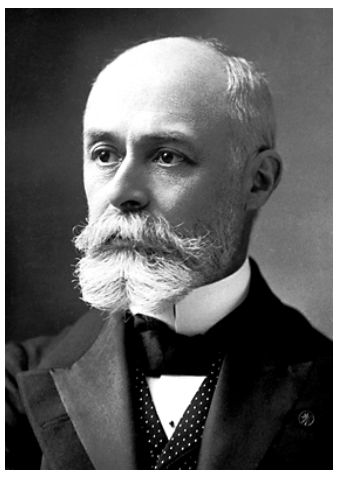

Henri Becquerel
Nobel foundation

2 *Radiation, which has enough energy to ionize an atom or molecule by removing bound electrons from them.*

3 *We call this product nuclide also a progeny or radiogenic nuclide.*

atomic number ($Z$) and mass number ($A$), typically denoted with the mass number as a superscript and the atomic number as a subscript in the notation ${}^{A}_{Z}\mathrm{X}$ [30, 31].

In alpha decay, the radionuclide emits an alpha particle, which is a helium nucleus consisting of two protons and two neutrons, thereby forming a new nuclide with the mass number $A$ reduced by four and the atomic number $Z$ reduced by two [30]:

$$ {}^{A}_{Z}\mathrm{X} \xrightarrow{\alpha} {}^{A-4}_{Z-2}\mathrm{Y}^{*} + {}^{4}_{2}\mathrm{He} \tag{2.1} $$

In beta decay, the radionuclide emits a beta particle, which is an electron or a positron. In the case of an electron ($e^-$), we call this decay mode also beta-minus decay ($\beta^-$) [30]:

$$ {}^{A}_{Z}\mathrm{X} \xrightarrow{\beta^-} {}_{Z+1}^{\phantom{Z}A}\mathrm{Y}^{*} + e^- + \overline{\nu}_e \tag{2.2} $$

and in case of a positron ($e^+$) we call it a beta-plus decay ($\beta^+$) [31]:

$$ {}^{A}_{Z}\mathrm{X} \xrightarrow{\beta^+} {}_{Z-1}^{\phantom{Z}A}\mathrm{Y}^{*} + e^+ + \nu_e \tag{2.3} $$

where $\nu_e$ and $\overline{\nu}_e$ are the electron neutrino and electron antineutrino, respectively. For radionuclides exhibiting a $\beta^+$ decay mode, the EC decay mode is frequently observed as well. In this decay mode, the radionuclide's nucleus captures an electron, predominantly from the inner electron shells, and emits an electron neutrino [31]:

$$ {}^{A}_{Z}\mathrm{X} + e^- \xrightarrow{\mathrm{EC}} {}_{Z-1}^{\phantom{Z}A}\mathrm{Y}^{*} + \nu_e \tag{2.4} $$

Ok, with that let us come back to the high-energy electromagnetic radiation, in which we are mainly interested in AGRS. In most cases, the decay modes discussed above leave the nucleus of the newly formed nuclide Y in an excited state Y*. De-excitation to the ground state may occur by the emission of a variable number $n$ of high-energy photons which we call gamma rays ($\gamma$) [30]:

$$ {}^{A}_{Z}\mathrm{Y}^{*} \xrightarrow{\gamma} {}^{A}_{Z}\mathrm{Y} + n\gamma \tag{2.5} $$

So what are these photons? Photons are massless and chargeless quanta of electromagnetic radiation, i.e. light. And here I have to introduce already one of the probably most unintuitive concepts, which physics has to offer: the wave-particle dualism of light. This concept is a cornerstone of quantum mechanics and states that light can behave both as a wave and as a particle, which we call in this

4 *Derived from the Greek word for light, φως.*

5 *Max Karl Ernst Ludwig Planck (*1858, †1947), a German theoretical physicist and father of quantum theory. He is best known for the theoretical prediction of the quantum nature of light introducing the famous quantum of action constant, which became later known as Planck's constant. For his discovery, he received the Nobel Prize in Physics in 1918.*

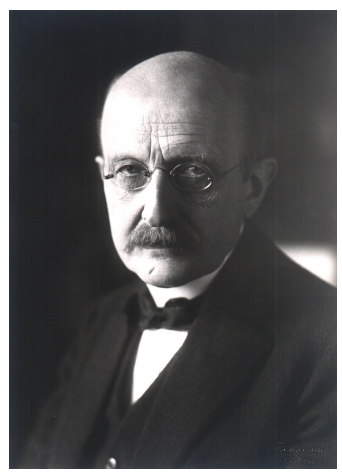

Max Planck
© Transocean Berlin

context a light quantum or a photon[4]. This revolutionary concept was first introduced by Max Planck[5] in 1900 [32, 33] and later refined by Albert Einstein[6] in 1905 [34]. After the theoretical predictions were confirmed experimentally by Robert Millikan[7] in 1916 [37], all three, Planck, Einstein and Millikan were eventually awarded the Nobel Prize in Physics for their discoveries in 1918, 1921 and 1923, respectively. According to this quantum theory, we may compute the energy of a photon $E_\gamma$ for a given electromagnetic wave with frequency $f$ using the Planck-Einstein relation:

$$E_\gamma = \mathrm{h}f \tag{2.6}$$

where:

| | | |
|---|---|---|
| $E_\gamma$ | photon energy | J |
| $f$ | electromagnetic wave frequency | Hz |
| h | Planck constant | J s |

The fundamental constant h called Planck's constant in honor of Max Planck is really tiny with a value of $6.626\,070\,15 \times 10^{-34}$ in units of J s [38]. As an example, for visible light with a frequency range of $(4.00–7.90) \times 10^{14}$ Hz, we get photon energies in the range of $(2.7–5.2) \times 10^{-19}$ J. From this we see already, that we have to deal with quite small amounts of energy for photons.[8] To avoid such small numbers, we therefore often use a different unit of energy in particle physics, the electronvolt. The electronvolt is defined as the energy gained by an electron when it is accelerated through an electric potential difference of 1 V. Consequently, one electronvolt is equivalent to the numerical value of the charge of an electron in units of coulomb, i.e. $1\,\mathrm{eV} = 1.602\,176\,634 \times 10^{-19}$ J. With this new unit of energy, we get much more convenient numbers for the energy of photons, e.g. for visible light we have photon energies of 1.7–3.3 eV.

Like the photons, the nucleus is a quantum system with well-defined energy states. Consequently, transitions between these excited states are also quantized, leading to the emission of photons with discrete energies. These energies correspond to the difference between the energy states and are characteristic of the specific radionuclide. Typical gamma-ray energies range from ~10 keV to multiple MeV.

In addition to gamma rays generated by energy-state transitions of the nucleus, high-energy photons may also be generated outside the nucleus in the electron shells. These photons are called characteristic

6 *Albert Einstein (*1879, †1955), a German-born theoretical physicist best known for developing the theory of relativity. Einstein made also important contributions to improve our understanding of the quantum mechanic nature of light, for which he eventually obtained the Nobel Prize in Physics in 1921.*

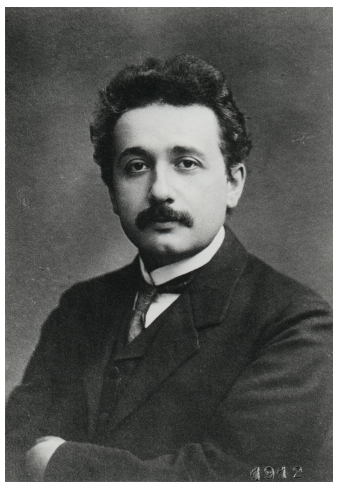

Albert Einstein
Ⓢ Jan Langhans

7 *Robert Andrews Millikan (*1868, †1953), a US experimental physicist. He is best known for the experimental studies on the elementary electric charge as well as the photoelectric effect [35–37]. For these contributions, he received the Nobel Prize in Physics in 1923.*

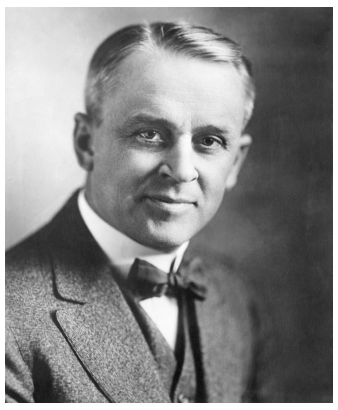

Robert Millikan
Ⓢ Nobel foundation

8 *Which can be challenging for both, our brains and our computers in numerical computations.*

X-rays and are emitted when an electron from a higher energy level fills a vacancy in a lower energy level.[9] These vacancies in the lower electron shells can be created in a EC decay as discussed before in Eq. 2.4 or by a process called internal conversion (IC).[10] In the internal conversion process, the energy of the excited nucleus is transferred not to gamma rays, but to a bound electron in the electron shell [40]:

$$ {}^{A}_{Z}\mathrm{Y}^{*} \xrightarrow{\mathrm{IC}} {}^{\phantom{-}A}_{Z-1}\mathrm{Y} + \mathrm{e}^{-} \tag{2.7} $$

The electron gets subsequently emitted from the atom leaving behind a vacancy in the electron shell [30]. Like the nucleus, electron shells are also a quantum system with well-defined energy states. Consequently, similar to the gamma rays, the characteristic X-ray photons have discrete energies, which correspond to the energy differences between the states and are also characteristic for the specific radionuclide. However, the production of these X-rays in radionuclide decay is somewhat more complex than gamma-ray emission. The filling of one electron vacancy in an inner shell will create an entire cascade of electron transitions in the outer shells, each eventually emitting characteristic X-rays or Auger electrons[11]. The energy of these photons is typically in the range of ~1 keV to ~100 keV.

There are two more processes associated with radionuclides, which can produce high-energy photons. However, in contrast to the already described sources, these processes take place outside the radionuclide atom in the surrounding matter. The first process is the so-called bremsstrahlung[12] or simply X-ray[13] radiation, which are high-energy photons produced when charged particles, such as electrons or positrons created in a beta-decay, are accelerated by electromagnetic interaction with nuclei or bound electrons.[14] Because of the continuous nature of the acceleration, the emitted photons have a continuous energy spectrum, which extends up to the kinetic energy of the incident particle. Therefore, we call bremsstrahlung also continuous X-rays in contrast to the characteristic X-rays with discrete photon energies. Typical kinetic energies for the emitted electrons and positrons in beta decay can reach >1 MeV, consequently extending the bremsstrahlung spectrum to these energies. A detailed quantum mechanical description of the bremsstrahlung process is quite involved and beyond the scope of this work. I recommend the monograph by Eberhard Haug and Werner Nakel [43] for a more comprehensive introduction to this topic.

A second source of high-energy photons outside the radionuclide atom is related to the $\beta^{+}$ decay mode, more specifically the positron

9 *The terminology regarding high-energy photons varies between scientific disciplines. In radiation detection and protection [30], we often refer to gamma rays as ionizing photons originating from energy-state transitions of the nucleus of an atom while X-rays arise from transitions outside the nucleus. In other fields such as astrophysics [39], gamma rays are often defined as photons having energies* >100 keV, *while below* 100 keV *(ionizing) photons are classified as X-rays. For the sake of convenience, I will follow the radiation detection terminology in this work.*

10 *These vacancies can not only be produced by nuclear decays but also by the interaction of external radiation with the atomic electrons. We will discuss this further in Chapter 3.*

11 *Like the internal conversion process in the de-excitation of a nucleus, the emission of Auger electrons is an alternative de-excitation mode in atomic electron transitions [30]. The energy of these Auger electrons is equal to the difference between the original atomic excitation energy and the binding energy of the energy state from which the electron was ejected.*

12 *From German "bremsen", which translates to "to brake". Thus, bremsstrahlung is typically associated with the negative acceleration of charged particles (deceleration), while the term "X-ray" is more generally used to describe the emission of high-energy photons resulting from either the positive or the negative acceleration of charged particles.*

emitted by radionuclides undergoing $\beta^+$ decay. After slowing down in the nearby matter, the emitted positron will eventually interact[15] with its antiparticle, an electron, and they will annihilate each other emitting two photons in opposite directions [30]. We can write this annihilation reaction as [40]:

$$e^+ + e^- = 2\gamma \tag{2.8}$$

The energy of the photons is equivalent to the rest mass of the electron and the positron, i.e. about 511 keV for each photon. The resulting photons are also known as annihilation radiation or annihilation photons.[16]

In summary, we have four main sources of high-energy photons associated with radionuclides:

1. Energy-state transitions of an excited nucleus after radioactive decay leading to gamma rays (discrete photon energies)
2. Atomic electron transitions in the electron shells outside the nucleus leading to characteristic X-rays (discrete photon energies)
3. Acceleration of charged particles emitted by the radionuclide in the surrounding matter leading to X-rays (continuous photon energies)
4. Annihilation of positrons emitted by the radionuclide with a $\beta^+$ decay mode in the surrounding matter leading to annihilation radiation (discrete photon energy of about 511 keV)

Although AGRS has the term "gamma-ray" in its name, it is important to note that AGRS systems are also capable of detecting characteristic X-rays, bremsstrahlung and annihilation photons.[17] That said, characteristic X-rays and bremsstrahlung related to radionuclides tend to possess photon energies that are lower than those of gamma rays or annihilation photons. Consequently, as we will see in Chapter 3, these X-rays are more easily absorbed by the atmosphere, which makes them harder for us to detect.

Another reason why we focus on gamma rays in AGRS is that, in contrast to annihilation photons or bremsstrahlung, they possess discrete energies characteristic for each individual radionuclide. Thanks to the dedicated work of many experimental physicists and chemists, comprehensive databases[18] containing detailed information on these gamma-ray energies are readily available today for hundreds of radionuclides. Using these databases, we have the means to fulfill one

13 *The term "X-rays" was established by Wilhelm Röntgen in 1985 after he discovered these mysterious rays generated by electrical discharges in a vacuum tube. He first observed them using a fluorescent screen coated with barium platinocyanide and subsequently termed these mysterious radiation X-rays using the mathematical designation "X" for something unknown [41]. For his discovery, he was awarded with the first Nobel Prize in Physics in 1901.*

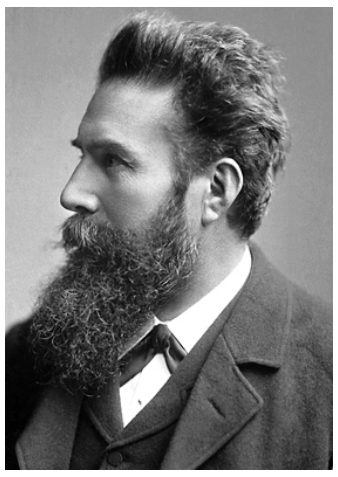

Wilhelm Conrad Röntgen (*1845, †1923)
© Nobel foundation

14 *In most cases, this interaction takes place outside the radionuclide, where the beta-decay took place. However, there is also the possibility that the beta particle gets accelerated by the nucleus, when it leaves the atom, leading to bremsstrahlung emitted by the radionuclide itself. In this case, we call the associated radiation also internal bremsstrahlung [42].*

15 *Either directly or by forming a quasi-stable combination, known as positronium [30].*

16 *From the Latin roots "ad" and "nihil", which translates to "to nothing".*

17 *I will demonstrate this in Part II.*

18 *Such as the ENDF/B-VIII.0 [44] or the JEFF-3.3 [45] nuclear data libraries.*

of the main tasks in AGRS, which is the identification of individual radionuclide sources by measuring the energies of the emitted photons with a spectrometer and comparing the detected photon energies with those from a nuclear data library.

### 2.1.2 Characteristics of Radionuclides

In the section before, we have seen how we may in principle be able to identify individual radionuclide sources by measuring the energies of the emitted gamma rays with a spectrometer and comparing the measured values with those tabulated in nuclear data libraries. But, once identified, how do we quantify the amount of a specific radionuclide? To answer this question, I have to introduce the concept of activity and related quantities.

The decay of radionuclides is a spontaneous stochastic process. However, we can characterize the probability for a radionuclide to decay within a differential time interval using the decay constant $\lambda$ in units of $s^{-1}$.[19] As a result, to quantify the decay rate of a given sample with $N$ identical radionuclides, we can define the activity $\mathcal{A}$ of this radionuclide as the differential number of decay events $-\mathrm{d}N$ per differential interval of time $\mathrm{d}t$ for a given time $t$ in units of Bq[20] as [30]:

$$\mathcal{A}(t) = -\left.\frac{\mathrm{d}N(t)}{\mathrm{d}t}\right|_{\text{decay}} = N\lambda \tag{2.9}$$

where:

| | | |
|---|---|---|
| $N$ | number of radionuclide atoms | |
| $t$ | time | s |
| $\lambda$ | decay constant | $s^{-1}$ |

The specific activity $a$ is then defined as the activity normalized to the mass of the specified radionuclide $i$ in units of $\mathrm{Bq\,kg^{-1}}$ [30]:

$$a_i = \frac{\lambda_i \mathrm{N_A}}{M_i} \tag{2.10}$$

where:

| | | |
|---|---|---|
| $M$ | molar mass | $\mathrm{kg\,mol^{-1}}$ |
| $\mathrm{N_A}$ | Avogadro constant (cf. Constants) | $\mathrm{mol^{-1}}$ |
| $\lambda$ | decay constant | $s^{-1}$ |

19 *Note that we can parametrize this probability also with the inverse $1/\lambda$, which is also known as the mean life $\tau$.*

20 *The SI unit "becquerel" corresponds to $1\,\mathrm{Bq} = 1\,s^{-1}$ and was named in honor of Henri Becquerel [38].*

To characterize radionuclides in the context of AGRS, depending on the specific radionuclide distribution in the environment, it is often more convenient to normalize the activity $\mathcal{A}$ to the mass of a given solid, to the volume of a liquid/gas or to a surface area. The corresponding quantities are defined as the activity mass concentration $a_{\mathrm{m},i}$, the activity volume concentration $a_{\mathrm{v},i}$ and the surface activity concentration $a_{S,i}$ for a given radionuclide $i$ in units of $\mathrm{Bq\,kg^{-1}}$, $\mathrm{Bq\,m^{-3}}$ and $\mathrm{Bq\,m^{-2}}$, respectively:

$$a_{\mathrm{m},i} = a_i w_i \tag{2.11a}$$

$$a_{\mathrm{v},i} = a_i w_i \rho \tag{2.11b}$$

$$a_{S,i} = a_i w_i \sigma_S \tag{2.11c}$$

where[21]:

| | | |
|---|---|---|
| $a_i$ | specific activity of the radionuclide $i$ | $\mathrm{Bq\,kg^{-1}}$ |
| $w_i$ | mass fraction of the radionuclide $i$ in given source matter | $\mathrm{kg\,kg^{-1}}$ |
| $\rho$ | mass density for a given source matter | $\mathrm{kg\,m^{-3}}$ |
| $\sigma_S$ | surface density for a given source matter | $\mathrm{kg\,m^{-2}}$ |

21 *Please note that, to avoid confusion with other unitless concentration quantities, I state here the unit of $w_i$ explicitly.*

In AGRS, there are instances where it is necessary to consider how the activity, as described in Eq. 2.9, evolves over time. There are three things to consider here. First, the progeny nuclide Y of a specific radioactive decay X → Y might also be radioactive, i.e. a radionuclide. In fact, a single radionuclide can form an entire decay chain of one radionuclide decaying into another radionuclide and so forth. We call such a chain also a decay series.

Second, a radionuclide might not only possess a single decay mode but multiple. In these cases, different progeny nuclides might be formed by the decay of this specific parent radionuclide. We describe this by the branching ratio $b_{i,j}$, which quantifies the fraction of decays of a specific radionuclide $i$ that occur by a specific decay mode resulting in a progeny nuclide $j$.

Third, in AGRS, as discussed before, we detect not the radionuclide decay itself but the subsequent emission of high-energy photons with characteristic energies. Consequently, we need to take into account the mean number of emitted photons for a specific characteristic energy per decay of a given radionuclide, which we call the relative photon intensity $I_\gamma$ often given in %. Similar to the photon

22 *Linear in the sense that each member nuclide of the chain has only one single parent nuclide. Nonlinear chains on the other hand may possess members in the chain with more than one parent nuclide.*

energies, this quantity is also characteristic for each radionuclide and readily available from nuclear data libraries [44, 45].

Based on these thoughts, we can write a general linear[22] decay series scheme for a radionuclide $X_1$ decaying into a sequence of progeny nuclides $X_2, X_3, \cdots, X_{n-1}$ eventually ending with the stable nuclide $X_n$:

$$X_1 \xrightarrow{b_{1,2}\lambda_1} X_2 \xrightarrow{b_{2,3}\lambda_2} \cdots \xrightarrow{b_{i-1,i}\lambda_{i-1}} X_i \xrightarrow{b_{i,i+1}\lambda_i} \cdots \xrightarrow{b_{n-1,n}\lambda_{n-1}} X_n \tag{2.12}$$

In 1910, Harry Bateman[23] published a general solution, which is today known as the Bateman equation, to describe the activity of an arbitrary progeny nuclide $X_i$ in such a linear decay series for $i < n$ [46, 47]:

23 *Harry Bateman (*1882, †1946), an English mathematician. His research was focused on general solutions to differential equations in the field of physics, e.g. nuclear physics or electrodynamics.*

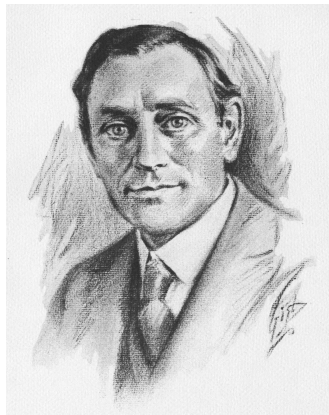

Harry Bateman
© Carl Rudolph Gist

$$\mathcal{A}_i(t) = \sum_{l=1}^{i} \left( \prod_{j=l}^{i-1} b_{j,j+1}\lambda_j \right) \sum_{k=l}^{i} \frac{\mathcal{A}_l(0)\,\mathrm{e}^{-\lambda_k t}}{\prod_{j=l,j\neq k}^{i} \lambda_j - \lambda_k} \frac{\lambda_i}{\lambda_l} \tag{2.13}$$

where:

| | | |
|---|---|---|
| $\mathcal{A}_l(0)$ | activity of the radionuclide $l$ for $t = 0$ | Bq |
| $b_{i,j}$ | branching ratio from the radionuclide $i$ to the nuclide $j$ | |
| $\lambda$ | decay constant | $s^{-1}$ |

Solutions for nonlinear decay chain problems, which also account for additional source or drain terms apart from nuclear decay, are readily available in the literature [48–50].

Let us consider now the simplest case of such a decay series, that is the decay of a single radionuclide X into a stable nuclide Y:[24]

24 *Other decay channels are accounted for by the indicated branching ratio $b_{X,Y}$.*

$$X \xrightarrow{b_{X,Y}\lambda_X} Y \tag{2.14}$$

Using Eq. 2.13, we get the activity of the radionuclide X at a given time $t$ as:

$$\mathcal{A}_X(t) = \mathcal{A}_X(0)\,\mathrm{e}^{-\lambda_X t} \tag{2.15}$$

from this we see that the activity of a radionuclide in such a scenario decays exponentially with time. The time it takes for the activity of a radionuclide to decay to half of its initial value is called the half-life $t_{1/2}$. From Eq. 2.15, it is easy to see that the half-life $t_{1/2}$ is related to the decay constant as follows:

$$t_{1/2} = \frac{\log(2)}{\lambda} \tag{2.16}$$

In summary, we can characterize the decay time dynamics of a radionuclide by employing one of the time-independent parameters: $\lambda$, $t_{1/2}$ or $\tau$. In most nuclear data libraries [44, 45, 51], the half-life $t_{1/2}$ is the preferred parameter. Therefore, I will adopt the half-life $t_{1/2}$ instead of the decay constant $\lambda$ or the mean life $\tau$ throughout this book.

### 2.1.3 Radionuclides in the Environment

In the last two sections, we analyzed what gamma rays are and how they are generated by radionuclides. We then continued to discuss how we can quantify these radionuclides using activity quantities and concluded with a brief discussion on how these activity quantities evolve over time. In this last subsection, I will give a brief overview which kind of radionuclides are present in the environment and what their typical activities are. For that purpose, I will limit the discussion to radionuclides relevant for AGRS applications, i.e. gamma-ray emitting radionuclides. We will use this information in Parts II and III to interpret measurement and simulation results as well as to provide the basis for data evaluation in Part IV.

In general, we can divide the radionuclides present in the environment in two broad categories: natural and anthropogenic[25,26] ones. Natural radionuclides were formed by natural processes whereas anthropogenic ones have been generated by mankind, i.e. they would not be present on Earth today in significant quantities without human intervention. We start our discussion with the natural radionuclides. As proposed by Kogan et al. [52], we may subdivide the natural radionuclides further into three subgroups based on their formation process:

1. Primordial radionuclides
2. Cosmogenic radionuclides
3. Radiogenic radionuclides

In the next subsections, I will explore the formation and characteristics of these groups of natural radionuclides in more detail before proceeding to discuss the anthropogenic ones.

25 *From the Ancient Greek roots $\alpha\nu\theta\rho\omega\pi\varsigma$ (anthropos) and $\gamma\epsilon\nu\eta\varsigma$ (genes) for "human" and "offspring".*

26 *Sometimes also referred to as man-made or artificial radionuclides.*

### 2.1.3.1 Primordial Radionuclides

So, how can natural radionuclides still be around, if they continuously decay, as we have learned in the previous subsections? The first option is that the associated radionuclide possess a sufficiently long half-life to be still present today since their formation. These radionuclides are called primordial[27] radionuclides, and they were formed mainly by stellar and supernova nucleosynthesis[28] processes, i.e. before the Earth was formed around $4.54 \times 10^9$ a ago [55]. Consequently, these primordial radionuclides require half-lives $t_{1/2} \geq \mathcal{O}(10^8)$ a.[29] In Table 2.1, I have listed all known gamma-ray emitting primordial radionuclides together with the corresponding half-life values and decay modes. In addition, I estimated the activity mass concentration in the Earth's crust ($a_{\text{crust}}$) and the activity volume concentration in the sea ($a_{\text{sea}}$) for these primordial radionuclides based on readily available natural isotopic and elemental abundance values [56, 57] (cf. Appendix A.1). From these estimates, we see that ${}^{40}_{19}$K shows the highest activity concentrations followed by ${}^{232}_{90}$Th and ${}^{238}_{92}$U. The remaining primordial radionuclides show all at least one order of magnitude smaller activity concentration values and are thereby in most circumstances negligible in environmental radiation measurements [58]. Furthermore, in general, we find that the activity concentrations for the primordial radionuclides are significantly higher (at least by a factor 50) in the Earth's crust than in the sea.[30] We can conclude that the three relevant primordial radionuclides for AGRS are ${}^{40}_{19}$K, ${}^{232}_{90}$Th and ${}^{238}_{92}$U and that they are mainly found in and on the Earth's crust.

In Table 2.2, I have compiled a more detailed list of the activity concentrations for various materials found in Earth's environment focusing on ${}^{40}_{19}$K, ${}^{232}_{90}$Th and ${}^{238}_{92}$U. Starting with rocks, the activity concentrations found for the Earth's crust listed in Table 2.1 are also reflected in the values for the listed rocks in Table 2.2. Magmatic rocks (granites and basalts) tend to possess higher concentrations of ${}^{40}_{19}$K, ${}^{232}_{90}$Th and ${}^{238}_{92}$U compared to sedimentary rocks (limestones, sandstones, shales and sands) [52]. For soils, there is significant dispersion in the activity mass concentrations locally but also globally [2]. As a rule of thumb, the higher the clay fraction in the specific soil is, the higher the activity mass concentrations for ${}^{40}_{19}$K, ${}^{232}_{90}$Th and ${}^{238}_{92}$U are [52]. This trend can be explained by the high sorption capacity of clays for these radionuclides. Regarding the biosphere, the total

*27 From the Latin word "primordialis" for "of the beginning".*

*28 Nucleosynthesis is the process of forming new nuclides from preexisting nucleons (protons and neutrons) and nuclei [53, 54]. The nucleosynthesis theory, by successfully explaining the abundances of the nuclides in the universe, marks one of the cornerstones of modern astrophysics.*

*29 Here, I assumed that for a primordial radionuclide to maintain a significant fraction of 0.1 % of its original amount today, its half-life must be at least a tenth of the Earth's age (cf. Eq. 2.15).*

*30 Considering that seawater has a mass density of $\sim 1\,\text{kg}\,\text{L}^{-1}$.*

**Table 2.1** List of gamma-ray emitting primordial radionuclides on Earth.

| Nuclide★ | Half-life [a] | Decay modes● | Decay products○ | $a_{crust}$† [Bq kg$^{-1}$] | $a_{sea}$‡ [Bq L$^{-1}$] | References |
|---|---|---|---|---|---|---|
| $^{40}_{19}$K | $1.2405(30) \times 10^{9}$ | $\beta^+, \beta^-, \gamma$, EC | $^{18}_{14}$Ar, $^{40}_{20}$Ca | $6.67 \times 10^{2}$ | $1.27 \times 10^{1}$ | [65] |
| $^{50}_{23}$V | $2.65(18) \times 10^{17}$ | $\beta^-, \gamma$, EC | $^{50}_{22}$Ti, $^{50}_{24}$Cr | $3 \times 10^{-7}$ | $6 \times 10^{-12}$ | [66] |
| $^{138}_{57}$La | $1.036(20) \times 10^{11}$ | $\beta^-, \gamma$, EC | $^{138}_{56}$Ba, $^{138}_{58}$Ce | $3 \times 10^{-2}$ | $3 \times 10^{-9}$ | [66] |
| $^{176}_{71}$Lu | $3.76(7) \times 10^{10}$ | $\beta^-, \gamma$ | $^{176}_{72}$Hf | $4 \times 10^{-2}$ | $8 \times 10^{-9}$ | [67] |
| $^{232}_{90}$Th | $1.402(6) \times 10^{10}$ | $\alpha, \gamma$ | thorium series | $4.0 \times 10^{1}$ | $4 \times 10^{-6}$ | [65] |
| $^{235}_{92}$U | $7.04(1) \times 10^{8}$ | $\alpha, \gamma$, SF | actinium series | 1.5 | $1.8 \times 10^{-3}$ | [65] |
| $^{238}_{92}$U | $4.468(5) \times 10^{9}$ | $\alpha, \gamma$, SF | uranium series | $3.3 \times 10^{1}$ | $4.0 \times 10^{-2}$ | [68] |

★ Only confirmed primordial radionuclides are listed. As an example, in 1971, Hoffman et al. [59] found evidence for primordial $^{244}_{94}$Pu in the rare-earth mineral bastnäsite and estimated the abundance in Earth's crust to be $2.8 \times 10^{-25}$ g g$^{-1}$. However, Lachner et al. [60] could not confirm the presence of $^{244}_{94}$Pu in bastnäsite in 2012. The evidence for primordial $^{244}_{94}$Pu remains therefore inconclusive and is not included in this list.

● $\alpha$: radioactive decay with alpha particle emission, $\beta^+$: radioactive beta-decay with positron emission, $\beta^-$: radioactive beta-decay with electron emission, $\gamma$: gamma-ray emission, EC: electron capture decay, SF: spontaneous fission decay (cf. Section 2.3.1).

○ Except the three main natural decay chains, i.e. the actinium, thorium and uranium series, all listed decay products are stable. $^{50}_{24}$Cr and $^{138}_{58}$Ce are potential candidates to be long-lived radionuclides($t_{1/2} > 1 \times 10^{17}$ a) exhibiting the rare double-beta decay mode, i.e. $^{50}_{24}\text{Cr} \xrightarrow{2\beta^+} {}^{50}_{22}\text{Ti}$ [61, 62] and $^{138}_{58}\text{Ce} \xrightarrow{2\beta^+} {}^{138}_{56}\text{Ba}$ [63, 64].

† Average activity mass concentration in the Earth's crust (cf. Appendix A.1).

‡ Average activity volume concentration in the sea (cf. Appendix A.1).

**Table 2.2** Terrestrial radionuclide activities in the environment.

| Sources | Activity mass concentration $a_m$ [Bq kg$^{-1}$] | | | References |
|---|---|---|---|---|
| | $^{40}_{19}$K | $^{232}_{90}$Th | $^{238}_{92}$U | |
| Rocks★ | | | | |
| Basalt | 300–900 | 10–15 | 7–10 | [69, 70] |
| Granite | 1000–1100 | 70–90 | 40–60 | [69, 70] |
| Limestones | 81–160 | 5 | 15 | [69, 70] |
| Sandstones | 330 | 24 | 15 | [69] |
| Shales | 810 | 41 | 15 | [69] |
| Sands | <300 | 25–30 | 40 | [69, 70] |
| Soil (World average)● | 400 (140–850) | 30 (11–64) | 35 (16–110) | [2] |
| Soil (Switzerland)● | 370 (40–1000) | 25 (4–70) | 40 (10–150) | [2] |
| Hydrosphere★○ | | | | |
| Sea water | 24–25 | (4–1000) × $10^{-6}$ | 0.49 | [52, 71] |
| Fresh water | 0.7–25 | (4–1000) × $10^{-6}$ | 0.16–0.21 | [52, 71] |
| Flora†‡ | | | | |
| Grasses and herbs | 163 (3–2576) | 2.2 (1 × $10^{-3}$–171) | 2.4 (1 × $10^{-3}$–609) | [2, 73, 75] |
| Lichens and bryophytes | 82 (4–867) | 6.1 (0.07–352) | 29 (0.23–1134) | [2, 73, 75] |
| Shrubs | 353 (6–1403) | 0.9 (2 × $10^{-3}$–86) | 1.4 (2 × $10^{-4}$–164) | [2, 73, 75] |
| Trees | 39 (0.2–564) | 0.04 (1 × $10^{-4}$–0.20) | 0.18 (2 × $10^{-4}$–3.5) | [2, 73, 75] |
| Fauna†‡ | | | | |
| Arthropods | 190 (25–695) | 0.09 (5 × $10^{-3}$–0.46) | 0.48 (0.01–2.2) | [2, 73, 75] |
| Birds$ | 1210 (36–3994) | 0.01 (3 × $10^{-3}$–0.03) | 0.50 (6 × $10^{-3}$–0.47) | [2, 73, 75] |
| Mammals | 511 (9–4592) | 4 × $10^{-3}$ (1 × $10^{-4}$–0.04) | 0.22 (2 × $10^{-4}$–2.3) | [2, 73, 75] |
| Human body§ | 63 | 4.2 × $10^{-3}$ | 1.2 × $10^{-2}$ | [76, 78] |
| Building materials‡ | | | | |
| Adobe | 300–2000 | 23–200 | 20–90 | [81] |
| Cement | 225 (135–380) | 15 (11–21) | 24 (19–40) | [82] |
| Concrete (normal) | 360 (130–560) | 20 (10–47) | 19 (13–24) | [82] |
| Concrete (lightweight) | 850 (710–950) | 70 (27–98) | 49 (22–83) | [82] |

*Continued on next page*

**Table 2.2** *Continued from previous page*

| Sources | Activity mass concentration $a_m$ [Bq kg$^{-1}$] | | | References |
|---|---|---|---|---|
| | $^{40}_{19}$K | $^{232}_{90}$Th | $^{238}_{92}$U | |
| Brick | 670 (470–1200) | 53 (37–98) | 50 (38–67) | [82] |
| Marble | 40–200 | 20 | 20–30 | [81] |
| Other materials ‡ | | | | |
| Coal ash | 44–1060 | 44–280 | 56–440 | [81] |
| Fertilizer | 40–8000 | 20–30 | 230–2300 | [81] |

★ The range of activity mass concentrations represent minimum and maximum values provided by the denoted references [52, 69–71].

• Mean and range (in parenthesis) provided by the indicated reference.

○ The molar fraction values for the radionuclide $i$ in sea water $n_{\text{sea},i}$ and fresh water $n_{\text{fresh},i}$ provided by Kogan et al. [52] as well as Smith [71] were converted to activity mass concentrations using the same methodology introduced in Appendix A.1, i.e. for sea water we have $a_m = \log(2) N_A n_{\text{sea},i} / (t_{1/2,i} M_{H_2O})$ with $N_A$, $t_{1/2,i}$ and $M_{H_2O}$ being the Avogadro constant, the half-life of the corresponding radionuclide $i$ and the molar mass for water $H_2O$ in units of mol$^{-1}$, s and kg mol$^{-1}$, respectively.

† Activity mass concentrations for flora and fauna data entries were computed using the radionuclide transfer methodology established and standardized by the International Atomic Energy Agency (IAEA) [72–74] and the International Commission on Radiological Protection (ICRP) [75] (cf. Appendix A.2).

‡ These numbers represent the arithmetic mean (fauna & flora) and median (building materials) in combination with minimum and maximum values (enclosed in parentheses).

§ These values represent arithmetic mean activity mass concentrations for the corresponding radionuclides and a reference male human body according to Snyder et al. [76]. The reference man is defined as being between 20–30 years of age, weighing 70 kg, is 170 cm in height, and lives in a climate with an average temperature of 10–20 °C. He is Caucasian and follows the lifestyle and customs typical of Western Europe or North America. Similar to Eq. A.1 in Appendix A.1, the activity mass concentration of $^{40}_{19}$K and $^{238}_{92}$U was computed based on elemental composition data of the human body from Snyder et al. [76], natural isotopic abundance data from Meija et al. [56] and atomic weight data from Prohaska et al. [77]. Adopted half-life values are indicated in Table 2.1. The activity mass concentration of $^{232}_{90}$Th was adopted from Pertsov [78]. It is worth adding that the elemental composition of the human body can vary significantly among individuals [79].

$ The arithmetic mean value of the $^{40}_{19}$K activity mass concentration in the bird biota should be interpreted with care. Literature sources indicate significantly lower $^{40}_{19}$K activity mass concentrations in birds compared to the arithmetic mean value derived from the radionuclide transfer methodology outlined in Appendix A.2 [80].

dry mass of living matter on Earth is estimated to be about $1.1 \times 10^{15}$ kg [83],[31] which results in an average surface density of about $0.2\,\mathrm{g\,cm^2}$ on Earth. Estimating the activity mass concentrations for the individual members of the flora and fauna is complex. I adopted the radionuclide transfer methodology established and standardized by the International Atomic Energy Agency (IAEA) [72–74] and the International Commission on Radiological Protection (ICRP) [75] to provided activity mass concentration estimates for the biota listed in Table 2.2. More information on this methodology can be found in Appendix A.2. In general, the most prominent radionuclide in the biosphere is $^{40}_{19}$K, with $^{232}_{90}$Th and $^{238}_{92}$U being almost insignificant for most biota. In general, members of the fauna tend to have higher concentrations of $^{40}_{19}$K but lower concentrations of $^{232}_{90}$Th and $^{238}_{92}$U compared to the members of the flora [52]. However, it is important to point out that the activity mass concentrations for both, fauna and flora, heavily depend on the local environment and can vary by multiple orders of magnitude as indicated in Table 2.2 [52]. As a last category, I have listed common building materials in Table 2.2. The activity mass concentrations for these materials are similar to those found in rocks and soil. However, similar to the biosphere, the activity mass concentrations for these materials can vary quite significantly depending on the origin of the raw materials.

31 *Estimated based on a conversion factor of 2 between carbon and dry mass as suggested by Bar-On et al. [83].*

#### 2.1.3.2 Radiogenic Radionuclides

The second option, how natural radionuclides in the environment can be formed, is by the decay of primordial radionuclides. These radionuclides are referred to as radiogenic [84]. As listed in Table 2.1, $^{232}_{90}$Th, $^{235}_{92}$U and $^{238}_{92}$U possess radiogenic radionuclides. These three primordial radionuclides form three natural decay chains, which are also known as the thorium, actinium and uranium series[32], as indicated in Table 2.1. In Fig. 2.1, all three series are shown with the associated radiogenic progeny nuclides and related decay modes. More details on the half-lives as well as the branching ratios can be found in the Tables C.1–C.3. From this, we find two important properties of these natural decay series.

First, the half-life of the parent primordial radionuclide is significantly longer than the ones of the radionuclide progeny, i.e. $t_{1/2,1} \gg t_{1/2,i>1}$. In this special case, a steady state equilibrium in the activities of all members of the decay chain might be eventually

32 *One reason, why we call these decay chains "series" is because the decay modes of these chains are limited to the $\alpha$ and $\beta^-$ decay. Consequently, the mass number A for each member of a given chain can be expressed by the relation $4n + k$ with $n \in \mathbb{N}$ and $k \in \{0, 2, 3\}$ for the thorium, uranium and actinium series, respectively. You might wonder, is there also a series for $k = 1$? The answer is yes. The decay series with $k = 1$ is called the neptunium series. However, this series does not occur naturally on Earth due to the relatively short half-life of its longest-lived member, $^{237}_{93}$Np, which is only $2.144(9) \times 10^6$ a [85], significantly smaller than the required $\mathcal{O}(10^8)$ a discussed above. We will meet the neptunium series again when we discuss anthropogenic radionuclides.*

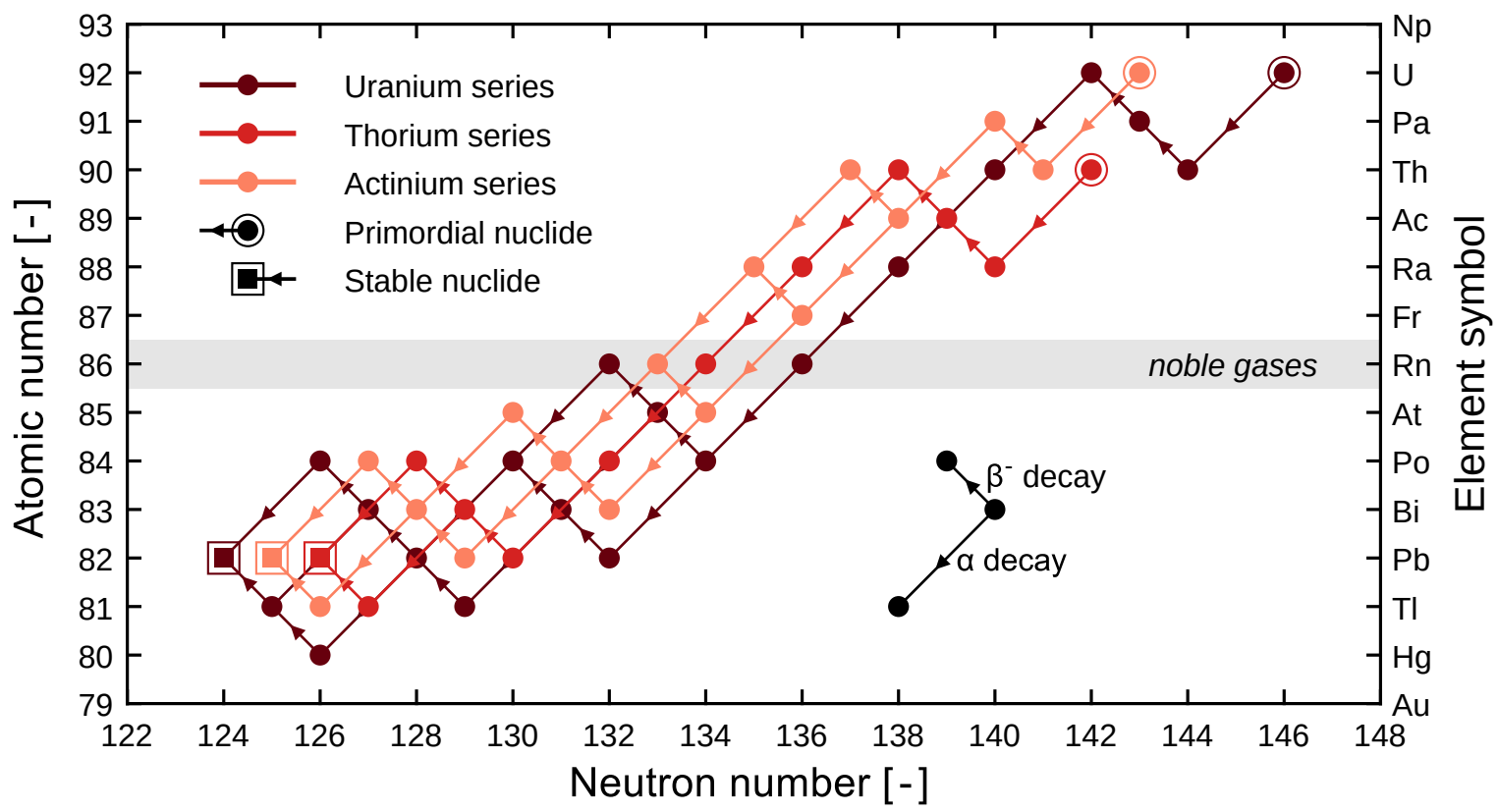

**Figure 2.1** Overview of the three natural decay series on Earth, i.e. the uranium, thorium and actinium series with the primordial radionuclides $^{238}_{92}$U, $^{232}_{90}$Th and $^{235}_{92}$U, as a function of the atomic number $Z$ and neutron number $N_n$. The stable nuclides forming the end points of the decay chains are $^{206}_{82}$Pb, $^{208}_{82}$Pb and $^{207}_{82}$Pb, respectively. Because of the inconclusive empirical results regarding its primordial status [59, 60], $^{244}_{94}$Pu and its decay products up to $^{236}_{92}$U are not included in this figure. Moreover, products from spontaneous fission decay (SF), heavy cluster Decay (HCD) and other rare decay modes are also not shown. Additional information on the half-lives, decay modes and associated branching ratios for each nuclide in the three series (including $^{244}_{94}$Pu and its decay products) can be found in the Tables C.1–C.3.

established, where the number of the progeny radionuclides decaying equals the number produced by the preceding parent nuclide per unit time. We call this steady state condition also a secular equilibrium.[33] If $t_{1/2,1}$ is significantly higher than the measurement time, we may approximate the activity of a member in a linear decay chain, having reached this secular equilibrium, by:

*33 First described by Ernest Rutherford and Frederick Soddy in their seminal study about the causes of natural radioactivity [86].*

$$\mathcal{A}_i \approx \left( \prod_{j=1}^{i-1} b_{j,j+1} \right) \mathcal{A}_1 \tag{2.17}$$

with $\mathcal{A}_1$ and $\mathcal{A}_i$ being the activity of the primordial and radiogenic radionuclides, respectively (cf. Eq. 2.13). This secular equilibrium might get disturbed by the selective removal of one or several members of a specific decay chain. For the natural decay series in the environment shown in Fig. 2.1, such a disequilibrium state might be the result of a variety of natural processes, e.g. selective leaching, fractionation or the removal of gaseous decay products [87]. After such a disruption in the series occurs, the time needed to restore again secular equilibrium is several times the half-life of the disturbed member [87].[34] In the environment, we observe that the uranium series is often in disequilibrium, while the thorium and actinium[35] series tend to maintain secular equilibrium [58, 87]. We will keep this in mind for the subsequent parts of this book when we discuss the interpretation of measurement and simulation results.

34 *Similar to Eq.* A.7 *in our discussion in Appendix A.3, the progeny approaches the primordial activity by the factor* $1 - \exp(-\lambda t)$*, given the half-life of the primordial radionuclide is significantly higher than the one of the progeny.*

35 *However, for members of the fauna and flora, secular equilibrium is always disrupted [52, 58].*

Having introduced the secular equilibrium, the second important property of the natural decay series is their photon emission. As listed in Tables C.1–C.3, most of the radiogenic members of the decay series emit gamma rays and characteristic X-rays. In Fig. 2.2, the relative photon intensity $I_\gamma$ for the three natural decay series (assuming secular equilibrium) as well as the most prominent primordial radionuclide in nature, $^{40}_{19}$K, is shown. In general, we find that the photons cover an energy range between about 10 keV and 3 MeV. We also see that the decay series, due to their radiogenic members, possess a much more complex emission spectrum compared to the primordial $^{40}_{19}$K. This is very attractive for radiation detection applications and especially AGRS, as we have an enhanced signal from these series. It is interesting to note that the emission of photons is not equally distributed among the decay series members, rather, a minority of members emit the majority of photons [52]. For the uranium series, about 80 % of the photons are emitted by $^{214}_{82}$Pb and $^{214}_{83}$Bi. For the actinium series, four members contribute to about 70 % of the photons, i.e. $^{211}_{82}$Pb, $^{227}_{90}$Th, $^{231}_{90}$Th and $^{235}_{92}$U. For the thorium series, the two members $^{208}_{81}$Tl, and $^{228}_{89}$Ac emit about 85 % of the photons. In the context of the AGRS, we identify and quantify predominantly these members in the decay series and not the associated primordial radionuclides, i.e. $^{232}_{90}$Th and $^{238}_{92}$U.[36] This is especially relevant for cases, where the uranium series is in disequilibrium, as we cannot infer the activity of $^{238}_{92}$U directly from the quantified activities of $^{214}_{82}$Pb and $^{214}_{83}$Bi.

36 *As discussed in the previous subsection,* $^{235}_{92}$U *is not relevant for AGRS applications. For completeness, I have kept it in the discussion of the radiogenic radionuclides here.*

Before we continue with the discussion of the third group of natural radionuclides, I would like to highlight a particular subset of radiogenic radionuclides. As discussed in the context of the secular equilibrium, due to their mobility, gaseous radiogenic nuclides might escape from the primordial source matrix into the Earth's atmosphere, where they are subjected to diffusion and convective mixing. In case of the natural decay series, these gaseous radionuclides are isotopes of the noble gas radon, i.e. ${}^{218}_{86}$Rn, ${}^{219}_{86}$Rn, ${}^{220}_{86}$Rn and ${}^{222}_{86}$Rn (cf. Fig. 2.1). Because of the short half-lives[37] of ${}^{218}_{86}$Rn, ${}^{219}_{86}$Rn and ${}^{220}_{86}$Rn, ${}^{222}_{86}$Rn from the uranium series is the main contributor in AGRS measurements [52]. That said, ${}^{222}_{86}$Rn shows no significant high-energy photon emission, but the ${}^{222}_{86}$Rn progeny ${}^{214}_{82}$Pb and ${}^{214}_{83}$Bi do. As a result, ${}^{222}_{86}$Rn leads to a significant background of atmospheric gamma rays by the decay of its ${}^{214}_{82}$Pb and ${}^{214}_{83}$Bi progeny. Activity volume concentrations of ${}^{222}_{86}$Rn in the lower Earth's atmosphere may vary significantly in time and space due to various meteorological (pressure, temperature, humidity, wind speed and direction, cloudiness, precipitation etc.) as well as geological (local geology and soil type, moisture content, porosity etc.) factors [52]. Radon concentrations in the lower atmosphere (<1000 m in altitude) typically vary between $1\,\mathrm{Bq\,m^{-3}}$ and $14\,\mathrm{Bq\,m^{-3}}$ with a mean value of $4(3)\,\mathrm{Bq\,m^{-3}}$ [88]. Above 1000 m, radon concentrations reduce to about $0.3(4)\,\mathrm{Bq\,m^{-3}}$ at 3000 m [88]. Activity volume concentrations in the lower atmosphere are significantly reduced over the sea compared to the land, typically by a factor of hundred [52]. More information including an in-depth discussion on the distribution and migration not only of ${}^{222}_{86}$Rn but also the progeny ${}^{214}_{82}$Pb and ${}^{214}_{83}$Bi in the Earth's atmosphere are provided by Jacobi et al. [89], Moore et al. [90], Beck [91], and Porstendörfer [92].

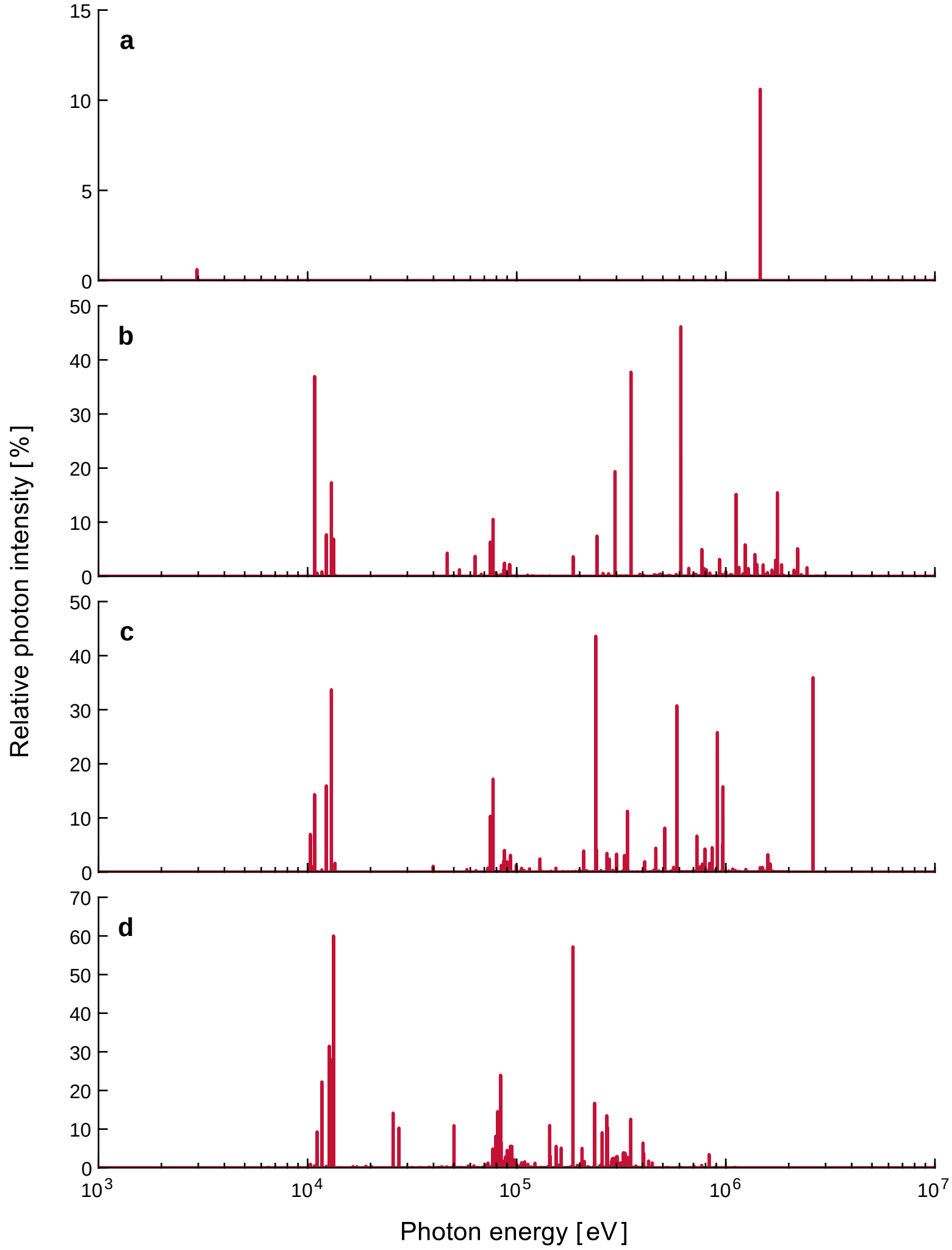

**Figure 2.2** Relative photon intensity $I_\gamma$ for selected primordial and radiogenic radionuclides as a function of the photon energy. **a** $^{40}_{19}K$. **b** $^{238}_{92}U$ including all associated radiogenic radionuclides in the uranium series. **c** $^{232}_{90}Th$ including all associated radiogenic radionuclides in the thorium series. **d** $^{235}_{92}U$ including all associated radiogenic radionuclides in the actinium series. For the decay series, secular equilibrium was assumed and $I_\gamma$ was computed with respect to the decay of the corresponding primordial radionuclide using the `FLUKA` code (version: `4-4.3`) [20].

37 *In addition to the low branching ratio leading to* ${}^{218}_{86}$Rn *and overall low primordial activities of* ${}^{235}_{92}$U *associated with* ${}^{219}_{86}$Rn.

#### 2.1.3.3 Cosmogenic Radionuclides

The third group of natural radionuclides on Earth are the cosmogenic radionuclides. These radionuclides are continuously formed by the interaction of cosmic rays with the Earth's atmosphere.[38] We will discuss cosmic rays in the next section. I have listed all known gamma-ray emitting cosmogenic radionuclides in Table 2.3 including the corresponding half-lives and production modes. Because of the low production rates, activity concentration measurements in the atmosphere for cosmogenic radionuclides are challenging. Consequently, empirical activity concentration data for corresponding radionuclides listed in Table 2.3 are not readily available. In contrast, the production rates of cosmogenic radionuclides in the atmosphere are well known [94–101]. Consequently, I have used best-estimate production rate values to infer an upper limit for the activity volume concentration values of the cosmogenic radionuclides listed in Table 2.3 (cf. Appendix A.3). The resulting concentration values are also included in Table 2.3. Comparing these limits, it is evident that

38 *Cosmogenic radionuclides are also produced in the hydrosphere, cryosphere or lithosphere, but at an even lower rate than in the atmosphere [93].*

**Table 2.3** List of cosmogenic radionuclides in the Earth's atmosphere showing gamma-ray or positron emission.

| Nuclide | Half-life | Decay modes★ | Main production mode | Activity• [Bq m$^{-3}$] | References |
|---|---|---|---|---|---|
| $^{7}_{4}$Be | $5.322(6) \times 10^{1}$ d | EC, $\gamma$ | spallation(N, O) | $<7.3 \times 10^{-2}$ | [97, 99, 102–104] |
| $^{11}_{6}$C | $2.0361(23) \times 10^{1}$ min | $\beta^{+}$, EC | spallation(N, O) | $2.3 \times 10^{-4}$ ○ | [96, 102] |
| $^{18}_{9}$F | 1.828 90(23) h | $\beta^{+}$, EC | spallation(Ar) | $<5.5 \times 10^{-5}$ | [102, 105] |
| $^{22}_{11}$Na | 2.6029(8) a | $\beta^{+}$, EC, $\gamma$ | spallation(Ar) | $<1.6 \times 10^{-4}$ | [65, 94, 103, 106] |
| $^{24}_{11}$Na | $1.4958(2) \times 10^{1}$ h | $\beta^{-}$, $\gamma$ | spallation(Ar) | $<2.2 \times 10^{-4}$ | [102, 103] |
| $^{28}_{12}$Mg | $2.0915(9) \times 10^{1}$ h | $\beta^{-}$, $\gamma$ | spallation(Ar) | $<9.4 \times 10^{-5}$ | [103, 107] |
| $^{26}_{13}$Al | $7.17(24) \times 10^{5}$ a ○ | $\beta^{+}$, EC, $\gamma$ | spallation(Ar) | $<2.5 \times 10^{-4}$ | [103, 108] |
| $^{28}_{13}$Al | 2.245(2) min | $\beta^{-}$, $\gamma$ | $^{28}_{12}\mathrm{Mg} \xrightarrow{\beta^{-}} {}^{28}_{13}\mathrm{Al}$ | $<9.4 \times 10^{-5}$ | [94, 103, 107] |
| $^{31}_{14}$Si | $1.5724(20) \times 10^{2}$ min | $\beta^{-}$, $\gamma$ | spallation(Ar) | $<8.0 \times 10^{-4}$ | [69, 94, 109] |
| $^{34m}_{17}$Cl | $3.199(3) \times 10^{1}$ min | EC, IC, $\gamma$ | spallation(Ar) | $<3.6 \times 10^{-4}$ | [94, 110, 111] |
| $^{34}_{17}$Cl | 1.5266(4) s | $\beta^{+}$, EC | $^{34m}_{17}\mathrm{Cl} \xrightarrow{\mathrm{IC}} {}^{34}_{17}\mathrm{Cl}$ | $<3.6 \times 10^{-4}$ | [94, 110, 111] |
| $^{38}_{17}$Cl | $3.7230(14) \times 10^{1}$ min | $\beta^{-}$, $\gamma$ | spallation(Ar) | $<3.6 \times 10^{-3}$ | [94, 112] |
| $^{39}_{17}$Cl | $5.62(6) \times 10^{1}$ min | $\beta^{-}$, $\gamma$ | spallation(Ar) | $<2.5 \times 10^{-3}$ | [94, 113] |
| $^{81}_{36}$Kr | $2.29(11) \times 10^{5}$ a | EC, $\gamma$ | spallation(Kr) | $<2.7 \times 10^{-5}$ | [94, 103, 105, 114, 115] |
| $^{129}_{53}$I | $1.61(7) \times 10^{7}$ a | $\beta^{-}$, $\gamma$ | spallation(Xe) | $<7.6 \times 10^{-6}$ | [102, 104, 115] |

★ $\beta^{+}$: radioactive beta-decay with positron emission, $\beta^{-}$: radioactive beta-decay with electron emission, $\gamma$: gamma-ray emission, EC: electron capture decay, IC: internal conversion.

• These values characterize temporal and spatial averaged activity volume concentration in the Earth's atmosphere. In cases, where no empirical or simulation data is available, I constrain the activity concentration by adopting known generation rates for the corresponding cosmogenic radionuclide as discussed in Appendix A.3. The resulting upper limits are denoted by the symbol " < ".

○ The half-life value was taken from the unpublished report vol. 9 by the Bureau International des Poids et Mesures (BIMP).

† Altitude average between 0–3 km [96].

the concentration values are orders of magnitude lower than those of radon.[39] As such, we can deduce that gamma rays emitted by cosmogenic radionuclides in the atmosphere are overshadowed by other atmospheric sources such as radon or cosmic rays[40] (factor $10^3$ or more) and thereby irrelevant for AGRS applications. However, if you are interested in the production, migration and deposition of cosmogenic radionuclides on Earth as well as their applications as tracers and for dating purposes, I recommend the comprehensive monograph by Beer et al. [104].

39 *Please note that the estimated threshold in Table 2.3 are conservative upper limits. The actual values in the atmosphere may be significantly lower. Please refer to Appendix A.3 for more information.*

40 *As we will see in the next section.*

#### 2.1.3.4 Anthropogenic Radionuclides

Moving on from the natural radionuclides to the anthropogenic ones. There are over 1300 artificially created radionuclides, mainly produced by particle accelerators and nuclear reactors [42]. A detailed discussion of these radionuclides as well as the involved nuclear reactions are beyond the scope of this work. Here, I will only provide a brief overview of the most prominent anthropogenic radionuclides relevant for AGRS applications, i.e. those emitting high-energy photons and which can be found in significant quantities in the environment. The most significant source of anthropogenic radionuclides in the environment in terms of global emissions are the nuclear weapons tests carried out in the Cold War, followed by major nuclear accidents such as the ones at the Chernobyl Nuclear Power Station (Unit 4) in 1986 [3] and the Fukushima Daiichi Nuclear Power Station (Units 1, 2 and 3) in 2011 [4]. In Table 2.4, I have listed anthropogenic radionuclides with high-energy photon emissions and half-lives $t_{1/2} > 1\,\mathrm{d}$ released during the nuclear weapons tests between 1945 and 2000 as well as during the nuclear accidents at Chernobyl and Fukushima together with estimates on the related activity values provided by the United Nations Scientific Committee on the Effects of Atomic Radiation (UNSCEAR) [2–4]. These estimates are not only interesting in reconstructing what could have been measured with AGRS systems, but they might also provide a blueprint for the potential radionuclide vector in future nuclear incidents.

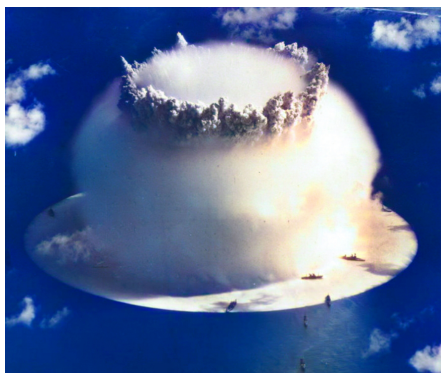

Operation Crossroads - Baker underwater test
U.S. Army Photographic Signal Corps

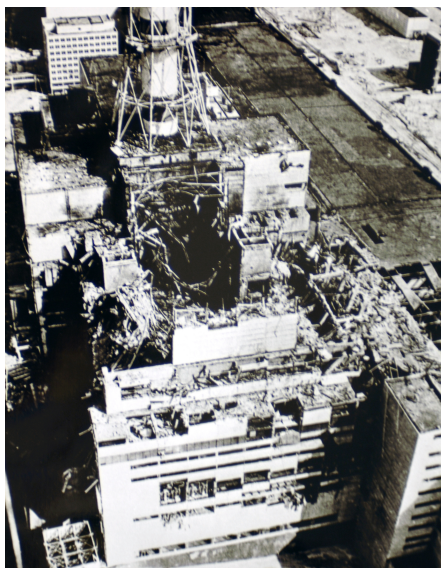

Chernobyl Unit 4 after the accident
IAEA Imagebank

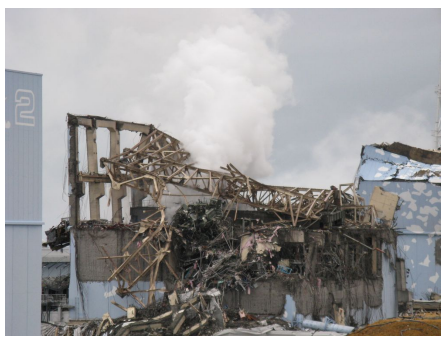

Fukushima Daiichi Unit 3 after the accident
Agency for Natural Resources and Energy

For major nuclear incidents, considering the half-lives and mobility in the environment [52, 70], the most important anthropogenic radionuclides in terms of activity releases and global distribution are the noble gases ($^{85}_{36}$Kr, $^{133}_{54}$Xe), the highly volatile iodides ($^{131}_{54}$I, $^{132}_{54}$I, $^{133}_{54}$I), tellurides ($^{129m}_{52}$Te, $^{132}_{52}$Te) as well as the cesium isotopes ($^{134}_{55}$Cs, $^{136}_{55}$Cs, $^{137}_{55}$Cs). In the immediate aftermath (first days after release), $^{132}_{52}$Te and the iodide radionuclides pose the most substantial threat by environmental contamination in terms of radiation dose. In contrast, over the long term (years following the incident), $^{137}_{55}$Cs becomes the predominant contributor to anthropogenic radiation in the environment for both, accidents at nuclear power plants and nuclear weapons explosions [2, 116, 117].[41] In fact, $^{137}_{55}$Cs from the Chernobyl accident can still be measured today in Switzerland using AGRS systems [118].

41 *At least outside the immediate vicinity of the incident site.*

Apart from nuclear incidents, there are also continuous gaseous radionuclide emissions from various facilities related to the industry, medicine and research, such as nuclear reactors or accelerators installations, some of which can emerge as an atmospheric gamma-ray source significant enough to be measurable with AGRS systems [119, 120]. For nuclear reactors, these are fission products, mainly $^{133}_{54}$Xe and $^{85}_{36}$Kr, but also activation products[42] from the coolant cycle, e.g. $^{16}_{7}$N or $^{41}_{18}$Ar in light-water reactors [121].[43] Accelerators on the other hand often produce proton-rich positron emitters such as $^{15}_{8}$O, $^{13}_{7}$N or $^{11}_{6}$C leading to annihilation photons as discussed in Section 2.1.1 [122]. In Figs. B.1–B.3, I have compiled selected graphs displaying the relative photon intensity $I_\gamma$ for the main anthropogenic radionuclides discussed above. From these, we see that the anthropogenic radionuclides emit photons in the energy range between about 10 keV and 1.5 MeV. There is one exception, $^{16}_{7}$N, which has main gamma-ray emissions at about 6129 keV and 7115 keV [123]. On the other end of the spectrum, $^{133}_{54}$Xe and $^{132}_{52}$Te possess comparably low maximum[44] photon emissions at 81 keV and 228 keV, respectively [85, 124].

42 *Please cf. Section 2.3.*

43 *Typically, $^{16}_{7}$N is emitted in significant quantities to the atmosphere only in boiling water reactors. This is attributed to the circulation of the coolant between the reactor and turbine building, a setup distinct from pressurized water reactors where the primary coolant circulates solely within the reactor building.*

44 *For $I_\gamma > 0.01\,\%$.*

**Table 2.4** Global radionuclide releases from nuclear accidents and nuclear weapons explosions.

| Nuclide★ | Half-life | Decay modes● | Decay products○ | Chernobyl activity releases [PBq]† | Fukushima activity releases [PBq]‡ | Nuclear weapons activity releases [PBq]§ | References |
|---|---|---|---|---|---|---|---|
| $^{54}_{25}$Mn | 312.19(3) d | EC, $\gamma$ | | | | 3980 | [102] |
| $^{55}_{26}$Fe | 2.747(8) a | EC, $\gamma$ | | | | 1530 | [68] |
| $^{85}_{36}$Kr | 10.752(23) a | $\beta^-, \gamma$ | | 33 | | | [102] |
| $^{89}_{38}$Sr | 50.57(3) d | $\beta^-, \gamma$ | | 115 | | 117 000 | [102] |
| $^{90}_{38}$Sr | 28.80(7) a | $\beta^-$ | $^{90}_{39}$Y | 10 | <0.01 | 622 | [68] |
| $^{91}_{39}$Y | 58.51(6) d | $\beta^-, \gamma$ | | | | 120 000 | [126] |
| $^{95}_{40}$Zr | 64.032(6) d | $\beta^-, \gamma$ | $^{95}_{41}$Nb | 84 | | 148 000 | [127] |
| $^{99}_{42}$Mo | 2.7479(6) d | $\beta^-, \gamma$ | $^{99}_{43}$Tc | >72 | | | [102] |
| $^{103}_{44}$Ru | 39.247(3) d | $\beta^-, \gamma$ | | >168 | | 247 000 | [128] |
| $^{106}_{44}$Ru | 371.5(21) d | $\beta^-$ | $^{106}_{45}$Rh | >73 | | 12 200 | [51] |
| $^{125}_{51}$Sb | 2.758 55(25) a | $\beta^-, \gamma$ | | | | 741 | [68] |
| $^{129m}_{52}$Te | 33.6(1) d | IC, $\gamma$ | $^{129}_{52}$Te | 240 | | | [129] |
| $^{132}_{52}$Te | 3.230(13) d | $\beta^-, \gamma$ | $^{132}_{53}$I | 1150 | | | [85] |
| $^{131}_{53}$I | 8.0233(19) d | $\beta^-, \gamma$ | | 1760 | 120 | 675 000 | [51] |
| $^{133}_{53}$I | 20.87(8) h | $\beta^-, \gamma$ | | 910 | | | [124] |
| $^{133}_{54}$Xe | 5.2474(5) d | $\beta^-, \gamma$ | | 6500 | 7300 | | [124] |
| $^{140}_{56}$Ba | 12.753(5) d | $\beta^-, \gamma$ | $^{140}_{57}$La | 240 | | 759 000 | [102] |
| $^{141}_{58}$Ce | 32.503(11) d | $\beta^-, \gamma$ | | 84 | | 263 000 | [108] |
| $^{144}_{58}$Ce | 284.89(6) d | $\beta^-, \gamma$ | $^{144}_{59}$Pr | 50 | | 30 700 | [51] |
| $^{134}_{55}$Cs | 2.0644(14) a | $\beta^-, \gamma$ | | 47 | 10 | | [108] |
| $^{136}_{55}$Cs | 13.01(5) d | $\beta^-, \gamma$ | | 36 | | | [130] |
| $^{137}_{55}$Cs | 30.018(22) a | $\beta^-, \gamma$ | | 85 | 10 | 948 | [68] |
| $^{239}_{93}$Np | 2.356(3) d | $\beta^-, \gamma$ | actinium series | 400 | | | [124] |
| $^{238}_{94}$Pu | 87.74(3) a | $\alpha, \gamma$ | uranium series | 0.015 | | | [65] |

*Continued on next page*

**Table 2.4** *Continued from previous page*

| Nuclide★ | Half-life | Decay modes● | Decay products○ | Chernobyl activity releases [PBq]† | Fukushima activity releases [PBq]‡ | Nuclear weapons activity releases [PBq]§ | References |
|---|---|---|---|---|---|---|---|
| $^{239}_{94}Pu$ | $2.4100(11) \times 10^4$ a | $\alpha, \gamma$ | actinium series | 0.013 | | 6.52 | [124] |
| $^{240}_{94}Pu$ | $6.561(7) \times 10^3$ a | SF, $\alpha, \gamma$ | thorium series | 0.018 | | 4.35 | [65] |
| $^{241}_{94}Pu$$ | 14.33(4) a | $\beta^-$ | neptunium series | 2.6 | | 142 | [124] |
| $^{242}_{96}Cm$ | 162.86(8) d | SF, $\alpha, \gamma$ | uranium series | 0.4 | | | [108] |
| | | | Radionuclide daughters# | | | | |
| $^{90}_{39}Y$ | 2.6684(13) d | $\beta^-, \gamma$ | | | | | [68] |
| $^{95}_{41}Nb$ | 34.991(6) d | $\beta^-, \gamma$ | | | | | [127] |
| $^{99}_{43}Tc$ | $2.115(11) \times 10^5$ a | $\beta^-, \gamma$ | | | | | [85] |
| $^{106}_{45}Rh$ | 30.1(3) s | $\beta^-, \gamma$ | | | | | [51] |
| $^{129}_{52}Te$ | 69.6(3) min | $\beta^-, \gamma$ | $^{129}_{53}I$ | | | | [129] |
| $^{129}_{53}I$ | $1.61(7) \times 10^7$ a | $\beta^-, \gamma$ | | | | | [102] |
| $^{132}_{53}I$ | 1.387(15) h | $\beta^-, \gamma$ | | | | | [131] |
| $^{140}_{57}La$ | 1.678 58(21) d | $\beta^-, \gamma$ | | | | | [102] |
| $^{144}_{59}Pr$ | 17.29(4) min | $\beta^-, \gamma$ | $^{144}_{60}Nd$●● | | | | [51] |

★ Only radionuclides are listed which exhibit gamma-ray emission directly or via daughter nuclides and have a half-life $t_{1/2} > 1$ d.
● $\alpha$: radioactive decay with alpha particle emission, $\beta^-$: radioactive beta-decay with electron emission, $\gamma$: gamma-ray emission, EC: electron capture decay, SF: spontaneous fission decay.
○ Only radioactive decay products are listed.
† Total activity released in the nuclear accident at the Chernobyl Nuclear Power Station (Unit 4) for each listed radionuclide [3].
‡ Total activity released in the nuclear accident at the Fukushima Daiichi Nuclear Power Station (Units 1, 2 and 3) for each listed radionuclide to the atmosphere [4].
§ Total activity released by nuclear weapons explosions for each listed radionuclide between 1945 and 2000 [2].
$ $^{241}_{94}Pu$ exhibits also a rare radioactive decay with alpha particle emission mode with a branching ratio of 0.002 44 % [124].
# For the actinium and thorium series, please refer to Tables C.2 and C.3. The radionuclide daughters associated with the neptunium series are readily available in the literature [42, 70].
●● $^{144}_{60}Nd$ has a half-life of $2.29(16) \times 10^{15}$ year and decays to the stable $^{140}_{58}Ce$ nuclide by alpha decay without gamma-ray emissions [125].

## 2.2 Cosmic Sources

### 2.2.1 Cosmic Rays

It was April 17th 1912, when Viktor Hess[45] together with the balloon pilot Hauptmann (major) Wilhelm Hoffory started with their balloon "Excelsior" to a scientific journey, which should change our understanding of ionizing radiation on Earth forever. During this balloon flight, Viktor Hess investigated the ionizing radiation in the atmosphere using three independent electroscopes. What he found was quite astonishing. Starting from an altitude of about 1500 m above ground, the ionizing radiation exponentially increased with altitude, which was a complete surprise to the scientific community at that time [132]. The prevailing opinion was that the ionizing radiation was generated solely by radionuclides on the ground as well as radon and associated progeny radionuclides in the air. The ionizing radiation was therefore expected to decrease with altitude. Moreover, to exclude the possibility that the ionizing radiation was generated by the Sun, the balloon flight was carefully timed in such a way, that Hess could perform the measurements during a partial solar eclipse, which took place on that day in Austria. Hess found no change in the ionizing radiation during the eclipse, which allowed him to exclude at least the possibility that ionizing radiation coming straight from the Sun was responsible for the observed trends in his measurements. Hess concluded that the ionizing radiation must come from outer space. This was the discovery of what we call today the "cosmic rays".[46]

Cosmic rays (CRs) are charged elementary particles and nuclei that pervade the universe and impact the top of Earth's atmosphere almost isotropically[47] with kinetic energies ($E_\mathrm{k}$) ranging from $\mathcal{O}(10^6)$ eV up to $\mathcal{O}(10^{20})$ eV [134].[48] The majority of CRs are protons (~80 %) and He nuclei (~15 %). The remaining part of the CRs consists of heavier nuclei, electrons, positrons and antiprotons. The CR composition changes significantly over the particle energy range, though [134].

At this point, I need also to clarify the terminology when it comes to CRs particles and gamma rays in space. By convention [134, 137], gamma rays coming from outer space are not counted as CR particles but as a separate category called "cosmic gamma rays".[49] The origin of both, CRs particles as well as cosmic gamma rays is still an active field of research [134, 137]. Possible sources for galactic cosmic gamma rays are pulsars, X-ray binaries and supernova remnants.

45 *Viktor Franz Hess (*1883, †1964), an Austrian experimental physicist. He is best known for the discovery of cosmic rays in 1912 using several balloon flights he performed himself together with Hauptmann Wilhelm Hoffory. For his discovery, he was awarded the Nobel Prize in Physics in 1936.*

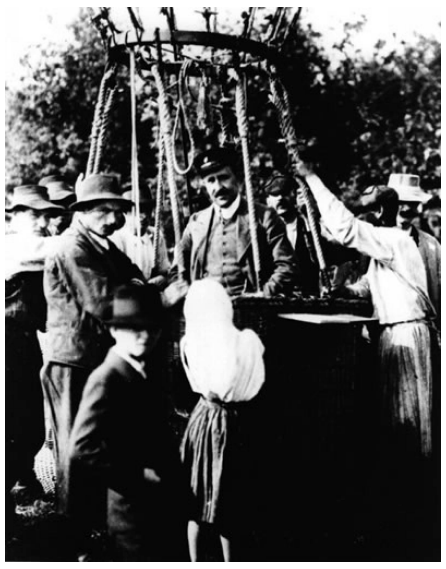

Viktor Hess back from his seventh balloon flight in August 1912 © American Physical Society

46 *This term was coined by Robert Millikan, who we have met already in Section 2.1.1 [133].*

47 *Angular anisotropies in the order of $\mathcal{O}(10^{-4})$ below ~$10^{15}$ eV and $\mathcal{O}(10^{-1})$ at the highest energies have been observed. The current understanding is that these anisotropies are due to galactic and extragalactic discrete CR sources and properties of the Galactic magnetic fields [134, 135].*

48 *The current record holder CR particle was detected in 1991 and had a kinetic energy of $3.2(9) \times 10^{20}$ eV [136]. That is roughly the energy equivalent of printing this entire book on paper and throwing it out of a first-floor window.*

49 *The same is true for cosmic neutrinos [137].*

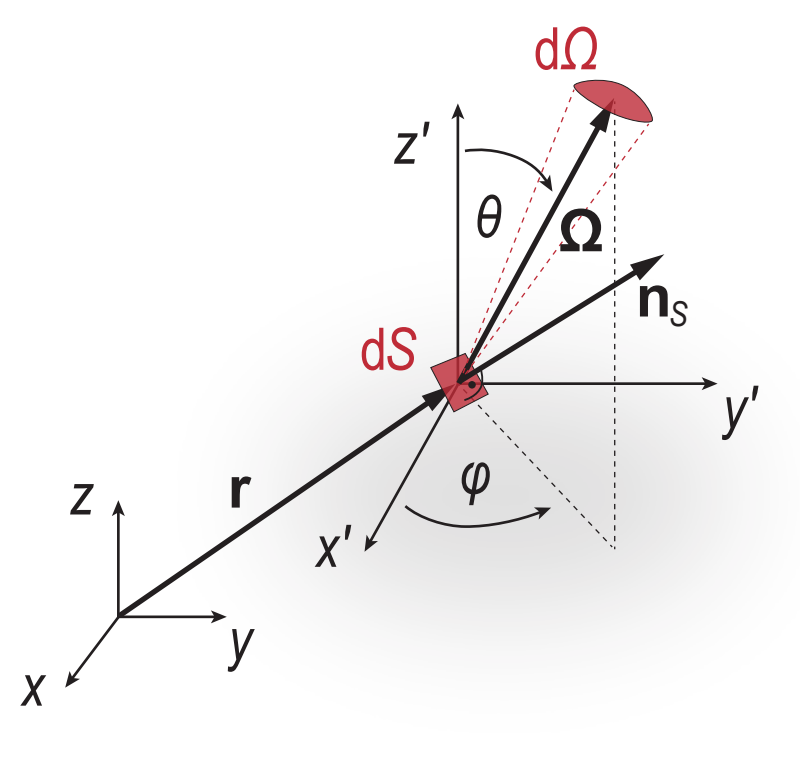


**Figure 2.3** Graphical depiction of position and direction related variables for radiation transport. These variables are the position vector $\mathbf{r}$ given in Cartesian coordinates $(x, y, z)$ and $(x', y', z')$, the direction vector $\mathbf{\Omega}$ parametrized by the polar angle $\theta$ and the azimuthal angle $\varphi$. In addition, the differential surface area $\mathrm{d}S$ together with its unit normal vector $\mathbf{n}_S$ as well as the differential solid angle $\mathrm{d}\Omega$ are displayed, too.

For extragalactic gamma rays, certain types of active galactic nuclei as well as gamma-ray bursts have been identified to contribute to the diffuse gamma-ray background [137]. Newest studies have shown that the interaction of CRs with the Sun and the Moon can generate gamma rays, too [138–142]. With various sources of ionizing radiation originating from outer space, the question arises: How do we accurately quantify these CR particles?

To answer this question, I need to introduce some more radiation transport terminology. In general, if we want to quantify the state of ionizing radiation particles of type $p$ for a given point in time and space, we may describe this state using the following seven independent variables: the position vector $\mathbf{r}$ (3 variables), the direction/direction-of-flight unit vector $\mathbf{\Omega}$ (2 variables with $|\mathbf{\Omega}| = 1$), the kinetic energy $E_\mathrm{k}$ (1 variable) and the time $t$ (1 variable). By definition, kinetic energy and time are scalar variables, whereas position and direction are vector variables defined in any coordinate system, which is convenient for the specific problem at hand. The direction vector is commonly parametrized using the polar angle $\theta$ and the azimuthal angle $\varphi$. In Cartesian coordinates, the direction vector $\mathbf{\Omega}$ is then defined as follows:

$$\mathbf{\Omega}\,(\theta, \varphi) = \begin{pmatrix} \sin\theta\cos\varphi \\ \sin\theta\sin\varphi \\ \cos\theta \end{pmatrix} \tag{2.18}$$

with:

| | | |
|---|---|---|
| $\theta$ | polar angle | rad |
| $\varphi$ | azimuthal angle | rad |

To describe an integral direction range, we may use the solid angle $\Omega$, which is defined in the context of this book as the integral of the differential solid angle $d\Omega$ over a predefined part of the unit sphere (given by the angular limits $[\theta_1, \theta_2]$ and $[\varphi_1, \varphi_2]$):

$$d\Omega = \sin(\theta)\, d\theta d\varphi \tag{2.19a}$$

$$\Omega = \int_{\varphi_1}^{\varphi_2} \int_{\theta_1}^{\theta_2} d\Omega \tag{2.19b}$$

$$= \left[\cos(\theta_1) - \cos(\theta_2)\right] \left[\varphi_2 - \varphi_1\right] \tag{2.19c}$$

where the associated unit of the solid angle is the steradian (sr) [38]. A graphical depiction of all the introduced position and direction-related variables is given in Fig. 2.3.

Using these seven variables, we may define the double differential[50] particle flux $\partial^2\phi_p/\partial E_k \partial\Omega$ for a given particle $p$ as the differential number $dN_p$ of particles $p$ with energies $[E_k, E_k + dE_k]$ crossing a differential surface area element $dS$ at a given position defined by the position vector $\mathbf{r}$ and coming from a solid angle $d\Omega$ in the time $dt$ in units of $s^{-1}\, m^{-2}\, eV^{-1}\, sr^{-1}$ as [134, 143, 144]:

50 *The "double" is referring to the kinetic energy $E_k$ and the solid angle $\Omega$.*

$$\frac{\partial^2\phi_p}{\partial E_k \partial\Omega}\left(\mathbf{r}, \boldsymbol{\Omega}, E_k, t\right) = \frac{dN_p}{dE_k d\Omega dS dt} \tag{2.20}$$

with:

| | | |
|---|---|---|
| $E_k$ | kinetic energy | eV |
| $N_p$ | number of particles | |
| $S$ | surface area | $m^2$ |
| $t$ | time | s |
| $\Omega$ | solid angle | sr |

and where we assume that $\boldsymbol{\Omega}$ is perpendicular to $dS$, i.e. $\boldsymbol{\Omega} \cdot \mathbf{n}_S = 1$ with $\mathbf{n}_S$ being the surface unit normal vector and "$\cdot$" the scalar product. The double differential flux $\phi_p$ may be integrated over a predefined range of solid angles $\Delta\Omega$ and kinetic energies $\Delta E_k$ separately to obtain the energy flux $d\phi_p/dE_k$ and the angular flux $d\phi_p/d\Omega$ as:

$$\frac{\mathrm{d}\phi_p}{\mathrm{d}E_\mathrm{k}}\left(\mathbf{r}, E_\mathrm{k}, t\right) = \int_{\Delta\Omega} \phi_p\left(\mathbf{r}, \mathbf{\Omega}, E_\mathrm{k}, t\right)\mathrm{d}\Omega \tag{2.21a}$$

$$\frac{\mathrm{d}\phi_p}{\mathrm{d}\Omega}\left(\mathbf{r}, \mathbf{\Omega}, t\right) = \int_{\Delta E_\mathrm{k}} \phi_p\left(\mathbf{r}, \mathbf{\Omega}, E_\mathrm{k}, t\right)\mathrm{d}E_\mathrm{k} \tag{2.21b}$$

in units of $\mathrm{s}^{-1}\,\mathrm{m}^{-2}\,\mathrm{eV}^{-1}$ and $\mathrm{s}^{-1}\,\mathrm{m}^{-2}\,\mathrm{sr}^{-1}$, respectively. Or we integrate over both variables combined to obtain the total flux[51] for a particle $p$ as a function of the position $\mathbf{r}$ and time $t$ in units of $\mathrm{s}^{-1}\,\mathrm{m}^{-2}$:

51 *Please note that the terminology, names, and symbols for these flux quantities vary significantly across scientific disciplines. Bell et al. [143] offers a comprehensive overview of the terminological diversity in various fields. The terminology introduced in this work loosely aligns with that commonly found in nuclear reactor and cosmic ray physics.*

$$\phi_p\left(\mathbf{r}, t\right) = \int_{\Delta\Omega}\int_{\Delta E_\mathrm{k}} \phi_p\left(\mathbf{r}, \mathbf{\Omega}, E_\mathrm{k}, t\right)\mathrm{d}E_\mathrm{k}\mathrm{d}\Omega \tag{2.22}$$

If we integrate not only over predefined ranges of the solid angles $\Delta\Omega$ and the kinetic energies $\Delta E_\mathrm{k}$ but also a given time interval $\Delta t$, we obtain the total fluence $\psi_p$ for a particle $p$ in units of $\mathrm{m}^{-2}$ as a function of the position $\mathbf{r}$:

$$\psi_p\left(\mathbf{r}\right) = \int_{\Delta t}\int_{\Delta\Omega}\int_{\Delta E_\mathrm{k}} \frac{\partial^2\phi_p}{\partial E_\mathrm{k}\partial\Omega}\left(\mathbf{r}, \mathbf{\Omega}, E_\mathrm{k}, t\right)\mathrm{d}E_\mathrm{k}\mathrm{d}\Omega\mathrm{d}t \tag{2.23}$$

Now, coming back to our CR particles, we can quantify CR nuclei in near-Earth space beyond the Earth's magnetosphere using one of the new quantities, we have learned. That is the isotropic[52] double differential flux defined in Eq. 2.20 at a fixed distance of 1 au[53] from the Sun, which we may estimate using the Matthiä model [145]:

52 *No explicit dependence on* $\mathbf{\Omega}$.

53 *The astronomical unit* (au) *is defined as* 1 au = 149 597 870 700 m, *which is approximately the mean distance between the Earth and the Sun [38].*

$$\frac{\partial^2\phi_p}{\partial E_\mathrm{k}\partial\Omega}\left(\mathbf{r}, E_\mathrm{k}, t\right) = \frac{a_{1,p}\beta^{a_{2,p}-1}}{\mathfrak{R}\left(E_\mathrm{k}\right)^{a_{3,p}}}\left\{\frac{\mathfrak{R}\left(E_\mathrm{k}\right)}{\mathfrak{R}\left(E_\mathrm{k}\right) + \left[c_1 + c_2 W\left(t\right)^{1.45}\right]}\right\}^{a_4 W(t) + a_5}\frac{A_p}{Z_p} \tag{2.24}$$

where:[54]

54 *Please note that I drop the* $|\cdot|$ *operator for* $Z_p$ *used by Matthiä et al., because in the context of this work, the atomic number* $Z_p$ *is always positive.*

| | | |
|---|---|---|
| $A_p$ | mass number | |
| $a_{1,p}$ | model parameter | $s^{-1}\,m^{-2}\,eV^{-1}\,sr^{-1}\,V^{a_{3,p}}$ |
| $a_{2,p}$ | model parameter | |
| $a_{3,p}$ | model parameter | |
| $a_3$ | model parameter | |
| $a_4$ | model parameter | |
| $c_1$ | model constant (= 0.37 V) | V |
| $c_2$ | model constant (= $3 \times 10^{-4}$ V) | V |
| $W$ | solar modulation index | |
| $Z_p$ | atomic number | |
| $\mathfrak{R}$ | rigidity | V |
| $\beta$ | ratio of the particle speed to the speed of light in vacuum | |

for a given CR nuclide $p$ with mass number $A_p$ and atomic number $Z_p$ as well as the nuclide-specific model parameters $a_{1,p}$ and $a_{2,p}$ provided by Matthiä et al. [145]. $\mathfrak{R}$ is the rigidity of the particle as a function of the kinetic energy $E_k$. I will discuss this quantity below. Last but not least, we have the solar modulation index $W$, which is a mechanistic parameter quantifying the solar activity as a function of time $t$ together with the two fixed model parameters $a_3$ and $a_4$.[55] Why do we have to consider solar activity, if Hess showed already more than a century ago, that the radiation coming straight from the Sun is not responsible for the cosmic rays?

Well, while the Sun does not directly produce cosmic rays, it does influence them nonetheless. There is a strong anti-correlation between solar activity and the CR flux near Earth. We can see this if we compare the number of sunspots, which are a measure of the solar activity[56], with the CR flux measured here at Earth as shown in Fig. 2.4. The reason for this anti-correlation is the solar wind[57], which correlates with the solar activity. When CR particles enter the heliosphere[58], they interact with the solar wind coming from the Sun by inelastic scattering, which decelerates and dampens the CR flux in the inner part of the solar system [39]. We call this effect of the Sun on the cosmic rays also "solar modulation".

But not only the solar wind has an impact on the CR flux. Earth's magnetic field, also known as the geomagnetic field (GMF), affects the CR flux near Earth, too. The GMF deflects charged particles, which are not coming from the direction of the poles, away from

55 *$a_3$ and $a_4$ depend on the detector system, which is used to derive W [145].*

56 *The solar activity has both periodic and aperiodic features as a function of time. Well-known periodic cycles are the 27-day cycle due to the Sun's rotation, the sunspot-related 11-year cycle and the 22-year Hale cycle [132]. Aperiodic features are related for example to solar flares or coronal mass ejections.*

57 *The solar wind is a continuous outflow of hot plasma from the corona of the Sun and consists mainly of highly ionized hydrogen and helium atoms and an approximately equal number of electrons [39, 137]. Typical kinetic energies for the protons are about* ~0.5 keV.

58 *Region in space, where the influence of the Sun by the solar wind dominates (up to* ~120 au *from the Sun)*

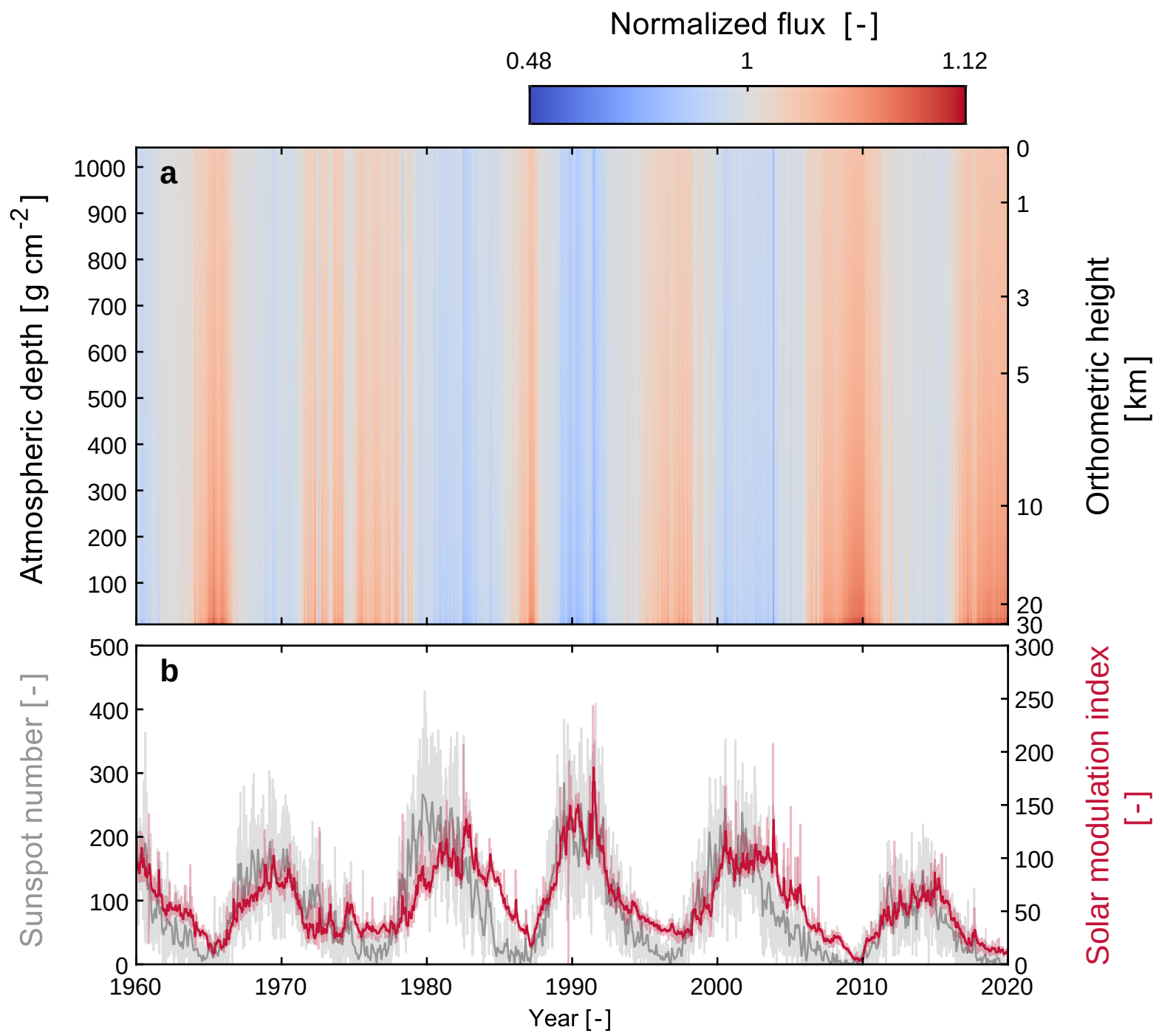


**Figure 2.4** Temporal evolution of the sunspot number and the cosmic-ray flux in the Earth's atmosphere between 1960 and 2020 (Gregorian calendar base). **a** The cosmic-ray flux was estimated using the PHITS-based Analytical Radiation Model in the Atmosphere (`PARMA`) code (version: `4.13`) [146] for a reference location at 45°N, 0°E. For all major cosmic-ray particles ($\gamma$, n, $e^{\pm}$, $\mu^{\pm}$, p, $\alpha$), the double differential flux was integrated over the full solid angle as well as the entire energy range available in `PARMA`, i.e. $10^{-2}$ eV to $10^{12}$ eV for neutrons and $10^{4}$ eV to $10^{12}$ eV for all remaining particles. The displayed total cosmic-ray flux represents the sum of all evaluated particle fluxes, screened with a three-day moving median filter and normalized to the temporal median cosmic-ray flux at each evaluated atmospheric depth. The maximum and minimum values are both observed at high altitudes ≥30 km and correspond to a 12 % increase and a 52 % decrease with respect to the temporal median flux at the corresponding altitude, respectively. The atmospheric depth ($d_{\text{atm}}$) was computed based on the NRLMSISE-00 atmospheric model [147]. **b** Sunspot number $S_{\text{N}}$ and solar modulation index $W$ time series (daily and monthly averaged datasets). $S_{\text{N}}$ was extracted from the World Data Center Sunspot Index and Long-term Solar Observations (WDC-SILSO) database (version: `2.0`) by the Royal Observatory of Belgium, Brussels [148, 149]. The solar modulation index $W$ was computed by the `PARMA` code (version: `4.13`).

the Earth's surface and thereby shields the Earth partly from the cosmic rays. To quantify the resistance of particles to this deflection by magnetic fields, we may define the rigidity $\mathfrak{R}$ as a function of the kinetic energy $E_\mathrm{k}$ in units of volt (V) as follows [137]:

$$\mathfrak{R}\left(E_\mathrm{k}\right) = \frac{E_\mathrm{k}}{Z\mathfrak{e}} \tag{2.25}$$

where:[59]

| | | |
|---|---|---|
| $E_\mathrm{k}$ | kinetic energy | J |
| $\mathfrak{e}$ | elementary charge (cf. Constants) | C |
| $Z$ | atomic number | |

The neat thing about this quantity is that different particles with the same rigidity have the same deflection, i.e. follow identical paths in a magnetic field.

To account for the GMF, we may therefore define an effective vertical cut-off rigidity[60] $\mathfrak{R}_\mathrm{c}$ as the minimum rigidity required for a particle to overcome trapping in the GMF and thereby eventually reach the top of the Earth's surface [137, 150]. Because the GMF is quite complex and evolves over time and space, estimating the vertical cut-off rigidity $\mathfrak{R}_\mathrm{c}$ is not trivial. Thanks to advanced numerical simulation studies adopting trajectory-tracing methods [150], today, world-grids for $\mathfrak{R}_\mathrm{c}$ with a spatial resolution of 5° in latitude, 15° in longitude and a temporal resolution of 5 a are available [150]. These simulations adopt high-fidelity GMF models such as the International Geomagnetic Reference Field (IGRF) [151]. In Fig. 2.5, the current best-estimate world map for $\mathfrak{R}_\mathrm{c}$ is shown. From this, we see that the vertical cut-off rigidity $\mathfrak{R}_\mathrm{c}$ varies significantly with the latitude and to a smaller degree with the longitude. The latitude anisotropy is mainly attributed to the core GMF, which may be approximated by a simple magnetic dipole. The longitude anisotropy can be explained by the fact that the geomagnetic dipole is asymmetrically located and tilted with respect to the Earth's rotation axis [137, 150]. In addition, magnetic anomalies as well as external field contributions by the ionosphere and the magnetosphere distort the GMF as well [150].

### 2.2.2 Extensive Air Shower

Once the CR particles reach the top of the Earth's atmosphere, they interact with the atmospheric constituents and produce a cascade

59 *Please note the similar symbols for the Euler's number* e *and the elementary charge* $\mathfrak{e}$.

60 *Often called simply cut-off rigidity [137].*

61 *Note that the flux of cosmic gamma rays is multiple orders of magnitude smaller than that of CR particles [134]. Consequently, cosmic gamma rays are in most cases neglected for the quantification of secondary particles in the Earth's atmosphere [146].*

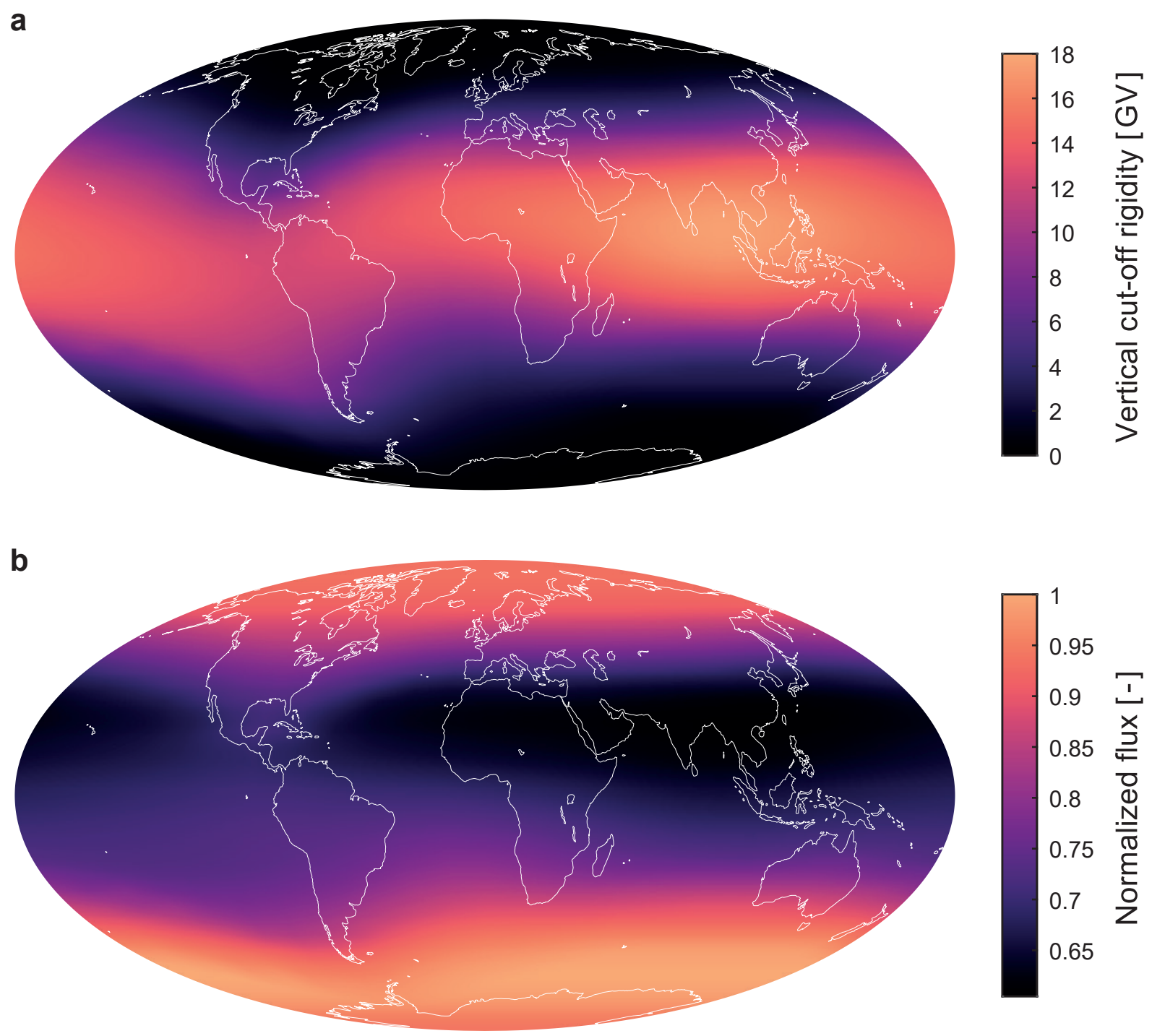


**Figure 2.5** World map (Mollweide projection) displaying CR related quantities. **a** Vertical cut-off rigidity ($\mathfrak{R}_c$) derived from a high-fidelity dataset based on the International Geomagnetic Reference Field (IGRF) for epoch 2020 [150]. **b** Normalized total cosmic ray induced ionizing particle flux ($\phi_{tot}$) ($\gamma$, n, $e^{\pm}$, $\mu^{\pm}$, p, $\alpha$) at sea level on 2022-05-04 estimated using the `PARMA` code (version: `4.13`). The double differential flux was integrated over the full solid angle as well as the entire energy range available in `PARMA`, i.e. $10^{-2}$ eV to $10^{12}$ eV for neutrons and $10^{4}$ eV to $10^{12}$ eV for all remaining particles. The resulting total flux was then normalized by the peak value ($\max(\phi_{tot}) = 0.19\,\mathrm{s^{-1}\,cm^{-2}}$). For the vertical cut-off rigidity $\mathfrak{R}_c$, the high-fidelity dataset displayed in **a** was used. Note that the current version of `PARMA` adopts a less accurate dataset for $\mathfrak{R}_c$ only accounting for the first order dipole moment of the geomagnetic field [146]. To derive the atmospheric depth ($d_{atm}$), the NRLMSISE-00 database was consulted [147].

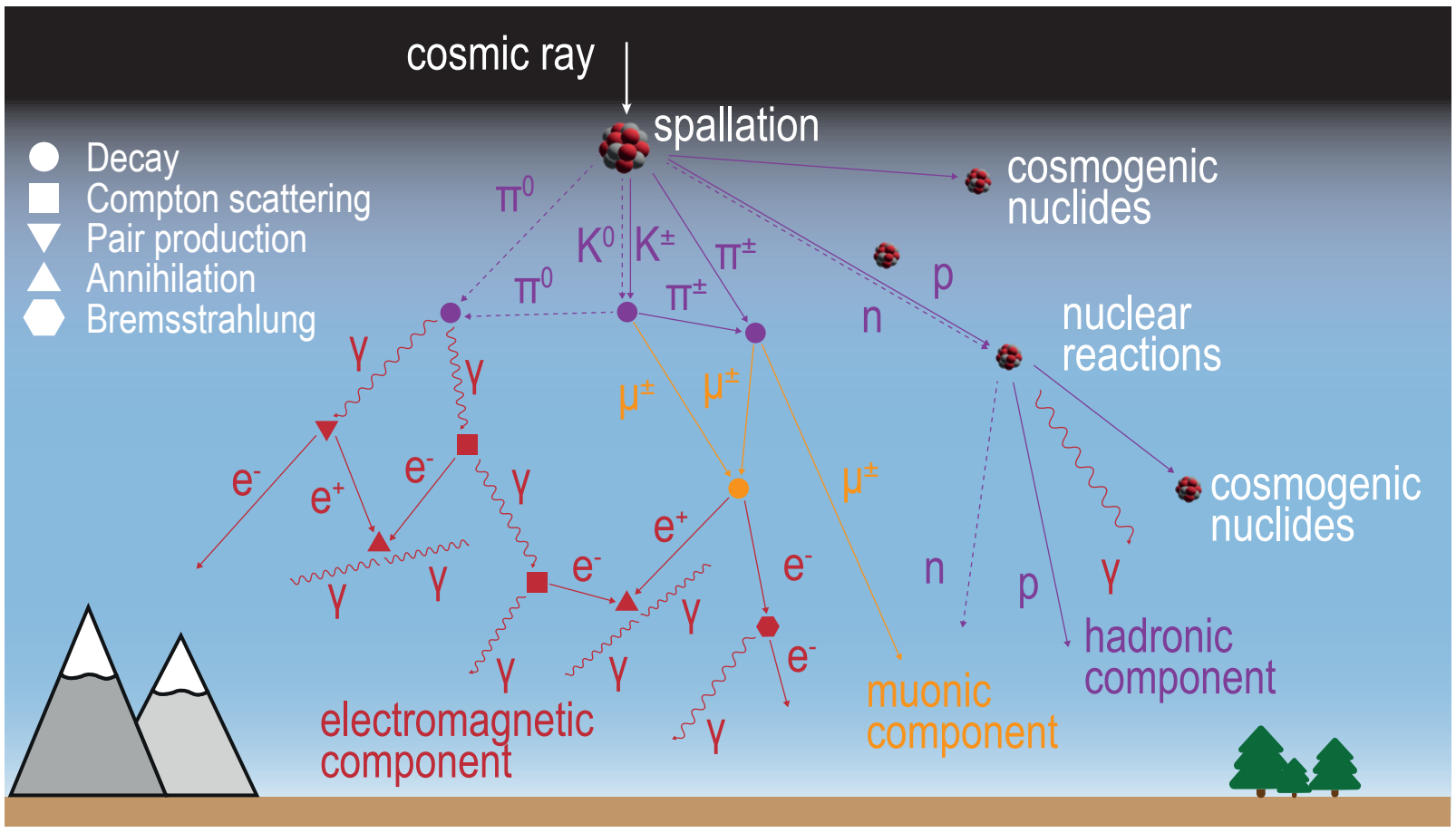


**Figure 2.6** Schematic depiction of an extensive air shower in Earth's atmosphere. This includes the three main components and associated particles: electromagnetic ($\gamma$: gamma ray, $e^-$: electron, $e^+$: positron), muonic ($\mu^-$: muon, $\mu^+$: antimuon) as well as the hadronic component (n: neutron, p: proton, $\pi^0$: neutral pion, $\pi^\pm$: charged pions, $K^0$: neutral kaon, $K^\pm$: charged kaons). In addition, main interaction modes are shown (see also Chapter 3 and Sections 2.1 and 2.3). This scheme was adapted from Grieder [137] and Zombeck [152].

of secondary particles as graphically depicted in Fig. 2.6.[61] A single CR particle may produce millions to billions of secondary particles, which eventually reach the ground with a lateral spread up to $\mathcal{O}(10)\,\mathrm{km}^2$ [134]. We call this cascade of secondary particles therefore also an extensive air shower (EAS). All EASs possess three distinct components: The "hadronic" component consists of high energetic heavy ions, protons and neutrons as well as some short-lived hadrons like neutral and charged pions or kaons generated by strong interactions.[62] This component includes also the cosmogenic radionuclides, which we have discussed in Section 2.1.3. The "muonic" component consists, as the name implies, of muons[63] ($\mu^\pm$), which are mainly generated by the decay of short-lived charged pions[64] ($\pi^\pm$). Last but not least, there is the "electromagnetic" component consisting of electrons, positrons and gamma rays. Electromagnetic shower components are mainly initialized by the decay of neutral pions[65] ($\pi^0$). The main processes leading to high-energy photon emission in an EAS are [137]:[66]

62 *We will discuss some of the related reactions later in this chapter.*

63 *An elementary particle, similar to electrons, with a rest mass of* $105.658\,375\,5(23)\,\mathrm{MeV/c^2}$, *a charge of* $1\mathrm{e}$ *and a mean life of* $2.196\,981\,1(22) \times 10^{-6}\,\mathrm{s}$ *[134]. There is also an antiparticle of opposite charge but equal mass called the "antimuon".*

64 *Subatomic hadron particles (meson) with a rest mass of* $139.570\,39(18)\,\mathrm{MeV/c^2}$, *a charge of* $\pm 1\mathrm{e}$ *and a mean life of* $2.6033(5) \times 10^{-8}\,\mathrm{s}$ *[134].*

1. Decay of neutral pions, i.e. $\pi^0 \rightarrow \gamma + \gamma$, creating two high-energy photons with a combined energy of at least 134 MeV [134].
2. Bremsstrahlung of electrons and positrons (cf. Section 2.1.1).
3. Electron-positron annihilation (cf. Section 2.1.1).
4. De-excitation of highly-excited nuclei after nuclear reactions such as spallation or radiative capture (cf. Section 2.3.1).

Due to the complex coupled interactions between the various secondary particles in EASs over multiple orders of magnitude in energy, estimating all the different secondary particle double differential fluxes in the Earth's atmosphere as a function of space $\mathbf{r}$, angular direction $\mathbf{\Omega}$, kinetic energy $E_\mathrm{k}$ and time $t$ is a challenging task. Advances in computational hardware made it possible to perform high-fidelity numerical simulations of the entire EASs transport in the Earth's atmosphere [99, 153–158]. One of the most accurate codes for this purpose is the `PARMA` (PHITS-based Analytical Radiation Model in the Atmosphere) code, which is a meta-model framework developed by Sato [146] and based on high-fidelity numerical simulations with the Monte Carlo code PHITS [23]. `PARMA` provides accurate estimates of the double differential flux $\partial^2\phi_p/\partial E_\mathrm{k}\partial\Omega(\mathbf{r}, \mathbf{\Omega}, E_\mathrm{k}, t)$ for electrons ($\mathrm{e}^-$), positrons ($\mathrm{e}^+$), muons ($\mu^-$), antimuons ($\mu^+$), gamma rays ($\gamma$) and heavy nuclei (H–Ni) globally, from the top of the atmosphere to the sea level, with an energy range of $10^4$ eV to $10^{12}$ eV[67] and with a time resolution of 1 d since 1614 to the present [146, 159]. The primary CR flux near Earth is estimated using the already presented Matthiä model (cf. Eq. 2.24). The solar modulation index $W$ is computed based on the methodology established by Matthiä et al. [145] using neutron monitor data, which is readily available from many stations all over the world since 1951 [146].[68] Before 1951, $W$ is estimated using a reconstruction method established by Usoskin et al. [160].

To get a feeling for the transport of primary CR particles through the atmosphere, I will discuss now some general properties of the flux field for selected secondary particles in the lower atmosphere using `PARMA` (version: `4.13`).[69] Note that not only the secondary gamma rays but also other CR induced secondary particles contribute to the measurement signal in AGRS systems, as I will demonstrate in Part IV. This is why I not only cover gamma rays here but extend the discussion also to other secondary particles relevant in AGRS. First, let us have a look at the lateral variation of the total CR induced

65 *Electrically neutral equivalent of $\pi^\pm$ with a lower rest mass* 134.9768(5) MeV/c$^2$ *and lower mean life of* $8.43(13) \times 10^{-17}$ s. *[134]*

66 *There are also various other nuclear interactions, which may generate gamma rays. Grieder [137] provides an in-depth analysis of gamma ray generation in EASs in Chapter 6. I will discuss some of those later in Chapter 3.*

67 *For neutrons, the lower kinetic energy threshold is* $10^{-2}$ eV. *For muons, the upper energy threshold is at* $10^{14}$ eV.

68 *If you want to use neutron monitor data yourself, have a look at the neutron monitor database (NMDB): https://www.nmdb.eu/.*

69 *Please note that I used the updated version with a patch to account for the production of annihilation photons. The missing annihilation photons were identified in preliminary simulations for this work and subsequently reported to the developers.*

ionizing particle flux at sea level shown in Fig. 2.5. As expected, the GMF through the vertical cut-off rigidity causes a significant angular anisotropy in the flux values, i.e. the total flux is reduced by more than 40 % close to the equator compared to the polar regions. The magnitude of this anisotropy is more pronounced at higher altitudes (cf. Fig. B.6). Due to the different rigidities, there is also a significant difference in this spatial anisotropy between the secondary particles (cf. Figs. B.7–B.14). As an example, the muons are much less affected by the GMF than the electrons are.

Going on, let us next investigate how the flux of secondary particles evolves across the Earth's atmosphere. For that purpose, first, we need to discuss briefly the different measures, which can be used to quantify the vertical position within the atmosphere. As shown already by Hess with his balloon flight, the CR induced ionizing radiation flux is a strong function of the altitude. The reason for this is simply that the atmosphere, similar to the GMF, acts as a shield against the primary CR particles. On the other hand, the more atmosphere the primary CR particles traverse, the more secondary particles are produced. So, there is a trade-off between attenuation and generation of secondary particles. To quantify the vertical column of atmosphere a primary CR particle traverses, in particle physics, we often use the so-called atmospheric depth $d_{\text{atm}}$[70] in units of $\text{kg}\,\text{m}^{-2}$ [134, 146] defined as:

70 *Sometimes also referred to as vertical column density [137].*

$$d_{\text{atm}}(h) = \int_{h}^{h_{\text{top}}} \rho_{\text{atm}}(h')\,\mathrm{d}h' \tag{2.26a}$$

$$\approx p_{\text{atm}}(h)/\text{g} \tag{2.26b}$$

where:

| | | |
|---|---|---|
| $\text{g}$ | standard gravity (cf. Constants) | $\text{m}\,\text{s}^{-2}$ |
| $h$, $h'$ | altitude | m |
| $h_{\text{top}}$ | top atmosphere altitude | m |
| $p_{\text{atm}}$ | atmospheric pressure | Pa |
| $\rho_{\text{atm}}$ | atmospheric mass density | $\text{kg}\,\text{m}^{-3}$ |

Depending on the selected datum for $h$, we distinguish[71] between the orthometric height ($h_{\text{ort}}$) with a reference geoid as datum and the geodetic height ($h_{\text{geo}}$) with a reference ellipsoid as datum. Atmospheric models to compute $\rho_{\text{atm}}(h)$ or $p_{\text{atm}}(h)$ are readily available,

71 *Meyer et al. provide a comprehensive summary about the different height systems and related datum quantities such as the geoid or reference ellipsoid [161–163].*

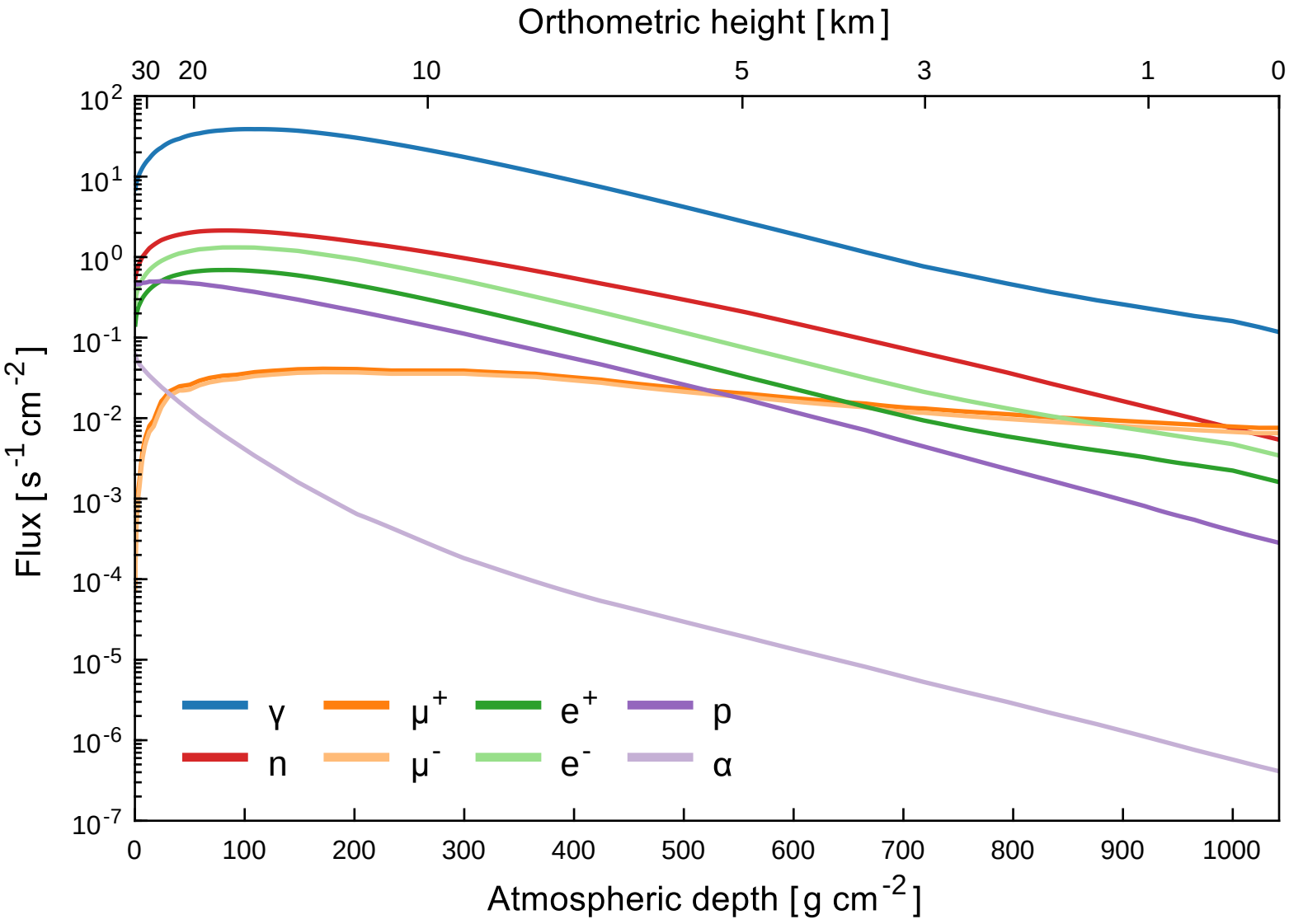


**Figure 2.7** Cosmic-ray induced total flux in the Earth's atmosphere for various secondary particles ($\gamma$: gamma ray, n: neutron, $e^-$: electron, $e^+$: positron, $\mu^-$: muon, $\mu^+$: antimuon, p: proton, $\alpha$: alpha particle) as a function of the atmospheric depth. The cosmic-ray flux was estimated using the `PARMA` code (version: `4.13`) for a reference location and date at 45°N, 0°E and 2022-03-21, respectively [146]. The cosmic-ray flux was integrated over the full solid angle as well as the entire energy range available in `PARMA` for each particle type, i.e. $10^{-2}$ eV to $10^{12}$ eV for neutrons and $10^4$ eV to $10^{12}$ eV for all remaining particles. The atmospheric depth ($d_{\text{atm}}$) was computed based on the NRLMSISE-00 atmospheric model [147].

such as the COESA, the NRLMSISE-00 or the ISA standard atmospheric models [147, 164, 165].

In Fig. 2.7, the CR induced total flux for selected secondary particles as a function of the atmospheric depth $d_{\text{atm}}$ and the orthometric height $h_{\text{ort}}$ is shown. As expected, we observe an exponential increase in the flux for $d_{\text{atm}} \gtrsim 300\,\text{g/cm}^2$ with a peak at $d_{\text{atm}} \sim 100\,\text{g/cm}^2$ for all particles, except for protons and helium ions due to the fact that these two ions are also part of the primary CR spectrum. Second, we find a significant difference in the magnitude of $\phi_p$ for the different secondary particles with the gamma rays having the highest values followed by the neutrons and all the other particles. In general, as

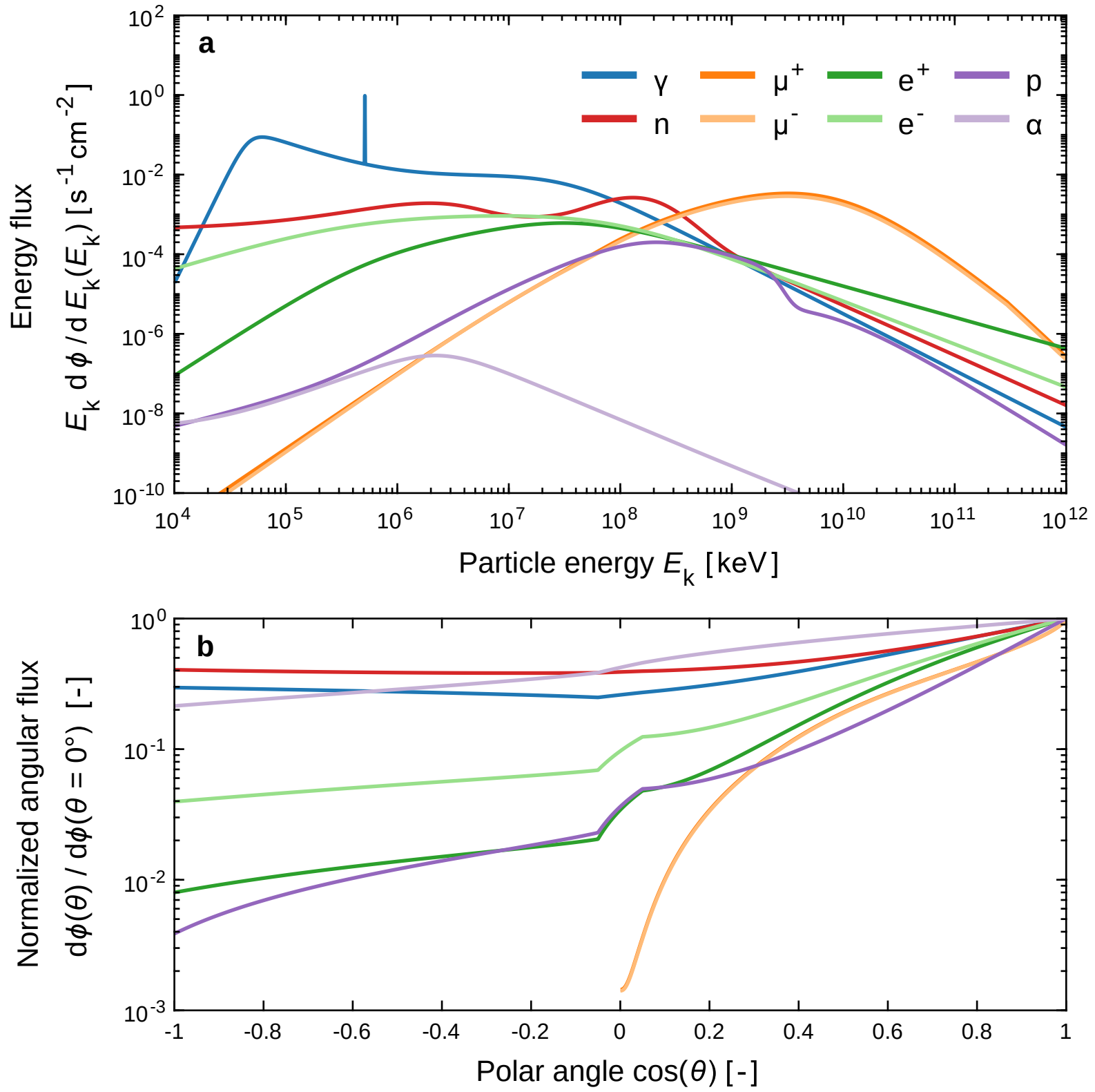


**Figure 2.8** Cosmic ray induced energy and angular flux for various particles ($\gamma$: gamma ray, n: neutron, $e^-$: electron, $e^+$: positron, $\mu^-$: muon, $\mu^+$: antimuon, p: proton, $\alpha$: alpha particle) in the lower atmosphere. All fluxes were estimated using the `PARMA` code (version: `4.13`) for a reference location, date and atmospheric depth of 46.75°N, 7.60°E, 2022-06-16 and 955 g cm$^{-2}$, respectively [146]. **a** Energy flux $\mathrm{d}\phi_p/\mathrm{d}E_\mathrm{k}(E_\mathrm{k})$ as a function of the kinetic energy $E_\mathrm{k}$. $\mathrm{d}\phi_p/\mathrm{d}E_\mathrm{k}$ was derived by integrating the double differential flux over the full solid angle. **b** Angular flux $\mathrm{d}\phi_p/\mathrm{d}\Omega(\theta)$ as a function of the polar angle $\theta$ and normalized by $\mathrm{d}\phi_p/\mathrm{d}\Omega(\theta = 0°)$. $\mathrm{d}\phi_p/\mathrm{d}\Omega$ was derived by integrating the double differential flux over the entire energy range available in `PARMA` for each particle type, i.e. $10^{-2}$ eV to $10^{12}$ eV for neutrons and $10^4$ eV to $10^{12}$ eV for all remaining particles.

the atomic number of the ions increases, the total flux of these ions tends to decrease [146].

Going back to Fig. 2.4, we see in subgraph a that not only the primary CR flux described by Eq. 2.24 evolves with time but also the secondary particle flux across the entire atmosphere shows a significant temporal variation. The maximum and minimum values are both observed at high altitudes ≥ 30 km and correspond to a 12 % increase and a 52 % decrease with respect to the temporal median flux at the corresponding altitude, respectively.[72]

72 *An interesting sidenote: the composition of the secondary particle flux, i.e. the relative contributions of the individual particles to the combined secondary particle flux, shows only minor variation over time <1 %, even during solar or geomagnetic storms (cf. Fig. B.15).*

As a last step, let us have a closer look at the different energies and directions of the secondary particles in the lower atmosphere. For that purpose, I have plotted the energy and angular flux in Fig. 2.8 for selected secondary particles. We find that gamma rays and neutrons dominate the lower part of the energy spectrum, whereas at higher energies, muons and electrons are the most abundant particles. The pronounced peak in the gamma-ray spectrum at 511 keV is a result of the electron-positron annihilation (cf. Eq. 2.8). The angular flux is also strongly anisotropic with a peak at the zenith ($\cos(\theta) = 1$) and a minimum at the nadir ($\cos(\theta) = -1$). The two main reasons for this are that the Earth blocks primary CR particles for $\cos(\theta) < 0$, and that primary CR particles with $\cos(\theta) > 0$ experience minimal attenuation as they traverse the atmosphere coming from the zenith direction.[73]

73 *Note that Sato assumes* $\phi_{\mu^\pm}(\cos(\theta) < 0) = 0$ *in the* `PARMA` *code [146]. Moreover, Sato adopts a four-component model approach to compute the polar angle dependence, which leads to the observed discontinuities around* $\cos(\theta) = 0$ *[159].*

## 2.3 Other Sources

Having discussed now in-depth radionuclides and CRs, in this last section, I would like to briefly highlight two other sources of gamma rays, which can play a role in some specific AGRS applications. These sources are nuclear reactions and terrestrial gamma-ray flashes (TGFs).

### 2.3.1 Nuclear Reactions

Let us start with nuclear reactions. As with the radioactive decay discussed in Section 2.1.1, nuclear reactions also create excited nuclei $Y^*$. However, in contrast to the radioactive decay, the excited nuclear state $Y^*$ is not induced by radioactive decay but by a nuclear reaction between the original nucleus X and some other high-energy particle:

$$X + x \rightarrow Y^* \tag{2.27}$$

The resulting excited nucleus $Y^*$ may then de-excite by emitting gamma rays as discussed in Eq. 2.5. Since this de-excitation typically lasts only a fraction of a second, we refer to this gamma-ray emission as "prompt" gamma rays. Furthermore, the resulting nucleus may also be radioactive and subsequently decay, emitting, what we call in the context of nuclear reactions "delayed" gamma rays.

As I will show in Part IV, prompt gamma rays generated by CR induced nuclear reactions are a common source of the background in AGRS. Therefore, I will provide here a short overview of four common nuclear reaction classes and their associated gamma-ray emissions. These reactions are:

1. radiative capture
2. fission
3. fusion
4. spallation

Let us start with the radiative capture reaction. In a radiative capture reaction[74], a nucleus absorbs a neutron forming a new excited nuclide with the mass number increased by one. The excited nucleus subsequently de-excites emitting a variable number $n$ of prompt gamma rays $\gamma$ [121]:

$$ {}^{A}_{Z}\mathrm{X} + {}^{1}_{0}\mathrm{n} \longrightarrow {}^{A+1}_{\phantom{A+}Z}\mathrm{X}^{*} \longrightarrow {}^{A+1}_{\phantom{A+}Z}\mathrm{X} + n\gamma \tag{2.28} $$

which we may also write in short form as ${}^{A}_{Z}\mathrm{X}(\mathrm{n},\gamma){}^{A+1}_{\phantom{A+}Z}\mathrm{X}$. If the resulting de-excited nuclide is radioactive, we call this nuclide also an activation product.[75] Like the gamma rays emitted by radionuclides, the prompt gamma rays emitted in radiative capture reactions are also discrete and characteristic for the involved nuclides.[76] Most of the prompt gamma rays emitted by radiative capture reactions cover an energy range between 10 keV and 10 MeV.

In contrast to radiative capture, fission and fusion reactions are rarely observed in the environment on Earth. In most cases, they are occurring under controlled conditions in nuclear fission and fusion reactors. In nuclear fission, a free neutron induces an atomic nucleus to split into two or more smaller nuclei and several free neutrons.[77] This is the core process in every nuclear reactor and nuclear fission weapon discussed in Section 2.1.3. Because of the statistical nature of the fission fragment distribution as well as continuum state transitions, the prompt gamma-ray spectrum related to fission reactions is a

74 *Sometimes also referred to as neutron capture reactions [166].*

75 *As an example, the anthropogenic ${}^{16}_{7}\mathrm{N}$ discussed in Section 2.1.3 is such an activation product.*

76 *A common example for a radiative capture reaction is ${}^{1}_{1}\mathrm{H}(\mathrm{n},\gamma){}^{2}_{1}\mathrm{H}$ emitting a single high-energy gamma-ray photon with an energy of about* 2223 keV *[167].*

77 *Please note that there is also a rare spontaneous fission (SF) decay mode, in which a heavy nucleus undergoes fission without a neutron inducing it [30].*

combination of a continuum component with discrete emission lines [168]. The total energy released by prompt gamma rays per neutron-induced fission reaction is typically in the range of 5 MeV to 7 MeV with a mean energy per gamma ray of ~0.7 MeV to ~1.0 MeV [169–173].[78]

78 *Note that the energy of the prompt gamma rays depends on the nuclide undergoing fission as well as the energy of the incident neutron [168, 174].*

In fusion reactions, instead of splitting a nucleus, two light nuclei are merged to form a heavier nucleus, in the process releasing nucleons and in some rare cases also prompt gamma rays. This is the core reaction in every fusion reactor or thermonuclear weapon. Like radiative capture, fusion reactions emit also prompt gamma rays with discrete energies characteristic of the resulting fused nucleus. For example, following gamma ray energies have been experimentally determined for three of the most intensively studied fusions reactions [175]: ~13.5 MeV and 16.75 MeV for ${}^{2}_{1}\mathrm{H}({}^{3}_{1}\mathrm{H},\gamma){}^{5}_{2}\mathrm{He}$ [176–184], 23.8 MeV for ${}^{2}_{1}\mathrm{H}({}^{2}_{1}\mathrm{H},\gamma){}^{4}_{2}\mathrm{He}$ [176, 185, 186], ~12 MeV and 16.66 MeV for ${}^{2}_{1}\mathrm{H}({}^{3}_{2}\mathrm{He},\gamma){}^{5}_{3}\mathrm{Li}$ [187], respectively. The number of gamma rays released per fusion reaction is typically in the range of $10^{-5}$ to $10^{-4}$, i.e. significantly lower than the gamma-ray yields achieved in fission reactions.

The spallation reaction is by far the most complex nuclear reaction. We have met this reaction already in Section 2.1.3, when we were discussing the interaction of CRs with the Earth's atmosphere. In a spallation reaction, a high-energy particle, in most cases a hadron, with a kinetic energy of at least 100 MeV per nucleon collides with an atomic nucleus [188, 189]. In the initial collision stage, the incident projectile interacts directly with individual nucleons in the target nucleus (intranuclear cascade) leading to the emission of various high-energy hadrons. In the second stage called the evaporation stage, the newly formed excited nucleus de-excites in the processing emitting nucleons and prompt gamma rays. The emitted secondary particles may decay, such as the $\pi^0$, or interact with the surrounding atmospheric molecules triggering an entire cascade of nuclear reactions, including fission and radiative capture, creating additional prompt gamma rays [190]. Consequently, the prompt gamma-ray spectrum from spallation reactions is typically a combination of a continuum attributed to fission and decay of mesons such as neutral pion ($\pi^0$) together with discrete emission lines related to the various excited nuclei formed during the spallation and subsequent radiative capture reactions [190]. If you want to learn more about this fascinating topic, I recommend the *Handbook of Spallation Research: Theory, Experiments and Applications* by Filges and Goldenbaum [190].

### 2.3.2 Terrestrial Gamma-Ray Flashes

The second gamma-ray source I would like to highlight here are terrestrial gamma-ray flashes (TGFs). TGFs are short, intense bursts of high-energy gamma rays generated in thunderstorms in Earth's atmosphere and were first discovered in 1991 by the Burst and Transient Source Experiment (BATSE) on board the Compton Gamma-Ray Observatory (CGRO) spacecraft [191]. Typical TGFs have an average pulse duration of ~0.5 ms [192] and a continuous[79] energy spectrum extending up to ~100 MeV [194]. The intensity of these TGFs is remarkably high reaching a gamma-ray fluence at low Earth orbit of $\psi_\gamma \sim 0.7\,\mathrm{cm}^{-2}$ with estimates for the number of photons released per flash being between $10^{17}$ and $10^{19}$ [195–197].

79 *The spectral index and high energy cut-off parameters are typically about* 0.45 *and* 8.8 MeV, *respectively [193].*

Recent spaceborne observations coupled with numerical simulations suggest that TGFs are produced in the electric fields associated with lightning leaders in the top layers of thunderclouds with source altitude estimates ranging from 12 km to 21 km [197–200]. The detailed production mechanisms are still not fully understood. The current best accepted theory is based on the relativistic runaway electron avalanche (RREA) process [193]. In this process, seed electrons[80] with kinetic energies above a few hundred eV are accelerated to relativistic energies by the strong electric fields in thunderstorm cells. Concurrently, inelastic collisions with atmospheric molecules lead to the generation of secondary electrons, triggering an avalanche of high-energy electrons. These high-energy electrons are then able to generate high-energy photons by bremsstrahlung as described in Section 2.1.[81]

80 *The origin of these seed electrons is still unclear. Possible options are cosmic ray interactions with the atmosphere or cold electrons in streamers [193, 198].*

81 *Please note that, despite being generated through bremsstrahlung, the emitted photons are referred to as gamma rays. This nomenclature persists because, at the time of the discovery of TGFs, the source of the observed high-energy gamma rays was still unclear. Considering the high energies produced by TGFs, I will relax the terminology introduced in Section 2.1 to align with the prevalent practice of calling the high-energy photons produced in TGFs gamma rays.*

Gamma rays produced by TGFs were not only observed by spaceborne instruments but also by aircraft flying at a cruising altitude of ~14 km [201] as well as by multiple ground-based detector arrays [202–205]. These results demonstrate that TGFs can not only be probed by spaceborne instruments but also by low and mid-altitude AGRS systems. However, with a global occurrence rate of only $220\,\mathrm{d}^{-1}$ to $570\,\mathrm{d}^{-1}$ [193], TGFs are rare events and thereby only relevant for AGRS in dedicated survey flights close to thunderstorm cells and not for typical AGRS survey flights in Switzerland.

To summarize all gamma-ray sources discussed in this chapter, a schematic overview of the main gamma-ray sources on Earth and in space is included in Fig. 2.9. The figure also highlights the different energy ranges of the gamma rays produced by the various sources. With that, I conclude the discussion of gamma-ray sources in the

environment. In the next chapter, we will discuss how the gamma rays interact with the environment and subsequently our detector systems in AGRS.

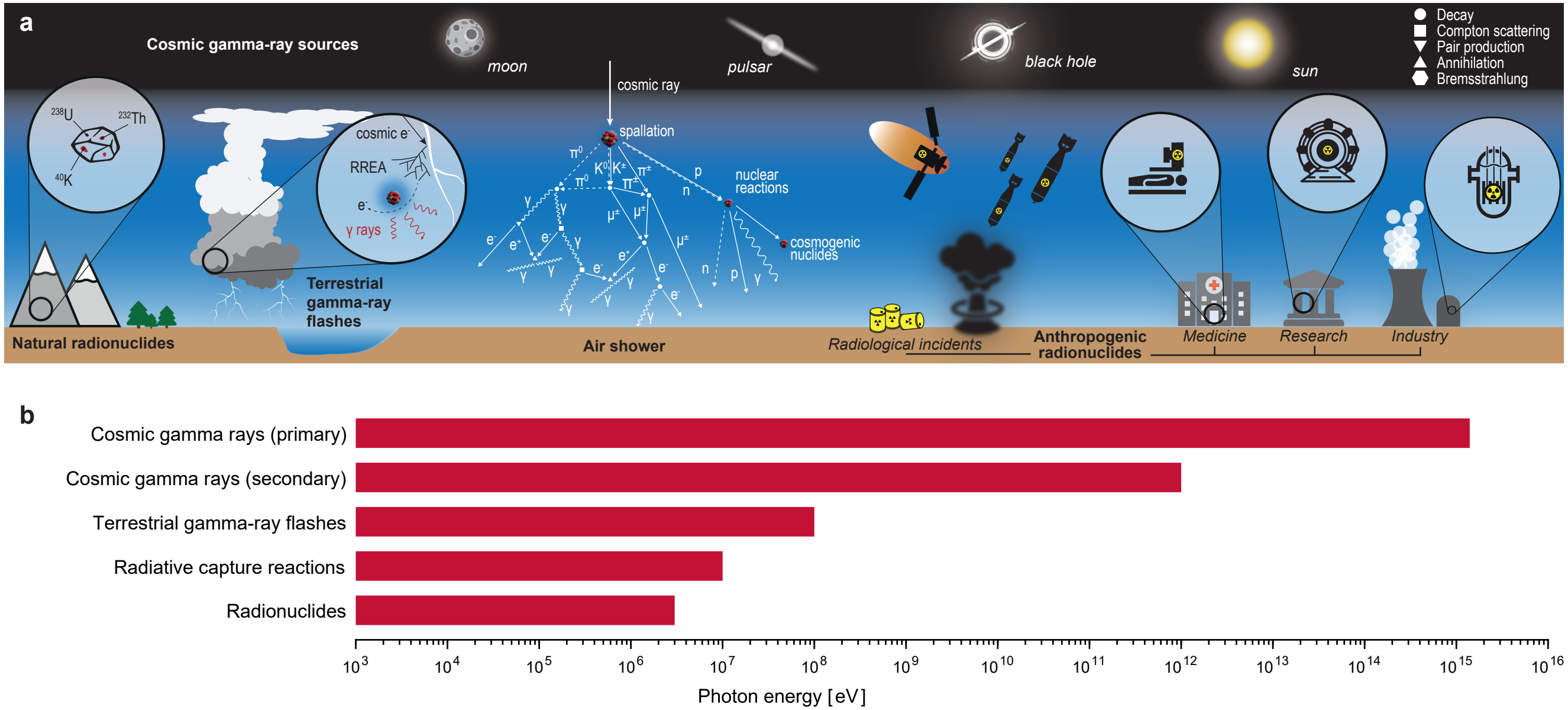


**Figure 2.9** Gamma rays in the sky. **a** Overview of gamma-ray sources on Earth and in space. This includes primary cosmic gamma rays in space, secondary cosmic gamma rays generated in EASs (γ: gamma ray, n: neutron, $e^-$: electron, $e^+$: positron, $\mu^-$: muon, $\mu^+$: antimuon, p: proton, $\pi^0$: neutral pion, $\pi^\pm$: charged pions, $K^0$: neutral kaon, $K^\pm$: charged kaons) in the Earth's atmosphere, gamma rays generated by terrestrial gamma-ray flashes (TGFs) in thunderstorms and natural as well as anthropogenic radionuclides. The air shower scheme was adapted from Grieder [137] and Zombeck [152]. **b** Spectral range of the main categories of gamma-ray sources in the environment on Earth and in space.

*"We must be clear that when it comes to atoms, language can be used only as in poetry."*

*— Niels Bohr*

# Chapter 3
# Interaction with Matter

**Contents**

In the chapter before, we have discussed how gamma rays are generated in the environment. In the first part of this chapter, I will give a brief overview of the various ways these generated gamma rays interact with matter. In the second part, I will present two approaches to predict the transport of gamma rays in the environment considering all the interaction modes presented in the first part.

The purpose of this chapter is not to provide a detailed discussion on the mathematical modeling of photon-matter interactions but rather to give a general overview of the main characteristics and properties of the high-energy photon transport relevant for AGRS. We will need the ideas and concepts introduced in this chapter in the next chapter when we will discuss gamma-ray spectrometry as well as later in this book to set up physics models and interpret measurement and simulation results.

## 3.1 Interaction of High-Energy Photons with Matter

As discussed in Section 2.1.1, high-energy photons can be described as point-like particles or quanta of light. But do they also behave as particles when they interact with matter? To answer this question, I have to introduce the wavelength of a photon $\lambda_\gamma$ in units of m [206]:

$$\lambda_\gamma = \frac{c}{f} \tag{3.1}$$

where:

| | | |
|---|---|---|
| c | speed of light in vacuum (cf. Constants) | $\mathrm{m\,s^{-1}}$ |
| $f$ | electromagnetic wave frequency | Hz |

using this definition, we can rewrite the Planck-Einstein relation in Eq. 2.6 as:

$$E_\gamma = \frac{hc}{\lambda_\gamma} \tag{3.2}$$

with:

| | | |
|---|---|---|
| c | speed of light in vacuum (cf. Constants) | $\mathrm{m\,s^{-1}}$ |
| $E_\gamma$ | photon energy | J |
| h | Planck constant (cf. Constants) | J s |
| $\lambda_\gamma$ | photon wavelength | m |

Now, when it comes to the interaction of photons with matter, or more specifically with the atoms in the matter, we may distinguish two main cases [206]: If the wavelength of the photon is much larger than the size of the atoms, i.e. $\lambda_\gamma \gg \mathcal{O}(10^{-10})$ m, the wave-character of the photon prevails, and it will interact with the matter as a whole. Under these circumstances, the interaction depends on various macroscopic properties of the material such as molecular or lattice structures. Conversely, when the photon wavelength is comparable or smaller than the size of individual atoms, i.e. $\lambda_\gamma \leq \mathcal{O}(10^{-10})$ m, the particle-character of the photon dominates. In this scenario, we may treat the interaction of photons with matter as an interaction between point-like particles with free atoms.

We can express the threshold between these two worlds not only in terms of the photon wavelength but also in terms of the photon energy using the adapted Planck-Einstein relation in Eq. 3.2. For an atom-size equivalent photon wavelength of $\lambda_\gamma \sim 10^{-10}$ m, we get a threshold energy of $E_\gamma \sim 10$ keV. As we have seen in the previous chapter, gamma rays typically have energies $E_\gamma \gtrsim 10$ keV. Consequently, we may describe the transport of gamma rays through matter as a series of particle-like interactions between point-like photons with free atoms, given the gamma-ray energy is not $\ll$10 keV. It is important to note that, considering the different sizes of atoms, 10 keV is not a threshold set in stone. As an example, nuclear models based on a particle-like approach typically provide accurate predictions down to a photon energy threshold of $E_\gamma \sim 1$ keV [207, 208]. As a result, I will implicitly assume $E_\gamma \geq 1$ keV in the remainder of this chapter.[1]

Before I continue presenting the main interaction mechanisms of gamma rays with matter, first I need to introduce a fundamental quantity in photon transport: the cross-section.

### 3.1.1 Cross-Section

The cross-section is the core quantity in radiation transport to quantify interactions of high-energy photons with matter. The following discussion is a summary of the monographs by Kogan et al. [52], Bell et al. [143], Prinja et al. [144], and Shultis et al. [209] as well as the thesis by Leppänen [210].[2] As discussed above, we assume here that

1 *So, if you find the term "high-energy photon" in the remainder of this chapter, this term denotes a photon with* $E_\gamma \geq 1$ keV.

2 *Similar to the flux quantities introduced in Section 2.2.1, the terminology for cross-sections is again not consistent across the different scientific fields. As a result, I adapt the notation and symbols in this section to align with that introduced in Section 2.2.1.*

the photon interacts with individual atoms as a point-like particle. Under this assumption, we may define the microscopic cross-section $\sigma_{i,j}$ as the differential number of interactions $\mathrm{d}N_{\mathrm{int},i,j}$ of type $i$ per differential time $\mathrm{d}t$ taking place between individual atoms of nuclide type $j$ and high-energy photons with a time-independent homogeneous[3] photon energy flux $\mathrm{d}\phi_\gamma/\mathrm{d}E_\gamma(E_\gamma)$ (cf. Eq. 2.21a) normalized by $\mathrm{d}\phi_\gamma/\mathrm{d}E_\gamma(E_\gamma)$ as well as the differential photon energy $\mathrm{d}E_\gamma$:

3 *No explicit dependence on the position* $\mathbf{r}$ *(cf. Eq. 2.21a).*

$$\sigma_{i,j}\left(E_\gamma\right) = \frac{\mathrm{d}N_{\mathrm{int},i,j}}{\frac{\mathrm{d}\phi_\gamma}{\mathrm{d}E_\gamma}\,\mathrm{d}E_\gamma\,\mathrm{d}t} \tag{3.3}$$

where:

| | | |
|---|---|---|
| $E_\gamma$ | photon energy | eV |
| $N_{\mathrm{int},i,j}$ | number of interactions of type $i$ with nuclide $j$ | |
| $t$ | time | s |
| $\frac{\mathrm{d}\phi_\gamma}{\mathrm{d}E_\gamma}$ | photon energy flux | $\mathrm{s}^{-1}\,\mathrm{m}^{-2}\,\mathrm{eV}^{-1}$ |

From Eq. 3.3, we see that the microscopic cross-section shares the same dimensional units as a geometric area, i.e. $\mathrm{m}^2$. As a result, we can interpret the microscopic cross-section either geometrically as an effective cross-sectional area or probabilistically as the probability of the interaction process $i$ to occur between an individual nuclide atom $j$ and an incoming high-energy photon with energy $E_\gamma$ normalized by the associated photon energy flux, photon energy and time. Similar to the small dimension of photon energies discussed in Section 2.1.1, the microscopic cross-section in $\mathrm{m}^2$ is typically also very small, e.g. for common photon interaction modes we have cross-sectional values in the order of $10^{-30}\,\mathrm{m}^2$ to $10^{-21}\,\mathrm{m}^2$. Therefore, we often use a different unit for microscopic cross-sections, the barn b, which is defined as $1\,\mathrm{b} = 10^{-28}\,\mathrm{m}^2$.[4]

4 *The barn was introduced 1941 by the nuclear physicists M. Holloway and C. Baker during the Manhattan Project to provide a more convenient unit for the cross-sections of nuclear reactions. According to Holloway and Baker, a cross-section of* $10^{-24}\,\mathrm{cm}^2$ *is really "as big as a barn" [211]. Note that, as of writing this book, the barn is not an official SI unit but is widely accepted both in nuclear and particle physics as well as in radiation measurement.*

There is one special photon interaction mechanism with matter, which requires an extension of the microscopic cross-section, that is scattering. In scattering, unlike absorption interactions, the photon does not get fully absorbed by the atom; instead, it gets scattered in a new direction. We distinguish two categories of scattering interactions: elastic and inelastic scattering. In elastic scattering, the photon retains its energy upon interaction with the atom, whereas in inelastic scattering, there is a change in the energy of the photon. Consequently, we need to describe not only the probability of the

interaction process to occur but also the properties, i.e. energy and direction, of the scattered photon, as the transport of the photon undergoing a scattering interaction does not end with the interaction. For that purpose, we can define the double differential cross-section for a scattering interaction in units of $m^2\,eV^{-1}\,sr^{-1}$ as the differential number of scattering events $dN_{int,s,j}$ per differential time $dt$, where a photon with energy $E_\gamma$ coming from a direction $\mathbf{\Omega}$ is scattered by an atom $j$ in a new direction $\mathbf{\Omega}'$ with energy $E'_\gamma$, normalized by the time-independent homogeneous double differential photon flux $\partial^2\phi_\gamma/\partial E_\gamma \partial\Omega(E_\gamma, \mathbf{\Omega})$ (cf. Eq. 2.20) and the differential photon energies $dE_\gamma$, $dE'_\gamma$ as well as the solid angles $d\Omega$, $d\Omega'$:

$$\frac{\partial^2\sigma_{s,j}}{\partial E'_\gamma \partial\Omega'}\left(E_\gamma, E'_\gamma, \mathbf{\Omega}\cdot\mathbf{\Omega}'\right) = \frac{dN_{int,s,j}\left(E_\gamma, E'_\gamma, \mathbf{\Omega}\cdot\mathbf{\Omega}'\right)}{\frac{\partial^2\phi_\gamma}{\partial E_\gamma \partial\Omega}\left(E_\gamma, \mathbf{\Omega}\right) dE_\gamma dE'_\gamma d\Omega d\Omega' dt} \tag{3.4}$$

where:

| | | |
|---|---|---|
| $E_\gamma$ | energy of incoming photon | eV |
| $E'_\gamma$ | energy of scattered photon | eV |
| $N_{int,s,j}$ | number of scattering events with nuclide $j$ | |
| $\frac{\partial^2\phi_\gamma}{\partial E_\gamma \partial\Omega}$ | double differential photon flux | $s^{-1}\,m^{-2}\,eV^{-1}\,sr^{-1}$ |
| $\Omega$ | solid angle of incoming photon | sr |
| $\Omega'$ | solid angle of scattered photon | sr |
| $\mathbf{\Omega}$ | direction vector of incoming photon | |
| $\mathbf{\Omega}'$ | direction vector of scattered photon | |

That is the general case for inelastic scattering interactions. For elastic scattering, we have by definition $E_\gamma = E'_\gamma$. Note also that, based on the assumption of isotropic media, I presume here rotational invariance for the scattering direction. Consequently, the probability of the scattering interaction only depends on the scattering angle $\theta_\gamma$ between $\mathbf{\Omega}$ and $\mathbf{\Omega}'$, which I indicate similar to Prinja et al. [144] by $\mathbf{\Omega}\cdot\mathbf{\Omega}' = \cos(\theta_\gamma)$.

In full analogy to the discussion on the flux quantity in Section 2.2.1, it is straightforward to introduce the single differential energy and angular cross-sections as the integrals over a predefined range of solid angles $\Delta\Omega'$ and energies $\Delta E'_\gamma$ for the scattered photon:

$$\frac{d\sigma_{s,j}}{dE'_\gamma}\left(E_\gamma, E'_\gamma\right) = \int_{\Delta\Omega'} \frac{\partial^2\sigma_{s,j}}{\partial E'_\gamma \partial\Omega'}\left(E_\gamma, E'_\gamma, \boldsymbol{\Omega}\cdot\boldsymbol{\Omega}'\right) d\Omega' \tag{3.5a}$$

$$\frac{d\sigma_{s,j}}{d\Omega'}\left(E_\gamma, \boldsymbol{\Omega}\cdot\boldsymbol{\Omega}'\right) = \int_{\Delta E'_\gamma} \frac{\partial^2\sigma_{s,j}}{\partial E'_\gamma \partial\Omega'}\left(E_\gamma, E'_\gamma, \boldsymbol{\Omega}\cdot\boldsymbol{\Omega}'\right) dE'_\gamma \tag{3.5b}$$

By integrating over both, $\Omega'$ and $E'_\gamma$, we obtain again the microscopic cross-section $\sigma_{i,j}$ defined in Eq. 3.3 with $i$ = s:

$$\sigma_{s,j}\left(E_\gamma\right) = \int\limits_{\Delta E'_\gamma}\int\limits_{\Delta\Omega'} \frac{\partial^2\sigma_{s,j}}{\partial E'_\gamma \partial\Omega'}\left(E_\gamma, E'_\gamma, \boldsymbol{\Omega}\cdot\boldsymbol{\Omega}'\right) d\Omega' dE'_\gamma \tag{3.6}$$

The cross-sections introduced so far only quantify the photon interaction with an individual atom. To describe the interaction with a macroscopic homogeneous[5] isotropic[6] material, I have to introduce another quantity, which is the attenuation coefficient $\mu$.[7] The attenuation coefficient $\mu_{i,j}$ is defined as the number of interactions $dN_{\text{int},i,j}$ of type $i$ between atoms of nuclide type $j$ and high-energy photons per differential time $dt$ taking place in a homogeneous material characterized by the mass density $\rho$ and a mass fraction $w_j$ normalized by the photon energy flux $d\phi_\gamma/dE_\gamma(E_\gamma)$ as well as the photon energy $dE_\gamma$:

$$\mu_{i,j}\left(E_\gamma\right) = \frac{dN_{\text{int},i,j}\left(E_\gamma\right) \mathrm{N_A} w_j \rho}{\frac{d\phi_\gamma}{dE_\gamma}\left(E_\gamma\right) dE_\gamma\, dt M_j} \tag{3.7a}$$

$$= \sigma_{i,j}\left(E_\gamma\right)\frac{\mathrm{N_A} w_j \rho}{M_j} \tag{3.7b}$$

where:

5 *Material whose properties do not depend on the position. From the Ancient Greek roots ομος (homos) and γενος (genos) translating to "of the same kind".*

6 *Material whose properties do not depend on the direction* $\boldsymbol{\Omega}$. *From the Ancient Greek roots ἴσος (isos) for "equal" and τροπικός (tropikos) for "pertaining to a turn".*

7 *This quantity is sometimes also called the macroscopic cross-section Σ, especially in the context of neutron transport [121, 143, 144].*

| | | |
|---|---|---|
| $M_j$ | molar mass of nuclide $j$ | $\mathrm{kg\,mol^{-1}}$ |
| $\mathrm{N_A}$ | Avogadro constant (cf. Constants) | $\mathrm{mol^{-1}}$ |
| $w_j$ | mass fraction of nuclide $j$ in matter | $\mathrm{kg\,kg^{-1}}$ |
| $\rho$ | mass density | $\mathrm{kg\,m^{-3}}$ |
| $\sigma_{i,j}$ | microscopic cross-section for interaction $i$ between high-energy photons and the nuclide $j$ | $\mathrm{m^2}$ |
| $\frac{\mathrm{d}\phi_\gamma}{\mathrm{d}E_\gamma}$ | photon energy flux | $\mathrm{s^{-1}\,m^{-2}\,eV^{-1}}$ |

In contrast to the microscopic cross-section, the attenuation coefficient has the dimension not of an area but of an inverse length in units of $\mathrm{m^{-1}}$. As a result, we may interpret the attenuation coefficient as the probability of the interaction process $i$ to occur between a high-energy photon and a nuclide $j$ per unit travel length of the photon in a homogeneous material characterized by the mass density $\rho$ and the mass fraction $w_j$. We can also define a total attenuation coefficient $\mu_{\mathrm{tot}}$[8] as the sum of the individual attenuation coefficients for all interaction processes $i$ and nuclides $j$ with the nuclide specific properties $M_j$ and $w_j$ for a given material as:

8 *Sometimes also referred to as linear attenuation coefficient [30, 206].*

$$\mu_{\mathrm{tot}}\left(E_\gamma\right) = \sum_i \sum_j \mu_{i,j}\left(E_\gamma\right) \tag{3.8a}$$

$$= \mathrm{N_A}\rho \sum_j \frac{w_j}{M_j}\left[\sum_i \sigma_{i,j}\left(E_\gamma\right)\right] \tag{3.8b}$$

There is one interesting relation between the total attenuation coefficient $\mu_{\mathrm{tot}}$ and the mean distance $\ell$ a photon travels in a given material before an interaction occurs. It can be shown that this distance $\ell$ is equivalent to the inverse of the total attenuation coefficient $\mu_{\mathrm{tot}}$ [30, 121, 144]:

$$\ell\left(E_\gamma\right) = \frac{1}{\mu_{\mathrm{tot}}\left(E_\gamma\right)} \tag{3.9}$$

We refer to this mean distance $\ell$ also as the mean free path. With these important quantities introduced, let us come back to the main interaction mechanisms of high-energy photons with matter.

### 3.1.2 Main Interaction Mechanisms

The interaction mechanisms between high-energy photons and matter can be classified into two main categories: scattering and absorption interactions. As discussed before, in scattering interactions the photon gets deflected whereas in absorption interactions the photon gets fully absorbed by the atom. There are five main interaction mechanisms of high-energy photons with matter: two modes of scattering interactions, Rayleigh and Compton scattering, and three modes of absorption interactions, photoelectric absorption, electron-positron pair production, and photonuclear reactions:

- I. Scattering interactions
  1. Rayleigh scattering
  2. Compton scattering
- II. Absorption interactions
  3. Photoelectric absorption
  4. Electron-positron pair production
  5. Photonuclear reactions

In the following, I will briefly highlight the main characteristics of these interaction mechanisms. The purpose of this section is not to provide a detailed discussion on the mathematical modeling of these interaction mechanisms but rather to give a general overview of the main properties of these mechanisms.

The most accurate framework to describe these interactions is quantum electrodynamics (QED) theory, which is beyond the scope of this work. Instead, I will limit the discussion on simple models to provide a qualitative discussion on the main trends in the interaction cross-sections with photon energy and material properties as well as the energies of the secondary particles produced during the interaction. For a more detailed discussion on the mathematical modeling of these interaction mechanisms, I recommend the monographs *An Introduction to the Passage of Energetic Particles through Matter* by Carron [206] and *The Atomic Nucleus* by Evans [212]. Furthermore, high-fidelity cross-section data for all interaction mechanisms described in this chapter can be found in open-access nuclear data libraries such as XCOM [208] or the Evaluated Photon Data Library, 1997 version (EPDL97) [207].

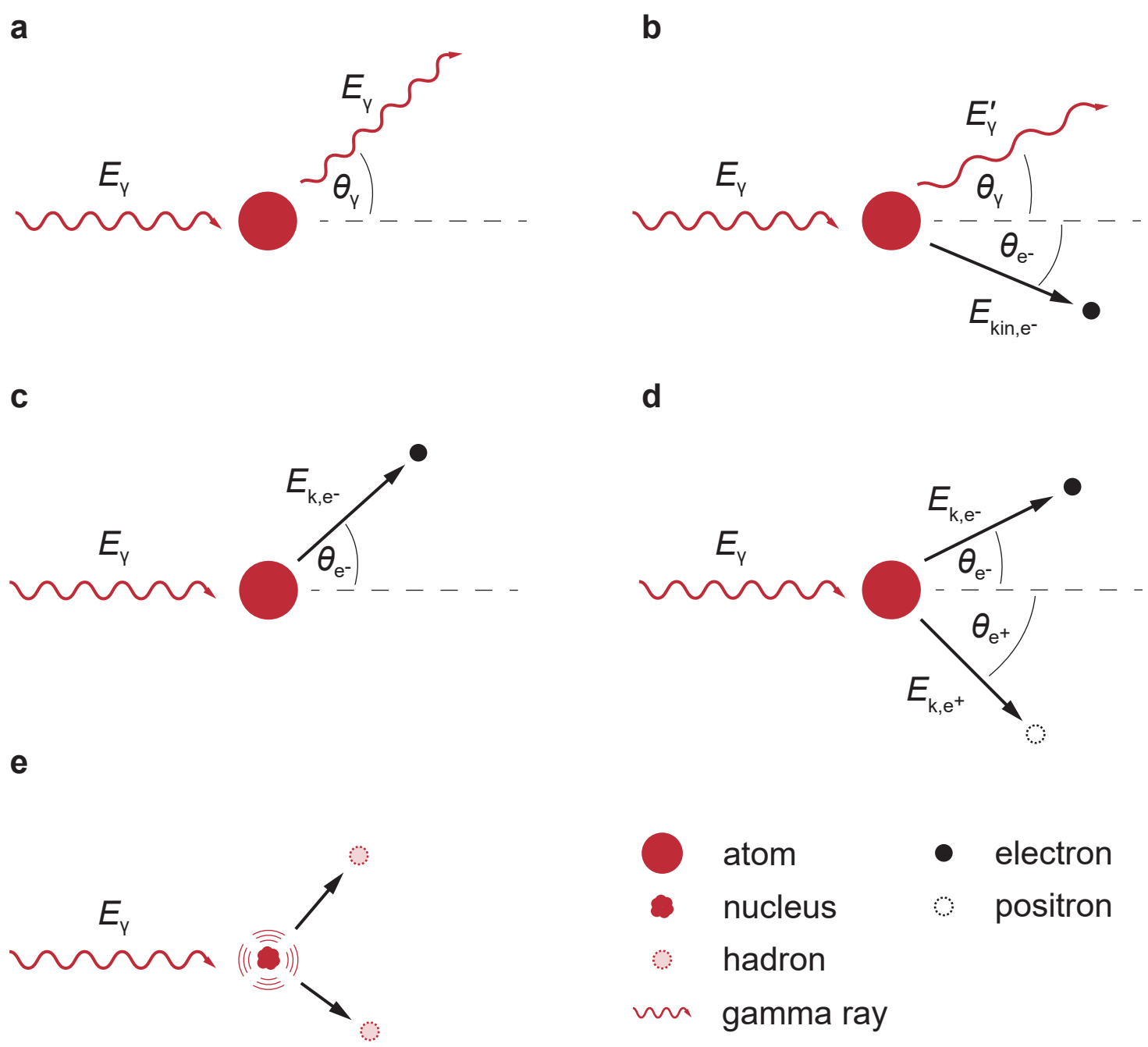


**Figure 3.1** Main interaction mechanisms of high-energy photons with matter. **a** Rayleigh scattering. **b** Compton scattering. **c** Photoelectric absorption. **d** Electron-positron pair production. **e** Photonuclear reaction.

### 3.1.2.1 Rayleigh Scattering

Let us start with probably the most simple interaction mechanism, that is Rayleigh scattering.[9] Rayleigh scattering describes an elastic scattering interaction of photons with an individual atom as schematically shown in Fig. 3.1 [206, 212]. In this process, the photon scatters coherently from all the electrons of the atom transferring momentum to the entire atom, but apart from that leaves the atom without excitation or ionization. Therefore, Rayleigh scattering is sometimes also referred to as coherent scattering. Due to the large mass of the atom, its recoil is negligible. Consequently, the photon scatters elastically from the atom and the photon energy remains nearly unchanged.

9 *Named after John William Strutt, 3rd Baron Rayleigh (*1842, †1919), a British mathematician and physicist. He made important contributions to fluid dynamics and the theory of elastic light scattering, in the process answering also a long-lasting question all parents know only too well from their children: why the sky is blue [213]. For his work, he was awarded the Nobel Prize in Physics in 1904.*

By approximating the atom as a point charge, the Rayleigh scattering cross-section may be approximated by the Thomson scattering angular cross-section for low photon energies [206]:

$$\frac{\mathrm{d}\sigma_{\mathrm{ray}}\left(\theta_{\gamma}\right)}{\mathrm{d}\Omega'} \approx \mathrm{r}_{\mathrm{e}}^{2} Z^{2} \frac{1+\cos^{2}\theta_{\gamma}}{2} \tag{3.10}$$

where:

| | | |
|---|---|---|
| $\mathrm{r_e}$ | classical electron radius (cf. Constants) | m |
| $Z$ | atomic number | |
| $\theta_\gamma$ | photon scattering angle | rad |

and which I will refer to as the Thomson limit throughout this book. For an accurate modeling of Rayleigh scattering interactions, we have to consider the charge distribution of the atom, which can be done for example by form factor approximations [207, 208, 214, 215].[10]

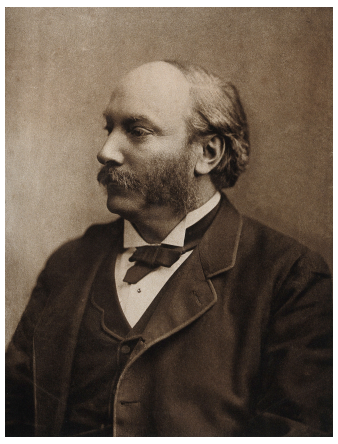

John William Strutt, 3rd Baron Rayleigh
Wellcome Collection

10 *The Monte Carlo code used in this book adopts high-fidelity models and atomic form factors from the EPDL97 nuclear data library [207, 216].*

#### 3.1.2.2 Compton Scattering

In contrast to Rayleigh scattering, Compton scattering is an inelastic incoherent scattering interaction mode of photons with matter.[11] In Compton scattering, as schematically displayed in Fig. 3.1, a high-energy photon scatters off a single electron, transferring part of its energy to the electron. The photon gets scattered in a new direction with a reduced energy.[12] In case the electron is bound in an atom and if the transferred energy is larger than the electron's binding energy, the electron gets ejected from the atom as a free electron leaving the atom in an ionized state, i.e. with a vacancy in the electron shell. Electron state transitions caused by this vacancy may then lead to the additional emission of characteristic X-rays and Auger electrons as discussed in Section 2.1.1. The ejected electron, which is also called a Compton electron, may create additional high-energy photons by bremsstrahlung interaction with the surrounding matter (cf. Section 2.1.1).

As we will see later in this chapter, Compton scattering emerges as one of the predominant interaction modes for the majority of primary gamma rays emitted by radionuclides, thus playing a crucial role in AGRS modeling. Given its significance in this context, I will discuss the Compton scattering interaction in somewhat more detail in the following paragraphs. Moreover, I will restrict the discussion not only to photons but will investigate also the main properties of the

11 *Named after Arthur Holly Compton (*1892, †1962), an American physicist. He is best known for the discovery and theoretical description of the Compton scattering effect confirming the quantum nature of photons [217]. For his work, he was awarded the Nobel Prize in Physics in 1927. Besides his influential work on the Compton effect, he was also a key figure during the Manhattan Project. As head of the Metallurgical Laboratory, Compton was responsible for the production and extraction of weapons-grade plutonium for the program.*

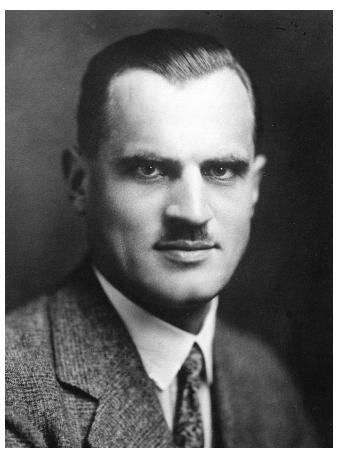

Arthur Compton
Mondadori Publishers

generated Compton electrons as they can play an important role in the generation of secondary photons by bremsstrahlung interactions.

For an accurate description of Compton scattering, the electron shell structure of the atom has to be taken into account. This can be achieved by incoherent scattering functions or relativistic impulse approximation methods [214, 218–220].[13] However, in this section, I will limit the discussion to a simplified model for Compton scattering of unpolarized photons from free electrons at rest, which is sufficient to provide a qualitative understanding of the main trends.

For Compton scattering from a free electron at rest, we can adopt energy and momentum conservation laws to derive the energy of the scattered photon $E'_\gamma$ as a function of the incoming photon energy $E_\gamma$ and the scattering angle $\theta_\gamma$ between the incoming and outgoing photon direction (cf. Fig. 3.1), as first demonstrated by Compton [217]:

$$E'_\gamma = \frac{E_\gamma}{1 + \alpha_\gamma \left(1 - \cos\theta_\gamma\right)} \tag{3.11}$$

where:

| | | |
|---|---|---|
| $E_\gamma$ | energy of incoming photon | eV |
| $E'_\gamma$ | energy of scattered photon | eV |
| $\theta_\gamma$ | photon scattering angle | rad |

and with $\alpha_\gamma$ being defined as:

$$\alpha_\gamma = \frac{E_\gamma}{\mathrm{m_e c^2}} \tag{3.12}$$

with:

| | | |
|---|---|---|
| c | speed of light in vacuum (cf. Constants) | $\mathrm{m\,s^{-1}}$ |
| $\mathrm{m_e}$ | electron mass (cf. Constants) | $\mathrm{eV\,s^2\,m^{-2}}$ |

The photon may scatter by any angle $\theta_\gamma \in [0, \pi]$. Consequently, we have the maximum energy transfer occurring for full backscattering events ($\theta_\gamma = \pi$) and minimum/no energy transfer for $\theta_\gamma = 0$:

$$E'_\gamma = \begin{cases} \dfrac{E_\gamma}{1 + 2\alpha_\gamma} & \text{for } \theta_\gamma = \pi \quad (3.13a) \\ E_\gamma & \text{for } \theta_\gamma = 0 \quad (3.13b) \end{cases}$$

12 *If the electron has enough kinetic energy before the interaction with the photon, the photon can gain energy from the scattering interaction. We call such an interaction also an inverse Compton scattering event [39]. However, inverse Compton scattering is only relevant in areas where high-energy electrons are abundant in significant quantities, e.g. in plasma clouds in astrophysics. As a result, I focus the discussion here on "classical"Exchanged the term "classic" with "classical". Compton scattering with bound electrons of sufficiently low kinetic energy to neglect inverse Compton scattering events.*

13 *The Monte Carlo code adopted in this work uses the relativistic impulse approximation method, which not only allows to account for electron shell but also Doppler shift effects due to the momentum distribution of the atomic electrons [219, 220].*

14 *It is worth mentioning that the kinetic energy of Compton electrons ejected from atoms are in addition reduced by the corresponding electron binding energy [206]. However, because we discuss here only Compton scattering of free electrons, the binding energy is not considered.*

The kinetic energy $E_{k,e^-}$ and emission angle $\theta_{e^-}$ of the Compton electron are also fully determined by the conservation of energy and momentum as [206]:[14]

$$E_{k,e^-} = E_\gamma - E'_\gamma \tag{3.14a}$$
$$= E_\gamma \frac{\alpha_\gamma \left(1 - \cos\theta_\gamma\right)}{1 + \alpha_\gamma \left(1 - \cos\theta_\gamma\right)} \tag{3.14b}$$
$$= E_\gamma \frac{2\alpha_\gamma \cos^2\theta_{e^-}}{\left(1 + \alpha_\gamma\right)^2 - \alpha_\gamma^2 \cos^2\theta_{e^-}} \tag{3.14c}$$

and

$$\cot\theta_{e^-} = \left(1 + \alpha_\gamma\right) \tan\frac{\theta_\gamma}{2} \tag{3.15a}$$
$$\cos\theta_\gamma = 1 - \frac{2}{1 + \left(1 + \alpha_\gamma\right)^2 \tan^2\theta_{e^-}} \tag{3.15b}$$

with:

| | | |
|---|---|---|
| $E_{k,e^-}$ | electron kinetic energy | eV |
| $\theta_{e^-}$ | electron emission angle | rad |

Having analyzed the kinematic variables of the scattered photon and Compton electron for a single scattering event, let us now turn our discussion to the distribution of these variables if we consider a large number of Compton scattering events. For this purpose, we can adopt again differential cross-sections. The differential angular cross-section for photons with energy $E_\gamma$ scattering through an angle $\theta_\gamma$ into a differential solid angle $d\Omega'$ is described by the Klein-Nishina cross-section [221]:[15]

15 *Derived by Oskar Benjamin Klein (*1894, †1977) and Yoshio Nishina (*1890, †1951) in 1929, which was one of the first successful applications of QED [221].*

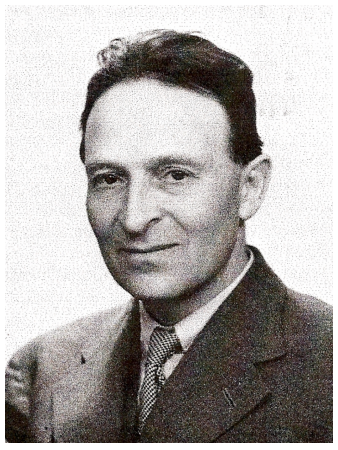

Oskar Klein


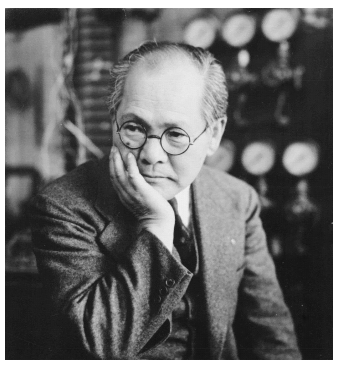

Yoshio Nishina

$$\frac{d\sigma_{com}}{d\Omega'}\left(E_\gamma, \theta_\gamma\right) = r_e^2 \frac{1}{\left[1+\alpha_\gamma\left(1-\cos\theta_\gamma\right)\right]^2} \frac{1+\cos^2\theta_\gamma}{2} \left\{1 + \frac{\alpha_\gamma^2\left(1-\cos\theta_\gamma\right)^2}{\left(1+\cos^2\theta_\gamma\right)\left[1+\alpha_\gamma\left(1-\cos\theta_\gamma\right)\right]}\right\} \tag{3.16}$$

where:

| | | |
|---|---|---|
| $r_e$ | classical electron radius (cf. Constants) | m |
| $\alpha_\gamma$ | ratio of the photon energy to the energy-equivalent electron rest mass | |
| $\theta_\gamma$ | photon scattering angle | rad |

from which also the differential angular cross-section for Compton electrons with energy $E_{k,e^-}$ emitted at an angle $\theta_{e^-}$ into a differential solid angle $d\Omega_e$ can be derived by combining Eq. 3.16 with Eqs. 3.15a and 3.15b [212]:

$$\frac{d\sigma_{com}}{d\Omega_e}\left(E_\gamma, \theta_{e^-}\right) = \frac{d\sigma_{com}}{d\Omega'}\left|\frac{1}{1+\alpha_\gamma}\frac{\left(1+\cos\theta_\gamma\right)\sin\theta_\gamma}{\sin^3\theta_{e^-}}\right| \tag{3.17}$$

Both angular cross-sections are plotted in Fig. 3.2 for selected photon energies $E_\gamma$. From this, we see that the angular distribution of scattered photons is increasingly peaked in the forward direction with increasing photon energy $E_\gamma$. In contrast, at low energies, the distribution shows a significant backscatter component and approaches the Thomson limit introduced, when we were discussing the Rayleigh scattering. On the other hand, the Compton electron angular distribution is kinematically always limited to the forward direction $\theta_{e^-} \in [0, \pi/2]$. It gets as well increasingly peaked in the forward direction with increasing photon energy $E_\gamma$ as displayed in Fig. 3.2.

Moving on to the energy distribution of the scattered photons and emitted Compton electrons: By combining the Klein-Nishina cross-section in Eq. 3.16 with Eq. 3.14b, one can show that the differential energy cross-section for the Compton electron is given by [206]:

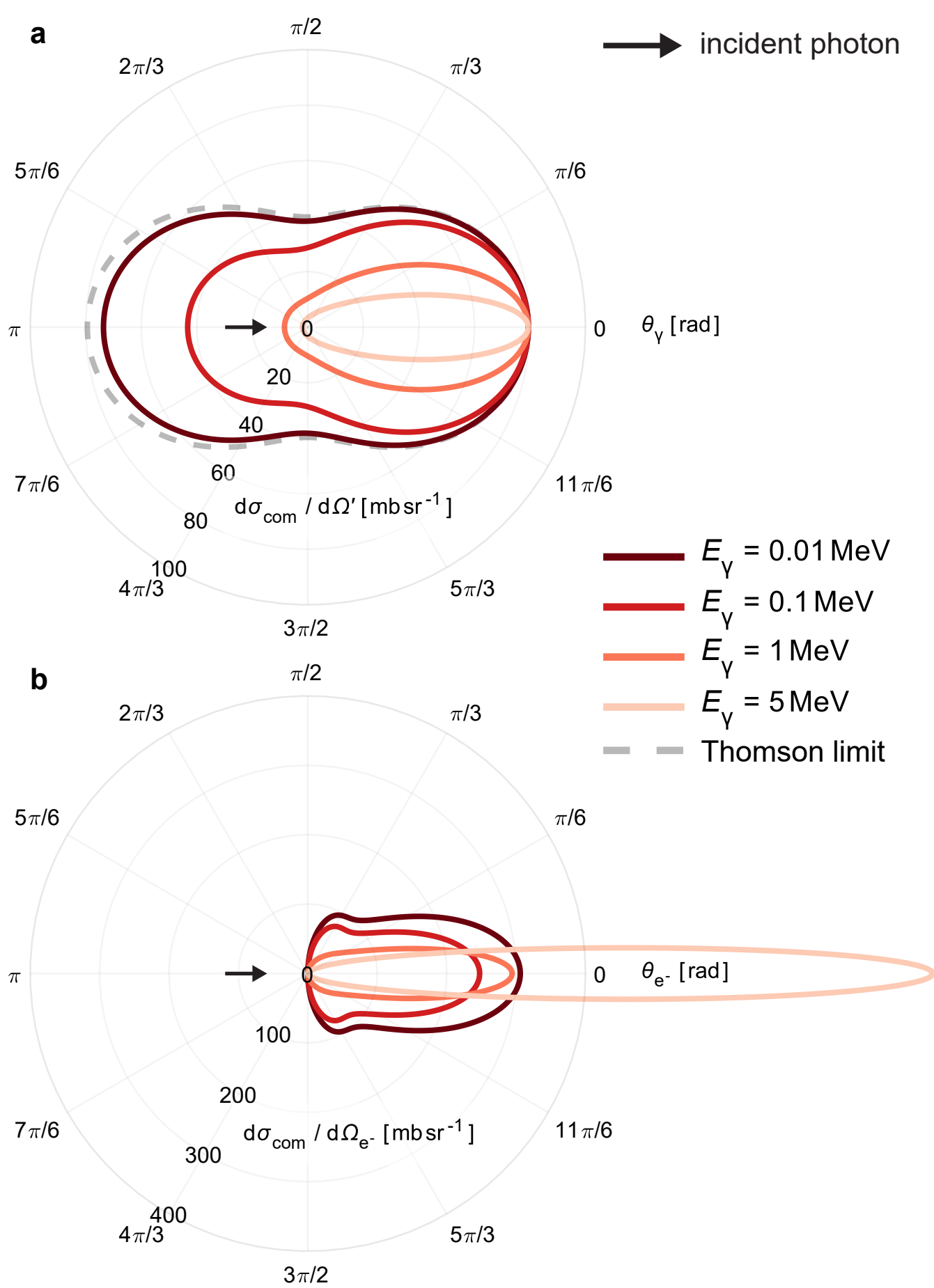


**Figure 3.2** Compton scattering differential angular cross-sections. **a** Compton scattering differential angular cross-section $d\sigma_{com}/d\Omega'$ for scattered photons as a function of the energy $E_\gamma$ of the incident photon and the scattering angle $\theta_\gamma$ (cf. Eq. 3.16). In addition, the Thomson limit is displayed as well (cf. Eq. 3.10). **b** Compton scattering differential angular cross-section $d\sigma_{com}/d\Omega_e$ for Compton electrons as a function of the energy $E_\gamma$ of the incident photon and the electron emission angle $\theta_{e^-}$ (cf. Eq. 3.17). For $E_\gamma = 10\,\mathrm{MeV}$, the angular cross-section peaks at about 900 mb sr$^{-1}$.

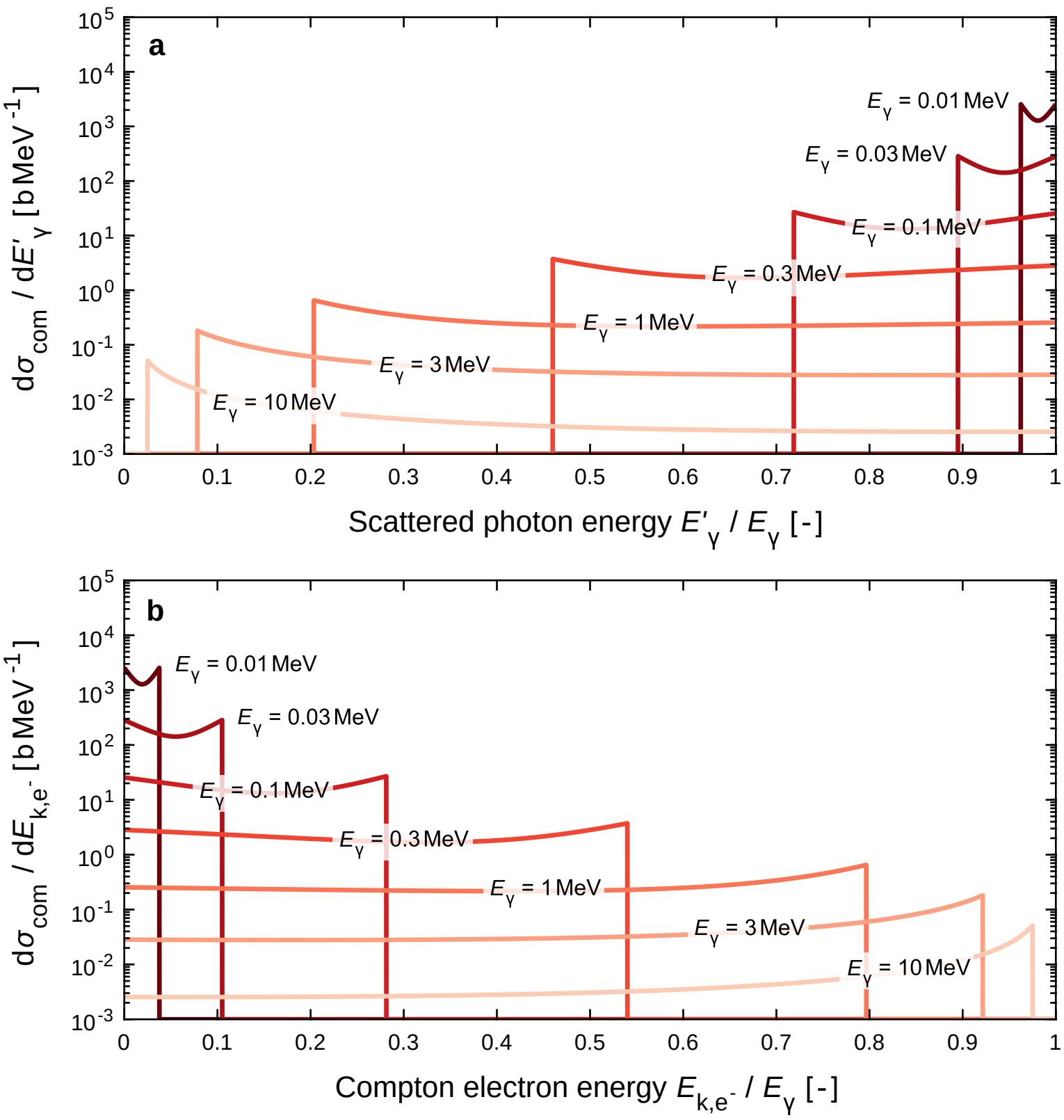


**Figure 3.3** Compton scattering differential energy cross-sections. **a** Compton scattering differential energy cross-section $d\sigma_{com} / dE'_\gamma$ for scattered photons as a function of the energies $E_\gamma$ and $E'_\gamma$ of the incoming and scattered photons, respectively (cf. Eq. 3.19). **b** Compton scattering differential energy cross-section $d\sigma_{com} / dE_{k,e^-}$ for Compton electrons as a function of the incoming photon energy $E_\gamma$ and kinetic electron energy $E_{k,e^-}$ (cf. Eq. 3.18). As indicated in the main text, $d\sigma_{com} / dE_{k,e^-}$ is mirroring $d\sigma_{com} / dE'_\gamma$.

$$\frac{d\sigma_{com}}{dE_{k,e^-}}\left(E_\gamma, E_{k,e^-}\right) = \frac{\pi r_e^2}{\alpha_\gamma E_\gamma}\left[1 + \left(1 - \frac{1}{\alpha_\gamma}\frac{E_{k,e^-}}{E_\gamma - E_{k,e^-}}\right)^2 + \frac{E_{k,e^-}^2}{E_\gamma\left(E_\gamma - E_{k,e^-}\right)}\right] \tag{3.18}$$

and by combining Eq. 3.18 with Eq. 3.14a, i.e. switching $E_{k,e^-}$ with $E_\gamma - E'_\gamma$, we obtain also the differential energy cross-section for the scattered photon [206]:

$$\frac{d\sigma_{com}}{dE'_\gamma}\left(E_\gamma, E'_\gamma\right) = \frac{\pi r_e^2}{\alpha_\gamma E_\gamma}\left[1 + \left(1 - \frac{1}{\alpha_\gamma}\frac{E_\gamma - E'_\gamma}{E'_\gamma}\right)^2 + \frac{\left(E_\gamma - E'_\gamma\right)^2}{E_\gamma E'_\gamma}\right] \tag{3.19}$$

which is simply mirroring Eq. 3.18. In Fig. 3.3, the differential energy cross-sections for scattered photons and Compton electrons are plotted for selected photon energies $E_\gamma$. As expected, the energy of the scattered photon is limited between the maximum energy transfer (given by Eq. 3.13a) and $E_\gamma$. The distribution is quite flat but peaks at the maximum energy transfer values and for lower photon energies also at $E_\gamma$. In line with Eq. 3.11, we find a larger relative shift in the photon energy $1 - E'_\gamma/E_\gamma$ as the photon energy $E_\gamma$ increases. The energy distribution of the Compton electrons is the perfect mirror of the photon energy distribution with the maximum kinetic energy determined by the maximum energy transfer in Eq. 3.13a.

#### 3.1.2.3 Photoelectric Absorption

As schematically displayed in Fig. 3.1, in photoelectric absorption interactions, a photon gets absorbed by an atomic electron.[16] The atomic electron gets subsequently ejected from the atom with a kinetic energy $E_{k,e^-}$ equal to the difference between the photon energy $E_\gamma$ and the binding energy $E_b$ of the corresponding electron along with the recoil energy of the atom. Because of the large mass of the

16 *First theoretically described by Albert Einstein in 1905 [34] by using the newly developed quantum mechanics theory and subsequently experimental confirmed by Robert Millikan [35–37], as already discussed in Section 2.1.1. The successful description of the photoelectric effect represents thereby an important milestone in the development of the quantum mechanics theory at the beginning of the 20th century.*

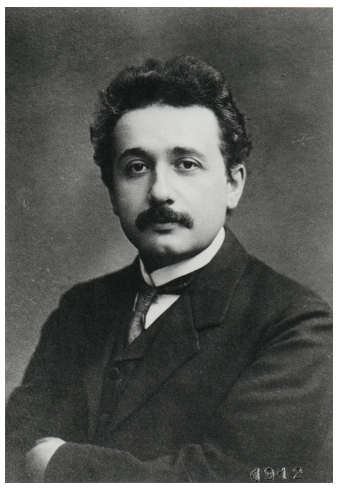

Albert Einstein


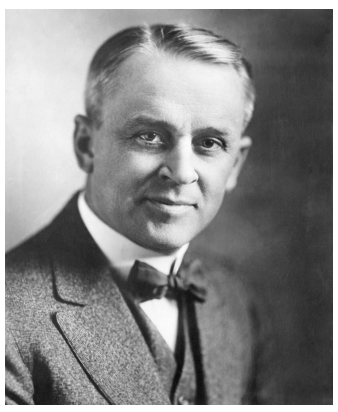

Robert Millikan

atom with respect to the electron mass, the recoil energy is usually small and can be neglected [206, 212]. We can therefore compute the discrete kinetic energy of the ejected electron as:

$$E_{\mathrm{k,e^-}} = E_\gamma - E_\mathrm{b} \tag{3.20}$$

where:

| | | |
|---|---|---|
| $E_\mathrm{b}$ | electron binding energy | eV |
| $E_\gamma$ | photon energy | eV |

As a result, the binding energies $E_\mathrm{b}$ for the associated atoms represent energy thresholds, under which no photoelectric absorption can occur with atomic electrons in the corresponding electron shells. We call the ejected electron also a photoelectron.

In contrast to Compton scattering, photoelectric absorption cannot occur with free electrons due to momentum conservation. However, photoelectric absorption can take place with atomic electrons as momentum is conserved by the recoil of the atom. In fact, the more tightly bound the electron is, the larger the microscopic cross-section for photoelectric absorption. As a general rule of thumb, about 80 % of the photoelectric absorption events occur in the K-shell of the atom, which is the innermost electron shell [212]. Because photoelectric absorption always involves the atomic shell configuration of the associated atom, the microscopic cross-section $\sigma_\mathrm{pe}$ for the photoelectric absorption is difficult to compute analytically. However, detailed numerical-based cross-section data are available in nuclear data libraries such as XCOM [208] or EPDL97 [207].[17]

17 *The Monte Carlo code used in this work adopts high-fidelity datasets and numerical models from the EPDL97 nuclear data library [207, 216].*

Similar to Compton scattering, photoelectric absorption creates also a vacancy in the electron shell of the atom. Because this vacancy is predominantly located in the inner electron shells, the resulting number of electron energy transitions and thereby also the emission of characteristic X-rays and Auger electrons is particularly high for photoelectric absorption events. As an example, for a vacancy in the K-shell of a uranium atom, 2772 different Auger electron as well as 154 characteristic X-ray emissions are possible [206, 207]. In addition, the ejected photoelectron may also create additional high-energy photons by bremsstrahlung interactions (cf. Section 2.1.1). Consequently, the angular distribution of the photoelectrons is important in photon transport as it influences the direction of secondary bremsstrahlung photons. In brief, the photoelectrons are emitted into the full solid angle with a distribution that is increasingly peaked into the forward

direction with increasing photon energy. A more detailed discussion about the differential angular cross-section of the photoelectrons including a graphical depiction of this distribution is provided in Appendix A.4.

#### 3.1.2.4 Electron-Positron Pair Production

When interacting with an electric field of a charged particle, a photon can be converted into a pair of a subatomic particle and its antiparticle. We call such a photon absorption process therefore also a pair production interaction.[18] As the energy has to be conserved, the energy $E_\gamma$ of the interacting photon has to be at least the total equivalent rest mass energy of the particle-antiparticle pair. With an equivalent rest mass energy $\mathrm{m_e c^2}$ of about 0.511 MeV, the electron and its antiparticle, the positron, are the lightest known subatomic particle-antiparticle pair [134]. The energy threshold for electron-positron pair production is thereby about 1.022 MeV. Other particle-antiparticle pairs such as the muon-antimuon ($\mu^\pm$) or pion-antipion ($\pi^\pm$) pairs possess significantly higher energy thresholds at about 211 MeV and 279 MeV, respectively [134]. Given the spectral range of the main gamma-ray sources discussed in the previous chapter, electron-positron pair production emerges as the dominant pair production mode for AGRS applications. Consequently, I will limit the discussion in this section here to this mode.

18 *Production of electron-positron pairs by high-energy photons were first described by Anderson and Neddermeyer in 1933 [222] shortly after the positron has been discovered by Anderson [223, 224]. For the discovery of the positron, Anderson was awarded the Nobel Prize in Physics in 1936.*

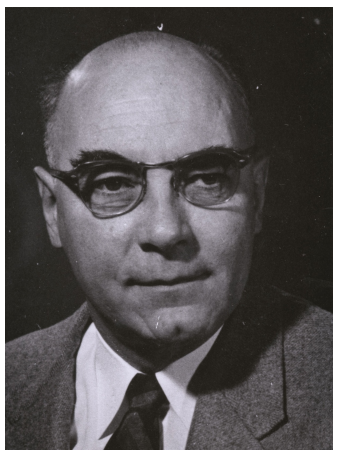

Carl David Anderson


As stated above, pair production interactions cannot occur in empty space but have to take place in the electric field of a charged particle. If the photon interacts with the electric field of a nucleus, almost all energy is transferred from the photon to the rest mass and kinetic energy of the newly created electron-positron pair. Because of the large mass difference between the electron/positron and the atomic nucleus, the recoil energy of the nucleus can be neglected for all practical purposes [206]. As a result, we can write:

$$E_\gamma = 2\mathrm{m_e c^2} + E_{\mathrm{k,e^-}} + E_{\mathrm{k,e^+}} \tag{3.21}$$

with:

| | | |
|---|---|---|
| $E_\gamma$ | photon energy | eV |
| $E_{\mathrm{k,e^+}}$ | positron kinetic energy | eV |
| $E_{\mathrm{k,e^-}}$ | electron kinetic energy | eV |
| $\mathrm{m_e}$ | electron mass (cf. Constants) | $\mathrm{eV\,s^2\,m^{-2}}$ |
| c | speed of light in vacuum (cf. Constants) | $\mathrm{m\,s^{-1}}$ |

This represents the energy constraint for the electron and positron emitted in an electron-positron pair production event with an energy threshold of $2m_ec^2 \sim 1.022$ MeV as stated before. A schematic depiction of a pair production event in the atomic nucleus field including all relevant kinematic variables is shown in Fig. 3.1.

Electron-positron pair production events cannot only occur in the electric field of atomic nuclei but also in that of electrons.[19] In the latter case, we cannot neglect the recoil energy of the particle associated with the electric field, i.e. the electron, as the mass of this electron is equal to the mass of the newly created electron/positron. We have therefore to adapt the energy conservation formula in Eq. 3.21 as follows:

19 *In fact, pair production events can occur in the electric field of any charged particle as stated before. However, atomic nuclei and electrons are by far the most abundant charged particles in the environment and thereby the dominant contributors to the overall pair production rate for AGRS applications.*

$$E_\gamma = 2m_ec^2 + E_{k,e^-} + E_{k,e^+} + E_{k,r} \tag{3.22}$$

with:

| | | |
|---|---|---|
| $E_{k,r}$ | recoil electron kinetic energy | eV |

Consequently, the energy threshold for such a pair production interaction in the electric field of an electron is higher than the 1.022 MeV threshold in the electric field of nuclei. For free electrons at rest, the energy threshold is equal to $4m_ec^2 \sim 2.044$ MeV for such pair production events [206, 212]. Additionally, if the interacting electron is bound in an atom, binding energies have to be considered as well. In case the kinetic energy of the recoil electron exceeds the binding energy, the electron is ejected from the atom. We call such an event also triplet production as three high-energy subatomic particles are emitted from the atom [206, 212].

Similar to the photoelectric effect, no simple analytical expressions exist to accurately model the differential cross-sections for electron-positron pair production. However, a broad variety of semi-analytical and numerical models with varying degrees of accuracy are readily available in nuclear data libraries and the literature [207, 208, 225–227].[20] In the following, I will limit the discussion to simplified models to highlight the main trends, with additional comprehensive information accessible in the literature [206, 212, 225, 228].

20 *The Monte Carlo code used in this work adopts high-fidelity models from the EPDL97 nuclear data library [207, 216].*

Both, the electron and the positron created in pair production events can subsequently emit high-energy photons by bremsstrahlung (cf. Section 2.1.1). In addition, the positron will eventually get annihilated in the surrounding matter giving rise to an additional annihilation radiation component accompanying each pair

production event. Moreover, in triplet production, the recoil electron can also create characteristic X-rays and Auger electrons.

Given the significance of emitted electrons and positrons in the production of secondary photons, I analyze the angular and energy distribution of these particles in Appendix A.5. In brief, we find a similar trend in the angular distribution as for the photoelectrons, i.e. emission into the full solid angle with an increasing peak in the forward direction with increasing photon energy. The kinetic energy for both, the electron and the positron, is symmetrically distributed around a mean of $E_\gamma/2 - \mathrm{m_e c^2}$ with an increasing dispersion with increasing photon energy. We conclude that pair production events can be a significant source of high-energy electrons and positrons reaching kinetic energies comparable to the energy of the photon inducing the pair production event.

#### 3.1.2.5 Photonuclear Reactions

In Section 2.3.1, we have learned already that nuclear reactions such as neutron capture or fission can create excited nuclei and thereby lead to the emission of high-energy photons. In these cases, neutrons were the main projectiles inducing the reaction. It comes without surprise that not only neutrons but also other subatomic particles including photons can induce nuclear reactions. In the case of photon projectiles, we call the corresponding reactions "photonuclear reactions".[21] Consequently, these photonuclear reactions can alter the photon field and, depending on the application, need to be considered therefore in photon transport.

Photonuclear reactions include neutron emission reactions, e.g. $(\gamma, \mathrm{n})$[22] or $(\gamma, 2\mathrm{n})$; charged particle emission reactions such as $(\gamma, \mathrm{p})$ or $(\gamma, \alpha)$; combined neutron-charged particle emission reactions, e.g. $(\gamma, \mathrm{pn})$; and photofission $(\gamma, \mathrm{f})$, among others [231]. Similar to the interaction processes discussed in the previous sections, we can describe the probability of a photonuclear reaction ${}^A_Z\mathrm{X}(\gamma, *)$ to occur for a photon energy $E_\gamma$ and target nuclide ${}^A_Z\mathrm{X}$ with the microscopic cross-section, which I denote in this book by $\sigma_{\mathrm{pn}}(E_\gamma)$.[23] Note that, because the photons interact directly with the nucleus (cf. Fig. 3.1), the photonuclear cross-section specifically depends on the individual nuclide and not only the element as for the previously discussed interaction modes. The absorption of high-energy photons by a nucleus in a photonuclear reaction is commonly described phenomenologically by the combination of the giant dipole resonance (GDR) and the quasi-deuteron (QD) mechanisms [231, 233]:

21 *In analogy to the photoelectric effect, we refer to these reactions also as photonuclear absorption in the context of photon transport [226].*

22 *This is exactly the opposite of the neutron capture reaction discussed in Section 2.3.1. This photonuclear reaction was famously utilized by James Chadwick and Maurice Goldhaber in 1934 to determine the mass of the neutron, which was discovered two years earlier [229, 230]. Chadwick and Goldhaber used gamma rays emitted from the* ${}^{208}_{82}\mathrm{Tl}$ *nuclide contained in natural thorium (cf. Section 2.1.3) with an energy of about* 2.62 MeV, *which is sufficient to overcome the energy threshold of* ~2.22 MeV *of a photodesintegration reaction in* ${}^2_1\mathrm{H}$ *(cf. Section 2.3.1).*

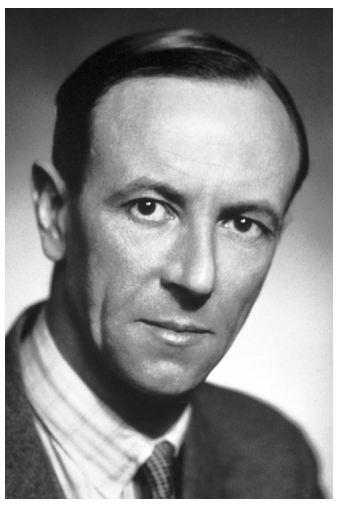

Sir James Chadwick
Nobel foundation

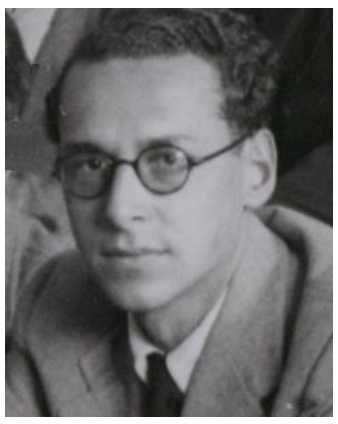

Maurice Goldhaber
GFHund

$$\sigma_{\text{pn}}(E_\gamma) = \sigma_{\text{GDR}}(E_\gamma) + \sigma_{\text{QD}}(E_\gamma) \tag{3.23}$$

The GDR describes the dominant excitation process at photon energies $E_\gamma \lesssim 30\,\text{MeV}$ in which a collective oscillation of the neutrons and protons in the nucleus against each other is induced [234–237]. Various parametric models are available to accurately predict $\sigma_{\text{GDR}}$ as a function of the photon energy and the corresponding target nuclide [237, 238]. At photon energies $E_\gamma \gtrsim 30\,\text{MeV}$, the QD mechanism becomes the dominant absorption process. The QD mechanism describes the excitation of a quasi-deuteron pair, i.e. a pair of a single neutron and a proton, within the nucleus [239–241]. For the QD mechanism, mainly empirical models are adopted in modern nuclear data libraries and nuclear reaction codes [231, 233, 239].[24]

In Fig. 3.4, I have used high-fidelity microscopic cross-section data provided by the TENDL nuclear library (version: 2023) [232] to compute $\sigma_{\text{pn}}$ as a function of the photon energy $E_\gamma$ for natural elements by weighting the individual nuclide cross-sections with the corresponding isotopic abundance values provided by Meija et al. [56]. We find a significant increase in $\sigma_{\text{pn}}$ around ~20 MeV for all elements corresponding to the peak of the GDR. The absolute microscopic cross-section generally increases with increasing atomic number reaching peak values of 0.50 b to 0.67 b for $Z \geq 80$. In addition, the GDR peak shifts from ~20 MeV for light elements to ~13 MeV for heavy elements. However, due to the variable isotopic abundances and nuclear properties associated with the different nuclides, there are some significant discontinuities in this general trends.[25] As an example, tin ($Z$=50) possesses the highest number of stable isotopes and shows a particularly high GDR peak compared to the neighboring elements of ~0.51 b. The photonuclear cross-sections of some major naturally occurring elements are also highlighted in Fig. 3.4, from which iron ($Z$=26) shows the highest cross-section values.

23 *Individual microscoscopic cross-sections for each photonuclear reaction channel are readily available in nuclear data libraries [231, 232]. I limit the discussion here to the combined effect of all these reactions together. As a consequence, only the absorption process is described using the microscopic cross-section. For the energy and angular distribution of the various secondary reaction products, you can find more information in the literature [231–233]. Although I have to emphasize that the experimental status of these secondary particle quantities is not satisfactory for accurate modeling purposes as of writing this book, and as a result, nuclear data libraries have to rely currently heavily on numerical models [231].*

24 *The Monte Carlo code adopted in this work uses GDR and QD models derived from experimental data obtained by the IAEA Photonuclear Data Library and the EXFOR database [231, 242–244]. Moreover, an adapted vector meson dominance model is included to accurately model photonuclear interactions in the GeV energy range and beyond [242, 245, 246].*

25 *Apart from the fact that several elements, especially for higher atomic numbers, are not naturally present in significant quantities in the environment [56].*

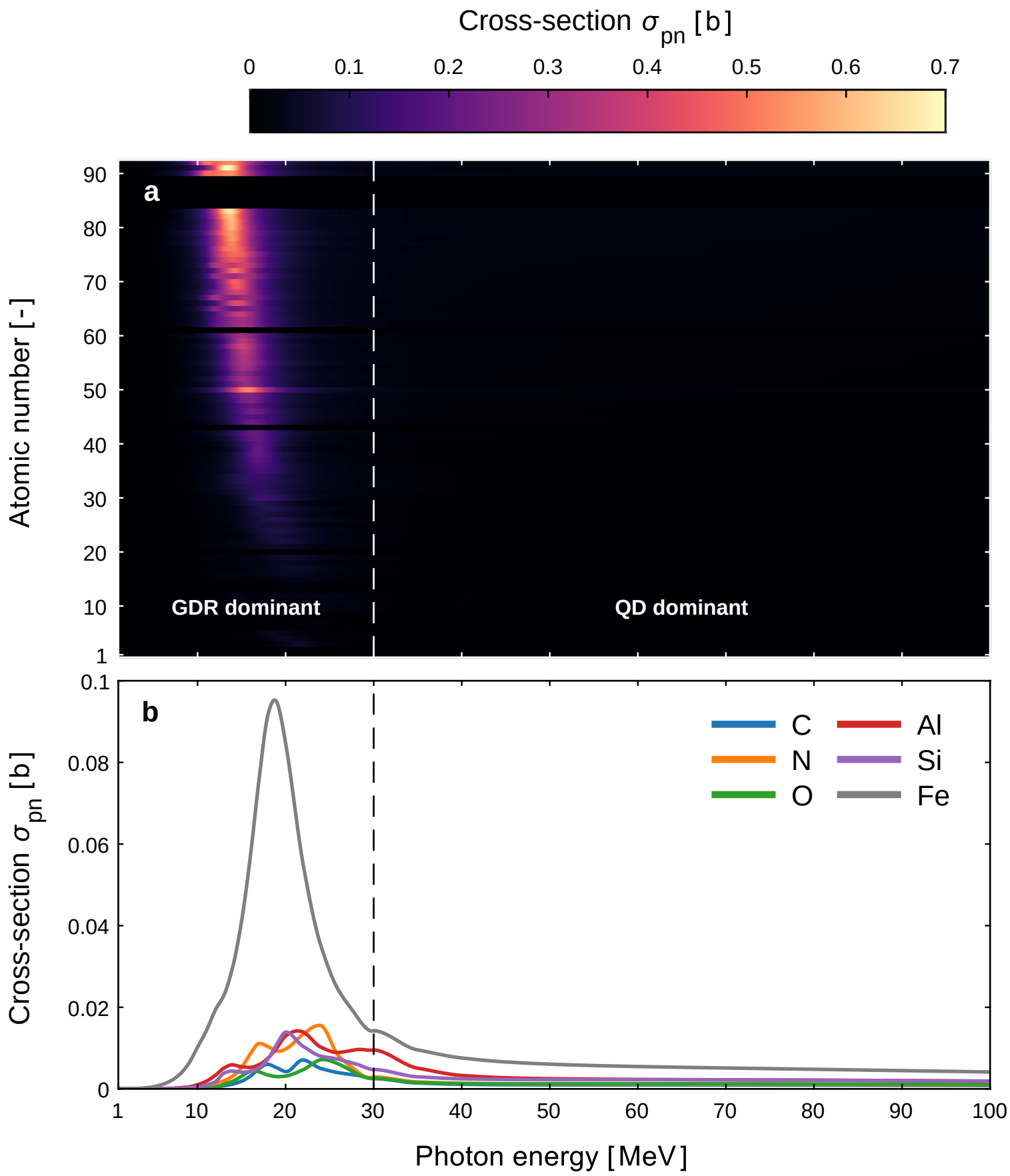


**Figure 3.4** Photonuclear microscopic cross-section $\sigma_{pn}$ for natural elements as a function of the photon energy $E_\gamma$ and the atomic number $Z$. $\sigma_{pn}$ was computed by weighting the individual nuclide cross-sections obtained by the TENDL nuclear data library (version: 2023) [232] with the corresponding isotopic abundance values provided by Meija et al. [56]. **a** Selected spectral window between 1 MeV and 100 MeV. The spectral regions, in which the GDR and QD mechanisms dominate, are highlighted. **b** Microscopic cross-section for selected elements with significant elemental abundance values in the environment (Earth's atmosphere & crust).

#### 3.1.2.6 Total Cross-Section & Mean Free Path

To describe the combined interaction probability of a high-energy photon with energy $E_\gamma$ with an individual nuclide $j$ considering all the interaction modes discussed in the previous paragraphs, we can define the total microscopic cross-section $\sigma_{\text{tot}}$ as:

$$\sigma_{\text{tot},j}(E_\gamma) = \sigma_{\text{ray},j}(E_\gamma) + \sigma_{\text{com},j}(E_\gamma) + \sigma_{\text{pe},j}(E_\gamma) + \sigma_{\text{pp,n},j}(E_\gamma) + \sigma_{\text{pp,e},j}(E_\gamma) + \sigma_{\text{pn},j}(E_\gamma) \quad (3.24)$$

with:

| | | |
|---|---|---|
| $\sigma_{\text{ray}}$ | microscopic Rayleigh scattering cross-section | $\text{m}^2$ |
| $\sigma_{\text{com}}$ | microscopic Compton scattering cross-section | $\text{m}^2$ |
| $\sigma_{\text{pe}}$ | microscopic photoelectric cross-section | $\text{m}^2$ |
| $\sigma_{\text{pp,n}}$ | microscopic electron-positron pair production cross-section in the electric field of the nucleus | $\text{m}^2$ |
| $\sigma_{\text{pp,e}}$ | microscopic electron-positron pair production cross-section in the electric field of the electron | $\text{m}^2$ |
| $\sigma_{\text{pn}}$ | microscopic photonuclear cross-section | $\text{m}^2$ |

This total cross-section is displayed in Fig. 3.5 for selected elements with atomic numbers ≤100 and photon energies from $10^3$ eV to $10^{11}$ eV. We can identify three distinct regions, in which one interaction mechanism dominates the total cross-section: photoelectric absorption at low energies, Compton scattering at intermediate energies and pair production at high energies (cf. also Fig. B.16).

In general, $\sigma_{\text{tot}}$ is highest in the photoelectric dominant region and decreases as photon energy increases towards the Compton dominant region. For $E_\gamma \ll \text{m}_\text{e}\text{c}^2$, we find $\sigma_{\text{pe}} \propto E_\gamma^{-7/3}$ [206, 248].[26] Because $\sigma_{\text{pe}}$ has a significantly increased dependence on the atomic number (between $Z^4$ and $Z^5$) compared to Compton scattering, the spectral range for which photoelectric absorption dominates, gets broader for heavier elements [206]. In this region, we find also some significant discontinuities, which can be attributed to the binding energies $E_\text{b}$ of the atomic electron shells. As discussed before, the binding energies $E_{\text{b},i}$ for the individual electron shells $i$ represent energy thresholds

26 *At higher energies $E_\gamma \gg \text{m}_\text{e}\text{c}^2$, we have $\sigma_{\text{pe}} \propto E_\gamma^{-1}$ [206, 248].*

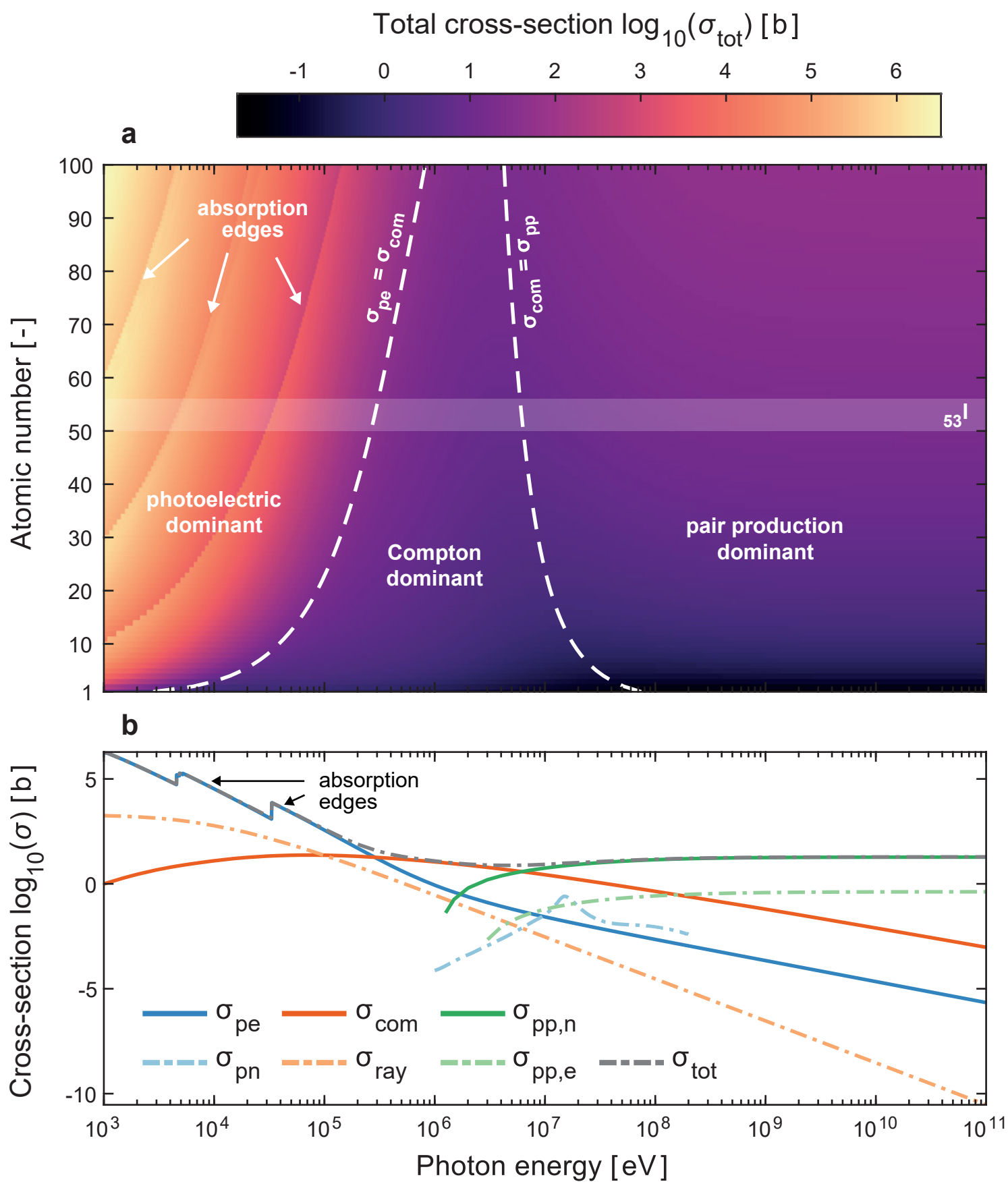


**Figure 3.5** Energy dependence of the total microscopic cross-section for photon-matter interactions. **a** Total microscopic cross-section $\sigma_{tot}$ as a function of the photon energy $E_\gamma$ and the atomic number $Z$ of the associated element. Regions with one dominant interaction process are highlighted with $\sigma_{pe}$, $\sigma_{com}$ and $\sigma_{pp}$ ($= \sigma_{pp,n} + \sigma_{pp,e}$) being the individual microscopic cross-sections for photoelectric, Compton scattering and pair production interactions, respectively. **b** Microscopic cross-section as a function of the photon energy for various interaction processes of photons with $_{53}$I ($Z = 53$), i.e. photoelectric absorption ($\sigma_{pe}$), Rayleigh scattering ($\sigma_{ray}$), Compton scattering ($\sigma_{com}$), pair production in the nuclear ($\sigma_{pp,n}$) and electron field ($\sigma_{pp,e}$) as well as the sum of these processes ($\sigma_{tot}$). The corresponding $\sigma_{tot}$ values are also highlighted in subfigure a. Cross-section data was obtained from the XCOM database by the National Institute of Standards and Technology (NIST) [208, 247].

for photoelectric absorption with atomic electrons of the associated shell or subshell. Because $\sigma_{pe}$ is the sum of all shell contributions, we find a sharp increase as the photon energy exceeds the corresponding binding energy of the atomic electron shell (or to a lesser degree also the subshell). As a result, these transitions are commonly referred to as absorption edges [206].

In the Compton dominant region, the total microscopic cross-section $\sigma_{tot}$ reaches a local minimum. Compton scattering dominates $\sigma_{tot}$ over a broad range of photon energies from a few keV up to several MeV for elements with low atomic numbers. This spectral range gets narrower for heavier elements as the microscopic cross-section of photoelectric absorption and pair production increases more rapidly with increasing atomic numbers.[27] Nonetheless, considering the abundance of low atomic number elements in the environment, Compton scattering remains the dominant interaction process for most primary gamma rays emitted by radionuclides.

27 *As discussed by Carron [206], the microscopic cross-section for Compton scattering is approximately proportional to the atomic number Z for photon energies significantly larger than the highest electron binding energy.*

At even higher photon energies, pair production becomes the dominant interaction process. The pair production dominant region starts for photon energies $\gtrsim 100\,\text{MeV}$ for low atomic number elements and only a few MeV for high atomic number elements. $\sigma_{pp,n}$ and $\sigma_{pp,e}$ are approximatly proportional to $Z^2$ and $Z$, respectively [206]. The ratio of the nuclear and electron field contributions can be estimated by $\sigma_{pp,n}/\sigma_{pp,e} = C(E_\gamma)Z$ with $C > 1$ for finite photon energies and $\lim_{E_\gamma \to \infty} C = 1$ [212]. Consequently, the combined cross-section $\sigma_{pp}$ is dominated by the nuclear electric field contribution $\sigma_{pp,n}$ except for light elements. Furthermore, at very high photon energies ($\gtrsim 10\,\text{GeV}$) $\sigma_{pp}$ approaches a constant value, which only depends on the atomic number $Z$ [212].

Rayleigh scattering and photonuclear reactions, in contrast to the other three interaction modes, never dominate the total cross-section $\sigma_{tot}$. Nevertheless, they can still contribute significantly to the total cross-section for some elements and photon energies. For Rayleigh scattering, the contribution is highest for light elements and photon energies below 100 keV (cf. Fig. B.17).[28] Photonuclear reactions on the other hand are most significant in the GDR domain with $10\,\text{MeV} \lesssim E_\gamma \lesssim 30\,\text{MeV}$.[29] As a result, in AGRS, photonuclear reactions should be considered for TGF and CR induced high-energy photons (cf. Chapter 2).[30]

28 *The highest contribution is found for helium at about 6 keV with ~31 % of the total cross-section.*

29 *Considering natural isotopic composition, it is the three light elements lithium, beryllium, and boron that show the highest contributions, i.e. ~30 %, ~20 % and ~14 %, respectively.*

30 *Apart from man-made artificial photon sources such as nuclear fusion induced prompt gamma rays (cf. Section 2.3.1).*

The total cross-section $\sigma_{tot}$ only describes the interaction of photons with an individual atom. To characterize the interaction of pho-

tons with a macroscopic material, we can adopt the total attenuation coefficient introduced in Eq. 3.8b as:

$$\mu_{\text{tot}}(E_\gamma) = \mathrm{N_A}\rho \sum_j \frac{w_j}{M_j}\sigma_{\text{tot},j}(E_\gamma) \tag{3.25a}$$

$$= \sum_j \mu_{\text{tot},j}(E_\gamma) \tag{3.25b}$$

with:

| | | |
|---|---|---|
| $M_j$ | molar mass of nuclide $j$ | $\mathrm{kg\,mol^{-1}}$ |
| $\mathrm{N_A}$ | Avogadro constant (cf. Constants) | $\mathrm{mol^{-1}}$ |
| $w_j$ | mass fraction of nuclide $j$ in matter | $\mathrm{kg\,kg^{-1}}$ |
| $\rho$ | mass density | $\mathrm{kg\,m^{-3}}$ |
| $\sigma_{\text{tot},j}$ | microscopic total interaction cross-section for nuclide $j$ | $\mathrm{m^2}$ |

and where $\mu_{\text{tot},j}$ represents the related partial attenuation coefficients for nuclide $j$. As expected, the total attenuation coefficient scales linearly with the mass fraction of the corresponding nuclide as well as the mass density of the material. The macroscopic cross-section is also inversely proportional to the mean free path $\ell$ of the photons in a given material (cf. Eq. 3.9). In Fig. 3.6, I have computed the mean free path for some common materials relevant in AGRS using again the XCOM database [208, 247] combined with material composition data provided by McConn et al. [249].[31] What immediately catches the eye in Fig. 3.6 is the fact that gases such as air have a significantly increased mean free path compared to solid or liquid materials, e.g. at a photon energy of 1 MeV the mean free path in air is about 128 m compared to 0.14 m in water or 0.02 m in steel. This difference has some important implications for AGRS, which I will discuss in the next section. In addition, we find a pronounced decrease in the mean free path for photon energies below about 100 keV for most materials plotted in Fig. 3.6. This can be explained by the pronounced increase in the photoelectric absorption cross-section at low photon energies. As we will discuss in the next chapter, the detector materials used in AGRS are typically selected to have a high mass density and contain heavy elements to maximize the total macroscopic cross-section or, conversely, minimize the mean free path of the photons in the corresponding material. This is why all displayed detector materials in Fig. 3.6 show a significantly lower mean free path than most other materials found in nature and aerospace engineering.

31 *As $E_\gamma < 10\,\mathrm{MeV}$, I neglected photonuclear reactions in these computations for simplicity (cf. Fig. 3.4).*

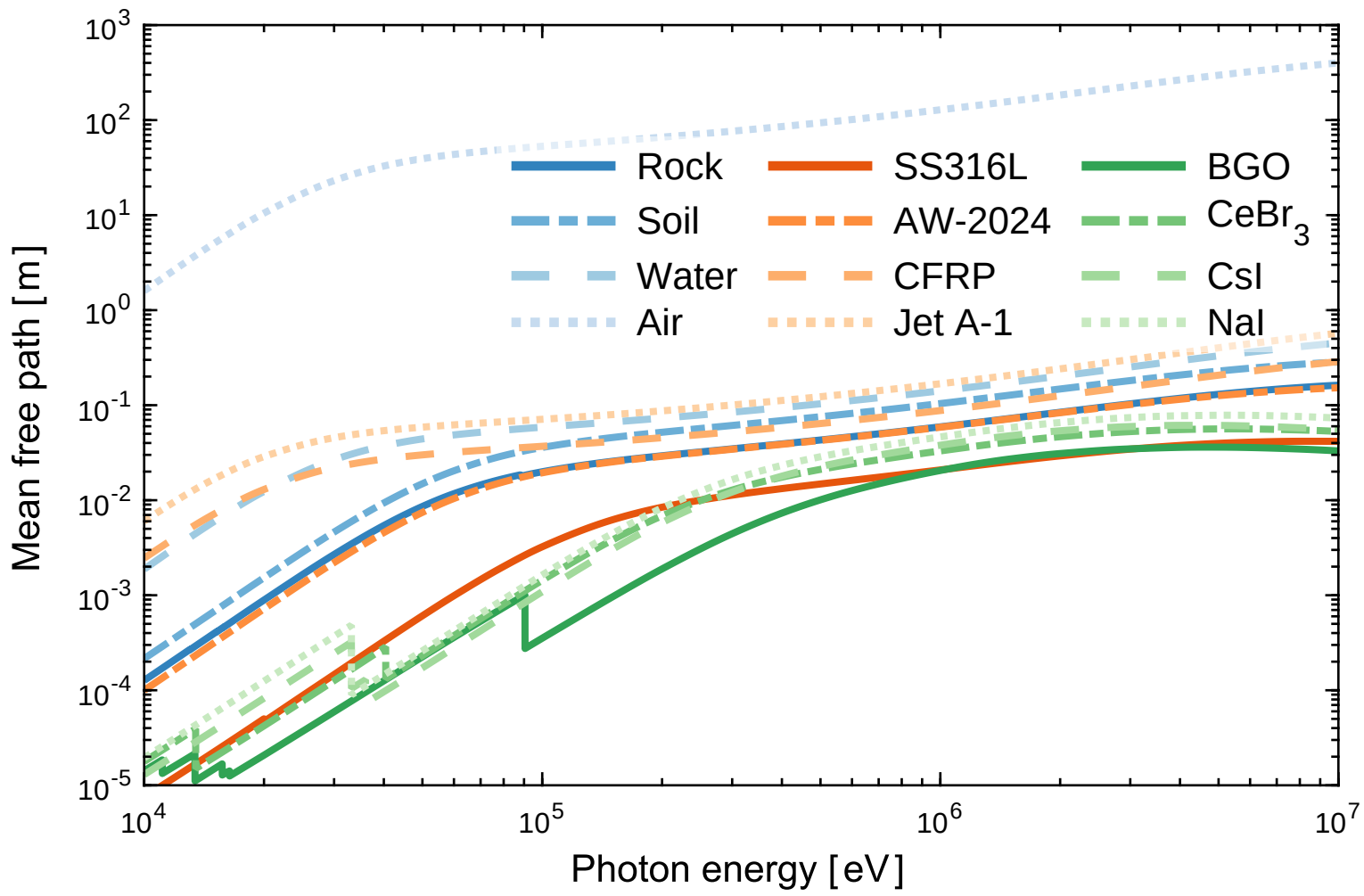


**Figure 3.6** Mean free path $\ell$ for selected materials as a function of the photon energy $E_\gamma$. Three material categories are shown, i.e. natural materials (rock mixture with equal weight fractions of basalt, granite, limestone, sandstone, and shale; soil mixture with 28 soils (dried) from throughout the U.S.; liquid water and dry air near sea level at $T = 15\,°\mathrm{C}$ and $p_{\mathrm{atm}} = 1.013\,25 \times 10^5\,\mathrm{Pa}$), materials used in aerospace engineering (stainless steel grade SS316L; aluminum alloy EN AW-2024 T3; carbon fiber-reinforced polymers (CFRP) with equal volume fractions of carbon fibers embedded in an epoxy matrix; aviation turbine fuel Jet A-1) as well as common inorganic scintillators applied in radiation detection (Bismuth germanate $Bi_4Ge_3O_{12}$ or BGO; Cerium bromide $CeBr_3$; Cesium iodide CsI and sodium iodide NaI). Material composition and cross-section data were adopted from McConn et al. [249] as well as the XCOM database by the NIST [208, 247], respectively.

## 3.2 Photon Transport Modeling

In the first part, we have analyzed in detail all the possible interaction processes of high-energy photons with matter. In this second part, we will discuss how we can use these interaction models presented in the previous paragraphs to predict the propagation of photons through the environment. The transport of photons in the environment can be described as a conservation or balance equation in phase space, that is for any given point in time $t$ and space $\mathbf{r}$, the density of photons with energy $E_\gamma$ and direction $\mathbf{\Omega}$ is the sum of "source" and "destruction"

processes. This balance equation can be written in the integral form of the linear Boltzmann transport equation (LBTE)[32] as [143]:

32 *Named after Ludwig Eduard Boltzmann (*1844, †1906), an Austrian theoretical physicist and philosopher. He is best known for his seminal work in thermodynamics and gas theory laying the ground for the field of statistical mechanics. In the context of gas theory, he derived a nonlinear version of the transport equation written in Eq. 3.26 [250].*

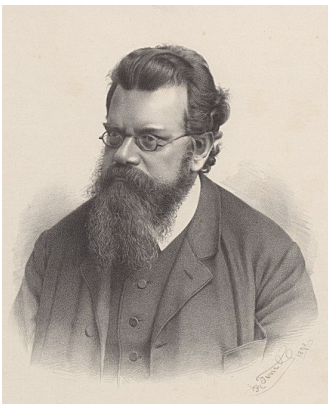

Ludwig Boltzmann
Ⓢ Rudolf Fenzl

$$\frac{\partial^2 \phi_\gamma}{\partial E_\gamma \partial \Omega}\left(\mathbf{r}, \mathbf{\Omega}, E_\gamma, t\right) = \int_0^\infty \underbrace{e^{-\int_0^s \mu_{\text{tot}}(\mathbf{r}'', E_\gamma)\, ds'}}_{\text{I}} \underbrace{Q_{\text{tot}}\left(\mathbf{r}', \mathbf{\Omega}, E_\gamma, t'\right)}_{\text{II}} ds \quad (3.26)$$

with:

| | | |
|---|---|---|
| $Q_{\text{tot}}$ | total source strength | $\text{s}^{-1}\,\text{m}^{-3}\,\text{sr}^{-1}\,\text{eV}^{-1}$ |
| $s$ | path length | m |
| $\mu_{\text{tot}}$ | total attenuation coefficient | $\text{m}^{-1}$ |
| $\partial^2\phi_\gamma/\partial E_\gamma \partial\Omega$ | double differential photon flux | $\text{s}^{-1}\,\text{m}^{-2}\,\text{sr}^{-1}\,\text{eV}^{-1}$ |

There are two distinct terms in Eq. 3.26. The first term (I) represents the probability for a given photon with energy $E_\gamma$ to propagate from a position $\mathbf{r}'$ over a distance $s = |\mathbf{r} - \mathbf{r}'|$ in the direction $\mathbf{\Omega}$ in time $\Delta t = t - t'$ without interacting with the surrounding matter. Consequently, we may interpret this term as an attenuation or loss factor. A graphical depiction of this propagation process is shown in Fig. 3.7. The second term (II) describes all source processes leading to the emission of photons with energy $E_\gamma$ at the position $\mathbf{r}'$ at time $t'$ in the direction $\mathbf{\Omega}$. This source term can be further divided into the

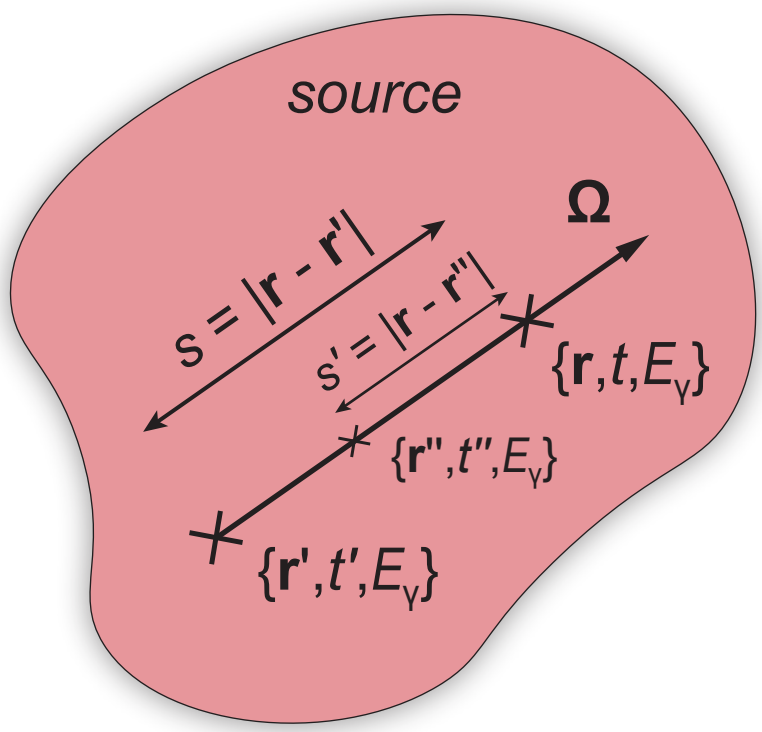


**Figure 3.7** Graphical depiction of phase-space variables adopted in the transport equation Eq. 3.26. These variables are the position vectors $\mathbf{r}$, $\mathbf{r}'$ and $\mathbf{r}''$, the direction vector $\mathbf{\Omega}$ as well as the time $t$, $t'$,$t''$ and photon energy $E_\gamma$. A similar graph can be found in the monograph by Prinja et al. [144].

sum of all external sources $Q_{\text{ext}}$, e.g. a radionuclide emitting gamma rays, and internal sources $Q_{\text{int}}$, which are defined as all interaction processes $i$ between a high-energy particle $p$ and matter leading to the emission of high-energy photons with energy $E_\gamma$ in the direction $\mathbf{\Omega}$ at a position $\mathbf{r}'$ and time $t'$:

$$Q_{\text{tot}}\left(\mathbf{r}', \mathbf{\Omega}, E_\gamma, t'\right) = Q_{\text{int}}\left(\mathbf{r}', \mathbf{\Omega}, E_\gamma, t'\right) + Q_{\text{ext}}\left(\mathbf{r}', \mathbf{\Omega}, E_\gamma, t'\right) \quad (3.27a)$$

$$Q_{\text{int}}\left(\mathbf{r}', \mathbf{\Omega}, E_\gamma, t'\right) = \sum_p \sum_i \int_0^\infty \int_0^{4\pi} \frac{\partial^2 \mu_{p,i}}{\partial E_\gamma \partial \Omega}\left(\mathbf{r}', \mathbf{\Omega}'_p \cdot \mathbf{\Omega}, E'_p, E_\gamma\right) \frac{\partial^2 \phi_p}{\partial E'_p \partial \Omega'_p}\left(\mathbf{r}', \mathbf{\Omega}'_p, E'_p, t'\right) \mathrm{d}\Omega'_p \, \mathrm{d}E'_p \quad (3.27b)$$

with:[33]

| | | |
|---|---|---|
| $E_p$ | particle energy | eV |
| $Q_{\text{ext}}$ | external source strength | $\text{s}^{-1}\,\text{m}^{-3}\,\text{sr}^{-1}\,\text{eV}^{-1}$ |
| $Q_{\text{int}}$ | internal source strength | $\text{s}^{-1}\,\text{m}^{-3}\,\text{sr}^{-1}\,\text{eV}^{-1}$ |
| $\partial^2\mu_{p,i}/\partial E_\gamma \partial\Omega$ | double differential attenuation coefficient | $\text{m}^{-1}\,\text{sr}^{-1}\,\text{eV}^{-1}$ |
| $\partial^2\phi_p/\partial E'_p \partial\Omega'_p$ | double differential particle flux | $\text{s}^{-1}\,\text{m}^{-2}\,\text{sr}^{-1}\,\text{eV}^{-1}$ |

The $Q_{\text{int}}$ source term in Eq. 3.27b includes not only direct photon interactions like Rayleigh or Compton scattering[34] but also all secondary photon emissions induced by secondary processes or secondary particles, e.g. characteristic X-ray emission, bremsstrahlung or gamma-ray emission from nuclear reactions. As a result, the photon transport described by the LBTE in Eq. 3.26 is a complex process coupled to the transport of other particles such as electrons or positrons. Thus, solving the transport equation presents a significant challenge, requiring accurate modeling not only of all photon interactions discussed in the first part of this chapter but also of all interactions involving other particles that can lead to the emission of high-energy photons. So, how do we solve this complex problem?

Before I answer this question, I would like to briefly highlight the main assumptions that are required for Eq. 3.26 to hold. Some of

33 *The double differential attenuation coefficient is obtained by replacing the microscopic cross-section in Eq. 3.7b with a generalized double differential cross-section $\sigma_{i,j,p}$ for interaction $i$ of particle $p$ with nuclide $j$. It quantifies the probability of a particle $p$ interacting with the corresponding matter, in the process emitting photons with energy $E_\gamma$ in the direction $\mathbf{\Omega}$.*

34 *It might seem confusing that scattering is a source of photons but remember that the LBTE describes the double differential flux. Consequently, all photons that get scattered from a direction $\mathbf{\Omega}'$ to the direction $\mathbf{\Omega}$ have to be considered in the source term $Q_{\text{int}}$. Of course, scattering always signifies also a loss term as photons get scattered from $\mathbf{\Omega}$ to $\mathbf{\Omega}'$, which is accounted for by $\mu_{\text{tot}}$ in the first term (I) in Eq. 3.26 (cf. Eqs. 3.24 and 3.25a).*

these assumptions may be relaxed but others are more fundamental and represent the main limits of the LBTE. These assumptions are [144]:

A. We assume that photons propagate through matter as a series of particle-like interactions between a point-like photon with free atoms. As discussed at the beginning of Section 3.1, this limits the applicability of the transport equation to high-energy photons with energies $\gtrsim 1\,\text{keV}$.

B. The propagation of photons between interactions can be described by classical mechanics and is fully determined by the seven phase-space variables $(\mathbf{r}, \boldsymbol{\Omega}, E_\gamma, t)$. Additional photon properties such as spin or polarization only affect the interaction processes. As such, we can write the position $\mathbf{r}'$ and time $t'$ in Eq. 3.26 as:[35]

$$\mathbf{r}' = \mathbf{r} - s\boldsymbol{\Omega} \tag{3.28a}$$

$$t' = t - \frac{s}{c} \tag{3.28b}$$

with:

| | | |
|---|---|---|
| c | speed of light in vacuum (cf. Constants) | $\mathrm{m\,s^{-1}}$ |
| $s$ | path length | m |

C. All media are considered isotropic. This implies rotational invariance for all interaction processes as already discussed in Section 3.1.1.

D. The attenuation coefficients do not depend on the photon flux. For this to hold, we need to introduce two additional assumptions:

   i. Photon-photon interactions must be negligible compared to other interaction modes. This is generally the case for photon transport in the environment as the density of high-energy photons and associated secondary particles is orders of magnitudes smaller than the number density of target nuclei.[36]

   ii. High-energy photons do not significantly alter the composition and structure of the matter within a time frame comparable to the transport of individual photons. Given the comparably short time periods for photons to propagate through matter, this assumption is generally fulfilled.

35 *In full analogy, we have of course also* $\mathbf{r}'' = \mathbf{r} - s'\boldsymbol{\Omega}$.

36 *In addition, photon-photon interaction processes such as electron-positron pair production by a photon pair or photon-photon scattering possess cross-sections only relevant over astronomical distances [251–254].*

This assumption is crucial as it ensures the linearity of the transport equation in Eq. 3.26. The linearity is a fundamental property of the LBTE that allows us to solve the equation by superposition of individual source terms. In general, if we can divide a complex source $Q_{\text{tot}}$ into a series of simpler sources $Q_{\text{tot}} = \sum_i Q_{\text{tot},i}$, then under certain boundary conditions [143], we can write the solution of the transport equation as $\phi_\gamma = \sum_i \phi_{\gamma,i}$ with $\phi_{\gamma,i}$ being the solutions for the individual transport problems with a $Q_{\text{tot},i}$ source term. The solutions $\phi_{\gamma_i}$ are also known as Green's functions. This is a significant simplification as it allows us to solve the transport equation for each source term separately and then sum up the results to obtain the total photon flux.

E. The propagation of the photon through matter is assumed to be Markovian, implying that only the photon's current state dictates its future interactions, independent of its prior history. This assumption holds naturally for most high-energy particles [255]. In a Markovian process, we can write the probability $\Pr(s)$ of an interaction to take place along a differential path length $\mathrm{d}s$ as:

$$\Pr(s) = \mu_{\text{tot}}\left(\mathbf{r}, E_\gamma\right) \mathrm{d}s \tag{3.29}$$

This is directly applied in the first term (I) in Eq. 3.26.

F. Due to the quantum mechanic nature of the interaction processes, photon transport is an inherently stochastic process. As a result, the photon flux fluctuates with time. This fluctuation is neglected and Eq. 3.26 thereby only describes the mean or expected value of $\phi_\gamma$.

With these assumptions and resulting limits in mind, let us come back to the question of how we can solve this transport equation, especially in the context of AGRS applications. There are three general approaches to solve the LBTE in Eq. 3.26:

**I. Analytical Methods** Analytical solutions for Eq. 3.26 are only possible for simple geometries and by applying additional assumptions regarding the source terms and time dynamics in Eq. 3.26. As a result, they have some limits in terms of accuracy and applicability. Furthermore, they are strictly limited to the flux as observable, i.e. detector response quantities are not accessible. Nevertheless, due to the simplicity of the solutions, the analytical approach is currently the main method adopted in AGRS in line with the current guidelines of the IAEA [8].

**II. Stochastic Methods** Stochastic methods, commonly known as Monte Carlo methods, approach the radiation transport problem by simulating the transport of photons on a particle-by-particle basis. Monte Carlo methods utilize the inherent statistical nature of the interaction of photons with matter by directly sampling from the continuous probability distributions described by the cross-sectional models.[37] In fact, Monte Carlo methods can simulate any high-energy particle transport through matter and thereby also account for coupled physics effects. As no approximations or discretizations are introduced in the underlying physical models, Monte Carlo methods are considered high-fidelity transport methods. They not only provide the highest accuracy but also the most flexibility in terms of geometry and source terms. As long as accurate physics models are available to describe the particle-matter interaction processes, arbitrary complex transport problems can be solved by simulating all relevant particle interactions. Furthermore, because each individual particle track is simulated, Monte Carlo methods are not restricted to photon flux estimates but can provide any observable quantity related to the particle transport including correlation and coincident measures between individual particles. However, due to the stochastic nature of the method, a large number of particles needs to be tracked to achieve accurate results for collective particle measures. This makes Monte Carlo simulations computationally expensive, especially for transport problems with extended simulation volumes and high-density materials. This might be one of the reasons why Monte Carlo simulations are not yet widely adopted in AGRS applications. Common Monte Carlo codes for radiation transport problems are `FLUKA` [20, 216], `Geant4` [21, 256, 257], `MCNP6` [22, 258] and `PHITS` [23]. I will discuss the Monte Carlo method in more detail later in this chapter.

37 *This sampling from a known probability distribution is also known as random sampling or just sampling.*

**III. Deterministic Methods** These numerical methods are based on the discretization of the phase-space variables in Eq. 3.26 and the subsequent solution of the resulting set of coupled differential equations. Deterministic codes provide global solutions for the double differential flux without statistical noise and with significantly lower computational effort compared to Monte Carlo methods. However, the discretization in the phase-space variables can lead to inaccuracies in the solution, especially in low-density materials relevant in radiation transport problems in the atmosphere [259].[38] Furthermore, similar to the analytical methods, deter-

38 *Especially the discretization in the angular variables can lead to significantly biased results, a phenomenon which is known as the ray effect [259].*

ministic methods are restricted to the photon flux as observable. Because of these shortcomings, they have rarely been adopted in AGRS, primarily in a hybrid approach together with Monte Carlo simulations [27]. Common deterministic codes used for radiation transport problems include `PARTISN` [260], `TORT` [261], `PENTRAN` [262] and `ATTILA` [263].

Monte Carlo simulations are currently the only available method that provides not only high-fidelity photon flux estimates for complex geometries but any physical quantity related to particle transport. Analytical and deterministic methods on the other hand are restricted to flux estimates. These methods require therefore additional experimental measurements or a hybrid approach combining the corresponding method with additional Monte Carlo simulations for AGRS applications [259, 264, 265]. Consequently, if the computational resources are available, Monte Carlo simulations are the method of choice for radiation transport problems, in particular for AGRS applications where the main radiation transport takes place in low-density materials, i.e. the atmosphere.

Before I introduce the Monte Carlo method in more detail, I will briefly highlight some analytical solutions for selected photon transport problems frequently adopted in AGRS. Although these solutions only partially represent the complex physics, they provide valuable insights into the main trends and properties of the photon flux for typical radiation transport problems in AGRS.

### 3.2.1 Analytical Methods

As indicated before, to be able to derive analytical solutions for common photon transport problems in AGRS, we need to introduce additional assumptions regarding the source term and flux in the LBTE in Eq. 3.26. These assumptions are:

G. Only external sources are considered, i.e. all internal sources, in particular Rayleigh and Compton scattering, are neglected. Furthermore, the external sources are assumed to be isotropic. Consequently, we can replace the source term $Q_{\text{tot}}(\mathbf{r}', \mathbf{\Omega}, E_\gamma, t')$ in Eq. 3.26 by $Q_{\text{iso}}(\mathbf{r}', E_\gamma, t')/(4\pi)$ with $Q_{\text{iso}}$ being the external isotropic source strength.

H. The time dynamics in the transport between interactions is neglected, i.e. $t \approx t'$.

I. Angular dependence of the photon flux is neglected by integrating Eq. 3.26 over all directions $\mathbf{\Omega}$. The resulting double integral with the differential element product $\mathrm{d}\Omega\mathrm{d}s$ can then be further simplified by replacing it with a single integral over the system volume $V$, as $\mathrm{d}\Omega\mathrm{d}s$ is equivalent to $\mathrm{d}V/s^2$ (cf. the solid angle definition discussed in Section 2.2.1).

Using these assumptions, we obtain a simplified transport equation for the energy flux instead of the double differential flux as a function of the position $\mathbf{r}$, photon energy $E_\gamma$ [143, 144] and time $t$:

$$\frac{\mathrm{d}\phi_\gamma}{\mathrm{d}E_\gamma}\left(\mathbf{r}, E_\gamma, t\right) = \int_V \underbrace{\frac{\mathrm{e}^{-\int_0^s \mu_{\mathrm{tot}}\left(\mathbf{r}'', E_\gamma\right)\mathrm{d}s'}}{4\pi s^2}}_{\mathrm{I}} \underbrace{Q_{\mathrm{iso}}\left(\mathbf{r}', E_\gamma, t\right)}_{\mathrm{II}} \mathrm{d}V \tag{3.30}$$

where:

| | | |
|---|---|---|
| $Q_{\mathrm{iso}}$ | external isotropic source strength | $\mathrm{s}^{-1}\,\mathrm{m}^{-3}\,\mathrm{eV}^{-1}$ |
| $s$ | path length | m |
| $V$ | system volume | $\mathrm{m}^3$ |
| $\mu_{\mathrm{tot}}$ | total attenuation coefficient | $\mathrm{m}^{-1}$ |

with the first term (I) representing again the attenuation of the photon flux due to the interaction with the surrounding matter as in Eq. 3.26 but this time also accounting for the geometric spreading of the isotropic source. This term is also referred to as the free flight kernel and is equivalent to the Green's function for a unit isotropic point source at the position $\mathbf{r}'$ with energy $E_\gamma$ [143]. The second term in Eq. 3.30 (II) represents the external isotropic source term. As all internal sources, in particular Rayleigh and Compton scattering, are neglected, Eq. 3.30 describes essentially the primary photon flux, that is the flux of photons emitted by the source without any interaction with the surrounding matter. Analytical methods based on Eq. 3.30 are therefore also referred to as primary quanta theory or monoenergetic transport theory [8, 52].

In the following paragraphs, I will briefly present four selected source geometries and resulting energy flux solutions using the simplified transport equation in Eq. 3.30. This discussion is based on the results presented in the monographs by Schwarz [10] and Kogan et al. [52]. The four source geometries are:

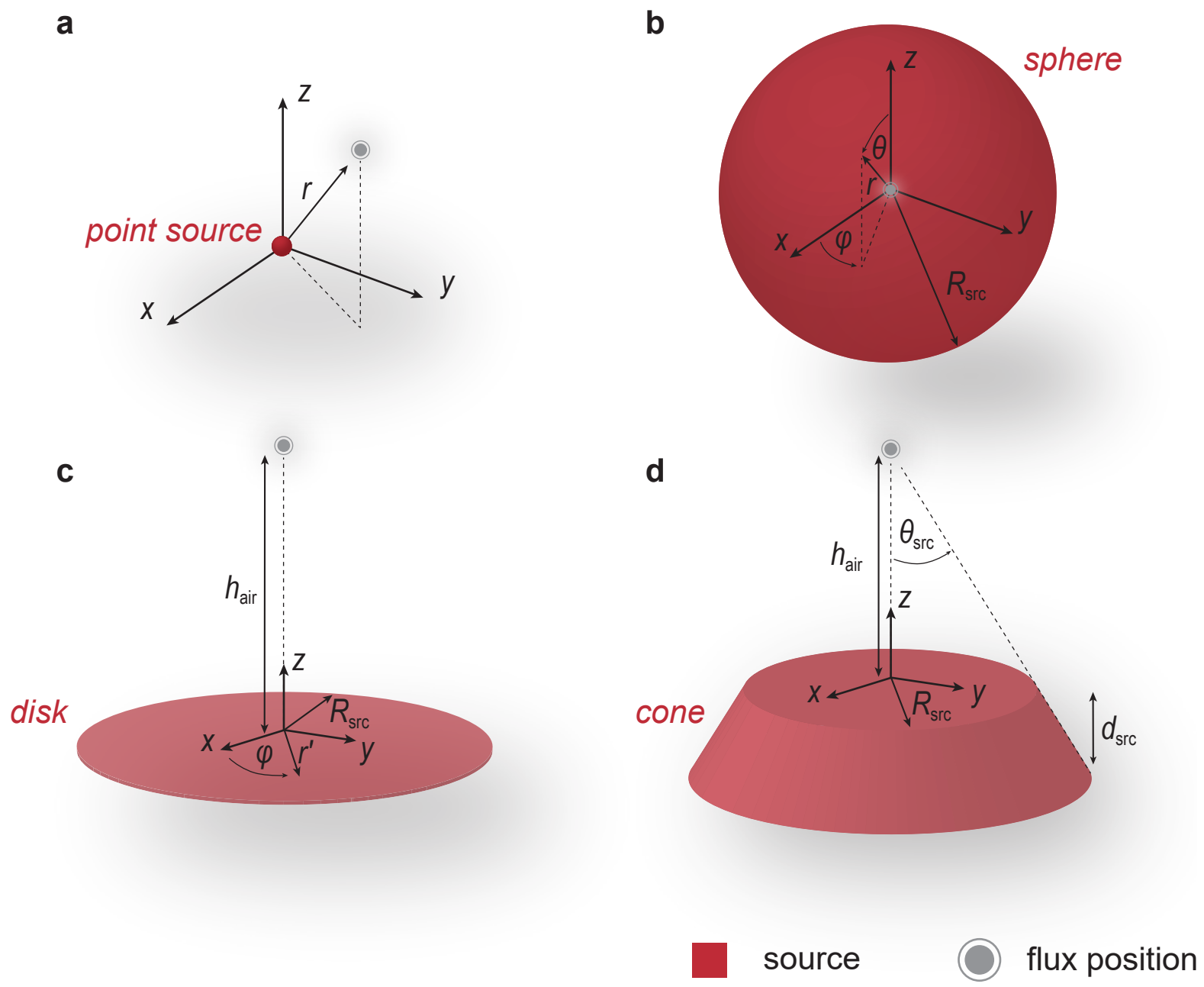


**Figure 3.8** Primary quanta source geometries. **a** Point source. **b** Sphere source. **c** Disk source. **d** Truncated cone source.

1. Point source
2. Sphere source
3. Disk source
4. Truncated cone source

A graphical illustration including the relevant geometrical parameters for these source geometries is shown in Fig. 3.8.

### 3.2.1.1 Point Source

Let us start our discussion with the simplest source: the point source. For this case, we assume an isotropic point source with source strength $Q_p$ is located in the air or on the ground surface at a position $\mathbf{r}'$ emitting high-energy photons with energy $E_\gamma$. The air is assumed to be a homogeneous and isotropic medium. The primary

energy flux in the air in a distance $r = |\mathbf{r} - \mathbf{r}'|$ from the source is then given simply by the free flight kernel discussed before combined with the external source strength:

$$\frac{\mathrm{d}\phi_\gamma}{\mathrm{d}E_\gamma}\left(r, E_\gamma, t\right) = \frac{Q_\mathrm{p}\left(E_\gamma, t\right)\mathrm{e}^{-\mu_\mathrm{air}(E_\gamma)r}}{4\pi r^2} \tag{3.31}$$

where:

| | | |
|---|---|---|
| $Q_\mathrm{p}$ | external isotropic point source strength | $\mathrm{s}^{-1}\,\mathrm{eV}^{-1}$ |
| $\mu_\mathrm{air}$ | total attenuation coefficient in air | $\mathrm{m}^{-1}$ |
| $r$ | distance to source | m |

Consequently, the primary energy flux for an isotropic point source in the homogeneous atmosphere is proportional to the point source strength in line with the linearity of the LBTE and exponentially decreases with the distance to the source. Possible applications of this source geometry in the context of AGRS include lost-source or nuclear-debris scenarios with man-made gamma-ray emitting radionuclide point sources in the environment. In such scenarios, the point source strength $Q_\mathrm{p}$ is equivalent to $\mathcal{A}(t)\mathrm{d}I_\gamma(E_\gamma)/\mathrm{d}E_\gamma$ with the source activity $\mathcal{A}$ and gamma-ray emission intensity $I_\gamma$ (cf. Section 2.1.2).

#### 3.2.1.2 Sphere Source

Next, let us consider the sphere source. In this geometry, an external isotropic sphere source with strength $Q_\mathrm{v}$ and radius $R_\mathrm{src}$ is located in an infinite homogeneous and isotropic air medium emitting high energy photons with energy $E_\gamma$. Possible applications of this source geometry in AGRS are gaseous gamma-ray emitting radionuclides discussed in Section 2.1.3, e.g. natural radon progeny radionuclides or anthropogenic radionuclides released by accelerators and nuclear reactors. We obtain the primary energy flux at the center of the sphere source by integrating the free flight kernel over the sphere volume in spherical coordinates with the volume element $\mathrm{d}V = r^2 \sin(\theta)\,\mathrm{d}r\mathrm{d}\theta\mathrm{d}\varphi$ (cf. Fig. 3.8):

$$\frac{\mathrm{d}\phi_\gamma}{\mathrm{d}E_\gamma}\left(R_{\mathrm{src}}, E_\gamma, t\right) = \int_0^{2\pi}\int_0^{\pi}\int_0^{R_{\mathrm{src}}} \frac{Q_{\mathrm{v}}\left(E_\gamma, t\right)\mathrm{e}^{-\mu_{\mathrm{air}}\left(E_\gamma\right)r}}{4\pi} \sin(\theta)\,\mathrm{d}r\,\mathrm{d}\theta\,\mathrm{d}\varphi \tag{3.32a}$$

$$= \frac{Q_{\mathrm{v}}\left(E_\gamma, t\right)}{\mu_{\mathrm{air}}\left(E_\gamma\right)}\left[1 - \mathrm{e}^{-\mu_{\mathrm{air}}\left(E_\gamma\right)R_{\mathrm{src}}}\right] \tag{3.32b}$$

with:

| | | |
|---|---|---|
| $Q_{\mathrm{v}}$ | external isotropic volume source strength | $\mathrm{s}^{-1}\,\mathrm{m}^{-3}\,\mathrm{eV}^{-1}$ |
| $R_{\mathrm{src}}$ | source radius | m |
| $\mu_{\mathrm{air}}$ | total attenuation coefficient in air | $\mathrm{m}^{-1}$ |

For airborne radionuclide sources, the source strength $Q_{\mathrm{v}}$ is equivalent to $a_{\mathrm{v}}(t)\mathrm{d}I_\gamma(E_\gamma)/\mathrm{d}E_\gamma$ with the activity volume concentration $a_{\mathrm{v}}$ and gamma-ray emission intensity $I_\gamma$ (cf. Section 2.1.2). From Eq. 3.32b, we can see that the primary energy flux for a sphere source in the homogeneous atmosphere is proportional to the source strength $Q_{\mathrm{v}}$ and inversely proportional to the total attenuation coefficient in air. Moreover, the primary energy flux approaches a maximum value of:

$$\frac{\mathrm{d}\phi_\gamma}{\mathrm{d}E_\gamma}\left(E_\gamma, t\right) = \frac{Q_{\mathrm{v}}\left(E_\gamma, t\right)}{\mu_{\mathrm{air}}\left(E_\gamma\right)} \tag{3.33}$$

for an infinite volume source as $R_{\mathrm{src}} \to \infty$. Consequently, the relative contribution of a finite sphere source to the primary energy flux of an infinite volume source is given by the term $1 - \exp\left(-\mu_{\mathrm{air}}R_{\mathrm{src}}\right)$. As an example, for a homogeneous air medium[39], 95 % of the primary energy flux is attributed to photons emitted within a distance of 5 m, 160 m and 385 m for photon energies of 10 keV, 100 keV and 1 MeV, respectively. This showcases the attenuation of the atmosphere on high-energy photons. As discussed in the first part, attenuation is exponentially increasing with decreasing photon energy which explains the significant reduction in the contributing volume for low photon energies. We will keep this interesting property in mind for later when we discuss Monte Carlo simulations of sphere volume sources.

39 *Dry air near sea level as defined by McConn et al. [249] with $\mu_{\mathrm{air}}$ adopted from the XCOM database by the NIST [208, 247].*

#### 3.2.1.3 Disk Source

The third source geometry displayed in Fig. 3.8 is the disk source. In this geometry, an external isotropic disk source with strength $Q_s$ and radius $R_{src}$ is located on the ground surface emitting high-energy photons with energy $E_\gamma$. Possible applications of the disk source geometry in AGRS are scenarios with deposition of airborne radionuclides on the ground surface, e.g. deposition of natural radon progeny radionuclides or anthropogenic radionuclides released in nuclear weapon explosions or nuclear accidents. The primary energy flux with a ground clearance of $h_{air}$ in an infinite homogeneous and isotropic air medium located above the disk source is given by:

$$\frac{\mathrm{d}\phi_\gamma}{\mathrm{d}E_\gamma}\left(R_{src}, h_{air}, E_\gamma, t\right) = \frac{Q_s(E_\gamma, t)}{2}\left\{\mathcal{E}_1\left[\mu_{air}(E_\gamma)h_{air}\right] - \mathcal{E}_1\left[\mu_{air}(E_\gamma)\sqrt{h_{air}^2 + R_{src}^2}\right]\right\} \quad (3.34)$$

with:

| | | |
|---|---|---|
| $h_{air}$ | ground clearance | m |
| $Q_s$ | external isotropic surface source strength | $s^{-1}\,m^{-2}\,eV^{-1}$ |
| $R_{src}$ | source radius | m |
| $\mu_{air}$ | total attenuation coefficient in air | $m^{-1}$ |

and where $\mathcal{E}_1$ denotes the exponential integral function of the first order. Selected properties of this function alongside a derivation of the above solution are discussed in Appendix A.6. For a gamma-ray emitting radionuclide surface source, the source strength $Q_s$ is equivalent to $a_S(t)\mathrm{d}I_\gamma(E_\gamma)/\mathrm{d}E_\gamma$ with the surface activity concentration $a_S$ and gamma-ray emission intensity $I_\gamma$ (cf. Section 2.1.2). If we consider an infinite disk source with $R_{src} \to \infty$, the primary energy flux simplifies to (cf. Eq. A.19a in Appendix A.6):

$$\frac{\mathrm{d}\phi_\gamma}{\mathrm{d}E_\gamma}\left(h_{air}, E_\gamma, t\right) = \frac{Q_s(E_\gamma, t)}{2}\mathcal{E}_1\left[\mu_{air}(E_\gamma)h_{air}\right] \quad (3.35)$$

For both, the finite and infinite disk source, the primary energy flux is proportional to the source strength $Q_s$. As expected, the flux decreases with the ground clearance $h_{air}$ but at a slower rate than for

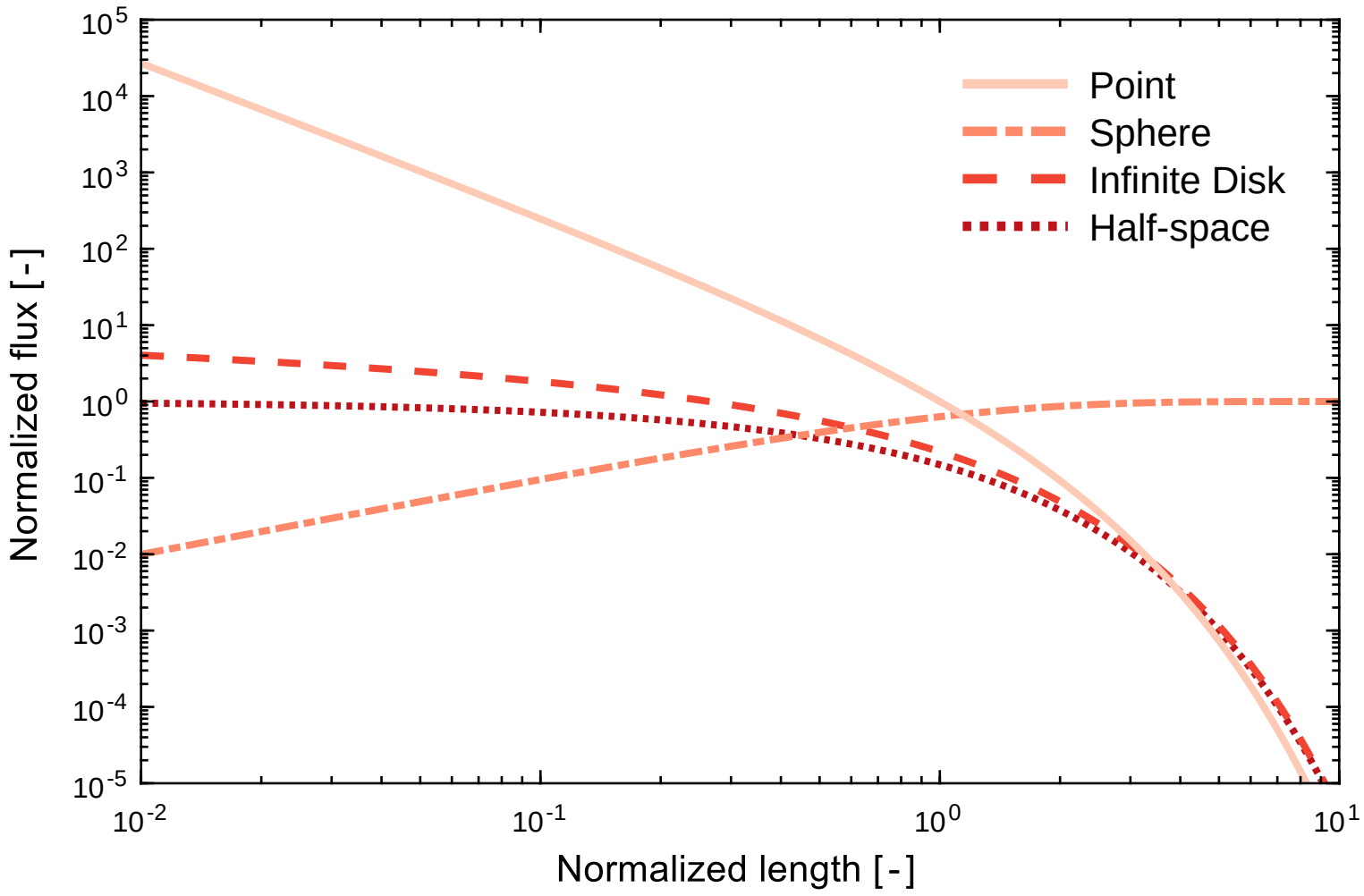


**Figure 3.9** Normalized primary quanta flux $(\mathrm{d}\phi_\gamma/\mathrm{d}E_\gamma)/(\mathrm{d}\phi_{\gamma,\mathrm{ref}}/\mathrm{d}E_\gamma)$ for selected source geometries as a function of the normalized length $l/\ell$ with $\ell$ being the mean free path (cf. Eq. 3.9). These source geometries are a point source with $l = r$, Eq. 3.31 for $\mathrm{d}\phi_\gamma/\mathrm{d}E_\gamma$ and $\mathrm{d}\phi_{\gamma,\mathrm{ref}}/\mathrm{d}E_\gamma = Q_\mathrm{p}\mu_\mathrm{air}^2/(4\pi\mathrm{e})$; a sphere source with $l = R_\mathrm{src}$, Eq. 3.32b for $\mathrm{d}\phi_\gamma/\mathrm{d}E_\gamma$ and Eq. 3.33 for $\mathrm{d}\phi_{\gamma,\mathrm{ref}}/\mathrm{d}E_\gamma$; a disk source with infinite extent with $l = h_\mathrm{air}$, Eq. 3.35 for $\mathrm{d}\phi_\gamma/\mathrm{d}E_\gamma$ and $\mathrm{d}\phi_{\gamma,\mathrm{ref}}/\mathrm{d}E_\gamma = Q_\mathrm{s}/2$; as well as a half-space source with $l = h_\mathrm{air}$, Eq. 3.38c for $\mathrm{d}\phi_\gamma/\mathrm{d}E_\gamma$ and $\mathrm{d}\phi_{\gamma,\mathrm{ref}}/\mathrm{d}E_\gamma = Q_\mathrm{v}/(2\mu_\mathrm{src})$.

the point source which can be seen in Fig. 3.9. Similar to the sphere source, we may also analyze the relative contribution of a finite disk source with source radius $R_\mathrm{src}$ to the energy flux of an infinite disk source by dividing Eq. 3.34 by Eq. 3.35:

$$\frac{\mathrm{d}\phi_\gamma/\mathrm{d}E_\gamma\left(R_\mathrm{src}, h_\mathrm{air}, E_\gamma\right)}{\mathrm{d}\phi_\gamma/\mathrm{d}E_\gamma\left(h_\mathrm{air}, E_\gamma\right)} = 1 - \frac{\mathscr{E}_1\left[\mu_\mathrm{air}\left(E_\gamma\right)\sqrt{h_\mathrm{air}^2 + R_\mathrm{src}^2}\right]}{\mathscr{E}_1\left[\mu_\mathrm{air}\left(E_\gamma\right)h_\mathrm{air}\right]} \tag{3.36}$$

This relative contribution is displayed in Fig. 3.10 as a function of the source radius $R_\mathrm{src}$ and the ground clearance $h_\mathrm{air}$ evaluated at three different photon energies $E_\gamma$, i.e. 10 keV (subfigure a), 100 keV (subfigure b) and 1 MeV (subfigure c) for dry air near sea level.[40,41] From that, we see that 95 % of the primary energy flux from an infinite disk source is attributed to photons emitted within a fairly restricted

*40 As defined by McConn et al. [249] with $\mu_\mathrm{air}$ adopted again from the XCOM database by the NIST [208, 247].*

*41 A word of warning if you would like to evaluate Eq. 3.36 yourself. As $\mathscr{E}_1$ rapidly converges to zero for fairly moderate argument values, evaluating ratios of exponential integral functions such as that in Eq. 3.36 can be problematic due to numerical underflow of $\mathscr{E}_1$. Therefore, common numerical software packages offer scaled exponential integral functions, e.g. $\mathscr{E}_1(x)\mathrm{e}^x$, to prevent numerical underflow.*

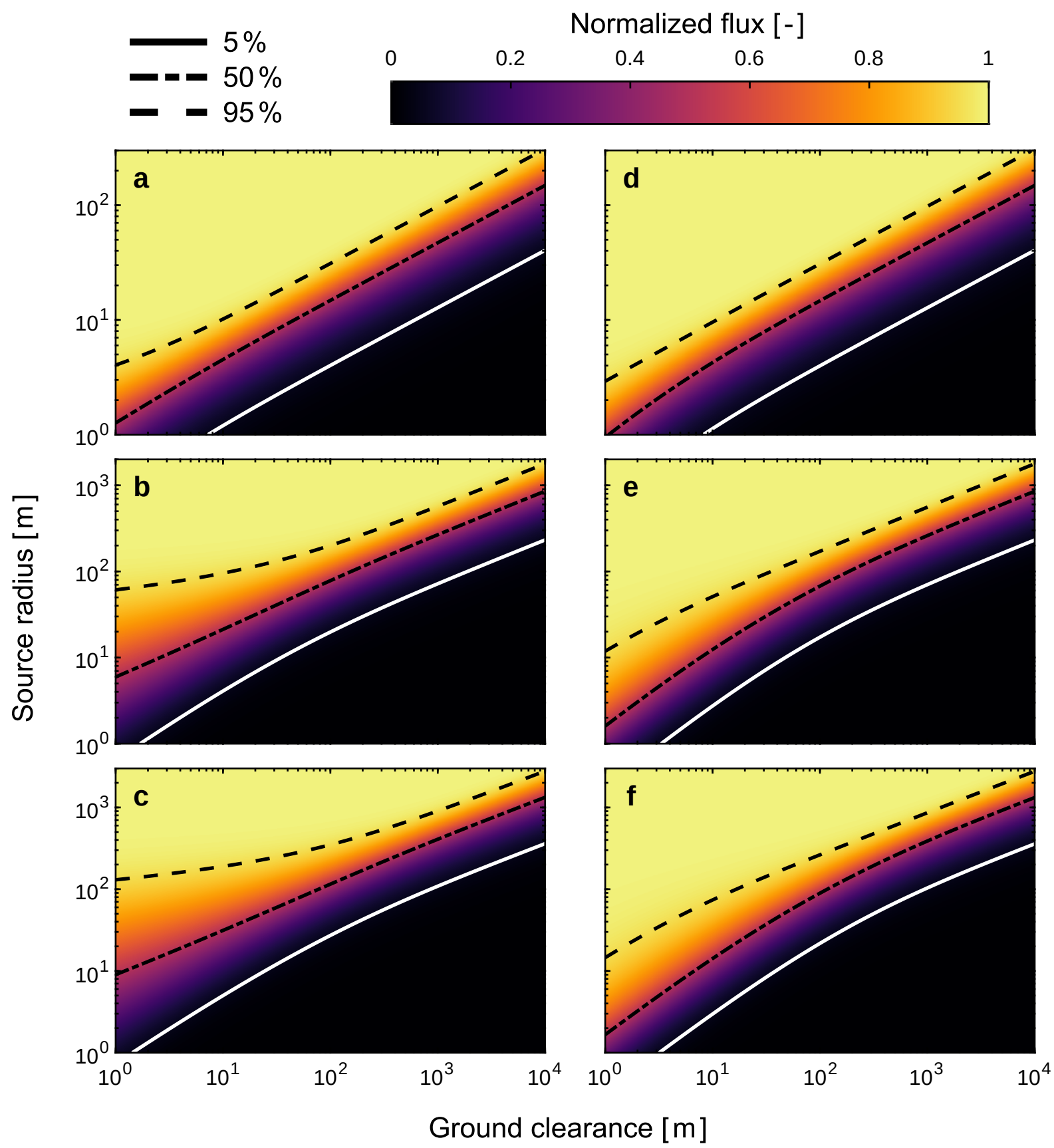


**Figure 3.10** Normalized primary quanta flux as a function of the ground clearance $h_{\mathrm{air}}$ and the source radius $R_{\mathrm{src}}$ in air. **a–c** Disk source with the normalized flux given by Eq. 3.36. **d–f** Solid truncated cone source ($d_{\mathrm{src}} \to \infty$) with the normalized flux given by Eq. 3.39b. For both source geometries, the total attenuation coefficient in air $\mu_{\mathrm{air}}$ was evaluated at three different photon energies $E_{\gamma}$, i.e. $10^4$ eV (**a**, **d**), $10^5$ eV (**b**, **e**) and $10^6$ eV (**c**, **f**). Material composition and cross-section data were adopted from McConn et al. [249] as well as the XCOM database by the NIST [208, 247], respectively.

area, e.g. with a typical ground clearance of 100 m for AGRS survey flights we have 95 % of the contributing primary photons emitted within a radius of about 30 m, 200 m and 350 m for the selected photon energies $E_\gamma$ of 10 keV, 100 keV and 1 MeV, respectively. This showcases again the attenuation of the atmosphere on high-energy photons and the significant reduction in the contributing source area for low photon energies.

#### 3.2.1.4 Truncated Cone Source

The last source geometry we will discuss is the truncated cone source. The truncated cone source consists of a homogeneous isotropic source medium characterized by the total attenuation coefficient $\mu_{\text{src}}$, the external source strength $Q_{\text{v}}$, the source depth $d_{\text{src}}$ and the source radius $R_{\text{src}}$ as displayed in Fig. 3.8. It is a prototypic geometry for terrestrial volume sources, e.g. natural radionuclides in the soil or rock such as ${}^{40}_{19}\text{K}$, ${}^{232}_{90}\text{Th}$ and ${}^{238}_{92}\text{U}$ with related progeny radionuclides as discussed in Section 2.1.3. The primary energy flux with a ground clearance $h_{\text{air}}$ in an infinite homogeneous isotropic air medium above the truncated cone source as a function of the photon energy is then given by:[42]

42 *Please note that for the sake of brevity, I do not explicitly denote the energy dependence of the attenuation coefficient variables in the subsequent expressions.*

$$\frac{\mathrm{d}\phi_\gamma}{\mathrm{d}E_\gamma}\left(d_{\text{src}}, R_{\text{src}}, h_{\text{air}}, E_\gamma, t\right) = \frac{Q_{\text{v}}(E_\gamma, t)}{2\mu_{\text{src}}}\Bigg\{ \mathscr{E}_2\left(\mu_{\text{air}}h_{\text{air}}\right) - \mathscr{E}_2\left(\mu_{\text{air}}h_{\text{air}} + \mu_{\text{src}}d_{\text{src}}\right) - \cos(\theta_{\text{src}})\left[\mathscr{E}_2\left(\frac{\mu_{\text{air}}h_{\text{air}}}{\cos(\theta_{\text{src}})}\right) - \mathscr{E}_2\left(\frac{\mu_{\text{air}}h_{\text{air}} + \mu_{\text{src}}d_{\text{src}}}{\cos(\theta_{\text{src}})}\right)\right]\Bigg\} \tag{3.37}$$

where:

| | | |
|---|---|---|
| $d_{\text{src}}$ | source depth | m |
| $h_{\text{air}}$ | ground clearance | m |
| $Q_{\text{v}}$ | external isotropic volume source strength | $\text{s}^{-1}\,\text{m}^{-3}\,\text{eV}^{-1}$ |
| $R_{\text{src}}$ | source radius | m |
| $\theta_{\text{src}}$ | view angle of the source | rad |
| $\mu_{\text{air}}$ | total attenuation coefficient in air | $\text{m}^{-1}$ |
| $\mu_{\text{src}}$ | total attenuation coefficient in the source | $\text{m}^{-1}$ |

and with $\cos(\theta_{\mathrm{src}}) = h_{\mathrm{air}}/(h_{\mathrm{air}}^2 + R_{\mathrm{src}}^2)^{1/2}$. As for the sphere source, for a gamma-ray emitting radionuclide volume source, the source strength $Q_{\mathrm{v}}$ is equivalent to $a_{\mathrm{v}}(t)\mathrm{d}I_\gamma(E_\gamma)/\mathrm{d}E_\gamma$ with the activity volume concentration $a_{\mathrm{v}}$ and gamma-ray emission intensity $I_\gamma$ (cf. Section 2.1.2). $\mathscr{E}_2$ in Eq. 3.37 denotes again the exponential integral function, but this time of the second order. Please refer to Appendix A.6 for selected properties of this function alongside a derivation of the above solution. Based on the properties listed in Appendix A.6, we can derive three special cases from Eq. 3.37:

$$\frac{\mathrm{d}\phi_\gamma}{\mathrm{d}E_\gamma}\left(d_{\mathrm{src}}, h_{\mathrm{air}}, E_\gamma, t\right) = \frac{Q_{\mathrm{v}}(E_\gamma, t)}{2\mu_{\mathrm{src}}}\left[\mathscr{E}_2\left(\mu_{\mathrm{air}} h_{\mathrm{air}}\right) - \mathscr{E}_2\left(\mu_{\mathrm{air}} h_{\mathrm{air}} + \mu_{\mathrm{src}} d_{\mathrm{src}}\right)\right] \tag{3.38a}$$

$$\frac{\mathrm{d}\phi_\gamma}{\mathrm{d}E_\gamma}\left(R_{\mathrm{src}}, h_{\mathrm{air}}, E_\gamma, t\right) = \frac{Q_{\mathrm{v}}(E_\gamma, t)}{2\mu_{\mathrm{src}}}\left[\mathscr{E}_2\left(\mu_{\mathrm{air}} h_{\mathrm{air}}\right) - \cos(\theta_{\mathrm{src}})\,\mathscr{E}_2\left(\frac{\mu_{\mathrm{air}} h_{\mathrm{air}}}{\cos(\theta_{\mathrm{src}})}\right)\right] \tag{3.38b}$$

$$\frac{\mathrm{d}\phi_\gamma}{\mathrm{d}E_\gamma}\left(h_{\mathrm{air}}, E_\gamma, t\right) = \frac{Q_{\mathrm{v}}(E_\gamma, t)}{2\mu_{\mathrm{src}}}\mathscr{E}_2\left(\mu_{\mathrm{air}} h_{\mathrm{air}}\right) \tag{3.38c}$$

That is a volume source with infinite radius $R_{\mathrm{src}} \to \infty$ but finite source depth in Eq. 3.38a, a volume source with infinite depth $d_{\mathrm{src}} \to \infty$ but finite radius in Eq. 3.38b, and a half-space source combining both, an infinite radius and depth $R_{\mathrm{src}}, d_{\mathrm{src}} \to \infty$ given by Eq. 3.38c. For the general case as well as the three special cases, we find again that the primary energy flux is proportional to the source strength $Q_{\mathrm{v}}$. Furthermore, the flux decreases with increasing ground clearance $h_{\mathrm{air}}$ at an even slower rate than for the disk source as illustrated in Fig. 3.9 for the half-space source geometry. Similar to the disk and sphere source, we can also compute the relative contribution of the sources with finite depth or radius to the energy flux of a reference source with infinite extent, i.e. the half-space source, by dividing Eq. 3.38a and Eq. 3.38b by Eq. 3.38c:

$$\frac{\mathrm{d}\phi_\gamma/\mathrm{d}E_\gamma\left(d_{\mathrm{src}}, h_{\mathrm{air}}, E_\gamma\right)}{\mathrm{d}\phi_\gamma/\mathrm{d}E_\gamma\left(h_{\mathrm{air}}, E_\gamma\right)} = 1 - \frac{\mathcal{E}_2\left(\mu_{\mathrm{air}} h_{\mathrm{air}} + \mu_{\mathrm{src}} d_{\mathrm{src}}\right)}{\mathcal{E}_2\left(\mu_{\mathrm{air}} h_{\mathrm{air}}\right)} \tag{3.39a}$$

$$\frac{\mathrm{d}\phi_\gamma/\mathrm{d}E_\gamma\left(R_{\mathrm{src}}, h_{\mathrm{air}}, E_\gamma\right)}{\mathrm{d}\phi_\gamma/\mathrm{d}E_\gamma\left(h_{\mathrm{air}}, E_\gamma\right)} = 1 - \frac{\cos(\theta_{\mathrm{src}})\,\mathcal{E}_2\left(\frac{\mu_{\mathrm{air}} h_{\mathrm{air}}}{\cos(\theta_{\mathrm{src}})}\right)}{\mathcal{E}_2\left(\mu_{\mathrm{air}} h_{\mathrm{air}}\right)} \tag{3.39b}$$

First, it is interesting to note that the relative contribution for a source with finite radius and infinite depth in Eq. 3.39b does not depend on the source medium, i.e. the total attenuation coefficient $\mu_{\mathrm{src}}$. Eq. 3.39b is displayed in the subfigures d–f in Fig. 3.10 as a function of the source radius $R_{\mathrm{src}}$ and the ground clearance $h_{\mathrm{air}}$ for dry air near sea level and the three photon energies $E_\gamma$ 10 keV (subfigure d), 100 keV (subfigure e) and 1 MeV (subfigure f).[43] It is evident that the contributing area enclosed by the source radius is smaller compared to the source disk (subfigures a–c in the same figure), in particular for higher photon energies and at lower ground clearance. For example, with a typical ground clearance of 100 m for AGRS survey flights, 95 % of the primary energy flux from a half-space source is attributed to photons emitted within a radius of about 30 m, 175 m and 265 m for the photon energies $E_\gamma$ 10 keV, 100 keV and 1 MeV, respectively.

Using Eq. 3.39a, we can also investigate the relative contribution of a volume source with finite depth and infinite radius to the primary energy flux of a half-space source. In Fig. 3.11, this relative contribution is displayed as a function of the source depth $d_{\mathrm{src}}$ and the ground clearance $h_{\mathrm{air}}$ again for dry air near sea level and the three photon energies $E_\gamma$ 10 keV, 100 keV and 1 MeV. As Eq. 3.39a not only depends on $\mu_{\mathrm{air}}$ but also $\mu_{\mathrm{src}}$, we need to select a source medium for the following discussion. For that purpose, I chose two representative source materials relevant in AGRS, that is a soil mixture[44] in the subfigures a–c and a rock mixture[45] in the subfigures d–f. There are two interesting points to note. First, the normalized energy flux only weakly depends on the ground clearance. This indicates that variations in the source depth primarily impact the attenuation within the source medium rather than in the surrounding air. Second, the majority of the photons contributing to the primary energy flux of a half-space source are emitted within a comparably shallow depth of $d_{\mathrm{src}} < 1$ m. This characteristic thickness decreases with decreasing photon energy and increasing attenuation coefficient of the source matter. For example, with a typical ground clearance of 100 m for

43 *Again, numerically evaluating ratios of exponential integral functions such as in Eq. 3.39a or Eq. 3.39b can be problematic due to numerical underflow. Therefore, consider using scaled exponential integral functions, e.g.* $\mathcal{E}_1(x)e^x$, *if you would like to evaluate Eqs. 3.39a and 3.39b yourself.*

44 *Defined as a mixture of 28 soils (dried) from throughout the U.S. with a mass density of* $\rho = 1.52\,\mathrm{g\,cm^{-3}}$ *[249]*

45 *Rock mixture with equal weight fractions of basalt, granite, limestone, sandstone, and shale with a mass density of* $\rho = 2.66\,\mathrm{g\,cm^{-3}}$ *[249]*

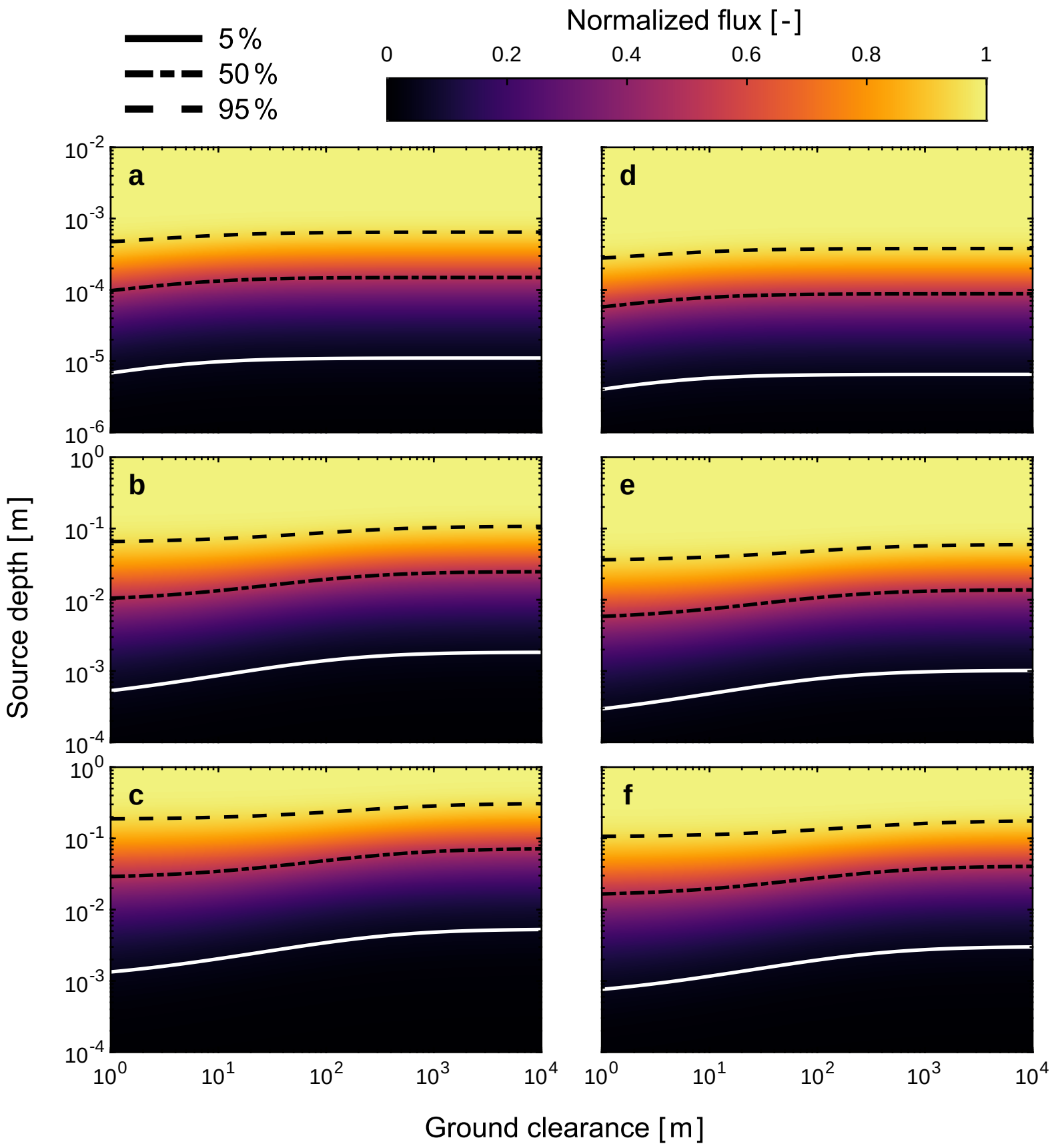


**Figure 3.11** Normalized primary quanta flux as a function of the ground clearance $h_{air}$ and the source depth $d_{src}$ in air. The normalized flux $\phi_\gamma$ was evaluated for a truncated cone source with finite depth and infinite radius ($R_{src} \rightarrow \infty$) using Eq. 3.39a. For the total attenuation coefficient in the source $\mu_{src}$, two different source materials were adopted: **a–c** Soil mixture (28 soils (dried) from throughout the U.S.). **d–f** Rock mixture (equal weight fractions of basalt, granite, limestone, sandstone, and shale). The total attenuation coefficient in the source $\mu_{src}$ as well as the total attenuation coefficient in air $\mu_{air}$ were evaluated as in Fig. 3.10 at three photon energies, i.e. $E_\gamma$ $10^4$ eV (**a, d**), $10^5$ eV (**b, e**) and $10^6$ eV (**c, f**). Material composition and cross-section data were adopted from McConn et al. [249] as well as the XCOM database by the NIST [208, 247], respectively.

AGRS survey flights, 95 % of the primary energy flux from a half-space source consisting of soil/rock is attributed to photons emitted within a depth of only 0.6 mm/0.4 mm, 9 cm/5 cm and 23 cm/13 cm for photon energies of 10 keV, 100 keV and 1 MeV, respectively. This highlights the fact that the primary energy flux emitted by terrestrial gamma-ray radionuclide volume sources is restricted to the first few centimeters of the ground surface.

As noted at the beginning of this section, based on the linearity of the LBTE, we can use the analytical solutions presented in this section also as building blocks to predict the primary energy flux of more complex source geometries. Here, I would like to give such an example, that is the primary energy flux of an atmospheric source such as gamma-ray emitting radon progeny radionuclides close to the ground. To estimate the primary energy flux for such a scenario, we can combine the sphere source and the half-space source by subtracting Eq. 3.38c from Eq. 3.32b, which results in:

$$\frac{\mathrm{d}\phi_\gamma}{\mathrm{d}E_\gamma}\left(h_{\mathrm{air}}, E_\gamma, t\right) = \frac{Q_{\mathrm{v}}(E_\gamma, t)}{\mu_{\mathrm{air}}}\left[1 - \frac{1}{2}\mathscr{E}_2\left(\mu_{\mathrm{air}} h_{\mathrm{air}}\right)\right] \tag{3.40}$$

With that, I conclude the discussion of the primary quanta theory. Let me summarize three main findings for AGRS, that we could derive from the analytical solutions presented in this section:

1. The primary energy flux $\mathrm{d}\phi_\gamma/\mathrm{d}E_\gamma$ of terrestrial radionuclide sources decreases with increasing ground clearance $h_{\mathrm{air}}$ with the highest rate observed for terrestrial point sources with $\mathrm{d}\phi_\gamma/\mathrm{d}E_\gamma \propto \exp(-\mu_{\mathrm{air}} h_{\mathrm{air}})/h_{\mathrm{air}}^2$, followed by terrestrial surface sources with $\mathrm{d}\phi_\gamma/\mathrm{d}E_\gamma \propto \mathscr{E}_1(\mu_{\mathrm{air}} h_{\mathrm{air}})$ and terrestrial volume sources with $\mathrm{d}\phi_\gamma/\mathrm{d}E_\gamma \propto \mathscr{E}_2(\mu_{\mathrm{air}} h_{\mathrm{air}})$.
2. The majority of the primary energy flux from terrestrial radionuclide volume sources is emitted within a comparably shallow depth of the ground surface $d_{\mathrm{src}} < 1$ m.
3. The majority of the primary energy flux from terrestrial radionuclide surface and volume sources is emitted within a comparably small area of the ground surface with the associated surface radius increasing approximately linearly with the ground clearance $h_{\mathrm{air}}$. As an example, for a maximum photon energy $E_\gamma$ of 3 MeV with a ground clearance $h_{\mathrm{air}}$ of 100 m, 99 % of the primary energy flux from a terrestrial surface/volume source is attributed to photons emitted within a radius of 822 m/558 m.

46 *The Monte Carlo method was originally developed by Stanisław Ulam and John von Neumann at the Los Alamos National Laboratory in the U.S. in the late 1940s to study neutron diffusion and criticality problems for nuclear weapon applications [266]. Inspired by the Monte Carlo Casino in Monaco, where Ulam's uncle used to gamble, one of Ulam's working colleagues, Nicholas Metropolis, named the newly developed method the "Monte Carlo method". It is worth adding that Enrico Fermi developed the same method already 15 years prior to the development at Los Alamos but never published his results [266].*

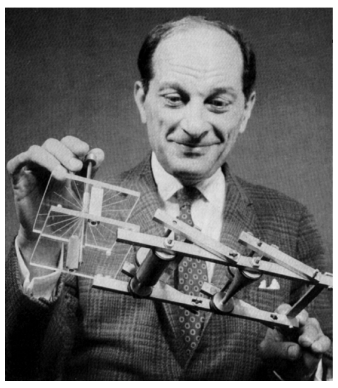

Stanisław Ulam
Ⓢ Unknown

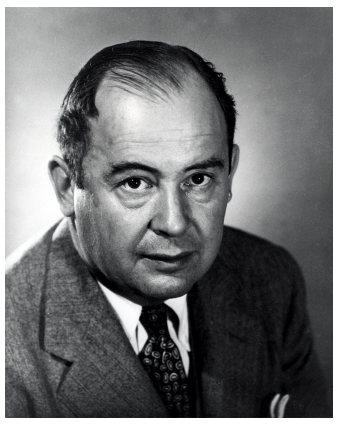

John von Neumann
Ⓢ U.S. Department of Energy

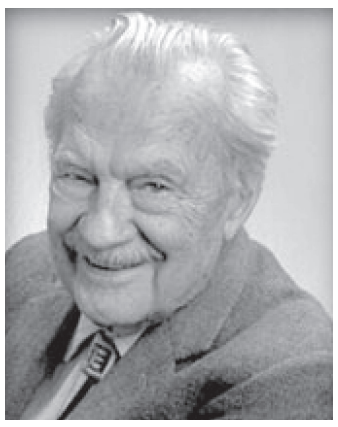

Nicholas Metropolis
Ⓢ Los Alamos National Laboratory

## 3.2.2 A Monte Carlo Approach

At the beginning of Section 3.2, we discussed how radiation transport problems can be tackled numerically using Monte Carlo methods.[46] In this section, I would like to provide a brief introduction to this interesting method. However, I will limit the discussion to the basic principles and concepts underlying Monte Carlo methods. For more detailed information on various aspects of Monte Carlo methods, I will refer to the relevant literature throughout the discussion.

### 3.2.2.1 Main Simulation Procedure

As outlined in Section 3.2, in Monte Carlo methods, the transport of photon transport through matter is simulated by independently following a large number of photons through a series of interactions. Each photon starts from a predefined state described by the position $\mathbf{r}$, the direction $\boldsymbol{\Omega}$ and the energy $E_\gamma$. These values could be deterministic, e.g. a perfect monoenergetic photon beam, or statistically sampled from a known distribution, e.g. a radionuclide source with known relative photon intensity $I_\gamma$.

The path length $s$ the photon travels until a first interaction with the matter occurs is stochastic and depends on the total attenuation coefficient $\mu_{\text{tot}}$. The normalized continuous probability density function (PDF) $\pi(s) := f_S(s)$,[47] which describes this stochastic process by modeling $s$ as a random variable $S$, is given by [255]:[48]

$$\pi(s) = \frac{e^{-\mu_{\text{tot}} s}}{\mu_{\text{tot}}} \tag{3.41}$$

Using the inverse transformation method, we can sample the path length $s$ from this PDF by inverting the corresponding cumulative distribution function (CDF) resulting in [255]:

$$s = -\frac{\log(\xi)}{\mu_{\text{tot}}} \tag{3.42}$$

with $\xi \sim \mathcal{U}(0,1)$ being a uniformly distributed (pseudo-)random[49] number between 0 and 1.

After the path length $s$ has been computed by Eq. 3.42, the new position of the photon $\mathbf{r}'$, where the next interaction will take place, is given by $\mathbf{r}' = \mathbf{r} + s\boldsymbol{\Omega}$. Next, the specific target nuclide $j$ as well as the interaction type $i$ have to be determined using again random

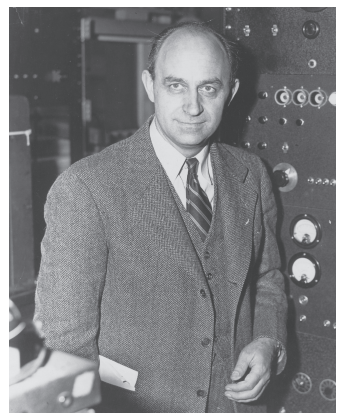

Enrico Fermi
Ⓢ U.S. Department of Energy

sampling. Both parameters are given by discrete probability distributions, i.e. $\Pr_j = \mu_{\text{tot},j}/\mu_{\text{tot}}$ for nuclide $j$ (cf. Eqs. 3.25a and 3.25b) and $\Pr_i = \sigma_{i,j}/\sigma_{\text{tot},j}$ (cf. Eqs. 3.3 and 3.24) for interaction type $i$. If the sampled interaction process $i$ represents an absorption, e.g. photoelectric absorption or pair production, the track of the specific photon ends at this point. On the other hand, if $i$ represents a scattering process such as Compton or Rayleigh scattering, a new direction and energy are sampled from the corresponding double differential cross-sections described in Section 3.1.2. As CDFs are often not available for complex differential cross-sections, the inverse transformation method cannot be applied. Consequently, other sampling techniques such as rejection or importance sampling have to be adopted. More information on this topic is provided by Rubinstein et al. [271] and Vassiliev [272].

The process described above is repeated until an absorption interaction occurs. The track might also be terminated if the photon energy drops below a predefined transport threshold as discussed at the beginning of Section 3.1 or if the photon leaves the simulation domain. In addition, depending on the interaction type, secondary particles such as characteristic X-rays are generated at the corresponding interaction point with their energy and direction sampled from distributions related to the corresponding differential cross-sections.

While I have focused solely on photon transport in this section, Monte Carlo methods are capable of simulating the transport of all types of high-energy radiation, including charged particles. The simulation of charged particles is more complex due to their interaction with matter through numerous discrete Coulomb interactions that ionize and excite the surrounding material. Simulating each of these Coulomb interactions would currently require prohibitively large computational resources, especially for particles with high kinetic energies [255]. As a result, charged particles are often simulated using a condensed-history approach, in which discrete interaction events are modeled as a continuous energy loss over a predefined step length together with an associated deflection angle [273]. More information on this topic is provided in Chapter 5 in the monograph by Vassiliev [272].

#### 3.2.2.2 Assumptions

Similar to the analytical solutions presented in the previous section, the Monte Carlo method is based on several assumptions. At this

47 *Note that in this book, I adopt the symbol $\pi(\cdot)$ to denote the PDF, which is the common notation in Bayesian statistics [267, 268].*

48 *Please note that I report here the normalized PDF in contrast to the unnormalized PDF reported by Fassò et al. [255].*

49 *Modern Monte Carlo codes in radiation transport do not apply true but pseudo-random numbers generated by so-called pseudorandom number generators (PRNGs). These pseudo-random numbers reproduce sequences of true random numbers but are completely determined by an initial value, called the "seed" [255]. There are two reasons why we apply pseudo- and not true random numbers. First, pseudo-random numbers are easier and faster to produce. Second and more importantly, PRNGs ensure the reproducibility of the conducted Monte Carlo simulations, given the seed value is known. This is an important feature for scientific applications as it allows to test and verify the results from such simulations. Common PRNGs applied in Monte Carlos codes are the Mersenne Twister by Matsumoto and Nishimura [269] and the RM64 PRNG by Marsaglia and Tsang [270], which is the one used by the Monte Carlo code adopted in this work.*

point, it is important to emphasize the fact that the Monte Carlo method is not a direct solution of the LBTE but rather a statistical approach to solve the overall photon transport problem behind the LBTE [274]. As such, only a subset of the assumptions discussed at the beginning of Section 3.2 is required for Monte Marlo simulations, specifically assumptions A–D. However, the Monte Carlo method introduces one more assumption:

J. In addition to the isotropic assumption (D), Monte Carlo methods usually require the matter to be defined as a subdivision of homogeneous media with well-defined boundaries. This allows the sampling of the path length $s$ and the interaction type $i$ as outlined above.

One important point I have not included in the description of the main simulation procedure above is the topic of boundary crossing. When a photon or any other high-energy particle for that matter crosses a boundary between two different media, the attenuation coefficient and consequently the mean free path changes. As a result, $\mu_{\text{tot}}$ in Eq. 3.41 has to be adapted accordingly. This can be achieved for example by ray-tracing algorithms or delta-tracking methods [210, 275].

#### 3.2.2.3 Mass Model

As a direct result of assumption J above, the Monte Carlo method requires the setup of a simulation domain composed of individual homogeneous sections with well-defined boundaries and material properties. This model is referred to as the mass model. Algorithms to build the geometry of such a mass model can be classified into three broad categories [276]:

1. **Combinatorial Solid Geometry (CSG)** In this approach, Boolean operations (intersection, union and subtraction) together with geometrical transformations (rotation and translation) are used to combine simple geometric solid shapes such as spheres, cylinders or boxes to form more complex shapes. Thanks to its simplicity and robustness it is still the most widely used approach in Monte Carlo codes [20, 22, 23, 216, 258].
2. **Voxelization** Here, the simulation domain is divided into a regular grid of voxels, i.e. tiny identical parallelepipeds forming a 3-dimensional grid [255]. For curved surfaces, voxel-based models

can only provide approximate descriptions of the corresponding shapes. Typically, it is only a valid option if related geometrical files, e.g. Digital Imaging and Communications in Medicine (DICOM) files generated by computed tomography (CT) scans, are available [216].

3. **CAD-based Modeling** Modern computer-aided design (CAD) software allows the creation of complex geometries using either tessellation or boundary representation (BREP) methods. The former is based on a 3D polygon mesh, while the latter uses a combination of topological instances (faces, edges and vertices) and geometric instances (points, lines and surfaces) to define the shape of 3D objects. Compared to CSG or voxel-based models, CAD-based modeling is more flexible but also more computationally expensive. It is generally only a viable option if accurate CAD models of the simulation domain or related files are available [277–280].

It is worth adding that modern Monte Carlo codes are not restricted to one of these approaches but often offer multiple options to build the geometric model [21, 22, 281]. In addition, there are various supporting codes that allow the conversion from one geometrical format to another [276, 282–286].

Once the geometric model has been built, assigning the material properties to the individual sections is straightforward. Required material properties for photon transport are the isotopic composition of each homogeneous medium together with the mass density. This allows the computation of the total and partial attenuation coefficients using Eqs. 3.7b and 3.8b as discussed in Section 3.1.

#### 3.2.2.4 Scoring

As noted already at the beginning of Section 3.2, Monte Carlo simulations are not limited to photon flux quantities but can record any quantity of interest or observable $y$ sampled during the simulation or derivatives thereof as an output result. This process is referred to as scoring. Like in empirical measurements, these scored quantities are subjected to statistical and systematic uncertainties [287].

The mathematical foundation of scoring observables $y$ in Monte Carlo simulations relies on the central limit theorem (CLT). Simply speaking, this theorem states that the distribution of averages or sums of large numbers $N$ of independent and identically distributed (i.i.d.)

random variables with true mean $\mu_y$ and variance $\sigma_y^2$ tend to a normal distribution with sample mean $\overline{y}$ and variance of the sample mean $\sigma_y^2/N$ as $N \to \infty$ assuming $\mu_y, \sigma_y^2 < \infty$. More rigorous mathematical formulations of the CLT are readily available in the literature [271, 288, 289].

As a direct result of the CLT, assuming we have simulated a large number of individual photon histories $N_{\mathrm{pr}} \gg 1$, we may estimate the true mean of the observable $y$ by the arithmetic mean of the sampled values $y_i$:

$$\overline{y} = \frac{1}{N_{\mathrm{pr}}} \sum_{i=1}^{N_{\mathrm{pr}}} y_i \tag{3.43}$$

We also commonly refer to these individual histories as primaries.[50] In the same way, we can estimate the true standard deviation of the sample mean $\sigma_y/(N)^{1/2}$ by the sample standard deviation $\sigma_{\overline{y}}$ [255, 288, 290]:

$$\sigma_{\overline{y}} = \sqrt{\frac{1}{N_{\mathrm{pr}}\left(N_{\mathrm{pr}}-1\right)} \sum_{i=1}^{N_{\mathrm{pr}}} \left(y_i - \overline{y}\right)^2} \tag{3.44a}$$

$$= \sqrt{\frac{1}{N_{\mathrm{pr}}\left(N_{\mathrm{pr}}-1\right)} \left[\sum_{i=1}^{N_{\mathrm{pr}}} y_i^2 - \frac{1}{N_{\mathrm{pr}}}\left(\sum_{i=1}^{N_{\mathrm{pr}}} y_i\right)^2\right]} \tag{3.44b}$$

where Eq. 3.44b is more practical for numerical computations than Eq. 3.44a as we do not need to store the individual values $y_i$ until the end of the simulation [290].[51] $\sigma_{\overline{y}}$ quantifies the statistical uncertainty of the computed arithmetic mean $\overline{y}$ and is sometimes referred to as the standard error of the mean.

To check the precision of the arithmetic mean we may compute the relative standard deviation or coefficient of variation CV as follows [255]:

$$\mathrm{CV}_{\overline{y}} = \frac{\sigma_{\overline{y}}}{\overline{y}} \tag{3.45}$$

For the sample standard deviation estimates on the other hand, we can use the relative variance of the variance VOV for the same purpose [291, 292]:

50 *In more rigorous terms, a primary is defined as the initial state of all high-energy particles released in an independent source emission event. In the simplest case, this is equivalent to the emission of a single particle such as a photon with a predefined position, energy and direction. However, more complex sources can be applied, e.g. the decay of a radionuclide and subsequent emission of associated high-energy particles, i.e. gamma-rays, characteristic X-rays and Auger electrons, among others.*

51 *Some observables $y$ are not sufficiently described as a single history value but may require multiple histories to be sampled. In this case, we can split the simulation into a number of cycles $N_{\mathrm{cl}}$, each containing a defined number of histories $N_{\mathrm{pr}}$. Mean and standard deviation are then estimated using the arithmetic means of each cycle instead of the individual scoring values [255].*

$$\mathrm{VOV}_{\overline{y}} = \frac{\mathrm{var}\left(\sigma_{\overline{y}}^2\right)}{\sigma_{\overline{y}}^4} \tag{3.46a}$$

$$\approx \frac{\sum_{i=1}^{N_{\mathrm{pr}}} \left(y_i - \overline{y}\right)^4}{\sum_{i=1}^{N_{\mathrm{pr}}} \left(y_i - \overline{y}\right)^2} - \frac{N_{\mathrm{pr}} - 3}{N_{\mathrm{pr}} - 1} \frac{1}{N_{\mathrm{pr}}} \tag{3.46b}$$

Typically, we aim for $\mathrm{CV}_{\overline{y}} < 10\,\%$ and $\mathrm{VOV}_{\overline{y}} < 10\,\%$ to ensure a reliable estimate of both, $\overline{y}$ and $\sigma_{\overline{y}}$ [255, 291].

In contrast to statistical uncertainties, systematic uncertainties in Monte Carlo simulations are more difficult to quantify. We may distinguish three main sources of systematic uncertainties in Monte Carlo simulations:

1. **Physics Models** These uncertainties arise from the fundamental limits in the accuracy of the adopted physics model in Monte Carlo simulations. Some of these systematic uncertainties are unavoidable as the models themselves are based on experimental data. Consequently, Monte Carlo simulations can never exceed the accuracy of the underlying experimental data used to build the physics models. Moreover, the physics models are often simplified to speed up the simulation process, which can introduce additional systematic uncertainties. An example of such a simplification is the condensed-history for charged particles discussed above.
2. **Numerical Models** Monte Carlo models require various numerical algorithms to perform the simulation, e.g. random sampling algorithms or scoring methods. These algorithms can be the source of additional systematic uncertainties in the simulation results.
3. **Mass Models** In most cases, the mass model used in Monte Carlo simulations is an approximation of the real-world geometry as the geometry and material composition of the various components in the simulation domain are either not well known or too complex to be modeled in every detail. These simplifications in the mass model can introduce additional systematic uncertainties in the simulation results.

Common techniques to quantify systematic uncertainties in Monte Carlo radiation transport simulations include brute-force propagation and surrogate modeling [293–295]. It is important to note that brute-force propagation is only feasible for small-scale problems involving a handful of model parameters subjected to systematic uncertainties. For large-scale Monte Carlo simulations involving hundreds or thousands of model parameters, the computational cost of brute-force methods is prohibitively high. In such cases, surrogate modeling, as advocated by Sudret [295], becomes the method of choice.

This concludes our discussion on the interaction of high-energy photons with matter. In the next chapter, we will use the theoretical foundations presented in this chapter to discuss how we can detect and quantify high-energy photons using gamma-ray spectrometry.

*"As long as there's light, we've got a chance."*

*— Poe Dameron, The Force Awakens*

# Chapter 4
# Gamma-Ray Spectrometry

**Contents**

In this chapter, I will discuss the fundamental principles of gamma-ray spectrometry. I will start by introducing the main components of the most common gamma-ray spectrometer type adopted in AGRS, which are inorganic scintillators coupled to photomultiplier tubes. In the second part, I will discuss the main features of the data product of gamma-ray spectrometry, the pulse-height spectrum, related calibration procedures and how it can be derived using detector response functions.

The purpose of this chapter is not to provide a detailed discussion on every aspect of gamma-ray spectrometry but rather to give a brief overview of the main aspects and physical processes involved in spectrometers based on inorganic scintillators coupled to photomultiplier tubes. We will use the ideas and concepts introduced in this chapter later to develop a numerical model of a gamma-ray spectrometer as well as to interpret the conducted radiation measurements and simulations.

As we will learn in the next chapter, inorganic scintillators coupled to photomultiplier tubes (PMTs) are by far the most common type of gamma-ray spectrometer used in AGRS. I will therefore focus in this chapter on this specific detector type. Furthermore, I will limit the discussion in the first part to the inorganic scintillator and the PMT and associated physical phenomena. Additional hardware components such as the electronics, power supply or the data acquisition system will not be analyzed in detail here. Comprehensive overviews on these components are readily available in the literature [30, 296, 297].

## 4.1 Inorganic Scintillators

Inorganic scintillators[1] are materials based on dielectric inorganic compounds that exhibit luminescence when exposed to ionizing radiation [298]. Luminescence[2] is defined as the "spontaneous emission of radiation from an electronically or vibrationally excited species not in thermal equilibrium with its environment" [299]. The electromagnetic radiation emitted in luminescence processes is typically in the visible or near visible spectral range [30, 298]. For gamma

1 *From the Latin verb "scintillare" which translates to "to sparkle".*

2 *From the Latin noun "lumen" for "light".*

rays as ionizing radiation source, scintillators can therefore be seen as wavelength shifters that convert high-energy photons with short wavelengths to low-energy photons with longer wavelengths [298].

The development of inorganic scintillators began over a century ago, playing a pivotal role in numerous scientific discoveries in nuclear and particle physics during the early 20th century. For instance, Wilhelm Röntgen, whom we previously encountered in Section 2.1.1, used the inorganic scintillator barium platinocyanide to detect the X-rays in his famous experiments in 1895 [41]. Similarly, zinc sulfide, another widely used inorganic scintillator, was employed by Ernest Rutherford[3] and his co-workers in their seminal gold foil experiments to investigate alpha particle scattering [300]. However, before the development of the PMT in the 1930s, scientists were limited to visual observation for detecting the scintillation light produced by the inorganic scintillators. As a result, these initial applications of inorganic scintillators were confined to qualitative counting experiments.

3 *Ernest Rutherford, 1st Baron Rutherford of Nelson (*1871, †1937), a New Zealand experimental physicist. Rutherford made groundbreaking contributions to our understanding of atomic structure and radioactivity. Together with his co-workers Hans Geiger and Ernest Marsden, he discovered the nucleus of the atom, revolutionizing the prevailing model of the atom. His work laid the foundation for the field of nuclear physics and earned him a Nobel Prize in Chemistry in 1908.*

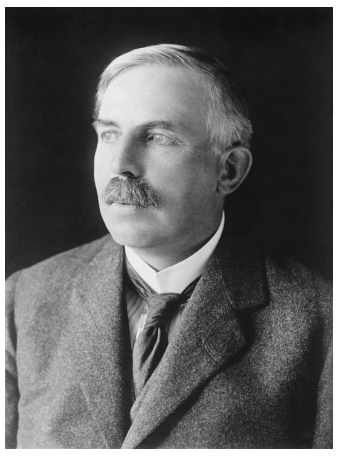

Ernest Rutherford
Library of Congress

It was Hartmut Kallmann[4] who discovered in 1947 together with his PhD student Immanuel Broser that the scintillation light emitted by inorganic scintillators is related to the deposited energy of the ionizing radiation [302]. This discovery paved the way for the development of quantitative gamma-ray spectrometry using inorganic scintillators coupled to photosensors such as PMTs [301]. Briefly after this discovery, Robert Hofstadter[5] developed the NaI(Tl) scintillator in 1948, which marks the beginning of the modern era of inorganic scintillators [298, 306]. In contrast to the inorganic scintillators used before, NaI(Tl) could be grown as transparent crystals forming large scintillator volumes ideal for gamma-ray spectrometry when coupled with PMTs [30, 306]. Since then, hundreds of new inorganic scintillators have been developed [298, 307–309]. However, due to its comparably high light yield, high density and availability in large crystal volumes, NaI(Tl) remains still one of the most widely used inorganic scintillators in gamma-ray spectrometry and in particular in AGRS, as we will see in the next chapter.

4 *Hartmut Paul Kallmann (*1896, †1978), a German physicist, best known for the first development of a scintillation spectrometer [301, 302]. As a German jew, he was actively hunted by the German Reich during the Second World War [303]. Shortly after the war, he emigrated to the U.S. [304].*

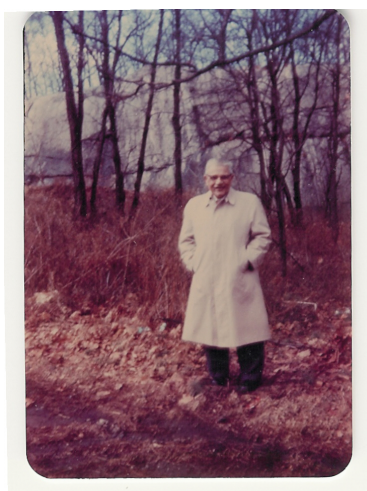

Hartmut Kallmann
Knox marsh

### 4.1.1 Scintillation Process

Let us now turn our discussion to the main physical processes related to scintillation events in inorganic scintillators. Scintillation is a complex process that is still not fully understood and thus continues to be the subject of ongoing research efforts [298, 310, 311]. Here, I only provide a brief phenomenological review of the main processes

involved in the scintillation process. For a detailed discussion on this topic including a thorough analysis of the various spectral and spatiotemporal aspects of scintillation, I recommend the monographs by Rodnyi [311] and Lecoq et al. [298].

The starting point of each scintillation event is the interaction of the ionizing radiation with the inorganic scintillator material. As we have seen in the last chapter, for high-energy photons, the primary interaction processes are the photoelectric effect, Compton scattering and pair production. What initiates the scintillation process are not the photons themselves but the charged particles that are produced as a result of the interaction event, i.e. photoelectrons, Compton electrons and electron-positron pairs, among others [30]. These charged particles are ejected with considerable kinetic energy (cf. Section 3.1.2) and are therefore capable of exciting and ionizing the scintillator material.[6]

We can describe this process using the energy band model for inorganic scintillator materials [312]. In this model, the electronic energy states of the atoms in the scintillator are described by a series of discrete energy bands separated by "forbidden" regions. In a normal/unexcited state, the lower energy bands are completely filled with electrons. The highest-filled energy band is also known as the valence band. The empty energy band above the valence band is the conduction band separated by the band gap energy $E_{\mathrm{gap}}$ from the valence band. A graphical depiction of a general energy band structure for an inorganic scintillator is shown in Fig. 4.1. Ionization can now be interpreted as the process of transferring electrons from the core bands to the conduction band leaving behind a hole in the core band. We can divide the subsequent scintillation processes into four steps [311]:

1. **Relaxation** As a result of the subsequent ionizations by the high-energy charged particles, a cascade of secondary electrons in the conduction band and holes in the core bands are produced in the vicinity of the primary interaction site. This avalanche process continues until the kinetic energy of the electrons drops below the mean excitation energy $I_0$ of the scintillator material. At this point, the electrons are further slowed down by electron-electron scattering until they reach a kinetic energy of $E_{\mathrm{k}} \leq 2E_{\mathrm{gap}}$ [298]. The characteristic time scale of the relaxation process is in the order of $\mathcal{O}(10^{-16})$ to $\mathcal{O}(10^{-14})$ s for common inorganic scintillators [298].

5 *Robert Hofstadter (*1915, †1990), an American physicist. Besides his important contributions to inorganic scintillation physics, Hofstadter is also known for his pioneering work in electron scattering experiments, which led to the discovery of the structure of the nucleons [305]. For his work, he was awarded the Nobel Prize in Physics in 1961.*

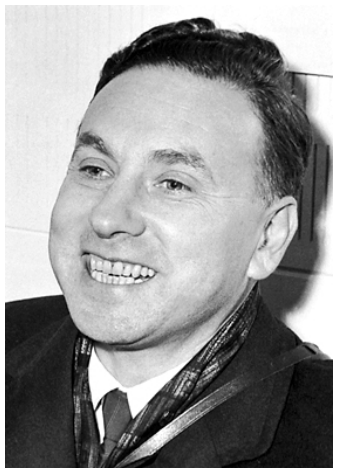

Robert Hofstadter
© Nobel foundation

6 *In common inorganic scintillator materials, the range of these electrons is typically a few millimeters [30]. Consequently, I will assume in the following discussion that the size of the scintillator material is sufficiently big to neglect the escape of these electrons from the scintillator volume.*

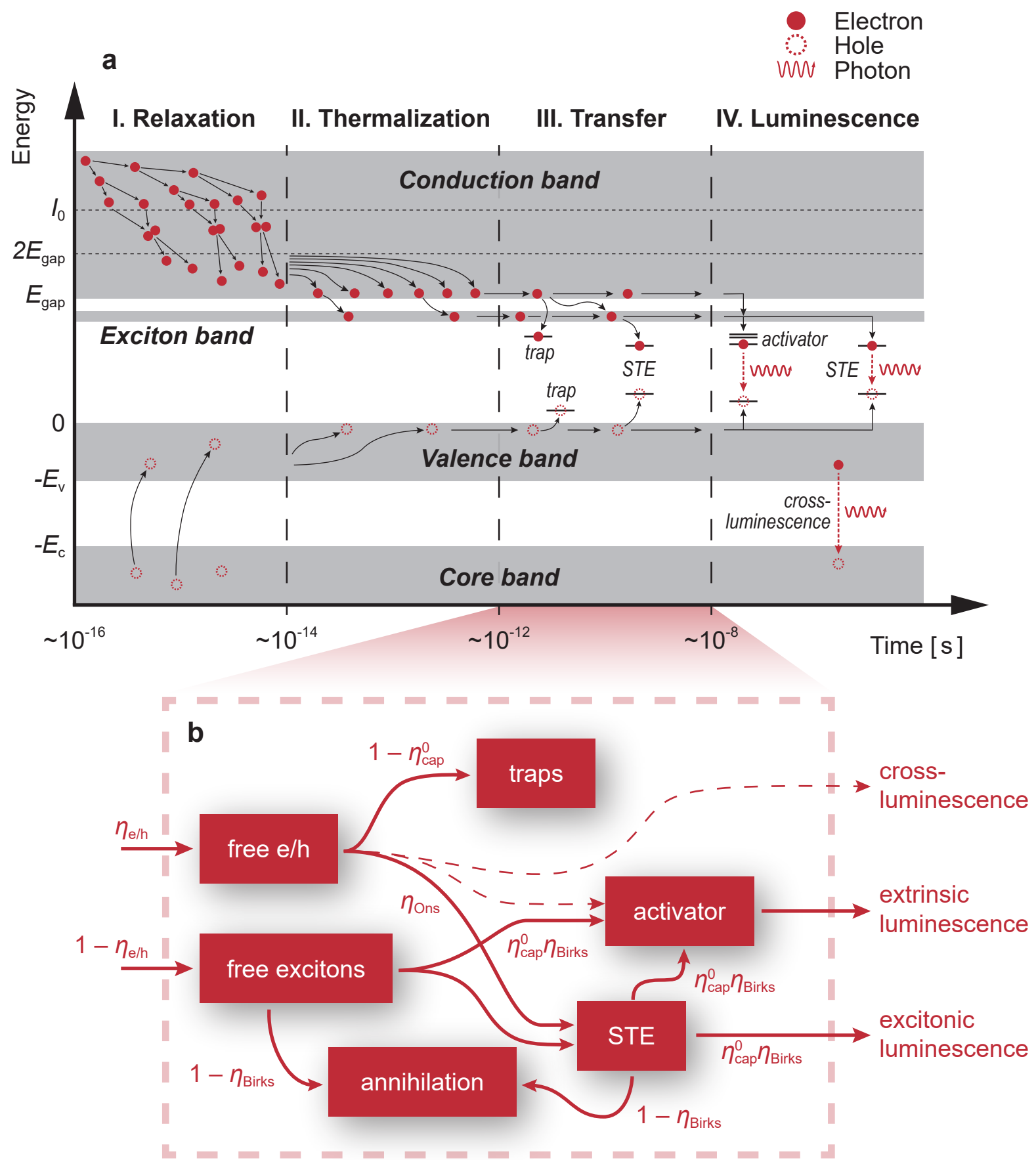


**Figure 4.1** Scintillation process in inorganic scintillators. **a** Electronic band structure of a generic inorganic scintillator according to Lecoq et al. [298] together with a graphical depiction of the four main stages in the scintillation process, i.e. relaxation, thermalization, energy transfer and luminescence. Characteristic energies in the band structure are highlighted such as the mean excitation energy $I_0$, the band gap energy $E_{gap}$ between the conduction and the valence band, the valence band energy $E_v$ and the upper core band energy $E_c$. Moreover, selected transfer and luminescence processes are shown including the self-trapped exciton (STE) mechanism. This subgraph was adapted from similar graphs shown by Lecoq et al. [298]. **b** Main mechanisms involved in the energy transfer and associated transfer efficiencies according to Payne et al. [313] (e/h: electron-hole pair, STE: self-trapped exciton). A detailed discussion on the physical meaning of the highlighted parameters is provided in Appendix A.7.

2. **Thermalization** In the second stage, the electrons in the conduction band are gradually further slowed down not by electron-electron interactions but by the interaction between the electrons and the vibrations of the crystal lattice, which we refer to also as electron-phonon[7] relaxation [311]. At the end of this thermalization process, the majority of the electrons are located at the bottom of the conduction band and the associated holes at the top of the valence band as shown in Fig. 4.1. These electrons and holes can move separately within the conduction and the valence band as free electron-hole pairs ($e^-$/h pairs). However, they can also form an electrical neutral quasiparticle known as exciton, in which case they remain associated with each other with an energy gap slightly smaller than the bandgap energy $E_{gap}$. The thermalization stage lasts about $\mathcal{O}(10^{-14})$ s to $\mathcal{O}(10^{-12})$ s for common inorganic scintillators [298].

3. **Energy transfer** In the third stage, the excitation carriers, i.e. excitons and free $e^-$/h pairs [314], migrate to the luminescent centers in the inorganic scintillator. On their way, they can get captured by local traps[8] such as impurities or lattice defects or get trapped themselves as STEs. A schematic overview of the different quenching and conversion processes involved in the energy transfer stage is shown in Fig. 4.1. In addition, a detailed discussion on these processes is also provided in Appendix A.7. The characteristic timescale of the energy transfer phase is in the order of $\mathcal{O}(10^{-12})$ s to $\mathcal{O}(10^{-10})$ s for common inorganic scintillators [298].

4. **Luminescence** The last step in the scintillation process is the actual luminescence, i.e. the excitation of the luminescent centers in the inorganic scintillator by the excitation carriers and their subsequent radiative relaxation. We refer to the emitted radiation also as scintillation photons . There are many different types of luminescent centers and thereby luminescence processes. However, they all share a common characteristic: they alter the electronic band structure in a way that the photon energy of the emitted light is smaller than the band gap energy $E_{gap}$. This is a fundamental condition to prevent reabsorption of the emitted photons in the host lattice [30].

   We distinguish between two broad categories of luminescent centers: intrinsic and extrinsic. Extrinsic luminescent centers are

7 *"Elementary excitation in the quantum mechanical treatment of vibrations in a crystal lattice" [299].*

8 *It is worth adding, that, depending on the trap depth in the electronic band structure, these trapped electrons or holes might escape back to the conduction or valence band either by intrinsic thermal excitation or by extrinsic stimulation. In fact, this phenomenon of metastable trapped excitation carriers is exploited in thermoluminescence (TL) and optically stimulated luminescence (OSL), where excitation is induced by external sources such as heating and optical stimulation, respectively. These techniques find extensive use in radiation protection, personal dosimetry or chronometric dating for archeological and geological applications [315, 316].*

impurities deliberately incorporated into the inorganic scintillator lattice. These impurities, which are also called activators [30], introduce excited levels within the forbidden zone between the valence and conduction band as highlighted in Fig. 4.1. These activators can either be excited sequentially by a free $e^-$/h pair or at once by an exciton [311].

Intrinsic luminescent centers are on the other hand, as the name implies, intrinsic to the inorganic scintillator and do not require activators. There are three subcategories of intrinsic luminescent centers. First, these intrinsic centers can be constituents of the host lattice itself. We call these scintillators therefore also "self-activated" [298]. Second, STEs can also form intrinsic luminescent centers in which case relaxation of the STEs results in what we call excitonic luminescence [311]. Last but not least, there is cross-luminescence[9] [298]. In this process, an electron from the valence band recombines radiatively with a deep hole in the core band emitting ultraviolet (UV) light with ultra-short decay time constants, typically <1 ns. Cross-luminescence can only take place in scintillators with special electronic shell configurations, i.e. $E_v - E_c < E_{gap}$, in order to prevent reabsorption of the emitted UV by the host lattice. (cf. Fig. 4.1).[10]

9 *Also known as core-valence luminescence [311].*

10 *A famous example for a scintillator exhibiting cross-luminescence is $BaF_2$, whose fast cross-luminescent emission component was discovered 1983 by Laval et al. [317] (cf. Table 4.1 at the end of this chapter).*

It is important to add that these different luminescence processes are not mutually exclusive but can occur in parallel in the same inorganic scintillator material [298, 311]. The potentially combined effects of the different luminescent centers determine also the characteristic time for the luminescence process. This brings us to the topic of the next subsection: the temporal evolution of a scintillation event in an inorganic scintillator.

### 4.1.2 Scintillation Pulse

Considering the time scales of the different scintillation stages, the time evolution of a scintillation pulse for many inorganic scintillators is characterized by an approximately instantaneous rise followed by a single exponential decay with the scintillator decay time constant $\tau_{sci,d}$ [311]:

$$\dot{N}_{sci}(t) = \dot{N}_{sci}(0)\exp\left(-\frac{t}{\tau_{sci,d}}\right) \tag{4.1}$$

where:

| | | |
|---|---|---|
| $\dot{N}_{sci}$ | scintillation photon intensity | $s^{-1}$ |
| $\tau_{sci,d}$ | scintillator decay time constant | s |

This model describes the first-order kinetic dynamics in the luminescence, i.e. luminescence when the concentration of luminescent centers is significantly bigger than that of the thermalized excitation carriers [311]. More complex models are required for scintillators with multiple characteristic decay times, rise times comparable to $\tau_{sci,d}$ or recombination mechanisms governed by second-order kinetics. Rodnyi [311] provides a comprehensive overview of those more complex models and associated scintillation pulse shapes. A list of experimentally determined decay time constants $\tau_{sci,d}$ for common inorganic scintillators is provided in Table 4.1 at the end of this chapter. From this table, we find that $\tau_{sci,d}$ for most[11] inorganic scintillators ranges between ∼$10^{-8}$ s and ∼$10^{-6}$ s [298].

11 *An exception at the lower end is the cross-luminescent* $BaF_2$ *with a decay time constant of only about* 0.8 ns. *On the other end of the temporal range, there is CWO with a decay constant reaching up to* 20 µs.

These characteristic pulse times of the scintillation process in inorganic scintillators have important implications for gamma-ray spectrometry. Consider an incoming high-energy photon that undergoes multiple Compton scatter events followed by photoelectric absorption in the scintillator material. Let us further assume that the total track length of the high-energy photon is $\mathcal{O}(0.1)$ m [30]. Given the speed of light, the characteristic time of the entire interaction series in the scintillator material is then $\mathcal{O}(0.1)$ ns. The time scale of the high-energy electrons and positrons produced in the interaction events is even shorter. As an example, a high-energy electron with a kinetic energy of $\mathcal{O}(1)$ MeV continuously slowing down in a common inorganic scintillator such as NaI(Tl) has a range of ∼2 mm [318]. This translates to a time scale of $\mathcal{O}(0.01)$ ns.[12] Consequently, the scintillation pulses of a series of interactions between a high-energy photon and the scintillator material will overlap in time for most inorganic scintillator materials and can therefore be considered as a single scintillation event [30, 319].

12 *Estimated assuming a continuous slow down of the electron resulting in* $\Delta t \approx 2\Delta x / v$ *with the range* $\Delta x \sim 2\,\mathrm{mm}$ *and the initial speed* $v$ *of the electron given by Eq.* A.12 *and* $E_{k,e^-} \sim 1\,\mathrm{MeV}$.

### 4.1.3 Light Yield & Non-Proportionality

Another core quantity to describe inorganic scintillators is the absolute light yield $Y_{sci,a}$, which is defined as the number of emitted scintillation photons per unit energy $E_{dep}$ deposited in the scintillator material by the ionizing radiation [298, 307]. Based on the scintillation processes discussed above, we may estimate the absolute light yield $Y_{sci,a}$ for inorganic scintillators using the so-called three-stage[13] formula [310, 320, 321]:

13 *The term "three-stage" refers to the consideration of the relaxation and thermalization phases as a single stage, with the other two stages being the energy transfer and the luminescence.*

$$Y_{\text{sci,a}} = \frac{\eta_{\text{gen}}\eta_{\text{cap}}\eta_{\text{lum}}}{E_{\text{dep}}} \tag{4.2}$$

where:

| | | |
|---|---|---|
| $E_{\text{dep}}$ | deposited energy | eV |
| $\eta_{\text{cap}}$ | energy transfer efficiency | |
| $\eta_{\text{gen}}$ | conversion efficiency | |
| $\eta_{\text{lum}}$ | luminescence quantum yield | |

and with $\eta_{\text{gen}}$, $\eta_{\text{cap}}$ and $\eta_{\text{lum}}$ representing the quantum efficiencies for the relaxation/thermalization, energy transfer and luminescence phases described above, respectively.

The conversion efficiency $\eta_{\text{gen}}$ describes how many thermalized excitation carriers are created on average for an energy $E_{\text{dep}}$ deposited into the scintillator material by the ionizing radiation:

$$\eta_{\text{gen}} = \frac{E_{\text{dep}}}{\beta_{\text{sci}} E_{\text{gap}}} \tag{4.3}$$

where:

| | | |
|---|---|---|
| $E_{\text{dep}}$ | deposited energy | eV |
| $E_{\text{gap}}$ | band gap energy (cf. Fig. 4.1) | eV |
| $\beta_{\text{sci}}$ | ratio between the energy needed to create an electron-hole pair and the bandgap energy | |

and $\beta_{\text{sci}}$, $E_{\text{gap}}$ being scintillator dependent material parameters[14] with $2.3 \lesssim \beta_{\text{sci}} \lesssim 8.0$ [320, 322] and $5\,\text{eV} \lesssim E_{\text{gap}} \lesssim 13\,\text{eV}$ [320] for common inorganic scintillators.[15]

The luminescence quantum yield $\eta_{\text{lum}}$ characterizes the last step in the scintillation process, i.e. the luminescence, by quantifying the efficiency for excited luminescence centers to relax radiatively and thereby emit scintillation photons. Similar to $\beta_{\text{sci}}$ and $E_{\text{gap}}$, $\eta_{\text{lum}}$ is a material/activator dependent parameter and can easily be determined using optical spectroscopy [320]. For most inorganic scintillators, $\eta_{\text{lum}}$ is close to one [320].[16].

In contrast to $\eta_{\text{gen}}$ and $\eta_{\text{lum}}$, estimating the energy transfer efficiency $\eta_{\text{cap}}$, which describes the energy transfer phase, is a much more challenging task as $\eta_{\text{cap}}$ is not a material constant but depends on the thermalized excitation carrier density during the energy transfer phase [314]. There are various methods by which we can theoretically predict $\eta_{\text{gen}}$ [314]. In this work, I focus on a thoroughly

14 *It is important to point out that these parameters can also depend on the activator type and the temperature, among others [298].*

15 *For the inorganic scintillator NaI(Tl), which was adopted in this work,* $\beta_{\text{sci}} \sim 2.6$ *and* $E_{\text{gap}} \sim 5.9\,\text{eV}$ *resulting in* $\frac{\eta_{\text{gen}}}{E_{\text{dep}}} \sim 6.5 \times 10^4\,\text{MeV}^{-1}$. *This translates to about* $65\,000\ \text{e}^-/\text{h}$ *pairs produced in a single photon interaction event depositing* $1\,\text{MeV}$ *energy into the NaI(Tl) scintillator [320].*

16 *Known exceptions are the BGO (*$Bi_4Ge_3O_{12}$*) inorganic scintillator with* $\eta_{\text{lum}} \sim 0.13$ *and pure CsI with* $\eta_{\text{lum}} \sim 0.10$ *to* $0.15$ *[320, 323].*

17 *Experimental validation was performed with an extensive database of measured scintillation light yields for inorganic scintillators, i.e. BGO,* $CaF_2$*(Eu),* $CeBr_3$*, CsI(Tl), CsI(Na),* $LaBr_3$*(Ce), LSO(Ce), NaI(Tl),* $SrI_2$*,* $SrI_2$*(Eu), YAP(Ce) and YAG(Ce), among others [313, 324, 325].*

18 *Which was termed the "Minimalistic Approach" by Moses et al. [314] as it only considers the main physical processes during the energy transfer phase and does not consider kinetic effects (cf. Appendix A.7).*

validated[17] mechanistic model framework[18] established by Stephen Payne and his co-workes [313, 324, 325]. In this framework, we may estimate the local energy transfer efficiency $\eta'_{\mathrm{cap}}$ as a function of the collisional stopping power of electrons $\mathcal{S}_{\mathrm{e,col}}$[19] in an inorganic scintillator as [325, 328]:[20]

$$\eta'_{\mathrm{cap}}\left(\mathcal{S}_{\mathrm{e,col}}\right) = \eta^{0}_{\mathrm{cap}} \frac{1 - \eta_{\mathrm{e/h}} \exp\left[-\frac{\mathcal{S}_{\mathrm{e,col}}}{\mathrm{d}E/\mathrm{d}x\,|_{\mathrm{Ons}}} \exp\left(-\frac{\mathrm{d}E/\mathrm{d}x\,|_{\mathrm{trap}}}{\mathcal{S}_{\mathrm{e,col}}}\right)\right]}{1 + \frac{\mathcal{S}_{\mathrm{e,col}}}{\mathrm{d}E/\mathrm{d}x\,|_{\mathrm{Birks}}}} \tag{4.4}$$

where:

| | | |
|---|---|---|
| $\mathcal{S}_{\mathrm{e,col}}$ | collisional stopping power of electrons | $\mathrm{eV\,m^{-1}}$ |
| $\mathrm{d}E/\mathrm{d}x\,\vert_{\mathrm{Birks}}$ | Birks stopping parameter | $\mathrm{eV\,m^{-1}}$ |
| $\mathrm{d}E/\mathrm{d}x\,\vert_{\mathrm{Ons}}$ | Onsager stopping parameter | $\mathrm{eV\,m^{-1}}$ |
| $\eta^{0}_{\mathrm{cap}}$ | first order non-radiative loss survival efficiency | |
| $\eta_{\mathrm{e/h}}$ | electron-hole pair fraction | |
| $\mathrm{d}E/\mathrm{d}x\,\vert_{\mathrm{trap}}$ | trapping stopping parameter | $\mathrm{eV\,m^{-1}}$ |

$\mathrm{d}E/\mathrm{d}x\,|_{\mathrm{Birks}}$, $\mathrm{d}E/\mathrm{d}x\,|_{\mathrm{Ons}}$, $\eta^{0}_{\mathrm{cap}}$, $\eta_{\mathrm{e/h}}$ and $\mathrm{d}E/\mathrm{d}x\,|_{\mathrm{trap}}$ are all scintillator dependent parameters, some of which are sensitive on the lattice structure including distribution and concentration of impurities, activators as well as lattice defects, among others. A detailed discussion on these dependencies, as well as the physical meaning of these parameters alongside a thorough derivation and interpretation of Eq. 4.4, is provided in Appendix A.7. In addition, a graphical depiction of the main processes and associated parameters considered in Eq. 4.4 is shown in the subgraph b in Fig. 4.1.

The collisional stopping power of electrons $\mathcal{S}_{\mathrm{e,col}}$[21] in Eq. 4.4 is defined as the mean energy loss per unit path length of electrons due to ionizations and excitations in matter [329]:

$$\mathcal{S}_{\mathrm{e,col}} = \left\langle - \left.\frac{\mathrm{d}E}{\mathrm{d}x}\right|_{\mathrm{e,col}} \right\rangle \tag{4.5}$$

where:

| | | |
|---|---|---|
| $\mathrm{d}E/\mathrm{d}x\,\vert_{\mathrm{e,col}}$ | energy loss per unit path length of electrons due to ionizations and excitations in matter | $\mathrm{eV\,m^{-1}}$ |

19 *The mechanistic model developed by Payne and his co-workes was derived for electrons as the primary ionizing radiation source (induced by photon interactions). That said, it is general consensus that the light yield as a function of the stopping power is, at least to a first approximation, independent of the ionizing particle type [326, 327]. Consequently, we may adapt this model for any charged particle type by exchanging the stopping power for electrons shown below with that of the corresponding particle type. For the sake of brevity, I will use the electron as a representative charged particle for the remainder of this section.*

20 *Please note that I denote the energy transfer efficiency here by $\eta'_{\mathrm{cap}}$ to avoid confusion with $\eta_{\mathrm{cap}}$ in Eq. 4.2, which is a function of $E_{\mathrm{dep}}$ and not $\mathcal{S}_{\mathrm{e,col}}$.*

21 *Sometimes also called linear ionization density [314].*

$\mathcal{S}_{e,col}$ itself is a function of the electron kinetic energy $E_{k,e^-}$ as well as material-specific properties. For high-energy electrons, the collisional stopping power in units of $eV\,m^{-1}$ for a homogeneous elemental medium can be estimated by an adapted Bethe formula[22] using a density correction term $\delta$ [332, 333] to account for the reduction in $\mathcal{S}_{e,col}$ due to the polarization of the medium (cf. Fig. B.18) [134]:

$$\mathcal{S}_{e,col} = \frac{KZ\rho}{2M\beta_{e^-}^2}\left\{\log\left[\frac{m_e^2c^4\beta_{e^-}^2\gamma^2(\gamma-1)}{2I_0^2}\right] + 1 - \beta_{e^-}^2 - \log(2)\,\frac{2\gamma-1}{\gamma^2} + \frac{1}{8}\left(\frac{\gamma-1}{\gamma}\right)^2 - \delta\right\} \quad (4.6)$$

where:

| | | |
|---|---|---|
| c | speed of light in vacuum (cf. Constants) | $m\,s^{-1}$ |
| $I_0$ | mean excitation energy | eV |
| K | stopping power coefficient ($= 4\pi N_A r_e^2 m_e c^2$) | $eV\,mol^{-1}\,m^2$ |
| $m_e$ | electron mass (cf. Constants) | $eV\,m^{-2}\,s^2$ |
| $M$ | molar mass | $kg\,mol^{-1}$ |
| $Z$ | atomic number | |
| $\beta_{e^-}$ | ratio of the electron speed to the speed of light in vacuum | |
| $\gamma$ | Lorentz factor | |
| $\delta$ | density effect parameter | |
| $\rho$ | mass density | $kg\,m^{-3}$ |

with the Lorentz factor being defined as $\gamma = (1-\beta_{e^-}^2)^{-1/2}$. $\beta_{e^-}$ is related to the electron kinetic energy $E_{k,e^-}$ as:

$$\beta_{e^-} = \sqrt{1-\left(\frac{E_{k,e^-}}{m_e c^2}+1\right)^{-2}} \quad (4.7)$$

From Eq. 4.7, it is easy to see that the collisional stopping power of electrons $\mathcal{S}_{e,col}$ can be written as a function of the kinetic electron energy $E_{k,e^-}$ and the material-specific parameters $I_0$, $M$, $Z$ and $\rho$. Models to predict the collisional stopping power for non-elemental media as well as for other charged particles than electrons are readily available, e.g. from the NIST ESTAR database [318].

22 *First derived by Hans Albrecht Bethe (*1906, †2005), a German-American theoretical physicist, whose scientific contributions spanned a wide variety of different fields including nuclear and particle physics, quantum mechanics and quantum electrodynamics [330]. For his discovery of the carbon-nitrogen-oxygen (CNO) cycle [331], a fundamental nuclear process in stellar energy generation, he was awarded the Nobel Prize in Physics in 1967.*

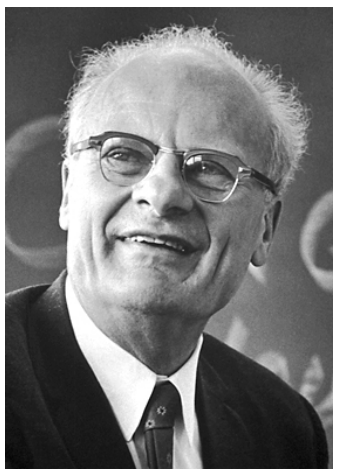

Hans Bethe
© Nobel Foundation

As discussed by Payne [313, 324], the adapted Bethe formula in Eq. 4.6 only provides accurate predictions of $\mathcal{S}_{\mathrm{e,col}}$ for high-energy electrons, i.e. $E_{\mathrm{k,e^-}} \gtrsim 10\,\mathrm{keV}$ [318]. For lower energies, we can use an empirical formula derived by Joy et al. [334] instead of Eq. 4.6 to estimate $\mathcal{S}_{\mathrm{e,col}}$ in units of eV Å$^{-1}$:

$$\mathcal{S}_{\mathrm{e,col}} = \frac{c_{\mathrm{J},1}\rho Z}{M E_{\mathrm{k,e^-}}} \log\left[\frac{1.166\left(E_{\mathrm{k,e^-}} + c_{\mathrm{J},2} I_0\right)}{I_0}\right] \tag{4.8}$$

where:[23]

23 *Note that for non-elemental media, we can in a first approximation exchange $Z$ with the effective atomic number $Z_{\mathrm{tot}}$ [313].*

| | | |
|---|---|---|
| $c_{\mathrm{J},1}$ | model constant (= 785) | $\mathrm{m^3\,eV^2\,mol^{-1}\,Å^{-1}}$ |
| $c_{\mathrm{J},2}$ | model parameter (= 2.8 for NaI(Tl) [313, 324]) | |
| $E_{\mathrm{k,e^-}}$ | electron kinetic energy | eV |
| $I_0$ | mean excitation energy | eV |
| $M$ | molar mass | $\mathrm{kg\,mol^{-1}}$ |
| $Z$ | atomic number | |
| $\rho$ | mass density | $\mathrm{kg\,m^{-3}}$ |

$\eta'_{\mathrm{cap}}$ in Eq. 4.4 represents the energy transfer efficiency for the local light yield, that is the instantaneous differential light yield at any given point along the electron track in the scintillator material [328, 335]. Assuming the scintillation event in an inorganic scintillator is induced by a single or multiple photon interaction events generating a set of high-energy electrons $\{E^{(i)}_{\mathrm{k,e^-}}\}$ with $i = \{1, 2, \ldots, N_{\mathrm{e^-}}\}$, we can compute the energy transfer efficiency $\eta_{\mathrm{cap}}$ in Eq. 4.2 by integrating $\eta'_{\mathrm{cap}}$ along each individual electron track and then sum the contributions together as follows [313]:

$$\eta_{\mathrm{cap}}\left(\{E^{(i)}_{\mathrm{k,e^-}}\}\right) = \sum_{i=1}^{N_{\mathrm{e^-}}} \frac{1}{E^{(i)}_{\mathrm{k,e^-}} - I_0} \int_{I_0}^{E^{(i)}_{\mathrm{k,e^-}}} \eta'_{\mathrm{cap}}\left(\mathcal{S}_{\mathrm{e,col}}\left(E'_{\mathrm{k,e^-}}\right)\right) \mathrm{d}E'_{\mathrm{k,e^-}} \tag{4.9}$$

where:

| | | |
|---|---|---|
| $E_{\mathrm{k,e^-}}$ | electron kinetic energy | eV |
| $I_0$ | mean excitation energy | eV |
| $N_{\mathrm{e^-}}$ | number of high-energy electrons | |
| $\mathcal{S}_{\mathrm{e,col}}$ | collisional stopping power of electrons | $\mathrm{eV\,m^{-1}}$ |
| $\eta'_{\mathrm{cap}}$ | local energy transfer efficiency | |

with $E_{dep} = \sum_{i=1}^{N_{e^-}} E_{k,e^-}^{(i)} - I_0$ being the total deposited energy introduced in Eq. 4.2. Consequently, as a direct result of the Birks and Onsager mechanisms governing the energy transfer stage (cf. Eq. 4.4 and Appendix A.7), we can conclude that the energy transfer efficiency $\eta_{cap}$ and thereby the absolute light yield $Y_{sci,a}$ (cf. Eq. 4.2) is a non-proportional function of the deposited energy $E_{dep}$, more specifically the kinetic energy distribution of the high-energy electrons created by photon interaction events [314]. We refer to this effect also as the scintillation non-proportionality [314, 327, 336]. The two interesting questions now are, how pronounced is this predited non-proportionality for common inorganic scintillators and how can we measure it?

As of writing this book, three main experimental techniques allow the direct measurement of the scintillation response to electrons in inorganic scintillators:

1. Compton coincidence technique In this technique, Compton electrons (cf. Section 3.1.2) induced by Compton scatter events inside the scintillator are used as electron source [337–343].
2. K-dip spectroscopy Here, photoelectrons (cf. Section 3.1.2) induced by photoelectric interaction events inside the scintillator are adopted as electron source [344–350].
3. External electron source As the name implies, in this technique the scintillator sample is irradiated by an external electron source [351].

As the performance of the external electron source technique is significantly deteriorated by surface effects, K-dip spectroscopy and Compton coincidence technique (CCT) are currently the preferred methods for accurate electron scintillation response measurements [314, 337, 352].

We can use these techniques to characterize the scintillation non-proportionality by computing the relative light yield $Y_{sci,r}$ as the ratio of the absolute light yield $Y_{sci,a}$ for a scintillation event induced by a single electron with kinetic energy $E_{k,e^-}$ to the absolute light yield at a reference kinetic energy $E_{k,e^-,ref}$ [313]:

$$Y_{\text{sci,r}}\left(E_{\text{k,e}^-}\right) = \frac{Y_{\text{sci,a}}\left(E_{\text{k,e}^-}\right)}{Y_{\text{sci,a}}\left(E_{\text{k,e}^-,\text{ref}}\right)} \quad (4.10a)$$

$$= \frac{\eta_{\text{cap}}\left(E_{\text{k,e}^-}\right)}{\eta_{\text{cap}}\left(E_{\text{k,e}^-,\text{ref}}\right)} \quad (4.10b)$$

The relation between the measured $Y_{\text{sci,r}}$ and $\eta_{\text{cap}}$ in Eq. 4.10b was used by Payne and his co-workers to determine the model parameters in Eq. 4.4, i.e. $\mathrm{d}E/\mathrm{d}x|_{\text{Birks}}$, $\mathrm{d}E/\mathrm{d}x|_{\text{Ons}}$, $\eta^0_{\text{cap}}$, $\eta_{\text{e/h}}$ and $\mathrm{d}E/\mathrm{d}x|_{\text{trap}}$, for various inorganic scintillators as already noted above. Using Payne's models and the experimentally determined model parameters from the literature (cf. Table C.4), I have compiled the relative light yield $Y_{\text{sci,r}}$ as a function of the electron kinetic energy $E_{\text{k,e}^-}$ for a selection of common inorganic scintillators in Fig. 4.2. A detailed discussion of the different trends in $Y_{\text{sci,r}}$ for the various inorganic scintillators displayed in Fig. 4.2 is provided in Appendix A.7. In addition, some of the model parameters are provided in Table C.4. In summary, the relative light yield response of iodide-based inorganic scintillators in Fig. 4.2 indicates both, a pronounced Onsager and Birks mechanism. On the other hand, oxides and bromides show a much less pronounced Onsager effect but a dominant Birks mechanism. In general, none of the inorganic scintillators exhibit a proportional scintillation response.

As already discussed above, the scintillation response of high-energy photons is always convoluted with the generated high-energy electrons in the photon-matter interactions.[24] This is especially true for photoelectric interactions as in this case, besides the photoelectron, various secondary electrons can be generated in the relaxation cascade following the emission of the photoelectron (cf. Section 3.1.2). As we have seen in Eq. 4.9, $\eta_{\text{cap}}$ depends on the detailed energy distribution of the generated electrons, which in turn is modulated by the rate and type of the individual photon-matter interactions. These interactions themselves depend again on various factors including the scintillator material properties, size and geometry, as explored in detail in Section 3.1.2. As a result, if we want to compute the relative light yield for high-energy photon interactions and a specific scintillator detector, we need to convolve the corresponding scintillation response to electrons given by Eqs. 4.9 and 4.10b with the characteristic energy distribution of the generated electrons in the photon-scintillator interactions. The energy distribution of the electrons can be estimated by Monte Carlo simulations [355–357].

24 *It is an interesting side note that scintillation non-proportionality of some of the first developed inorganic scintillators such as NaI(Tl) was experimentally observed before the underlying scintillation physics, in particular the relation to the electron scintillation response was fully understood [353, 354].*

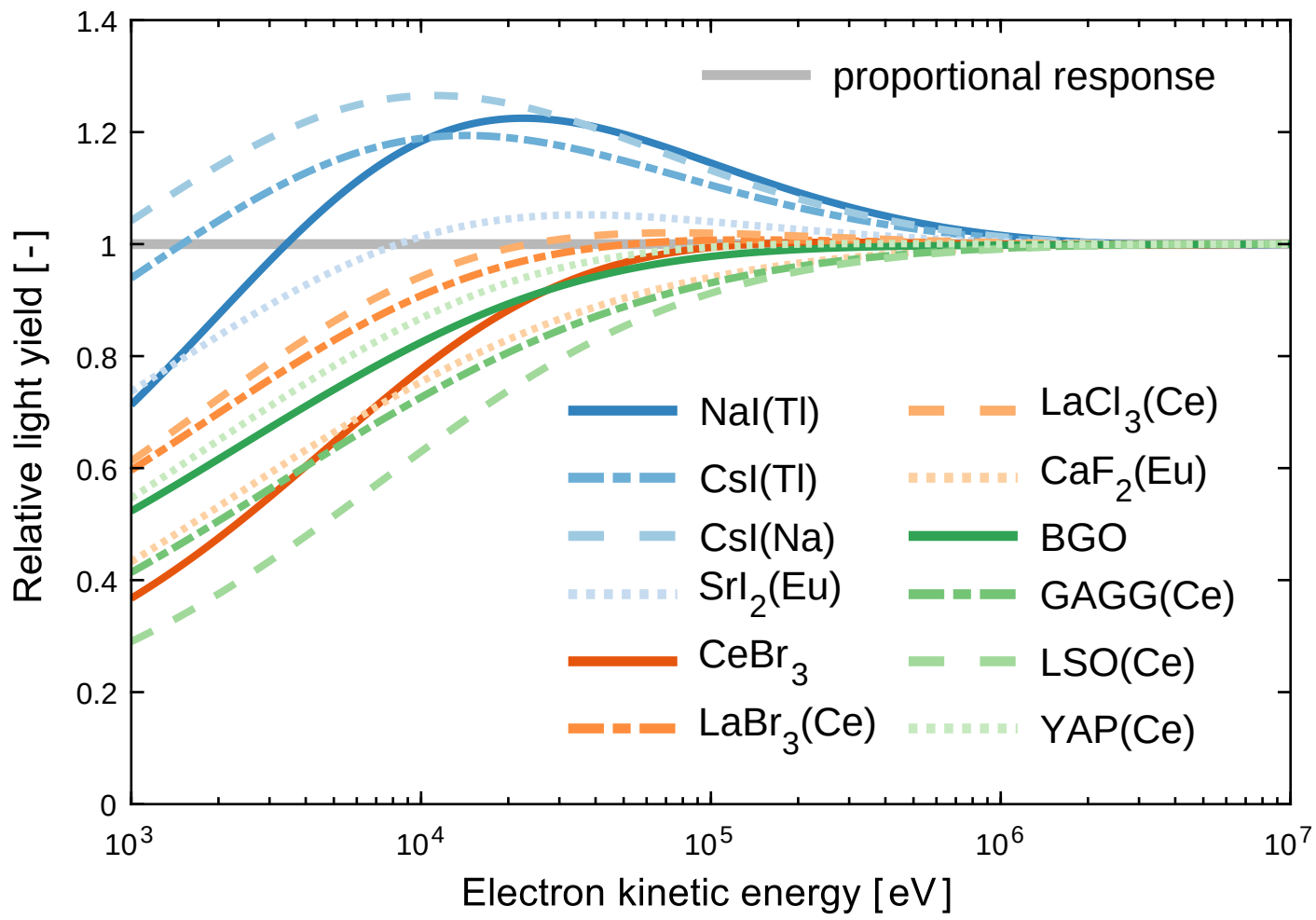


**Figure 4.2** Relative light yield $Y_{\mathrm{sci,r}}$ as a function of the electron kinetic energy $E_{\mathrm{k,e^-}}$ for selected inorganic scintillators. The reference energy for $Y_{\mathrm{sci,r}}$ was set to 10 MeV. The $Y_{\mathrm{sci,r}}$ was computed using the models described by Eqs. 4.9 and A.32b (cf. Appendix A.7). A combined stopping power model for electrons was used to perform the integration in Eq. 4.9 (cf. Fig. B.18). Model parameters ( $\mathrm{d}E/\mathrm{d}x|_{\mathrm{Birks}}$, $\mathrm{d}E/\mathrm{d}x|_{\mathrm{Ons}}$, $\eta_{\mathrm{e/h}}$) were adopted from the literature (cf. Table C.4) for all scintillators except GAGG(Ce), for which these parameters were estimated using experimental data provided by Kaewkhao et al. [342].

To simplify the comparison of the scintillation non-proportionality between different scintillation detectors, Khodyuk et al. [358] introduced a performance metric, the degree of non-proportionality to photons $\sigma_{\mathrm{nPR}}$, which quantifies the scintillation non-proportionality for scintillation events induced by individual high-energy photons with a single number:

$$\sigma_{\mathrm{nPR}} = \frac{1}{E_{\gamma,\mathrm{max}} - E_{\gamma,\mathrm{min}}} \int_{E_{\gamma,\mathrm{min}}}^{E_{\gamma,\mathrm{max}}} \left| 1 - \frac{Y_{\mathrm{sci,a}}\left(E_\gamma\right)}{Y_{\mathrm{sci,a}}\left(E_{\gamma,\mathrm{max}}\right)} \right| \mathrm{d}E_\gamma \qquad (4.11)$$

where:

| | | |
|---|---|---|
| $E_\gamma$ | photon energy | eV |
| $Y_{\mathrm{sci,a}}$ | absolute light yield | $\mathrm{eV}^{-1}$ |

This metric can be computed numerically by conducting light yield measurements with monoenergetic photon sources over a predefined spectral range $E_{\gamma,\mathrm{min}} \leq E_\gamma \leq E_{\gamma,\mathrm{max}}$ [358, 359]. Alternatively, we could estimate $\sigma_{\mathrm{nPR}}$ also by convolving the relative light yield $Y_{\mathrm{sci,r}}$ for electrons with energy distributions obtained by Monte Carlo simulation as discussed above [343, 355–357]. $\sigma_{\mathrm{nPR}}$ together with the absolute light yield $Y_{\mathrm{sci,a}}$ at a reference photon energy of 662 keV is compiled in Table 4.1 at the end of this chapter for common inorganic scintillators in gamma-ray spectrometry.

Besides the scintillator parameters discussed in the sections above, i.e. absolute light yield, degree of non-proportionality and decay time constant, there are various other factors affecting the performance of inorganic scintillators in gamma-ray spectrometry. These include the density, the composition, the intrinsic background due to natural radionuclides incorporated into the lattice[25], the scintillation photon spectrum, the temperature dependence, the hygroscopicity, the radiation hardness, the geometry and size, the cost and the commercial availability, among others. A discussion of all these parameters is beyond the scope of this work. Comprehensive overviews of these properties are provided in the literature [30, 298, 307]. For the reader's convenience, some of these additional properties are included in Table 4.1 at the end of this chapter, too.

25 *As an example, inorganic scintillators incorporating La such as $LaBr_3$(Ce) or $LaCl_3$(Ce) show an increased intrinsic background due to the associated primordial radionuclide $^{138}_{57}$La (cf. Table 2.1) [298].*

## 4.2 Photomultiplier Tube

As already discussed at the beginning of this chapter, we need to couple the inorganic scintillators to photosensors to convert the scintillation light into an electrical signal, which then can be further processed by the electronic hardware to obtain the energy spectrum of the incident radiation [30]. In AGRS, PMTs are the prevalent type of photosensor used for this purpose (cf. Chapter 5). I will therefore focus in this section on this specific photosensor type. If you are interested in other photosensors such as silicon photomultipliers (SiPMs) or avalanch photodiodes (APDs), I recommend the comprehensive reviews by Renker et al. [360], Vinogradov et al. [361], and Gundacker et al. [362]. These reviews provide an in-depth analysis of each photosensor's properties and their respective advantages and disadvantages when compared to PMTs.

The basic components of a PMT are schematically depicted in Fig. 4.3. The scintillator is optically coupled to the PMT by a light window, which is usually made of a transparent material such as silica

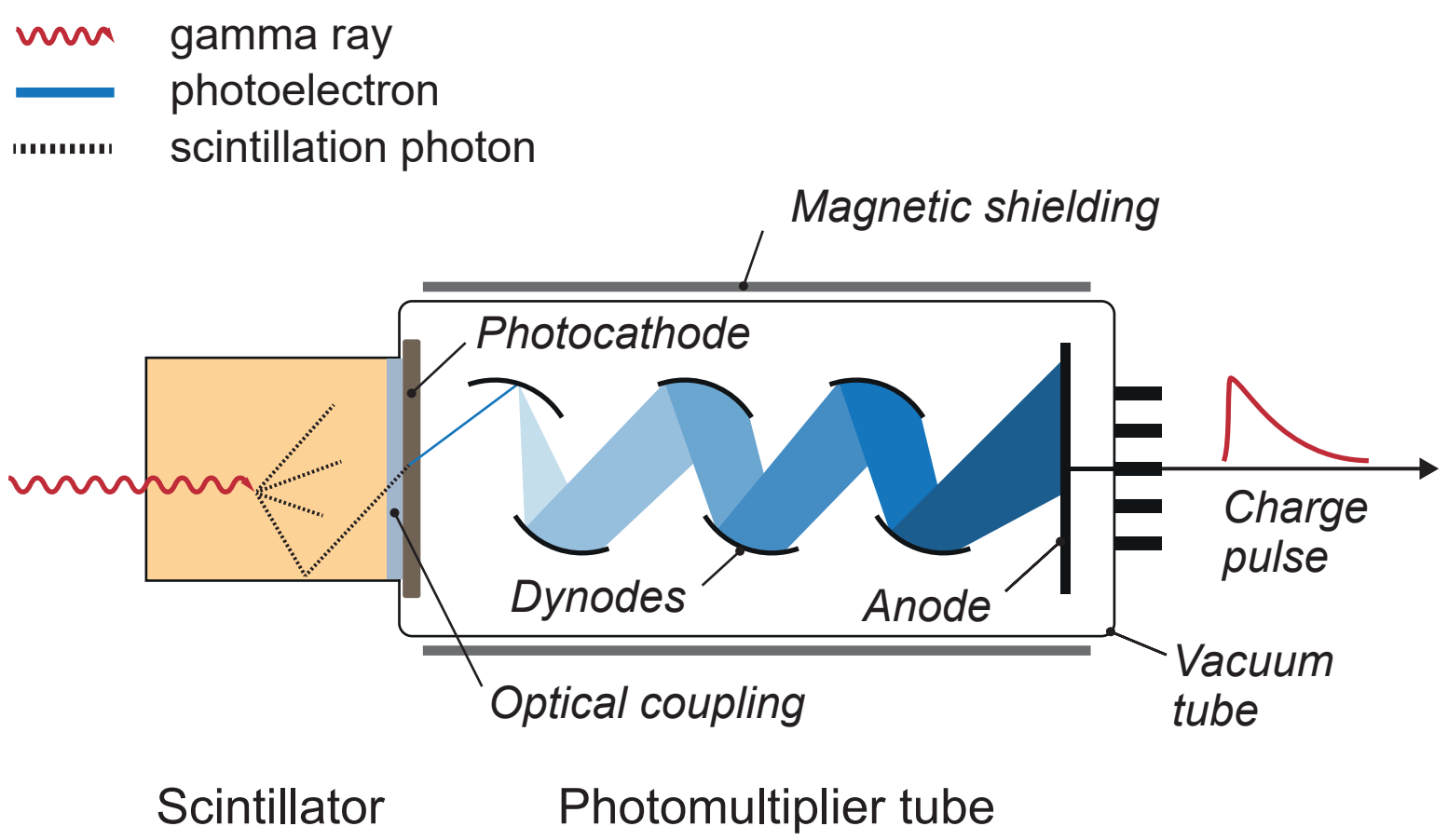


**Figure 4.3** Schematic depiction of a generic photomultiplier tube coupled to an inorganic scintillator for gamma-ray spectrometry applications.

or borosilicate glass [363]. When scintillation photons are generated in the scintillator, they are transmitted[26] through the light window and impinge on a photocathode[27] converting the light into photoelectrons. A typical scintillation pulse in common inorganic scintillator materials releases only a few hundred photoelectrons, a signal too weak for the subsequent electronic hardware [30]. Therefore, the signal has to be amplified, which is achieved by a series of electrodes called dynodes inside a vacuum tube. The photoelectrons are accelerated towards the first dynode by an external electric field. When the photoelectrons strike the first dynode, they release secondary electrons in a cascade-like process. These secondary electrons are then accelerated towards the next dynode, where they release further secondary electrons. This process is repeated over multiple stages of dynodes amplifying the number of electrons from a few hundred to about $10^7$ to $10^{10}$ electrons [30]. This amplified charge signal is then collected at the anode located at the end of the vacuum tube and further processed by the detector electronics [30].

The total charge $Q_{\mathrm{PMT}}$ released in a PMT for a scintillation pulse induced by ionizing radiation depositing an energy $E_{\mathrm{dep}}$ into the scintillator and subsequently emitting $N_{\mathrm{sci}}$ scintillation photons can be estimated as [307]:

26 *To maximize the light collection and transmission from the scintillator to the light window, it is customary to cover all faces of the scintillator, except the one adjacent to the PMT, with a reflector material such as magnesium oxide, aluminum foil, or polytetrafluoroethylene (PTFE) tape [30, 312].*

27 *Typically made out of alkali metals [307, 363].*

$$Q_{\mathrm{PMT}}(E_{\mathrm{dep}}) = \mathfrak{e}N_{\mathrm{sci}}(E_{\mathrm{dep}})\eta_{\mathrm{PMT}}G_{\mathrm{PMT}} \tag{4.12}$$

where:

| | | |
|---|---|---|
| $\mathfrak{e}$ | elementary charge (cf. Constants) | C |
| $E_{\mathrm{dep}}$ | deposited energy | eV |
| $G_{\mathrm{PMT}}$ | photomultiplier total gain | |
| $N_{\mathrm{sci}}$ | number of emitted scintillation photons | |
| $\eta_{\mathrm{PMT}}$ | photomultiplier transfer efficiency | |

The number of scintillation photons $N_{\mathrm{sci}}$ as a function of the deposited energy $E_{\mathrm{dep}}$ can be computed using the absolute light yield $Y_{\mathrm{sci,a}}$ discussed in the previous section, i.e. $N_{\mathrm{sci}}(E_{\mathrm{dep}}) = E_{\mathrm{dep}}Y_{\mathrm{sci,a}}(E_{\mathrm{dep}})$ (cf. Eq. 4.2).[28] The photomultiplier total gain $G_{\mathrm{PMT}}$ quantifies the amplification factor of the PMT, i.e. the ratio between the number of electrons collected at the anode and the number of photoelectrons released in the photocathode [307]. It is not only a function of the number and properties of the adopted dynodes but also of the applied voltage between the dynodes [30]. Last but not least, the photomultiplier transfer efficiency $\eta_{\mathrm{PMT}}$ characterizes both the efficiency with which scintillation photons are collected and converted into photoelectrons by the photocathode, and the transfer efficiency of these generated photoelectrons to the first dynode [312]:

28 *As we have discussed in Section 4.1.3, for scintillators with a non-proportional scintillation response, the energy deposition alone is not sufficient to compute $N_{\mathrm{sci}}$. For that purpose, the detailed distribution of high-energy electrons released in the scintillation pulse needs to be known (cf. Eq. 4.9).*

$$\eta_{\mathrm{PMT}} = \eta_{\mathrm{ph,col}}\eta_{\mathrm{spec}}\eta_{\mathrm{pe,col}} \tag{4.13}$$

where:

| | |
|---|---|
| $\eta_{\mathrm{spec}}$ | degree of spectral match |
| $\eta_{\mathrm{pe,col}}$ | photoelectron collection efficiency at the first dynode |
| $\eta_{\mathrm{ph,col}}$ | light collection efficiency |

The individual terms in Eq. 4.13 quantify the subsequent stages in the conversion of scintillation photons into photoelectrons by the PMT, i.e. the efficiency with which the scintillation photons are collected at the photocathode ($\eta_{\mathrm{ph,col}}$), the conversion efficiency from photons to photoelectrons at the photocathode ($\eta_{\mathrm{spec}}$) and the transfer efficiency from the photocathode to the first dynode ($\eta_{\mathrm{pe,col}}$). For an ideal scintillation spectrometer, we have $\eta_{\mathrm{ph,col}}\eta_{\mathrm{pe,col}} = 1$ [312]. The degree of spectral match $\eta_{\mathrm{spec}}$ on the other hand is limited by the spectral

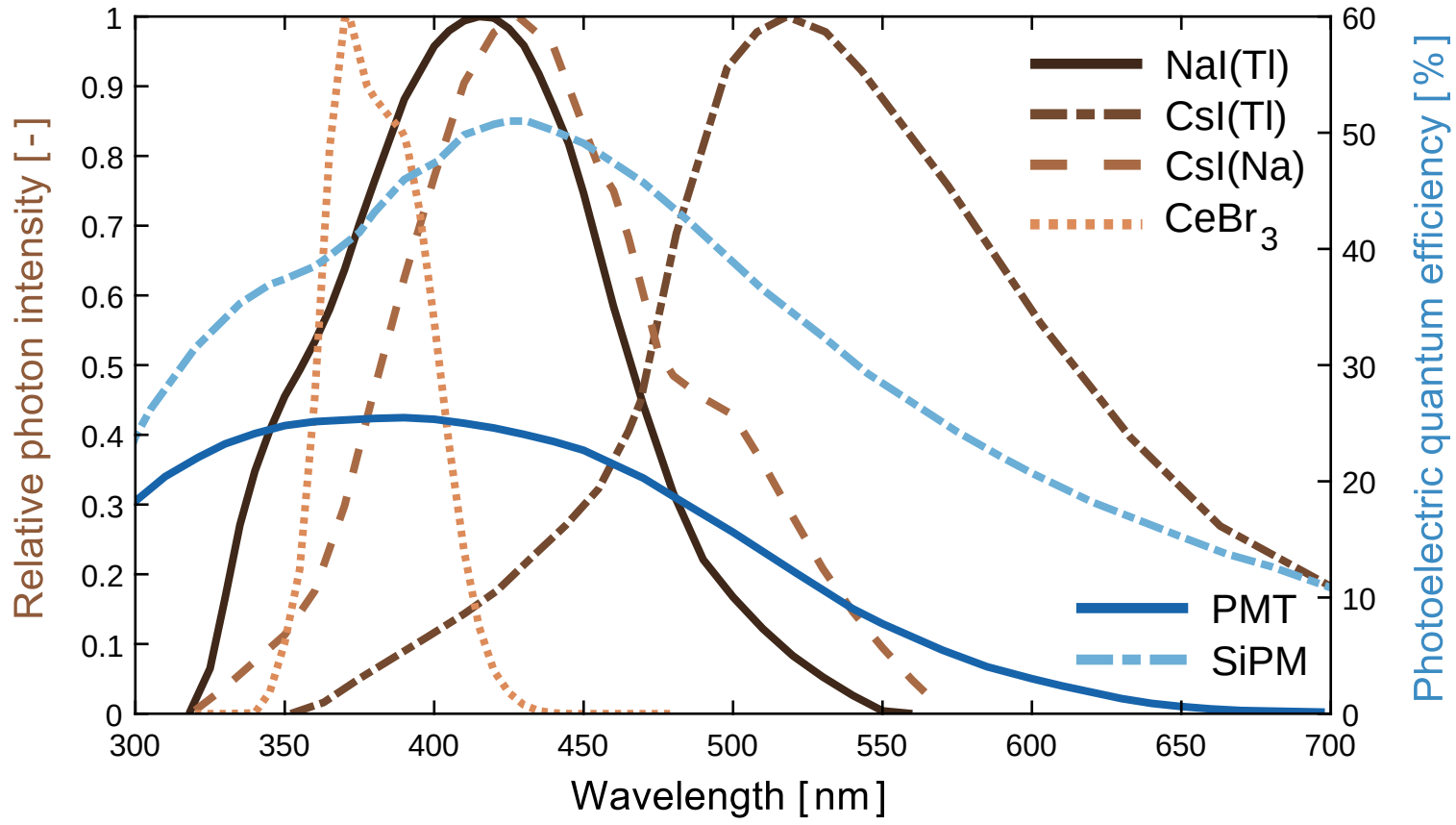


**Figure 4.4** This graph shows the relative scintillation photon spectrum $I_{sci}$ as a function of the scintillator emission wavelength $\lambda_{sci}$ for selected inorganic scintillators. The normalization of $I_{sci}$ was performed by the peak intensity, i.e. $I_{sci}(\lambda_{sci,max})$. In addition, the photoelectric quantum efficiency $\eta_{pe}$ is presented for two generic photodetectors, i.e. a PMT (bialkali photocathode) and a SiPM. All displayed data was provided by Schotanus [364]. Additional scintillator properties including mean values for $\lambda_{sci,max}$ can be found in Table 4.1 at the end of this chapter.

match between the quantum efficiency of the photocathode and the scintillation photon spectrum [312]:

$$\eta_{spec} = \frac{\int_0^\infty I_{sci}\left(\lambda_{sci}\right) \eta_{pe}\left(\lambda_{sci}\right) \mathrm{d}\lambda_{sci}}{\int_0^\infty I_{sci}\left(\lambda_{sci}\right) \mathrm{d}\lambda_{sci}} \tag{4.14}$$

where:

| | | |
|---|---|---|
| $I_{sci}$ | relative scintillation photon spectrum | |
| $\lambda_{sci}$ | scintillator emission wavelength | m |
| $\eta_{pe}$ | photoelectric quantum efficiency | |

and $\eta_{spec} \in [0, 1]$. Typical quantum efficiencies of the photocathodes in modern PMTs can reach values up to 37 % [363]. The relative scintillation photon spectrum for common inorganic scintillators in gamma-ray spectrometry alongside the wavelength-dependent photoelectric quantum efficiency for a typical PMT is shown in Fig. 4.4. In addition, characteristic scintillator peak emission wavelengths

$\lambda_{sci,max}$ for additional inorganic scintillators are provided in Table 4.1 at the end of this chapter. Fig. 4.4 effectively illustrates the spectral match between the PMT and the various scintillators. For instance, it is easy to see that the spectral match between $CeBr_3$ or NaI(Tl) and the selected PMT is significantly better compared to CsI(Tl).

In summary, the overall performance of a gamma-ray spectrometer based on inorganic scintillators coupled to PMTs is not only determined by the standalone performance of each component but also by their mutual compatibility and integration.

## 4.3 Pulse-Height Spectra

As we have discussed in the previous sections, the individual scintillation pulses carry information about the deposited energy of the incident radiation. After the scintillation light is converted into electrical signals by the PMT, these signals are further processed by the detector electronics[29]. The main product of this processing is a histogram of the number of scintillation events $C$ as a function of the pulse height $\mathcal{H}$, i.e. the amplitude of the voltage pulse generated by the detector electronics from the scintillation pulse discussed in Section 4.1.2. We call such a histogram a differential pulse-height spectrum $\mathrm{d}C/\mathrm{d}\mathcal{H}$ [30]. This differential pulse-height spectrum is approximated by a discrete histogram with a finite number of bins, which we refer to as pulse-height channels $n$, normalized by the bin width $\Delta\mathcal{H}$ [30]. But how should we choose the widths of these channels? The most straightforward solution is to divide the pulse-height range $\mathcal{H}_{min} \leq \mathcal{H} \leq \mathcal{H}_{max}$ into $N_{ch}$ equal intervals resulting in a constant channel width of:

*29 As already noted at the beginning of this chapter, a detailed discussion of the detector electronics and multichannel pulse analysis is beyond the scope of this work. For a comprehensive understanding of these subjects, I recommend consulting the monographs authored by Knoll [30], Gilmore [296], and Tsoulfanidis et al. [297].*

$$\Delta\mathcal{H} = \frac{\mathcal{H}_{max} - \mathcal{H}_{min}}{N_{ch}} \tag{4.15}$$

where:

| | | |
|---|---|---|
| $\mathcal{H}_{min}$ | lower-level discriminator (LLD) | V |
| $\mathcal{H}_{max}$ | upper-level discriminator (ULD) | V |
| $N_{ch}$ | number of pulse-height channels | |

As we will see later, it might be beneficial to choose a variable channel width instead of a constant one to account for the scintillation non-proportionality discussed in Section 4.1.3. However, for the sake of simplicity, let us assume a constant channel width

for now. The number of pulses in each pulse-height channel with $\{n \in \mathbb{N}_+ \mid n \leq N_{\mathrm{ch}}, N_{\mathrm{ch}} \in \mathbb{N}_+\}$ is then equivalent to [30, 365]:

$$C(n) = \int_{\mathcal{H}_{\min}+(n-1)\Delta\mathcal{H}}^{\mathcal{H}_{\min}+n\Delta\mathcal{H}} \frac{\mathrm{d}C}{\mathrm{d}\mathcal{H}}\,\mathrm{d}\mathcal{H} \tag{4.16}$$

where:

| | | |
|---|---|---|
| $\mathrm{d}C/\mathrm{d}\mathcal{H}$ | differential pulse-height spectrum | $\mathrm{V}^{-1}$ |
| $\Delta\mathcal{H}$ | pulse-height channel width | V |
| $n$ | pulse-height channel number | |

Although the pulse-height channel number $n$ is an integer quantity, to characterize the location of spectral features within a pulse-height channel, it is often advantageous to define a continuous channel number $\tilde{n}$ [365]. By definition, this parameter ranges continuously from 0.5 to $N_{\mathrm{ch}} + 0.5$ to align the integer channel numbers with the center of the pulse-height channels $\mathcal{H}_{\min} + (\tilde{n} - \frac{1}{2})\Delta\mathcal{H}$ and cover the equivalent pulse-height range between $\mathcal{H}_{\min}$ and $\mathcal{H}_{\max}$. This is often more convenient to work with, especially if a variable pulse-height channel width is adopted.

But how can we relate these pulse-height channels $\tilde{n}$ or the pulse-height $\mathcal{H}$ to the deposited energy $E_{\mathrm{dep}}$ of the incident radiation? Before I answer this question, we need to discuss first the main spectral features in a pulse-height spectrum associated with high-energy photon sources. Understanding these features is crucial for all subsequent data analysis methods as well as for the interpretation of measurement and simulation results throughout this book.

### 4.3.1 Spectral Features

For the purposes of this discussion, let us assume that there is a proportional relationship between the pulse-height $\mathcal{H}$ and the deposited energy $E_{\mathrm{dep}}$ of the incident radiation, i.e. $\mathcal{H} \propto E_{\mathrm{dep}}$. In Fig. 4.5, a typical pulse-height spectrum is depicted for a generic inorganic scintillator spectrometer exposed to a single monoenergetic gamma-ray source.[30] As you can see in the bottom subfigure of Fig. 4.5, there is a variety of very distinct spectral features present in the pulse-height spectrum. These features are the result of a series of complex photon-matter interactions with the scintillator material as well as

30 *To highlight the different spectral features, we assume here that no other ionizing radiation sources are present, i.e. a perfectly shielded spectrometer without any radiation background. In addition, no secondary radiation is emitted by the source such as electrons, positrons or characteristic X-rays.*

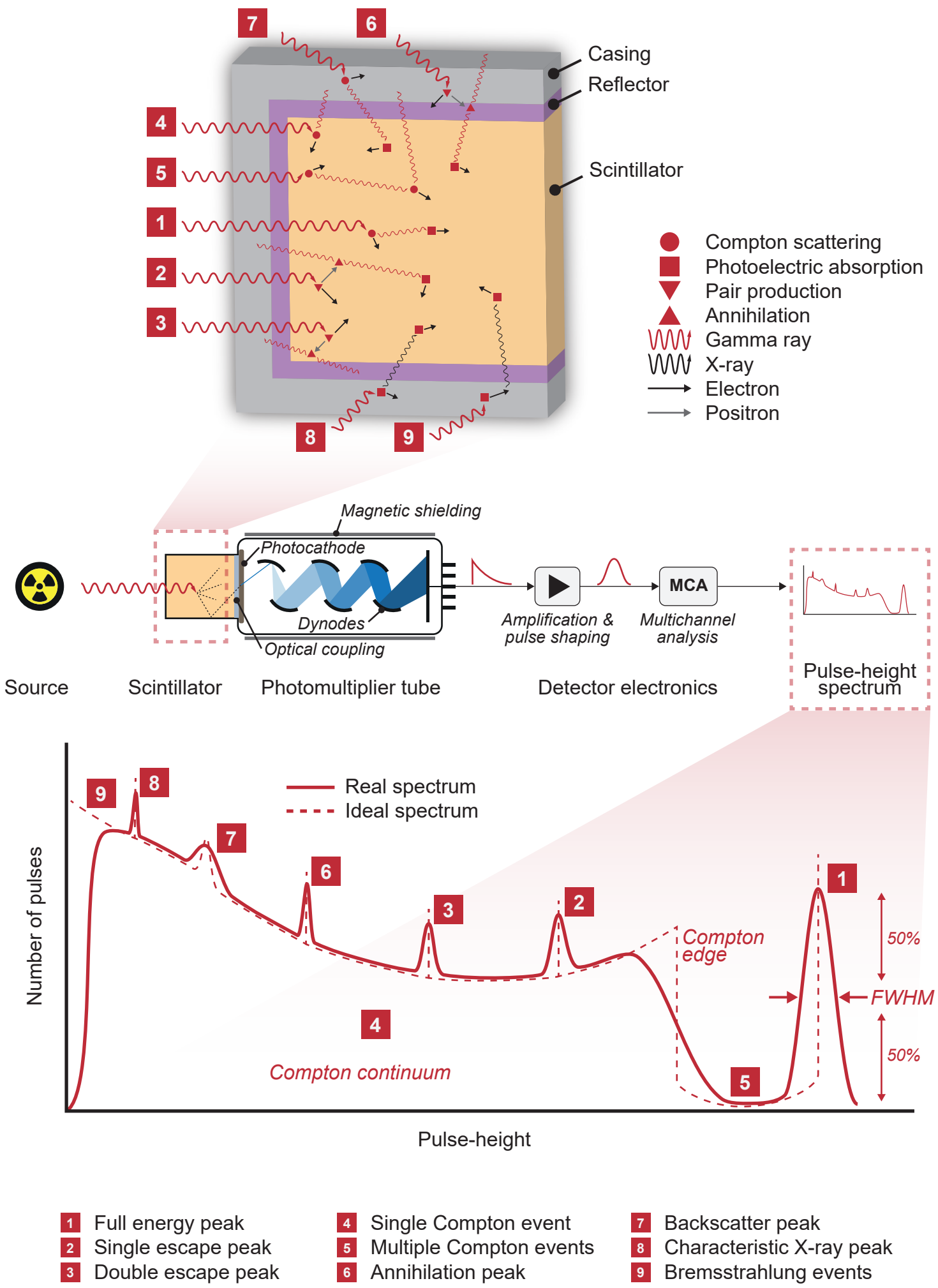


**Figure 4.5** Gamma-ray spectrometry overview highlighting gamma-ray interactions with the scintillator (top), a generic detector configuration scheme (middle) as well as the main features in a pulse-height spectrum associated with the highlighted gamma-ray interactions for an ideal (perfect resolution, $\mathcal{H}_{min} = 0\,V$) and real (finite resolution, $\mathcal{H}_{min} > 0\,V$) detector (bottom). For convenience and ease of interpretation, the energy scale of the pulse-height spectrum has been adjusted to a non-linear format.

with other detector components and the environment in the vicinity of the spectrometer. These interactions are indicated in the top subfigure of Fig. 4.5. For a typical inorganic scintillator spectrometer used in AGRS, the main spectral features in a pulse-height spectrum associated with gamma-ray interactions are the following [30]:

1. **Full energy peak (FEP)** One of the most prominent features in the pulse-height spectrum is the FEP. It corresponds to the deposition of all the energy of a single gamma ray emitted by the monoenergetic source into the scintillator. Such a deposition can be the result of a single photoelectric absorption of the gamma-ray photon or a series of photon-matter interactions[31] ending with the absorption of the gamma-ray photon, all taking place inside the scintillator (cf. also the discussion about the time resolution of inorganic scintillators in Section 4.1.2). Apart from the primary photon, all secondary radiation created by the primary interactions (photoelectrons, Compton electrons, characteristic X-rays, ...) needs to be fully absorbed by the scintillator, too.[32] Consequently, FEPs are a prominent spectral feature that allow us to directly infer the energy of the incident gamma-ray photon $E_\gamma$ as:

$$E_{\mathrm{dep}} = E_\gamma \tag{4.17}$$

   Let us denote this energy in the pulse-height spectrum by $E'_{\mathrm{FEP}}$.

2. **Single escape peak (SEP)** This spectral feature is the result of a pair production event inside the scintillator with one of the annihilation photons (created by the annihilation of the positron as discussed in Section 3.1.2) escaping from the scintillator volume without interaction. All other secondary radiation is completely absorbed by the scintillator. As a result, the deposited energy is not the full gamma-ray energy but is reduced by the energy of the one escaping annihilation photon (cf. Eq. 2.8):

$$E_{\mathrm{dep}} = E_\gamma - \mathrm{m_e c^2} \tag{4.18}$$

   where:

| | | |
|---|---|---|
| $\mathrm{m_e}$ | electron mass (cf. Constants) | $\mathrm{eV\,m^{-2}\,s^2}$ |
| c | speed of light in vacuum (cf. Constants) | $\mathrm{m\,s^{-1}}$ |

31 *For example, multiple Compton scattering events followed by photoelectric absorption or a pair production event with subsequent photoelectric absorption of the two secondary annihilation photons (cf. Section 2.1.1).*

32 *Note that I have not included characteristic X-ray escape peaks in the pulse-height spectrum. These peaks are the result of an interaction series in the scintillator where the entire energy of the gamma ray is deposited except the energy of a single characteristic X-ray created during the photoelectric absorption, which leaves the scintillator volume without interaction [30]. For typical NaI(Tl) based spectrometers used in AGRS, because of the finite resolution, characteristic X-ray escape peaks are in most cases obscured by the FEPs. Moreover, because of the comparably large scintillator volumes adopted in AGRS, the relative rate of such escape events is low [30].*

Consequently, we expect to find the SEP about 511 keV below the FEP in the pulse-height spectrum as indicated in the bottom subfigure of Fig. 4.5. Let us denote this energy in the pulse-height spectrum by $E'_{\text{SEP}}$. Note also that for a SEP, the initial photon energy $E_\gamma$ needs to be $E_\gamma \geq 2\text{m}_\text{e}\text{c}^2$, i.e. ~1.022 MeV or higher, to allow for the creation of a positron-electron pair in a pair production event (cf. Section 3.1.2).

3. **Double escape peak (DEP)** This feature is very similar to the SEP. The DEP is again the result of a pair production event inside the scintillator, but this time not with one but both of the annihilation photons escaping from the scintillator volume without interaction. Consequently, the deposited energy is reduced by the energy of both escaping annihilation photons:

$$E_{\text{dep}} = E_\gamma - 2\text{m}_\text{e}\text{c}^2 \tag{4.19}$$

which translates to a spectral peak about 1022 keV below the FEP. Let us denote this energy in the pulse-height spectrum by $E'_{\text{DEP}}$. Note again that the initial photon energy needs to be $E_\gamma \geq 2\text{m}_\text{e}\text{c}^2$ in order for the DEP to occur.

4. **Compton continuum (CC)** In contrast to the already discussed features, the Compton continuum is not a discrete but a continuous part of the pulse-height spectrum as indicated in Fig. 4.5. It is the result of a single Compton scattering event with the scattered gamma ray escaping from the scintillator volume and the secondary radiation (Compton electron, bremsstrahlung, . . . ) being fully absorbed by the scintillator. Consequently, we have an energy deposition equivalent to the initial energy of the Compton electron created by the Compton scattering of the gamma ray with energy $E_\gamma$ by a scattering angle of $\theta_\gamma$ (cf. Eq. 3.14b):

$$E_{\text{dep}} = E_\gamma \frac{\alpha_\gamma \left(1 - \cos\theta_\gamma\right)}{1 + \alpha_\gamma \left(1 - \cos\theta_\gamma\right)} \tag{4.20}$$

where:

| | | |
|---|---|---|
| $\alpha_\gamma$ | ratio of the photon energy to the energy-equivalent electron rest mass (cf. Eq. 3.12) | |
| $\theta_\gamma$ | photon scattering angle | rad |

The characteristic shape of the Compton continuum is determined by the differential energy cross-section for Compton electrons given by Eq. 3.18 and displayed in Fig. 3.3 for selected photon energies. As we have discussed in Section 3.1.2, there is an upper threshold for the kinetic energy of the Compton electron, which translates to a Compton scatter event with $\theta_\gamma = \pi$:

$$E_{\text{dep}} = E_\gamma \left(1 - \frac{1}{1 + 2\alpha_\gamma}\right) \tag{4.21}$$

This upper limit is a distinct feature in the pulse-height spectrum where the Compton continuum abruptly ends. This spectral feature is also known as the **Compton edge (CE)** [30]. Let us denote this energy in the pulse-height spectrum by $E'_{\text{CE}}$.

5. **Compton gap (CG)** As a direct result of the upper limit for the kinetic energy of the Compton electron, there is a spectral region between the CE and the FEP where no single Compton scattering events are possible. As a result, the spectrum shows a characteristic gap or valley in this region. The width of this gap can be computed by simply subtracting Eq. 4.21 from $E_\gamma$ resulting in [30]:

$$\Delta E_{\text{dep}} = \frac{E_\gamma}{1 + 2\alpha_\gamma} \tag{4.22}$$

which has an upper limit of [30]:

$$\lim_{E_\gamma \to \infty} \Delta E_{\text{dep}} = \frac{\mathrm{m_e c^2}}{2} \tag{4.23}$$

and translates to about 0.256 MeV. Consequently, the CE is always located above the SEP or the DEP in the pulse-height spectrum. It is also important to add that the spectrum is not suppressed to zero pulses, as there are still other interactions contributing to the pulse-height spectrum in this region. These are mainly multiple Compton scattering events [30], i.e. a gamma ray that undergoes more than one Compton scattering event inside the scintillator and with the scattered gamma ray escaping in the end from the scintillator volume as indicated in Fig. 4.5.

33 *Note that for radionuclide sources with a $\beta^+$ decay mode (cf. Eq. 2.3), the AP can also be the result of the annihilation of the positron emitted by the source itself.*

6. **Annihilation peak (AP)** Similar to SEPs and DEPs, the AP is also the result of a pair production event.[33] Specifically, given the time resolution of the gamma-ray spectrometer discussed in Section 4.1.2, the AP results from a pair production event occurring in the surrounding material, where one of the annihilation photons from the subsequent positron annihilation interacts with the scintillator material, undergoing complete absorption. This absorption process is similar to the FEP and can occur through either a single photoelectric absorption event or a series of photon-matter interactions, as depicted in Fig. 4.5. The deposited energy is thus given by the photon energy of the annihilation photon (Eq. 2.8):

$$E_{\text{dep}} = \mathrm{m_e c^2} \tag{4.24}$$

which translates to a characteristic spectral peak at about 511 keV in the pulse-height spectrum. Let us denote this energy in the pulse-height spectrum by $E'_{\text{AP}}$. Similar to the SEP and the DEP, the initial photon energy needs to be $E_\gamma \geq 2\mathrm{m_e c^2}$ in order for the AP to occur.

7. **Backscatter peak (BSP)** Like the AP, the BSP is also the result of a photon-matter interaction event taking place in the surrounding matter. In this case however, it is not a pair production event but a single or multiple Compton scattering events with the scattered gamma ray getting fully absorbed by the scintillator. For a gamma ray absorbed by the scintillator which was undergoing one Compton scattering event in the surrounding matter, the deposited energy is equivalent to:

$$E_{\text{dep}} = \frac{E_\gamma}{1 + \alpha_\gamma \left(1 - \cos\theta_\gamma\right)} \tag{4.25}$$

which is simply mirroring the Compton continuum in Eq. 4.20. Consequently, like the Compton continuum, it is a continuous part of the pulse-height spectrum. But why do we call it a peak? Well, for that we have to take a look at the distribution of the scattered photon energies $E'_\gamma$ and thereby the deposited energy $E_{\text{dep}}$ as a function of the scattering angle $\theta_\gamma$ which is displayed in Fig. 4.6 for selected initial photon energies $E_\gamma$. We can easily see that for backscattered photons ($\theta_\gamma > \pi/2$), the deposited energy is

fairly restricted to $E_{dep} \lesssim 0.5\,\text{MeV}$. In fact, for fully backscattered photons ($\theta_\gamma = \pi$), the deposited energy reduces to:

$$E_{dep}\Big|_{\theta_\gamma=\pi} = \frac{E_\gamma}{1 + 2\alpha_\gamma} \tag{4.26}$$

Let us denote this energy in the pulse-height spectrum by $E'_{BSP}$. Eq. 4.26 is equivalent to the CG[34] in Eq. 4.22 and has also an upper limit of [30]:

34 *As for both cases, this energy quantifies the energy of the photon after a Compton scattering event occurred with maximum energy loss for the photon, i.e. full backscattering with $\theta_\gamma = \pi$.*

$$\lim_{E_\gamma \to \infty} E_{dep}\Big|_{\theta_\gamma=\pi} = \frac{m_e c^2}{2} \tag{4.27}$$

which translates again to about 0.256 MeV. Consequently, a large number of backscattered photons possess energies in the range of ~0.200 MeV to 0.256 MeV resulting in a peak-like feature in this spectral region in the pulse-height spectrum [30]. This is true for a large variety of initial photon energies with $E_\gamma \gtrsim 0.3\,\text{MeV}$ which can also be seen in Fig. 4.6. It is interesting to add that there is a photon energy $E_\gamma$ for which the BSP energy given by Eq. 4.26 is equal to the CE energy given by Eq. 4.21. We can obtain this photon energy by setting Eq. 4.26 equal to Eq. 4.21 and solving for $E_\gamma$ which gives us a photon energy of $E_\gamma = m_e c^2/2$, i.e. about 0.256 MeV.

8. **Characteristic X-ray peak (CXP)** Similar to the AP or the BSP, the CXP is again a result of a photon-matter interaction event taking place in the surrounding matter. This time, it is a photoelectric absorption event induced by the gamma ray close to the scintillator volume which is the source of the CXP. As we have learned in Section 3.1.2, photoelectric absorption gives rise to secondary characteristic X-ray photons. The energy of these characteristic X-ray photons is typically $\lesssim 100\,\text{keV}$ resulting in a relatively limited range for these photons. However, if the photoelectric absorption takes place close to the scintillator volume, e.g. in the surrounding detector material, the characteristic X-ray photon may reach the scintillator material and get subsequently absorbed by an additional photoelectric absorption event. This leads to a characteristic spectral peak in the pulse-height spectrum at the corresponding photon energy as indicated in Fig. 4.5 [30].[35]

35 *Note that most radionuclide sources also emit characteristic X-rays as a result of the decay process. However, these characteristic X-rays are typically absorbed in the surrounding matter before they can reach the scintillator material.*

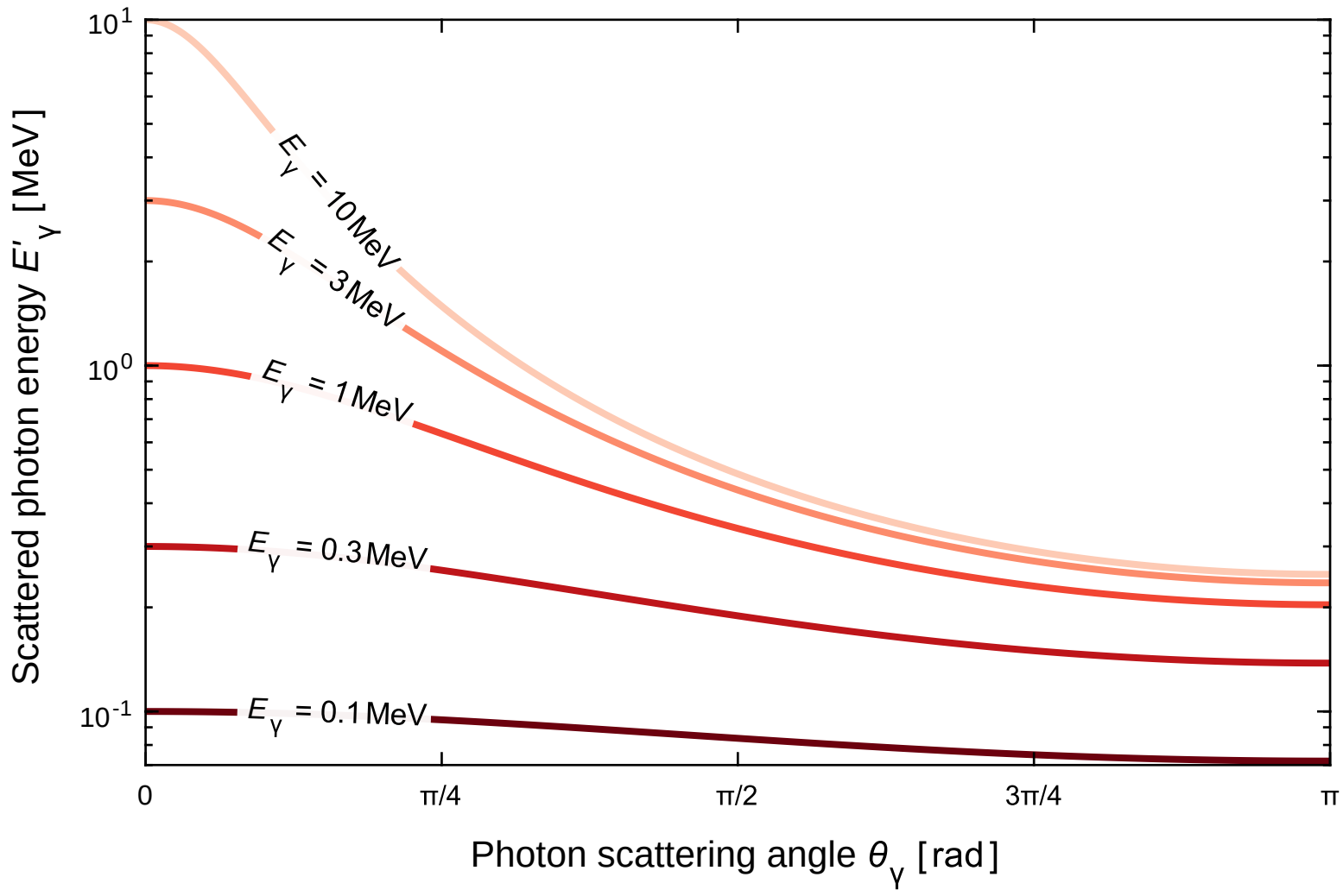


**Figure 4.6** Compton scattering photon energy $E'_\gamma$ as a function of the photon scattering angle $\theta_\gamma$ and initial photon energies $E_\gamma$.

9. **Bremsstrahlung continuum** Last but not least, the bremsstrahlung continuum is another continuous part of the pulse-height spectrum. It is the result of high-energy secondary electrons emitting bremsstrahlung in the surrounding matter (cf. Section 2.1.1). The high-energy secondary electrons or positrons can be created by a variety of photon-matter interactions, e.g. Compton scattering, pair production, or photoelectric absorption events. The bremsstrahlung emission spectrum typically peaks at the very low end of the pulse-height spectrum as indicated in Fig. 4.5 [30].

As already indicated above for some cases, the number and order of the spectral features in a pulse-height spectrum are not arbitrary but follow a specific spectral scheme. We can distinguish in general between five different such schemes:

$$[E'_{\mathrm{BSP}}, E'_{\mathrm{AP}}, E'_{\mathrm{DEP}}, E'_{\mathrm{SEP}}, E'_{\mathrm{CE}}, E'_{\mathrm{FEP}}] \quad E_\gamma > 3\mathrm{m_e c^2} \tag{4.28a}$$

$$[E'_{\mathrm{BSP}}, E'_{\mathrm{DEP}}, E'_{\mathrm{AP}}, E'_{\mathrm{SEP}}, E'_{\mathrm{CE}}, E'_{\mathrm{FEP}}] \quad E_\gamma < 3\mathrm{m_e c^2} \tag{4.28b}$$

$$E_\gamma > (1+\sqrt{2})\mathrm{m_e c^2}$$

$$[E'_{\mathrm{DEP}}, E'_{\mathrm{BSP}}, E'_{\mathrm{AP}}, E'_{\mathrm{SEP}}, E'_{\mathrm{CE}}, E'_{\mathrm{FEP}}] \quad E_\gamma < (1+\sqrt{2})\mathrm{m_e c^2} \tag{4.28c}$$

$$E_\gamma > 2\mathrm{m_e c^2}$$

$$[E'_{\mathrm{BSP}}, E'_{\mathrm{CE}}, E'_{\mathrm{FEP}}] \quad E_\gamma < 2\mathrm{m_e c^2} \tag{4.28d}$$

$$E_\gamma > \mathrm{m_e c^2}/2$$

$$[E'_{\mathrm{CE}}, E'_{\mathrm{BSP}}, E'_{\mathrm{FEP}}] \quad E_\gamma < \mathrm{m_e c^2}/2 \tag{4.28e}$$

with:[36]

| | | |
|---|---|---|
| $E'_{\mathrm{AP}}$ | spectral energy of the annihilation peak | eV |
| $E'_{\mathrm{BSP}}$ | spectral energy of the backscatter peak | eV |
| $E'_{\mathrm{CE}}$ | spectral energy of the Compton edge | eV |
| $E'_{\mathrm{FEP}}$ | spectral energy of the full energy peak | eV |
| $E'_{\mathrm{SEP}}$ | spectral energy of the single escape peak | eV |
| $E'_{\mathrm{DEP}}$ | spectral energy of the double escape peak | eV |

where I sorted the corresponding characteristic energies of the spectral features in the pulse-height spectrum in ascending order, e.g. $[E'_1, E'_2, E'_3]$ is equivalent to $E'_1 < E'_2 < E'_3$. The specific type of spectral scheme is determined by the initial photon energy $E_\gamma$ according to the four transition energies $\mathrm{m_e c^2}/2$, $2\mathrm{m_e c^2}$, $(1+\sqrt{2})\mathrm{m_e c^2}$ and $3\mathrm{m_e c^2}$. These schemes will become important in Chapter 9.

There is a second important remark in regard to the pulse-height spectrum. As indicated by the solid and dashed lines in the bottom subfigure of Fig. 4.5, there are some differences between the idealized pulse-height spectrum discussed so far and a pulse-height spectrum recorded in practice. There are four main differences:

I. **Lower-level discriminator (LLD)** As indicated at the beginning of this section, the pulse-height spectrum is obtained by performing multichannel pulse analysis. During this analysis, pulses that are smaller than a certain threshold are rejected to suppress electronic noise registered at low pulse heights (cf. Eqs. 4.15 and 4.16) [30, 296]. This threshold is referred to as the lower-level discriminator (LLD) and is typically set to a value that corresponds to

36 *Note that I have not included CXPs as they depend on the specific elements in the materials surrounding the scintillator. Furthermore, they are typically significantly suppressed compared to other spectral features, especially in the case of large scintillator volumes such as the ones used in AGRS.*

a spectral energy of about 10 keV to 50 keV for inorganic scintillator spectrometers used in AGRS. The result from the LLD is a complete suppression of the pulses in the low energy region of the pulse-height spectrum as indicated in the bottom subfigure of Fig. 4.5.

II. **Summation effects** Due to the finite time resolution of inorganic scintillator spectrometers (mainly determined by the scintillator decay time constant $\tau_{sci,d}$ and the detector electronics as discussed in Section 4.1.2), there is a probability for two or more pulses to occur within a time scale that is comparable to or smaller than the detector time resolution. In this case, the two pulses are summed together which can lead to characteristic sum coincidence continua and sum coincidence peaks in the pulse-height spectrum [30, 296]. For low-count rate applications, these coincidence summing effects are typically overshadowed by the various background sources, except at the upper end of the pulse-height spectrum where these backgrounds are low [30]. In general, we distinguish two main categories of coincidence summing: true coincidence and chance coincidence [30].

True coincidence is the result of a radionuclide emitting two or more gamma rays during its radioactive decay [30].[37] These gamma rays are emitted typically with a time difference in the order of picoseconds or less.[38] Consequently, if both emitted gamma rays interact with the scintillator, they will be detected as a single scintillation event. The rate of these true coincidence events to occur is only a function of the radionuclide decay scheme and the source-detector geometry.

37 *An example for such a radionuclide is* $^{60}_{27}$Co *emitting two gamma rays with energies of about* 1.173 MeV *and* 1.332 MeV *during its decay.*

38 *For* $^{60}_{27}$Co*, the time difference is about* 0.7 ps *[68].*

Chance coincidence on the other hand is the accidental summing of scintillation pulses induced by high-energy photons emitted by independent source events, e.g. gamma rays emitted by independent radionuclide decays. Because the time intervals separating scintillation events induced by such independent gamma rays are randomly distributed, some time intervals will be smaller than the inherent resolving time of the scintillator or pulse-processing system resulting in a chance coincidence summing of the corresponding pulses [30]. This phenomenon is also known as random summing, random coincidence or pile-up [296]. In contrast to true coincidence summing, the rate of chance coincidence events increases with the count rate of the spectrometer [30]. Consequently, it is mainly relevant for high-count rate applications and

is of lesser importance for low-count rate applications like AGRS. Moreover, most modern spectrometer systems are equipped with pulse-processing systems that include active pile-up rejection to reduce the effect of chance coincidence events [30].

**III. Dead time** This phenomenon is somewhat related to the summation effect, i.e. pile-up. Dead time is the time interval during which the scintillation spectrometer is not able to detect a new scintillation event because it is still processing a previous event [366]. The basic difference between dead time and pile-up is that in the case of dead time, additional scintillation pulses have no effect on the pulse-height spectrum, i.e. these additional events are automatically rejected. Consequently, if we compute the count rate from the recorded number of pulses $C$ in a pulse-height channel, we need to account for these dead time losses. In case the dead time loss events do not affect the dead time itself,[39] we can compute the true count rate $c$ from the recorded number of counts $C$ by [8, 10, 30]:

$$c(n) = \frac{C(n)}{t_{\text{meas}} - C(n)\,\tau_{\text{dead}}} \tag{4.29}$$

with:

| | | |
|---|---|---|
| $C$ | number of pulses in pulse-height channel $n$ | |
| $n$ | pulse-height channel number | |
| $t_{\text{meas}}$ | measurement time | s |
| $\tau_{\text{dead}}$ | dead time constant | s |

The dead time constant $\tau_{\text{dead}}$ quantifies the mean dead time per scintillation pulse [52]. For typical inorganic scintillator spectrometers, $\tau_{\text{dead}}$ is in the order of 1 µs to 15 µs [8, 10, 297]. The corrected measurement time $t_{\text{meas}} - C\tau_{\text{dead}}$ in Eq. 4.29 is also called the live time $t_{\text{live}}$. Modern scintillation spectrometers automatically record this live time to allow for the computation of the true count rate $c$ [8]. If not otherwise stated, the count rate $c$ is assumed to be dead-time corrected in the remainder of this book.

**IV. Spectral resolution** Probably the most evident difference between the idealized and real pulse-height spectrum shown in the bottom subfigure of Fig. 4.5 is the finite spectral resolution of the detector system. The spectral resolution is a measure of the minimal difference in the deposited energy that can be distinguished

39 *Which is true for most inorganic scintillator spectrometers [8, 10, 52]. We call such systems also nonparalyzable and the related dead time "dead time of type I" [366].*

in the pulse-height spectrum [30]. For a spectrometer with perfect spectral resolution, the FEP associated with a monoenergetic gamma-ray source would thus be a Dirac delta function resulting in a peak contained in a single channel in the pulse-height spectrum as indicated in Fig. 4.5. In a spectrometer with finite resolution, the FEP is broadened and the peak is spread over several channels. But what is the reason for this spread? That is an interesting question. The fundamental origin of the finite spectral resolution is the inherent fluctuation in the detector response signal, i.e. the electric charge pulses discussed in Section 4.2 [30]. Because the spectral resolution has important implications not only for radiation measurements but also for detector response simulations, I will provide a more formal definition of this quantity in the next subsection alongside a detailed discussion about its origin and methods for its characterization.

### 4.3.2 Spectral Resolution

The spectral resolution $R_E$ for an inorganic scintillator spectrometer is defined as the full width at half maximum (FWHM) of a FEP in the pulse-height spectrum divided by the centroid[40] of the FEP $\mu_{\mathcal{H}}$ [30]:[41]

$$R_E\left(\mu_{\mathcal{H}}\right) = \frac{\text{FWHM}\left(\mu_{\mathcal{H}}\right)}{\mu_{\mathcal{H}}} \tag{4.30}$$

where:

| | |
|---|---|
| FWHM | full width at half maximum of the FEP |
| $\mu_{\mathcal{H}}$ | centroid of the FEP |

Thus, the spectral resolution quantifies the finite spectral width of a monoenergetic photon response in the pulse-height spectrum. As indicated in Eq. 4.30, it is a function of the centroid of the FEP $\mu_{\mathcal{H}}$ in the pulse-height spectrum and thereby the deposited energy $E_{\text{dep}}$.

By adopting generating functions [367] and the PMT model introduced in Section 4.2, Breitenberger derived a model to compute the spectral resolution $R_E$ as a function of the total charge released in a photomultiplier tube $Q_{\text{PMT}}$ [368]:[42]

40 *We can perform this characterization either on a pulse-height $\mathcal{H}$, spectral energy $E'$ or continuous pulse-height channel number $\tilde{n}$ basis. By definition, I will use the continuous pulse-height channel number $\tilde{n}$ as the basis for the spectral resolution characterization in this work.*

41 *Please note that in this definition, we assume the baseline, onto which the peak may be superimposed, is negligible or has already been subtracted [30].*

42 *Note that we assume here that $Q_{\text{PMT}}$ is sufficiently large so that we may approximate the* FWHM *assuming a Gaussian distribution, i.e.* $\text{FWHM} = 2\sqrt{2\log 2}\sqrt{\text{var}(Q_{\text{PMT}})}$ *[312, 368].*

$$R_E = 2\sqrt{2\log 2}\sqrt{\frac{\mathrm{var}\left(Q_{\mathrm{PMT}}\right)}{Q_{\mathrm{PMT}}^2}} \tag{4.31a}$$

$$= 2\sqrt{2\log 2}\left\{\frac{\mathrm{var}\left(\eta_{\mathrm{PMT}}\right)}{\eta_{\mathrm{PMT}}^2} + \frac{1+\mathrm{var}\left(G_{\mathrm{PMT}}\right)/G_{\mathrm{PMT}}^2}{N_{\mathrm{sci}}\eta_{\mathrm{PMT}}} + \left[1+\frac{\mathrm{var}\left(\eta_{\mathrm{PMT}}\right)}{\eta_{\mathrm{PMT}}^2}\right]\left[\frac{\mathrm{var}\left(N_{\mathrm{sci}}\right)}{N_{\mathrm{sci}}^2} - \frac{1}{N_{\mathrm{sci}}}\right] + \frac{R_{E,\mathrm{elec}}^2}{8\log 2}\right\}^{\frac{1}{2}} \tag{4.31b}$$

with:[43]

| | | |
|---|---|---|
| $G_{\mathrm{PMT}}$ | photomultiplier total gain | |
| $N_{\mathrm{sci}}$ | number of emitted scintillation photons | |
| $Q_{\mathrm{PMT}}$ | total charge released in a photomultiplier tube | C |
| $R_{E,\mathrm{elec}}$ | spectral resolution induced by electronics | |
| $\eta_{\mathrm{PMT}}$ | photomultiplier transfer efficiency | |

If we assume $\mathrm{var}\left(\eta_{\mathrm{PMT}}\right)/\eta_{\mathrm{PMT}}^2 \ll 1$[44], we can simplify Eq. 4.31b to [368, 370]:

$$R_E \approx 2\sqrt{2\log 2}\left\{\overbrace{\frac{1}{N_{\mathrm{sci}}\eta_{\mathrm{PMT}}}}^{R_{E,\mathrm{stat}}^2} + \overbrace{\left[\frac{\mathrm{var}\left(N_{\mathrm{sci}}\right)}{N_{\mathrm{sci}}^2} - \frac{1}{N_{\mathrm{sci}}}\right]}^{R_{E,\mathrm{intr}}^2} + \underbrace{\frac{\mathrm{var}\left(\eta_{\mathrm{PMT}}\right)}{\eta_{\mathrm{PMT}}^2}}_{R_{E,\mathrm{trans}}^2} + \underbrace{\frac{\mathrm{var}\left(G_{\mathrm{PMT}}\right)}{N_{\mathrm{sci}}\eta_{\mathrm{PMT}}G_{\mathrm{PMT}}^2}}_{R_{E,\mathrm{pmt}}^2} + \frac{R_{E,\mathrm{elec}}^2}{8\log 2}\right\}^{\frac{1}{2}} \tag{4.32}$$

From Eq. 4.32 we see that $R_E^2$ may be written as a linear combination of the following terms:

$$R_E^2 = R_{E,\mathrm{stat}}^2 + R_{E,\mathrm{intr}}^2 + R_{E,\mathrm{trans}}^2 + R_{E,\mathrm{pmt}}^2 + R_{E,\mathrm{elec}}^2 \tag{4.33}$$

43 *Note that I adapted the notation introduced by Birks [312] and Breitenberger [368] to align with that introduced in Section 4.2. Moreover, for completeness, I added an additional term $R_{E,\mathrm{elec}}$ in Eqs. 4.31b and 4.32, which Breitenberger has not included yet. This term characterizes the contribution of the detector electronics to $R_E$ as discussed by Knoll [30] and Rasco et al. [369].*

44 $\mathrm{var}\left(\eta_{\mathrm{PMT}}\right)/\eta_{\mathrm{PMT}}^2$ *is typically in the order of* $10^{-3}$ *or smaller [312, 368].*

which are also indicated already in Eq. 4.32. Each of these terms corresponds to a distinct physical process in the signal processing chain of a scintillator spectrometer using a PMT. As a result, they represent the individual contributions of these processes to the spectral resolution $R_E$. Let us break down the individual terms one by one.

The first term $R^2_{E,\text{stat}} = 1/(N_{\text{sci}}\eta_{\text{PMT}})$ quantifies the intrinsic quantum fluctuation in the number of photoelectrons reaching the first dynode in the PMT for a scintillation pulse comprising $N_{\text{sci}}$ scintillation photons [30]. It is inversely proportional to the product of the number of scintillation photons $N_{\text{sci}}$ and the photomultiplier transfer efficiency $\eta_{\text{PMT}}$ [30]. From this, we see already that to reduce this term, we aim to increase the number of scintillation photons $N_{\text{sci}}$ and the photomultiplier transfer efficiency $\eta_{\text{PMT}}$, i.e. use a scintillator with high absolute light yield $Y_{\text{sci,a}}$ coupled to a PMT with high photomultiplier transfer efficiency $\eta_{\text{PMT}}$.

The second term $R^2_{E,\text{intr}} = \text{var}(N_{\text{sci}})/N^2_{\text{sci}} - 1/N_{\text{sci}}$ accounts for the spectral resolution degradation due to the scintillation non-proportionality as well as non-uniformities[45] in the scintillator. Because of its intrinsic nature, we call $R_{E,\text{intr}}$ also the intrinsic resolution of a scintillator. $R_{E,\text{intr}}$ can also be interpreted as a measure of the deviation from a pure Poisson process statistic [370], as for a perfectly proportional scintillator without any inhomogeneities in the crystal lattice, we have $\text{var}(N_{\text{sci}})/N^2_{\text{sci}} = 1/N_{\text{sci}}$ and this second term vanishes [370]. The intrinsic spectral resolution $R_{E,\text{intr}}$ can only be improved by reducing the inhomogeneities in the scintillator crystal structure. Hence, the non-proportional scintillation response contribution to $R_{E,\text{intr}}$ represents a fundamental limit to the overall spectral resolution of a scintillator spectrometer [359, 370, 372]. Depending on the degree of non-proportionality for a given scintillator, the intrinsic resolution $R_{E,\text{intr}}$ can be of similar magnitude or even higher than the statistical contribution $R_{E,\text{stat}}$ over a large part of the pulse-height spectrum [298, 370, 373].

The third term $R^2_{E,\text{trans}} = \text{var}(\eta_{\text{PMT}})/\eta^2_{\text{PMT}}$ quantifies the statistical variation in the photomultiplier transfer efficiency $\eta_{\text{PMT}}$ itself. According to cf. Eq. 4.13, this variance may be further decomposed into three different contributions, which are the variation in the collection efficiency $\eta_{\text{ph,col}}$ of the scintillation photons at the photocathode, the variation in the conversion of the scintillation photons into photoelectrons at the photocathode $\eta_{\text{spec}}$ as well as variations in the photoelectron collection efficiency at the first dynode $\eta_{\text{pe,col}}$. As this term

45 *Non-uniformities may include inhomogeneous distribution of (intentional) activators or (unintentional) impurities as well as defects in the structure of the scintillator such as vacancy or interstitial defects in the crystal lattice [359, 370, 371].*

is inversely proportional to $\eta_{\mathrm{PMT}}^2$, it can be reduced by improving the overall photomultiplier transfer efficiency $\eta_{\mathrm{PMT}}$, e.g. by improving the light collection efficiency or the spectral match between the scintillator and the PMT (cf. Section 4.2), among others. On the other hand, one can reduce the variance in $\eta_{\mathrm{PMT}}$ by reducing for example inhomogeneities in the light coupling or photocathode material. In modern scintillation spectrometers using PMTs, this term is often negligible compared to the other terms [359].

The fourth term $R_{E,\mathrm{pmt}}^2 = \mathrm{var}(G_{\mathrm{PMT}})/(G_{\mathrm{PMT}}^2 N_{\mathrm{sci}} \eta_{\mathrm{PMT}})$ quantifies the inherent statistical variation in the photomultiplier total gain $G_{\mathrm{PMT}}$ due to fluctuations in the supply voltage [368]. It is a function of the individual design of the PMT and is typically in the range of 10 % to 20 % [307, 374].

The last term $R_{E,\mathrm{elec}}^2$ was only included for completeness in Eqs. 4.31b and 4.32. This term quantifies the contribution of the detector electronics to the spectral resolution $R_E$. It is in most cases negligible for modern scintillator spectrometers [30, 296, 369].

To estimate the limit in the spectral resolution $R_E$ for an ideal scintillator spectrometer with a proportional scintillation response and a homogeneous crystal lattice, we can simplify Eq. 4.32 further by assuming $R_{E,\mathrm{intr}}^2 = 0$, $R_{E,\mathrm{trans}}^2 = 0$, $R_{E,\mathrm{elec}}^2 = 0$ resulting in [298, 307, 312, 359, 368, 370, 375]:

$$R_E = 2\sqrt{2\log 2}\sqrt{\frac{1 + \mathrm{var}\left(G_{\mathrm{PMT}}\right)/G_{\mathrm{PMT}}^2}{N_{\mathrm{sci}}\eta_{\mathrm{PMT}}}} \tag{4.34}$$

In Fig. 4.7, the ultimate statistical limit in the spectral resolution $R_E$ for an ideal detector with $\mathrm{var}(G_{\mathrm{PMT}})/G_{\mathrm{PMT}}^2 = 0$ and $\eta_{\mathrm{PMT}} = 1$ is shown alongside a more realistic limit for a state-of-the-art detector system with $\mathrm{var}(G_{\mathrm{PMT}})/G_{\mathrm{PMT}}^2 = 0.1$ and $\eta_{\mathrm{PMT}} = 0.3$ as a function of the number of emitted scintillation photons $N_{\mathrm{sci}}$ using Eq. 4.34. In the same graph, the empirically determined overall spectral resolution for selected inorganic scintillator materials is shown for a reference photon energy of 662 keV (cf. Table 4.1). There are two interesting trends to observe in this graph. First, as expected, the spectral resolution $R_E$ of the plotted scintillator materials tends to improve with increasing number of emitted scintillation photons $N_{\mathrm{sci}}$, i.e. absolute light yield. Second, by comparing the difference between the empirically observed spectral resolution and the resolution limits according to Eq. 4.34, there is a clear trend that the difference increases for scintillators with a high degree of non-proportionality to photons $\sigma_{\mathrm{nPR}}$.

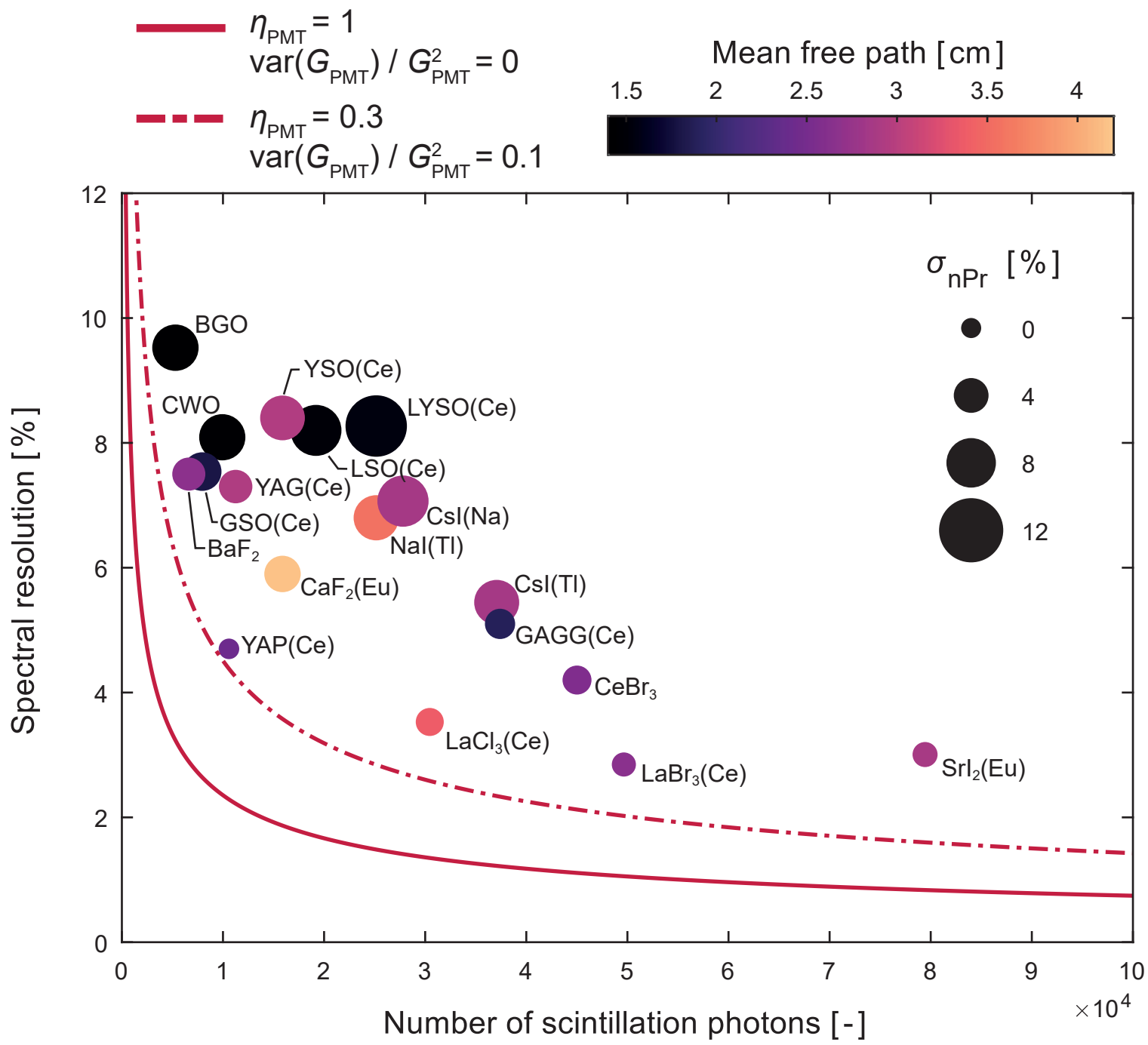


**Figure 4.7** Scintillator properties. Spectral resolution $R_E$ as a function of the number of emitted scintillation photons $N_{sci}$ at a reference photon energy of 662 keV for selected inorganic scintillator materials. Color and size-coding were used for the mean free path $\ell$ at 662 keV and the degree of non-proportionality to photons $\sigma_{nPR}$, respectively. Spectral resolution limits according to Eq. 4.34 are shown as straight and dashed lines for different combinations of the photomultiplier transfer efficiency $\eta_{PMT}$ and the relative variance of the photomultiplier total gain $var(G_{PMT})/G^2_{PMT}$. The raw data used for plotting together with detailed references and discussions on the applied data reduction algorithms are included in Table 4.1 at the end of this chapter.

We find that the spectral resolution of most inorganic scintillators is far from the ultimate statistical limit. However, those scintillator materials with low degree of non-proportionality to photons $\sigma_{\mathrm{nPR}}$ are closer to the limit. This indicates that the intrinsic resolution is a significant factor limiting the spectral resolution of most inorganic scintillators, as we have discussed already above.

### 4.3.3 Spectral Calibration

After discussing the spectral features and the spectral resolution, let us come back to the question I posed at the beginning of this section: How can we relate the pulse-height $\mathcal{H}$ to the deposited energy $E_{\mathrm{dep}}$ of the incident radiation in a given pulse-height spectrum?

#### 4.3.3.1 Energy Calibration

The answer to this question is energy calibration. As we have learned in the Section 4.3.1, FEPs are the result of scintillation events which deposit the full energy of the incident gamma ray into the scintillator. Consequently, the centroid of the FEPs $\mu_{\mathcal{H}}$, is equivalent to the gamma-ray energy $E_{\gamma}$. So, by determining the centroid of a series of FEPs in a pulse-height spectrum for known monoenergetic photon sources, a functional relationship can be derived between the pulse-height $\mathcal{H}$ and the corresponding deposited energy $E_{\mathrm{dep}}$ using regression analysis [365]:

$$E_{\mathrm{dep}} = f\,(\mathcal{H}) \tag{4.35}$$

where:

| | | |
|---|---|---|
| $E_{\mathrm{dep}}$ | deposited energy | eV |
| $\mathcal{H}$ | pulse-height | V |

The function $f$ in Eq. 4.35 may be an arbitrarily complex parametric or non-parametric function [365]. From our discussion in Section 4.1.3, we expect $f$ to be non-proportional. That said, modern scintillation spectrometers often provide automatic spectrum linearization to simplify the energy calibration in Eq. 4.35 and to allow the combination of multiple spectra from different scintillation spectrometers. In this linearization, the pulse-height channel width $\Delta\mathcal{H}$ is not held constant but varied in order to obtain a constant energy width $\Delta E'$ for each pulse-height channel $n$. For such detectors, the functional

relationship between the pulse-height channel $\tilde{n}$ and the energy associated with this channel, which I will denote in this work as the spectral energy $E'$, is then given by a simple polynomial of degree one [10, 365]:

$$E'(\tilde{n}) = E'_0 + \Delta E' \tilde{n} \tag{4.36}$$

where:

| | | |
|---|---|---|
| $E'_0$ | spectral energy offset | eV |
| $\Delta E'$ | spectral energy bin width | eV |
| $\tilde{n}$ | continuous pulse-height channel number | |

The centroid of the full energy peak $\mu_{\mathcal{H}}$ can be determined by another regression analysis[46] of given singlet or multiplet FEPs in measured pulse-height spectra [30, 296, 365]. For inorganic scintillator spectrometers, a simple Gaussian model approach is often sufficient to characterize $\mu_{\mathcal{H}}$ for a fixed number of $N_{\text{FEP}}$ FEPs in a given pulse-height spectrum [10, 30, 365]:

46 *This regression analysis is often termed peak fitting [30].*

$$C(\tilde{n}) = C_{\text{b}}(\tilde{n}) + \sum_{i=1}^{N_{\text{FEP}}} a_i \mathrm{e}^{-\frac{\left(\tilde{n} - \mu_{\mathcal{H},i}\right)^2}{2\sigma^2_{\mathcal{H},i}}} \tag{4.37}$$

where:

| | |
|---|---|
| $a_i$ | normalization parameter |
| $C$ | number of counts |
| $C_{\text{b}}$ | baseline counts |
| $N_{\text{FEP}}$ | number of full energy peaks |
| $\mu_{\mathcal{H}}$ | mean of the FEP |
| $\sigma_{\mathcal{H}}$ | standard deviation of the FEP |

To estimate the baseline $C_{\text{b}}$, simple parametric models [30, 297, 376–378] or more sophisticated non-parametric models [379–385] can be used.

#### 4.3.3.2 Resolution Calibration

A regression analysis based on the Gaussian model in Eq. 4.37 provides not only an estimate of the centroids $\mu_{\mathcal{H}}$ but at the same time quantifies also the dispersion of the FEPs with the standard deviations $\sigma_{\mathcal{H}}$. This dispersion data together with the centroid data can

47 *For Gaussian distributions, we can compute* FWHM *in Eq.* 4.30 *analytically as* $\text{FWHM} = 2\sqrt{2\log 2}\,\sigma_{\mathcal{H}}$ *[30].*

then be used to characterize the spectral resolution of a given inorganic scintillator spectrometer according to Eq. 4.30.[47]

In Section 4.3.2, we have seen that the spectral resolution is a function of the number of scintillation photons $N_{sci}$. Hence, the spectral resolution is also a function of the pulse-height channel $\tilde{n}$. Similar to Eq. 4.35, we can derive a functional relationship $f$ between the spectral resolution $R_E$ and the pulse-height channel $\tilde{n}$ using again regression analysis [365]:

$$R_E = f(\tilde{n}) \tag{4.38}$$

where:

| | |
|---|---|
| $\tilde{n}$ | continuous pulse-height channel number |
| $R_E$ | spectral resolution |

In contrast to the energy calibration, selecting a function $f$ is non-trivial. In principle, the model derived by Breitenberger in Eq. 4.31b may provide the basis to define $f$ in Eq. 4.38. However, in order to use Eq. 4.31b, detailed knowledge of the individual variance terms is required. This is often not the case, in particular the intrinsic term in Eq. 4.31b is not known in most applications. This is why in practice, we use empirical models to describe the relationship between $n$ and $R_E$ in Eq. 4.38 [386]. Considering the dependence of intrinsic resolution on the scintillator type, no single empirical model universally applies to all inorganic scintillators. That said, for NaI(Tl) used in this work, power-law models such as:

$$R_E(\tilde{n}) = a_1 \tilde{n}^{a_2} \tag{4.39}$$

with:

| | |
|---|---|
| $a_1$ | scale coefficient |
| $a_2$ | power coefficient |

have been proven to provide satisfactory results [387–391].

### 4.3.4 Detector Response Modeling

In the previous sections, we have discussed in detail how high-energy photons interact with inorganic scintillators and how the scintillation light generated in these interactions can be used to reconstruct the deposited energy of the incident photons in a pulse-height spectrum.

We can describe this process in a more general way as a convolution between the incoming continuous double differential photon flux $\partial^2\phi_\gamma/\partial E_\gamma \partial \Omega'$ (cf. Eq. 2.20) and the so-called detector response function (DRF) $R$[48] to obtain the count rate $c$ as a function of the spectral energy $E'$ and time $t$ [30, 365]:[49]

$$c(E',t) = \int_0^\infty \int_0^{4\pi} R\left(E', E_\gamma, \mathbf{\Omega}', t\right) \frac{\partial^2 \phi_\gamma}{\partial E_\gamma \partial \Omega'} \left(E_\gamma, \mathbf{\Omega}', t\right) \mathrm{d}\Omega' \mathrm{d}E_\gamma \quad (4.40)$$

where:

| | | |
|---|---|---|
| $E'$ | spectral energy | eV |
| $E_\gamma$ | photon energy | eV |
| $R$ | detector response function | $\mathrm{m}^2$ |
| $\Omega'$ | solid angle in the detector frame | sr |
| $\mathbf{\Omega}'$ | direction unit vector in the detector frame | |
| $\frac{\partial^2 \phi_\gamma}{\partial E_\gamma \partial \Omega'}$ | double differential photon flux | $\mathrm{s}^{-1}\,\mathrm{m}^{-2}\,\mathrm{eV}^{-1}\,\mathrm{sr}^{-1}$ |

The DRF $R$ in Eq. 4.40 characterizes the response of the detector system to the incoming photons, i.e. it predicts the number of pulses per time in a pulse-height channel with associated spectral energy $E'$ for a monoenergetic monodirectional photon flux incident from a direction $\mathbf{\Omega}'$ (measured in the detector reference frame) with photon energy $E_\gamma$[50]. It has the dimensional units of a geometric area and can therefore be interpreted similarly to the microscopic cross-section in Section 3.1.1 as an effective cross-sectional area.

The number of pulses $C$ introduced in Eq. 4.16 as a function of the spectral energy $E'$ in a pulse-height spectrum is then obtained by integrating Eq. 4.40 over a given measurement time interval $t_1 \leq t \leq t_2$:

$$C(E') = \int_{t_1}^{t_2} c(E',t)\,\mathrm{d}t \quad (4.41)$$

For Eqs. 4.40 and 4.41 to be valid, we made two important assumptions:

48 *Sometimes also termed as the instrument response function [392], the efficiency-area product [393] or area-efficiency response [394].*

49 *Note that the angular dependence of the detector response function (DRF) is often neglected in the literature [30, 365, 395–398]. As we will see later in this book, for anisotropic DRFs, it is crucial to account for the angular dependence. I have therefore extended the convolution integral here to account also for angular dependence in both, the DRF and the photon flux. Similar approaches have been used in astrophysical applications [392–394]. For completeness, I have also included the time dependence in both, the detector response and the double differential flux, to account for temporal changes in these quantities.*

50 *Note that we could write Eq. 4.40 also as a function of the continuous pulse-height channel number $\tilde{n}$ or the pulse-height $\mathcal{H}$ instead of the spectral energy $E'$ (cf. Eq. 4.36). Moreover, we could conduct the convolution also in a global static reference frame with directions $\mathbf{\Omega}$ and solid angle $\Omega$. The selection of the specific set of variables and frame of reference mainly depends on the specific application.*

A. Similar to the assumption D for the radiation transport modeling in Section 3.2, we assume that the DRF $R$ is independent of the double differential photon flux $\partial^2\phi_\gamma/\partial E_\gamma \partial\Omega'$. For this to hold, we again need to introduce the two additional assumptions D.i. and D.ii. already discussed in Section 3.2. For your convenience, I repeat them here briefly:
   i. Photon-photon interactions must be negligible compared to other interaction modes.
   ii. High-energy photons do not significantly alter the composition and structure of the matter within a time frame comparable to the transport of individual photons.

   This first assumption is crucial as it ensures the linearity of the convolution equation in Eq. 4.40. This allows us to compute the count rate as the product of the DRF and the photon flux. The resulting linear homogeneous integral equation in Eq. 4.40 is also known as the Fredholm integral equation of the first kind [399, 400].[51]

B. We assume that the double differential photon flux $\partial^2\phi_\gamma/\partial E_\gamma \partial\Omega'$ in Eq. 4.40 is homogeneous within the finite extent of the detector volume. This allows us to neglect the position dependence $\mathbf{r}$ of the photon flux (cf. Eq. 2.20) [30, 365, 392–394].

51 *Erik Ivar Fredholm (*1866, †1927), a Swedish mathematician. Fredholm is best known for his influential work on integral equations, which laid the groundwork for what is today known as the Fredholm theory [401]. Both a lunar crater (18.4°N, 4.65°E) and a small asteroid in the asteroid belt ((21659) Fredholm) are named in his honor.*

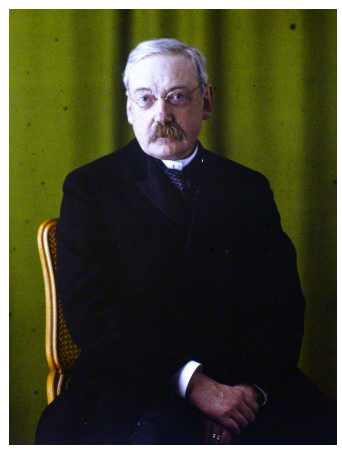

Erik Fredholm
Musée Départemental Albert-Kahn

Both, the DRF and the double differential photon flux in Eq. 4.40 are continuous functions. In most cases, we are not able to predict these functions analytically [30]. Instead, as we have seen in Section 3.2, we can use deterministic or Monte Carlo methods to estimate the double differential flux. The DRF $R$ can be estimated using radiation measurements or Monte Carlo simulations as we will see later in this book [30, 365]. Consequently, we have two main approaches to solve the detector response problem described by Eq. 4.40:

I. **Full Simulation** We can adopt a full simulation approach by performing either a single Monte Carlo simulation of the entire radiation transport and detector response process or alternatively a hybrid simulation, where deterministic simulations of the double differential photon flux are used as input to a Monte Carlo simulation (cf. Section 3.2). In both cases, an estimate for the count rate $c$ in a finite number of pulse-height channels is obtained directly from the simulations, without the need to solve any integral equations.

**II. Numerical Integration** Instead of a full simulation approach, we can derive the DRF $R$ and the double differential photon flux $\partial^2\phi_\gamma/\partial E_\gamma \partial\Omega'$ separately as indicated above. We can then solve the integral equation in Eq. 4.40 numerically using numerical integration techniques, e.g. numerical quadrature [400].

I will discuss both approaches in more detail later in this book. The fundamental principles of deterministic and Monte Carlo simulations were already introduced in Section 3.2. Therefore, in the remainder of this section, I will briefly outline the key aspects of the numerical integration approach.

Following a numerical quadrature approach, we evaluate the DRF $R$ on a discrete set of spectral energies $\{E'_i \mid i \in \mathbb{N}_+, i \leq N_{\text{ch}}\}$ corresponding to the individual pulse-height channels $n$, a series of photon energies $\{E_{\gamma,j} \mid j \in \mathbb{N}_+, j \leq N_{E_\gamma}\}$, a series of directions $\{\boldsymbol{\Omega}'_k \mid k \in \mathbb{N}_+, k \leq N_{\boldsymbol{\Omega}'}\}$ as well as a series of time instances $\{t_l \mid l \in \mathbb{N}_+, l \leq N_t\}$ with $N_{\text{ch}}$, $N_{E_\gamma}$, $N_{\boldsymbol{\Omega}'}$ and $N_t$ being the number of pulse-height channels, number of evaluated photon energies, number of evaluated directions and number of evaluated time instances, respectively. At the same time, we evaluate the double differential photon flux $\partial^2\phi_\gamma/\partial E_\gamma \partial\Omega'$ on the same discrete set of photon energies, directions and time instances as the DRF $R$. Using a general binning scheme with variable bin widths, we can rewrite Eq. 4.40 in its discretized form as follows [30, 365]:[52]

52 *Please note that I have again extended the discretized convolution equation here to account for angular dependence in both, the DRF and the photon flux. Similar approaches have been used in astrophysical applications [393].*

$$c(E'_i, t_l) \approx \sum_{j=1}^{N_{E_\gamma}} \sum_{k=1}^{N_{\boldsymbol{\Omega}'}} R\left(E'_i, E_{\gamma,j}, \boldsymbol{\Omega}'_k, t_l\right) \frac{\partial^2\phi_\gamma}{\partial E_\gamma \partial\Omega'}\left(E_{\gamma,j}, \boldsymbol{\Omega}'_k, t_l\right) \Delta E_{\gamma,j} \Delta\Omega'_k \quad (4.42)$$

where:

| | | |
|---|---|---|
| $N_{E_\gamma}$ | number of evaluated photon energies | |
| $\Delta E_{\gamma,j}$ | photon energy bin width | eV |
| $N_{\boldsymbol{\Omega}'}$ | number of evaluated directions | |
| $\Delta\Omega'_k$ | solid angle bin width | sr |

We can further simplify Eq. 4.42 by appling matrix notation with the count rate vector $\mathbf{c} \in \mathbb{R}_+^{N_{\text{ch}} \times 1}$ and a series of $N_{\boldsymbol{\Omega}'}$ different detector response matrices $\mathbf{R} \in \mathbb{R}_+^{N_{\text{ch}} \times N_{E_\gamma}}$ and photon flux vectors $\boldsymbol{\phi}_\gamma \in \mathbb{R}_+^{N_{E_\gamma} \times 1}$ [30, 365]:

$$\mathbf{c}\left(t_l\right) \approx \sum_{k=1}^{N_{\Omega'}} \mathbf{R}\left(\boldsymbol{\Omega}'_k, t_l\right) \boldsymbol{\phi}_\gamma\left(\boldsymbol{\Omega}'_k, t_l\right) \tag{4.43}$$

where:

| | | |
|---|---|---|
| $\mathbf{R}$ | detector response matrix | $\mathrm{m}^2$ |
| $\boldsymbol{\phi}_\gamma$ | photon flux vector | $\mathrm{s}^{-1}\,\mathrm{m}^{-2}$ |

Note that the individual elements of the photon flux vector are computed by multiplying the double differential photon flux elements with the corresponding energy and solid angle bin widths, i.e. $\phi_\gamma(E_{\gamma,j}, \boldsymbol{\Omega}'_k) = \partial^2\phi_\gamma/\partial E_\gamma \partial\Omega'(E_{\gamma,j}, \boldsymbol{\Omega}'_k)\Delta E_{\gamma,j}\Delta\Omega'_k$. The number of counts per pulse-height channel $\mathbf{C} \in \mathbb{N}^{N_{\mathrm{ch}}\times 1}$ can then be computed by summing simply over all evaluated time instances:

$$\mathbf{C} \approx \sum_{l=1}^{N_t} \sum_{k=1}^{N_{\Omega'}} \mathbf{R}\left(\boldsymbol{\Omega}'_k, t_l\right) \boldsymbol{\phi}_\gamma\left(\boldsymbol{\Omega}'_k, t_l\right) \Delta t_l \tag{4.44}$$

with:

| | | |
|---|---|---|
| $N_t$ | number of time bins | |
| $\Delta t_l$ | time bin width | s |

The matrix versions in Eqs. 4.43 and 4.44 are particularly useful for numerical computations as they allow the parallel evaluation of the individual elements of the count and count rate vector $\mathbf{C}$ and $\mathbf{c}$, respectively.[53]

This concludes this chapter on gamma-ray spectrometry. In the next chapter, we will discuss the application of gamma-ray spectrometry in the context of AGRS. We will start with a brief overview of the history of AGRS and the current state of the art in the field. I will then continue with a discussion of current AGRS detector systems with a special focus on the Swiss AGRS system. In the last part, I will analyze key challenges in AGRS and how advanced simulation tools such as Monte Carlo simulations can help to address these challenges.

53 *Modern numerical codes such as* `Python/NumPy` *with* `numpy.matmul` *or MATLAB with* `pagemtimes` *provide efficient algorithms for such multi-dimensional matrix multiplication operations.*

**Table 4.1** Properties of common inorganic scintillators used in gamma-ray spectrometry.

| Material★ | $\rho$ [g cm$^{-3}$]● | $\tau_{sci,d}$ [ns]○ | $\lambda_{sci,max}$ [nm]† | $Y_{sci,a}$ [keV$^{-1}$]‡ | $R_E$ [%]§ | $\ell$ [cm]$ | $\sigma_{nPR}$ [%]# | Cost [USD cm$^{-3}$]●● | References |
|---|---|---|---|---|---|---|---|---|---|
| | | | | | **I. Iodides** | | | | |
| NaI(Tl) | 3.67 | 230 | 415 | 38–45 | 6.5–7.1 | 3.55 | 6.83 | 4–8 | [347, 358, 370, 371, 373, 410–413] |
| CsI(Tl) | 4.51 | 600/3400 | 550 | 57–65 | 4.3–6.6○○ | 2.86 | 6.78 | 7–12 | [30, 343, 358, 370, 371, 373, 408, 411, 414] |
| CsI(Na) | 4.51 | 630 | 420 | 39–49 | 6.7–7.4 | 2.86 | 8.46 | 7–10 | [358, 370, 371, 410, 411, 415] |
| $SrI_2$(Eu)†† | 4.59 | 1200 | 435 | 68–120 | 2.8–3.2 | 2.88 | 1.35 | 150–300 | [313, 348, 358, 371, 402, 416–418] |
| | | | | | **II. Bromides** | | | | |
| $CeBr_3$ | 5.20 | 17 | 370 | 43–68 | 3.7–4.3‡‡ | 2.57 | 2.42 | 140–170 | [325, 358, 371, 419, 420] |
| $LaBr_3$(Ce) | 5.08 | 16 | 380 | 48–75 | 2.7–3.3 | 2.66 | 1.10 | 135–175 | [358, 371, 373, 375, 386, 421–424] |
| | | | | | **III. Oxides** | | | | |
| BGO | 7.13 | 300 | 480 | 8–9 | 9.05–10.0 | 1.40 | 7.15 | 13–30 | [30, 358, 370, 371, 411, 425] |
| CWO | 7.90 | 5000/20 000 | 540/470 | 15–28 | 6.8–9.5 | 1.45 | 7.11 | 13 | [358, 370, 371, 411, 426, 427] |
| GAGG(Ce)§§ | 6.63 | 121/510 | 520/550 | 32–57 | 5.1–6.1○○ | 1.89 | 2.38–3.01$$ | 200 | [371, 405–407, 409, 428, 429] |
| GSO(Ce) | 6.71 | 56/600 | 440 | 9–13 | 6.7–8.1 | 1.79 | 5.03 | 400 | [346, 358, 371, 411, 430–432] |
| LSO(Ce) | 7.40 | 40 | 420 | 23–30 | 7.9–8.5 | 1.50 | 8.40 | 70 | [355, 358, 370, 371, 422, 433–435] |
| LYSO(Ce) | 7.10 | 40 | 420 | 32 | 7.84–8.7 | 1.54 | 11.24 | 70 | [371, 436, 437] |
| YAG(Ce) | 4.57 | 70/300 | 550 | 14–20 | 7.0–7.6 | 2.93 | 3.64 | 125 | [358, 371, 436, 438, 439] |
| YAP(Ce) | 5.55 | 27 | 350 | 13–18 | 4.4–5.0 | 2.42 | 0.13 | 125 | [349, 358, 371, 440, 441] |
| YSO(Ce) | 4.55 | 37/82 | 420 | 23–45 | 8.3–8.5 | 2.95 | 6.71 | 40 | [349, 358, 371, 434, 442] |

*Continued on next page*

**Table 4.1** *Continued from previous page*

| Material★ | $\rho$ [g cm$^{-3}$]● | $\tau_{\mathrm{sci,d}}$ [ns]○ | $\lambda_{\mathrm{sci,max}}$ [nm]† | $Y_{\mathrm{sci,a}}$ [keV$^{-1}$]‡ | $R_E$ [%]§ | $\ell$ [cm]$ | $\sigma_{\mathrm{nPR}}$ [%]# | Cost [USD cm$^{-3}$]●● | References |
|---|---|---|---|---|---|---|---|---|---|
| | | | | | **IV. Chlorides & Fluorides** | | | | |
| $BaF_2$ | 4.88 | 0.8/630 | 220/315 | 1.4/10–11 | 7–8 | 2.67 | 3.54 | 45 | [358, 370, 371, 443] |
| $CaF_2$(Eu) | 3.18 | 840 | 435 | 24 | 5.7–6.1 | 4.16 | 4.45 | 10–35 | [355, 358, 370, 371, 410, 433, 444] |
| $LaCl_3$(Ce) | 3.85 | 28 | 350 | 40–49 | 3.0–4.2 | 3.37 | 2.16 | 135–175 | [356, 358, 371, 373, 445–449] |

★ For the sake of brevity, if available, common scintillator abbreviations are listed. The corresponding chemical formulas are as follows: BGO: $Bi_4Ge_3O_{12}$, CWO: $CdWO_4$, GAGG(Ce): $Gd_3Al_2Ga_3O_{12}$(Ce), GSO(Ce): $Gd_2SiO_5$(Ce), LSO(Ce): $Lu_2SiO_5$(Ce), LYSO(Ce): $Lu_{1.8}Y_{0.2}SiO_2$(Ce), YAG(Ce): $Y_3Al_5O_{12}$(Ce), YAP(Ce): $YAlO_3$(Ce), YSO(Ce): $Y_2SiO_5$(Ce).
● Scintillator mass density.
○ Scintillator decay time constant (cf. Eq. 4.1).
† Scintillator peak emission wavelength (cf. Section 4.2).
‡ Absolute light yield for high-energy photons with $E_\gamma = 662$ keV (cf. Eq. 4.2).
§ Spectral resolution at a spectral energy $E'$ of 662 keV (cf. Eq. 4.30).
$ Mean free path at 662 keV using the XCOM database by the National Institute of Standards and Technology (NIST).
# Degree of non-proportionality to photons (cf. Eq. 4.11).
●● Cost estimates are adopted from Bell [371] and are based on a standard cylindrical crystal size of ⌀2.54 × 2.54 cm as of 2018. The cost for CSO(Ce) is based on an informal inquiry to Advatech UK Limited.
†† Scintillation properties of $SrI_2$(Eu) strongly depend on the temperature and dopant concentration, among other factors [402]. The listed values are for a reference dopant concentration of 5 % at room temperature.
‡‡ Quarati et al. [403] and Guss et al. [404] reported an improvement in the spectral resolution by adopting co-doping achieving $R_E \approx 3$ %.
§§ Because GAGG(Ce) scintillators are still in a comparably early stage in their development, crystal growth and packaging are not yet sufficiently standardized to ensure reproducibility in the scintillation properties [405–407].
○○ Due to the comparably large scintillator peak emission wavelength $\lambda_{\mathrm{sci,max}}$ of ≈520 nm and ≈550 nm, GAGG(Ce) and CsI(Tl) scintillators are ideal candidates to match with SiPMs. Using SiPMs instead of PMTs results in an absolute reduction of the spectral resolution by ≈1 % at 662 keV [405, 408].
$$ Because the degree of non-proportionality to photons $\sigma_{\mathrm{nPR}}$ for GAGG(Ce) is not readily available from the literature, $\sigma_{\mathrm{nPR}}$ was estimated based on experimental results obtained by Campana et al. [409].

*"Flying is for droids."*

*— Obi-Wan Kenobi, Revenge of the Sith*

# Chapter 5
# Airborne Gamma-Ray Spectrometry

**Contents**

After covering the fundamentals of gamma-ray sources, their interaction with matter, and the principles of gamma-ray spectrometry, this final chapter of the first part offers a concise overview of the historical evolution and current status of AGRS worldwide. Subsequently, a detailed review of the Swiss AGRS system, which constitutes the primary focus of this book, will be presented.

In the second part, I will discuss one of the core tasks of AGRS, the quantification of gamma-ray sources. For this purpose, I will formally analyze the quantification problem in AGRS as an ill-posed linear discrete inverse problem after Hadamard alongside a comprehensive review of the current standard methodologies to solve this challenging problem. This will include a detailed discussion of standard calibration techniques to derive the required model parameters for the individual quantification methodologies.

The chapter will conclude with a comprehensive analysis of the current limitations in the quantification and calibration methodologies and outline alternative approaches to overcome these limitations motivating the scope of this work.

## 5.1 A Brief History of AGRS

### 5.1.1 The Geophysical Roots

The success story of AGRS started as early as 1949 when Pringle et al.[1] mounted for the first time a gamma-ray spectrometer consisting of a small NaI(Tl) scintillator with a mass of ~100 g coupled to a PMT into an aircraft to localize and map uranium deposits in Canada [450, 452]. These first AGRS flights took place less than a year after Robert Hofstadter developed the NaI(Tl) scintillator in 1948 (cf. Section 4.1) [306]. After the successful demonstration by Pringle et al., AGRS established itself over the following years as a powerful tool for geological mapping and the exploration of raw resources. Compared to the prototype developed by Pringle et al., these standardized systems used significantly larger NaI(Tl) scintillator crystals with masses >1 kg to increase sensitivity to the natural

1 *Robert William Pringle (*1920, †1996) was a Scottish experimental physicist, father of AGRS and a passionate rugby player [451].*

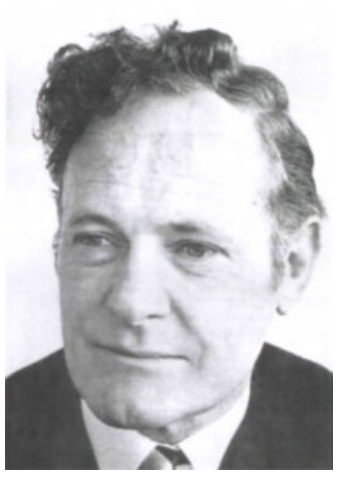

Robert Pringle
© Reproduced with permission from *Physics in Canada*, Vol. 52 (6), p. 303 (1996)

terrestrial gamma-ray sources of interest, i.e. ${}^{40}_{19}$K, ${}^{232}_{90}$Th and ${}^{238}_{92}$U with related progeny radionuclides as discussed in Section 2.1.3 [453–459]. Today, AGRS systems are routinely used in geophysical surveys by both, governmental and non-governmental organizations across the globe [8, 9]. Apart from geological mapping and raw resource exploration[2], AGRS has been successfully applied in a variety of other fields associated with the mapping of natural radionuclides, i.e. soil and peat mapping [469–476], soil moisture quantification [477, 478], landslide mapping [479, 480], snow cover quantification [481–487], radon risk mapping [488–499], forest management [500, 501] and high-precision agriculture [502–504], among others.

2 *It is worth adding that AGRS can not only be used to explore natural uranium, thorium or potassium deposits but also various raw resources associated with these elements such as zinc, sulfides, salt, gold, copper, rare earth elements or petroleum and gas deposits, among others [8, 52, 460–468].*

### 5.1.2 Crashing Satellites and Lost Missiles

Shortly after its establishment in geophysics at the beginning of the 1950s, the potential of AGRS to quantify not only natural but also anthropogenic radionuclides was recognized. This extended the application range of AGRS to nuclear safety and security applications. Within this domain, applications are broadly categorized into two main areas: emergency response, on the one hand, and routine surveys of nuclear facilities and their surrounding areas, on the other hand.

Surveys of nuclear facilities with AGRS systems began shortly after the first commercial nuclear power plants started their operation in the U.S. in the late 1950s [505–508] and have continued in various countries to the present day [118–120, 508–536]. These surveys are typically conducted to monitor the environmental impact of nuclear facilities and to verify the compliance with environmental regulations. In addition to nuclear facilities, AGRS systems have also been used to survey specific areas with known or suspected long-term anthropogenic radionuclide contamination, such as nuclear weapon test sites or regions affected by nuclear accidents [118, 537–542].

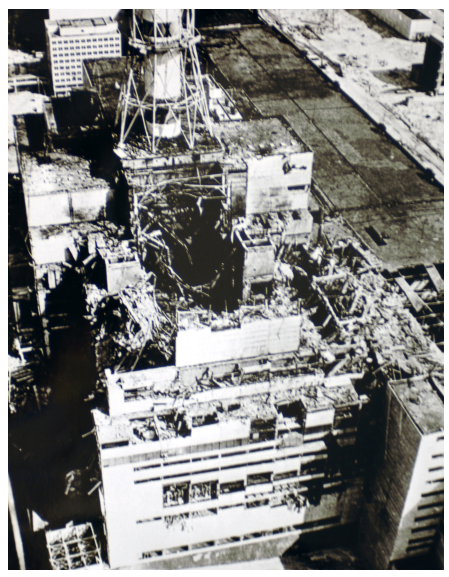

Chernobyl Unit 4 after the accident


In addition to these routine surveys, AGRS systems have been successfully deployed in several nuclear emergency response scenarios in the past. One of the first successful deployments of AGRS systems in such a scenario was at the Windscale nuclear accident in the U.K. in 1957 [543]. Over the following decades, AGRS has proven to be an essential tool in nuclear emergency response with its unique characteristics to rapidly identify and quantify radionuclide contaminations in severe nuclear accidents over large areas [8, 9]. As such it has become an integral part of the emergency response

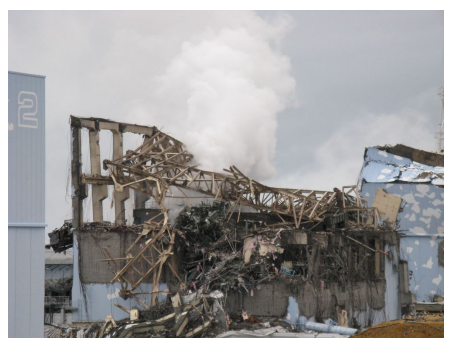

Fukushima Daiichi Unit 3 after the accident

toolset of many countries, providing decision-makers with the necessary information to make informed decisions regarding radiation protection measures for the public in case of a nuclear emergency [8, 9, 544, 545]. In addition to the Windscale accident, AGRS systems have been instrumental in most major radiological incidents in the 20th and 21st century. These incidents include the severe nuclear accidents at the Three Mile Island Nuclear Generating Station (Unit 2) in the U.S. in 1979 [546], the Chernobyl Nuclear Power Station (Unit 4) in Ukraine in 1986 [547–549] and the Fukushima Daiichi Nuclear Power Station (Units 1, 2 and 3) in Japan in 2011 [6, 28, 550] as well as the radiological incident in Goiânia in Brazil in 1987 [551, 552]. In some of these accidents, AGRS systems were not only used to quantify surface contamination but also airborne radionuclides released into the Earth's atmosphere [24, 553–555].

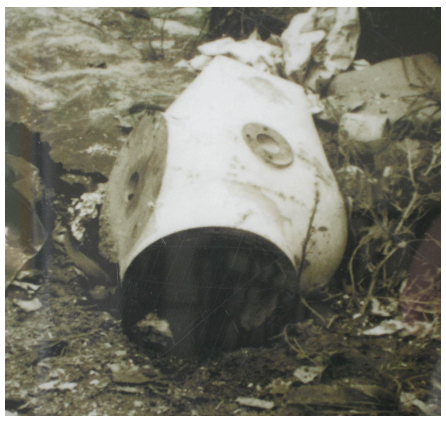

${}^{137}_{55}$Cs source involved in the Goiânia incident (cropped)


A second important application of AGRS in radiological emergency response was demonstrated shortly after the Windscale nuclear accident in 1957. In the summer of 1968, a ${}^{60}_{27}$Co point source with an activity of ~12 GBq was lost during a truck transport along the ~1800 km route between Salt Lake City and Kansas City in the U.S. [556]. To localize the lost source, an AGRS system was deployed and within two days, the ${}^{60}_{27}$Co source could be successfully located and retrieved [556]. This incident was one of the first reported applications of AGRS to locate lost or stolen radioactive point sources. Since then, AGRS systems have been effectively used in several orphan source incidents. A famous example is the retrieval of the highly radioactive remains of the nuclear reactor core of the Soviet Cosmos 954 reconnaissance satellite, which reentered Earth's atmosphere over the subarctic tundra in Canada in 1978 and subsequently dispersed radioactive materials over a large area ~24 000 km$^2$ [557–560]. Another example, where AGRS was successfully applied in an orphan source scenario, was the Athena radiological incident. In this incident, two ${}^{57}_{27}$Co sources with a total activity of ~35 GBq mounted in an Athena missile were lost north of Torreon, Mexico during a missile test conducted by the U.S. Air Force from Green River, Utah in 1972 [561]. In a span of two days, the ${}^{57}_{27}$Co sources were successfully located and retrieved within the 60 km$^2$ search area, employing once again an AGRS system [561].

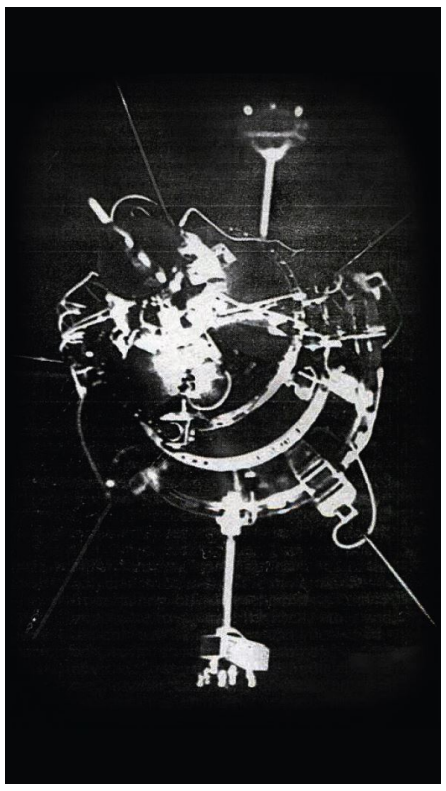

Cosmos 954 satellite


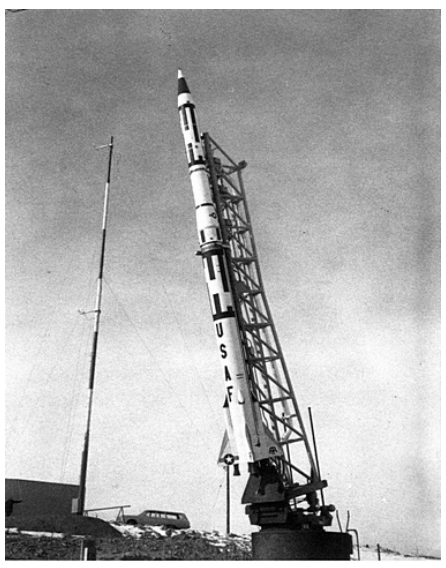



Athena missile


### 5.1.3 The Advent of UAV-based AGRS Systems

So far, the discussed AGRS systems were all operated using crewed aircraft. With the advent of unmanned aerial vehicles (UAVs) at the

beginning of the 21st century, we have also seen the development of AGRS systems mounted on UAVs. First unmanned AGRS prototypes were developed already since 1993 [562–568]. In the aftermath of the nuclear accident at the Fukushima Daiichi Nuclear Power Station, in addition to the established manned systems, unmanned AGRS systems were successfully deployed for the first time in a nuclear emergency response scenario [569–573]. The successful demonstration at Fukushima prompted new research efforts in the field of unmanned AGRS systems, a field that continues to evolve and innovate to this day [504, 574–603].

In general, UAV-based AGRS systems tend to possess a shorter endurance and smaller payload capacities compared to more traditional manned AGRS systems [571, 604–606]. Consequently, manned and unmanned systems are not mutually exclusive but rather complement each other in their operational capabilities, i.e. manned systems can provide the rapid coverage of large areas from $\mathcal{O}(10^2)\,\mathrm{m}^2$ up to $\mathcal{O}(10^8)\,\mathrm{m}^2$, while unmanned systems are a cost-effective way to improve the spatial resolution in mapping, localization and identification in smaller regions of interest with areas between $\mathcal{O}(10^1)\,\mathrm{m}^2$ and $\mathcal{O}(10^4)\,\mathrm{m}^2$ [536].

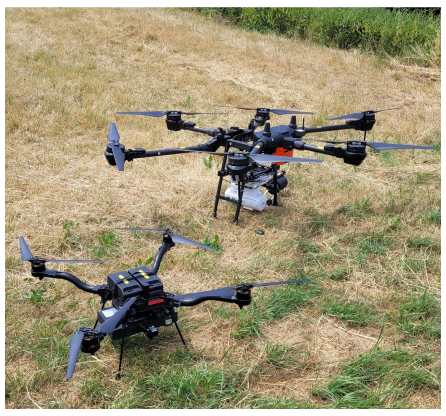

UAV-based AGRS systems during the ARM23 exercise [536]
© David Breitenmoser

## 5.2 Current Status

After this short review about the origins and evolution of AGRS, let us have now a closer look at the technical parameters of AGRS systems currently operational as of writing this book. Because of the broad range of applications, AGRS systems are quite diverse in terms of detector characteristics, adopted aircraft systems and overall operational capabilities. Given the scope of this work, I limit the discussion here to manned AGRS systems operated by non-commercial organizations such as governmental agencies, research institutions or universities for the localization, identification and quantification of terrestrial gamma-ray sources.

In Table 5.1, I have compiled a list of currently operational manned AGRS systems alongside their main characteristics such as aircraft type, survey flight parameters and secondary sensors other than the primary gamma-ray spectrometer. Details about the adopted primary gamma-ray spectrometer are provided in a separate table, Table 5.2. The majority of the data presented in these tables was collected in a systematic survey conducted during the 11th International Airborne Radiometry Technical Exchange Meeting held on September 18th–22nd 2023 in Spiez, Switzerland.

**Table 5.1** Overview about manned AGRS systems affiliated with non-commercial organizations and operational as of writing this book. The majority of the data presented in this table was collected in a systematic survey conducted during the 11th International Airborne Radiometry Technical Exchange Meeting held on September 18th–22nd 2023 in Spiez, Switzerland. The remainder of the data was retrieved from the literature as indicated in the corresponding data entries. A graphical size comparison of the listed aircraft can be found in Fig. B.19.

| ID★ | Country• | Organization◦ | Aircraft† | Survey flight parameters | | Secondary radiation detectors$ | Secondary sensors# |
|---|---|---|---|---|---|---|---|
| | | | | $h_{air}$ [m]‡ | $v_g$ [m s$^{-1}$]§ | | |
| AT1•• | AT | GSA | B212(H) | 80 | 28 | | APS, ATS, DCAM, GNSSr, HYG, LA, RA |
| CA1 | CA | GSC | AS50(H), B06(H), B412(H) | 15–40 | 25–50 | | GNSSr, IMU, LA |
| CH1 | CH | NBC-EOD, NEOC, PSI | AS32(H) | 90 | 28–42 | GMT | APS, ATS, GNSSr, RA |
| CN1◦◦ | CN | CAGS | AS50(H) | 150 | 42 | | GNSSr, RA |
| CN2†† | CN | CAGRC, CMGB | AN2(F), Y12(F) | 120–200 | 39 | | GNSSr, RA |
| CZ1 | CZ | AČR, SÚRO | B412(H), MI8(H) | 100 | 28 | GMT, HPGe | APS, ATS, GNSSr, HYG, RA |
| DE1 | DE | BfS | EC35(H) | 90 | 33 | GMT, HPGe | GNSSr, RA |
| DE2‡‡ | DE | BGR | S76(H) | 70–100 | 39 | | GNSSr, LA |
| FR1 | FR | IRSN | AS50(H) | 50–150 | 22–33 | GMT | GNSSr |
| FR2§§ | FR | CEA | AS55(H), EC45(H) | 40 | 19 | HPGe | GNSSr,RA |
| IT1 | IT | INFN | Radgyro(G)$$ | 40–200 | 19–39 | | APS, ATS, DCAM, FLP, GNSSr, HYG, IMU, NIRCAM, RA, |
| NO1 | NO | NGU | AS50(H) | 60–80 | 25–33 | HPGe | APS, ATS, RA |
| NO2 | NO | DSA | EH10(H) | 60–90 | 28 | | GNSSr |
| | | | B350(F) | 100–150 | 69 | | GNSSr |

*Continued on next page*

**Table 5.1** *Continued from previous page*

| ID★ | Country• | Organization○ | Aircraft† | Survey flight parameters $h_{air}$ [m]‡ | $v_g$ [m s$^{-1}$]§ | Secondary radiation detectors$ | Secondary sensors# |
|---|---|---|---|---|---|---|---|
| SE1 | SE | SGU | AC50(F), AC90(F) | 60–150 | 69 | HPGe | APS, ATS, GNSSr, RA |
| TW1 | TW | RMC | AS65(H), H60(H) | 150–200 | 36 | | GNSSr |
| US1 | US | NNSA | B412(H) | 15–150 | 36 | HPGe | GNSSr, RA, |
| | | | B350(F) | 300 | 133 | GMT, HPGe | GNSSr, RA |

★ Custom identifier.
• Country code defined by the ISO 3166-1 alpha-2 standard, i.e. AT: Austria, CA: Canada, CN: People's Republic of China, CZ: Czech Republic, FR: France, DE: Germany, IT: Italy, NO: Norway, SE: Sweden, CH: Switzerland, TW: Taiwan, US: United States of America.
○ AČR: Armáda České republiky (Army of the Czech Republic), BfS: Bundesamt für Strahlenschutz (Federal Office for Radiation Protection), BGR: Bundesanstalt für Geowissenschaften und Rohstoffe (Federal Institute for Geosciences and Natural Resources), CAGRC: China Aero Geophysical Survey & Remote Sensing Center for Land and Resources, CAGS: Chinese Academy of Geological Sciences, CEA: Commissariat à l'énergie atomique et aux énergies alternatives (French Alternative Energies and Atomic Energy Commission), CMGB: China Metallurgical Geology Bureau, DSA: Direktoratet for strålevern og atomsikkerhet (Norwegian Radiation and Nuclear Safety Authority), GSA: GeoSphere Austria, GSC: Geological Survey of Canada, INFN: Istituto Nazionale di Fisica Nucleare (National Institute for Nuclear Physics), IRSN: Institut de Radioprotection et de Sûreté Nucléaire (Radioprotection and Nuclear Safety Institute), NBC-EOD: Nuclear, Biological, Chemical, Explosive Ordnance Disposal and Mine Action Centre of Competence, NEOC: National Emergency Operations Centre, NGU: Norges Geologiske Undersøkelse (Geological Survey of Norway), NNSA: National Nuclear Security Administration, PSI: Paul Scherrer Institute, RMC: Radiation Monitoring Center, SGU: Sveriges Geologiska Undersökning (Geological Survey of Sweden), SÚRO: Státní Ústav Radiační Ochrany (National Radiation Protection Institute).
† Aircraft type designator according to the International Civil Aviation Organization (ICAO) together with a custom designator in parentheses (F: Fixed-wing aircraft, G: Gyrocopter, H: Helicopter).
‡ Ground clearance.
§ Ground speed.
$ Secondary radiation detectors in addition to primary scintillation detector (GMT: Geiger-Müller tube, HPGe: high-purity germanium detector).
# Secondary aircraft sensors accessible to the AGRS system (APS: air pressure sensor, ATS: air temperature sensor, DCAM: downlooking camera, FLP: fuel level probe, GNSSr: global navigation satellite system receiver, HYG: hygrometer, IMU: inertial measurement unit, LA: laser altimeter, NIRCAM: near infrared camera, RA: radar altimeter).
•• At the time of writing, not operational since 2015.
○○ Reference: Zhang et al. [607], Xiong [608], and Liao et al. [609].
†† Reference: Zhang et al. [607], Xiong [608], Xiong et al. [610], and Xu et al. [611].
‡‡ Reference: Siemon et al. [476], [612].
§§ Reference: Bucher et al. [613] and Wilhelm et al. [614].
$$ Experimental gyrocopter based on a Pagotto Brakogyro, cf. Albéri et al. [615].

**Table 5.2** Detailed information on the scintillation spectrometers adopted by the manned AGRS systems listed in Table 5.1. The majority of the data presented in this table was collected in a systematic survey conducted during the 11th International Airborne Radiometry Technical Exchange Meeting held on September 18th–22nd 2023 in Spiez, Switzerland. The remainder of the data was retrieved from the literature as indicated in the corresponding data entries. For more information about the general specifications of the corresponding AGRS systems please refer to Table 5.1.

| ID★ | Scintillator | $V_{sci}$ [dm$^3$]● | Manufacturer○ | Position† | DAM‡ | $N_{ch}$ [−]§ | $t_s$ [s]$ | DEA# | Calibration●● |
|---|---|---|---|---|---|---|---|---|---|
| AT1 | NaI(Tl) | 37.8 | Picodas | cabin | BM | 256 | 1 | PF, WM | CPm, CAR |
| CA1 | NaI(Tl) | 16.8 | RSI | basket or cabin | BM | 1024 | 1 | WM | CPf, PS, CAR |
| CH1 | NaI(Tl) | 16.8 | Mirion | fuselage | BM | 1024 | 1○○ | WM | PS, CAR |
| CN1†† | NaI(Tl) | 37.8 | RSI | cabin | BM | 256 | 1 | WM | CPf |
| CN2‡‡ | NaI(Tl) | 37.8 | Nuvia | cabin | BM | 256 | 1 | WM | CPf, CAR |
| CZ1 | NaI(Tl) | 16.8 | Nuvia | cabin | BM | 512 | 1 | FSA, WM | CPf, CPm, MC, CAR |
| DE1 | NaI(Tl) | 16.8 | Envinet | cabin | BM | 2048 | 2 | WM | CPf, PS, CAR |
| DE2§§ | NaI(Tl) | 21.0 | RSI | cabin | BM | 256 | 1 | WM | CPm, CAR |
| FR1 | NaI(Tl) | 16.8 | Mirion | basket | BM | 1024 | 1 | PF, WM | CPf, MC, PS, CAR |
| FR2$$ | NaI(Tl) | 16.8 | RSI | basket | BM | 1024 | 2 | PF, WM | MC, PS |
| IT1 | NaI(Tl) | 16.8 | Scionix | fuselage | LM | | 1 | FSA, PF, WM | CPf, MC, CAR |
| NO1 | NaI(Tl) | 21.0 | RSI | basket | BM | 1024 | 1 | FSA, WM | CPm, PS, CAR |
| NO2 | NaI(Tl) | 16.8/21.0 | RSI | cabin/cabin | BM | 1024 | 1 | FSA | |

*Continued on next page*

**Table 5.2** *Continued from previous page*

| ID★ | Scintillator | $V_{sci}$ [dm$^3$]● | Manufacturer○ | Position† | DAM‡ | $N_{ch}$ [−]§ | $t_s$ [s]$ | DEA# | Calibration●● |
|---|---|---|---|---|---|---|---|---|---|
| SE1 | NaI(Tl) | 21.0 | RSI | cabin | BM | 1024 | 1 | FSA, WM | CPf, PS, CAR |
| TW1 | NaI(Tl) | 8.4 | Mirion | cabin | BM | 1024 | 1 | WM | |
| US1 | NaI(Tl) | 25.2/12.6 | RSI | basket/cabin | BM | 1024 | 1 | FSA, PF, WM | CPf, PS, CAR |

★ Custom identifier, please refer to Table 5.1.
● Scintillator total volume (helicopter / fixed wing aircraft). The total scintillator total volume comprises a varying number of identical 10.2 cm × 10.2 cm × 40.6 cm prismatic NaI(Tl) scintillation crystals for all listed systems, except for US1 and TW1, which use 5.1 cm × 10.2 cm × 40.6 cm prismatic crystals instead.
○ Mirion Technologies Inc. (Mirion), NUVIA Instruments GmbH (Nuvia), Picodas Group Inc. (Picodas), Radiation Solutions Inc. (RSI), Scienta Envinet (Envinet), SCIONIX Holland B.V. (Scionix).
† Scintillation detector location (helicopter / fixed-wing aircraft): incorporated into the aircraft's fuselage (fuselage), inside transport baskets on the side or below the aircraft (basket), inside the aircraft's crew cabin (cabin).
‡ Data acquisition mode (BM: bin-mode, LM: list-mode).
§ Number of pulse-height channels.
$ Sampling time.
# Data evaluation approaches introduced in Sections 5.4.3–5.4.5 (FSA: full spectrum analysis, PF: peak fitting, WM: window method).
●● Calibration methods introduced in Section 5.5 (CPf: calibration pad (fixed), CPm: calibration pad (mobile), MC: Monte Carlo simulation, PS: point source, CAR: calibration range).
○○ The NBC-EOD adopts a sliding 5 s binning of the raw 1 s accumulated spectra in their data postprocessing routines [533]. The NEOC applies directly the raw 1 s accumulated spectra in their data postprocessing routines.
†† Reference: Zhang et al. [607], Xiong [608], and Liao et al. [609].
‡‡ Reference: Zhang et al. [607], Xiong [608], Xiong et al. [610], and Xu et al. [611].
§§ Reference: Siemon et al. [476], [612].
$$ Reference: Bucher et al. [613] and Wilhelm et al. [614].

The remainder of the data was retrieved from the literature as indicated in the corresponding data entries. Although significant effort was put into the compilation of these tables, they are not exhaustive and do not claim to be complete but rather provide a representative overview of the technical parameters of currently operational manned AGRS systems.

Interestingly, when comparing the overall diversity of both manned and unmanned AGRS systems, including those designed for TGF and atmospheric or cosmic gamma-ray surveys, manned systems tend to exhibit greater uniformity in their technical parameters. Most manned AGRS systems utilize helicopters, with a minority deployed on small fixed-wing aircraft or gyrocopters. The maximum extent of these aircraft can vary between 5 m and 25 m (cf. Fig. B.19). Typical survey ground clearances $h_{\mathrm{air}}$ and survey speeds $v_{\mathrm{g}}$ tend to be smaller for helicopter-based systems, i.e. $15\,\mathrm{m} \leq h_{\mathrm{air}} \leq 200\,\mathrm{m}$ and $19\,\mathrm{m\,s^{-1}} \leq v_{\mathrm{g}} \leq 50\,\mathrm{m\,s^{-1}}$ compared to $60\,\mathrm{m} \leq h_{\mathrm{air}} \leq 300\,\mathrm{m}$ and $39\,\mathrm{m\,s^{-1}} \leq v_{\mathrm{g}} \leq 133\,\mathrm{m\,s^{-1}}$ for fixed-wing aircraft systems.

Most AGRS systems are equipped with secondary sensors or have access to equivalent sensor data provided by the aircraft, such as air pressure sensors (APSs), air temperature sensors (ATSs), global navigation satellite system receivers (GNSSrs), hygrometers (HYGs), inertial measurement units (IMUs), laser altimeters (LAs) or radar altimeters (RAs). The GNSSr is essential for positioning and navigation while the aircraft's ground clearance $h_{\mathrm{air}}$ can be derived from GNSSr, APSs, LAs or RAs.[3] The orientation of the aircraft can be determined using an IMU. In addition, atmospheric sensors such as APSs, ATSs, and HYGs provide detailed information about the atmospheric conditions during the survey.[4]

Let us have now a closer look at the primary gamma-ray spectrometer systems listed in Table 5.2. What immediately stands out is the fact that all systems adopt the same inorganic scintillator, NaI(Tl), as the primary gamma-ray detector. The total scintillator volume $V_{\mathrm{sci}}$ varies between $\sim 8\,\mathrm{dm}^3$ and $\sim 38\,\mathrm{dm}^3$. Most systems apply a bin-mode (BM) for data acquisition, i.e. the scintillation events are recorded as separated pulse-height spectra with a fixed sampling time $t_{\mathrm{s}}$ and a fixed number of pulse-height channels $N_{\mathrm{ch}}$.[5] For these systems, the sampling time $t_{\mathrm{s}}$ varies between 1 s and 2 s while the number of pulse-height channels $N_{\mathrm{ch}}$ ranges from 256 to 2048. As indicated in Table 5.1, some AGRS systems adopt also secondary radiation detectors such as high-purity germanium detectors (HPGes) or Geiger-Müller tubes (GMTs) to complement the primary NaI(Tl)

3 *A detailed discussion of the various techniques to derive $h_{\mathrm{air}}$ with their respective merits and drawbacks can be found in the thesis by Bucher [11].*

4 *Air pressure, temperature and humidity affect the air density in the atmosphere. As we have seen Section 3.1.2, in particular Eq. 3.25a, the interaction rate between gamma rays and air molecules is proportional to the air density. Consequently, changes in the aforementioned atmospheric parameters can affect the gamma-ray transport in the atmosphere and therefore need to be considered in the data evaluation pipelines.*

5 *There is an alternative acquisition mode, the list-mode (LM). In this mode, which is adopted by the team IT1, each scintillation event is recorded separately in an event-by-event manner [616].*

spectrometer. These secondary detectors are typically used to identify radionuclides in a given area of interest and for high dose-rate measurements, respectively [120, 522, 533, 617].

In Table 5.2, I listed also the data evaluation techniques and related calibration methods adopted by the different teams to quantify terrestrial radionuclides. But before we discuss these techniques in more detail, I will provide some more information about the Swiss AGRS system, which is the main system used in the work presented in this book.

## 5.3 The Swiss AGRS System

### 5.3.1 Background

Initiated by the Swiss Geophysical Commission (SGPK), the first Swiss AGRS system was developed in 1985 at the Institute of Geophysics of the Swiss Federal Institute of Technology Zurich (ETHZ) by Professor Ladislaus Rybach[6] and his PhD student Georg Schwarz[7] [10, 618]. In the following years, the system was used in geophysical surveys to systematically map natural terrestrial radionuclides ($^{40}_{19}$K, $^{232}_{90}$Th and $^{238}_{92}$U with related progeny radionuclides as discussed in Section 2.1.3) in Switzerland [618]. In 1989, funded by the Swiss Federal Nuclear Safety Inspectorate (ENSI), the AGRS system was used for the first time for surveys of Swiss nuclear facilities and their surrounding areas. These survey flights were repeated annually between 1989 and 1993 [10, 619–622]. In 1993, the Expert Group for Aeroradiometrics (FAR) was established to coordinate the scientific aspects of the Swiss AGRS system [623].

In 1994, the system was integrated into the Emergency Organization Radioactivity (EOR) of the Federal Office for Civil Protection (FOCP) and the survey flights were subsequently reduced to biennial intervals [623]. Since then, the organization and deployment of the Swiss AGRS system are managed by the National Emergency Operations Centre (NEOC) of the FOCP. Alongside this reorganization of AGRS in Switzerland came also a change in the aircraft system, transitioning from helicopters commercially operated by Heliswiss AG to the Aérospatiale AS332M1 Super Puma (TH06) helicopters operated by the Swiss Air Force [623]. The TH06 helicopter has excellent flight performance capabilities, enabling it to effectively conduct emergency operations even in challenging weather conditions and during nighttime. Scientific support and further development of the

6 *Ladislaus Rybach (*1935) is Professor Emeritus in the Department of Earth Sciences at ETHZ. Professor Rybach is a renowned geophysicist with a focus on AGRS, uranium exploration and applied geothermics, among others. He has served on numerous federal commissions, including the Federal Nuclear Safety Commission (NSC) (formerly KSA), and held the presidency of the Expert Group for Aeroradiometrics (FAR) from 1993 to 2013. Throughout his decades of dedicated research, Professor Rybach has authored over 400 scientific articles and several textbooks. He continues to publish actively and participates annually in the Swiss AGRS exercises.*

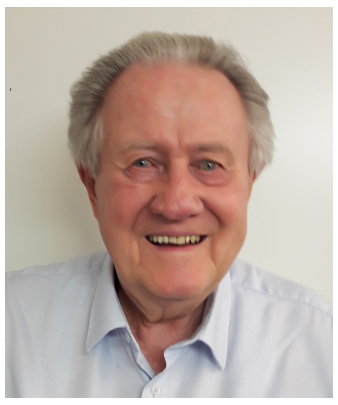

Ladislaus Rybach


7 *Georg Schwarz (*1959) studied geophysics at ETHZ. After his PhD in AGRS [10], he joined the ENSI as head of IT Systems, eventually becoming Deputy Director of the ENSI. On an international level, Schwarz served as the official representative of Switzerland in the Committee on Nuclear Regulatory Activities of the Nuclear Energy Agency (NEA), Vice President of the Seventh Review Conference of the Convention on Nuclear Safety, and led five international review missions of the IAEA. Since his retirement, he has worked as an independent consultant.*

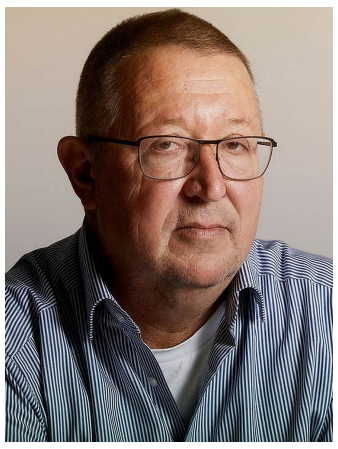

Georg Schwarz
© Georg Schwarz

AGRS system was continued by the Institute of Geophysics of the ETHZ until 2003 when these tasks were transferred to the Radiation Metrology Section of the Paul Scherrer Institute (PSI) in cooperation with the ENSI [624]. As the result of a second PhD project, conducted by the PhD student Benno Bucher[8] under the supervision of Professor Ladislaus Rybach between 1997 and 2001, significant improvements in the AGRS system capabilities could be achieved. These improvements included the introduction of a differential global positioning system (DGPS) for the survey flights, the development of a new data processing software enabling real-time data evaluation as well as the application of novel ground clearance correction methods, among others [11, 625].

*8 Benno Kurt Bucher (*1970) studied environmental sciences at ETHZ. After his PhD [11], Bucher joined the PSI as a research associate in the Department of Radiation Safety and Security. In 2005, he transitioned to the ENSI in the section for radiation measurement. Since 2013, he has been president of the FAR as well as deputy chairman of the Environmental Monitoring Working Group (AKU) of the German-Swiss Association for Radiation Protection (FS).*

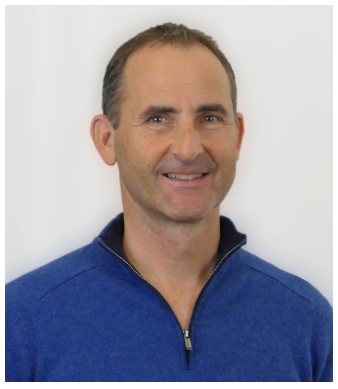

Benno Bucher
© Benno Bucher

In 2018, the AGRS system used by the NEOC in the past exercises was replaced by a new system, the Radiometrie Land-Luft (RLL) AGRS system manufactured by Mirion Technologies Ltd. and integrated into the TH06 helicopter in collaboration with RUAG AG [532–534]. Four identical systems were acquired by the Swiss Armed Forces and are now stationed at the military airfields of Dübendorf (ICAO: LSMD) and Payerne (ICAO: LSMP) [536]. There is a formal service agreement between the civil NEOC and the Swiss Armed Forces which allows the deployment of the RLL system for regular civil surveys and the emergency response to radiological incidents. Of the four systems available, two systems are operated by the staff of the Nuclear, Biological, Chemical, Explosive Ordnance Disposal and Mine Action Centre of Competence (NBC-EOD) for measurement tasks with military character and two systems are assigned to the NEOC [536]. Each system can be fully operative and airborne within four hours [536]. All performed survey flights since 1989 are documented in annual scientific reports [10, 118–120, 526–536, 613, 619–624, 626–638] and are freely accessible to the public on the FAR website: https://far.ensi.ch/.

### 5.3.2 Technical Specifications

Here, I only review the technical specifications of the currently operational Swiss AGRS system, that is the RLL system discussed above and summarized in the Tables 5.1 and 5.2. This system was used for all measurements presented in this book. More information on previous systems used in Switzerland is readily available in the literature [533, 618, 619, 639].

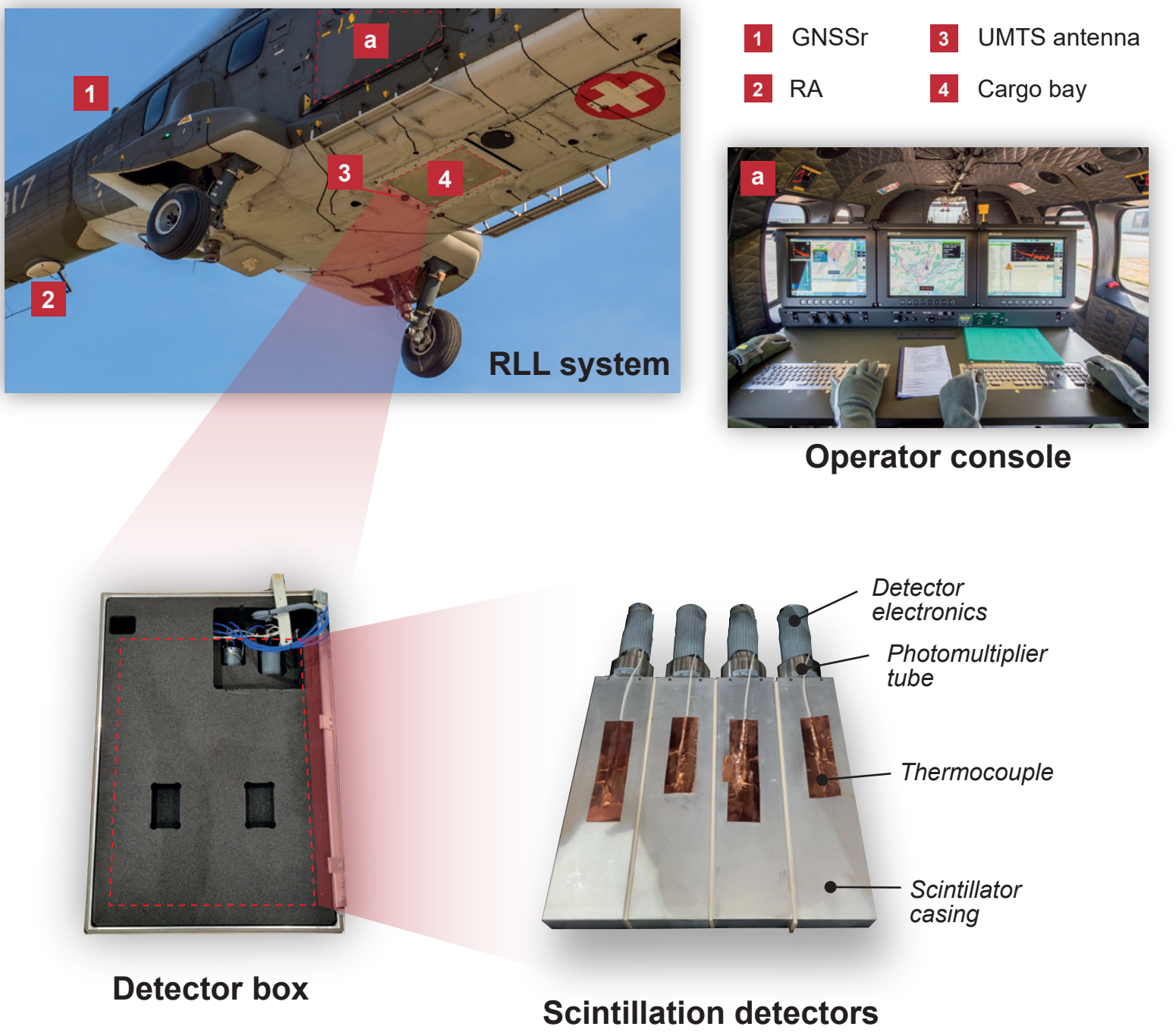


**Figure 5.1** The Swiss AGRS system RLL: Overview. This illustration highlights a close-up view of the detector location within the helicopter's fuselage, the detector box and scintillation detectors as well as the operator console. In addition, secondary helicopter sensors accessible to the RLL system (GNSSr: global navigation satellite system receiver, RA: radar altimeter) together with the UMTS antenna are marked.

As illustrated in Fig. 5.1, the RLL AGRS system comprises four identical 10.2 cm × 10.2 cm × 40.6 cm prismatic NaI(Tl) scintillation crystals (Saint-Gobain 4*4H16/3.5-X) protected by individual aluminum casings. Each crystal is coupled to a separate PMT (Hamamatsu R10755) with associated detector electronic components (preamplifier and multichannel analyzer (MCA)) allowing individual read-out for each scintillation crystal as discussed in Chapter 4. The four scintillation spectrometer assemblies are embedded in a thermal-insulating and vibration-damping polyethylene foam, further protected by a rugged aluminum box with outer dimensions of 90 cm × 64 cm × 35 cm. The total mass of the detector box is ~90 kg. The spectrometer uses a BM scheme for data acquisition (cf. Section 5.2) with a sampling time of 1 s.[9] It comprises 1024 pulse-height channels, with the LLD and upper-level discriminator (ULD) configured to encompass a spectral range $E'$ between ~30 keV and ~3.072 MeV. Furthermore, the spectrometer features automatic gain stabilization[10], spectrum linearization with offset correction as well as live time recording, significantly simplifying the data postprocessing as discussed in Section 4.3. In addition to the primary gamma-ray spectrometer, the system is also equipped with a GMT (Centronic ZP1202) for high dose-rate measurements up to 40 mSv h$^{-1}$.

*9 The NBC-EOD adopts a sliding 5 s binning of the raw 1 s accumulated spectra in their data postprocessing routines [533]. The NEOC applies directly the raw 1 s accumulated spectra in their data postprocessing routines.*

*10 Using known FEPs related to the natural radionuclides $^{40}_{19}$K and $^{208}_{81}$Tl (cf. Section 2.1.3). In addition, the crystal temperature is measured using thermocouples during start-up to preset the corresponding regions of interest [640].*

The detector box is mounted in the cargo bay of the TH06 helicopter, with the bottom of the detector box aligned with the lower helicopter surface for optimal sensitivity to terrestrial gamma-ray sources, as illustrated in Fig. 5.1. A rugged computer functions as the system's data server, while two additional rugged and redundant client computers serve as interfaces for online evaluation, data mapping and communication. These computers are housed within an equipment rack situated inside the crew cabin of the TH06 helicopter. The total mass of the system components within the crew cabin is ~290 kg.[11] During regular surveys, two members of the NEOC or the NBC-EOD operate the system with their associated client computer, display, keyboard and trackball as highlighted in Fig. 5.1. The operator console features a third central display which is mirrored on a screen in the cockpit positioned between the pilot and co-pilot. This setup facilitates efficient information exchange between the pilots and the operators [536]. In addition, the system includes also an universal mobile telecommunications system (UMTS) modem and related antenna for real-time data transmission to the ground station. Using an internal ARINC 429 avionics data bus, the RLL system can access secondary aircraft sensor data, i.e. data from

*11 Excluding the ~40.6 kg steel plates at the back of the helicopter cabin to balance the additional mass of the RLL system.*

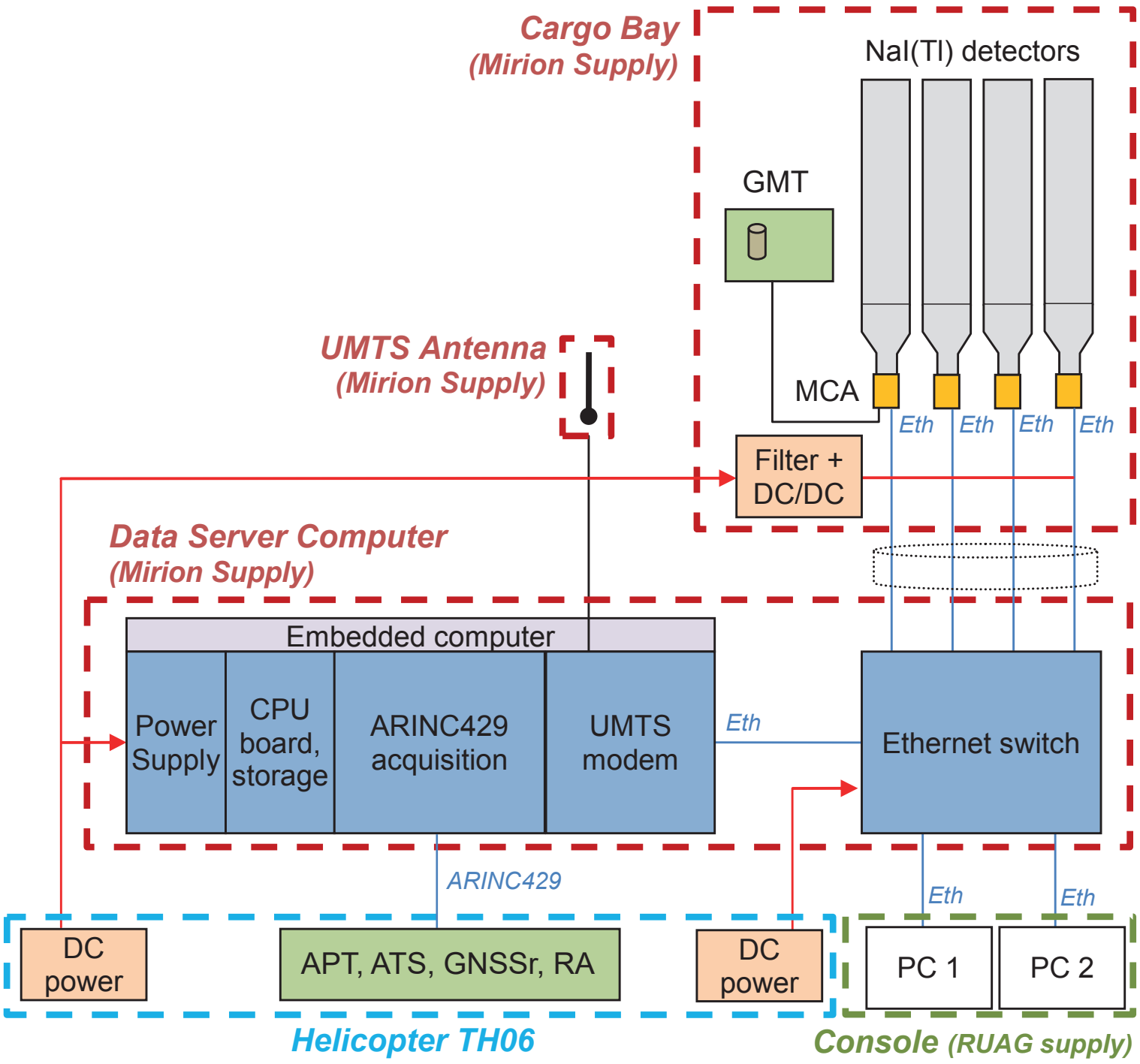


**Figure 5.2** The Swiss AGRS system RLL: System architecture. This scheme highlights the radiation detectors (NaI(Tl) scintillators and GMT: Geiger-Müller tube), detector electronics (MCA: multichannel analyzer), secondary helicopter sensors (APS: air pressure sensor, ATS: air temperature sensor, GNSSr: global navigation satellite system receiver, RA: radar altimeter), the rugged computer server, the operator consoles and the data transmission system (UMTS: universal mobile telecommunications system). In addition, the manufacturers of the individual system components are indicated, too (Mirion Technologies Inc. & RUAG AG). Adapted from Perez et al. [640].

the APS, ATS, GNSSr and RA as already indicated in Table 5.1. A schematic overview of the RLL system architecture including also information about the different manufacturers of the individual system components is provided in Fig. 5.2.

Data acquisition and evaluation of the survey flights are performed using a proprietary software suite (`SpirIDENT`) developed and maintained by Mirion Technologies Inc. [640]. Additionally, an independent offline data evaluation software (`AGS_CH`) developed at

PSI has been used to produce the results presented in the scientific reports since 2020 [118, 120, 536]. This offline software is based on the preceding data evaluation software `MGS32` developed at the Institute of Geophysics of the ETHZ by Bucher et al. [11, 625]. A detailed description of both codes, `SpirIDENT` and `AGS_CH`, is provided by Butterweck et al. [533].

## 5.4 Quantification

After reviewing the history and current status of AGRS in Switzerland and abroad, let us now focus on the data evaluation and calibration routines adopted in AGRS. As discussed in Chapter 1, in this work, I focus on the quantification task of AGRS systems, i.e. the determination of the individual source strengths associated with the various high-energy photon sources discussed in Chapter 2 given a set of recorded pulse-height spectra. This inference task belongs to the class of inverse problems [400]. However, before we analyze this inverse problem and explore potential solution strategies in detail, we must first consider its counterpart, i.e. the computation of the expected pulse-height spectra given a set of known high-energy photon sources and their associated source strengths. This problem is referred to as the forward problem [400].

### 5.4.1 The Forward Problem

Given the scope of this work and based on the available source information reviewed in Chapter 2, I will assume the following in this and the subsequent chapters:

A. The number and type of all high-energy photon sources affecting the measurement response of the AGRS system are known (identification task).

B. The location and distribution of all high-energy photon sources in the environment affecting the measurement response of the AGRS system are known (location task).

In addition, I will continue to assume that the detector response function (DRF) is independent of the double differential photon flux (discussed in Section 4.3.4). Using these assumptions, we may write the detector response of the AGRS system, i.e. the pulse-height spectrum characterized by the count rate vector **c** (cf. Section 4.3.4), as

a linear combination of the individual source contributions $\hat{\mathbf{c}}_m$ with the corresponding source strengths $\xi_m$ [641–646]:

$$\mathbf{c}\,(t) = \sum_{m=1}^{N_{\mathrm{src}}} \hat{\mathbf{c}}_m\,(\mathfrak{d}(t))\,\xi_m\,(t) \tag{5.1}$$

where:

| | | |
|---|---|---|
| $\mathbf{c}$ | count rate vector (cf. Section 4.3.4) | $\mathrm{s}^{-1}$ |
| $\hat{\mathbf{c}}$ | spectral signature | $\mathrm{s}^{-1}\,[\xi]^{-1}$ |
| $\mathfrak{d}$ | experimental condition | $[\mathfrak{d}]$ |
| $N_{\mathrm{src}}$ | number of sources | |
| $t$ | time | s |
| $\xi$ | source strength | $[\xi]$ |

The coefficients $\xi_m$ in Eq. 5.1 represent the source strength of the individual high-energy photon sources. For radionuclide sources, $\xi$ may be represented by the activity mass concentration $a_{\mathrm{m}}$ (e.g. terrestrial volume sources), the activity volume concentration $a_{\mathrm{v}}$ (e.g. atmospheric volume sources), the surface activity concentration $a_S$ (e.g. terrestrial surface source) or simply the activity $\mathcal{A}$ (e.g. point sources) in units of $\mathrm{Bq\,kg^{-1}}$, $\mathrm{Bq\,m^{-3}}$, $\mathrm{Bq\,m^{-2}}$ and Bq, respectively (cf. Section 2.1.2). For sources not associated with radionuclides, e.g. the cosmic rays discussed in Section 2.2, the source strength may be characterized by the photon flux $\phi_\gamma$ in units of $\mathrm{s^{-1}\,m^{-2}}$. Both, for radionuclide and non-radionuclide sources, the source strength $\xi$ may vary over time $t$ (cf. Chapter 2) as indicated in Eq. 5.1. Because of the different units of the source strengths, I will continue to represent the source strength of all high-energy photon sources as $\xi$ and denote its physical unit with $[\xi]$.

The individual source contributions $\hat{\mathbf{c}}_m$ in Eq. 5.1 are the detector response for the corresponding source $m$, i.e. the part of the pulse-height spectrum $\mathbf{c}$ related to the source $m$. There is no consistent terminology for naming these source contribution spectra in the literature. Depending on the field, this quantity is named standard spectrum [16, 647, 648], (unit) component spectrum [645, 646], fundamental spectrum [616, 649], elemental library spectrum [641] or spectral signature [643, 644], among others. I will use the term "spectral signature" for the remainder of this book. The spectral signature $\hat{\mathbf{c}}$ is a function of the detailed source-detector geometry, the environment as well as the detector system, among others. We can summarize this complex dependency by writing the spectral signature $\hat{\mathbf{c}}$ as

a function of the experimental condition $\mathfrak{d} \in \mathbb{R}_+^{N_{\mathfrak{d}} \times 1}$, a generic vector containing all relevant experimental parameters affecting the detector response, e.g. the position and orientation of the AGRS system and the atmospheric and ground conditions.[12] I will assume that these parameters are deterministic and well-known.[13] As some of these parameters in $\mathfrak{d}$ are subject to temporal variations, the spectral signature $\hat{\mathbf{c}}$ is also a function of the time $t$ as indicated in Eq. 5.1. Please note also that by definition, the spectral signature $\hat{\mathbf{c}}$ is normalized by the source strength $\xi$.

12 *Please note that I indicate the unit of the experimental condition $\mathfrak{d}$ with $[\mathfrak{d}]$ as the physical unit are not necessarily the same for all the individual parameters in $\mathfrak{d}$.*

13 *With other words, I assume that these parameters are known a priori or can be measured/predicted with sufficient accuracy to neglect any statistical or systematic uncertainty [650].*

The forward problem, as described by Eq. 5.1, shares similarities with the detector response problem detailed in Section 4.3.4. However, the key distinction lies in the quantity of interest: rather than the double differential flux $\partial^2\phi_\gamma/\partial E_\gamma \partial \Omega'$, here we focus on the source strength $\xi$ of the individual sources. Similar to Eq. 4.43 in Section 4.3.4, we can simplify Eq. 5.1 by applying matrix notation with the spectral signature matrix $\boldsymbol{\mathcal{M}} \in \mathbb{R}_+^{N_{\mathrm{ch}} \times N_{\mathrm{src}}}$ and source strength vector $\boldsymbol{\xi} \in \mathbb{R}_+^{N_{\mathrm{src}} \times 1}$ [641, 642, 645, 646]:

$$\mathbf{c}\,(t) = \boldsymbol{\mathcal{M}}\,(\mathfrak{d}(t))\,\boldsymbol{\xi}\,(t) \tag{5.2}$$

with:

| | | |
|---|---|---|
| $\boldsymbol{\mathcal{M}}$ | spectral signature matrix | $\mathrm{s}^{-1}\,[\xi]^{-1}$ |
| $\boldsymbol{\xi}$ | source strength vector | $[\xi]$ |

The spectral signature matrix $\boldsymbol{\mathcal{M}}$ is simply a block matrix with each column being a separate spectral signature $\hat{\mathbf{c}}_m \in \mathbb{R}_+^{N_{\mathrm{ch}} \times 1}$, i.e. $\boldsymbol{\mathcal{M}} = [\hat{\mathbf{c}}_1 \cdots \hat{\mathbf{c}}_m \cdots \hat{\mathbf{c}}_{N_{\mathrm{src}}}]$. Similarly, the source strength vector $\boldsymbol{\xi}$ is a vector with each element being the source strength $\xi_m$ of the individual high-energy photon source, i.e. $\boldsymbol{\xi} = [\xi_1 \cdots \xi_m \cdots \xi_{N_{\mathrm{src}}}]^\top$.

Assuming the pulse-height spectra are recorded over a series of discrete time intervals $\Delta t_l$ with $\{l \in \mathbb{N}_+ \mid l \leq N_t\}$ and $N_t \in \mathbb{N}_+$, the number of counts per pulse-height channel $\mathbf{C} \in \mathbb{N}^{N_{\mathrm{ch}} \times 1}$ can then be computed by summing over all evaluated time instances $l$:

$$\mathbf{C} = \sum_{l=1}^{N_t} \boldsymbol{\mathcal{M}}\left(\mathfrak{d}(t_l)\right)\,\boldsymbol{\xi}\left(t_l\right)\Delta t_l \tag{5.3}$$

with:

| | | |
|---|---|---|
| $\mathbf{C}$ | count vector | |
| $N_t$ | number of time instances | |
| $\Delta t_l$ | time interval | s |

The matrix versions in Eqs. 5.2 and 5.3 are particularly useful for numerical computations as they allow the parallel evaluation of the individual elements of the count and count rate vector $\mathbf{C}$ and $\mathbf{c}$, respectively.[14]

### 5.4.2 The Inverse Problem

Using the quantities introduced in the forward problem, we can define the quantification task in AGRS now more formally as follows:

**Definition 5.1 ▶ Quantification Task in AGRS**

Inference problem to derive the unknown source strength vectors $\xi_l \in \mathbb{R}_+^{N_{\mathrm{src}}\times 1}$ for a given set of measured pulse-height spectra $\mathbf{c}_l \in \mathbb{R}_+^{N_{\mathrm{ch}}\times 1}$ recorded over discrete time intervals $\Delta t_l$ under known experimental conditions $\mathfrak{d}_l \in \mathbb{R}_+^{N_{\mathfrak{d}}\times 1}$ with $\{l \in \mathbb{N}_+ \mid l \leq N_t, N_t \in \mathbb{N}_+\}$, $N_{\mathrm{ch}} \in \mathbb{N}_+$, $N_{\mathrm{src}} \in \mathbb{N}_+$ and $N_{\mathfrak{d}} \in \mathbb{N}_+$.

Mathematically, this inference problem in AGRS belongs to the class of linear discrete inverse problems, that is a problem where quantities that are not accessible by direct observations are determined based on measurements of related quantities [400].[15] Consequently, in inverse problems, we try to estimate the unknown causes (e.g. source strength vectors $\xi_l$) from observed consequences (e.g. pulse-height spectra $\mathbf{c}_l$) against the natural causal direction [652]. The opposite, that is the inference problem of predicting the consequences from the causes is also referred to as the forward problem, which we have discussed in the previous subsection [653]. Inverse problems are ubiquitous in various scientific fields such as astrophysics and cosmology [654–657], particle physics [654, 658], geophysics [659–661], imaging [662–664] or engineering [665–668]. Most of these inverse problems, including the one described above for AGRS, are ill-posed after Hadamard[16] [669], meaning they violate at least one of the following three conditions [400]:

1. **Existence** The problem has a solution.
2. **Uniqueness** The problem has a unique solution.
3. **Stability** The solution depends continuously on the data.

This makes inverse problems and consequently the quantification task in AGRS particularly challenging. There are various reasons

14 *Modern numerical codes such as `Python/NumPy` with `numpy.matmul` or MATLAB with `pagemtimes` provide efficient algorithms for single and multi-dimensional matrix multiplication operations.*

15 *A more formal definition of inverse problems is provided by Nakamura et al. [651].*

16 *Jacques Salomon Hadamard (*1865, †1963) was a French mathematician renowned for his contributions to number theory, complex function theory and partial differential equations, among others.*

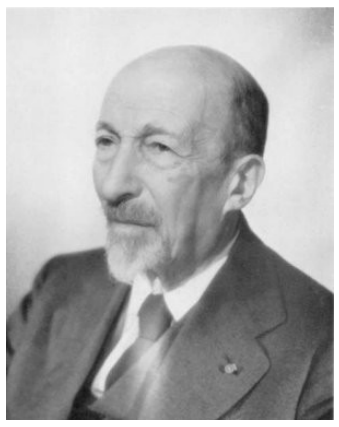

Jacques Hadamard
Wikimedia Commons

why the quantification task in AGRS tends to violate the conditions stated above [670].

First, regarding the existence of a solution to the quantification task in AGRS, let us consider the scenario where we would like to estimate the source strength vector $\boldsymbol{\xi}$ in Eq. 5.2 for a given pulse-height spectrum $\mathbf{c}$ and spectral signature matrix $\boldsymbol{\mathcal{M}}$. In this case, Eq. 5.2 represents a system of $N_{\mathrm{ch}}$ linear equations and $N_{\mathrm{src}}$ unknowns. Assuming $\mathrm{rank}(\boldsymbol{\mathcal{M}}) = N_{\mathrm{ch}}$, this linear system of equations has at least one solution if $N_{\mathrm{ch}} \leq N_{\mathrm{src}}$ [671]. In AGRS, typically we have $256 \leq N_{\mathrm{ch}} \leq 2048$ (cf. Table 5.2) and $N_{\mathrm{src}} \lesssim 10$ considering a single pulse-height spectrum measurement (cf. Chapter 2) and thereby $N_{\mathrm{ch}} \gg N_{\mathrm{src}}$. Consequently, the quantification task is an overdetermined problem without a direct solution [672]. There is one other factor that can lead to the non-existence of a solution to inverse problems. The data ($\mathbf{c}$) and model ($\boldsymbol{\mathcal{M}}$) are in most cases subject to statistical and systematic uncertainties. As a result, the model may not be able to reproduce the measured data exactly but only within the related uncertainty boundaries [670]. This is especially true for low-counting applications such as AGRS surveys.

Regarding the uniqueness of a solution to the quantification task in AGRS, there are again several reasons why the solution is non-unique. First, there is an inherent physical ambiguity that cannot be avoided [672]. To see this, consider an AGRS survey that takes place right after heavy rainfall has occurred. In this scenario, the radon progeny $^{214}_{82}\mathrm{Pb}$ and $^{214}_{83}\mathrm{Bi}$ are washed out from the atmosphere and deposited on the ground. The spectral signature $\hat{\mathbf{c}}_{\mathrm{Rn}}$ from these deposited radionuclides will be almost identical to the spectral signature $\hat{\mathbf{c}}_{\mathrm{U}}$ of a shallow uranium deposit on the ground [87, 673, 674].[17] In the extreme case where the spectral signatures are identical, $\boldsymbol{\mathcal{M}}$ will be rank deficient, i.e. $\mathrm{rank}(\boldsymbol{\mathcal{M}}) < N_{\mathrm{src}}$, moving the problem from an overdetermined toward an underdetermined problem [670, 671]. A second reason for the non-uniqueness of the solution to inverse problems is again the statistical and systematic uncertainty in the data ($\mathbf{c}$) and the model ($\boldsymbol{\mathcal{M}}$). As stated above, the model may not be able to reproduce the measured data exactly but only within the related uncertainty boundaries. If we allow solutions within these boundaries, in most cases, there will be many solutions, i.e. source strength vectors $\boldsymbol{\xi}$, which are consistent with the measured data $\mathbf{c}$ [672]. Again, this is especially true for low-counting applications such as AGRS surveys.

17 *This is one of the reasons why geophysical AGRS surveys should not be conducted during or shortly after rainfall [87]. As a rule of thumb, ~3 h after the rainfall stopped, the deposited radon progeny $^{214}_{82}\mathrm{Pb}$ and $^{214}_{83}\mathrm{Bi}$ have sufficiently decayed to resume AGRS surveys (cf. Table C.1) [87].*

Last but not least, most inverse problems including the quantification task in AGRS are ill-conditioned, i.e. small changes in the data

can lead to tremendous changes in the estimated source strength vector $\xi$ violating the stability condition after Hadamard [670]. We can quantify the ill-conditioning of discrete linear inverse problems by the so-called condition number cond($\mathcal{M}$), which is the ratio of the largest to the smallest singular value of $\mathcal{M}$ [672]:

$$\mathrm{cond}\left(\mathcal{M}\right) = \frac{\nu_{\mathrm{s,max}}}{\nu_{\mathrm{s,min}}} \tag{5.4}$$

where:

| | |
|---|---|
| $\nu_{\mathrm{s,max}}$ | largest singular value of $\mathcal{M}$ |
| $\nu_{\mathrm{s,min}}$ | smallest singular value of $\mathcal{M}$ |

The condition number quantifies the maximum relative change in the source strength vector $\xi$ with respect to a relative change in $\mathcal{M}^{\top}\mathbf{c}$ [672]. If $\mathrm{cond}\left(\mathcal{M}\right) \gg 1$, we call the inverse problem ill-conditioned [670, 672]. For rank-deficient matrices, we have $\nu_{\mathrm{s,min}} = 0$ and consequently $\mathrm{cond}(\mathcal{M}) \rightarrow \infty$ [672]. The fundamental reason for ill-conditioning in inverse problems is the fact that as a direct consequence of the second law of thermodynamics, natural physical processes tend to damp high-frequency signals resulting in smoothed data and an increase in entropy [400, 652]. Consequently, the reversal of this physical process in inverse problems is inherently unstable as high-frequency noise in the data will inadvertently get amplified.

So, how do we solve this challenging quantification task outlined in Def. 5.1? Since its inception at the beginning of the 1960s, three main approaches to the quantification problem in AGRS have been established. These methods are:

1. **Spectral window approach**
2. **Full spectrum approach**
3. **Peak fitting approach**

and they are also included in Table 5.2 indicating the specific method adopted for the the quantification task by the various organizations affiliated with the listed AGRS systems. Each of these methods has its advantages and disadvantages, and the choice of the method depends on the specific requirements of the quantification task in AGRS. In the following three subsections, I will introduce each of these approaches in more detail highlighting their respective merits and drawbacks when compared to the other two methods.

### 5.4.3 Spectral Window Approach

The spectral window approach, often denoted simply as window method (WM) [16, 17], is as of writing this book the most commonly adopted method for the quantification task in AGRS (cf. Table 5.2) and is also used in the data evaluation codes (`SpirIDENT` and `AGS_-CH`) in Switzerland (cf. Section 5.3.2) [533, 640]. Additionally, it is the method recommended by the IAEA in its guidelines published in 1991 and 2003 [8, 9].

The spectral window approach is based on the idea of defining a wide spectral domain comprising several pulse-height channels for each high-energy photon source, where the spectral signature of the primary gamma rays from the source is particularly pronounced compared to the background. As a result, these domains are typically defined to encompass single or multiple FEPs with high photon intensity $I_\gamma$ of the related radionuclide (cf. Sections 2.1 and 4.3). These spectral domains are referred to as spectral windows or simply windows [8, 9]. Currently adopted spectral windows by the NEOC and the NBC-EOD in Switzerland are displayed in Fig. 5.3. Using these spectral windows, the count rate vector $\mathbf{c} \in \mathbb{R}_+^{N_{\text{ch}} \times 1}$ from Eq. 5.2 is converted into a spectral window count rate vector $\tilde{\mathbf{c}} \in \mathbb{R}_+^{N_{\text{w}} \times 1}$ with $N_{\text{w}} = N_{\text{src}} \ll N_{\text{ch}}$ by summing over the pulse-height channels in the related spectral windows. Similarly, the rectangular spectral signature matrix $\boldsymbol{\mathcal{M}} \in \mathbb{R}_+^{N_{\text{ch}} \times N_{\text{src}}}$ from Eq. 5.2 is reduced to a square matrix $\tilde{\boldsymbol{\mathcal{M}}} \in \mathbb{R}_+^{N_{\text{src}} \times N_{\text{src}}}$, which is often referred to as the sensitivity matrix [8, 9], by summing over the pulse-height channels in the related spectral windows for each column in $\boldsymbol{\mathcal{M}}$.

By focusing on the highest relative spectral signature imprint of the primary gamma rays from the corresponding source in the pulse-height spectrum, the signal-to-noise ratio is maximized and at the same time, by combining the pulses over multiple pulse-height channels, the statistical uncertainty is reduced. This in turn reduces the ill-conditioning of the quantification problem leading to a more well-posed problem after Hadamard. Moreover, by constraining the number of spectral windows $N_{\text{w}}$ to the number of sources $N_{\text{src}}$, i.e. $N_{\text{w}} = N_{\text{src}}$, the overdetermined inference problem according to Def. 5.1 is turned into a determined problem with a unique solution given $\text{rank}(\tilde{\boldsymbol{\mathcal{M}}}) = N_{\text{src}}$. To further improve the conditioning of the quantification problem, nuisance background sources are excluded from the quantification task. Typically, the following three nuisance sources are declared as background in AGRS [8, 9]:

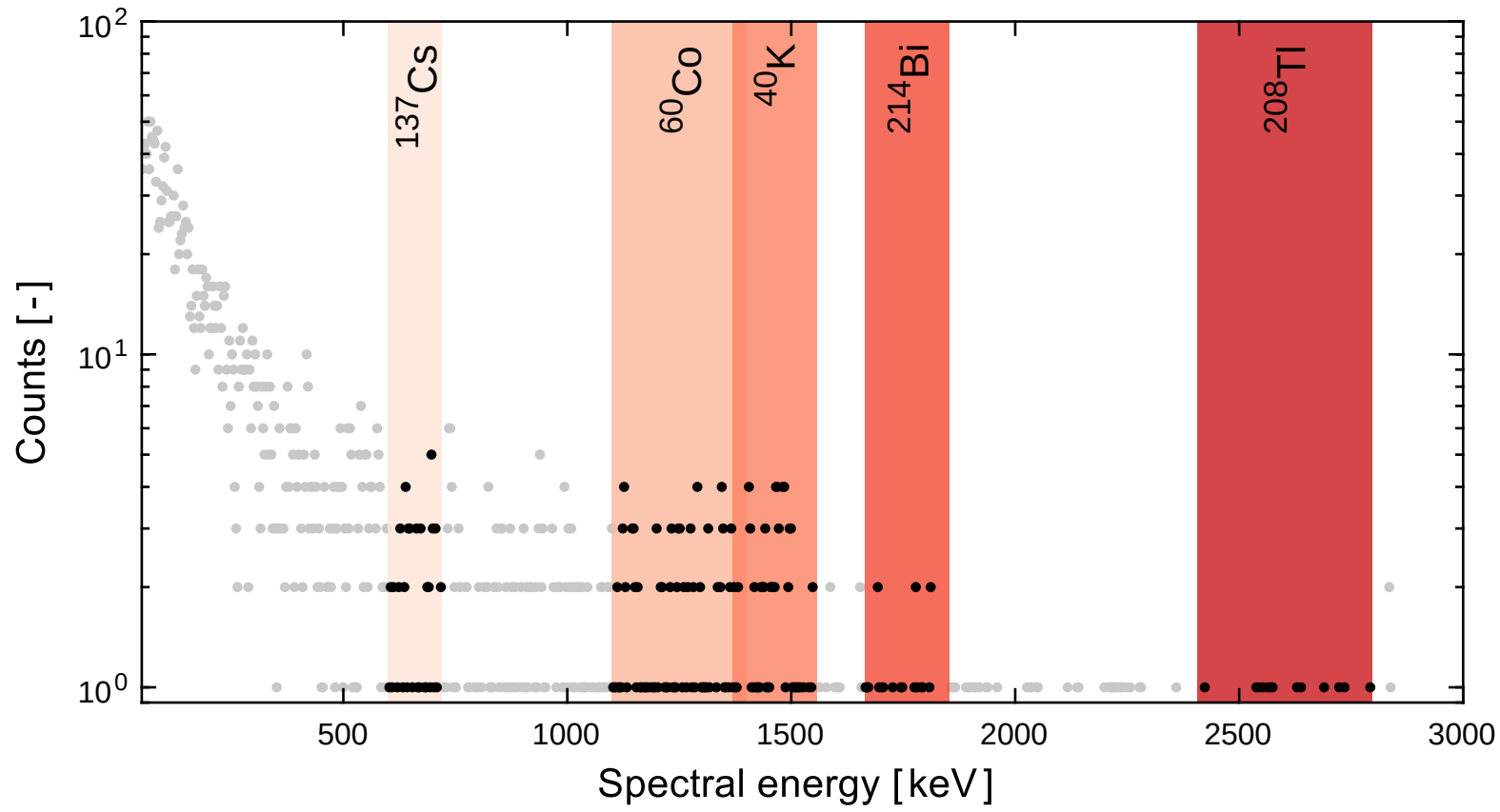


**Figure 5.3** Spectral windows currently applied by the the NEOC and NBC-EOD for quantification in AGRS [120]. The spectral windows are centered around the characteristic gamma-ray energies of the radionuclides of interest, i.e. 661.657(3) keV for $^{137}_{55}$Cs [68], 1173.228(3) keV and 1332.492(4) keV for $^{60}_{27}$Co [68], 1460.822(6) keV for $^{40}_{19}$K [65], 1764.494(14) keV for $^{214}_{83}$Bi [124] associated with the uranium decay series (cf. Table C.1) and 2416.511(10) keV for $^{208}_{81}$Tl [108] associated with the thorium decay series (cf. Table C.2). The displayed net experimental spectrum with a spectral energy bin width of $\Delta E' \sim 3$ keV is representative of a pulse-height spectrum **C** recorded with a manned AGRS system ($V_{sci}$ = 16.8 dm$^3$) at a ground clearance of ~90 m ( orthometric height of ~660 m, vertical datum reference: `EGM2008` [675]) and an acquisition time of 1 s over flat grassland terrain. It was recorded with the Swiss RLL system in a hover-flight mode during the exercise ARM22 on the military training ground in Thun, Switzerland (46.753°N, 7.596°E) on 2022-06-16 [120]. Background correction was applied using survey flight data over Lake Thun. More information on this measurement and the data processing can be found in Section 8.3.2.1.

1. **Intrinsic Background** Natural radionuclides ($^{40}_{19}$K, $^{232}_{90}$Th and $^{238}_{92}$U with related progeny radionuclides) incorporated in the AGRS system (detector system, aircraft and crew) as discussed in Section 2.1.3.
2. **Cosmic Background** Cosmic ray induced background as discussed in Section 2.2.
3. **Radon Background** Background induced by airborne radon progeny (mainly $^{214}_{82}$Pb and $^{214}_{83}$Bi) as discussed in Section 2.1.3.

Of course, by excluding these sources from the quantification problem, their contributions in the individual spectral windows, denoted here by the count rate vector $\tilde{\mathfrak{c}}_{\mathrm{b}} \in \mathbb{R}_{+}^{N_{\mathrm{w}} \times 1}$, must be determined by other means, typically experimentally by additional background flights over water bodies combined with simplified analytical physics models [8, 9].

Using the spectral window approach outlined above, we can rewrite the forward problem in Eq. 5.2 as follows:

$$\tilde{\mathfrak{c}}(t) = \tilde{\mathcal{M}}(\mathfrak{d}(t))\, \xi(t) + \sum_{n=1}^{N_{\mathrm{bg}}} \tilde{\mathfrak{c}}_{\mathrm{b},n}(\mathfrak{d}(t), t) \tag{5.5}$$

where:

| | | |
|---|---|---|
| $\tilde{\mathfrak{c}}$ | spectral window count rate vector | $\mathrm{s}^{-1}$ |
| $\tilde{\mathfrak{c}}_{\mathrm{b}}$ | spectral window background count rate vector | $\mathrm{s}^{-1}$ |
| $\tilde{\mathcal{M}}$ | sensitivity matrix | $\mathrm{s}^{-1}\,[\xi]^{-1}$ |
| $N_{\mathrm{bg}}$ | number of background sources | |

Given $\mathrm{rank}(\tilde{\mathcal{M}}) = N_{\mathrm{src}}$, $N_{\mathrm{w}} = N_{\mathrm{src}}$ and by excluding the three background sources discussed above from the source strength vector, the quantification problem becomes well-conditioned with a unique source strength vector $\xi$ solution for a given spectral window count rate vector $\tilde{\mathfrak{c}}$ and associated experimental condition $\mathfrak{d}$ [8, 9]:

$$\xi = \tilde{\mathcal{M}}^{-1}(\mathfrak{d}) \left[ \tilde{\mathfrak{c}} - \sum_{n=1}^{N_{\mathrm{bg}}} \tilde{\mathfrak{c}}_{\mathrm{b},n}(\mathfrak{d}) \right] \tag{5.6}$$

The advantages of the spectral window approach are evident. Thanks to the compression of the pulse-height spectrum into a set

of spectral windows, the quantification problem can be solved in a simple and robust way using a single matrix inversion operation outlined in Eq. 5.6. This simplicity and robustness in combination with the low computer memory requirements made this approach especially attractive before large computer memory capacities were available. Moreover, by limiting the spectral windows to spectral domains where primary gamma rays dominate the pulse-height spectrum, simplified physics models for calibration purposes such as the monoenergetic transport theory reviewed in Section 3.2.1 can be adopted. If combined with experimental calibration measurements, this eliminates the need for complex radiation transport simulations discussed in Section 3.2. [8, 9, 52, 87, 618].

However, compressing the pulse-height spectrum into a small set of spectral windows with dominant primary gamma-ray signatures comes also with several drawbacks. First, all the information outside the spectral windows is lost in the analysis, in particular all the Compton scattered photons from the source but also primary gamma rays with energies outside the adopted spectral windows. As a direct result, the sensitivity to the various terrestrial sources is significantly reduced [16].[18] Second, as the conditioning number tends to increase with an increase in the number of sources, the spectral window approach is limited to a small number of spectral windows and thereby to a small number of selected sources, typically $N_{\mathrm{src}} \lesssim 5$, to ensure rank($\tilde{\boldsymbol{\mathcal{M}}}$)=$N_{\mathrm{src}}$. Given the omnipresence of the three natural terrestrial radionuclides ${}^{40}_{19}\mathrm{K}$, ${}^{232}_{90}\mathrm{Th}$ and ${}^{238}_{92}\mathrm{U}$ with related progeny radionuclides in the environment (cf. Section 2.1.3), only two additional sources can be quantified in the spectral window approach. Considering the large number of anthropogenic radionuclides that could be emitted in a single nuclear incident (cf. Section 2.1.3), this is a significant limitation for nuclear emergency response applications. Third, the required background measurements can be challenging in practice, especially the radon background, which, depending on the location, can be highly variable and difficult to quantify experimentally [120, 642, 676].

18 *A reduction in the sensitivity by a factor of* ∼1.5 *to* >3 *compared to the full spectrum approach was reported in the literature [16, 17].*

### 5.4.4 Full Spectrum Approach

The full spectrum approach, often termed full spectrum analysis (FSA) [16, 649, 676], is fundamentally different from the spectral window approach. In the full spectrum approach, we aim to incorporate the inherent statistical noise into our forward model rather than suppressing it, as done in the spectral window approach. This allows

us to keep the full pulse-height spectrum as the input data for the quantification problem instead of a reduced set of spectral windows.

Let us consider a set of measured pulse-height spectra $\mathcal{Y} = \{\mathbf{y}_k \in \mathbb{R}_+^{N_{\text{ch}} \times 1} \mid k \in \mathbb{N}_+, k \leq N_{\mathcal{Y}}, N_{\mathcal{Y}} \in \mathbb{N}_+\}$ recorded under known experimental conditions $\mathfrak{D} = \{\mathfrak{d}_k \in \mathbb{R}^{N_{\mathfrak{d}} \times 1} \mid k \in \mathbb{N}_+, k \leq N_{\mathcal{Y}}\}$ and constant[19] sources with unknown source strengths $\boldsymbol{\xi} \in \mathbb{R}_+^{N_{\text{src}} \times 1}$. Formally, we can interpret the individual pulse-height spectra in the dataset $\mathcal{Y}$ as independent realizations of an underlying random vector $\boldsymbol{Y}$ following an associated conditional PDF given the set of known experimental conditions $\mathfrak{D}$ under which the dataset was recorded, as well as constant but unknown model parameters $\mathbf{x}$ with $\boldsymbol{\xi} \subseteq \mathbf{x}$ [267]:

19 *In other words, we assume that the source strength vector $\boldsymbol{\xi}$ is time independent for all recorded pulse-height spectra within $\mathcal{Y}$. The methodology introduced here can be easily generalized to time-varying sources by repeating the full spectrum approach for a series of datasets $\mathcal{Y}$, each representing a time instance with constant source strength vector $\boldsymbol{\xi}$.*

$$\boldsymbol{Y} \sim \pi\left(\mathbf{y} \mid \mathbf{x}, \mathfrak{d}\right) \tag{5.7}$$

where:

| | | |
|---|---|---|
| $\mathfrak{d}$ | experimental condition | $[\mathfrak{d}]$ |
| $\mathbf{x}$ | model parameter vector | $[x]$ |
| $\mathbf{y}$ | data vector | $[y]$ |

Note that I generalized the notation for the parameters of the forward model from the source strength vector $\boldsymbol{\xi}$ to $\mathbf{x}$. This adjustment reflects the fact that the full spectrum approach is not limited to quantifying source strength vectors but can also naturally incorporate nuisance parameters, e.g. parameters quantifying noise dispersion. Similarly, I denote the pulse-height spectra within the dataset $\mathcal{Y}$ by $\mathbf{y}$ and the associate unit by $[y]$ to account for the fact that the dataset $\mathcal{Y}$ can be defined using count or count rate based pulse-height spectra.

If we consider the dataset $\mathcal{Y}$ to be fixed and to consist of independently recorded measurements under known deterministic measurement conditions $\mathfrak{D}$, we may define the likelihood function $\mathcal{L} : \mathcal{D}_{\mathbf{X}} \mapsto \mathbb{R}_+$ as [267]:[20]

20 *Note that, as $\mathcal{L}$ is defined as a function of $\mathbf{x}$ and in general $\int_{\mathcal{D}_{\mathbf{X}}} \mathcal{L}(\mathbf{x})\,\mathrm{d}\mathbf{x} \neq 1$, it is not a PDF [267].*

$$\mathcal{L}\left(\mathbf{x}; \mathcal{Y}, \mathfrak{D}\right) := \prod_{k=1}^{N_{\mathcal{Y}}} \pi\left(\mathbf{y}_k \mid \mathbf{x}, \mathfrak{d}_k\right) \tag{5.8}$$

Intuitively, the likelihood function quantifies how well the statistical model describes the dataset $\mathcal{Y}$ as a function of the model parameters $\mathbf{x}$. It is the core quantity in statistical inference and incorporates the complete physical forward model $\mathcal{M}(\boldsymbol{\xi}, \mathfrak{d})$ combined with a statistical model to account for the statistical variation in the data. The likelihood function allows us to find the most likely model parameters

$\mathbf{x}_{\text{MLE}}$ given the dataset $\mathcal{Y}$ and set of experimental conditions $\mathfrak{D}$ by maximizing $\mathcal{L}(\mathbf{x})$ over the parameter space $\mathcal{D}_X$:

$$\mathbf{x}_{\text{MLE}} = \underset{\mathbf{x}\in\mathcal{D}_X}{\arg\max}\ \mathcal{L}(\mathbf{x}; \mathcal{Y}, \mathfrak{D}) \tag{5.9}$$

This inference approach is known as maximum likelihood estimation (MLE) and is a widely used method in statistics and machine learning [677]. For general likelihood functions and forward models, the MLE is often analytically intractable and requires numerical optimization methods to find the maximum [677]. However, for specific cases, the MLE problem in Eq. 5.9 can be solved analytically. As this approach was applied in the past for the quantification problem in AGRS, I will briefly highlight such a solution in the following paragraphs.

If we assume that the statistical variation in the dataset $\mathcal{Y}$ is generated by additive zero-mean Gaussian[21] noise[22] with covariance matrix $\mathbf{K}$ on top of the linear forward model introduced in Section 5.4.1:

$$\boldsymbol{Y} = \boldsymbol{\mathcal{M}}(\boldsymbol{\mathfrak{d}})\,\boldsymbol{\xi} + \boldsymbol{E} \tag{5.10a}$$

$$\boldsymbol{E} \sim \mathcal{N}\left(\mathbf{0}, \mathbf{K}(\mathbf{x}_\varepsilon)\right) \tag{5.10b}$$

where:[23]

| | | |
|---|---|---|
| $\boldsymbol{E}$ | random discrepancy vector | $[y]$ |
| $\boldsymbol{\mathfrak{d}}$ | experimental condition | $[\boldsymbol{\mathfrak{d}}]$ |
| $\boldsymbol{\mathcal{M}}$ | spectral signature matrix | $[y][\xi]^{-1}$ |
| $\mathbf{K}$ | covariance matrix | $[y]^2$ |
| $\mathbf{x}_\varepsilon$ | discrepancy parameter vector | $[x]$ |
| $\boldsymbol{\xi}$ | source strength vector | $[\xi]$ |

and with $\mathcal{N}(\cdot)$ denoting the multivariate Gaussian or normal distribution, we can write the likelihood function as [268]:

$$\mathcal{L}(\mathbf{x}; \mathcal{Y}, \mathfrak{D}) = \prod_{k=1}^{N_{\mathcal{Y}}} \mathcal{N}\left(\mathbf{y}_k \mid \boldsymbol{\mathcal{M}}(\boldsymbol{\mathfrak{d}}_k)\,\boldsymbol{\xi}, \mathbf{K}(\mathbf{x}_\varepsilon)\right) \tag{5.11a}$$

$$= \prod_{k=1}^{N_{\mathcal{Y}}} \frac{\exp\left(-\frac{1}{2}\left[\mathbf{y}_k - \boldsymbol{\mathcal{M}}(\boldsymbol{\mathfrak{d}}_k)\,\boldsymbol{\xi}\right]^\top \mathbf{K}(\mathbf{x}_\varepsilon)^{-1}\left[\mathbf{y}_k - \boldsymbol{\mathcal{M}}(\boldsymbol{\mathfrak{d}}_k)\,\boldsymbol{\xi}\right]\right)}{\sqrt{(2\pi)^{N_{\text{ch}}}\det\left(\mathbf{K}(\mathbf{x}_\varepsilon)\right)}} \tag{5.11b}$$

21 *Named after Johann Carl Friedrich Gauss (*1777, †1855), a German mathematician and physicist who made significant contributions to many fields, including number theory, algebra, statistics, and astronomy. Known as the "Princeps mathematicorum", the Prince of Mathematicians, his work laid the foundation for modern mathematics, and his discovery of the Gaussian distribution is fundamental to probability theory and statistics.*

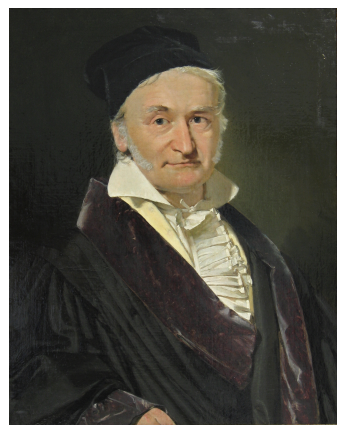

Carl Friedrich Gauss
© Christian Albrecht Jensen

22 *There are several reasons, why this can be a good assumption for a wide range of applications. Murphy [677] and Jaynes [678] provide an in-depth discussion on the applicability of the normal distribution for modeling statistical noise.*

23 *Please note again that, as the dataset $\mathcal{Y}$ can be defined using count or count rate based pulse-height spectra, I denote the unit of $\boldsymbol{\mathcal{M}}$, $\mathbf{K}$ and $\boldsymbol{E}$ by $[y][\xi]^{-1}$, $[y]^2$ and $[y]$, respectively.*

with $\mathbf{x} = [\boldsymbol{\xi}, \mathbf{x}_\varepsilon]^\top$ being the parameter vector consisting of the source strength vector $\boldsymbol{\xi}$ and discrepancy parameters $\mathbf{x}_\varepsilon$ modeling the covariance matrix $\mathbf{K}$. For simplicity, let us assume $N_{\mathcal{Y}} = 1$, specifically $\mathcal{Y} = \{\mathbf{c}\}$. If we know the spectral signature matrix $\boldsymbol{\mathcal{M}}$ as well as the discrepancy parameters $\mathbf{x}_\varepsilon$ and thereby the covariance matrix $\mathbf{K}$, we can solve the MLE problem in Eq. 5.9 analytically resulting in [672]:

$$\boldsymbol{\xi} = \left[\boldsymbol{\mathcal{M}}^\top(\boldsymbol{\mathfrak{d}})\,\mathbf{K}^{-1}\left(\mathbf{x}_\varepsilon\right)\boldsymbol{\mathcal{M}}(\boldsymbol{\mathfrak{d}})\right]^{-1}\boldsymbol{\mathcal{M}}^\top(\boldsymbol{\mathfrak{d}})\,\mathbf{K}^{-1}\left(\mathbf{x}_\varepsilon\right)\mathbf{c} \tag{5.12}$$

Eq. 5.12 is known as the generalized least-squares (GLS) solution. Following a full spectrum approach, simplified versions of the GLS solution in Eq. 5.12 with $\mathbf{K} = \mathrm{diag}(\mathbf{c})$ as well as $\mathbf{K} = \mathbb{I}_{N_{\mathrm{ch}}}$ known as weighted least-squares (WLS) and ordinary least-squares (OLS), respectively, were successfully applied in AGRS in previous studies [16, 17, 642, 645, 646, 676].

By keeping the full pulse-height spectrum as input data for the quantification problem, the full spectrum approach overcomes two main limitations of the spectral window approach. First, the sensitivity to the various terrestrial sources is significantly increased compared to the spectral window approach [16].[24] Second, by utilizing the entire spectral signature of the different sources, the background sources such as radon or cosmic rays can be included in the quantification problem, thereby eliminating the need for separate experimental background measurements [642, 645, 676, 679, 680]. This can also allow for a more accurate estimation of the source strength vector $\boldsymbol{\xi}$.

24 *As noted already above, an improvement in the sensitivity by a factor of ~1.5 to >3 compared to the spectral window approach was reported in the literature [16, 17].*

Although the full spectrum approach using MLE successfully overcomes two main limitations of the spectral window approach, it still has several drawbacks. First and foremost, by adopting the MLE approach, only the overdeterminism of the quantification problem described in Section 5.4.2 is addressed (existence & uniqueness conditions), but not the ill-conditioning (stability condition). As a result, like the spectral window approach, a full spectrum approach using MLE alone is limited to a small number of sources that can be quantified, typically $N_{\mathrm{src}} \lesssim 5$ [16, 17, 642, 645, 646, 676].

Second, at low numbers of counts per pulse-height channel $C \lesssim 20$, a Gaussian likelihood function is not appropriate to model the statistical variation. In low-count regimes, Poisson or negative binomial likelihood functions should be used instead to avoid systematic errors in the MLE estimates [30, 681–683]. Unfortunately, for

these likelihood functions, no analytical solutions exist, and the MLE problem in Eq. 5.9 must be solved numerically using iterative optimization methods such as the expectation-maximization algorithm (EM) [684–686]. As a result, for the same number of sources $N_{\mathrm{src}}$, the full spectrum approach requires in general more computational resources compared to the spectral window approach.

Last but not least, as the entire pulse-height spectrum must be reproduced with high accuracy, derivation and calibration of the spectral signature matrix $\boldsymbol{\mathcal{M}}$ is considerably more complex for the full spectrum approach compared to the spectral window approach. This includes in particular the lower part of the spectrum where photons scattered in the aircraft and the environment significantly alter the spectral signatures.

### 5.4.5 Peak Fitting Approach

The last of the three quantification techniques in AGRS is the peak fitting (PF) method. This is a method that is widely used in laboratory-based gamma-ray spectrometry [30, 296, 297, 365]. Similar to the spectral window approach, the peak fitting method also focuses on the FEPs. However, instead of combining pulse-height channels into spectral windows, the peak fitting method aims to fit analytical and/or numerical models to the FEPs to quantify the net area under the peaks $\check{\mathbf{c}} \in \mathbb{R}_{+}^{N_{\mathrm{FEP}} \times 1}$. Assuming that for each source a single unique FEP is fitted, i.e. $N_{\mathrm{FEP}} = N_{\mathrm{src}}$, and that contributions from other sources to the specific FEP are fully corrected by the baseline model (cf. Section 4.3.3), the associated sensitivity matrix $\check{\boldsymbol{\mathcal{M}}}$ is diagonal, i.e. $\check{\boldsymbol{\mathcal{M}}} = \mathrm{diag}(\epsilon_{\mathrm{det},1}, \ldots, \epsilon_{\mathrm{det},N_{\mathrm{src}}})$ with $\epsilon_{\mathrm{det},l}$ being the detector sensitivity coefficient at $E' = E_{\gamma,l}$ for a specific source-detector configuration in units of $\mathrm{s}^{-1}\,[\xi]^{-1}$. Consequently, the ill-posed quantification problem is turned into a well-posed problem, which can be solved by a simple matrix inversion with $\check{\boldsymbol{\mathcal{M}}}^{-1} = \mathrm{diag}(\epsilon_{\mathrm{det},1}^{-1}, \ldots, \epsilon_{\mathrm{det},N_{\mathrm{src}}}^{-1})$:

$$\boldsymbol{\xi} = \check{\boldsymbol{\mathcal{M}}}^{-1}(\boldsymbol{\mathfrak{d}}) \left[ \check{\mathbf{c}} - \sum_{n=1}^{N_{\mathrm{bg}}} \check{\mathbf{c}}_{\mathrm{b},n}(\boldsymbol{\mathfrak{d}}) \right] \tag{5.13}$$

where:

| | | |
|---|---|---|
| $\check{\mathbf{c}}$ | net peak area count rate vector | $\mathrm{s}^{-1}$ |
| $\check{\mathbf{c}}_\mathrm{b}$ | net peak area background count rate vector | $\mathrm{s}^{-1}$ |
| $\check{\mathcal{M}}$ | sensitivity matrix | $\mathrm{s}^{-1}\,[\xi]^{-1}$ |
| $N_\mathrm{bg}$ | number of background sources | |

Note that, similar to the spectral window approach, background sources with identical FEPs as terrestrial sources, e.g. radon and intrinsic background discussed in Section 5.4.3, need to be corrected separately using related background count rate vectors $\check{\mathbf{c}}_\mathrm{b}$, as indicated in Eq. 5.13.

There are two main advantages when it comes to the peak fitting approach. First, thanks to the baseline correction and specific selection of FEPs, the sensitivity matrix is diagonal and no complex deconvolution operations are required. As a result, the quantification problem is well-posed and can be solved for a large number of sources $N_\mathrm{src} \gg 5$. Second, as only FEP events are used for the quantification and scattering events are numerically corrected by the baseline model, calibration is considerably simplified only requiring to model the transport of primary gamma rays.

However, the peak fitting method also has several shortcomings. First, the method requires good statistics in each pulse-height channel to perform the peak fitting operation. For typical pulse-height spectra recorded in AGRS with $t_\mathrm{s} \leq 2\,\mathrm{s}$, necessary statistics are only achieved in rare occasions, e.g. close to strong anthropogenic point sources or in emergency response to severe nuclear accidents with a significant release and subsequent deposition of anthropogenic radionuclides in the environment [28].

Moreover, as we have discussed in Section 4.3.2, typical primary gamma-ray spectrometers adopted in AGRS such as NaI(Tl) scintillators possess a quite poor energy resolution $R_E \gtrsim 6.5\,\%$ (cf. Table 4.1). As a result, for surveys with NaI(Tl), FEPs associated with weak anthropogenic sources are in most cases heavily obscured by other peaks associated with natural terrestrial sources (in particular $^{232}_{90}\mathrm{Th}$ and $^{238}_{92}\mathrm{U}$ with related progeny radionuclides) and therefore cannot be quantified by peak fitting methods. Last but not least, similar to the spectral window approach, background sources with identical FEPs as terrestrial sources need to be corrected separately which can lead to additional problems discussed already above regarding background estimation.

Considering all these drawbacks, if used on pulse-height spectra obtained with NaI(Tl) scintillators, the peak fitting method is only applicable in special situations with strong anthropogenic sources

and needs to be combined with other quantification methods to ensure complete quantification of all terrestrial sources. Consequently, it is not suitable as a stand-alone method for general quantification problems in AGRS according to Def. 5.1.[25]

25 *However, it is worth adding that peak fitting methods can still be useful in AGRS if combined with high-resolution secondary spectrometers such as HPGe detectors. As discussed in Section 5.2, provided that several pulse-height spectra from HPGe spectrometers are combined to ensure good statistics, the peak fitting method can be applied to identify radionuclides in selected areas of interest [522, 533, 617].*

## 5.5 Calibration

In the previous subsections, we have discussed in detail how we can quantify the source strength vector $\boldsymbol{\xi}$ for a given pulse-height spectrum $\mathbf{c}$ recorded under known experimental condition $\boldsymbol{\mathfrak{d}}$ using a linear forward model of the form $\mathbf{c} \propto \boldsymbol{\mathcal{M}}\boldsymbol{\xi}$. As defined in Section 5.4.1, the experimental condition $\boldsymbol{\mathfrak{d}}$ is assumed to be well known, typically obtained from secondary measurement sensors or other reliable sources. However, I have not yet discussed how to derive the spectral signature matrix $\boldsymbol{\mathcal{M}}$. This derivation process, which I refer to in this book as the calibration task in AGRS, can be formally defined as follows:

**Definition 5.2 ▶ Calibration Task in AGRS**

All operations required to derive the matrix $\boldsymbol{\mathcal{M}} \in \mathbb{R}_{+}^{N_y \times N_{\text{src}}}$ linearly relating the source strength vector $\boldsymbol{\xi} \in \mathbb{R}_{+}^{N_{\text{src}} \times 1}$ to the pulse-height spectrum derived data vector $\mathbf{y} \in \mathbb{R}_{+}^{N_y \times 1}$ by $\mathbf{y} = \boldsymbol{\mathcal{M}}(\boldsymbol{\mathfrak{d}})\boldsymbol{\xi}$ as a function of the experimental condition $\boldsymbol{\mathfrak{d}} \in \mathbb{R}_{+}^{N_{\mathfrak{d}} \times 1}$ with $N_y \in \mathbb{N}_{+}$, $N_{\text{src}} \in \mathbb{N}_{+}$ and $N_{\mathfrak{d}} \in \mathbb{N}_{+}$.

Depending on the selected quantification method, the data vector $\mathbf{y}$ is given as a count rate vector $\mathbf{c}$ with $N_y = N_{\text{ch}}$ representing the raw pulse-height spectrum (full spectrum approach), $\tilde{\mathbf{c}}$ with $N_y = N_{\text{src}}$ representing the spectral window count rates (spectral window approach) or $\check{\mathbf{c}}$ with $N_y = N_{\text{src}}$ representing the net FEP count rates (peak fitting approach). Naturally, this translates to the three matrices $\boldsymbol{\mathcal{M}} \in \mathbb{R}_{+}^{N_{\text{ch}} \times N_{\text{src}}}$, $\tilde{\boldsymbol{\mathcal{M}}} \in \mathbb{R}_{+}^{N_{\text{src}} \times N_{\text{src}}}$ and $\check{\boldsymbol{\mathcal{M}}} \in \mathbb{R}_{+}^{N_{\text{src}} \times N_{\text{src}}}$ introduced in the previous subsections relating the corresponding count rate vectors to the source strength vector $\boldsymbol{\xi}$.

Deriving these matrices is challenging. They incorporate all the complex physics involved in the generation of the high-energy photons and secondary particles discussed in Chapter 2, their transport through the environment analyzed in Chapter 3 and the subsequent

energy deposition generating in the end the pulse-height spectrum through a series of energy conversion processes and signal processing steps covered in Chapter 4. In essence, one has to be able to reliably predict the spectral signatures $\mathbf{c}$, $\tilde{\mathbf{c}}$ or $\check{\mathbf{c}}$ separately for all relevant sources in the environment as a function of the experimental conditions $\mathfrak{d}$, e.g. the ground clearance or the detector orientation relative to the source.

### 5.5.1 Empirical Calibration

Because of the complex physical processes involved in the radiation transport and detector response, standard calibration methods proposed in the guidelines for AGRS by the IAEA are all based on experimental measurements combined with simplified physics models using the monoenergetic transport theory introduced in Section 3.2.1 [8, 9, 687]. Reconstruction of the sensitivity matrix is typically performed by combining separately derived stripping and sensitivity factors [10]. The most common radiation sources adopted for calibration measurements in AGRS are:

1. **Fixed Calibration Pads** Large concrete pads with extensions of ~8 m × 8 m × 0.5 m doped with known natural radionuclide concentrations ($^{40}_{19}$K, $^{232}_{90}$Th and $^{238}_{92}$U with related progeny radionuclides) and located typically on airfields for ground calibration [8, 9, 14, 687].
2. **Mobile Calibration Pads:** Smaller mobile versions of the fixed calibration pads with extensions of ~1 m × 1 m × 0.3 m or disks with a diameter of ~0.7 m and a thickness of ~0.35 m [8, 12, 15].
3. **Point Sources** Calibration point sources [9–11, 13, 618].
4. **Calibration ranges** Reference areas with an extension of about 1 km × 8 km, whose mean radionuclide content was estimated by ground-based analysis techniques [8, 9, 687].

All these empirical methods suffer from three main drawbacks. First, only a small number of radionuclides can be calibrated, in general, natural volume sources ($^{40}_{19}$K, $^{232}_{90}$Th and $^{238}_{92}$U with related progeny radionuclides) and selected anthropogenic point sources. Second, some of the methods (mobile calibration pads & point sources) require a considerable amount of calibration time of up to 3 d to 5 d to cover all relevant source-detector configurations [12, 13], while

the others suffer from high costs in the preparation (fixed calibration pads & calibration ranges) and/or decommissioning (fixed calibration pads) of the calibration sources. Last but not least, all these methods are subject to significant systematic uncertainties, primarily originating from the adopted monoenergetic transport theory, along with its simplifications and approximations discussed in Section 3.2.1.

### 5.5.2 Numerical Calibration

As already discussed in Sections 3.2.2 and 4.3.4, modern high-fidelity Monte Carlo simulation codes such as `FLUKA` [20, 216] or `MCNP` [22, 258] can be used to predict the spectral signatures with high accuracy.[26] As a result, Monte Carlo based calibration approaches are today commonly used for laboratory, in-situ and proximal gamma-ray spectrometry [616, 648, 688], in the marine environment [689–692], in nuclear and particle physics [369, 693], in astrophysics [392, 694–699] or in space exploration [700–704], among others. Despite the apparent advantages, i.e. unrestricted capabilities in terms of source type and source-detector geometry as well as cost-effective and fast evaluation, simulation-based calibration has not yet established itself as an equivalent methodology for manned AGRS systems.

26 *It is worth adding that numerical calibration methods also require some radiation measurements to determine hyperparameters in the numerical models. I will discuss this in more detail in the Chapters 6 and 7.*

First attempts to simulate the spectral signatures for manned AGRS systems were performed by Allyson et al. [18] in 1998 using an in-house Monte Carlo code and a simplified mass model, i.e. the multi-component detector system was reduced to a single scintillation crystal together with an equivalent aluminum casing and the entire aircraft system was excluded from the simulation. Allyson et al. adopted a full simulation approach combined with spectral window analysis. The code systematically underestimated the measured pulse-height spectra, in particular at the lower end of the spectral range with relative errors >200 %. In addition, significant deviations between measured and simulated pulse-height spectra were found around the CE and the CG (cf. Section 4.3.1).

Billings et al. [19] adopted a modified version of the Monte Carlo code developed by Allyson et al. and performed extensive simulations to characterize the angular detector response of an AGRS system. Again, the AGRS system was simplified to a single crystal and all casing and aircraft components were neglected.

In the aftermath of the nuclear accident at the Fukushima Daiichi Nuclear Power Station, Torii et al. [28] adopted a hybrid model combining the `EGS5` Monte Carlo code [705] with analytical flux models

to calibrate their AGRS system for ${}^{131}_{54}$I and ${}^{134}_{55}$Cs surface contamination sources following a peak fitting approach. In contrast to the earlier studies, Torii et al. included the multi-component crystals and detector components such as PMTs and detector electronics in their mass model. However, the aircraft system was still excluded from the simulation.

At the same time, Sinclair et al. [24] used the `EGSnrc` Monte Carlo code [706] in a full simulation approach to calibrate their AGRS system for the quantification of airborne ${}^{133}_{54}$Xe released during the accident at the Fukushima Daiichi Nuclear Power Station. Similar to Torii et al., Sinclair et al. followed a spectral window approach for the quantification and included the multi-component crystals and additional detector components in their mass model. In addition, a simplified aircraft system represented by a primitive spherical shell was incorporated into the mass model.

Sinclair et al. [25] used the same Monte Carlo code and a similar mass model in a second study to calibrate their AGRS system for the quantification of surface contamination by the radionuclide ${}^{140}_{57}$La released in a series of experiments conducted at the Defence Research and Development Canada's Suffield Research Centre in 2012. In contrast to the previous study, Sinclair and her co-workers adopted a hybrid simulation approach combining the Monte Carlo simulations with analytical flux models. The aircraft system was excluded from the simulation.

In their study from 2018, Zhang et al. [26] adopted the `MCNP` Monte Carlo code [22, 258] combined with analytical flux models to calibrate their AGRS system following again a spectral window approach. Similar to Sinclair et al. [24], Zhang et al. combined the multi-component crystals and additional detector components in their mass model with a simplified aircraft model based on a primitive cubic shell.

One of the most advanced Monte Carlo models was developed by Kulisek et al. [27] who adopted a hybrid simulation approach combining the deterministic `ATTILA` [263] code with the Monte Carlo code `MCNP` [22, 258]. The mass model included the multi-component crystals and detector components as well as a simplified aircraft model. In contrast to the previous studies, Kulisek et al. applied Monte Carlo calibration following a full spectrum approach. However, as Allyson et al. [18], Kulisek et al. found large deviations between the simulated and measured pulse-height spectra, overestimating the measured count rates at the lower end of the spectrum $E' \lesssim 100\,\mathrm{keV}$ by a factor of >40.

## 5.6 Current Limitations & Scope

Let us conclude this chapter by summarizing the main limitations of manned AGRS systems when it comes to the quantification of terrestrial high-energy photon sources and what is needed to overcome these limitations.

1. **Calibration** As we have discussed in Section 5.5.1, current empirical calibration methods are severely limited in the type and number of sources that can be calibrated. In addition, they suffer from significant systematic uncertainties, mainly due to the adopted monoenergetic transport theory with its simplifications and approximations. In contrast, simulation-based calibration approaches using high-fidelity Monte Carlo codes have the potential to overcome these limitations. However, as discussed in Section 5.5.2, the current Monte Carlo models for manned AGRS systems are still in their infancy. This can be attributed to two main challenges when it comes to the calibration of AGRS systems using brute-force Monte Carlo simulations: model complexity and computational cost.

   The model complexity arises from the large simulation domain featuring a mobile aircraft platform with a characteristic extension of $\mathcal{O}(10^1)$ m operating at a ground clearance of $\mathcal{O}(10^2)$ m in a survey area of up to $\mathcal{O}(10^2)$ km$^2$ with a varying radiation field of primary and secondary high-energy particles interacting in the environment and the aircraft. To reduce the complexity, previous studies have adopted Monte Carlo models, which either completely neglected or significantly simplified the aircraft system. These simplifications have proven to introduce significant systematic errors in the simulations, in particular at low energies with relative errors >200 %. In addition, most studies have not aimed to reproduce the full spectrum response of the AGRS system to high-energy photons. Instead, they focused on the spectral window response following a hybrid simulation approach that combines analytical monoenergetic flux models with Monte Carlo simulations. Consequently, to minimize the systematic errors and allow for the accurate calibration of the AGRS system in a full spectrum analysis approach, more advanced Monte Carlo models are required that include the full aircraft system and high-fidelity physics models accounting for all relevant physics processes discussed in the Chapters 2–4.

Computational cost is the second challenge. As discussed in Section 3.2, Monte Carlo simulations are computationally expensive due to their stochastic nature. To achieve the necessary precision, a typical simulation for AGRS requires a characteristic computation time of $\mathcal{O}(10^1)$ h on a standard[27] computer cluster. For a typical AGRS survey flight, the number of required evaluations of the adopted physics models can easily exceed $\mathcal{O}(10^5)$. Given these numbers, it is easy to see that the calibration of AGRS systems using brute-force Monte Carlo simulations as proposed in previous studies becomes computationally prohibitively expensive in practical applications with computation times of $\mathcal{O}(10^6)$ h or $\mathcal{O}(10^2)$ a. This computational limitation of Monte Carlo simulations is not unique to AGRS systems but is common across various scientific disciplines, where radiation spectrometers with complex spectral and angular dispersion characteristics are applied to varying source-detector configurations. In the field of gamma-ray and neutron spectrometry, this limitation was addressed by the development of the detector response modeling approach introduced in Section 4.3.4. This method has already been successfully applied in astrophysics [392, 395, 694, 707, 708] and planetary science [393, 394, 396]. Given the similarities in the detector-response configurations, this approach also shows promise for overcoming the limitations of brute-force Monte Carlo simulations for manned AGRS systems.

27 500 *cores at a nominal clock speed of* 2.6 GHz

2. **Inversion** As we have seen in Section 5.4.2, the quantification problem in AGRS is in general ill-posed. I presented three different approaches for solving this ill-posed problem, i.e. the spectral window approach, the full spectrum approach and the peak fitting approach. Considering that the spectral window approach has significantly reduced sensitivities compared to the other two methods and the peak fitting approach is only applicable on rare occasions, full spectrum analysis is currently the most promising quantification method in AGRS. Apart from the increased sensitivity and versatility, the full spectrum approach has the additional benefit of including nuisance sources such as radon or cosmic backgrounds in the quantification problem. However, as discussed in Section 5.4.4, the MLE-based algorithms using Gaussian likelihood functions, which have been employed previously in AGRS for full spectrum analysis, exhibit significant drawbacks. Specifically, they are limited to quantifying only a small number of sources, typically $N_{\text{src}} \lesssim 5$, and they introduce systematic errors

in the source strength vector estimates at low count rates. More advanced inversion methods such as regularized MLE algorithms and Bayesian inversion methods using Poisson or negative binomial likelihood functions could overcome these limitations [400, 651, 658, 670, 677, 709]. However, these methods have not yet been applied to quantification problems in AGRS.

3. **Background** It has been recognized in the past that variable nuisance sources such as the radon or cosmic background can introduce significant systematic errors in the quantification of terrestrial sources [120, 642, 676]. As discussed in Section 5.5, the current methods to estimate these backgrounds are all based on empirical measurements and suffer from large systematic uncertainties. In principle, the full spectrum approach has the potential to include these background sources in the quantification problem and thereby reduce the systematic errors. However, this requires an accurate estimate of the spectral signatures of these background sources, which is difficult to obtain in practice using empirical measurements [642, 676, 679]. Monte Carlo simulations could offer a solution to this problem, but they would require in turn high-fidelity mass models of the entire aircraft system to accurately quantify the attenuation and modulation of the double-differential photon flux over the full solid angle. Additionally, regarding the cosmic background, advanced physics models are necessary to account for nuclear reactions induced by secondary high-energy cosmic ray particles generated by EASs in the Earth's atmosphere (cf. Sections 2.2.2 and 2.3.1). Consequently, due to the complexity of creating such detailed Monte Carlo models and the significant unresolved systematic errors in pulse-height spectra estimates reported in previous studies, there have been no attempts so far to estimate the spectral signatures of these background sources using Monte Carlo simulations.

These limitations in quantification for current AGRS systems motivate the scope of this work introduced already in Chapter 1, which is rephrased here using the various new terms introduced in this and in the previous chapters:

**► Scope of this Work**

1. **AGRS Monte Carlo** Development and validation of a high-fidelity Monte Carlo model of the Swiss AGRS system including the aircraft and advanced physics models to accurately simulate the full spectrum response of the gamma-ray spectrometer for arbitrary source-detector configurations.
2. **Generalization** Derivation and validation of detector response functions and double-differential photon flux models to emulate the Monte Carlo model for arbitrary source-detector configurations allowing efficient and rapid computations of spectral signature matrices.
3. **Quantification** Development and validation of a full spectrum Bayesian inversion methodology allowing to accurately quantify any number of high-energy photon sources including cosmic and radon backgrounds by AGRS both at low and high count rates.

This concludes the first part of this book introducing the fundamentals of AGRS. Following a bottom-up approach, I will start in the next part by discussing the development and validation of the high-fidelity Monte Carlo model of the Swiss RLL spectrometer under laboratory conditions.

# Part II
# Detector Modeling

> *"Your theory is crazy, but it's not crazy enough to be true."*
>
> — *Niels Bohr*

# Chapter 6 Proportional Scintillation Monte Carlo

**Contents**

High-fidelity Monte Carlo codes have the potential to overcome main limitations in the experimental-based calibration and spectral response inversion of AGRS detector systems. However, as analyzed in Section 5.5.2, Monte Carlo modeling of manned AGRS systems is still in its infancy and previously developed models showed significant deviations >200 % between the simulated and measured pulse-height spectra.

As a first step towards accurate Monte Carlo based full spectrum modeling of AGRS detector systems, I will present in this chapter the derivation and validation of a high-fidelity Monte Carlo radiation transport model of the Swiss RLL AGRS detector system under laboratory conditions using the multi-purpose Monte Carlo code `FLUKA`. Consistent with previous studies, I will simplify the complex detector response physics to an energy deposition problem, assuming a proportional scintillation response of the RLL spectrometer. To validate the developed model, radiation measurements were conducted under laboratory conditions using seven different calibration point sources with known activities and source-detector geometries.

The simulation results show superior accuracy compared to previous AGRS simulation models with a median relative deviation <10 % for the majority of the evaluated radiation sources. Yet thorough statistical analysis revealed statistically significant deviations between the simulated and measured pulse-height spectra around the Compton edge, which can be attributed to the scintillation non-proportionality.

For full spectrum AGRS detector modeling, these findings imply that the common energy deposition model assuming proportional scintillation response should be revised and considered to be replaced by more accurate non-proportional models.

Parts of this chapter were reproduced from the following study published by the author in the context of the present work:

D. Breitenmoser et al. "Experimental and Simulated Spectral Gamma-Ray Response of a NaI(Tl) Scintillation Detector Used in Airborne Gamma-Ray Spectrometry". *Advances in Geosciences* **57** 10.5194/ADGEO-57-89-2022 (2022).

## 6.1 Introduction

To fulfill the first objective of this work, that is the development and validation of a high-fidelity Monte Carlo model of the entire Swiss RLL AGRS detector system including the aircraft (cf. Section 5.6), I will follow a bottom-up modeling approach. In the following two chapters, I will present the derivation and validation of two Monte Carlo models of the RLL spectrometer, excluding all aircraft components. This allows me to validate the developed physics and mass models under controlled laboratory conditions with high-counting statistics. The extension of the Monte Carlo model to the entire Swiss RLL AGRS detector system including the aircraft will be presented in Part III.

As discussed in Section 5.6, Monte Carlo based full spectrum modeling of AGRS systems is challenging. It encompasses all the complex physics involved in the generation of the high-energy photons and secondary particles discussed in Chapter 2, their transport through the environment analyzed in Chapter 3 and the subsequent energy deposition generating in the end the pulse-height spectrum through a series of energy conversion processes and signal processing steps covered in Chapter 4. In particular, the last step, i.e. the generation and transport of the scintillation photons, is a complex multiscale problem, whose modeling is still in its infancy and not well understood as discussed in Section 4.1.1 [298, 310, 311]. Furthermore, as we have seen in Section 4.1.3, a single energy deposition event can generate thousands of scintillation photons resulting in a dramatic increase in the number of particles to be simulated. This makes the problem computationally prohibitively expensive, in particular for large-scale simulation problems such as the one in AGRS [369]. Additionally, scintillation processes depend on various detector-specific material and geometrical parameters, which are often not known with sufficient accuracy for these modeling purposes.

To address these challenges and make the problem tractable for large-scale simulations, a simplified scintillation physics model is traditionally adopted in Monte Carlo based full spectrum modeling. In this approach, a proportional relationship between the energy $E_{\text{dep}}$ deposited in the scintillation crystal and the number of scintillation photons produced during the energy deposition event is assumed (cf. Eq. 4.2 in Section 4.1.3):

$$Y_{\text{sci,a}}(E_{\text{dep}}) = \text{const.} \tag{6.1}$$

where:

| | | |
|---|---|---|
| $E_{\text{dep}}$ | deposited energy | eV |
| $Y_{\text{sci,a}}$ | absolute light yield | $\text{eV}^{-1}$ |

With this proportional scintillation physics model, the detector response characterization is reduced to a comparably simple energy deposition problem without the need for explicit simulation of the scintillation processes discussed above. Reconstruction of the pulse-height spectrum is achieved by convolving the simulated energy deposition events with empirically derived spectral resolution and spectral energy functions discussed in Section 4.3.3. I will refer to this Monte Carlo approach as proportional scintillation Monte Carlo (PSMC) for the remainder of this book. PSMC is currently the standard approach for Monte Carlo based full spectrum modeling of scintillation spectrometers and was successfully applied in various fields including laboratory gamma-ray spectrometry [387, 648], proximal gamma-ray spectrometry [391, 616, 711], in-situ gamma-ray spectrometry, both on land [648, 688] and in the marine environment [689–692, 712], remote sensing in space exploration [700, 701, 713], gamma-ray astronomy [694, 695, 697, 699] as well as in AGRS in all the previous studies reviewed in Section 5.5.2.

Given the long history of successful applications of PSMC across various fields, including all previous Monte Carlo studies in AGRS, PSMC also forms the basis for the Monte Carlo model derived in this chapter. The main goal of this chapter is the derivation and validation of a high-fidelity PSMC model of the Swiss RLL AGRS detector system under laboratory conditions using the multi-purpose Monte Carlo code `FLUKA` [20, 216, 281]. The developed model was validated using benchmark radiation measurements conducted under laboratory conditions.

In this chapter and the following one, the discussion will be confined to centered source-detector configurations, with a primary emphasis on physics modeling and the analysis of spectral detector response. The angular detector response will be investigated later in Part III.

## 6.2 Methods

Since all the PSMC simulations in this chapter aim to reproduce the conducted radiation measurements, I will first present the experimental setup and postprocessing pipeline for creating spectral signatures and performing meta-model calibration, followed by the PSMC model and its postprocessing pipeline.

### 6.2.1 Radiation Measurements

All radiation measurements presented in this chapter were performed in the calibration laboratory at the Paul Scherrer Institute in Switzerland using the RLL spectrometer discussed in detail in Section 5.3.2 irradiated by seven different calibration radionuclide point sources ($^{57}_{27}$Co, $^{60}_{27}$Co, $^{88}_{39}$Y, $^{109}_{48}$Cd, $^{133}_{56}$Ba, $^{137}_{55}$Cs and $^{152}_{63}$Eu) manufactured by Eckert & Ziegler Nuclitec GmbH. These sources consist of a radionuclide carrying ion exchange sphere (diameter 1 mm) embedded in a 3 mm thick solid plastic disk with a diameter of 25 mm. Specific source properties and measurement specifications including the activity at the corresponding measurement date, the half-life and the measurement live time are listed in Table 6.1. The relative photon intensities $I_\gamma$ of these sources are compiled in the Figs. B.3–B.5 in Appendix B.

#### 6.2.1.1 Measurement Setup

The adopted experimental setup is visualized in Fig. 6.1. The calibration point sources were inserted in a low absorption source holder made of polylactide polymer (PLA) and mounted on a tripod. The source holder was specifically developed and constructed for these measurements using additive manufacturing techniques. The ion exchange resin sphere inside the calibration disk source was then aligned on the detector z-axis and placed at a fixed distance of 1 m to the detector box mounted on an aluminum frame. The reference datum was defined as the center of mass of the four scintillation crystals, projected to the outer surface of the detector box (cf. Fig. 6.1). To measure the source-detector distances and to position the sources accurately, distance as well as positioning laser systems were used.

Data acquisition was performed with the proprietary software suite `SpirIDENT` (cf. Section 5.3.2). Between the gross radiation measurements, background measurements were performed regularly for background correction and gain stability checks.

**Table 6.1** Physical properties of the calibration radionuclide point sources adopted for the radiation measurements.

| Nuclide | Half-life | Gross live time [h]★ | Background live time [h]• | Activity [Bq]○ | References |
|---|---|---|---|---|---|
| $^{57}_{27}$Co | $2.7181(4) \times 10^2$ d | 7.4 | 15 | $1.113(18) \times 10^5$ | [102] |
| $^{60}_{27}$Co | 5.2711(8) a | 3.7 | 61 | $3.08(5) \times 10^5$ | [68] |
| $^{88}_{39}$Y | $1.0663(5) \times 10^2$ d | 4.2 | 15 | $6.83(14) \times 10^5$ | [51] |
| $^{109}_{48}$Cd | $4.619(4) \times 10^2$ d | 15 | 16 | $7.38(15) \times 10^4$ | [102] |
| $^{133}_{56}$Ba | $1.0539(6) \times 10^1$ a | 6.3 | 61 | $2.152(32) \times 10^5$ | [102] |
| $^{137}_{55}$Cs | $3.0018(22) \times 10^1$ a | 4.0 | 14 | $2.266(34) \times 10^5$ | [68] |
| $^{152}_{63}$Eu | $1.3522(16) \times 10^1$ a | 3.8 | 16 | $1.973(30) \times 10^4$ | [714] |

★ Gross measurement live time $t_{gr}$ for the corresponding source rounded to two significant digits (cf. Section 4.3.1).
• Background measurement live time $t_b$ for the corresponding source rounded to two significant digits (cf. Section 4.3.1).
○ Source activity $\mathcal{A}$ computed at the corresponding measurement start date (cf. Eqs. 2.15 and 2.16) using the information provided by the calibration certificate as well as the half-life $t_{1/2}$ indicated in this table. More information on the uncertainty quantification is provided in Appendix A.8.1.

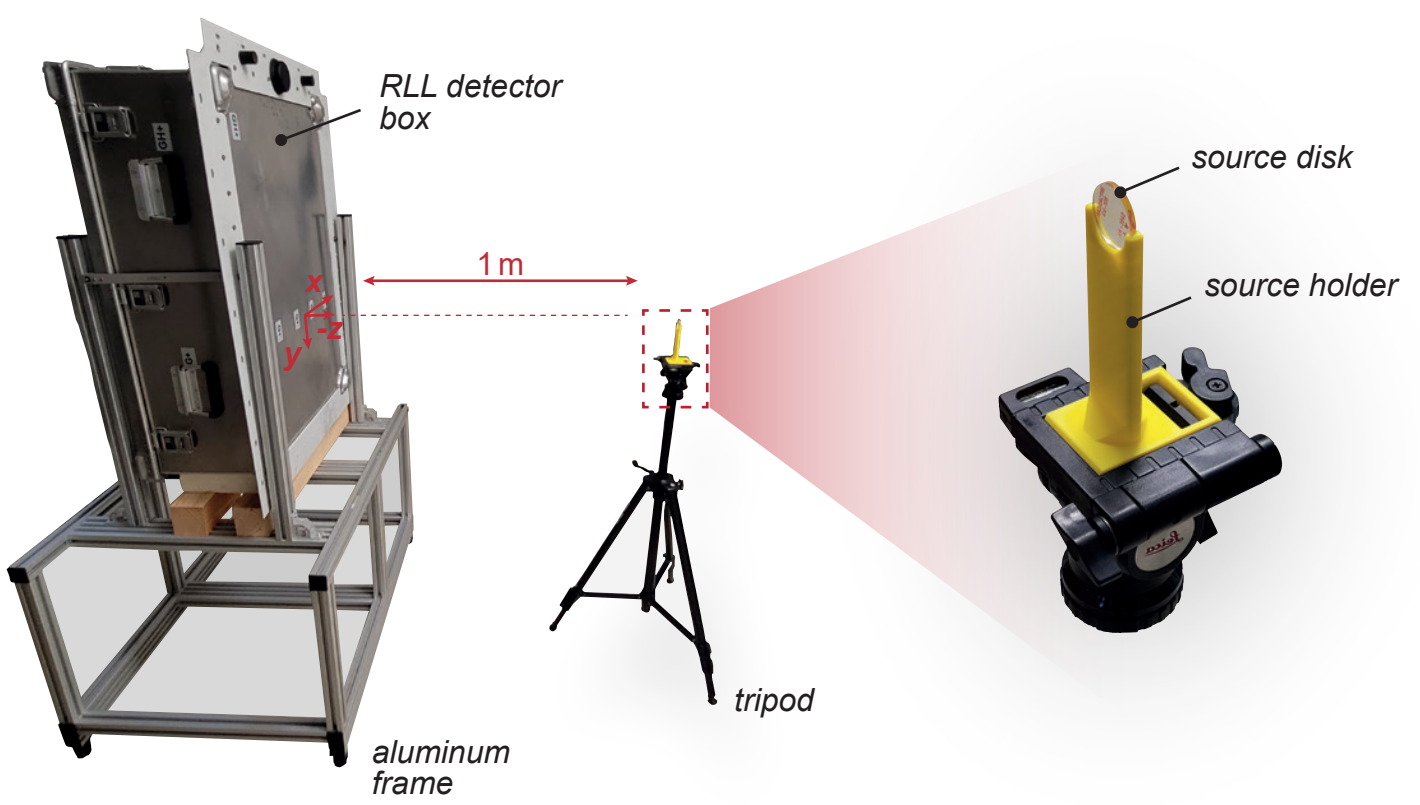


**Figure 6.1** Experimental setup of the laboratory-based radiation measurements. The main elements of the setup are the RLL detector box (90 cm × 64 cm × 35 cm, cf. also Fig. 5.1) mounted on an aluminum frame and the calibration point source placed on a tripod at a distance of 1 m on the central detector $z$-axis. The source consists of a radionuclide carrying ion exchange sphere (diameter 1 mm) embedded in a 25 mm × 3 mm solid plastic disk and inserted into a custom low absorption source holder made out of polylactide polymer (PLA) which itself is positioned on the tripod.

For all measurements, the air temperature in the calibration laboratory was controlled by an air conditioning unit and maintained at 18.8(6) °C. The relative humidity and the ambient air pressure were continuously monitored by external sensors, with mean values of 42(4) % and 982(6) hPa, respectively.

#### 6.2.1.2 Postprocessing Pipeline: `RLLSpec`

To postprocess the pulse-height spectra measured by the RLL spectrometer, a dedicated pipeline named `RLLSpec` was developed using the `MATLAB` code. As reviewed in Section 5.3.2, the pulse-height spectra are recorded with a sampling time of 1 s for each of the four scintillation crystals. These four acquisition channels will be referred to as detector channels `#1` through `#4`. In addition to these four single channels, there is an additional fifth channel `#SUM`, which is simply the sum of the channels `#1` through `#4`. The postprocessing steps of the `RLLSpec` pipeline, which I will describe in the following paragraphs, were conducted for each of the five detector channels separately. For the sake of brevity, I will not distinguish between the individual detector channels in the notation of the postprocessed quantities. Depending on the available measurement data, the `RLLSpec` pipeline executes the following four steps or a subset thereof:

1. **Count Vector Aggregation** As a first step, the set of gross pulse-height spectra $\mathbf{C}_{\mathrm{gr},k} \in \mathbb{N}^{N_{\mathrm{ch}} \times 1}$ (cf. Section 4.3) with $\mathcal{D}_{\mathrm{gr}} = \{k \in \mathbb{N}_+ \mid k \leq N_{\mathrm{gr}}, N_{\mathrm{gr}} \in \mathbb{N}_+\}$ and (if available) background pulse-height spectra $\mathbf{C}_{\mathrm{b},k} \in \mathbb{N}^{N_{\mathrm{ch}} \times 1}$ with $\mathcal{D}_{\mathrm{b}} = \{k \in \mathbb{N}_+ \mid k \leq N_{\mathrm{b}}, N_{\mathrm{b}} \in \mathbb{N}_+\}$ are summed over the subset of recorded spectra $\mathcal{D}_{\mathrm{gr}*} \subseteq \mathcal{D}_{\mathrm{gr}}$ an $\mathcal{D}_{\mathrm{b}*} \subseteq \mathcal{D}_{\mathrm{b}}$ for which we may assume no or negligible variation in the experimental condition $\mathfrak{d}$ discussed in Section 5.4. In this chapter, the conducted radiation measurements were designed in such a way that $\mathcal{D}_{\mathrm{gr}*} = \mathcal{D}_{\mathrm{gr}}$ an $\mathcal{D}_{\mathrm{b}*} = \mathcal{D}_{\mathrm{b}}$. Consequently, the gross and background count vectors associated with the conducted radiation measurements are given by:

$$\mathbf{C}_{\mathrm{gr}} = \sum_{k=1}^{N_{\mathrm{gr}}} \mathbf{C}_{\mathrm{gr},k} \tag{6.2a}$$

$$\mathbf{C}_\mathrm{b} = \sum_{k=1}^{N_\mathrm{b}} \mathbf{C}_{\mathrm{b},k} \tag{6.2b}$$

where:

| | |
|---|---|
| $\mathbf{C}_\mathrm{b}$ | background count vector |
| $\mathbf{C}_\mathrm{gr}$ | gross count vector |

2. **Count Rate Vector Computation** In a second step, dead-time corrected mean gross and background count rate pulse-height spectra are computed using the recorded live times $t_\mathrm{gr} = \sum_{k=1}^{N_\mathrm{gr}} t_{\mathrm{gr},k}$ and $t_\mathrm{b} = \sum_{k=1}^{N_\mathrm{b}} t_{\mathrm{b},k}$ (cf. also Eq. 4.29):

$$\mathbf{c}_\mathrm{gr} = \mathbf{C}_\mathrm{gr}/t_\mathrm{gr} \tag{6.3a}$$

$$\mathbf{c}_\mathrm{b} = \mathbf{C}_\mathrm{b}/t_\mathrm{b} \tag{6.3b}$$

where:

| | | |
|---|---|---|
| $\mathbf{c}_\mathrm{b}$ | background count rate vector | $\mathrm{s}^{-1}$ |
| $\mathbf{c}_\mathrm{gr}$ | gross count rate vector | $\mathrm{s}^{-1}$ |
| $t_\mathrm{b}$ | background measurement live time | s |
| $t_\mathrm{gr}$ | gross measurement live time | s |

The individual live times of the gross and background measurements conducted in this chapter are provided in Table 6.1.

3. **Background Correction** In a third step, given background measurements are available, dead-time and background corrected mean net count rate pulse-height spectra $\mathbf{c}_\mathrm{net} \in \mathbb{R}_+^{N_\mathrm{ch}\times 1}$ can be estimated:

$$\mathbf{c}_\mathrm{net} = \mathbf{c}_\mathrm{gr} - \mathbf{c}_\mathrm{b} \tag{6.4}$$

4. **Spectral Signature Computation** If the source strength $\xi$ of the adopted source[1] during the gross measurement is known, the mean spectral signature can be computed by normalizing the mean net count rate pulse-height spectra in Eq. 6.4 by the corresponding source strength $\xi$ (cf. Section 5.4.1):[2]

*1 Note that I adopt again the notation of a generic source with associated source strength $\xi$ to account for the different source scenarios detailed in Section 5.4.1.*

*2 Here, we implicitly assume that the source strength remains constant throughout the measurement. For radionuclide sources, the source strength varies over time, as detailed in Section 2.1.2. However, if the measurement duration is short relative to the source's half-life, the variation in the source strength during the measurements can be considered negligible (cf. Appendix A.8.1.4).*

$$\hat{\mathbf{c}}_{\mathrm{exp}} = \mathbf{c}_{\mathrm{net}} \xi^{-1} \tag{6.5}$$

with:

| | | |
|---|---|---|
| $\hat{\mathbf{c}}_{\mathrm{exp}}$ | measured mean spectral signature | $\mathrm{s}^{-1}\,[\xi]^{-1}$ |
| $\xi$ | source strength | $[\xi]$ |

For the radionuclide sources adopted in this chapter, the source strength $\xi$ is equivalent to the source activity $\mathcal{A}$ (cf. Eq. 2.9), resulting in the measured mean spectral signature $\hat{\mathbf{c}}_{\mathrm{exp}}$ being expressed in units of $\mathrm{s}^{-1}\,\mathrm{Bq}^{-1}$. The activity $\mathcal{A}$ for the adopted radionuclide sources can be easily estimated using the source calibration data and the model provided in Eq. 2.15 (cf. also to Section 2.1.2). Resulting mean activities alongside uncertainty estimates are provided in Table 6.1.

For all quantities discussed in this subsection, the `RLLSpec` pipeline provides uncertainty estimates characterizing both statistical and systematic contributions. More information on these uncertainty estimates is provided in Appendix A.8.1.

#### 6.2.1.3 Calibration Pipeline: `RLLCal`

As already indicated in Section 5.5.2, no numerical calibration method is completely free of the need for some radiation measurements to determine hyperparameters in the numerical models. For the PSMC model adopted in this chapter, there are two main meta-models with associated hyperparameters, which need to be derived from radiation measurements:

I. **Spectral energy** Model relating the continuous pulse-height channel number $\tilde{n}$ (definition given in Section 4.3) to the spectral energy $E'$ reviewed in Section 4.3.3.1.

II. **Spectral resolution** A model to characterize the spectral resolution $R_E$ analyzed in the Sections 4.3.2 and 4.3.3.2 as a function of the continuous pulse-height channel number $\tilde{n}$.

A dedicated pipeline, named `RLLCal`, was developed using the `MATLAB` code to derive these meta-models from radiation measurements. This pipeline is applied separately to each of the four single

3 *As I will discuss later, only the calibration models for the four single detector channels are used in the postprocessing of the simulation data. The calibration models for the detector channel `#SUM` are solely used for visualization purposes.*

detector channels `#1` through `#4` and to the detector channel `#SUM`.[3] Since the general spectral calibration methodology has already been outlined in Section 4.3.3, this section will focus on the specific modeling details employed by the `RLLCal` pipeline. For a comprehensive overview of the spectral calibration methodology, please refer to Section 4.3.3. The `RLLCal` pipeline executes the following three steps:

1. **Peak Fitting** The first step in the `RLLCal` pipeline is the extraction of the centroid and FWHM dispersion parameters from a set of selected FEPs in measured net pulse-height spectra. In this work, I have adopted Gaussian singlet and multiplet peak models (cf. Eq. 4.37) combined with a physics-oriented empirical baseline correction [380, 381] to extract these parameters. The regression analysis is performed by weighted non-linear least-squares (WNLLS) regression using the interior-reflective Newton method[4] [715].[5] The selected FEPs listed in Table C.5 cover a spectral range between ~88 keV and ~2615 keV. Peak fit results are attached in the Figs. B.20–B.30 in Appendix B for all five detector channels.

2. **Energy Calibration** As discussed in Section 5.3.2, the RLL spectrometer features automatic spectrum linearization with offset correction. As a result, a simple proportional energy calibration model has proven sufficient to characterize the spectral energy $E'$ as a function of the continuous pulse-height channel number $\tilde{n}$ for the specific detector channel:

$$E' = \Delta E' \tilde{n} \tag{6.6}$$

where:

| | | |
|---|---|---|
| $E'$ | spectral energy | eV |
| $\Delta E'$ | spectral energy bin width | eV |
| $\tilde{n}$ | continuous pulse-height channel number | |

Weighted linear least-squares (WLLS) regression is applied to derive the spectral energy bin width $\Delta E'$ for each of the five detector channels using the centroid data from the first step.[6] The derived energy calibration models are displayed in Fig. 6.2. The associated model parameters $\Delta E'$ are summarized in Table C.6.

4 *Named after Sir Isaac Newton (*1642, †1727), an English physicist and mathematician who revolutionized science with his laws of motion and universal gravitation, laying the foundation for classical mechanics. He also made significant contributions to optics and co-invented calculus.*

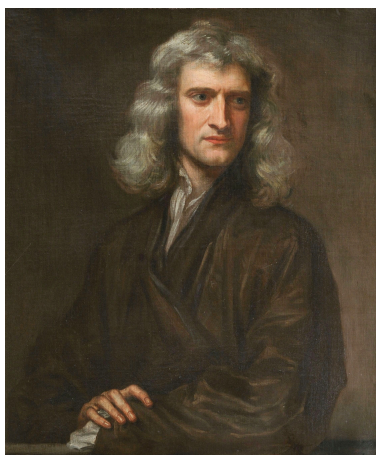

Sir Isaac Newton
Ⓢ Godfrey Kneller

5 *I applied the `fitnlm` function from the MATLAB code to perform this regression analysis.*

6 *I applied the `fitlm` function from the MATLAB code to perform this regression analysis.*

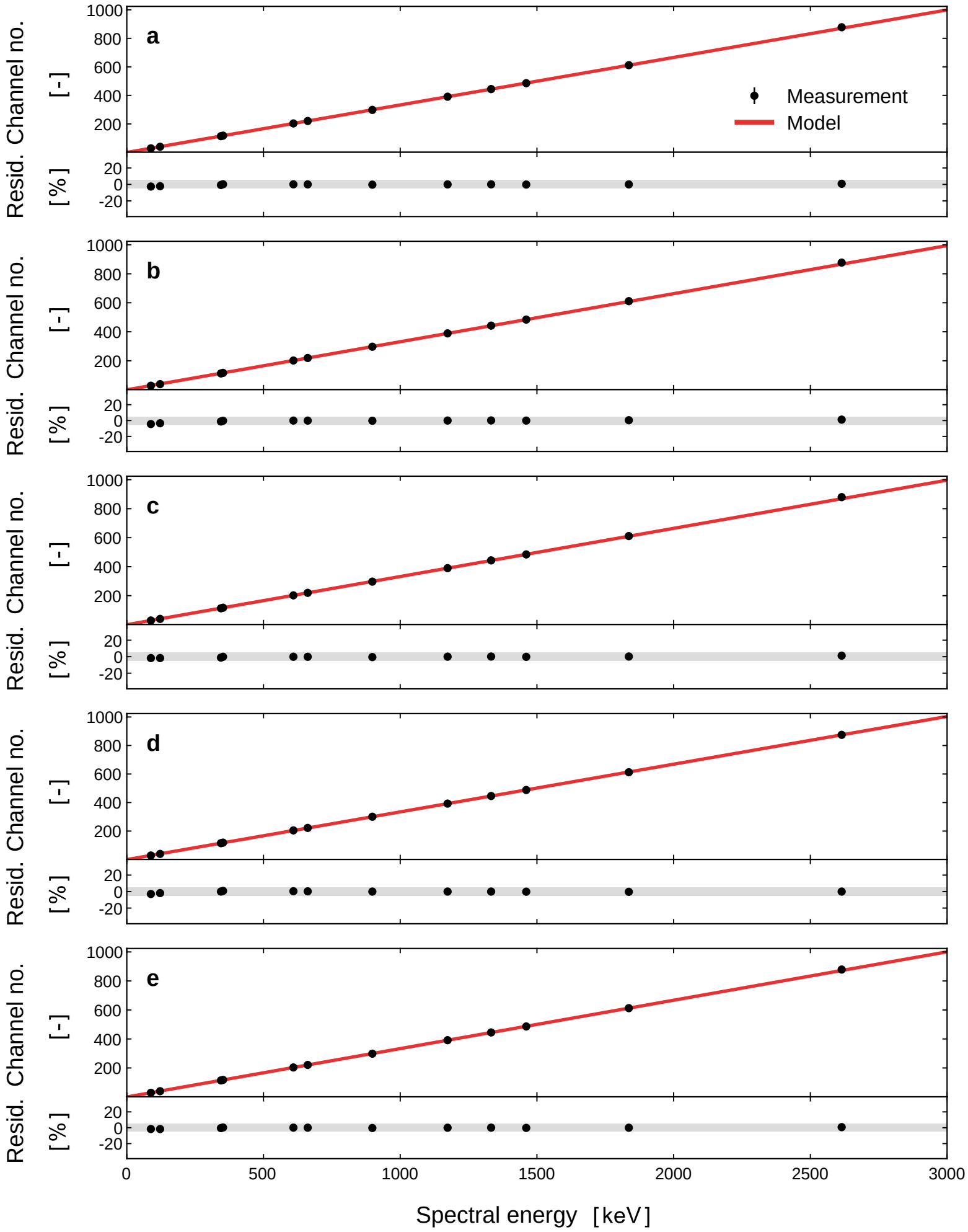


**Figure 6.2** Energy calibration models relating the spectral energy $E'$ to the continuous pulse-height channel number $\tilde{n}$ for the four single detector channels #1 through #4 (**a–d**) and the detector channel #SUM (**e**). Measurement uncertainties are provided as 1 standard deviation (SD) error bars (hidden by the marker size). The model uncertainty is characterized by 99 % prediction intervals displayed as red-shaded areas (hidden by the line width). In addition, the relative residuals (Resid., normalized by the mean measured values) are provided for each detector channel with the 5 % band highlighted.

3. **Resolution Calibration** As discussed in Section 4.3.3.2, selecting a spectral resolution model for inorganic scintillator spectrometers is nontrivial. Detailed knowledge of the individual variance terms discussed in Section 4.3.2 is required to derive a physics-informed model. As this knowledge is often not available in practice, empirical models are used instead to describe the relationship between the continuous pulse-height channel number $\tilde{n}$ and the spectral resolution $R_E$ [386]. I used the predicted residual error sum of squares (PRESS) statistic [716, 717] to compare several empirical models adopted in the past to describe the spectral resolution of inorganic scintillators [386]. The power-law model (Eq. 4.39) introduced in Section 4.3.3.2, similar to previous studies [387–391], has proven to be the best-performing model based on the PRESS statistic (cf. Table C.7). Therefore, this model has been selected to characterize the spectral resolution in this work. I apply again WNLLS regression to derive the parameters of the spectral resolution model, that is the scale coefficient $a_1$ and the power coefficient $a_2$, for each of the five detector channels using the FWHM dispersion data from the first step of the `RLLCal` pipeline (cf. Section 4.3.3.2).[7] The resulting calibrated models are displayed in Fig. 6.3. Derived model parameters are summarized in Table C.6.

7 *I applied again the `fitnlm` function from the `MATLAB` code to perform this regression analysis.*

### 6.2.2 Monte Carlo Simulations

Monte Carlo simulations in this chapter are conducted using the multi-purpose `FLUKA` code[8] maintained by the FLUKA.CERN Collaboration [20, 216, 281] together with the graphical interface `FLAIR`[9] [718]. The `FLUKA` code adopts high-fidelity and thoroughly benchmarked physics models to accurately simulate the interaction and coupled transport of over 60 different particles, including photons, electrons, muons, hadrons and heavy ions in arbitrary materials from 1 keV ($10^{-5}$ eV for neutrons) up to ultrarelativistic energies (20 TeV for hadrons, 1 PeV for electrons and muons, 10 PeV for heavy ions[10] and photons) [216]. This allows me to perform not only high-fidelity simulations of the transport of photons coupled with electrons and positrons but also the transport of hadrons and heavy ions, which will become important later in this book. All Monte Carlo simulations were performed on a local computer cluster (7 nodes with a total number of 520 cores at a nominal clock speed of 2.6 GHz) at the Paul Scherrer Institute (PSI) utilizing parallel computing.

8 *Version `4-2.1` for the simulations conducted in this and in the next chapter.*

9 *Version `3.1-15.1` for the simulations conducted in this and in the next chapter.*

10 *per nucleon*

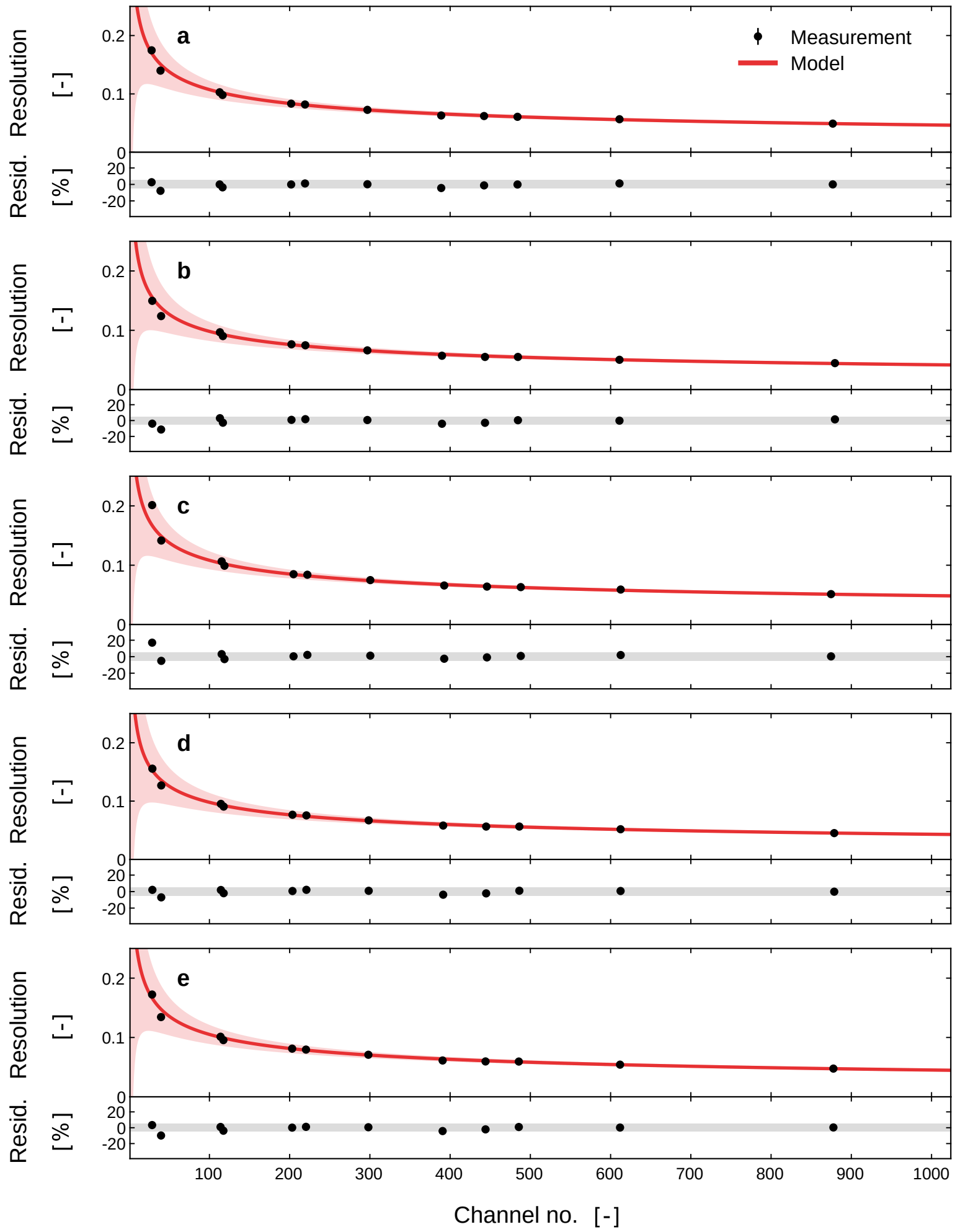


**Figure 6.3** Spectral resolution models relating the continuous pulse-height channel number $\tilde{n}$ to the spectral resolution $R_E$ for the four single detector channels `#1` through `#4` (**a**–**d**) and the detector channel `#SUM` (**e**). Measurement uncertainties are provided as 1 standard deviation (SD) error bars (for most data points hidden by the marker size). The model uncertainty is characterized by 99 % prediction intervals displayed as red-shaded areas. In addition, the relative residuals (Resid., normalized by the mean measured values) are provided for each detector channel with the 5 % band highlighted.

General principles of the Monte Carlo method were reviewed already in Section 3.2.2. Consequently, I focus in this section on the setup for the conducted simulations and the pipeline developed to postprocess the simulation data. For a comprehensive overview of the Monte Carlo simulation principles, please refer to Section 4.3.3.

#### 6.2.2.1 Physics Models

The most accurate physics mode available in `FLUKA`, `precisio`, was used featuring a fully coupled photon, electron and positron radiation transport. Lower transport thresholds of 1 keV were applied to the scintillation crystals and their immediate surroundings, including reflectors, optical windows and aluminum casings, for all four crystals. For other components in the model, the transport threshold was set to 10 keV to optimize computational efficiency while maintaining accurate transport simulation within the scintillation crystals (cf. Section 3.2.2.1). For the adopted calibration radionuclide point sources in the radiation measurements discussed in Section 6.2.1, detailed photon, electron and positron emissions are simulated using the `raddecay` card in a semi-analogue mode according to the corresponding decay schemes (cf. Section 2.1.1).[11]

11 *It is important to add that in the adopted `FLUKA` code version `3.1-15.1`, an erroneous implementation of the photon emission spectrum of* $^{109}_{48}$Cd *was detected and subsequently confirmed by the developers. Therefore, a custom source user routine was written using the emission data from the ENDF/B-VIII.0 nuclear data file library [44] to simulate the* $^{109}_{48}$Cd *source adequately [719].*

#### 6.2.2.2 Scoring

As discussed at the beginning of this chapter, for a traditional PSMC model, the detector response simulation of inorganic scintillators reduces to an energy deposition problem. A custom user routine `usreou` was developed and combined with the `detect` card to score the energy deposition events in each of the four scintillation crystals individually on an event-by-event basis [719]. The number of primaries $N_{\text{pr}}$ was set to $10^7$ for all simulations, which guarantees a median coefficient of variation med($\mathbf{CV}_{\text{stat}}$) < 1 % and a median relative variance of the variance med($\mathbf{VOV}$) < 0.01 % over the spectral domain of interest (SDOI) for all five detector channels (`#1` through `#4` and `#SUM`) (cf. Appendix A.8). This domain is defined as follows:

**Definition 6.1 ► Spectral Domain of Interest (SDOI)**

The spectral domain of interest (SDOI) $\mathcal{D}_{\text{SDOI}}$ for a given spectral signature $\hat{\mathbf{c}}$ is defined as the spectral band with continuous pulse-height channel numbers $\{\tilde{n} \in \mathbb{R}_+ \mid \mu_{\text{LLD}} + 2\sigma_{\text{LLD}} \leq \tilde{n} \leq \max(E_\gamma)/\Delta E' + 2\sigma_{\mathcal{H}}\}$. Here, $\max(E_\gamma)$ represents the maximum photon source energy, $\Delta E'$ the constant spectral energy bin width and $\sigma_{\mathcal{H}}$ the standard deviation of the associated FEP in the pulse-height spectrum. For practical reasons, only emissions with a relative photon intensity $I_\gamma$ over 1 % are considered for radionuclide sources. The lower limit of $\mathcal{D}_{\text{SDOI}}$ is defined by the mean continuous pulse-height channel number $\mu_{\text{LLD}}$ and standard deviation $\sigma_{\text{LLD}}$ associated with the lower-level discriminator (LLD) in the given pulse-height spectrum.

#### 6.2.2.3 Mass Model

A detailed mass model of the experimental setup described in Section 6.2.1.1 was developed using `FLAIR` and `FLUKA`'s core Combinatorial Solid Geometry (CSG) builder. The simulation model illustrated in Fig. 6.4 includes all relevant detector components such as scintillation crystals, reflectors, aluminum casings, insulation materials, PMTs and detector electronics in high detail. Geometrical as well as material properties were provided by the manufacturer of the Swiss AGRS system. The calibration disk sources together with the source holder are modeled in high detail with known material and geometrical properties. On the other hand, the laboratory room together with additional instruments and equipment are modeled in less detail. For these simplifications, care was taken to preserve the overall opacity as well as the mass density. The composition and mass density of the air in the laboratory was approximated as dry air near sea level [249].[12]

12 *As I will demonstrate later in this chapter, the expected changes in air density resulting from the laboratory's elevation, as well as the temporal variations in air temperature, pressure, and humidity, have a negligible impact on the spectral signature.*

#### 6.2.2.4 Postprocessing Pipeline: `PScinMC`

Similar to the postprocessing of the measurement data, a dedicated pipeline named `PScinMC` was developed using the `MATLAB` code to postprocess the Monte Carlo simulation data, that is the set of energy deposition events $\mathbf{E}_{\text{dep}} \in \mathbb{R}_+^{N_{\text{dep}} \times 1}$ for each of the four scintillation crystals with $N_{\text{dep}}$ being the number of scored energy deposition

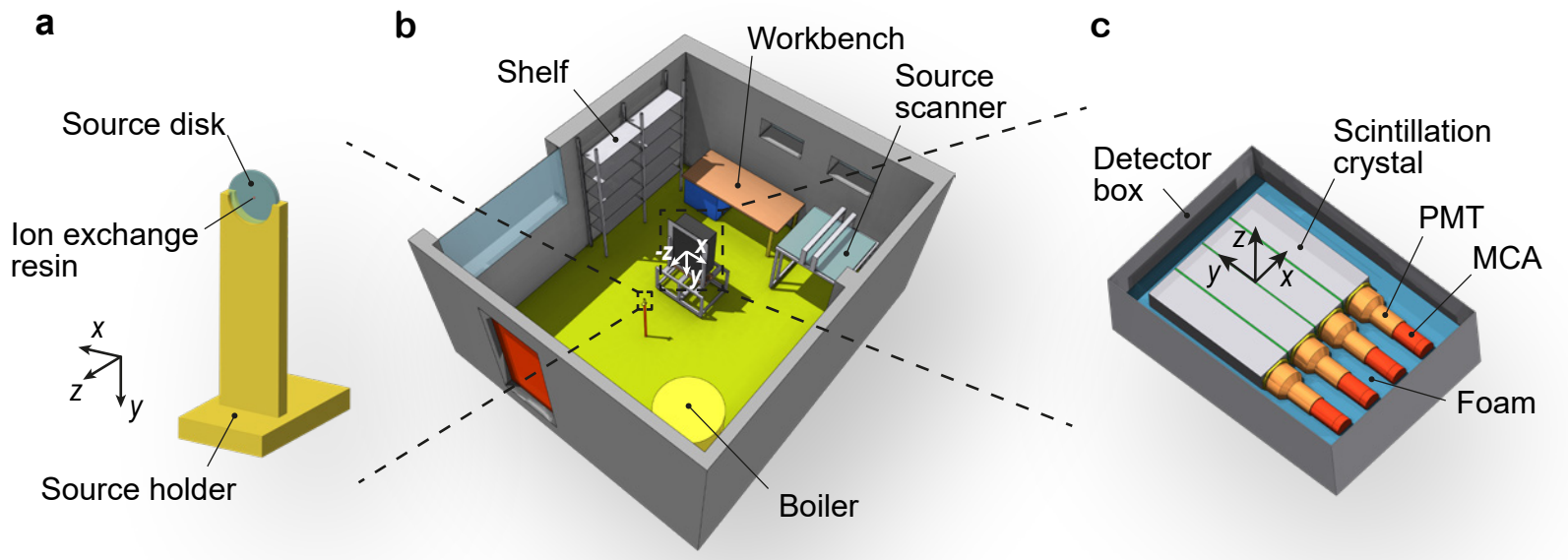


**Figure 6.4** Monte Carlo mass model of the laboratory-based experimental setup. **a** Zoomed view of the adopted radiation sources consisting of a radionuclide carrying ion exchange sphere (diameter 1 mm) embedded in a 25 mm × 3 mm solid plastic disc and inserted into a custom low absorption source holder made out of PLA. **b** Cut view of the calibration laboratory room (inner room dimensions 5.3 m × 4.5 m × 3 m) at the PSI with the detector mounted on an aluminum frame and the source holder placed on a tripod in a fixed distance of 1 m to the detector front on the central detector $x$-axis. **c** Cut view of the RLL detector box (90 cm × 64 cm × 35 cm) including the individual detector components such as the scintillation crystals, photomultiplier tubes (PMT) or multi-channel analyzers (MCA) embedded in a thermal-insulating and vibration-damping polyethylene (PE) foam protected by a rugged aluminum detector box (cf. Section 5.3.2). The mass model figures were created using the graphical interface `FLAIR` [718]. For better visibility and interpretability, false colors were applied.

events. The `PScinMC` pipeline is applied to each of the four single detector channels `#1` through `#4` separately. The `#SUM` detector channel is obtained by summing the derived pulse-height spectra of the four individual single detector channels. Consequently, as noted before, only the calibration models for the four single detector channels are used in the postprocessing of the simulation data (cf. Section 6.2.1.3). The `PScinMC` pipeline reconstructs the spectral signature for a given set of energy deposition events $\mathbf{E}_{\text{dep}}$ in four distinct steps:

1. **Pulse-Height Channel Conversion** In the first step, the scored deposited energies $E_{\text{dep},i}$ with $\{i \in \mathbb{N}_+ \mid i \leq N_{\text{dep}}, N_{\text{dep}} \in \mathbb{N}_+\}$ are transformed to continuous pulse-height channel numbers $\tilde{n}^*_{\text{dep}} \in \mathbb{R}_+$ using the energy calibration models (Eq. 6.6) for the

corresponding detector channel derived by the `RLLCal` pipeline (cf. Section 6.2.1.3):

$$\tilde{n}^*_{\text{dep},i} = E_{\text{dep},i} \Delta E'^{-1} \tag{6.7}$$

where:

| | | |
|---|---|---|
| $E_{\text{dep}}$ | deposited energy | eV |
| $\Delta E'$ | spectral energy bin width | eV |
| $\tilde{n}^*_{\text{dep}}$ | continuous raw pulse-height channel number of an energy deposition event | |

2. **Broadening & Binning** In the second step, to account for the finite detector resolution, the converted individual energy deposition events $\tilde{n}^*_{\text{dep},i}$ are convoluted with the spectral resolution models derived by the `RLLCal` pipeline. As noted above, this is done for each detector channel individually with the related spectral resolution model displayed in Fig. 6.3. Like in previous studies [18, 27, 388, 389, 391, 616, 648, 689, 700, 701, 711–713, 720], the energy deposition events $\tilde{n}_{\text{dep},i}$ are assumed to be normally distributed:

$$\tilde{\mathfrak{N}}_{\text{dep}} \sim \mathcal{N}\left(\tilde{n}_{\text{dep}} \mid \tilde{n}^*_{\text{dep}}, \sigma^2_E(\tilde{n}^*_{\text{dep}})\right) \tag{6.8}$$

with a mean equal to the simulated raw energy deposition $\tilde{n}^*_{\text{dep}}$ given by Eq. 6.7 and a standard deviation $\sigma_E$ given by the spectral resolution model Eq. 4.39 [30]:

$$\sigma_E(\tilde{n}^*_{\text{dep}}) = \frac{R_E(\tilde{n}^*_{\text{dep}})\, \tilde{n}^*_{\text{dep}}}{2\sqrt{2\log 2}} \tag{6.9a}$$

$$= \frac{a_1\, (\tilde{n}^*_{\text{dep}})^{a_2+1}}{2\sqrt{2\log 2}} \tag{6.9b}$$

where:

| | |
|---|---|
| $a_1$ | scale coefficient |
| $a_2$ | power coefficient |
| $R_E$ | spectral resolution |

The convolution, sometimes also referred to as Gaussian broadening [27, 648, 689, 712, 720], is performed in parallel with the binning of the simulated energy deposition events $\tilde{n}^*_{\text{dep},i}$ to generate the simulated mean number of scintillation pulses $C_n$ in pulse-height channel $n$ with $\{n \in \mathbb{N}_+ \mid n \leq N_{\text{ch}}, N_{\text{ch}} \in \mathbb{N}_+\}$:

$$C_n = \sum_{i=1}^{N_{\text{dep}}} \int_{n-\frac{1}{2}}^{n+\frac{1}{2}} \frac{1}{\sigma_E(\tilde{n}^*_{\text{dep},i})} \varphi\left(\frac{\tilde{n} - \tilde{n}^*_{\text{dep},i}}{\sigma_E(\tilde{n}^*_{\text{dep},i})}\right) \mathrm{d}\tilde{n} \tag{6.10a}$$

$$= \sum_{i=1}^{N_{\text{dep}}} \Phi\left(\frac{n + \frac{1}{2} - \tilde{n}^*_{\text{dep},i}}{\sigma_E(\tilde{n}^*_{\text{dep},i})}\right) - \Phi\left(\frac{n - \frac{1}{2} - \tilde{n}^*_{\text{dep},i}}{\sigma_E(\tilde{n}^*_{\text{dep},i})}\right) \tag{6.10b}$$

with the standard normal probability density function $\varphi(\cdot)$ and standard normal cumulative distribution function $\Phi(\cdot)$. Considering that typical high-fidelity simulations require $N_{\text{dep}} \geq 10^5$, sequential summation of the individual energy deposition events in Eqs. 6.10a and 6.10b is computationally inefficient, in particular, if the convolution is performed multiple times for identical energy deposition events such as FEP or AP events (cf. Section 4.3.1). Therefore, I apply a modified discrete convolution operation in `PScinMC` by discretizing the continuous $\tilde{n}$ domain introducing refined pulse-height channel numbers $\tilde{n}^\dagger_j = (j-1)\Delta\tilde{n}^\dagger$ with refined bin width $\Delta\tilde{n}^\dagger$ and index $\{j \in \mathbb{N}_+ \mid j \leq N^\dagger_{\text{ch}}, N^\dagger_{\text{ch}} \in \mathbb{N}_+\}$. To cover the full spectral domain, the number of refined bins $N^\dagger_{\text{ch}}$ should be $N^\dagger_{\text{ch}} > N_{\text{ch}}/\Delta\tilde{n}^\dagger$. In this work, a refined bin width of $\Delta\tilde{n}^\dagger = 10^{-1}$ and $N^\dagger_{\text{ch}} = (N_{\text{ch}} + 30)/\Delta\tilde{n}^\dagger$ was applied. Using this discretization scheme, we can rewrite Eq. 6.10b in matrix form as follows:

$$\mathbf{C} = \mathbf{G}\mathbf{C}^\dagger \tag{6.11}$$

where:

| | |
|---|---|
| $\mathbf{C}$ | mean count vector |
| $\mathbf{C}^\dagger$ | prebinned count vector |
| $\mathbf{G}$ | Gaussian weight matrix |

As the name implies, $\mathbf{C} \in \mathbb{R}_{+}^{N_{\text{ch}} \times 1}$ represents the mean number of energy deposition events in the individual pulse-height channels. In contrast, $\mathbf{C}^{\dagger} \in \mathbb{N}^{N_{\text{ch}}^{\dagger} \times 1}$ denotes the prebinned count vector of the simulated energy deposition events using the refined pulse-height channel numbers $\tilde{n}^{\dagger}$, i.e. the number of simulated energy deposition events in each refined pulse-height channel $j$ binned by $\tilde{n}_j^{\dagger} - \Delta\tilde{n}^{\dagger}/2 \leq \tilde{n}_{\text{dep}}^{*} \leq \tilde{n}_j^{\dagger} + \Delta\tilde{n}^{\dagger}/2$. The elements of the Gaussian weight matrix $\mathbf{G}$ are computed as follows:

$$G_{n,j} = \Phi\left(\frac{n + \frac{1}{2} - \tilde{n}_j^{\dagger}}{\sigma_E(\tilde{n}_j^{\dagger})}\right) - \Phi\left(\frac{n - \frac{1}{2} - \tilde{n}_j^{\dagger}}{\sigma_E(\tilde{n}_j^{\dagger})}\right) \tag{6.12}$$

with the index $\{j \in \mathbb{N}_{+} \mid j \leq N_{\text{ch}}^{\dagger}, N_{\text{ch}}^{\dagger} \in \mathbb{N}_{+}\}$ of the newly introduced refined pulse-height channel number $\tilde{n}^{\dagger}$ and the (original) pulse-height channel number $\{n \in \mathbb{N}_{+} \mid n \leq N_{\text{ch}}, N_{\text{ch}} \in \mathbb{N}_{+}\}$ giving $\mathbf{G} \in \mathbb{R}_{+}^{N_{\text{ch}} \times N_{\text{ch}}^{\dagger}}$.

Using the matrix form in Eq. 6.11 and the outlined discretization scheme, the computational efficiency of the numerical convolution operation can be improved by a factor $> \mathcal{O}(10^2)$ with discretization errors kept below <1 % in the relevant spectral domain $\mathcal{D}_{\text{SDOI}}$ (cf. Fig. B.31). This substantial acceleration will become important in the subsequent chapters of this book.

3. **Normalization** To obtain spectral signature estimates, the mean count vector $\mathbf{C}$ of the previous step needs to be normalized by the integrated source strength $\Xi$:[13]

$$\hat{\mathbf{c}}_{\text{sim}} = \mathbf{C}\,\Xi^{-1} \tag{6.13}$$

where:

| | | |
|---|---|---|
| $\Xi$ | integrated source strength | $[\xi]\,\text{s}$ |
| $\mathbf{C}$ | mean count vector | |
| $\hat{\mathbf{c}}_{\text{sim}}$ | simulated mean spectral signature | $\text{s}^{-1}\,[\xi]^{-1}$ |

The integrated source strength $\Xi$ is defined as:

$$\Xi = \int_{\Delta t} \xi \, \mathrm{d}t \tag{6.14}$$

13 *Similar to Eq. 6.5 in Section 6.2.1.2, I adopt again the notation of a generic source with associated integrated source strength $\Xi$ to account for the different source scenarios detailed in Section 5.4.1.*

with:

| | | |
|---|---|---|
| $\xi$ | source strength | $[\xi]$ |

In Monte Carlo radiation transport simulations, $\Xi$ is equivalent to the number of primaries $N_{pr}$, that is the number of simulated source histories (cf. Section 3.2.2.4), normalized by a source specific factor $F_{src}$ characterizing the detailed source geometry and/or the source matter:

$$\Xi = N_{pr} F_{src}^{-1} \tag{6.15}$$

with:

| | | |
|---|---|---|
| $F_{src}$ | source geometry-matter factor | $[F_{src}]$ |
| $N_{pr}$ | number of primaries | |

Here, I list factors for the most prominent radionuclide sources adopted in Monte Carlo simulations and resulting units for the simulated mean spectral signature:

$$F_{src} = \begin{cases} 1 & [\hat{\mathbf{c}}_{sim}] = \frac{1}{\mathrm{s\,Bq}} & \text{point source} \quad (6.16a) \\ A_{src} & [\hat{\mathbf{c}}_{sim}] = \frac{\mathrm{m}^2}{\mathrm{s\,Bq}} & \text{surface source} \quad (6.16b) \\ V_{src} & [\hat{\mathbf{c}}_{sim}] = \frac{\mathrm{m}^3}{\mathrm{s\,Bq}} & \text{volume source} \quad (6.16c) \\ V_{src}\rho_{src} & [\hat{\mathbf{c}}_{sim}] = \frac{\mathrm{kg}}{\mathrm{s\,Bq}} & \text{volume source} \quad (6.16d) \end{cases}$$

where:

| | | |
|---|---|---|
| $A_{src}$ | source area | $\mathrm{m}^2$ |
| $V_{src}$ | source volume | $\mathrm{m}^3$ |
| $\rho_{src}$ | source mass density | $\mathrm{kg\,m}^{-3}$ |

As we adopt calibration radionuclide point sources in this chapter, we have $F_{src} = 1$ and $[\hat{\mathbf{c}}_{sim}] = \mathrm{s}^{-1}\,\mathrm{Bq}^{-1}$ inline with the dimension of the measured mean spectral signature $\hat{\mathbf{c}}_{exp}$ estimate obtained from the `RLLSpec` pipeline (cf. Section 6.2.1.2).

4. **LLD Correction** As a last step in the PScinMC pipeline, the simulated mean spectral signature $\hat{\mathbf{c}}_{\text{sim}}$ estimated by Eq. 6.13 is corrected for the lower-level discriminator (LLD) effect discussed in Section 4.3 using a heuristic Gaussian model:

$$\hat{c}_{\text{sim,LLD},n} = \hat{c}_{\text{sim},n}\,\Phi\left(\frac{n - \mu_{\text{LLD}}}{\sigma_{\text{LLD}}}\right) \tag{6.17}$$

where:

| | | |
|---|---|---|
| $\hat{c}_{\text{sim}}$ | simulated mean spectral signature element | $\text{s}^{-1}\,[\xi]^{-1}$ |
| $n$ | pulse-height channel number | |
| $\mu_{\text{LLD}}$ | mean lower-level discriminator | |
| $\sigma_{\text{LLD}}$ | lower-level discriminator standard deviation | |

and with $\Phi(\cdot)$ denoting again the standard normal cumulative distribution function. This heuristic model is motivated by the idea that the fixed lower level threshold $\mathcal{H}_{\text{min}}$ (cf. Section 4.3) is fluctuating due to inherent electronic noise in the MCAs and associated electronics resulting in an approximately normally distributed cut-off with the mean $\mu_{\text{LLD}}$ corresponding to the predefined LLD threshold and a standard deviation $\sigma_{\text{LLD}}$ determined by the level of electronic noise. The model-specific parameters, i.e. $\mu_{\text{LLD}}$ and $\sigma_{\text{LLD}}$ in Eq. 6.17, are calibrated for the four detector channels #1 through #4 using global optimization techniques (cf. Appendix A.9). Depending on the detector channel, the mean lower-level discriminator is found to be between ~10 keV and ~20 keV with a standard deviation between ~3 keV and ~6 keV (cf. Fig. B.32).[14] For notational convenience, the corrected mean spectral signature $\hat{\mathbf{c}}_{\text{sim,LLD}}$ will be denoted by $\hat{\mathbf{c}}_{\text{sim}}$ in the remainder of this book.

Like for the RLLSpec pipeline, the PScinMC pipeline provides also uncertainty estimates for the derived quantities presented in this subsection characterizing both statistical and systematic contributions. More information on these uncertainty estimates is provided in Appendix A.8.2.

14 *As indicated in Eq. 6.17, $\mu_{\text{LLD}}$ and $\sigma_{\text{LLD}}$ are expressed and applied as continuous pulse-height channel numbers. Here, to simplify the interpretation, I report these values as spectral energies using the energy calibration models derived by the* RLLCal *pipeline (cf. Section 6.2.1.3).*

## 6.3 Results & Discussion

As outlined in Section 6.2.1.2, the RLL spectrometer features five different detector channels, i.e. four single channels #1 through #4 corresponding to the individual scintillation detectors and the sum channel #SUM representing the combined signal of all four single channels. The developed PSMC model and associated postprocessing pipeline PScinMC presented in Section 6.2.2 provide spectral signature estimates for all five detector channels.

In this section, I present the measured and simulated results specifically for the detector channel #SUM. However, it is important to note that all four single detector channels were also carefully evaluated in parallel. Corresponding results for these single channels are provided in Appendix B. Where applicable, I will cross-reference relevant findings from these single detector channels.

### 6.3.1 Spectral Signature Analysis

Seven different calibration radionuclide point sources ($^{57}_{27}$Co, $^{60}_{27}$Co, $^{88}_{39}$Y, $^{109}_{48}$Cd, $^{133}_{56}$Ba, $^{137}_{55}$Cs and $^{152}_{63}$Eu) were used to validate the PSMC model of the Swiss RLL AGRS detector system under laboratory conditions. Measurement and model setup as well as the postprocessing pipelines RLLSpec and PScinMC were described in Section 6.2.1 and Section 6.2.2, respectively.

The resulting measured ($\hat{c}_{\mathrm{exp}}$) and simulated ($\hat{c}_{\mathrm{sim}}$) spectral signatures for all seven sources are presented in the Figs. 6.5–6.8 alongside uncertainty estimates (cf. Appendix A.8) and relative deviations computed as $|\hat{c}_{\mathrm{sim}} - \hat{c}_{\mathrm{exp}}|/\hat{c}_{\mathrm{exp}}$. To quantify the statistical distribution of the relative deviations, adjusted box plots [721, 722] were computed for the different sources and $n \in \mathcal{D}_{\mathrm{SDOI}}$ (cf. Def. 6.1) using the Library for Robust Analysis (LIBRA) code [723, 724]. These results are shown in Fig. 6.9.

Given the substantial deviations in previous studies [18, 27], the obtained spectral results show a significantly improved agreement between measured and simulated spectral signatures. The median relative deviation for the majority of the sources is <10 %. However, a comparison of the spectral signatures reveals three systematic deviations between the simulation and measurement results, which require further discussion:

1. I find increased relative deviations between the measured and the simulated spectral signatures at the low end of the spectral

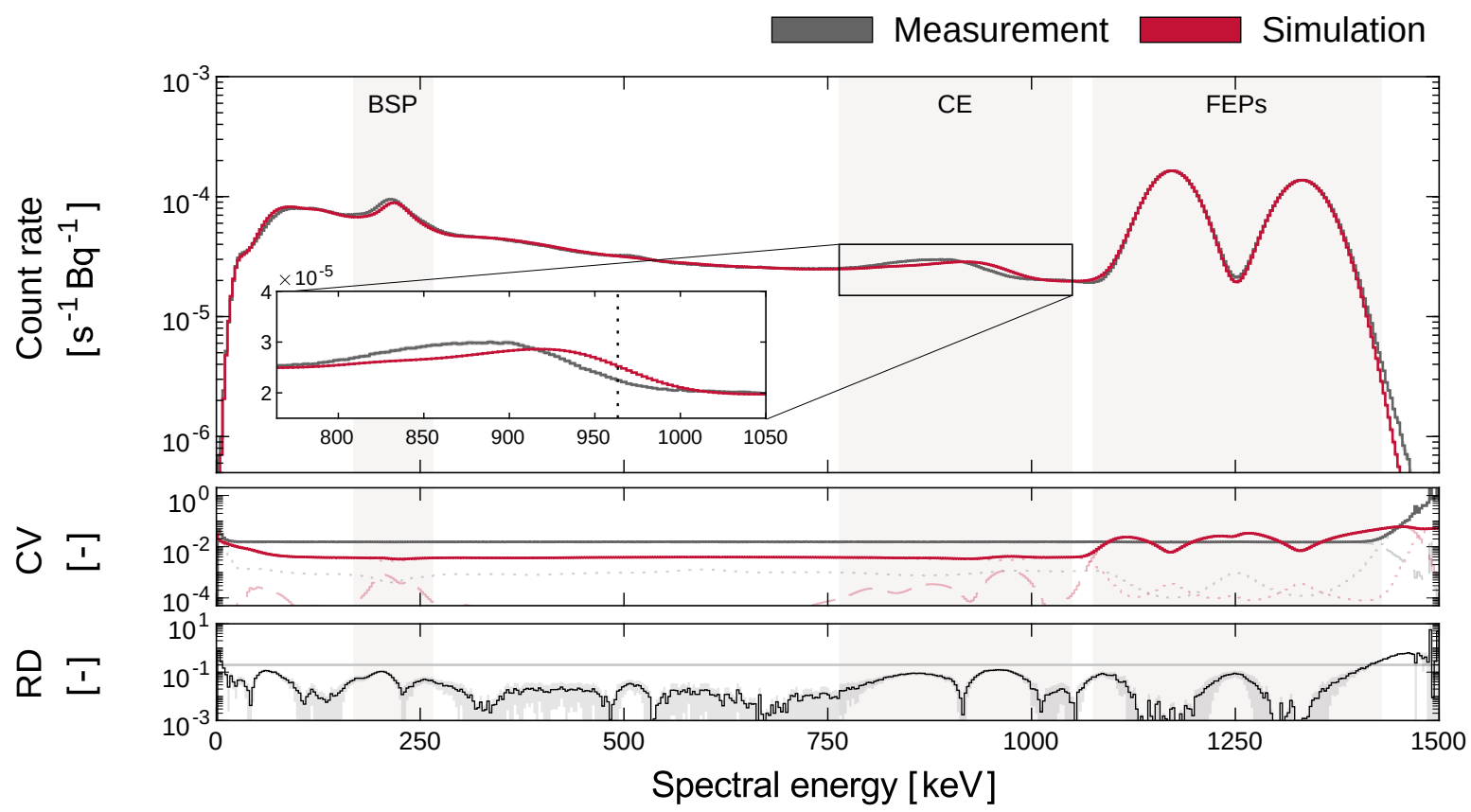


**Figure 6.5** The measured ($\hat{c}_{\text{exp}}$) and simulated ($\hat{c}_{\text{sim}}$) mean spectral signatures for a $^{60}_{27}$Co source ($\mathcal{A}$ = 3.08(5) × $10^5$ Bq) are shown for the detector channel `#SUM` as a function of the spectral energy $E'$ with a spectral energy bin width of $\Delta E' \sim 3$ keV. Distinct spectral regions, i.e. the backscatter peak (BSP), the domain around the Compton edge (CE) as well as the full energy peaks (FEPs) are marked. The zoomed-in subfigure highlights the spectral region around the Compton edge marked by the vertical dotted line and associated with the photon emission line at 1173.228(3) keV [68] (cf. Eq. 4.21). Uncertainties ($\hat{\sigma}_{\text{exp}}$, $\hat{\sigma}_{\text{sim}}$) are provided as 1 standard deviation (SD) shaded areas. In addition, the coefficient of variation (CV) for the measured and simulated signatures (statistical and systematic contributions are indicated by shaded dotted and dashed lines, respectively, cf. Appendix A.8) as well as the relative deviation (RD) computed as $|\hat{c}_{\text{sim}} - \hat{c}_{\text{exp}}|/\hat{c}_{\text{exp}}$ (20 % mark highlighted by a horizontal grey line) are provided.

domain $E' \lesssim 50$ keV. This is particularly evident in the spectral signatures featuring CXPs ($^{133}_{56}$Ba, $^{109}_{48}$Cd, $^{137}_{55}$Cs) as well as in the adjusted box plots for the low-energy emitting radionuclides $^{57}_{27}$Co and $^{109}_{48}$Cd with an increased median relative deviation of ~12 % and ~26 %, respectively.

This deviation can be mainly explained by the increased systematic uncertainty in the empirical energy calibration and spectral resolution models adopted by the developed PSMC model. As discussed in Section 6.2.1.3, these models have been derived based on peak fit analysis of FEPs with photon energies between ~88 keV and ~2615 keV. As a result, we expect that the accuracy of the

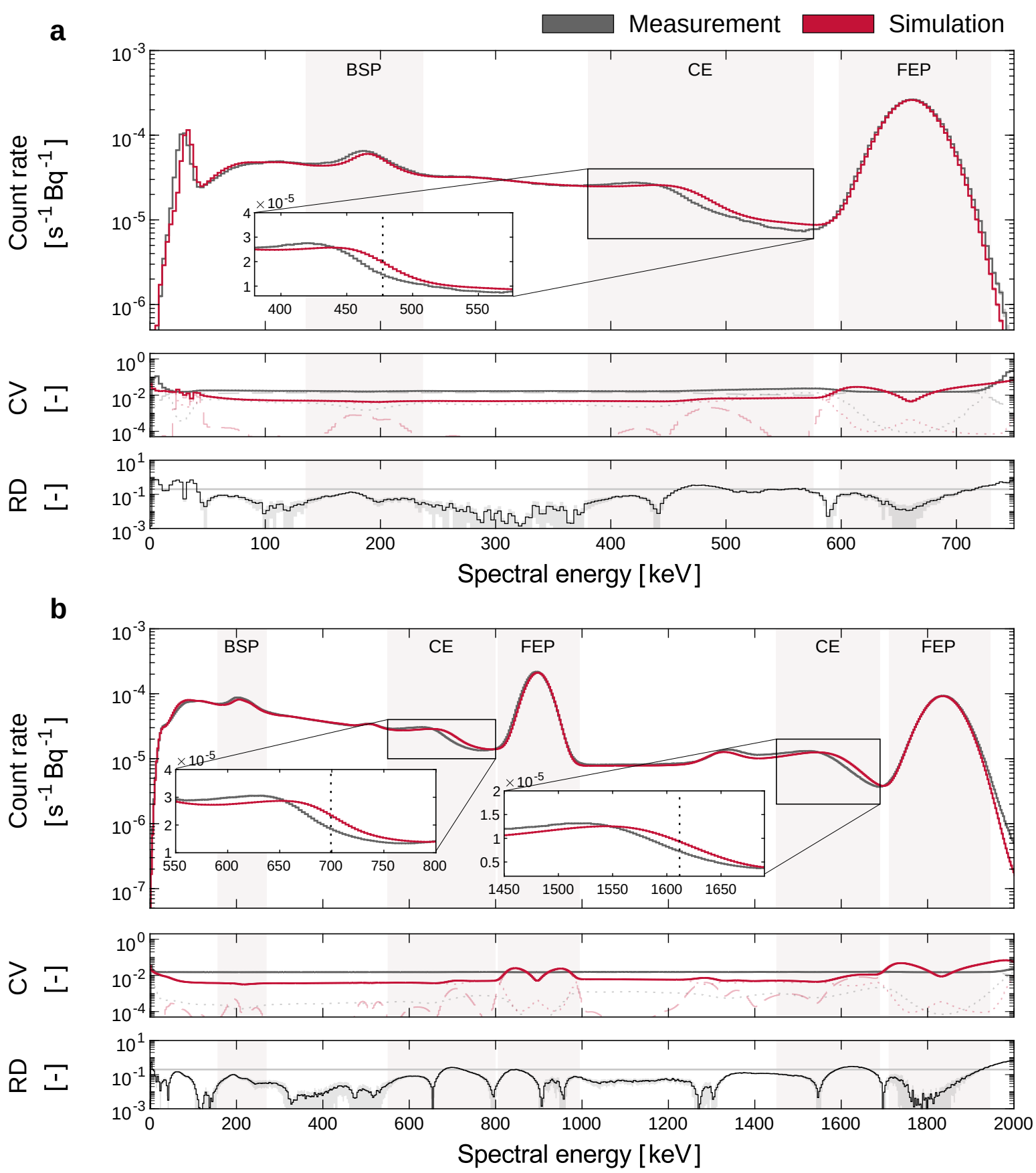


**Figure 6.6** The measured ($\hat{c}_{\mathrm{exp}}$) and simulated ($\hat{c}_{\mathrm{sim}}$) mean spectral signatures for two radionuclides sources and the detector channel #SUM as a function of the spectral energy $E'$ with a spectral energy bin width of $\Delta E' \sim 3\,\mathrm{keV}$ are displayed: **a** $^{137}_{55}\mathrm{Cs}$ ($\mathcal{A} = 2.266(34) \times 10^5\,\mathrm{Bq}$). **b** $^{88}_{39}\mathrm{Y}$ ($\mathcal{A} = 6.83(14) \times 10^5\,\mathrm{Bq}$). Distinct spectral regions, i.e. the backscatter peak (BSP), the domain around the Compton edge (CE) as well as the full energy peak (FEP) are marked. The zoomed-in subfigures highlight the spectral regions around the Compton edge marked by the vertical dotted line and associated with the photon emission lines at 661.657(3) keV for $^{137}_{55}\mathrm{Cs}$ [68] as well as 898.042(11) keV and 1836.070(8) keV for $^{88}_{39}\mathrm{Y}$ [51]. Uncertainties ($\hat{\sigma}_{\mathrm{exp}}$, $\hat{\sigma}_{\mathrm{sim}}$) are provided as 1 standard deviation (SD) shaded areas. In addition, the coefficient of variation (CV) for the measured and simulated signatures (statistical and systematic contributions are indicated by shaded dotted and dashed lines, respectively, cf. Appendix A.8) as well as the relative deviation (RD) computed as $|\hat{c}_{\mathrm{sim}} - \hat{c}_{\mathrm{exp}}|/\hat{c}_{\mathrm{exp}}$ (20 % mark highlighted by a horizontal grey line) are provided.

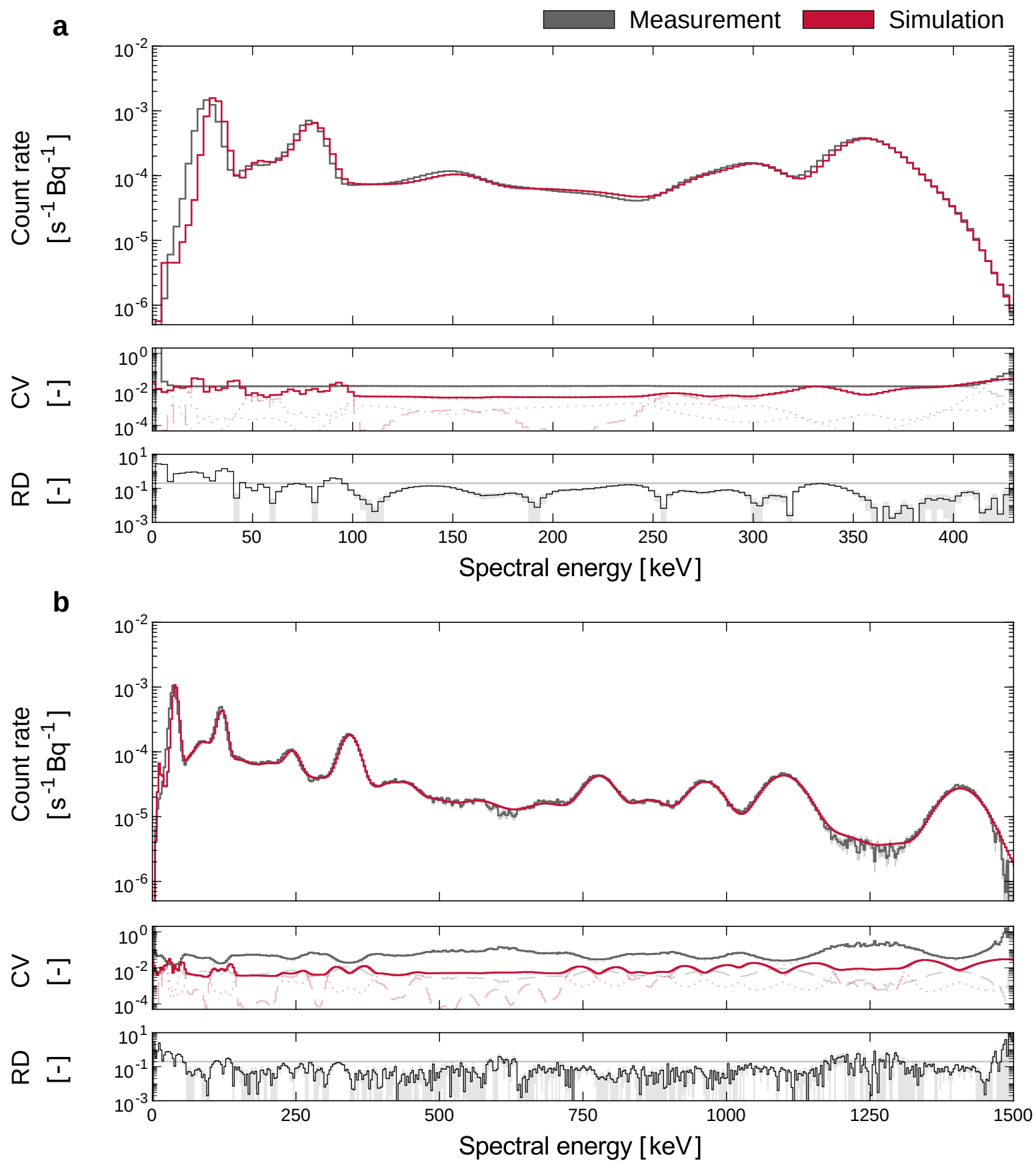


**Figure 6.7** The measured ($\hat{c}_{\mathrm{exp}}$) and simulated ($\hat{c}_{\mathrm{sim}}$) mean spectral signatures for two radionuclides sources and the detector channel `#SUM` as a function of the spectral energy $E'$ with a spectral energy bin width of $\Delta E' \sim 3\,\mathrm{keV}$ are displayed: **a** $^{133}_{56}\mathrm{Ba}$ ($\mathcal{A} = 2.152(32) \times 10^5\,\mathrm{Bq}$). **b** $^{152}_{63}\mathrm{Eu}$ ($\mathcal{A} = 1.973(30) \times 10^4\,\mathrm{Bq}$). Uncertainties ($\hat{\sigma}_{\mathrm{exp}}$, $\hat{\sigma}_{\mathrm{sim}}$) are provided as 1 standard deviation (SD) shaded areas. In addition, the coefficient of variation (CV) for the measured and simulated signatures (statistical and systematic contributions are indicated by shaded dotted and dashed lines, respectively, cf. Appendix A.8) as well as the relative deviation (RD) computed as $|\hat{c}_{\mathrm{sim}} - \hat{c}_{\mathrm{exp}}|/\hat{c}_{\mathrm{exp}}$ (20 % mark highlighted by a horizontal grey line) are provided.

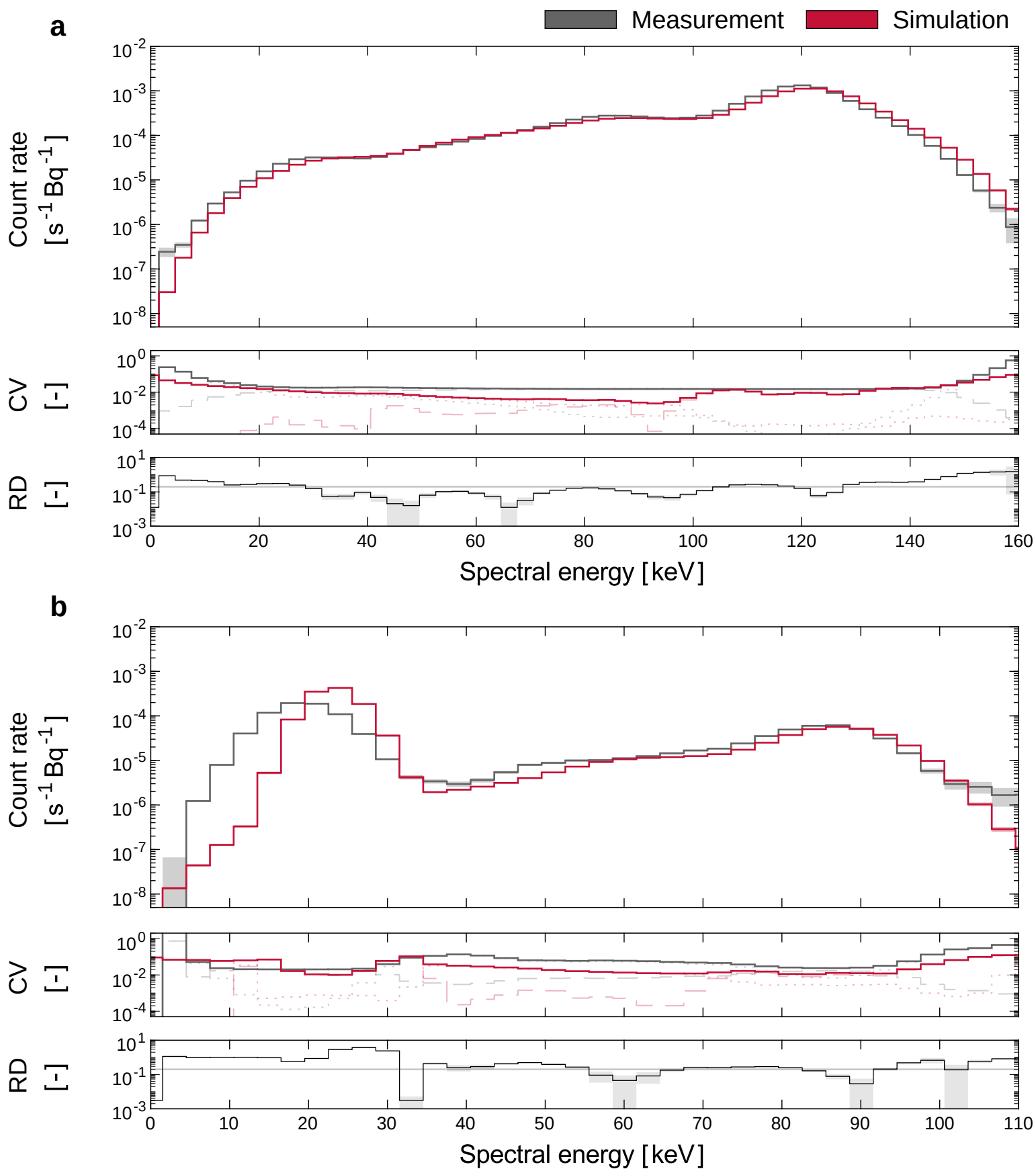


**Figure 6.8** The measured ($\hat{c}_{\text{exp}}$) and simulated ($\hat{c}_{\text{sim}}$) mean spectral signatures for two radionuclides sources and the detector channel `#SUM` as a function of the spectral energy $E'$ with a spectral energy bin width of $\Delta E' \sim 3$ keV are displayed: **a** $^{57}_{27}$Co ($\mathcal{A} = 1.113(18) \times 10^5$ Bq). **b** $^{109}_{48}$Cd ($\mathcal{A} = 7.38(15) \times 10^4$ Bq). Uncertainties ($\hat{\sigma}_{\text{exp}}$, $\hat{\sigma}_{\text{sim}}$) are provided as 1 standard deviation (SD) shaded areas. In addition, the coefficient of variation (CV) for the measured and simulated signatures (statistical and systematic contributions are indicated by shaded dotted and dashed lines, respectively, cf. Appendix A.8) as well as the relative deviation (RD) computed as $|\hat{c}_{\text{sim}} - \hat{c}_{\text{exp}}|/\hat{c}_{\text{exp}}$ (20 % mark highlighted by a horizontal grey line) are provided.

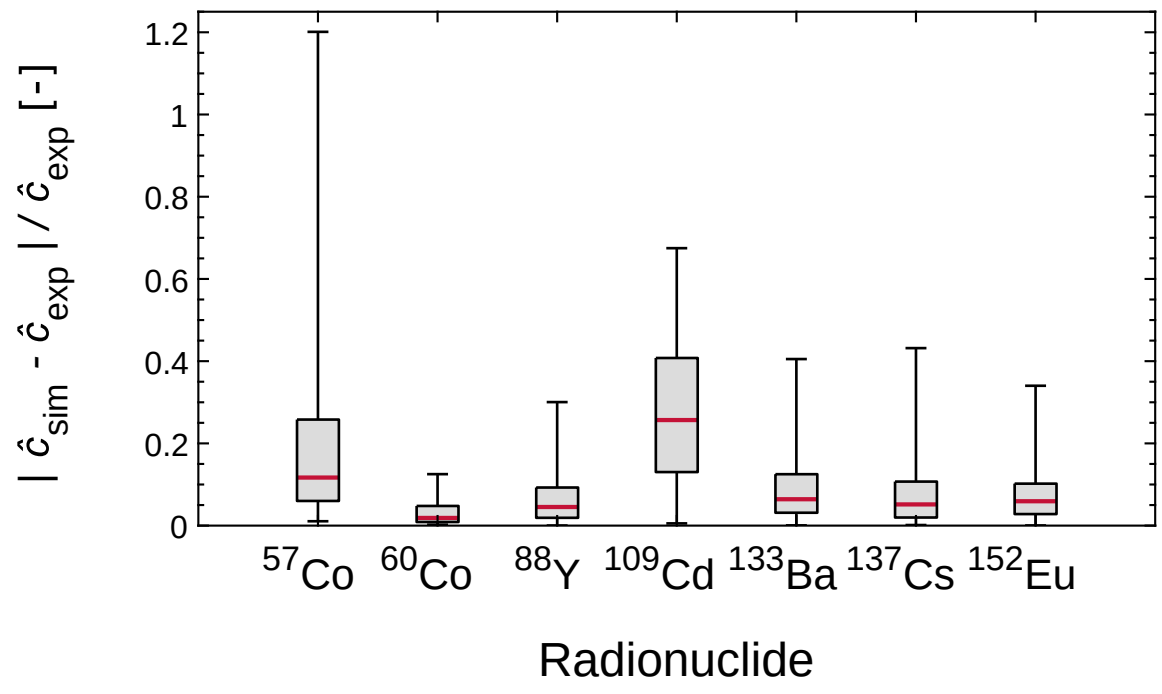


**Figure 6.9** Adjusted box plot [721, 722] are displayed for several radionuclides for the detector channel `#SUM` characterizing the statistical distribution of the related relative deviations $|\hat{c}_{\text{sim}} - \hat{c}_{\text{exp}}|/\hat{c}_{\text{exp}}$ between the experimental ($\hat{c}_{\text{exp}}$) and the simulated spectral signatures ($\hat{c}_{\text{sim}}$). The spectral evaluation domain was limited to $n \in \mathcal{D}_{\text{SDOI}}$ (cf. Def. 6.1). The statistical analysis was performed using the Library for Robust Analysis (`LIBRA`) code [723, 724].

derived models deteriorates outside the covered spectral range, in particular below $\sim 88\,\text{keV}$ due to discontinuities in the scintillation response as a result of the absorption edges of NaI(Tl) [697].

Other factors contributing to the increased deviation at $E' \lesssim 50\,\text{keV}$ are systematic uncertainties in the applied LLD correction model (cf. Section 6.2.2.4) and in the adopted mass model (cf. Section 6.2.2.3). The susceptibility of the detector response to the individual mass model elements will be further investigated in the next section.

Last but not least, increased systematic deviations at low energies can also be attributed to the adopted physics models in the Monte Carlo code, e.g. the condensed history approach for the electron transport [273]. This is particularly true for the $^{109}_{48}$Cd source, as in this case, a custom user routine had to be used (cf. Section 6.2.2.1), which only incorporates photon but no electron emission and related bremsstrahlung.

2. A second systematic deviation between the measured and simulated spectral signatures is found around the BSP highlighted in Figs. 6.5 and 6.6 for the $^{60}_{27}$Co, $^{88}_{39}$Y and $^{137}_{55}$Cs sources. Here, I

observe a significant deviation in the lower energy region of the BSP. This discrepancy may again be attributed to systematic uncertainties in the mass model used in the simulation, which will be investigated further in the next section.

3. A systematic shift of the measured CE toward lower spectral energies compared to the simulated ones can be observed for the $^{60}_{27}$Co, $^{88}_{39}$Y and $^{137}_{55}$Cs sources highlighted in Figs. 6.5 and 6.6. Similar deviations for NaI(Tl) have been reported in the literature [18, 387, 388, 725, 726]. Based on dedicated studies using electron spectroscopy [351] and Compton coincidence measurements [337, 338, 341], this shift can be attributed to the scintillation non-proportionality of the NaI(Tl) scintillator discussed in Section 4.1.3 [387, 388, 725, 726].

It is worth adding that these findings for the detector channel `#SUM` are in line with the results obtained for the single detector channels `#1` through `#4` (cf. Figs. B.36–B.51 and B.54–B.57).

### 6.3.2 Mass Model Sensitivity Analysis

To assess the sensitivity of the spectral detector response on the individual mass model components, a sensitivity analysis was performed changing one set of model elements at the time. More specifically, starting with the high-fidelity model illustrated in Fig. 6.4 as the reference mass model (model A), specific element sets were excluded from this model and replaced with air. These additional mass models are:

B. **No Laboratory Room** Mass model without laboratory walls, floor and ceiling.

C. **No Detector Components** Mass model without detector box components, i.e. no PMTs, no detector electronics, no protective foam and no detector box casing. Only the scintillation crystals with their individual aluminum casings remain.

D. **No Source Components** Mass model without source components, i.e. no source disk, no source holder and no tripod.

E. **No Laboratory Equipment** Mass model with additional laboratory equipment removed, i.e. no shelves, no workbench, no source scanner and no boiler.

As already noted in Section 6.2.2.3, the air atmosphere in the laboratory room is modeled as dry air near sea level with a mass density of $1.205 \times 10^{-3}\,\mathrm{g\,cm^{-3}}$ [249]. To assess the sensitivity of the spectral signature on the air temperature, pressure and humidity, an additional mass model (model F) was created with the dry air exchanged by humid air for the mean measurement conditions reported in Section 6.2.1.1, i.e. a mean air temperature of $T = 18.8\,°\mathrm{C}$, a mean air pressure of $p = 982\,\mathrm{hPa}$ and a mean relative humidity of RH = 42 % resulting in an equivalent mass density of $1.168 \times 10^{-3}\,\mathrm{g\,cm^{-3}}$ [727]:[15]

**F. Humid Air** Mass model with the dry air atmosphere ($\rho = 1.205 \times 10^{-3}\,\mathrm{g/cm^3}$, RH = 0 %) exchanged by humid air ($\rho = 1.168 \times 10^{-3}\,\mathrm{g/cm^3}$, RH = 41.6 %).

Furthermore, there are some uncertainties regarding the elemental composition and mass density of the reflector, which is located between the scintillation crystals and their individual aluminum casings (cf. Section 4.2). Based on publicly available data, the reflector is modeled as a PTFE foil with an equivalent mass density of $\rho = 2.25\,\mathrm{g/cm^3}$. In contrast, most of the other studies adopted MgO for the reflector with an equivalent mass density of $2\,\mathrm{g/cm^3}$ [387, 729] based on an early study by Saito et al. [725]. To quantify the sensitivity of the reflector on the detector response, an additional mass model (model G) was created using MgO instead of PTFE for the reflector:

**G. MgO Reflector** Mass model with the PTFE scintillator reflector ($\rho = 2.25\,\mathrm{g/cm^3}$) exchanged by MgO ($\rho = 2\,\mathrm{g/cm^3}$).

The resulting simulated $^{60}_{27}\mathrm{Co}$ spectral signatures for these additional mass models B–G are presented in Fig. 6.10 together with the signature of the reference model A and the related measurement.[16] To quantify the statistical distribution of the relative deviations $|\hat{c}_{\mathrm{sim}} - \hat{c}_{\mathrm{exp}}|/\hat{c}_{\mathrm{exp}}$, adjusted box plots [721, 722] were computed for different spectral windows characterizing specific spectral features highlighted in Fig. 6.10. The statistical analysis was performed using again the Library for Robust Analysis (`LIBRA`) code [723, 724]. The results from this sensitivity analysis can be summarized as follows:

- The laboratory room (model B) has by far the highest impact on the spectral signature for $E' \lesssim 250\,\mathrm{keV}$. This is particularly

15 *To compute the equivalent mass density and water mass fraction, I adapted the `AirProperties` code by Fitzgerald [728]. The composition of the dry air was modeled based on data provided by McConn et al. [249].*

16 $^{60}_{27}\mathrm{Co}$ *was selected for this analysis as it contains all relevant spectral features to be tested and at the same time does not feature any significant CXPs overshadowing the low-energy part of the spectrum.*

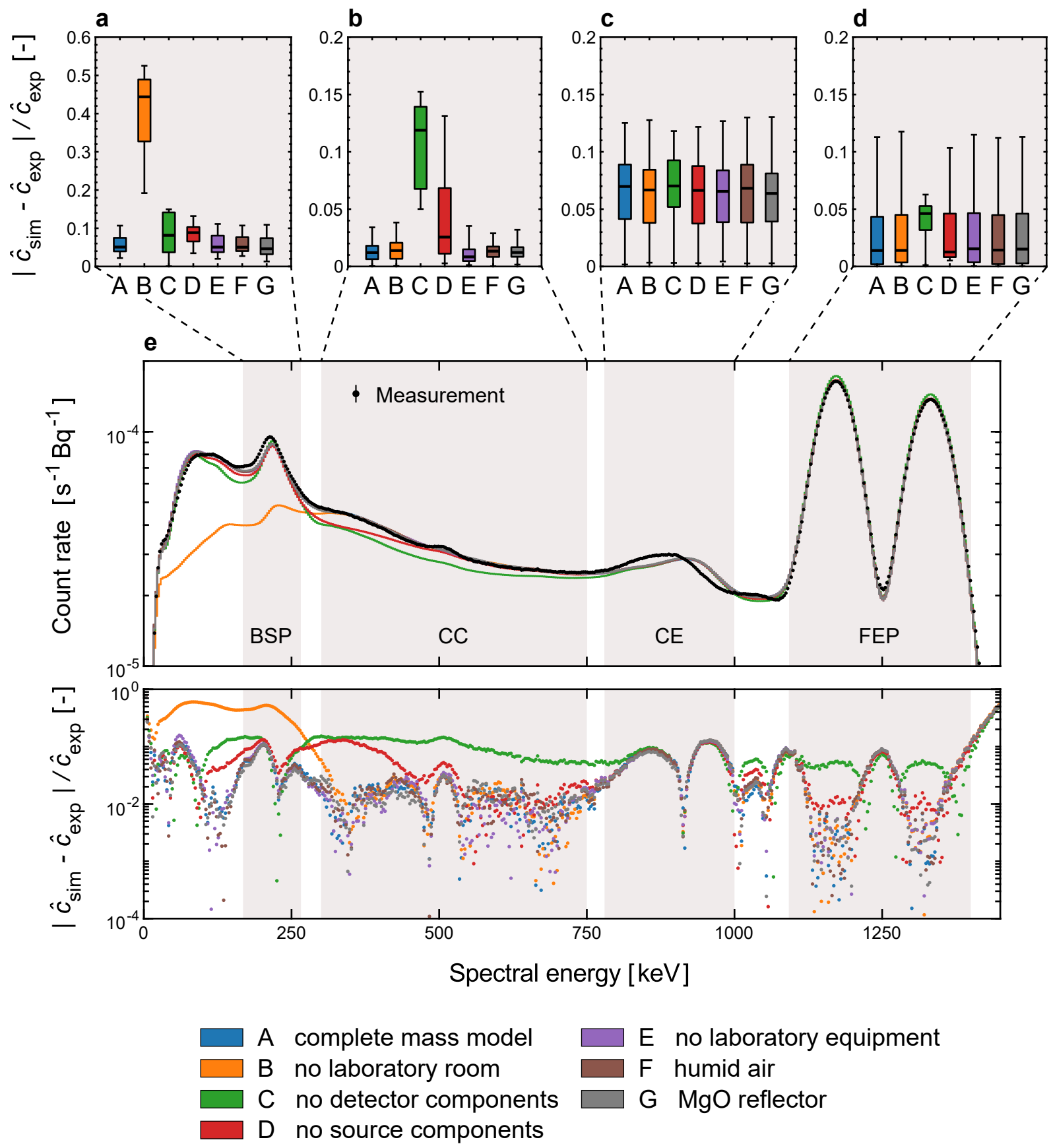


**Figure 6.10** Mass model sensitivity analysis for the $^{60}_{27}$Co spectral signature and the detector channel `#SUM`. **a–d** Adjusted box plots [721, 722] characterizing the statistical distribution of the relative deviations $|\hat{c}_{\text{sim}} - \hat{c}_{\text{exp}}|/\hat{c}_{\text{exp}}$ between the experimental ($\hat{c}_{\text{exp}}$) and the simulated spectral signatures ($\hat{c}_{\text{sim}}$) are shown for four different spectral domains (BSP: backscatter peak, CC: medium part of the Compton continuum, CE: Compton edge, FEP: full energy peaks, cf. also Section 4.3.1) and six different mass models A–G (cf. Section 6.3.2). The statistical analysis was performed using the Library for Robust Analysis (`LIBRA`) [723, 724]. **e** The measured ($\hat{c}_{\text{exp}}$) and simulated ($\hat{c}_{\text{sim}}$) mean spectral signatures for $^{60}_{27}$Co are displayed as a function of the spectral energy $E'$ with a spectral energy bin width of $\Delta E' \sim 3\,\text{keV}$. Spectral domains evaluated in **a–d** are highlighted. Uncertainties ($\hat{\sigma}_{\text{exp}}$, $\hat{\sigma}_{\text{sim}}$) are provided as 1 standard deviation (SD) shaded areas (coverage factor $k = 1$, cf. Appendix A.8). In addition, the relative deviation $|\hat{c}_{\text{sim}} - \hat{c}_{\text{exp}}|/\hat{c}_{\text{exp}}$ as a function of the spectral energy $E'$ is displayed for all six mass models A–G.

evident in the BSP. As discussed in Section 4.3.1, these BSP events in the spectral response are attributed to gamma rays undergoing single or multiple Compton scattering in the surrounding matter with a significant change in direction ($\theta_\gamma \gg 0$) before ultimately being absorbed in the scintillation crystal.

- At medium spectral energies between the BSP and the CE, the detector box components (model C) and the source materials (model D) affect the spectral signature the most. These events can be related to single Compton scatter events in the corresponding materials with $\theta_\gamma \lesssim \pi/2$ before they get absorbed in the scintillation crystals (cf. Section 4.3.1).
- The detector box components (model C) are also those that show the highest sensitivity on the FEPs. The observed increase in the FEP can be attributed to the reduction in the attenuation of the primary gamma rays in the corresponding detector box materials (cf. Section 4.3.1).
- None of the changes in the mass model significantly alter the spectral signature around the CE. This result suggests that the observed systematic shift in the CE discussed before is not susceptible to the systematic uncertainties in the mass model components.
- The performed mass model changes in the laboratory equipment (model E), air atmosphere in the laboratory room (model F) and the reflector (model G) show no significant impact on any part of the spectral signature.

These results hold also for the single detector channels `#1` through `#4` (cf. Figs. B.58–B.61). It is worth noting that the calibration sources in this study exhibit comparably low attenuation and self-shielding. Consequently, for sources with metal shields or bigger active volume sources, the sensitivity of the source elements on the spectral detector response can be expected to be higher.

## 6.4 Conclusion

In this chapter, I presented the development and validation of a high-fidelity PSMC model of the Swiss RLL AGRS detector system under laboratory conditions using the multi-purpose Monte Carlo code `FLUKA` [20, 216, 281]. The simulation results show good agreement

with the measurements with a median relative deviation <10 % for the majority of the evaluated radiation sources ($^{60}_{27}$Co, $^{88}_{39}$Y, $^{133}_{56}$Ba, $^{137}_{55}$Cs and $^{152}_{63}$Eu). This is a significant improvement over previously developed AGRS Monte Carlo models, which showed relative spectral deviations >200 % at low spectral energies [18, 27]. This superiority in model accuracy can be explained by the high-fidelity physics models adopted in the `FLUKA` code [20, 216, 281] compared to the simplified physics models applied in the in-house codes used in some of the previous studies [18, 19], the careful and detailed design of the mass model incorporating the entire laboratory room as well as the implementation of a LLD model in the simulation postprocessing pipeline `PScinMC`.

A sensitivity analysis confirmed the conjecture that the laboratory room, the source and the detector box components are essential parts of the mass model for an accurate detector response simulation under laboratory conditions. In consequence, neglecting the detector box components surrounding the scintillation crystals such as aluminum casings, PMTs or insulating foam as it was done by Allyson et al. [18] and Billings et al. [19], appears to be questionable.

A careful investigation of the spectral signatures revealed statistically significant systematic deviations between measurements and simulations at the very low end of the spectral range ($E' \lesssim 50$ keV), in the BSP as well as around the CE. The deviations at $E' \lesssim 50$ keV and in the BSP can be attributed to the systematic uncertainties in the empirical energy calibration and spectral resolution models adopted by the developed PSMC model as well as the simplifications in the mass model of the laboratory room, among other factors.

On the other hand, the systematic deviations around the CE are a known bias of the PSMC approach and can be attributed to the scintillation non-proportionality of the NaI(Tl) scintillator [387, 388, 725, 726]. This deviation is substantial with a spectral shift of $\gtrsim 20$ keV between the measured and simulated spectral signature. The performed mass model sensitivity analysis revealed that the observed systematic deviations around the CE are not susceptible to the systematic uncertainties in the mass model components. I conclude therefore that the observed systematic spectral shift of the CE is conclusively related to the scintillation non-proportionality, consistent with the results reported in previous studies [387, 388, 725, 726].

Given the significance of the CE shift, these findings imply that the traditional PSMC approach assuming a proportional scintillation response should be revised and considered to be replaced by more

accurate non-proportional models to improve the accuracy of the spectral detector response predictions. This is particularly important for full spectrum AGRS detector modeling, where the entire spectral domain including the CE needs to be reproduced with high accuracy as discussed in Section 5.4.4. The development of such a non-proportional scintillation Monte Carlo approach will be addressed in the next chapter.

> *"This is where the fun begins."*
>
> — *Anakin Skywalker, Revenge of the Sith*

# Chapter 7
# Non-Proportional Scintillation Monte Carlo

**Contents**

In the previous chapter, I found systematic errors in the simulated spectral signature near the Compton edges. These Compton edge shifts can be attributed to the scintillation non-proportionality, which is neglected in the traditional proportional scintillation Monte Carlo (PSMC) method.

To address these systematic errors, I propose a new non-proportional scintillation Monte Carlo (NPSMC) approach utilizing online evaluation of a parametric non-proportional scintillation model (NPSM). The model parameters of the NPSM are calibrated by a novel calibration method, Compton edge probing, combining Bayesian inference with machine learning trained vector-valued surrogate models to emulate the simulated detector response obtained from Monte Carlo simulations. Validation of the new methodology was performed with dedicated laboratory-based radiation measurements.

The results demonstrate that the proposed NPSMC method successfully corrects the systematic errors in the simulated spectral signature around the Compton edges. Moreover, the newly developed calibration method offers a simple way to infer NPSMs for any inorganic scintillator without the need for additional electron response measurements.

Parts of this chapter were reproduced from the following study published by the author in the context of the present work:

D. Breitenmoser et al. "Emulator-Based Bayesian Inference on Non-Proportional Scintillation Models by Compton-Edge Probing". *Nature Communications* **14** 10.1038/s41467-023-42574-y (2023).

## 7.1 Introduction

In the previous chapter, I have presented a high-fidelity PSMC model to simulate the spectral signature of the RLL spectrometer under laboratory conditions. By utilizing the advanced physics models in the `FLUKA` Monte Carlo code, designing a detailed mass model of the entire laboratory room and incorporating a LLD model into the simulation postprocessing pipeline, I achieved a significant

improvement in the accuracy of the simulated spectral signatures compared to previous studies [18, 27].

However, thorough validation of the developed PSMC model with dedicated radiation measurements revealed systematic spectral deviations between the measurements and simulations around the Compton edges. These deviations have been reported in previous studies and can be attributed to the neglected scintillation non-proportionality in PSMC [387, 388, 725, 726]. We refer to the scintillation non-proportionality as the physical phenomenon of inorganic scintillators to exhibit a non-proportional relation $f$ between the deposited energy[1] $E_{\text{dep}}$ and the absolute light yield $Y_{\text{sci,a}}$:

$$Y_{\text{sci,a}} = f\left(E_{\text{dep}}\right) \tag{7.1}$$

where:

| | | |
|---|---|---|
| $E_{\text{dep}}$ | deposited energy | eV |
| $Y_{\text{sci,a}}$ | absolute light yield | $\text{eV}^{-1}$ |

1 *More specifically, the kinetic energy distribution of the high-energy electrons created by ionization events in the scintillator (cf. Section 4.1.3).*

As discussed in detail in Section 4.1.3 and Appendix A.7, this non-proportional scintillation response is a direct result of the Birks and Onsager mechanisms governing the energy transfer stage in the scintillation process of inorganic scintillators. Nonetheless, our understanding of scintillation physics in inorganic scintillators remains incomplete and continues to be an active field of research [308–310, 313, 314, 324, 325, 327, 328, 357, 358].

To account for the scintillation non-proportionality, previous studies proposed to extend traditional PSMC to non-proportional scintillation Monte Carlo (NPSMC) by incorporating dedicated non-proportional scintillation models (NPSMs) [341, 355, 369, 412, 693]. However, the implementation of NPSMC methods is challenging, mainly due to the computational cost and the required calibration of the NPSMs:

1. **Computational Cost** As discussed in Section 4.1.3, implementing NPSMs in Monte Carlo simulations involves the convolution of a dedicated NPSM with the characteristic energy distribution of the generated electrons in the photon-scintillator interactions. Furthermore, a full detector response simulation requires the generation and transport of the scintillation photons. For large-scale simulation models like an AGRS system, this is in most cases computationally prohibitively expensive [369]. Additionally, accurate optical photon transport simulations depend on detailed knowledge of the various material and geometrical parameters of the detector system, which are in practice often not known with sufficient accuracy for these modeling purposes. To address these challenges, previous studies proposed a hybrid approach, where the PSMC method is extended by the convolution of a NPSM with the energy distribution of the generated electrons in the photon-scintillator interactions, while excluding at the same time the generation and transport of the scintillation photons from the Monte Carlo simulation [341, 355, 369, 412, 693]. This hybrid approach has the major advantage of being computationally tractable for large-scale simulation models, while also enabling an accurate description of scintillation non-proportionality effects. However, this advantage comes at the cost of requiring additional offline convolution operations to account for the excluded signal processing steps, similar to PSMC (cf. Section 6.2.2.4).

2. **Calibration** As the scintillation non-proportionality depends on various detector-specific properties, in particular the activator concentration and impurities, NPSMs adopted in NPSMC need to be calibrated for each individual detector system [413]. Currently available calibration methods [337–351] reviewed in Section 4.1.3 require specialized equipment and are not readily available for extensive calibration campaigns of custom detectors. Moreover, they cannot be applied during detector deployment, which might be important for certain applications where the properties of the scintillator change over time, e.g. due to radiation damage or temperature changes.

The scope of this chapter is to develop and validate a NPSMC model of the RLL spectrometer under laboratory conditions. Two new methods will be presented in the following section, which addresses some of the limitations in the implementation and calibration of NPSMs discussed above.

## 7.2 Methods

In this methods section, I will focus on the newly developed NPSMC model as well as associated calibration and postprocessing pipelines. Validation measurements and general setup of the developed Monte Carlo model of the RLL spectrometer (physics models, mass model & scoring) were discussed already in detail in the previous chapter. For a comprehensive overview of these topics, please refer to the Sections 6.2.1 and 6.2.2.1–6.2.2.3.

### 7.2.1 Online NPSMC with FLUKA

To describe the scintillation non-proportionality in the RLL spectrometer, I adopt a thoroughly validated[2] mechanistic NPSM recently published by Stephen Payne and his co-workes [313, 324, 325]. This model quantifies the scintillation non-proportionality by characterizing the local energy transfer efficiency $\eta'_{\mathrm{cap}}$ of the energy transfer stage in the scintillation process of inorganic scintillators. I have discussed this model and the related scintillation mechanisms in detail in Section 4.1.3 and Appendix A.7. The basic NPSM is given in Eq. 4.4.

2 Experimental validation was performed with an extensive database of measured scintillation light yields for inorganic scintillators, i.e. BGO, $CaF_2$(Eu), $CeBr_3$, CsI(Tl), CsI(Na), $LaBr_3$(Ce), LSO(Ce), NaI(Tl), $SrI_2$, $SrI_2$(Eu), YAP(Ce) and YAG(Ce), among others [313, 324, 325].

To save computation time, I performed the convolution of this NPSM with the characteristic energy distribution of the generated electrons in the photon-scintillator interactions online during the Monte Carlo simulations. For this purpose, I developed a custom user routine, `COMSCW`, using the multi-purpose Monte Carlo code `FLUKA` [20, 216, 281].[3] The primary steps of the implemented routine are outlined in Algorithm 7.1.

3 The code was developed with the support of the FLUKA.CERN Collaboration, specifically, Dr. Francesco Cerutti.

**Algorithm 7.1 ▶ Online Non-Proportional Scintillation Monte Carlo with FLUKA**

```
load EventID                                   ▷ EventID = {"continuous", "local"}
load ParticleID                                ▷ Particle type
load dE                                        ▷ Deposited energy
if EventID = "continuous" AND
( ParticleID = "electron" OR
  ParticleID = "positron" ) then
    load ds                                    ▷ Curved particle path
    S ← dE/ds                                  ▷ Estimate stopping power S
else if EventID = "local" then
    load E_k                                   ▷ Kinetic particle energy
    if ParticleID = "electron" then
        S ← EDEDXT(E_k)                        ▷ Call stopping power function EDEDXT
    else if ParticleID = "positron" then
        S ← PDEDXT(E_k)                        ▷ Call stopping power function PDEDXT
    end if
end if
return dE × η'_cap(S)                          ▷ Scale deposited energy by NPSM η'_cap
```

The routine is invoked at each energy deposition event within the scintillation crystals resulting in an integrated deposited energy, scaled by the NPSM, i.e. $E_{\text{dep}}\eta_{\text{cap}}$ (cf. Eq. 4.9).

As described in Section 6.2.2.1, I apply a lower transport threshold of 1 keV below which the electrons and positrons as well as particles generated by atomic de-excitation are no longer transported and their energy is deposited on the spot. I refer to these events as "local" energy deposition events. Conversely, for energy deposition events above this threshold, ionization losses are distributed uniformly along the particle's path [216, 731]. These are termed "continuous" energy deposition events. The algorithm accounts for both local and continuous energy deposition events. The pseudocode included in Algorithm 7.1 is a simplified version of the one implemented in `FLUKA`. For more details, I kindly refer to the actual routine deposited on the ETH Research Collection repository under accession code https://doi.org/10.3929/ethz-b-000595727 [732].

### 7.2.2 Calibration Pipeline: `NPScinCal`

Apart from the meta-models obtained through the `RLLCal` pipeline presented in Section 6.2.1.3, the online NPSMC approach requires the calibration of three additional meta-models. These are:

I. **Non-proportional scintillation scaling** Model relating the scaled energy deposition events $E_{\text{dep}}\eta_{\text{cap}}$ to the continuous pulse-height channel number $\tilde{n}$ (definition given in Section 4.3).

II. **Intrinsic spectral resolution** In contrast to unscaled energy deposition events obtained by PSMC, the scaled energy deposition events resulting from online NPSMC are intrinsically broadened due to the scintillation non-proportionality. This broadening is quantified by the intrinsic spectral resolution $R_{E,\text{intr}}$ discussed in Section 4.3.2 as a function of the continuous pulse-height channel number $\tilde{n}$.

III. **Non-proportional scintillation** As discussed in the introduction to this chapter, due to the sensitivity to the activator concentration and impurities, NPSMs adopted in NPSMC such as the one in Eq. 4.4 need to be calibrated themselves for each individual detector system.

Considering the complexity of the third calibration task, that is the calibration of the NPSM, I will discuss this topic in a separate sub-section 7.2.4. For now, let us assume we possess a calibrated NPSM for each scintillation crystal.

Here, I will present the calibration of the first two meta-models, the non-proportional scintillation scaling and the intrinsic spectral resolution models. For this purpose, an additional calibration pipeline, `NPScinCal`, was developed using the `MATLAB` code. Similar to the `RLLCal` pipeline, this `NPScinCal` pipeline is applied to each single detector channel and consists of the following three steps:

1. **Peak fitting** The first step of the `NPScinCal` pipeline is the quantification of the centroid and FWHM dispersion parameters of intrinsic FEP events simulated by the NPSMC model. The peak fitting is performed with a peak fitting algorithm adapted from the `RLLCal` pipeline (cf. Section 6.2.1.3). The intrinsic FEP events were simulated for a generic 10.2 cm × 10.2 cm × 40.6 cm prismatic NaI(Tl) scintillation crystal exposed to an isotropic and homogeneous monoenergetic photon flux (10 keV $\leq E_\gamma \leq$ 3.2 MeV) using online NPSMC described above. To account for the different spectral scales, a series of photon energies $E_\gamma$ was simulated with a 2 keV spacing below 110 keV and 100 keV spacing above.
2. **Non-proportional scintillation scaling** The centroid data from the previous peak fitting step is then used to derive a model $f_{\mathrm{np}}$ to relate the scaled energy deposition events $E_{\mathrm{dep}}\eta_{\mathrm{cap}}$ to the continuous pulse-height channel number $\tilde{n}$:

$$\tilde{n} = \frac{f_{\mathrm{np}}(E_{\mathrm{dep}}\eta_{\mathrm{cap}})}{\Delta E'} \tag{7.2}$$

where:

| | | |
|---|---|---|
| $E_{\mathrm{dep}}$ | deposited energy | eV |
| $\Delta E'$ | spectral energy bin width | eV |
| $\tilde{n}$ | continuous pulse-height channel number | |
| $\eta_{\mathrm{cap}}$ | energy transfer efficiency | |

with the spectral energy bin width $\Delta E'$ being derived by the `RLLCal` pipeline (cf. Section 6.2.1.3). Given the narrow spacing in $E_\gamma$ and the close-to-linear trend in $f_{\mathrm{np}}$, a modified Akima spline interpolation model [733] has proven to be sufficient to quantify the scaling of the adopted NPSM (cf. Fig. B.62).[4]

[4] *I adopted the function `makima` from the `MATLAB` code to perform this interpolation.*

3. **Intrinsic spectral resolution** In the third step of the NPScinCal pipeline, the dispersion data obtained from the peak fitting step is used to derive a model $g$ relating the intrinsic spectral resolution $R_{E,\text{intr}}$ to the continuous pulse-height channel number $\tilde{n}$:

$$R_{E,\text{intr}} = g\left(\tilde{n}\right) \tag{7.3}$$

To model the complex non-linear behavior of the intrinsic spectral resolution, I adopted a Gaussian process (GP) regression model [734]:[5]

$$R_{E,\text{intr}}\left(\tilde{n}\right) \sim \mathcal{GP}\big(\mathbf{f}(\tilde{n})\boldsymbol{\beta}_{\mathcal{GP}}, \text{Cov}(\tilde{n},\tilde{n}') + \sigma^2_{\mathcal{GP}}\delta_{\tilde{n},\tilde{n}'}\big) \tag{7.4}$$

with:

| | |
|---|---|
| $\boldsymbol{\beta}_{\mathcal{GP}}$ | Gaussian process trend function parameters |
| $\sigma^2_{\mathcal{GP}}$ | Gaussian process noise variance |

I applied a polynomial trend function of the second order, i.e. $\mathbf{f}(\tilde{n}) := (1, \tilde{n}, \tilde{n}^2)$ and $\boldsymbol{\beta}_{\mathcal{GP}} := (\beta_0, \beta_1, \beta_2)^\top$, a homoscedastic[6] noise model with the noise variance $\sigma^2_{\mathcal{GP}}$ and Kronecker delta function $\delta_{\tilde{n},\tilde{n}'}$, as well as a Matérn-3/2 covariance function $\text{Cov}(\tilde{n},\tilde{n}') := (1+\sqrt{3}|\tilde{n}-\tilde{n}'|/\theta_{\mathcal{GP}})\exp\left(-\sqrt{3}|\tilde{n}-\tilde{n}'|/\theta_{\mathcal{GP}}\right)$ with the kernel scale $\theta_{\mathcal{GP}}$ [734]. To derive the GP hyperparameters $\boldsymbol{\beta}_{\mathcal{GP}}$, $\sigma^2_{\mathcal{GP}}$ and $\theta_{\mathcal{GP}}$, I used the `fitrgp` model from the MATLAB code with a Bayesian optimization solver combined with a 5-fold cross-validation scheme. I trained two GP models, one for $E_\gamma \leq 90\,\text{keV}$ and one for $E_\gamma > 90\,\text{keV}$ to account for the different spectral scales. The results of the GP regression model will be discussed later in this chapter in Section 7.3.3.

5 *Note that for brevity, I write the GP regression model here as a function of the continuous pulse-height channel number $\tilde{n}$. As the dispersion is quantified based on the scaled energy deposition $E_{\text{dep}}\eta_{\text{cap}}$, the non-proportional scintillation scale models derived in the second step of the NPScinCal pipeline (cf. Eq. 7.2) are required to convert $E_{\text{dep}}\eta_{\text{cap}}$ to $\tilde{n}$.*

6 *From Ancient Greek skedastós, which translates to "to scatter". Homoscedasticity is a property of a set of random variables where each variable has the same finite variance.*

### 7.2.3 Postprocessing Pipeline: NPScinMC

To account for the NPSM adopted in the developed online NPSMC framework, the postprocessing pipeline PScinMC for the simulation data presented in Section 6.2.2.4 was adapted. This adapted pipeline will be referred to as NPScinMC. Given that only minor modifications were required, this subsection will focus on the specific changes made to the pipeline. For a comprehensive overview of the original postprocessing pipeline, please refer to Section 6.2.2.4.

Two main modifications were made to the `PScinMC` pipeline. First, in the pulse-height channel conversion step, Eq. 6.7 was exchanged with the non-proportional scintillation scaling model Eq. 7.2. Second, given that the scaled energy deposition events resulting from online NPSMC are intrinsically broadened, the spectral resolution standard deviation $\sigma_E$ adopted in the broadening step (Eqs. 6.8, 6.10a, 6.10b and 6.12) needs to be replaced with an adapted standard deviation $\sigma'_E$ [693] (cf. also to Eq. 4.33):

$$\sigma'_E(\tilde{n}^*_{\text{dep}}) = \sqrt{\sigma^2_E(\tilde{n}^*_{\text{dep}}) - \frac{R^2_{E,\text{intr}}(\tilde{n}^*_{\text{dep}})}{2\sqrt{2\log 2}}} \tag{7.5}$$

where:

| | |
|---|---|
| $\tilde{n}^*_{\text{dep}}$ | continuous raw pulse-height channel number of an energy deposition event |
| $R_{E,\text{intr}}$ | intrinsic spectral resolution |
| $\sigma_E$ | spectral resolution standard deviation |

and with $\sigma_E$ and $R_{E,\text{intr}}$ being computed by the Eqs. 6.9b and 7.4, respectively.[7]

Like for the `PScinMC` pipeline, the `NPScinMC` pipeline provides also uncertainty estimates for the derived quantities characterizing both statistical and systematic contributions. More information on these uncertainty estimates is provided in Appendix A.8.2.

7 For $R_{E,\text{intr}}$, the predicted mean of the GP regression model was used [734].

### 7.2.4 Compton Edge Probing

As discussed in the introduction to this chapter, due to the sensitivity to the activator concentration and impurities, NPSMs adopted in NPSMC such as the one in Eq. 4.4 need to be calibrated for each individual detector system. Currently available calibration methods [337–351] reviewed in Section 4.1.3 require specialized equipment and were not available to calibrate the NPSM parameters for the RLL spectrometer. Therefore, a novel calibration method was developed to infer the parameters of the adopted NPSM directly from the pulse-height spectra. The model parameters $\mathbf{x}_{\mathcal{M}} := (\,\mathrm{d}E/\mathrm{d}x\,|_{\text{Birks}}, \eta_{\text{e/h}}, \,\mathrm{d}E/\mathrm{d}x\,|_{\text{trap}})^\top$ are the Birks stopping parameter $\mathrm{d}E/\mathrm{d}x\,|_{\text{Birks}}$, the electron-hole pair fraction $\eta_{\text{e/h}}$ and the trapping stopping parameter $\mathrm{d}E/\mathrm{d}x\,|_{\text{trap}}$ discussed in Section 4.1.3 and Appendix A.7. As suggested by previous investigators [313, 325], I fixed

*8 As discussed in Appendix A.7, in contrast to the other model parameters, $\mathrm{d}E/\mathrm{d}x\,|_{\text{Ons}}$ does only depend on fundamental material properties. More specifically, $\mathrm{d}E/\mathrm{d}x\,|_{\text{Ons}}$ does not depend on the concentration and distribution of impurities or lattice defects in the scintillator and is therefore believed to be constant for different samples of the same scintillator material (under the same laboratory conditions).*

the Onsager stopping parameter $\mathrm{d}E/\mathrm{d}x\,|_{\mathrm{Ons}}$ to 36.4 MeV cm$^{-1}$.[8] Note also that, as it scales proportionally with the deposited energy, the first order non-radiative loss survival efficiency $\eta^0_{\mathrm{cap}}$ in Eq. 4.4 does not need to be calibrated.

This calibration problem represents an inverse problem like the one discussed in Section 5.4.2 with the forward model $\mathcal{M}(\mathbf{x}_{\mathcal{M}})$ being equivalent to the online NPSMC model combined with the postprocessing pipeline `NPScinMC` described above. Given the non-linear behavior of the NPSM in Eq. 4.4 convolved with the complex emission and transport of electrons in the scintillator, the calibration of the model parameters is an ill-posed problem after Hadamard (cf. Section 5.4.2). To solve this inverse problem, I applied Bayesian inversion with a state-of-the-art Markov chain Monte Carlo (MCMC) algorithm [735]. In contrast to traditional frequentist methods like MLE reviewed in Section 5.4.4 or simple data-driven optimization algorithms, a Bayesian approach offers a natural, consistent and transparent way of combining existing information from the literature and physics constraints with empirical data to solve complex inverse problems using a solid probabilistic decision theory framework [658, 736–739].

As we have seen in Section 6.3.1, spectral distortions due to the scintillation non-proportionality are most pronounced around the Compton edges (cf. also to Section 4.3.1). Thus, to further improve the conditioning of the inverse problem, I restrict the spectral domain for NPSM calibration to the Compton edge domain $\mathcal{D}_{\mathrm{CE}}$ defined in Def. 7.1. Motivated by this specific spectral constraint, I call this new calibration method of NPSMs "Compton edge probing".

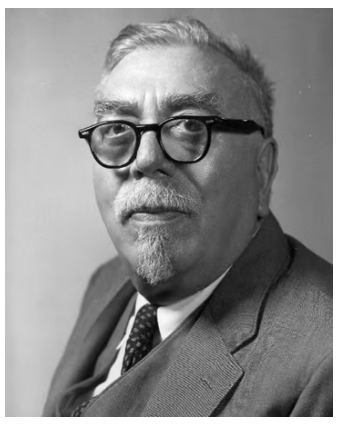

Norbert Wiener

However, there is one major challenge when it comes to the application of the Bayesian inversion methodology to complex computational forward models such as NPSMC presented above. Bayesian inversion requires a large number of model evaluations $\geq \mathcal{O}(10^4)$. Because the forward model encompasses the entire NPSMC simulation with a characteristic computation time of $\mathcal{O}(10^3)$ s to achieve the required precision on the available computer cluster (cf. Section 6.2.2), this calibration task is computationally prohibitively expensive.

To address this challenge, I adopted machine learning to train a custom vector-valued polynomial chaos expansion (PCE) surrogate model[9] $\hat{\mathcal{M}}(\mathbf{x}_{\mathcal{M}})$ emulating the expensive-to-evaluate forward model $\mathcal{M}(\mathbf{x}_{\mathcal{M}})$ [741]. This surrogate model has an excellent evaluation time of $\mathcal{O}(10^{-4})$ s on a local workstation compared to $\mathcal{O}(10^3)$ s for the original NPSMC model evaluated on a computer cluster. Correspondingly, the surrogate model provides a significant acceleration of the

9 *Also known as Wiener chaos expansion after Norbert Wiener (*1894, †1964), an American computer scientist, mathematician and philosopher, who first introduced this method in 1938 [740].*

**Definition 7.1 ▶ Compton edge domain $\mathcal{D}_{\mathrm{CE}}$**

I define the Compton edge domain $\mathcal{D}_{\mathrm{CE}}$ as the spectral band $\mathcal{D}_{\mathrm{CE}} = \{\tilde{n} \in \mathbb{R}_{+} \mid \tilde{n}_{\mathrm{min}} \leq \tilde{n} \leq \tilde{n}_{\mathrm{max}}\}$ with the lower limit $\tilde{n}_{\mathrm{min}} = E'_{\mathrm{CE}}/\Delta E' - 3\sigma_{E}(E'_{\mathrm{CE}}/\Delta E')$ and upper limit $\tilde{n}_{\mathrm{max}} = E_{\gamma}/\Delta E' - 2\sigma_{E}(E_{\gamma}/\Delta E')$. Here, $E'_{\mathrm{CE}}$ represents the spectral energy of the Compton edge given by Eq. 4.21 and associated with the photon emission line with energy $E_{\gamma}$. The spectral energy bin width $\Delta E'$ and the spectral resolution standard deviation $\sigma_{E}$ are obtained by the `RLLCal` pipeline (cf. Section 6.2.1.3) for each of the five detector channels.

Bayesian inversion computations, reducing their processing time by a factor of $10^7$ from about $\mathcal{O}(10^7)$ s to $\mathcal{O}(1)$ s.[10]

As calibration data, I selected the $^{60}_{27}$Co spectral dataset presented in Chapter 6 leaving the remaining measurements ($^{57}_{27}$Co, $^{88}_{39}$Y, $^{109}_{48}$Cd, $^{133}_{56}$Ba, $^{137}_{55}$Cs and $^{152}_{63}$Eu) for validation.[11] $^{60}_{27}$Co possesses two main photon emission lines at 1173.228(3) keV and 1332.492(4) keV [68] with corresponding Compton edges (cf. Eq. 4.21) at 963.419(3) keV and 1118.101(4) keV, respectively. However, in this work, I will focus on the lower Compton edge at 963.419(3) keV, because the upper edge is heavily obscured by the FEP at 1173.228(3) keV (cf. Fig. 6.5). The Bayesian inversion methodology will be independently applied to all five detector channels, i.e. the four single detector channels `#1` through `#4` and the detector channel `#SUM`.

In the following two subsections, I will provide a concise overview of the Bayesian inversion methodology and the surrogate modeling approach that was adopted in this chapter. I will begin by discussing the surrogate modeling approach.

#### 7.2.4.1 Surrogate Modeling

To emulate the expensive-to-evaluate NPSMC forward model within the selected $\mathcal{D}_{\mathrm{CE}}$, I adopted a custom machine learning trained vector-valued PCE surrogate model. PCE surrogate models are ideal candidates to emulate expensive-to-evaluate vector-valued computational models [741]. In addition, PCE surrogate models allow the computation of Sobol' indices[12] analytically, enabling global sensitivity analysis of the emulated model with minimal additional computational cost [742].

10 *Of course, the surrogate model is only an approximation of the original forward model. However, as detailed in Section 7.2.4.1 and Appendix A.11, thanks to the adoption of machine learning training [741], the induced generalization error is kept <2 % for the four single detector channels #1 through #4 and <1 % for the detector channel #SUM, respectively.*

11 *Note that only $^{88}_{39}$Y, $^{60}_{27}$Co and $^{137}_{55}$Cs possess resolved Compton edges. For the other sources, the Compton edges are obscured by other FEPs (cf. Section 6.3.1).*

12 *First derived by Ilya Meyerovich Sobol' (*1926), a Russian mathematician, best known for his influential work on Monte Carlo methods and global sensitivity analysis.*

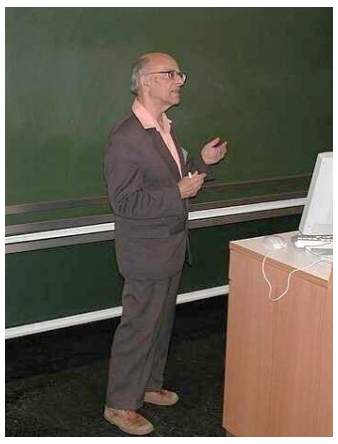

Ilya M. Sobol'

As shown in previous studies [743–745], any function $\boldsymbol{Y} = \mathcal{M}(\boldsymbol{X})$ with the random input vector $\boldsymbol{X} \in \mathbb{R}^{M\times 1}$ and random response vector $\boldsymbol{Y} \in \mathbb{R}^{N\times 1}$ can be expanded as a so-called polynomial chaos expansion provided that $\mathbb{E}[|\boldsymbol{Y}|^2] < \infty$:

$$\boldsymbol{Y} = \hat{\mathcal{M}}(\boldsymbol{X}) = \sum_{\boldsymbol{\alpha}\in\mathbb{N}^{M\times 1}} \mathbf{a}_{\boldsymbol{\alpha}} \Psi_{\boldsymbol{\alpha}}(\boldsymbol{X}) \tag{7.6}$$

with

| | | |
|---|---|---|
| $\mathbf{a}_{\boldsymbol{\alpha}}$ | expansion coefficient vector | $[y]$ |
| $\boldsymbol{\alpha}$ | multi-index vector | |
| $\Psi_{\boldsymbol{\alpha}}$ | multivariate polynomial basis function | |

and where $\mathbf{a}_{\boldsymbol{\alpha}} = \left(a_{1,\boldsymbol{\alpha}}, \ldots, a_{N,\boldsymbol{\alpha}}\right)^\top \in \mathbb{R}^{N\times 1}$ are the deterministic expansion coefficients and $\boldsymbol{\alpha} = \left(\alpha_1, \ldots, \alpha_M\right)^\top \in \mathbb{N}^{M\times 1}$ the multi-indices storing the degrees of the univariate polynomials $\psi_{\alpha}$, which in turn define the multivariate polynomial functions $\Psi_{\boldsymbol{\alpha}}(\boldsymbol{X})$ [741]:

$$\Psi_{\boldsymbol{\alpha}}(\boldsymbol{X}) = \prod_{i=1}^{M} \psi_{\alpha_i}^{(i)}(X_i) \tag{7.7}$$

assuming independent input components in $\boldsymbol{X}$. In this work, the random input vector $\boldsymbol{X} \in \mathbb{R}^{M\times 1}$ corresponds to the $M = 3$ input parameters of the NPSM. Conversely, the random response vector $\boldsymbol{Y} \in \mathbb{R}^{N\times 1}$ represents the variable number of pulse-height channels $N$ within the Compton edge domain $\mathcal{D}_{\mathrm{CE}}$, which depends on the detector channel and the selected Compton edge.

By choosing the set of multivariate polynomial functions that satisfy [746]:

$$\langle \Psi_{\boldsymbol{\alpha}}, \Psi_{\boldsymbol{\beta}} \rangle_{\pi_X} = \int_{\mathcal{D}_X} \Psi_{\boldsymbol{\alpha}}(\mathbf{x}) \Psi_{\boldsymbol{\beta}}(\mathbf{x}) \pi(\mathbf{x}) \, \mathrm{d}\mathbf{x} = \delta_{\boldsymbol{\alpha},\boldsymbol{\beta}} \tag{7.8}$$

they build an orthonormal basis of the space of square-integrable functions with respect to the joint probability density function $\pi_X := \pi(\mathbf{x})$ [746].[13] Provided certain mild conditions hold, a set of multivariate polynomial functions fulfilling Eq. 7.8 can be found for arbitrary joint marginal probability density functions [745].[14] For common parametric probability density functions such as uniform or normal distributions, there exist well-known families of univariate polynomial functions that can be used to build the multivariate polynomial basis functions $\Psi_{\boldsymbol{\alpha}}$ [748].

13 *Please note again that in this book, I use the symbol $\pi(\cdot)$ to denote the PDF, which is the common notation in Bayesian statistics [267, 268].*

14 *However, in practice, the analytical form of such arbitrary orthonormal polynomials is often not known and needs to be constructed numerically using Stieltjes procedure or Gram-Schmidt orthogonalization [267, 747].*

To exploit the correlation between the pulse-height channels within the Compton edge domain $\mathcal{D}_{\mathrm{CE}}$, I combine the PCE model with principal component analysis (PCA) which allows me to characterize the model response by means of a significantly reduced number $N'$ of output variables compared to the original $N$ pulse-height channels, i.e. $N' \ll N$ [746, 749].[15] The emulated computational model response $\hat{\mathcal{M}}(\boldsymbol{X})$ may then be approximated in matrix form as [746]:[16]

$$\boldsymbol{Y} \approx \hat{\mathcal{M}}(\boldsymbol{X}) = \boldsymbol{\mu}_{\boldsymbol{Y}} + \mathrm{diag}(\boldsymbol{\sigma}_{\boldsymbol{Y}})\, \boldsymbol{\Phi}' \mathbf{A} \boldsymbol{\Psi}(\boldsymbol{X}) \tag{7.9}$$

where:

| | | |
|---|---|---|
| $\mathbf{A}$ | expansion coefficient matrix | |
| $\boldsymbol{\mu}_{\boldsymbol{Y}}$ | mean of the random response vector $\boldsymbol{Y}$ | $[y]$ |
| $\boldsymbol{\sigma}_{\boldsymbol{Y}}$ | standard deviation of the random response vector $\boldsymbol{Y}$ | $[y]$ |
| $\boldsymbol{\Phi}'$ | matrix of retained eigenvectors | |
| $\boldsymbol{\Psi}$ | multivariate polynomial basis function vector | |

The matrix $\boldsymbol{\Phi}' \in \mathbb{R}^{N \times N'}$ contains the retained eigenvectors $\boldsymbol{\Phi}' = (\boldsymbol{\phi}_1, \ldots, \boldsymbol{\phi}_{N'})$ of the PCA. On the other hand, the vector $\boldsymbol{\Psi}(\boldsymbol{X}) \in \mathbb{R}^{\mathrm{card}(\mathcal{A}^\star) \times 1}$ and matrix $\mathbf{A} \in \mathbb{R}^{N' \times \mathrm{card}(\mathcal{A}^\star)}$ store the multivariate orthonormal polynomials and corresponding PCE coefficients, respectively. The union set $\mathcal{A}^\star = \bigcup_{j=1}^{N'} \mathcal{A}_j$ includes the finite sets of multi-indices $\mathcal{A}_j$ for the $N'$ output variables following a given PCE truncation scheme. In this work, I adopted a hyperbolic truncation scheme $\mathcal{A}_j = \{\boldsymbol{\alpha} \in \mathbb{N}^M \mid (\sum_{i=1}^{M} \alpha_i^q)^{1/q} \leq p\}$ with $p$ and $q$ being hyperparameters defining the maximum degree for the associated polynomial and the q-norm, respectively [741].

To compute the PCE coefficient matrix $\mathbf{A}$, I applied adaptive least angle regression [750] and optimized the hyperparameters $p := \{1, 2, \ldots, 7\}$ and $q := \{0.5, 0.6, \ldots, 1\}$ using machine learning with a holdout partition of 80 % and 20 % for the training and test set, respectively. The training and test set were derived based on a Latin hypercube experimental design $\boldsymbol{\mathcal{X}} \in \mathbb{R}_+^{M \times K}$ [751, 752] with $K = 200$ instances sampled from a predefined probabilistic model (cf. Table C.8). This probabilistic model defines also the related univariate polynomial functions $\psi_\alpha$, which in this work are equivalent to Legendre[17] polynomials $\mathcal{P}_\alpha$ [748]. The model response $\boldsymbol{\mathcal{Y}} \in \mathbb{R}_+^{N \times K}$, that is the spectral signature within the predefined Compton edge domain $\mathcal{D}_{\mathrm{CE}}$, was then evaluated using the NPSMC forward model described in the previous subsections with the model parameters defined by $\boldsymbol{\mathcal{X}}$.[18] For

15 *A compression factor of 8 to 9 was obtained in this work.*

16 *Note that I standardize the response data before performing PCA (cf. Appendix A.11). As a result, the second term in Eq. 7.9 is, in contrast to Wagner et al. [746], scaled by the standard deviation of the random response vector $\boldsymbol{Y}$ and the expansion coefficient matrix $\mathbf{A}$ is unitless.*

17 *First described by Adrien-Marie Legendre (*1752, †1833), a prominent French mathematician who made significant contributions to number theory, algebra and statistics.*

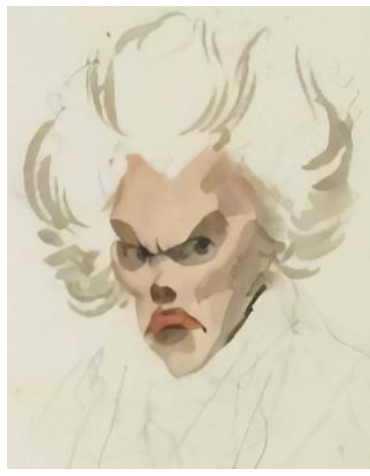

Adrien-Marie Legendre
Julien-Léopold Boilly

18 *Note that this involves the complete Monte Carlo simulation (Section 7.2.1), calibration (Section 7.2.2) and post-processing (Section 7.2.3) for each individual instance.*

the PCA truncation, I adopted a relative PCA truncation error $\varepsilon_{\text{PCA}}$ of 0.1 %, i.e. $N' = \min\{S \in \mathbb{N}_+ \mid S \leq N, \sum_{j=1}^{S} \Lambda_j / \sum_{j=1}^{N} \Lambda_j \geq 1 - \varepsilon_{\text{PCA}}\}$ with $\Lambda$ being the eigenvalues from the PCA [749].

The resulting generalization error of the trained surrogate models, characterized by the relative mean squared error over the test set [267], is <2 % for the four single detector channels `#1` through `#4` and <1 % for the detector channel `#SUM`, respectively. All PCE computations were performed with the `UQLab` code [753] in combination with custom `MATLAB` scripts to perform the PCA.[19] More information on the custom PCE-PCA model including detailed derivations is provided in Appendix A.11. In addition, a discussion on the computation of Sobol′ indices using the PCE surrogate model is included in Appendix A.12.

19 *I used the built-in* `pca` *function in* `MATLAB`*, which adopts a singular value decomposition (SVD) algorithm to perform the PCA.*

#### 7.2.4.2 Bayesian Inference

Bayesian inference is a powerful statistical tool that establishes a probabilistic framework for combining existing information from the literature and physics constraints with newly obtained empirical data to solve complex inference problems. It has been successfully applied in a wide variety of fields, including physics [658, 737, 754], engineering [755, 756], earth sciences [661, 672, 757, 758], biology and medicine [759, 760], machine learning [677, 761, 762] or social sciences [763, 764]. Bayesian inference encompasses a whole range of philosophical attitudes including the interpretation of probability as a measure of belief instead of a frequency like in frequentist statistics. There a several reasons, both from a statistical and practical standpoint, why a Bayesian interpretation of probability is superior to a frequentist one, in particular when it comes to solving ill-posed inverse problems [539, 677, 678, 737, 765]. Here, I will not delve into the philosophical debate regarding the merits and drawbacks of the Bayesian paradigm. Detailed discussions on this topic are available in the cited literature. Instead, given its importance not only for this but also for later chapters in this book, this subsection will focus on a comprehensive introduction to the parametric Bayesian inference methodology and its application to solving inverse problems. For a more thorough general introduction, I recommend the excellent textbooks by Gelman et al. [738], McElreath [766], and Sivia et al. [767]. Introductions with a special focus on applications in physics are also readily available in the literature [658, 737, 754, 768–770]. More rigorous mathematical treatments are provided in the monographs by Shao [771] and Keener [772]. Although they lack discussions on modern computational aspects, the classic literature on the topic can provide additional valuable perspectives [773, 774]. Lastly, I recommend the theory sections of the two dissertations by Wagner [267] and Nagel [268]. Both provide a concise overview of the Bayesian inference methodology and its application to solving inverse problems in physics and engineering applications. My discussion in this subsection is largely adapted from these last two references.

**Likelihood Function** In parametric Bayesian inference, we aim to infer some unknown parameters $\mathbf{x} = (x_1, \ldots, x_M)^\top \in \mathbb{R}^{M\times 1}$ from a set of vector-valued measurements $\mathcal{Y} = \{\mathbf{y}_i \in \mathbb{R}^{N\times 1} \mid i \in \mathbb{N}_+, i \leq N_\mathcal{Y}\}$ obtained under known[20] deterministic experimental conditions $\mathfrak{D} = \{\mathfrak{d}_k \in \mathbb{R}^{N_\mathfrak{d}\times 1} \mid k \in \mathbb{N}\}$ with $M, N, N_\mathfrak{d}$ and $N_\mathcal{Y}$ being the number of unknown parameters, number of observables, number of experimental condition parameters and the number of independent measurements, respectively. Following a probabilistic modeling approach, similar to the MLE frequentist method discussed in Section 5.4.4, we interpret the individual recorded observables $\mathbf{y}$ in the dataset $\mathcal{Y}$ as independent realizations of an underlying random vector $\boldsymbol{Y}$ following an associated conditional PDF:

20 *Following the Bayesian paradigm, unknown experimental conditions can always be incorporated as additional model parameters $x$.*

$$\boldsymbol{Y} \sim \pi(\mathbf{y} \mid \mathbf{x}, \mathfrak{d}) \tag{7.10}$$

where:

| | | |
|---|---|---|
| $\mathfrak{d}$ | experimental condition | $[\mathfrak{d}]$ |
| $\mathbf{x}$ | model parameter vector | $[x]$ |
| $\boldsymbol{Y}$ | random response vector | $[y]$ |
| $\mathbf{y}$ | data vector | $[y]$ |

which I will refer to as the probabilistic data model [268]. If we consider the dataset $\mathcal{Y}$ to be fixed and to consist of independently recorded measurements under known deterministic measurement conditions $\mathfrak{D}$, we may define the likelihood function $\mathcal{L} : \mathcal{D}_X \mapsto \mathbb{R}_+$ as [267]:[21]

21 *Please note that, as $\mathcal{L}$ is defined as a function of $\mathbf{x}$ and in general $\int_{\mathcal{D}_X} \mathcal{L}(\mathbf{x})\,\mathrm{d}\mathbf{x} \neq 1$, it is not a PDF [267].*

$$\mathcal{L}(\mathbf{x}; \mathcal{Y}, \mathfrak{D}) := \prod_{k=1}^{N_\mathcal{Y}} \pi\left(\mathbf{y}_k \mid \mathbf{x}, \mathfrak{d}_k\right) \tag{7.11}$$

Intuitively, the likelihood function quantifies how well the statistical model describes the dataset $\mathcal{Y}$ as a function of the model parameters $\mathbf{x}$. It is not only the core quantity in Bayesian but also in frequentist statistical inference and incorporates the complete physical forward model $\mathcal{M}(\mathbf{x})$ combined with a statistical model to account for the statistical variation in the data.

**Prior Distribution** Where the frequentist and Bayesian approaches differ is in the interpretation of the model parameters $\mathbf{x}$. In the Bayesian framework, we treat the model parameters not as fixed but as random variables with associated random vector $\boldsymbol{X}$ and assign a prior probability distribution $\pi(\mathbf{x})$ to them:

$$\boldsymbol{X} \sim \pi(\mathbf{x}) \tag{7.12}$$

We may interpret the prior distribution $\pi(\mathbf{x})$ as a measure of the level of information existing on the parameters before the set of measurements $\mathcal{Y}$ is considered. Possible sources of such information include physics constraints, expert knowledge and results from previous measurements, either conducted by the investigators themselves or retrieved from the literature.

Generally, priors fall into two categories: subjective and objective. Subjective priors are derived from personal or expert beliefs, incorporating prior knowledge or experience into the model [775, 776]. These priors are often termed informative, as they reflect specific information or assumptions about the parameters.

On the other hand, objective priors are constructed according to some formal rules designed to minimize the influence of subjective judgments [777–780]. Examples include priors based on the maximum entropy principle, which ensures that the prior is as noncommittal as possible given the known constraints [781–786]. These objective priors are also often referred to as uninformative or noninformative because they aim to provide a neutral starting point that relies minimally on prior beliefs.

In practice, prior distributions often fall somewhere between these two extremes. A prior can be more subjective or informative if it incorporates some prior knowledge, or it can be more objective or uninformative if it adheres, at least partly, to some formal rules designed to minimize subjective input.

Historically, the choice of the prior distribution has been a subject of much debate in Bayesian statistics and has been criticized for its subjectivity. However, today, many Bayesian statisticians argue that the prior distribution is not a limitation or drawback but rather a feature and a benefit of Bayesian statistics [658, 737, 766]. No statistical inference can be made without assumptions, and the prior distribution is a way to explicitly state these assumptions and incorporate them into the analysis in a transparent and consistent way. In contrast, frequentist methods often rely on implicit assumptions that are not always clearly stated. As an example, the choice of the loss function in a frequentist setting can be translated to a specific pair of likelihood functions and prior distributions in a Bayesian setting [267, 677].

Furthermore, the prior distribution is inherently flexible and can accommodate complex physics constraints and external information thereby constraining and regularizing the solution space of inverse problems. This is particularly important in cases where the data are sparse and the related inverse problem is ill-posed like the one formulated in Def. 5.1.

**Posterior Distribution** The core of Bayesian inference is the posterior distribution, a conditional PDF which combines the likelihood function $\mathcal{L}(\mathbf{x}; \mathcal{Y}, \mathfrak{D})$ with the prior distribution $\pi(\mathbf{x})$ to provide a complete probabilistic description of the model parameters $\mathbf{x}$ given the newly obtained data $\mathcal{Y}$ and experimental conditions $\mathfrak{D}$. The posterior distribution can be computed by applying Bayes' theorem[22] [267]:

$$\pi(\mathbf{x} \mid \mathcal{Y}, \mathfrak{D}) = \frac{\overbrace{\mathcal{L}(\mathbf{x}; \mathcal{Y}, \mathfrak{D})}^{\text{likelihood}} \overbrace{\pi(\mathbf{x})}^{\text{prior}}}{\underbrace{\int_{\mathcal{D}_X} \mathcal{L}(\mathbf{x}; \mathcal{Y}, \mathfrak{D}) \, \pi(\mathbf{x}) \, \mathrm{d}\mathbf{x}}_{\text{evidence}}} \tag{7.13}$$

The posterior distribution $\pi(\mathbf{x} \mid \mathcal{Y}, \mathfrak{D})$ represents the complete probabilistic solution to an inverse problem with the forward model being described by the likelihood function $\mathcal{L}(\mathbf{x}; \mathcal{Y}, \mathfrak{D})$ and the prior information being encoded in the prior distribution $\pi(\mathbf{x})$. As a result, we may interpret the prior and the posterior as the state of knowledge or level of epistemic uncertainty before and after incorporating the newly obtained data $\mathcal{Y}$. The computation of the posterior distribution can therefore also be seen as an update of our prior beliefs in the model parameters based on the newly obtained data, reflecting the learning process in Bayesian inference.

To ensure that the posterior distribution in Eq. 7.13 integrates to one, it is normalized by a constant known as the evidence or marginal

22 *Derived by Thomas Bayes (*1701, †1761), an 18th-century English statistician, philosopher, and clergyman. His work, published posthumously by his friend Richard Price [787], laid the groundwork for modern Bayesian probability theory and has wide applications in various scientific fields.*

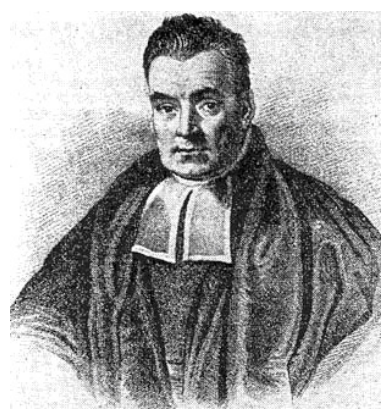

Thomas Bayes

23 *This evidence can be interpreted as the probability of the probabilistic data model itself and as such is central for model comparison [268, 658, 737].*

likelihood.[23] The evidence is the integral of the likelihood function over the entire parameter space weighted by the prior distributions and it is this integral which makes the computation of the posterior distribution challenging. Unfortunately, for general likelihood and prior distributions, no closed-form expression for the posterior distribution exists [738].[24] As a result, numerical methods are required to compute the posterior distribution in Eq. 7.13. Over the decades, an entire arsenal of methods has been developed to efficiently perform the Bayesian update in Eq. 7.13. Common methods are approximate Bayesian computations (ABC) [788–791], variational Bayesian methods (VB) [792, 793], Laplace approximations [677, 762, 794], spectral likelihood expansions (SLE) [795] and nested sampling [654, 796–799].

24 *Exceptions are likelihood functions combined with conjugate prior distributions [738]. Given the limited applicability of these models, I will not discuss them in this work and instead focus on general probabilistic models without any constraints in selecting likelihood functions and prior distributions.*

However, the most widely used method remains Markov chain Monte Carlo (MCMC). This class of algorithms generates samples $\hat{\mathcal{X}} = \{\mathbf{x}^{(i)} \in \mathbb{R}^{M\times 1} \mid i \in \mathbb{N}_+, i \leq N_{\hat{\mathcal{X}}}, N_{\hat{\mathcal{X}}} \in \mathbb{N}_+\}$ from the posterior distribution $\pi(\mathbf{x} \mid \mathcal{Y}, \mathfrak{D})$ by constructing a Markov chain that converges to the target posterior distribution [800–804]. One can distinguish these MCMC algorithms again in many subclasses such as the classic Metropolis-Hastings (MH) algorithm[25] [805, 806], Hamiltonian Monte Carlo (HMC) [807, 808] or affine invariant ensemble sampler (AIES) [735, 809]. Given its simplicity, versatility and robustness for complex posterior distributions, I selected a state-of-the-art AIES MCMC algorithm [735] implemented in the `UQLab` code [753] to perform all Bayesian computations in this book. A detailed discussion of the MCMC algorithms is beyond the scope of this work. For a comprehensive overview, I recommend the excellent textbooks by Brooks et al. [801] and Liu [803] as well as Chapter 1 of Wagner's dissertation [267].

25 *First published by Nicholas Metropolis (*1915, †1999) in 1953 [805] and later generalized by Wilfred Keith Hastings (*1930, †2016) [806].*

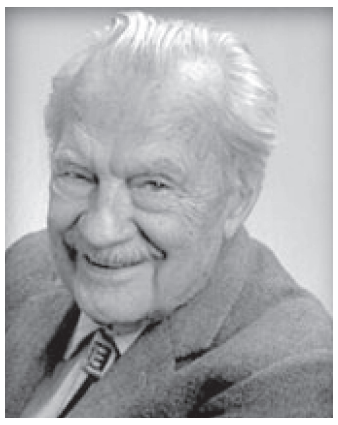

Nicholas Metropolis
Ⓢ Los Alamos National Laboratory

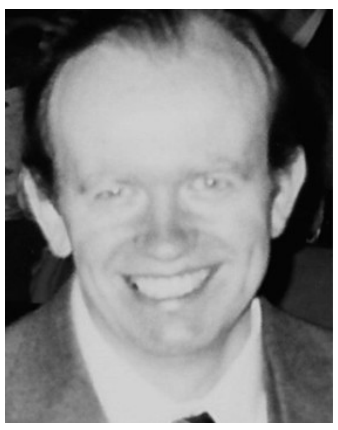

Wilfred K. Hastings
Ⓢ unknown

**Posterior Summaries** The posterior distribution discussed above encodes the complete probabilistic solution to our inverse problem. However, in practice, interpreting a set of posterior samples $\hat{\mathcal{X}}$ can be challenging, especially for high-dimensional parameter spaces. As a result, we often adopt posterior summary statistics to simplify the reporting and interpretation of the Bayesian inference results. These statistics can be broadly categorized into two types: point estimates, which, simply speaking, characterize the most likely parameter values; and dispersion measures, which quantify the variability or uncertainty of single and combined parameters. Common point estimates are the posterior mean [267, 268]:

$$
\begin{aligned}
\mathbf{x}_{\text{Mean}} &= \mathbb{E}\left(\boldsymbol{X} \mid \mathcal{Y}, \mathfrak{D}\right) && \text{(7.14a)}\\
&= \int_{\mathcal{D}_X} \mathbf{x}\, \pi\left(\mathbf{x} \mid \mathcal{Y}, \mathfrak{D}\right) \mathrm{d}\mathbf{x} && \text{(7.14b)}\\
&\approx \frac{1}{N_{\hat{\mathcal{X}}}} \sum_{\substack{i=1\\ \mathbf{x}\in\hat{\mathcal{X}}}}^{N_{\hat{\mathcal{X}}}} \mathbf{x}^{(i)} && \text{(7.14c)}
\end{aligned}
$$

or alternatively the statistically more robust posterior median $\mathbf{x}_{\text{Median}}$ [738, 766, 770]. A third point estimate is the maximum a posteriori (MAP) estimate [267, 268], which is equivalent to the mode of the posterior distribution [267, 268]:

$$
\begin{aligned}
\mathbf{x}_{\text{MAP}} &= \underset{\mathbf{x}\in\mathcal{D}_X}{\arg\max}\ \pi\left(\mathbf{x} \mid \mathcal{Y}, \mathfrak{D}\right) && \text{(7.15a)}\\
&= \underset{\mathbf{x}\in\mathcal{D}_X}{\arg\max}\ \mathcal{L}\left(\mathbf{x}; \mathcal{Y}, \mathfrak{D}\right) \pi\left(\mathbf{x}\right) && \text{(7.15b)}\\
&\approx \underset{\mathbf{x}\in\hat{\mathcal{X}}}{\arg\max}\ \mathcal{L}\left(\mathbf{x}; \mathcal{Y}, \mathfrak{D}\right) \pi\left(\mathbf{x}\right) && \text{(7.15c)}
\end{aligned}
$$

The MAP point estimate is closely related to the MLE $\mathbf{x}_{\text{MLE}}$ discussed in Section 5.4.4 (cf. Eq. 5.9). Unlike the MLE, the MAP estimate incorporates the prior information imposed by the prior distribution. As a result, the difference between the MAP and MLE estimates tends to be more pronounced with sparser datasets. When using uniform priors, the two are equivalent, as long as the MLE point lies within the support of the prior distribution [267].

To quantify the dispersion of the parameters, we may compute the full posterior covariance matrix of the posterior distribution as follows [267, 268]:

$$
\begin{aligned}
&\operatorname{Cov}\left(\boldsymbol{X} \mid \mathcal{Y}, \mathfrak{D}\right) =\\
&= \int_{\mathcal{D}_X} \left[\mathbf{x} - \mathbb{E}\left(\boldsymbol{X} \mid \mathcal{Y}, \mathfrak{D}\right)\right] \left[\mathbf{x} - \mathbb{E}\left(\boldsymbol{X} \mid \mathcal{Y}, \mathfrak{D}\right)\right]^{\top} \pi\left(\mathbf{x} \mid \mathcal{Y}, \mathfrak{D}\right) \mathrm{d}\mathbf{x} && \text{(7.16a)}\\
&\approx \frac{1}{N_{\hat{\mathcal{X}}} - 1} \sum_{\substack{i=1\\ \mathbf{x}\in\hat{\mathcal{X}}}}^{N_{\hat{\mathcal{X}}}} \left[\mathbf{x}^{(i)} - \mathbf{x}_{\text{Mean}}\right] \left[\mathbf{x}^{(i)} - \mathbf{x}_{\text{Mean}}\right]^{\top} && \text{(7.16b)}
\end{aligned}
$$

with the posterior standard deviation $\boldsymbol{\sigma}_\mathbf{x}$ being the square root of the diagonal elements of the covariance matrix. Instead of the covariance matrix, we may characterize the joint variability of the parameters also by computing the posterior correlation matrix using for example Spearman's rank correlation coefficients[26] [810].

In addition to covariance and correlation matrices, we can compute also the Bayesian analogue of frequentist confidence intervals, the credible region $\mathcal{C}_X$, to quantify the spread of the individual parameters [658, 770]. The credible region is defined as the region in parameter space that contains a predefined amount of the posterior probability mass $\mathcal{B} \in [0,1]$ [658, 770]:

$$\int_{\mathcal{C}_X} \pi(\mathbf{x} \mid \mathcal{Y}, \mathfrak{D})\, \mathrm{d}\mathbf{x} = \mathcal{B} \tag{7.17}$$

If we refer to an individual parameter, we call this region also the credible interval. In general, such credible regions are not unique, i.e. there may be many regions that fulfill the condition in Eq. 7.17 [677, 811]. As a result, we need to add further constraints to find a unique credible region. One of the most common constraints is the one where we choose the region in such a way that the probability being below the individual credible intervals is as likely as being above them, i.e. the posterior probability mass $1 - \mathcal{B}$ is equally distributed on both sides of the intervals. The resulting credible region is also known as the central credible region/interval [677] or percentile region/interval [766]. If MCMC samples are available, these credible intervals can be easily estimated for each parameter by computing the quantiles of the posterior samples $\hat{\mathcal{X}}$ for the probabilities $(1 - \mathcal{B})/2$ and $(1 + \mathcal{B})/2$ [766].[27]

It is important to note that, despite their similar intended purposes, Bayesian credible intervals and frequentist confidence intervals differ significantly in their interpretation [677]. The interpretation of a credible interval is straightforward: it represents a range within which the unknown parameter lies with a certain predefined probability, given the data and prior information. In contrast, frequentist confidence intervals provide a range of values that would contain the true parameter value in a predefined percentage of hypothetical repeated experiments [677].

Often, not all parameters included in Bayesian inference are of equal interest. Sometimes, there are parameters that are necessary to construct a realistic model but do not provide any direct information about the quantities of interest. These parameters are often called

26 *Spearman's rank correlation coefficient quantifies the strength and direction of association between two variables. It ranges from* $-1$ *to* $+1$*, where* $+1$ *indicates a perfect positive monotonic relationship (as one variable increases, so does the other),* $-1$ *indicates a perfect negative monotonic relationship (as one variable increases, the other decreases) and 0 indicates the absence of such monotonic relationships between the variables.*

27 *Corresponding functions are readily available in common statistics codes, e.g.* `quantile` *in MATLAB and* `numpy.quantile` *in Python.*

nuisance parameters [738, 770]. Classic examples are discrepancy variables which quantify the variability in the data. In such cases, we can summarize the posterior distribution by integrating over these nuisance parameters. Specifically, we split the random parameter vector $\boldsymbol{X}$ into two vectors: the parameters of interest $\boldsymbol{X}_{\mathcal{U}}$ and the nuisance parameters $\boldsymbol{X}_{\mathcal{V}}$ with components $\{X_i \mid i \in \mathcal{U}\} \in \mathcal{D}_{x_{\mathcal{U}}} \subseteq \mathbb{R}^{M_{\mathcal{U}} \times 1}$ and $\{X_i \mid i \in \mathcal{V}\} \in \mathcal{D}_{x_{\mathcal{V}}} \subseteq \mathbb{R}^{M_{\mathcal{V}} \times 1}$ using two non-empty disjoint index sets $\mathcal{U}$ and $\mathcal{V}$ such that $\mathcal{U} \cup \mathcal{V} = \{1, \ldots, M\}$ and $M_{\mathcal{U}} + M_{\mathcal{V}} = M$. The integration then reads [812]:[28]

$$\pi\left(\mathbf{x}_{\mathcal{U}} \mid \mathcal{Y}, \mathfrak{D}\right) = \int_{\mathcal{D}_{X_{\mathcal{V}}}} \pi\left(\mathbf{x} \mid \mathcal{Y}, \mathfrak{D}\right) \mathrm{d}\mathbf{x}_{\mathcal{V}} \tag{7.18}$$

28 *Note that for MCMC samples, the marginalization is equivalent to only considering the parameters of interest in the samples [766].*

This integration over selected parameters is also known as marginalization and the resulting distributions are commonly referred to as the posterior marginals [267, 268] or marginal posterior distributions [766, 770]. As discussed by Trotta [737] and Gregory [770] in more detail, marginalization is one of the many technical advantages that Bayesian inference holds over standard frequentist statistics.

**Predictive Distributions** In the previous subsections, I have outlined the process for computing the posterior distribution of the model parameters $\mathbf{x}$ given the data $\mathcal{Y}$ and experimental conditions $\mathfrak{D}$, as well as methods for summarizing this distribution. However, in many cases, we are not only interested in the parameters themselves but also in the predictions of the model response $\boldsymbol{Y}$ based on the inferred parameters.

One approach to estimate the model response is to use the point estimates discussed earlier and incorporate them into the probabilistic data model in Eq. 7.10. However, this straightforward approach does not account for the uncertainty in the parameters and may lead to overly confident predictions [267, 268]. Following the Bayesian paradigm, we can instead compute predictive distributions defined in the data space by marginalizing over the full model parameter space. Depending if this marginalization is performed before or after the Bayesian update, we distinguish between the prior and posterior predictive distributions [267]. The prior predictive distribution is obtained by averaging the probabilistic data model in Eq. 7.10 over the prior distribution given in Eq. 7.12:

$$\pi\left(\mathbf{y}^* \mid \mathfrak{D}^*\right) = \int_{\mathcal{D}_X} \pi\left(\mathbf{y}^* \mid \mathbf{x}, \mathfrak{D}^*\right) \pi\left(\mathbf{x}\right) \mathrm{d}\mathbf{x} \tag{7.19}$$

which quantifies our expectations about future data $\mathbf{y}^*$ recorded under predefined experimental conditions $\mathfrak{D}^*$ without considering the dataset $\mathcal{Y}$. Samples from this predictive distribution can be obtained by first sampling from the prior distribution in Eq. 7.12[29] and then propagating those parameter samples through the probabilistic data model in Eq. 7.10 [812].

29 *Using standard statistical sampling techniques like the inverse transformation method or rejection sampling [800, 802, 803].*

The posterior predictive distribution, on the other hand, is obtained by averaging the probabilistic data model in Eq. 7.10 over the posterior distribution given in Eq. 7.13:

$$\pi(\mathbf{y}^* \mid \mathcal{Y}, \mathfrak{D}, \mathfrak{D}^*) = \int_{\mathcal{D}_X} \pi(\mathbf{y}^* \mid \mathbf{x}, \mathcal{Y}, \mathfrak{D}, \mathfrak{D}^*)\, \pi(\mathbf{x} \mid \mathcal{Y}, \mathfrak{D})\, \mathrm{d}\mathbf{x} \tag{7.20a}$$

$$= \int_{\mathcal{D}_X} \pi(\mathbf{y}^* \mid \mathbf{x}, \mathfrak{D}^*)\, \pi(\mathbf{x} \mid \mathcal{Y}, \mathfrak{D})\, \mathrm{d}\mathbf{x} \tag{7.20b}$$

where I assumed conditional independence between the obtained data $\mathcal{Y}$ and the future data $\mathbf{y}^*$ to simplify Eq. 7.20a to Eq. 7.20b [268]. The posterior predictive distribution quantifies our updated expectation about future data $\mathbf{y}^*$ recorded under predefined experimental conditions $\mathfrak{D}^*$ after performing a Bayesian update with the dataset $\mathcal{Y}$ using Bayes' theorem in Eq. 7.13. Samples from Eq. 7.20b can be obtained by first sampling from the posterior distribution Eq. 7.13, e.g. by MCMC, and then propagating those parameter samples through the probabilistic data model in Eq. 7.10 [812].

Similar to the posterior distribution, we may summarize the spread of the prior and posterior predictive distributions by computing (predictive) credible regions $\mathcal{C}_Y$ as [811]:

$$\int_{\mathcal{C}_Y} \pi(\mathbf{y}^* \mid \mathfrak{D}^*)\, \mathrm{d}\mathbf{y} = \mathcal{B} \tag{7.21a}$$

$$\int_{\mathcal{C}_Y} \pi(\mathbf{y}^* \mid \mathcal{Y}, \mathfrak{D}, \mathfrak{D}^*)\, \mathrm{d}\mathbf{y} = \mathcal{B} \tag{7.21b}$$

which I will refer to as prediction regions or intervals to distinguish them from the credible regions associated with the model parameters (cf. Eq. 7.17). As for the credible regions defined in Eq. 7.17, we need to add further constraints to find unique prediction regions such as the central prediction regions with the probability mass $1 - \mathcal{B}$ equally distributed on both sides of the intervals.

30 *A probabilistic model is identifiable if there is a one-to-one mapping $X \mapsto \pi_X$ [766, 772, 813].*

Given the probabilistic data model is identifiable[30], we expect the dispersion of the posterior predictive distribution to be smaller than the one of the prior predictive distribution, reflecting the information gain from the newly obtained data [267]. The predictive distributions are essential for quantifying the uncertainty in the model predictions and are often used to assess the model's predictive performance [267].

**Implementation Details** Given the high-counting statistics obtained in the laboratory-based measurements to probe the Compton edge (cf. Section 6.3.1) [30], I assume that the statistical variation in the spectral signature $\mathbf{y} := \mathbf{c} \in \mathbb{R}_{+}^{N\times 1}$ within the Compton edge domain $\mathcal{D}_{\mathrm{CE}}$ is generated by additive zero-mean Gaussian noise with a covariance matrix $\sigma_{\varepsilon}^{2}\mathbb{I}_{N}$ and an unknown constant discrepancy model variance $\sigma_{\varepsilon}^{2}$ on top of the deterministic NPSMC forward model $\mathcal{M}$ discussed in the Sections 7.2.1–7.2.3:

$$\boldsymbol{Y} = \mathcal{M}\left(\mathbf{x}_{\mathcal{M}}, \mathfrak{d}\right) + \boldsymbol{E} \tag{7.22a}$$

$$\boldsymbol{E} \sim \mathcal{N}\left(\mathbf{0}, \sigma_{\varepsilon}^{2}\mathbb{I}_{N}\right) \tag{7.22b}$$

where:

| | | |
|---|---|---|
| $\mathfrak{d}$ | experimental condition | $[\mathfrak{d}]$ |
| $\boldsymbol{E}$ | random discrepancy vector | $\mathrm{s}^{-1}\,\mathrm{Bq}^{-1}$ |
| $\mathbb{I}_{N}$ | $N \times N$ identity matrix | |
| $\mathcal{M}$ | forward model | $\mathrm{s}^{-1}\,\mathrm{Bq}^{-1}$ |
| $\mathbf{x}_{\mathcal{M}}$ | forward model parameter vector | $[x]$ |
| $\sigma_{\varepsilon}^{2}$ | discrepancy model variance | $\mathrm{s}^{-2}\,\mathrm{Bq}^{-2}$ |

and $\mathcal{N}(\cdot)$ denoting the multivariate Gaussian or normal distribution detailed in Section 5.4.4. The forward model parameter vector is equivalent to $\mathbf{x}_{\mathcal{M}} := (\,\mathrm{d}E/\mathrm{d}x|_{\mathrm{Birks}}, \eta_{\mathrm{e/h}},\, \mathrm{d}E/\mathrm{d}x|_{\mathrm{trap}})^{\top}$ as already defined at the beginning of Section 7.2.4. Similar to our discussion in Section 5.4.4, we can rewrite this probabilistic data model as:

$$\pi\left(\mathbf{y} \mid \mathbf{x}, \mathfrak{d}\right) = \mathcal{N}\left(\mathbf{y} \mid \hat{\mathcal{M}}\left(\mathbf{x}_{\mathcal{M}}, \mathfrak{d}\right), \sigma_{\varepsilon}^{2}\mathbb{I}_{N}\right) \tag{7.23}$$

with $\mathbf{x} := [\mathbf{x}_{\mathcal{M}}, \sigma_{\varepsilon}^{2}]^{\top}$. Note also that I have replaced the expensive-to-evaluate NPSMC forward model $\mathcal{M}$ with the trained surrogate model $\hat{\mathcal{M}}$ discussed in Section 7.2.4.1. As the available data is limited to a single spectral signature vector in this chapter ($N_{\mathcal{Y}} = 1$, $\mathcal{Y} = \{\mathbf{y}\}$ and $\mathfrak{D} = \{\mathfrak{d}\}$), the associated likelihood function is equivalent to Eq. 7.23, i.e. $\mathcal{L}(\mathbf{x}; \mathbf{y}, \mathfrak{d}) = \mathcal{N}(\mathbf{y} \mid \hat{\mathcal{M}}(\mathbf{x}_{\mathcal{M}}, \mathfrak{d}), \sigma_\varepsilon^2 \mathbb{I}_N)$.

Regarding the prior distribution, I adopted statistically independent marginal priors motivated by the maximum entropy principle [781, 782, 786] as well as results from previous studies [313, 324, 325]:

$$\pi(\mathbf{x}) = \prod_{i=1}^{M} \pi(x_i) \tag{7.24}$$

It should be emphasized that the Bayesian computations were performed first for the detector channel `#SUM`, followed by the four single detector channels `#1` through `#4`. Consistent with the Bayesian paradigm, the posterior results obtained for the detector channel `#SUM` were subsequently incorporated into the marginal priors used for the four individual detector channels. A full list of all adopted marginal priors is provided in Table C.9.

All Bayesian computations were performed using the `UQLab` code [753]. As stated already above, I applied an AIES algorithm [735] to perform MCMC sampling from the posterior distribution in Eq. 7.13. I used ten parallel chains with $2 \times 10^4$ MCMC iterations per chain together with a 50 % burn-in resulting in a total of $N_{\hat{\mathcal{X}}} = 10^5$ posterior samples. The convergence and precision of the MCMC simulations were carefully assessed using standard diagnostics tools [738, 814], showing a potential scale reduction factor $\hat{R} < 1.02$ and effective sample size ESS > 400 across all MCMC runs. Additional trace and convergence plots for the individual parameters and point estimates are provided in Figs. B.63–B.67.

## 7.3 Results

Similar to the previous chapter, I will focus here on the various results obtained for the detector channel `#SUM`. However, it is important to note that all four single detector channels were also carefully evaluated in parallel. Corresponding results for these single channels are provided in Appendix B. Where applicable, I will cross-reference relevant findings from these single detector channels.

Due to its importance for all subsequent findings, I will begin by presenting the Bayesian inversion results for the selected NPSM acquired via Compton edge probing and discuss the implications for the Compton edge shift. Following this, I will explore the intrinsic resolution models resulting from the calibrated NPSM. Finally, I will conclude by comparing spectral signatures of the newly introduced online NPSMC model with the PSMC model presented in the previous chapter.

### 7.3.1 Bayesian Inference on NPSM

In Fig. 7.1, I present the solution to the NPSM inversion problem outlined in Section 7.2.4 as a multivariate posterior distribution estimate for the detector channel `#SUM`. I find a unimodal posterior distribution with a MAP point estimate given by $\eta_{\mathrm{e/h}} = 5.96^{+0.08}_{-0.14} \times 10^{-1}$, $\mathrm{d}E/\mathrm{d}x\,|_{\mathrm{trap}} = 1.46^{+0.02}_{-0.17} \times 10^{1}\,\mathrm{MeV\,cm^{-1}}$ and $\mathrm{d}E/\mathrm{d}x\,|_{\mathrm{Birks}} = 3.22^{+0.38}_{-0.36} \times 10^{2}\,\mathrm{MeV\,cm^{-1}}$. The indicated uncertainties were estimated using the central credible intervals with a probability mass $\mathcal{B}$ of 95 %. Combining the individual posterior of the four single detector channels,[31] I obtain statistically consistent estimates, i.e. $\eta_{\mathrm{e/h}} = 5.75^{+0.36}_{-0.08} \times 10^{-1}$, $\mathrm{d}E/\mathrm{d}x\,|_{\mathrm{trap}} = 1.41^{+0.17}_{-0.15} \times 10^{1}\,\mathrm{MeV\,cm^{-1}}$ and $\mathrm{d}E/\mathrm{d}x\,|_{\mathrm{Birks}} = 2.64^{+1.65}_{-0.29} \times 10^{2}\,\mathrm{MeV\,cm^{-1}}$.

31 *By combining the individual posterior samples from the four single detector channels together and performing the posterior analysis on this aggregated sample set.*

However, as expected, I find statistically significant differences between the posterior point estimates for the four individual scintillation crystals (cf. Table C.10). Furthermore, the results significantly differ from best-estimate literature values provided by Payne and his co-workers [325]. With a marginal adaption to the laboratory temperature of $T = 18.8\,°\mathrm{C}$ (cf. Section 6.2.1.1) using linear interpolation, these values read $\eta_{\mathrm{e/h}} = 4.53 \times 10^{-1}$, $\mathrm{d}E/\mathrm{d}x\,|_{\mathrm{trap}} = 1.2 \times 10^{1}\,\mathrm{MeV\,cm^{-1}}$ and $\mathrm{d}E/\mathrm{d}x\,|_{\mathrm{Birks}} = 1.85 \times 10^{2}\,\mathrm{MeV\,cm^{-1}}$.

From a modeling perspective, it is interesting to note that, considering the uncertainty estimates, only minor differences among the different posterior point estimates for the individual detector channels can be observed. Furthermore, to investigate the sensitivity of the predefined Compton edge domain $\mathcal{D}_{\mathrm{CE}}$ according to Def. 5.1 on the Bayesian inversion results, I performed a sensitivity analysis on $\mathcal{D}_{\mathrm{CE}}$ by altering the domain boundaries of $\mathcal{D}_{\mathrm{CE}}$. Within the uncertainty bounds, the inversion results have proven to be insensitive to these changes (cf. Table C.11). A full list of all Bayesian results is summarized in Table C.10. Corner plots for the four individual scintillation crystals are provided in Figs. B.68–B.71.

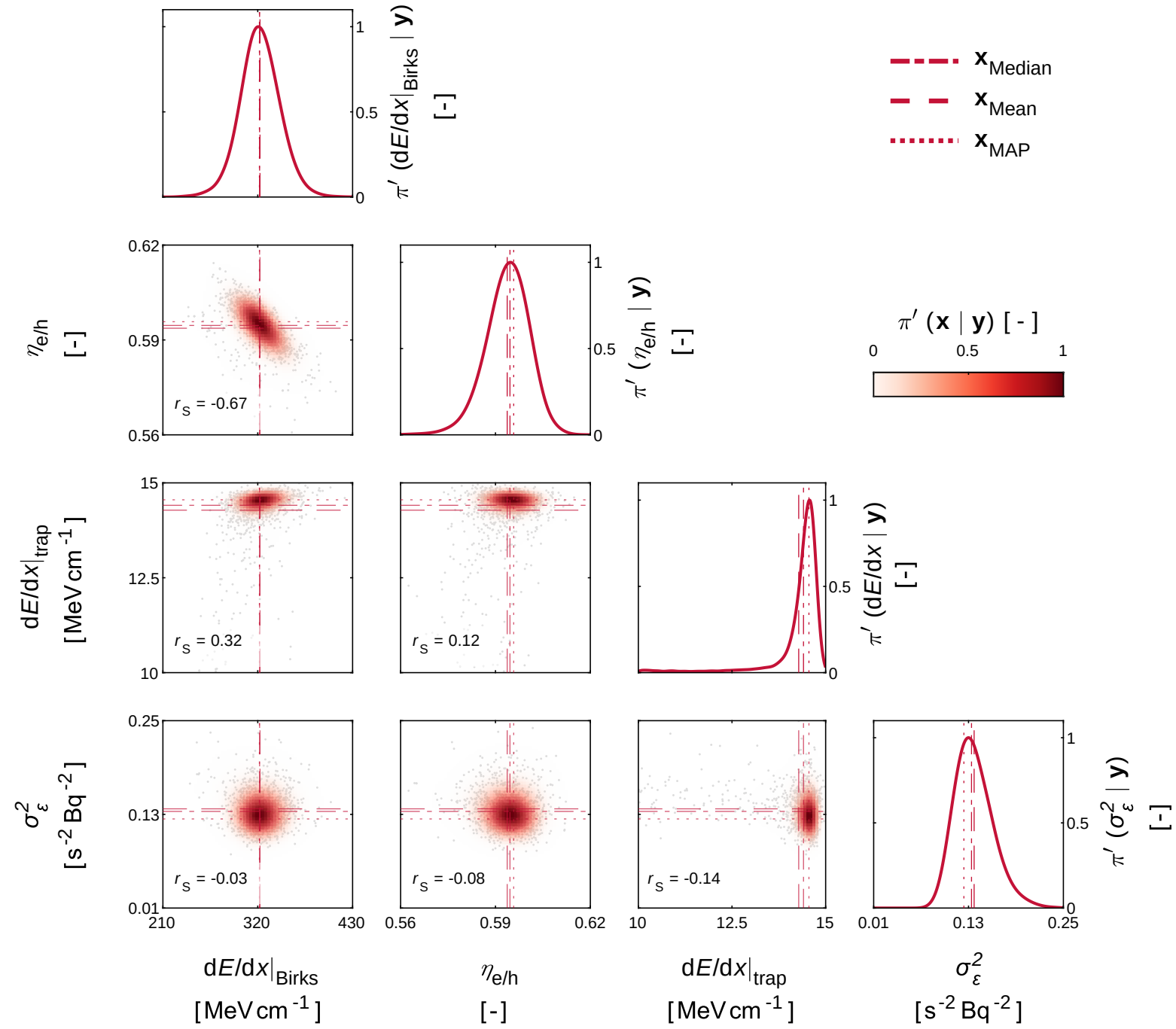

**Figure 7.1** Here, I present the Bayesian inversion results for the detector channel `#SUM` in the form of a corner plot. The off-diagonal subfigures display the normalized bivariate posterior marginal estimates (color-encoded) along with a subset of the first $10^3$ MCMC samples (in gray) for the model parameters $\mathbf{x} := (\,\mathrm{d}E/\mathrm{d}x\,|_{\mathrm{Birks}}, \eta_{\mathrm{e/h}}, \,\mathrm{d}E/\mathrm{d}x\,|_{\mathrm{trap}}, \sigma_\varepsilon^2)^\top$ and experimental data $\mathbf{y}$. In addition, Spearman's rank correlation coefficient $r_S$ is provided for the model parameters in the corresponding off-diagonal subfigures (cf. Section 7.2.4.2). The subfigures on the diagonal axis highlight the normalized univariate posterior marginals for the corresponding model parameter. Both the univariate and multivariate marginals were normalized by their corresponding global maxima. Derived posterior point estimates, i.e. the maximum a posteriori (MAP) probability estimate $\mathbf{x}_{\mathrm{MAP}}$, the posterior mean $\mathbf{x}_{\mathrm{Mean}}$ and the posterior median $\mathbf{x}_{\mathrm{Median}}$, are indicated as well in each subfigure.

### 7.3.2 Compton Edge Predictions

As discussed in Section 7.2.4.2, we can leverage the Bayesian probabilistic framework to predict the model response as prior and posterior predictive distributions. In Fig. 7.2, I present these distributions for all five detector channels as prediction intervals with a probability mass of 99 % within the Compton edge domain $\mathcal{D}_{\mathrm{CE}}$. A comparison of the prior and posterior prediction intervals indicates that the newly developed methodology successfully constrains the adopted NPSM.[32] From a modeling perspective, it is interesting to add that no significant difference for Compton edge predictions using the various point estimates discussed in Section 7.2.4.2 can be observed. Furthermore, I find no statistically significant difference in the posterior predictive distribution between the detector channel `#SUM` and the aggregated single detector channel response (cf. Fig. B.72).

32 *Note that the prior predictive distribution spans the entire display area in Fig. 7.2 for each detector channel.*

In addition to the Bayesian posterior probabilistic predictions, we can also use the trained PCE surrogate models $\hat{\mathcal{M}}(\mathbf{x}_{\mathcal{M}})$ to examine the spectral Compton edge response as a function of the NPSM parameters $\mathbf{x}_{\mathcal{M}}$. In Fig. 7.3, I present the spectral response of the PCE surrogate model for the detector channel `#SUM` as a function of the Birks stopping parameter $\mathrm{d}E/\mathrm{d}x\,|_{\mathrm{Birks}}$, the electron-hole pair fraction $\eta_{\mathrm{e/h}}$ and the trapping stopping parameter $\mathrm{d}E/\mathrm{d}x\,|_{\mathrm{trap}}$ discussed in Section 4.1.3 and Appendix A.7. We can observe a consistent shift of the Compton edge toward smaller spectral energies with an increase in $\mathrm{d}E/\mathrm{d}x\,|_{\mathrm{Birks}}$ and $\eta_{\mathrm{e/h}}$, and, to a lesser degree, with a decrease in $\mathrm{d}E/\mathrm{d}x\,|_{\mathrm{trap}}$. Consistent results are obtained for the individual single detector channels `#1` through `#4` (cf. Figs. B.73–B.76).

As discussed in Section 7.2.4.1, we can leverage the analytical relation between the polynomial chaos expansion and the Hoeffding-Sobol decomposition [742] to perform a global sensitivity analysis of the NPSM. In Fig. 7.3, I also include total Sobol' indices[33] $S^{\mathrm{T}}$ for the NPSM model parameters as a function of the spectral energy $E'$. From Fig. 7.3, I find that the total Sobol' indices can be ordered as $S^{\mathrm{T}}(\eta_{\mathrm{e/h}}) > S^{\mathrm{T}}(\mathrm{d}E/\mathrm{d}x\,|_{\mathrm{Birks}}) > S^{\mathrm{T}}(\mathrm{d}E/\mathrm{d}x\,|_{\mathrm{trap}})$ over the entire spectral Compton edge domain $\mathcal{D}_{\mathrm{CE}}$ indicating a corresponding relative contribution to the total model response variance. Again, consistent results are obtained for the individual single detector channels (cf. Figs. B.73–B.76).

33 *Total Sobol' indices quantify the total effect of an input parameter $X$ on the model response $\mathcal{M}(X)$ (large values correspond to a large effect and vice versa). More details on Hoeffding-Sobol decomposition and associated sensitivity indices can be found in Appendix A.12.*

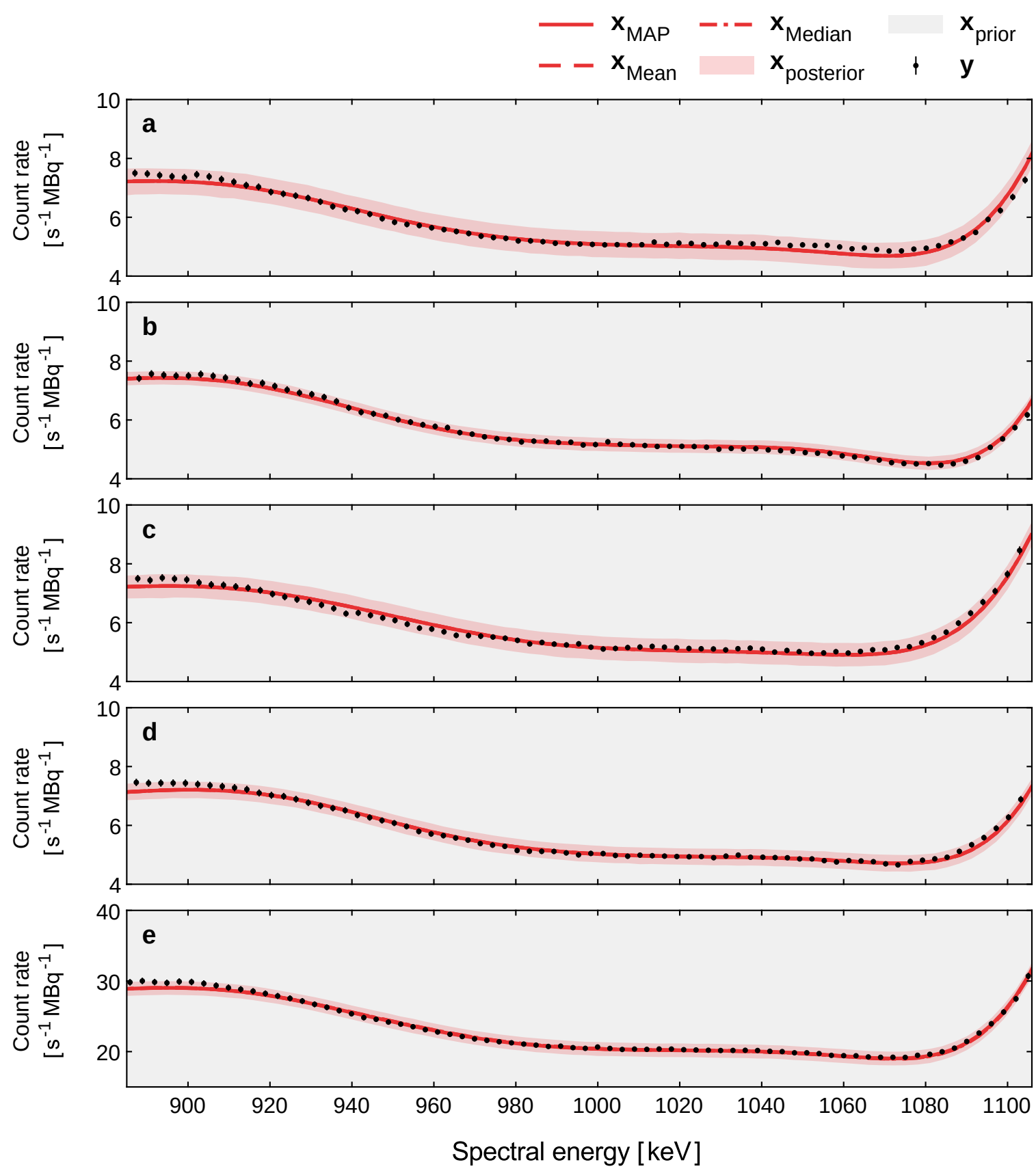

**Figure 7.2** In these graphs, I present the prior and posterior predictive distributions using 99 % central credible intervals for the single detector channels `#1` through `#4` (a–d) as well as the detector channel `#SUM` (e). In addition, the corresponding experimental data $\mathbf{y}$ together with the derived posterior predictions using point estimates, i.e. the maximum a posteriori (MAP) probability estimate $\mathbf{x}_{\mathrm{MAP}}$, the posterior mean $\mathbf{x}_{\mathrm{Mean}}$ and the posterior median $\mathbf{x}_{\mathrm{Median}}$, are indicated as well in each subfigure.

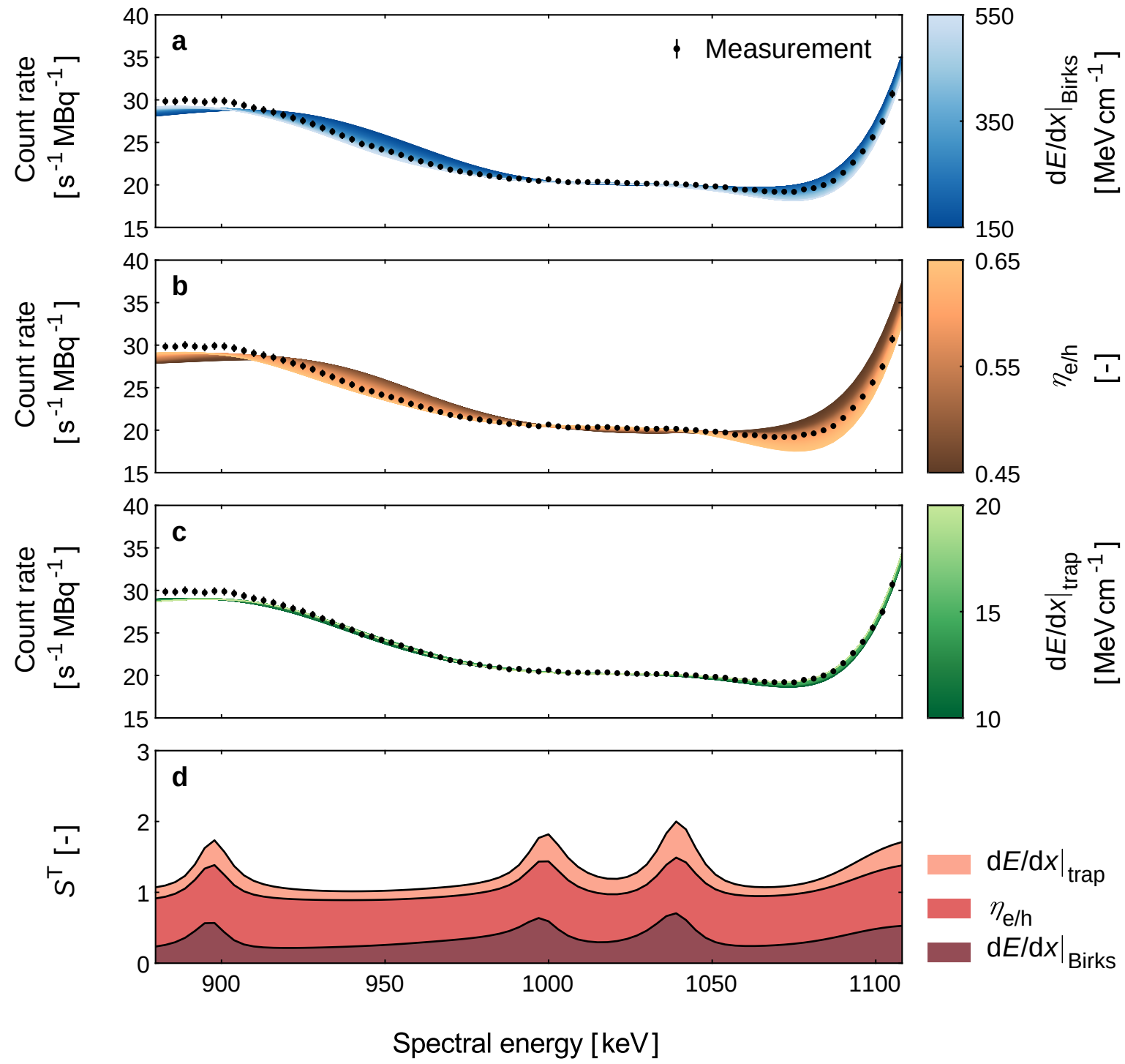

**Figure 7.3** Compton edge dynamics characterized by the trained PCE surrogate model for the detector channel `#SUM`. **a–c** In these subgraphs, the PCE surrogate model predictions are displayed as a function of the spectral energy $E'$ and the individual NPSM parameters, i.e. the Birks stopping parameter $\mathrm{d}E/\mathrm{d}x|_{\mathrm{Birks}}$, the electron-hole pair fraction $\eta_{\mathrm{e/h}}$ as well as the trapping stopping parameter $\mathrm{d}E/\mathrm{d}x|_{\mathrm{trap}}$. The remaining parameters are fixed at the corresponding maximum a posteriori (MAP) probability estimates. The measured $^{60}_{27}\mathrm{Co}$ spectral signature for the detector channel `#SUM` (cf. Chapter 6) is indicated as well as a reference. **d** In this subgraph, I show stacked total Sobol′ indices $S^{\mathrm{T}}$ computed by the trained PCE surrogate model [742] as a function of the spectral energy $E'$ and the individual NPSM parameters.

### 7.3.3 Intrinsic Resolution

The intrinsic resolution discussed in detail in Section 4.3.2 is of great importance for gamma-ray spectrometry for two key reasons. First, it sets a fundamental lower limit on the achievable spectral resolution for a given scintillator material, making it a crucial factor in the development of new scintillators (cf. Section 4.3.2). Second, from a more technical perspective, the intrinsic resolution is a key meta-model required to perform NPSMC (cf. Section 7.2.2). As a result, I will discuss the intrinsic resolution models derived by the `NPScinCal` pipeline here in more detail.

In Fig. 7.4, I present the intrinsic resolution models for the four single detector channels `#1` through `#4` and the detector channel `#SUM` as a function of the spectral energy $E'$. In addition, I have included the total spectral resolution models derived by the `RLLCal` pipeline together with the empirical dataset discussed in Chapter 6 (cf. also to Fig. 6.3).

Comparing the intrinsic and total spectral resolution, I find an almost constant ratio $R^2_{E,\mathrm{intr}}/R^2_E$ between 0.2 and 0.6, depending on the scintillation crystal, for $E' \gtrsim 1.0\,\mathrm{MeV}$. Around $E' \approx 440\,\mathrm{keV}$, there is a pronounced peak with a relative increase with respect to the values at higher spectral energies of ~15 % to ~40 %, depending on the scintillation crystal. For $E' \lesssim 440\,\mathrm{keV}$, I observe a significant decrease in $R^2_{E,\mathrm{intr}}/R^2_E$ with decreasing photon energy $E'$. Moreover, for $E' \lesssim 100\,\mathrm{keV}$, I find a more complex behaviour in the $R_{E,\mathrm{intr}}$ with a pronounced discontinuity at the iodine K-absorption edge at $E' = 33.1694(4)\,\mathrm{keV}$ [815]. On the other hand, at even smaller spectral energies, there is again a marked increase in $R_{E,\mathrm{intr}}$ with decreasing spectral energy compared to the mere moderate increase for $60\,\mathrm{keV} \lesssim E' \lesssim 100\,\mathrm{keV}$.

From a modeling perspective, it is important to add that the complex trend in $R_{E,\mathrm{intr}}$ at lower spectral energies requires careful consideration in the modeling of the intrinsic resolution for NPSMC applications. Here, I have adopted a non-parametric GP regression model [734] to accurately quantify the intrinsic resolution at low spectral energies.

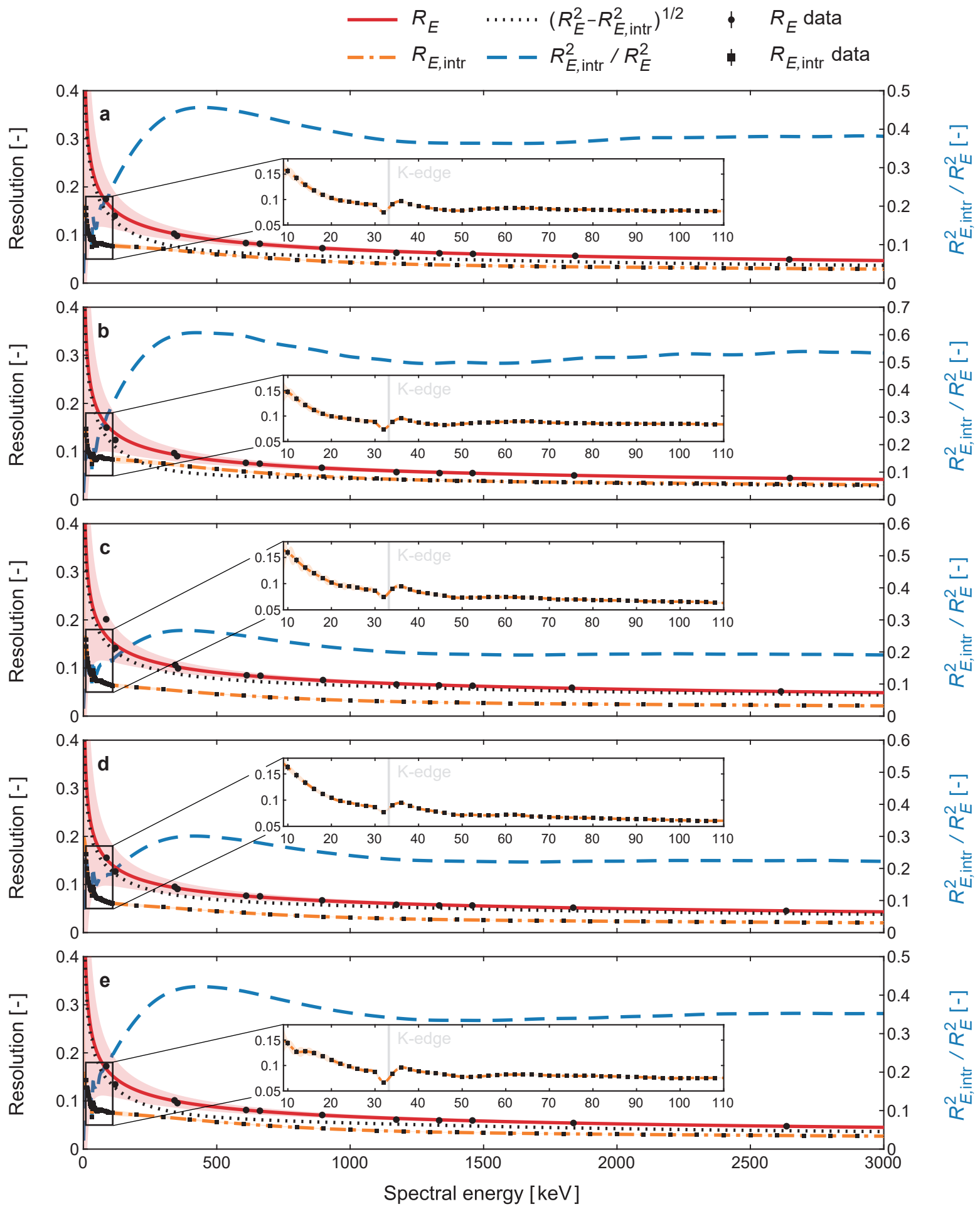

**Figure 7.4** This graph presents the total ($R_E$) and the intrinsic ($R_{E,\mathrm{intr}}$) spectral resolution models for the four single detector channels `#1` through `#4` (**a–d**) and the detector channel `#SUM` (**e**) as a function of the spectral energy $E'$ (for the reader's convenience, the continuous pulse-height channel number $\tilde{n}$ was converted to the spectral energy $E'$ using the energy calibration models, cf. Fig. 6.2). The total spectral resolution models derived by the `RLLCal` were discussed already in Chapter 6 (cf. also Fig. 6.3). The intrinsic spectral resolution models were derived by the `NPScinCal` pipeline discussed in Section 7.2.2. For the zoomed insets with $E' \leq 110\,\mathrm{keV}$, the K-absorption edge for iodine is highlighted [815]. Uncertainties are provided as 1 standard deviation (SD) values for all subgraphs presented in this figure.

### 7.3.4 Spectral Signature

To validate the newly developed NPSMC methodology, I utilized the measured ${}^{57}_{27}$Co, ${}^{88}_{39}$Y, ${}^{109}_{48}$Cd, ${}^{133}_{56}$Ba, ${}^{137}_{55}$Cs and ${}^{152}_{63}$Eu spectral signatures presented in Chapter 6. For that purpose, I performed online NPSMC as described in Section 7.2.1 with meta-models derived by the `NPScinCal` pipeline (cf. Section 7.2.3) and spectral postprocessing conducted by the `NPScinMC` pipeline (cf. Section 7.2.2). Motivated by the results discussed in Sections 7.3.1 and 7.3.2 as well as previous results from the literature [369], I applied the Bayesian calibrated NPSM of the detector channel `#SUM` to all four scintillation crystals using the corresponding posterior point estimates $\mathbf{x}_{\mathrm{MAP}}$ (cf. Table C.10) together with the related meta-models, i.e. the non-proportional scintillation scaling and intrinsic resolution model discussed in Section 7.2.2. Because the Compton edges are obscured by other FEPs for the ${}^{57}_{27}$Co, ${}^{109}_{48}$Cd, ${}^{133}_{56}$Ba and ${}^{152}_{63}$Eu spectral signatures, I will focus here on the ${}^{88}_{39}$Y and ${}^{137}_{55}$Cs measurements.

In Figs. 7.5 and 7.6, I present the measured and simulated spectral signatures of the detector channel `#SUM` for ${}^{88}_{39}$Y and ${}^{137}_{55}$Cs as well as ${}^{60}_{27}$Co, whose Compton edge domain was used to perform the calibration of the adopted NPSM. I display both, simulation results obtained by PSMC discussed in Chapter 6 (in red) as well as the newly introduced NPSMC methodology from this chapter (in blue).

For all three measurements, I observe a significant improvement in the Compton edge prediction for the NPSMC simulations compared to the standard PSMC method. However, there are still some discrepancies at the lower end of the Compton edge domain. Moreover, I find also some deviations at the Compton gap between the Compton edge and the related FEP for ${}^{88}_{39}$Y and ${}^{137}_{55}$Cs. It is important to note that these discrepancies are smaller or at least of similar size for the NPSMC simulations compared to the proportional approach indicating that the former performs significantly better for all examined spectral signatures. Consistent results are obtained for the single detector channels `#1` through `#4` (cf. Figs. B.36–B.43). Additional validation results for the spectral signatures of ${}^{57}_{27}$Co, ${}^{109}_{48}$Cd, ${}^{133}_{56}$Ba and ${}^{152}_{63}$Eu are provided in the Figs. B.44–B.53.

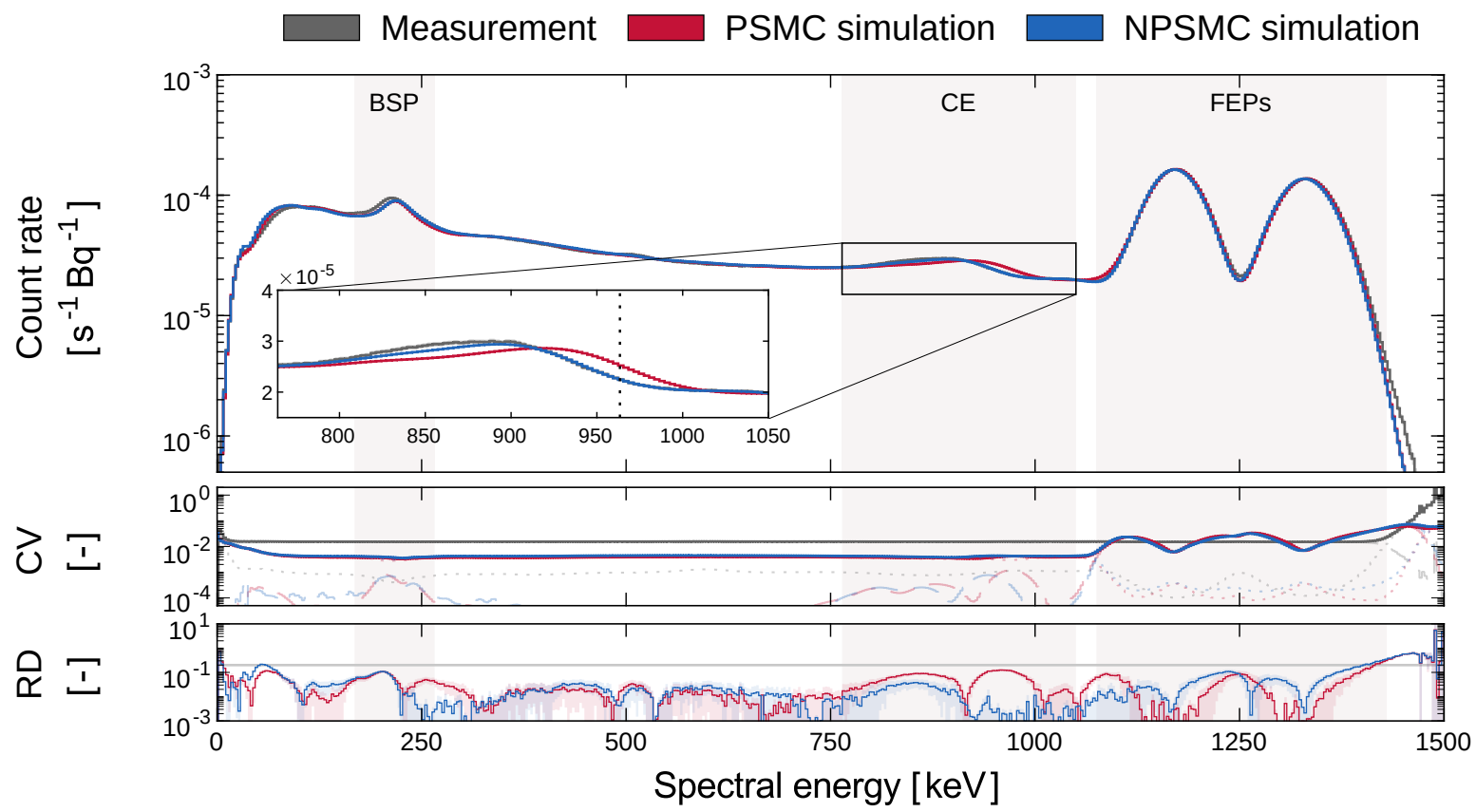

**Figure 7.5** The measured ($\hat{c}_{exp}$) and simulated ($\hat{c}_{sim}$) mean spectral signatures for a $^{60}_{27}$Co source ($\mathcal{A}$ = 3.08(5) × 10$^5$ Bq) are shown for the detector channel `#SUM` as a function of the spectral energy $E'$ with a spectral energy bin width of $\Delta E' \sim 3$ keV. The simulated spectral signatures were obtained by PSMC (in red, cf. Chapter 6) and NPSMC (in blue). Distinct spectral regions, i.e. the backscatter peak (BSP), the domain around the Compton edge (CE) as well as the full energy peaks (FEPs) are marked. The zoomed-in subfigure highlights the spectral region around the Compton edge marked by the vertical dotted line and associated with the photon emission line at 1173.228(3) keV [68] (cf. Eq. 4.21). Uncertainties ($\hat{\sigma}_{exp}$, $\hat{\sigma}_{sim}$) are provided as 1 standard deviation (SD) shaded areas. In addition, the coefficient of variation (CV) for the measured and simulated signatures (statistical and systematic contributions are indicated by shaded dotted and dashed lines, respectively, cf. Appendix A.8) as well as the relative deviation (RD) computed as $|\hat{c}_{sim} - \hat{c}_{exp}|/\hat{c}_{exp}$ (20 % mark highlighted by a horizontal grey line) are provided.

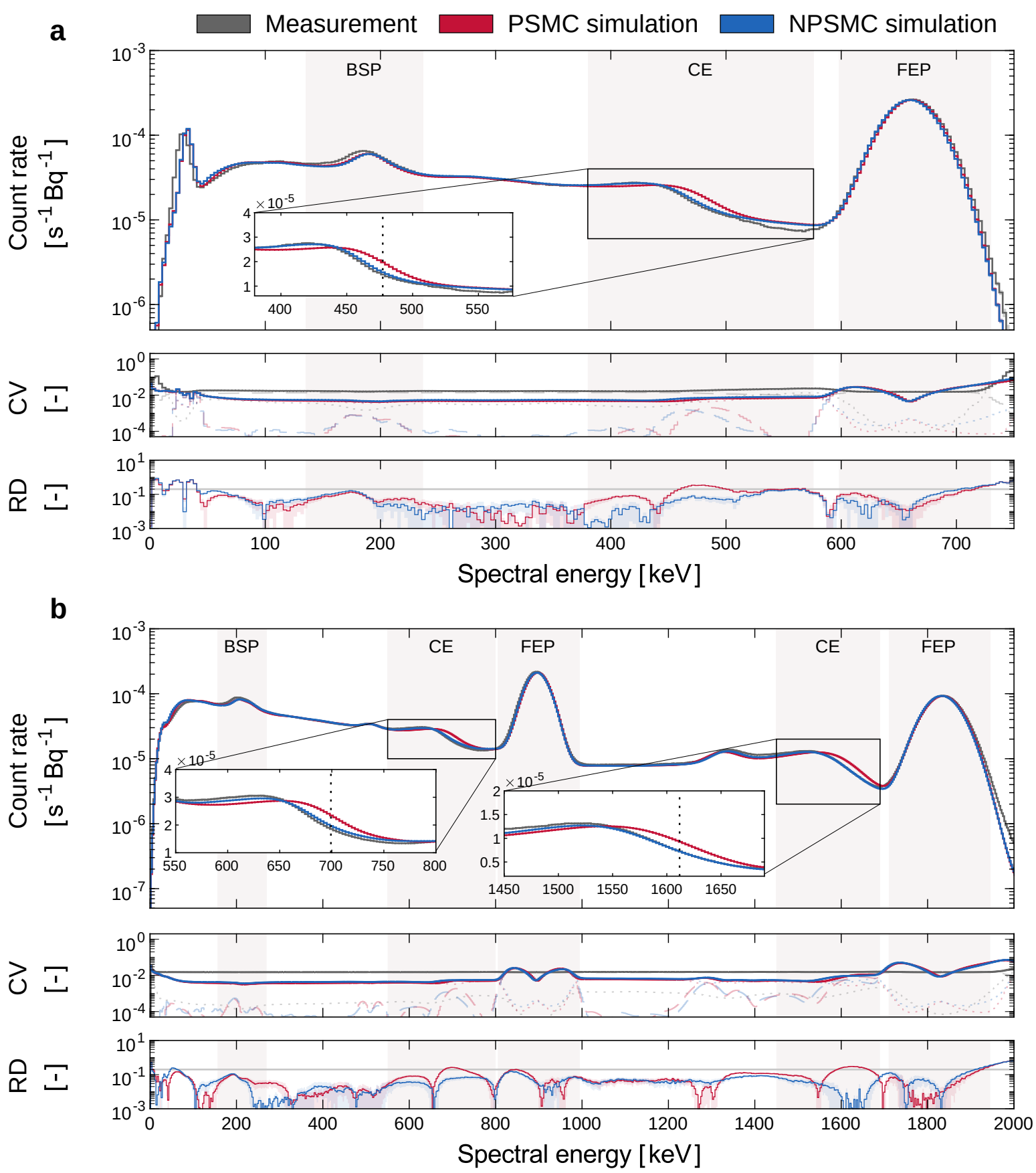

**Figure 7.6** The measured ($\hat{c}_{\mathrm{exp}}$) and simulated ($\hat{c}_{\mathrm{sim}}$) mean spectral signatures for two radionuclides sources and the detector channel `#SUM` as a function of the spectral energy $E'$ with a spectral energy bin width of $\Delta E' \sim 3\,\mathrm{keV}$ are displayed: **a** $^{137}_{55}\mathrm{Cs}$ ($\mathcal{A} = 2.266(34) \times 10^5\,\mathrm{Bq}$). **b** $^{88}_{39}\mathrm{Y}$ ($\mathcal{A} = 6.83(14) \times 10^5\,\mathrm{Bq}$). The simulated spectral signatures were obtained by PSMC (in red, cf. Chapter 6) and NPSMC (in blue). Distinct spectral regions, i.e. the backscatter peak (BSP), the domain around the Compton edge (CE) as well as the full energy peak (FEP) are marked. The zoomed-in subfigures highlight the spectral regions around the Compton edge marked by the vertical dotted line and associated with the photon emission lines at 661.657(3) keV for $^{137}_{55}\mathrm{Cs}$ [68] as well as 898.042(11) keV and 1836.070(8) keV for $^{88}_{39}\mathrm{Y}$ [51]. Uncertainties ($\hat{\sigma}_{\mathrm{exp}}$, $\hat{\sigma}_{\mathrm{sim}}$) are provided as 1 standard deviation (SD) shaded areas. In addition, the coefficient of variation (CV) for the measured and simulated signatures (statistical and systematic contributions are indicated by shaded dotted and dashed lines, respectively, cf. Appendix A.8) as well as the relative deviation (RD) computed as $|\hat{c}_{\mathrm{sim}} - \hat{c}_{\mathrm{exp}}|/\hat{c}_{\mathrm{exp}}$ (20 % mark highlighted by a horizontal grey line) are provided.

## 7.4 Conclusion

In this chapter, I have developed and validated a NPSMC model of the RLL spectrometer under laboratory conditions. In contrast to previous studies [341, 355, 412], the calibrated NPSM is evaluated online during the Monte Carlo simulation, thereby saving considerable computation time and avoiding the need to store and analyze large files with secondary particle data.

The calibration of the NPSM was performed using a novel calibration method, Compton edge probing, which provides a reliable and cost-effective alternative to existing NPSM calibration methods, eliminating the need for additional measurement equipment and facilitating online calibration during detector deployment. Comparing the calibrated NPSM parameters with best-estimate literature values [325], I found statistically significant differences, which may be attributed to the sensitivity of these parameters to the activator concentration and impurities in the scintillation crystals. This conclusion is supported by the Bayesian calibrated NPSM for the individual scintillation crystals of the RLL spectrometer, which showed statistically significant differences between the individual NPSM parameters, too. These findings corroborate the experimental results of Hull and his co-workers [413] and underscore the criticality of the NPSM calibration for NPSMC.

Validation of the newly developed NPSMC methodology was performed with spectral signatures of two radionuclide sources under laboratory conditions. A comparison between the simulated spectral signatures and the measured data indicated a significant improvement in the prediction accuracy of NPSMC relative to the standard PSMC method, particularly in the Compton edge domain. Nonetheless, some discrepancies persist. Apart from the already discussed discrepancies at $E' \lesssim 50\,\mathrm{keV}$ and in the BSPs (cf. Sections 6.3.1 and 6.4), some deviations were observed at the lower and higher ends of the Compton edge domain. These discrepancies might be attributed to systematic uncertainties in the Monte Carlo mass model or deficiencies in the adopted NPSM.

Using the derived PCE-PCA surrogate model, predictions of the spectral Compton edge shift as a function of the NPSM parameters can be made. I observed a Compton edge shift toward smaller spectral energies for an increase in $\mathrm{d}E/\mathrm{d}x|_{\mathrm{Birks}}$ and $\eta_{\mathrm{e/h}}$ as well as a decrease in $\mathrm{d}E/\mathrm{d}x|_{\mathrm{trap}}$. These results imply that the Compton edge shift is enhanced for an increase in the electron-hole pair fraction as well

as an increase in the Onsager mechanism relative to the Birks mechanism, which in turn promotes the scintillation non-proportionality (cf. Appendix A.7).

Further, levering PCE-PCA-based Sobol′ indices, I quantified the sensitivity of the spectral response in the Compton edge domain on the individual NPSM parameters. The sensitivity results indicate that $\eta_{\mathrm{e/h}}$ has the highest sensitivity, followed by $\mathrm{d}E/\mathrm{d}x|_{\mathrm{Birks}}$ and $\mathrm{d}E/\mathrm{d}x|_{\mathrm{trap}}$. However, the obtained sensitivity results should be interpreted with care as the relative importance of the NPSM parameters might change due to changes in the scintillator properties [325].

The proposed NPSMC method can not only be used to accurately predict the full spectrum response of inorganic scintillators but it can also be utilized to investigate various scintillator properties related to the scintillation non-proportionality, such as the Compton edge shift discussed above or the intrinsic resolution. A detailed analysis of the intrinsic resolution for the RLL spectrometer revealed a significant contribution of the scintillation non-proportionality to the total spectral resolution with some dispersion between the individual scintillation crystals. Similar variability has been found in previous studies for NaI(Tl) with $R_{E,\mathrm{intr}}^2/R_E^2$ ranging from ~30 % [324] up to ~58 % [373] at $E' = 662\,\mathrm{keV}$ and $R_E = 8\,\%$[34] (cf. also to [341, 359, 372, 375, 412, 816]).

34 *Which is the spectral resolution for the detector channel* ***#SUM*** *of the RLL spectrometer at* $E' = 662\,\mathrm{keV}$ *(cf. Fig. 6.3).*

While I focused in this chapter on NaI(Tl) in electron and gamma-ray fields, the presented NPSMC methodology can easily be extended to a much broader range of applications. First, it is consensus that the light yield as a function of the stopping power is, at least to a first approximation, independent of the ionizing particle type [326, 327]. Second, the adopted NPSM was validated with an extensive database of measured scintillation light yields for inorganic scintillators, i.e. BGO, $CaF_2$(Eu), $CeBr_3$, CsI(Tl), CsI(Na), $LaBr_3$(Ce), LSO(Ce), NaI(Tl), $SrI_2$, $SrI_2$(Eu), YAP(Ce) and YAG(Ce), among others [313, 324, 325]. From this, it follows that the proposed NPSMC methodology may in principle be applied to any combination of inorganic scintillator and ionizing radiation field, including protons, $\alpha$-particles and heavy ions.

Regarding the newly developed NPSM calibration method, Compton edge probing, it is important to note that it relies on the probing of Compton edge shifts with a sufficient signal-to-noise ratio (SNR). I have shown that these shifts are influenced by the strength of the non-proportional scintillation response of a given scintillator.

Furthermore, we expect a decrease in the Compton edge shift for a decrease in the scintillator volume as well as a decrease in the Compton edge energy itself (cf. Appendix A.10). As a result, applications with small scintillator volumes, low energy Compton edges or scintillator materials that exhibit only a mild non-proportional scintillation response, e.g. $LaBr_3$(Ce) or YAP(Ce) (cf. Table 4.1), may present challenges for Compton edge probing. That said, the presented calibration method can be readily adapted using Bayes's theorem to address low SNR cases more effectively by combining multiple Compton edge domains or by probing additional spectral features distorted by the non-proportional scintillation response.

In summary, the developed NPSMC model allows for accurate predictions of the full spectrum response of the RLL spectrometer, successfully correcting the observed discrepancies in the Compton edge domain of the standard PSMC method presented in the previous chapter. These discrepancies are expected to increase with larger scintillator volumes, making NPSMC particularly valuable for applications involving large crystal scintillators such as AGRS, but also in dark matter research, total absorption spectroscopy or remote sensing [693, 817–820].

# Part III

# Integrated System Modeling

> *"This is the Way."*
>
> — *Paz Vizsla, The Mandalorian*

Chapter 8

# AGRS Monte Carlo

**Contents**

In the previous part, I focused on the development and validation of the physics models of the RLL AGRS detector system using high-fidelity Monte Carlo simulations. Following a bottom-up approach, all aircraft components were excluded, which allowed me to validate the developed physics models under controlled laboratory conditions with high-counting statistics.

As a next step towards accurate Monte Carlo based full spectrum modeling of AGRS detector systems, I extend in this chapter the developed mass model of the RLL AGRS detector system to include the entire aircraft as well as the environment. For all performed simulations, the validated NPSMC physics model from the previous chapter was adopted using again the multi-purpose Monte Carlo code `FLUKA`.

To validate the extended Monte Carlo model, several field measurement campaigns were performed using natural and man-made radionuclides. The results demonstrate that the developed Monte Carlo model accurately reproduces the measured pulse-height spectra with a median relative deviation of less than 8 % in all performed radiation measurements.

The validated AGRS Monte Carlo model allows for the computation of spectral signatures for arbitrary source-detector configurations, marking a crucial step toward full spectrum simulation-based calibration of AGRS systems.

## 8.1 Introduction

In the previous part, I have focused on the development and validation of the physics models of the RLL detector system under laboratory conditions, specifically excluding all aircraft components. This allowed me to validate the developed physics models under controlled laboratory conditions with high-counting statistics.

From our discussions in Chapters 4 and 5, given the substantial size and mass of the aircraft systems used in AGRS (cf. Section 5.2), we can anticipate that the aircraft system has a significant impact on the measured pulse-height spectra, in particular in the Compton continuum at lower spectral energies. Incorporating the aircraft in the PSMC or NPSMC mass model is therefore essential for accurate full spectrum modeling of AGRS systems.

However, the literature review in Section 5.5.2 revealed that fully integrated Monte Carlo models of AGRS detector systems are scarce. In most of the studies, aircraft components were completely neglected or represented by primitive geometric shapes such as spheres or cubes leading to large systematic errors at low spectral energies. Moreover, scintillation non-proportionality effects discussed in Section 4.1.3 were not considered in any of the studies reviewed in Section 5.5.2.

As a next step towards accurate Monte Carlo based full spectrum modeling of AGRS detector systems, the scope of this chapter is the derivation and validation of a fully integrated NPSMC model of the Swiss AGRS system combining the already validated mass model of the RLL spectrometer with a detailed mass model of the aircraft system, an Aérospatiale AS332M1 Super Puma (TH06) helicopter (cf. Section 5.3.2). A comprehensive analysis of the derived AGRS Monte Carlo model's response to external photon fields will be presented in the subsequent Chapter 9.

## 8.2 Monte Carlo Model

Similar to the previous two chapters, I adopted the `FLUKA` code[1] [20, 216, 281] maintained by the FLUKA.CERN Collaboration [20, 216, 281] together with the graphical interface `FLAIR`[2] [718] to perform the Monte Carlo simulations presented in this chapter. All Monte Carlo simulations were conducted on a local computer cluster (7 nodes with a total number of 520 cores at a nominal clock speed of 2.6 GHz) at the Paul Scherrer Institute (PSI) utilizing parallel computing.

1 *Version* `4-2.2` *for the simulations presented in this chapter.*

2 *Version* `3.2-4.5` *for the simulations presented in this chapter.*

Following the bottom-up modeling approach outlined in Chapter 6, I utilized the already validated NPSMC model of the RLL spectrometer presented in Chapter 7. Consequently, the derivation of the AGRS Monte Carlo model primarily involved extending the mass model of the RLL AGRS detector system to include the aircraft, while all physics models and postprocessing routines remain unchanged. In this section, I limit the discussion therefore to the extension of the mass model. Comprehensive information on the physics models, scoring and postprocessing routines is available in the Sections 6.2.2.1, 6.2.2.2, 7.2.1 and 7.2.3.

Similar to Chapter 6, I used `FLAIR` and `FLUKA`'s core Combinatorial Solid Geometry (CSG) builder to extend the already validated RLL

spectrometer mass model derived in Section 6.2.2.3 with the aircraft system components. Geometrical as well as material properties of the various aircraft components were obtained from a combination of sources including expert interviews, technical schemes and CAD files provided by RUAG AG.[3] Due to their size and mass, the RLL supporting systems[4] installed in the crew cabin during survey flights were also included in the mass model (Section 5.3.2). Given the complexity of the individual subsystems, the mass model was built in a modular fashion, with the individual components being modeled as separate homogeneous objects. For these simplifications, care was taken to preserve the overall geometry, the opacity as well as the mass density.[5] Complex geometries of some aircraft components like the fuselage were modeled by fitting quadrics to the corresponding aircraft surfaces.[6] The atmosphere around the TH06 was represented as homogeneous humid air with adjustable air temperature $T$, air pressure $p$ and relative humidity RH.[7] The resulting mass model of the AGRS detector system is illustrated in Fig. 8.1. It features over $10^3$ bodies and $10^2$ different materials.

There is another important aspect in the derivation of AGRS mass models, which I would like to highlight here. Previous studies reviewed in Section 5.5.2 either completely neglected the aircraft system or represented it as a static object using primitive geometric shapes. However, aircraft systems employed in AGRS are inherently dynamic, with many components changing position, composition or orientation during or between survey flights. As I will demonstrate in this and in the next chapter, some of this dynamic behavior has a significant impact on the measured pulse-height spectra. Therefore, in contrast to previous studies, I explicitly account for the following dynamic changes in the aircraft system:[8]

1. **Fuel** The TH06 possesses six fuel tanks with a total capacity of ~2 m$^3$ holding about $1.6 \times 10^3$ kg of jet fuel A-1. As illustrated in Fig. 8.1, these tanks are arranged around the RLL spectrometer within the helicopter's fuselage. To maintain the aircraft's center of gravity, these fuel tanks are emptied in a predetermined sequence during the course of a survey flight. To account for these fuel level changes, I modeled the fuel level in each tank separately as a dynamic parameter following the aircraft's fuel depletion during the course of a survey flight (cf. Fig. B.81).[9]
2. **Crew** As discussed in Section 5.3.2, the core personnel for the Swiss AGRS system consists of two pilots operating the aircraft

3 *The available CAD files were incomplete, missing many key subsystems due to the fact that the TH06 was designed in 1974 based on technical drawings. Therefore, I did not consider CAD-based mass model builders in this work (cf. Section 3.2.2.1).*

4 *This includes the operator console, the operator seats and an equipment rack, among others.*

5 *There is one exception. Given the small longitudinal cross-section, a large mass of $\gtrsim 300$ kg and a distance of >4 m from the RLL spectrometer, the tail of the TH06 helicopter was simplified as a perfect black body absorber. This simplification ensured that the maximum extension of the AGRS Monte Carlo model in the longitudinal direction was limited to 10 m (neglecting the rotor blades).*

6 *Using the geometry card `QUA` in `FLUKA` combined with dimensional data from technical schemes and CAD files.*

7 *To compute the equivalent mass density and water mass fraction, I adapted the `AirProperties` code by Fitzgerald [728]. The composition of the dry air was modeled based on data provided by McConn et al. [249].*

8 *Note that the landing gears are always extended during AGRS surveys. Consequently, even though a dynamic model for the landing gears was developed, their position in the mass model remained fixed in the extended position for all simulations presented in this book.*

9 *Note that the fuel surface is modeled as a plane parallel to the principal axes $x'$-$y'$.*

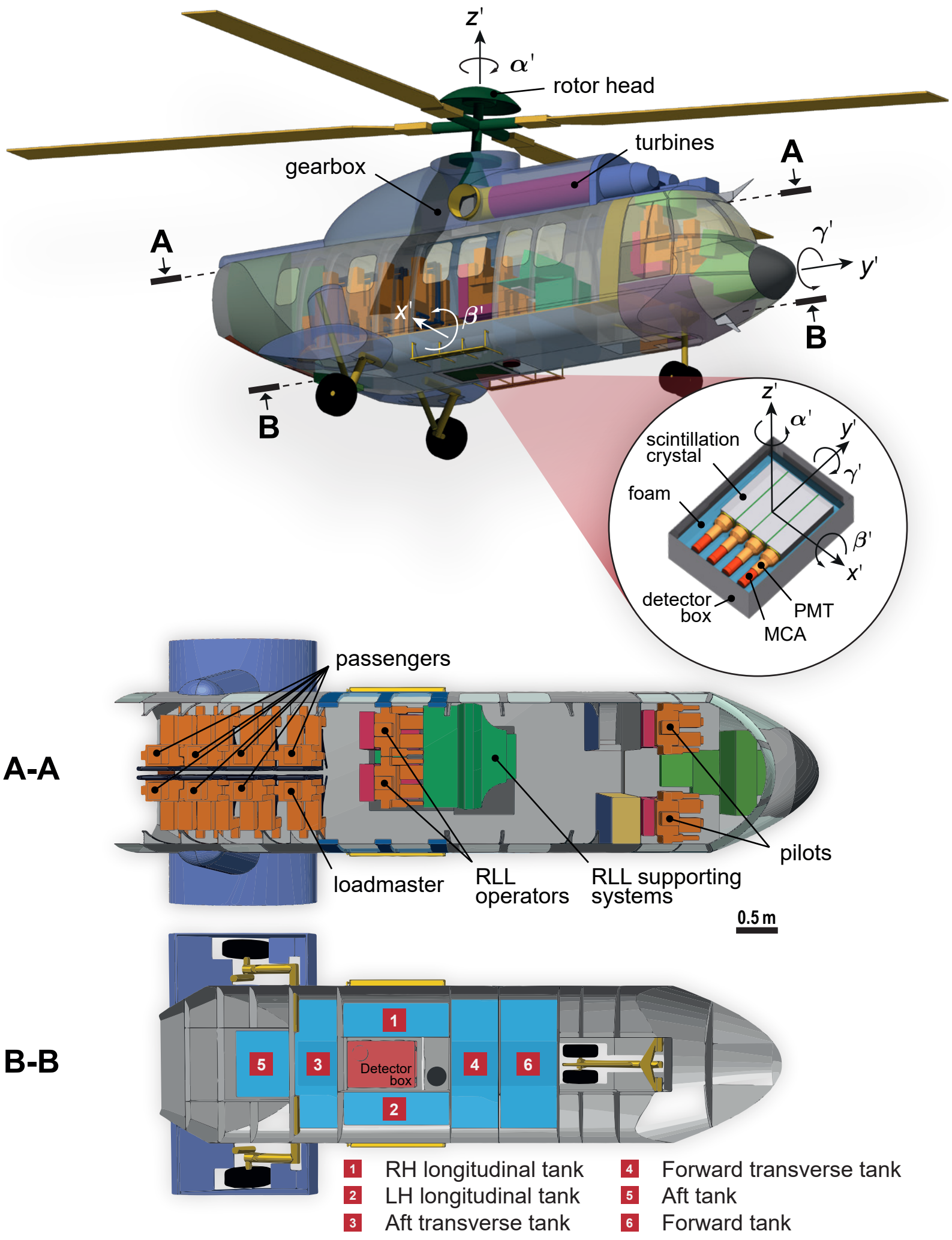

**Figure 8.1** Here, I illustrate the derived Monte Carlo mass model of the Swiss AGRS system including the entire TH06 aircraft and the RLL detector system. For reference, I also indicate the adopted principal axes $x'$-$y'$-$z'$ and corresponding Tait-Bryan angles, i.e. yaw $\alpha'$, pitch $\beta'$ and roll $\gamma'$ quantifying the orientation of the mass model with respect to the world frame. **A-A** Cross-section view of the aircraft cockpit and the cabin including the RLL supporting systems and the crew (max. passenger capacity). Note that for visualization purposes, the rear of the cabin was shortened. **B-B** Cross-section view of the fuselage highlighting the position of the six fuel tanks relative to the RLL detector box (landing gear retracted). All mass model figures displayed here were created using the graphical interface `FLAIR` [718]. For better visibility and interpretability, semi-transparent false colors were applied.

in the cockpit in the front as well as a loadmaster and two RLL operators situated in the cabin. In addition to this core crew, a variable number of up to seven passengers can be seated in the rear of the aircraft. An illustration of the cockpit and cabin layout is provided in Fig. 8.1. To account for changes in the number of passengers across several flights, the presence or absence of the individual passengers was modeled as a dynamic parameter in the derived AGRS mass model. The crew members were modeled using mass, elemental composition and anthropometric dimensional data for a reference human male provided by the ICRP [76] and the National Aeronautics and Space Administration (NASA) [821].

3. **Orientation** In order to account for the orientation of the aircraft system with respect to the world frame, I applied rotation transformations using the `ROT-DEFIni` card provided by `FLUKA`. For this purpose, three intrinsic rotations about the aircraft principal axes in the sequence $z'$-$x'$-$y'$ were used with the corresponding Tait-Bryan angles[10] $\alpha'$, $\beta'$ and $\gamma'$, as indicated in Fig. 8.1. In aerospace engineering, these angles are also known as yaw ($\alpha'$), pitch ($\beta'$) and roll ($\gamma'$). Note that there is no standard convention on how to orient the principal axes for Tait-Bryan rotations. In this work, I orient the principal axes of the aircraft system as indicated in Fig. 8.1: $z'$ points upward parallel to the fuselage station, $y'$ aligns with the longitudinal axis pointing forward and $x'$ aligns with the transverse axis pointing to starboard. The origin of the aircraft reference system is defined as the center of gravity of the four scintillation crystals.

10 *Named after Peter Guthrie Tait (*1831, †1901), a Scottish mathematical physicist and George Hartley Bryan (*1864, †1928), an English applied mathematician.*

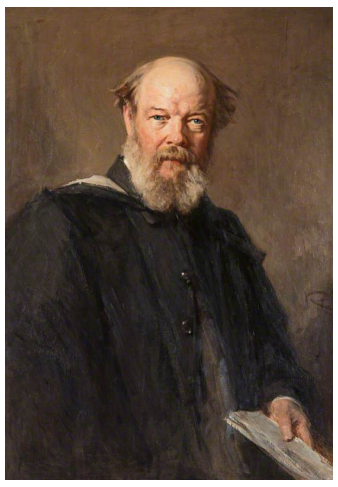

Peter G. Tait

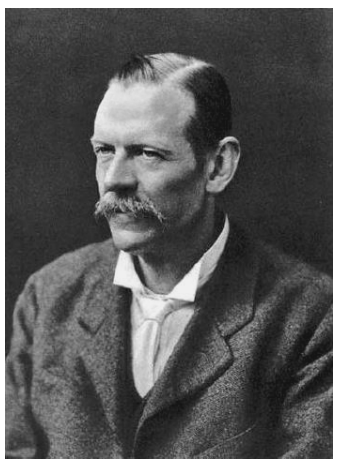

George H. Bryan

## 8.3 Model Validation

To validate the AGRS Monte Carlo model described above, I performed two field measurement campaigns:

1. Dübendorf Validation Campaign
2. ARM22 Validation Campaign

In the following two subsections, I will present the measurement setup and the results from these two validation campaigns. Given the large number of measurements and simulations performed, I will limit the discussion here to the detector channel `#SUM` (cf. Section 6.2.1.2).

### 8.3.1 Dübendorf Validation Campaign

The scope of this field measurement campaign was the spectral and angular validation of the derived AGRS Monte Carlo model presented in the previous section for extended radionuclide sources positioned below the TH06 aircraft in the near field, i.e. with the source-detector distance being smaller or of comparable size to the characteristic extension of the aircraft (~10 m).

#### 8.3.1.1 Radiation Measurements

All measurements were conducted at the Dübendorf Airfield, specifically in hangar No. 12 of the Swiss Air Force (47.405°N, 8.643°E), over an extension of four days between 2021-12-14 and 2021-12-17. As illustrated in Fig. 8.2, the measurements were performed with the fully integrated Swiss AGRS system[11] located inside hangar No. 12.

Custom-made $K_{nat}$ radionuclide sources were utilized for the measurements. As illustrated in Fig. 8.2, these sources consist of granular KCl sealed into aluminum casings. In total, seven sources have been employed for each measurement with a total activity of 456.6(26) kBq. Details on the KCl mass and resulting $^{40}_{19}$K activities of the individual sources are provided in Table C.13. Technical drawings illustrating the design of the custom-made $K_{nat}$ radionuclide sources can be found in Fig. B.82.

In total, ten radiation measurements have been performed. For each measurement, the seven $K_{nat}$ radionuclide sources have been placed on a predefined position on the hangar floor covering an area of 1.5 m × 0.65 m as indicated in Fig. 8.2. The selected source positions cover a solid angle between $\theta' = 180°$ (0C) and $\theta' = 110°$ (2E, 2N, 2S) with $\theta'$ being the polar angle with respect to the principal axis $z'$ (cf. Fig. 8.1). To position the sources accurately, distance as well as positioning laser systems were used.

Data acquisition was performed with the proprietary software suite `SpirIDENT` (cf. Section 5.3.2). Between the gross radiation measurements, background measurements were performed regularly for background correction and gain stability checks. Measurement live times for both gross and background measurements are provided in Table C.14. At the beginning and end of each measurement, air temperature, pressure and humidity were recorded using an external sensor (HM30 Meteo Station).[12]

Postprocessing of the measured pulse-height spectra was performed using the `RLLSpec` pipeline described in Section 6.2.1.2. The

11 *Including the RLL supporting systems in the cabin of the TH06 helicopter (cf. Section 5.3.2). No personnel was present inside the aircraft during the measurements.*

12 *Note that the TH06 helicopter was offline during the measurement campaign and thus, no external sensor data of the TH06 helicopter were available.*

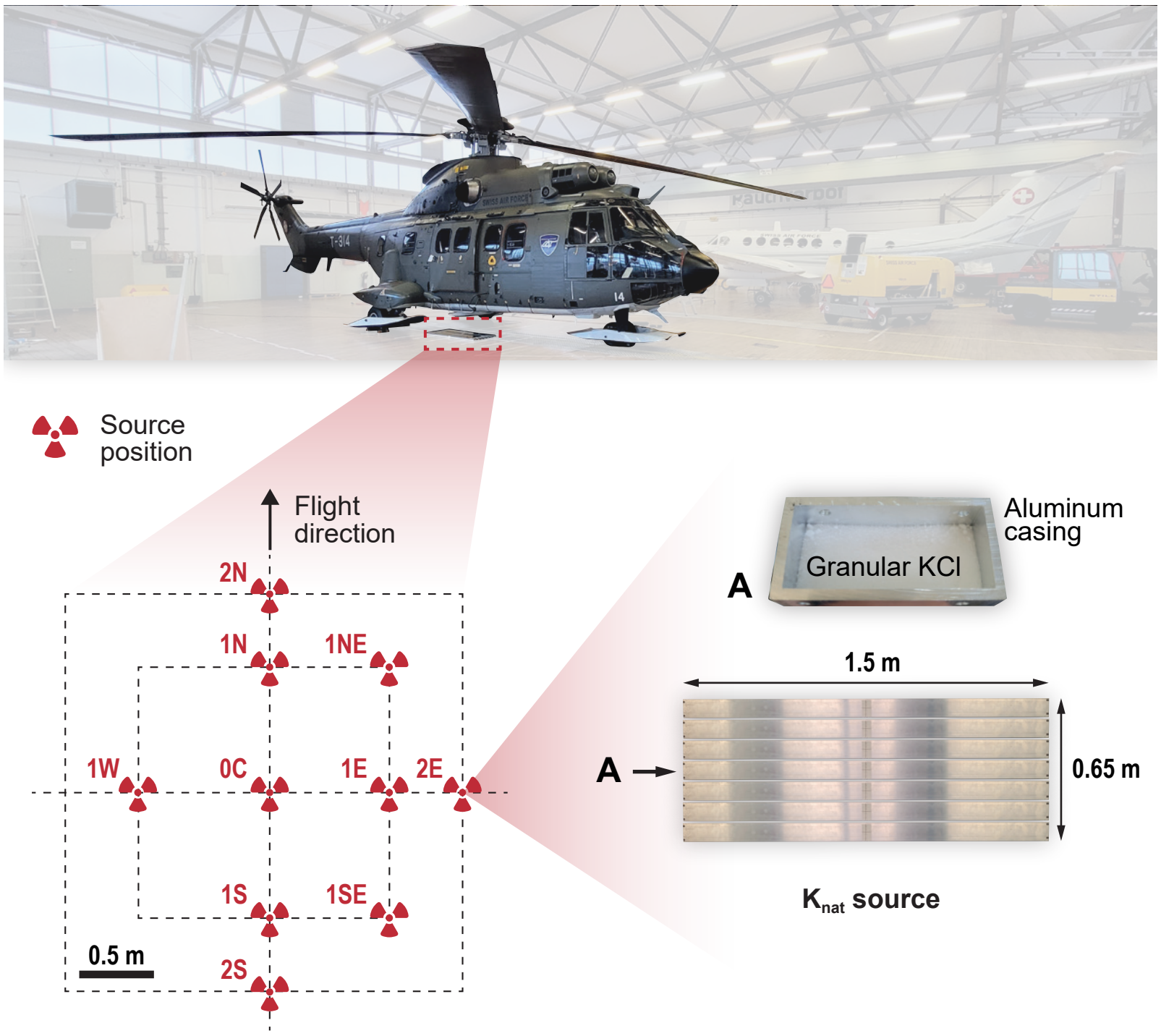

**Figure 8.2** Experimental setup of the field measurements conducted during the Dübendorf validation campaign. The main elements of the setup are the fully integrated Swiss AGRS system located in hangar No. 12 of the Dübendorf Airfield (47.405°N, 8.643°E) and the custom-made $K_{nat}$ radionuclide sources consisting of granular KCl sealed into aluminum casings. These sources were placed on ten different predefined positions on the hangar floor covering an area of 1.5 m × 0.65 m. Note that for visualization purposes, the top cover of the aluminum casing in the zoomed-in view of the $K_{nat}$ radionuclide source is removed (cf. Fig. B.82).

spectral energy and the spectral resolution were calibrated using the RLLCal pipeline detailed in Section 6.2.1.3. This calibration was necessary because a different RLL spectrometer was deployed during the Dübendorf validation campaign compared to the one used for the measurements described in Chapter 6. Except for $^{88}_{39}Y$ and $^{109}_{48}Cd$, which were not available, the same radionuclides have been adopted for the calibration as listed in Table C.5. Calibrated models for the spectral energy, the spectral resolution and the LLD[13] are provided in Figs. B.83–B.85 and Table C.15.

13 *For the LLD, the methods described in Appendix A.9 were adopted to calibrate its model parameters.*

#### 8.3.1.2 Monte Carlo Simulations

As already indicated in Section 8.2, all Monte Carlo simulations in this chapter were performed using NPSMC discussed in Chapter 7. The general mass model of the AGRS Monte Carlo model was described already in Section 8.2. Therefore, I limit the discussion here to the specific simulation setup for the Dübendorf validation campaign.

Similar to the mass model derived for the simulations presented in the Chapters 6 and 7, I extended the AGRS mass model described in Section 8.2 to include the hangar floor, walls, doors and ceiling. The atmosphere in the hangar was modeled as homogeneous humid air using the measured air temperature, pressure and humidity (cf. Section 8.3.1.1).[14] The fuel level in each of the six fuel tanks was set by the corresponding fuel volume fractions obtained from the fuel tank management panel in the cockpit.[15]

14 *As only small fluctuations were observed in these parameters during the individual measurements, the arithmetic mean of the recorded values was used for the corresponding simulations.*

15 *Note that the fuel level in the tanks can only be accessed via this panel. The TH06 helicopter does not possess an option for digital logging of these parameters.*

The seven custom-made $K_{nat}$ radionuclide sources were modeled as individual radionuclide volume sources using the `raddecay` card in a semi-analogue mode. Detailed mass models of the individual sources were derived using the structural information from Fig. B.82 and mass data from Table C.13.

The number of primaries $N_{pr}$ was set to $7 \times 10^8$ for all simulations, which guarantees a median coefficient of variation $\mathrm{med}(\mathbf{CV}_{stat}) < 7\,\%$ and a median relative variance of the variance $\mathrm{med}(\mathbf{VOV}) < 0.1\,\%$ over the SDOI defined in Def. 6.1 in the detector channel `#SUM` (cf. also to Appendix A.8).

Postprocessing of the obtained simulation data was performed using the `NPScinMC` pipeline detailed in Section 7.2.3. Motivated by the observed prediction power for the single detector channels, I adopted the NPSM derived in the previous chapter for the detector channel `#SUM`, although a different RLL spectrometer was used during the Dübendorf validation campaign. As I will demonstrate in the next subsection, the Compton edges are well reproduced by the derived NPSM, indicating that the calibrated NPSM parameters for the detector channel `#SUM` are transferable between the two deployed RLL spectrometers.

#### 8.3.1.3 Results

In the Figs. 8.3–8.7, I present the measured ($\hat{c}_{\text{exp}}$) and simulated ($\hat{c}_{\text{sim}}$) spectral signatures for all ten source positions evaluated during the Dübenrdorf validation campaign alongside uncertainty estimates (cf. Appendix A.8) and relative deviations computed as $|\hat{c}_{\text{sim}} - \hat{c}_{\text{exp}}|/\hat{c}_{\text{exp}}$.

Similar to the NPSMC results obtained in Chapter 7, good agreement between the measured and simulated spectral signatures is observed for all ten source positions with a median relative deviation <8 % within the SDOI. As already indicated in Section 8.3.1.2, thanks to NPSMC, I find also an excellent agreement at the Compton edge for all ten source positions.

However, there are three systematic deviations between the measured and simulated spectral signatures which require further discussion:

1. One of the most evident deviations is observed around the 609.312(7) keV emission line of $^{214}_{83}$Bi [124] and, to a lesser degree, around the 351.932(2) keV emission line of $^{214}_{82}$Pb [124]. I highlighted the corresponding spectral range of the FEPs associated with these emission lines for the measurement 2N in Fig. 8.7, where the deviations are most pronounced. Considering that the large doors of the aircraft hangar were regularly opened during the field measurement campaign to allow for aircraft to be moved in and out, these deviations can be attributed to the changing background levels of the radon progeny $^{214}_{83}$Bi and $^{214}_{82}$Pb in the hangar atmosphere between the gross and background measurements. (cf. also to the discussion in Section 2.1.3.2).

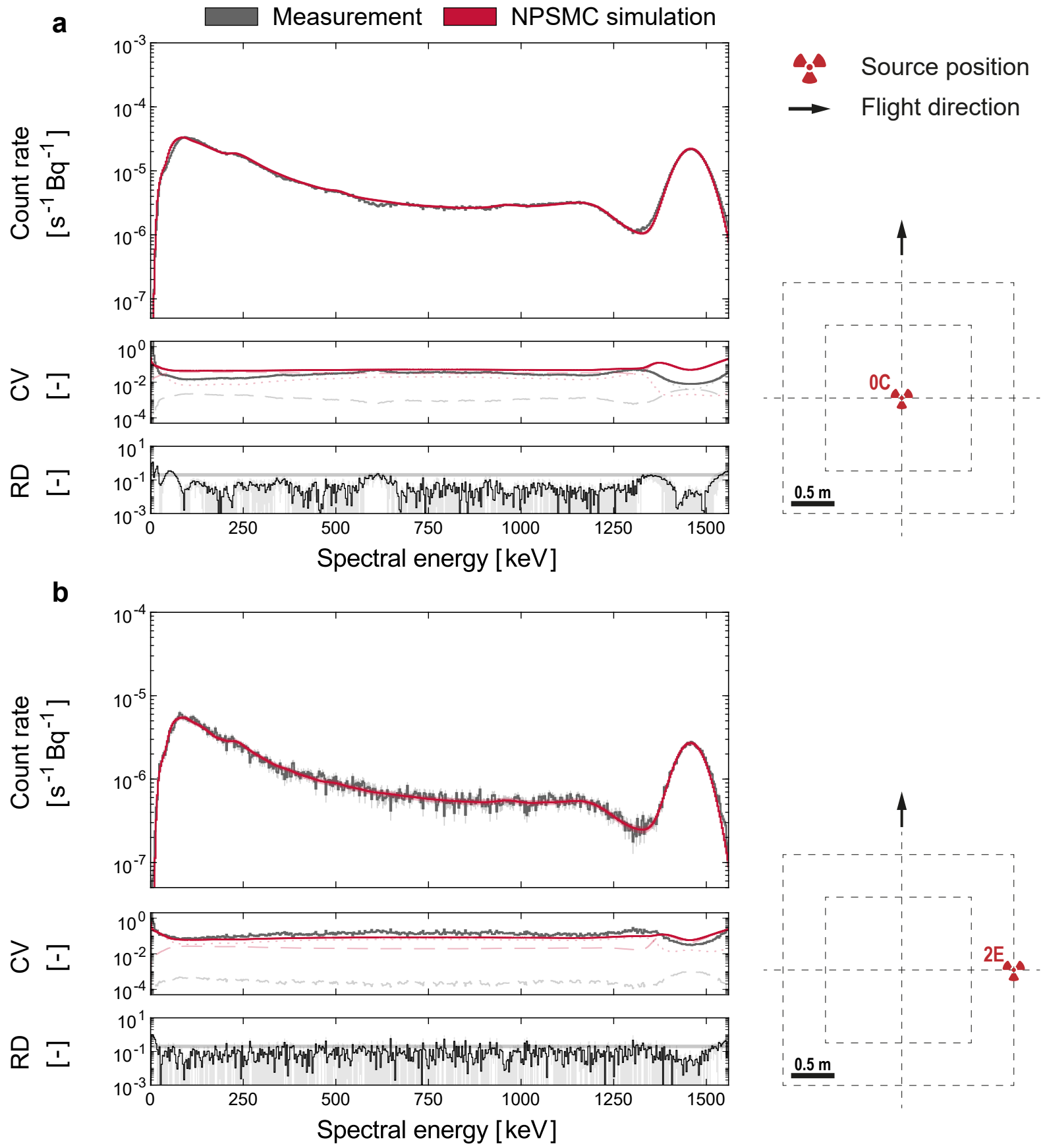

**Figure 8.3** The measured ($\hat{c}_{exp}$) and simulated ($\hat{c}_{sim}$) mean spectral signatures in the detector channel `#SUM` are shown as a function of the spectral energy $E'$ with a spectral energy bin width of $\Delta E' \sim 3$ keV for two measurement positions: **a** 0C. **b** 2E. For both measurements, a custom-made $K_{nat}$ radionuclide source was deployed (total activity $\mathcal{A} = 456.6(26)$ kBq). The simulated spectral signatures were obtained by NPSMC. Uncertainties ($\hat{\sigma}_{exp}$, $\hat{\sigma}_{sim}$) are provided as 1 standard deviation (SD) shaded areas. In addition, the coefficient of variation (CV) for the measured and simulated signatures as well as the relative deviation (RD) computed as $|\hat{c}_{sim} - \hat{c}_{exp}|/\hat{c}_{exp}$ (20 % mark highlighted by a horizontal grey line) are provided. Statistical and systematic contributions to the CV are indicated by shaded dotted and dashed lines, respectively (cf. Appendix A.8).

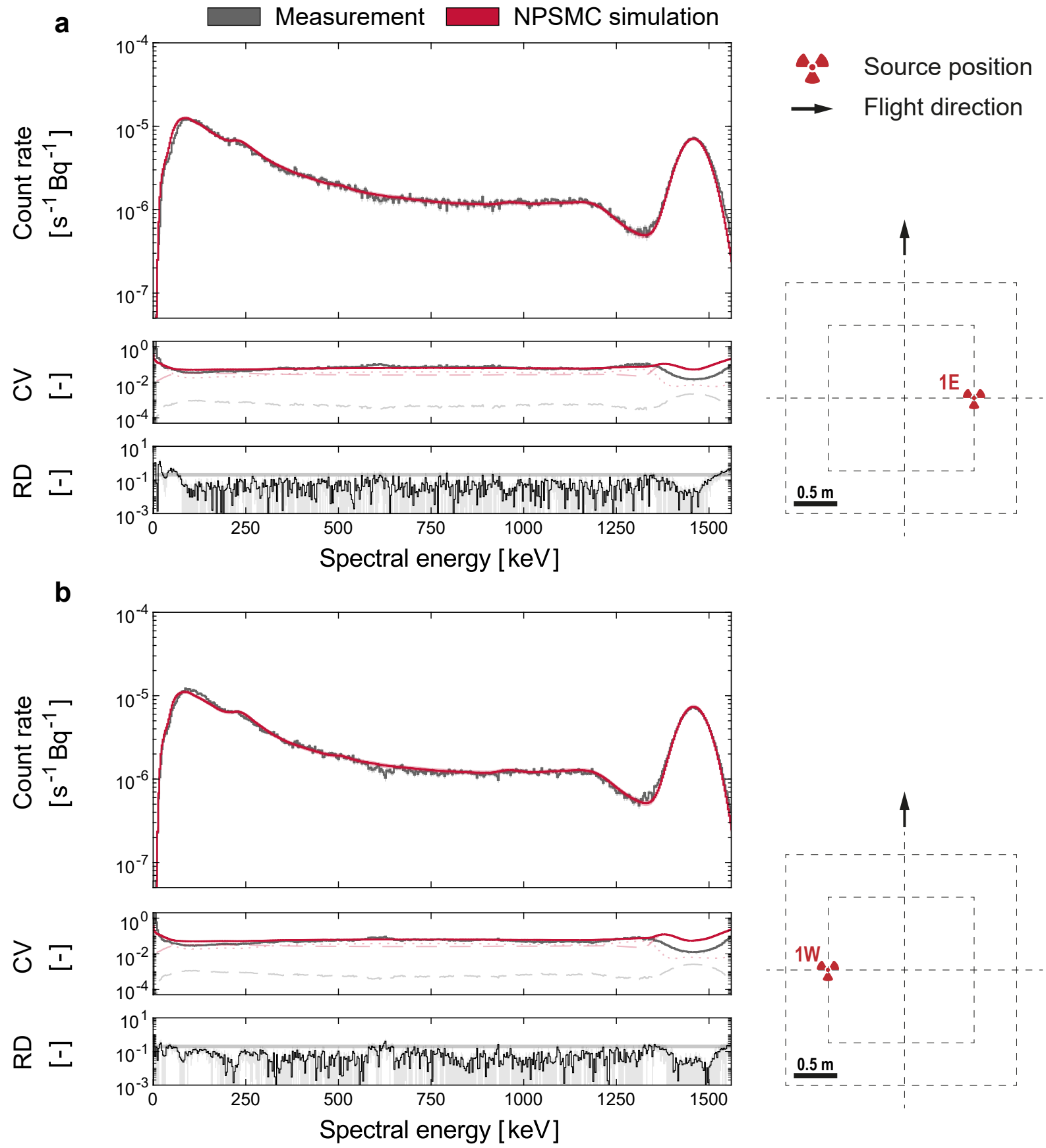

**Figure 8.4** The measured ($\hat{c}_{exp}$) and simulated ($\hat{c}_{sim}$) mean spectral signatures in the detector channel `#SUM` are shown as a function of the spectral energy $E'$ with a spectral energy bin width of $\Delta E' \sim 3$ keV for two measurement positions: **a** 1E. **b** 1W. For both measurements, a custom-made $K_{nat}$ radionuclide source was deployed (total activity $\mathcal{A} = 456.6(26)$ kBq). The simulated spectral signatures were obtained by NPSMC. Uncertainties ($\hat{\sigma}_{exp}$, $\hat{\sigma}_{sim}$) are provided as 1 standard deviation (SD) shaded areas. In addition, the coefficient of variation (CV) for the measured and simulated signatures as well as the relative deviation (RD) computed as $|\hat{c}_{sim} - \hat{c}_{exp}|/\hat{c}_{exp}$ (20 % mark highlighted by a horizontal grey line) are provided. Statistical and systematic contributions to the CV are indicated by shaded dotted and dashed lines, respectively (cf. Appendix A.8).

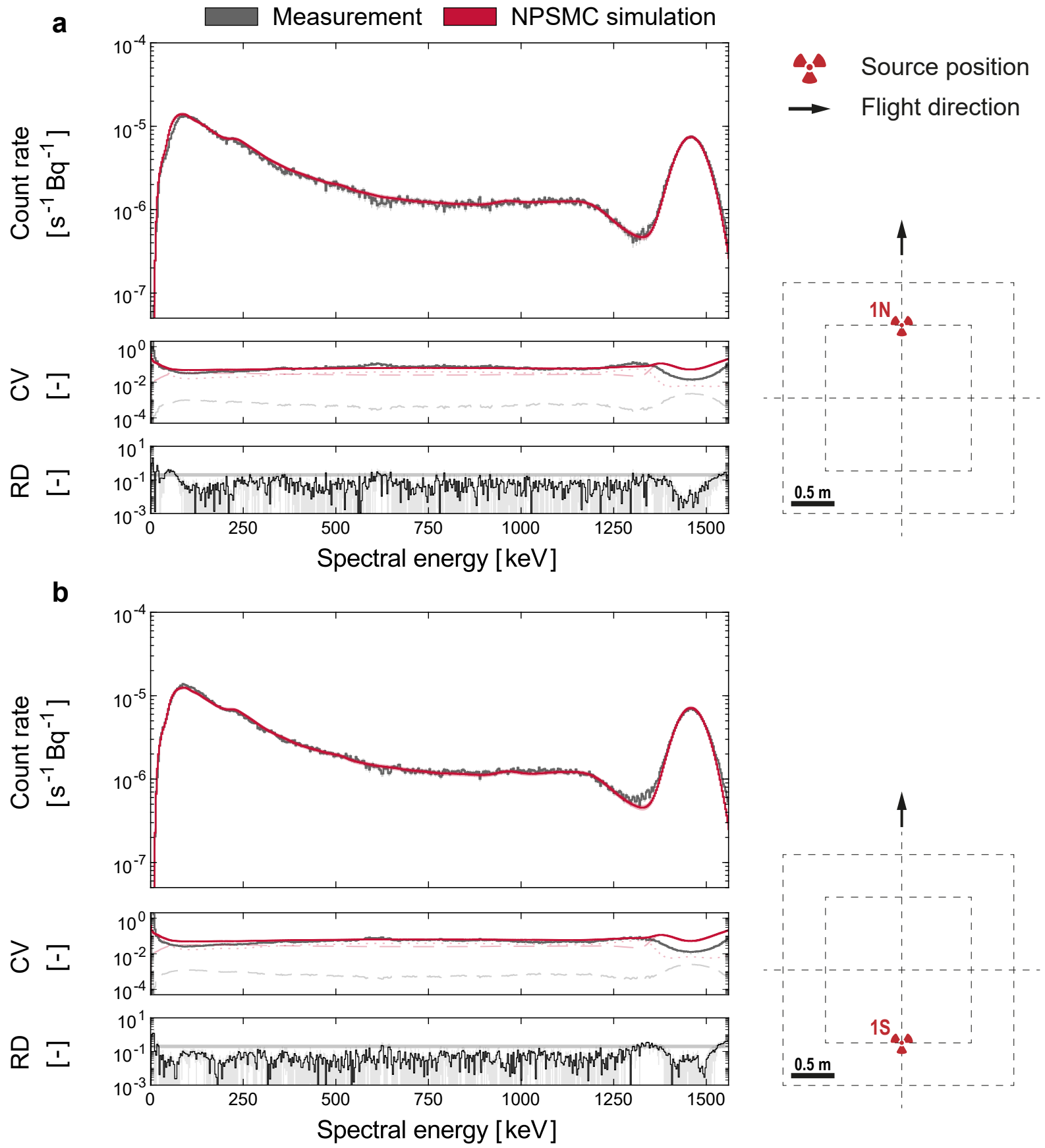

**Figure 8.5** The measured ($\hat{c}_{\mathrm{exp}}$) and simulated ($\hat{c}_{\mathrm{sim}}$) mean spectral signatures in the detector channel #SUM are shown as a function of the spectral energy $E'$ with a spectral energy bin width of $\Delta E' \sim 3$ keV for two measurement positions: **a** 1N. **b** 1S. For both measurements, a custom-made $K_{\mathrm{nat}}$ radionuclide source was deployed (total activity $\mathcal{A} = 456.6(26)$ kBq). The simulated spectral signatures were obtained by NPSMC. Uncertainties ($\hat{\sigma}_{\mathrm{exp}}$, $\hat{\sigma}_{\mathrm{sim}}$) are provided as 1 standard deviation (SD) shaded areas. In addition, the coefficient of variation (CV) for the measured and simulated signatures as well as the relative deviation (RD) computed as $|\hat{c}_{\mathrm{sim}} - \hat{c}_{\mathrm{exp}}|/\hat{c}_{\mathrm{exp}}$ (20 % mark highlighted by a horizontal grey line) are provided. Statistical and systematic contributions to the CV are indicated by shaded dotted and dashed lines, respectively (cf. Appendix A.8).

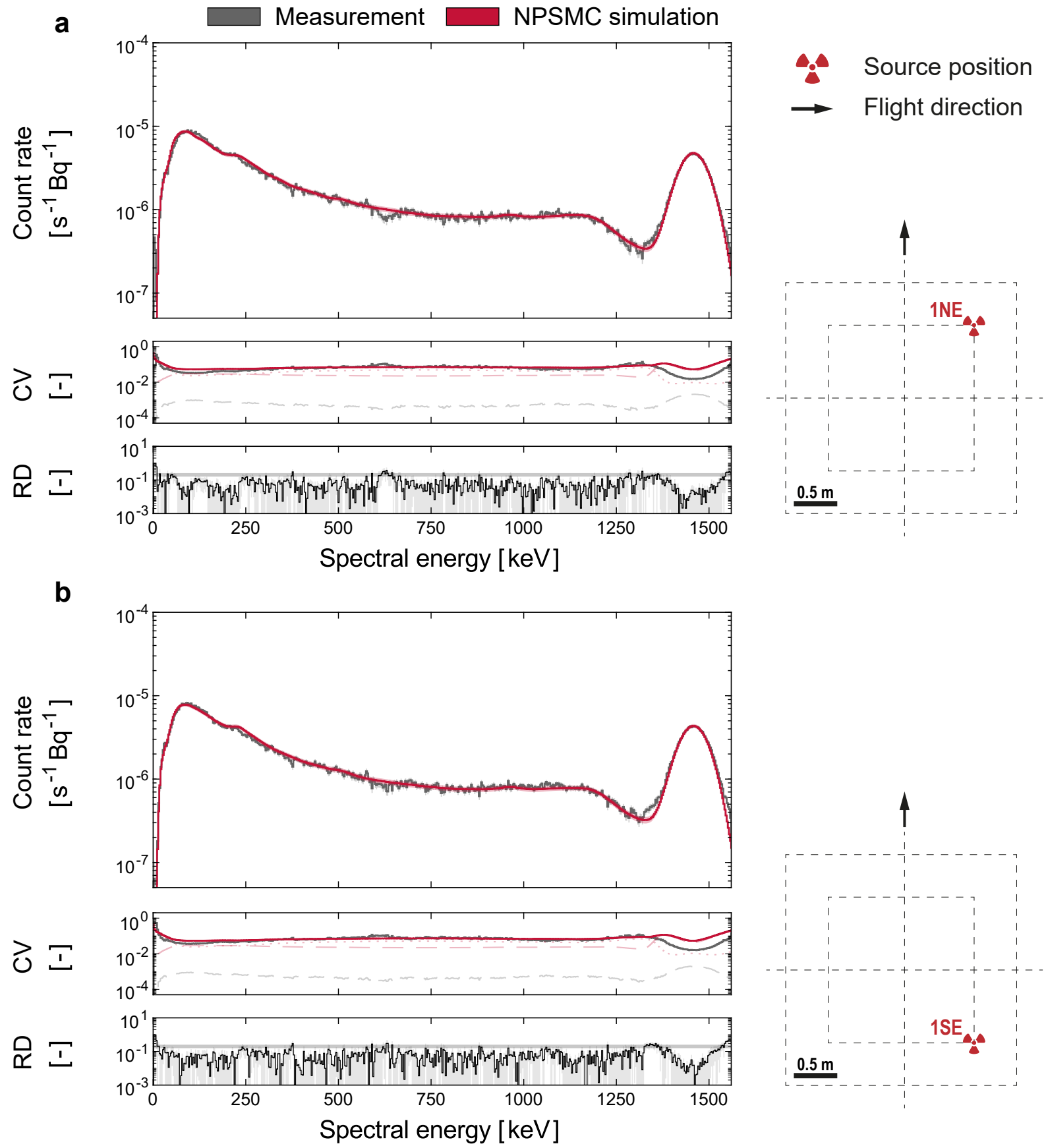

**Figure 8.6** The measured ($\hat{c}_{exp}$) and simulated ($\hat{c}_{sim}$) mean spectral signatures in the detector channel #SUM are shown as a function of the spectral energy $E'$ with a spectral energy bin width of $\Delta E' \sim 3$ keV for two measurement positions: **a** 1NE. **b** 1SE. For both measurements, a custom-made $K_{nat}$ radionuclide source was deployed (total activity $\mathcal{A} = 456.6(26)$ kBq). The simulated spectral signatures were obtained by NPSMC. Uncertainties ($\hat{\sigma}_{exp}$, $\hat{\sigma}_{sim}$) are provided as 1 standard deviation (SD) shaded areas. In addition, the coefficient of variation (CV) for the measured and simulated signatures as well as the relative deviation (RD) computed as $|\hat{c}_{sim} - \hat{c}_{exp}|/\hat{c}_{exp}$ (20 % mark highlighted by a horizontal grey line) are provided. Statistical and systematic contributions to the CV are indicated by shaded dotted and dashed lines, respectively (cf. Appendix A.8).

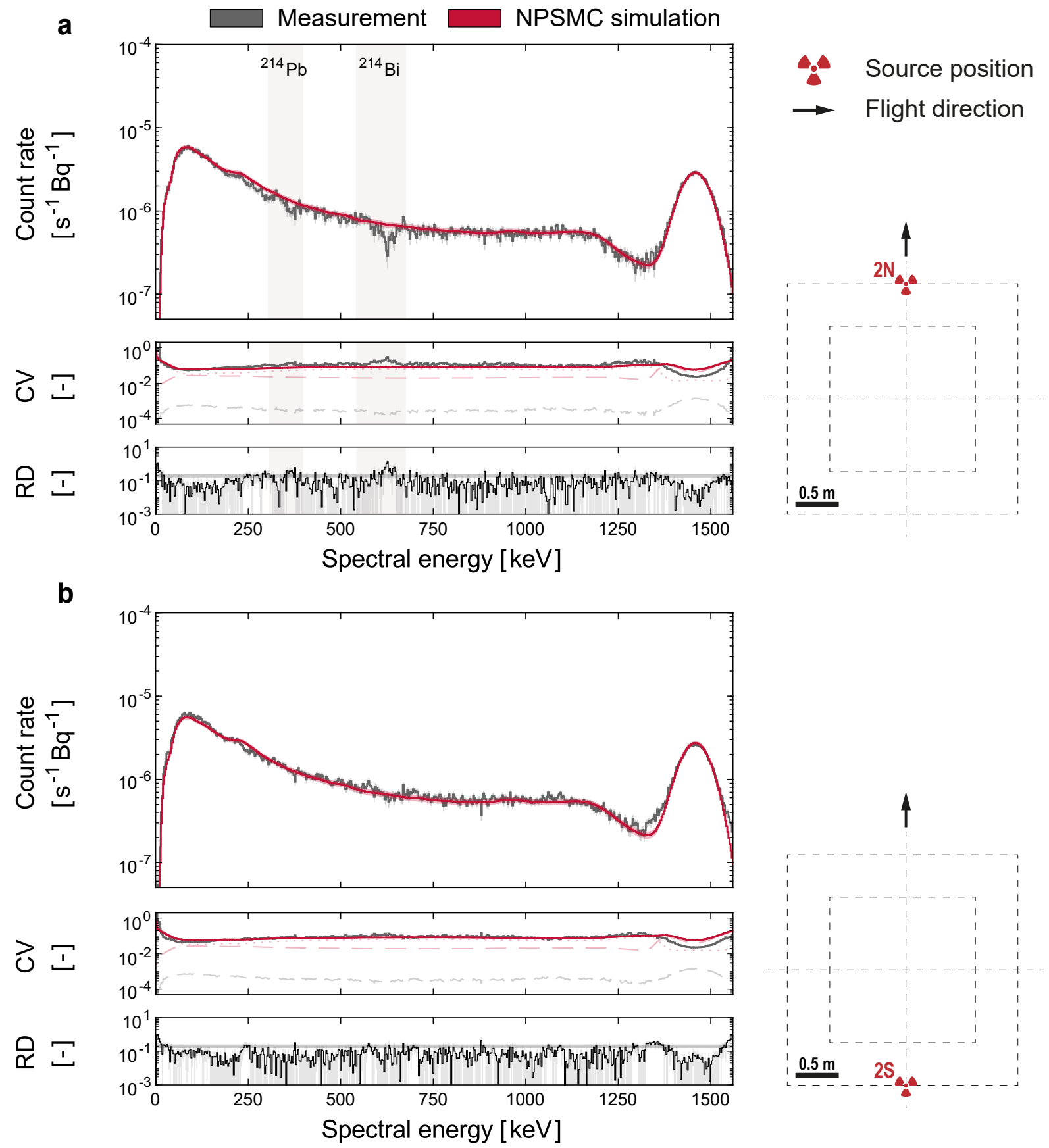

**Figure 8.7** The measured ($\hat{c}_{exp}$) and simulated ($\hat{c}_{sim}$) mean spectral signatures in the detector channel #SUM are shown as a function of the spectral energy $E'$ with a spectral energy bin width of $\Delta E' \sim 3$ keV for two measurement positions: **a** 2N. **b** 2S. For both measurements, a custom-made $K_{nat}$ radionuclide source was deployed (total activity $\mathcal{A} = 456.6(26)$ kBq). The simulated spectral signatures were obtained by NPSMC. Uncertainties ($\hat{\sigma}_{exp}$, $\hat{\sigma}_{sim}$) are provided as 1 standard deviation (SD) shaded areas. In addition, the coefficient of variation (CV) for the measured and simulated signatures as well as the relative deviation (RD) computed as $|\hat{c}_{sim} - \hat{c}_{exp}|/\hat{c}_{exp}$ (20 % mark highlighted by a horizontal grey line) are provided. Statistical and systematic contributions to the CV are indicated by shaded dotted and dashed lines, respectively (cf. Appendix A.8). For the measurement 2N, the spectral range of the FEPs associated with the 351.932(2) keV and 609.312(7) keV emission lines of $^{214}_{82}$Pb and $^{214}_{83}$Bi [124] is highlighted. The spectral range is defined as $E' = E_\gamma \pm 3\sigma_E(E_\gamma/\Delta E')\Delta E'$ with $E_\gamma$, $\sigma_E$ and $\Delta E'$ being the photon energy of the emission line, the spectral resolution and the spectral energy bin width, respectively (cf. Section 6.2.1.3).

2. A second systematic deviation can be observed at the low end of the spectral domain $E' \lesssim 100\,\text{keV}$. Similar deviations have been observed already in Chapters 6 and 7. As discussed in Section 6.3.1, these deviations can mainly be attributed to the increased systematic uncertainty at low spectral energies in the calibration models (spectral energy, spectral resolution and LLD) as well as in the physics models adopted by `FLUKA`.
3. A third deviation is found in the Compton gap between the FEP and the Compton edge. Similar deviations were noted in the previous chapter in Section 7.3.4. The consistent magnitude of this deviation across all ten source positions supports the hypothesis proposed in Section 7.4 that the observed Compton gap deviations may be attributed to deficiencies in the adopted NPSM (cf. also to Section 7.2.1).

### 8.3.2 ARM22 Validation Campaign

In addition to the Dübendorf validation campaign discussed in the previous subsection, a second field measurement campaign was performed as part of the ARM22 exercise [120]. The scope of this second validation campaign was twofold:

1. The first objective was to validate the AGRS Monte Carlo model under a realistic source-detector scenario for AGRS in the far field, i.e. with the source-detector distance being significantly larger than the characteristic extension of the aircraft (~10 m), thereby not only assessing the mass model of the AGRS system but also the increased effect of the environment, particularly the atmosphere.
2. The second goal was to validate the AGRS Monte Carlo model under a source-detector scenario where a significant part of the aircraft's fuselage is attenuating the primary gamma-ray field, thereby assessing the mass model of the AGRS in a source-detector scenario where the effect of the aircraft on the detector response is maximized.

To meet these objectives, two dedicated series of field measurements have been performed. In the following subsections, I will discuss the measurement setup, the Monte Carlo simulations performed and the results obtained for both series of measurements.

#### 8.3.2.1 Radiation Measurements

All field measurements of the ARM22 validation campaign were conducted at the Thun military training ground (46.753°N, 7.596°E) on the 2022-06-16 as part of the of the civil phase of the ARM22 exercise (ARM22c) [120]. As indicated above, two series of field measurements were performed within the scope of the validation campaign:[16]

**Hover Flight** To meet the first objective of the ARM22 validation campaign, the Swiss AGRS system was operated in a hover flight mode with the TH06 helicopter positioned at a predefined ground clearance above a sealed radionuclide point source positioned on the ground in a custom source holder aligned with the principal axis $z'$ as indicated in Fig. 8.8. The hover flight was performed at three different ground clearances: ~30 m, ~60 m and ~90 m with a measurement live time of ~5 min[17] for each measurement. As illustrated in Fig. 8.8, the source holder consisted of a polyvinyl chloride (PVC) base plate with two custom source holders made from PLA and constructed using additive manufacturing techniques. The base plate was secured to the ground using four lead bricks, each with a mass of 11.3 kg.

Given the exponential decrease in the photon flux with an increase in the source-detector distance (cf. Eq. 3.31), radionuclide sources with significant activities $\geq \mathcal{O}(10^8)$ Bq had to be deployed to obtain sufficient counting statistics within the available measurement time. For that purpose, two sealed radionuclide sources, a $^{137}_{55}$Cs source ($\mathcal{A} = 9.0(5) \times 10^9$ Bq) and a $^{133}_{56}$Ba source ($\mathcal{A} = 4.7(2) \times 10^8$ Bq), were provided by the NBC-EOD and deployed separately for each hover flight by placing them in the corresponding source holder.[18] To correct for the background radiation, additional background measurements were performed for each gross measurement using the same source-detector configuration but with the sources removed.[19]

Data acquisition was performed with the proprietary software suite `SpirIDENT` by the RLL operators on board the TH06 aircraft (cf. Section 5.3.2). At the beginning and end of each measurement, the relative fuel fraction was recorded manually by the author.[20] Ground clearances were determined using the radar altimeter of the TH06 helicopter accessed by the ARINC 429 avionics data bus (cf. Section 5.3.2). In addition, orientation data (yaw $\alpha'$, pitch $\beta'$ and roll $\gamma'$) of the helicopter was obtained by the flight data recorder.

16 *The validation campaign on the 2022-06-16 was co-organized by the members of the NEOC and the author of this book. Aircraft operation was performed by members of the Swiss Air Force. A detailed list of all involved personnel contributing to the validation campaign can be found in the Acknowledgements at the end of this book.*

17 *Which was mainly constrained by the endurance of the TH06 helicopter and its flight crew. Hover flights are both fuel-consuming and challenging for the pilots to maintain.*

18 *Activity values were computed for the start time of the measurement using Eqs. 2.15 and 2.16 and the information provided by the calibration certificates as well as the half-life $t_{1/2}$ indicated in Table 6.1. Note that given the short separation in time between the individual measurements, there is no statistically significant difference between the activities for the individual gross measurements. Furthermore, it is worth adding that no uncertainty values were provided in the certificates. Therefore, as a reference, $\sigma_{\mathcal{A}_0}/\mathcal{A}_0 = 5\,\%$ was assumed and propagated according to Eq. A.41.*

19 *Note that since the hover flights for the $^{137}_{55}$Cs source were performed in the morning and those for $^{133}_{56}$Ba in the afternoon, background measurements were repeated for both sources to account for changes in the background radiation.*

20 *As noted already in Section 8.3.1.2, the fuel level in the tanks can only be accessed via the fuel tank management panel in the cockpit of the TH06 helicopter. The TH06 helicopter does not possess an option for digital logging of these parameters.*

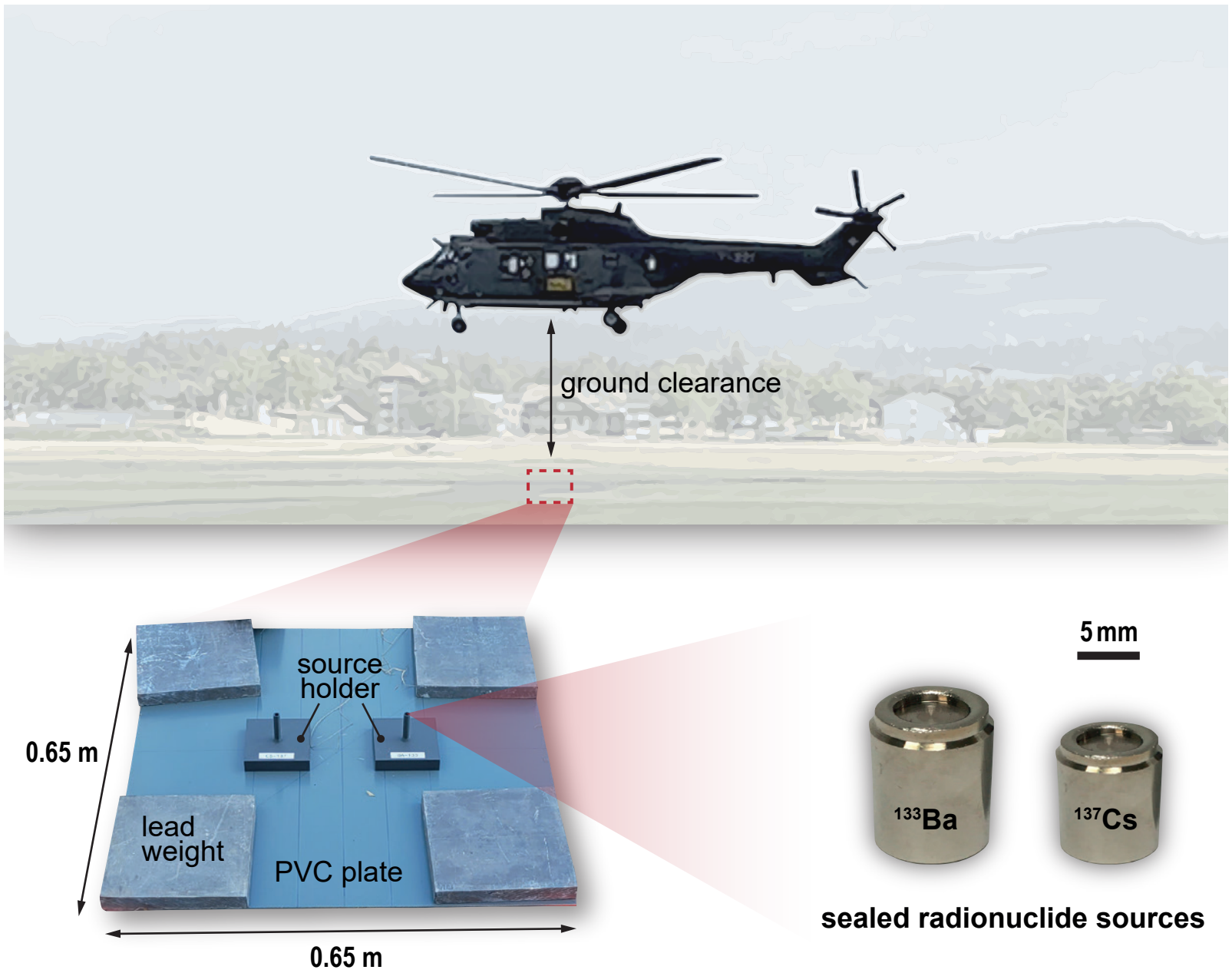

**Figure 8.8** Experimental setup of the hover flight measurements conducted during the ARM22 validation campaign. The main elements of the setup are the fully integrated Swiss AGRS system operated in hover flight mode at a predefined ground clearance above a sealed radionuclide point source positioned on the ground in a custom source holder. Two different sources have been deployed separately: a $^{137}_{55}\mathrm{Cs}$ source ($\mathcal{A} = 9.0(5) \times 10^{9}$ Bq) and a $^{133}_{56}\mathrm{Ba}$ source ($\mathcal{A} = 4.7(2) \times 10^{8}$ Bq). The source holder consisted of a PVC base plate with two custom source holders made from PLA and constructed using additive manufacturing techniques. The base plate was secured to the ground using four lead bricks, each with a mass of 11.3 kg. Note that the adopted zoom-in figures of the radionuclide sources are only representative of their size and geometry and do not depict the actual sources deployed during the measurements.

A list of all performed measurements with associated gross and background measurement live times, mean ground clearances and orientation angles of the helicopter is provided in Table C.16.

**Ground Measurement** To meet the second objective of the ARM22 validation campaign, the Swiss AGRS system was parked on the ground on the Thun military training ground as displayed in Fig. 8.9. A sealed $^{137}_{55}$Cs radionuclide point source ($\mathcal{A} = 7.7(4) \times 10^8$ Bq)[21], provided again by the NBC-EOD, was positioned at a distance of 11 m with respect to the center of gravity of the scintillation crystals on the principal axis $x'$ on the starboard side of the TH06 helicopter using a tripod and a custom source holder made from PLA using additive manufacturing techniques. To position the sources accurately, distance as well as positioning laser systems were used. Given the position of the fuel tanks discussed in Section 8.2 (cf. also to Fig. 8.1), the line of sight between the source and the detector was obstructed not only by the fuselage of the TH06 helicopter but also by the fuel tank #2. This source-detector configuration enables validation of the mass model for both the fuselage and the adopted dynamic fuel level model of the TH06 helicopter detailed in Section 8.2.

To assess the effect of the aircraft fuel on the detector response in this specific source-detector configuration, measurements were repeated three times with three different fuel volume fractions: 9.5 %, 43.7 % and 92.2 %. For that purpose, the helicopter was gradually refueled in the field between the individual measurements using a tanker truck provided by the Swiss Air Force. After each refueling, the fuel volume fractions in the six fuel tanks were manually logged using the fuel tank management panel in the cockpit. To correct for background radiation, additional background measurements were conducted for each gross measurement using the same source-detector configuration but with the source removed. Data acquisition was performed by the author using the proprietary software suite `SpirIDENT`. A list of all performed measurements with associated gross and background measurement live times is provided in Table C.17.

Postprocessing of the measured pulse-height spectra for both measurement series was performed using the `RLLSpec` pipeline described in Section 6.2.1.2.[22] The spectral energy and the spectral resolution were calibrated using the `RLLCal` pipeline detailed in Section 6.2.1.3.[23]

21 *The activity was computed for the start time of the measurement using Eqs. 2.15 and 2.16 and the information provided by the calibration certificates as well as the half-life $t_{1/2}$ indicated in Table 6.1. Note that given the short separation in time between the individual measurements, there is no statistically significant difference between the activities for the individual gross measurements.*

22 *It is worth adding that, in the case of the hover flights, the set of recorded pulse-height spectra was filtered to only include those instances in which the helicopter was aligned with the source on the ground. This analysis was performed by Dr. Gernot Butterweck using the data obtained via the ARINC 429 avionics data bus from the GNSSr (cf. also to Section 5.3.2).*

23 *The same RLL spectrometer used in the Dübendorf validation campaign was employed here, thus the same calibration models for spectral energy, spectral resolution and LLD were adopted (cf. Figs. B.83–B.85). For the LLD, the methods described in Appendix A.9 were adopted to calibrate its model parameters.*

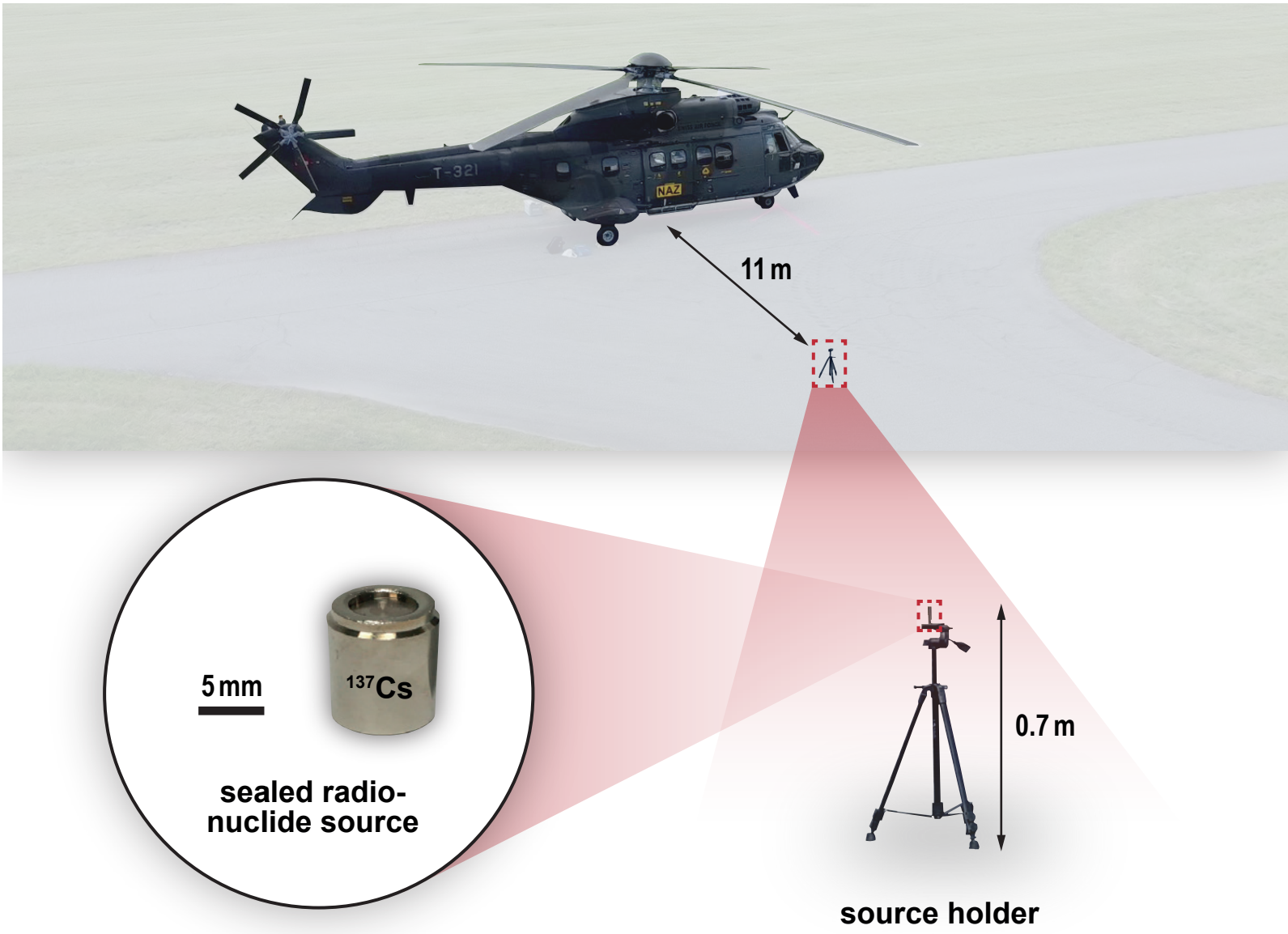

**Figure 8.9** Experimental setup of the ground measurements conducted during the ARM22 validation campaign. The main elements of the setup are the fully integrated Swiss AGRS system parked on the ground at the Thun military training ground (46.753°N, 7.596°E) and a sealed $^{137}_{55}$Cs radionuclide point source ($\mathcal{A}$ = 7.7(4) × $10^8$ Bq). This point source was inserted in a custom source holder made from PLA, mounted on a tripod and positioned on the principal axis $x'$ on the starboard side of the TH06 helicopter in a distance of 11.0(5) m with respect to the center of gravity of the scintillation crystals. Note that the adopted zoom-in figure of the radionuclide source is only representative of its size and geometry and does not depict the actual source deployed during the measurements.

#### 8.3.2.2 Monte Carlo Simulations

As already indicated in Section 8.2, all Monte Carlo simulations in this chapter were performed using NPSMC discussed in Chapter 7. The general mass model of the AGRS Monte Carlo model was described already in Section 8.2. Therefore, I limit the discussion here to the specific simulation setups adopted for the ARM22 validation campaign.

Similar to the mass model derived for the Dübendorf validation campaign discussed in Section 8.3.1.2, I extended the AGRS mass model described in Section 8.2 to encompass all major parts of the environment in which the field measurements described above were conducted. This includes the atmosphere modeled as homogeneous humid air as described in Section 8.2 using the measured air temperature, pressure and humidity obtained by the automatic weather station THU operated by MeteoSwiss (WIGOS-ID: 0-20000-0-06731) and located on the Thun military training ground (46.749 853°N, 7.585 222°E).[24] The deployed radionuclide sources and all associated source holder elements (tripod, lead bricks, ...) described above were reproduced in detail in the Monte Carlo mass model.[25] The ground was modeled as a homogeneous mixture of 28 soils with the compositional data provided by McConn et al. [249]. The street where the source holders were positioned was incorporated into the mass model as homogeneous asphalt pavement, with compositional data sourced again from McConn et al. [249]. Motivated by the mean free path of the primary photons in air (cf. Fig. 3.6), the atmosphere was confined to a sphere with a radius of 200 m, centered at the RLL spectrometer's location. Similarly, the ground was limited to a vertical extension of 10 m. The effect of these environmental elements on the detector response will be discussed in the next subsection.

For all performed simulations, the fuel level in the individual tanks was set by the corresponding fuel volume fractions obtained from the fuel tank management panel in the cockpit. In addition, the position and orientation of the aircraft were accurately reproduced in the Monte Carlo model using the data obtained via the ARINC 429 avionics data bus as well as the flight data recorder (cf. Table C.16).[26]

As for the Dübendorf validation campaign, the three radionuclide sources were modeled using the `raddecay` card from `FLUKA` in a semi-analogue mode. The number of primaries $N_{\mathrm{pr}}$ was adjusted for each simulation to ensure a median coefficient of variation $\mathrm{med}(\mathbf{CV}_{\mathrm{stat}}) < 2\,\%$ and a median relative variance of the variance

24 *Note that for the ground measurements, the TH06 helicopter was offline and thus, no external sensor data of the TH06 helicopter were available. For consistency reasons, the same automatic weather station was also used to quantify the atmospheric conditions for the hover flights.*

25 *Structural and compositional data for the* $^{133}_{56}\mathrm{Ba}$ *source was provided by the NBC-EOD. For the two* $^{137}_{55}\mathrm{Cs}$ *sources used for the ground and hover flight gross measurements, no detailed data was available. For these sources, best-estimate values were derived based on the information obtained for the* $^{133}_{56}\mathrm{Ba}$ *source.*

26 *For these parameters, arithmetic mean values computed over the course of the individual gross measurements were applied.*

med(**VOV**) < 0.1 % over the SDOI defined in Def. 6.1 in the detector channel `#SUM` (cf. also to Appendix A.8).[27] Postprocessing of the obtained simulation data was performed using again the `NPScinMC` pipeline detailed in Section 7.2.3.[28]

27 *The resulting number of primaries ranged between* $5 \times 10^9$ *and* $4 \times 10^{10}$.

28 *Building upon the findings from the Dübendorf validation campaign detailed in Section 8.3.1.3, the same NPSM derived in the previous chapter for the detector channel* `#SUM` *was adopted for the ARM22 validation campaign.*

#### 8.3.2.3 Hover Flight Results

In the Figs. 8.10 and 8.11, I present the measured ($\hat{c}_{\mathrm{exp}}$) and simulated ($\hat{c}_{\mathrm{sim}}$) spectral signatures for all hover flight measurements conducted during the ARM22 validation campaign alongside uncertainty estimates (cf. Appendix A.8) and relative deviations computed as $|\hat{c}_{\mathrm{sim}} - \hat{c}_{\mathrm{exp}}|/\hat{c}_{\mathrm{exp}}$.

As predicted by the monoenergetic transport theory (cf. Section 3.2.1), I find an exponential decrease in the count rate in the FEPs with increasing ground clearance. In general, good agreement between the measured and simulated spectral signatures is observed for both radionuclides at all three ground clearances with a median relative deviation <5 % within the SDOI. Specifically, I would like to highlight again the excellent agreement at the Compton edge in Fig. 8.11 for $^{137}_{55}\mathrm{Cs}$ at all three ground clearances.

However, similar to the Dübendorf validation campaign, there are also some systematic deviations between the measured and simulated spectral signatures which require further discussion:

1. One of the most evident deviations is observed at low spectral energies in the backscatter peak (BSP) highlighted in Fig. 8.11 for the $^{137}_{55}\mathrm{Cs}$ source and, to a lesser degree, around ~150 keV for the $^{133}_{56}\mathrm{Ba}$ source. The fact that the magnitude of the relative deviation tends to decrease with increasing ground clearance indicates that it is likely related to systematic uncertainties in mass model elements close to the ground, i.e. the source holder components, street and soil.
2. A second deviation is found in the Compton gap between the FEP and the Compton edge for the $^{137}_{55}\mathrm{Cs}$ source. Similar deviations were noted already in Sections 7.3.4 and 8.3.1.3. As with the deviations observed at the BSP, the magnitude of the relative deviation tends to decrease with increasing ground clearance. Consequently, these deviations may be attributed to systematic uncertainties in the mass model elements close to the ground, too. Alternatively, as discussed in Sections 7.3.4 and 8.3.1.3, these deviations could also stem, at least partly, from deficiencies in the adopted NPSM.

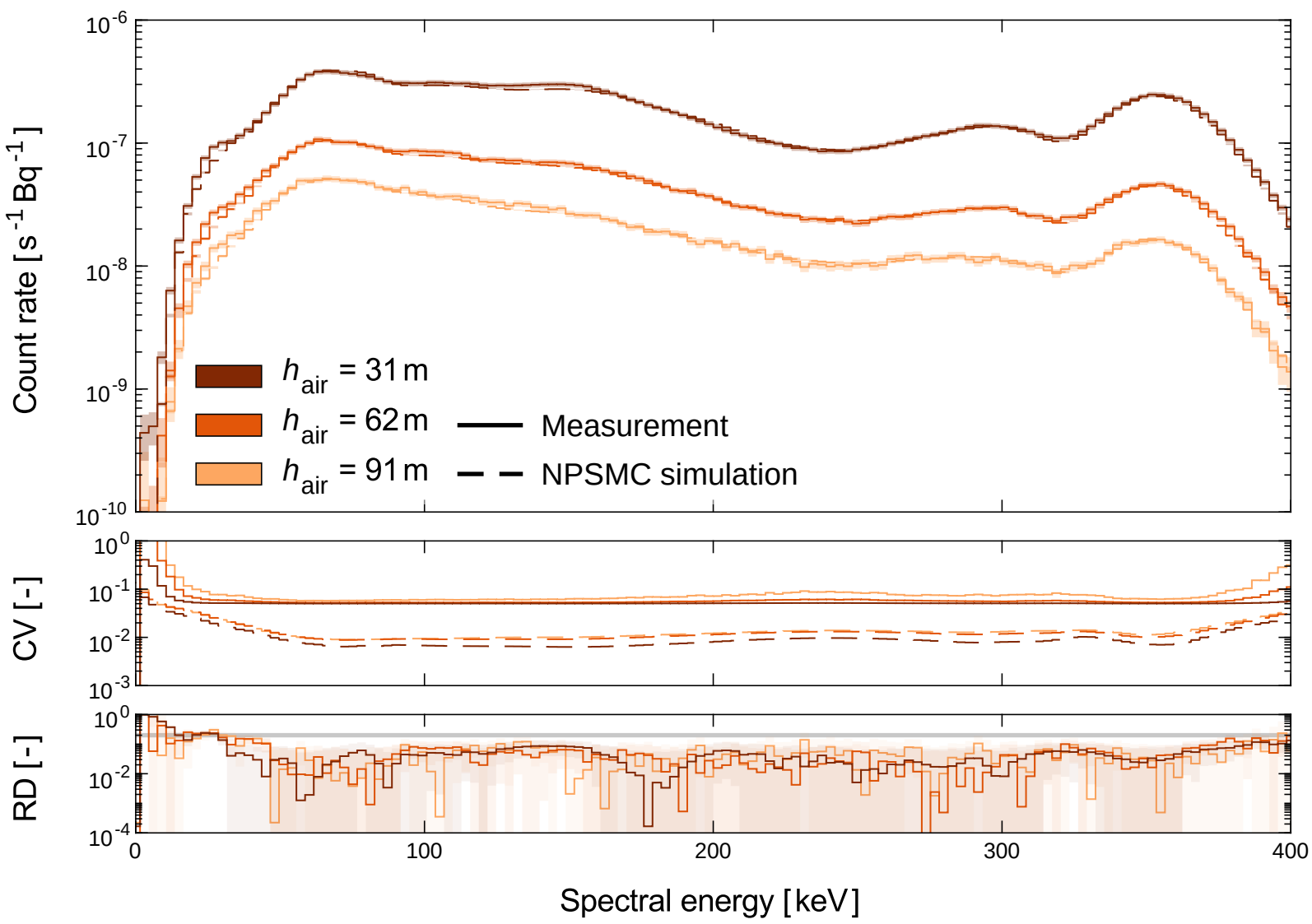

**Figure 8.10** The measured ($\hat{c}_{exp}$) and simulated ($\hat{c}_{sim}$) mean spectral signatures in the detector channel `#SUM` are shown as a function of the spectral energy $E'$ with a spectral energy bin width of $\Delta E' \sim 3$ keV for the hover flight measurements conducted during the ARM22 validation campaign using the $^{133}_{56}$Ba point source ($\mathcal{A} = 4.7(2) \times 10^8$ Bq). The hover flights were performed at three different altitudes with mean ground clearances ($h_{air}$): **a** 30.7(1) m. **b** 61.7(1) m. **c** 90.8(2) m. The simulated spectral signatures were obtained by NPSMC. Uncertainties ($\hat{\sigma}_{exp}$, $\hat{\sigma}_{sim}$) are provided as 1 standard deviation (SD) shaded areas. In addition, the coefficient of variation (CV) for the measured and simulated signatures (cf. Appendix A.8) as well as the relative deviation (RD) computed as $|\hat{c}_{sim} - \hat{c}_{exp}|/\hat{c}_{exp}$ (20 % mark highlighted by a horizontal grey line) are provided.

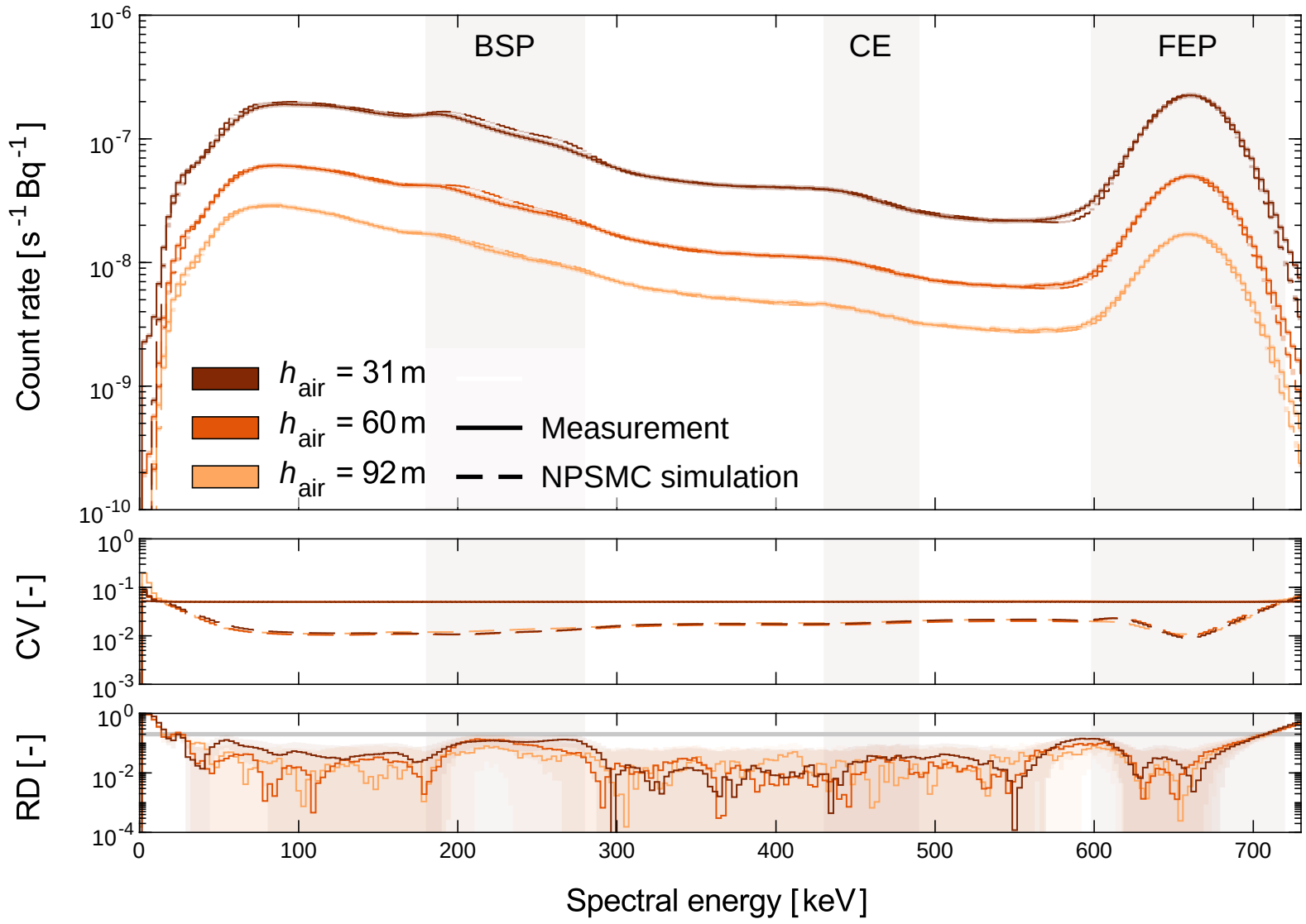

**Figure 8.11** The measured ($\hat{c}_{exp}$) and simulated ($\hat{c}_{sim}$) mean spectral signatures in the detector channel `#SUM` are shown as a function of the spectral energy $E'$ with a spectral energy bin width of $\Delta E' \sim 3$ keV for the hover flight measurements conducted during the ARM22 validation campaign using the $^{137}_{55}$Cs point source ($\mathcal{A} = 9.0(5) \times 10^9$ Bq). The hover flights were performed at three different altitudes with mean ground clearances ($h_{air}$): **a** 30.9(1) m. **b** 60.2(1) m. **c** 91.6(2) m. The simulated spectral signatures were obtained by NPSMC. Distinct spectral regions, i.e. the backscatter peak (BSP), the domain around the Compton edge (CE) as well as the full energy peak (FEP) are marked. Uncertainties ($\hat{\sigma}_{exp}$, $\hat{\sigma}_{sim}$) are provided as 1 standard deviation (SD) shaded areas. In addition, the coefficient of variation (CV) for the measured and simulated signatures (cf. Appendix A.8) as well as the relative deviation (RD) computed as $|\hat{c}_{sim} - \hat{c}_{exp}|/\hat{c}_{exp}$ (20 % mark highlighted by a horizontal grey line) are provided.

To demonstrate the sensitivity of the environment mass model elements on the detector response, similarly to Section 6.3.2, I performed additional NPSMC simulations selectively excluding various elements of the environment from the mass model. The results presented in Fig. B.86 revealed a pronounced sensitivity of the mass model components for the ground, atmosphere as well as the source-related elements on the spectral detector response. This underscores the importance of accurate modeling of the environment in which the AGRS surveys are conducted, in particular at low ground clearances.

#### 8.3.2.4 Ground Measurement Results

In Fig. 8.12, I present the measured ($\hat{c}_{\mathrm{exp}}$) and simulated ($\hat{c}_{\mathrm{sim}}$) spectral signatures for all ground measurements performed during the ARM22 validation campaign alongside uncertainty estimates (cf. Appendix A.8) and relative deviations computed as $|\hat{c}_{\mathrm{sim}} - \hat{c}_{\mathrm{exp}}|/\hat{c}_{\mathrm{exp}}$.

Before I compare the measured and simulated spectral signatures, I would like to discuss two interesting trends in the measured spectral signatures in Fig. 8.12:

1. The count rate in the FEP for the lowest fuel volume fraction of 9.5 % is significantly increased, while only minimal differences can be observed in the count rate of the FEPs for the two higher fuel volume fractions.
2. The count rate in the CC is continuously decreasing with increasing fuel volume fraction.

Both trends can be explained by the increasing fuel levels in the six fuel tanks of the TH06 helicopter and the resulting attenuation by the aircraft fuel of the primary photons and Compton scattered photons. As noted in Section 8.3.2.1, the fuel tank #2 is the main tank that obstructed the line of sight for the primary photons in the adopted source-detector configuration. It is also one of the tanks which is filled first during the refueling process (cf. Fig. B.81). As a result, at a fuel volume fraction of 9.5 %, only the lower part of the fuel tank #2 is filled, leading to a reduced attenuation of the primary photons in the fuel and thus a higher count rate in the FEPs.[29] In contrast, at a fuel volume fraction of >20 %, the fuel tank #2 is completely filled (cf. Fig. B.81), causing a maximized attenuation of the primary photons in the fuel. Therefore, the attenuation of the primary photons remains constant for the two higher fuel volume fractions, resulting in nearly identical FEPs.

29 *It is worth noting that we can observe a significant increase in the energy deposition events in the Compton gap and a shift of the FEP mode towards lower spectral energies for all fuel volume fractions in Fig. 8.12. This indicates a substantial contribution of photons to these spectral regions that underwent Compton scattering in the helicopter's fuselage with small deflection angles, a process known as buildup [30]. This explains why an increased count rate in the FEP is observed at a fuel volume fraction of 9.5 %, even though the scintillation crystals are already fully obstructed by the fuel in tank #2 (cf. Fig. B.81).*

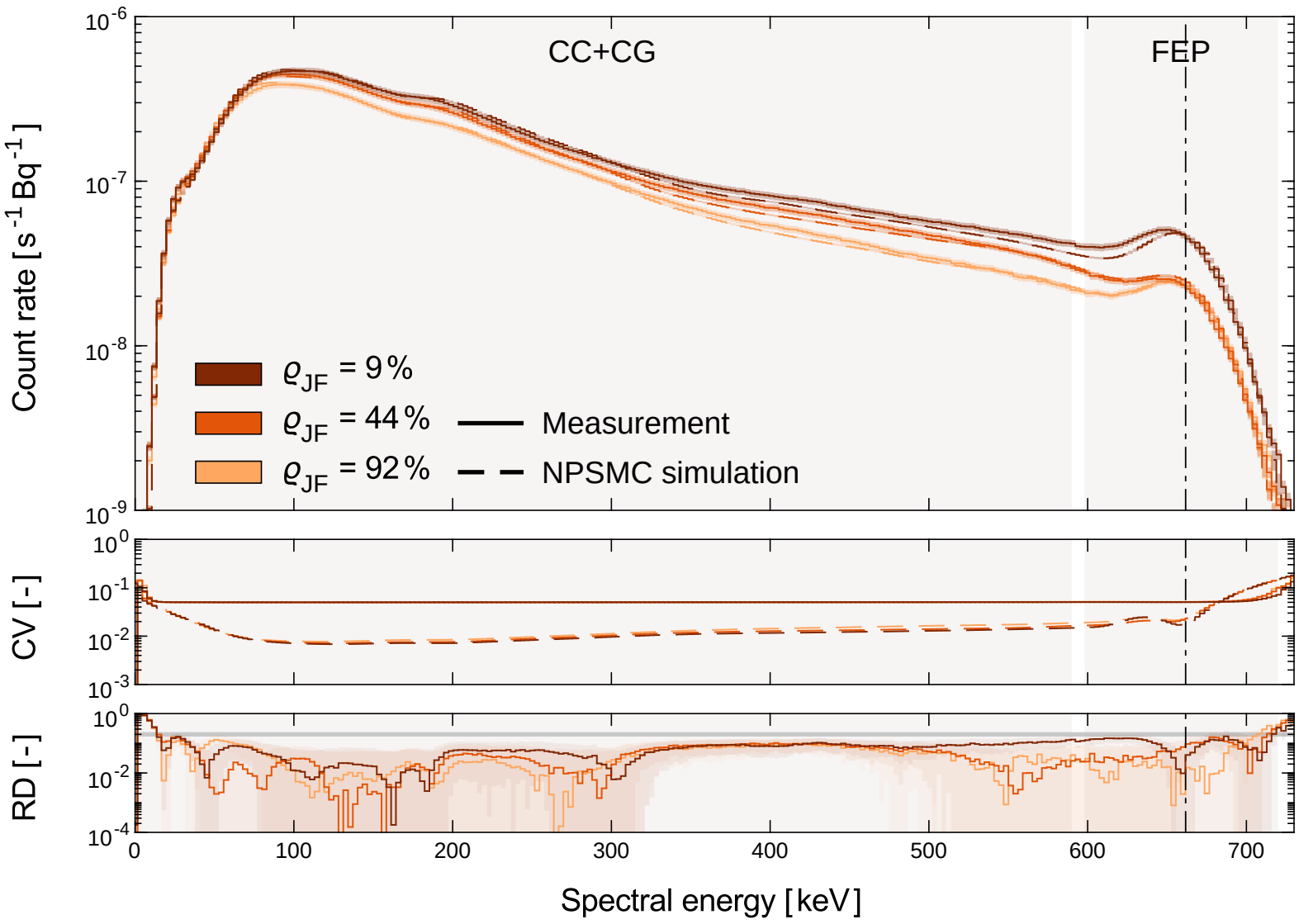

**Figure 8.12** The measured ($\hat{c}_{\mathrm{exp}}$) and simulated ($\hat{c}_{\mathrm{sim}}$) mean spectral signatures in the detector channel `#SUM` are shown as a function of the spectral energy $E'$ with a spectral energy bin width of $\Delta E' \sim 3\,\mathrm{keV}$ for the ground measurements conducted during the ARM22 validation campaign using a $^{137}_{55}\mathrm{Cs}$ point source ($\mathcal{A} = 7.7(4) \times 10^{8}\,\mathrm{Bq}$). The ground measurements were performed at three different fuel volume fractions ($\varrho_{\mathrm{JF}}$): **a** 9.5 %. **b** 43.7 %. **c** 92.2 %. The simulated spectral signatures were obtained by NPSMC. Distinct spectral regions, i.e. the Compton continuum (CC) together with the Compton gap (CG) as well as the full energy peak (FEP) with the 661.657(3) keV emission line [68] (black dash-dotted line) are marked. Uncertainties ($\hat{\sigma}_{\mathrm{exp}}, \hat{\sigma}_{\mathrm{sim}}$) are provided as 1 standard deviation (SD) shaded areas. In addition, the coefficient of variation (CV) for the measured and simulated signatures (cf. Appendix A.8) as well as the relative deviation (RD) computed as $|\hat{c}_{\mathrm{sim}} - \hat{c}_{\mathrm{exp}}|/\hat{c}_{\mathrm{exp}}$ (20 % mark highlighted by a horizontal grey line) are provided.

On the other hand, the continuous decrease in the CC with increasing fuel volume fraction is attributed to the combined attenuation of the secondary photons by all six fuel tanks. As illustrated in Fig. 8.1, the fuel tanks are all arranged around the RLL spectrometer, acting as a shield for the photons undergoing Compton scattering in the environment and the helicopter's fuselage. As a result, the count rate in the CC is continuously decreasing with increasing fuel volume fraction as observed in Fig. 8.12.

The simulated spectral signatures reproduce both these trends well. In general, I find again a good agreement between the measured and simulated spectral signatures for all three fuel volume fractions with median relative deviations of 7.5 %, 3.5 % and 2.8 % within the SDOI for the fuel volume fractions of 9.5 %, 43.7 % and 92.2 %, respectively.

However, I observe a systematic deviation between the measured and simulated spectral signatures starting at the lower end of the BSP at $E' \sim 200\,\mathrm{keV}$ and extending up to the FEP for all three fuel volume fractions. The magnitude of the relative deviation increases as the fuel volume fraction decreases, suggesting that this deviation is likely due to systematic uncertainties in the fuselage mass model. Specifically, the approximation of the TH06 aircraft's complex semi-monocoque fuselage design to a homogeneous equivalent metal sheet is expected to introduce a systematic error in the spectral detector response. In contrast, the aircraft fuel, being a liquid, is well approximated as a homogeneous mass model element. Consequently, as the fuel level decreases, the relative influence of the fuel diminishes while the effect of the fuselage becomes more pronounced, leading to increased deviations in the Compton continuum, Compton gap and the FEP. It is worth noting that the relative deviation is restricted to <20 % over the entire SDOI, even for the lowest fuel volume fraction of 9.5 %.

## 8.4 Conclusion

In this chapter, I presented the derivation and validation of a high-fidelity AGRS Monte Carlo model including the already validated NPSMC model of the RLL spectrometer presented in the previous chapter as well as a newly introduced detailed mass model of the entire TH06 aircraft system. In contrast to previous studies reviewed in Section 5.5.2, care was taken to accurately model all major parts of the aircraft system by preserving the geometry, the opacity as well as the mass density of the individual aircraft components. Moreover, the aircraft is not assumed to be static like in previous studies but is modeled as a dynamic system, specifically accounting for changes in the orientation of the aircraft and the depletion of the fuel tanks, among others.

The AGRS Monte Carlo model was validated by performing a series of field measurements using the fully integrated Swiss AGRS system under various source-detector configurations adopting custom-made $K_{nat}$ volume sources as well as $^{133}_{56}Ba$ and $^{137}_{55}Cs$ point sources. Compared to previous studies reviewed in Section 5.5.2, excellent agreement between the measured and simulated spectral signatures was obtained with a median relative deviation of less than 8 % in the SDOI for all performed radiation measurements. This includes radiation measurements performed under source-detector configuration with maximized attenuation by the aircraft fuselage. Furthermore, leveraging the use of a calibrated NPSM combined with NPSMC presented in the previous chapter, I found an excellent agreement between the measured and simulated spectral signatures in the Compton edge domain. This is particularly noteworthy as significant systematic deviations were previously observed in this domain using the standard PSMC method (cf. Chapter 6).

However, a careful investigation of the spectral signatures revealed statistically significant systematic deviations between the measured and simulated spectral signatures. Apart from the already observed discrepancies in previous chapters (cf. Sections 6.4 and 7.4), I observed systematic deviations at ~610 keV and, to a lesser degree, also at ~350 keV in the spectral signatures of some of the performed field measurements conducted in a hangar at the Dübendorf airfield. Given the prolonged measurement times and changing atmosphere under which the specific field measurements were conducted, the observed deviations are likely related to changing background levels of the radon progeny $^{214}_{82}Pb$ (351.932(2) keV) and $^{214}_{84}Po$ (609.312(7) keV).

A second systematic deviation was observed at higher spectral energies, starting at the BSP and extending to the FEP, in the spectral signatures of the ground measurements performed during the ARM22 validation campaign. The magnitude of the relative deviation increases as the fuel volume fraction decreases, suggesting that these deviations are likely attributed to the level of detail used in the mass model of the aircraft's fuselage. It is worth adding that these deviations are restricted to <20 % over the entire SDOI, even for the lowest fuel volume fraction of 9.5 % tested during the ARM22 validation campaign.

In summary, the developed AGRS Monte Carlo model demonstrated excellent agreement with the measured spectral signatures for a series of field measurements. Systematic deviations related to the AGRS model remained below 20 % across the entire SDOI, even under worst-case source-detector scenarios involving significant attenuation by the aircraft fuselage, suggesting a satisfactory degree of detail used in defining the mass model in the simulation. The detailed modeling of the entire aircraft, including dynamic models for its orientation and fuel depletion, enables accurate predictions of spectral signatures for arbitrary source-detector configurations. As such, the developed AGRS Monte Carlo model represents a critical step towards full spectrum simulation-based calibration of AGRS systems and serves as the foundation of all subsequent work presented in this book.

> *"Divide each of the difficulties under examination into as many parts as possible, and as might be necessary for its adequate solution."*
>
> — *René Descartes, Discourse on Method*

# Chapter 9
# Detector Response Model

**Contents**

In the previous chapter, I derived and validated a high-fidelity Monte Carlo model of the Swiss AGRS system, allowing for the accurate simulation of spectral signatures for arbitrary source-detector configurations. While Monte Carlo simulations are the method of choice for accurate gamma-ray spectrometry modeling, their computational expense, with typical evaluation times of $\mathcal{O}(10^1)$ h per run on a computer cluster, renders them impractical for the $\mathcal{O}(10^5)$ evaluations needed for AGRS system calibration.

To overcome this limitation, I developed a detector response model that emulates the AGRS Monte Carlo model. This model integrates numerically derived detector response functions and photon fluxes, enabling rapid computation of spectral signatures with evaluation times of $\mathcal{O}(1)$ s per signature on a local workstation. Unlike previous studies, this model incorporates the angular dependence of both the detector response function and photon fluxes, derived using the `FLUKA` Monte Carlo code.

Verification against the AGRS Monte Carlo model showed good agreement, with a median relative deviation <6 % for all tested source-detector configurations and photon energies. In contrast, previously proposed isotropic approaches deviated by >250 %, underscoring the significance of the anisotropic response of AGRS systems.

The detector response model presented in this chapter marks not only an essential step towards full spectrum simulation-based calibration of AGRS systems, but also allows for detailed analyses of the spectral and the angular dispersion in the AGRS system for varying experimental conditions.

Parts of this chapter were reproduced from the following study published by the author in the context of the present work:

D. Breitenmoser et al. "Numerical Derivation of High-Resolution Detector Response Matrices for Airborne Gamma-Ray Spectrometry Systems". *2022 IEEE Nuclear Science Symposium and Medical Imaging Conference (NSS/MIC)* 10.1109/NSS/MIC44845.2022.10399024 (2022).

## 9.1 Introduction

As discussed in Section 3.2.2, Monte Carlo simulations are the method of choice for accurate full spectrum modeling of gamma-ray spectrometry systems. However, due to the stochastic nature of the method, a large number of particles needs to be tracked to achieve accurate results for collective particle measures. This makes Monte Carlo simulations computationally expensive, especially for transport problems with extended simulation volumes like AGRS systems. To achieve the required precision, a typical NPSMC simulation of a single spectral signature requires a characteristic computation time of $\Delta t_{\mathrm{MC}} = \mathcal{O}(10^1)\,\mathrm{h}$ on a typical computer cluster.[1]

*1 Like the one at the PSI used for this work, offering 520 cores at a nominal clock speed of* 2.6 GHz.

Considering now that the calibration task in AGRS outlined in Def. 5.2 requires the computation of spectral signature matrices $\boldsymbol{\mathcal{M}} \in \mathbb{R}_+^{N_{\mathrm{ch}} \times N_{\mathrm{src}}}$ for each individual recorded pulse-height spectrum, the necessary number of model evaluations for a typical AGRS survey flight with $N_{\mathcal{Y}} = \mathcal{O}(10^4)$ recorded spectra and $N_{\mathrm{src}} = \mathcal{O}(10^1)$ sources of interest is in the order of $N_{\mathrm{src}} N_{\mathcal{Y}} \Delta t_{\mathrm{MC}} = \mathcal{O}(10^6)\,\mathrm{h} = \mathcal{O}(10^2)\,\mathrm{a}$. Given these numbers, it is easy to see that the calibration of AGRS systems using brute-force PSMC or NPSMC simulations becomes computationally prohibitively expensive in practical applications.

This limitation of Monte Carlo simulations is not unique to AGRS systems but is common across various scientific disciplines, where radiation spectrometers with complex spectral and angular dispersion characteristics are applied to varying source-detector configurations. In the field of gamma-ray and neutron spectrometry, this limitation was addressed by the development of the detector response modeling approach, which was already successfully applied in a variety of different fields, including astrophysics [392, 395, 694, 707, 708], planetary science [393, 394, 396] and in-situ gamma-ray spectrometry [823].[2]

*2 One of the first documented applications of detector response modeling in gamma-ray spectrometry was performed by Reedy and collaborators in the context of the Apollo 15 and 16 missions [396, 824].*

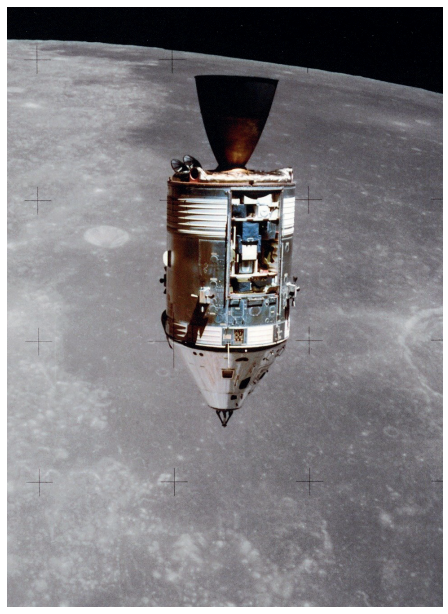

Apollo 15 Service Module in lunar orbit which included a gamma-ray spectrometer for elemental mapping of the lunar surface 

In this approach, which I have reviewed already in detail in Section 4.3.4, the pulse-height spectra are reconstructed by splitting the radiation transport and detector response simulation into two parts: the detector response function (DRF) $R$ characterizing the detector system, and the double differential photon flux $\partial^2\phi_\gamma/\partial E_\gamma \partial\Omega$ (cf. Eq. 2.20) quantifying the source and the photon transport in the environment.[3] Both parts are then combined in the Fredholm integral equation of the first kind (cf. Eq. 4.40) to obtain the pulse-height spectra characterized by the count rate vector $\mathbf{c} \in \mathbb{R}_+^{N_{\mathrm{ch}} \times 1}$ with $N_{\mathrm{ch}}$ being

*3 Note that the methodology is not limited to photons but can, in principle, be extended to any type of radiation particle or a combination thereof. However, as stated below, the scope of this chapter is limited to photon sources.*

the number of pulse-height channels. I will refer to the methodology encompassing the DRF, the double differential photon flux and the techniques for integrating the Fredholm integral equation of the first kind as the detector response model (DRM). The computational advantage of this approach is based on four main factors:

1. Both the DRF and the double differential photon flux can be separately computed using Monte Carlo simulations on a modern computer cluster with significantly reduced characteristic computation times. Depending on the specific source-detector configuration, a reduction in the computation time by a factor five or more can be achieved compared to a brute-force full simulation.
2. As the DRF is typically constant or only slowly changing[4], it can be precomputed using a comparable small number of Monte Carlo simulations $\mathcal{O}(10^3)$.
3. Using the double differential photon flux, on the other hand, has the advantage that different source-detector configurations can be computed in a single Monte Carlo simulation, thereby reducing the number of required simulations essentially to the number of $N_{\mathrm{src}}$ characteristic sources which may vary between $\mathcal{O}(10^1)$ and $\mathcal{O}(10^3)$ for typical AGRS surveys.[5] As a result, the double differential photon flux can also be precomputed.
4. The Fredholm integral equation of the first kind can be solved efficiently using a variety of numerical methods with a characteristic computation time $\Delta t_{\mathrm{DRM}}$ of $\mathcal{O}(1)$ s on a local workstation.

Consequently, once the DRF and the double differential photon flux are precomputed and stored in a database, the spectral detector response can be predicted for arbitrary source-detector configurations with a characteristic computation time $\Delta t_{\mathrm{DRM}}$ of $\mathcal{O}(1)$ s on a local workstation. This represents a dramatic reduction in the computation time by a factor of at least $\mathcal{O}(10^4)$ compared to brute-force full simulations performed on a computer cluster. If we account for the number of cores available on the computer cluster and the local workstation,[6] the factor increases to $\mathcal{O}(10^6)$.

However, in AGRS, despite its computational advantages, detector response modeling is still in its infancy and has only been utilized in a few studies by Klusoň and his co-workers [397, 825]. In these studies, Klusoň proposed a simplified isotropic approach, neglecting the angular dependence of both the DRF and the photon flux [397,

4 *For example due to changes in the fuel level.*

5 *Assuming that the experimental conditions are constant or only slowly varying.*

6 *As noted in Section 6.2.2, ∼500 cores were available on the utilized computer cluster in this work. The local workstation possessed an eight-core central processing unit.*

825]. Considering the mass and the extension of the aircraft systems typically adopted in AGRS surveys, as well as the characteristic photon transport lengths for the source-detector configurations encountered in these surveys (cf. Section 5.2), a significant anisotropy both in the DRF and in the photon flux is expected. Consequently, the isotropic approach proposed by Klusoň may lead to significant deviations in the predicted spectral detector response, especially for AGRS surveys with varying source-detector configurations like those typically encountered in alpine environments.

Motivated by this analysis, the scope of this chapter is the development and verification of a DRM for the Swiss AGRS system to accurately emulate the spectral detector response for arbitrary photon source fields in AGRS surveys. For this purpose, instead of the isotropic approach proposed by Klusoň, I will adopt a more realistic anisotropic approach accounting for the angular dependence in the DRF and the double differential photon flux using methods originally developed for astrophysics [392, 694] as well as planetary science [393, 394] applications.

## 9.2 Implementation

A detailed discussion of the anisotropic detector response modeling approach is provided in Section 4.3.4. Therefore, I limit the discussion here to the specific implementation of the models outlined in Section 4.3.4, specifically, the precomputations of the DRF and the double differential photon flux, as well as the numerical integration of the Fredholm integral equation of the first kind.

### 9.2.1 Double Differential Photon Flux

As discussed already in detail in Section 3.2, computing the double differential photon flux for a given source-detector configuration is a standard task in radiation transport and can be achieved by various numerical codes. For this purpose, I applied again the multi-purpose Monte Carlo code `FLUKA` [20, 216, 281] with the built-in `USRYIELD` card, adopting a custom binning scheme with a bin width of 3 keV for the photon energy $E_\gamma$ and 5° for the polar angle $\theta$.[7] To obtain spectral signature estimates, in full analogy to the normalization applied in the `PScinMC` and `NPScinMC` pipelines, the double differential photon flux $\partial^2\phi_\gamma/\partial E_\gamma \partial \Omega$ is additionally scaled by the source geometry-matter factor $F_{\mathrm{src}}$ introduced in Section 6.2.2.4:[8]

7 *Note that I assume azimuthal symmetry with respect to the source-detector axis for all computed double differential photon fluxes.*

8 *Note that the double differential photon flux signature is expressed in the global reference frame x-y-z with the corresponding direction unit vector* $\mathbf{\Omega}$.

$$\hat{\phi}_\gamma \left(E_\gamma, \mathbf{\Omega}, \mathfrak{d}\right) := \frac{\partial^2 \phi_\gamma}{\partial E_\gamma \partial \Omega} \left(E_\gamma, \mathbf{\Omega}, \mathfrak{d}\right) F_{\mathrm{src}} \tag{9.1}$$

where:

| | | |
|---|---|---|
| $\mathfrak{d}$ | experimental condition | $[\mathfrak{d}]$ |
| $F_{\mathrm{src}}$ | source geometry-matter factor | $[F_{\mathrm{src}}]$ |
| $E_\gamma$ | photon energy | eV |
| $\mathbf{\Omega}$ | direction unit vector in the global reference frame | |

and which I will refer to as the double differential photon flux signature. Given that the double differential photon flux signature characterizes the source and the environment, the set of required simulations depends on the sources of interest as well as the specific environmental conditions under which the AGRS survey takes place. A detailed discussion of the modeling of sources relevant for AGRS surveys will be provided in the next chapter. In this chapter, I limit the discussion to simplified source-detector configurations used for verification purposes.

### 9.2.2 Detector Response Function

As already detailed in Section 4.3.4, the continuous DRF can be approximated by a set of discrete detector response evaluations using again Monte Carlo simulations. For that purpose, similar to previous studies [394], I simulated an irradiation of the AGRS Monte Carlo mass model[9] presented in Section 8.2, with a monoenergetic plane-wave photon source at a discrete set of 30 different photon energies $E_\gamma$ (between 50 keV and 3000 keV)[10] from 134 directions $\mathbf{\Omega}'$ (covering the full $4\pi$ solid angle)[11] resulting in a set of 4020 individual NPSMC simulations. A graphical depiction of this simulation setup is provided in Fig. 9.1. All Monte Carlo simulations were executed on the same local computer cluster that was used for the simulations detailed in Section 8.3, with each simulation running a fixed number of primaries of $N_{\mathrm{pr}} = 10^7$. The total computation time for the DRF was approximately 225 h on the adopted computer cluster[12].

Having obtained the counts $C$ as a function of the spectral energy $E'$ for each of the 4020 NPSMC simulations using the `NPScinMC` pipeline (cf. Section 7.2.3), the DRF for a given photon energy $E_\gamma$, direction $\mathbf{\Omega}'$ and experimental condition $\mathfrak{d}$ can be approximated as:[13]

9 *Embedded in a vacuum environment. The atmosphere inside the aircraft is assumed to be homogeneous humid air with an air temperature* $T = 20\,°\mathrm{C}$, *an air pressure* $p = 1013.25\,\mathrm{hPa}$ *and a relative humidity* $\mathrm{RH} = 50\,\%$.

10 *With increasing spacing starting with 25 keV at low photon energies and reaching up to 500 keV spacing at the highest photon energies. In addition, the three main characteristic threshold energies obtained in Eqs. 4.28a–4.28e are included as well, i.e.* $\mathrm{m_e c^2}/2$, $2\mathrm{m_e c^2}$ *and* $3\mathrm{m_e c^2}$. *Given the significant suppression of DEPs close to* $E_\gamma = 2\mathrm{m_e c^2}$, *the threshold* $(1+\sqrt{2})\mathrm{m_e c^2}$ *was not included.*

11 $15°$ *spacing for the polar angle and* $30°$ *spacing for the azimuth.*

12 *7 nodes with a total number of 520 cores at a nominal clock speed of 2.6 GHz.*

13 *Please note that to be consistent with the notation used in Section 4.3.4, I adopt here again the spectral energy* $E'$ *instead of the pulse-height channel number* $\mathcal{n}$. *Furthermore, instead of the time variable* $t$, *I use the experimental condition vector* $\mathfrak{d}$ *to implicitly quantify the time dependence of the DRF. Last but not least, in line with the methodology established in Section 4.3.4, I derive the DRF in the detector frame* $x'$-$y'$-$z'$ *characterized by the Tait-Bryan angles (cf. Section 8.2) instead of the global reference frame* $x$-$y$-$z$.

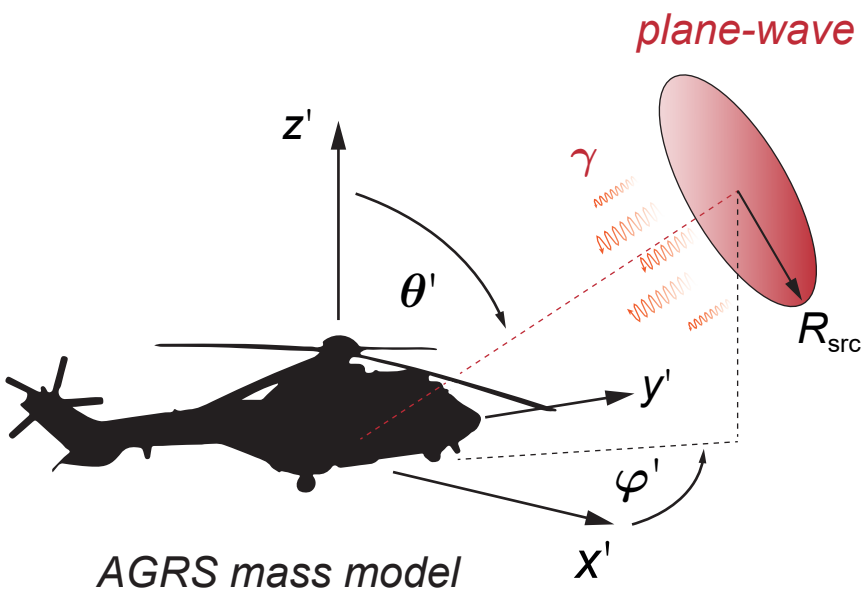

**Figure 9.1** Graphical depiction of the simulation setup for Monte Carlo based detector response function estimation. This includes the AGRS mass model presented in Section 8.2 exposed to a monoenergetic plane-wave photon source with radius $R_{src}$ from a direction $\mathbf{\Omega}'$ characterized by the azimuthal angle $\varphi'$ and the polar angle $\theta'$ in the detector frame $x'$-$y'$-$z'$.

$$R\left(E', E_\gamma, \mathbf{\Omega}', \mathfrak{d}\right) \approx \frac{C\,(E')\,A_{\text{src}}}{N_{\text{pr}}} \tag{9.2}$$

where:

| | | |
|---|---|---|
| $A_{\text{src}}$ | plane-wave source area | $\text{m}^2$ |
| $C$ | counts | |
| $\mathfrak{d}$ | experimental condition | $[\mathfrak{d}]$ |
| $E'$ | spectral energy | eV |
| $E_\gamma$ | photon energy | eV |
| $R$ | detector response function | $\text{m}^2$ |
| $N_{\text{pr}}$ | number of primaries | |
| $\mathbf{\Omega}'$ | direction unit vector in the detector frame | |

and with $A_{\text{src}}$ being the source area of the plane-wave photon source and $N_{\text{pr}}$ the number of primaries adopted in the NPSMC simulation. As discussed in Section 4.3.4, the detector response model approach is based on the assumption that the double differential photon flux is homogeneous within the finite extent of the detector system.[14] Consequently, care must be taken to ensure that the source area of the plane-wave photon source encompasses the full extent of the detector system, or at the very least, is sufficiently large so that the DRF remains unaffected by the simulated source size.[15] In this work, I adopted a circular plane-wave source with a radius of 2 m which has proven to be sufficient for the AGRS mass model described in Section 8.2.[16]

*14 The detector system equates here to all the system components which modulate the photon field.*

*15 Note that the statistical uncertainty in the count vector **C** for a given number of primaries $N_{\text{pr}}$ is proportional to the source area $A_{\text{src}}$. Therefore, to minimize the simulation time, we are interested in minimizing the source area.*

*16 As demonstrated in Fig. B.87, where significant deviations are appreciated only for a radius <1.5 m.*

As indicated in Eq. 9.2, the detector response is also a function of the experimental condition vector $\mathfrak{d}$, encoding all parameters which may change the AGRS mass model and thereby the detector response during or between AGRS surveys. These are mainly the fuel dynamics and the crew composition, as discussed in Section 8.2.[17] To account for these changes, multiple DRFs may be precomputed for different experimental condition vectors $\mathfrak{d}$. I will investigate the impact of the noted parameters on the DRF in Section 9.4. For now, I will assume a standard mass model with a fuel volume fraction of 50 % and a standard crew composition of two pilots, two RLL operators and one loadmaster (cf. Fig. 8.1).

Furthermore, as the DRF quantifies the spectral detector response, it encompasses all the meta-models discussed in the Sections 6.2.1.3 and 7.2.2, too. Consequently, if the meta-models are updated, the DRF must be recomputed.[18] In this chapter, I adopted the meta-models that were used for the Dübendorf and the ARM22 validation campaigns (cf. Section 8.3).

### 9.2.3 Numerical Integration

To perform the numerical integration of the Fredholm integral equation of the first kind (cf. Eq. 4.40), I have implemented the Eqs. 4.43 and 4.44 in `MATLAB` using the `pagemtimes` function allowing for efficient multi-dimensional matrix multiplication operation by applying array programming.[19] As discussed in Section 4.3.4, the numerical integration requires a matching discretization grid between the DRF and the double differential photon flux signature.

Regarding the photon energy, given the coarse discretization scheme used to approximate the DRF, I interpolate the DRF to the photon energy grid employed for the double differential photon flux signature. This interpolation is performed using a spectral interpolation method developed by Jandel et al. [826].[20]

For the angular variables, as the double differential photon flux signature is derived in the global reference frame $x$-$y$-$z$ ($\theta$ and $\varphi$) and the DRF in the detector frame with the aircraft principal axes $x'$-$y'$-$z'$ ($\theta'$ and $\varphi'$), a coordinate transformation needs to be performed for one of the two quantities. In this work, I transform the angular grid employed for the double differential photon flux signature to the detector frame using rotation matrices defined by the Tait-Bryan angles (cf. Appendix A.13).[21] The DRF is then interpolated for all energies on the obtained transformed angular grid.[22]

17 *Note that by deriving the DRF in the detector frame $x'$-$y'$-$z'$, changes in the orientation of the aircraft can be easily accounted for through coordinate transformations of the double differential photon flux signature or the DRF.*

18 *For changes in the spectral energy, the spectral resolution or the LLD models, it is sufficient to repeat the postprocessing using the `NPScinMC` pipeline with the revised meta-models. However, any modifications to the NPSM require a complete rerun of the NPSMC simulations, followed by the postprocessing with the `NPScinMC` pipeline.*

19 *Note that the double differential photon flux $\partial^2\phi_\gamma/\partial E_\gamma \partial\Omega'$ adopted in Eqs. 4.43 and 4.44 needs to be replaced by the double differential photon flux signature $\hat{\phi}_\gamma$ in order to obtain the spectral signature $\hat{\mathbf{c}}$.*

20 *Note that I have slightly adapted Jandel's method by applying the spectral schemes derived in Eqs. 4.28a–4.28e instead of the ones proposed by Jandel et al. to perform the spectral interpolation. Given the significant suppression of DEPs close to $E_\gamma = 2\mathrm{m_e c^2}$, the threshold $(1+\sqrt{2})\mathrm{m_e c^2}$ was again not included.*

21 *It is worth adding that, by definition, $\mathbf{\Omega}$ in the double differential photon flux signature quantifies the direction-of-flight of the photon in the global reference frame (cf. Eq. 2.18), whereas $\mathbf{\Omega}'$ in the DRF characterizes the opposite of the direction-of-flight, i.e. the direction from which the photons are arriving with respect to the detector reference frame (cf. Section 4.3.4).*

Like for PSMC and NPSMC, uncertainty estimates are provided for the derived pulse-height spectra using the detector response model by applying standard error propagation techniques. More information on these uncertainty estimates is provided in Appendix A.8.3.

22 *To account for the anisotropy in the transformed azimuth $\varphi'$, I enrich the angular grid of the double differential photon flux signature by adding additional azimuthal angles at 30° spacing. As noted above, due to the azimuthal symmetry, the azimuth is marginalized in the binning of the double differential flux in the global reference frame. Interpolation is performed using a modified Akima spline interpolation model [733].*

## 9.3 Verification

To assess the accuracy of the derived DRM, I conducted a verification study by comparing its spectral signature predictions to those from the validated AGRS Monte Carlo model presented in the previous chapter.

### 9.3.1 Verification Methods

Two different source-detector configurations have been adopted for the verification study:

1. A monoenergetic isotropic photon point source located 30 m below the Swiss AGRS system on the normal $z'$ axis ($\theta' = 180°$).
2. A monoenergetic isotropic photon point source located in a distance of 30 m on the starboard side of the Swiss AGRS system ($\theta' = 45°$, $\varphi' = 0°$).

For both configurations, the AGRS mass model[23] detailed in Section 8.2 was embedded in homogeneous humid air[24]. Two different photon source energies $E_\gamma$ were considered for each configuration: 120 keV and 2700 keV.

The spectral signature for the two source-detector configurations was computed using the DRM detailed in the previous subsection as well as the validated AGRS Monte Carlo model from Chapter 8 using brute-force NPSMC[25]. The monoenergetic photon point source was modeled using `FLUKA`'s `beam` card for both the full NPSMC and the double differential photon flux signature simulations.[26]

To assess the validity of the simplified isotropic approach proposed by Klusoň [397] (cf. Section 9.1), I also computed the spectral signature using an isotropic DRM. For that purpose, I applied an isotropic DRF $R(E', E_\gamma, \mathfrak{d}) := R(E', E_\gamma, \theta' = 180°, \varphi' = 0°, \mathfrak{d})$ as well as an energy photon flux signature $F_{\mathrm{src}} \mathrm{d}\phi_\gamma \,/\, \mathrm{d}E_\gamma$ (cf. Eq. 2.21a). I will term these two DRMs the isotropic and anisotropic DRM.

23 *Using the reference conditions introduced in Section 9.2, i.e. a fuel volume fraction of 50 % and a standard crew composition of two pilots, two RLL operators and one loadmaster (cf. Fig. 8.1).*

24 *With an air temperature $T = 20\,°\mathrm{C}$, an air pressure $p = 1013.25\,\mathrm{hPa}$ and a relative humidity $\mathrm{RH} = 50\,\%$. The atmosphere was confined to a sphere with a radius of 200 m, centered at the RLL spectrometer's location.*

25 *The adopted number of primaries varied between $9 \times 10^9$ and $3 \times 10^{10}$. The accumulated computation time on the utilized cluster for all four NPSMC simulations was ~10 d.*

26 *Note that given the rotational symmetry of the source-detector configuration, only one Monte Carlo simulation had to be performed per photon energy to derive the double differential photon flux signature. The accumulated computation time for the double differential photon flux signature derivation was less than 30 h with the number of primaries being $8 \times 10^{10}$.*

### 9.3.2 Verification Results

In the Figs. 9.2 and 9.3, I present the spectral signatures obtained by NPSMC ($\hat{c}_{\text{sim}}$) as well as by the isotropic and anisotropic DRMs ($\hat{c}_{\text{DRM}}$) for the two source-detector configurations detailed in Section 9.3.1 alongside uncertainty estimates (cf. Appendices A.8.2 and A.8.3) and relative deviations computed as $|\hat{c}_{\text{DRM}} - \hat{c}_{\text{sim}}|/\hat{c}_{\text{sim}}$.[27]

I find an excellent agreement in the spectral signature between the NPSMC and the anisotropic DRM for all tested source-detector configurations with a median relative deviation <6 % within the SDOI $\mathcal{D}_{\text{SDOI}}$ (cf. Def. 6.1). In contrast, the isotropic DRM shows significant systematic deviations already in the best-case scenario[28] represented in Fig. 9.2 with a median relative deviation exceeding 50 % for the low energy case ($E_\gamma = 120\,\text{keV}$). This deviation can mainly be attributed to the anisotropic nature of the photon flux with a significant fraction of low-energy photons impinging on the detector off-axis due to Compton scattering in the air (cf. Fig. B.88). As expected, due to the increased attenuation by the aircraft structure, the deviations dramatically increase for the source-detector configuration in Fig. 9.3 with a median relative deviation >250 % ($E_\gamma = 120\,\text{keV}$) and >40 % ($E_\gamma = 2700\,\text{keV}$).

27 *Note that I use the unit* Bq *to denote the source strength of the point source, implying that the source consists of a radionuclide with a single emission line at the given photon source energy $E_\gamma$ with a relative photon intensity of $I_\gamma$=100 %.*

28 *Considering that the isotropic DRF was evaluated for the source location at $\theta' = 180°$.*

## 9.4 Detector Response Results

As discussed at the beginning of Section 9.1, the DRF quantifies the spectral response of the AGRS system to incoming photons as a function of the spectral energy $E'$, the photon energy $E_\gamma$ as well as the direction $\mathbf{\Omega}'$ from which the photons are arriving. It has the dimensional units of a geometric area and can therefore be interpreted similarly to the microscopic cross-section in Section 3.1.1 as an effective cross-sectional area. As such, it is not only a key property in the DRM but also a fundamental quantity that fully characterizes the spectral response of the AGRS system to incoming high-energy photons. On the other hand, the double differential photon flux, which is the second part of the DRM is more generic as it quantifies the source as well as the interaction of the photons in the environment for a specific source-detector configuration.

In this last section, I will focus on the DRF to explore its fundamental properties in detail. The response of the system to the various sources relevant in AGRS will be discussed later in Part IV.

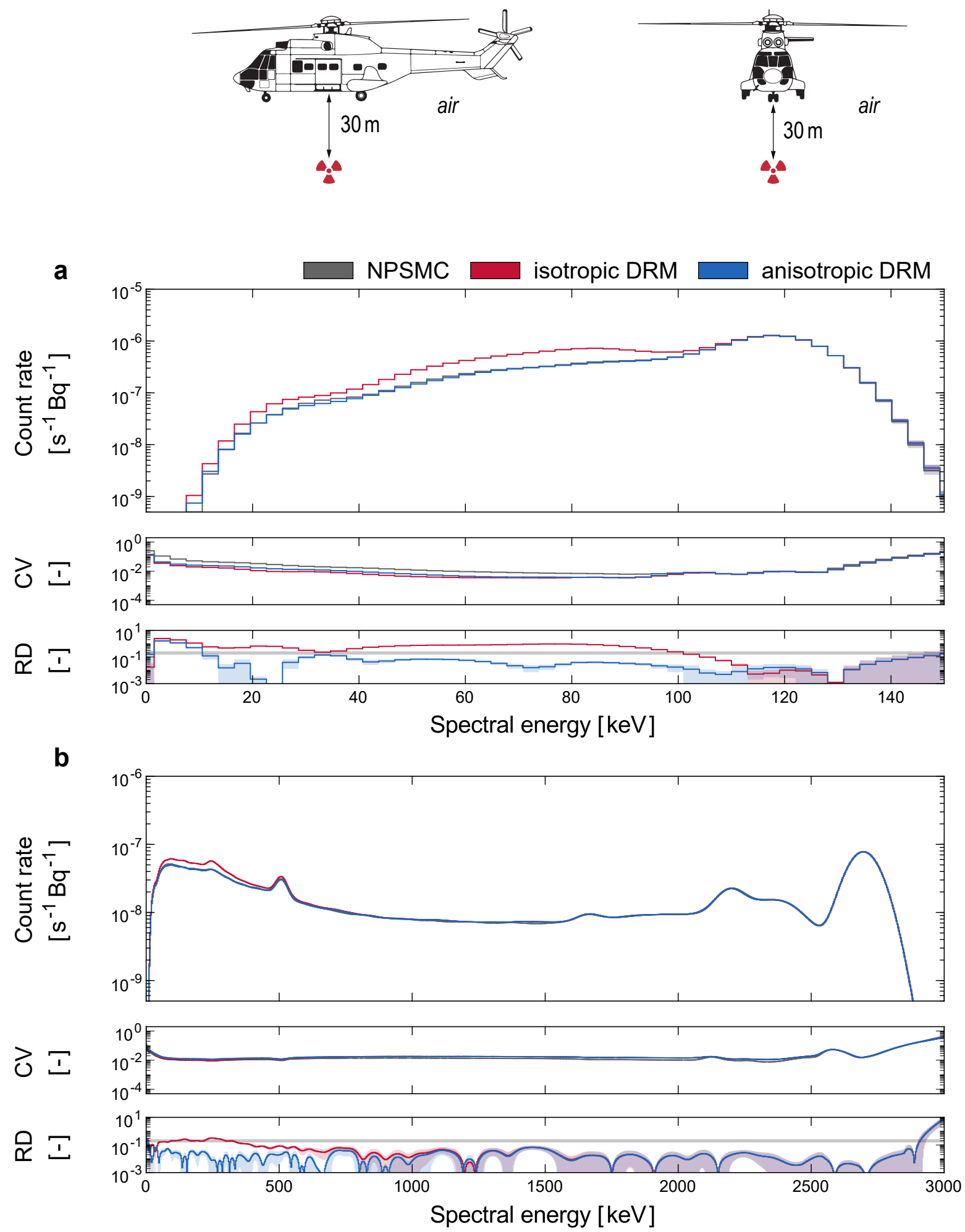

**Figure 9.2** Here, I present the verification results of the DRM using a source-detector configuration depicted at the top of the figure. For that purpose, the spectral signature of the Swiss AGRS system was computed using NPSMC ($\hat{c}_{\mathrm{sim}}$) as well as the isotropic and anisotropic DRMs ($\hat{c}_{\mathrm{DRM}}$) for two different photon source energies: **a** $E_{\gamma} = 120\,\mathrm{keV}$. **b** $E_{\gamma} = 2700\,\mathrm{keV}$. For each photon energy, the mean spectral signature in the detector channel `#SUM` is displayed as a function of the spectral energy $E'$ with a spectral energy bin width of $\Delta E' \sim 3\,\mathrm{keV}$. Uncertainties are provided as 1 standard deviation (SD) shaded areas. In addition, the coefficient of variation (CV) (cf. Appendix A.8) as well as the relative deviation (RD) computed as $|\hat{c}_{\mathrm{DRM}} - \hat{c}_{\mathrm{sim}}|/\hat{c}_{\mathrm{sim}}$ (20 % mark highlighted by a horizontal grey line) are provided.

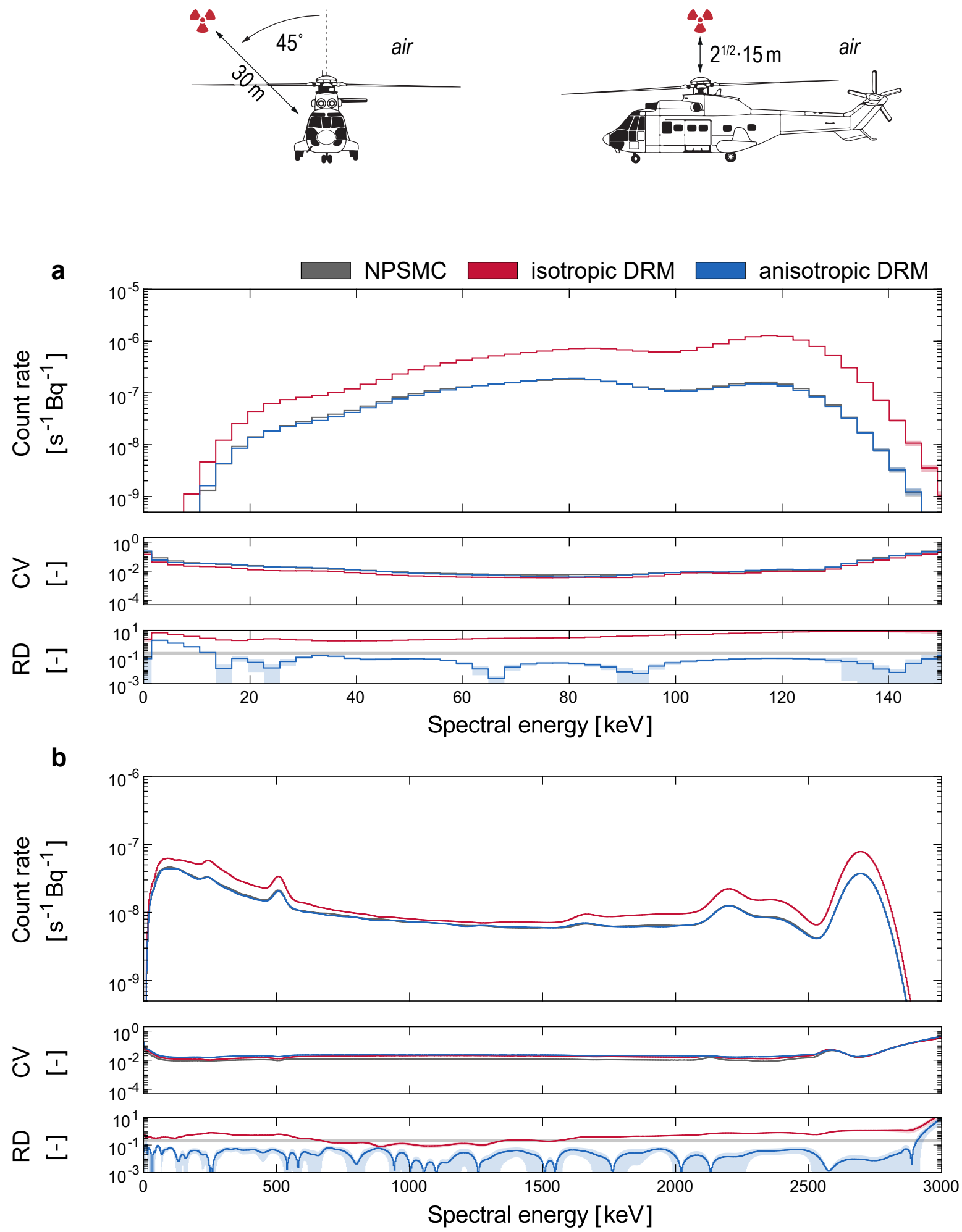

**Figure 9.3** Here, I present the verification results of the DRM using a source-detector configuration depicted at the top of the figure. For that purpose, the spectral signature of the Swiss AGRS system was computed using NPSMC ($\hat{c}_{\text{sim}}$) as well as the isotropic and anisotropic DRMs ($\hat{c}_{\text{DRM}}$) for two different photon source energies: **a** $E_{\gamma} = 120$ keV. **b** $E_{\gamma} = 2700$ keV. For each photon energy, the mean spectral signature in the detector channel `#SUM` is displayed as a function of the spectral energy $E'$ with a spectral energy bin width of $\Delta E' \sim 3$ keV. Uncertainties are provided as 1 standard deviation (SD) shaded areas. In addition, the coefficient of variation (CV) (cf. Appendix A.8) as well as the relative deviation (RD) computed as $|\hat{c}_{\text{DRM}} - \hat{c}_{\text{sim}}|/\hat{c}_{\text{sim}}$ (20 % mark highlighted by a horizontal grey line) are provided.

As discussed already Section 9.2, the TH06 aircraft is an inherently dynamic system with many components changing position, composition and orientation during or between survey flights. To assess the effect of changes in the various parameters on the DRF, I introduce the following reference mass model:

**I. Reference mass model ($R_{\mathrm{ref}}$)** Complete AGRS Monte Carlo mass model presented in Section 8.2 with a fuel volume fraction of 50 % and a standard crew composition of two pilots, two RLL operators and one loadmaster (cf. also to Fig. 8.1).

This reference mass model, which was already used for the verification study in Section 9.3.1, will serve as a basis for the subsequent analyses. The corresponding DRF, denoted as $R_{\mathrm{ref}}$, was computed using the methodology outlined in Section 9.2.

To investigate the effect of specific mass model components on the DRF, I derived additional DRFs by selectively changing the parameters of interest in the reference mass model. Specifically, I assessed the effect of the aircraft, fuel and the crew using the following scenarios and associated DRFs:

**II. Aircraft effect**

1. **No aircraft ($R_{\mathrm{det}}$)** Reference mass model with all aircraft structures removed, i.e. only the RLL detector box and its internals remain.
2. **No RLL supporting systems ($R_{\mathrm{RLL}}$)** Reference mass model with all RLL supporting systems located in the aircraft cabin removed, i.e. the operator console, the operator seats, the equipment rack as well as the two operators (cf. Section 5.3.2).

**III. Fuel effect**

1. **Empty fuel tank ($R_{\mathrm{JF0}}$)** Reference mass model with the relative fuel volume reduced to 0 %.
2. **Quarter-filled fuel tank ($R_{\mathrm{JF25}}$)** Reference mass model with the relative fuel volume reduced to 25 %.
3. **Filled fuel tank ($R_{\mathrm{JF100}}$)** Reference mass model with the relative fuel volume increased to 100 %.

**IV. Crew effect**

1. **Full capacity ($R_{FC}$)** Reference mass model with personnel on board increased to full capacity (2 pilots, 2 operators, 1 loadmaster, 7 passengers, cf. Fig. 8.1).
2. **No personnel ($R_{NP}$)** Reference mass model with all personnel removed.

In the following subsections, I will start by discussing the DRF for the reference mass model and then proceed to analyze the effect of the aircraft, the fuel and the crew on the DRF using the additional scenarios introduced above.

### 9.4.1 Reference Model

In this first subsection, I will analyze the DRF for the reference mass model outlined above. Since the DRF is a function of four fundamental variables,[29] I will investigate its characteristics using two-dimensional projections: one to the spectral space $(E', E_\gamma)$ denoted as the spectral dispersion projection and one to the angular space $(\theta', \varphi')$ named angular dispersion projection.

29 *Assuming the experimental conditions to be fixed.*

**Spectral Dispersion** The spectral dispersion projection is performed by keeping the direction of the incoming photons fixed and by varying the spectral energy $E'$ and the photon energy $E_\gamma$. In the Figs. 9.4–9.6, I present the resulting projections for the six cardinal directions along the three aircraft principal axes $x'$, $y'$ and $z'$. There are two interesting trends in the Figs. 9.4–9.6 that I would like to highlight.

First, as expected from our discussion in Chapter 3, the absolute detector response is highest for the normal direction ($\theta' = 180°$) and significantly reduced by a factor $\mathcal{O}(10^2)$ for all other cardinal directions with the lowest response observed in the forward direction ($\theta' = 90°$, $\varphi' = 90°$). This reduction is mainly due to the increased attenuation by the aircraft structure obstructing the incoming photons for the corresponding directions, as well as the positioning and arrangement of the scintillation crystals within the aircraft.

Second, regarding the spectral dispersion, I observe a significant variation in the spectral response both across the photon energy $E_\gamma$ range as well as among the different directions. The most pronounced features are highlighted in the individual graphs.

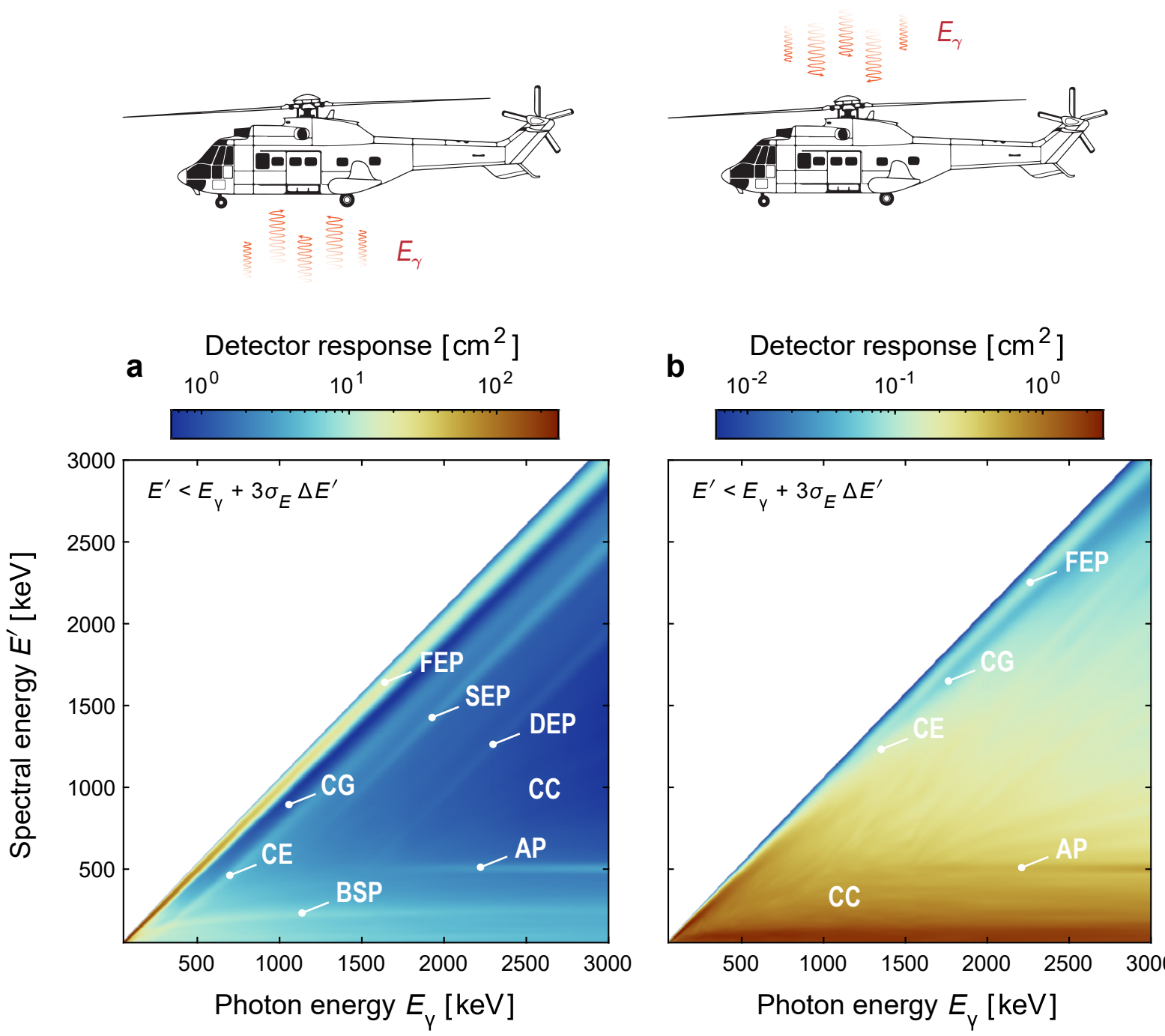

**Figure 9.4** Spectral dispersion of the Swiss AGRS system's detector response $R_{\mathrm{ref}}$ in the detector channel `#SUM` as a function of the spectral energy $E'$ and the photon energy $E_\gamma$ with a spectral energy bin width of $\Delta E' \sim 3$ keV. The DRF was computed for two different directions $\mathbf{\Omega}'$: **a** Normal ($\theta' = 180°$). **b** Anti-normal ($\theta' = 0°$). For visualization purposes, the DRF is limited to the spectral energy range $E' < E_\gamma + 3\sigma_E \Delta E'$ with $\sigma_E$ being the spectral resolution standard deviation (measured in the pulse-height channel number space, cf. Section 4.3.2) at $E_\gamma / \Delta E'$ and $\Delta E'$ the spectral energy bin width (cf. Section 6.2.1.3). Characteristic spectral features discussed in Section 4.3.1 are marked, i.e. the full energy peak (FEP), the single escape peak (SEP), the double escape peak (DEP), the annihilation peak (AP), the backscatter peak (BSP), the Compton edge (CE), the Compton gap (CG) as well as the Compton continuum (CC).

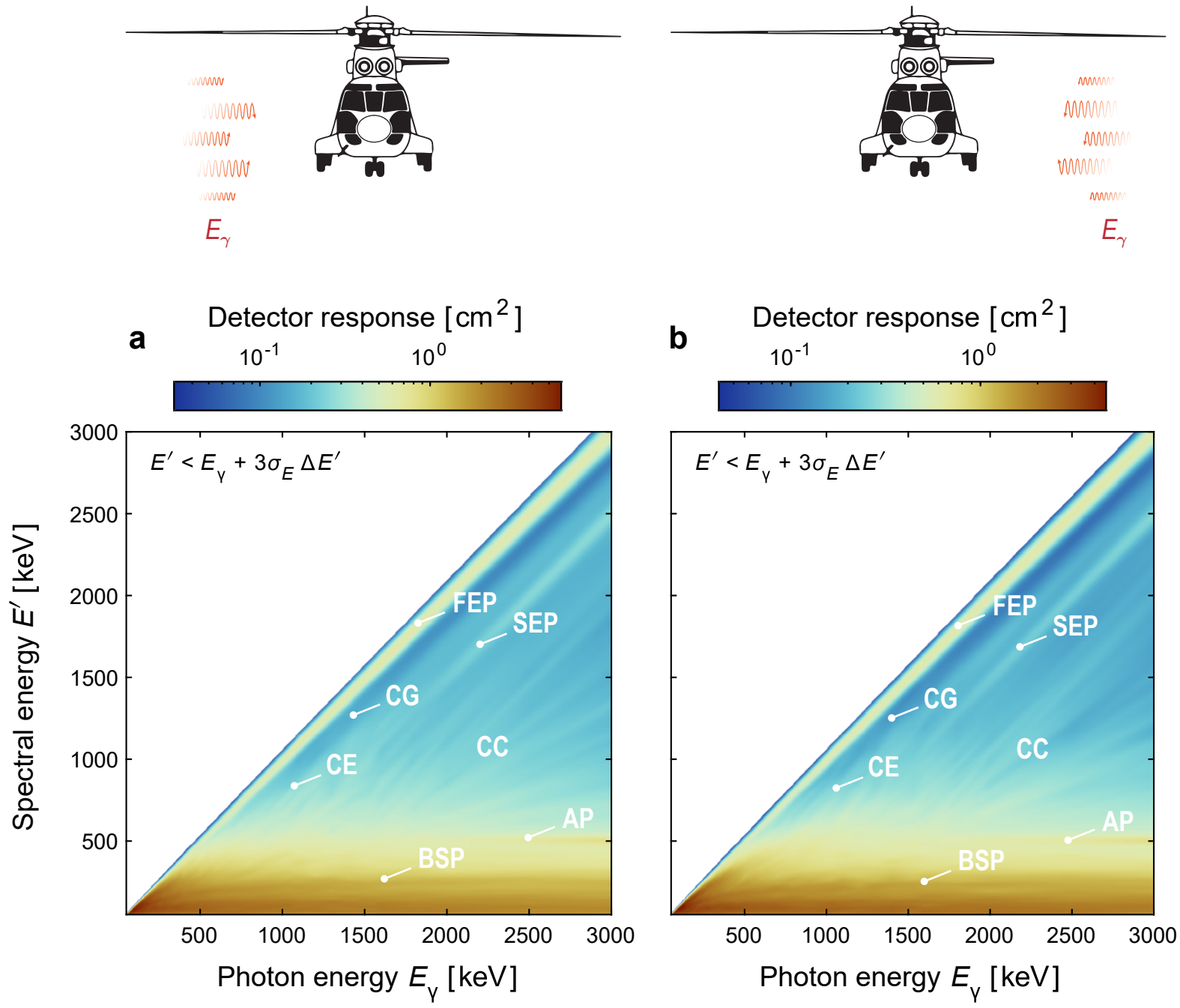

**Figure 9.5** Spectral dispersion of the Swiss AGRS system's detector response $R_{\text{ref}}$ in the detector channel `#SUM` as a function of the spectral energy $E'$ and the photon energy $E_\gamma$ with a spectral energy bin width of $\Delta E' \sim 3\,\text{keV}$. The DRF was computed for two different directions $\mathbf{\Omega}'$: **a** Starboard ($\theta' = 90°$, $\varphi' = 0°$). **b** Port ($\theta' = 90°$, $\varphi' = 180°$). For visualization purposes, the DRF is limited to the spectral energy range $E' < E_\gamma + 3\sigma_E \Delta E'$ with $\sigma_E$ being the spectral resolution standard deviation (measured in the pulse-height channel number space, cf. Section 4.3.2) at $E_\gamma / \Delta E'$ and $\Delta E'$ the spectral energy bin width (cf. Section 6.2.1.3). Characteristic spectral features discussed in Section 4.3.1 are marked, i.e. the full energy peak (FEP), the single escape peak (SEP), the annihilation peak (AP), the backscatter peak (BSP), the Compton edge (CE), the Compton gap (CG) as well as the Compton continuum (CC).

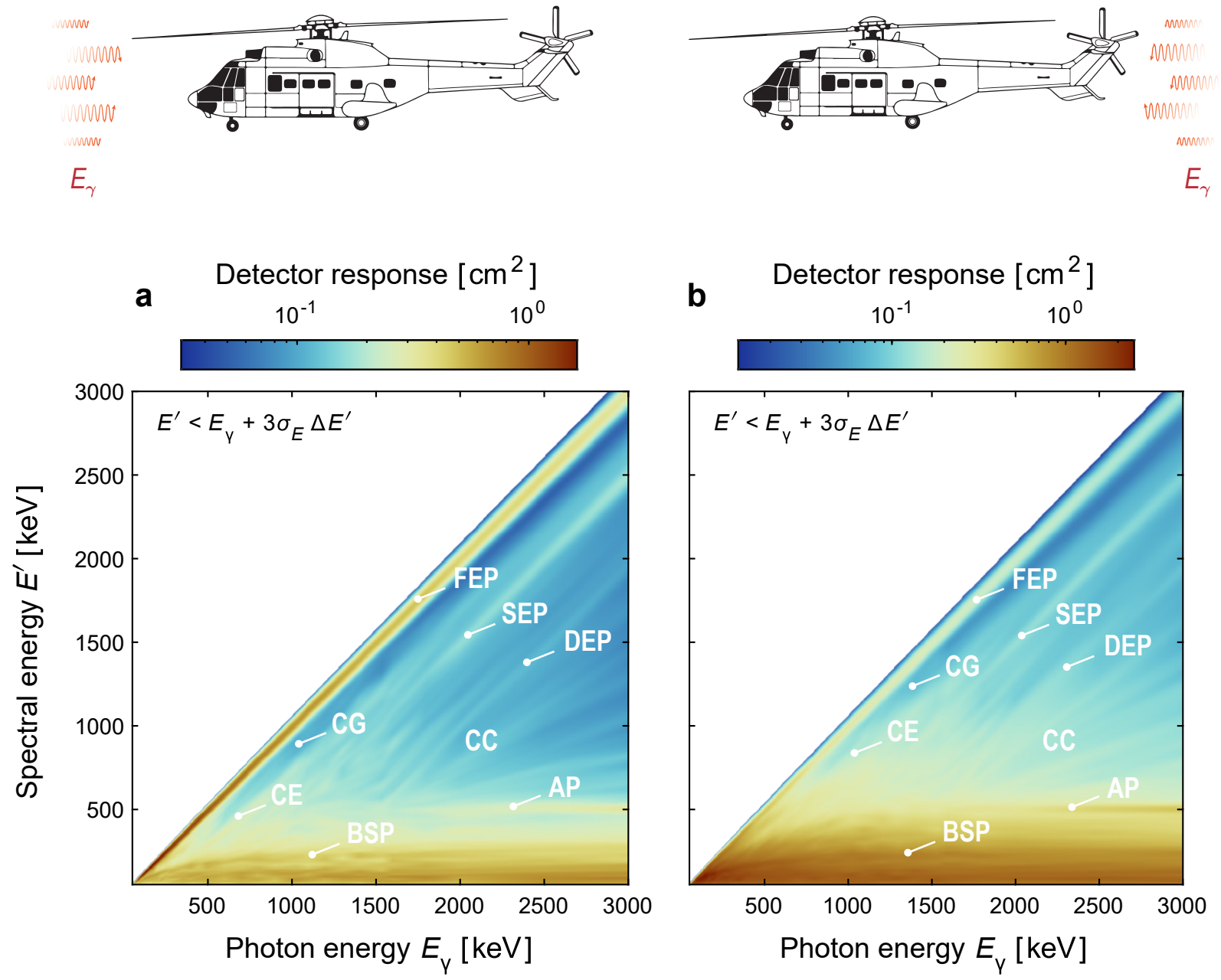

**Figure 9.6** Spectral dispersion of the Swiss AGRS system's detector response $R_{\text{ref}}$ in the detector channel `#SUM` as a function of the spectral energy $E'$ and the photon energy $E_\gamma$ with a spectral energy bin width of $\Delta E' \sim 3\,\text{keV}$. The DRF was computed for two different directions $\boldsymbol{\Omega}'$: **a** Backward ($\theta' = 90°$, $\varphi' = -90°$). **b** Forward ($\theta' = 90°$, $\varphi' = 90°$). For visualization purposes, the DRF is limited to the spectral energy range $E' < E_\gamma + 3\sigma_E \Delta E'$ with $\sigma_E$ being the spectral resolution standard deviation (measured in the pulse-height channel number space, cf. Section 4.3.2) at $E_\gamma / \Delta E'$ and $\Delta E'$ the spectral energy bin width (cf. Section 6.2.1.3). Characteristic spectral features discussed in Section 4.3.1 are marked, i.e. the full energy peak (FEP), the single escape peak (SEP), the double escape peak (DEP), the annihilation peak (AP), the backscatter peak (BSP), the Compton edge (CE), the Compton gap (CG) as well as the Compton continuum (CC).

This includes the full energy peak (FEP), the single escape peak (SEP), the double escape peak (DEP), the annihilation peak (AP), the backscatter peak (BSP), the Compton edge (CE), the Compton gap (CG) as well as the Compton continuum (CC) (cf. also to Section 4.3.1). As a general trend, the ratio between the FEP and the CC is maximized for cases where the attenuation of the primary photons is minimal and the attenuation of photons undergoing Compton scattering in the aircraft structure is maximized. This is the case for the normal direction. In contrast, due to the high-density rotor head (~472 kg without the rotor blades) and the associated gearbox (~350 kg) obstructing the trajectory of the primary photons combined with the low-attenuating cabin structure,[30] the ratio between the FEP and the CC is minimized for the anti-normal direction (cf. Fig. 8.1). The remaining directions represent intermediate cases with a varying ratio between the FEP and the CC depending on the specific aircraft structure obstructing the trajectories of the primary photons.

30 *The position of these aircraft components is highlighted in Fig. 8.1.*

It is interesting to note that, given the almost symmetric design of the longitudinal fuel tanks (cf. Figs. 8.1 and B.81), only minimal differences in the spectral response are observed between the starboard and port directions. Conversely, there are notable differences between the forward and backward directions. The reduced ratio between the FEP and the CC in the backward direction can be mainly attributed to the increased attenuation of the primary photons by the detector electronics and the PMTs, all positioned at the rear of the RLL spectrometer (cf. Fig. 8.1). Additionally, for the reference fuel volume fraction of 50 %, the fuel level in the aft fuel tank (tank #5) is slightly higher than in the forward fuel tanks, covering a larger part of the scintillation crystals' cross-sectional area. This also contributes to the increased attenuation of the primary photons in the backward direction.[31]

31 *Note that the tail of the TH06 aircraft does not affect the spectral response in the backward direction, as it is vertically displaced by >1 m from the detector position. This was confirmed in a sensitivity study (cf. Fig. B.89).*

**Angular Dispersion** The angular dispersion projection is performed by integrating[32] the DRF over three prefined spectral bands for a fixed photon energy $E_\gamma$. These three spectral bands are defined as follows:

32 *As the spectral energy represents discrete pulse-height channels with a fixed spectral energy bin width $\Delta E'$, the integration operation translates to summing the pulse-height channels within the predefined spectral bands. Consequently, the unit of the detector response remains unchanged for the projected detector response.*

1. **Full spectrum domain** Spectral band to quantify the full spectrum response, defined as $\mathcal{D}_{\text{tot}} = \{E' \in \mathbb{R}_+ \mid E' \leq N_{\text{ch}}\Delta E'\}$ with $N_{\text{ch}}$ being the number of pulse-height channels and $\Delta E'$ the spectral energy bin width (cf. Sections 4.3.3.1 and 6.2.1.3).
2. **Full energy peak domain** Spectral band to quantify the FEP, defined as $\mathcal{D}_{\text{FEP}} = \{E' \in \mathbb{R}_+ \mid E_\gamma - 2\sigma_E\Delta E' \leq E' \leq E_\gamma + 2\sigma_E\Delta E'\}$ with $\sigma_E$ being the spectral resolution standard deviation (measured in the pulse-height channel number space, cf. Section 4.3.2) at $E_\gamma/\Delta E'$.

3. **Compton continuum domain** Spectral band to quantify the CC, defined as $\mathcal{D}_{\text{CC}} = \{E' \in \mathbb{R}_+ \mid E' \leq E'_{\text{CE}} - 2\sigma_E \Delta E'\}$ with $E'_{\text{CE}}$ being the spectral energy of the Compton edge (cf. Eq. 4.21) associated with the photon energy $E_\gamma$ and $\sigma_E$ denoting the spectral resolution at $E'_{\text{CE}}/\Delta E'$.

The resulting angular dispersion projections are displayed in the Figs. 9.7–9.9 for three selected photon energies $E_\gamma$: 88 keV, 662 keV and 2615 keV.[33] Four interesting trends can be observed in these graphs.

First, as anticipated in Section 9.3.2, we can identify a pronounced angular anisotropy for all photon energies and all spectral bands with the maximum response being observed consistently for the normal direction and the lowest response in the directions where the attenuation by the aircraft structure is maximized. Given the position of the scintillation crystals in the cargo bay of the TH06 aircraft (cf. Section 5.3.2), the aircraft acts as a collimator for the incoming primary photons, leading to a significant attenuation of the response for $\theta' \lesssim 120°$. However, due to the higher penetration power, i.e. larger mean free paths and lower attenuation coefficients (cf. Section 3.1.2.6), this collimation effect is reduced for higher photon energies, resulting in a more homogeneous angular response across all spectral bands. This is particularly true for the directions crossing the cabin walls and cabin doors at $30° \lesssim \theta' \lesssim 60°$ and $180°k - 60° \lesssim \varphi' \lesssim 180°k + 60°$ with $k = 0$ for the starboard side and $k = 1$ for the port side.[34]

Second, regarding the full energy peak domain $\mathcal{D}_{\text{FEP}}$, I observe a consistent decrease in the response for an increase in the photon energy across the entire angular space. This trend can be explained by the decrease in the microscopic cross-section of the photoelectric absorption interaction with increasing photon energy (cf. Fig. B.16 and Section 3.1.2.3). The reduced photoelectric cross-section results in a reduction of the photoelectric absorption events in the scintillation crystals and thereby a decrease in the FEP.

33 *These photon energies were arbitrarily selected to represent generic low, intermediate and high-energy photon sources that may be encountered in AGRS surveys.*

34 *The notable asymmetry in the angular response between the starboard and port side will be discussed in the next subsection.*

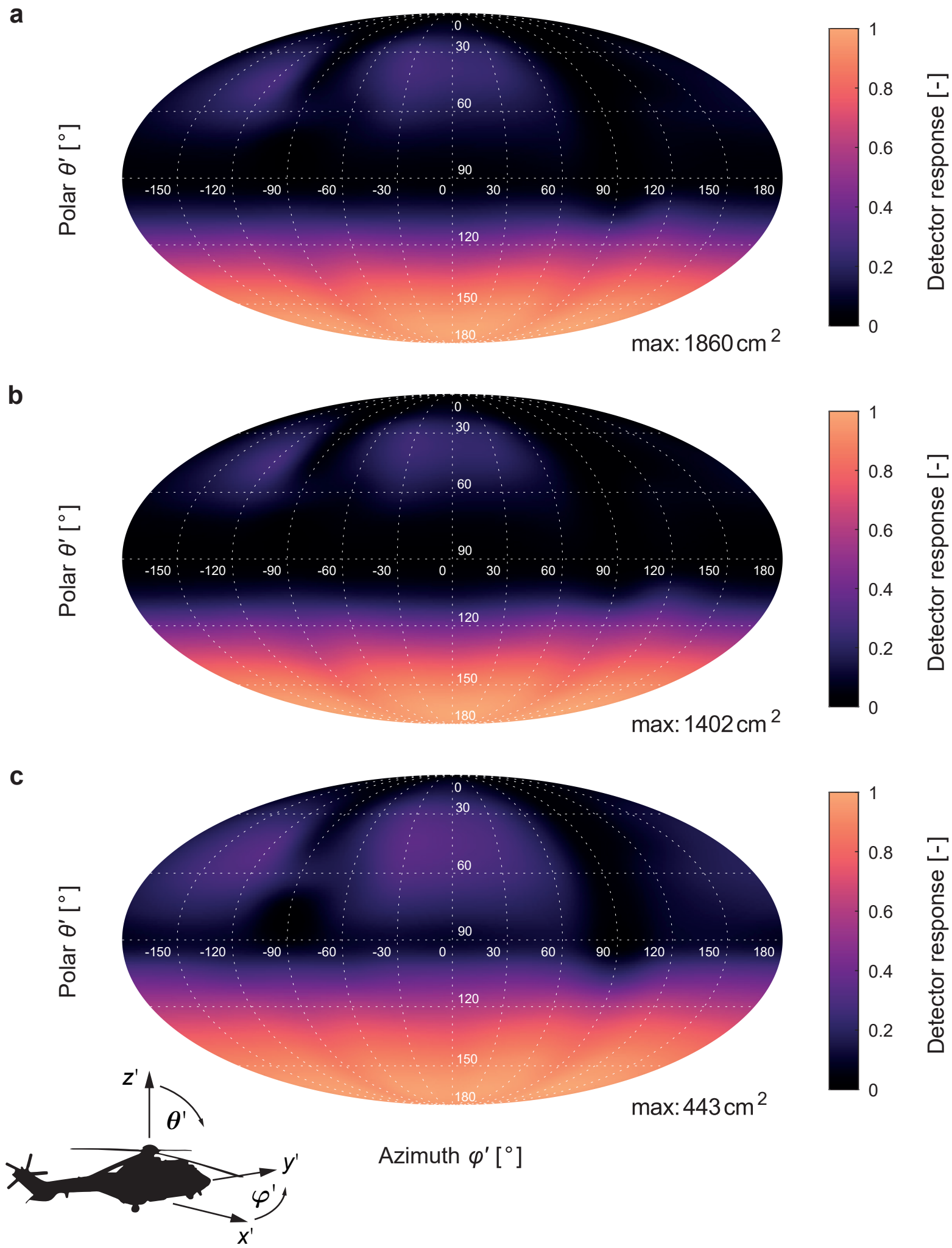

**Figure 9.7** Angular dispersion of the Swiss AGRS system's detector response $R_{\text{ref}}$ in the detector channel `#SUM` as a function of the polar angle $\theta'$ and the azimuthal angle $\varphi'$ with respect to the detector reference frame $x'$-$y'$-$z'$. The angular dispersion was computed for a photon energy $E_\gamma$ = 88 keV and normalized by its maximum, i.e. $R_{\text{ref}}/\max(R_{\text{ref}})$ with $\max(R_{\text{ref}})$ indicated in each graph. Three different spectral domains were evaluated: **a** Full spectrum domain $\mathcal{D}_{\text{tot}}$. **b** Full energy peak domain $\mathcal{D}_{\text{FEP}}$. **c** Compton continuum domain $\mathcal{D}_{\text{CC}}$. The angular dispersion graphs are interpolated on a 1° × 1° angular grid (modified Akima spline interpolation [733]) and displayed using the Mollweide projection.

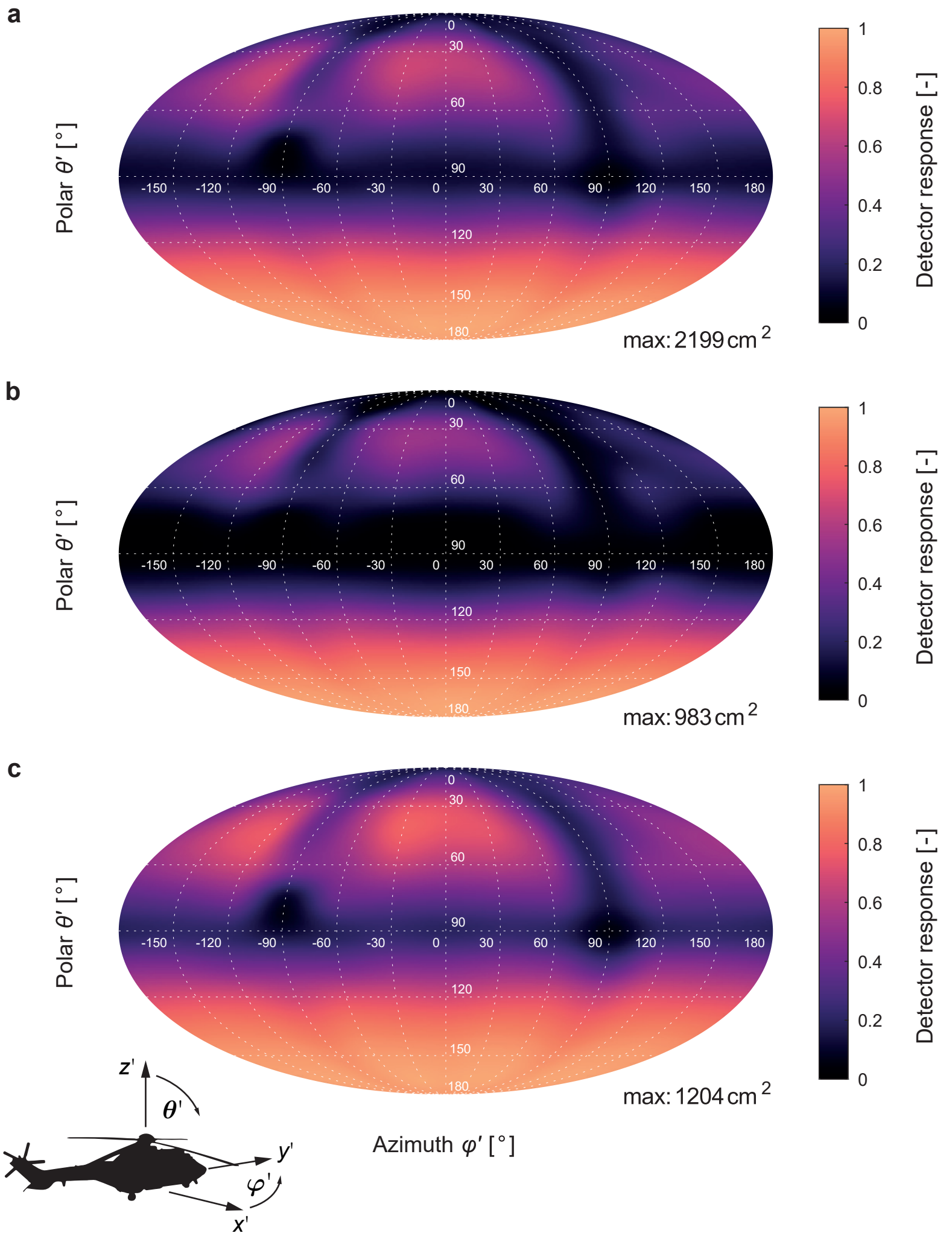

**Figure 9.8** Angular dispersion of the Swiss AGRS system's detector response $R_{\text{ref}}$ in the detector channel `#SUM` as a function of the polar angle $\theta'$ and the azimuthal angle $\varphi'$ with respect to the detector reference frame $x'$-$y'$-$z'$. The angular dispersion was computed for a photon energy $E_\gamma = 662\,\text{keV}$ and normalized by its maximum, i.e. $R_{\text{ref}}/\max(R_{\text{ref}})$ with $\max(R_{\text{ref}})$ indicated in each graph. Three different spectral domains were evaluated: **a** Full spectrum domain $\mathcal{D}_{\text{tot}}$. **b** Full energy peak domain $\mathcal{D}_{\text{FEP}}$. **c** Compton continuum domain $\mathcal{D}_{\text{CC}}$. The angular dispersion graphs are interpolated on a $1° \times 1°$ angular grid (modified Akima spline interpolation [733]) and displayed using the Mollweide projection.

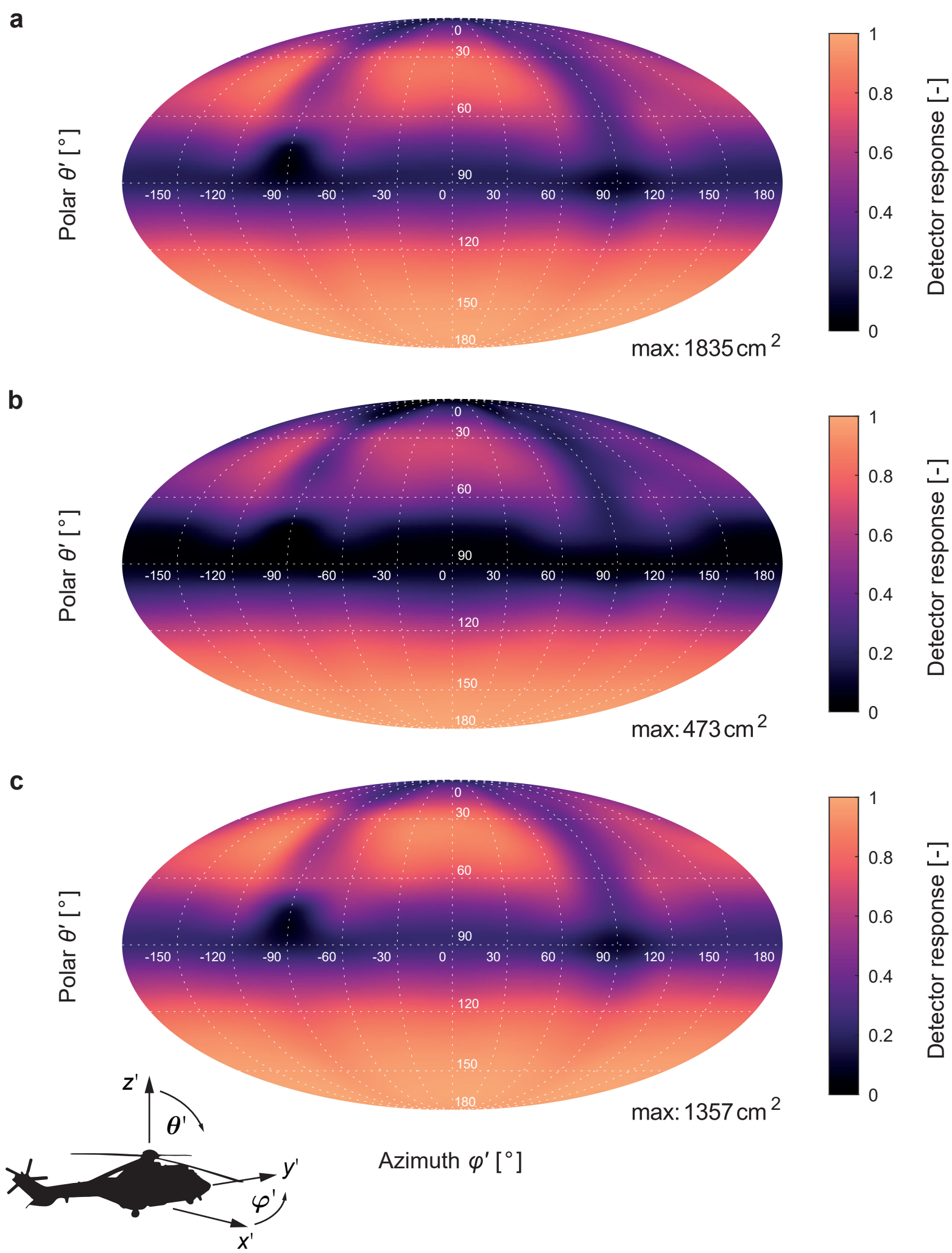

**Figure 9.9** Angular dispersion of the Swiss AGRS system's detector response $R_{\text{ref}}$ in the detector channel `#SUM` as a function of the polar angle $\theta'$ and the azimuthal angle $\varphi'$ with respect to the detector reference frame $x'$-$y'$-$z'$. The angular dispersion was computed for a photon energy $E_\gamma$ = 2615 keV and normalized by its maximum, i.e. $R_{\text{ref}}/\max(R_{\text{ref}})$ with $\max(R_{\text{ref}})$ indicated in each graph. Three different spectral domains were evaluated: **a** Full spectrum domain $\mathcal{D}_{\text{tot}}$. **b** Full energy peak domain $\mathcal{D}_{\text{FEP}}$. **c** Compton continuum domain $\mathcal{D}_{\text{CC}}$. The angular dispersion graphs are interpolated on a 1° × 1° angular grid (modified Akima spline interpolation [733]) and displayed using the Mollweide projection.

Third, regarding the Compton continuum domain $\mathcal{D}_{\mathrm{CC}}$, I find a consistent increase in the response for an increase in the photon energy across the entire angular space. This trend can be attributed to the increase in the microscopic cross-section of the Compton scattering interaction with increasing photon energy (cf. Fig. B.16 and Section 3.1.2.2). The increase in the Compton cross-section results in an increase of Compton scattering events both in the aircraft structure and in the scintillation crystals, leading to an increase in the CC.

Fourth, regarding the full spectrum domain $\mathcal{D}_{\mathrm{tot}}$, I find an interesting trend with the highest response occurring at the photon energy of 662 keV, while reduced responses are observed at 88 keV and 2615 keV. This observation can be explained by the combination of the two opposing trends discussed above for the $\mathcal{D}_{\mathrm{FEP}}$ and the $\mathcal{D}_{\mathrm{CC}}$ spectral bands. Specifically, the reduction in FEP events, coupled with an increase in Compton scattering events for increasing photon energies results in a maximum response in the full spectrum band $\mathcal{D}_{\mathrm{CC}}$ for intermediate photon energies.

In summary, the DRF of the Swiss AGRS system exhibits a pronounced anisotropic response with complex spectral and angular dispersion. The observed anisotropic response is attributed mainly to the aircraft structure, as well as the positioning and arrangement of the scintillation crystals within the aircraft. To investigate the effect of the aircraft structure in more detail, I performed additional detector response simulations which I will present in the next subsection.

### 9.4.2 Aircraft Effect

In the previous subsection, I found a pronounced anisotropic response of the Swiss AGRS system to incoming photons. To assess the effect of the aircraft on the detector response, I compared the DRF for the reference mass model ($R_{\mathrm{ref}}$) presented in the previous subsection with a DRF that includes only the RLL detector box and its internals ($R_{\mathrm{det}}$), i.e. all aircraft structures and all personnel were removed. Specifically, I computed the angular dispersion[35] as detailed in the previous subsection using the three spectral bands $\mathcal{D}_{\mathrm{tot}}$, $\mathcal{D}_{\mathrm{FEP}}$ and $\mathcal{D}_{\mathrm{CC}}$ for both models and then calculated the relative deviation as $(R_{\mathrm{det}} - R_{\mathrm{ref}})/\max(R_{\mathrm{ref}})$ with $\max(R_{\mathrm{ref}})$ being equivalent to $R_{\mathrm{ref}}$ evaluated in the normal direction ($\theta'$=180°). These computations were repeated for the same photon energies $E_\gamma$ as in the previous subsection, i.e. 88 keV, 662 keV and 2615 keV.

35 *Note that by evaluating the angular dispersion not only for the full spectrum band but also for the FEP and the CC, trends in the major two spectral features are characterized. Therefore, I focus on the angular dispersion in the succeeding analyses. A supplementary analysis of the effect of the NPSM adopted in NPSMC on the spectral dispersion in the Compton edge (CE) compared to PSMC is provided in Fig. B.90.*

The resulting angular dispersion graphs are presented in the Figs. 9.10–9.12. As expected, there is a pronounced difference between the two models for all energies and all spectral bands with maximum relative deviations varying between ~40 % and ~100 %. In line with our previous discussion, the effects are highest at 88 keV and decrease for higher photon energies. With increasing photon energy, specific high-density aircraft components such as the rotor head (~472 kg without the rotor blades), the gearbox (~350 kg) and the turbines (two times ~232 kg) start to form distinct features in the angular response, as indicated in the corresponding graphs.[36] There are three other interesting trends.

36 *The position of these aircraft components is highlighted in Fig. 8.1.*

First, in line with the observations from the previous subsection, I find that the attenuation effect of the primary photons quantified by the angular dispersion in the full energy peak domain $\mathcal{D}_{\text{FEP}}$ is restricted to $\theta' \lesssim 120°$ for all photon energies. As discussed previously, this trend can be attributed to the position of the scintillation crystals in the cargo bay of the TH06 aircraft with no aircraft structures obstructing the trajectories of the primary photons for $\theta' \gtrsim 120°$.

Second, from the Compton continuum domain $\mathcal{D}_{\text{CC}}$ graphs it is evident that the aircraft structure does not only attenuate primary and secondary photons for $\theta' \lesssim 120°$, but also results in a significant increase in the Compton scattering events for $\theta' \gtrsim 120°$ and thereby an increase in the response in $\mathcal{D}_{\text{CC}}$ (~35 % at $E_\gamma = 88$ keV, ~32 % at $E_\gamma = 662$ keV and ~20 % at $E_\gamma = 2615$ keV). We can therefore conclude that the aircraft does not only act as a collimator but also as a scatterer for the primary photons, significantly increasing the Compton continuum response for photons arriving at a direction with $\theta' \gtrsim 120°$. At higher energies, this effect is not restricted to $\theta' \gtrsim 120°$ but occurs also in the directions where only low absorption aircraft structures obstruct the trajectories of the primary photons. This is particularly true for the directions crossing the cabin walls and the cabin doors at $30° \lesssim \theta' \lesssim 60°$ and $180°k - 60° \lesssim \varphi' \lesssim 180°k + 60°$ with $k = 0$ for the starboard side and $k = 1$ for the port side.

Third, as already indicated in the previous subsection, there are notable asymmetries in the angular response between the starboard and port side. This is most pronounced between $30° \lesssim \varphi' \lesssim 60°$ and $120° \lesssim \varphi' \lesssim 150°$ for $30° \lesssim \theta' \lesssim 60°$. Given the bilateral symmetry in the design of the TH06 aircraft, this trend can be attributed to the asymmetric placement of the supporting systems for the RLL spectrometer in the aircraft cabin, i.e. the rack, console, operator seats as well as the operators themselves. As indicated in Fig. 8.1, these systems are positioned with an offset of approximately 20 cm towards the port side, equating to roughly 20 % of the aircraft cabin's half-width.

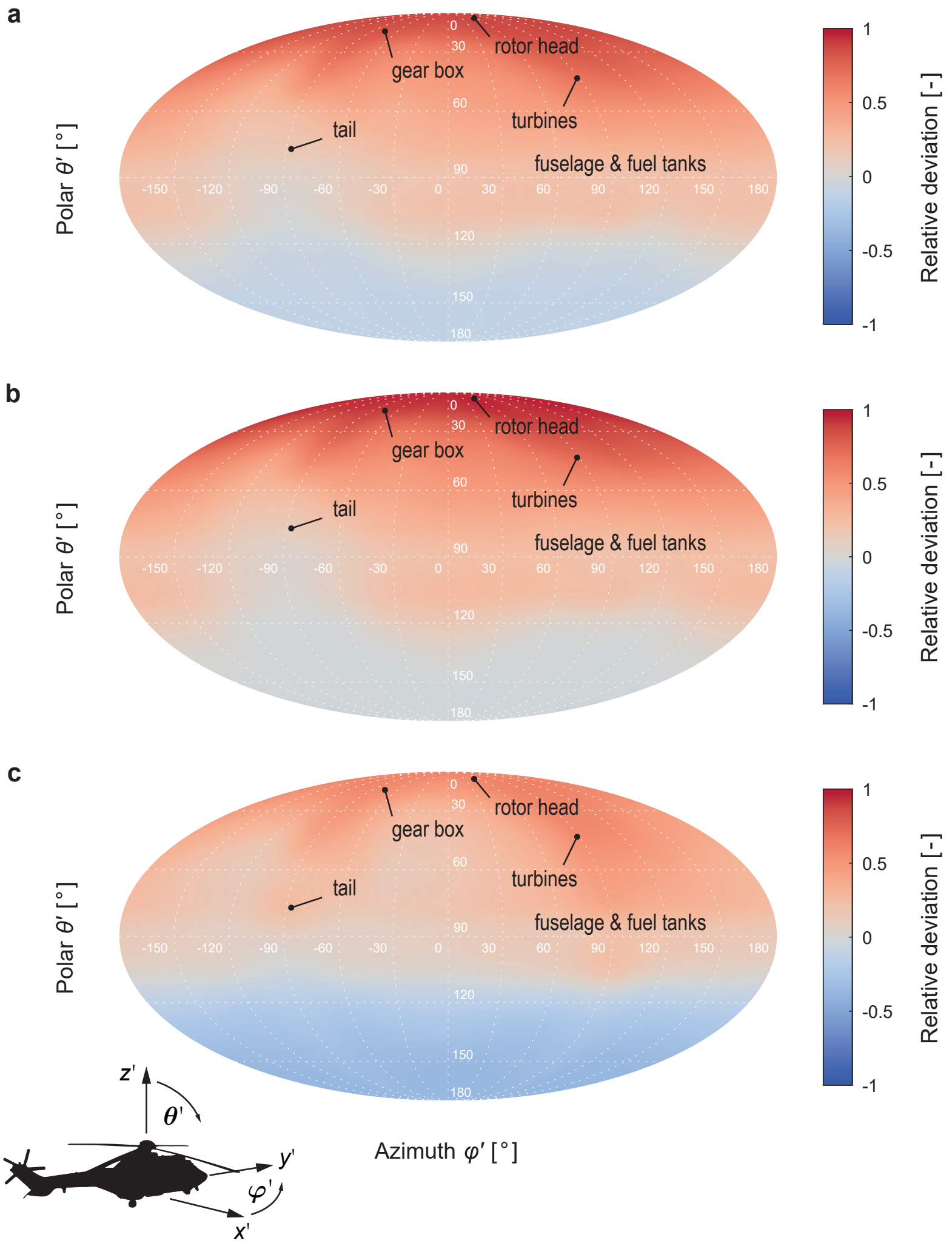

**Figure 9.10** Relative deviation in the angular dispersion between the reference DRF ($R_{\text{ref}}$) and a DRF including only the RLL spectrometer ($R_{\text{det}}$) at a photon energy of $E_\gamma = 88\,\text{keV}$. The relative deviation was computed as $(R_{\text{det}} - R_{\text{ref}})/\max(R_{\text{ref}})$ in the detector channel `#SUM` as a function of the polar angle $\theta'$ and the azimuthal angle $\varphi'$ with respect to the detector reference frame $x'$-$y'$-$z'$. Three different spectral domains were evaluated: **a** Full spectrum domain $\mathcal{D}_{\text{tot}}$. **b** Full energy peak domain $\mathcal{D}_{\text{FEP}}$. **c** Compton continuum domain $\mathcal{D}_{\text{CC}}$. The graphs are interpolated on a $1° \times 1°$ angular grid (modified Akima spline interpolation [733]) and displayed using the Mollweide projection.

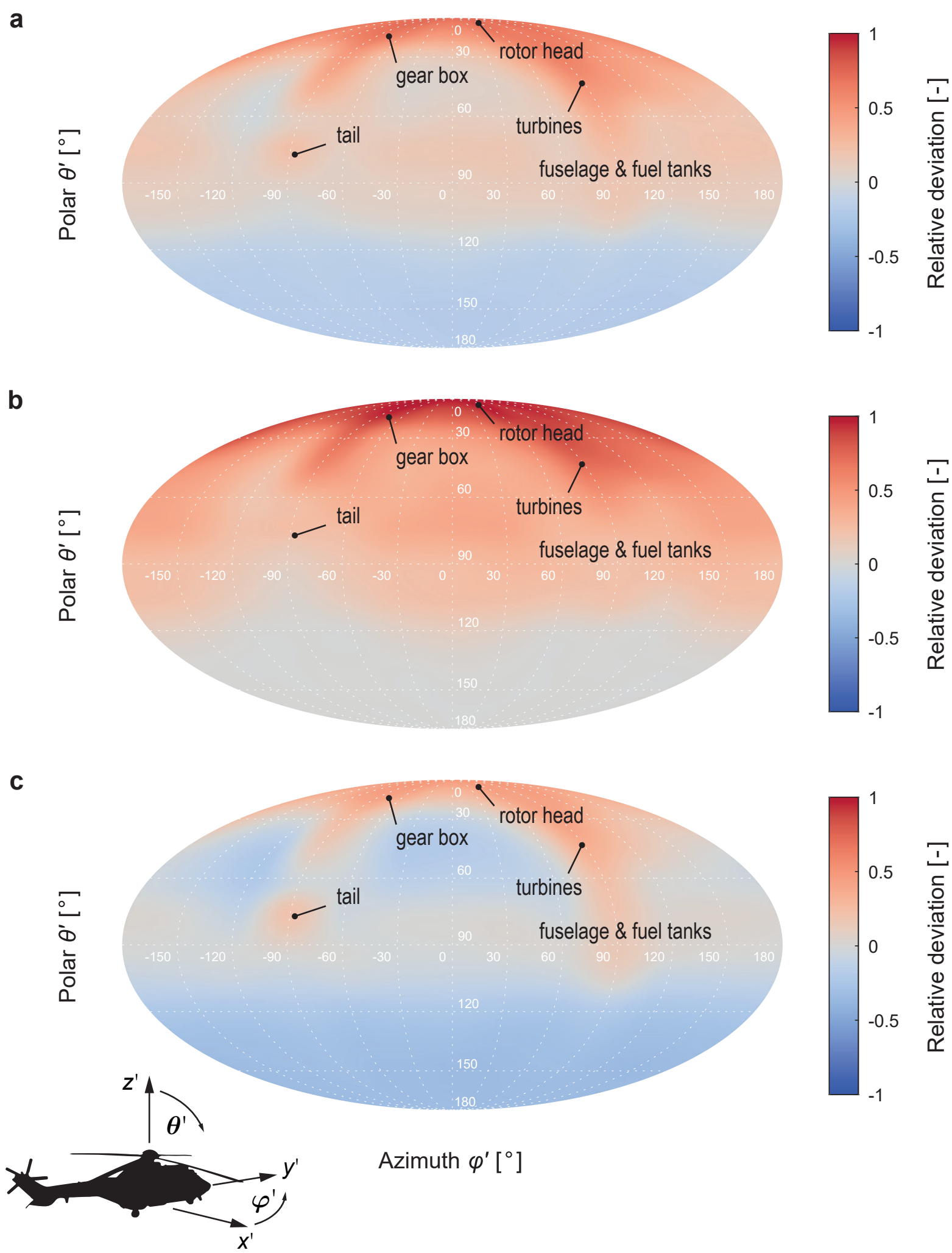

**Figure 9.11** Relative deviation in the angular dispersion between the reference DRF ($R_{\text{ref}}$) and a DRF including only the RLL spectrometer ($R_{\text{det}}$) at a photon energy of $E_\gamma = 662\,\text{keV}$. The relative deviation was computed as $(R_{\text{det}} - R_{\text{ref}})/\max(R_{\text{ref}})$ in the detector channel `#SUM` as a function of the polar angle $\theta'$ and the azimuthal angle $\varphi'$ with respect to the detector reference frame $x'$-$y'$-$z'$. Three different spectral domains were evaluated: **a** Full spectrum domain $\mathcal{D}_{\text{tot}}$. **b** Full energy peak domain $\mathcal{D}_{\text{FEP}}$. **c** Compton continuum domain $\mathcal{D}_{\text{CC}}$. The graphs are interpolated on a $1° \times 1°$ angular grid (modified Akima spline interpolation [733]) and displayed using the Mollweide projection.

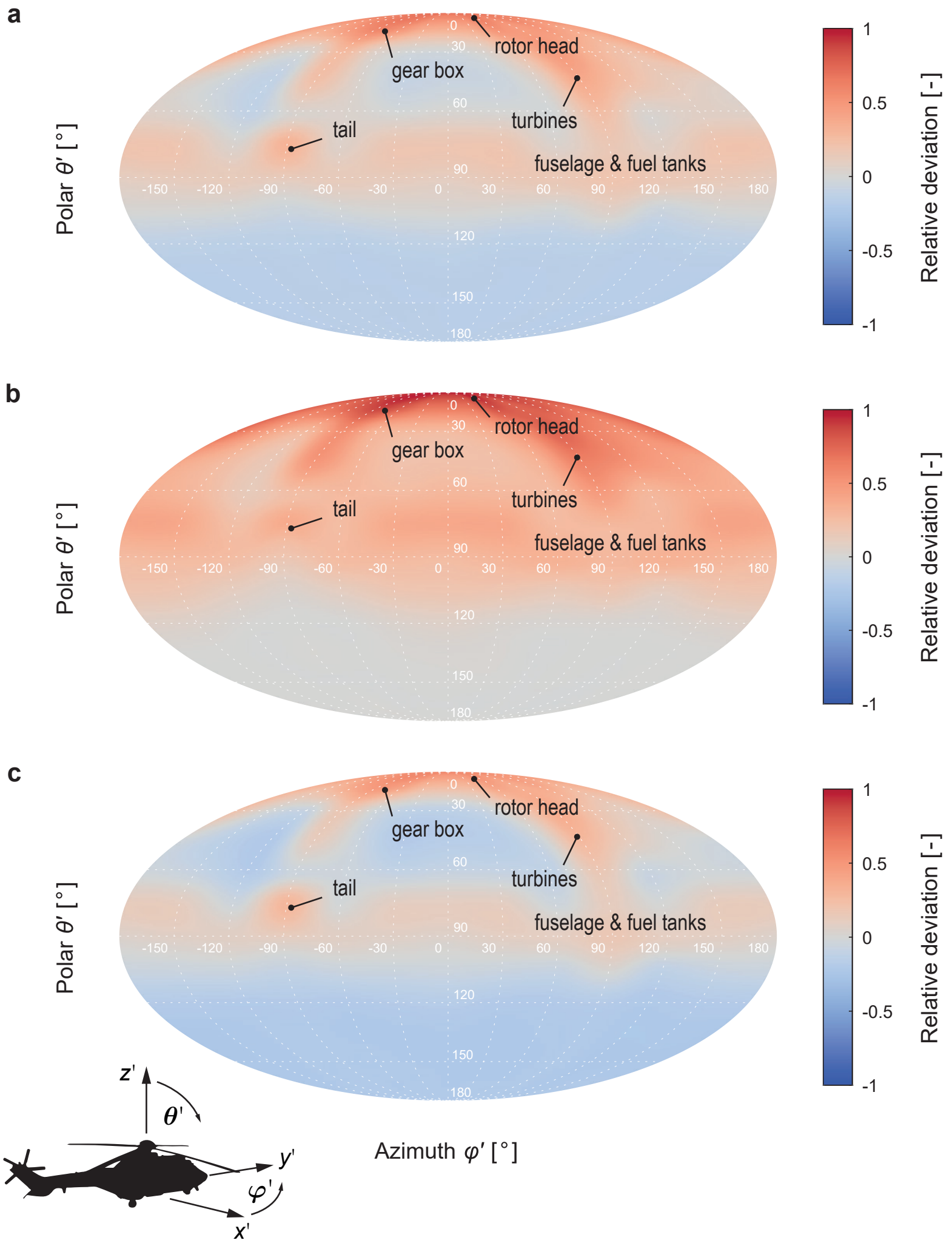

**Figure 9.12** Relative deviation in the angular dispersion between the reference DRF ($R_{\text{ref}}$) and a DRF including only the RLL spectrometer ($R_{\text{det}}$) at a photon energy of $E_\gamma = 2615\,\text{keV}$. The relative deviation was computed as $(R_{\text{det}} - R_{\text{ref}})/\max(R_{\text{ref}})$ in the detector channel `#SUM` as a function of the polar angle $\theta'$ and the azimuthal angle $\varphi'$ with respect to the detector reference frame $x'$-$y'$-$z'$. Three different spectral domains were evaluated: **a** Full spectrum domain $\mathcal{D}_{\text{tot}}$. **b** Full energy peak domain $\mathcal{D}_{\text{FEP}}$. **c** Compton continuum domain $\mathcal{D}_{\text{CC}}$. The graphs are interpolated on a $1° \times 1°$ angular grid (modified Akima spline interpolation [733]) and displayed using the Mollweide projection.

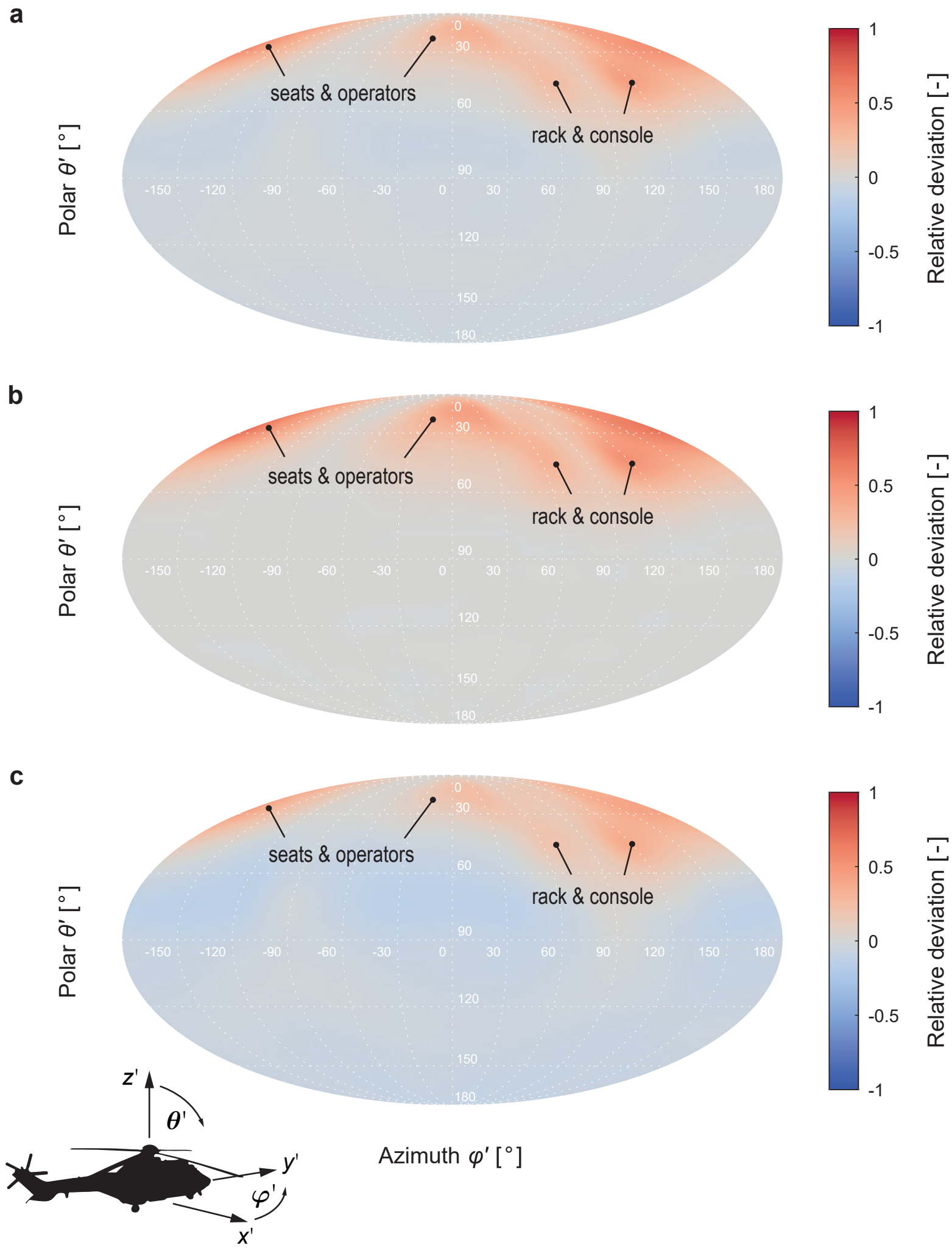

**Figure 9.13** Relative deviation in the angular dispersion between the reference DRF ($R_{\mathrm{ref}}$) and a DRF excluding the RLL supporting systems ($R_{\mathrm{RLL}}$) at a photon energy of $E_\gamma = 662\,\mathrm{keV}$. The relative deviation was computed as $(R_{\mathrm{RLL}} - R_{\mathrm{ref}})/\max(R_{\mathrm{ref}})$ in the detector channel `#SUM` as a function of the polar angle $\theta'$ and the azimuthal angle $\varphi'$ with respect to the detector reference frame $x'$-$y'$-$z'$. Three different spectral domains were evaluated: **a** Full spectrum domain $\mathcal{D}_{\mathrm{tot}}$. **b** Full energy peak domain $\mathcal{D}_{\mathrm{FEP}}$. **c** Compton continuum domain $\mathcal{D}_{\mathrm{CC}}$. The graphs are interpolated on a $1° \times 1°$ angular grid (modified Akima spline interpolation [733]) and displayed using the Mollweide projection.

To assess the effect of the RLL supporting systems, I compared the DRF for the reference mass model ($R_{\text{ref}}$) with a DRF that selectively excludes only the RLL supporting systems as well as the two operators from the full aircraft model ($R_{\text{RLL}}$). The resulting relative deviation in the angular dispersion is presented in Fig. 9.13 for a photon energy of 662 keV. Additional graphs for 88 keV and 2615 keV are provided in the Figs. B.91 and B.92.

As anticipated given their mass (~290 kg without the operators, cf. Section 5.3.2), the RLL supporting systems show a notable effect on the DRF with a maximum relative deviation reaching ~40 % in the full spectrum band, thereby explaining the observed asymmetry between the starboard and port side. It is interesting to note that the effect is less pronounced for $\theta' \leq 90°$ and $\varphi' = \pm 90°$ due to the significant attenuation of the primary photons by the rotor head, the gearbox and the turbines for these directions (cf. Fig. 8.1).

### 9.4.3 Fuel Effect

To quantify the maximum effect of the fuel on the detector response, I computed the relative deviation in the angular dispersion between the DRF with the empty fuel tanks ($R_{\text{JF0}}$) and the full fuel tanks ($R_{\text{JF100}}$) as $(R_{\text{JF0}} - R_{\text{JF100}})/\max(R_{\text{JF100}})$ for the three spectral bands $\mathcal{D}_{\text{tot}}$, $\mathcal{D}_{\text{FEP}}$ and $\mathcal{D}_{\text{CC}}$.

The resulting angular dispersion graphs for $E_\gamma = 662$ keV are presented in Fig. 9.14. As anticipated, the fuel has a significant effect on the DRF with maximum relative deviations ~30 % in the full spectrum band. The largest attenuation effect in the primary photons is observed for the directions obstructed by the longitudinal fuel tanks #1 and #2. Note that, due to the shape of the helicopter fuselage, the cross-sectional area covered by the fuel in the longitudinal fuel tanks in the $x'$-$y'$ plane continuously expands as the fuel level increases. This explains the increased effect of the fuel at $\theta' \sim 75°$. Furthermore, similar to the aircraft structure, the fuel acts as a scatterer for the primary photons, increasing the Compton continuum response by ~10 % for photons arriving at a direction with $\theta' \gtrsim 120°$ and $\theta' \lesssim 60°$. As expected, these effects are reduced at lower photon energies and increased at higher ones (cf. Figs. B.93 and B.94).

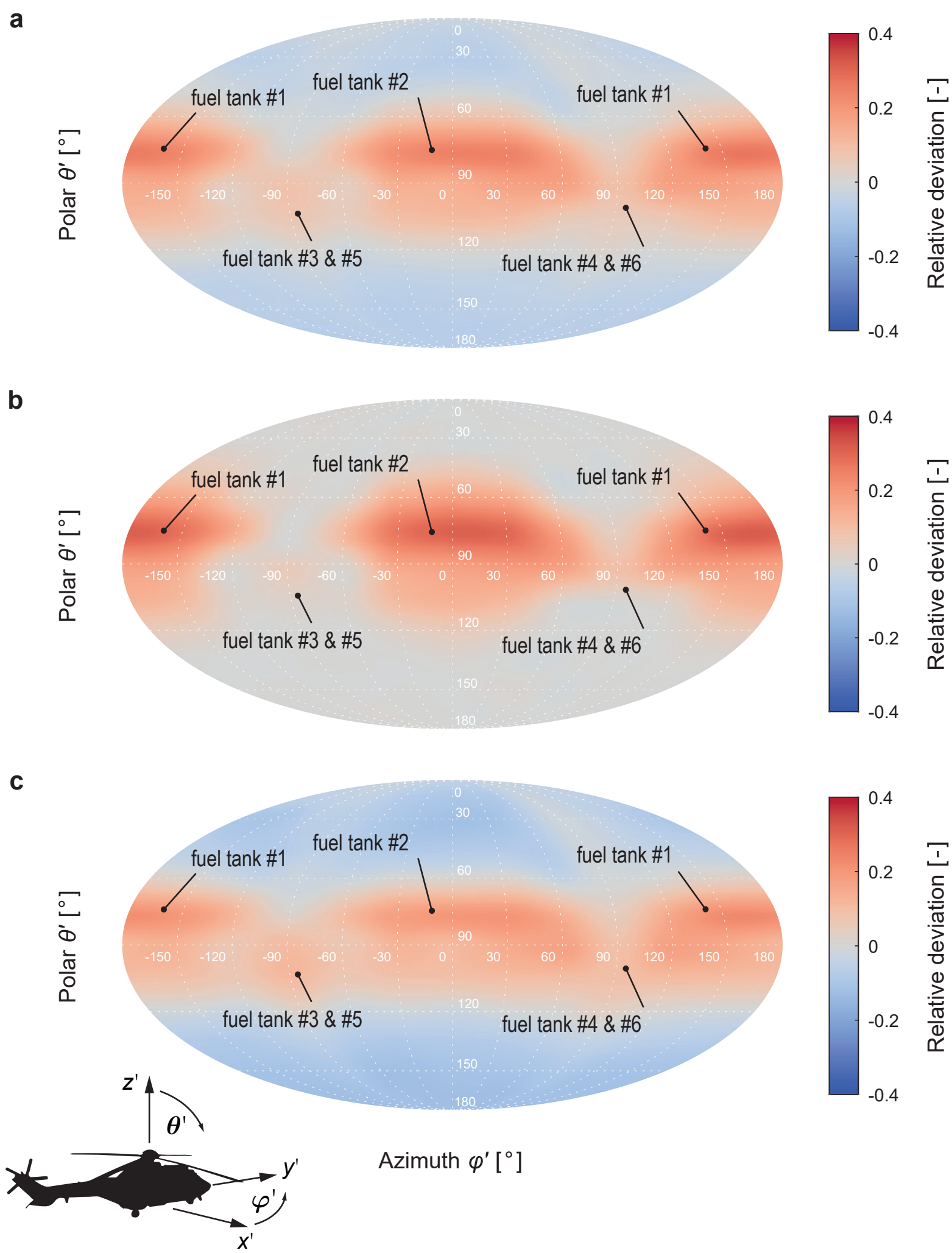

**Figure 9.14** Relative deviation in the angular dispersion between the DRF with the empty fuel tanks ($R_{\text{JF0}}$) and the full fuel tanks ($R_{\text{JF100}}$) at a photon energy of $E_\gamma = 662\,\text{keV}$. The relative deviation was computed as $(R_{\text{JF0}} - R_{\text{JF100}})/\max(R_{\text{JF100}})$ in the detector channel `#SUM` as a function of the polar angle $\theta'$ and the azimuthal angle $\varphi'$ with respect to the detector reference frame $x'$-$y'$-$z'$. Three different spectral domains were evaluated: **a** Full spectrum domain $\mathcal{D}_{\text{tot}}$. **b** Full energy peak domain $\mathcal{D}_{\text{FEP}}$. **c** Compton continuum domain $\mathcal{D}_{\text{CC}}$. The graphs are interpolated on a $1° \times 1°$ angular grid (modified Akima spline interpolation [733]) and displayed using the Mollweide projection.

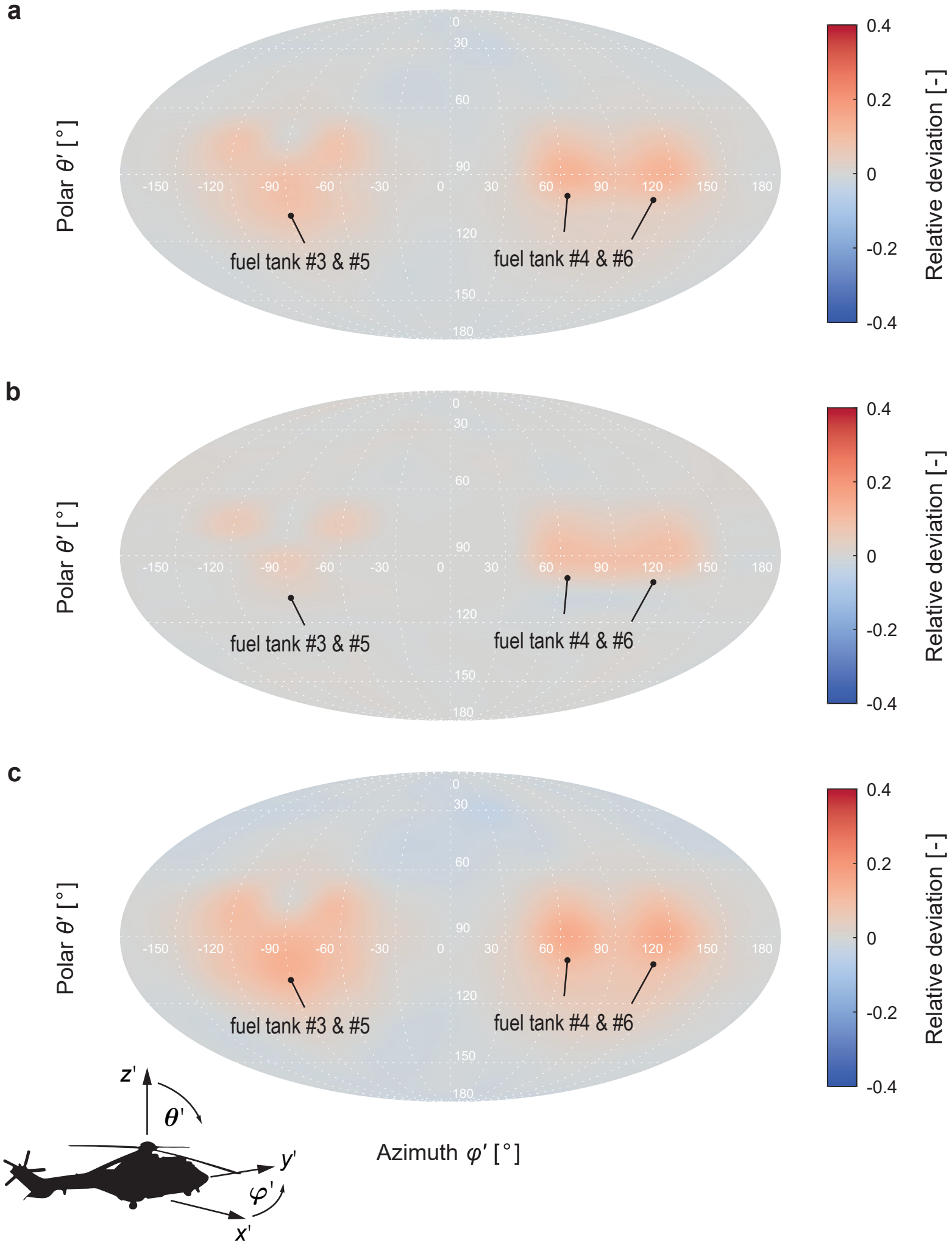

**Figure 9.15** Relative deviation in the angular dispersion between the DRF with the quarter-filled fuel tanks ($R_{\mathrm{JF25}}$) and the full fuel tanks ($R_{\mathrm{JF100}}$) at a photon energy of $E_\gamma = 662\,\mathrm{keV}$. The relative deviation was computed as $(R_{\mathrm{JF25}} - R_{\mathrm{JF100}})/\max(R_{\mathrm{JF100}})$ in the detector channel `#SUM` as a function of the polar angle $\theta'$ and the azimuthal angle $\varphi'$ with respect to the detector reference frame $x'$-$y'$-$z'$. Three different spectral domains were evaluated: **a** Full spectrum domain $\mathcal{D}_{\mathrm{tot}}$. **b** Full energy peak domain $\mathcal{D}_{\mathrm{FEP}}$. **c** Compton continuum domain $\mathcal{D}_{\mathrm{CC}}$. The graphs are interpolated on a $1° \times 1°$ angular grid (modified Akima spline interpolation [733]) and displayed using the Mollweide projection.

In the TH06 aircraft, fuel levels in the longitudinal tanks #1 and #2 begin to decrease only when the total fuel volume fraction falls below 25 %, due to the predetermined sequence of tank depletion (cf. Fig. B.81). Consequently, reduced effects in the detector response are expected for changes in the fuel volume fraction between 25 % and 100 %. To confirm this conjecture, I computed the relative deviation in the angular dispersion between the DRF with a fuel volume fraction of 25 % ($R_{JF25}$) and 100 % ($R_{JF100}$) as $(R_{JF25} - R_{JF100})/\max(R_{JF100})$ using again the three spectral bands $\mathcal{D}_{tot}$, $\mathcal{D}_{FEP}$ and $\mathcal{D}_{CC}$. The resulting angular dispersion graphs for $E_\gamma = 662$ keV are presented in Fig. 9.15. As expected, I find reduced relative deviations ≤15 % for all three spectral bands. The effects are restricted almost entirely to the directions obstructed by the fuel tanks #3 to #6, i.e. $60° \lesssim \theta' \lesssim 120°$ and $\pm 90° - 60° \lesssim \varphi' \lesssim \pm 90° + 60°$ for the starboard and port sides, respectively.[37] In particular, there is no significant increase in the Compton continuum response for photons arriving at a direction with $\theta' \gtrsim 120°$ and $\theta' \lesssim 60°$, like it was observed in Fig. 9.14. Consistent results are obtained at $E_\gamma = 88$ keV and $E_\gamma = 2615$ keV (cf. Figs. B.95 and B.96).

*37 Note also, that similar to to Fig. 9.13, the effects of the fuel changes are reduced for $\varphi' = \pm 90°$ as in this directions, the attenuation of the primary photons by the aircraft fuselage is maximized.*

### 9.4.4 Crew Effect

To assess the maximum effect of the aircraft personnel (pilots, operators, loadmaster and passengers) on the DRF, I computed the relative deviation in the angular dispersion between the DRF with no personnel on board ($R_{NP}$) and the full crew capacity ($R_{FC}$) as $(R_{NP} - R_{FC})/\max(R_{FC})$ using again the three spectral bands $\mathcal{D}_{tot}$, $\mathcal{D}_{FEP}$ and $\mathcal{D}_{CC}$.

The resulting angular dispersion graphs for $E_\gamma = 662$ keV are presented in Fig. 9.16. Similar to the fuel dynamics, I find a significant effect of the aircraft personnel on the detector response with maximum relative deviations ∼25 % in the full spectrum band. These effects are reduced at lower photon energies and increased at higher ones (cf. Figs. B.97 and B.98). Three additional trends in Fig. 9.16 are worth highlighting.

First, consistent with the results in Fig. 9.13, I observe again an asymmetry in the detector response between the starboard and port side. As discussed above, this can be attributed to the offset of the operator seats and consequently the operators themselves towards the port side (cf. Fig. 8.1).

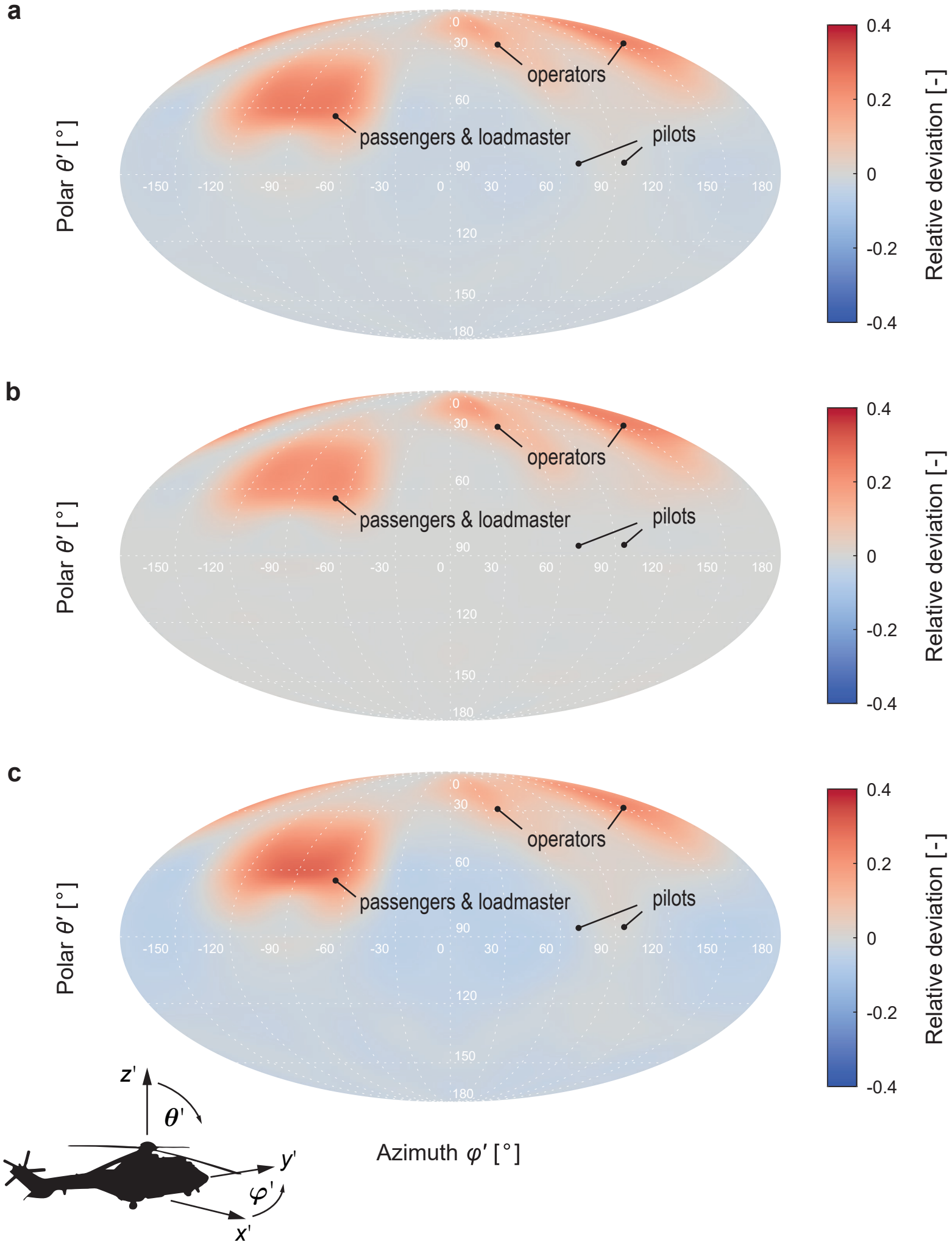

**Figure 9.16** Relative deviation in the angular dispersion between the DRF with no personnel on board ($R_{\text{NP}}$) and the full crew capacity ($R_{\text{FC}}$) at a photon energy of $E_\gamma = 662\,\text{keV}$. The relative deviation was computed as $(R_{\text{NP}} - R_{\text{FC}})/\max(R_{\text{FC}})$ in the detector channel `#SUM` as a function of the polar angle $\theta'$ and the azimuthal angle $\varphi'$ with respect to the detector reference frame $x'$-$y'$-$z'$. Three different spectral domains were evaluated: **a** Full spectrum domain $\mathcal{D}_{\text{tot}}$. **b** Full energy peak domain $\mathcal{D}_{\text{FEP}}$. **c** Compton continuum domain $\mathcal{D}_{\text{CC}}$. The graphs are interpolated on a $1° \times 1°$ angular grid (modified Akima spline interpolation [733]) and displayed using the Mollweide projection.

Second, no significant effect is found for the pilots. This can be attributed to the fact that the cockpit is separated from the cabin by high-density aircraft structures (hydraulic and electronic systems), thereby reducing changes in the detector response by the pilots to negligible levels.

Third, like in Fig. 9.15, no significant increase in the Compton continuum response due to the presence of personnel is observed. As a result, the effect of the passengers, operators and the loadmaster on the detector response is limited to photon directions with $\theta' \lesssim 75°$.

## 9.5 Conclusion

In this chapter, I presented the derivation and verification of an anisotropic DRM for the Swiss AGRS system. The DRM was developed using high-fidelity NPSMC simulations adopting the AGRS Monte Carlo model detailed in the previous chapter. By leveraging array programming, the derived DRM enables rapid computation of spectral signatures, with evaluation times of $\mathcal{O}(1)$ s per signature on a local workstation. This represents a significant reduction in the computation time by a factor of at least $\mathcal{O}(10^4)$ compared to brute-force Monte Carlo simulations performed on a computer cluster.

To assess the accuracy of the derived DRM, a detailed verification study was conducted by comparing its spectral signature predictions to those from the validated AGRS Monte Carlo model for four different source scenarios. Good agreement was found between the two models with a maximum median relative deviation <6 %. In contrast, previously proposed isotropic approaches [397] deviated by >250 %. These results highlight the significance of anisotropy in the detector response function of the Swiss AGRS system, underscoring the need for an anisotropic DRM to achieve accurate spectral signature predictions.

The DRM presented in this chapter marks not only an essential step towards full spectrum simulation-based calibration of AGRS systems, but also allows for detailed analyses of the AGRS system's spectral response to external photon fields under varying experimental conditions. I have leveraged the developed DRM to investigate the fundamental angular and spectral detector response of the AGRS system over the full $4\pi$ solid angle and for photon energies ranging from 50 keV to 3000 keV. In line with the verification results, I found a significant spectral and angular dispersion with the maximum detector response varying by a factor of up to $\mathcal{O}(10^2)$ in the angular space.

This spectral and angular dispersion can be primarily attributed to the interactions of the primary photons with the aircraft structure, as well as the positioning and arrangement of the scintillation crystals within the aircraft.

Additional DRM computations revealed that the aircraft acts as a collimator for the incoming primary photons, leading to a significant attenuation of the response for $\theta' \lesssim 120°$. However, due to the higher penetration power, this collimation effect is reduced at higher photon energies, resulting in a more homogeneous angular response and thereby a reduction in the angular dispersion.

Moreover, it was found that the aircraft does not only act as a collimator but also as a scatterer for the primary photons, increasing the Compton continuum response for photons arriving at a direction with $\theta' \gtrsim 120°$ significantly (~30 % at $E_\gamma$ = 662 keV). At higher photon energies, this effect is not restricted to $\theta' \gtrsim 120°$, but also occurs in directions where only low absorption aircraft structures obstruct the primary photons' trajectories. This indicates that the effect of the aircraft structure on the detector response is not limited to a certain subset of photon directions but instead affects the full $4\pi$ solid angle, including photon trajectories with $\theta' \gtrsim 120°$. These results underscore the need for detailed modeling of the AGRS system's aircraft structure to achieve accurate spectral signature predictions.

Given the similar design of the detector systems as well as the comparable mass and size of the adopted aircraft for other AGRS systems reviewed in Section 5.2, similar effects of the aircraft structure on the detector response can be expected across these systems. However, the magnitude of these effects may vary depending on the positioning of the scintillation crystals (cf. Table 5.2). Specifically, based on the results presented in this chapter, I anticipate a significant reduction in the full energy peak response for systems with scintillation crystals positioned inside the aircraft cabin. Conversely, there will likely be a significant increase in the Compton continuum response for systems with scintillation crystals mounted in baskets on the aircraft's exterior.

As discussed in the previous chapter, the TH06 aircraft is an inherently dynamic system with many components changing position, composition and orientation during or between survey flights. By adopting dynamic modeling, the DRM allowed me to investigate the effect of these changes on the detector response, specifically, changes in the fuel level as well as the number of crew members. The maximum effect of these changes in the full spectrum band was found to be

**Table 9.1** Summary table of the impact of individual system components on the detector response function.

| Component | Effect [%]★ | Polar angle $\theta'$ [°] |
|---|---|---|
| Aircraft• | 0 – 100 | 0 – 180 |
| Fuel | 0 – 30 | 0 – 180 |
| Crew | 0 – 25 | $\lesssim 75$ |
| RLL supporting systems∘ | 0 – 40 | $\lesssim 75$ |

★ Relative deviation in the full spectrum band $\mathcal{D}_{\text{tot}}$ with respect to the reference model defined in Section 9.4.1 at $\theta' = 180°$.

• Including the RLL supporting systems, a fuel volume fraction of 50 % and a standard crew of two pilots, two RLL operators and one loadmaster.

∘ This refers to all systems associated with the RLL spectrometer located in the aircraft cabin, i.e. the operator console, the operator seats and the equipment rack (cf. Section 5.3.2), as well as the two operators.

~30 % (fuel) and ~25 % (crew). These maximum effects are restricted to specific photon directions, i.e. $60° \lesssim \theta' \lesssim 120°$ and $\theta' \lesssim 75°$ for the fuel and the crew, respectively.[38] Furthermore, due to the specific arrangements of the fuel tanks as well as the fuel depletion sequence in the TH06 aircraft, the effect of fuel level changes on the detector response is significantly reduced for fuel volume fractions >25 %. These results are consistent with the experimental data presented in Section 8.3.2.4. A summary of the effects of the aircraft, the RLL supporting systems, the fuel and the crew on the detector response is provided in Table 9.1.

38 *Unlike the crew, the fuel significantly affects the Compton continuum response. Consequently, the overall influence of the fuel is not limited to specific photon directions.*

The developed DRM is a powerful tool that has proven to provide accurate spectral signature predictions at a fraction of the computational cost of brute-force Monte Carlo simulations. However, it is based on some simplifications and assumptions that may limit its application in certain source scenarios. First, I have limited the DRM to quantify the response of the Swiss AGRS system specifically to photons. Although the DRM can be extended to include responses to other radiation particles like neutrons or muons, given the focus on gamma-ray spectrometry in this work, this was not considered. Second, the DRM is based on the fundamental assumption of a homogeneous photon flux within the aircraft's extent. As a result, the DRM is not suitable to quantify spectral signatures for point sources in the near field of the aircraft, i.e. sources placed at a characteristic distance $\lesssim$10 m to the RLL spectrometer. This includes also radionuclide sources within the aircraft itself.

That said, considering that typical AGRS surveys are performed at a ground clearance $\geq$40 m by most of the AGRS systems reviewed in Section 5.2,[39] the DRM approach is well-suited to model the spectral signature for all relevant terrestrial gamma-ray source scenarios encountered in AGRS surveys. In conclusion, the developed DRM represents a major advancement in full spectrum simulation-based calibration of AGRS systems, enabling rapid computation of spectral signature matrices for arbitrary terrestrial source scenarios. This capability will be demonstrated in the next chapter.

39 *In Switzerland, the surveys are typically conducted at a mean ground clearance of* 90 m *(cf. Table 5.1).*

# Part IV

# Spectral Inversion

*"Inside every Non-Bayesian, there is a Bayesian struggling to get out."*

— *Dennis V. Lindley*

# Chapter 10
# Full Spectrum Bayesian Inversion

**Contents**

One of the fundamental challenges in AGRS is accurately quantifying terrestrial radionuclide activities from measured pulse-height spectra. Full spectrum analysis (FSA), due to its increased sensitivity and versatility compared to other methods, stands as the most promising approach to address this ill-posed quantification problem in AGRS. However, current FSA algorithms exhibit significant drawbacks, including limitations on the number of quantifiable sources and systematic errors in activity estimates at low count rates. Additionally, these algorithms require accurate spectral signatures for each source, which are difficult to derive using existing empirical calibration methods. As a result, FSA algorithms have not been widely adopted in AGRS.

To address these limitations, I propose a novel FSA approach that combines Bayesian inversion with numerically derived spectral signatures from high-fidelity NPSMC and DRM models presented in previous chapters.

The proposed method was validated with the Swiss AGRS system using radiation measurements of $^{133}_{56}Ba$ and $^{137}_{55}Cs$ point sources. The results showed excellent accuracy and precision in quantifying the activities of the employed point sources, with relative deviations <2 % for a measurement time of 1 s.

To further demonstrate the capabilities of the methodology, I applied the Bayesian inversion method to quantify the cosmic and radon background in the atmosphere during a series of AGRS measurement flights conducted over Lake Thun and the North Sea.

The full spectrum Bayesian inversion method presented in this chapter not only offers a powerful tool to accurately quantify arbitrarily complex terrestrial radionuclide sources at low count rates, but also provides a pathway to probe the cosmic flux and radon activity volume concentrations in the atmosphere, thereby opening new possibilities for real-time background correction and broader environmental monitoring applications of AGRS systems.

## 10.1 Introduction

In the previous four chapters, I have focused on the two first objectives of this work, namely the derivation and validation of Monte Carlo and detector response forward models to accurately compute spectral signatures of arbitrarily complex gamma-ray sources for the Swiss AGRS system. In this chapter, I will address the third objective, the implementation of these forward models in a new framework to solve one of the three fundamental tasks in AGRS: the quantification of gamma-ray sources from measured pulse-height spectra.

As analyzed in detail in Section 5.4.2, the quantification task in AGRS defined in Def. 5.1 is an ill-posed problem after Hadamard, making it challenging to solve in practice. In the Sections 5.4.3–5.4.5, I reviewed three different approaches to solving this ill-posed problem: the spectral window approach, the full spectrum approach and the peak fitting approach. Considering that the spectral window approach has reduced sensitivities and the peak fitting approach is only applicable on rare occasions (cf. Section 5.4.2), full spectrum analysis is currently the most promising quantification method in AGRS. Apart from the increased sensitivity and versatility, the full spectrum approach has the additional benefit of including nuisance sources such as radon or cosmic backgrounds in the quantification problem.

Nevertheless, as discussed in Section 5.6, full spectrum analysis (FSA) faces three fundamental unresolved problems that have prevented its widespread application in AGRS:

1. **Calibration** In contrast to other methods, FSA algorithms require accurate spectral signatures over the full spectral range of the employed AGRS spectrometer to solve the quantification problem. As discussed in Section 5.5, deriving these signatures is challenging with existing empirical calibration techniques. Simulation-based calibration approaches, leveraging high-fidelity Monte Carlo codes, offer a promising alternative to overcome these challenges. Yet, as reviewed in Section 5.5.2, the current Monte Carlo models for manned AGRS systems are still in their infancy and, due to various simplifications, suffer from large deviations between the simulated and measured pulse-height spectra, severely limiting the applicability of these models for calibration purposes in a full spectrum approach in AGRS.

2. **Inversion** Previously employed FSA algorithms in AGRS exclusively rely on MLE methods combined with Gaussian likelihood functions [16, 17, 642, 645, 646, 676]. These methods are limited to quantifying only a small number of sources, typically $\lesssim$5, and in addition suffer from significant systematic errors in the source strength vector estimates at low count rates.
3. **Background** It has been recognized in the past that variable nuisance sources such as the radon or cosmic background can introduce significant systematic errors in the quantification of terrestrial sources [120, 642, 676]. Current empirical methods for estimating these backgrounds suffer from large systematic uncertainties, as discussed in Section 5.5. In principle, the full spectrum approach has the potential to include these background sources in the quantification problem and thereby reduce the systematic errors. However, this requires an accurate estimate of the spectral signatures of these background sources, which is difficult to obtain in practice using empirical measurements [642, 676, 679]. Monte Carlo simulations could offer a solution to this problem, but they require in turn high-fidelity mass models of the entire aircraft system and advanced physics models to accurately quantify the attenuation and modulation of the complex ionizing radiation field of primary and secondary high-energy particles in the atmosphere and the aircraft. This complexity in modeling has prevented any attempts so far to quantify radon and cosmic backgrounds using Monte Carlo based FSA in AGRS.

To address these three unresolved problems, I propose a novel FSA approach by combining two key concepts:

1. **Bayesian Inversion** I propose to replace the frequentist MLE methods employed in previous studies with a Bayesian inversion approach using advanced Markov chain Monte Carlo (MCMC) algorithms. Bayesian inversion offers a natural, consistent and transparent way of combining existing information from the literature and physics constraints with empirical data to solve complex inverse problems using a solid probabilistic decision theory framework [658, 736–739]. By combining the available prior knowledge and physics constraints on gamma-ray sources reviewed in Chapter 2 with dedicated likelihood functions for low-counting statistics, Bayesian inversion can overcome the main limitations of the current frequentist MLE methods, such as the limited number of quantifiable sources and the systematic errors in the source strength estimates at low count rates. Furthermore, Bayesian inversion provides a robust and reliable way to quantify uncertainties in the source strength estimates, which is essential for the reliable interpretation of the quantification results.

2. **Numerical Calibration** I propose replacing the empirical calibration methods reviewed in Section 5.5.1 with a fully numerical approach using the high-fidelity NPSMC and DRM models presented in the previous chapters. These models have demonstrated the capability to provide accurate spectral signature predictions for arbitrary gamma-ray sources and varying experimental conditions. Furthermore, due to the detailed modeling of the entire aircraft and the integration of advanced physics models in the adopted Monte Carlo code `FLUKA` [20, 216, 281], the developed NPSMC and DRM models can be leveraged to derive spectral signatures of cosmic and radon radiation fields. Consequently, numerical calibration offers a promising alternative to overcome the limitations of the current empirical calibration methods and provides a pathway to accurately quantify not only the terrestrial gamma-ray sources but also the cosmic and radon backgrounds in the atmosphere during AGRS measurement flights.

The scope of this chapter is twofold: In the first part, I will present the derivation and validation of the proposed FSA approach, which I will refer to as full spectrum Bayesian inversion (FSBI). In the second part, I will illustrate the potential of this new method to address one of the persistent challenges in AGRS: the quantification of cosmic and radon backgrounds in the atmosphere. This will be shown using data from measurement flights over Lake Thun and the North Sea performed by the Swiss AGRS system.

## 10.2 A Bayesian Approach

Bayesian inference is a powerful statistical tool that establishes a probabilistic framework for combining existing information from the literature and physics constraints with newly obtained empirical data to solve complex ill-posed inverse problems [539, 677, 678, 737, 765]. It has been successfully applied in a wide variety of fields, including physics [658, 737, 754], engineering [755, 756], earth sciences [661, 672, 757, 758], biology and medicine [759, 760], machine learning [677, 761, 762] or social sciences [763, 764]. As reviewed in Section 7.2.4.2,

Bayesian inversion involves three main steps: (1) the definition of the likelihood function, (2) the specification of the prior distribution and (3) the computation of the posterior distribution using Bayes' theorem in Eq. 7.13. In the following paragraphs, I will discuss the implementation of these three steps within a FSA framework to solve the quantification problem in AGRS. A more general introduction to the core concepts of Bayesian inference and related probabilistic quantities is provided in Section 7.2.4.2. The general theory of FSA has been introduced earlier in Section 5.4.4.

**Likelihood Function** Consistent with the terminology introduced in Section 5.4.4, let us consider datasets $\mathcal{Y} = \{\mathbf{y}_k \in \mathbb{R}_+^{N_\mathrm{ch}\times 1} \mid k \in \mathbb{N}, k \leq N_\mathcal{Y}\}$ containing a varying number of $N_\mathcal{Y} \in \mathbb{N}_+$ pulse-height spectra $\mathbf{y}$ with $N_\mathrm{ch}$ channels recorded during an AGRS survey over local areas with already identified and localized constant sources but unknown source strengths $\boldsymbol{\xi} \in \mathbb{R}_+^{N_\mathrm{src}\times 1}$.[1] In FSA, we formally interpret the individual pulse-height spectra within a dataset $\mathcal{Y}$ as independent realizations of an underlying random vector $\boldsymbol{Y}$ following an associated conditional PDF given the corresponding sets of known experimental conditions $\mathfrak{D} = \{\mathfrak{d}_k \in \mathbb{R}^{N_\mathfrak{d}\times 1} \mid k \in \mathbb{N}, k \leq N_\mathcal{Y}\}$:

$$\boldsymbol{Y} \sim \pi(\mathbf{y} \mid \mathbf{x}, \mathfrak{d}) \tag{10.1}$$

where:

| | | |
|---|---|---|
| $\mathfrak{d}$ | experimental condition | $[\mathfrak{d}]$ |
| $\mathbf{x}$ | model parameter vector | $[x]$ |
| $\mathbf{y}$ | data vector | $[y]$ |

and with $\mathbf{x} = (\mathbf{x}_\mathcal{M}, \mathbf{x}_\nu)^\top$ encoding the model parameters of interest $\mathbf{x}_\mathcal{M}$, i.e. the source strength vector $\mathbf{x}_\mathcal{M} = \boldsymbol{\xi}$, as well as additional nuisance parameters $\mathbf{x}_\nu$ that may contain unknown experimental conditions or meta-parameters for the probabilistic model.

As noted already in the introduction to this chapter, because of its simplicity and the availability of closed-form solutions, the PDF in Eq. 10.1 is typically assumed to be Gaussian in the current frequentist MLE methods employed in AGRS [16, 17, 642, 645, 646, 676] (cf. Eqs. 5.11a and 5.11b). However, as discussed by Knoll [30], Gilmore [296], and Kirkpatrick et al. [827], the Gaussian probabilistic model approximates the underlying count statistics of the pulse-height spectra only in the limit of large counts per channel $C \rightarrow \infty$.[2]

1 *In other words, we assume that the source strength vector $\boldsymbol{\xi}$ is time independent for all recorded pulse-height spectra within $\mathcal{Y}$. The methodology introduced here can be easily generalized to time-varying sources by repeating the full spectrum approach for a series of datasets $\mathcal{Y}$, each representing a time instance with constant source strength vector $\boldsymbol{\xi}$.*

2 *As a rule of thumb, for $C \gtrsim 20$, a Gaussian probabilistic model starts to provide less biased results and may be considered to approximate the underlying count statistics [30].*

Considering the short sampling times of $t_s \sim 1\,\text{s}$ and low count rates (cf. Section 5.2), the number of counts is typically $C < 20$ for the majority of the pulse-height channels in AGRS (cf. Fig. 5.3).

To prevent systematic errors for low-count data in AGRS, Poisson probabilistic models[3] were proposed in the past [684, 828]. These models have also been successfully applied in related fields such as particle physics [829–831], astrophysics [832, 833], nuclear security [643, 644, 834] and gamma-ray imaging [835, 836], among others. In contrast to the Gaussian probabilistic model, the Poisson model provides a discrete statistical description of the number of counts $C \in \mathbb{N}$ in a pulse-height channel with the PDF defined as [30]:[4]

$$\pi(C \mid \mathbf{x}, \boldsymbol{\mathfrak{d}}) = \frac{\lambda_{\text{Pois}}^{C} \exp\left(-\lambda_{\text{Pois}}\right)}{\Gamma(C+1)} \tag{10.2}$$

where:

| | |
|---|---|
| $C$ | number of counts in a pulse-height channel |
| $\lambda_{\text{Pois}}$ | mean parameter |

and with $\lambda_{\text{Pois}} \in \mathbb{R}_+$, while $\Gamma(\cdot)$ denotes the gamma function.

However, it has been recognized that Poisson probabilistic models often underestimate the dispersion in count data, i.e. the variance in the obtained data is larger than predicted by the Poisson model [681, 837, 838]. This discrepancy, known as overdispersion, arises because the experimental conditions $\boldsymbol{\mathfrak{d}}$ typically fluctuate during the counting process, introducing additional variability in the data. To address this, the Poisson model can be generalized to a gamma-Poisson mixture model, where the Poisson mean parameter ($\lambda_{\text{Pois}}$) is not fixed but instead follows a gamma distribution. The PDF of this generalized model can be formulated as:[5]

$$\pi(C \mid \mathbf{x}, \boldsymbol{\mathfrak{d}}) = \frac{\Gamma\left(C + \frac{1}{\alpha_{\text{NB}}}\right)}{\Gamma\left(\frac{1}{\alpha_{\text{NB}}}\right)\Gamma(C+1)} \left(\frac{1}{1+\alpha_{\text{NB}}\lambda_{\text{NB}}}\right)^{\frac{1}{\alpha_{\text{NB}}}} \left(\frac{\alpha_{\text{NB}}\lambda_{\text{NB}}}{1+\alpha_{\text{NB}}\lambda_{\text{NB}}}\right)^{C} \tag{10.3}$$

where:

| | |
|---|---|
| $\alpha_{\text{NB}}$ | dispersion parameter |
| $\lambda_{\text{NB}}$ | mean parameter |

In contrast to the Poisson distribution, the gamma-Poisson mixture model introduces a second parameter $\alpha_{\text{NB}} \in \mathbb{R}_+$ in addition to the

3 *Named after Siméon Denis Poisson (*1781, †1840), a French mathematician and physicist known for his significant contributions to the fields of probability theory, partial differential equations, electricity and magnetism, optics, fluid dynamics and mechanics, among others. Poisson is also one of the 72 French scientists honored by having their names inscribed on the Eiffel Tower in Paris.*

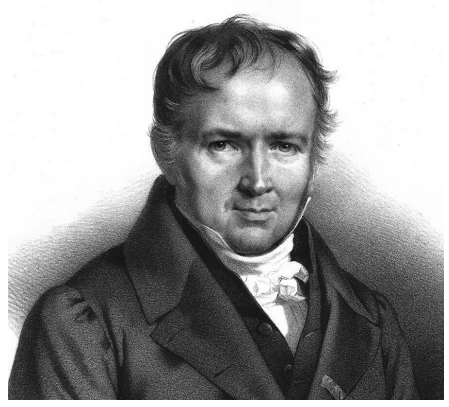

Siméon D. Poisson
Ⓢ François-Séraphin Delpech

4 *Please note that I replaced the more commonly used factorial expression in the denominator, $C!$, with the equivalent expression, $\Gamma(C+1)$, using the gamma function to facilitate interpretation alongside the gamma-Poisson mixture distribution, which I will introduce in the next paragraph.*

5 *Note that there are different conventions in the literature on the parametrization of the gamma-Poisson mixture distribution. I adopt here the parametrization used by Hünnefeld et al. [681] and Salinas et al. [682] because of its preferable properties in inference [837].*

mean $\lambda_{\text{NB}} \in \mathbb{R}_+$. This second parameter, commonly referred to as the dispersion, shape or clustering parameter [681, 682, 837], controls the variance $\sigma^2$ of the gamma-Poisson mixture distribution and thus the overdispersion with $\sigma^2 = \lambda_{\text{NB}} + \alpha_{\text{NB}}\lambda_{\text{NB}}^2 \in \mathbb{R}_+$ [682].[6] In the special case of $\alpha_{\text{NB}} \in \mathbb{N}$, the gamma-Poisson mixture model is also referred to as the negative binomial distribution [772, 839]. The gamma-Poisson mixture model has been successfully applied in the past to accurately quantify overdispersed count data in various disciplines, including neutrino physics [681], machine learning [682], positron emission tomography [840], weather forecast [841–843], ecology [844] and disease control [837, 845], among others.

*6 For $\alpha_{\text{NB}} \to 0$, the gamma-Poisson mixture model approaches the Poisson distribution and $\alpha_{\text{NB}} = 1$ is equivalent to the geometric distribution [837].*

Considering the inherent variability in the experimental conditions during the surveys, the gamma-Poisson mixture model is a natural choice to describe the statistical variation in pulse-height spectra obtained in AGRS. To integrate the gamma-Poisson mixture model in the Bayesian inversion framework, I replaced the Gaussian likelihood function in Eq. 5.11a with a gamma-Poisson mixture likelihood function by inserting Eq. 10.3 in Eq. 5.8 and exchanging $\lambda_{\text{NB}}$ with the forward model $\mathcal{M}(\boldsymbol{\xi}, \mathbf{\mathfrak{d}})$[7], resulting in:

*7 Here, I implicitly assume that all relevant experimental conditions are known.*

$$\mathcal{L}(\mathbf{x}; \mathcal{Y}, \mathfrak{D}) = \prod_{k=1}^{N_{\mathcal{Y}}} \prod_{j=1}^{N_{\text{ch}}} \frac{\Gamma\left(C_{k,j} + \frac{1}{\alpha_{\text{NB}}}\right)}{\Gamma\left(\frac{1}{\alpha_{\text{NB}}}\right)\Gamma\left(C_{k,j}+1\right)} \left(\frac{1}{1 + \alpha_{\text{NB}}\mathcal{M}(\boldsymbol{\xi}, \mathbf{\mathfrak{d}}_k)}\right)^{\frac{1}{\alpha_{\text{NB}}}} \cdot \left(\frac{1}{1 + [\alpha_{\text{NB}}\mathcal{M}(\boldsymbol{\xi}, \mathbf{\mathfrak{d}}_k)]^{-1}}\right)^{C_{k,j}} \tag{10.4}$$

with:

| | | |
|---|---|---|
| $\mathfrak{D}$ | set of experimental conditions | [$\mathfrak{d}$] |
| $\mathcal{M}$ | forward model | |
| $N_{\text{ch}}$ | number of pulse-height channels | |
| $N_{\mathcal{Y}}$ | number of pulse-height spectra within the dataset | |
| $\mathcal{Y}$ | dataset | |
| $\xi$ | source strength vector | [$\xi$] |

and where $C_{k,j}$ denotes the number of counts in channel $j$ of the pulse-height spectrum $k$ in $\mathcal{Y}$. Following the Bayesian paradigm, I integrated the dispersion parameter $\alpha_{\text{NB}}$ as a nuisance parameter in the Bayesian inversion framework, i.e. $\mathbf{x} = (\xi, \alpha_{\text{NB}})^\top$. As the forward model $\mathcal{M}(\boldsymbol{\xi}, \mathfrak{d})$ in Eq. 10.4 depends on the specific AGRS survey, I will specify it in more detail in the following sections.

Because the gamma function terms in Eq. 10.4 grow rapidly for moderately large arguments,[8] special care is required for numerical evaluation. To avoid under- or overflow, it is advisable to compute the logarithm of the likelihood function, known as the log-likelihood, instead of directly evaluating Eq. 10.4 [766]:[9]

8 *As an example,* $\Gamma(20)$ *is already exceeding* $10^{17}$.

9 *Modern numerical codes offer specialized functions for computing* $\log\Gamma(\cdot)$*, e.g.* `gammaln` *in MATLAB and* `scipy.special.gammaln` *in Python, to prevent under- or overflow.*

$$\begin{aligned}\log \mathcal{L}(\mathbf{x}; \mathcal{Y}, \mathfrak{D}) = \sum_{k=1}^{N_{\mathcal{Y}}} \sum_{j=1}^{N_{\mathrm{ch}}} \log\Gamma\left(C_{k,j} + \frac{1}{\alpha_{\mathrm{NB}}}\right) - \log\Gamma\left(\frac{1}{\alpha_{\mathrm{NB}}}\right) \\ - \log\Gamma\left(C_{k,j} + 1\right) - \frac{1}{\alpha_{\mathrm{NB}}} \log\left(1 + \alpha_{\mathrm{NB}} \mathcal{M}(\boldsymbol{\xi}, \mathfrak{d}_k)\right) \\ - C_{k,j} \log\left(1 + \frac{1}{\alpha_{\mathrm{NB}} \mathcal{M}(\boldsymbol{\xi}, \mathfrak{d}_k)}\right) \end{aligned} \tag{10.5}$$

**Prior Distribution** As reviewed in Section 7.2.4.2, the prior distribution quantifies the level of information existing on the parameters before the set of pulse-height spectra $\mathcal{Y}$ is considered. Possible sources of such information include physics constraints, expert knowledge and results from previous measurements, either conducted by the investigators themselves or retrieved from the literature.

In this chapter, I adopted weakly informative statistically independent marginal priors using Eq. 7.24 for all parameters $\mathbf{x}$. For the model parameters of interest, i.e. the source strengths $\boldsymbol{\xi}$, I defined the specific marginal prior distributions based on the available information compiled in Chapter 2. As the prior distributions depend on the AGRS survey as well as the adopted forward model, I will specify them in more detail in the following sections.

**Posterior Distribution** The core of Bayesian inversion is the posterior distribution, a conditional PDF which combines the likelihood function $\mathcal{L}(\mathbf{x}; \mathcal{Y}, \mathfrak{D})$ with the prior distribution $\pi(\mathbf{x})$ via Bayes' theorem, as detailed in Eq. 7.13. This posterior distribution provides a complete probabilistic description of the model parameters $\mathbf{x}$ given the newly obtained data $\mathcal{Y}$ and experimental conditions $\mathfrak{D}$. As such it represents a comprehensive probabilistic solution to the quantification problem in AGRS, as defined in Def. 5.1, encoding the statistically consistent update of our prior beliefs in the model parameters based on the newly obtained data.

For the gamma-Poisson likelihood and prior distributions adopted in this chapter, no closed-form expression for the posterior distribution exists [738]. As a result, numerical methods are required to compute the posterior distribution in Eq. 7.13. Given its simplicity, versatility and robustness for complex posterior distributions, I selected a state-of-the-art affine invariant ensemble sampler (AIES) MCMC algorithm [735] implemented in the `UQLab` code [753] to perform all Bayesian computations in this chapter, combined with a custom `MATLAB` script to evaluate the log-likelihood function in Eq. 10.5.[10] I will provide additional details on the performed MCMC simulations in the following sections.

10 *A detailed discussion of the MCMC algorithms is beyond the scope of this work. For a comprehensive overview, I recommend the excellent textbooks by Brooks et al. [801] and Liu [803] as well as Chapter 1 of Wagner's dissertation [267]. Additional information on numerical algorithms for Bayesian inference is also provided in Section 7.2.4.2.*

## 10.3 Validation

The scope of the validation study presented in this section is to demonstrate the capability of the FSBI method to accurately quantify the source strengths of multiple gamma-ray sources both for high and low-count data in a source-detector configuration that is both reproducible and representative of a typical AGRS survey in Switzerland. For that purpose, I applied FSBI on a selected subset of the experimental datasets obtained during the ARM22 validation campaign discussed in Chapter 8. By comparing the derived source strengths with known reference values obtained from calibration certificates and in-situ gamma-ray spectrometry measurements, the accuracy and precision of the proposed FSBI method presented above was assessed. In the following subsections, I will detail the selected experimental datasets, the inversion setup and the validation results obtained with the FSBI method.

### 10.3.1 Radiation Measurements

To validate the FSBI method for a reproducible source-detector configuration that is most similar to a AGRS survey in Switzerland, I used datasets obtained from the hover flight radiation measurements conducted at the Thun military training ground (46.753°N, 7.596°E) on the 2022-06-16 (cf. Section 8.3.2.1) as part of the ARM22 validation campaign. In these radiation measurements, the Swiss AGRS system was operated in a hover flight mode for ~5 min with the TH06 helicopter positioned at a ground clearance of ~90 m above two separately deployed sealed radionuclide point sources, a $^{137}_{55}Cs$ source and a $^{133}_{56}Ba$ source, which were positioned on the ground and aligned with the aircraft principal $z'$ axis (cf. Figs. 8.1 and 8.8).[11]

**Table 10.1** Overview of the experimental datasets considered for the validation of the FSBI method. These datasets were derived from two hover flight measurements conducted during the ARM22 validation campaign. More information on these measurements is provided in Section 8.3.2.1.

| Identifier | Point source | Ground clearance [m]★ | Gross measurement live time [s]• |
|---|---|---|---|
| Cs_I | $^{137}_{55}Cs$ | 88 | 1 |
| Cs_II | $^{137}_{55}Cs$ | 92 | 5 |
| Cs_III | $^{137}_{55}Cs$ | 92 | 232 |
| Ba_I | $^{133}_{56}Ba$ | 93 | 1 |
| Ba_II | $^{133}_{56}Ba$ | 92 | 5 |
| Ba_III | $^{133}_{56}Ba$ | 91 | 296 |

★ Mean ground clearance $h_{air}$ rounded to two significant digits.
• Gross measurement live time $t_{gr}$ rounded to seconds.

To investigate the effect of data sparsity on the Bayesian inversion results, I summed the counts from individual pulse-height spectra in the detector channel #SUM recorded with a 1 s sampling time into three different gross count spectra $\mathbf{C}_{gr} \in \mathbb{N}^{N_{ch} \times 1}$.[12] These spectra correspond to gross measurement live times of 1 s, 5 s and the full measurement time of ~5 min. This was repeated for both the $^{137}_{55}Cs$ and the $^{133}_{56}Ba$ source, resulting in a total of six different datasets. Each dataset is characterized by the gross count spectrum $\mathbf{C}_{gr}$, the gross measurement live time $t_{gr}$ and the mean experimental conditions $\mathfrak{d}$ under which the aggregated pulse-height spectra in $\mathbf{C}_{gr}$ were recorded. An overview of the datasets is provided in Table 10.1. More information on the measurement setup, data acquisition and data processing can be found in Section 8.3.2.1.

11 *Here, I only considered the hover flights conducted at ~90 m ground clearance, which is the regular ground clearance for AGRS surveys in Switzerland (cf. Table 5.1).*

12 *As for the previous two chapters, I limit the discussion here to the detector channel #SUM.*

### 10.3.2 Inversion Setup

The general FSBI methodology was outlined already in Section 10.2. In the following subsections, I focus on the specific setup of the FSBI method for the selected measurement datasets listed in Table 10.1. Specifically, I will provide details on the forward model and its calibration as well as the Bayesian inversion computations performed for the validation study.

#### 10.3.2.1 Forward Model

The forward model $\mathcal{M}(\boldsymbol{\xi}, \boldsymbol{\mathfrak{d}})$ employed in Eq. 10.5 is the core quantity in FSBI and incorporates the complete physical model to predict the expected mean number of counts $C$ in each pulse-height channel as a function of the source strengths $\boldsymbol{\xi}$ and the experimental conditions $\boldsymbol{\mathfrak{d}}$ for the given radiation measurement.

As discussed in detail in Section 5.4.1, in this book, I assume that both the number/type and the location of all high-energy photon sources affecting the detector response of the AGRS system are known. Under these assumptions, we can write the forward model for the hover flight measurements as a simple linear combination of the spectral signatures encoded in the spectral signature matrix $\boldsymbol{\mathcal{M}} \in \mathbb{R}_+^{N_{\text{ch}} \times N_{\text{src}}}$ scaled by their respective source strengths $\boldsymbol{\xi} \in \mathbb{R}_+^{N_{\text{src}} \times 1}$ (cf. Eq. 5.3):

$$\mathbf{C} = \left[\boldsymbol{\mathcal{M}}\left(\boldsymbol{\mathfrak{d}}\right)\boldsymbol{\xi} + \mathbf{c}_{\text{b}}\right] t_{\text{gr}} \tag{10.6}$$

where:

| | | |
|---|---|---|
| $\mathbf{C}$ | count vector | |
| $\mathbf{c}_{\text{b}}$ | background count rate vector | $\text{s}^{-1}$ |
| $\boldsymbol{\mathfrak{d}}$ | experimental condition | $[\boldsymbol{\mathfrak{d}}]$ |
| $\boldsymbol{\mathcal{M}}$ | spectral signature matrix | $\text{s}^{-1}\,[\xi]^{-1}$ |
| $t_{\text{gr}}$ | gross measurement live time | s |
| $\boldsymbol{\xi}$ | source strength vector | $[\xi]$ |

Note that in this validation study, I distinguish between sources of interest, for which reference values are available and nuisance sources, for which no reference information is available. The sources of interest are explicitly considered in the source strength vector $\boldsymbol{\xi}$, while the nuisance sources are represented as a fixed background count vector $\mathbf{c}_{\text{b}}$ scaled by the gross measurement time $t_{\text{gr}}$.

The following sources of interest were explicitly considered in the forward model for the hover flight measurements:

1. **Point Sources** Sealed $^{133}_{56}\text{Ba}$ and $^{137}_{55}\text{Cs}$ point sources. Reference activity values for these two sources could be derived from calibration certificates (cf. Section 8.3.2.1). The source strengths $\xi_{\text{Ba-133}}$ and $\xi_{\text{Cs-137}}$ were parametrized by the associated absolute activities $\mathcal{A}$ in Bq (cf. Eq. 2.9).

2. **Terrestrial Primordial Sources** Terrestrial primordial radionuclides $^{40}_{19}K$, $^{232}_{90}Th$ and $^{238}_{92}U$ with all associated progeny assumed to be in secular equilibrium (cf. Tables C.1 and C.2) and uniformly distributed in the soil of the Thun military training ground. I will refer to these sources as $K_{nat}$, $Th_{nat}$ and $U_{nat}$ in this section. Average activity mass concentration estimates for $K_{nat}$, $Th_{nat}$ and $U_{nat}$ were available from a series of in-situ spectrometry measurements conducted during the ARM20 Exercise [119]. The source strengths $\xi_{\text{K-nat}}$, $\xi_{\text{Th-nat}}$ and $\xi_{\text{U-nat}}$ were parametrized by the associated activity mass concentrations $a_m$ (specified for soil in natural condition) in Bq kg$^{-1}$ (cf. Eq. 2.11a).

All other sources were treated as nuisance sources, and their combined contribution to the detector response was modeled by the background count vector $\mathbf{c}_b$ in Eq. 10.6. These nuisance sources are assumed to be fixed and encompass the intrinsic, the cosmic and the radon backgrounds discussed in Section 5.4.3. However, as the radon background can vary considerably over short temporal and spatial scales (cf. Section 2.1.3.2), I included an additional radon source term in the forward model to explicitly account for changes in the radon background:

3. **Radon** Atmospheric radon $^{222}_{86}Rn$ with the progeny up to but not including $^{210}_{82}Pb$ assumed to be in secular equilibrium (cf. Table C.1) and uniformly distributed in the lower atmosphere. The source strength $\Delta\xi_{\text{Rn-222}}$ was parametrized by the associated activity volume concentration $a_v$ in Bq m$^{-3}$ (cf. Eq. 2.11b).[13] I will refer to this source as $\Delta^{222}_{86}Rn$ in this section.

13 *Note that I denote the source strength of this radon term with a $\Delta$ to emphasize that it is a correction term relative to the combined background encoded in $\mathbf{c}_b$, rather than a measure of the absolute radon source strength.*

In summary, five different sources were considered in the forward model ($N_{src}$=5), i.e. the deployed point source (either $^{133}_{56}Ba$ or $^{137}_{55}Cs$), the three terrestrial primordial radionuclides in the soil ($K_{nat}$, $Th_{nat}$, $U_{nat}$) and the radon bias term ($\Delta^{222}_{86}Rn$).

#### 10.3.2.2 Calibration

The forward model in Eq. 10.6 required the derivation of the spectral signature matrix $\boldsymbol{\mathcal{M}}$ as well as the background count vector $\mathbf{c}_b$.

The background count vector $\mathbf{c}_b$ was determined experimentally using standard background flights over Lake Thun performed at orthometric heights equivalent to the ones at which the hover flight

14 *A general introduction to the experimental measurement of background radiation in AGRS are provided by Erdi-Krausz et al. [8] and Smith et al. [9]* measurements were performed (cf. Fig. B.99).[14] Since these background flights were conducted at the same orthometric height close to hover flight locations (<10 km) and with a small temporal separation (<1 h), the background measurements were considered to be representative for the cosmic and intrinsic background radiation during the hover flights. Temporal and spatial variation in the radon background was explicitly accounted for by the additional radon source term $\Delta\xi_{\text{Rn-222}}$ in the forward model discussed above.

The spectral signatures for each of the five sources encoded in the spectral signature matrix $\boldsymbol{\mathcal{M}}$ were derived using the DRM method discussed in Chapter 9.[15] Specifically, I derived a DRF for each of two source cases listed in Table 10.1 using the methodology detailed in Section 9.2.2.[16] The double differential photon flux signatures were computed with the methods discussed in Section 9.2.1. The environmental mass model was adapted from Section 8.3.2.2.[17] Based on the assumptions on the source locations and distributions stated above, the terrestrial primordial radionuclides and the radon source were modeled as uniformly distributed volume sources within the soil and the air, respectively. All progeny for $\text{Th}_{\text{nat}}$ and $\text{U}_{\text{nat}}$ are assumed to be in secular equilibrium (cf. Tables C.1 and C.2). For the radon source, only the progeny up to but not including ${}^{210}_{82}\text{Pb}$ were considered (cf. Table C.1). The mass model of the point sources was transferred from Section 8.3.2.2 without changes.[18] The resulting spectral signatures scaled with the derived source strengths will be presented in Section 10.3.3.

15 *Using mean experimental conditions* $\mathfrak{d}$ *during the individual measurements. This includes crew composition, fuel levels in the individual fuel tanks, aircraft orientation, ground clearance and atmospheric condition (temperature, pressure and humidity) (cf. Section 8.3.2.1).*

16 *In other words, I assumed a constant fuel level among the three datasets associated with a given source.*

17 *To account for the extension of the volume sources, the radius of the simulation volume was increased from* 200 m *to* 3 km *based on the findings obtained in Section 3.2.1 and considering the photon energies (cf. Fig. 2.2) and resulting mean free paths of the primary photons in the atmosphere (cf. Fig. 3.6).*

18 *It is worth adding that I exploited the translation invariance in the geometry for the terrestrial primordial radionuclides and the radon source using the reciprocal transform method [616, 846–852].*

#### 10.3.2.3 Bayesian Computations

The quantification problem is solved for each of the six datasets listed in Table 10.1 independently using FSBI by combining the likelihood function $\mathcal{L}(\mathbf{x}; \mathbf{C}_{\text{gr}}, \mathfrak{d})$ and the prior distribution $\pi(\mathbf{x})$ via Bayes' theorem (Eq. 7.13) to compute the posterior distribution $\pi(\mathbf{x} \mid \mathbf{C}_{\text{gr}}, \mathfrak{d})$.[19] The likelihood function is fully defined by Eq. 10.4, the calibrated forward model in Eq. 10.6 and the model parameter vector $\mathbf{x} = (\boldsymbol{\xi}, \alpha_{\text{NB}})^\top$. To prevent overly restrictive assumptions on the prior distribution, I adopted weakly informative statistically independent marginal priors for all model parameters $\mathbf{x}$. A full list of all marginal prior distributions is provided in Table C.19.

As noted already in Section 10.2, I used an AIES MCMC algorithm [735] implemented in the `UQLab` code [753] to compute the posterior distribution.[20] The MCMC algorithm was executed with 24 paral-

19 *Note that, since each dataset defined in Section 10.3.1 contains only a single pulse-height spectrum* ($\mathbf{C}_{\text{gr}}$)*, I specified the dataset* $\mathcal{Y}$ *as* $\mathbf{C}_{\text{gr}}$ *and the set of experimental conditions* $\mathfrak{D}$ *as* $\mathfrak{d}$.

20 *It is worth adding that to account for the systematic uncertainties at low spectral energies observed in Chapters 7 and 8, the FSBI was limited to spectral energies* $E' \geq 50$ keV.

lel chains, each running $5 \times 10^3$ iterations. The burn-in phase was set to 20 % resulting in a total of $N_{\hat{\mathcal{X}}} = 9.6 \times 10^4$ posterior samples. The convergence and precision of the MCMC simulations were carefully assessed using standard diagnostics tools [738, 814], showing a potential scale reduction factor $\hat{R} < 1.02$ and effective sample size ESS > 400 across all MCMC runs. Additional trace and convergence plots for the individual parameters and point estimates are provided in the Figs. B.100–B.105.

### 10.3.3 Results

In this section, I present the validation results obtained with the FSBI method for the selected hover flight datasets listed in Table 10.1. I will provide a detailed discussion of the posterior distribution estimates for the source strengths of the deployed point sources and the terrestrial primordial radionuclides. The results will be compared with the known reference values obtained from calibration certificates and in-situ gamma-ray spectrometry measurements.

A summary of the Bayesian inversion results for each of the six analyzed datasets is provided in Table 10.2 using maximum a posteriori (MAP) point estimators with uncertainties indicated as central credible intervals with a probability mass of 95 % alongside the reference values for the source strengths of the deployed point sources and the terrestrial primordial radionuclides. Full posterior point and dispersion estimate results for all six datasets are provided in the Tables C.21 and C.22 together with univariate and bivariate marginal posterior distributions displayed in the Figs. B.106–B.111.

Comparing the predicted MAP with the reference activities for the two point sources $^{133}_{56}$Ba and $^{137}_{55}$Cs in Table 10.2, I find excellent agreement for both sources with relative deviations <2 % for the low-count datasets `Cs_I`, `Cs_II`, `Ba_I` and `Ba_II`. The high-count dataset `Ba_III` shows a moderately increased relative deviation of 5.7 %, but this remains statistically insignificant within the credible intervals.

This increased deviation is likely due to the greater variability in experimental conditions over the extended measurement time of ~5 min compared to the shorter measurement times of 1 s and 5 s for the low-count datasets. This is further supported by the posterior results for the dispersion parameter of the gamma-Poisson mixture distribution, which is highest with $\alpha_{\text{NB}} = 3.1^{+0.6}_{-0.4} \times 10^{-2}$ for the high-count dataset `Ba_III` (cf. Table C.22 and Fig. B.111), suggesting an

**Table 10.2** FSBI validation results. More detailed Bayesian posterior inversion results are provided in the Tables C.21 and C.22.

| **Sources**★ | **Reference**● | **Prediction**○ | | | | | | **Unit** |
|---|---|---|---|---|---|---|---|---|
| | | `Ba_I` | `Ba_II` | `Ba_III` | `Cs_I` | `Cs_II` | `Cs_III` | |
| $^{133}_{56}$Ba | $4.7 \times 10^8$ | $4.7^{+0.8}_{-1.1} \times 10^8$ | $4.7^{+0.4}_{-0.5} \times 10^8$ | $4.4^{+0.4}_{-0.5} \times 10^8$ | | | | Bq |
| $^{137}_{55}$Cs | $9.0 \times 10^9$ | | | | $9.1^{+0.2}_{-0.2} \times 10^9$ | $9.1^{+0.1}_{-0.1} \times 10^9$ | $9.1^{+0.1}_{-0.1} \times 10^9$ | Bq |
| $K_{nat}$ | $1.9(6) \times 10^2$ | $2.0^{+1.0}_{-0.9} \times 10^2$ | $2.3^{+0.4}_{-0.4} \times 10^2$ | $2.4^{+0.2}_{-0.2} \times 10^2$ | $2.7^{+1.1}_{-0.8} \times 10^2$ | $2.5^{+0.5}_{-0.4} \times 10^2$ | $2.4^{+0.1}_{-0.1} \times 10^2$ | $\mathrm{Bq\,kg^{-1}}$ |
| $Th_{nat}$ | $1.7(6) \times 10^1$ | $2.1^{+1.0}_{-0.7} \times 10^1$ | $2.5^{+0.4}_{-0.4} \times 10^1$ | $2.8^{+0.1}_{-0.1} \times 10^1$ | $2.0^{+1.1}_{-0.7} \times 10^1$ | $2.3^{+0.4}_{-0.4} \times 10^1$ | $2.4^{+0.1}_{-0.1} \times 10^1$ | $\mathrm{Bq\,kg^{-1}}$ |
| $U_{nat}$ | $1.7(4) \times 10^1$ | $2.1^{+1.5}_{-1.5} \times 10^1$ | $1.8^{+0.7}_{-0.9} \times 10^1$ | $1.4^{+0.4}_{-0.4} \times 10^1$ | $3.6^{+1.5}_{-1.8} \times 10^1$ | $2.4^{+0.8}_{-0.9} \times 10^1$ | $2.8^{+0.4}_{-0.4} \times 10^1$ | $\mathrm{Bq\,kg^{-1}}$ |

★ Sources defined in Section 10.3.2.

● The activity values $\mathcal{A}$ for the two point sources $^{133}_{56}$Ba and $^{137}_{55}$Cs were computed for the start time of the corresponding measurement using Eqs. 2.15 and 2.16 and the information provided by the calibration certificates as well as the half-life $t_{1/2}$ indicated in Table 6.1. Note that no uncertainty values were provided in the certificates. Assuming $\sigma_{\mathcal{A}_0}/\mathcal{A}_0 = 5\,\%$ and propagating the uncertainties according to Eq. A.41, the reference activities with uncertainties (coverage factor $k = 1$) are $4.7(2) \times 10^8$ and $9.0(5) \times 10^9$ for $^{133}_{56}$Ba and $^{137}_{55}$Cs, respectively. The activity mass concentrations $a_m$ (specified for soil in natural condition) were sourced from a series of in-situ spectrometry measurements conducted during the ARM20 Exercise [119] using the average values for all measurement positions on the Thun military training ground (46.753°N, 7.596°E). Uncertainties for these activity mass concentrations are reported for a coverage factor $k = 1$.

○ Maximum a posteriori point estimators (cf. Eq. 7.15a) with uncertainties indicated as central credible intervals with a probability mass of 95 % (cf. Eq. 7.17) for each of the six validation datasets listed in Table 10.1.

overdispersion in the data as discussed in Section 10.2. Consistent results are obtained for `Cs_III`, with a slightly reduced dispersion parameter of $\alpha_{\text{NB}} = 6.8^{+0.8}_{-0.6} \times 10^{-3}$, attributed to the shorter measurement live time of 232 s compared to 296 s for the `Ba_III` dataset (cf. Table 10.1). In contrast, the low-count datasets `Cs_I`, `Cs_II`, `Ba_I` and `Ba_II` show a reduced dispersion parameter $\alpha_{\text{NB}}$ with a MAP $<10^{-3}$, indicating a near Poisson statistic in the count data for these datasets.

The predicted MAP estimates for the terrestrial primordial radionuclides $K_{\text{nat}}$, $Th_{\text{nat}}$ and $U_{\text{nat}}$ are more difficult to assess given the increased magnitude of the uncertainties in both the posterior predictions and the reference values. Nevertheless, the predicted MAP estimates are statistically consistent with the reference values for all six datasets, confirming the robustness of the predictions within the given uncertainty ranges.

As discussed in Section 7.2.4.2, we can leverage the Bayesian probabilistic framework not only to quantify the source strengths but also to provide probabilistic predictions for the model response by computing prior and posterior predictive distributions (cf. Eqs. 7.19 and 7.20b). In the Figs. 10.1 and 10.2, I present these distributions[21] alongside measured pulse-height spectra for the six different datasets as well as point posterior predictions using the MAP estimates (red line).

Overall, the posterior predictive distributions consistently align with the measured pulse-height spectra across all datasets. Furthermore, as expected for an identifiable probabilistic data model (cf. Section 7.2.4.2), there is a significant reduction in the dispersion of the posterior predictive distribution compared to the prior predictive distribution, reflecting the information gain from the newly obtained data. This reduction is particularly evident for the high-count datasets `Cs_III` and `Ba_III`. For the low-count datasets `Cs_I`, `Cs_II`, `Ba_I`, and `Ba_II`, the increased sparsity in the data results in a less pronounced reduction in the dispersion of the posterior predictive distribution, particularly at higher spectral energies.[22]

In the Figs. 10.1 and 10.2, in addition to the prior and predictive distributions, I also display the spectral signatures for the five sources considered in the forward model in Eq. 10.6 scaled by the MAP source strengths and the measurement live times as $\hat{\mathbf{c}}\xi t_{\text{gr}}$.[23] These scaled spectral signatures represent the absolute contributions of the individual sources to the predicted pulse-height spectra (red line). As expected from the photon emission spectra (cf. Figs. 2.2,

21 *Color-coded using the `ksdensity` kernel density estimator function provided by the `MATLAB` code.*

22 *The overdispersion discussed above moderates this trend to some extent. This effect is most apparent in the increased dispersion of the posterior predictive distribution at low spectral energies for the high-count dataset `Ba_III` when compared to `Ba_II`. A similar trend can be observed for `Cs_III`, albeit less pronounced.*

23 *Note that for $\Delta^{222}_{86}$Rn, I scaled the related spectral signature with the magnitude $|\Delta\xi_{Rn\text{-}222}|$ in order to display the resulting spectral contribution estimate in the log-plot.*

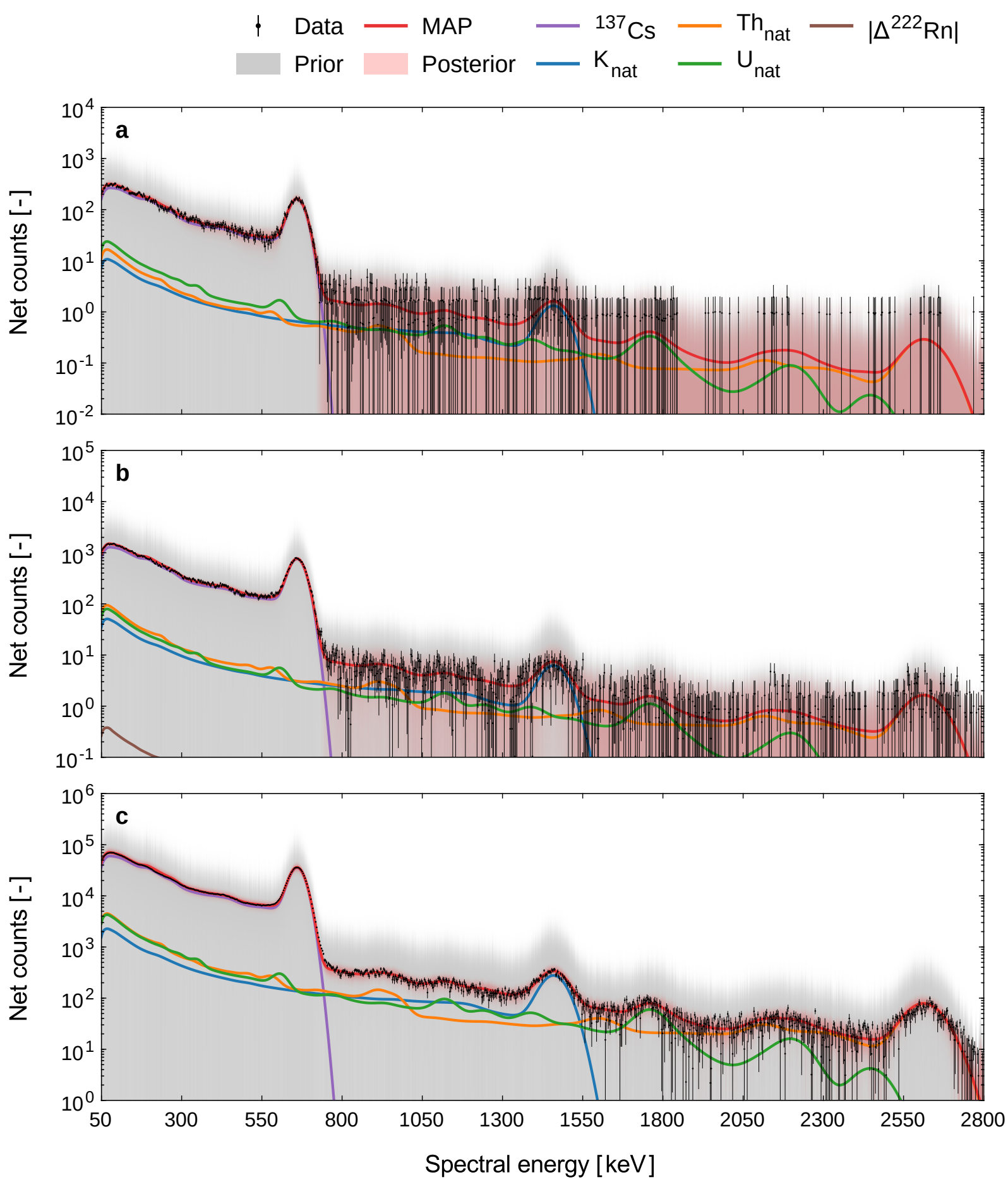

**Figure 10.1** In these graphs, I present the prior and posterior predictive distributions (color-coded) obtained by the FSBI alongside the measured pulse-height spectra $\mathbf{C}_{gr}$ for the three datasets: **a** `Cs_I` ($t_{gr} = 1\,s$). **b** `Cs_II` ($t_{gr} = 5\,s$). **c** `Cs_III` ($t_{gr} = 232\,s$). For the measured spectra, the uncertainties are indicated as 1 standard deviation (SD) errorbars (cf. Appendix A.8). In addition to the posterior predictive distributions, I also display the point posterior predictions derived by the maximum a posteriori (MAP) estimates (red line). Moreover, the spectral signatures scaled by the MAP source strengths and the measurement live time ($\hat{\mathbf{c}}\xi t_{gr}$) for all five sources considered in the forward model in Eq. 10.6 are indciated as well, i.e. the sealed $^{137}_{55}$Cs point source ($\xi_{\text{Cs-137}}$), the three natural terrestrial radionuclides $K_{nat}$, $Th_{nat}$ and $U_{nat}$ ($\xi_{\text{K-nat}}$, $\xi_{\text{Th-nat}}$, $\xi_{\text{U-nat}}$) and the radon source term $\Delta^{222}_{86}$Rn ($\Delta\xi_{\text{Rn-222}}$, only partly visible in **b**). For better interpretability, all spectral quantities were corrected for the background count rate vector $\mathbf{c}_b$ (cf. Eq. 10.6).

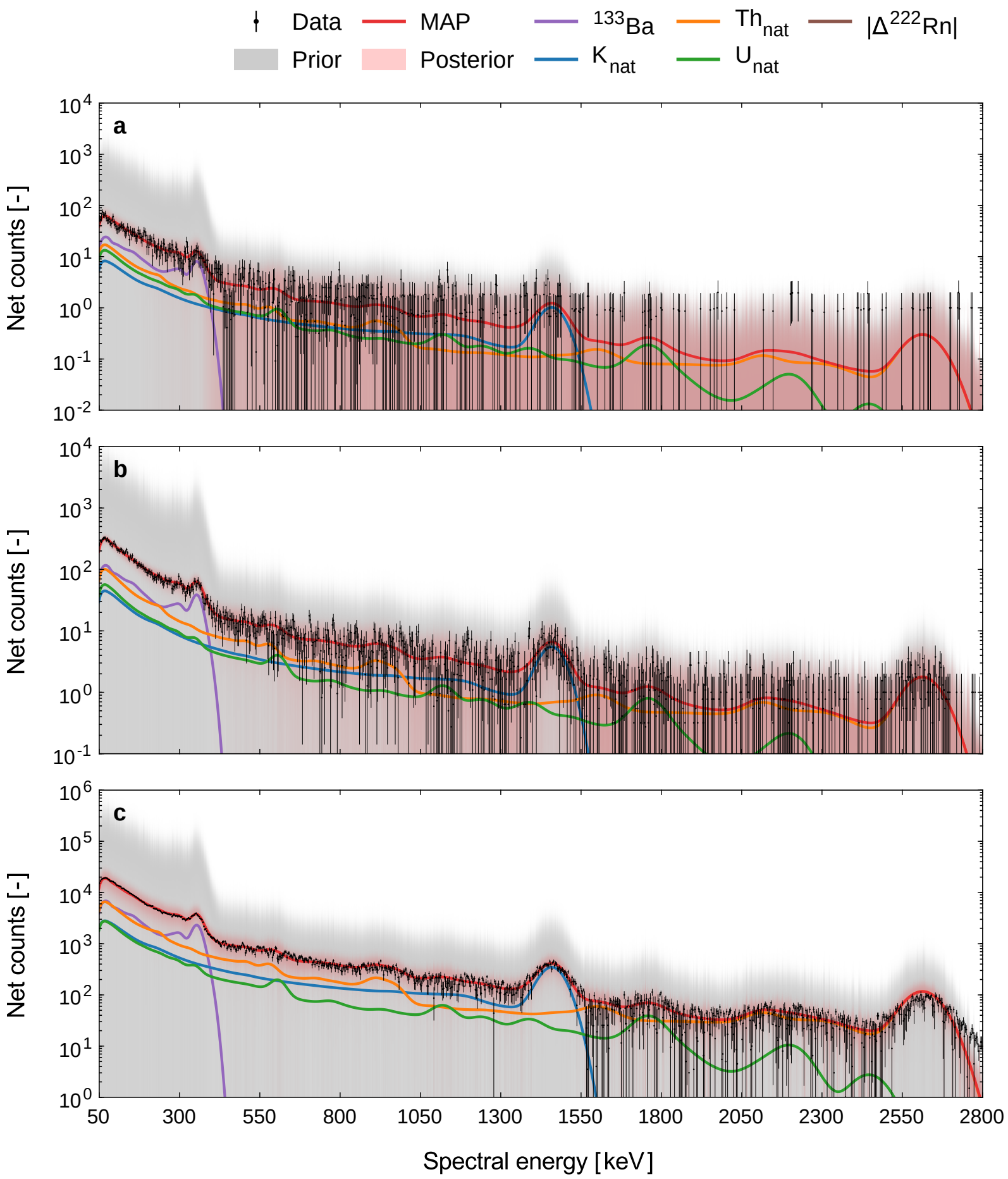

**Figure 10.2** In these graphs, I present the prior and posterior predictive distributions (color-coded) obtained by the FSBI alongside the measured pulse-height spectra $\mathbf{C}_{gr}$ for the three datasets: **a** `Ba_I` ($t_{gr} = 1$ s). **b** `Ba_II` ($t_{gr} = 5$ s). **c** `Ba_III` ($t_{gr} = 296$ s). For the measured spectra, the uncertainties are indicated as 1 standard deviation (SD) errorbars (cf. Appendix A.8). In addition to the posterior predictive distributions, I also display the point posterior predictions derived by the maximum a posteriori (MAP) estimates (red line). Moreover, the spectral signatures scaled by the MAP source strengths and the measurement live time ($\hat{\mathbf{c}}\xi t_{gr}$) for all five sources considered in the forward model in Eq. 10.6 are indciated as well, i.e. the sealed $^{133}_{56}$Ba point source ($\xi_{\text{Cs-137}}$), the three natural terrestrial radionuclides $K_{nat}$, $Th_{nat}$ and $U_{nat}$ ($\xi_{\text{K-nat}}$, $\xi_{\text{Th-nat}}$, $\xi_{\text{U-nat}}$) and the radon source term $\Delta^{222}_{86}$Rn ($\Delta\xi_{\text{Rn-222}}$, too small to be visible). For better interpretability, all spectral quantities were corrected for the background count rate vector $\mathbf{c}_b$ (cf. Eq. 10.6).

B.3 and B.5), I find that the spectral signatures for the point sources $^{133}_{56}$Ba and $^{137}_{55}$Cs are the dominant contributors to the measured pulse-height spectra at low spectral energies, while the natural terrestrial radionuclides $K_{nat}$, $Th_{nat}$ and $U_{nat}$ become the dominant sources at higher spectral energies. It is interesting to observe that the radon source term $\Delta^{222}_{86}$Rn does not significantly contribute to the predicted pulse-height spectra across all datasets. This observation is corroborated by the posterior results in the Tables C.21 and C.22, which show a statistically insignificant source strength for the radon term in all cases. This implies that, for the considered datasets listed in Table 10.1, there were no statistically significant variations in the radon background between the background flights over Lake Thun and the hover flights. That said, the credible intervals for the radon source term are relatively broad, mainly due to the similarity of the spectral signatures between $\Delta^{222}_{86}$Rn and $U_{nat}$. Therefore, it is worth highlighting that variations in the radon source term could potentially be masked by this increased uncertainty.

## 10.4 Background Quantification

As discussed in the introduction to this chapter as well as in Section 5.6, it has been recognized in the past that variable nuisance sources such as radon or cosmic backgrounds can introduce significant systematic errors in the quantification of terrestrial sources [120, 642, 676]. The full spectrum approach has the potential to include these background sources in the quantification problem, allowing for real-time background correction and thereby reducing systematic errors. However, this requires accurate spectral signature estimates for these background sources, which are difficult to obtain in practice using empirical measurements [642, 676, 679].

In the previous section, for validation purposes, I quantified the combined intrinsic, cosmic and radon background effect on the pulse-height spectra empirically using standard background flights over Lake Thun combined with a radon correction term to account for changes in the radon background between the background and hover flights.

Here, I will demonstrate the capability of the new methodology to accurately quantify the cosmic and radon background in the atmosphere by applying the FSBI method to a series of AGRS flights conducted over Lake Thun and the North Sea. Spectral signature calibration was performed numerically using the high-fidelity NPSMC and DRM models presented in the previous chapters. In the following subsections, I will detail the selected experimental datasets, the Bayesian inversion setup and the results obtained from the FSBI method.

### 10.4.1 Radiation Measurements

To minimize the effect of natural terrestrial gamma-ray sources on the detector response at low ground clearances, all AGRS flights considered in this section were conducted over extended bodies of water (cf. Section 2.1.3.1). In total three different datasets were derived from three independent flights conducted by the Swiss AGRS system:

1. `Lake_I` Flight conducted over Lake Thun in Switzerland on the morning of 2022-06-16 (beginning of the recording at 06:08:38 UTC) as part of the ARM22 validation campaign [120]. During this flight, the AGRS system was flown in a continuous descent mode, beginning at a ground clearance of 2249 m and descending to an altitude of 32 m above the water surface. A map of the flight path is shown in Fig. B.112.
2. `Lake_II` Flight conducted over Lake Thun in Switzerland on the afternoon of 2022-06-16 (beginning of the recording at 13:53:27 UTC) as part of the ARM22 validation campaign [120]. During this flight, the AGRS system was flown in a continuous climb mode, beginning at a ground clearance of 30 m and ascending to an altitude of 2260 m above the water surface. A map of the flight path is provided in Fig. B.112.
3. `Sea_I` Flight conducted over the North Sea at the west coast of Denmark close to the city Søndervig in the morning of 2018-06-19 (beginning of the recording at 09:58:05 UTC) as part of the International Exercise CONTEX 2018 [534]. The AGRS system was operated in a series of six horizontal level flights at increasing altitudes ranging from a ground clearance of 87 m up to an altitude of 1236 m above the water surface. A map of the flight path is displayed in Fig. B.113.

A summary of the main experimental conditions including measurement live times and orthometric heights for these datasets is provided in Table 10.3.

To minimize systematic uncertainties in the quantification of the cosmic and radon background, the datasets were filtered to exclude data recorded during flight maneuvers deviating from horizontal or near-horizontal orientations, such as turns or rapid ascent/descent corrections. Furthermore, to account for the significantly reduced count statistics compared to regular AGRS survey flights over land, the filtered data was aggregated into gross count spectra $\mathbf{C}_{\mathrm{gr}}$ by summing the counts from the individual pulse-height spectra recorded in the detector channel `#SUM`[24] with a sampling time of 1 s over predefined time intervals.[25] Each dataset is characterized by a finite set of gross count spectra $\mathcal{Y} = \{\mathbf{C}_{\mathrm{gr},k} \in \mathbb{R}_+^{N_{\mathrm{ch}}\times 1} \mid k \in \mathbb{N}, k \leq N_{\mathcal{Y}}, N_{\mathcal{Y}} \in \mathbb{N}_+\}$, the associated gross measurement live times $t_{\mathrm{gr},k}$ and the mean experimental conditions $\mathfrak{d}_k$ under which the individual aggregated pulse-height spectra in $\mathbf{C}_{\mathrm{gr},k}$ were recorded. More information on the conducted measurements and data acquisition can be found in [120, 534].

**Table 10.3** Overview of the experimental datasets considered for the quantification of intrinsic, cosmic and radon backgrounds using FSBI. These datasets were derived from two survey flights conducted over Lake Thun (`Lake_I`, `Lake_II`) and one over the North Sea (`Sea_I`) using the Swiss AGRS system.

| Identifier | Site | Ground clearance [m]★ | Orthometric height [m]• | Gross measurement live time [s]◦ | Reference |
|---|---|---|---|---|---|
| `Lake_I` | Lake Thun | [32, 2249] | [591, 2803] | 669 | [120] |
| `Lake_II` | Lake Thun | [30, 2260] | [588, 2820] | 454 | [120] |
| `Sea_I` | North Sea | [87, 1236] | [87, 1236] | 592 | [534] |

★ Minimum and maximum ground clearance $h_{\mathrm{air}}$ rounded to meters.
• Minimum and maximum orthometric height $h_{\mathrm{ort}}$ based on the `EGM2008` earth gravitational model [675] and rounded to meters.
◦ Total gross measurement live time $t_{\mathrm{gr}}$ rounded to seconds.

24 *As for the previous two chapters, I limit the discussion here to the detector channel `#SUM`.*

25 *Fixed* ~60 s *time intervals for the `Lake_I` and `Lake_II` datasets while for the `Sea_I` dataset, the pulse-height spectra were aggregated over the six flight levels resulting in varying measurement live times between* 87 s *and* 112 s.

### 10.4.2 Inversion Setup

The general FSBI methodology was outlined already in Section 10.2. In the following subsections, I focus on the specific setup of the FSBI method for the selected measurement datasets listed in Table 10.3. Specifically, I will provide details on the forward model and its calibration as well as the Bayesian inversion computations performed to estimate the source strengths of the cosmic and radon background as well as the intrinsic background of the Swiss AGRS system.

#### 10.4.2.1 Forward Model

As discussed in detail in Section 5.4.1, in this book, I assume that both the number/type and the location of all high-energy photon sources affecting the detector response of the AGRS system are known. Under these assumptions, we can write the forward model for the AGRS measurements listed in Table 10.3 as a simple linear combination of the spectral signatures encoded in the spectral signature matrix $\boldsymbol{\mathcal{M}} \in \mathbb{R}_+^{N_{\mathrm{ch}} \times N_{\mathrm{src}}}$ scaled by their respective source strengths $\boldsymbol{\xi} \in \mathbb{R}_+^{N_{\mathrm{src}} \times 1}$ (cf. Eq. 5.3):

$$\mathbf{C} = \boldsymbol{\mathcal{M}}\,(\boldsymbol{\mathfrak{d}})\,\boldsymbol{\xi} t_{\mathrm{gr}} \tag{10.7}$$

where:

| | | |
|---|---|---|
| $\mathbf{C}$ | count vector | |
| $\boldsymbol{\mathfrak{d}}$ | experimental condition | $[\boldsymbol{\mathfrak{d}}]$ |
| $\boldsymbol{\mathcal{M}}$ | spectral signature matrix | $\mathrm{s}^{-1}\,[\xi]^{-1}$ |
| $t_{\mathrm{gr}}$ | gross measurement live time | s |
| $\boldsymbol{\xi}$ | source strength vector | $[\xi]$ |

The following sources were considered in the forward model for the hover flight measurements:

1. **Cosmic Background** Cosmic ray induced background in the Earth's atmosphere as discussed in Section 2.2. The source strength $\xi_{\mathrm{cos}}$ of this background was parametrized by the total cosmic ray induced ionizing particle flux $\phi_{\mathrm{tot}}$ for a given location in the Earth's atmosphere, that is defined as the sum of cosmic particle fluxes $\phi_p$ (cf. Eq. 2.22) for a selected set of ionizing particles ($\gamma$, n, $\mathrm{e}^{\pm}$, $\mu^{\pm}$, p)[26] in units of $\mathrm{s}^{-1}\,\mathrm{m}^{-2}$ (cf. Section 2.2).
2. **Radon Background** Atmospheric radon $^{222}_{86}\mathrm{Rn}$ with the progeny up to but not including $^{210}_{82}\mathrm{Pb}$ assumed to be in secular equilibrium (cf. Table C.1) and uniformly distributed in the vicinity of the aircraft system. The source strength $\xi_{\mathrm{Rn\text{-}222}}$ was parametrized by the associated activity volume concentration $a_{\mathrm{v}}$ in $\mathrm{Bq\,m^{-3}}$ (cf. Eq. 2.11b).
3. **Intrinsic Background** Natural radionuclides $^{40}_{19}\mathrm{K}$, $^{232}_{90}\mathrm{Th}$ and $^{238}_{92}\mathrm{U}$ with all associated progeny assumed to be in secular equilibrium (cf. Tables C.1 and C.2) and uniformly distributed in the AGRS system (cf. Section 2.1.3). I will refer to these sources as $\mathrm{K_{nat}}$,

26 *As discussed in Section 2.2, these are the main ionizing particles generated by the cosmic background in the lower Earth's atmosphere. Heavier hadrons show a negligible contribution to the total flux in the lower atmosphere.*

$Th_{nat}$ and $U_{nat}$ in this section. The source strengths $\xi_{\text{K-nat}}$, $\xi_{\text{Th-nat}}$ and $\xi_{\text{U-nat}}$ were parametrized by the associated absolute activities $\mathcal{A}$ in Bq (cf. Eq. 2.9).

In summary, five different sources were considered in the forward model ($N_{\text{src}}$=5) with the source strength vector being defined as $\boldsymbol{\xi} = \{\xi_{\text{cos}}, \xi_{\text{Rn-222}}, \xi_{\text{K-nat}}, \xi_{\text{Th-nat}}, \xi_{\text{U-nat}}\}^\top$.

#### 10.4.2.2 Calibration

The forward model in Eq. 10.7 required the derivation of the spectral signatures for each of the five sources encoded in the spectral signature matrix $\boldsymbol{\mathcal{M}}$ as a function of the experimental condition $\boldsymbol{\mathfrak{d}}$.

**Cosmic Background** The most complex part of the forward model calibration was the derivation of the spectral signatures for the cosmic background. As discussed in detail in Section 2.2, the cosmic background in the Earth's atmosphere generated by various primary and secondary high-energy ionizing particles is a complex function of both time and space. To derive the spectral signatures for the cosmic background as a function of the experimental conditions $\boldsymbol{\mathfrak{d}}$ under which the individual datasets in Table 10.3 were obtained, I combined the `PARMA` code (version: `4.13`)[27] [146, 159] presented in Section 2.2.2 with the `FLUKA` Monte Carlo code (version: `4-2.2`) [20, 216, 281] using the high-fidelity mass model of the Swiss AGRS system presented in Section 8.2.

Specifically, in a first step, I generated a set of phase space files for the double differential flux of the main cosmic ray induced ionizing particles in the lower atmosphere ($\gamma$, n, $e^{\pm}$, $\mu^{\pm}$, p, cf. Section 2.2.2) using the `PARMA` code.[28] These phase space files were generated for the location and time of the AGRS flights over Lake Thun and the North Sea, considering the full energy range available in `PARMA`, i.e. $10^4$ eV to $10^{12}$ eV,[29] and the full solid angle for all particles.

In a second step, I used these phase space files as input for the AGRS Monte Carlo model presented in Section 8.2 to simulate the spectral signatures of the cosmic background as a function of the experimental condition $\boldsymbol{\mathfrak{d}}$.[30] For that purpose, I developed a custom script based on the `source` routine provided by `FLUKA`. Given that the AGRS Monte Carlo model was developed for the transport of electrons, positrons and photons with kinetic energies $\leq \mathcal{O}(10^6)$ eV, the physics models had to be extended to account for the increased

27 *Please note that I used the updated version with a patch to account for the production of annihilation photons. The missing annihilation photons were identified in preliminary simulations for this work and subsequently reported to the developers.*

28 *I need to point out that I specifically accounted for the angular dependence of the double differential flux for all particles discussed in Section 2.2.2.*

29 *For neutrons, the lower energy threshold is $10^{-2}$ eV. For muons, the upper energy threshold is $10^{14}$ eV [159].*

30 *It is worth adding that the PSMC physics model was used together with the `PScinMC` postprocessing pipeline since the NPSM presented in Section 7.2.1 was only calibrated for photons and electrons.*

spectral range as well as the various additional secondary particles generated by the cosmic background in the Earth's atmosphere. In addition to the high-fidelity physics model actived by the `precisio` mode (cf. Section 6.2.2.1), I incorporated the `PHOTONUC` and `MUPHOTON` cards to account for photonuclear, electronuclear and muon-nuclear interactions in the Monte Carlo simulations.[31] For the neutrons, I applied the pointwise low energy neutron treatment activated by the `LOW-PWXS` card combined with the JEFF-3.3 nuclear data library [45]. A lower transport threshold of $10^3$ eV was applied for all particles, except for the neutrons, where a lower transport threshold of $10^{-5}$ eV was used.

To account for the variation in the double differential flux of the cosmic background as a function of the orthometric height, I generated a series of spectral signatures using the two-step procedure detailed above for different orthometric heights with a step size of 500 m ranging from 0 m to 3000 m/1500 m for the Lake Thun/North Sea datasets, respectively. The spectral signature for the cosmic background at a given orthometric height was then interpolated based on the generated spectral signatures using a modified Akima spline interpolation model [733]. Apart from the orthometric heights, I assumed constant experimental conditions $\mathfrak{d}$ during the three individual flights listed in Table 10.3.[32]

In general, I found only small variations in the spectral signatures for the cosmic background as a function of the orthometric height, location and time (cf. Fig. B.114). The spectral signature for the cosmic background shows two distinct spectral features: an annihilation peak at ~511 keV attributed to electron-positron annihilation events (cf. Eq. 2.8) as well as a less pronounced FEP at ~2223 keV related to ${}^1_1\mathrm{H}(\mathrm{n},\gamma){}^2_1\mathrm{H}$ nuclear reaction events (cf. Section 2.3.1).[33,34]

A detailed study on the response of individual secondary particles to the detector channel `#SUM` of the Swiss AGRS system revealed significant effects from all secondary particles except protons (cf. Fig. B.115). The relative contributions of the individual secondary particles to the total cosmic background were found to be as follows:[35]

| | |
|---|---|
| **Photons** | 77 % |
| **Neutrons** | 9 % |
| **Electrons** | 9 % |
| **Muons** | 4 % |
| **Protons** | <1 % |

31 *All settings were kept consistent for both the particles and the corresponding antiparticles.*

32 *This includes crew composition and fuel levels in the individual fuel tanks. The aircraft orientation was assumed to be horizontal with respect to the Earth's surface. Note that the `PARMA` code is based on the U.S. Standard Atmosphere 1976 model [164]. For the `Sea_I` dataset, no information on the crew composition and fuel levels was available. Therefore, best estimate values were assumed based on the information provided in [534].*

33 *Note also that not only the neutrons but also the protons showed a FEP at ~2223 keV which was attributed to a nuclear reaction sequence of a* $(\mathrm{p},\mathrm{n})$ *nuclear reaction followed by the radiative capture reaction* ${}^1_1\mathrm{H}(\mathrm{n},\gamma){}^2_1\mathrm{H}$ *in the aircraft structure and the detector system (cf. Fig. B.115).*

34 *Given the high hydrogen content of Jet A-1 fuel, the radiative capture reaction events primarily occur in the aircraft's fuel tanks. To assess the effect of the aircraft fuel on the cosmic background, a sensitivity study was performed by varying the fuel volume fractions in the aircraft fuel tanks. In line with the results presented in Chapter 9, no significant effect was found in the FEP at ~2223 keV between fuel volume fractions of 50 % and 100 % (cf. Fig. B.116), implying that it is the fuel tanks #1 and #2 that affect the FEP at ~2223 keV in the AGRS measurements (cf. Fig. 8.1).*

**Radon Background** The spectral signature of the radon background was derived by the DRM method discussed in Chapter 9. For that purpose, radon with the progeny up to but not including $^{210}_{82}$Pb was assumed to be in secular equilibrium (cf. Table C.1) and uniformly distributed around the aircraft.[36] Similar to Section 10.3.2.2, I derived three individual DRFs for each of the three datasets listed in Table 10.3 using the methodology detailed in Section 9.2.2 and accounting for the individual experimental conditions.[37] The double differential photon flux signatures were computed with the methods discussed in Section 9.2.1. The environmental mass model was adopted from Section 8.3.2.2 with the ground exchanged by water.[38]

To account for the variation in the double differential flux of the radon background as a function of the ground clearance[39], I generated a series of spectral signatures for different ground clearances ranging from 0 m to 2500 m/1500 m for the Lake Thun/North Sea datasets, respectively.[40] The spectral signature for the radon background at a given orthometric height was then interpolated based on the generated spectral signatures using a modified Akima spline interpolation model [733]. An overview of the resulting spectral signatures for the radon background as a function of the ground clearance is provided in Fig. B.117 for all three datasets.

**Intrinsic Background** Only limited information was available on the distribution of natural radionuclides in the TH06 aircraft. A laboratory-based gamma-ray spectrometry analysis[41] of a Jet A-1 fuel sample showed only insignificant intrinsic activities of the natural radionuclides $^{40}_{19}$K, $^{232}_{90}$Th, $^{238}_{92}$U and their progeny in the aircraft fuel (cf. Table C.18).[42] Therefore, the intrinsic background was modeled as a combination of the natural radionuclides $^{40}_{19}$K, $^{232}_{90}$Th and $^{238}_{92}$U with all associated progeny assumed to be in secular equilibrium (cf. Tables C.1 and C.2) and uniformly distributed in the AGRS system excluding the fuel.[43]

Given the near field character of the intrinsic sources, I used the AGRS Monte Carlo model presented in Chapter 8 to compute the spectral signature for the three natural radionuclides $K_{nat}$, $Th_{nat}$ and $U_{nat}$ as a function of the experimental conditions $\mathfrak{d}$ for each dataset listed in Table 10.3.[44] To accurately simulate the intrinsic volume sources, I developed a custom script based on the `source` routine provided by `FLUKA` that implements a region sampling algorithm to account for the spatial distribution of the natural radionuclides in the

35 *Computed by integrating the individual spectral signatures over all pulse-height channels for a reference orthometric height of 659 m, location: 46.753°N, 7.596°E and date: 2022-06-16. Note that the relative contributions are given in percent of the total cosmic background for the combined effect of the particles together with their antiparticles.*

36 *A similar modeling approache was used by Sinclair et al. to investigate radioxenon concentrations in the atmosphere using AGRS [24].*

37 *This includes crew composition, fuel levels in the individual fuel tanks and ground clearance. Standard atmospheric conditions were assumed for all three datasets, i.e. dry air with $T = 15\,°C$ and $p = 1013.25\,hPa$. Considering the obtained spectral signatures presented in Fig. B.117, I find only minor effects of the different experimental conditions on the spectral signatures with a maximum median relative deviation of ~4 % over the entire spectral range of the detector.*

38 *To account for the extension of the radon volume source, the radius of the simulation volume was increased from 200 m to 3 km based on the findings obtained in Section 3.2.1 and considering the photon energies (cf. Fig. 2.2) and resulting mean free paths of the primary photons in the atmosphere (cf. Fig. 3.6).*

39 *Note that, since the radon spectral signature is independent of the atmospheric depth in the adopted model, I chose here the ground clearance instead of the orthometric height as the reference parameter for the altitude.*

40 *Varying step size of 100 m below and 500 m above a ground clearance of $h_{air} = 500\,m$.*

complex aircraft structure. The resulting spectral signatures scaled with the derived source strengths will be presented in Section 10.4.3.

#### 10.4.2.3 Bayesian Computations

Given the expected variation in the source strengths of the cosmic and radon backgrounds as a function of the altitude, the quantification problem is solved independently for each of the gross count spectra $\mathbf{C}_{\mathrm{gr}}$ contained in the three datasets listed in Table 10.3 using FSBI by combining the likelihood function $\mathcal{L}(\mathbf{x}; \mathbf{C}_{\mathrm{gr}}, \mathfrak{d})$ and the prior distribution $\pi(\mathbf{x})$ via Bayes' theorem (Eq. 7.13) to compute the posterior distribution $\pi(\mathbf{x} \mid \mathbf{C}_{\mathrm{gr}}, \mathfrak{d})$.[45] The likelihood function is fully defined by Eq. 10.4, the calibrated forward model in Eq. 10.7 and the model parameter vector $\mathbf{x} = (\boldsymbol{\xi}, \alpha_{\mathrm{NB}})^\top$. To prevent overly restrictive assumptions on the prior distribution, I adopted weakly informative statistically independent marginal priors for all model parameters $\mathbf{x}$. A full list of all marginal prior distributions is provided in Table C.20.

As noted already in Section 10.2, I used an AIES MCMC algorithm [735] implemented in the `UQLab` code [753] to compute the posterior distribution.[46] In total, 25 independent FSBI computations were performed. The MCMC algorithm was executed with 20 parallel chains, each running $5 \times 10^3$ iterations. The burn-in phase was set 40 % resulting in a total of $N_{\hat{\mathcal{X}}} = 6 \times 10^4$ posterior samples. The convergence and precision of the MCMC simulations were carefully assessed using standard diagnostics tools [738, 814], showing a potential scale reduction factor $\hat{R} < 1.04$ and effective sample size ESS > 300 across all MCMC runs.

### 10.4.3 Results

In this section, I present the results of the FSBI computations for the three datasets listed in Table 10.3. Given the thorough validation of the FSBI method in Section 10.3, I limit the discussion here to the estimated source strengths and their evolution as a function of the altitude for the cosmic and radon background as well as the intrinsic background of the Swiss AGRS system.

In the Figs. 10.3–10.5, I show the estimated source strengths of the cosmic and radon background as well as the intrinsic background of the Swiss AGRS system as a function of the orthometric height and ground clearance for the datasets `Lake_I`, `Lake_II` and `Sea_I`, respectively.

41 *Performed by the Radioanalytics Group at the PSI using a HPGe detector (cf. Table C.18).*

42 *Considering the results obtained in Section 10.4.3, the fuel only contributes* $\lesssim$1 % *to the* $^{232}_{90}$Th *and* $^{238}_{92}$U *intrinsic backgrounds at a fuel volume fraction of* 50 %.

43 *Motivated by the results obtained in Fig. B.118, I explicitly considered the aircraft personnel in the simulations of the spectral signature for the intrinsic* $\mathrm{K}_{\mathrm{nat}}$ *background.*

44 *This includes again the crew composition and the mean fuel levels during the individual flights.*

45 *Note that, to assess the variability of the intrinsic background, the source strengths of the natural radionuclides* $\mathrm{K}_{\mathrm{nat}}$, $\mathrm{Th}_{\mathrm{nat}}$ *and* $\mathrm{U}_{\mathrm{nat}}$ *were allowed to vary across the individual pulse-height spectra within the three datasets.*

46 *It is worth adding that, to account for the systematic uncertainties at low spectral energies observed in Chapters 7 and 8, the FSBI was again limited to spectral energies* $E' \geq 50$ keV.

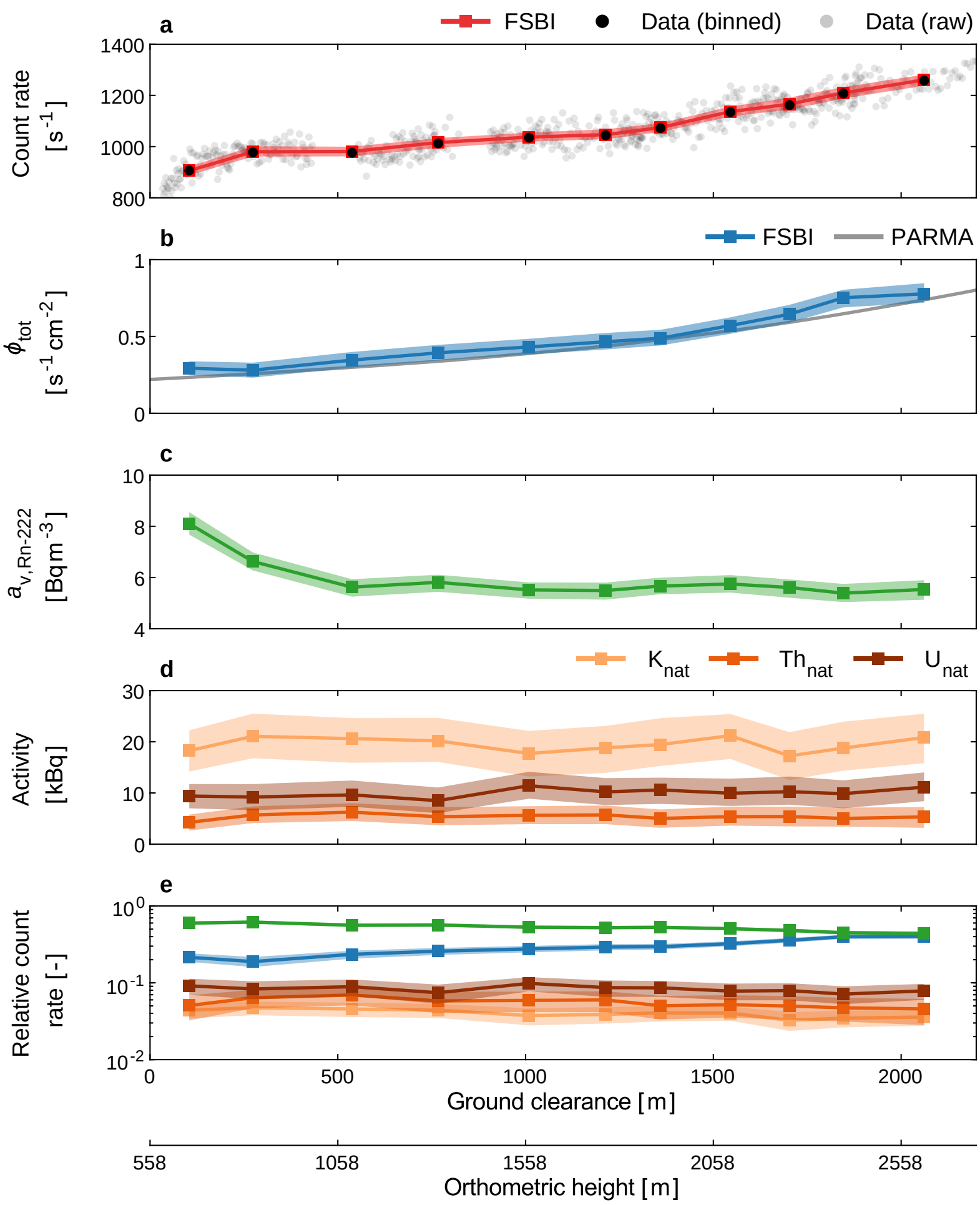

**Figure 10.3** In these graphs, I present the FSBI results for the dataset `Lake_I` as a function of the ground clearance and the orthometric height (datum reference: `EGM2008` [675]). The displayed posterior point and dispersion estimators correspond to the maximum a posteriori (MAP) probability estimates and central credible/predictive intervals with a probability mass of 99 %. **a** Measured and predicted total count rate in the detector channel `#SUM` of the Swiss AGRS system. For the measured data, both the raw and aggregated data described in Section 10.4.1 are displayed. **b** Cosmic background source strength $\xi_{\text{cos}}$, i.e. the total cosmic ray induced ionizing particle flux $\phi_{\text{tot}}$ predicted by FSBI and `PARMA`. **c** Radon background source strength $\xi_{\text{Rn-222}}$, i.e. the activity volume concentration $a_{\text{V,Rn-222}}$ predicted by FSBI. **d** Intrinsic background source strengths $\xi_{\text{K-nat}}$, $\xi_{\text{Th-nat}}$ and $\xi_{\text{U-nat}}$, i.e. the absolute activities $\mathcal{A}$ of $K_{\text{nat}}$, $Th_{\text{nat}}$ and $U_{\text{nat}}$ predicted by FSBI. **e** Relative contribution of the individual background sources to the total count rate in the detector channel `#SUM` of the Swiss AGRS system.

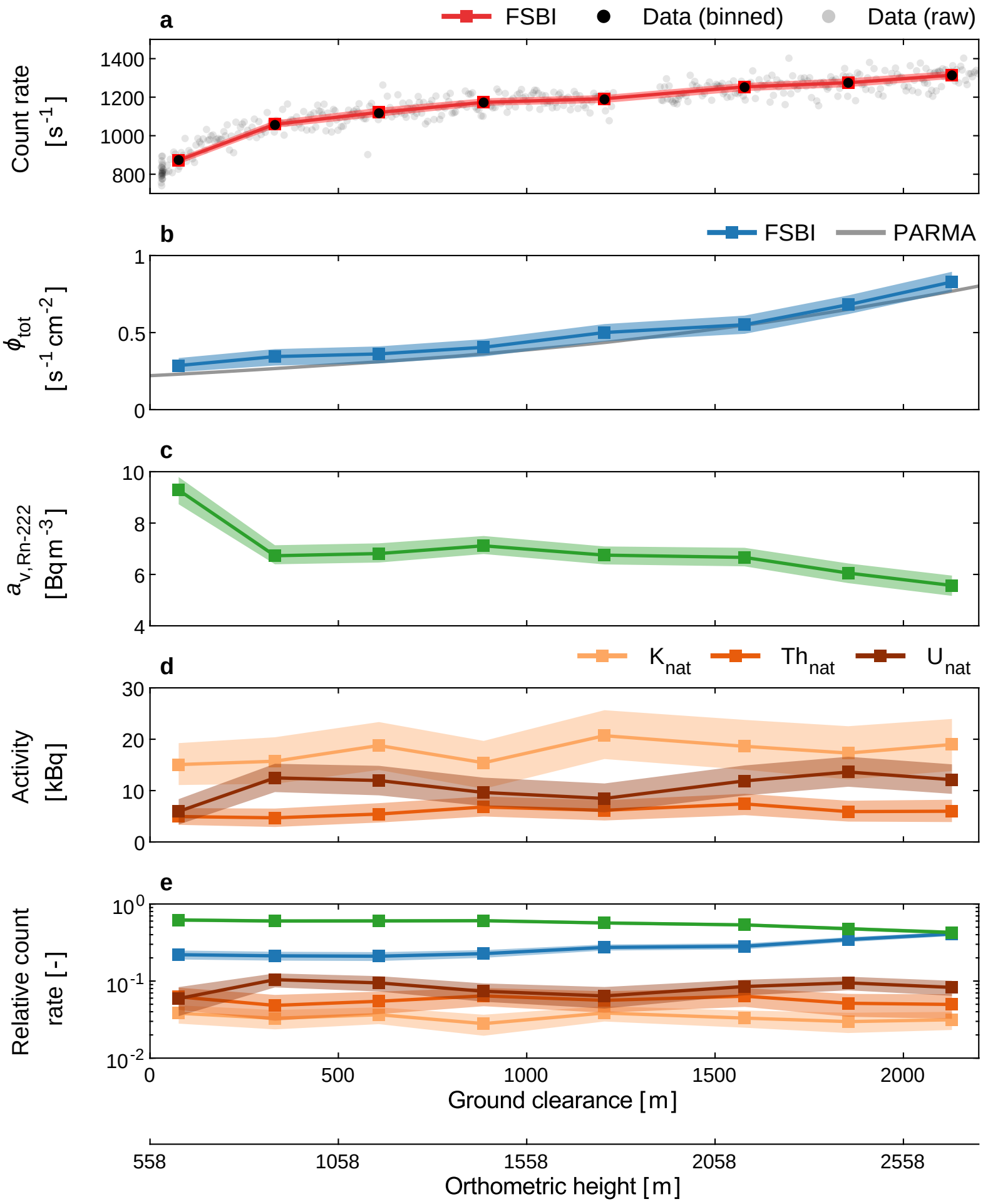

**Figure 10.4** In these graphs, I present the FSBI results for the dataset `Lake_II` as a function of the ground clearance and the orthometric height (datum reference: `EGM2008` [675]). The displayed posterior point and dispersion estimators correspond to the maximum a posteriori (MAP) probability estimates and central credible/predictive intervals with a probability mass of 99 %. **a** Measured and predicted total count rate in the detector channel `#SUM` of the Swiss AGRS system. For the measured data, both the raw and aggregated data described in Section 10.4.1 are displayed. **b** Cosmic background source strength $\xi_{\mathrm{cos}}$, i.e. the total cosmic ray induced ionizing particle flux $\phi_{\mathrm{tot}}$ predicted by FSBI and `PARMA`. **c** Radon background source strength $\xi_{\text{Rn-222}}$, i.e. the activity volume concentration $a_{\mathrm{v,Rn-222}}$ predicted by FSBI. **d** Intrinsic background source strengths $\xi_{\text{K-nat}}$, $\xi_{\text{Th-nat}}$ and $\xi_{\text{U-nat}}$, i.e. the absolute activities $\mathcal{A}$ of $K_{\mathrm{nat}}$, $Th_{\mathrm{nat}}$ and $U_{\mathrm{nat}}$ predicted by FSBI. **e** Relative contribution of the individual background sources to the total count rate in the detector channel `#SUM` of the Swiss AGRS system.

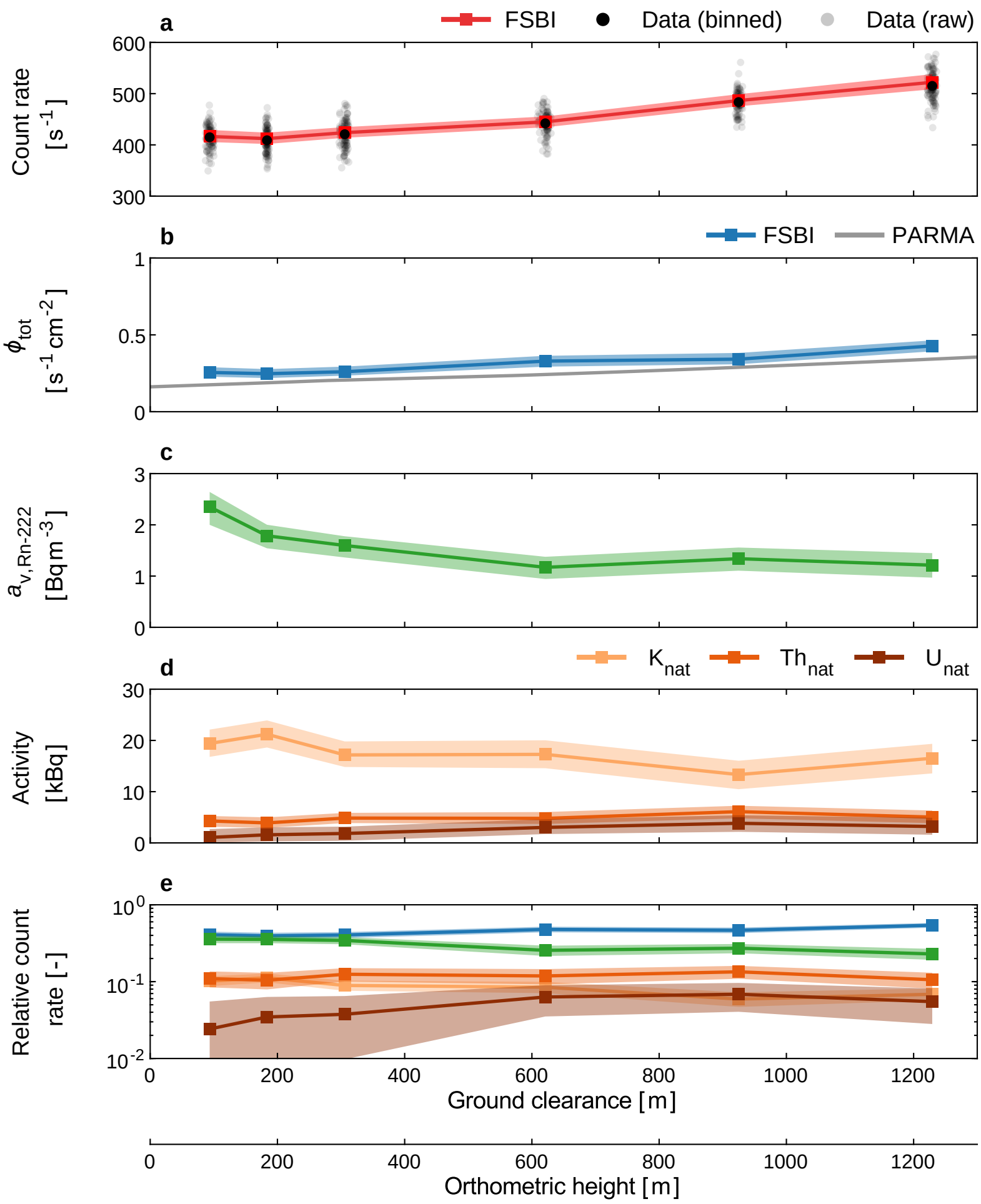

**Figure 10.5** In these graphs, I present the FSBI results for the dataset `Sea_I` as a function of the ground clearance and the orthometric height (datum reference: `EGM2008` [675]). The displayed posterior point and dispersion estimators correspond to the maximum a posteriori (MAP) probability estimates and central credible/predictive intervals with a probability mass of 99 %. **a** Measured and predicted total count rate in the detector channel `#SUM` of the Swiss AGRS system. For the measured data, both the raw and aggregated data described in Section 10.4.1 are displayed. **b** Cosmic background source strength $\xi_{\mathrm{cos}}$, i.e. the total cosmic ray induced ionizing particle flux $\phi_{\mathrm{tot}}$ predicted by FSBI and `PARMA`. **c** Radon background source strength $\xi_{\mathrm{Rn\text{-}222}}$, i.e. the activity volume concentration $a_{\mathrm{V,Rn-222}}$ predicted by FSBI. **d** Intrinsic background source strengths $\xi_{\mathrm{K\text{-}nat}}$, $\xi_{\mathrm{Th\text{-}nat}}$ and $\xi_{\mathrm{U\text{-}nat}}$, i.e. the absolute activities $\mathcal{A}$ of $\mathrm{K_{nat}}$, $\mathrm{Th_{nat}}$ and $\mathrm{U_{nat}}$ predicted by FSBI. **e** Relative contribution of the individual background sources to the total count rate in the detector channel `#SUM` of the Swiss AGRS system.

Starting with the cosmic background, as expected from our discussion in Section 2.2.2, the estimated total cosmic ray induced ionizing particle flux $\phi_{\mathrm{tot}}$ shows a clear increase with the orthometric height for all three datasets. Furthermore, a comparison with the absolute flux predictions obtained by the PARMA code shows good agreement for the datasets Lake_I and Lake_II while for the dataset Sea_I, the FSBI predictions show a slight overestimation of the flux $\phi_{\mathrm{tot}}$.[47] Consistent deviations in the cosmic ray induced photon flux in the lower atmosphere were reported by the developers of the PARMA code and were attributed to the neglection of primary cosmic rays with kinetic energies >1 TeV per nucleon in the PARMA code [159]. These discrepancies are expected to increase with decreasing orthometric height as well as increasing latitude.

47 *Note that for the calibration of the forward model in Section 10.4.2.2, I did not use any absolute flux quantities from PARMA.*

Continuing with the radon background, as anticipated from the discussion in Section 2.1.3.2, I find a significant drop in the radon activity volume concentration $a_{\mathrm{v,Rn-222}}$ between a ground clearance of 0 m and ~500 m. This reduction is attributed to the distribution and transport of the radon progeny $^{214}_{82}\mathrm{Pb}$ and $^{214}_{83}\mathrm{Bi}$ in the Earth's atmosphere [89–92]. On the other hand, based on findings from earlier studies [88, 853], a rapid decrease in the radon activity volume concentration is expected at ground clearances above ~1000 to ~1500 m. The almost constant radon activity volume concentration between a ground clearance of ~500 m and ~2000 m for the datasets Lake_I and Lake_II suggests a deviation from these previous findings. These discrepancies in the radon levels may be attributed to the unique topographical and geological features of the Swiss Alps around Lake Thun (cf. Fig. B.112), potentially causing elevated radon activity volume concentrations at higher altitudes.

It is also interesting to observe that the radon activity volume concentration in the Sea_I dataset is three to four times lower than in the Lake_I and Lake_II datasets. This substantial reduction is consistent with the expected decrease in radon concentration over marine environments [52].[48]

48 *It is important to note that because the AGRS flight of the dataset Sea_I was performed near the shore (cf. Fig. B.113), the reduction in radon activity volume concentration is less pronounced than what Kogan et al. [52] reported for open sea conditions.*

Regarding the intrinsic background characterized by the natural radionuclides $\mathrm{K_{nat}}$, $\mathrm{Th_{nat}}$ and $\mathrm{U_{nat}}$, I find only minor variations over the course of the three individual measurement flights, when accounting for the credible intervals. On the other hand, a significant reduction in the $\mathrm{U_{nat}}$ activity is observed for the dataset Sea_I compared to the datasets Lake_I and Lake_II. This reduction can be attributed to the fact that a different TH06 aircraft was used to record the dataset Sea_I, while the datasets Lake_I and Lake_II were recorded with the same aircraft.

In the Figs. 10.3–10.5, in addition to the individual source strengths, I display also the total count rate in the detector channel #SUM of the Swiss AGRS system to characterize the combined effect of all three background sources during the individual flights. There are two interesting trends to observe in these graphs. First, there is a significant reduction in the total count rate by a factor of two to three for the dataset Sea_I compared to the datasets Lake_I and Lake_II for the same ground clearance. Considering the only minor differences in the cosmic background, this reduction can be mainly attributed to the significant reduction in the radon activity volume concentration discussed before. Second, there is a rapid increase in the total count rate within the first ~200 m above the water surface for the datasets Lake_I and Lake_II, while for the dataset Sea_I, no significant increase can be observed. This trend is the result of three competing effects as a function of the altitude: (1) the increase in the cosmic background, (2) the reduction in the radon activity volume concentration and (3) the increase in the effective cross-section for the radon background due to the extended distribution of the radon progeny in the lower atmosphere combined with the increased sensitivity of the Swiss AGRS system for photons coming from directions with $\theta' \gtrsim 120°$ (cf. Section 9.4).

To quantify the relative contribution of the background sources to the total count rate, I summed the spectral signatures of the intrinsic, cosmic and radon backgrounds, scaled with their estimated source strengths, over all pulse-height channels for all three datasets. The results are included in the Figs. 10.3–10.5, too. As anticipated from the results presented before, there are significant differences between the datasets Lake_I and Lake_II compared to the dataset Sea_I. For a ground clearance of ~100 m, the approximate contributions of the individual background sources to the total count rate in the detector channel #SUM of the Swiss AGRS system are as follows for the datasets Lake_I and Sea_I (Lake Thun / North Sea):

| | |
|---|---|
| **Cosmic** | 20 / 40 % |
| **Radon** | 60 / 35 % |
| **Intrinsic** | 20 / 25 % |

These differences in the relative contribution to the total background are also reflected in the corresponding pulse-height spectra displayed in Figs. 10.6–10.8. Specifically, the relative intensities of the cosmic and radon backgrounds can be assessed by comparing the annihilation peak at ~511 keV induced by the cosmic background to the FEP at ~609 keV related to the corresponding emission line of the radon progeny $^{214}_{83}$Bi [124].

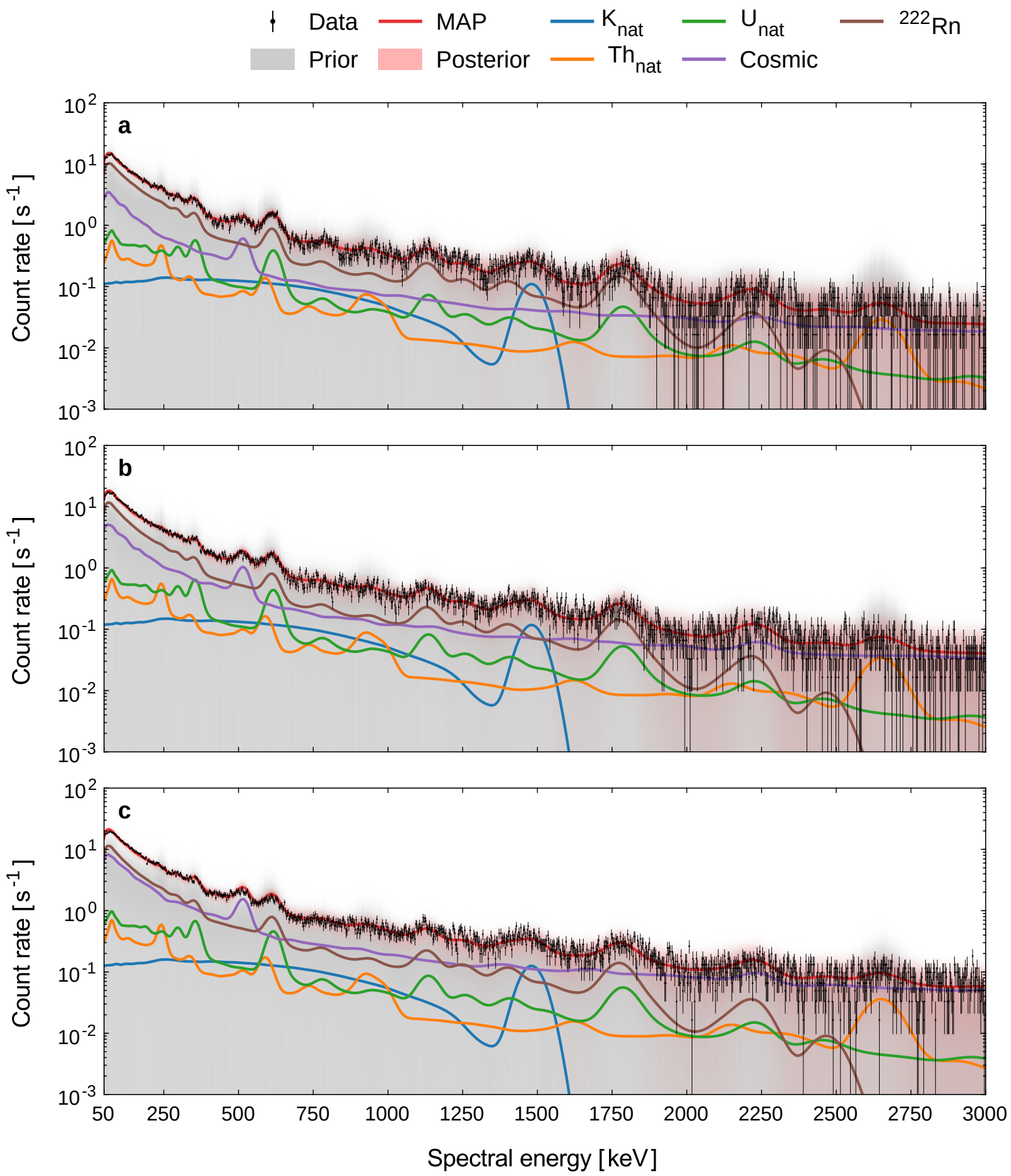

**Figure 10.6** In these graphs, I present the prior and posterior predictive distributions (color-coded) obtained by FSBI of the dataset `Lake_I` alongside the measured pulse-height spectra $\mathbf{C}_{gr}$ for three selected mean ground clearances: **a** $h_{air}$ = 104 m. **b** $h_{air}$ = 1359 m. **c** $h_{air}$ = 2061 m. For the measured spectra, the uncertainties are indicated as 1 standard deviation (SD) errorbars (cf. Appendix A.8). In addition to the posterior predictive distributions, I also display the point posterior predictions derived by the maximum a posteriori (MAP) estimates (red line). Moreover, the spectral signatures scaled by the MAP source strengths ($\hat{\mathbf{c}}\xi$) for all five sources considered in the forward model in Eq. 10.7 are indicated as well, i.e. the natural sources $K_{nat}$, $Th_{nat}$ and $U_{nat}$ associated with the intrinsic background ($\xi_{K\text{-nat}}$, $\xi_{Th\text{-nat}}$, $\xi_{U\text{-nat}}$), the cosmic background ($\xi_{cos}$) as well as the radon background ($\xi_{Rn\text{-}222}$).

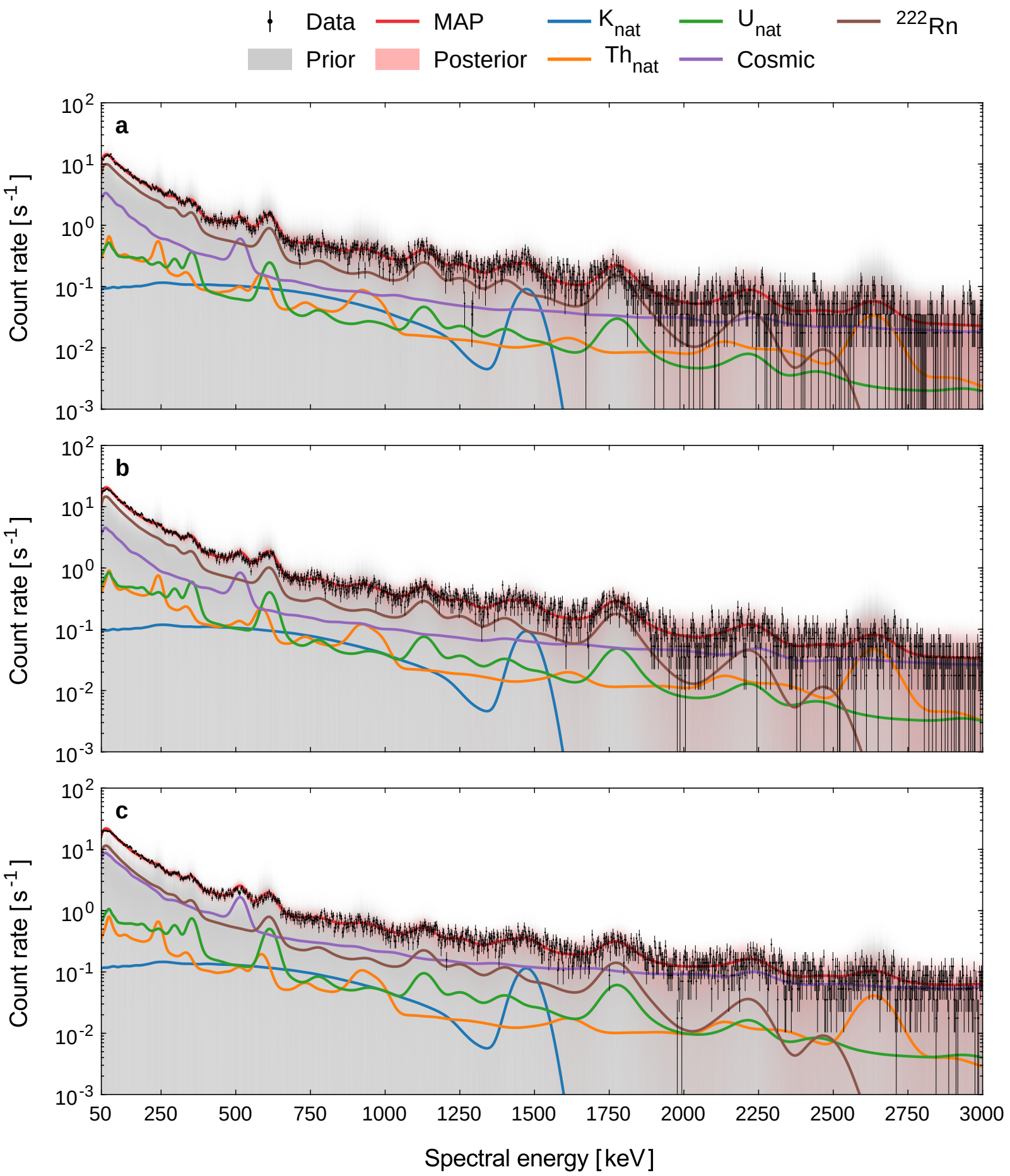

**Figure 10.7** In these graphs, I present the prior and posterior predictive distributions (color-coded) obtained by FSBI of the dataset `Lake_II` alongside the measured pulse-height spectra $\mathbf{C}_{gr}$ for three selected mean ground clearances: **a** $h_{air} = 76$ m. **b** $h_{air} = 886$ m. **c** $h_{air} = 2129$ m. For the measured spectra, the uncertainties are indicated as 1 standard deviation (SD) errorbars (cf. Appendix A.8). In addition to the posterior predictive distributions, I also display the point posterior predictions derived by the maximum a posteriori (MAP) estimates (red line). Moreover, the spectral signatures scaled by the MAP source strengths ($\hat{\mathbf{c}}\xi$) for all five sources considered in the forward model in Eq. 10.7 are indicated as well, i.e. the natural sources $K_{nat}$, $Th_{nat}$ and $U_{nat}$ associated with the intrinsic background ($\xi_{K\text{-nat}}$, $\xi_{Th\text{-nat}}$, $\xi_{U\text{-nat}}$), the cosmic background ($\xi_{cos}$) as well as the radon background ($\xi_{Rn\text{-}222}$).

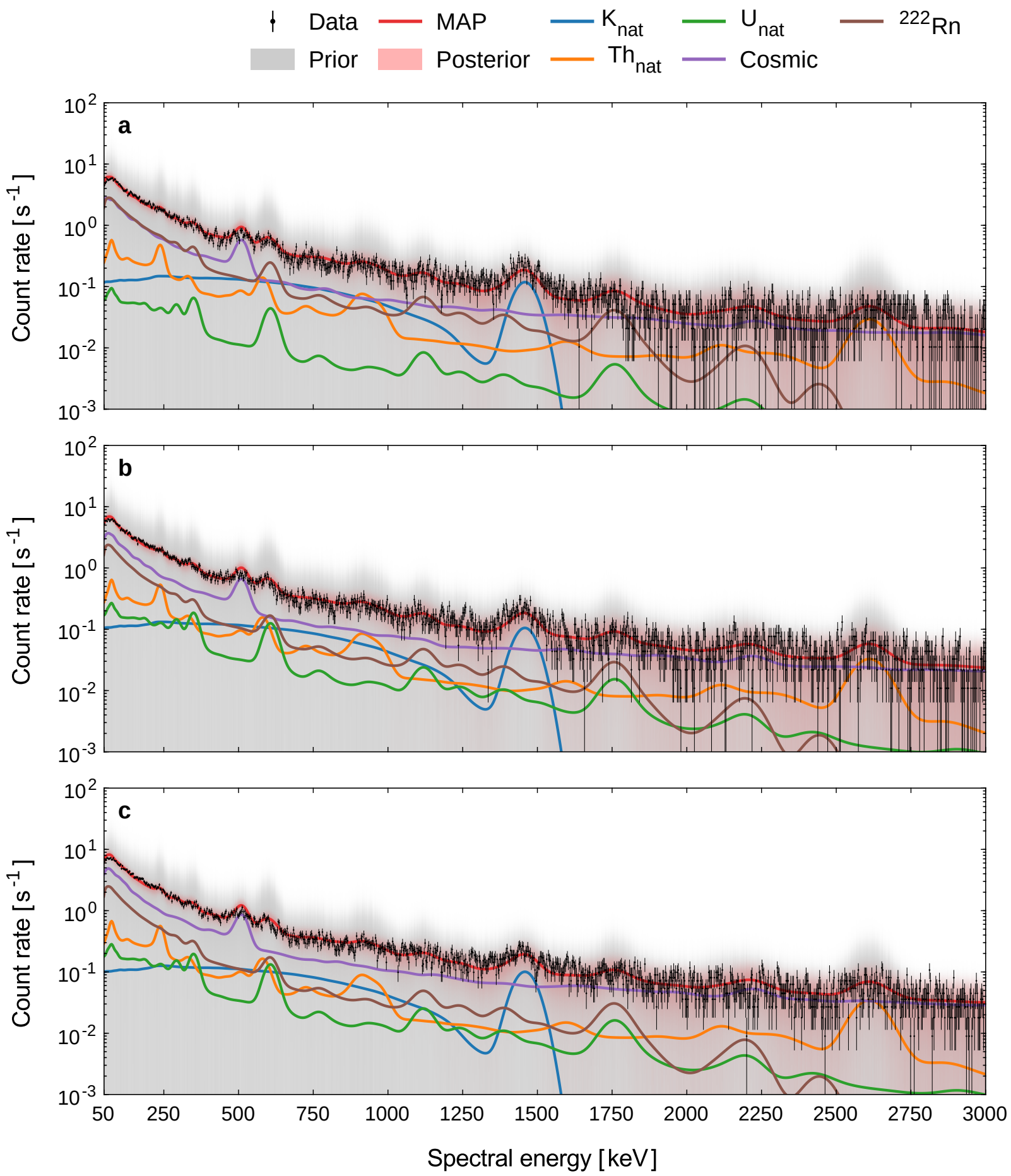

**Figure 10.8** In these graphs, I present the prior and posterior predictive distributions (color-coded) obtained by FSBI of the dataset `Sea_I` alongside the measured pulse-height spectra $\mathbf{C}_{gr}$ for three selected mean ground clearances: **a** $h_{air} = 93\,\text{m}$. **b** $h_{air} = 621\,\text{m}$. **c** $h_{air} = 1229\,\text{m}$. For the measured spectra, the uncertainties are indicated as 1 standard deviation (SD) errorbars (cf. Appendix A.8). In addition to the posterior predictive distributions, I also display the point posterior predictions derived by the maximum a posteriori (MAP) estimates (red line). Moreover, the spectral signatures scaled by the MAP source strengths ($\hat{\mathbf{c}}\xi$) for all five sources considered in the forward model in Eq. 10.7 are indicated as well, i.e. the natural sources $K_{nat}$, $Th_{nat}$ and $U_{nat}$ associated with the intrinsic background ($\xi_{K\text{-nat}}$, $\xi_{Th\text{-nat}}$, $\xi_{U\text{-nat}}$), the cosmic background ($\xi_{cos}$) as well as the radon background ($\xi_{Rn\text{-}222}$).

## 10.5 Conclusion

In this chapter, I presented a novel FSA approach that combines Bayesian inversion with numerically derived spectral signatures from high-fidelity NPSMC and DRM models presented in previous chapters. In contrast to existing FSA algorithms, the proposed method allows for the quantification of arbitrarily complex gamma-ray sources both for low and high-count data, while also effectively accounting for overdispersion effects.

The proposed method was validated with the Swiss AGRS system using radiation measurements of $^{133}_{56}Ba$ and $^{137}_{55}Cs$ point sources. The results showed excellent accuracy and precision in quantifying the activities of the employed point sources, with relative deviations <2 % for a measurement time of 1 s for both sources.

To further demonstrate the capabilities of the new method, I applied it to quantify the cosmic and radon background in the atmosphere as well as the intrinsic background of the Swiss AGRS system during a series of AGRS measurement flights conducted over Lake Thun and the North Sea. The estimated cosmic flux was in good agreement with predictions obtained by the `PARMA` code. The general trend in the radon activity volume concentration profiles was consistent with previous studies at lower altitudes.

In contrast, unlike the results reported in previous studies [88, 853], the radon activity volume concentration over Lake Thun remained almost constant between a ground clearance of ~500 m and ~2000 m, indicating a significant enhancement of the radon activity volume concentration at these altitudes. The enhanced radon levels may be attributed to the unique topographical and geological features of the Swiss Alps around Lake Thun (cf. Fig. B.112). In addition, I found also a significant reduction in the radon activity volume concentration by a factor of three to four over the North Sea compared to the measurements over Lake Thun. Further studies are required to verify the observed trends.

The developed FSBI method is a powerful tool that has proven to accurately quantify arbitrarily complex terrestrial radionuclide sources at low count rates. Furthermore, I have demonstrated that the method can be used to probe the cosmic flux and radon activity volume concentrations in the atmosphere. This offers a new pathway for real-time background correction in AGRS systems. Real-time background correction with FSBI would not only save time and costs, but also reduce systematic errors introduced by the empirical calibration methods [120, 642, 676], making current empirical background correction methods obsolete. In addition, probing the cosmic flux and radon activity volume concentrations in the atmosphere is not only of interest to the AGRS community, but also opens new possibilities for broader environmental monitoring applications.

That said, it is important to acknowledge that the Bayesian inversion method presented in this chapter is computationally intensive with a single inversion taking $\mathcal{O}(10^1)$ s to $\mathcal{O}(10^3)$ s on a local workstation, depending on the complexity of the quantification problem. As a result, this method is currently only suitable for offline applications. For online applications, a computationally more efficient approach could involve simplified MLE methods combined with numerically derived spectral signatures from high-fidelity NPSMC and DRM models. Meanwhile, offline postprocessing would be performed with the more accurate Bayesian inversion method presented in this chapter. Further research is needed to explore the viability and effectiveness of such a hybrid approach for AGRS systems.

# Part V

# Conclusion & Outlook

> "*One never notices what has been done; one can only see what remains to be done.*
>
> — *Marie Curie*

# Chapter 11
# Conclusion & Outlook

THE main objective of this work was to develop and validate a Monte Carlo based full spectrum modeling approach for the Swiss airborne gamma-ray spectrometry (AGRS) system. In this last chapter, I will reflect on the obtained results in light of this main objective, discuss the performance and limitations of the derived methodology and outline implications for the establishment of a new standard in calibration and data evaluation of AGRS systems.

In the second part, I will discuss the necessary steps for integrating the developed methods into the routine calibration and data evaluation protocols of the Swiss AGRS system. This section will also include a comprehensive analysis of open questions and challenges that need to be addressed in the future to further improve the accuracy and applicability of the developed Monte Carlo based full spectrum modeling approach not only for the Swiss AGRS system but also for other AGRS systems worldwide.

## 11.1 Conclusion

As reviewed in Chapter 5, Monte Carlo based full spectrum modeling has the potential to overcome the main limitations of current calibration and data evaluation methods for AGRS systems proposed by the International Atomic Energy Agency (IAEA). Successfully implementing this approach would achieve an extended range of applicability for AGRS systems with enhanced accuracy and increased sensitivity, particularly for detecting radionuclides with low-energy photon emissions expected to be released in radiological incidents like severe nuclear accidents or nuclear weapon explosions. Given the importance of these applications in the context of emergency response, the development of such a new standard for AGRS data analysis is of great interest not only to the scientific community in AGRS but also to the relevant governmental and regulatory authorities worldwide.

To address the shortcomings of the previous studies discussed in Section 5.5.2, this work proposes a novel Monte Carlo based full spectrum modeling approach, incorporating two major advancements:

1. The development and validation of a Monte Carlo model that combines the highest fidelity physics models available with a detailed mass model of the entire aircraft to accurately simulate the full spectrum response of the Swiss AGRS system to arbitrary gamma-ray source fields.
2. The derivation and validation of a detector response model that accurately emulates the simulated full spectrum response of the derived Monte Carlo model with an evaluation time of $\mathcal{O}(1)$ s.

Following a bottom-up modeling approach, I demonstrated in Chapter 6 that combining a high-fidelity multi-purpose Monte Carlo code with a detailed mass model of the gamma-ray spectrometer can reproduce the measured pulse-height spectra under laboratory conditions with a median relative deviation <10 % over the entire energy range of the detector, particularly at low energies. Yet statistical analysis revealed significant deviations between the simulated and measured pulse-height spectra around the Compton edge, with spectral shifts ≥20 keV, attributed to scintillation non-proportionality of the detector material.

To address this deficiency, a novel non-proportional scintillation Monte Carlo (NPSMC) model was derived in Chapter 7. Given that the scintillation non-proportionality of each detector material is unique [413], I developed a novel calibration method, Compton edge probing, that combines Bayesian inference with a custom machine learning trained vector-valued emulator to accurately infer the non-proportional scintillation model (NPSM) parameters of the detector material. Unlike previous methods, Compton edge probing eliminates the need for additional equipment by relying solely on measured and simulated pulse-height spectra. Laboratory-based valida-tion measurements confirmed that the calibrated NPSMC model successfully corrects the spectral shifts around the Compton edges observed with standard proportional scintillation Monte Carlo (PSMC) models. As discussed in Section 7.4, thanks to the general nature of the adopted NPSM, the proposed NPSMC methodology, while developed for gamma-ray spectrometry, is broadly applicable to any combination of inorganic scintillator and ionizing radiation field. This extends its utility to areas such as dark matter research [817, 818], total absorption spectroscopy [693] and remote sensing [819, 820].

After successfully developing and validating the necessary physics models to accurately simulate the full spectrum response of the gamma-ray spectrometer, I extended the model scope to the entire environment including the aircraft system in Part III. For that purpose, I combined the validated NPSMC methodology with a detailed mass model of the entire Aérospatiale AS332M1 Super Puma (TH06) helicopter. Extensive field measurements with the Swiss AGRS system showed that the developed Monte Carlo model accurately reproduces measured pulse-height spectra, achieving a median relative deviation <8 %, even for worst-case source-detector scenarios involving significant attenuation from the aircraft fuselage and fuel tanks. With these results, I demonstrated that Monte Carlo based full spectrum modeling, by combining high-fidelity physics models and detailed mass models of the entire aircraft, provides superior accuracy compared to current empirical-based calibration methods discussed in Section 5.5.1.

With the first objective fulfilled, what remained was to find a way to increase the computational performance of the simulations to make the developed Monte Carlo model applicable for routine calibration and data evaluation tasks of AGRS systems. For that purpose, I adopted a surrogate modeling approach in Chapter 9 originally developed for astrophysics and planetary science applications that emulates the full spectrum response predicted by NPSMC. The derived detector response model was benchmarked with the validated AGRS Monte Carlo model, showing good accuracy with a maximum median relative deviation <6 %. By adopting array programming, the derived detector response model enables rapid computation of the full spectrum detector response with evaluation times of $\mathcal{O}(1)$ s per spectrum on a local workstation. This represents a significant reduction in the computation time by a factor of at least $\mathcal{O}(10^4)$ compared to a brute-force Monte Carlo simulation performed on a computer cluster. With this, I have demonstrated that Monte Carlo based full spectrum modeling can not only achieve the necessary level of accuracy but also the computational performance to challenge current empirical-based calibration methods discussed in Section 5.5.1.

The developed detector response model marks not only an essential step towards full spectrum simulation-based calibration of AGRS systems, but also allows for detailed analyses of the AGRS system's spectral response to external photon fields. For the Swiss AGRS system, I found a significant anisotropy both in the full energy peak and the Compton continuum spectral band, suggesting that the influence of the aircraft structure extends beyond a specific range of photon directions and instead affects the detector's response across the entire $4\pi$ solid angle. Due to the similar design and aircraft specifications among various AGRS systems, similar impacts of the aircraft structure on the detector response can be expected. The magnitude of these effects may vary based on scintillation crystal positioning, with greater reductions in full energy peak response for crystals inside the cabin and increased Compton continuum response for crystals mounted externally.

Since no data evaluation framework exists to incorporate the newly developed surrogate models for AGRS applications, it became clear that a new data evaluation approach is required to fully exploit the capabilities of the Monte Carlo based full spectrum modeling approach. Therefore, motivated by the promising results obtained in the previous parts, I extended the scope of this work to include a proof-of-concept of a new data evaluation approach that incorporates the derived numerical models, marking a third key innovation:

3. The development and validation of a data evaluation methodology for AGRS to accurately quantify arbitrarily complex radioactive sources using a Monte Carlo based full spectrum modeling approach.

To achieve this, I developed a novel spectral inversion methodology that combines Bayesian inference with numerically calibrated spectral signature matrices derived by the detector response model. The new quantification algorithm was validated with the Swiss AGRS system using a series of field measurements, showing excellent accuracy and precision in quantifying the activities of the deployed radionuclide sources with relative deviations <2 % for a measurement time of 1 s. This could be demonstrated for both high-energy and low-energy photon sources, highlighting the versatility and robustness of the developed methodology.

To further demonstrate the capabilities and potential of the methodology, I applied the Bayesian inversion method to quantify the cosmic and radon background in the atmosphere during a series of measurement flights conducted over Lake Thun and the North Sea with the Swiss AGRS system. The estimated cosmic flux was in good agreement with predictions obtained by the PHITS-based Analytical Radiation Model in the Atmosphere (`PARMA`) code [146]. The general magnitude and trend in the radon activity volume concentration profiles were consistent with previous studies.

In conclusion, with the results presented in this work, I have demonstrated that a Monte Carlo based full spectrum modeling approach does improve both accuracy and range of applications of AGRS systems, thereby overcoming the main limitations of the current empirical-based calibration and data evaluation methods discussed in the introduction at the beginning of this document:

I. **Calibration** In contrast to existing empirical-based calibration methods, the developed Monte Carlo based full spectrum modeling approach enables the calibration for any arbitrary source geometry and energy distribution, thereby significantly extending the range of sources that can be calibrated and ultimately quantified using AGRS. In addition, once established, Monte Carlo based calibration offers a substantial reduction in both the cost and time required for calibration, while also minimizing systematic errors associated with varying radiation backgrounds and analytical corrections.

II. **Physics Modeling** By exploiting all the information contained in the measured pulse-height spectra, the developed Monte Carlo based full spectrum modeling approach significantly increases the sensitivity and accuracy in the data evaluation. This offers the potential to quantify arbitrarily complex radioactive sources over the entire spectral range of the detector, including low-energy terrestrial sources or cosmic and radon backgrounds in the atmosphere.

Given these benefits and considering the results obtained in this work, Monte Carlo based full spectrum modeling has the potential to not only replace the current calibration and data evaluation methods adopted for the Swiss AGRS system but also become a new standard worldwide for AGRS.

## 11.2 Outlook

As outlined in Chapter 1, this work marks only the first step in integrating a Monte Carlo based full spectrum modeling approach in the routine calibration and data evaluation protocols of the Swiss AGRS system. Further steps are needed to achieve full integration and operational implementation.

First, regarding the detector response models derived in Chapter 9, the current implementation is limited to simplified source-detector configurations. The existing database of double-differential photon flux models and detector response functions (DRFs) needs to be extended to account for the full range of sources and source-detector configurations encountered in practice. As part of this work, I developed a software toolbox containing the various data processing pipelines to facilitate this process.

Second, regarding the data evaluation, I limited the scope of this work to the quantification task in a proof-of-concept study. To fully integrate the developed Monte Carlo based full spectrum modeling approach into the routine data evaluation protocols of the Swiss AGRS system, the developed Bayesian inversion methods need to be extended to address the full range of data evaluation tasks required in practice for AGRS, including localization and identification of the sources. While Bayesian inference is powerful for these tasks [654, 658, 737, 798, 854], it is computationally intensive, making it suitable for offline applications only. For online use, a simplified approach combining maximum likelihood estimation with Monte Carlo-derived spectral signatures could be explored, with more accurate Bayesian methods reserved for offline postprocessing.

Third, regarding background correction, I have demonstrated in Chapter 10 that Bayesian inversion methods can be used to quantify the cosmic and radon background in the atmosphere. However, the results are based on a limited number of flights, all performed over water bodies. To fully exploit the Monter Carlo based full spectrum modeling approach, one may consider developing a background correction method that quantifies the cosmic and radon background in the atmosphere during survey flights over land in real-time. This would need however further investigations to determine the accuracy and applicability of such an approach.

Although this work focused on the Swiss AGRS system, given the similar design of the gamma-ray spectrometers and aircraft platforms among various AGRS systems, the developed Monte Carlo based full spectrum modeling approach is broadly applicable to other AGRS systems worldwide. The main challenge in transferring the developed methodology to other AGRS systems is the considerable effort required to build the Monte Carlo models for the specific system configurations and source-detector scenarios encountered in practice. This is particularly true for the detailed mass models of the aircraft system, which are essential for accurately simulating the full spectrum response of the gamma-ray spectrometer. To address this challenge, the adoption of CAD-based mass models could be considered in the future to streamline the process and minimize development time [277–280].

Furthermore, the developed Monte Carlo based full spectrum modeling approach is not restricted to AGRS systems mounted on manned aircraft but may also be applied to unmanned aerial vehicle (UAV)-based systems. Thanks to the reduced size and complexity of UAV platforms compared to manned aircraft, detailed mass models of the entire UAV system can be developed with reduced modeling effort and time compared to manned aircraft systems, making the Monte Carlo based full spectrum modeling approach especially attractive for these type of systems.

Last but not least, several critical questions and potential research directions emerge that extend beyond the immediate focus on the Swiss AGRS system but are of broader scientific importance. First, while this work addresses major systematic uncertainties, a more comprehensive uncertainty quantification could be achieved using surrogate modeling or Total Monte Carlo (TMC) sampling techniques, given the complexity and computational demands of the derived numerical models [293, 295]. Second, while this work has focused on the data obtained from the detector channel `#SUM`, there is potential to leverage data obtained from the four individual detector channels for enhanced data analysis, including directional gamma-ray imaging methods to improve source localization in the future [855, 856]. Lastly, the Monte Carlo based full spectrum modeling approach offers significant promise not only for cosmic flux and radon quantification but also for broader environmental monitoring applications. Future work could explore the extension of this capability to detect other atmospheric radiation phenomena, such as terrestrial gamma-ray flashes (TGFs) [201] or trace airborne radionuclides [24]. Further investigation is needed to validate and expand these applications effectively.

# Appendices

# Appendix A
# Supplementary Information

## Contents

## A.1 Primordial Activity Concentrations on Earth

To assess the significance of primordial radionuclides as gamma-ray emitters, I provided activity concentration values for the Earth's crust and sea in Table 2.1. In this section, I will briefly discuss, how these values can be derived based on best-estimate isotopic and elemental abundance data.

Combining the definitions for the specific activity $a$ (cf. Eq. 2.10), the activity mass concentration $a_{\mathrm{m}}$ (cf. Eq. 2.11a) and the half-life $t_{1/2}$ (cf. Eq. 2.16) for a given radionuclide $i$, we can compute the activity mass concentration for the corresponding element with atomic number $Z$ as follows:

$$a_{\mathrm{m},i} = a_i x_{\mathrm{iso},i} \frac{M_i}{M_Z} = \frac{\log(2) \mathrm{N_A} x_{\mathrm{iso},i}}{t_{1/2,i} M_Z} \tag{A.1}$$

where[1]:

| | | |
|---|---|---|
| $a_i$ | specific activity of radionuclide $i$ | $\mathrm{Bq\,kg^{-1}}$ |
| $M_i$ | molar mass for given radionuclide $i$ | $\mathrm{kg\,mol^{-1}}$ |
| $M_Z$ | molar mass for the element $Z$ associated with the radionuclide $i$ | $\mathrm{kg\,mol^{-1}}$ |
| $\mathrm{N_A}$ | Avogadro constant (cf. Constants) | $\mathrm{mol^{-1}}$ |
| $t_{1/2,i}$ | half-life for given radionuclide $i$ | s |
| $x_{\mathrm{iso},i}$ | natural isotopic abundance of radionuclide $i$ | $\mathrm{mol\,mol^{-1}}$ |

By adopting the natural elemental abundance in the Earth's crust $w_{\mathrm{crust},Z}$ and the natural elemental abundance in the sea $c_{\mathrm{sea},Z}$ for the associated element, we can compute the average activity mass concentration in the Earth's crust $a_{\mathrm{crust},i}$ and the average activity volume concentration in the sea $a_{\mathrm{sea},i}$ for a given radionuclide $i$ as follows:

$$a_{\mathrm{crust},i} = \frac{\log(2) \mathrm{N_A} x_{\mathrm{iso},i} w_{\mathrm{crust},Z}}{t_{1/2,i} M_Z} \tag{A.2a}$$

$$a_{\mathrm{sea},i} = \frac{\log(2) \mathrm{N_A} x_{\mathrm{iso},i} c_{\mathrm{sea},Z}}{t_{1/2,i} M_Z} \tag{A.2b}$$

where[2]:

1 *Please note that, to avoid confusion with other unitless concentration quantities, I state the unit of $x_{\mathrm{iso},i}$ explicitly.*

2 *Please note that, to avoid confusion with other unitless concentration quantities, I state the unit of $w_{\mathrm{crust},Z}$ explicitly.*

| | | |
|---|---|---|
| $c_{\text{sea},Z}$ | natural elemental abundance in the sea for the element $Z$ associated with the radionuclide $i$ | $\text{kg}\,\text{L}^{-1}$ |
| $w_{\text{crust},Z}$ | natural elemental abundance in the Earth's crust for the element $Z$ associated with the radionuclide $i$ | $\text{kg}\,\text{kg}^{-1}$ |

Using Eqs. A.2a and A.2b, we can estimate the activity concentration values in the Earth's crust and the sea for the primordial radionuclides listed in Table 2.1. For $x_{\text{iso},i}$, $M_Z$, $w_{\text{crust},Z}$ and $c_{\text{sea},Z}$, I adopted best-estimate values provided by Meija et al. [56], Prohaska et al. [77] and Rumble [57], respectively. Nuclear data sources for the half-life values $t_{1/2,i}$ are indicated in Table 2.1.

## A.2 Natural Radionuclide Activity in the Biosphere

To assess the activity concentration in biota, i.e. plants and animals, I provided activity mass concentration values for the biosphere in Table 2.2. In this section, we will briefly discuss, how these values can be derived based on the radionuclide transfer methodology established and standardized by the IAEA [72–74] and the International Commission on Radiological Protection (ICRP) [75].

I start by introducing $CR_{b,i}$ as the concentration ratio for biota b in a terrestrial ecosystem for a radionuclide $i$:

$$CR_{b,i} = \frac{a_{m,b,i}}{a_{m,soil,i}} \tag{A.3}$$

where:

| | | |
|---|---|---|
| $a_{m,b,i}$ | activity mass concentration in the entire biota b (per fresh mass) | $Bq\,kg^{-1}$ |
| $a_{m,soil,i}$ | activity mass concentration in the soil (per fresh mass) | $Bq\,kg^{-1}$ |

This standardized quantity characterizes the steady-state transfer of a radionuclide $i$ from the soil to the biota b. It was intensively studied in the past and is readily available for all relevant radionuclides via open-access databases [72, 73, 75]. It is important to point out that the concentration ratio approach is not applicable to non-equilibrium conditions that may occur after short-term radionuclide releases to the environment, such as nuclear weapons tests or radiological accident scenarios [74]. Using $CR_{b,i}$, we can compute $a_{m,b,i}$ for known $a_{m,soil,i}$ values as follows:

$$a_{m,b,i} = CR_{b,i} a_{m,soil,i} \tag{A.4}$$

To compute $a_{m,b,i}$ for all biota b and the primordial radionuclides $^{40}_{19}K$, $^{232}_{90}Th$ and $^{238}_{92}U$ listed in Table 2.2, I used minimum and maximum $a_{m,soil,i}$ values for the soil (global estimate provided in Table 2.2) and combined those with concentration ratios obtained from the Wildlife Transfer Parameter Database (WTPD) (https://www.wildlifetransferdatabase.org/, version: 2015-04-20) [73]. To ensure conservative estimates in Table 2.2, I paired the minimum/maximum $a_{m,soil,i}$ values together with the corresponding minimum/maximum $CR_{b,i}$ values. The mean point estimates were derived by combining mean $a_{m,soil,i}$ and $CR_{b,i}$ values.

## A.3 Upper Limit on Cosmogenic Radionuclide Activity Concentrations in the Atmosphere

As discussed in Section 2.1.3.3, activity concentration measurements in the atmosphere for cosmogenic radionuclides with low production rates and short half-lives are challenging. Consequently, empirical activity concentration data for corresponding radionuclides listed in Table 2.3 are not readily available. In contrast, the production rates of cosmogenic radionuclides in the atmosphere are well known [94–101]. Consequently, to estimate an upper limit on the activity concentration for cosmogenic radionuclides in the atmosphere, I formulate the production and kinetics of cosmogenic radionuclides in the atmosphere as an initial value problem using known production rates averaged over time and space.

Similar to Lal et al. [857] or Beno et al. [101], I start modeling the change in the activity volume concentration $a_{\mathrm{v},i}$ of a cosmogenic radionuclide $i$ in the atmosphere as a function of time $t$ by applying an adapted Bateman model [46] to our initial value problem as follows (cf. Section 2.1.2):

$$\frac{1}{\lambda_i}\frac{\mathrm{d}a_{\mathrm{v},i}}{\mathrm{d}t}(t) = Q_{\mathrm{cosm},i} - a_{\mathrm{v},i}(t) \tag{A.5}$$

where:

| | | |
|---|---|---|
| $a_\mathrm{v}$ | activity volume concentration | $\mathrm{Bq\,m^{-3}}$ |
| $Q_\mathrm{cosm}$ | cosmogenic radionuclide generation rate | $\mathrm{m^{-3}\,s^{-1}}$ |
| $\lambda$ | decay constant | $\mathrm{s^{-1}}$ |

In accordance with our aim to provide an upper limit on the activity volume concentration $a_{\mathrm{v},i}$, we consider only nuclear decay, neglecting all other loss terms such as settling or spallation. We continue by solving the ordinary differential equation (ODE) in Eq. A.5 for $a_{\mathrm{v},i}(t)$ using a separation of variables approach:

$$a_{\mathrm{v},i}(t) = C' \mathrm{e}^{-\lambda_i t} + Q_{\mathrm{cosm},i} \tag{A.6}$$

where $C'$ is a constant of integration. Using the initial condition $a_{\mathrm{v},i}(t = 0) = 0$, we get $C' = -Q_{\mathrm{cosm},i}$ and as a result:

$$a_{\mathrm{v},i}(t) = Q_{\mathrm{cosm},i}\left(1 - \mathrm{e}^{-\lambda_i t}\right) \tag{A.7}$$

We see from Eq. A.7 that the activity volume concentration $a_{\mathrm{v},i}(t)$ increases with time and approaches an equilibrium activity concentration $a_{\mathrm{v},i}^{\mathrm{eq}}$. We can calculate this equilibrium value by taking the limit of Eq. A.7 as $t$ goes to infinity:

$$a_{\mathrm{v},i}^{\mathrm{eq}} = \lim_{t\to\infty} a_{\mathrm{v},i}(t) = Q_{\mathrm{cosm},i} \tag{A.8}$$

From Eq. A.8 we see that the equilibrium activity concentration $a_{\mathrm{v},i}^{\mathrm{eq}}$ is equal to the cosmogenic radionuclide generation rate $Q_{\mathrm{cosm},i}$. In other words, the activity volume concentration $a_{\mathrm{v},i}$ of a cosmogenic radionuclide $i$ approaches the cosmogenic radionuclide generation rate $Q_{\mathrm{cosm},i}$ for $t \gg 1/\lambda_i$. It is interesting to note that $a_{\mathrm{v},i}^{\mathrm{eq}}$ is independent of the decay constant $\lambda_i$, and therefore also of the half-life $t_{1/2,i} = \log(2)/\lambda_i$ (cf. Section 2.1.2).

In the literature, the generation of cosmogenic radionuclides in the atmosphere is often quantified not by the cosmogenic radionuclide generation rate $Q_{\mathrm{cosm},i}$ but by the cosmogenic radionuclide generation rate density $q_{\mathrm{cosm},i}$ in $\mathrm{m^{-2}\,s^{-1}}$ [94, 97, 99, 100, 103–105, 115, 857–859]. Similar to Masarik et al. [97], I convert the cosmogenic radionuclide generation rate density $q_{\mathrm{cosm},i}$ into a cosmogenic radionuclide generation rate $Q_{\mathrm{cosm},i}$ as follows:

$$Q_{\mathrm{cosm},i} = \frac{q_{\mathrm{cosm},i}\,\rho_{\mathrm{atm}}}{d_{\mathrm{atm}}} \tag{A.9}$$

where:

| | | |
|---|---|---|
| $d_{\mathrm{atm}}$ | atmospheric depth (cf. Eq. 2.26a) | $\mathrm{kg\,m^{-2}}$ |
| $q_{\mathrm{cosm}}$ | cosmogenic radionuclide generation rate density | $\mathrm{m^{-2}\,s^{-1}}$ |
| $\rho_{\mathrm{atm}}$ | atmospheric mass density | $\mathrm{kg\,m^{-3}}$ |

Combining Eqs. 2.26b and A.9, I reformulate Eq. A.8 as follows:

$$a_{\mathrm{v},i}^{\mathrm{eq}} = \frac{q_{\mathrm{cosm},i}\,\rho_{\mathrm{atm}}\,g}{p_{\mathrm{atm}}} \tag{A.10}$$

where:

| | | |
|---|---|---|
| $g$ | standard gravity (cf. Constants) | $\mathrm{m\,s^{-2}}$ |
| $p_{\mathrm{atm}}$ | atmospheric pressure | Pa |

I use this result to estimate an upper limit on the activity concentration of the cosmogenic radionuclides $^{7}_{4}\mathrm{Be}$, $^{18}_{9}\mathrm{F}$, $^{22}_{11}\mathrm{Na}$, $^{24}_{11}\mathrm{Na}$, $^{28}_{12}\mathrm{Mg}$, $^{26}_{13}\mathrm{Al}$,

$^{28}_{13}$Al, $^{31}_{14}$Si, $^{34m}_{17}$Cl, $^{34}_{17}$Cl, $^{38}_{17}$Cl, $^{39}_{17}$Cl, $^{81}_{36}$Kr and $^{129}_{53}$I in Table 2.3 based on best-estimate cosmogenic radionuclide generation rate density values $q_{\mathrm{cosm},i}$ provided by Ferronsky et al. [103]. In line with our aim to provide an upper limit on the activity volume concentration $a_{\mathrm{v},i}$ and the reference models used in the literature [97, 103], I select the standard atmospheric mass density at sea level, i.e. $\rho_{\mathrm{atm}} = 1.2985\,\mathrm{kg\,m^{-3}}$, and a related atmospheric pressure value $p_{\mathrm{atm}}$ of $1.013\,25 \times 10^5\,\mathrm{Pa}$, respectively [38].

## A.4 Angular Distribution of Photoelectrons

As discussed in Section 3.1.2.3, photoelectrons emitted in a photoelectric absorption event can lead to secondary photon emission by bremsstrahlung. The angular distribution of these photoelectrons is of particular interest, as it determines the direction of the bremsstrahlung photons. Furthermore, photoelectrons play also an important role in the interaction with the detector material, as discussed in Chapter 4. Therefore, I briefly highlight in this section the angular distribution of photoelectrons.

For this purpose, we can use again the differential angular cross-section $\mathrm{d}\sigma_{\mathrm{pe}}/\mathrm{d}\Omega_{\mathrm{e}}$ for electrons emitted with a kinetic energy $E_{\mathrm{k}}$ at an emission angle $\theta_{\mathrm{e}^-}$ with respect to the incoming photon direction $\mathbf{\Omega}$. The most commonly used model for the differential angular cross-section of photoelectrons was derived by Fritz Sauter[3] in 1931 for light elements[4] ($Z \ll 137$) [861]:

3 *Fritz Sauter (*1906, †1983) an Austrian theoretical physicist. During his career, he made important contributions to the quantum electrodynamics (QED) by applying the Dirac theory to study the Klein paradox, photoelectric effect or Mott polarization, among others.*

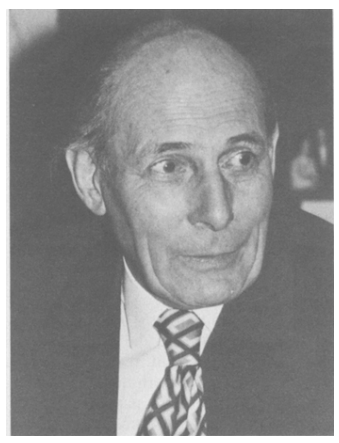

Fritz Sauter, used with permission of WILEY-VCH Verlag GmbH & Co, from [860]; permission conveyed through Copyright Clearance Center, Inc. © 1983 WILEY-VCH Verlag GmbH & Co. KGaA, Weinheim

4 *For heavier elements, more accurate numerical models are available [862–864].*

$$\frac{\mathrm{d}\sigma_{\mathrm{pe}}}{\mathrm{d}\Omega_{\mathrm{e}}}\left(E_{\mathrm{k,e}^-},\theta_{\mathrm{e}^-}\right) \propto \frac{\sin^2\theta_{\mathrm{e}^-}}{\left(1-\beta_{\mathrm{e}^-}\cos\theta_{\mathrm{e}^-}\right)^4}\Bigg[1 + \frac{1}{2}\gamma\left(\gamma-1\right)\left(\gamma-2\right)\left(1-\beta_{\mathrm{e}^-}\cos\theta_{\mathrm{e}^-}\right)\Bigg] \quad \text{(A.11)}$$

where:

| | | |
|---|---|---|
| $E_{\mathrm{k,e}^-}$ | electron kinetic energy | eV |
| $\beta_{\mathrm{e}^-}$ | ratio of the electron speed to the speed of light in vacuum | |
| $\gamma$ | Lorentz factor | |
| $\theta_{\mathrm{e}^-}$ | electron emission angle | rad |

with the Lorentz factor being defined as $\gamma = (1-\beta_{\mathrm{e}^-}^2)^{-1/2}$. $\beta_{\mathrm{e}^-}$ is related to the electron kinetic energy $E_{\mathrm{k,e}^-}$ as:

$$\beta_{\mathrm{e}^-} = \sqrt{1-\left(\frac{E_{\mathrm{k,e}^-}}{\mathrm{m_e c^2}}+1\right)^{-2}} \quad \text{(A.12)}$$

where:

| | | |
|---|---|---|
| $\mathrm{m_e}$ | electron mass (cf. Constants) | $\mathrm{eV\,s^2\,m^{-2}}$ |
| c | speed of light in vacuum (cf. Constants) | $\mathrm{m\,s^{-1}}$ |

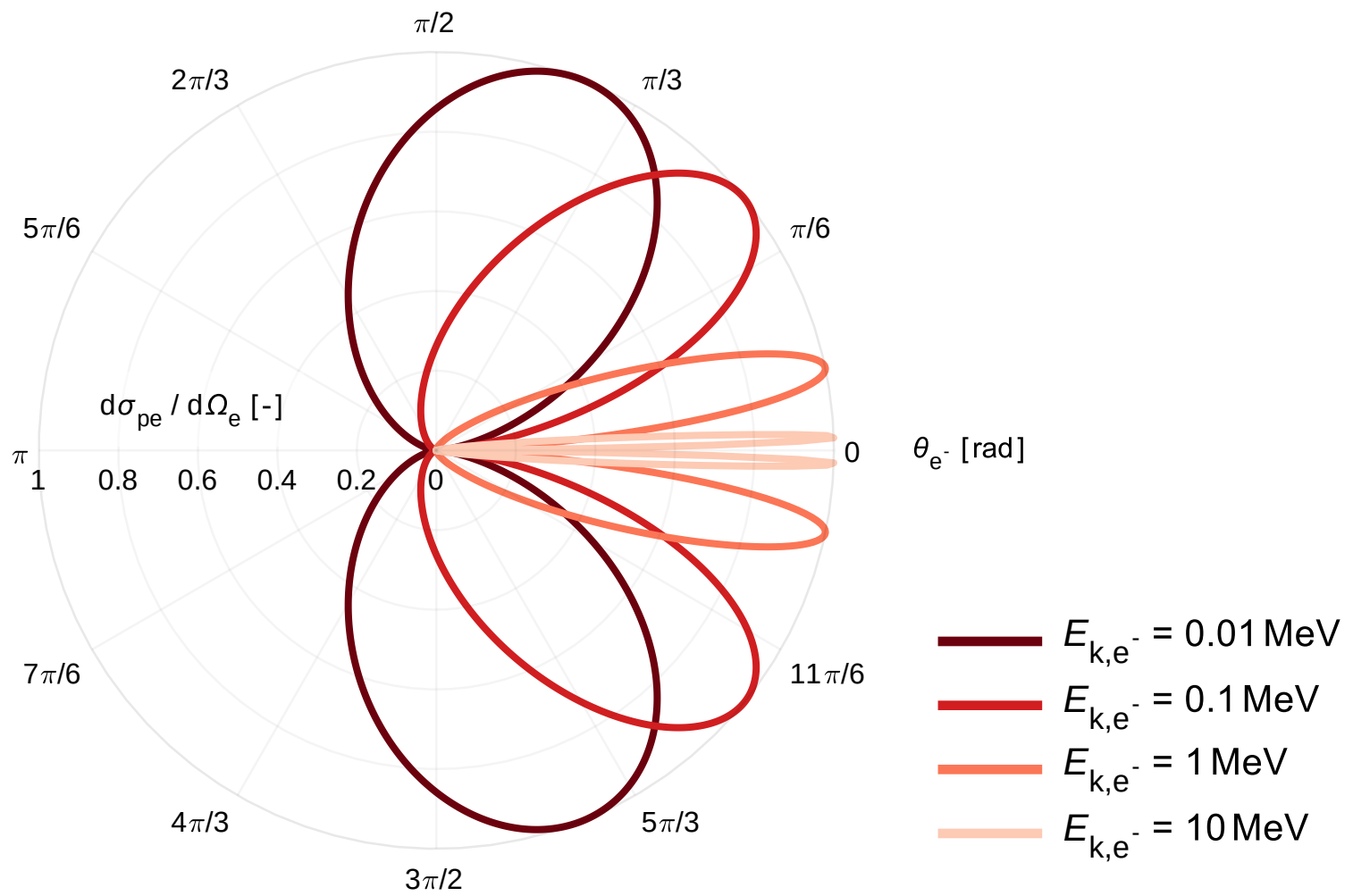

**Figure A.1** Normalized differential angular cross-section for photoelectron emission $d\sigma_{pe}/d\Omega_e$ as a function of the electron kinetic energy $E_{k,e^-}$ and the emission angle $\theta_{e^-}$ (cf. Eq. A.11). $d\sigma_{pe}/d\Omega_e$ was normalized by the corresponding peak value.

The angular cross-section in Eq. A.11 is plotted in Fig. A.1 for selected electron kinetic energies $E_{k,e^-}$. We find a similar trend as for the Compton electrons, i.e. with increasing electron kinetic energy and thereby also photon energy, the angular distribution of the photoelectrons is increasingly peaked in the forward direction. However, the angular cross-section for $\theta_{e^-} = 0$ is always zero and the most probable emission direction at low energies is perpendicular to the photon direction [206]. Moreover, due to the recoil of the atom, photoelectron emission with $\theta_{e^-} < \pi/2$ is kinematically possible, e.g. for 0.01 MeV about 34 % of the photoelectrons are emitted into the backward hemisphere with respect to the photon direction.

## A.5 Angular and Energy Distribution of the Electron-Positron Pair

As discussed in Section 3.1.2.4, electron-positron pairs emitted in pair production events can lead to secondary photon emission by bremsstrahlung, among other processes. The angular and energy distributions of these electrons and positrons are of particular interest, as they determine the direction and intensity of the bremsstrahlung photons. Furthermore, electrons and positrons produced in pair production events play also an important role in the interaction with the detector material, as discussed in Chapter 4. Therefore, I briefly highlight in this section the main properties of the emitted electron-positron pair, in particular the angular and energy distribution.

To investigate the main trend in the angular distribution of the electron/positron emission, we can use the leading term of the Sauter-Gluckstern-Hull formula describing the emission of the electron/positron in a pair production event with a (total) energy $E_{e^-} = E_{k,e^-} + m_e c^2$ and an emission angle $\theta_{e^-}$ with respect to the incoming photon direction $\mathbf{\Omega}$ into a differential solid angle $d\Omega_{e^-}$ as follows [225, 865]:

$$\frac{\partial^2 \sigma_{pp,n}}{\partial E_{e^-} \partial \Omega_{e^-}} \left(E_{k,e^-}, \theta_{e^-}\right) \propto \frac{1}{\left(1 - \beta_{e^-} \cos \theta_{e^-}\right)^2} \tag{A.13}$$

with:

| | | |
|---|---|---|
| $E_{e^-}$ | electron energy | eV |
| $\beta_{e^-}$ | ratio of the electron speed to the speed of light in vacuum (cf. Eq. 4.7) | |
| $\theta_{e^-}$ | electron emission angle | rad |

The corresponding distribution for selected electron kinetic energies $E_{k,e^-}$ is plotted in Fig. A.2. Note that for the positron emission, we can simply exchange the set of variables $\{E_{e^-}, \Omega_{e^-}, E_{k,e^-}, \theta_{e^-}, \beta_{e^-}\}$ in Eq. A.13 with $\{E_{e^+}, \Omega_{e^+}, E_{k,e^+}, \theta_{e^+}, \beta_{e^+}\}$, respectively.[5] Due to the fact that the photon momentum can be shared between the electron-positron pair as well as the interacting atomic nucleus, the electron/positron can be emitted into the full solid angle and is not restricted to the forward hemisphere, unlike the Compton electron. Similar to the photoelectron, with increasing electron/positron kinetic energy and thereby also photon energy, the angular distribution of the emitted electron/positron is increasingly peaked in the forward direction.

5 *Please note that for an accurate description of the combined angular distributions of the electron-positron pair, one has to consider the correlation between the electron and positron emission angles, e.g. by using higher order differential cross sections such as* $\partial^3\sigma_{pp,n}/\partial E_{e^-}\partial\Omega_{e^-}\partial\Omega_{e^+}$ *[225].*

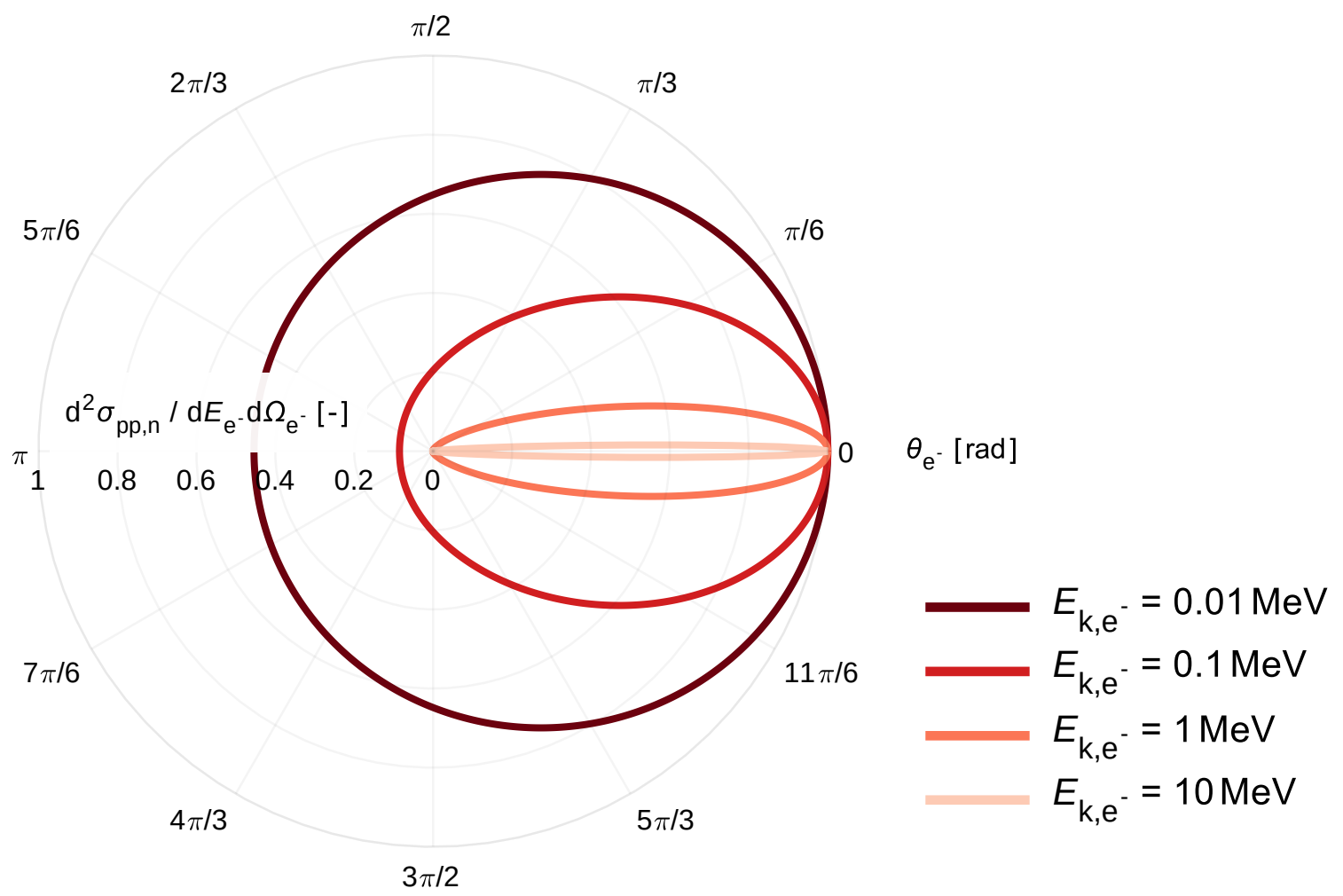

**Figure A.2** Normalized double differential cross-section for electron emission in pair production interactions in the nuclear field $\partial^2\sigma_{\mathrm{pp,n}}/\partial E_{\mathrm{e^-}}\partial\Omega_{\mathrm{e^-}}$ as a function of the electron kinetic energy $E_{\mathrm{k,e^-}}$ and the emission angle $\theta_{\mathrm{e^-}}$ (cf. Eq. A.11). $\partial^2\sigma_{\mathrm{pp,n}}/\partial E_{\mathrm{e^-}}\partial\Omega_{\mathrm{e^-}}$ was normalized by the corresponding peak value.

Not only the photon momentum but also the photon energy is shared between the electron-positron pair as well as the interacting particle as discussed in Eqs. 3.21 and 3.22. This has the important implication that the kinetic energy of the emitted electron/positron is not fixed but distributed over the transferred photon energy range. Because of the low contribution of the pair production events in the electron field for the majority of the elements, I will limit the discussion here again to the pair production events in the electric field of nuclei. To investigate the energy distribution of the emitted electron-positron pair, it is convenient to define a new energy quantity, the reduced electron/positron energy $\varepsilon_{\mathrm{e^-}}/\varepsilon_{\mathrm{e^+}}$:

$$\varepsilon_{\mathrm{e^-}} = \frac{E_{\mathrm{k,e^-}} + \mathrm{m_e c^2}}{E_\gamma} = 1 - \varepsilon_{\mathrm{e^+}} \tag{A.14a}$$

$$\varepsilon_{\mathrm{e^+}} = \frac{E_{\mathrm{k,e^+}} + \mathrm{m_e c^2}}{E_\gamma} = 1 - \varepsilon_{\mathrm{e^-}} \tag{A.14b}$$

which is simply the fraction of the photon energy $E_\gamma$ transferred to the electron and the positron, respectively (cf. Eq. 3.21). Using this new quantity and assuming $\alpha Z(1-4\alpha_\gamma^{-2})^{-1/2} \ll 1$ with $\alpha$ being the fine structure constant (cf. Constants), the differential energy cross-section $\mathrm{d}\sigma_{\mathrm{pp,n}}/\mathrm{d}E_{\mathrm{e}^-}$ of the emitted electron in a pair production event for high particle energies ($E_{\mathrm{e}^-}, E_{\mathrm{e}^+}, E_\gamma \gg \mathrm{m_e c^2}$) can be approximated as [225, 866]:[6]

6 *In addition, screening effects, the finite size of the nucleus as well as the nucleus recoil energy are neglected [225].*

$$\frac{\mathrm{d}\sigma_{\mathrm{pp,n}}}{\mathrm{d}E_{\mathrm{e}^-}}\left(\alpha_\gamma, \varepsilon_{\mathrm{e}^-}\right) \propto \frac{Z^2\left(5-2\varepsilon_{\mathrm{e}^-}\right)}{3\alpha_\gamma}\left\{\log\left[2\varepsilon_{\mathrm{e}^-}\left(1-\varepsilon_{\mathrm{e}^-}\right)\alpha_\gamma\right]-\frac{1}{2}\right\} \quad \text{(A.15)}$$

with:[7]

7 *Please note that I have rewritten the original expression in [225] by exchanging the variables $E_{\mathrm{e}^-}$ with $\varepsilon_{\mathrm{e}^-}$.*

| | | |
|---|---|---|
| $E_{\mathrm{e}^-}$ | electron energy | eV |
| $Z$ | atomic number | |
| $\alpha_\gamma$ | ratio of the photon energy to the energy-equivalent electron rest mass (cf. Eq. 3.12) | |
| $\varepsilon_{\mathrm{e}^-}$ | reduced electron energy | |

For the positron emission, we can simply exchange the set of variables $\{E_{\mathrm{e}^-}, \varepsilon_{\mathrm{e}^-}\}$ in Eq. A.15 with $\{E_{\mathrm{e}^+}, \varepsilon_{\mathrm{e}^+}\}$ [225, 866]. The distribution given by Eq. A.15 is displayed in Fig. A.3 as a function of the reduced electron energy $\varepsilon_{\mathrm{e}^-}$ for selected photon energies $E_\gamma$. We find a reduced mean electron energy of 0.5 for the entire range of photon energies, which is equivalent to mean kinetic energies of $E_\gamma/2 - \mathrm{m_e c^2}$ for both, the electron and the positron. However, the energy distribution shows increasing dispersion with increasing photon energies. It is easy to see from Eq. 3.21, that the reduced electron energy is kinematically constraint by [206]:

$$\varepsilon_{\mathrm{e}^-,\mathrm{min}} = \frac{\mathrm{m_e c^2}}{E_\gamma} \quad \text{(A.16a)}$$

$$\varepsilon_{\mathrm{e}^-,\mathrm{max}} = \frac{E_\gamma - \mathrm{m_e c^2}}{E_\gamma} = 1 - \varepsilon_{\mathrm{e}^-,\mathrm{min}} \quad \text{(A.16b)}$$

The same holds for the reduced positron energy $\varepsilon_{\mathrm{e}^+}$. Consequently, pair production events can be a significant source of high-energy electrons and positrons reaching kinetic energies in the same order as the energy of the photon inducing the pair production event. This has important implications for the photon transport as these produced particles can in turn create high-energy photons by bremsstrahlung radiation as discussed in Section 2.1.1.

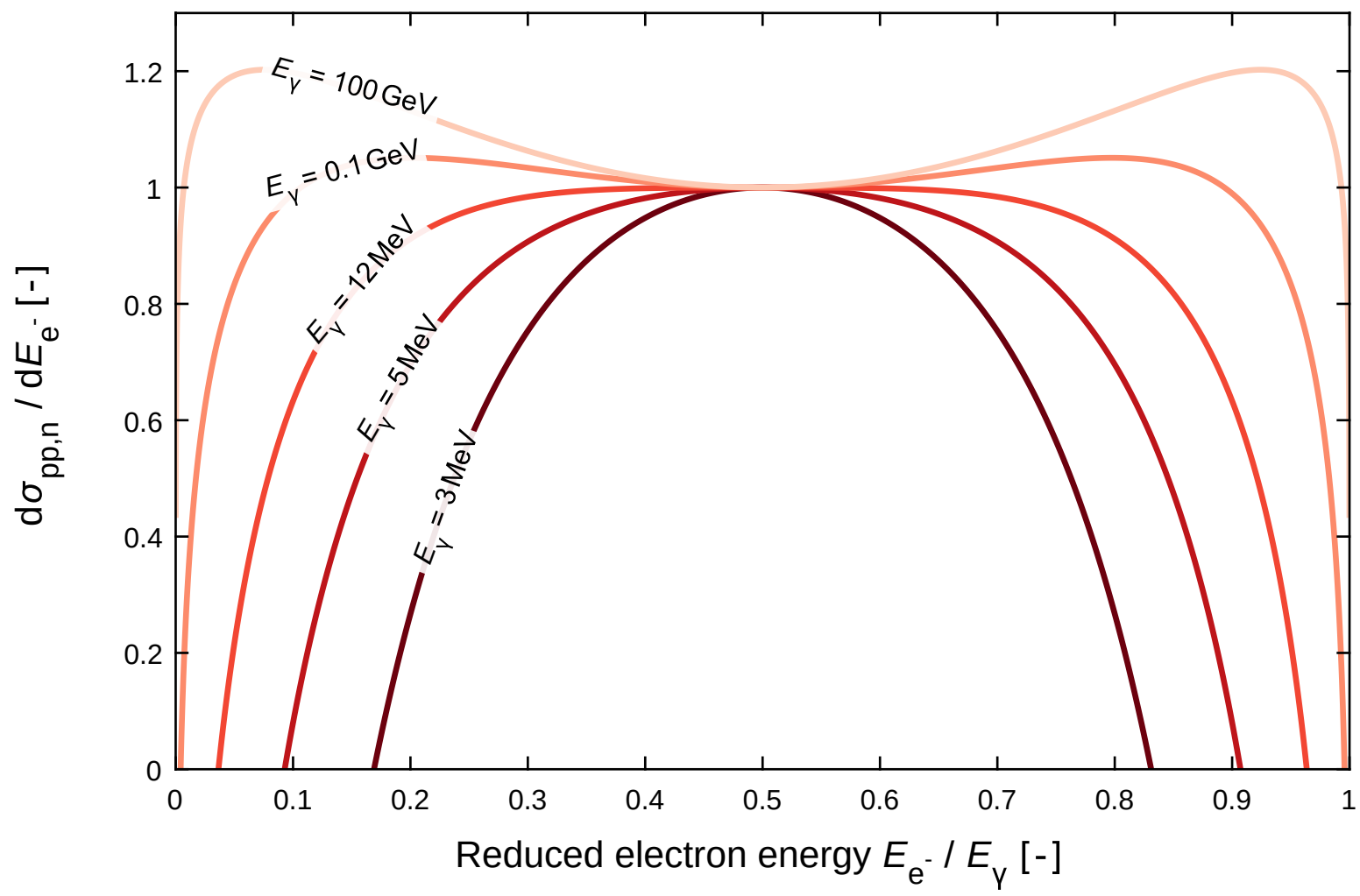

**Figure A.3** Normalized differential energy cross-section for electron emission in pair production interactions in the nuclear field $\mathrm{d}\sigma_{\mathrm{pp,n}}/\mathrm{d}E_{\mathrm{e}^-}$ as a function of the reduced electron energy $\varepsilon_{\mathrm{e}^-} = E_{\mathrm{e}^-}/E_\gamma$ and the photon energy $E_\gamma$ (cf. Eq. A.15). The energy cross-section $\mathrm{d}\sigma_{\mathrm{pp,n}}/\mathrm{d}E_{\mathrm{e}^-}(E_\gamma, \varepsilon_{\mathrm{e}^-})$ was normalized by $\mathrm{d}\sigma_{\mathrm{pp,n}}/\mathrm{d}E_{\mathrm{e}^-}(E_\gamma, \varepsilon_{\mathrm{e}^-} = 0.5)$.

## A.6 Primary Quanta Theory and the Exponential Integral Function

Based on the monoenergetic transport theory introduced in Section 3.2.1, we have discussed analytical solutions to the simplified linear Boltzmann transport equation (LBTE) given in Eq. 3.30 for selected source geometries relevant in AGRS. For the disk and the truncated cone geometries, the analytical solution for the primary energy flux was provided using the exponential integral function of the first $\mathcal{E}_1$ and second $\mathcal{E}_2$ order. In the first part of this section, I will provide a brief summary of the definition and main properties of the generalized exponential integral function. In the second part, using the definitions introduced in the first part, I will derive the primary energy flux for the disk and the truncated cone source geometries presented in Eq. 3.30. Similar derivations are readily available in the literature [10, 52].

But first, let me introduce the exponential integral function $\mathcal{E}_n$ and its properties. The exponential integral function $\mathcal{E}_n$ of the $n$th order is defined as the definite integral of the ratio between an exponential function and its argument as follows:

$$\mathcal{E}_n(x) = x^{n-1} \int_x^{\infty} \frac{\mathrm{e}^{-\tau}}{\tau^n} \,\mathrm{d}\tau \tag{A.17}$$

with $x \in \mathbb{R}$ and $n \in \mathbb{N}$.[8] Although originally introduced by Schlömilch in 1846 [867], it is often referred to as the King function [868] named after Louis Vessot King who was one of the first adopting it for calculations in the context of radiation transport [869]. The exponential integral function can also be expressed in terms of the more common upper incomplete gamma function $\Gamma$ as follows:

*8 More general definitions on the complex plane exist, i.e. $x, n \in \mathbb{C}$. However, for the solutions outlined in Section 3.2.1, the indicated definition is sufficient.*

$$\mathcal{E}_n(x) = x^{n-1}\Gamma(1-n, x) \tag{A.18a}$$

$$\Gamma(s, x) = \int_x^{\infty} \frac{\mathrm{e}^{-\tau}}{\tau^{1-s}} \,\mathrm{d}\tau \tag{A.18b}$$

with $s \in \mathbb{R}_+$. Some useful properties of $\mathcal{E}_n$ in the context of the monoenergetic transport theory are listed here [52]:

$$\lim_{x\to\infty} \mathscr{E}_n(x) = 0 \tag{A.19a}$$

$$\mathscr{E}_n(0) = \frac{1}{n-1} \qquad n > 1 \tag{A.19b}$$

$$\mathscr{E}_0(x) = \frac{e^{-x}}{x} \tag{A.19c}$$

$$\frac{d\mathscr{E}_n}{dx}(x) = -\mathscr{E}_{n-1}(x) \qquad n > 0 \tag{A.19d}$$

$$\int_x^\infty \mathscr{E}_n(\tau)\, d\tau = \mathscr{E}_{n+1}(x) \tag{A.19e}$$

$$\mathscr{E}_{n+1}(x) = \frac{1}{n}\left[e^{-x} - x\mathscr{E}_n(x)\right] \tag{A.19f}$$

More properties can be found in the monograph by Kogan et al. [52].

Let us continue with the derivation of the primary energy flux for the disk and the truncated cone source geometries presented in Section 3.2.1. The primary energy flux $d\phi_\gamma/dE_\gamma$ in a homogeneous isotropic air medium in an altitude $h_{\text{air}}$ above a disk source with source radius $R_{\text{src}}$ and source strength $Q_s$ emitting high-energy photons with energy $E_\gamma$ is obtained by solving the LBTE in Eq. 3.30 for the disk source geometry displayed in Fig. 3.8. For this purpose, it is convenient to perform the integration in cylindrical coordinates with the radius $r'$ and the azimuthal angle $\varphi$. Given that the source is reduced to a surface, the volume integral in Eq. 3.30 simplifies to a surface integral with the surface element being $r'\,dr'd\varphi$. The path length $s$ in Eq. 3.30 on the other hand is equivalent to $(h_{\text{air}}^2 + r'^2)^{1/2}$ resulting in the following surface integral:[9]

9 *Please note that for the sake of brevity, I do not explicitly denote the energy and time dependence of the source strength and attenuation coefficient variables in the subsequent expressions within this section.*

$$\frac{d\phi_\gamma}{dE_\gamma}\left(R_{\text{src}}, h_{\text{air}}, E_\gamma, t\right) = \int_0^{2\pi}\int_0^{R_{\text{src}}} \frac{Q_s e^{-\mu_{\text{air}}\sqrt{h_{\text{air}}^2 + r'^2}}}{4\pi\left(h_{\text{air}}^2 + r'^2\right)} r'\, dr' d\varphi \tag{A.20a}$$

$$= \frac{Q_s}{2}\int_0^{R_{\text{src}}} \frac{e^{-\mu_{\text{air}}\sqrt{h_{\text{air}}^2 + r'^2}}}{h_{\text{air}}^2 + r'^2} r'\, dr' \tag{A.20b}$$

with:

| | | |
|---|---|---|
| $h_{air}$ | ground clearance | m |
| $Q_s$ | external isotropic surface source strength | $s^{-1}\,m^{-2}\,eV^{-1}$ |
| $R_{src}$ | source radius | m |
| $\mu_{air}$ | total attenuation coefficient in air | $m^{-1}$ |

We can further simplify the integral in Eq. A.20b by substituting $\tau = \mu_{air}(h_{air}^2 + r'^2)^{1/2}$ and consequently $d\tau = \mu_{air}^2 r'\,dr'/\tau$ resulting in:

$$\frac{d\phi_\gamma}{dE_\gamma}\left(R_{src}, h_{air}, E_\gamma, t\right) = \frac{Q_s}{2} \underbrace{\int\limits_{\mu_{air}h_{air}}^{\mu_{air}\sqrt{h_{air}^2+R_{src}^2}} \frac{e^{-\tau}}{\tau}\,d\tau}_{\mathscr{E}_1[\mu_{air}h_{air}]-\mathscr{E}_1\left[\mu_{air}\sqrt{h_{air}^2+R_{src}^2}\right]} \tag{A.21}$$

It is easy to see that the highlighted integral term in Eq. A.21 can be expressed with the exponential integral function of the first order $\mathscr{E}_1$ as indicated, which leads to the final solution for the primary energy flux of a disk source given in Eq. 3.34. Based on Eq. 3.34, we can compute the primary energy flux for a disk source geometry with infinite extent ($R_{src} \to \infty$) using Eq. A.19a as discussed already in Section 3.2.1.

Moving on to the truncated cone source geometry displayed in Fig. 3.8. Based on the linearity of the LBTE (cf. Section 3.2), we can derive the primary energy flux for a truncated cone source with source depth $d_{src}$, source radius $R_{src}$ and source strength $Q_v$ by integrating a corresponding disk source with source radius $R_{src}$ and source strength $Q_s$ over the source depth $d_{src}$ while keeping the view angle of the source $\theta_{src}$ constant as follows:[10]

10 *It should be pointed out that in the original derivation by Kogan et al. [52], there are two typographical errors in the related equation 2.58, i.e. the second plus sign should be a minus sign and the upper integral limit should be denoted as* $h_3$.

$$\frac{d\phi_\gamma}{dE_\gamma}\left(d_{src}, R_{src}, h_{air}, E_\gamma, t\right) = \int\limits_0^{d_{src}} \frac{Q_v}{2}\Bigg[ \underbrace{\mathscr{E}_1\left(\mu_{air}h_{air} + \mu_{src}z\right)}_{\text{I}} - \underbrace{\mathscr{E}_1\left(\frac{\mu_{air}h_{air} + \mu_{src}z}{\cos(\theta_{src})}\right)}_{\text{II}}\Bigg] dz \tag{A.22}$$

where:

| | | |
|---|---|---|
| $d_{\mathrm{src}}$ | source depth | m |
| $h_{\mathrm{air}}$ | ground clearance | m |
| $Q_{\mathrm{v}}$ | external isotropic volume source strength | $\mathrm{s^{-1}\,m^{-3}\,eV^{-1}}$ |
| $R_{\mathrm{src}}$ | source radius | m |
| $\theta_{\mathrm{src}}$ | view angle of the source | rad |
| $\mu_{\mathrm{air}}$ | total attenuation coefficient in air | $\mathrm{m^{-1}}$ |
| $\mu_{\mathrm{src}}$ | total attenuation coefficient in the source | $\mathrm{m^{-1}}$ |

and with $\cos(\theta_{\mathrm{src}}) = h_{\mathrm{air}}/(h_{\mathrm{air}}^2 + R_{\mathrm{src}}^2)^{1/2}$ as well as $Q_{\mathrm{s}} = Q_{\mathrm{v}}\,\mathrm{d}z$. We can solve the integral in Eq. A.22 by substituting $\tau = \mu_{\mathrm{air}} h_{\mathrm{air}} + \mu_{\mathrm{src}} z$ and consequently $\mathrm{d}\tau = \mu_{\mathrm{src}}\,\mathrm{d}z$ in the first term (I) in Eq. A.22 as well as $\tau' = (\mu_{\mathrm{air}} h_{\mathrm{air}} + \mu_{\mathrm{src}} z)/\cos(\theta_{\mathrm{src}})$ and $\mathrm{d}\tau' = \mu_{\mathrm{src}}\,\mathrm{d}z/\cos(\theta_{\mathrm{src}})$ in the second term (II) resulting in:

$$\frac{\mathrm{d}\phi_\gamma}{\mathrm{d}E_\gamma}\left(d_{\mathrm{src}}, R_{\mathrm{src}}, h_{\mathrm{air}}, E_\gamma, t\right) = \frac{Q_{\mathrm{v}}}{2\mu_{\mathrm{src}}}\Bigg[ \underbrace{\int_{\mu_{\mathrm{air}} h_{\mathrm{air}}}^{\mu_{\mathrm{air}} h_{\mathrm{air}} + \mu_{\mathrm{src}} d_{\mathrm{src}}} \mathscr{E}_1(\tau)\,\mathrm{d}\tau}_{\mathscr{E}_2(\mu_{\mathrm{air}} h_{\mathrm{air}}) - \mathscr{E}_2(\mu_{\mathrm{air}} h_{\mathrm{air}} + \mu_{\mathrm{src}} d_{\mathrm{src}})} - \cos\left(\theta_{\mathrm{src}}\right) \underbrace{\int_{\frac{\mu_{\mathrm{air}} h_{\mathrm{air}}}{\cos(\theta_{\mathrm{src}})}}^{\frac{\mu_{\mathrm{air}} h_{\mathrm{air}} + \mu_{\mathrm{src}} d_{\mathrm{src}}}{\cos(\theta_{\mathrm{src}})}} \mathscr{E}_1(\tau')\,\mathrm{d}\tau'}_{\mathscr{E}_2\left(\frac{\mu_{\mathrm{air}} h_{\mathrm{air}}}{\cos(\theta_{\mathrm{src}})}\right) - \mathscr{E}_2\left(\frac{\mu_{\mathrm{air}} h_{\mathrm{air}} + \mu_{\mathrm{src}} d_{\mathrm{src}}}{\cos(\theta_{\mathrm{src}})}\right)} \Bigg] \tag{A.23}$$

By applying the integral identity for exponential integral functions given in Eq. A.19e to the two highlighted integrals in Eq. A.23, we obtain the final solution for the primary energy flux of a truncated cone source given in Eq. 3.37, as indicated. The three special cases in Eqs. 3.38a–3.38c can be readily obtained by applying Eq. A.19a on the corresponding terms in Eq. 3.37.

## A.7 Mechanistic Model to Describe Scintillation Non-Proportionality

As we saw in Section 4.1.3, the non-proportionality in the scintillation light yield of inorganic scintillators is governed by the dependence of the local energy transfer efficiency $\eta'_{\mathrm{cap}}$ on the collisional stopping power of electrons $\mathcal{S}_{\mathrm{e,col}}$. In Eq. 4.4, I presented a mechanistic model published by Payne and his co-workers [313, 324, 325, 328, 357] to describe $\eta'_{\mathrm{cap}}$ as a function of $\mathcal{S}_{\mathrm{e,col}}$. In this section, I will briefly summarize the main steps in the derivation of this model as well as discuss some of its assumptions and limitations. I will also provide a brief overview of the model parameters and their physical meaning.

Let us start by recalling the meaning of $\eta_{\mathrm{cap}}$ as defined by the three-stage formula (Eq. 4.2) [310, 320, 321]. The energy transfer efficiency $\eta_{\mathrm{cap}}$ quantifies the efficiency with which excitation carriers, i.e. free electron-hole pairs ($e^-$/h pairs) as well as excitons created at the end of the thermalization stage, migrate to and subsequently get captured by the luminescent centers. In other words, $\eta_{\mathrm{cap}}$ describes everything that happens to the excitation carriers during the energy transfer phase discussed in Section 4.1.1. A graphical depiction of all these processes is shown in Fig. 4.1. At the end of the thermalization stage, we end up with two different types of excitation carriers: free $e^-$/h pairs and free excitons. The fraction of free $e^-$/h pairs at the end of the thermalization phase is given by the electron-hole pair fraction $\eta_{\mathrm{e/h}}$, whereas the complementary fraction $1 - \eta_{\mathrm{e/h}}$ denotes the excitons share.

Let us discuss first the transfer of free excitons. Three main things can happen to these excitons: (1) they can migrate to the luminescent centers and recombine radiatively, i.e. trigger luminescence; (2) they can get trapped and form so-called self-trapped excitons (STEs) [311, 313, 870]; or (3) they collide with each other in a process which is known as exciton-exciton annihilation [312, 314]. In this annihilation process, one of the excitons dissociates non-radiatively giving all its energy to the other exciton. This other exciton then loses the gained excess energy by phonon interactions. Hence, this annihilation process reduces the number of excitons from two to one. We may characterize the survival probability of an exciton in an annihilation process $\eta_{\mathrm{Birks}}$ using Birk's law, which was first introduced by Birks[11] in 1951 to describe scintillation non-proportionality in anthracene, an organic scintillator[12] [30, 872]:

11 *Jon Betteley Birks (*1920, †1979) was a reader in physics at Manchester University and a leading expert in organic scintillators. He started his career as a scientist during the Second World War at the Telecommunications Research Establishment in the UK working on secret radar technologies. After his PhD in microwave absorption in ferrites, he was appointed as a professor of physics at Rhodes University, South Africa, where he quickly established an active research group focusing on organic scintillators. However, he strongly opposed the developing apartheid regime in South Africa at that time and decided to move back to the UK in 1954. After some years in industry, he joined Manchester University as a reader in physics, where he continued his research until his passing in 1979 [871].*

12 *Note that most of the organic scintillators exhibit even more pronounced scintillation non-proportionality than inorganics do.*

$$\eta_{\text{Birks}} = \left(1 + \frac{\mathcal{S}_{\text{e,col}}}{\mathrm{d}E/\mathrm{d}x\,|_{\text{Birks}}}\right)^{-1} \tag{A.24}$$

where:

| | | |
|---|---|---|
| $\mathrm{d}E/\mathrm{d}x\,|_{\text{Birks}}$ | Birks stopping parameter | $\mathrm{eV\,m^{-1}}$ |
| $\mathcal{S}_{\text{e,col}}$ | collisional stopping power of electrons (cf. Eq. 4.5) | $\mathrm{eV\,m^{-1}}$ |

The $\mathrm{d}E/\mathrm{d}x\,|_{\text{Birks}}$ Birks stopping parameter is an empirical parameter controlling the strength of the Birks mechanism in a particular scintillator.[13] From Eq. A.24, we see that the survival probability of an exciton in an annihilation process increases for a decrease in $\mathcal{S}_{\text{e,col}}$ as well as an increase in $\mathrm{d}E/\mathrm{d}x\,|_{\text{Birks}}$. This aligns intuitively, as a decrease in $\mathcal{S}_{\text{e,col}}$ reduces the exciton density and thereby the chance of annihilation.

13 *Birks adopted a slightly different notation using* $kB := \mathrm{d}E/\mathrm{d}x\,|_{\text{Birks}}^{-1}$ *instead of* $\mathrm{d}E/\mathrm{d}x\,|_{\text{Birks}}$, *which was introduced later by Payne and his co-workers [324]. For consistency reasons, I use Payne's notation in this work.*

Moving onward to the STEs, they possess lower mobility but increased lifetime compared to the free excitons [311]. Similar to free excitons, they can migrate to the luminescent centers and undergo radiative recombination through a hopping-type transport mechanism, or they may annihilate in exciton-exciton annihilation events described by $\eta_{\text{Birks}}$ [313]. However, they can also deexcite radiatively themselves, resulting in what is known as excitonic luminescence, a form of intrinsic luminescence that does not require activators [311].

Free $e^-$/h pairs are subjected to somewhat more complex transfer processes. First, electrons and holes can migrate independently to the luminescent centers and by subsequent capture lead to luminescence. However, on their way to the luminescent centers, they can get trapped by impurities or lattice defects and are thereby lost for the scintillation process. On the other hand, free $e^-$/h pairs can also form STEs by the classic Onsager mechanism [873]. As proposed by Onsager[14], the probability for a free $e^-$/h pair to form a STE may be computed as follows:

14 *Lars Onsager (**1903, †1976*) was a Norwegian physical chemist and theoretical physicist. He was awarded the Nobel Prize in Chemistry in 1968 for his influential work on reciprocal relations in thermodynamics [874].*

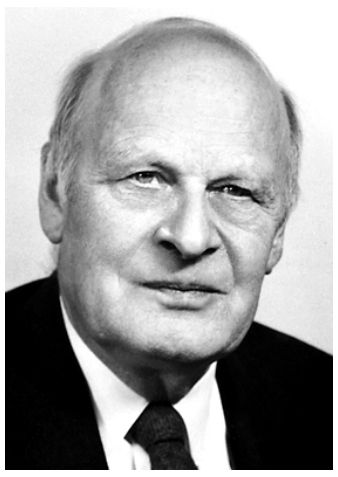

Lars Onsager

$$\eta_{\text{Ons}} = 1 - \exp\left(-\frac{r_{\text{Ons}}}{r_{\text{e/h}}}\right) \tag{A.25}$$

where:

| | | |
|---|---|---|
| $r_{\text{e/h}}$ | electron-hole separation length | m |
| $r_{\text{Ons}}$ | Onsager radius | m |

The electron-hole separation length $r_{e/h}$ represents the initial separation distance between the electron and the hole. The Onsager radius $r_{Ons}$ is defined as follows [873]:

$$r_{Ons} = \frac{e^2}{4\pi\varepsilon k_B T} \tag{A.26}$$

where:

| | | |
|---|---|---|
| $e$ | elementary charge (cf. Constants) | C |
| $k_B$ | Boltzmann constant (cf. Constants) | $\mathrm{J\,K^{-1}}$ |
| $T$ | temperature of the scintillator | K |
| $\varepsilon$ | permittivity of the scintillator | $\mathrm{F\,m^{-1}}$ |

As shown by Payne et al. [313], we may estimate the average electron-hole separation distance to a first order as follows:

$$\langle r_{e/h} \rangle = \frac{\beta_{sci} E_{gap}}{\mathcal{S}_{e,col}} \tag{A.27}$$

where:

| | | |
|---|---|---|
| $E_{gap}$ | band gap energy (cf. Eq. 4.3) | eV |
| $\mathcal{S}_{e,col}$ | collisional stopping power of electrons (cf. Eq. 4.5) | $\mathrm{eV\,m^{-1}}$ |
| $\beta_{sci}$ | ratio between the energy needed to create an electron-hole pair and the bandgap energy (cf. Eq. 4.3) | |

Combining Eqs. A.26 and A.27, we can rewrite Eq. A.25 as:

$$\eta_{Ons} = 1 - \exp\left(-\frac{\mathcal{S}_{e,col}}{\mathrm{d}E/\mathrm{d}x\,|_{Ons}}\right) \tag{A.28}$$

with

$$\mathrm{d}E/\mathrm{d}x\,|_{Ons} = \frac{4\pi\varepsilon k_B T \beta_{sci} E_{gap}}{e^2} \tag{A.29}$$

being the Onsager stopping parameter $\mathrm{d}E/\mathrm{d}x\,|_{Ons}$. From Eq. A.28, we see that $\eta_{Ons}$ decreases with decreasing $\mathcal{S}_{e,col}$ and increasing $\mathrm{d}E/\mathrm{d}x\,|_{Ons}$. As $\mathcal{S}_{e,col}$ decreases, so thus the number of free $e^-$/h pairs. Consequently, $\langle r_{e/h} \rangle$ increases and $\eta_{Ons}$ is suppressed.

Based on all the discussed transfer mechanisms, Payne and his co-workers developed a simplified mechanistic model[15] to describe $\eta'_{cap}$

15 *Sometimes also referred to as the minimalistic approach or model [314].*

as a function of $\mathcal{S}_{e,col}$ for extrinsic and excitonic luminescent inorganic scintillators [313, 324, 325, 328, 357]. Two simplifying assumptions were made by Payne et al. First, because excitons are more mobile and possess longer lifetimes than free $e^-/h$ pairs [314], Payne assumes in his model that only excitons directly contribute to luminescence. Second, steady-state conditions are assumed during the energy transfer stage, i.e. no kinetic dynamics is considered here. Using these two assumptions, we may calculate the local energy transfer efficiency $\eta'_{cap}$ as the sum of the free exciton and free $e^-/h$ pair contributions to luminescence, where only the part of free $e^-/h$ pairs is considered, which successfully convert to excitons.

The luminescent contribution from free excitons is simply the fraction of excitation carriers created as free excitons $1 - \eta_{e/h}$, which survive exciton-exciton annihilation described by the Birks mechanism $\eta_{Birks}$ and in the end get captured by the luminescent centers or radiatively recombine as STEs:

$$\eta'_{cap}\big|_{exc} = \eta^0_{cap}\left(1 - \eta_{e/h}\right)\eta_{Birks} \tag{A.30}$$

where Payne et al. introduced an additional term $\eta^0_{cap}$ to quantify first order non-radiative losses, e.g. non-radiative recombination in STEs.

On the other hand, the luminescent contribution from free $e^-/h$ pairs may be computed as the fraction of excitation carriers created as free $e^-/h$ pairs $\eta_{e/h}$, which get successfully converted to excitons by the Onsager mechanism $\eta_{Ons}$, survive exciton-exciton annihilation described by the Birks mechanism $\eta_{Birks}$ and in the end get captured by the luminescent centers or radiatively recombine as STEs:

$$\eta'_{cap}\big|_{e/h} = \eta^0_{cap}\eta_{e/h}\eta_{Ons}\eta_{Birks} \tag{A.31}$$

where we used again $\eta^0_{cap}$ to quantify first-order non-radiative losses. Combining Eqs. A.30 and A.31, we obtain [313, 324]:

$$\eta'_{cap} = \eta^0_{cap}\left[\eta_{e/h}\eta_{Ons} + \left(1 - \eta_{e/h}\right)\right]\eta_{Birks} \tag{A.32a}$$

$$= \eta^0_{cap}\frac{1 - \eta_{e/h}\exp\left(-\frac{\mathcal{S}_{e,col}}{dE/dx|_{Ons}}\right)}{1 + \frac{\mathcal{S}_{e,col}}{dE/dx|_{Birks}}} \tag{A.32b}$$

16 *Screening is a common concept in physics and describes the phenomenon, where an electric field is damped by the presence of charge carriers or charge centers.*

After investigating temperature effects on the scintillation non-proportionality for various inorganic scintillators [325], Payne et al. extended this model by accounting for screening effects[16] in the Onsager mechanism based on a study by Vasil'Ev et al. [310]. The Onsager mechanism in Eq. A.25 was subsequently modified as follows:

$$\eta_{\mathrm{Ons}} = 1 - \exp\left[-\frac{r_{\mathrm{Ons}}}{r_{\mathrm{e/h}}}\exp\left(-\frac{r_{\mathrm{e/h}}}{\lambda_{\mathrm{D}}}\right)\right] \tag{A.33}$$

with $\lambda_{\mathrm{D}}$ being the Debye length[17] defined as [875]:

$$\lambda_{\mathrm{D}} = \sqrt{\frac{\varepsilon \mathrm{k_B} T}{\mathrm{e}^2\left(n_{\mathrm{ch}} + n_{\mathrm{A}}\right)}} \tag{A.34}$$

where:

| | | |
|---|---|---|
| $n_{\mathrm{ch}}$ | charge carrier density | $\mathrm{m}^{-3}$ |
| $n_{\mathrm{A}}$ | charge center density | $\mathrm{m}^{-3}$ |

$n_{\mathrm{ch}}$ and $n_{\mathrm{A}}$ quantify the screening effect caused by free $\mathrm{e}^-/\mathrm{h}$ pairs as well as excitation centers and traps, respectively. Inserting Eqs. A.26, A.27 and A.34 in Eq. A.33, we obtain a modified version of Eq. A.28:

$$\eta_{\mathrm{Ons}} = 1 - \exp\left[-\frac{\mathcal{S}_{\mathrm{e,col}}}{\mathrm{d}E/\mathrm{d}x\,|_{\mathrm{Ons}}}\exp\left(-\frac{\mathrm{d}E/\mathrm{d}x\,|_{\mathrm{trap}}}{\mathcal{S}_{\mathrm{e,col}}}\right)\right] \tag{A.35}$$

with the trapping stopping parameter $\mathrm{d}E/\mathrm{d}x\,|_{\mathrm{trap}}$ defined as:

$$\mathrm{d}E/\mathrm{d}x\,|_{\mathrm{trap}} = \frac{\beta_{\mathrm{sci}} E_{\mathrm{gap}}}{\lambda_{\mathrm{D}}} \tag{A.36a}$$

$$= \beta_{\mathrm{sci}} E_{\mathrm{gap}} \mathrm{e} \sqrt{\frac{n_{\mathrm{ch}} + n_{\mathrm{A}}}{\varepsilon \mathrm{k_B} T}} \tag{A.36b}$$

From Eq. A.35, we see that a higher trapping stopping parameter $\mathrm{d}E/\mathrm{d}x\,|_{\mathrm{trap}}$ suppresses the Onsager mechanism. By inserting Eq. A.35 instead of Eq. A.28 in Eq. A.32a, we obtain a modified model for $\eta'_{\mathrm{cap}}$, which was presented in the main text in Eq. 4.4, accounting know also for Debye screening, i.e. higher order loss processes in the Onsager mechanism [325]:

*17 Named after Peter Joseph William Debye (*1884, †1966), a Dutch physicist and physical chemist. Debye made significant contributions to our understanding of molecular structures through investigations on dipole moments and on the diffraction of X-rays and electrons in gases. For these achievements, he received the Nobel Prize in Chemistry in 1936.*

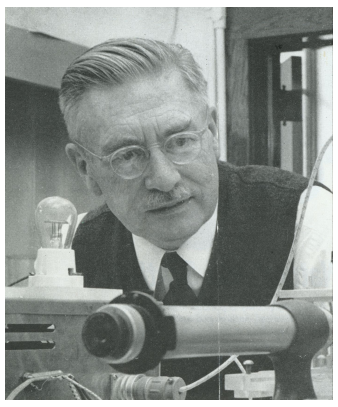

Peter Debye
unknown

$$\eta'_{\text{cap}} = \eta^0_{\text{cap}} \frac{1 - \eta_{\text{e/h}} \exp\left[-\frac{\mathcal{S}_{\text{e,col}}}{\text{d}E/\text{d}x\,|_{\text{Ons}}} \exp\left(-\frac{\text{d}E/\text{d}x\,|_{\text{trap}}}{\mathcal{S}_{\text{e,col}}}\right)\right]}{1 + \frac{\mathcal{S}_{\text{e,col}}}{\text{d}E/\text{d}x\,|_{\text{Birks}}}} \tag{A.37}$$

Let us conclude our discussion on Payne's mechanistic model with two interesting observations. First, as already highlighted by Payne et al. [313, 325], we see from Eqs. A.24, A.29 and A.36b that $\text{d}E/\text{d}x\,|_{\text{Ons}}$, in contrast to the remaining model parameters, does only depend on fundamental material properties. More specifically, $\text{d}E/\text{d}x\,|_{\text{Ons}}$ does not depend on the concentration and distribution of impurities or lattice defects in the scintillator and is therefore believed to be constant for different samples of the same scintillator material (under the same laboratory conditions). In general, $\text{d}E/\text{d}x\,|_{\text{Ons}}$ has proven to only weakly vary for different scintillator materials and is therefore often treated as a fundamental constant with a value of 36.4 MeV cm$^{-1}$ in regression analysis studies [313, 325]. On the other hand, the remaining parameters, i.e. $\text{d}E/\text{d}x\,|_{\text{Birks}}$, $\eta_{\text{e/h}}$ and $\text{d}E/\text{d}x\,|_{\text{trap}}$, are a function of the detailed lattice structure including distribution and concentration of impurities, activators as well as lattice defects. Consequently, these parameters are expected to vary for different samples of the same scintillator material. This was confirmed by Hull et al. in their experimental study on the scintillation non-proportionality across various NaI(Tl) samples [413].

Second, based on these findings, we see from Eq. A.37 that the non-proportional relation between $\eta'_{\text{cap}}$ and $\mathcal{S}_{\text{e,col}}$ is mainly governed by the Birks and Onsager mechanisms. These two mechanisms exhibit opposite trends in $\mathcal{S}_{\text{e,col}}$, i.e. at low $\mathcal{S}_{\text{e,col}}$ values, the Onsager mechanism suppresses the creation of excitons from free $\text{e}^-$/h pairs and thereby the scintillation light yield. On the other hand, at high $\mathcal{S}_{\text{e,col}}$ values, the Birks mechanism reduces the number of excitons forming from free $\text{e}^-$/h pairs, which suppresses the light yield, too. We conclude that depending on the relative strength of these two mechanisms controlled by the $\text{d}E/\text{d}x\,|_{\text{Birks}}$ and $\eta_{\text{e/h}}$ parameters (see discussion above), the light yield may be suppressed at lower, higher, or both extremes of the $\mathcal{S}_{\text{e,col}}$ range. In the latter scenario, we anticipate observing a local maximum in the light yield for intermediate $\mathcal{S}_{\text{e,col}}$ values.

## A.8 Uncertainty Analysis

The spectral quantities derived by the postprocessing pipelines `RLLSpec` and `PScinMC/NPScinMC` discussed in Sections 6.2.1.2, 6.2.2.4 and 7.2.3 as well as the ones obtained by the detector response model detailed in Section 9.2 are inherently subject to uncertainties. In this section, I will briefly discuss the uncertainty analysis for these spectral quantities. In line with the standard established by the Joint Committee for Guides in Metrology [287], I will distinguish between two main categories of uncertainties. The first category is associated with all uncertainties that are estimated by an observed frequency distribution (simulated or measured). This is known as type A [287, 296], aleatory [268, 876, 877] or statistical uncertainty [134, 878, 879]. The second category encompasses all uncertainties derived from sources other than observed frequency distributions, such as literature sources, calibration certificates or expert knowledge. This is known as type B [287, 296], epistemic [268, 876, 877] or systematic uncertainty [134, 878, 879]. I will use the terms "statistical" and "systematic" in this work to refer to these two categories of uncertainty, respectively.

### A.8.1 Measurement Uncertainty

Uncertainty quantification in radiation measurements is extensively addressed in the literature [30, 296, 297] and standardized by the International Organization for Standardization (ISO) in ISO 11929. In this work, I adopt the uncertainty analysis methodology for radiation measurements detailed by Knoll [30].

#### A.8.1.1 Statistical Measurement Uncertainty

For the radiation measurements, the statistical uncertainty of the spectral quantities discussed in Section 6.2.1.2 are characterized by the standard deviation and estimated by a probabilistic Poisson model according to Knoll [30]:

$$\boldsymbol{\sigma}_{C_{\mathrm{gr}}} = (\mathbf{C}_{\mathrm{gr}})^{\circ\frac{1}{2}} \tag{A.38a}$$

$$\boldsymbol{\sigma}_{C_{\mathrm{b}}} = (\mathbf{C}_{\mathrm{b}})^{\circ\frac{1}{2}} \tag{A.38b}$$

$$\boldsymbol{\sigma}_{c_{\mathrm{gr}}} = (\mathbf{C}_{\mathrm{gr}})^{\circ\frac{1}{2}} / t_{\mathrm{gr}} \tag{A.38c}$$

$$\boldsymbol{\sigma}_{c_b} = (\mathbf{C}_b)^{\circ\frac{1}{2}} / t_b \tag{A.38d}$$

$$\boldsymbol{\sigma}_{c_{net}} = \left(\mathbf{C}_{gr}/t_{gr}^2 + \mathbf{C}_b/t_b^2\right)^{\circ\frac{1}{2}} \tag{A.38e}$$

where:

| | | |
|---|---|---|
| $\mathbf{C}_b$ | background count vector | |
| $\mathbf{C}_{gr}$ | gross count vector | |
| $t_b$ | background measurement live time | s |
| $t_{gr}$ | gross measurement live time | s |

and with $\circ$ being defined as the Hadamard[18] , i.e. element-wise, power. By combining Eqs. 6.5 and A.38e, we may estimate the statistical uncertainty in the measured mean spectral signature $\hat{\mathbf{c}}_{exp}$ as [30]:[19]

18 *Named after Jacques Salomon Hadamard (*1865, †1963), a French mathematician renowned for his contributions to number theory, complex function theory, and partial differential equations, among others. We met him already in Section 5.4.2 when we were discussing ill-posed problems.*

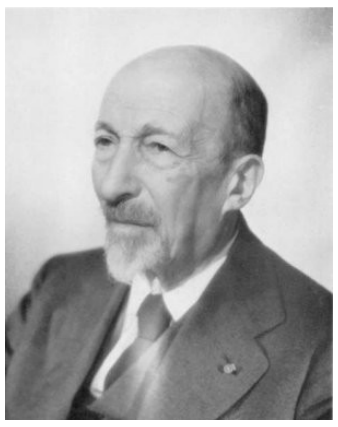

Jacques Hadamard
Ⓢ Wikimedia Commons

19 *In this work, only point source radionuclides were used to derive the measured spectral signatures $\hat{\mathbf{c}}_{exp}$. Consequently, I limit the discussion here to the case where the source strength $\xi$ is equivalent to the source activity $\mathcal{A}$.*

$$\hat{\boldsymbol{\sigma}}_{exp,stat} = \mathcal{A}^{-1}\boldsymbol{\sigma}_{c_{net}} \tag{A.39a}$$

$$= \mathcal{A}^{-1}\left(\mathbf{C}_{gr}/t_{gr}^2 + \mathbf{C}_b/t_b^2\right)^{\circ\frac{1}{2}} \tag{A.39b}$$

where:

| | | |
|---|---|---|
| $\mathcal{A}$ | source activity at measurement starting time | Bq |
| $\boldsymbol{\sigma}_{c_{net}}$ | net standard deviation count rate vector | $s^{-1}$ |

#### A.8.1.2 Systematic Measurement Uncertainty

The measured mean spectral signature $\hat{\mathbf{c}}_{exp}$ is subject to both statistical and systematic uncertainties. Here, I limit the quantification of the systematic uncertainty to that introduced by scaling $\hat{\mathbf{c}}_{exp}$ with the source strength $\xi$, which is anticipated to be the largest source of systematic uncertainty for the radiation measurements conducted in this work.[20] In the case of a calibration radionuclide source, we may estimate the systematic uncertainty in the measured mean spectral signature $\hat{\mathbf{c}}_{exp}$ by applying the standard error propagation methodology for independent variables [287, 878, 881] to Eq. 6.5 resulting in:

20 *Other sources such as the constant activity bias (cf. Appendix A.8.1.4) will be discussed on a case-by-case basis.*

$$\hat{\boldsymbol{\sigma}}_{exp,sys} = \mathbf{c}_{net}\sigma_{\mathcal{A}}\mathcal{A}^{-2} \tag{A.40}$$

where:

| | | |
|---|---|---|
| $\mathcal{A}$ | source activity at measurement starting time $t$ | Bq |
| $\sigma_{\mathcal{A}}$ | activity standard deviation | Bq |
| $\mathbf{c}_{\text{net}}$ | net count rate vector | $s^{-1}$ |

The activity standard deviation $\sigma_{\mathcal{A}}$ in turn can be estimated using the standard error propagation methodology for independent variables [287, 878, 881] applied on Eq. 2.15 with the three independent variables, that are the reference activity $\mathcal{A}_0$ at $t_0$, the half-life $t_{1/2}$ as well as the time difference $\Delta t = t - t_0$ between the measurement starting time $t$ and the reference date $t_0$:

$$\sigma_{\mathcal{A}} = \mathcal{A}\sqrt{\left(\frac{\sigma_{\mathcal{A}_0}}{\mathcal{A}_0}\right)^2 + \left(\frac{\log(2)\Delta t\,\sigma_{t_{1/2}}}{t_{1/2}^2}\right)^2 + \left(\frac{\log(2)\sigma_{\Delta t}}{t_{1/2}}\right)^2} \tag{A.41}$$

where:

| | | |
|---|---|---|
| $\mathcal{A}$ | source activity at measurement starting time $t$ | Bq |
| $\mathcal{A}_0$ | source activity at reference date $t_0$ | Bq |
| $t_{1/2}$ | half-life | s |
| $\sigma_{\mathcal{A}_0}$ | reference activity standard deviation | Bq |
| $\sigma_{t_{1/2}}$ | half-life standard deviation | s |
| $\sigma_{\Delta t}$ | standard deviation in the time difference between the reference date $t_0$ and the starting time $t$ of the measurement | s |

For all radiation sources used in this work, the relative contribution of the uncertainty in $\sigma_{t_{1/2}}$ and $\sigma_{\Delta t}$ to the total uncertainty in $\mathcal{A}$ is found to be smaller than 1 %.

### A.8.1.3 Combined Measurement Uncertainty

The combined uncertainty in the measured spectral signature $\hat{\mathbf{c}}_{\text{exp}}$ is estimated by combining the related statistical and systematic uncertainties in quadrature as suggested by the Particle Data Group et al. [134] and the Joint Committee for Guides in Metrology [287]:

$$\hat{\boldsymbol{\sigma}}_{\text{exp}} = \left(\hat{\boldsymbol{\sigma}}_{\text{exp,stat}}^{\circ 2} + \hat{\boldsymbol{\sigma}}_{\text{exp,sys}}^{\circ 2}\right)^{\circ \frac{1}{2}} \tag{A.42a}$$

$$= \hat{\mathbf{c}}_{\text{exp}} \odot \left[\left(\boldsymbol{\sigma}_{c_{\text{net}}} \oslash \mathbf{c}_{\text{net}}\right)^{\circ 2} + \left(\frac{\sigma_{\mathcal{A}}}{\mathcal{A}}\right)^2 \mathbf{J}_{N_{\text{ch}},1}\right]^{\circ \frac{1}{2}} \tag{A.42b}$$

where:

| | | |
|---|---|---|
| $\hat{\boldsymbol{\sigma}}_{\text{exp,stat}}$ | measured statistical spectral signature standard deviation | $\text{s}^{-1}\,\text{Bq}^{-1}$ |
| $\hat{\boldsymbol{\sigma}}_{\text{exp,sys}}$ | measured systematic spectral signature standard deviation | $\text{s}^{-1}\,\text{Bq}^{-1}$ |

and with $\odot$ and $\oslash$ denoting the Hadamard, i.e. element-wise, product and division, respectively. The all-ones matrix $\mathbf{J}_{m,n}$ is being defined as a matrix of ones with dimension $m \times n$.

By default, $\hat{\boldsymbol{\sigma}}_{\text{exp}}$ is indicated in each figure presenting measured spectral signatures $\hat{\mathbf{c}}_{\text{exp}}$. If applicable, the coefficient of variation vector $\mathbf{CV}_{\text{exp}} \in \mathbb{R}_{+}^{N_{\text{ch}}\times 1}$ defined as:

$$\mathbf{CV}_{\text{exp}} = \hat{\boldsymbol{\sigma}}_{\text{exp}} \oslash \hat{\mathbf{c}}_{\text{exp}} \tag{A.43}$$

together with the relative statistical and systematic contributions:

$$\hat{\boldsymbol{\sigma}}^{\circ 2}_{\text{exp,stat}} \oslash \hat{\boldsymbol{\sigma}}^{\circ 2}_{\text{exp}} \odot \mathbf{CV}_{\text{exp}} = \hat{\boldsymbol{\sigma}}^{\circ 2}_{\text{exp,stat}} \oslash \left(\hat{\boldsymbol{\sigma}}_{\text{exp}} \odot \hat{\mathbf{c}}_{\text{exp}}\right) \tag{A.44a}$$

$$\hat{\boldsymbol{\sigma}}^{\circ 2}_{\text{exp,sys}} \oslash \hat{\boldsymbol{\sigma}}^{\circ 2}_{\text{exp}} \odot \mathbf{CV}_{\text{exp}} = \hat{\boldsymbol{\sigma}}^{\circ 2}_{\text{exp,sys}} \oslash \left(\hat{\boldsymbol{\sigma}}_{\text{exp}} \odot \hat{\mathbf{c}}_{\text{exp}}\right) \tag{A.44b}$$

are provided as well. For typical laboratory-based radiation measurements such as those presented in Chapters 6 and 7, the relative contributions of the systematic uncertainty in the activity $\mathcal{A}$ to the total uncertainty can be substantial and even exceed the statistical contribution over large parts of the spectral domain of interest (SDOI) $\mathcal{D}_{\text{SDOI}}$ (cf. Section 6.3.1).

#### A.8.1.4 Constant Activity Bias

As already noted in Section 6.2.1.2, for the radiation measurements conducted in this work, I assume that the activity of the radionuclides remains constant during the measurement time $\Delta t$. However, in reality, the activity of the radionuclide source changes with time as discussed in Section 2.1.2. Consequently, this assumption of a constant activity introduces a systematic bias in the spectral signature $\hat{\mathbf{c}}_{\text{exp}}$ computation in Eq. 6.5. We can quantify this bias by comparing the number of decays $N_{\text{const}}$ occurring during the measurement time $\Delta t$ for a hypothetical radionuclide source with constant activity $\mathcal{A}$:

$$N_{\text{const}} = \mathcal{A}\,\Delta t \tag{A.45}$$

where:

| | | |
|---|---|---|
| $\mathcal{A}$ | source activity | Bq |
| $\Delta t$ | measurement time | s |

with the number of decays $N_t$ occurring during the measurement time $\Delta t$ for a real radionuclide source that decays according to the exponential law described by Eq. 2.15 into a stable nuclide (cf. Section 2.1.2):

$$N_t = \int_0^{\Delta t} \mathcal{A} \exp\left[-\frac{\log(2)t}{t_{1/2}}\right] \mathrm{d}t \tag{A.46a}$$

$$= \frac{\mathcal{A} t_{1/2}}{\log(2)} \left(1 - \exp\left[-\frac{\log(2)\Delta t}{t_{1/2}}\right]\right) \tag{A.46b}$$

where:

| | | |
|---|---|---|
| $\mathcal{A}$ | source activity at the start of the measurement ($t = 0$) | Bq |
| $t_{1/2}$ | half-life | s |
| $\Delta t$ | measurement time | s |

Combining Eqs. A.45 and A.46b, we may compute the relative error introduced by the constant activity assumption in the spectral signature as:

$$\frac{N_{\mathrm{const}} - N_t}{N_t} = \frac{\log(2)\Delta t}{t_{1/2}} \left(1 - \exp\left[-\frac{\log(2)\Delta t}{t_{1/2}}\right]\right)^{-1} - 1 \tag{A.47}$$

From Eq. A.47, two interesting observations can be made. First, the relative error is found to only depend on the half-life of the source and the measurement time $\Delta t$. In particular, it is found to be independent of the activity $\mathcal{A}$ and consequently the measurement start time $t$. Second, for $\Delta t \ll t_{1/2}$, the relative error tends to $\log(2)\Delta t / t_{1/2} - 1$, which may provide a good approximation to estimate the relative error for long-lived radionuclides.

For the radiation measurements conducted in the context of this work, relative errors introduced by the constant activity assumption are typically ≪1 %. The largest relative error, ~0.19 %, is found for the radiation measurement using $^{57}_{27}Co$ reported in Chapters 6 and 7 (cf. also Table 6.1).

### A.8.2 Simulation Uncertainty

A general discussion on uncertainty estimation for Monte Carlo derived quantities was already provided in Section 3.2.2.4. In this subsection, I focus on uncertainty estimates for the primary quantity of interest derived by the `PScinMC` and the `NPScinMC` pipelines, that is the simulated spectral signature $\hat{\mathbf{c}}_{\text{sim}} \in \mathbb{R}_{+}^{N_{\text{ch}} \times 1}$ (cf. Sections 6.2.2.4 and 7.2.3).

#### A.8.2.1 Statistical Simulation Uncertainty

Building on the quantities introduced in the Sections 3.2.2.4 and 6.2.2.4, the main multivariate observable $\mathbf{y} \in \mathbb{R}_{+}^{N_{\text{ch}} \times 1}$ derived from a simulated energy deposition event $\tilde{n}^*_{\text{dep}}$ is equivalent to:

$$y_n = \begin{cases} \Phi\left(\frac{n+\frac{1}{2}-\tilde{n}^*_{\text{dep}}}{\sigma_E(\tilde{n}^*_{\text{dep}})}\right) - \Phi\left(\frac{n-\frac{1}{2}-\tilde{n}^*_{\text{dep}}}{\sigma_E(\tilde{n}^*_{\text{dep}})}\right) & \tilde{n}^*_{\text{dep}} > 0 \quad \text{(A.48a)} \\ 0 & \tilde{n}^*_{\text{dep}} = 0 \quad \text{(A.48b)} \end{cases}$$

where:

| | |
|---|---|
| $n$ | pulse-height channel number |
| $\tilde{n}^*_{\text{dep}}$ | continuous raw pulse-height channel number of an energy deposition event |
| $\sigma_E$ | spectral resolution standard deviation |

and with the pulse-height channel number being defined as $\{n \in \mathbb{N}_+ \mid n \leq N_{\text{ch}}, N_{\text{ch}} \in \mathbb{N}_+\}$ (cf. Section 4.3). Note that I explicitly distinguish between the case of $\tilde{n}^*_{\text{dep}} > 0$ and $\tilde{n}^*_{\text{dep}} = 0$ in Eqs. A.48a and A.48b to emphasize that not all simulated primaries will result in an energy deposition event in one of the scintillation crystals. The number of recorded energy deposition events $N_{\text{dep}}$ for a given detector channel is defined as $N_{\text{dep}} = |\{\tilde{n}^*_{\text{dep}} \in \mathbb{R} \mid \tilde{n}^*_{\text{dep}} > 0\}|$. Consequently, the number of simulated primaries not resulting in an energy deposition event in a given detector channel is given by $N_0 = |\{\tilde{n}^*_{\text{dep}} \in \mathbb{R} \mid \tilde{n}^*_{\text{dep}} = 0\}|$. Although we could record primary histories without energy depositions as zero-events in our Monte Carlo simulations, i.e. $\{\mathbf{y}_i \in \mathbb{R}_{+}^{N_{\text{ch}} \times 1} \mid i \in \mathbb{N}_+, i \leq N_{\text{pr}}\}$, this is not done in practice. Instead, we only record the energy deposition events $\tilde{n}^*_{\text{dep}} > 0$ as $\{\mathbf{y}_j \in \mathbb{R}_+^{N_{\text{ch}} \times 1} \mid j \in \mathbb{N}_+, j \leq N_{\text{dep}}\}$ and reconstruct the number of zero-events by $N_0 = N_{\text{pr}} - N_{\text{dep}}$. In line with Eq. 3.43, the observable mean $\overline{\mathbf{y}} \in \mathbb{R}_+^{N_{\text{ch}} \times 1}$ is then estimated as:

$$\overline{\mathbf{y}} = \frac{1}{N_{\text{pr}}} \sum_{i=1}^{N_{\text{pr}}} \mathbf{y}_i \tag{A.49a}$$

$$= \frac{1}{N_{\text{pr}}} \sum_{j=1}^{N_{\text{dep}}} \mathbf{y}_j \tag{A.49b}$$

$$\approx \frac{1}{N_{\text{pr}}} \mathbf{G} \mathbf{C}^{\dagger} \tag{A.49c}$$

where:

| | |
|---|---|
| $\mathbf{C}^{\dagger}$ | prebinned count vector |
| $\mathbf{G}$ | Gaussian weight matrix |

In the last observable mean estimate Eq. A.49c, I have utilized the Gaussian weight matrix $\mathbf{G} \in \mathbb{R}_+^{N_{\text{ch}} \times N^{\dagger}_{\text{ch}}}$ and the prebinned count vector $\mathbf{C}^{\dagger} \in \mathbb{N}^{N^{\dagger}_{\text{ch}} \times 1}$ introduced in Section 6.2.2.4.

In the same way, we can estimate the standard deviation of the observable mean $\boldsymbol{\sigma}_{\overline{y}} \in \mathbb{R}_+^{N_{\text{ch}} \times 1}$ by applying Eq. 3.44a introduced in Section 3.2.2.4 to the observable vector $\mathbf{y} \in \mathbb{R}_+^{N_{\text{ch}} \times 1}$ as follows:

$$\boldsymbol{\sigma}_{\overline{y}} = \left\{ \frac{\sum_{i=1}^{N_{\text{pr}}} \left(\mathbf{y}_i - \overline{\mathbf{y}}\right)^{\circ 2}}{N_{\text{pr}} \left(N_{\text{pr}} - 1\right)} \right\}^{\circ \frac{1}{2}} \tag{A.50a}$$

$$= \left\{ \frac{\left(N_{\text{pr}} - N_{\text{dep}}\right) \overline{\mathbf{y}}^{\circ 2} + \sum_{j=1}^{N_{\text{dep}}} \left(\mathbf{y}_j - \overline{\mathbf{y}}\right)^{\circ 2}}{N_{\text{pr}} \left(N_{\text{pr}} - 1\right)} \right\}^{\circ \frac{1}{2}} \tag{A.50b}$$

$$\approx \left\{ \frac{\left(N_{\text{pr}} - N_{\text{dep}}\right) \left(\mathbf{G}\mathbf{C}^{\dagger}\right)^{\circ 2} + \left(N_{\text{pr}} \mathbf{G} - \mathbf{G}\mathbf{C}^{\dagger} \mathbf{J}_{1, N^{\dagger}_{\text{ch}}}\right)^{\circ 2} \mathbf{C}^{\dagger}}{N_{\text{pr}}^3 \left(N_{\text{pr}} - 1\right)} \right\}^{\circ \frac{1}{2}} \tag{A.50c}$$

where I used again the matrix quantities introduced in Section 6.2.2.4 to approximate Eq. A.50b with Eq. A.50c. With $\circ$, I denote again the Hadamard, i.e. element-wise, power and $\mathbf{J}_{m,n}$ is defined as a matrix of ones with dimension $m \times n$. Eqs. A.50a–A.50c quantify the statistical dispersion of the observable mean $\overline{\mathbf{y}}$. Consequently, we may estimate the statistical uncertainty of the simulated spectral signature $\hat{\mathbf{c}}_{\text{sim}}$ by scaling the standard deviation of the observable mean $\boldsymbol{\sigma}_{\overline{y}}$ given by Eqs. A.50a–A.50c with the source geometry-matter factor $F_{\text{src}}$ introduced in Section 6.2.2.4:

$$\hat{\boldsymbol{\sigma}}_{\text{sim,stat}} = F_{\text{src}} \boldsymbol{\sigma}_{\overline{y}} \tag{A.51}$$

where:

| | | |
|---|---|---|
| $F_{\text{src}}$ | source geometry-matter factor | $[F_{\text{src}}]$ |

As discussed in the related method sections, I select the number of primaries $N_{\text{pr}}$ in such a way that $\text{med}(\mathbf{CV}_{\text{stat}}) = \text{med}(\hat{\boldsymbol{\sigma}}_{\text{sim,stat}} \oslash \hat{\mathbf{c}}_{\text{sim}}) < 1\,\%$ with $\text{med}(\cdot)$ being the median operator and $\mathbf{CV}_{\text{stat}}$ the coefficient of variation vector with respect to the statistical simulation uncertainties.

The statistical uncertainty in Eq. A.51 can in turn be quantified by the relative variance of the variance VOV as discussed in Section 3.2.2.4. By applying Eq. 3.46b to the observable vector $\mathbf{y} \in \mathbb{R}_{+}^{N_{\text{ch}} \times 1}$, we obtain the relative variance of the variance vector $\mathbf{VOV}_{\overline{y}} \in \mathbb{R}_{+}^{N_{\text{ch}} \times 1}$ as:

$$\mathbf{VOV}_{\overline{y}} \approx \frac{\sum_{i=1}^{N_{\text{pr}}} \left(\mathbf{y}_i - \overline{\mathbf{y}}\right)^{\circ 4}}{\sum_{i=1}^{N_{\text{pr}}} \left(\mathbf{y}_i - \overline{\mathbf{y}}\right)^{\circ 2}} - \frac{N_{\text{pr}} - 3}{N_{\text{pr}} - 1} \frac{1}{N_{\text{pr}}} \tag{A.52a}$$

$$= \frac{\left(N_{\text{pr}} - N_{\text{dep}}\right) \overline{\mathbf{y}}^{\circ 4} + \sum_{j=1}^{N_{\text{dep}}} \left(\mathbf{y}_j - \overline{\mathbf{y}}\right)^{\circ 4}}{\left(N_{\text{pr}} - N_{\text{dep}}\right) \overline{\mathbf{y}}^{\circ 2} + \sum_{j=1}^{N_{\text{dep}}} \left(\mathbf{y}_j - \overline{\mathbf{y}}\right)^{\circ 2}} - \frac{N_{\text{pr}} - 3}{N_{\text{pr}} - 1} \frac{1}{N_{\text{pr}}} \tag{A.52b}$$

$$\approx \frac{\left(N_{\text{pr}} - N_{\text{dep}}\right) \left(\mathbf{G}\mathbf{C}^{\dagger}\right)^{\circ 4} + \left(N_{\text{pr}}\mathbf{G} - \mathbf{G}\mathbf{C}^{\dagger}\mathbf{J}_{1,N_{\text{ch}}^{\dagger}}\right)^{\circ 4} \mathbf{C}^{\dagger}}{N_{\text{pr}}^{2} \left[\left(N_{\text{pr}} - N_{\text{dep}}\right) \left(\mathbf{G}\mathbf{C}^{\dagger}\right)^{\circ 2} + \left(N_{\text{pr}}\mathbf{G} - \mathbf{G}\mathbf{C}^{\dagger}\mathbf{J}_{1,N_{\text{ch}}^{\dagger}}\right)^{\circ 2} \mathbf{C}^{\dagger}\right]}$$

$$- \frac{N_{\mathrm{pr}} - 3}{N_{\mathrm{pr}} - 1} \frac{1}{N_{\mathrm{pr}}} \tag{A.52c}$$

Given the definition of the relative variance of the variance (VOV) in Eq. 3.46a, $\mathrm{VOV}_{\overline{y}}$ is the same for the simulated spectral signature $\hat{\mathbf{c}}_{\mathrm{sim}}$, i.e. $\mathrm{VOV}_{\overline{y}} = \mathrm{VOV}_{\hat{c}_{\mathrm{sim}}}$. As discussed in the related method sections, I select the number of primaries $N_{\mathrm{pr}}$ in such a way that $\mathrm{med}(\mathbf{VOV}_{\hat{c}_{\mathrm{sim}}}) < 0.01\,\%$.

### A.8.2.2 Systematic Simulation Uncertainty

Similar to the measurement quantities discussed in the previous section, the simulated quantities are also subject to systematic uncertainties. As discussed in Section 3.2.2.4, systematic uncertainty estimation in Monte Carlo simulations is challenging. In this work, I limit the systematic uncertainty analysis to the meta-models applied in the postprocessing pipelines `PScinMC` and `NPScinMC` using the Monte Carlo sampling technique [293–295]. These models are the spectral energy and spectral resolution models with the related model parameters, i.e. the spectral energy bin width $\Delta E'$, the scale coefficient $a_1$ and the power coefficient $a_2$ derived by the `RLLCal` pipeline (cf. Section 6.2.1.3). In case of the `NPScinMC` pipeline, there is additionally the spectral intrinsic resolution standard deviation $\sigma_{\mathrm{intr}}$ derived by the `NPScinCal` pipeline (cf. Section 7.2.2). In line with the uncertainty quantification methodology [295], I interpret these model parameters as a random vector $\boldsymbol{X} = (\Delta E', A_1^*, A_2)^\top \in \mathbb{R}^{3\times1}$ for `PScinMC` and $\boldsymbol{X} = (\Delta E', A_1^*, A_2, \sigma_{\mathrm{intr}})^\top \in \mathbb{R}^{4\times1}$ for `NPScinMC` with the log-transformed variable $A_1^* := \log A_1$ and related marginal distribution:

$$\boldsymbol{X} \sim \mathcal{N}\left(\mathbf{x} \mid \boldsymbol{\mu}, \mathrm{diag}\left(\boldsymbol{\sigma}^{\circ 2}\right), \mathbf{x}_l, \mathbf{x}_u\right) \tag{A.53}$$

being the truncated multivariate normal distribution with mean $\boldsymbol{\mu}$ and standard deviation $\boldsymbol{\sigma}$ defined by the regression results of the `RLLCal` and `NPScinCal` pipelines. The truncation $\mathbf{x}_l < \mathbf{x} < \mathbf{x}_u$ is set by $\mathbf{x}_l = (-\infty, -\infty, 0)^\top$ and $\mathbf{x}_u = (\infty, \infty, \infty)^\top$ for `PScinMC` and $\mathbf{x}_l = (-\infty, -\infty, 0, 0)^\top$ and $\mathbf{x}_u = (\infty, \infty, \infty, \infty)^\top$ for `NPScinMC`. To account for the statistical dependence between the scale coefficient $a_1$ and the power coefficient $a_2$, I apply the copula theory to perform correlated sampling of these two parameters using the Gaussian copula $\mathcal{C}_{\mathcal{N}}$ [882–884]:

$$\{A_1^*, A_2\} \sim \mathcal{C}_{\mathcal{N}}\left(F_{A_1^*}\left(a_1^*\right), F_{A_2}\left(a_2\right); \boldsymbol{\mathcal{R}}\right) \tag{A.54a}$$

$$= \Phi_2\left(\Phi^{-1}\left(F_{A_1^*}\left(a_1^*\right)\right), \Phi^{-1}\left(F_{A_2}\left(a_2\right)\right); \boldsymbol{\mathcal{R}}\right) \tag{A.54b}$$

with the marginal cumulative distribution functions $F$, the inverse cumulative distribution function of the standard normal distribution $\Phi^{-1}$ as well as the bivariate Gaussian cumulative distribution function $\Phi_2$ with zero mean $\mathbf{0}$ and correlation matrix $\boldsymbol{\mathcal{R}} \in \mathbb{R}^{3\times3}$ defined as:

$$\boldsymbol{\mathcal{R}} = \begin{pmatrix} 1 & \mathrm{corr}\left(a_1^*, a_2\right) \\ \mathrm{corr}\left(a_2, a_1^*\right) & 1 \end{pmatrix} \tag{A.55}$$

The individual correlation coefficients $\mathrm{corr}\left(a_1^*, a_2\right) = \mathrm{corr}\left(a_2, a_1^*\right)$ are provided by the regression analysis of the `RLLCal` pipeline (cf. as an example to Table C.6). In contrast to the scale coefficient $a_1$ and the power coefficient $a_2$, the spectral energy bin width $\Delta E'$ and the spectral intrinsic resolution standard deviation $\sigma_{\mathrm{intr}}$ are sampled independently from the other parameters according to the corresponding marginal distribution given in Eq. A.53. For more details on the copula theory, I recommend the monographs by Joe [883] and Nelsen [884].

Following the Monte Carlo approach reviewed in Section 3.2.2.4, I draw a fixed number of $N_{\mathrm{sys}}$ independent samples $\mathcal{X} = \{\mathbf{x}_i \in \mathbb{R}^{3\times1} \mid i \in \mathbb{N}_+, i \leq N_{\mathrm{sys}}\}$ for `PScinMC` and $\mathcal{X} = \{\mathbf{x}_i \in \mathbb{R}^{4\times1} \mid i \in \mathbb{N}_+, i \leq N_{\mathrm{sys}}\}$ for `NPScinMC` from the probabilistic input model detailed above. These samples are then propagated independently through the postprocessing pipelines `PScinMC` or `NPScinMC` discussed in Sections 6.2.2.4 and 7.2.3, respectively. The resulting set of simulated spectral signatures $\mathcal{Y} = \{\hat{\mathbf{c}}_{\mathrm{sim},i} \in \mathbb{R}_+^{N_{\mathrm{ch}}\times1} \mid i \in \mathbb{N}_+, i \leq N_{\mathrm{sys}}\}$ can then be used to estimate the systematic uncertainty by computing the sample standard deviation $\hat{\boldsymbol{\sigma}}_{\mathrm{sim,sys}} \in \mathbb{R}_+^{N_{\mathrm{ch}}\times1}$ as:

$$\hat{\boldsymbol{\sigma}}_{\mathrm{sim,sys}} = \left[\frac{N_{\mathrm{sys}} \sum_{i=1}^{N_{\mathrm{sys}}} \hat{\mathbf{c}}_{\mathrm{sim},i}^{\circ 2} - \left(\sum_{i=1}^{N_{\mathrm{sys}}} \hat{\mathbf{c}}_{\mathrm{sim},i}\right)^{\circ 2}}{N_{\mathrm{sys}}^2\left(N_{\mathrm{sys}} - 1\right)}\right]^{\circ \frac{1}{2}} \tag{A.56}$$

#### A.8.2.3 Combined Simulation Uncertainty

In full analogy to the combined uncertainty estimation for the experimental spectral signature $\hat{\mathbf{c}}_{\text{exp}}$ in Eq. A.42a, the total or combined uncertainty $\hat{\boldsymbol{\sigma}}_{\text{sim}} \in \mathbb{R}_{+}^{N_{\text{ch}} \times 1}$ in the simulated spectral signature $\hat{\mathbf{c}}_{\text{sim}}$ is estimated by combining the statistical and systematic uncertainties computed by Eqs. A.51 and A.56 in quadrature [134, 287]:

$$\hat{\boldsymbol{\sigma}}_{\text{sim}} = \left(\hat{\boldsymbol{\sigma}}_{\text{sim,stat}}^{\circ 2} + \hat{\boldsymbol{\sigma}}_{\text{sim,sys}}^{\circ 2}\right)^{\circ \frac{1}{2}} \tag{A.57}$$

where:

| | | |
|---|---|---|
| $\hat{\boldsymbol{\sigma}}_{\text{sim,stat}}$ | simulated statistical spectral signature standard deviation | $\text{s}^{-1}\,[\xi]^{-1}$ |
| $\hat{\boldsymbol{\sigma}}_{\text{sim,sys}}$ | simulated systematic spectral signature standard deviation | $\text{s}^{-1}\,[\xi]^{-1}$ |

By default, $\hat{\boldsymbol{\sigma}}_{\text{sim}}$ is indicated in each figure presenting the simulated spectral signature $\hat{\mathbf{c}}_{\text{sim}}$. If applicable, the coefficient of variation vector $\mathbf{CV}_{\text{sim}} \in \mathbb{R}_{+}^{N_{\text{ch}} \times 1}$ defined as:

$$\mathbf{CV}_{\text{sim}} = \hat{\boldsymbol{\sigma}}_{\text{sim}} \oslash \hat{\mathbf{c}}_{\text{sim}} \tag{A.58}$$

together with the relative statistical and systematic contributions:

$$\hat{\boldsymbol{\sigma}}_{\text{sim,stat}}^{\circ 2} \oslash \hat{\boldsymbol{\sigma}}_{\text{sim}}^{\circ 2} \odot \mathbf{CV}_{\text{sim}} = \hat{\boldsymbol{\sigma}}_{\text{sim,stat}}^{\circ 2} \oslash (\hat{\boldsymbol{\sigma}}_{\text{sim}} \odot \hat{\mathbf{c}}_{\text{sim}}) \tag{A.59a}$$

$$\hat{\boldsymbol{\sigma}}_{\text{sim,sys}}^{\circ 2} \oslash \hat{\boldsymbol{\sigma}}_{\text{sim}}^{\circ 2} \odot \mathbf{CV}_{\text{sim}} = \hat{\boldsymbol{\sigma}}_{\text{sim,sys}}^{\circ 2} \oslash (\hat{\boldsymbol{\sigma}}_{\text{sim}} \odot \hat{\mathbf{c}}_{\text{sim}}) \tag{A.59b}$$

are provided as well. In general, considering the SDOI $\mathcal{D}_{\text{SDOI}}$ and the selected number of primaries, the systematic uncertainty is found to be dominant in distinct spectral features, particularly the full energy peaks (FEPs), whereas the statistical uncertainty mainly affects the spectral continua of $\hat{\mathbf{c}}_{\text{sim}}$.

### A.8.3 Detector Response Model Uncertainty

As discussed in detail in Section 9.2, the detector response model (DRM) is based on the combination of two Monte Carlo derived quantities, i.e. the detector response function $R$ and the double differential photon flux signature $\hat{\phi}_{\gamma}$. As a result, the uncertainty in the obtained mean spectral signature is directly influenced by the uncertainties in these two quantities.

The uncertainty in the mean spectral signature obtained by the DRM can be estimated by propagating the uncertainties in the detector response function and the double differential photon flux signature through the DRM in Eq. 4.43 using the standard error propagation methodology for independent variables [287, 878, 881]:

$$\sigma_{\mathrm{DRM}}\left(E'_i, \mathfrak{d}_l\right) = \left[\sum_j \sum_k \left(\hat{\phi}_{\gamma,jkl}\sigma_{R_{ijkl}}\Delta\Omega'_k\Delta E_{\gamma,j}\right)^2 + \sum_j \sum_k \left(R_{ijkl}\sigma_{\hat{\phi}_{\gamma,jkl}}\Delta\Omega'_k\Delta E_{\gamma,j}\right)^2\right]^{\frac{1}{2}} \tag{A.60}$$

where:

| | | |
|---|---|---|
| $E'$ | spectral energy | eV |
| $\mathfrak{d}$ | experimental condition | $[\mathfrak{d}]$ |
| $R$ | mean detector response function | $\mathrm{m}^2$ |
| $\hat{\phi}_\gamma$ | mean double differential photon flux signature | $\mathrm{s}^{-1}\,\mathrm{m}^{-2}\,\mathrm{eV}^{-1}\,\mathrm{sr}^{-1}\,[\xi]^{-1}$ |
| $\Delta E_\gamma$ | photon energy bin width | eV |
| $\Delta\Omega'$ | solid angle bin width | sr |
| $\sigma_R$ | standard deviation of the mean detector response function | $\mathrm{m}^2$ |
| $\sigma_{\hat{\phi}_\gamma}$ | standard deviation of the mean double differential photon flux signature | $\mathrm{s}^{-1}\,\mathrm{m}^{-2}\,\mathrm{eV}^{-1}\,\mathrm{sr}^{-1}\,[\xi]^{-1}$ |

and with $R_{ijkl}$ and $\hat{\phi}_{\gamma,jkl}$ denoting the mean detector response function and the mean double differential photon flux signature evaluated at a given spectral energy $E'_i$, photon energy $E_{\gamma,j}$, direction unit vector $\mathbf{\Omega}'_k$ and experimental condition vector $\mathfrak{d}_l$:

$$R_{ijkl} \coloneqq R\left(E'_i, E_{\gamma,j}, \mathbf{\Omega}'_k, \mathfrak{d}_l\right) \tag{A.61a}$$

$$\hat{\phi}_{\gamma,jkl} \coloneqq \hat{\phi}_\gamma\left(E_{\gamma,j}, \mathbf{\Omega}'_k, \mathfrak{d}_l\right) \tag{A.61b}$$

Given their derivation using Monte Carlo simulations, the individual uncertainties in the detector response function $\sigma_{R_{ijkl}}$ and the double differential photon flux signature $\sigma_{\hat{\phi}_{\gamma,jkl}}$ are estimated using the methods detailed in Appendix A.8.2. This includes both statistical and systematic contributions as well as the computation of the corresponding coefficient of variation vectors (cf. Appendix A.8.2.3). In full analogy to Eq. A.51, the statistical uncertainty can be estimated as:

$$\sigma_{R_{ijkl},\text{stat}} = A_{\text{src}} \boldsymbol{\sigma}_{\overline{y}} \tag{A.62a}$$

$$\sigma_{\hat{\phi}_{\gamma,jkl},\text{stat}} = F_{\text{src}} \boldsymbol{\sigma}_{\overline{y}} \tag{A.62b}$$

where:

| | | |
|---|---|---|
| $A_{\text{src}}$ | plane-wave source area | $\text{m}^2$ |
| $F_{\text{src}}$ | source geometry-matter factor | $[F_{\text{src}}]$ |

and with $\boldsymbol{\sigma}_{\overline{y}}$ being the standard deviation of the observable mean (cf. Eq. A.50c for $\sigma_{R_{ijkl},\text{stat}}$ and Eq. 3.44b for $\sigma_{\hat{\phi}_{\gamma,jkl},\text{stat}}$). The systematic uncertainty for the detector response function $\sigma_{R_{ijkl},\text{sys}}$ is estimated by the Monte Carlo sampling technique detailed in Appendix A.8.2.2. For the double differential photon flux signature, the systematic uncertainty is assumed to be negligible. Consequently, in full analogy to Eq. A.57, I estimate the combined uncertainties $\sigma_{R_{ijkl}}$ and $\sigma_{\hat{\phi}_{\gamma,jkl}}$ applied in Eq. A.60 as:

$$\sigma_{R_{ijkl}} = \left[ \sigma^2_{R_{ijkl},\text{stat}} + \sigma^2_{R_{ijkl},\text{sys}} \right]^{\frac{1}{2}} \tag{A.63a}$$

$$\sigma_{\hat{\phi}_{\gamma,jkl}} = \sigma_{\hat{\phi}_{\gamma,jkl},\text{stat}} \tag{A.63b}$$

## A.9 Lower-Level Discriminator Calibration

The PScinMC and NPScinMC pipelines developed in this study to postprocess the Monte Carlo simulation data include also a unique correction method for the lower-level discriminator (LLD) effect (cf. Section 4.3). In this section, I briefly discuss the calibration of the associated model parameters for this lower-level discriminator (LLD) correction.

As detailed in Section 6.2.2.4, I apply a heuristic Gaussian model to correct for the LLD effect in the simulated spectral signatures. For completeness, I repeat this model here again:[21]

21 *Please note that, as the calibration of the LLD model parameters is based on radiation measurements with calibration radionuclide point sources, I express the spectral signatures here in units of* $\mathrm{s^{-1}\,Bq^{-1}}$.

$$\hat{c}_{\mathrm{sim,LLD},n} = \hat{c}_{\mathrm{sim},n}\,\Phi\left(\frac{n-\mu_{\mathrm{LLD}}}{\sigma_{\mathrm{LLD}}}\right) \tag{A.64}$$

where:

| | | |
|---|---|---|
| $\hat{c}_{\mathrm{sim}}$ | simulated mean spectral signature without LLD correction | $\mathrm{s^{-1}\,Bq^{-1}}$ |
| $\hat{c}_{\mathrm{sim,LLD}}$ | simulated mean spectral signature with LLD correction | $\mathrm{s^{-1}\,Bq^{-1}}$ |
| $n$ | pulse-height channel number | |
| $\mu_{\mathrm{LLD}}$ | mean lower-level discriminator | |
| $\sigma_{\mathrm{LLD}}$ | lower-level discriminator standard deviation | |

and with $\Phi(\cdot)$ denoting the standard normal cumulative distribution function. The model parameters, $\mu_{\mathrm{LLD}}$ and $\sigma_{\mathrm{LLD}}$, are calibrated for each of the four detector channels #1 through #4 individually using a numerical optimization approach. Specifically, the model parameters are optimized based on the $\chi^2$ objective function to align the simulated spectral signatures with the measured ones:

$$\left(\mathring{\mu}_{\mathrm{LLD}}, \mathring{\sigma}_{\mathrm{LLD}}\right) = \underset{\mu_{\mathrm{LLD}},\sigma_{\mathrm{LLD}}}{\arg\min}\ \chi^2 \tag{A.65}$$

with $\mathring{\mu}_{\mathrm{LLD}}$ and $\mathring{\sigma}_{\mathrm{LLD}}$ representing the arguments of the global $\chi^2$ minimum. The $\chi^2$ objective function is defined as:

$$\chi^2 = \frac{1}{N^*_{\mathrm{ch}}-2}\sum_{n=1}^{N^*_{\mathrm{ch}}}\frac{\left(\hat{c}_{\mathrm{exp},n} - \hat{c}_{\mathrm{sim,LLD},n}(\mu_{\mathrm{LLD}},\sigma_{\mathrm{LLD}})\right)^2}{\hat{\sigma}^2_{\mathrm{exp},n} + \hat{\sigma}^2_{\mathrm{sim},n}} \tag{A.66}$$

where:

| | | |
|---|---|---|
| $\hat{c}_{\mathrm{exp}}$ | measured mean spectral signature | $\mathrm{s}^{-1}\,\mathrm{Bq}^{-1}$ |
| $\hat{c}_{\mathrm{sim,LLD}}$ | simulated mean spectral signature with LLD correction | $\mathrm{s}^{-1}\,\mathrm{Bq}^{-1}$ |
| $N^{*}_{\mathrm{ch}}$ | reduced number of pulse-height channels | |
| $\hat{\sigma}_{\mathrm{exp}}$ | measured spectral signature standard deviation | $\mathrm{s}^{-1}\,\mathrm{Bq}^{-1}$ |
| $\hat{\sigma}_{\mathrm{sim}}$ | simulated spectral signature standard deviation | $\mathrm{s}^{-1}\,\mathrm{Bq}^{-1}$ |

with $\hat{\mathbf{c}}_{\mathrm{exp}}$ and $\hat{\boldsymbol{\sigma}}_{\mathrm{exp}}$ obtained by the `RLLSpec` pipeline (cf. Section 6.2.1.2) and $\hat{\mathbf{c}}_{\mathrm{sim}}$ and $\hat{\boldsymbol{\sigma}}_{\mathrm{sim}}$ obtained by the `PScinMC` or the `NPScinMC` pipeline. The global minimization problem in Eq. A.65 is solved by the radial basis surrogate solver `surrogateopt` [885, 886] using the `MATLAB` code. The spectral range was restricted to $n \leq \lfloor 50\,\mathrm{keV}/\Delta E' \rfloor = N^{*}_{\mathrm{ch}}$ with $\{\mu_{\mathrm{LLD}} \in \mathbb{R}_{+} \mid \mu_{\mathrm{LLD}} \leq 30\,\mathrm{keV}/\Delta E'\}$ and $\{\sigma_{\mathrm{LLD}} \in \mathbb{R}_{+} \mid 0.1\,\mathrm{keV}/\Delta E' \leq \sigma_{\mathrm{LLD}} \leq 20\,\mathrm{keV}/\Delta E'\}$. The spectral signature of the $^{60}_{27}\mathrm{Co}$ calibration point source was selected for the calibration of the LLD model parameters (cf. Section 6.2) as this source does not show any characteristic X-ray emission above 10 keV, resulting in reduced systematic uncertainties in this low spectral domain compared to other radionuclide source presented in Chapters 6 and 7.

The resulting $\chi^2$ contour maps for the `PScinMC` and `PScinMC` pipeline are presented in Figs. B.32 and B.33, respectively. For each detector channel, the radial basis surrogate solver converged to a global minimum $\chi^2_{\mathrm{min}}$ indicated by a plus sign in the individual contour maps. Confidence limits are illustrated adopting 1 and 2 standard deviation contour lines [887]. Depending on the detector channel, the mean lower-level discriminator is found to be between ~10 keV and ~20 keV with a standard deviation between ~3 keV and ~6 keV (cf. Fig. B.32).[22] No significant difference between the model parameters derived by the `PScinMC` and the `NPScinMC` could be found. A comparison of the LLD corrected and uncorrected simulated spectral signatures for the $^{60}_{27}\mathrm{Co}$ calibration point source is presented in Figs. B.34 and B.35. The results demonstrate a significant improvement in the agreement between the simulated and measured spectral signatures after applying the calibrated LLD correction model. Furthermore, the spectral results presented in Chapters 6 and 7 confirm

22 *As indicated in Figs. B.32 and B.33, $\mu_{\mathrm{LLD}}$ and $\sigma_{\mathrm{LLD}}$ are expressed and applied as continuous pulse-height channel numbers. Here, to simplify the interpretation, I report these values as spectral energies using the energy calibration models derived by the `RLLCal` pipeline (cf. Section 6.2.1.3).*

that the calibrated LLD model corrects not only the LLD effect for the $^{60}_{27}$Co source adequately but also for the other sources, which were not used in the LLD model calibration. For simplicity, the optimized LLD model parameters $\mathring{\mu}_{\mathrm{LLD}}$ and $\mathring{\sigma}_{\mathrm{LLD}}$ will be denoted by $\mu_{\mathrm{LLD}}$ and $\sigma_{\mathrm{LLD}}$ in the main part of this work.

## A.10 Compton Edge Shift Analysis

To gain a deeper understanding of the spectral Compton edge shift observed in Chapter 6 and its application in Compton edge probing for NPSM calibration discussed in Chapter 7, I present here an expanded analysis of this phenomenon. In particular, I quantify the magnitude of the spectral Compton edge shift empirically for different radionuclides using radiation measurements obtained in Chapter 6. Following this, I explore how the Compton edge shift varies with photon energy and scintillator volume through a simplified semi-analytical approach using the physics models introduced in Chapters 3 and 4.

### A.10.1 Empirical Analysis

Because the measured Compton edges are obscured by the finite spectral resolution, quantification of the exact position in the measured pulse-height spectrum would require additional coincidence measurements [888, 889]. Here, similar to a previous study [890], I use an alternative approach by quantifying the spectral shift between the already available PSMC simulation results and the measured spectral signature presented in Chapter 6. Furthermore, as the Compton edges are obscured by the related FEPs for the $^{57}_{27}$Co, $^{109}_{48}$Cd, $^{133}_{56}$Ba and $^{152}_{63}$Eu spectral signatures, I will focus here on the $^{60}_{27}$Co, $^{88}_{39}$Y and $^{137}_{55}$Cs measurements.

In a first step, I determine the inflection points as a characteristic measure of the corresponding Compton edges using spline regression [891]. I then compute the Compton edge shift as the spectral difference between the determined inflection points for the measured and simulated spectral signatures. I apply this method to different Compton edges, i.e. [477.334(3), 699.133(3), 963.419(3), 1611.77(1)] keV associated with the photon emission lines of the radionuclides $\{^{137}_{55}\text{Cs}, ^{88}_{39}\text{Y}, ^{60}_{27}\text{Co}, ^{88}_{39}\text{Y}\}$ at $E_\gamma$ = [661.657(3), 898.042(11), 1173.228(3), 1836.070(8)] keV [51, 68], respectively. Moreover, like in Appendix A.8.2.2, I perform Monte Carlo based uncertainty quantification by systematically propagating the uncertainty in the hyperparameters for the individual spline regression models.

In Fig. A.4, I present the results of the empirical Compton edge shift analysis for the detector channel `#SUM`. In general, we can identify a consistent trend in the Compton edge shift, i.e. an increase in the magnitude of the shift as the associated photon energy $E_\gamma^0$ increases. As already reported in Chapter 6, the magnitude of the shift is substantial with ~20 keV for $E_\gamma = 661.657(3)$ keV increasing up to ~30 keV for $E_\gamma = 1836.070(8)$ keV. Additional results for the other detector channels are presented in the Figs. B.77–B.80.

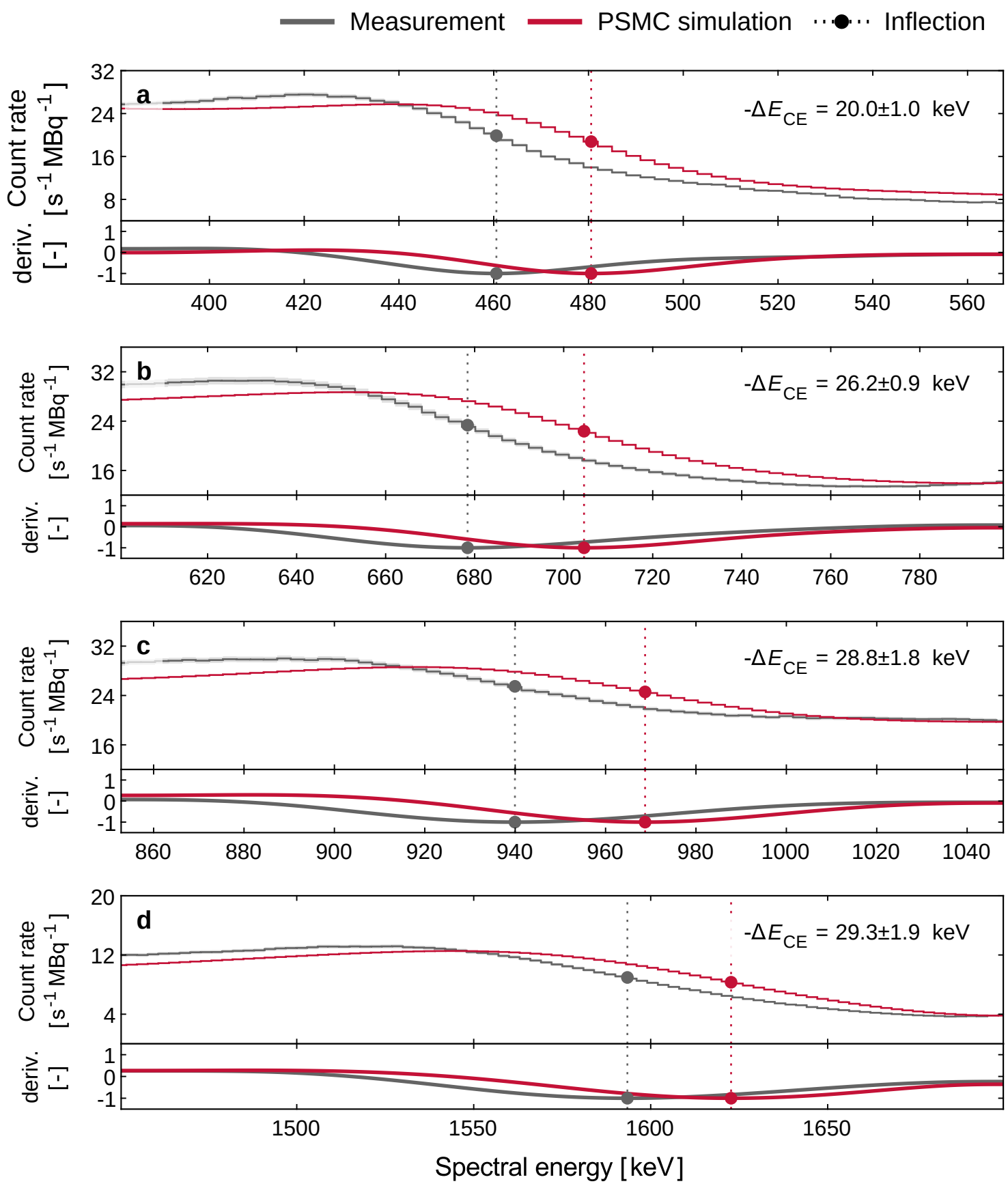

**Figure A.4** Here, the results from the Compton edge shift analysis of four different Compton edges are presented for the detector channel `#SUM`. These Compton edges are: **a** 477.334(3) keV associated with the $^{137}_{55}Cs$ emission line at 661.657(3) keV [68]. **b** 699.133(3) keV associated with the $^{88}_{39}Y$ emission line at 898.042(11) keV [51]. **c** 963.419(3) keV associated with the $^{60}_{27}Co$ emission line at 1173.228(3) keV [68]. **d** 1611.77(1) keV associated with the $^{88}_{39}Y$ emission line at 1836.070(8) keV [51]. The top panels present the measured ($\hat{c}_{exp}$) and simulated ($\hat{c}_{sim}$, derived by PSMC discussed in Chapter 6) mean spectral signatures for the corresponding sources around the Compton edges of interest. The bottom panels show the first derivative (deriv.) of the measured and simulated mean spectral signatures estimated by spline regression [891]. For visualization purposes, the first derivatives were normalized by their corresponding global minima. Uncertainties are provided as 1 standard deviation (SD) shaded areas (spectral signatures) or numerical values (Compton edge shift).

### A.10.2 Photon Energy Dependence

To better understand how the Compton edge shift varies with photon energy $E_\gamma$, I develop here a simplified semi-analytical model using the physics models introduced in Chapters 3 and 4.

In essence, the observed Compton edge shift is a spectral bias introduced by calibrating the spectral energy scale of an inorganic scintillator using full energy peaks (FEPs) as detailed in Section 6.2.1.3. To see this, let us consider an inorganic scintillator with a non-proportional scintillation response to electrons described by an energy-dependent absolute light yield function $Y_{\mathrm{sci,a}}(\mathcal{E}_\mathrm{k})$, where $\mathcal{E}_\mathrm{k}$ denotes the set of initial electron kinetic energies released in a given energy deposition process (cf. Section 4.1.3). Based on these definitions, we can compute the spectral bias for a Compton edge energy deposition event as the difference between the spectral energy $E'$ assigned to the Compton edge using a spectral model calibrated by FEPs and the true deposited energy $E_\mathrm{dep}$:

$$\Delta E_\mathrm{CE} = E' - E_\mathrm{dep} \tag{A.67a}$$

$$= E_\mathrm{dep}\left(\frac{Y_{\mathrm{sci,a}}\left(\mathcal{E}_\mathrm{k}^\mathrm{CE}\right)}{Y_{\mathrm{sci,a}}\left(\mathcal{E}_\mathrm{k}^\mathrm{FEP}\right)} - 1\right) \tag{A.67b}$$

with:

| | | |
|---|---|---|
| $E_\mathrm{dep}$ | deposited energy | eV |
| $E'$ | spectral energy | eV |
| $\mathcal{E}_\mathrm{k}^\mathrm{CE}$ | set of initial kinetic energies of the electrons released in a Compton edge event | |
| $\mathcal{E}_\mathrm{k}^\mathrm{FEP}$ | set of initial kinetic energies of the electrons released in a full energy peak event | |
| $Y_{\mathrm{sci,a}}$ | absolute light yield | $\mathrm{eV}^{-1}$ |

and where the sets of initial kinetic energies $E_{\mathrm{k,e^-}}$ of the electrons released in a Compton edge (CE) event $\mathcal{E}_\mathrm{k}^\mathrm{CE}$ and full energy peak (FEP) event $\mathcal{E}_\mathrm{k}^\mathrm{FEP}$ satisfy:

$$E_{\mathrm{dep}} = \sum_{i \in \mathcal{E}_{\mathrm{k}}^{\mathrm{CE}}} E_{\mathrm{k,e^-}}^{(i)} = \sum_{i \in \mathcal{E}_{\mathrm{k}}^{\mathrm{FEP}}} E_{\mathrm{k,e^-}}^{(i)} \tag{A.68}$$

We can rewrite Eq. A.67b using the energy transfer efficiency $\eta_{\mathrm{cap}}$ introduced in Eq. 4.2 and assuming that the conversion efficiency $\eta_{\mathrm{gen}}$ and the luminescence quantum yield $\eta_{\mathrm{lum}}$ are energy independent (cf. Section 4.1.3):

$$\Delta E_{\mathrm{CE}} = E_{\mathrm{dep}} \left( \frac{\eta_{\mathrm{cap}}\left(\mathcal{E}_{\mathrm{k}}^{\mathrm{CE}}\right)}{\eta_{\mathrm{cap}}\left(\mathcal{E}_{\mathrm{k}}^{\mathrm{FEP}}\right)} - 1 \right) \tag{A.69}$$

where:

| | |
|---|---|
| $\eta_{\mathrm{cap}}$ | energy transfer efficiency |

In order to compute the Compton edge shift $\Delta E_{\mathrm{CE}}$ using Eq. A.69, we need to determine the sets of initial kinetic energies of electrons $\mathcal{E}_{\mathrm{k}}^{\mathrm{CE}}$ and $\mathcal{E}_{\mathrm{k}}^{\mathrm{FEP}}$. For that purpose, we need to consider the energy deposition processes in the scintillator for Compton edge (CE) and FEP events.

As already discussed in Section 4.3.1, in a CE event, a photon enters the scintillator, undergoes a single Compton scattering (COM) event with a deflection angle of $\theta_\gamma = \pi$, i.e. full back-scattering, and subsequently escapes the scintillator. During this interaction, some of the photon's energy gets transferred to a single atomic electron. Neglecting Doppler broadening and atomic shell effects [892, 893], the transferred energy $E_{\mathrm{k,e^-}}^{\mathrm{CE}}$ is equivalent to the Compton edge energy discussed in Eq. 4.21:

$$E_{\mathrm{k,e^-}}^{\mathrm{CE}} = E_\gamma \left(1 - \frac{1}{1 + 2\alpha_\gamma}\right) \tag{A.70}$$

where:

| | | |
|---|---|---|
| $E_\gamma$ | photon energy | eV |
| $\alpha_\gamma$ | ratio of the photon energy to the energy-equivalent electron rest mass (cf. Eq. 3.12) | |

and consequently $\mathcal{E}_{\mathrm{k}}^{\mathrm{CE}} = \{E_{\mathrm{k,e^-}}^{\mathrm{CE}}\}$ and $E_{\mathrm{dep}} = E_{\mathrm{k,e^-}}^{\mathrm{CE}}$.

FEP events, on the other hand, are more complex because they involve a variable number of Compton scattering (COM) events with a subsequent photoelectric absorption (PE) of the photon in the scintillator, i.e. $\mathcal{E}_{\text{k}}^{\text{FEP}} = \{E_{\text{k,e}^-}^{\text{PE}}\} \cup \{E_{\text{k,e}^-,i}^{\text{COM}}\}_{i\in\{1,\dots,N_{\text{COM}}\}}$ with $E_{\text{k,e}^-}^{\text{PE}}$ and $E_{\text{k,e}^-}^{\text{COM}}$, being the initial kinetic energies of the photoelectron and a sequence of $N_{\text{COM}}$ Compton electrons, respectively.[23] For simplicity, I neglect again Doppler broadening as well as atomic shell effects and consider only secondary electrons generated during COM and PE events. In particular, I neglect fluorescence photons and Auger electrons. Using these simplifications, we can calculate the light yield for a FEP event as a sequence of $N_{\text{COM}}$ COM events followed by a single PE event:

23 *Note that I neglect here also FEP events caused by pair production interactions for which all secondary radiation gets absorbed in the scintillator (cf. Section 4.3).*

$$\eta_{\text{cap}}\left(\mathcal{E}_{\text{k}}^{\text{FEP}}\right) = \eta_{\text{cap}}\left(E_{\text{k,e}^-}^{\text{PE}}\right) + \sum_{i=1}^{N_{\text{COM}}} \eta_{\text{cap}}\left(E_{\text{k,e}^-,i}^{\text{COM}}\right) \tag{A.71a}$$

$$\approx \eta_{\text{cap}}\left(E_{\gamma}^{(N_{\text{COM}})}\right) + \sum_{i=1}^{N_{\text{COM}}} \eta_{\text{cap}}\left(E_{\gamma}^{(i-1)} - E_{\gamma}^{(i)}\right) \tag{A.71b}$$

where I denote with $E_{\gamma}^{(i)}$ the photon energy after $i$ subsequent COM events. Using the assumptions made to derive Eq. A.71b, we can also simplify the condition in Eq. A.68 to:

$$E_{\text{dep}} = E_{\text{k,e}^-}^{\text{CE}} = E_{\gamma}^{0} \tag{A.72}$$

with $E_{\gamma}^{0}$ being the initial photon energy, which induces the FEP event.

Using these results, we can rewrite Eq. A.69 as a function of the initial photon energy $E_{\gamma}^{0}$ and the number of COM events $N_{\text{COM}}$ associated with a FEP event in a given inorganic scintillator:

$$\Delta E_{\text{CE}} \approx E_{\gamma}^{0}\left(\frac{\eta_{\text{cap}}\left(E_{\gamma}^{0}\right)}{\eta_{\text{cap}}\left(E_{\gamma}^{(N_{\text{COM}})}\right) + \sum_{i=1}^{N_{\text{COM}}} \eta_{\text{cap}}\left(E_{\gamma}^{(i-1)} - E_{\gamma}^{(i)}\right)} - 1\right) \tag{A.73}$$

Both, the number of COM events ($N_{\text{COM}}$) as well as the transferred energy in these COM events ($E_{\gamma}^{(i-1)} - E_{\gamma}^{(i)}$) are linked in a complex stochastic process and depend on the photon energy as well as the properties of the scintillator [221, 892, 893]. Computing the Compton edge shift $\Delta E_{\text{CE}}$ by the simplified model in Eq. A.73 requires therefore two main steps:

1. **COM sequence** In the first step, we need to estimate the number of COM events ($N_{\text{COM}}$) as well as the transferred energy in these COM events ($E_\gamma^{(i-1)} - E_\gamma^{(i)}$). For that purpose, I performed Monte Carlo simulations using the `FLUKA` code [20, 216, 281] with the same physics settings as described in Section 6.2.2.1. As a mass model, I used a prismatic NaI(Tl) scintillator with dimensions 10.2 cm × 10.2 cm × 40.6 cm, i.e. the same dimension as the individual crystals in the Radiometrie Land-Luft (RLL) spectrometer (cf. Section 5.3.2), embedded in a vacuum environment. To estimate the scintillator response, I irradiated the mass model with an isotropic and homogeneous monoenergetic photon flux of energy $E_\gamma^0$ using the `FLOOD` mode with the `BEAMPOSit` card. I repeated these simulations for 33 different photon energies $E_\gamma^0$ in the spectral range [400–2000 keV] with a spacing of 50 keV. To score $N_{\text{COM}}$ as well as $E_\gamma^{(i)}$, I applied the user routine `mgdraw`.

   In Fig. A.5, I present the probability density for the scored number of COM events before absorption $N_{\text{COM}}$ as a function of the photon energy $E_\gamma^0$. In line with previous results [693], I find a moderate increase in the mean number of COM events with increasing photon energy ranging from ~1.1 at 400 keV up to ~2.3 at 2000 keV.

2. **Light yield model** In the second step, the obtained results from the first step need to be convolved with a light yield model quantifying the energy transfer efficiency $\eta_{\text{cap}}$ as a function of the initial electron kinetic energy. For that purpose, I adopted the light yield model derived by Payne and his co-workers [313, 324, 325] and discussed in detail in Section 4.1.3 and Appendix A.7. Specifically, I performed numerical integration of Eq. 4.9, which requires two additional physics models:

   i. **Stopping power model** To quantify the collisional stopping power of electrons $\mathcal{S}_{\text{e,col}}$ as a function of the electron kinetic energy $E_{\text{k,e}^-}$, similar to Payne and his co-workers [313, 324], I applied an empirical model derived by Joy and Luo [334] for $E_{\text{k,e}^-} < 10$ keV (cf. Eq. 4.8) and an adapted Bethe formula for $E_{\text{k,e}^-} \geq 10$ keV (cf. Eq. 4.6). I consulted the ESTAR database by the National Institute of Standards and Technology (NIST) [318] to evaluate Eq. 4.6 and to compute associated material properties such as the density effect parameter $\delta$ or mean excitation energy $I_0$. All adopted material parameters for NaI(Tl) are listed in Table C.12. The resulting collisional stopping power model is displayed in Fig. B.18 for NaI(Tl).

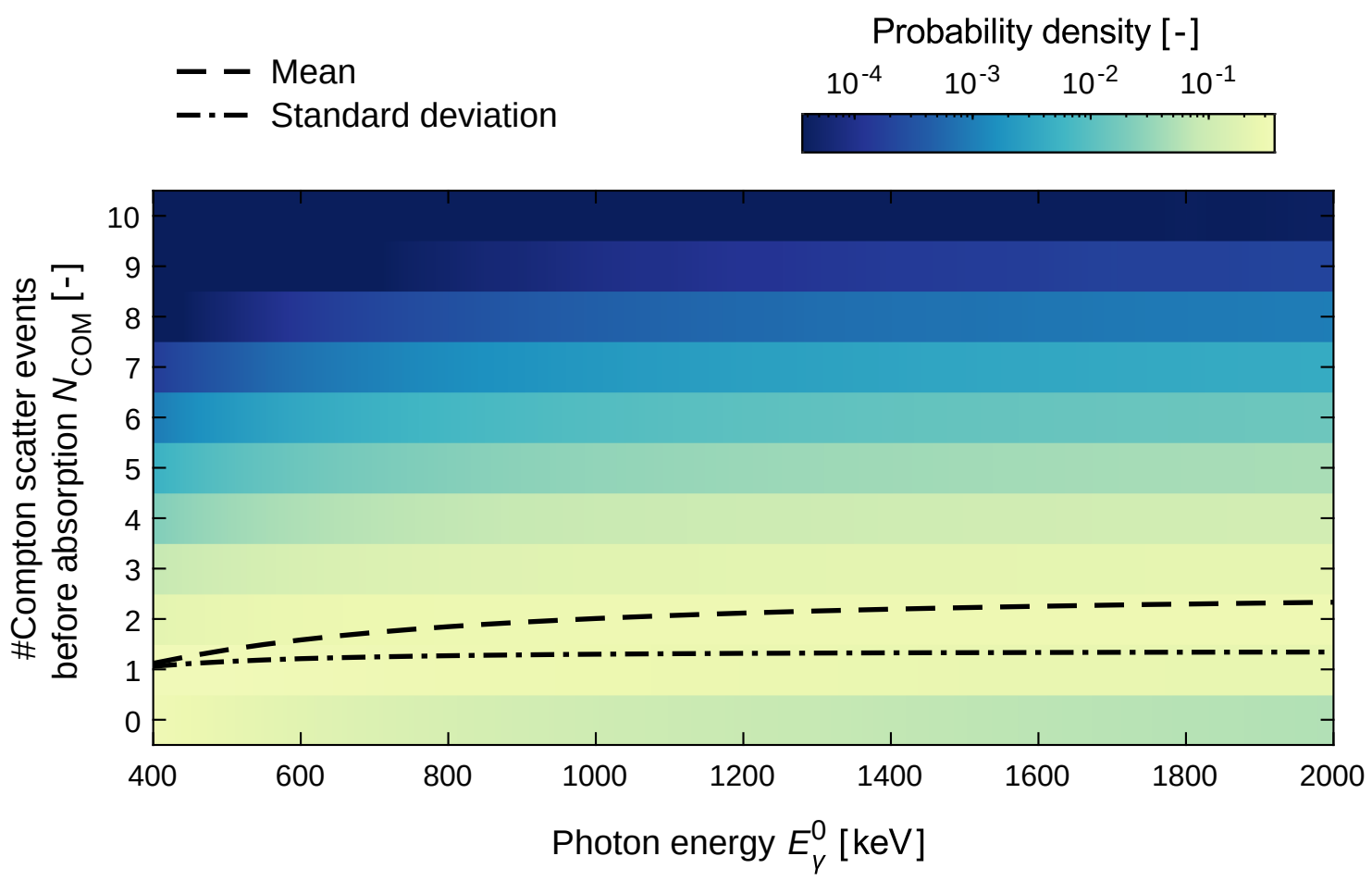

**Figure A.5** Here, I show the probability density for the number of Compton scatter events before absorption $N_{\mathrm{COM}}$ as a function of the initial photon energy $E_{\gamma}^{0}$ for a generic 10.2 cm × 10.2 cm × 40.6 cm prismatic NaI(Tl) scintillation crystal (cf. Table C.12) exposed to a homogeneous isotropic monoenergetic photon flux with photon energy $E_{\gamma}^{0}$. The probability density was estimated by Monte Carlo simulations using the multi-purpose code `FLUKA` [20, 216, 281].

ii. **Differential light yield** To quantify the differential light yield $\propto \eta'_{\mathrm{cap}}$ as a function of the electron kinetic energy $E_{\mathrm{k,e^-}}$, I adopted the mechanistic model in Eq. 4.4 derived by Payne and his co-workers [313, 324, 325], which is also the main scintillation physics model adopted in Chapter 7 to perform NPSMC. For the model parameters in Eq. 4.4, i.e. the Birks stopping parameter $\mathrm{d}E/\mathrm{d}x\,|_{\mathrm{Birks}}$, the electron-hole pair fraction $\eta_{\mathrm{e/h}}$ and the trapping stopping parameter $\mathrm{d}E/\mathrm{d}x\,|_{\mathrm{trap}}$, I applied the maximum a posteriori (MAP) point estimates obtained by Compton edge probing (cf. Section 7.2.4) as well as the full set of posterior samples to derive the corresponding prediction intervals for all five detector channels. As suggested by previous investigators [313, 325], I fixed the Onsager stopping parameter $\mathrm{d}E/\mathrm{d}x\,|_{\mathrm{Ons}}$ to 36.4 MeV cm$^{-1}$.

In Fig. A.6, I show the relative light yield $Y_{\mathrm{sci,r}}$ as a function of the electron kinetic energy $E_{\mathrm{k,e^-}}$ (cf. Eq. 4.10b) for the four individual scintillation crystals `#1` through `#4` and the detector channel `#SUM` of the RLL spectrometer. As a reference, the relative light yield obtained by model parameters retrieved from the literature [313] is shown as well (cf. also to Table C.4). The characteristic shape of these relative light yield curves in Fig. A.6 has been extensively documented by numerous previous empirical studies [313, 324, 345, 347, 355, 413], illustrating an increase in light yield with increasing energy for $E_{\mathrm{k,e^-}} < 7\,\mathrm{keV}$, a prominent peak around ~7 keV, followed by a subsequent decrease in yield for higher energies. As expected from the discussion in Section 7.3.1 and in line with previous experimental findings [413], I find statistically significant differences between the relative light yield for the individual scintillation crystals and the relative light yield obtained by the literature model parameters.

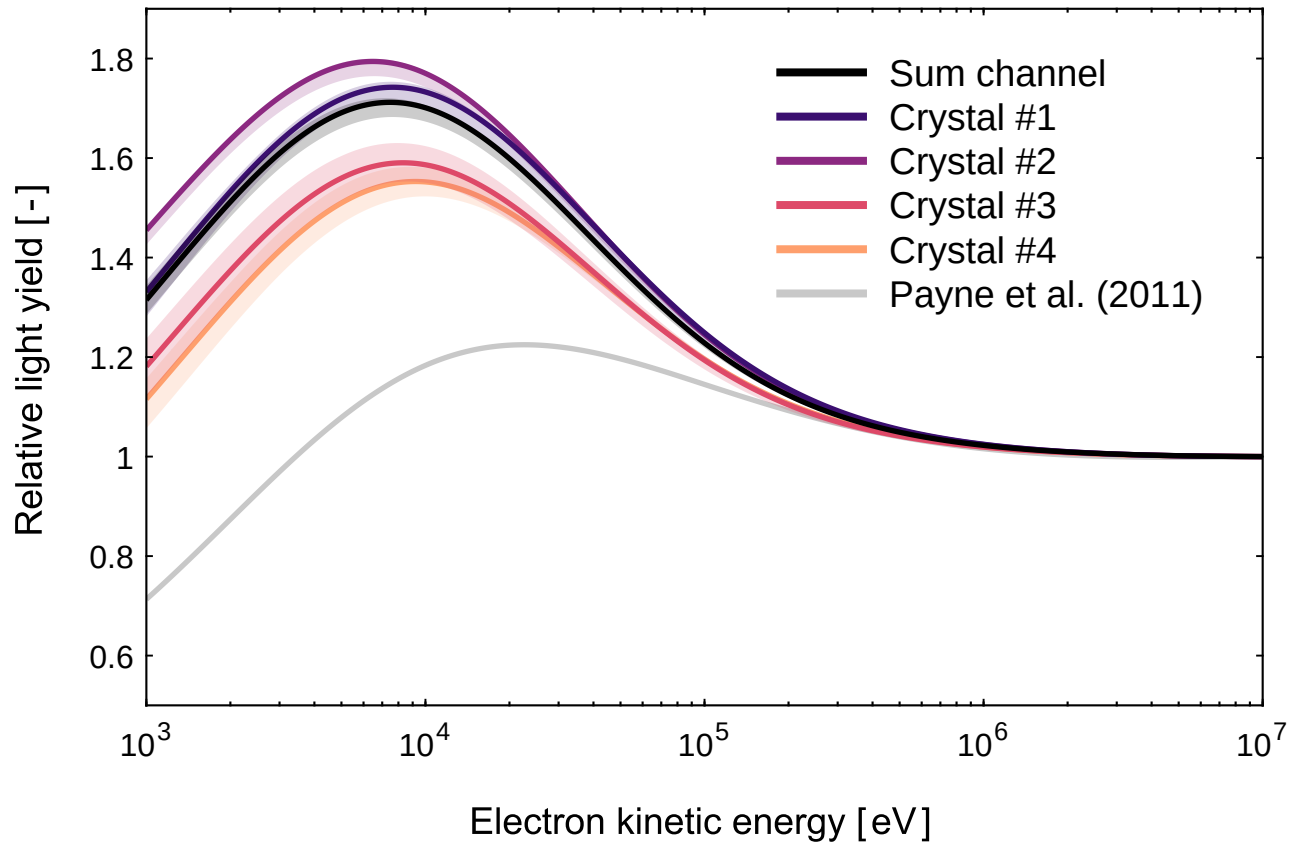

**Figure A.6** This graph shows the relative light yield $Y_{\mathrm{sci,r}}$ as a function of the electron kinetic energy $E_{\mathrm{k,e^-}}$ for the four individual scintillation crystals `#1` through `#4` (**a–d**) and the detector channel `#SUM` (**e**) of the RLL spectrometer. The reference energy for $Y_{\mathrm{sci,r}}$ was set to 10 MeV. The $Y_{\mathrm{sci,r}}$ was computed using the models described by Eqs. 4.4 and 4.9 (cf. Section 4.1.3). A combined stopping power model for electrons was used to perform the integration in Eq. 4.9 (cf. Fig. B.18). For the model parameters in Eq. 4.4 ( $\mathrm{d}E/\mathrm{d}x|_{\mathrm{Birks}}$, $\mathrm{d}E/\mathrm{d}x|_{\mathrm{Ons}}$, $\eta_{\mathrm{e/h}}$, $\mathrm{d}E/\mathrm{d}x|_{\mathrm{trap}}$), Bayesian calibrated maximum a posteriori (MAP) probability point estimates have been used. In addition, 99 % central prediction intervals (PIs) for each individual relative light yield function are displayed, derived by the full set of posterior samples. As a reference, the relative light yield obtained by model parameters retrieved from the literature [313] is shown as well (cf. also to Table C.4).

In Fig. A.7, I show the resulting median Compton edge shift $\Delta E_{\mathrm{CE}}$ predicted by Eq. A.73 together with the 95 % prediction interval as a function of the initial photon energy $E_\gamma^0$. In addition, I present also the predicted Compton edge shift for the discriminated number of COM events $N_{\mathrm{COM}}$. From Fig. A.7, it is evident that the magnitude of the Compton edge shift $|\Delta E_{\mathrm{CE}}|$ increases with increasing photon energy $E_\gamma^0$ as well as the number of COM events $N_{\mathrm{COM}}$. Both trends can be explained by the decreasing $\eta_{\mathrm{cap}}$ for higher energies (cf. Fig. A.6). Furthermore, this trend results also in $\eta_{\mathrm{cap}}(\mathcal{E}_{\mathrm{k}}^{\mathrm{CE}}) < \eta_{\mathrm{cap}}(\mathcal{E}_{\mathrm{k}}^{\mathrm{FEP}})$ and subsequently in a negative Compton edge shift $\Delta E_{\mathrm{CE}} < 0$ in line with the empirical results discussed in Appendix A.10.1.

From these findings, we can also conclude that the spectral bias introduced by the scintillation non-proportionality is in general maximized for energy deposition events releasing a single high-energy electron. Given that Compton edge events release a Compton electron with a particularly high kinetic energy (cf. Section 4.3.1), this explains why Compton edge events are particularly sensitive to the scintillation non-proportionality.

In Fig. A.7, I have also included the experimentally determined Compton edge shift results discussed in Appendix A.10.1. In general, the simplified model in Eq. A.73 provides a moderate agreement with the empirical results. Deviations can be attributed to the various simplifications and assumptions made during model derivation, e.g. neglection of pair production interactions, Landau fluctuations, Doppler broadening, atomic shell effects, detector cross-talk, secondary particles such as fluorescence photons and Auger electrons or escaping electrons.

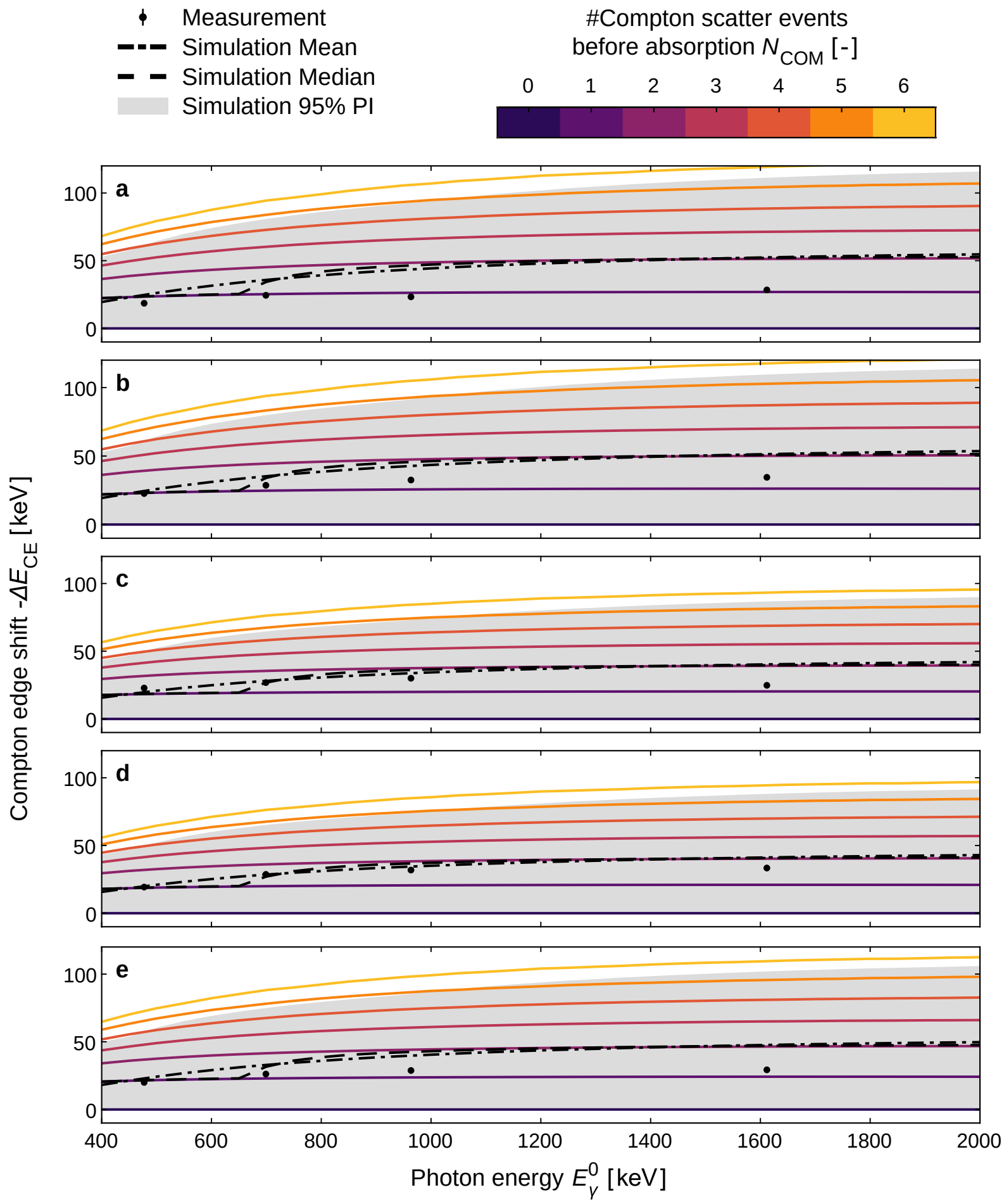

**Figure A.7** Measured and predicted mean and median Compton edge shift $\Delta E_{CE}$ are shown together with the 95 % prediction interval (PI) for the four single detector channels `#1` through `#4` (**a–d**) and the detector channel `#SUM` (**e**) of the RLL spectrometer. In addition, the predicted Compton edge shift $\Delta E_{CE}$ discriminated for individual number of Compton scatter events before absorption $N_{COM}$ are indicated as well for each detector channel. The measured Compton edge shifts were obtained by Compton edge shift analysis of four different Compton edges, i.e. [477.334(3), 699.133(3), 963.419(3), 1611.77(1)] keV associated with the photon emission lines of the radionuclides $\{^{137}_{55}Cs, ^{88}_{39}Y, ^{60}_{27}Co, ^{88}_{39}Y\}$ at [661.657(3), 898.042(11), 1173.228(3), 1836.070(8)] keV [51, 68], respectively. Measurement uncertainties are provided as 1 standard deviation (SD) error bars (hidden by the marker size).

### A.10.3 Scintillator Volume Dependence

The Compton edge shift model derived in the previous subsection can also be leveraged to investigate the relation between the Compton edge shift and the size of a scintillation crystal. For that purpose, I evaluated Eq. A.73 using the same method as described in Appendix A.10.2 for three additional equilateral cylindrical NaI(Tl) crystals with characteristic lengths of [2.54, 5.08, 7.62] cm and associated volumes of [4, 33, 111] $cm^3$. For all crystals, I applied the NPSM parameters of the detector channel `#SUM` obtained by Compton edge probing (cf. Table C.10).

In Fig. A.8, I present the predicted mean Compton edge shift $\Delta E_{CE}$ as a function of the photon energy $E_{\gamma}^{0}$ for the four different scintillator crystals alongside the relation between $N_{COM}$ and $E_{\gamma}^{0}$. I find a pronounced and consistent increase in the magnitude of the Compton edge shift $|\Delta E_{CE}|$ for an increase in crystal size over the entire spectral domain [400–2000 keV]. From our discussion in the previous subsection and the results displayed in Fig. A.8a, it is evident that this trend can be explained by the increase in $N_{COM}$ for bigger scintillation crystals [30, 693].

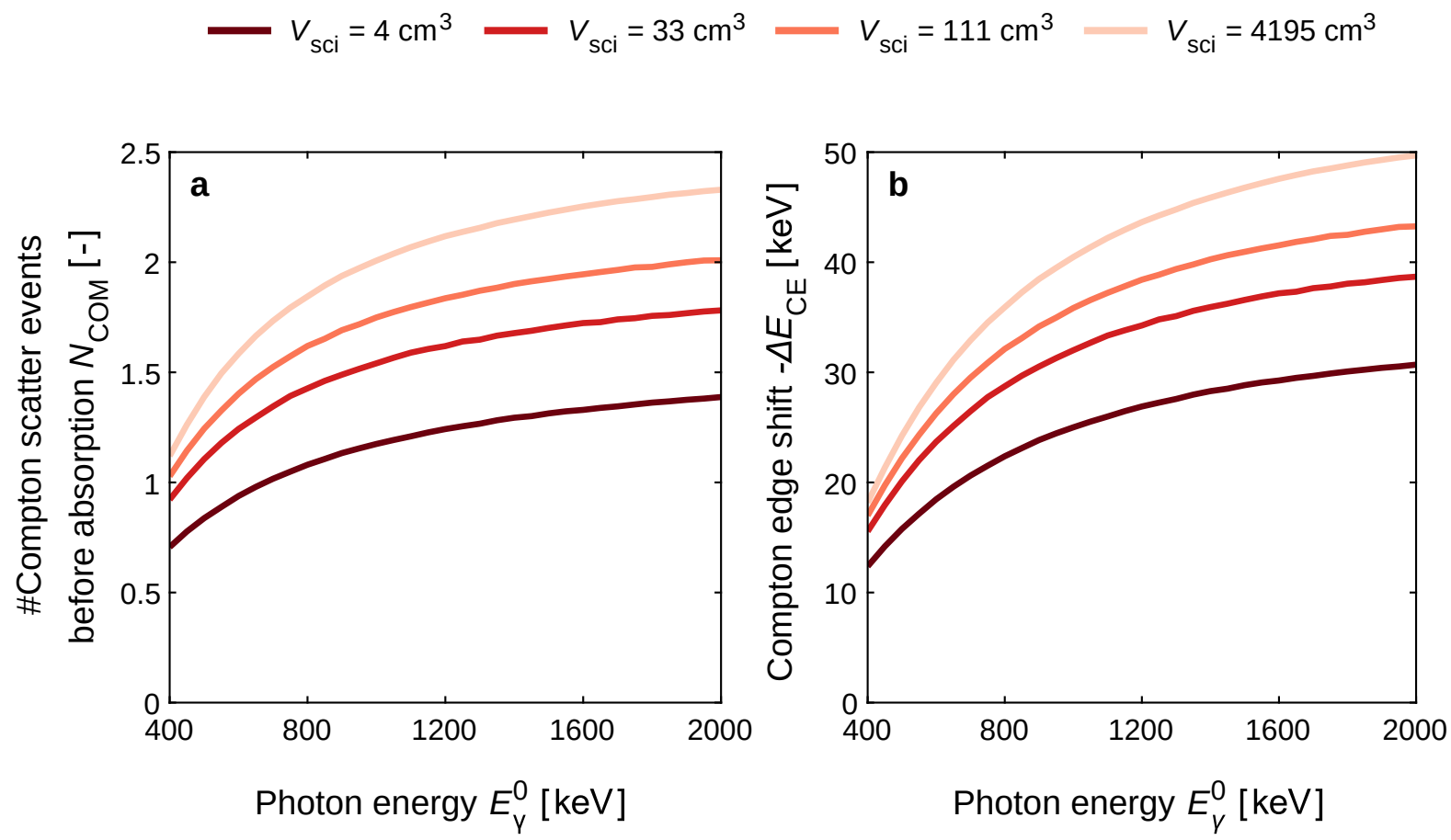

**Figure A.8** Here, I present Monte Carlo based estimates of the mean number of Compton scatter events before absorption $N_{COM}$ (**a**) together with mean Compton edge shift $\Delta E_{CE}$ predictions (**b**) as a function of the initial photon energy $E_\gamma^0$ for four different NaI(Tl) scintillation crystals, i.e. three equilateral cylindrical crystals with characteristic lengths [2.54, 5.08, 7.62] cm and associated volumes ($V_{sci}$) [4, 33, 111] cm$^3$ as well as a 10.2 cm × 10.2 cm × 40.6 cm prismatic crystal with a volume of 4195 cm$^3$.

## A.11 Adaptive Sparse PCE-PCA Surrogate Model

Here, based on previous work [746, 749, 894], I derive the custom vector-valued adaptive sparse polynomial chaos expansion (PCE)-principal component analysis (PCA) surrogate model, which I have applied in Section 7.2.4.1 to emulate the expensive-to-evaluate NPSMC forward model. The model consists of two parts: the PCA model and the PCE model. The PCA model reduces the dimensionality of the model response by transforming the response variables into a new space spanned by the principal components. The PCE model, on the other hand, approximates the model response in the new space using a truncated polynomial chaos expansion. The combination of both models allows for a more efficient emulation of the model response as detailed by Blatman et al. [749].

### A.11.1 Principal Component Analysis

Let us start with the PCA model part. Consider the vector-valued model response as a random vector $\boldsymbol{Y} \in \mathbb{R}^{N\times 1}$ with mean $\boldsymbol{\mu}_{\boldsymbol{Y}}$, standard deviation $\boldsymbol{\sigma}_{\boldsymbol{Y}}$ and correlation matrix $\boldsymbol{\mathcal{R}}_{\boldsymbol{Y}} := \operatorname{corr}(\boldsymbol{Y}) = \mathbb{E}[\boldsymbol{Y}^*(\boldsymbol{Y}^*)^\top]$.

Note that, in contrast to previous studies [746, 749, 894], I standardize the model response $\boldsymbol{Y}$ with $\boldsymbol{Y}^* := \operatorname{diag}(\boldsymbol{\sigma}_{\boldsymbol{Y}})^{-1}(\boldsymbol{Y} - \boldsymbol{\mu}_{\boldsymbol{Y}})$ to account for the differences in the variance of the individual response variables [677]. Using this standardized response, we can then perform an eigenvalue decomposition of the correlation matrix $\boldsymbol{\mathcal{R}}_{\boldsymbol{Y}}$ with the eigenvalues $\Lambda_j$ and eigenvectors $\boldsymbol{\phi}_j := (\phi_1, \ldots, \phi_N)^\top$ satisfying $\boldsymbol{\mathcal{R}}_{\boldsymbol{Y}}\boldsymbol{\phi}_j = \Lambda_j \boldsymbol{\phi}_j$ for $\{j \in \mathbb{N}_+ \mid j \leq N\}$. Since $\boldsymbol{\mathcal{R}}_{\boldsymbol{Y}}$ is symmetric and positive definite, the eigenvectors define an orthonormal basis $\mathbb{R}^N = \operatorname{span}(\{\boldsymbol{\phi}_j\}_{j=1}^N)$ and we can perform an orthogonal transformation of our random vectors $\boldsymbol{Y}^*$ as follows:

$$\boldsymbol{Z} = \boldsymbol{\Phi}^\top \boldsymbol{Y}^* \tag{A.74}$$

with the orthonormal matrix $\boldsymbol{\Phi} := (\boldsymbol{\phi}_1, \ldots, \boldsymbol{\phi}_N) \in \mathbb{R}^{N\times N}$, where $\Lambda_1 \geq \Lambda_2 \geq \ldots \geq \Lambda_N$. We call the eigenvectors $\boldsymbol{\phi}_j$ also the principal components or principal directions of the model response $\boldsymbol{Y}$ [677]. The original response vector $\boldsymbol{Y}$ can be retrieved from the transformed vectors $\boldsymbol{Z} = \left(Z_1, \ldots, Z_N\right)^\top$ by inverting Eq. A.74:

$$\boldsymbol{Y} = \boldsymbol{\mu}_{\boldsymbol{Y}} + \operatorname{diag}(\boldsymbol{\sigma}_{\boldsymbol{Y}}) \sum_{j=1}^{N} Z_j \boldsymbol{\phi}_j \tag{A.75}$$

To reduce the dimensions of our problem, we can now retain only the first $N'$ principal components with the highest variance thereby approximating the random vector $\boldsymbol{Y}$ as:

$$\boldsymbol{Y} \approx \boldsymbol{\mu}_{\boldsymbol{Y}} + \operatorname{diag}(\boldsymbol{\sigma}_{\boldsymbol{Y}}) \sum_{j=1}^{N'} Z_j \boldsymbol{\phi}_j \tag{A.76}$$

where I choose:

$$N' = \min\{S \in \mathbb{N}_+ \mid S \leq N, \sum_{j=1}^{S} \Lambda_j / \sum_{j=1}^{N} \Lambda_j \geq 1 - \varepsilon_{\text{PCA}}\} \tag{A.77}$$

with a prescribed relative truncation error $\varepsilon_{\text{PCA}}$.[24] This PCA transformation with a reduced number of principal components $N'$ is also known as principal component approximation [749] and is a prevalent tool in data analysis for dimensionality reduction applications [677].

24 *Set to* 0.1 % *in this work.*

### A.11.2 Polynomial Chaos Expansion

For the PCE model part, let us start with the polynomial chaos expansion of the model response $\mathcal{M}(\boldsymbol{X})$ with the random input vector $\boldsymbol{X} \in \mathbb{R}^{M\times 1}$ as described in Eq. 7.6:

$$\boldsymbol{Y} = \hat{\mathcal{M}}(\boldsymbol{X}) = \sum_{\boldsymbol{\alpha} \in \mathbb{N}^{M\times 1}} \mathbf{a}_{\boldsymbol{\alpha}} \Psi_{\boldsymbol{\alpha}}(\boldsymbol{X}) \tag{A.78}$$

where $\mathbf{a}_{\boldsymbol{\alpha}} = (a_{1,\boldsymbol{\alpha}}, \ldots, a_{N,\boldsymbol{\alpha}})^\top \in \mathbb{R}^{N\times 1}$ are the deterministic expansion coefficients, $\boldsymbol{\alpha} = (\alpha_1, \ldots, \alpha_M)^\top \in \mathbb{N}^{M\times 1}$ the multi-indices storing the degrees of the univariate polynomials $\psi_\alpha$ and $\Psi_{\boldsymbol{\alpha}}(\boldsymbol{X}) = \prod_{i=1}^{M} \psi_{\alpha_i}^{(i)}(X_i)$ the multivariate polynomial basis functions, which are orthonormal with respect to the joint probability density function $\pi_{\boldsymbol{X}}$, i.e. $\langle \Psi_{\boldsymbol{\alpha}}, \Psi_{\boldsymbol{\beta}} \rangle_{\pi_{\boldsymbol{X}}} = \delta_{\boldsymbol{\alpha},\boldsymbol{\beta}}$ (cf. Eq. 7.8). In this work, given the predefined probabilistic model (cf. Table C.8), the univariate polynomial functions $\psi_\alpha$ are equivalent to the Legendre[25] polynomials $\mathcal{P}_\alpha$ [748].

25 *First described by Adrien-Marie Legendre (*1752, †1833), a prominent French mathematician who made significant contributions to number theory, algebra and statistics.*

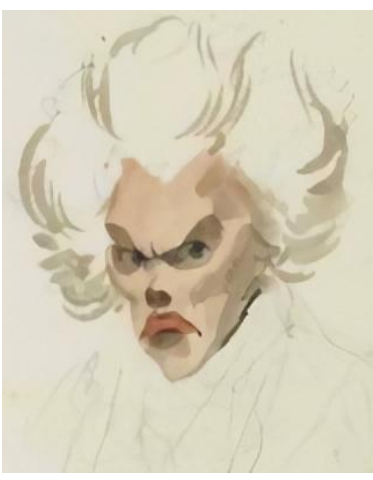

Adrien-Marie Legendre
Ⓢ Julien-Léopold Boilly

The sum in Eq. A.78 involves an infinite number of terms. As a result, for practical applications, we need to truncate the expansion to a finite number of terms by adopting a truncation set $\mathcal{A}_j$ for the multi-index $\boldsymbol{\alpha}$ of each individual response variable $\{j \in \mathbb{N}_+ \mid j \leq N\}$ resulting in:

$$Y_j \approx \hat{\mathcal{M}}(\boldsymbol{X}) = \sum_{\boldsymbol{\alpha}\in\mathcal{A}_j} a_{j,\boldsymbol{\alpha}} \Psi_{\boldsymbol{\alpha}}(\boldsymbol{X}) \tag{A.79}$$

As discussed in Section 7.2.4.1, I apply a hyperbolic truncation scheme defining the multi-index set as $\mathcal{A}_j = \{\boldsymbol{\alpha} \in \mathbb{N}^M \mid (\sum_{i=1}^{M} \alpha_i^q)^{1/q} \leq p\}$ with $p$ and $q$ being hyperparameters defining the maximum degree for the associated polynomial and the q-norm, respectively [750].

### A.11.3 PCE-PCA Surrogate Model

To reduce the computational burden, we can now combine these results and perform the PCE not in the original response variable space but in the truncated principal component space $\mathbb{R}^{N'} = \mathrm{span}(\{\boldsymbol{\phi}_j\}_{j=1}^{N'})$. For that, we insert Eq. A.79 into Eq. A.76:

$$\boldsymbol{Y} \approx \hat{\mathcal{M}}(\boldsymbol{X}) = \boldsymbol{\mu}_{\boldsymbol{Y}} + \mathrm{diag}(\boldsymbol{\sigma}_{\boldsymbol{Y}}) \sum_{j=1}^{N'} \left( \sum_{\boldsymbol{\alpha}\in\mathcal{A}_j} a_{j,\boldsymbol{\alpha}} \Psi_{\boldsymbol{\alpha}}(\boldsymbol{X}) \right) \boldsymbol{\phi}_j \tag{A.80}$$

which we can rearrange by introducing the union set $\mathcal{A}^\star = \bigcup_{j=1}^{N'} \mathcal{A}_j$ to:

$$\boldsymbol{Y} \approx \hat{\mathcal{M}}(\boldsymbol{X}) = \boldsymbol{\mu}_{\boldsymbol{Y}} + \mathrm{diag}(\boldsymbol{\sigma}_{\boldsymbol{Y}}) \sum_{\boldsymbol{\alpha}\in\mathcal{A}^\star} \sum_{j=1}^{N'} a_{j,\boldsymbol{\alpha}} \Psi_{\boldsymbol{\alpha}}(\boldsymbol{X}) \boldsymbol{\phi}_j \tag{A.81}$$

or expressed in a more compact matrix form:

$$\boldsymbol{Y} \approx \hat{\mathcal{M}}(\boldsymbol{X}) = \boldsymbol{\mu}_{\boldsymbol{Y}} + \mathrm{diag}(\boldsymbol{\sigma}_{\boldsymbol{Y}}) \boldsymbol{\Phi}' \mathbf{A} \boldsymbol{\Psi}(\boldsymbol{X}) \tag{A.82}$$

with the vector $\boldsymbol{\Psi}(\boldsymbol{X}) \in \mathbb{R}^{\mathrm{card}(\mathcal{A}^\star)\times 1}$ as well as the two matrices $\boldsymbol{\Phi}' \in \mathbb{R}^{N\times N'}$ and $\mathbf{A} \in \mathbb{R}^{N'\times \mathrm{card}(\mathcal{A}^\star)}$ storing the multivariate orthonormal polynomials $\Psi_{\boldsymbol{\alpha}}$, the retained eigenvectors $\boldsymbol{\phi}_j$ and the PCE coefficients $a_{j,\boldsymbol{\alpha}}$, respectively.

### A.11.4 Model Training

For training and validation of the PCE-PCA surrogate model, a set of model input parameters $\mathcal{X} = \{\mathbf{x}_{\mathcal{M}}^{(i)} \in \mathbb{R}^{M\times 1} \mid i \in \mathbb{N}_+, i \leq K\}$ together with the evaluated vector-valued model responses $\mathcal{Y} = \{\mathbf{y}^{(i)} \in \mathbb{R}^{N\times 1} \mid i \in \mathbb{N}_+, i \leq K\}$ is required. To define the set of input parameters, I adopt a Latin hypercube sampling scheme [751, 752] with $K = 200$ instances using a predefined probabilistic model (cf. Table C.8). These $K$ instances need then to be propagated through the original forward model $\mathcal{M}(\boldsymbol{X})$, i.e. the NPSMC model described in Sections 7.2.1–7.2.3, to obtain the related model response vectors $\mathbf{y} \in \mathbb{R}_+^{N\times 1}$, representing the spectral signature within the Compton edge domain $\mathcal{D}_{\text{CE}}$. These $K$ model response vectors are then arranged in a response matrix $\boldsymbol{\mathcal{Y}} = (\mathbf{y}^{(1)}, \ldots, \mathbf{y}^{(K)}) \in \mathbb{R}_+^{N\times K}$. Similarly, we may arrange also the sampled model input parameters in an experimental design or input matrix $\boldsymbol{\mathcal{X}} = (\mathbf{x}_{\mathcal{M}}^{(1)}, \ldots, \mathbf{x}_{\mathcal{M}}^{(K)}) \in \mathbb{R}_+^{M\times K}$.

For the PCA model, we can use the obtained set of model response vectors $\mathbf{y}$ to estimate $\boldsymbol{\mu}_Y$, $\boldsymbol{\sigma}_Y$ as well as $\boldsymbol{\mathcal{R}}_Y$:

$$\boldsymbol{\mu}_Y \approx \frac{1}{K}\sum_{k=1}^{K} \mathbf{y}^{(k)} \tag{A.83a}$$

$$\boldsymbol{\sigma}_Y \approx \left[\frac{1}{K-1}\sum_{k=1}^{K}\left(\mathbf{y}^{(k)} - \boldsymbol{\mu}_Y\right)^{\circ 2}\right]^{\circ\frac{1}{2}} \tag{A.83b}$$

$$\boldsymbol{\mathcal{R}}_Y \approx \frac{1}{K-1}\boldsymbol{\mathcal{Y}}^* \left(\boldsymbol{\mathcal{Y}}^*\right)^\top \tag{A.83c}$$

with $\circ$ denoting the Hadamard, i.e. element-wise, power and $\boldsymbol{\mathcal{Y}}^*$ the standardized response matrix storing the standardized response variables $\mathbf{y}^* := \operatorname{diag}(\boldsymbol{\sigma}_Y)^{-1}(\mathbf{y} - \boldsymbol{\mu}_Y)$, i.e. $\boldsymbol{\mathcal{Y}}^* = (\mathbf{y}^{*(1)}, \ldots, \mathbf{y}^{*(K)}) \in \mathbb{R}_+^{N\times K}$. To perform the PCA, i.e. compute the matrix $\boldsymbol{\Phi}'$, I developed a custom `MATLAB` script using the built-in `pca` function.

For PCE, on the other hand, the model parameters contained in the PCE coefficient matrix $\mathbf{A}$ are derived from both the input matrix $\boldsymbol{\mathcal{X}}$ and the response matrix $\boldsymbol{\mathcal{Y}}$. Because the PCE model is evaluated in the reduced principal component space $\mathbb{R}^{N'} = \operatorname{span}(\{\boldsymbol{\phi}_j\}_{j=1}^{N'})$, for PCE model training, we need to transform the response matrix $\boldsymbol{\mathcal{Y}}$ to this space using Eq. A.74:

$$\boldsymbol{\mathcal{Z}} = \boldsymbol{\Phi}'^\top \boldsymbol{\mathcal{Y}}^* \tag{A.84}$$

with the transformed response matrix $\boldsymbol{\mathcal{Z}} \in \mathbb{R}^{N'\times K}$.

A rich variety of non-intrusive and sparse methods exist to estimate the PCE coefficient matrix $\mathbf{A}$ [895, 896]. For the NPSMC

surrogate model in Chapter 7, I applied the least angle regression algorithm [750] due to its high evaluation speed and its high accuracy even for very small experimental designs. To estimate the PCE hyperparameters, i.e. $p$ and $q$ in Eq. A.77, I applied machine learning for $p \coloneqq \{1, 2, \ldots, 7\}$ and $q \coloneqq \{0.5, 0.6, \ldots, 1\}$ with a holdout partition of 80 % and 20 % for the training and test set, respectively. The resulting generalization error of the surrogate models, characterized by the relative mean squared error over the test set [267], is <2 % for the four single detector channels `#1` through `#4` and <1 % for the detector channel `#SUM`, respectively. All PCE computations were performed with the `UQLab` code [753]. Because the `UQLab` code [753] does not offer a direct implementation of the PCE-PCA surrogate model, custom `MATLAB` scripts were developed to combine the PCE and PCA models as described in Eq. A.82.

## A.12 PCE-PCA based Hoeffding-Sobol Decomposition

One benefit of using PCE surrogate models to emulate computationally expensive simulations is the analytical relation between PCE and Sobol′ indices, enabling global sensitivity analysis of the emulated model with minimal additional computational cost [742]. For completeness, I repeat here some of the theory already discussed elsewhere [742, 746, 897, 898] and derive the PCE-PCA based Sobol′ indices accounting for the custom standardization in the PCA discussed in the previous subsection.

### A.12.1 Hoeffding-Sobol Decomposition

Let me start with the global variance decomposition theory derived by Sobol′ [898]. It can be shown that any univariate integrable function $\mathcal{M}(\boldsymbol{X})$ with $M$ mutually independent random input variables $\boldsymbol{X} \in \mathcal{D}_{\boldsymbol{X}} \subseteq \mathbb{R}^{M\times 1}$ can be uniquely decomposed into a sum of orthogonal terms, often referred to as the Hoeffding-Sobol decomposition [898]:

$$\mathcal{M}(\boldsymbol{X}) = \mathcal{M}_0 + \sum_{i=1}^{M} \mathcal{M}_i(X_i) + \sum_{1\le i<j\le M} \mathcal{M}_{i,j}\left(X_i, X_j\right) + \dots + \mathcal{M}_{1,2,\dots,M}\left(X_1,\dots,X_M\right) \quad \text{(A.85)}$$

which can be written in a more compact form by introducing an index set $\mathcal{U}$ as [267]:

$$\mathcal{M}(\boldsymbol{X}) = \mathcal{M}_0 + \sum_{\mathcal{U}\in\mathcal{P}^+(\{1,\dots,M\})} \mathcal{M}_{\mathcal{U}}\left(\boldsymbol{X}_{\mathcal{U}}\right) \quad \text{(A.86)}$$

where $\mathcal{P}^+$ denotes the non-empty power set. The uniqueness of this Hoeffding-Sobol decomposition holds under the following two conditions:

1. The first term $\mathcal{M}_0$ is constant and equal to the expected value of $\mathcal{M}(\boldsymbol{X})$:

$$\mathcal{M}_0 = \mathbb{E}\left[\mathcal{M}(\boldsymbol{X})\right] = \int_{\mathcal{D}_{\boldsymbol{X}}} \mathcal{M}(\mathbf{x})\,\mathrm{d}\mathbf{x} \quad \text{(A.87)}$$

2. All the terms in the functional decomposition are orthogonal:

$$\int_{\mathcal{D}_X} \mathcal{M}_{\mathcal{U}}(\mathbf{x}_{\mathcal{U}})\, \mathcal{M}_{\mathcal{V}}(\mathbf{x}_{\mathcal{V}})\, \mathrm{d}\mathbf{x} = \delta_{\mathcal{U},\mathcal{V}} \tag{A.88}$$

for $\mathcal{U}, \mathcal{V} \in \{0\} \cup \mathcal{P}^{+}(\{1, \ldots, M\})$ and with $\delta_{\mathcal{U},\mathcal{V}}$ being the Kronecker delta function.

Further assuming that the function $\mathcal{M}(\boldsymbol{X})$ is square-integrable, the functional decomposition in Eq. A.85 may be squared and integrated to provide the variance decomposition:

$$\mathrm{var}[\mathcal{M}(\boldsymbol{X})] = \sum_{\mathcal{U} \in \mathcal{P}^{+}(\{1,\ldots,M\})}^{M} \mathrm{var}[\mathcal{M}_{\mathcal{U}}(\boldsymbol{X}_{\mathcal{U}})] \tag{A.89}$$

with the total variance $\mathrm{var}[\mathcal{M}(\boldsymbol{X})]$ and partial variances $\mathrm{var}[\mathcal{M}_{\mathcal{U}}(\boldsymbol{X}_{\mathcal{U}})]$ being defined as:

$$\mathrm{var}[\mathcal{M}(\boldsymbol{X})] = \int_{\mathcal{D}_X} \mathcal{M}^2(\mathbf{x})\, \mathrm{d}\mathbf{x} - \mathcal{M}_0^2 \tag{A.90a}$$

$$\mathrm{var}[\mathcal{M}_{\mathcal{U}}(\boldsymbol{X}_{\mathcal{U}})] = \int_{\mathcal{D}_{X_{\mathcal{U}}}} \mathcal{M}_{\mathcal{U}}^2(\mathbf{x}_{\mathcal{U}})\, \mathrm{d}\mathbf{x}_{\mathcal{U}} \tag{A.90b}$$

### A.12.2 Sobol' Indices

Based on these results, Sobol' indices $S_{\mathcal{U}}$ can be defined as a natural global sensitivity measure of $\mathcal{M}(\boldsymbol{X})$ on the input variables $\boldsymbol{X}_{\mathcal{U}}$:

$$S_{\mathcal{U}} := \frac{\mathrm{var}[\mathcal{M}_{\mathcal{U}}(\boldsymbol{X}_{\mathcal{U}})]}{\mathrm{var}[\mathcal{M}(\boldsymbol{X})]} \tag{A.91}$$

Consequently, $S_{\mathcal{U}}$ represents the relative contribution of the set of variables $\{X_i\}_{i \in \mathcal{U}}$ to the total variance $\mathrm{var}[\mathcal{M}(\boldsymbol{X})]$. First order indices $S_i$ with $i \in \{1, \ldots, M\}$ indicate the influence of $X_i$ alone, whereas the higher order indices quantify possible interactions or mixed influences between multiple variables. In addition, we can also define the $i$-th total Sobol' index $S_i^{\mathrm{T}}$ to evaluate the total effect of an input parameter $X_i$ on $\mathcal{M}(\boldsymbol{X})$:

$$S_i^{\mathrm{T}} := \frac{\sum_{\mathcal{U} \in \mathcal{I}_i} \mathrm{var}[\mathcal{M}_{\mathcal{U}}(\boldsymbol{X}_{\mathcal{U}})]}{\mathrm{var}[\mathcal{M}(\boldsymbol{X})]} \tag{A.92}$$

with $\mathcal{I}_i := \{\mathcal{U} \in \mathcal{P}^{+}(\{1, \ldots, M\} \mid i \in \mathcal{U})\}$ being the set of all indices $\mathcal{U}$ that contain the index $i \in \{1, \ldots, M\}$. As a result, $S_i^{\mathrm{T}}$ includes not only the effect of $X_i$ alone but in addition the effect induced by all interactions between $X_i$ and the other variables. This is also the reason, why the sum of the total Sobol' indices $\sum_{i \in \{1,\ldots,M\}} S_i^{\mathrm{T}}$ can in fact exceed 1. As an example, if we have an interaction between the variables $X_1$ and $X_2$, their interaction effect on $\mathcal{M}$ is counted twice, once in $S_1^{\mathrm{T}}$ and another time in $S_2^{\mathrm{T}}$. This example in mind, it is easy to see that the peaks in $S_i^{\mathrm{T}}$ in Figs. 7.3 and B.73–B.76 highlight spectral regions, where the interaction terms between the individual variables significantly contribute to the total effect on $\mathcal{M}$. It is important to add that to draw conclusions about the relative importance of the corresponding variables $X_i$ for the model response $\mathcal{M}(\boldsymbol{X})$, only the relative magnitudes of $S_i^{\mathrm{T}}$ matter.

### A.12.3 PCE-PCA based Total Sobol' Indices

Estimating the Sobol' indices $S_{\mathcal{U}}$ in Eq. A.91 and the total Sobol' indices $S_i^{\mathrm{T}}$ in Eq. A.92 requires the computation of the partial variance terms in Eq. A.90b. These terms can be computed analytically using recursion formulas or numerically using Monte Carlo methods [267]. Both these methods are computationally expensive [742].

PCE discussed in Section 7.2.4.1 and Appendix A.11 offers a computationally efficient alternative to estimate the partial variance terms in Eq. A.90b and in extension also the Sobol' indices $S_{\mathcal{U}}$ and the total Sobol' indices $S_{\mathcal{U}}^{\mathrm{T}}$. Here, I focus on the total Sobol' indices $S_i^{\mathrm{T}}$ for the custom PCE-PCA surrogate model discussed in Section 7.2.4.1 and Appendix A.11.

As shown by Sudret [742], the expression in Eq. A.92 is equivalent to:

$$S_i^{\mathrm{T}} = 1 - S_{\sim i} \tag{A.93a}$$

$$= 1 - \frac{\mathrm{var}_{X_{\sim i}}\left[\mathbb{E}_{X_i}\left[\mathcal{M}\left(\boldsymbol{X}\right)\right]\right]}{\mathrm{var}[\mathcal{M}\left(\boldsymbol{X}\right)]} \tag{A.93b}$$

where $\sim i$ denotes a set of indices, which does not include $i$. The subscripts in the expectation and variance operators $\mathrm{var}_{X_i}$ and $\mathbb{E}_{X_i}$ on the other hand indicate the variable(s) with respect to which the variance and expectation are taken, e.g. $\mathbb{E}_{X_i}[\cdot] := \int (\cdot)\pi(x_i)\,\mathrm{d}x_i$.

Suppose now that we have a PCE-PCA surrogate model to emulate the vector-valued model response $\boldsymbol{Y} = \mathcal{M}(\boldsymbol{X})$ with random input vector $\boldsymbol{X} \in \mathbb{R}^{M\times 1}$ and random response vector $\boldsymbol{Y} \in \mathbb{R}^{N\times 1}$. To derive the total Sobol' indices $S^{\mathrm{T}}_{i,k}$ for each response variable $k \in \{1, 2, \ldots, N\}$ and model parameter $i \in \{1, 2, \ldots, M\}$, I start with $\mathrm{var}_{X_{\sim i}}[\mathbb{E}_{X_i}[Y_k]]$ from Eq. A.93b by replacing $Y_k$ with the $k$-th component of Eq. 7.9 [746]:

$$\mathrm{var}_{X_{\sim i}}\left[\mathbb{E}_{X_i}\left[Y_k\right]\right] = \mathbb{E}_{X_{\sim i}}\left[\left(\mathbb{E}_{X_i}\left[Y_k\right]\right)^2\right] - \left(\mathbb{E}_{X}\left[Y_k\right]\right)^2 \tag{A.94a}$$

$$= \mathbb{E}_{X_{\sim i}}\left[\left(\mathbb{E}_{X_i}\left[\mu_{Y_k} + \sigma_{Y_k}\boldsymbol{\Phi}'_k\mathbf{A}\boldsymbol{\Psi}(\boldsymbol{X})\right]\right)^2\right] - \mu^2_{Y_k} \tag{A.94b}$$

where I used $\boldsymbol{\Phi}'_k := (\phi_{k,1}, \ldots, \phi_{k,N'})$ and with $\mu_{Y_k}$ and $\sigma_{Y_k}$ denoting the $k$-th component of the mean $\boldsymbol{\mu}_{\boldsymbol{Y}}$ and standard deviation $\boldsymbol{\sigma}_{\boldsymbol{Y}}$ of $\boldsymbol{Y}$ introduced in Eq. 7.9. We can simplify this expression by expanding the first term and considering that the expectation vanishes for all principal components, i.e. $\mathbb{E}[\mathbf{A}\boldsymbol{\Psi}(\boldsymbol{X})] = 0$ [746]:

$$\mathrm{var}_{X_{\sim i}}\left[\mathbb{E}_{X_i}\left[Y_k\right]\right] = \mathbb{E}_{X_{\sim i}}\left[\left(\sigma_{Y_k}\boldsymbol{\Phi}'_k\mathbf{A}\mathbb{E}_{X_i}\left[\boldsymbol{\Psi}(\boldsymbol{X})\right]\right)^2\right] \tag{A.95a}$$

$$= \mathbb{E}_{X_{\sim i}}\left[\left(\sum_{\boldsymbol{\alpha}\in\mathcal{A}^\star}\sum_{j=1}^{N'}\sigma_{Y_k}\phi_{k,j}a_{j,\boldsymbol{\alpha}}\mathbb{E}_{X_i}\left[\Psi_{\boldsymbol{\alpha}}(\boldsymbol{X})\right]\right)^2\right] \tag{A.95b}$$

where I used the summation notation detailed in Appendix A.11.3 to rewrite Eq. A.95a into Eq. A.95b with eigenvector entries $\phi_{k,j}$ and PCE coefficients $a_{j,\boldsymbol{\alpha}}$.

As shown by Wagner et al. [746], due to the orthonormality of the polynomial basis $\{\Psi_{\boldsymbol{\alpha}}\}_{\boldsymbol{\alpha}\in\mathcal{A}^\star}$, we can further simplify Eq. A.95b resulting in:

$$\mathrm{var}_{X_{\sim i}}\left[\mathbb{E}_{X_i}\left[Y_k\right]\right] = \sigma^2_{Y_k}\sum_{\boldsymbol{\alpha}\in\mathcal{A}^\star_{i=0}}\left(\sum_{j=1}^{N'}\phi_{k,j}a_{j,\boldsymbol{\alpha}}\right)^2 \tag{A.96}$$

with the subset $\mathcal{A}^\star_{i=0} := \{\boldsymbol{\alpha} \in \mathcal{A}^\star \mid \alpha_i = 0\}$. Using these results, we can compute the total variance with:

$$\operatorname{var}\left[Y_k\right] = \sigma_{Y_k}^2 \sum_{\boldsymbol{\alpha} \in \mathcal{A}^\star} \left( \sum_{j=1}^{N'} \phi_{k,j} a_{j,\boldsymbol{\alpha}} \right)^2 \tag{A.97}$$

In the end, we get the total PCE-PCA based Sobol′ indices $S_{i,k}^{\mathrm{T}}$ for the input variable $i$ and the response variable $k$ by inserting Eqs. A.96 and A.97 into Eq. A.93b:

$$S_{i,k}^{\mathrm{T}} = 1 - \frac{\sum\limits_{\boldsymbol{\alpha} \in \mathcal{A}_{i=0}^\star} \left( \sum\limits_{j=1}^{N'} \phi_{k,j}\, a_{j,\boldsymbol{\alpha}} \right)^2}{\sum\limits_{\boldsymbol{\alpha} \in \mathcal{A}^\star} \left( \sum\limits_{j=1}^{N'} \phi_{k,j} a_{j,\boldsymbol{\alpha}} \right)^2} \tag{A.98}$$

which is equivalent to the formula derived by Wagner et al. [746]. Note that the expression in Eq. A.98 only depends on the PCE coefficients and the eigenvector entries. As the `UQLab` code does not support PCE-PCA surrogate models [753], a custom `MATLAB` script was developed to compute the total Sobol′ indices $S_{i,k}^{\mathrm{T}}$ using the expression in Eq. A.98.

## A.13 Global-to-Detector Coordinate Transformation

In this section, I will derive the coordinate transformation to convert the direction unit vector $\mathbf{\Omega}$ in the global coordinate system to the one in the detector coordinate system $\mathbf{\Omega}'$.

Let us start by repeating the definition of the direction unit vector $\mathbf{\Omega}$ in the global coordinate system, which is parametrized by the azimuthal angle $\varphi$ and the polar angle $\theta$ in cartesian coordinates as (cf. Eq. 2.18):

$$\mathbf{\Omega}\left(\theta,\varphi\right) = \begin{pmatrix} \sin\theta\cos\varphi \\ \sin\theta\sin\varphi \\ \cos\theta \end{pmatrix} \tag{A.99}$$

where:

| | | |
|---|---|---|
| $\theta$ | polar angle | rad |
| $\varphi$ | azimuthal angle | rad |

Given that the orientation of the detector coordinate system is parametrized by the Tait-Bryan angles $\alpha'$ (yaw), $\beta'$ (pitch) and $\gamma'$ (roll) (cf. Fig. 8.1), we can write the coordinate transformation as a linear transformation mapping [899]:

$$\mathbf{\Omega}'\left(\theta',\varphi'\right) = \mathcal{R}\left(\alpha',\beta',\gamma'\right)\mathbf{\Omega}\left(\theta,\varphi\right) \tag{A.100}$$

with the rotation matrix $\mathcal{R} \in \mathbb{R}^{3\times3}$. The rotation matrix $\mathcal{R}$ can be expressed as the product of three elemental rotation matrices representing a sequence $z'$-$x'$-$y'$ of three intrinsic rotations about the principal aircraft axis $z'$ (yaw), $x'$ (pitch) and $y'$-axis (roll) [900] (cf. Section 8.2):

$$\mathcal{R} = \begin{pmatrix} c_{\gamma'} & 0 & -s_{\gamma'} \\ 0 & 1 & 0 \\ s_{\gamma'} & 0 & c_{\gamma'} \end{pmatrix} \begin{pmatrix} 1 & 0 & 0 \\ 0 & c_{\beta'} & s_{\beta'} \\ 0 & -s_{\beta'} & c_{\beta'} \end{pmatrix} \begin{pmatrix} c_{\alpha'} & s_{\alpha'} & 0 \\ -s_{\alpha'} & c_{\alpha'} & 0 \\ 0 & 0 & 1 \end{pmatrix} \tag{A.101a}$$

$$= \begin{pmatrix} c_{\alpha'}c_{\gamma'} - s_{\alpha'}s_{\beta'}s_{\gamma'} & c_{\gamma'}s_{\alpha'} + c_{\alpha'}s_{\beta'}s_{\gamma'} & -c_{\beta'}s_{\gamma'} \\ -c_{\beta'}s_{\alpha'} & c_{\alpha'}c_{\beta'} & s_{\beta'} \\ c_{\alpha'}s_{\gamma'} + c_{\gamma'}s_{\alpha'}s_{\beta'} & s_{\alpha'}s_{\gamma'} - c_{\alpha'}c_{\gamma'}s_{\beta'} & c_{\beta'}c_{\gamma'} \end{pmatrix} \tag{A.101b}$$

with $s_*$ and $c_*$ being a short notation for the sine $\sin(*)$ and cosine $\cos(*)$, respectively. Inserting Eq. A.101b and Eq. A.99 into Eq. A.100, the direction unit vector $\mathbf{\Omega}'$ in the detector coordinate system can be computed by performing a simple matrix multiplication operation as:

$$\mathbf{\Omega}' = \begin{pmatrix} \Omega'_x \\ \Omega'_y \\ \Omega'_z \end{pmatrix} = \begin{pmatrix} [\mathrm{c}_{\alpha'}\mathrm{c}_{\gamma'} - \mathrm{s}_{\alpha'}\mathrm{s}_{\beta'}\mathrm{s}_{\gamma'}]c_{\varphi}s_{\theta} + [\mathrm{c}_{\gamma'}\mathrm{s}_{\alpha'} + \mathrm{c}_{\alpha'}\mathrm{s}_{\beta'}\mathrm{s}_{\gamma'}]s_{\varphi}s_{\theta} - \mathrm{c}_{\beta'}\mathrm{c}_{\theta}\mathrm{s}_{\gamma'} \\ -\mathrm{c}_{\beta'}\mathrm{c}_{\varphi}\mathrm{s}_{\alpha'}\mathrm{s}_{\theta} + \mathrm{c}_{\alpha'}\mathrm{c}_{\beta'}s_{\varphi}s_{\theta} + \mathrm{c}_{\theta}\mathrm{s}_{\beta'} \\ [\mathrm{c}_{\alpha'}\mathrm{s}_{\gamma'} + \mathrm{c}_{\gamma'}\mathrm{s}_{\alpha'}\mathrm{s}_{\beta'}]\mathrm{c}_{\varphi}\mathrm{s}_{\theta} + [\mathrm{s}_{\alpha'}\mathrm{s}_{\gamma'} - \mathrm{c}_{\alpha'}\mathrm{c}_{\gamma'}\mathrm{s}_{\beta'}]\mathrm{s}_{\varphi}\mathrm{s}_{\theta} + \mathrm{c}_{\beta'}\mathrm{c}_{\gamma'}\mathrm{c}_{\theta} \end{pmatrix} \tag{A.102}$$

As a last step, the azimuthal angle $\varphi'$ and the polar angle $\theta'$ of the direction unit vector $\mathbf{\Omega}'$ in the detector coordinate system can be obtained by transforming the cartesian coordinates derived in Eq. A.102 into spherical coordinates as [399]:

$$\begin{pmatrix} \varphi' \\ \theta' \end{pmatrix} = \begin{pmatrix} \arctan2\left[\Omega'_y, \Omega'_x\right] \\ \arccos\left[\Omega'_z\right] \end{pmatrix} \tag{A.103}$$

where $\Omega'_x$, $\Omega'_y$ and $\Omega'_z$ represent the cartesian coordinates of Eq. A.102 and $\arctan2[\cdot,\cdot]$ denotes the two-argument arctangent.

It is worth noting that, as already discussed in Section 9.2, care must be taken when applying these coordinate transformations to align the angular grid in the double differential photon flux signature with the one of the detector response function. The direction $\mathbf{\Omega}$ in the double differential photon flux signature quantifies the direction-of-flight of the photon in the global reference frame (cf. Eq. 2.18), while $\mathbf{\Omega}'$ in the detector response function characterizes the opposite of the direction-of-flight, i.e. the direction from which the photons are arriving with respect to the detector reference frame (cf. Section 4.3.4). Consequently, if the direction unit vector $\mathbf{\Omega}$ in the global reference frame from the double differential photon flux signature is transformed into the detector reference frame to align with the detector response function angular grid, the sign of the transformed direction unit vector needs to be changed in Eq. A.102, i.e. $-\mathbf{\Omega}'$ before Eq. A.103 is applied.

# Appendix B
# Supplementary Figures

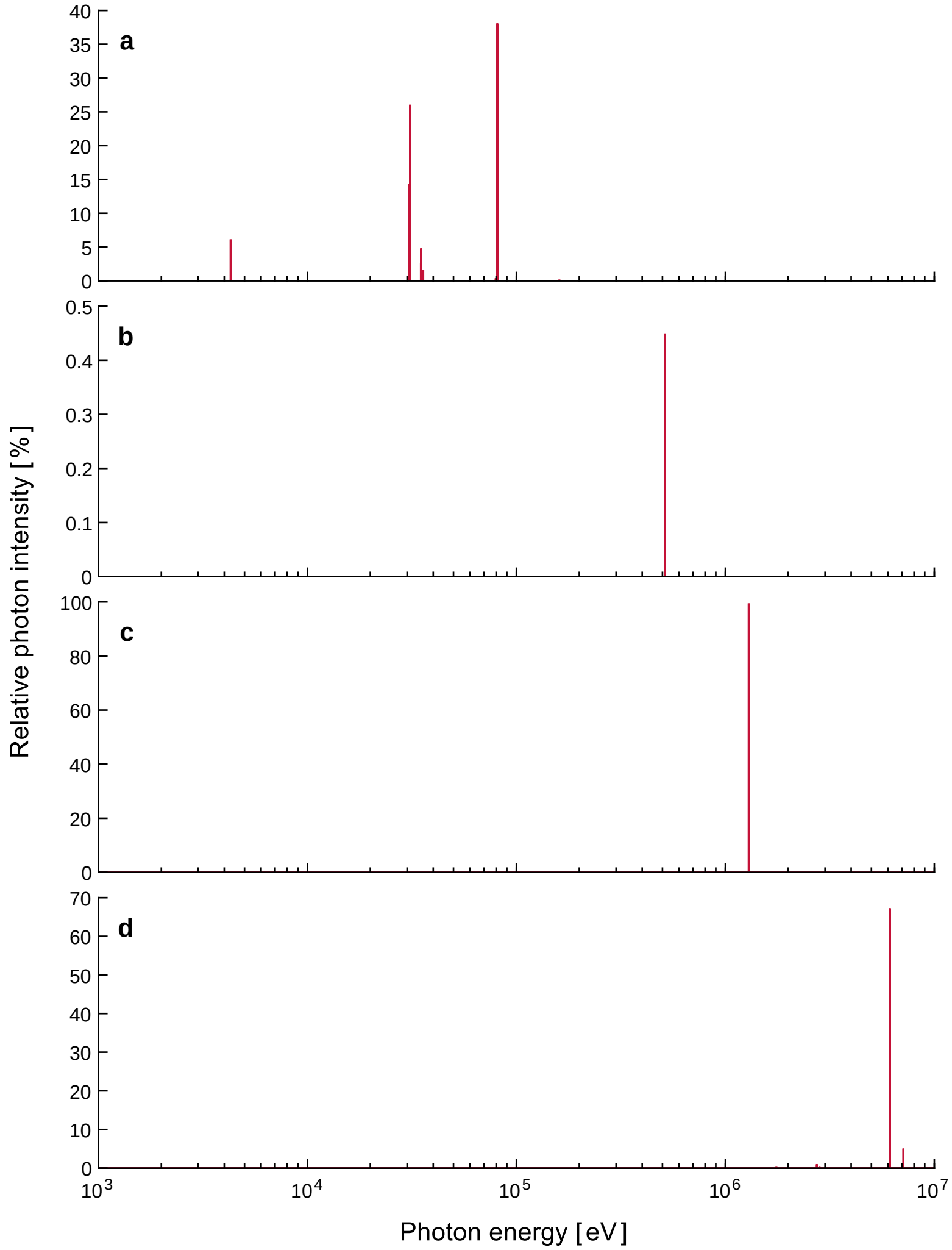

**Figure B.1** Relative photon intensity $I_\gamma$ for selected gaseous anthropogenic radionuclides as a function of the photon energy $E_\gamma$. **a** $^{133}_{54}$Xe. **b** $^{85}_{36}$Kr. **c** $^{41}_{18}$Ar. **d** $^{16}_{7}$N. $I_\gamma$ was computed using the `FLUKA` code (version: `4-4.3`) [20].

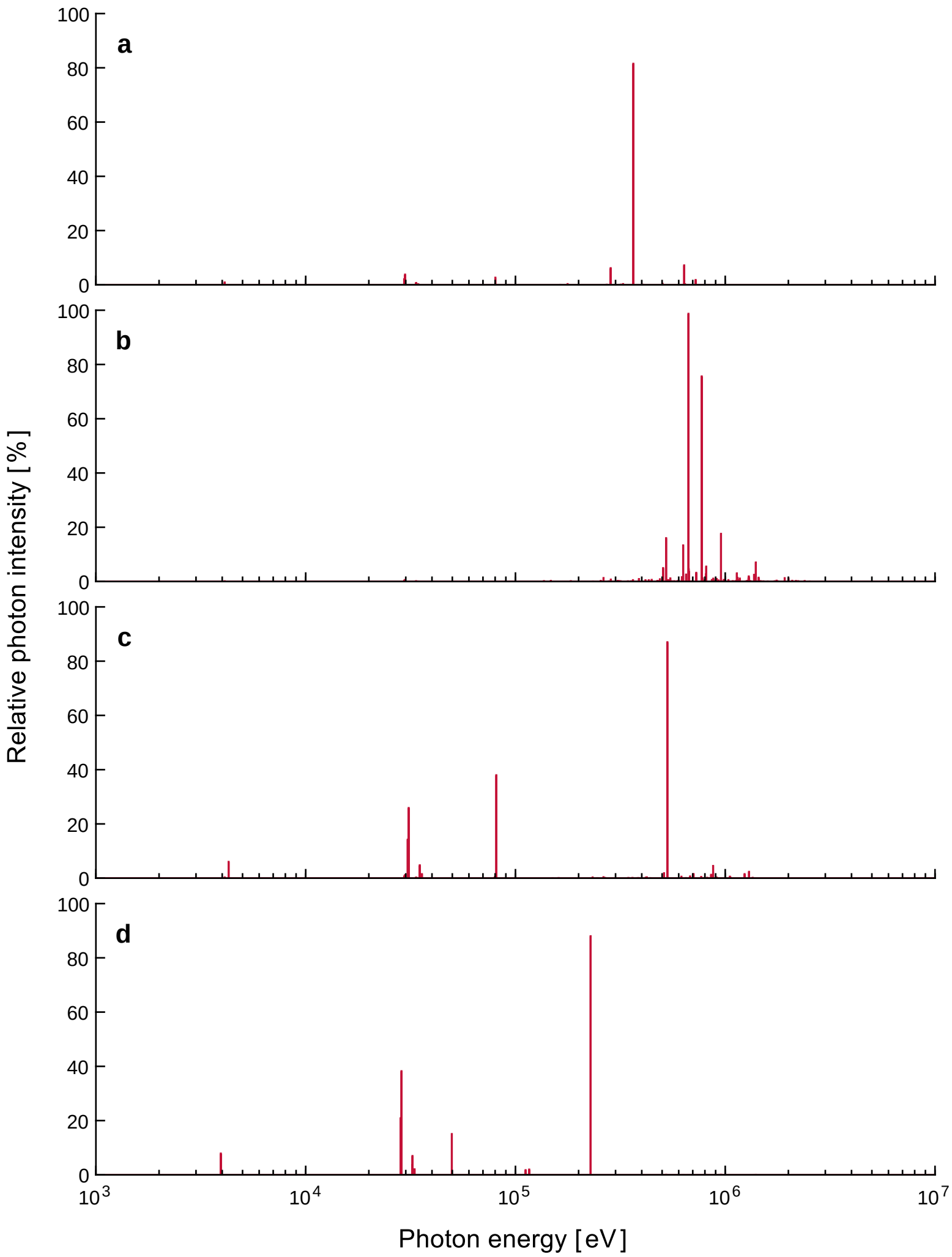

**Figure B.2** Relative photon intensity $I_\gamma$ for selected anthropogenic iodide and telluride radionuclides as a function of the photon energy $E_\gamma$. **a** $^{131}_{53}$I. **b** $^{132}_{53}$I. **c** $^{133}_{53}$I. **d** $^{132}_{52}$Te. $I_\gamma$ was computed using the `FLUKA` code (version: `4-4.3`) [20].

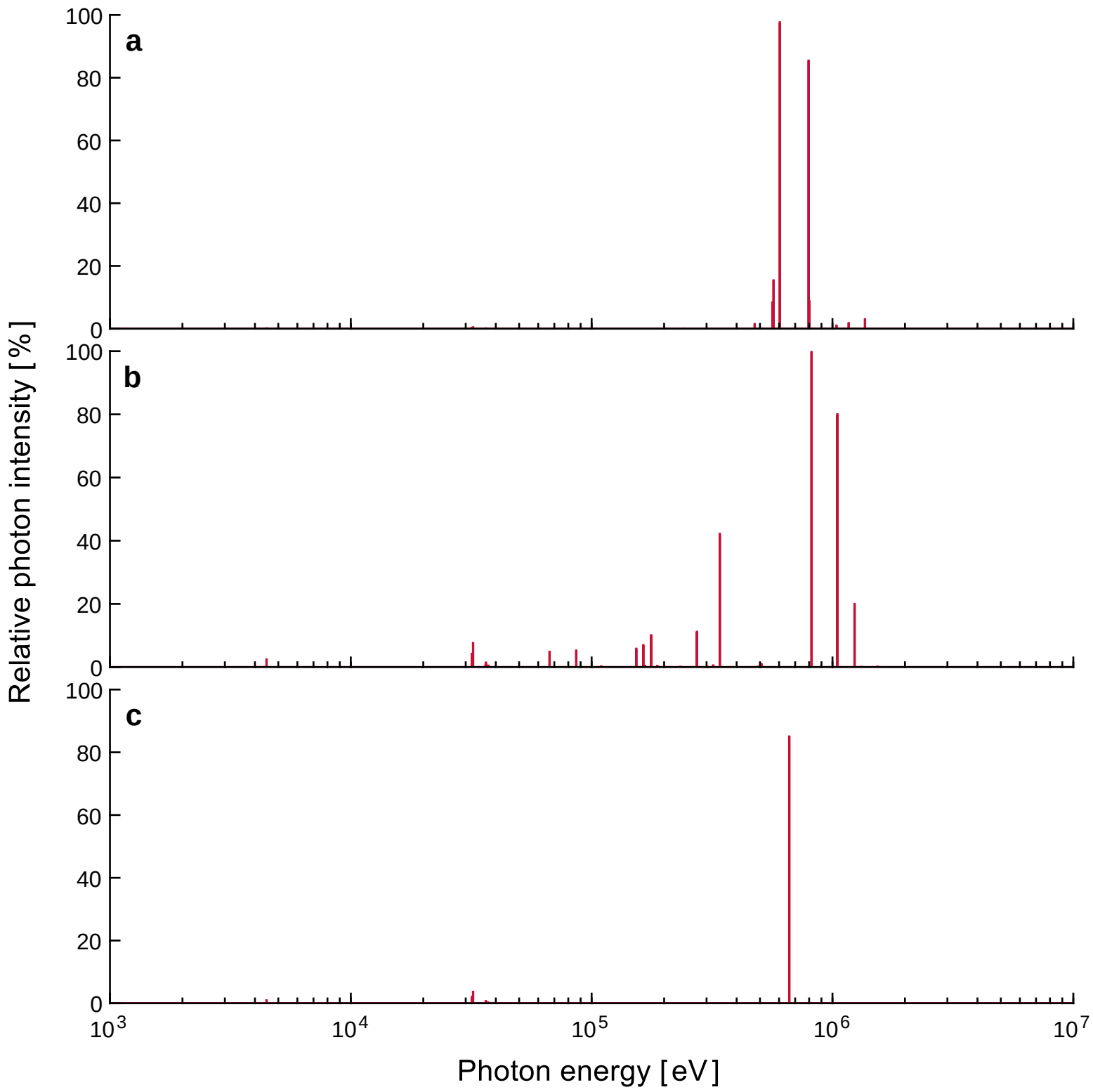

**Figure B.3** Relative photon intensity $I_\gamma$ for selected anthropogenic cesium radionuclides as a function of the photon energy $E_\gamma$. **a** $^{134}_{55}Cs$. **b** $^{136}_{55}Cs$. **c** $^{137}_{55}Cs$. $I_\gamma$ was computed using the `FLUKA` code (version: `4-4.3`) [20].

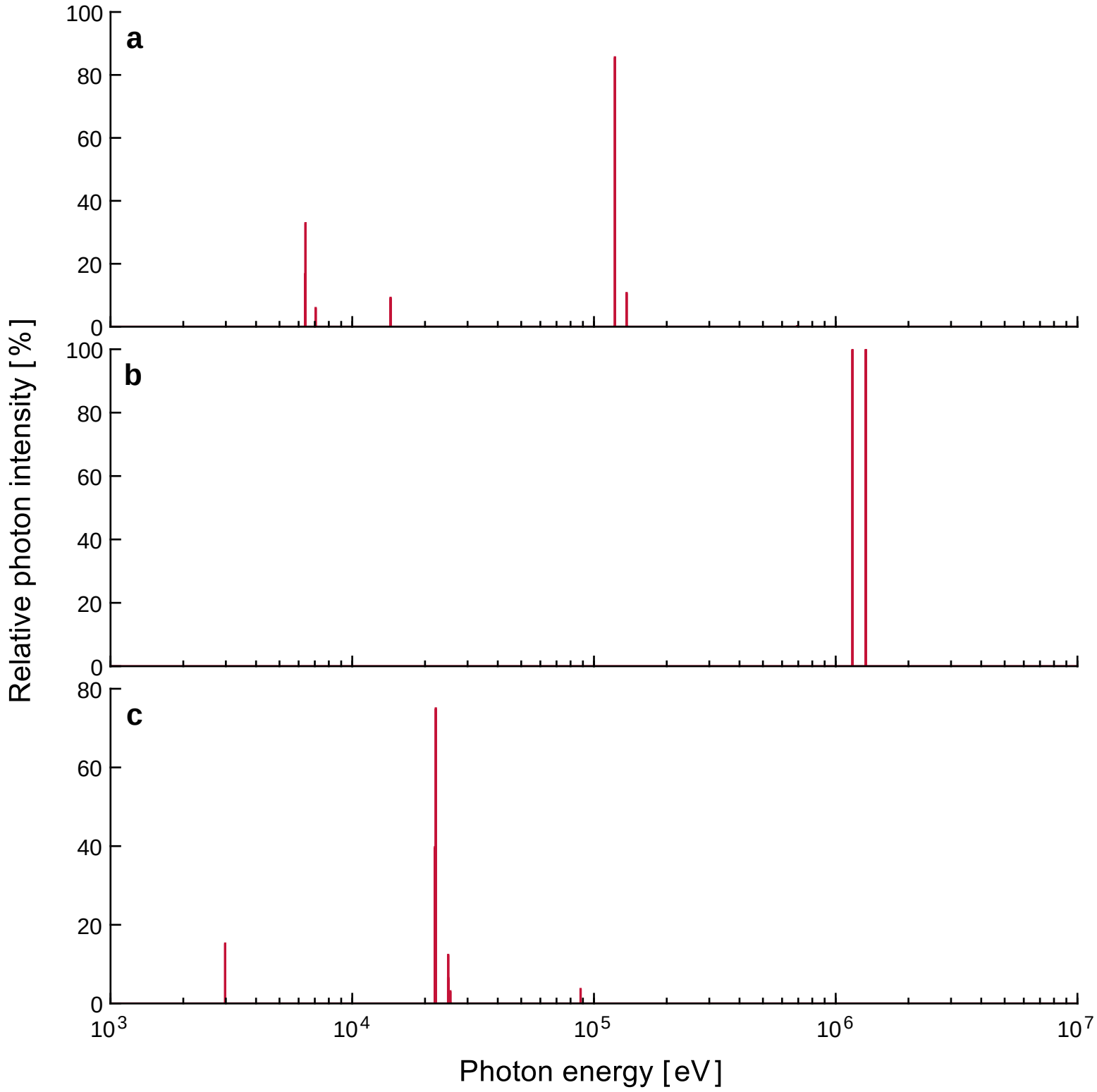

**Figure B.4** Relative photon intensity $I_\gamma$ for selected anthropogenic cobalt and cadmium radionuclides as a function of the photon energy $E_\gamma$. **a** $^{57}_{27}$Co. **b** $^{60}_{27}$Co. **c** $^{109}_{48}$Cd. $I_\gamma$ was computed using the `FLUKA` code (version: `4-4.3`) [20].

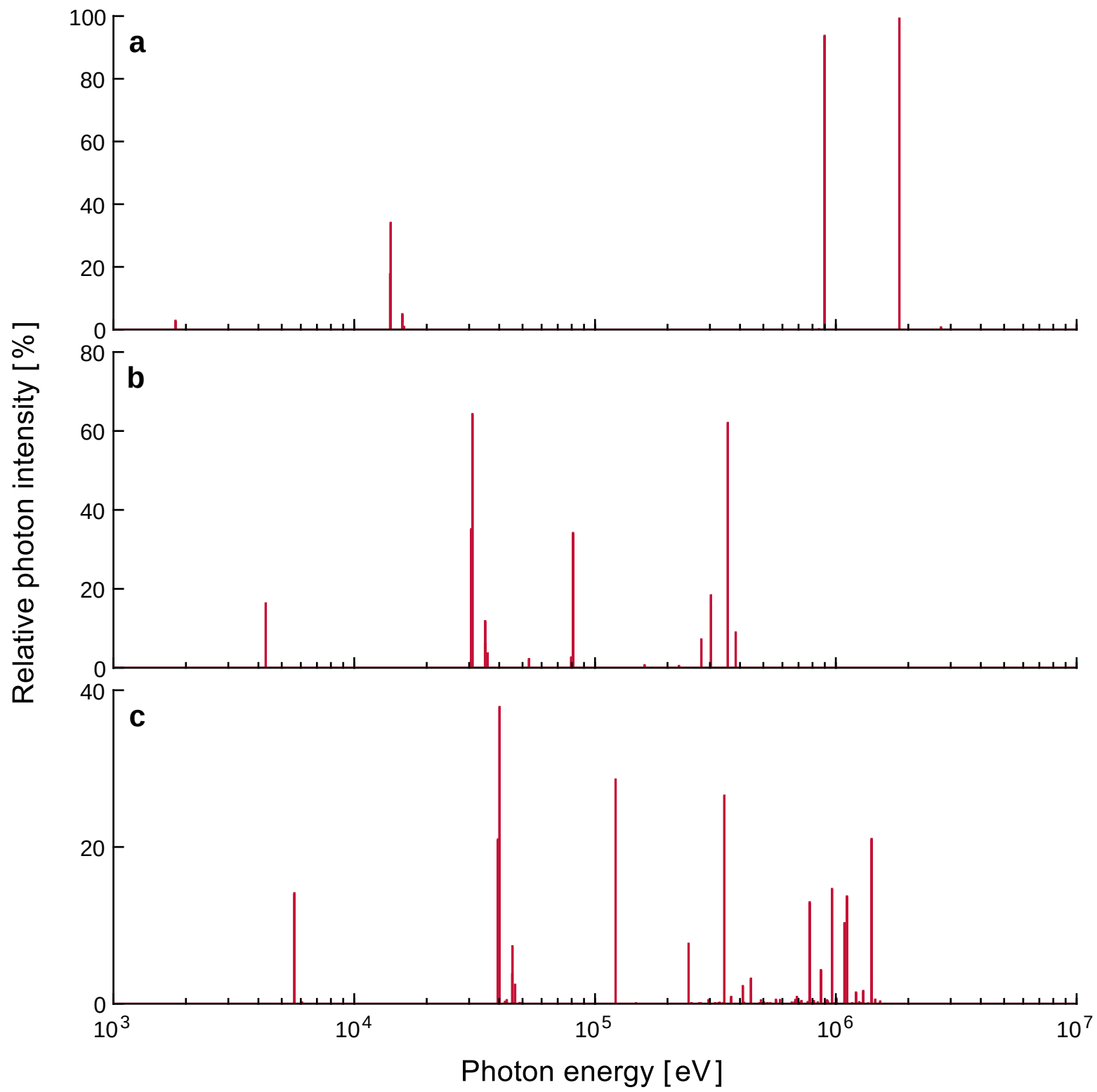

**Figure B.5** Relative photon intensity $I_\gamma$ for selected anthropogenic yttrium, barium and europium radionuclides as a function of the photon energy $E_\gamma$. **a** $^{88}_{39}$Y. **b** $^{133}_{56}$Ba. **c** $^{152}_{63}$Eu. $I_\gamma$ was computed using the `FLUKA` code (version: `4-4.3`) [20].

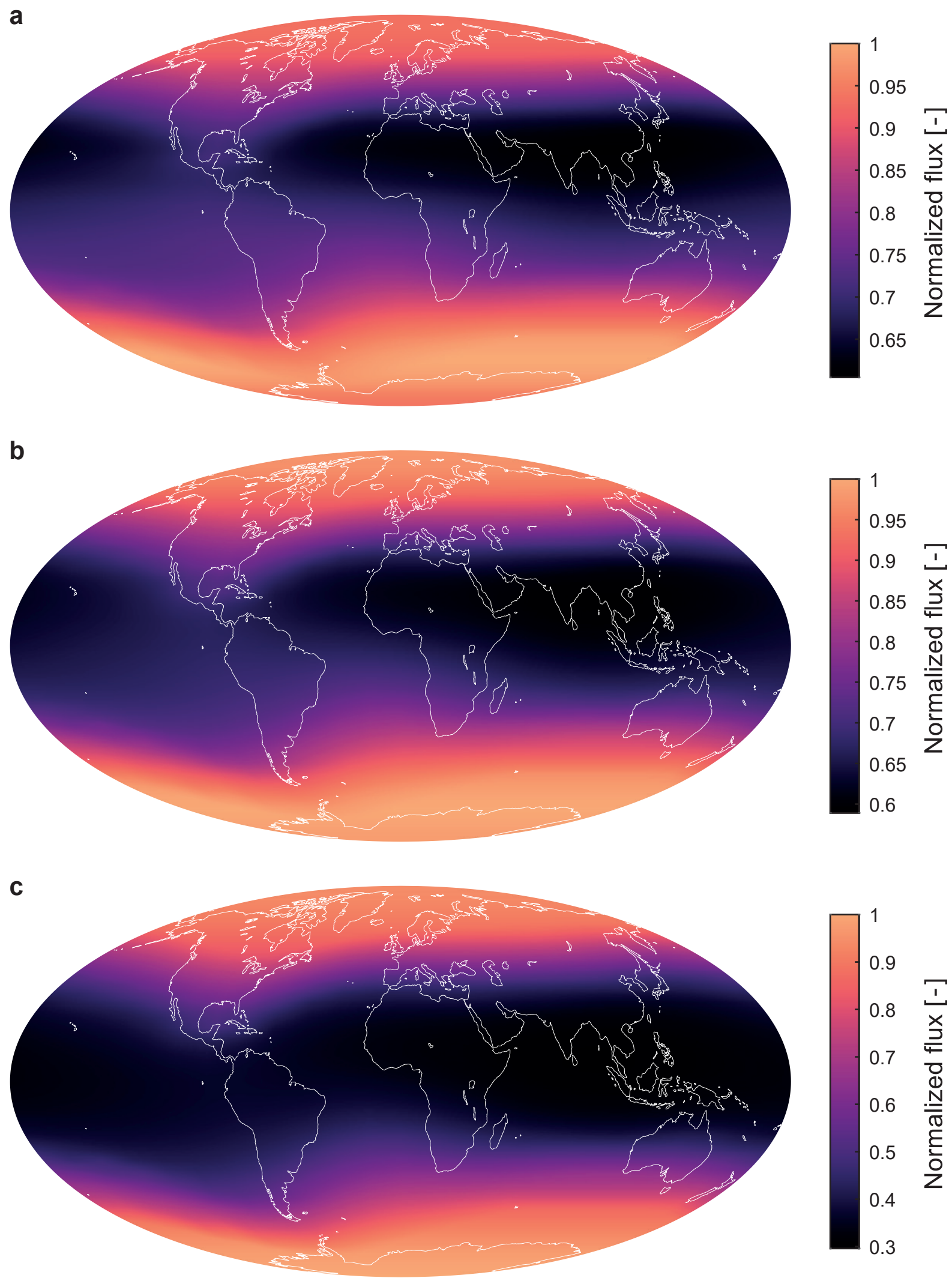

**Figure B.6** World map (Mollweide projection) displaying the normalized total cosmic rays (CR) induced ionizing particle flux ($\phi_{tot}$) ($\gamma$, n, $e^{\pm}$, $\mu^{\pm}$, p, $\alpha$) for different geodetic altitudes ($h_{geo}$) on 2022-05-04 estimated using the `PARMA` code (version: `4.13`). The total ionizing particle flux was normalized by the peak value of the corresponding maps as listed below. A more detailed description of the data processing steps is provided in Fig. 2.5. **a** $h_{geo} = 0$ km, $\max(\phi_{tot}) = 0.19\,s^{-1}\,cm^{-2}$. **b** $h_{geo} = 1$ km, $\max(\phi_{tot}) = 0.35\,s^{-1}\,cm^{-2}$. **c** $h_{geo} = 10$ km, $\max(\phi_{tot}) = 37.00\,s^{-1}\,cm^{-2}$.

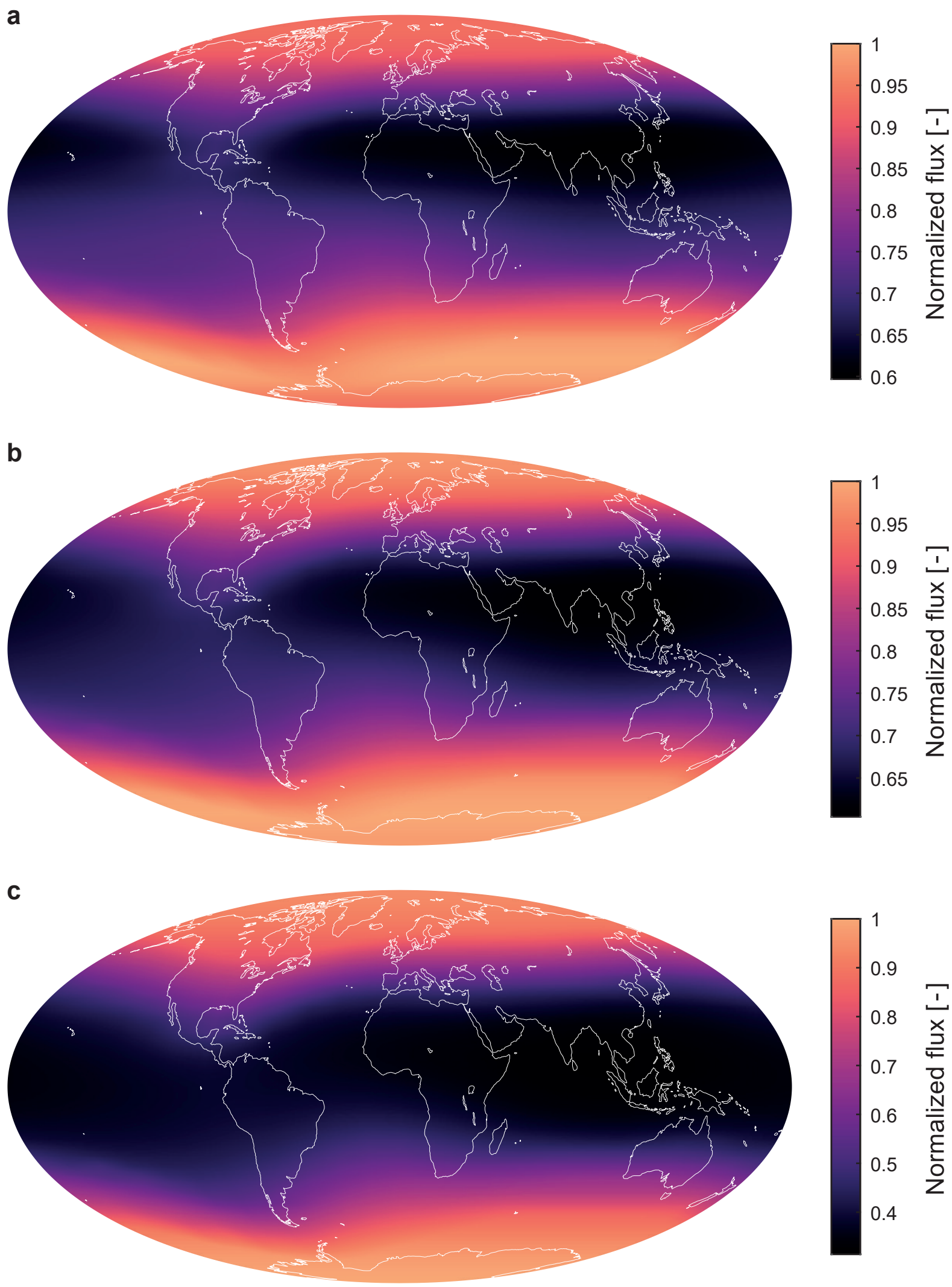

**Figure B.7** World map (Mollweide projection) displaying the normalized total CR induced gamma-ray flux ($\phi_\gamma$) for different geodetic altitudes ($h_{geo}$) on 2022-05-04 estimated using the `PARMA` code (version: `4.13`). The total gamma-ray flux was normalized by the peak value of the corresponding maps as listed below. A more detailed description of the data processing steps is provided in Fig. 2.5. **a** $h_{geo} = 0\,\mathrm{km}$, $\max(\phi_\gamma) = 0.16\,\mathrm{s^{-1}\,cm^{-2}}$. **b** $h_{geo} = 1\,\mathrm{km}$, $\max(\phi_\gamma) = 0.29\,\mathrm{s^{-1}\,cm^{-2}}$. **c** $h_{geo} = 10\,\mathrm{km}$, $\max(\phi_\gamma) = 32.08\,\mathrm{s^{-1}\,cm^{-2}}$.

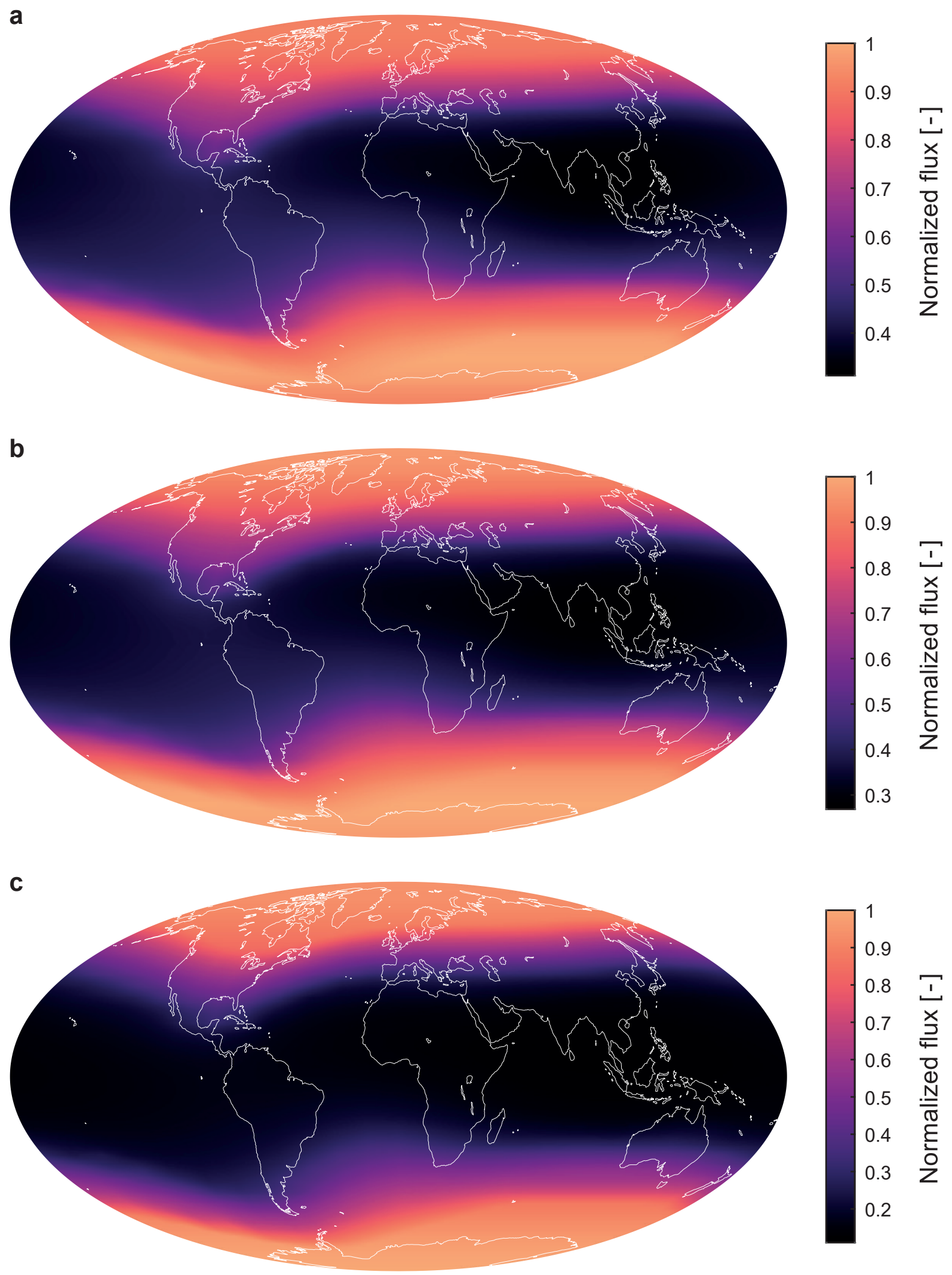

**Figure B.8** World map (Mollweide projection) displaying the normalized total CR induced neutron flux ($\phi_n$) for different geodetic altitudes ($h_{geo}$) on 2022-05-04 estimated using the `PARMA` code (version: `4.13`). The total neutron flux was normalized by the peak value of the corresponding maps as listed below. A more detailed description of the data processing steps is provided in Fig. 2.5. **a** $h_{geo} = 0\,\text{km}$, $\max(\phi_n) = 0.01\,\text{s}^{-1}\,\text{cm}^{-2}$. **b** $h_{geo} = 1\,\text{km}$, $\max(\phi_n) = 0.03\,\text{s}^{-1}\,\text{cm}^{-2}$. **c** $h_{geo} = 10\,\text{km}$, $\max(\phi_n) = 3.19\,\text{s}^{-1}\,\text{cm}^{-2}$.

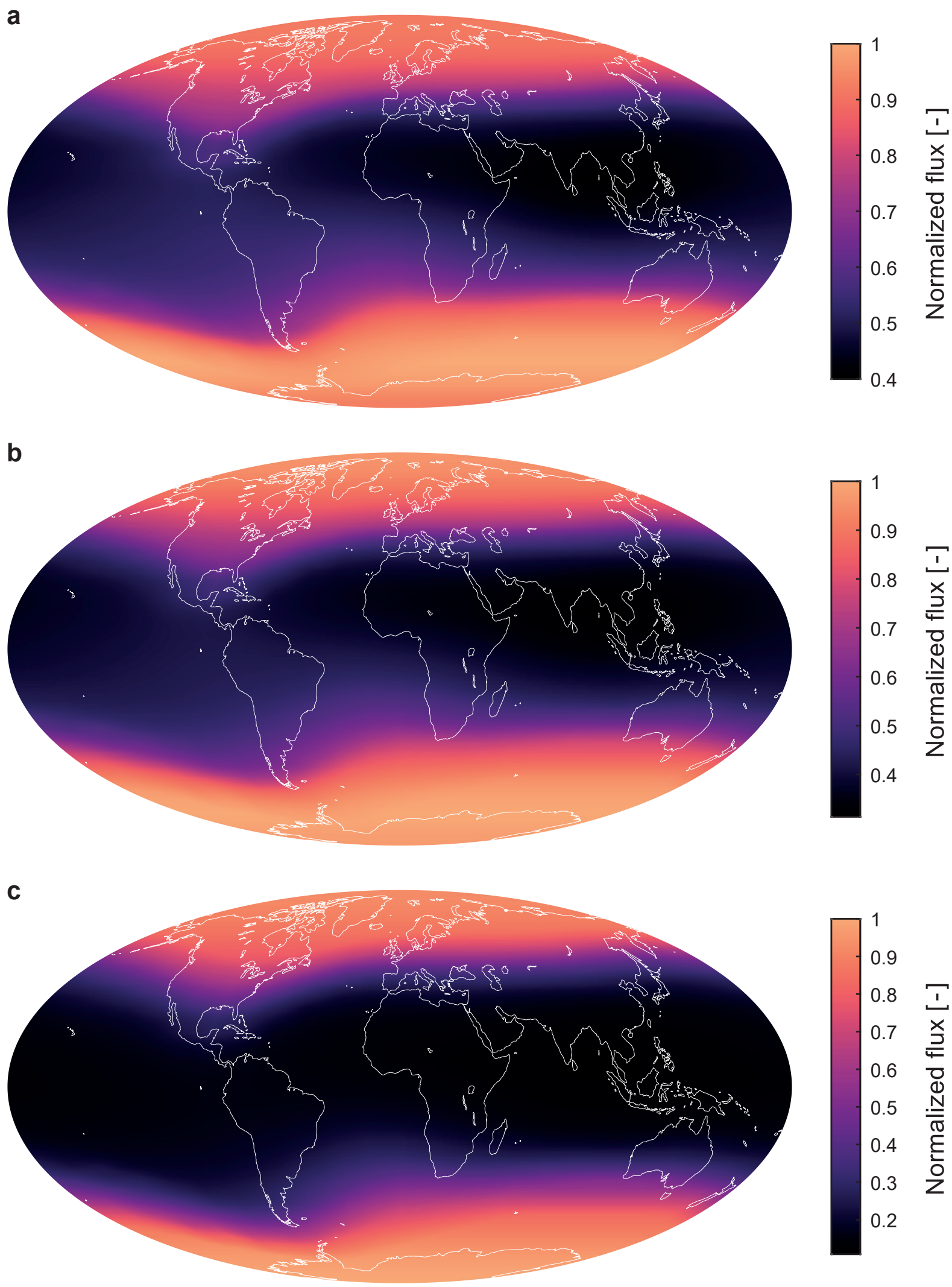

**Figure B.9** World map (Mollweide projection) displaying the normalized total CR induced proton flux ($\phi_p$) for different geodetic altitudes ($h_{geo}$) on 2022-05-04 estimated using the `PARMA` code (version: `4.13`). The total proton flux was normalized by the peak value of the corresponding maps as listed below. A more detailed description of the data processing steps is provided in Fig. 2.5. **a** $h_{geo} = 0\,\text{km}$, $\max(\phi_p) = 4.2 \times 10^{-4}\,\text{s}^{-1}\,\text{cm}^{-2}$. **b** $h_{geo} = 1\,\text{km}$, $\max(\phi_p) = 1.3 \times 10^{-3}\,\text{s}^{-1}\,\text{cm}^{-2}$. **c** $h_{geo} = 10\,\text{km}$, $\max(\phi_p) = 0.33\,\text{s}^{-1}\,\text{cm}^{-2}$.

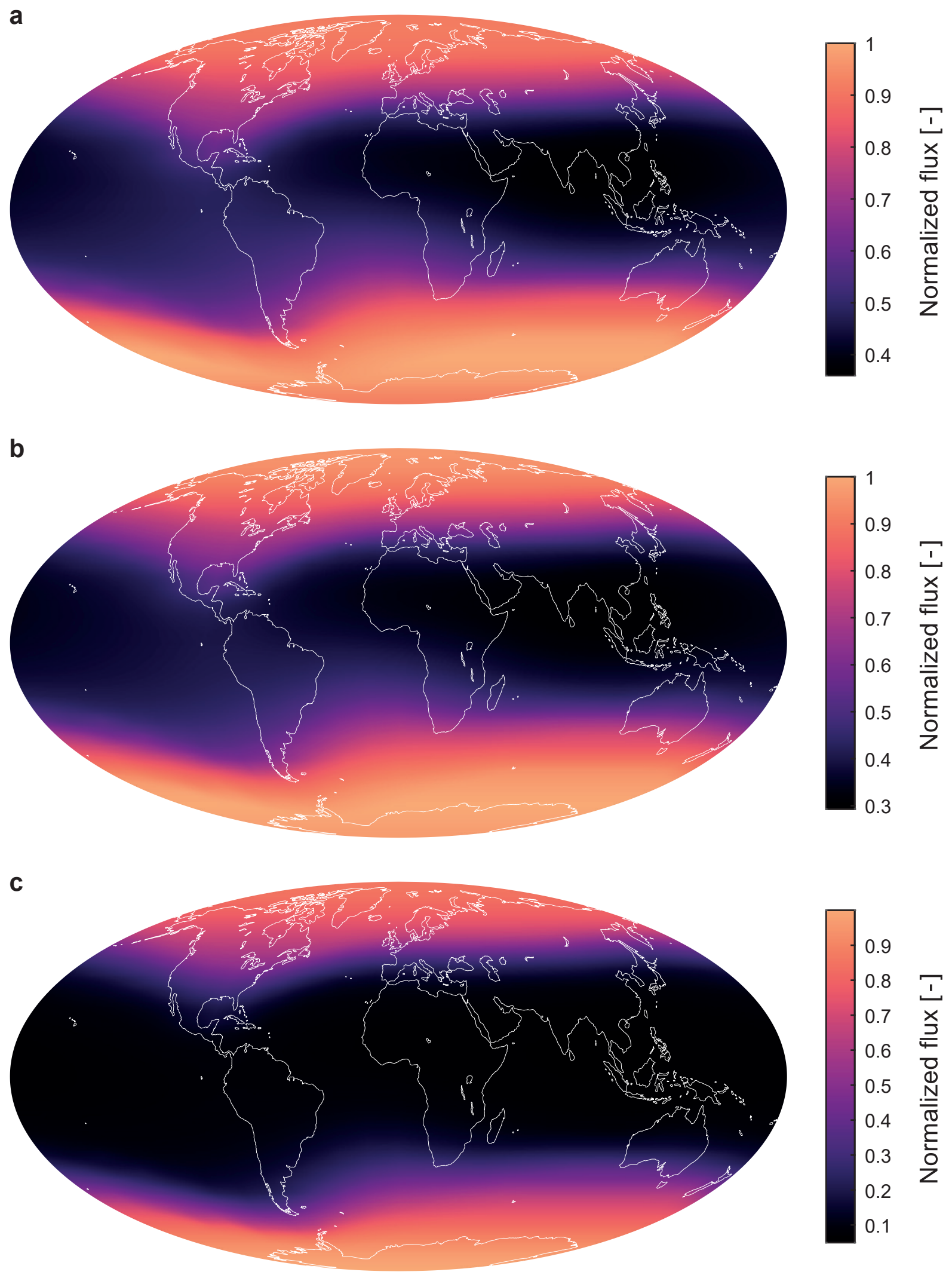

**Figure B.10** World map (Mollweide projection) displaying the normalized total CR induced alpha particle flux ($\phi_\alpha$) for different geodetic altitudes ($h_{\text{geo}}$) on 2022-05-04 estimated using the `PARMA` code (version: `4.13`). The total alpha particle flux was normalized by the peak value of the corresponding maps as listed below. A more detailed description of the data processing steps is provided in Fig. 2.5. **a** $h_{\text{geo}} = 0\,\text{km}$, $\max(\phi_\alpha) = 6.0 \times 10^{-7}\,\text{s}^{-1}\,\text{cm}^{-2}$. **b** $h_{\text{geo}} = 1\,\text{km}$, $\max(\phi_\alpha) = 1.7 \times 10^{-6}\,\text{s}^{-1}\,\text{cm}^{-2}$. **c** $h_{\text{geo}} = 10\,\text{km}$, $\max(\phi_\alpha) = 8.4 \times 10^{-4}\,\text{s}^{-1}\,\text{cm}^{-2}$.

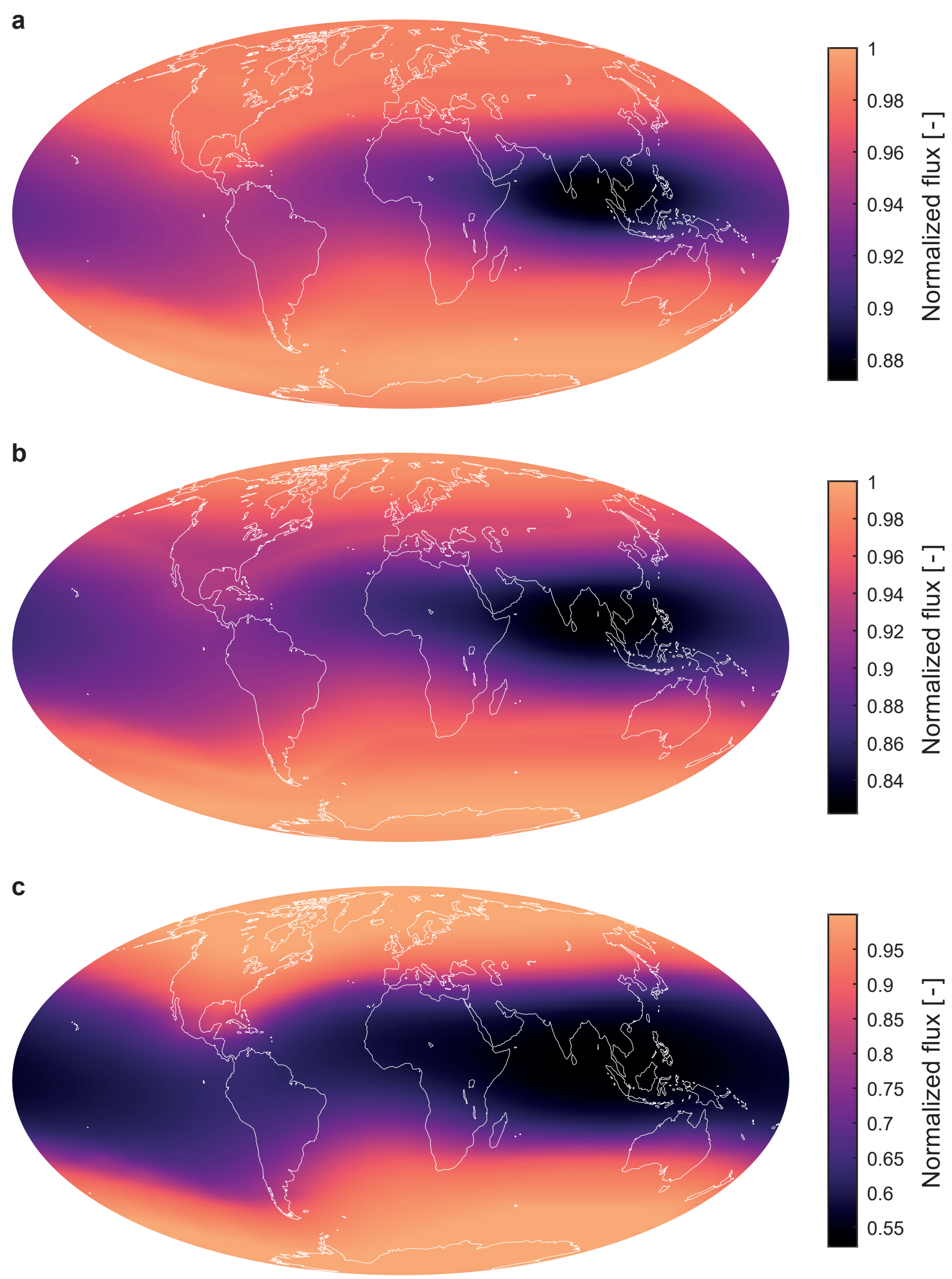

**Figure B.11** World map (Mollweide projection) displaying the normalized total CR induced muon flux ($\phi_{\mu^-}$) for different geodetic altitudes ($h_{\text{geo}}$) on 2022-05-04 estimated using the `PARMA` code (version: `4.13`). The tota muon flux was normalized by the peak value of the corresponding maps as listed below. A more detailed description of the data processing steps is provided in Fig. 2.5. **a** $h_{\text{geo}} = 0\,\text{km}$, $\max(\phi_{\mu^-}) = 6.6 \times 10^{-3}\,\text{s}^{-1}\,\text{cm}^{-2}$. **b** $h_{\text{geo}} = 1\,\text{km}$, $\max(\phi_{\mu^-}) = 8.1 \times 10^{-3}\,\text{s}^{-1}\,\text{cm}^{-2}$. **c** $h_{\text{geo}} = 10\,\text{km}$, $\max(\phi_{\mu^-}) = 3.9 \times 10^{-2}\,\text{s}^{-1}\,\text{cm}^{-2}$.

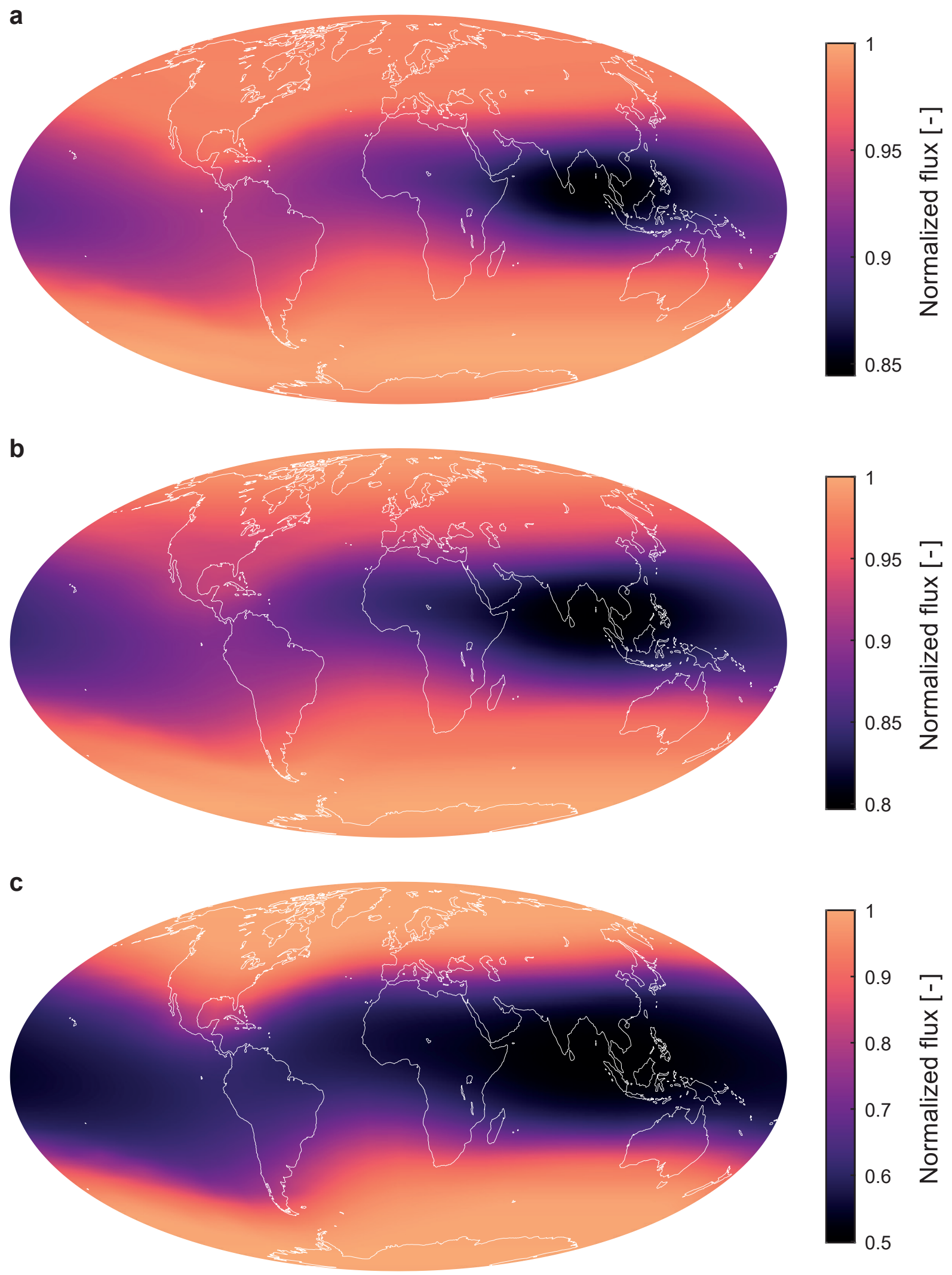

**Figure B.12** World map (Mollweide projection) displaying the normalized total CR induced antimuon flux ($\phi_{\mu^+}$) for different geodetic altitudes ($h_{\mathrm{geo}}$) on 2022-05-04 estimated using the `PARMA` code (version: `4.13`). The total antimuon flux was normalized by the peak value of the corresponding maps as listed below. A more detailed description of the data processing steps is provided in Fig. 2.5. **a** $h_{\mathrm{geo}} = 0\,\mathrm{km}$, $\max(\phi_{\mu^+}) = 7.7 \times 10^{-3}\,\mathrm{s}^{-1}\,\mathrm{cm}^{-2}$. **b** $h_{\mathrm{geo}} = 1\,\mathrm{km}$, $\max(\phi_{\mu^+}) = 9.3 \times 10^{-3}\,\mathrm{s}^{-1}\,\mathrm{cm}^{-2}$. **c** $h_{\mathrm{geo}} = 10\,\mathrm{km}$, $\max(\phi_{\mu^+}) = 4.3 \times 10^{-2}\,\mathrm{s}^{-1}\,\mathrm{cm}^{-2}$.

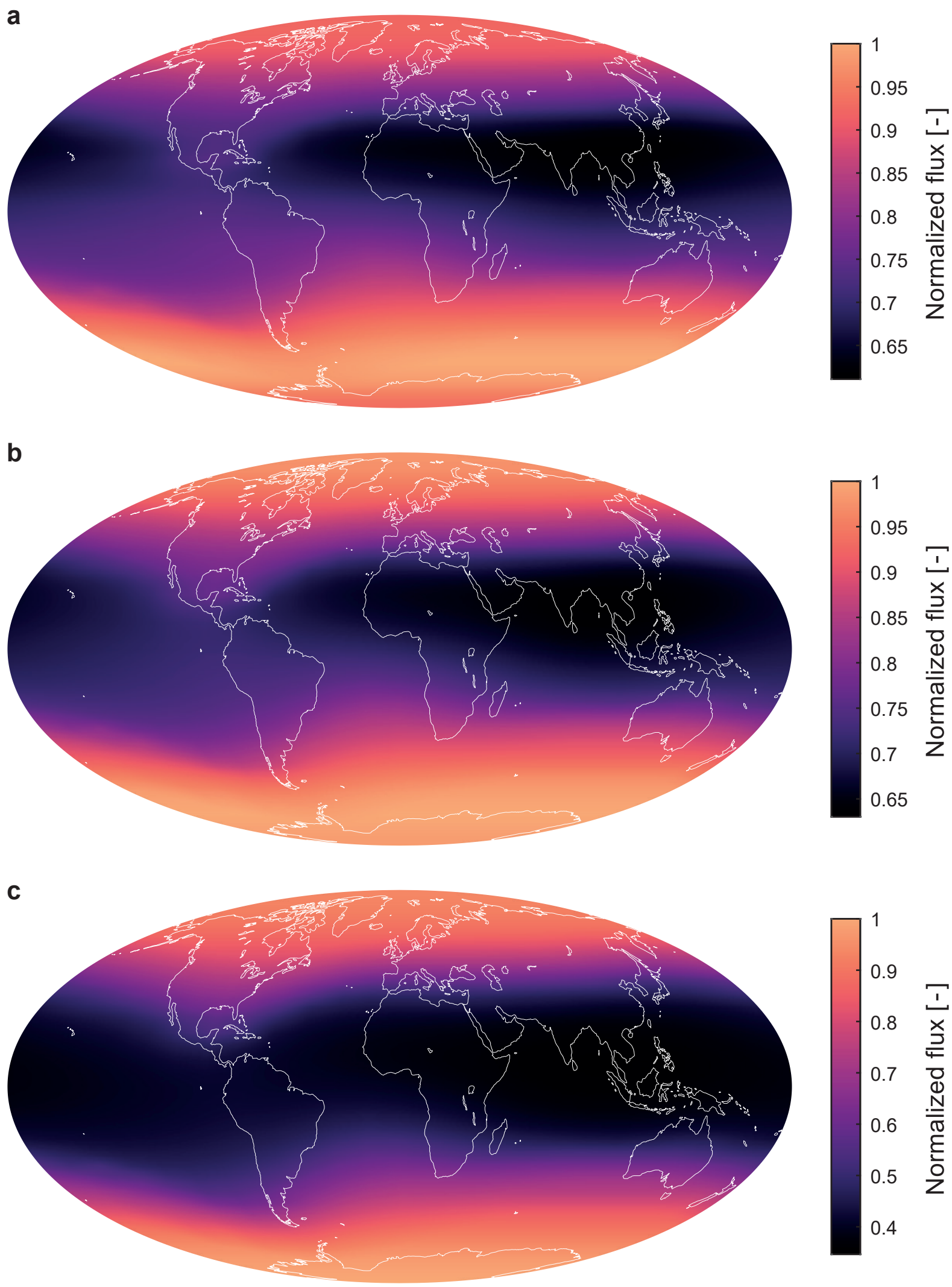

**Figure B.13** World map (Mollweide projection) displaying the normalized total CR induced electron flux ($\phi_{e^-}$) for different geodetic altitudes ($h_{geo}$) on 2022-05-04 estimated using the `PARMA` code (version: `4.13`). The total electron flux was normalized by the peak value of the corresponding maps as listed below. A more detailed description of the data processing steps is provided in Fig. 2.5. **a** $h_{geo} = 0\,\mathrm{km}$, $\max(\phi_{e^-}) = 4.5 \times 10^{-3}\,\mathrm{s^{-1}\,cm^{-2}}$. **b** $h_{geo} = 1\,\mathrm{km}$, $\max(\phi_{e^-}) = 8.3 \times 10^{-3}\,\mathrm{s^{-1}\,cm^{-2}}$. **c** $h_{geo} = 10\,\mathrm{km}$, $\max(\phi_{e^-}) = 8.9 \times 10^{-1}\,\mathrm{s^{-1}\,cm^{-2}}$.

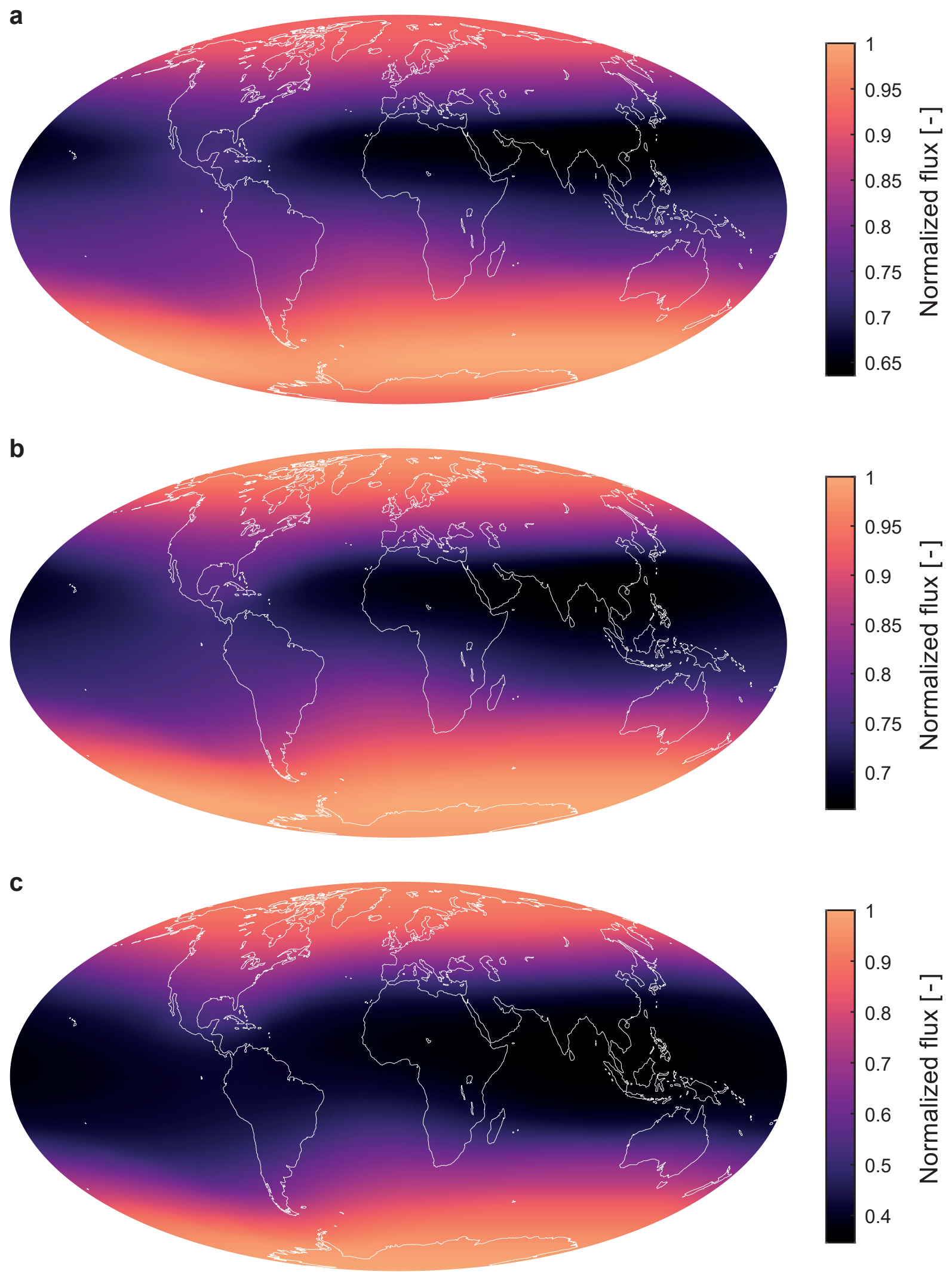

**Figure B.14** World map (Mollweide projection) displaying the normalized total CR induced positron flux ($\phi_{e^+}$) for different geodetic altitudes ($h_{geo}$) on 2022-05-04 estimated using the `PARMA` code (version: `4.13`). The total positron flux was normalized by the peak value of the corresponding maps as listed below. A more detailed description of the data processing steps is provided in Fig. 2.5. **a** $h_{geo} = 0\,\text{km}$, $\max(\phi_{e^+}) = 2.1 \times 10^{-3}\,\text{s}^{-1}\,\text{cm}^{-2}$. **b** $h_{geo} = 1\,\text{km}$, $\max(\phi_{e^+}) = 3.8 \times 10^{-3}\,\text{s}^{-1}\,\text{cm}^{-2}$. **c** $h_{geo} = 10\,\text{km}$, $\max(\phi_{e^+}) = 4.2 \times 10^{-1}\,\text{s}^{-1}\,\text{cm}^{-2}$.

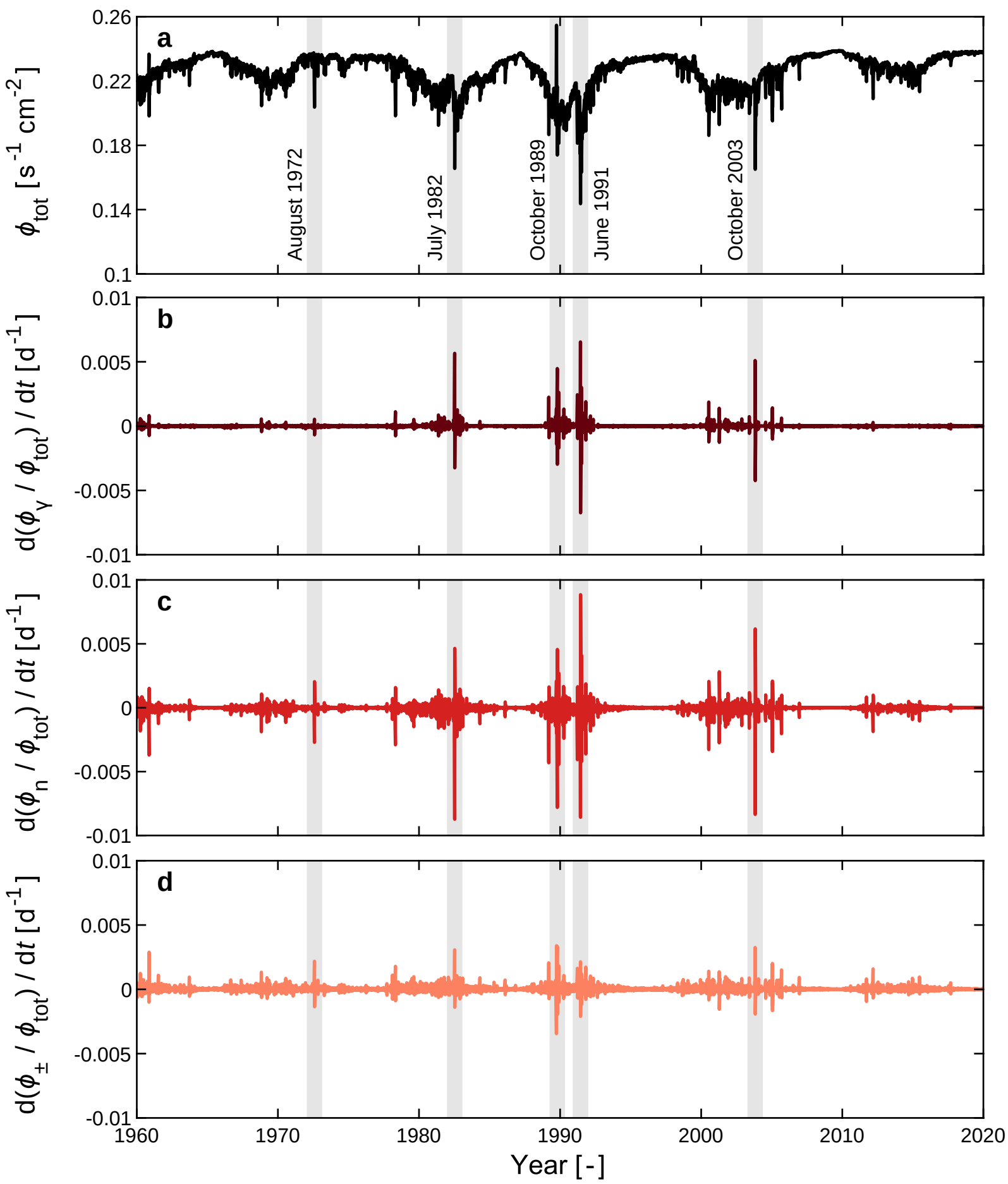

**Figure B.15** Temporal evolution of various secondary cosmic-ray particle fluences in the Earth's atmosphere between 1960 and 2020 (Gregorian calendar base). **a** The CR induced ionizing particle flux $\phi_{tot}$ was estimated using the `PARMA` code (version: `4.13`) for a reference location and altitude of 46.7°N, 7.6°E and 659 m, respectively. The same methodology to estimate $\phi_{tot}$ as described in Fig. 2.4 was applied. **b–d** Time derivatives of the normalized fluences for gamma rays ($\phi_\gamma/\phi_{tot}$), neutrons ($\phi_n/\phi_{tot}$), and charged particles ($\phi_\pm/\phi_{tot}$), i.e. $e^\pm$, $\mu^\pm$, p and $\alpha$ combined, are displayed. Major solar and geomagnetic storms, i.e. the solar flares on August 1972 [901, 902] and July 1982 [903–905], the geomagnetic storm on October 1989 [906–908], the solar flare on June 1991 [909–911] and the Halloween solar storm on October 2003 [912–918], are highlighted together with the corresponding dates of the highest activity in all subfigures.

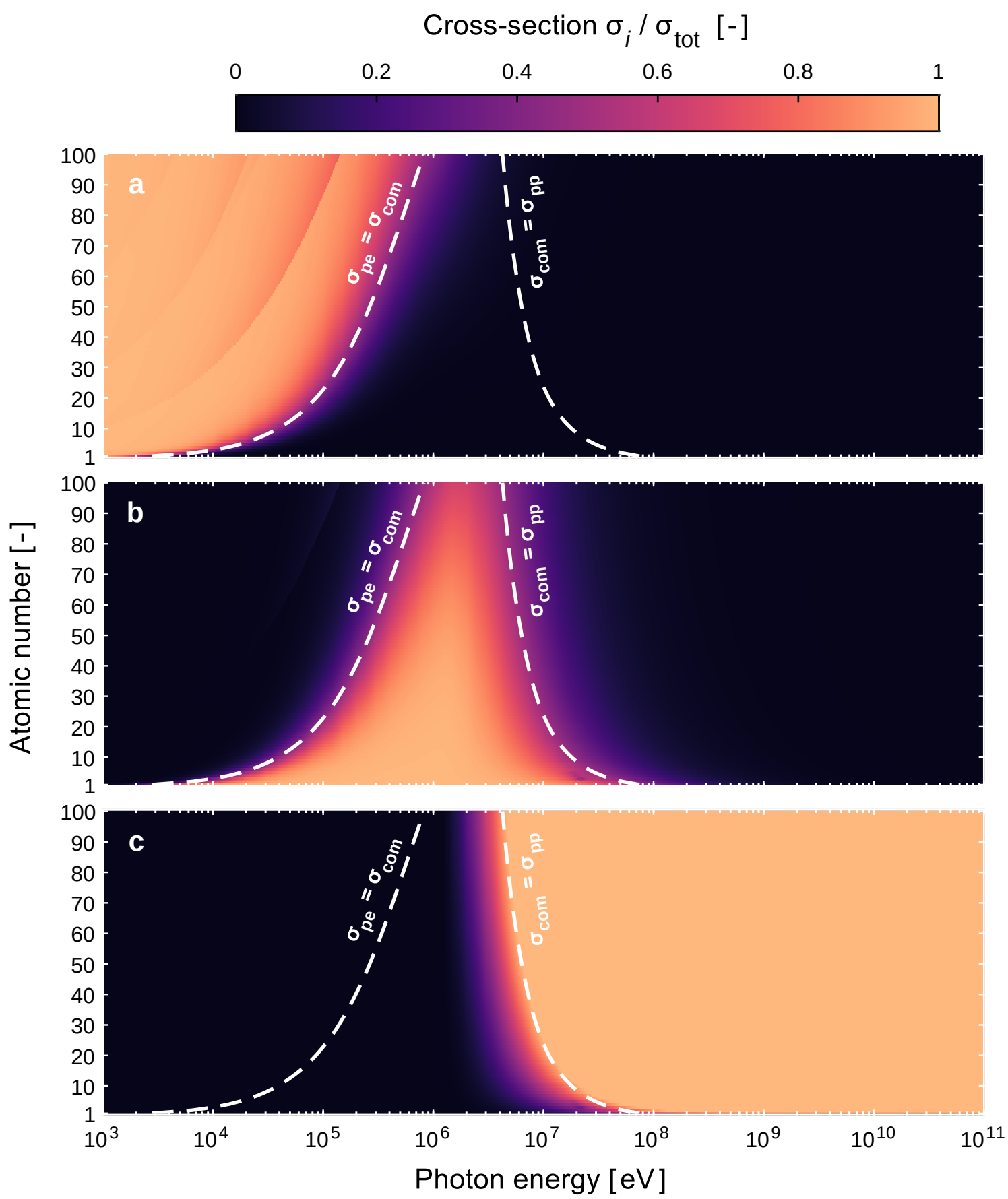

**Figure B.16** Relative microscopic cross-section for high-energy photon interaction processes $i$ with matter as a function of the photon energy and the atomic number of the associated matter. **a** Relative photoelectric cross-section $\sigma_{pe}/\sigma_{tot}$. **b** Relative Compton scattering cross-section $\sigma_{com}/\sigma_{tot}$. **c** Relative electron-positron pair production cross-section $\sigma_{pp}/\sigma_{tot}$ with $\sigma_{pp}$=$\sigma_{pp,n}$+$\sigma_{pp,e}$. The total interaction cross-section $\sigma_{tot}$ is presented in Fig. 3.5. Cross-section data was obtained from the XCOM database by the National Institute of Standards and Technology (NIST) [208, 247].

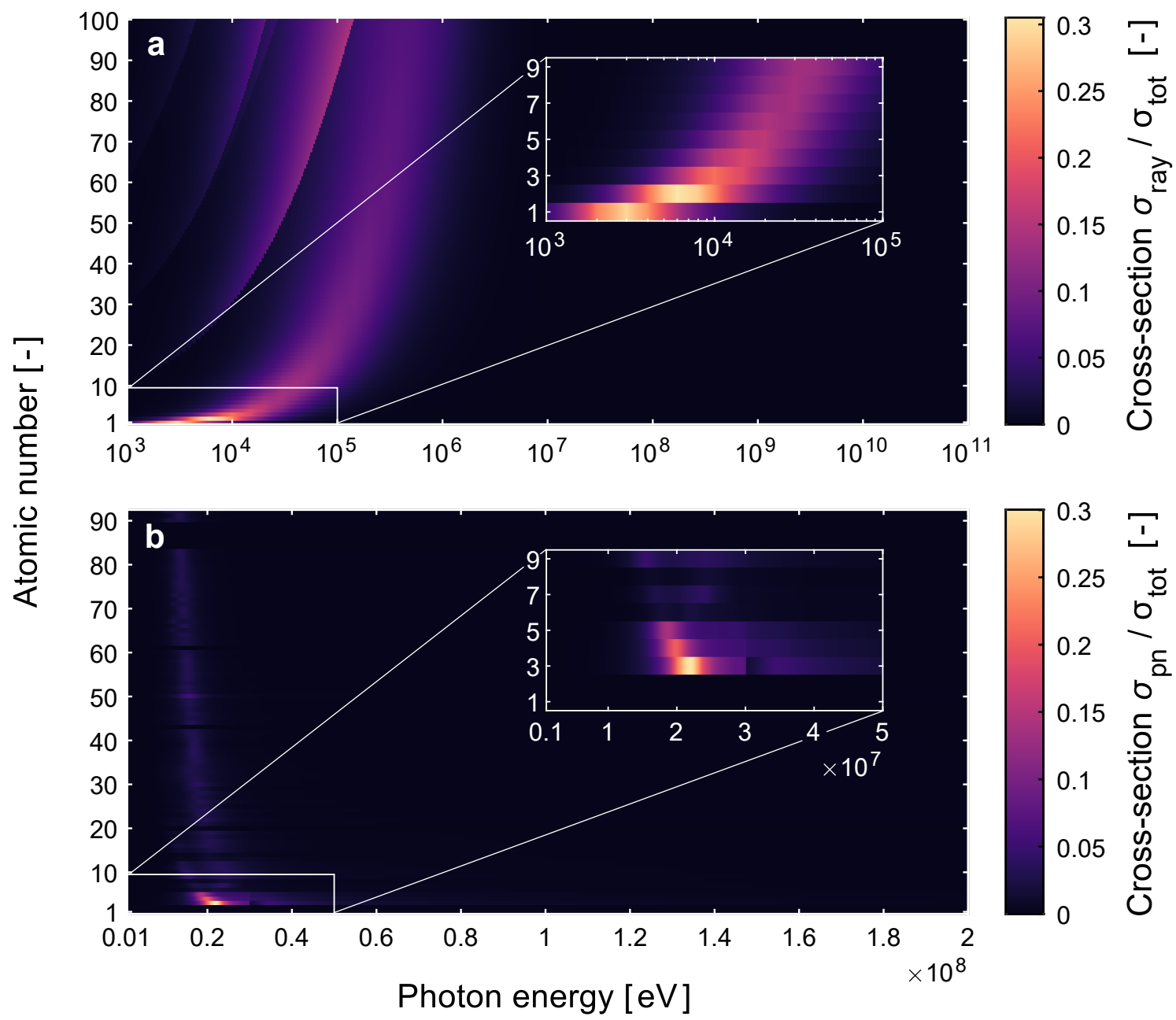

**Figure B.17** Relative microscopic cross-section for high-energy photon interaction processes with matter as a function of the photon energy and the atomic number of the associated matter. **a** Relative Rayleigh scattering cross-section $\sigma_{ray}/\sigma_{tot}$. Cross-section data was obtained from the XCOM database by the National Institute of Standards and Technology (NIST) [208, 247]. **b** Relative photonuclear cross-section $\sigma_{pn}/\sigma_{tot}$. $\sigma_{pn}$ was computed by weighting the individual nuclide cross-sections obtained by the TENDL nuclear data library (version: 2023) [232] with the corresponding isotopic abundance values provided by Meija et al. [56] over the entire spectral range provided by the TENDL nuclear data library, i.e. 1–200 MeV. The total interaction cross-section $\sigma_{tot}$ is presented in Fig. 3.5.

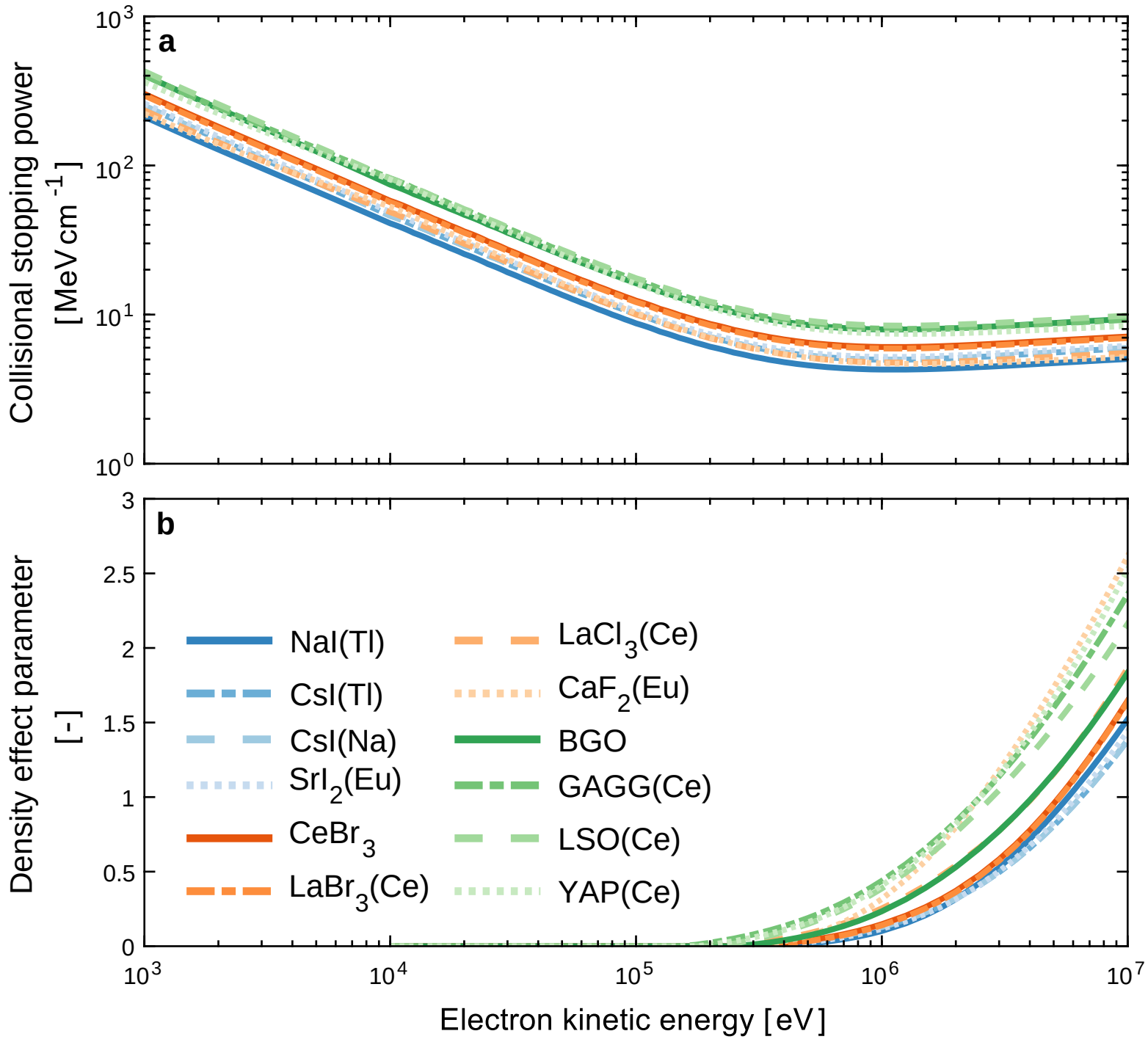

**Figure B.18** Collisional stopping power of electrons & density effect in selected inorganic scintillators. **a** Collisional stopping power of electrons $\mathcal{S}_{e,col}$ as function of the electron kinetic energy $E_{k,e^-}$ for selected inorganic scintillators. $\mathcal{S}_{e,col}$ was computed using Eq. 4.6 for $E_{k,e^-} > 10$ keV and Eq. 4.8 for $E_{k,e^-} < 10$ keV. **b** Density effect parameter $\delta$ (cf. Eq. 4.6) as function of the electron kinetic energy $E_{k,e^-}$ for selected inorganic scintillators. The ESTAR database by the NIST [318] was adopted to evaluate Eq. 4.6 and to compute associated material properties such as the density effect parameter $\delta$ or mean excitation energy $I_0$.

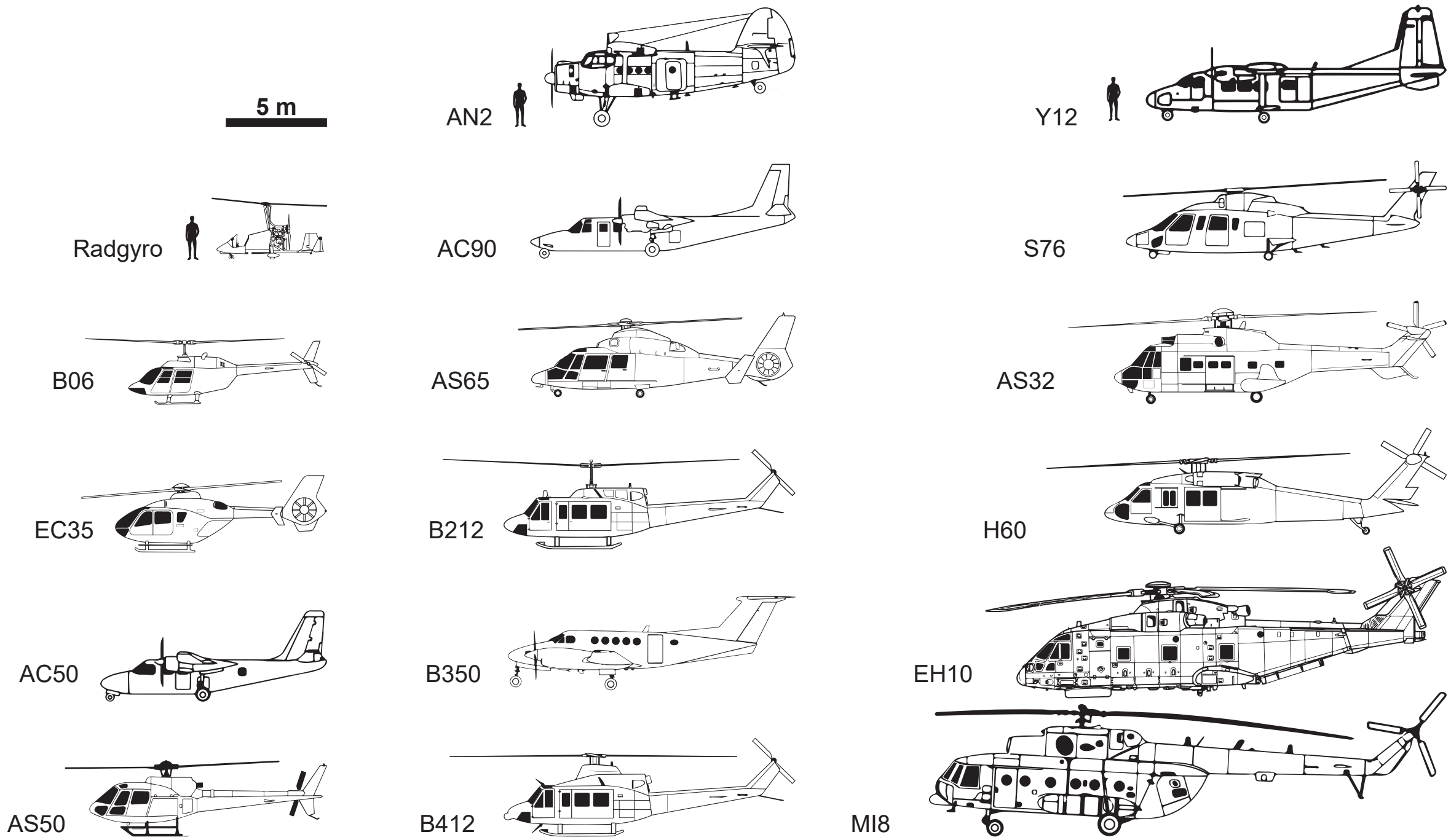

**Figure B.19** Common manned aircraft systems used for non-commercial AGRS at the time of writing (cf. Table 5.1). Aircraft type designators are in accordance with the International Civil Aviation Organization (ICAO) code for aircraft designators.

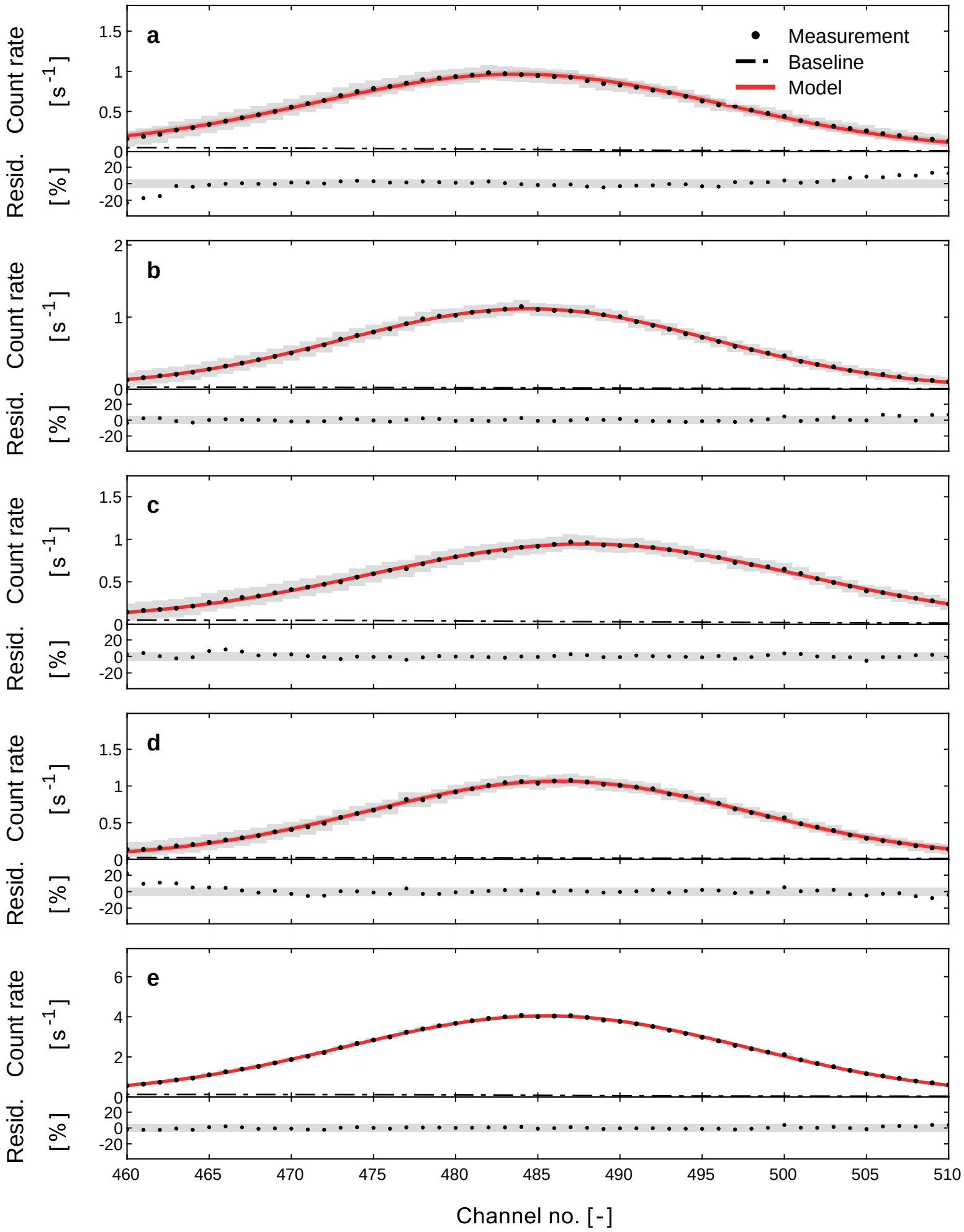

**Figure B.20** The measured and modeled net pulse-height spectra for a $K_{nat}$ source around the FEP related to the photon emission line at 1460.822(6) keV of $^{40}_{19}K$ [65] are displayed for the four single detector channels `#1` through `#4` (**a–d**) and the detector channel `#SUM` (**e**). The displayed part of the pulse-height spectrum was modeled as a Gaussian singlet combined with a numerical baseline correction [380, 381] (cf. Section 6.2.1.3). Measurement uncertainties are provided as 1 standard deviation (SD) gray-shaded areas. The model uncertainty is characterized by 99 % prediction intervals displayed as red-shaded areas. In addition, the relative residuals (Resid., normalized by the mean measured values) are provided for each detector channel with the 5 % band highlighted.

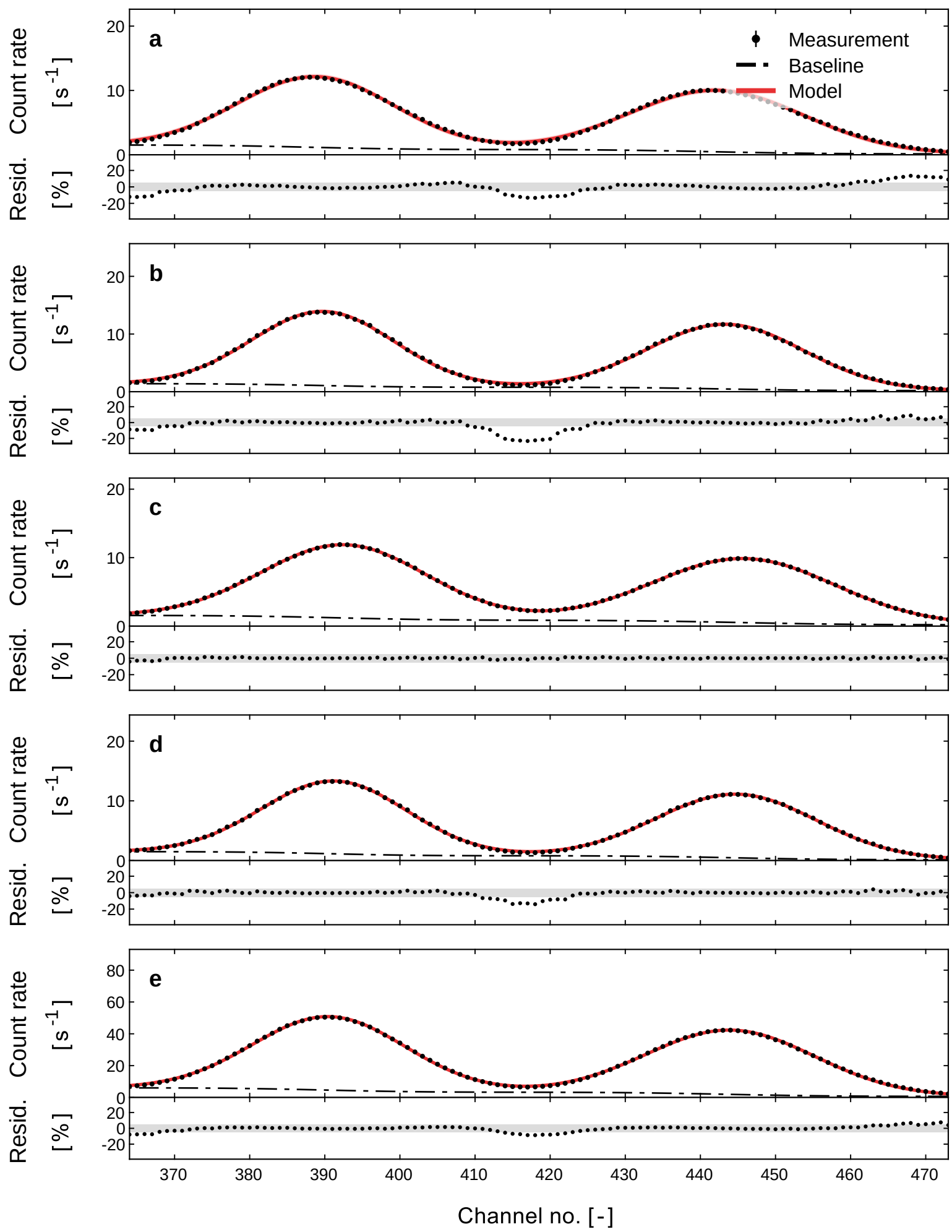

**Figure B.21** The measured and modeled net pulse-height spectra for a $^{60}_{27}$Co source around the FEP related to the photon emission lines at 1173.228(3) keV and 1332.492(4) keV of $^{60}_{27}$Co [68] are displayed for the four single detector channels `#1` through `#4` (**a–d**) and the detector channel `#SUM` (**e**). The displayed part of the pulse-height spectrum was modeled as a Gaussian doublet combined with a numerical baseline correction [380, 381] (cf. Section 6.2.1.3). Measurement uncertainties are provided as 1 standard deviation (SD) error bars (hidden by the marker size). The model uncertainty is characterized by 99 % prediction intervals displayed as red-shaded areas. In addition, the relative residuals (Resid., normalized by the mean measured values) are provided for each detector channel with the 5 % band highlighted.

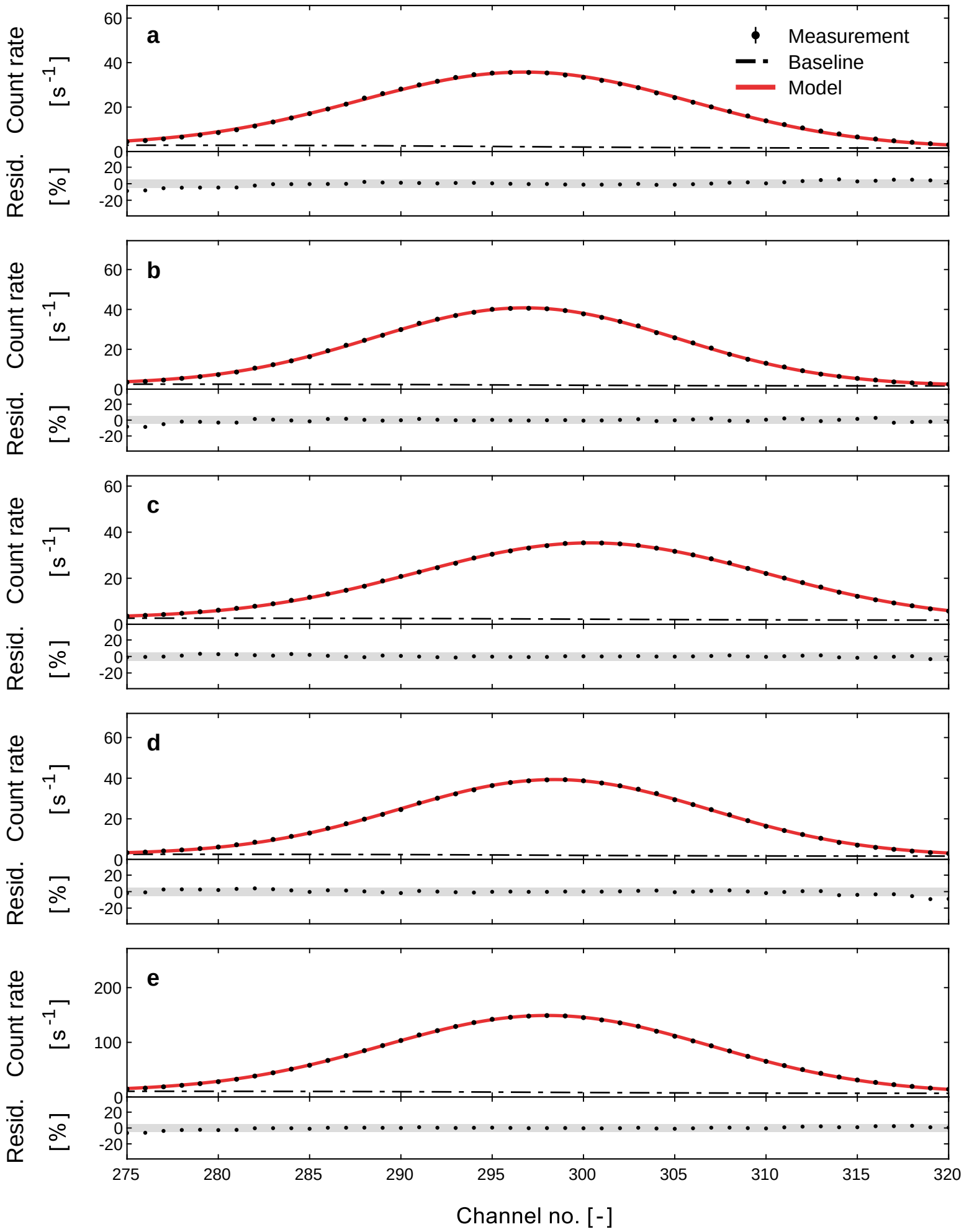

**Figure B.22** The measured and modeled net pulse-height spectra for a $^{88}_{39}$Y source around the FEP related to the photon emission line at 898.042(11) keV of $^{88}_{39}$Y [51] are displayed for the four single detector channels `#1` through `#4` (**a–d**) and the detector channel `#SUM` (**e**). The displayed part of the pulse-height spectrum was modeled as a Gaussian singlet combined with a numerical baseline correction [380, 381] (cf. Section 6.2.1.3). Measurement uncertainties are provided as 1 standard deviation (SD) error bars (hidden by the marker size). The model uncertainty is characterized by 99 % prediction intervals displayed as red-shaded areas (hidden by the line width). In addition, the relative residuals (Resid., normalized by the mean measured values) are provided for each detector channel with the 5 % band highlighted.

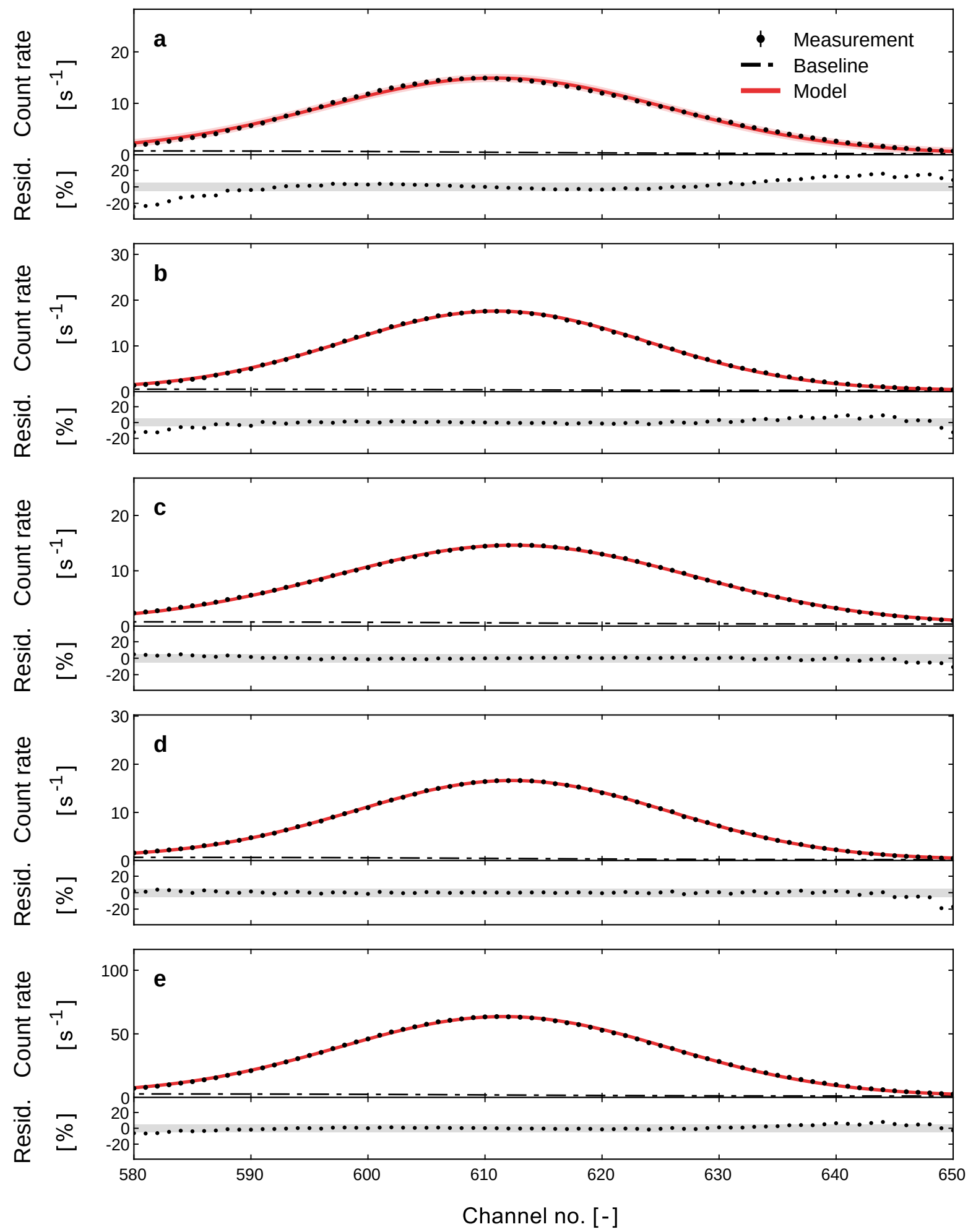

**Figure B.23** The measured and modeled net pulse-height spectra for a $^{88}_{39}$Y source around the FEP related to the photon emission line at 1836.070(8) keV of $^{88}_{39}$Y [51] are displayed for the four single detector channels `#1` through `#4` (**a–d**) and the detector channel `#SUM` (**e**). The displayed part of the pulse-height spectrum was modeled as a Gaussian singlet combined with a numerical baseline correction [380, 381] (cf. Section 6.2.1.3). Measurement uncertainties are provided as 1 standard deviation (SD) error bars (hidden by the marker size). The model uncertainty is characterized by 99 % prediction intervals displayed as red-shaded areas (hidden by the line width). In addition, the relative residuals (Resid., normalized by the mean measured values) are provided for each detector channel with the 5 % band highlighted.

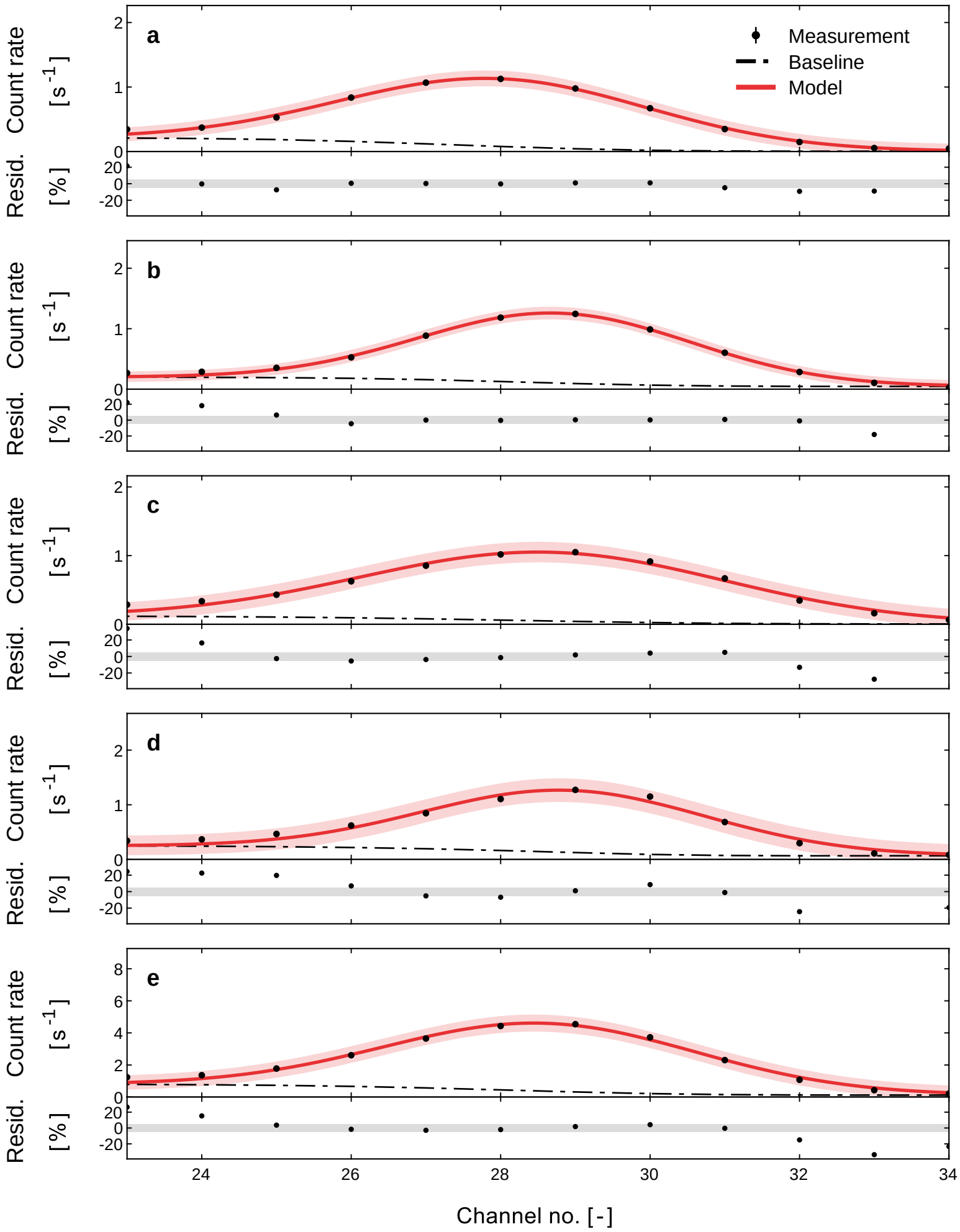

**Figure B.24** The measured and modeled net pulse-height spectra for a $^{109}_{48}$Cd source around the FEP related to the photon emission line at 88.0336(10) keV of $^{109}_{48}$Cd [102] are displayed for the four single detector channels `#1` through `#4` (**a–d**) and the detector channel `#SUM` (**e**). The displayed part of the pulse-height spectrum was modeled as a Gaussian singlet combined with a numerical baseline correction [380, 381] (cf. Section 6.2.1.3). Measurement uncertainties are provided as 1 standard deviation (SD) error bars (hidden by the marker size). The model uncertainty is characterized by 99 % prediction intervals displayed as red-shaded areas. In addition, the relative residuals (Resid., normalized by the mean measured values) are provided for each detector channel with the 5 % band highlighted.

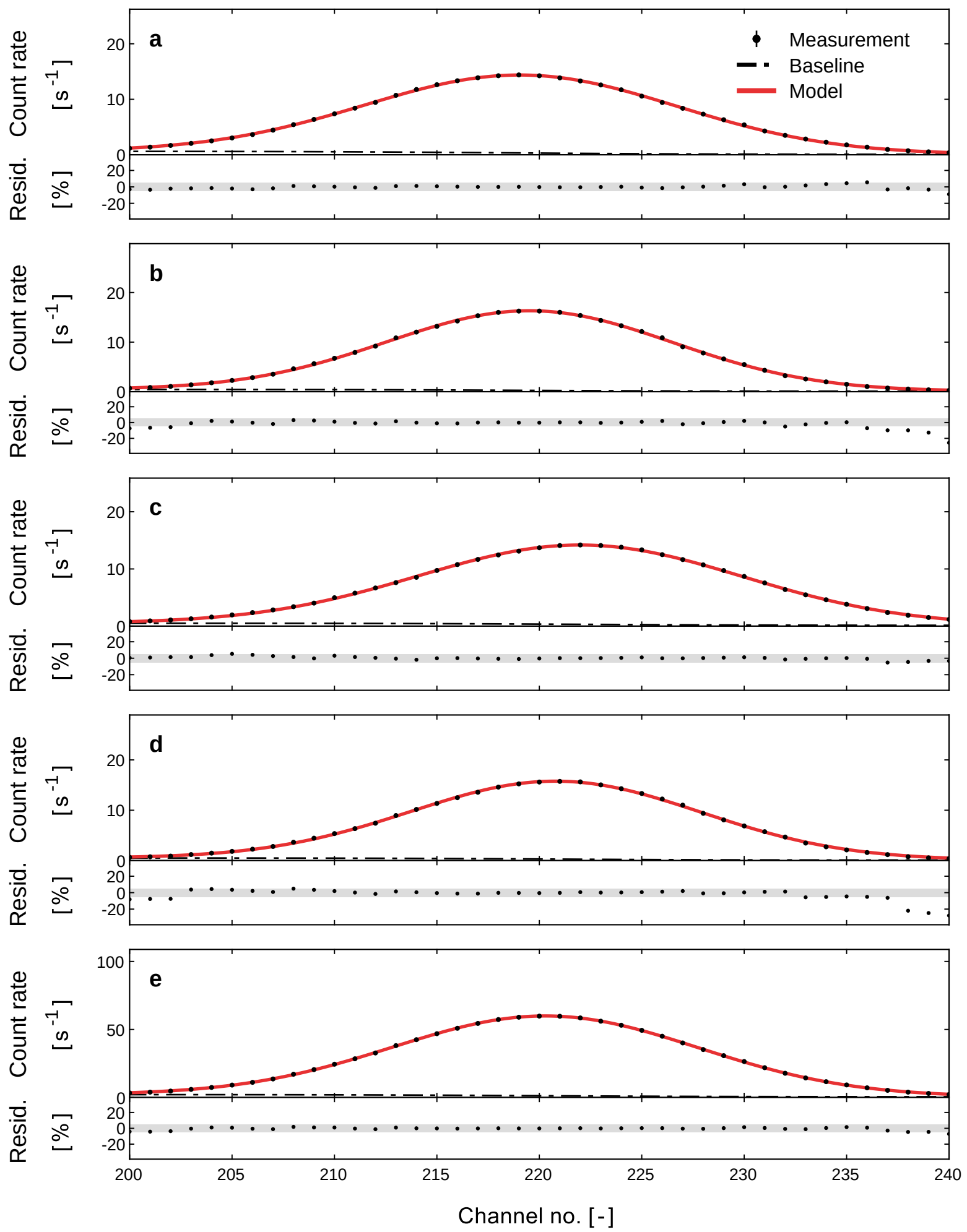

**Figure B.25** The measured and modeled net pulse-height spectra for a $^{137}_{55}Cs$ source around the FEP related to the photon emission line at 661.657(3) keV of $^{137}_{55}Cs$ [68] are displayed for the four single detector channels `#1` through `#4` (**a–d**) and the detector channel `#SUM` (**e**). The displayed part of the pulse-height spectrum was modeled as a Gaussian singlet combined with a numerical baseline correction [380, 381] (cf. Section 6.2.1.3). Measurement uncertainties are provided as 1 standard deviation (SD) error bars (hidden by the marker size). The model uncertainty is characterized by 99 % prediction intervals displayed as red-shaded areas (hidden by the line width). In addition, the relative residuals (Resid., normalized by the mean measured values) are provided for each detector channel with the 5 % band highlighted.

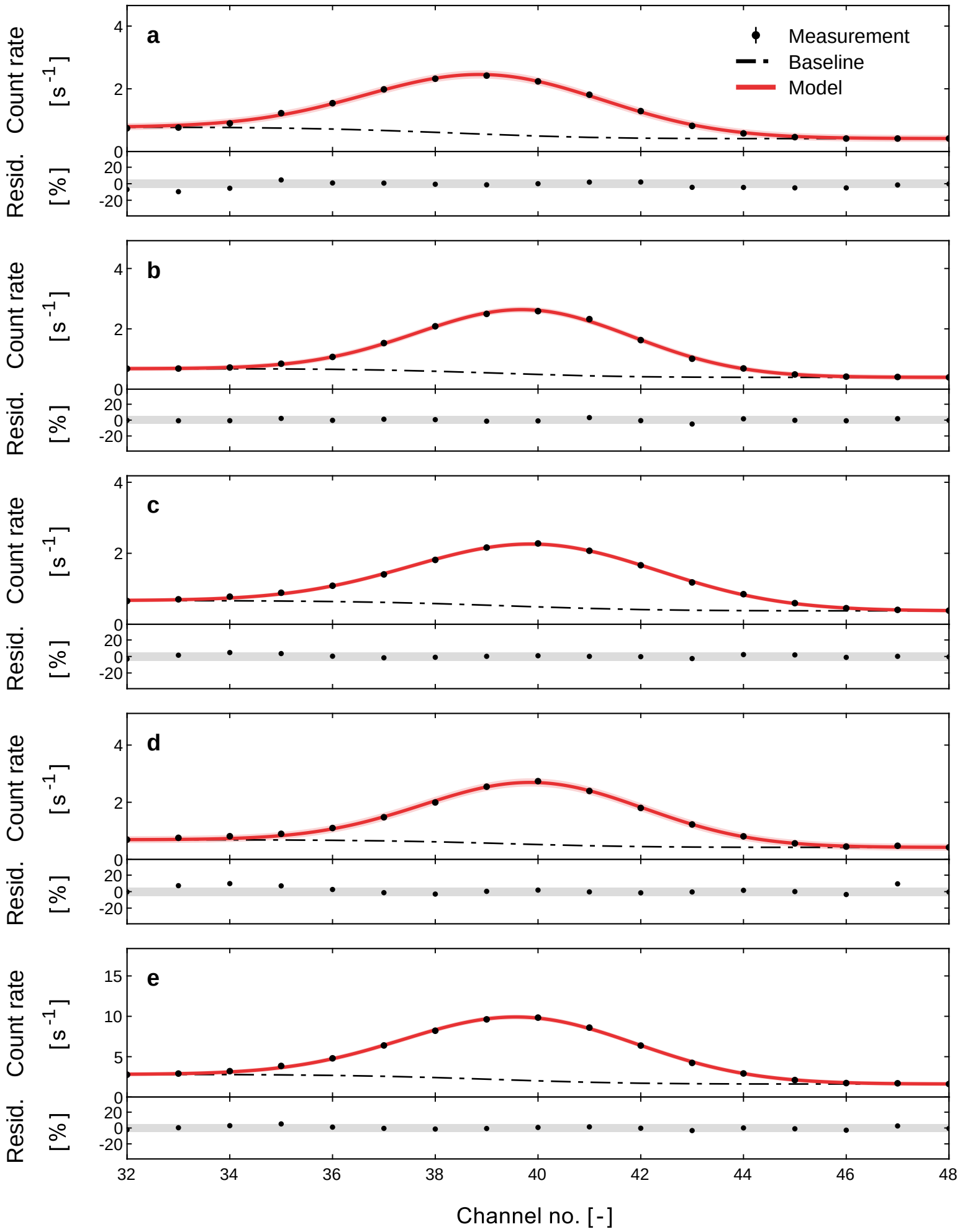

**Figure B.26** The measured and modeled net pulse-height spectra for a $^{152}_{63}$Eu source around the FEP related to the photon emission line at 121.7817(3) keV of $^{152}_{63}$Eu [714] are displayed for the four single detector channels `#1` through `#4` (a–d) and the detector channel `#SUM` (e). The displayed part of the pulse-height spectrum was modeled as a Gaussian singlet combined with a numerical baseline correction [380, 381] (cf. Section 6.2.1.3). Measurement uncertainties are provided as 1 standard deviation (SD) error bars (hidden by the marker size). The model uncertainty is characterized by 99 % prediction intervals displayed as red-shaded areas. In addition, the relative residuals (Resid., normalized by the mean measured values) are provided for each detector channel with the 5 % band highlighted.

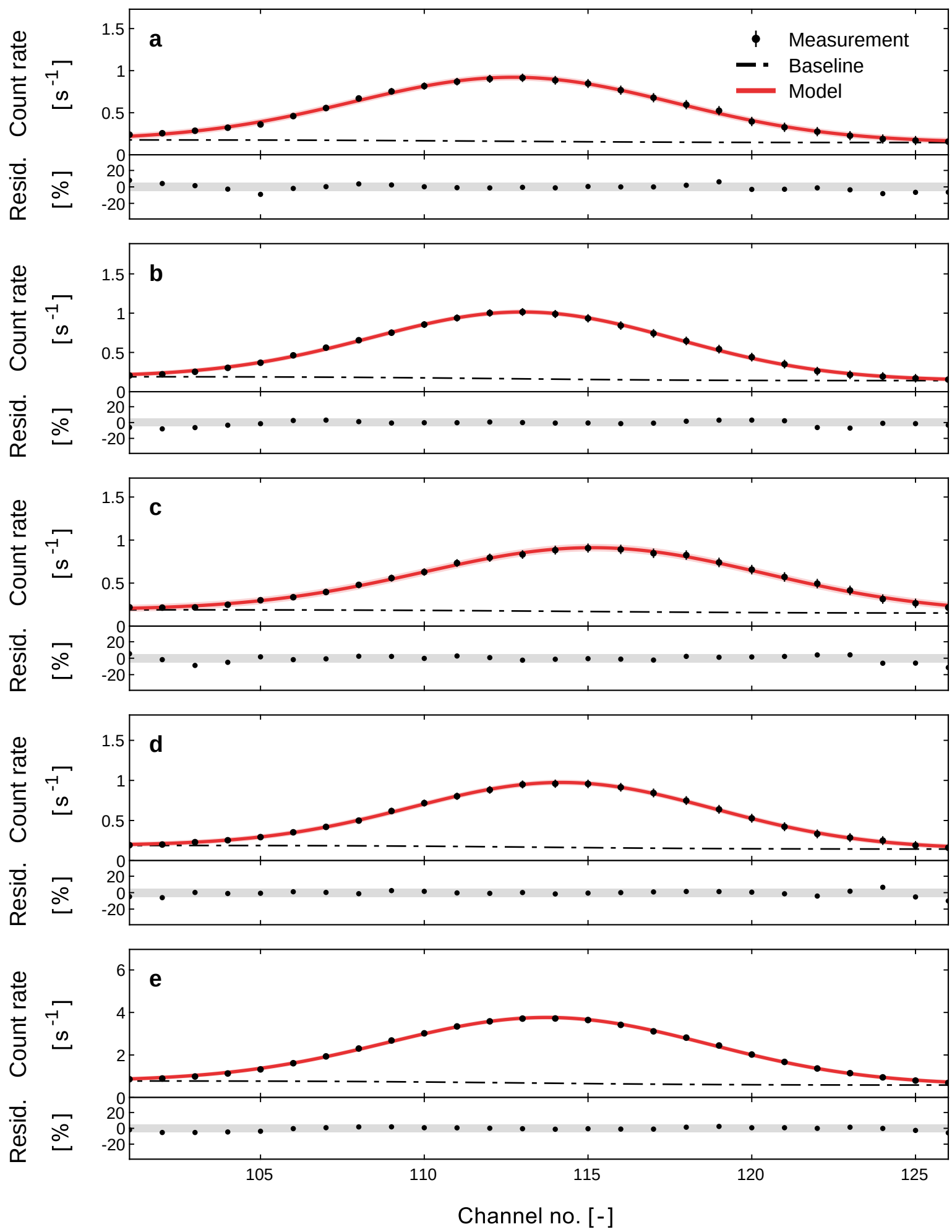

**Figure B.27** The measured and modeled net pulse-height spectra for a $^{152}_{63}$Eu source around the FEP related to the photon emission line at 344.2785(12) keV of $^{152}_{63}$Eu [714] are displayed for the four single detector channels `#1` through `#4` (**a–d**) and the detector channel `#SUM` (**e**). The displayed part of the pulse-height spectrum was modeled as a Gaussian singlet combined with a numerical baseline correction [380, 381] (cf. Section 6.2.1.3). Measurement uncertainties are provided as 1 standard deviation (SD) error bars. The model uncertainty is characterized by 99 % prediction intervals displayed as red-shaded areas. In addition, the relative residuals (Resid., normalized by the mean measured values) are provided for each detector channel with the 5 % band highlighted.

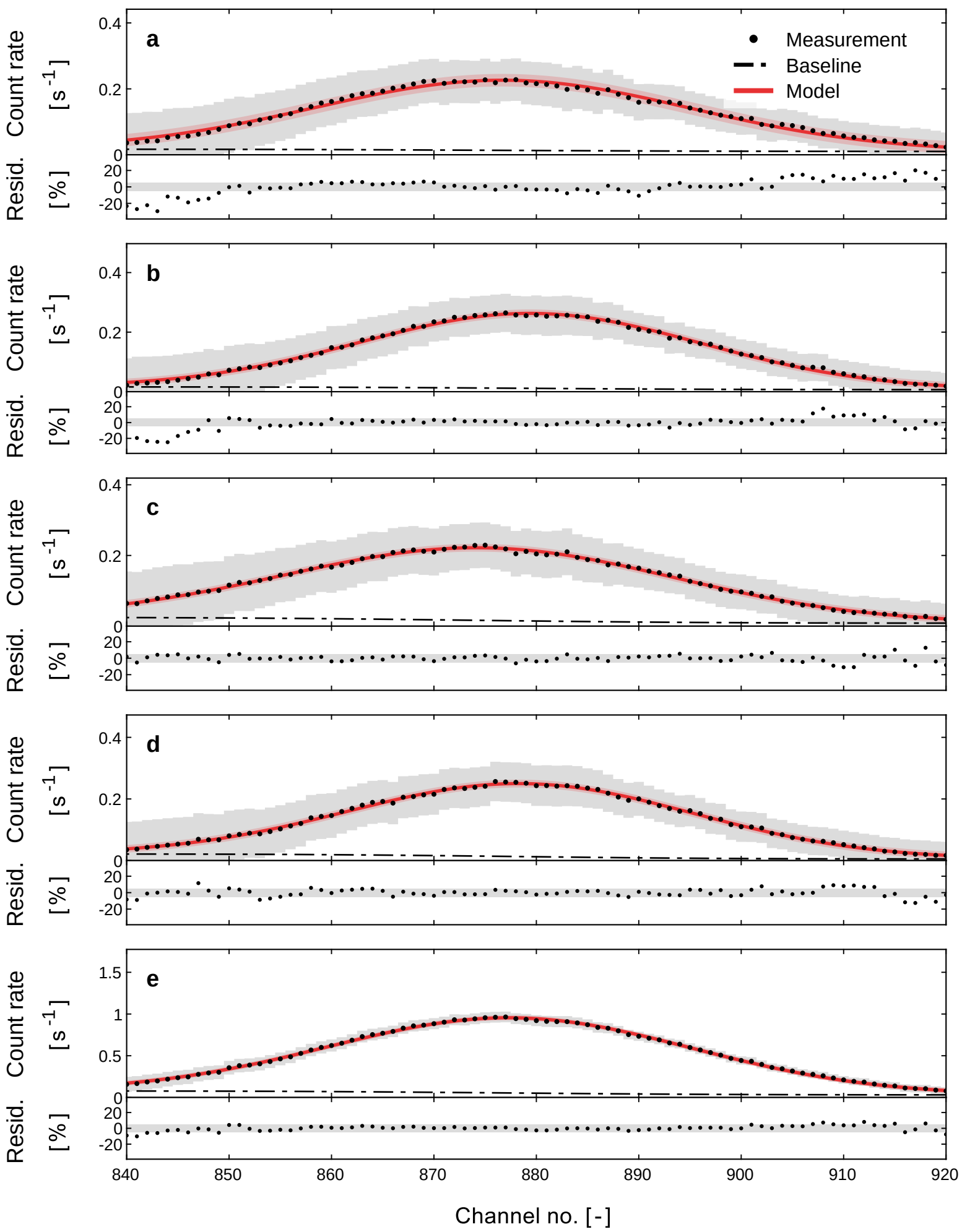

**Figure B.28** The measured and modeled net pulse-height spectra for a $Th_{nat}$ source around the FEP related to the photon emission line at 2614.511(10) keV of ${}^{208}_{81}Tl$ [108] are displayed for the four single detector channels `#1` through `#4` (**a–d**) and the detector channel `#SUM` (**e**). The displayed part of the pulse-height spectrum was modeled as a Gaussian singlet combined with a numerical baseline correction [380, 381] (cf. Section 6.2.1.3). Measurement uncertainties are provided as 1 standard deviation (SD) gray-shaded areas. The model uncertainty is characterized by 99 % prediction intervals displayed as red-shaded areas. In addition, the relative residuals (Resid., normalized by the mean measured values) are provided for each detector channel with the 5 % band highlighted.

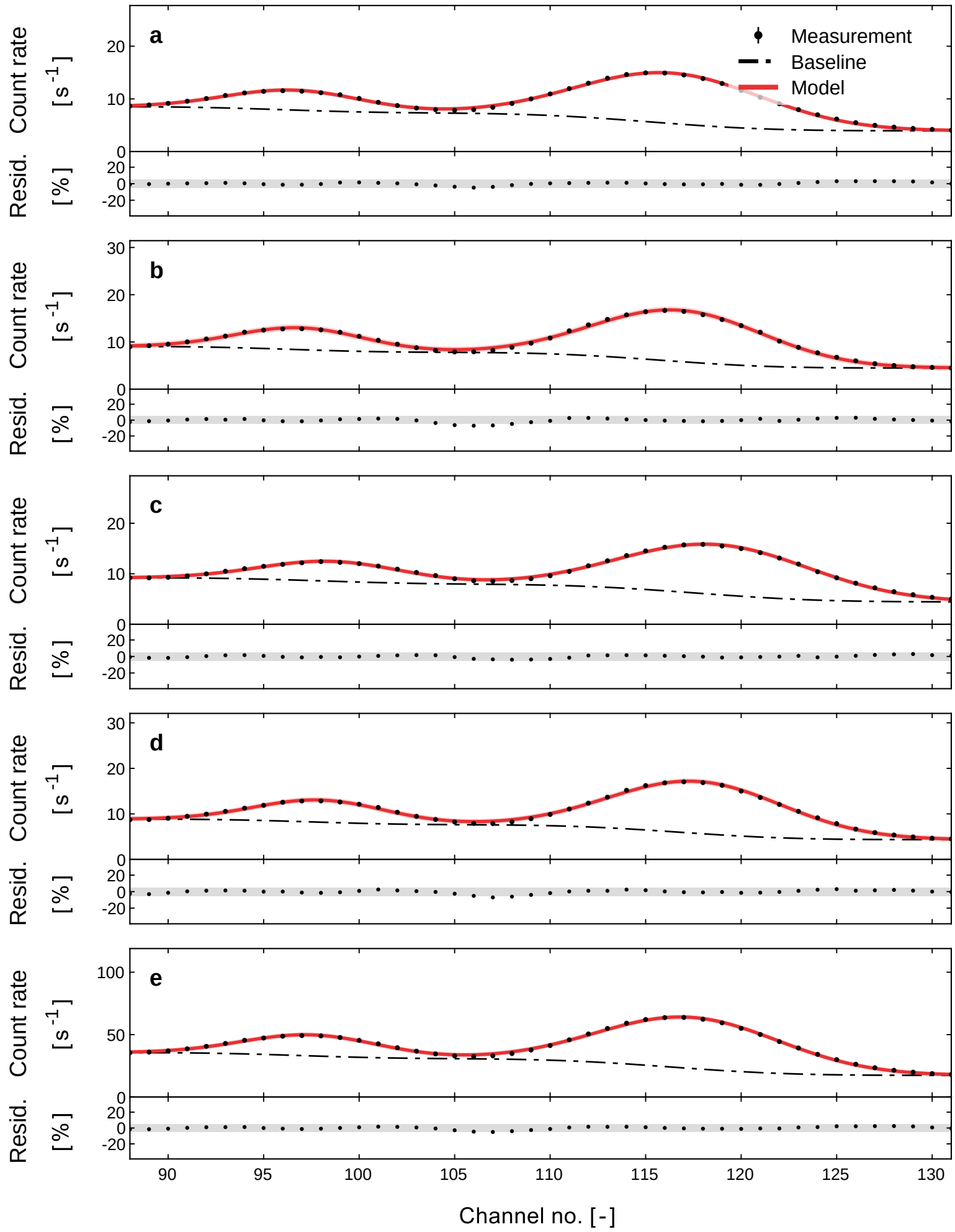

**Figure B.29** The measured and modeled net pulse-height spectra for a $U_{nat}$ source around the FEP related to the photon emission lines at 295.224(2) keV and 351.932(2) keV of $^{214}_{82}Pb$ [124] are displayed for the four single detector channels `#1` through `#4` (**a–d**) and the detector channel `#SUM` (**e**). The displayed part of the pulse-height spectrum was modeled as a Gaussian doublet combined with a numerical baseline correction [380, 381] (cf. Section 6.2.1.3). Measurement uncertainties are provided as 1 standard deviation (SD) error bars (hidden by the marker size). The model uncertainty is characterized by 99 % prediction intervals displayed as red-shaded areas. In addition, the relative residuals (Resid., normalized by the mean measured values) are provided for each detector channel with the 5 % band highlighted.

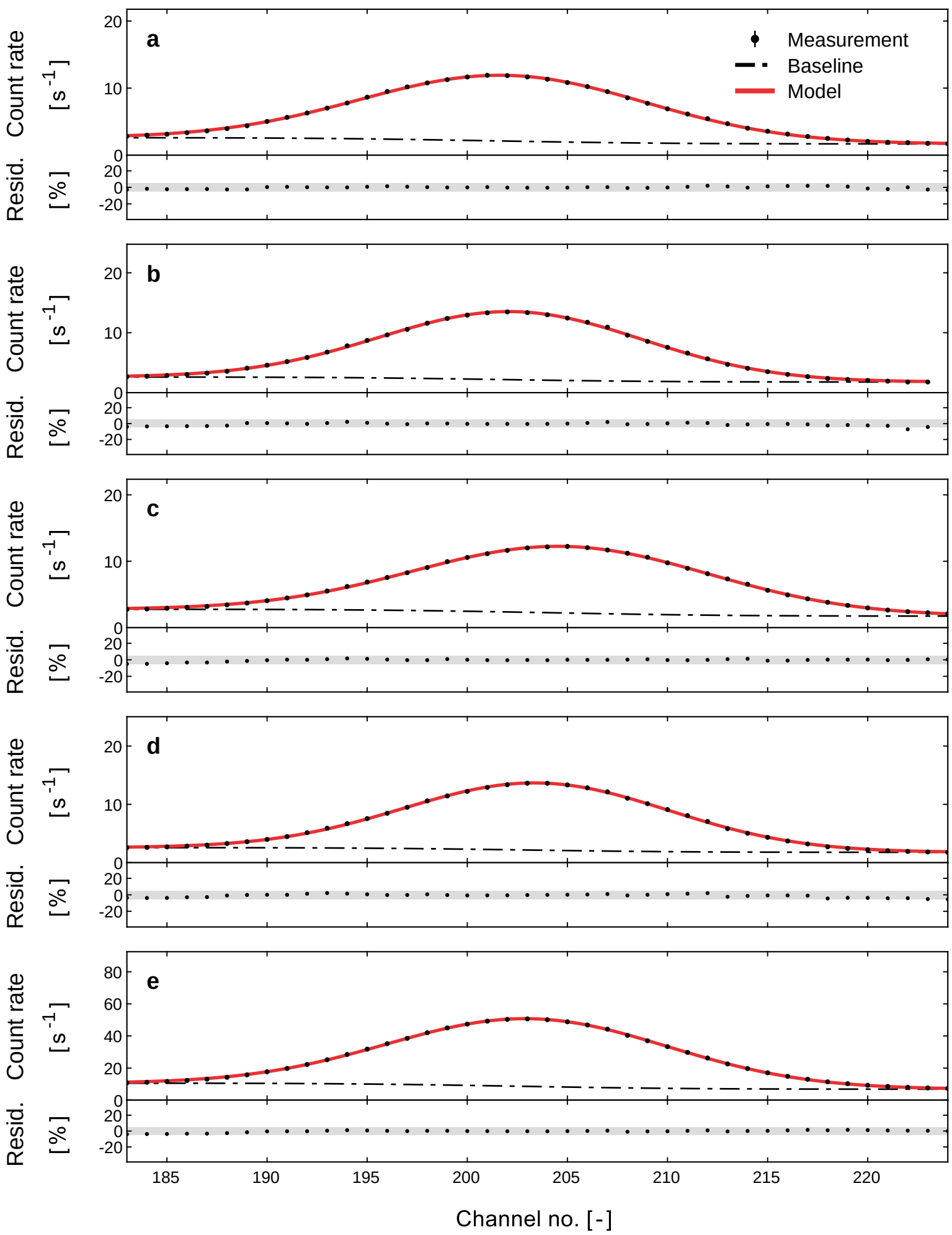

**Figure B.30** The measured and modeled net pulse-height spectra for a $U_{nat}$ source around the FEP related to the photon emission line at 609.312(7) keV of ${}^{214}_{83}$Bi [124] are displayed for the four single detector channels `#1` through `#4` (**a–d**) and the detector channel `#SUM` (**e**). The displayed part of the pulse-height spectrum was modeled as a Gaussian singlet combined with a numerical baseline correction [380, 381] (cf. Section 6.2.1.3). Measurement uncertainties are provided as 1 standard deviation (SD) error bars (hidden by the marker size). The model uncertainty is characterized by 99 % prediction intervals displayed as red-shaded areas (hidden by the line width). In addition, the relative residuals (Resid., normalized by the mean measured values) are provided for each detector channel with the 5 % band highlighted.

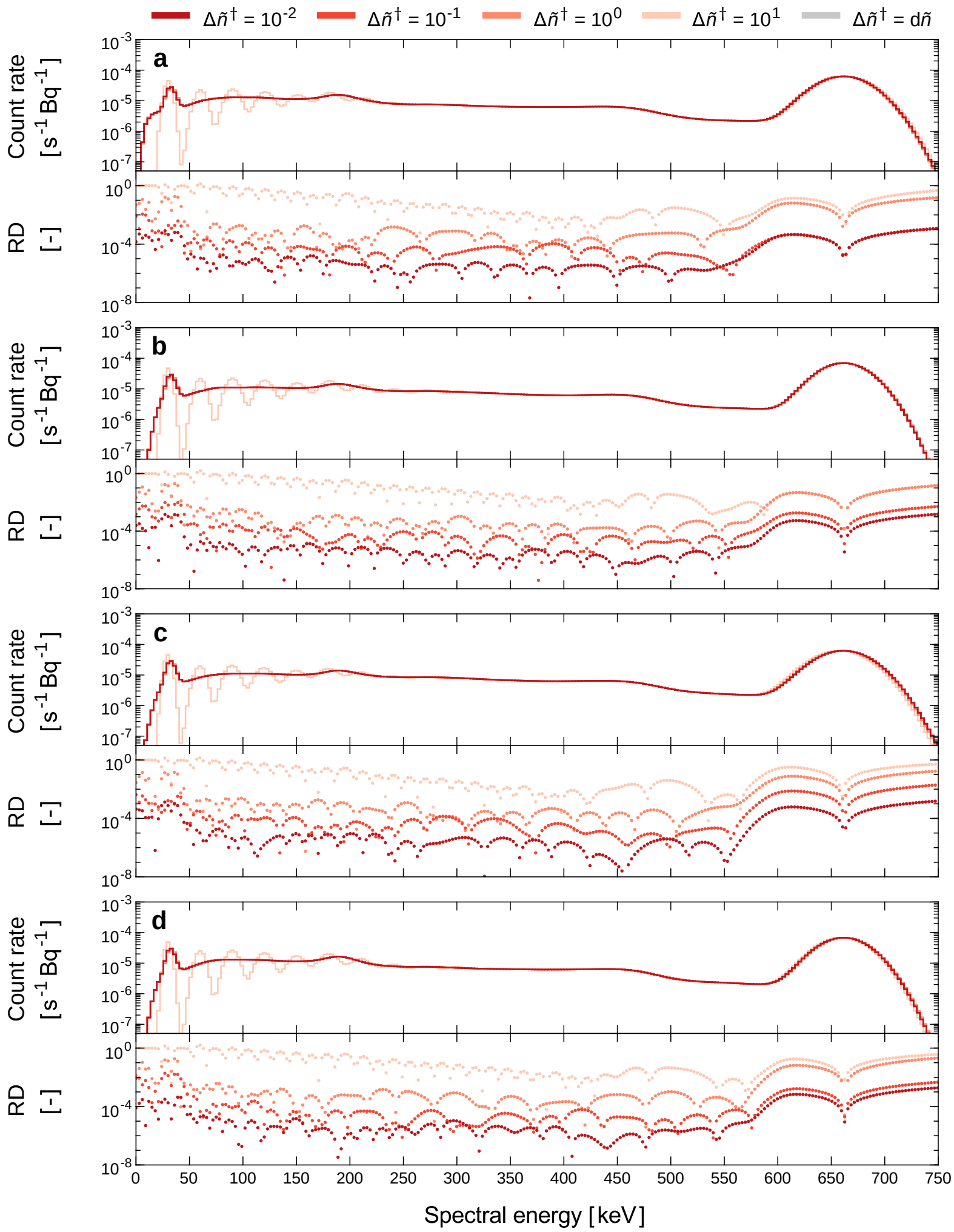

**Figure B.31** These graphs present the results from a convergence study for the bin width $\Delta\tilde{n}^\dagger$ of the refined pulse-height channel number $\tilde{n}^\dagger$ introduced in the `PScinMC` pipeline (cf. Section 6.2.2.4). The simulated mean spectral signatures $\hat{c}_{\text{sim}}$, corresponding to four distinct bin widths $\Delta\tilde{n}^\dagger = \{10^{-2}, 10^{-1}, 1, 10^1\}$ alongside the continuously convolved signature $\Delta\tilde{n}^\dagger = \mathrm{d}n$ (hidden by the line width), are depicted as a function of the spectral energy $E'$ with $\Delta E' \sim 3\,\text{keV}$ for the four detector channels `#1` through `#4` (**a–d**). These simulations are based on the $^{137}_{55}$Cs calibration source and simulation setup outlined in Section 6.2.2. In addition, the relative deviation (RD) between the continuously and discrete convolved signatures computed as $|\hat{c}_{\text{sim},\Delta\tilde{n}^\dagger} - \hat{c}_{\text{sim},\mathrm{d}n}|/\hat{c}_{\text{sim},\mathrm{d}n}$ are provided.

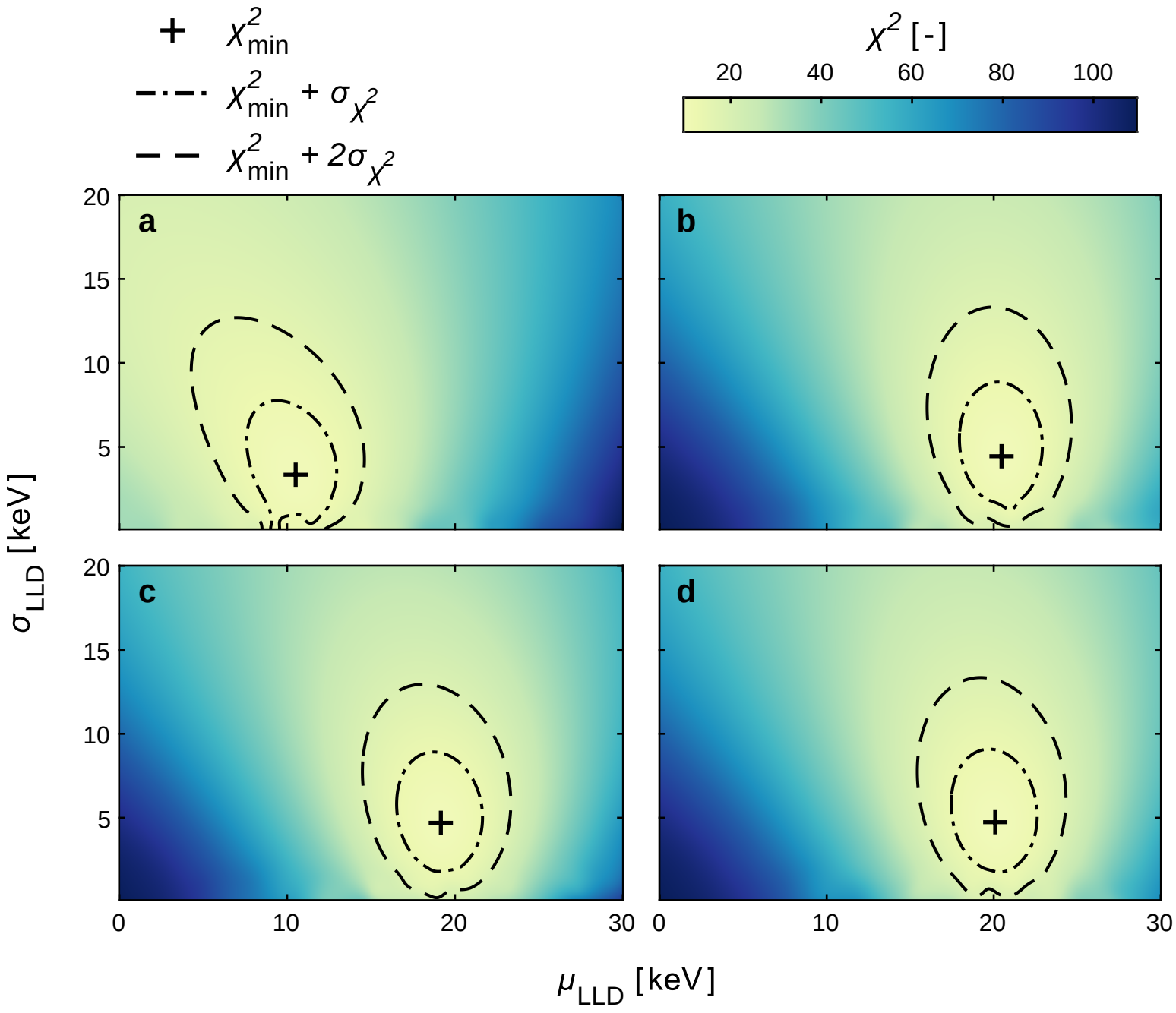

**Figure B.32** Here, the results of the LLD calibration described in Appendix A.9 are presented as chi-squared $\chi^2$ contour maps of the lower level discriminator model parameters $\mu_{LLD}$ and $\sigma_{LLD}$ for the four detector channels: **a** Detector channel `#1`. **b** Detector channel `#2`. **c** Detector channel `#3`. **d** Detector channel `#4`. The global minima $\chi^2_{min}$ are marked with a plus sign together with the corresponding 1-sigma and 2-sigma confidence regions indicated by the dashed-dotted and dashed contour lines, respectively. For the reader's convenience, the unit of the lower level discriminator model parameters $\mu_{LLD}$ and $\sigma_{LLD}$ was converted from continuous pulse-height channel numbers to spectral energies using the energy calibration models derived by the `RLLCal` pipeline (cf. Section 6.2.1.3). For the calibration, the laboratory-based radiation measurement with a ${}^{60}_{27}Co$ calibration point source described in Section 6.2.1 was adopted. The measured and simulated spectral signatures were derived by the `RLLSpec` and `PScinMC` pipelines, respectively.

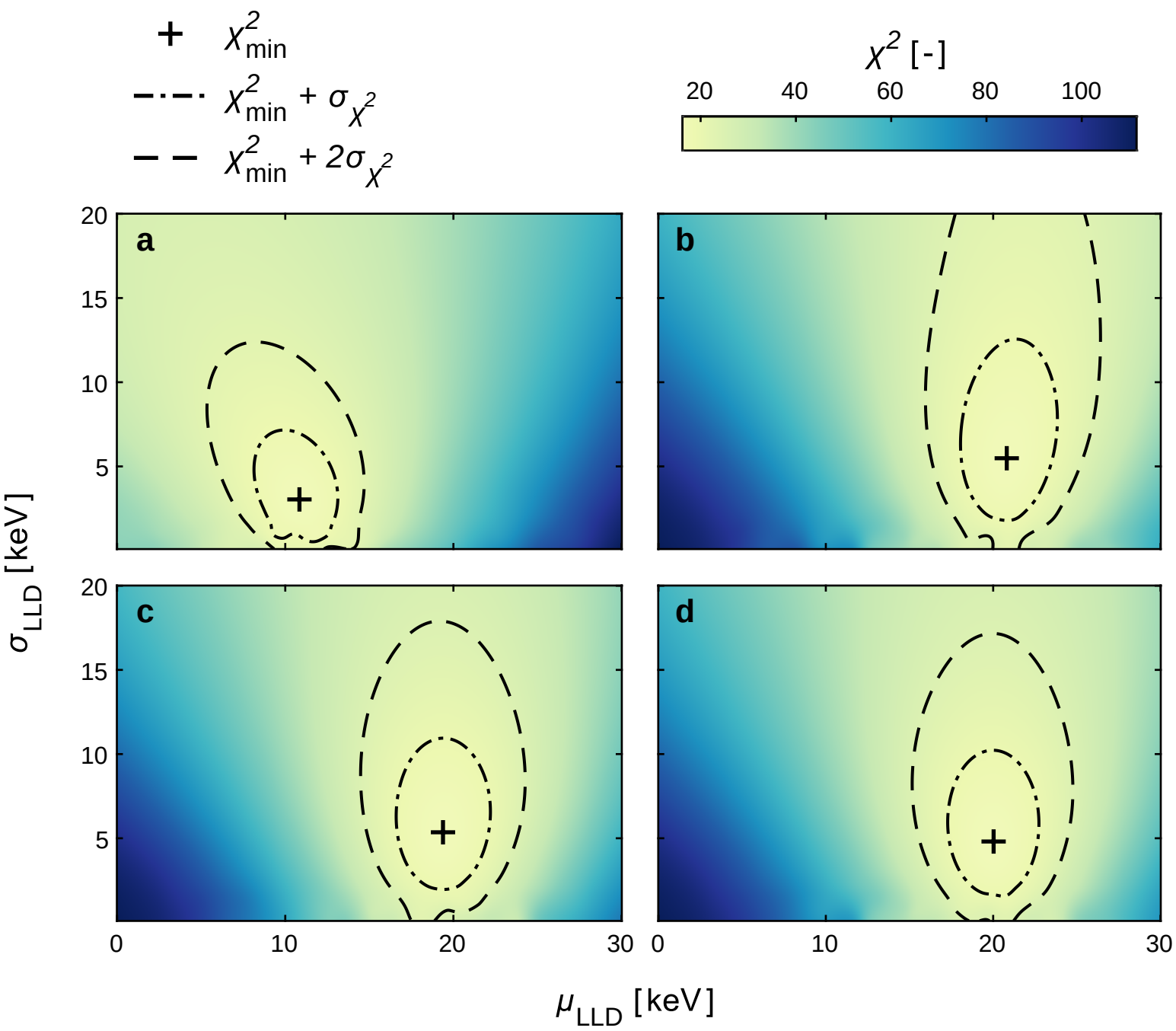

**Figure B.33** Here, the results of the LLD calibration described in Appendix A.9 are presented as chi-squared $\chi^2$ contour maps of the lower level discriminator model parameters $\mu_{LLD}$ and $\sigma_{LLD}$ for the four detector channels: **a** Detector channel `#1`. **b** Detector channel `#2`. **c** Detector channel `#3`. **d** Detector channel `#4`. The global minima $\chi^2_{min}$ are marked with a plus sign together with the corresponding 1-sigma and 2-sigma confidence regions indicated by the dashed-dotted and dashed contour lines, respectively. For the reader's convenience, the unit of the lower level discriminator model parameters $\mu_{LLD}$ and $\sigma_{LLD}$ was converted from continuous pulse-height channel numbers to spectral energies using the energy calibration models derived by the `RLLCal` pipeline (cf. Section 6.2.1.3). For the calibration, the laboratory-based radiation measurement with a $^{60}_{27}Co$ calibration point source described in Section 6.2.1 was adopted. The measured and simulated spectral signatures were derived by the `RLLSpec` and `NPScinMC` pipelines, respectively.

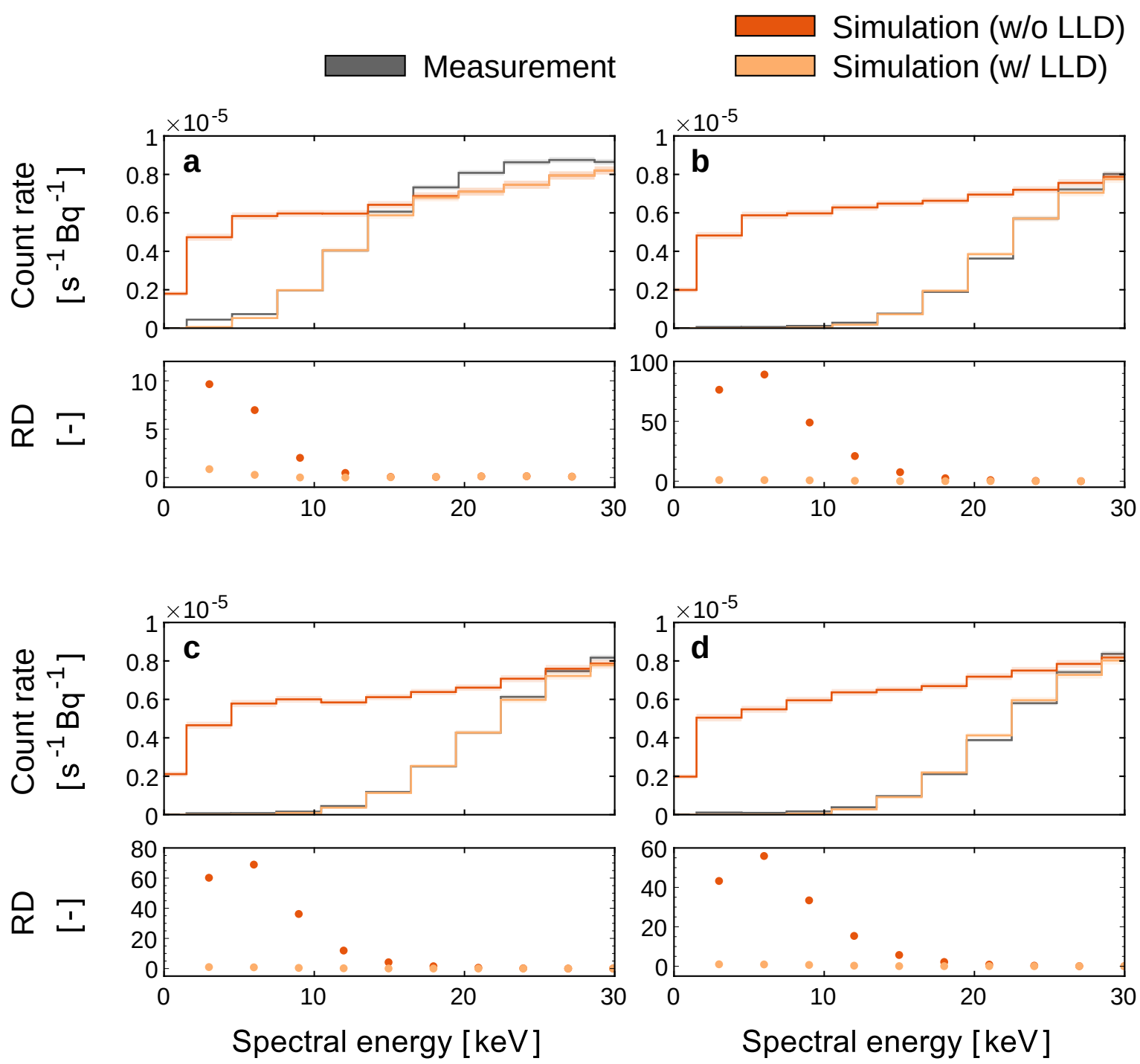

**Figure B.34** These graphs display the measured and simulated mean spectral signatures obtained by the `RLLSpec` and `PScinMC` pipeline for $^{60}_{27}$Co (cf. Appendix A.9) with (w/) and without (w/o) a calibrated lower-level discriminator (LLD) model as a function of the spectral energy $E'$ with a spectral energy bin width of $\Delta E' \sim 3\,\text{keV}$. **a** Detector channel `#1`. **b** Detector channel `#2`. **c** Detector channel `#3`. **d** Detector channel `#4`. Uncertainties are provided as 1 standard deviation (SD) shaded areas. In addition, the relative deviation (RD) computed as $|\hat{c}_{\text{sim}} - \hat{c}_{\text{exp}}|/\hat{c}_{\text{exp}}$ is provided for each detector channel.

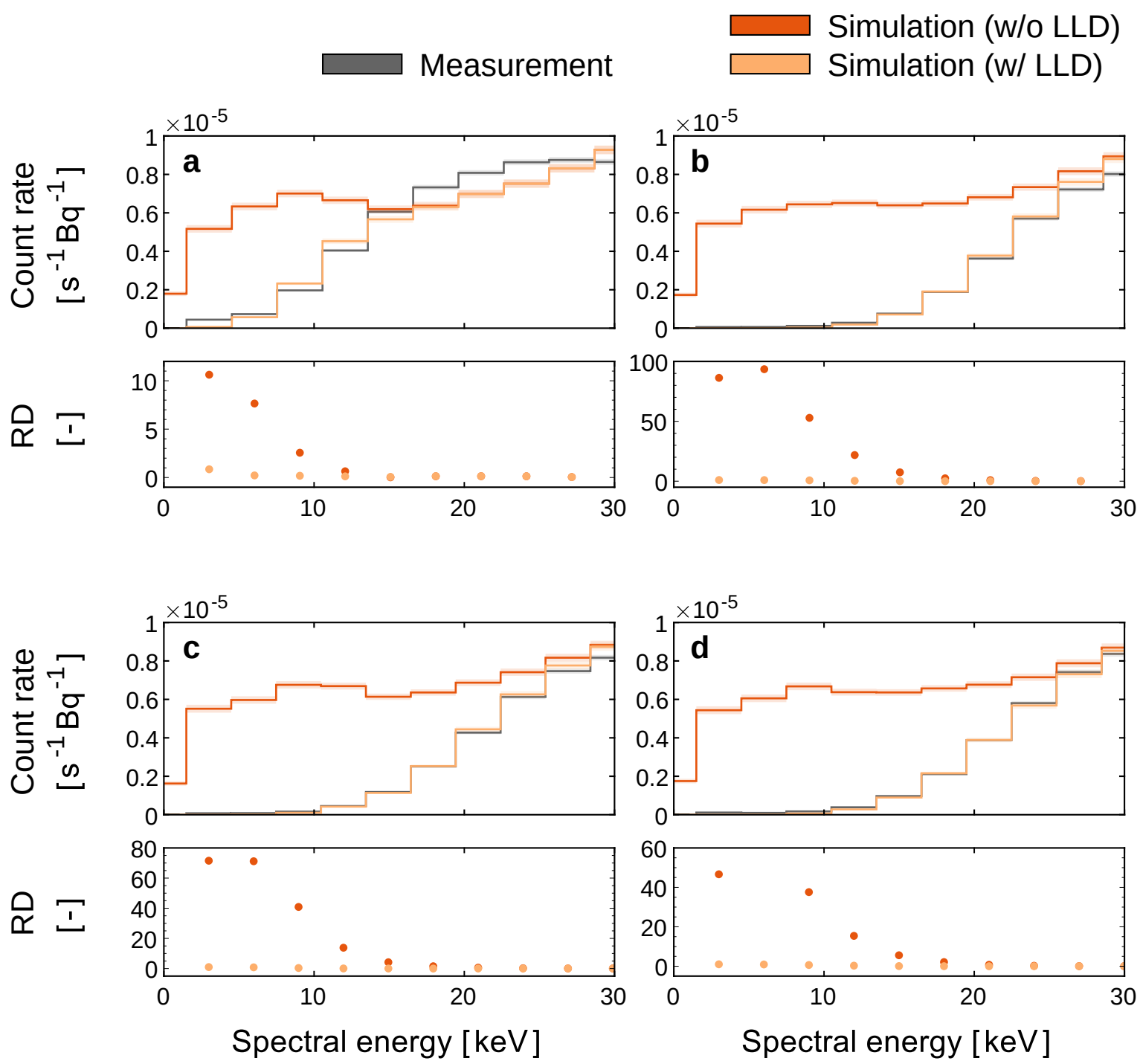

**Figure B.35** These graphs display the measured and simulated mean spectral signatures obtained by the `RLLSpec` and `NPScinMC` pipeline for $^{60}_{27}$Co (cf. Appendix A.9) with (w/) and without (w/o) a calibrated lower-level discriminator (LLD) model as a function of the spectral energy $E'$ with a spectral energy bin width of $\Delta E' \sim 3$ keV. **a** Detector channel #1. **b** Detector channel #2. **c** Detector channel #3. **d** Detector channel #4. Uncertainties are provided as 1 standard deviation (SD) shaded areas. In addition, the relative deviation (RD) computed as $|\hat{c}_{\text{sim}} - \hat{c}_{\text{exp}}|/\hat{c}_{\text{exp}}$ is provided for each detector channel.

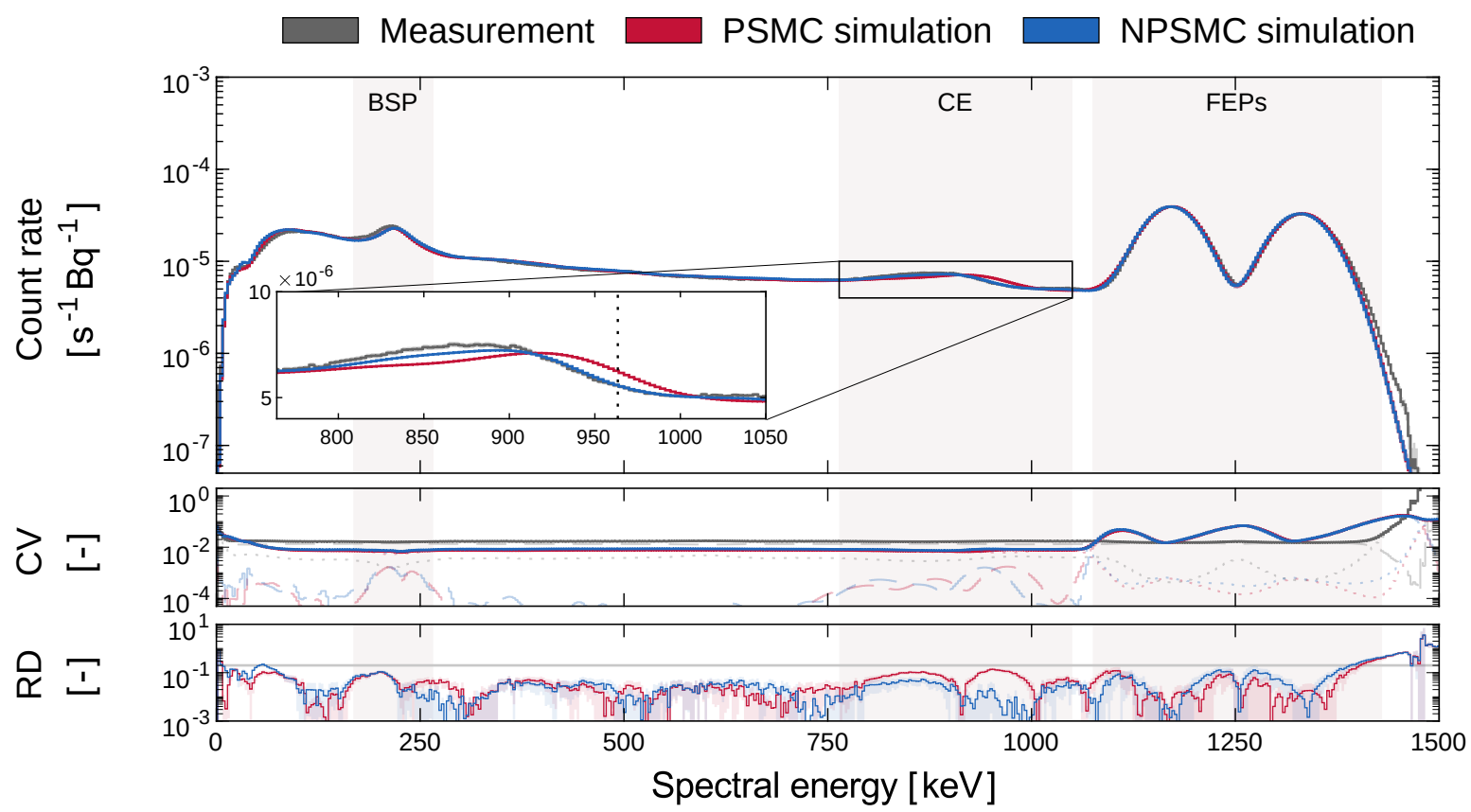

**Figure B.36** The measured ($\hat{c}_{\text{exp}}$) and simulated ($\hat{c}_{\text{sim}}$) mean spectral signatures for a $^{60}_{27}$Co source ($\mathcal{A} = 3.08(5) \times 10^5$ Bq) are shown for the detector channel #1 as a function of the spectral energy $E'$ with a spectral energy bin width of $\Delta E' \sim 3$ keV. The simulated spectral signatures were obtained by PSMC (in red, cf. Chapter 6) and NPSMC (in blue, using a NPSM derived by Compton edge probing applied to the detector channel #SUM as discussed in Chapter 7). Distinct spectral regions, i.e. the backscatter peak (BSP), the domain around the Compton edge (CE) as well as the full energy peaks (FEPs) are marked. The zoomed-in subfigure highlights the spectral region around the Compton edge marked by the vertical dotted line and associated with the photon emission line at 1173.228(3) keV [68] (cf. Eq. 4.21). Uncertainties ($\hat{\sigma}_{\text{exp}}$, $\hat{\sigma}_{\text{sim}}$) are provided as 1 standard deviation (SD) shaded areas. In addition, the coefficient of variation (CV) for the measured and simulated signatures (statistical and systematic contributions are indicated by shaded dotted and dashed lines, respectively, cf. Appendix A.8) as well as the relative deviation (RD) computed as $|\hat{c}_{\text{sim}} - \hat{c}_{\text{exp}}|/\hat{c}_{\text{exp}}$ (20 % mark highlighted by a horizontal grey line) are provided.

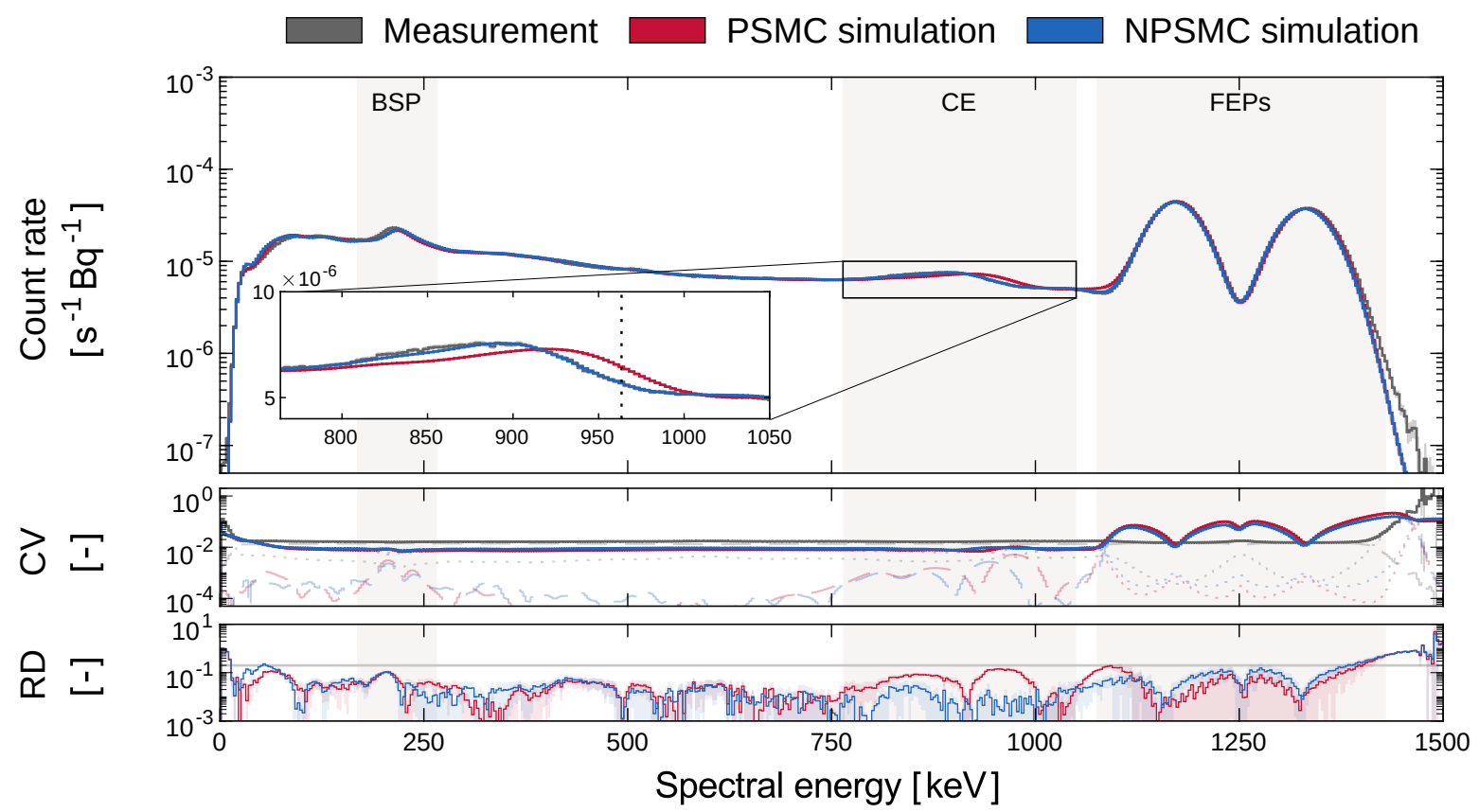

**Figure B.37** The measured ($\hat{c}_{\text{exp}}$) and simulated ($\hat{c}_{\text{sim}}$) mean spectral signatures for a $^{60}_{27}$Co source ($\mathcal{A} = 3.08(5) \times 10^5$ Bq) are shown for the detector channel `#2` as a function of the spectral energy $E'$ with a spectral energy bin width of $\Delta E' \sim 3$ keV. The simulated spectral signatures were obtained by PSMC (in red, cf. Chapter 6) and NPSMC (in blue, using a NPSM derived by Compton edge probing applied to the detector channel `#SUM` as discussed in Chapter 7). Distinct spectral regions, i.e. the backscatter peak (BSP), the domain around the Compton edge (CE) as well as the full energy peaks (FEPs) are marked. The zoomed-in subfigure highlights the spectral region around the Compton edge marked by the vertical dotted line and associated with the photon emission line at 1173.228(3) keV [68] (cf. Eq. 4.21). Uncertainties ($\hat{\sigma}_{\text{exp}}$, $\hat{\sigma}_{\text{sim}}$) are provided as 1 standard deviation (SD) shaded areas. In addition, the coefficient of variation (CV) for the measured and simulated signatures (statistical and systematic contributions are indicated by shaded dotted and dashed lines, respectively, cf. Appendix A.8) as well as the relative deviation (RD) computed as $|\hat{c}_{\text{sim}} - \hat{c}_{\text{exp}}|/\hat{c}_{\text{exp}}$ (20 % mark highlighted by a horizontal grey line) are provided.

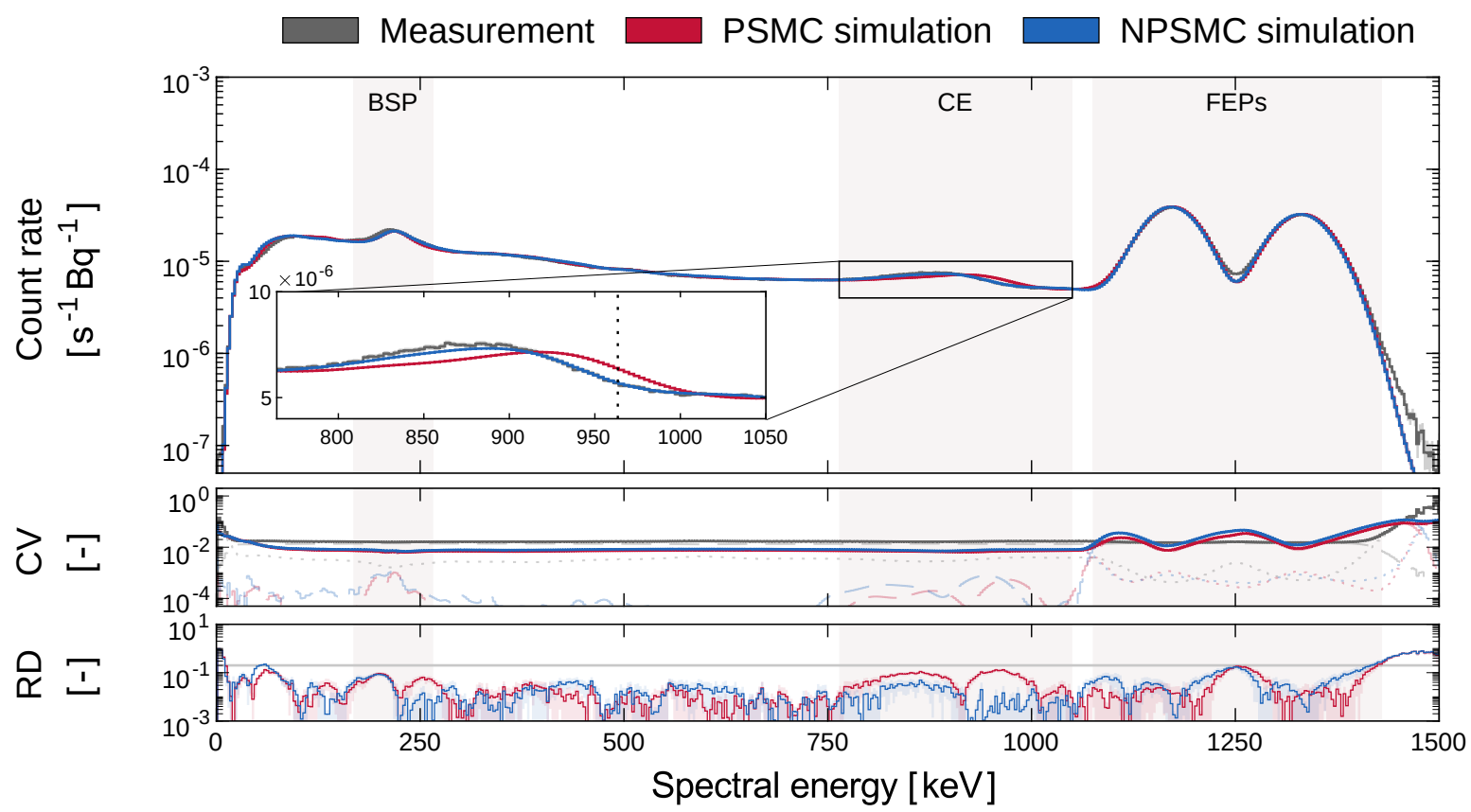

**Figure B.38** The measured ($\hat{c}_{\mathrm{exp}}$) and simulated ($\hat{c}_{\mathrm{sim}}$) mean spectral signatures for a ${}^{60}_{27}$Co source ($\mathcal{A} = 3.08(5) \times 10^5$ Bq) are shown for the detector channel #3 as a function of the spectral energy $E'$ with a spectral energy bin width of $\Delta E' \sim 3$ keV. The simulated spectral signatures were obtained by PSMC (in red, cf. Chapter 6) and NPSMC (in blue, using a NPSM derived by Compton edge probing applied to the detector channel #SUM as discussed in Chapter 7). Distinct spectral regions, i.e. the backscatter peak (BSP), the domain around the Compton edge (CE) as well as the full energy peaks (FEPs) are marked. The zoomed-in subfigure highlights the spectral region around the Compton edge marked by the vertical dotted line and associated with the photon emission line at 1173.228(3) keV [68] (cf. Eq. 4.21). Uncertainties ($\hat{\sigma}_{\mathrm{exp}}$, $\hat{\sigma}_{\mathrm{sim}}$) are provided as 1 standard deviation (SD) shaded areas. In addition, the coefficient of variation (CV) for the measured and simulated signatures (statistical and systematic contributions are indicated by shaded dotted and dashed lines, respectively, cf. Appendix A.8) as well as the relative deviation (RD) computed as $|\hat{c}_{\mathrm{sim}} - \hat{c}_{\mathrm{exp}}|/\hat{c}_{\mathrm{exp}}$ (20 % mark highlighted by a horizontal grey line) are provided.

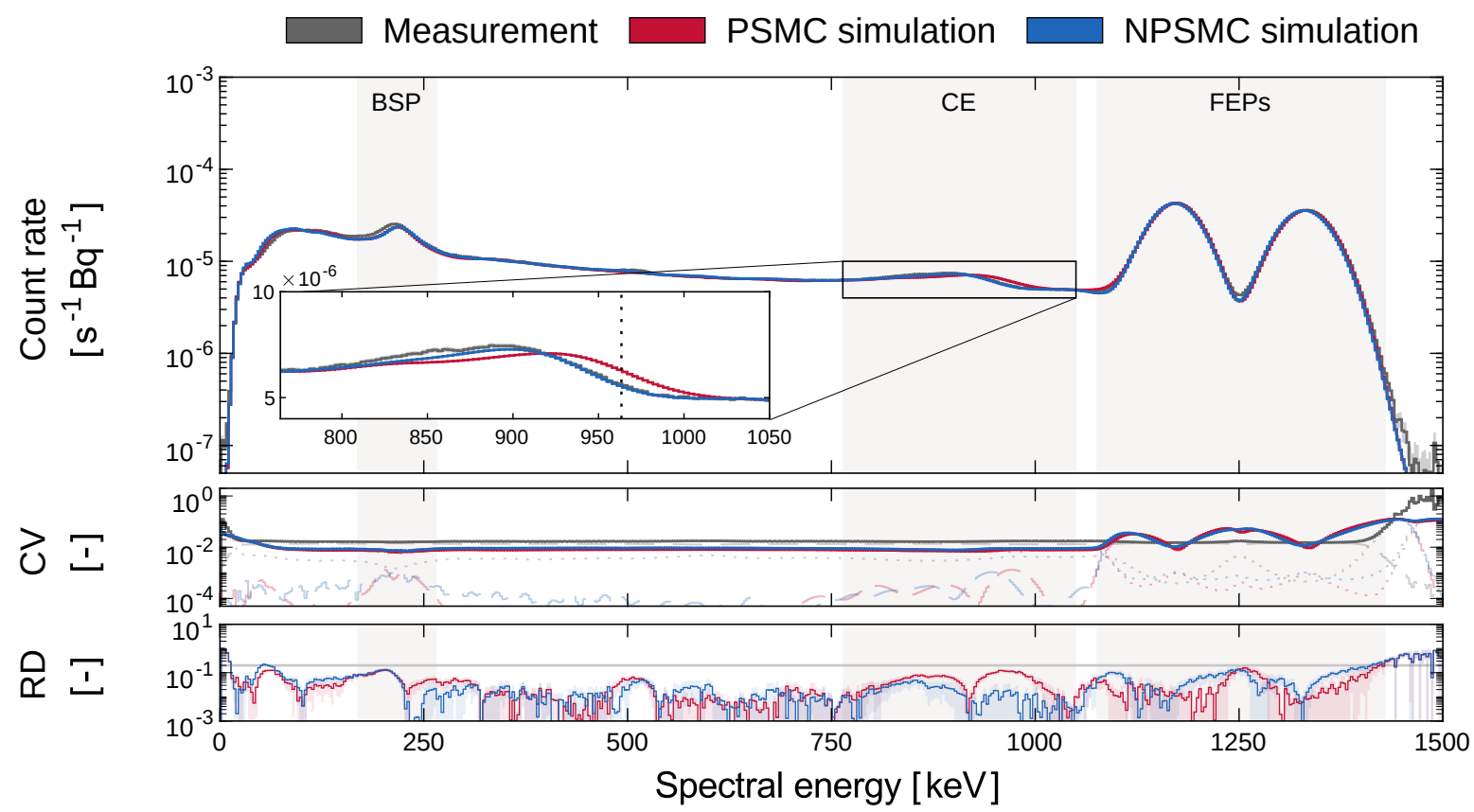

**Figure B.39** The measured ($\hat{c}_{\mathrm{exp}}$) and simulated ($\hat{c}_{\mathrm{sim}}$) mean spectral signatures for a ${}^{60}_{27}$Co source ($\mathcal{A} = 3.08(5) \times 10^5$ Bq) are shown for the detector channel `#4` as a function of the spectral energy $E'$ with a spectral energy bin width of $\Delta E' \sim 3$ keV. The simulated spectral signatures were obtained by PSMC (in red, cf. Chapter 6) and NPSMC (in blue, using a NPSM derived by Compton edge probing applied to the detector channel `#SUM` as discussed in Chapter 7). Distinct spectral regions, i.e. the backscatter peak (BSP), the domain around the Compton edge (CE) as well as the full energy peaks (FEPs) are marked. The zoomed-in subfigure highlights the spectral region around the Compton edge marked by the vertical dotted line and associated with the photon emission line at 1173.228(3) keV [68] (cf. Eq. 4.21). Uncertainties ($\hat{\sigma}_{\mathrm{exp}}$, $\hat{\sigma}_{\mathrm{sim}}$) are provided as 1 standard deviation (SD) shaded areas. In addition, the coefficient of variation (CV) for the measured and simulated signatures (statistical and systematic contributions are indicated by shaded dotted and dashed lines, respectively, cf. Appendix A.8) as well as the relative deviation (RD) computed as $|\hat{c}_{\mathrm{sim}} - \hat{c}_{\mathrm{exp}}|/\hat{c}_{\mathrm{exp}}$ (20 % mark highlighted by a horizontal grey line) are provided.

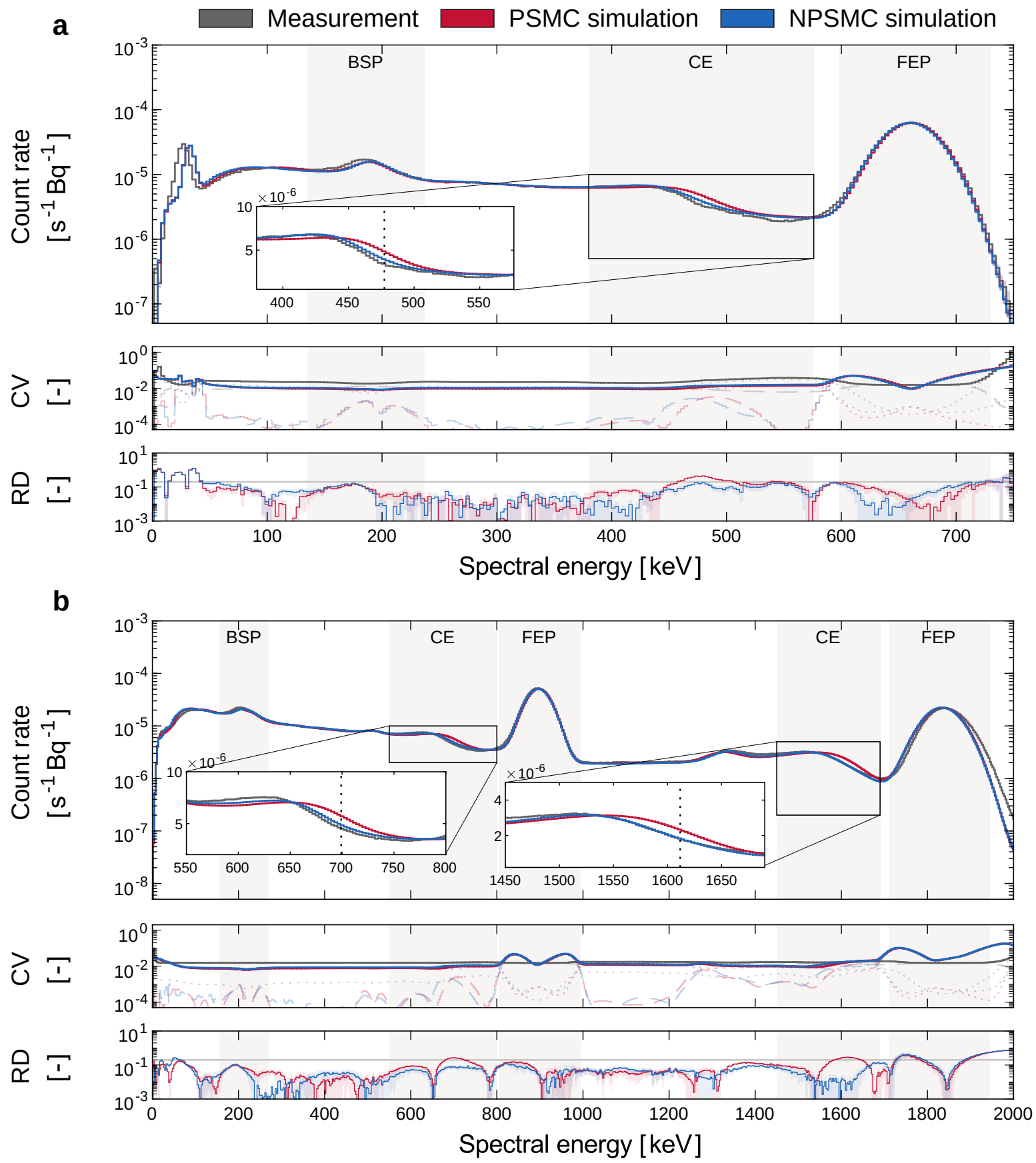

**Figure B.40** The measured ($\hat{c}_{\text{exp}}$) and simulated ($\hat{c}_{\text{sim}}$) mean spectral signatures for two radionuclides sources and the detector channel `#1` as a function of the spectral energy $E'$ with a spectral energy bin width of $\Delta E' \sim 3\,\text{keV}$ are displayed: **a** $^{137}_{55}$Cs ($\mathcal{A} = 2.266(34) \times 10^5\,\text{Bq}$). **b** $^{88}_{39}$Y ($\mathcal{A} = 6.83(14) \times 10^5\,\text{Bq}$). The simulated spectral signatures were obtained by PSMC (in red, cf. Chapter 6) and NPSMC (in blue, using a NPSM derived by Compton edge probing applied to the detector channel `#SUM` as discussed in Chapter 7). Distinct spectral regions, i.e. the backscatter peak (BSP), the domain around the Compton edge (CE) as well as the full energy peak (FEP) are marked. The zoomed-in subfigures highlight the spectral regions around the Compton edge marked by the vertical dotted line and associated with the photon emission lines at 661.657(3) keV for $^{137}_{55}$Cs [68] as well as 898.042(11) keV and 1836.070(8) keV for $^{88}_{39}$Y [51]. Uncertainties ($\hat{\sigma}_{\text{exp}}$, $\hat{\sigma}_{\text{sim}}$) are provided as 1 standard deviation (SD) shaded areas. In addition, the coefficient of variation (CV) for the measured and simulated signatures (statistical and systematic contributions are indicated by shaded dotted and dashed lines, respectively, cf. Appendix A.8) as well as the relative deviation (RD) computed as $|\hat{c}_{\text{sim}} - \hat{c}_{\text{exp}}|/\hat{c}_{\text{exp}}$ (20 % mark highlighted by a horizontal grey line) are provided.

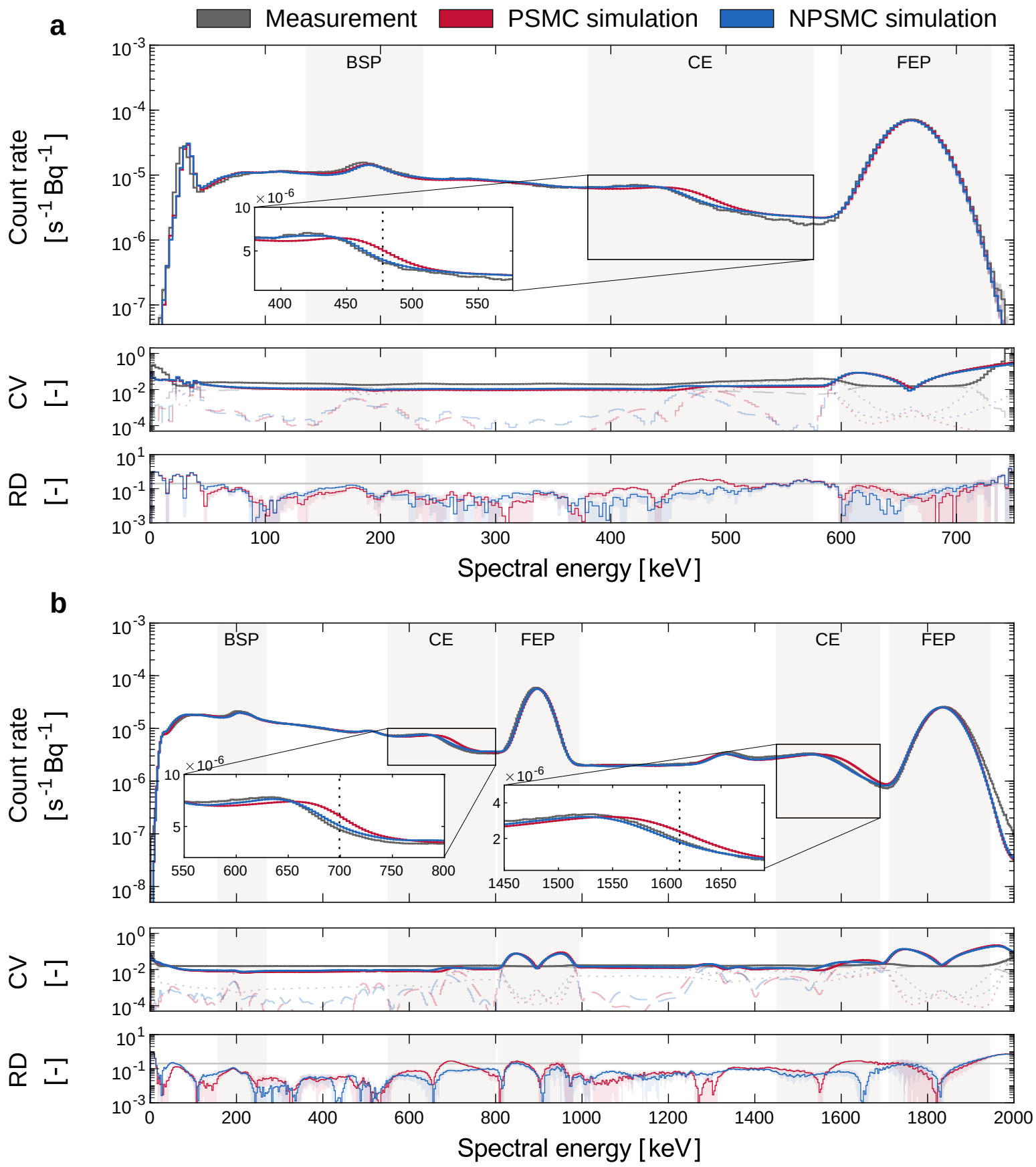

**Figure B.41** The measured ($\hat{c}_{\text{exp}}$) and simulated ($\hat{c}_{\text{sim}}$) mean spectral signatures for two radionuclides sources and the detector channel #2 as a function of the spectral energy $E'$ with a spectral energy bin width of $\Delta E' \sim 3$ keV are displayed: **a** $^{137}_{55}$Cs ($\mathcal{A} = 2.266(34) \times 10^5$ Bq). **b** $^{88}_{39}$Y ($\mathcal{A} = 6.83(14) \times 10^5$ Bq). The simulated spectral signatures were obtained by PSMC (in red, cf. Chapter 6) and NPSMC (in blue, using a NPSM derived by Compton edge probing applied to the detector channel #SUM as discussed in Chapter 7). Distinct spectral regions, i.e. the backscatter peak (BSP), the domain around the Compton edge (CE) as well as the full energy peak (FEP) are marked. The zoomed-in subfigures highlight the spectral regions around the Compton edge marked by the vertical dotted line and associated with the photon emission lines at 661.657(3) keV for $^{137}_{55}$Cs [68] as well as 898.042(11) keV and 1836.070(8) keV for $^{88}_{39}$Y [51]. Uncertainties ($\hat{\sigma}_{\text{exp}}$, $\hat{\sigma}_{\text{sim}}$) are provided as 1 standard deviation (SD) shaded areas. In addition, the coefficient of variation (CV) for the measured and simulated signatures (statistical and systematic contributions are indicated by shaded dotted and dashed lines, respectively, cf. Appendix A.8) as well as the relative deviation (RD) computed as $|\hat{c}_{\text{sim}} - \hat{c}_{\text{exp}}|/\hat{c}_{\text{exp}}$ (20 % mark highlighted by a horizontal grey line) are provided.

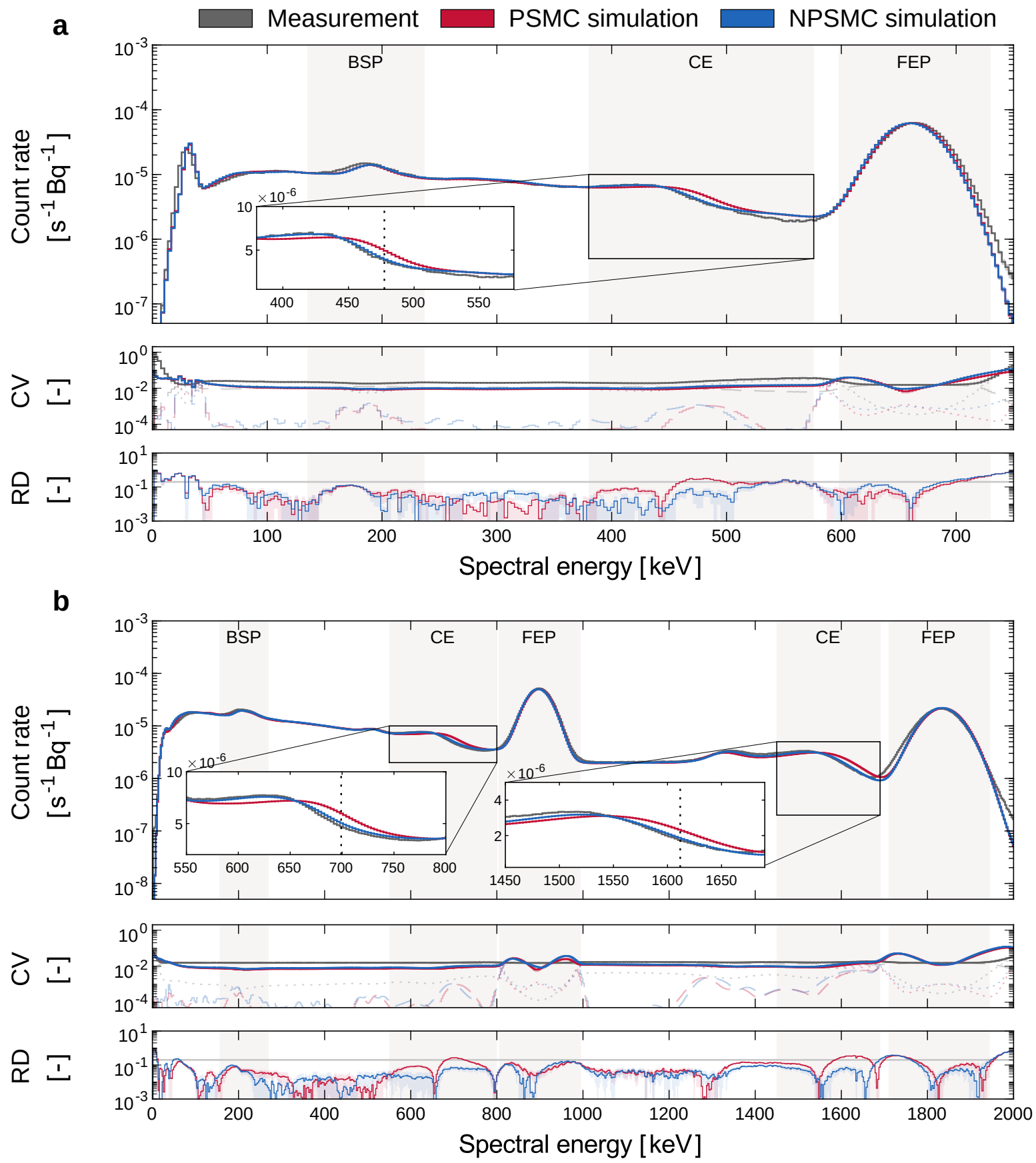

**Figure B.42** The measured ($\hat{c}_{\text{exp}}$) and simulated ($\hat{c}_{\text{sim}}$) mean spectral signatures for two radionuclides sources and the detector channel #3 as a function of the spectral energy $E'$ with a spectral energy bin width of $\Delta E' \sim 3\,\text{keV}$ are displayed: **a** $^{137}_{55}\text{Cs}$ ($\mathcal{A} = 2.266(34) \times 10^5\,\text{Bq}$). **b** $^{88}_{39}\text{Y}$ ($\mathcal{A} = 6.83(14) \times 10^5\,\text{Bq}$). The simulated spectral signatures were obtained by PSMC (in red, cf. Chapter 6) and NPSMC (in blue, using a NPSM derived by Compton edge probing applied to the detector channel #SUM as discussed in Chapter 7). Distinct spectral regions, i.e. the backscatter peak (BSP), the domain around the Compton edge (CE) as well as the full energy peak (FEP) are marked. The zoomed-in subfigures highlight the spectral regions around the Compton edge marked by the vertical dotted line and associated with the photon emission lines at 661.657(3) keV for $^{137}_{55}\text{Cs}$ [68] as well as 898.042(11) keV and 1836.070(8) keV for $^{88}_{39}\text{Y}$ [51]. Uncertainties ($\hat{\sigma}_{\text{exp}}$, $\hat{\sigma}_{\text{sim}}$) are provided as 1 standard deviation (SD) shaded areas. In addition, the coefficient of variation (CV) for the measured and simulated signatures (statistical and systematic contributions are indicated by shaded dotted and dashed lines, respectively, cf. Appendix A.8) as well as the relative deviation (RD) computed as $|\hat{c}_{\text{sim}} - \hat{c}_{\text{exp}}|/\hat{c}_{\text{exp}}$ (20 % mark highlighted by a horizontal grey line) are provided.

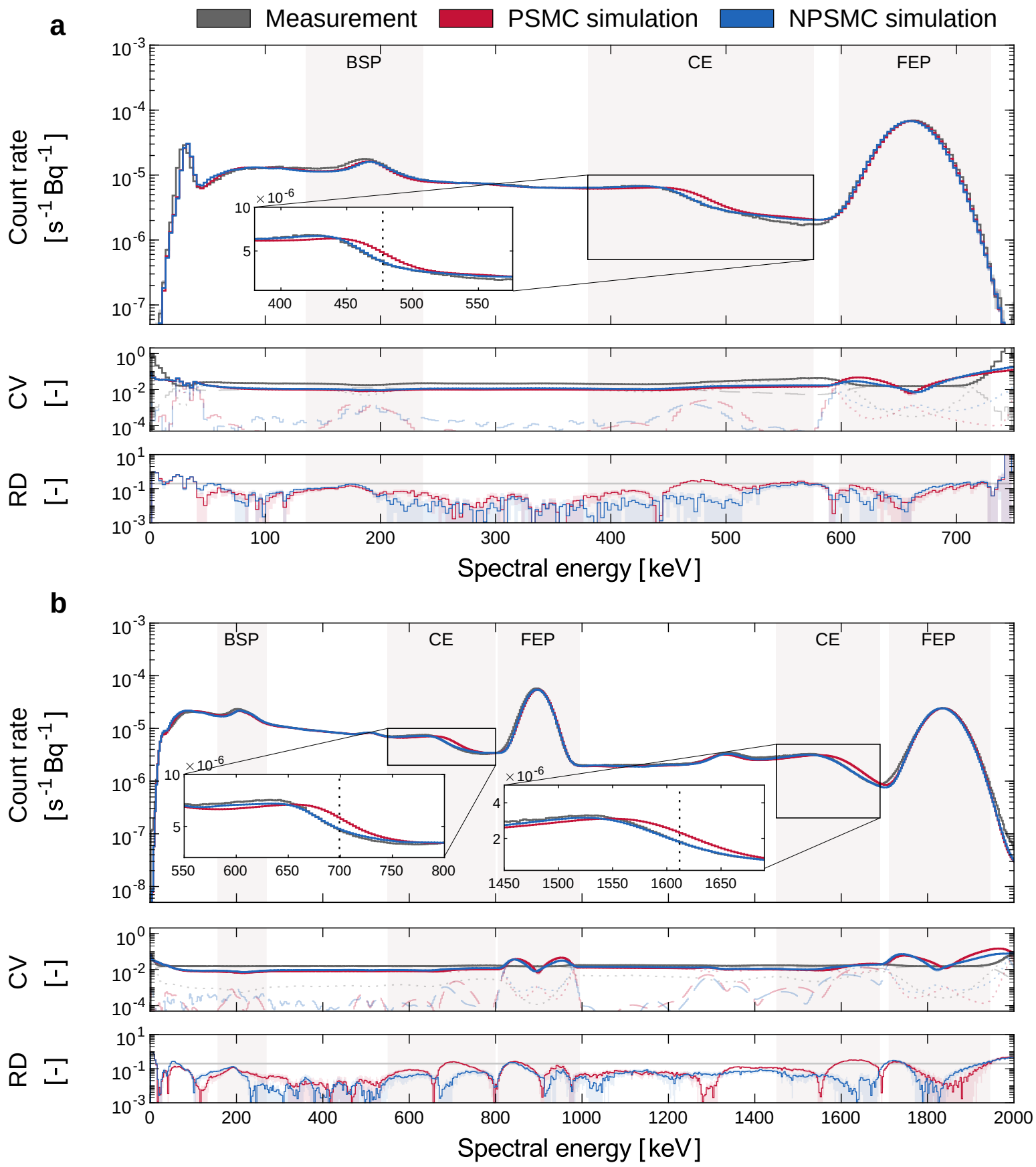

**Figure B.43** The measured ($\hat{c}_{\text{exp}}$) and simulated ($\hat{c}_{\text{sim}}$) mean spectral signatures for two radionuclides sources and the detector channel #4 as a function of the spectral energy $E'$ with a spectral energy bin width of $\Delta E' \sim 3\,\text{keV}$ are displayed: **a** $^{137}_{55}$Cs ($\mathcal{A} = 2.266(34) \times 10^5\,\text{Bq}$). **b** $^{88}_{39}$Y ($\mathcal{A} = 6.83(14) \times 10^5\,\text{Bq}$). The simulated spectral signatures were obtained by PSMC (in red, cf. Chapter 6) and NPSMC (in blue, using a NPSM derived by Compton edge probing applied to the detector channel #SUM as discussed in Chapter 7). Distinct spectral regions, i.e. the backscatter peak (BSP), the domain around the Compton edge (CE) as well as the full energy peak (FEP) are marked. The zoomed-in subfigures highlight the spectral regions around the Compton edge marked by the vertical dotted line and associated with the photon emission lines at 661.657(3) keV for $^{137}_{55}$Cs [68] as well as 898.042(11) keV and 1836.070(8) keV for $^{88}_{39}$Y [51]. Uncertainties ($\hat{\sigma}_{\text{exp}}$, $\hat{\sigma}_{\text{sim}}$) are provided as 1 standard deviation (SD) shaded areas. In addition, the coefficient of variation (CV) for the measured and simulated signatures (statistical and systematic contributions are indicated by shaded dotted and dashed lines, respectively, cf. Appendix A.8) as well as the relative deviation (RD) computed as $|\hat{c}_{\text{sim}} - \hat{c}_{\text{exp}}|/\hat{c}_{\text{exp}}$ (20 % mark highlighted by a horizontal grey line) are provided.

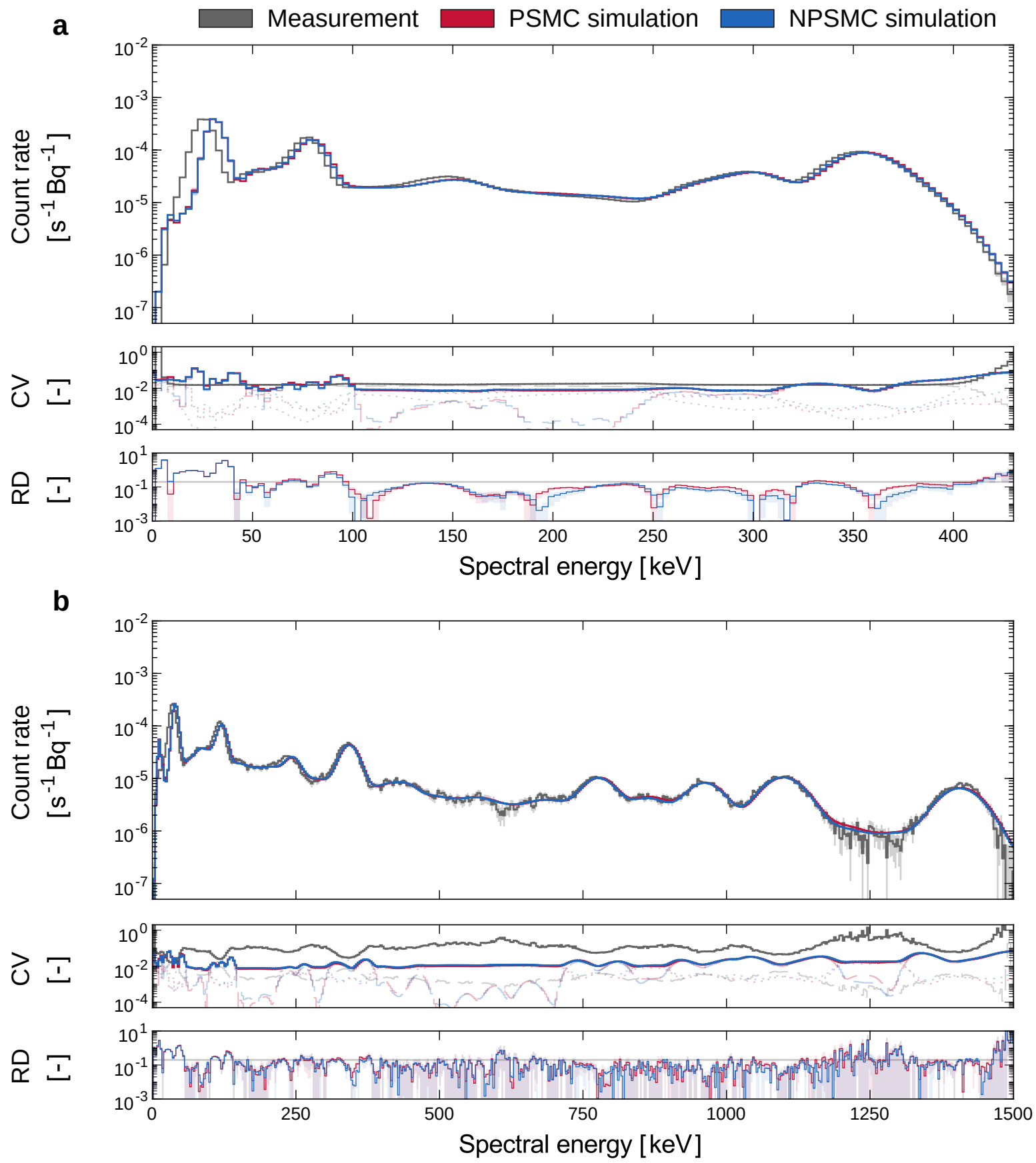

**Figure B.44** The measured ($\hat{c}_{\text{exp}}$) and simulated ($\hat{c}_{\text{sim}}$) mean spectral signatures for two radionuclides sources and the detector channel `#1` as a function of the spectral energy $E'$ with a spectral energy bin width of $\Delta E' \sim 3\,\text{keV}$ are displayed: **a** $^{133}_{56}\text{Ba}$ ($\mathcal{A} = 2.152(32) \times 10^5\,\text{Bq}$). **b** $^{152}_{63}\text{Eu}$ ($\mathcal{A} = 1.973(30) \times 10^4\,\text{Bq}$). The simulated spectral signatures were obtained by PSMC (in red, cf. Chapter 6) and NPSMC (in blue, using a NPSM derived by Compton edge probing applied to the detector channel `#SUM` as discussed in Chapter 7). Uncertainties ($\hat{\sigma}_{\text{exp}}$, $\hat{\sigma}_{\text{sim}}$) are provided as 1 standard deviation (SD) shaded areas. In addition, the coefficient of variation (CV) for the measured and simulated signatures (statistical and systematic contributions are indicated by shaded dotted and dashed lines, respectively, cf. Appendix A.8) as well as the relative deviation (RD) computed as $|\hat{c}_{\text{sim}} - \hat{c}_{\text{exp}}|/\hat{c}_{\text{exp}}$ (20 % mark highlighted by a horizontal grey line) are provided.

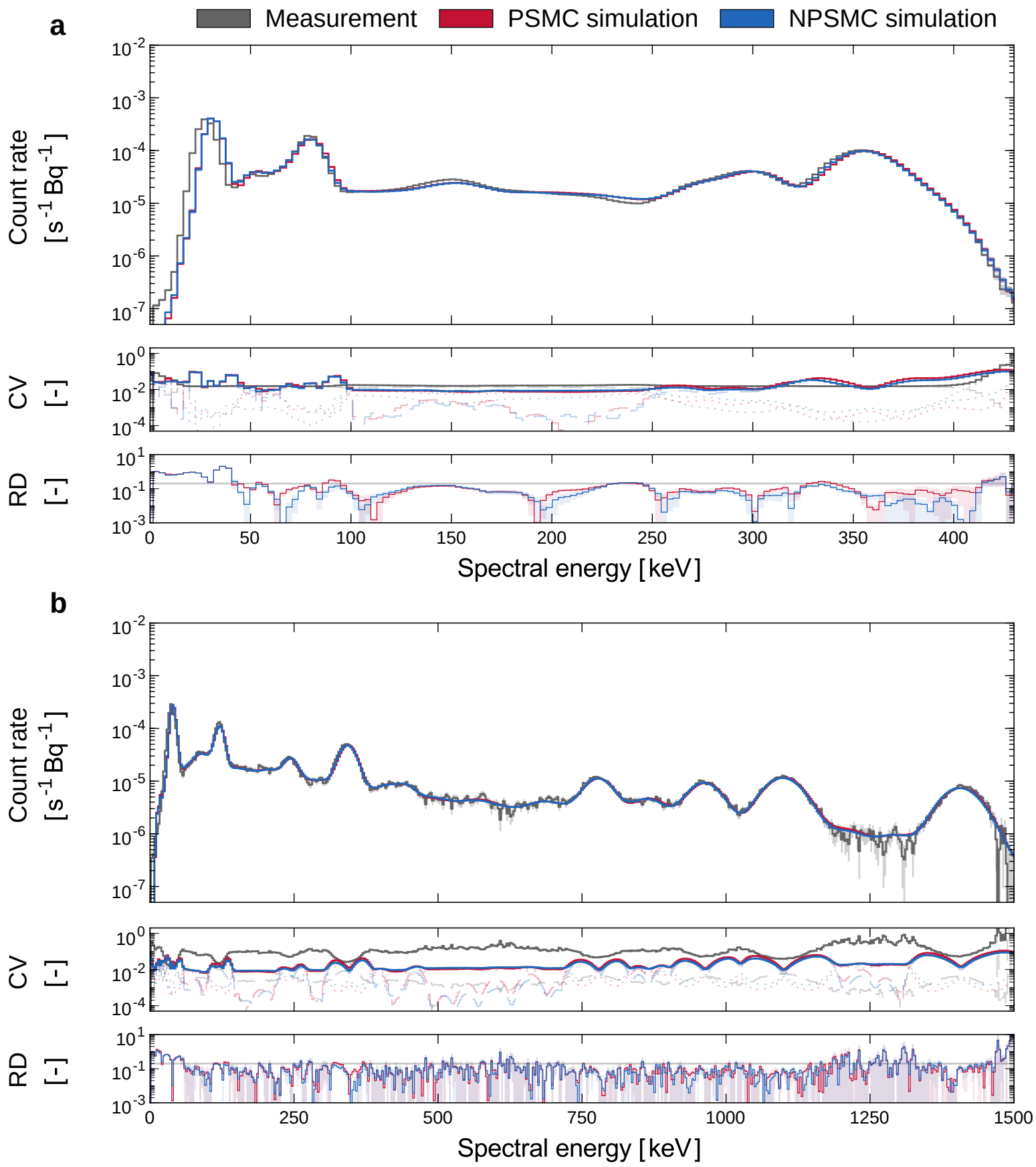

**Figure B.45** The measured ($\hat{c}_{\mathrm{exp}}$) and simulated ($\hat{c}_{\mathrm{sim}}$) mean spectral signatures for two radionuclides sources and the detector channel #2 as a function of the spectral energy $E'$ with a spectral energy bin width of $\Delta E' \sim 3\,\mathrm{keV}$ are displayed: **a** $^{133}_{56}\mathrm{Ba}$ ($\mathcal{A} = 2.152(32) \times 10^5\,\mathrm{Bq}$). **b** $^{152}_{63}\mathrm{Eu}$ ($\mathcal{A} = 1.973(30) \times 10^4\,\mathrm{Bq}$). The simulated spectral signatures were obtained by PSMC (in red, cf. Chapter 6) and NPSMC (in blue, using a NPSM derived by Compton edge probing applied to the detector channel `#SUM` as discussed in Chapter 7). Uncertainties ($\hat{\sigma}_{\mathrm{exp}}, \hat{\sigma}_{\mathrm{sim}}$) are provided as 1 standard deviation (SD) shaded areas. In addition, the coefficient of variation (CV) for the measured and simulated signatures (statistical and systematic contributions are indicated by shaded dotted and dashed lines, respectively, cf. Appendix A.8) as well as the relative deviation (RD) computed as $|\hat{c}_{\mathrm{sim}} - \hat{c}_{\mathrm{exp}}|/\hat{c}_{\mathrm{exp}}$ (20 % mark highlighted by a horizontal grey line) are provided.

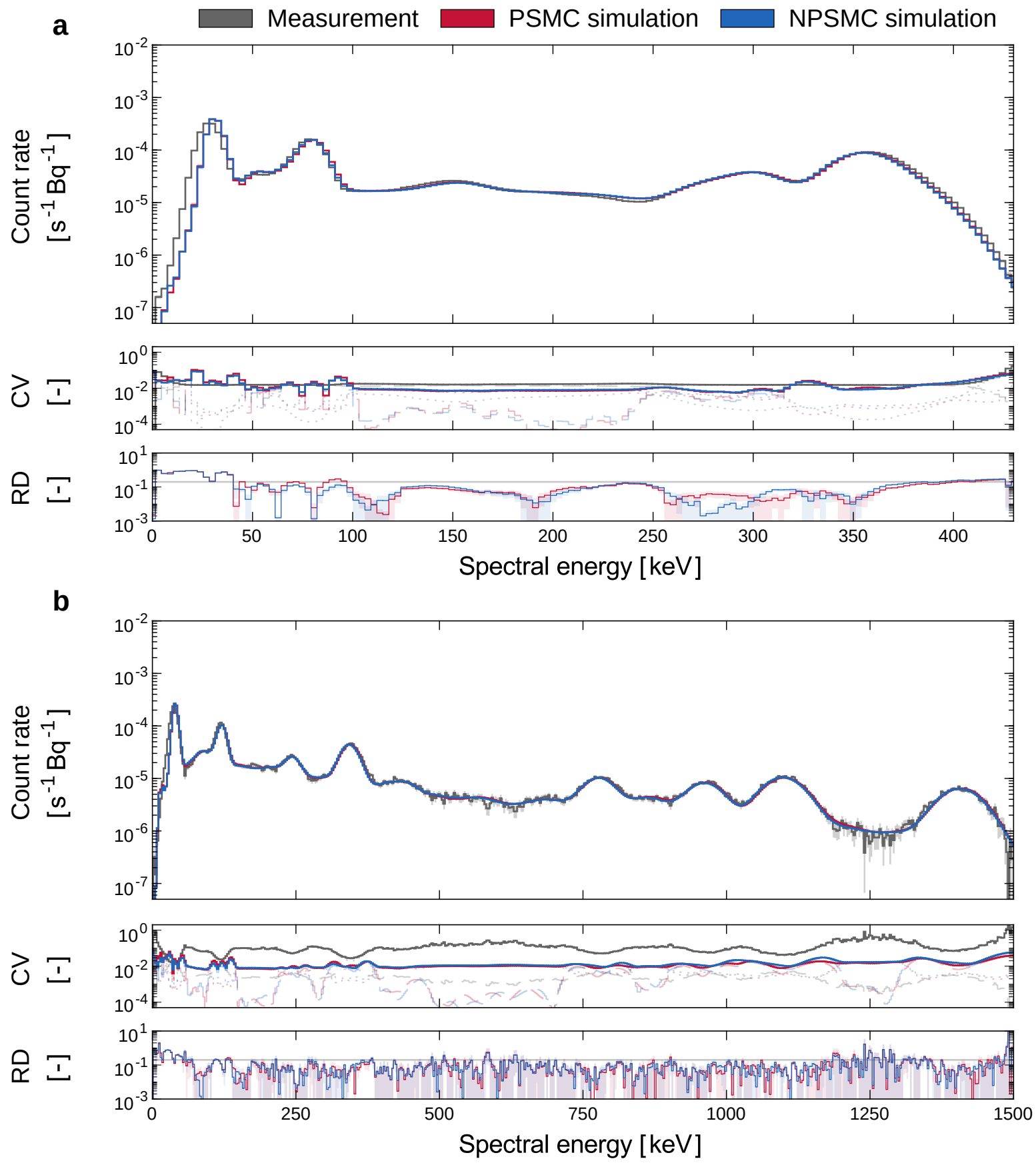

**Figure B.46** The measured ($\hat{c}_{\text{exp}}$) and simulated ($\hat{c}_{\text{sim}}$) mean spectral signatures for two radionuclides sources and the detector channel `#3` as a function of the spectral energy $E'$ with a spectral energy bin width of $\Delta E' \sim 3\,\text{keV}$ are displayed: **a** $^{133}_{56}$Ba ($\mathcal{A} = 2.152(32) \times 10^5\,\text{Bq}$). **b** $^{152}_{63}$Eu ($\mathcal{A} = 1.973(30) \times 10^4\,\text{Bq}$). The simulated spectral signatures were obtained by PSMC (in red, cf. Chapter 6) and NPSMC (in blue, using a NPSM derived by Compton edge probing applied to the detector channel `#SUM` as discussed in Chapter 7). Uncertainties ($\hat{\sigma}_{\text{exp}}, \hat{\sigma}_{\text{sim}}$) are provided as 1 standard deviation (SD) shaded areas. In addition, the coefficient of variation (CV) for the measured and simulated signatures (statistical and systematic contributions are indicated by shaded dotted and dashed lines, respectively, cf. Appendix A.8) as well as the relative deviation (RD) computed as $|\hat{c}_{\text{sim}} - \hat{c}_{\text{exp}}|/\hat{c}_{\text{exp}}$ (20 % mark highlighted by a horizontal grey line) are provided.

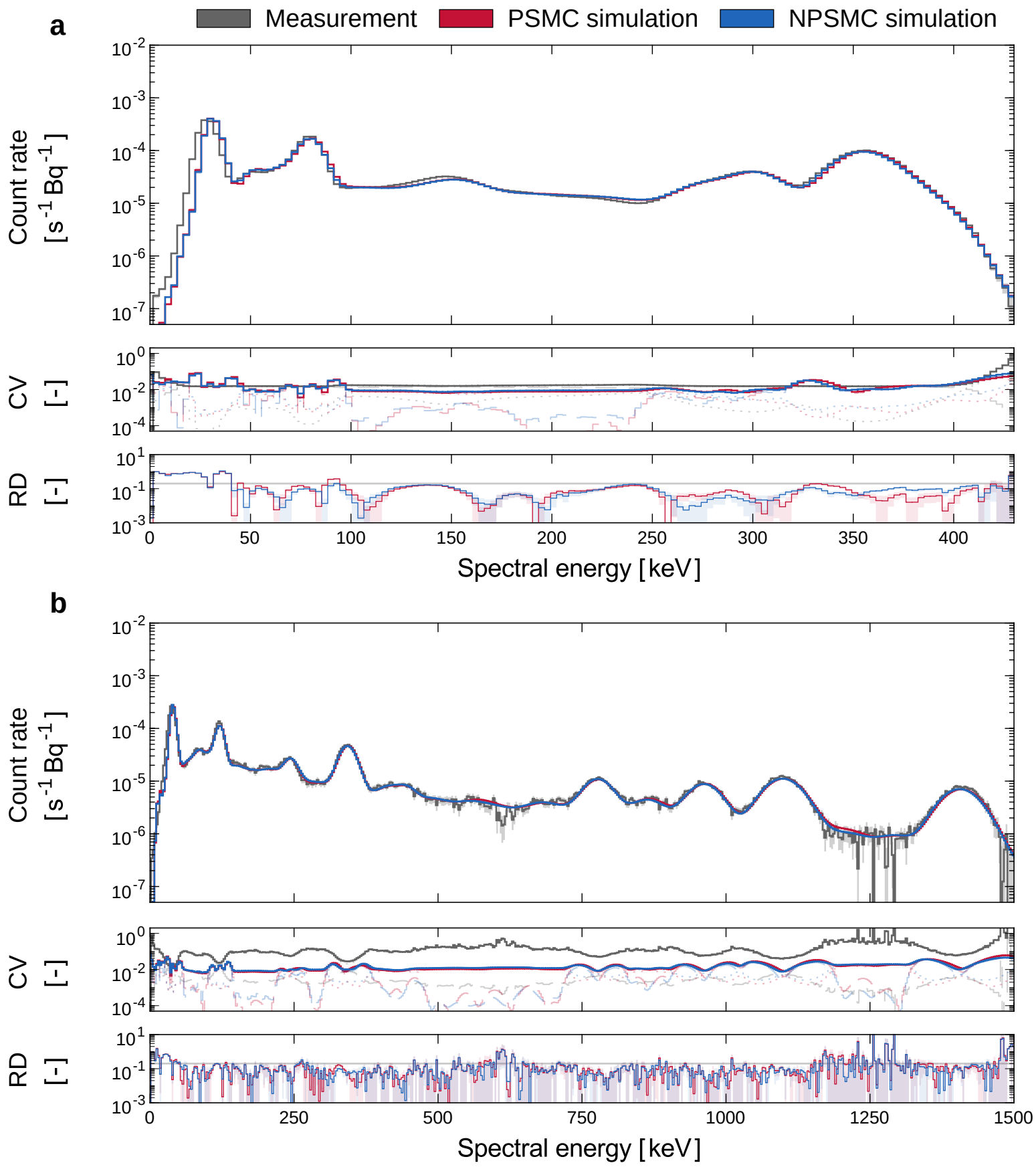

**Figure B.47** The measured ($\hat{c}_{\text{exp}}$) and simulated ($\hat{c}_{\text{sim}}$) mean spectral signatures for two radionuclides sources and the detector channel #4 as a function of the spectral energy $E'$ with a spectral energy bin width of $\Delta E' \sim 3\,\text{keV}$ are displayed: **a** $^{133}_{56}$Ba ($\mathcal{A} = 2.152(32) \times 10^5$ Bq). **b** $^{152}_{63}$Eu ($\mathcal{A} = 1.973(30) \times 10^4$ Bq). The simulated spectral signatures were obtained by PSMC (in red, cf. Chapter 6) and NPSMC (in blue, using a NPSM derived by Compton edge probing applied to the detector channel #SUM as discussed in Chapter 7). Uncertainties ($\hat{\sigma}_{\text{exp}}$, $\hat{\sigma}_{\text{sim}}$) are provided as 1 standard deviation (SD) shaded areas. In addition, the coefficient of variation (CV) for the measured and simulated signatures (statistical and systematic contributions are indicated by shaded dotted and dashed lines, respectively, cf. Appendix A.8) as well as the relative deviation (RD) computed as $|\hat{c}_{\text{sim}} - \hat{c}_{\text{exp}}|/\hat{c}_{\text{exp}}$ (20 % mark highlighted by a horizontal grey line) are provided.

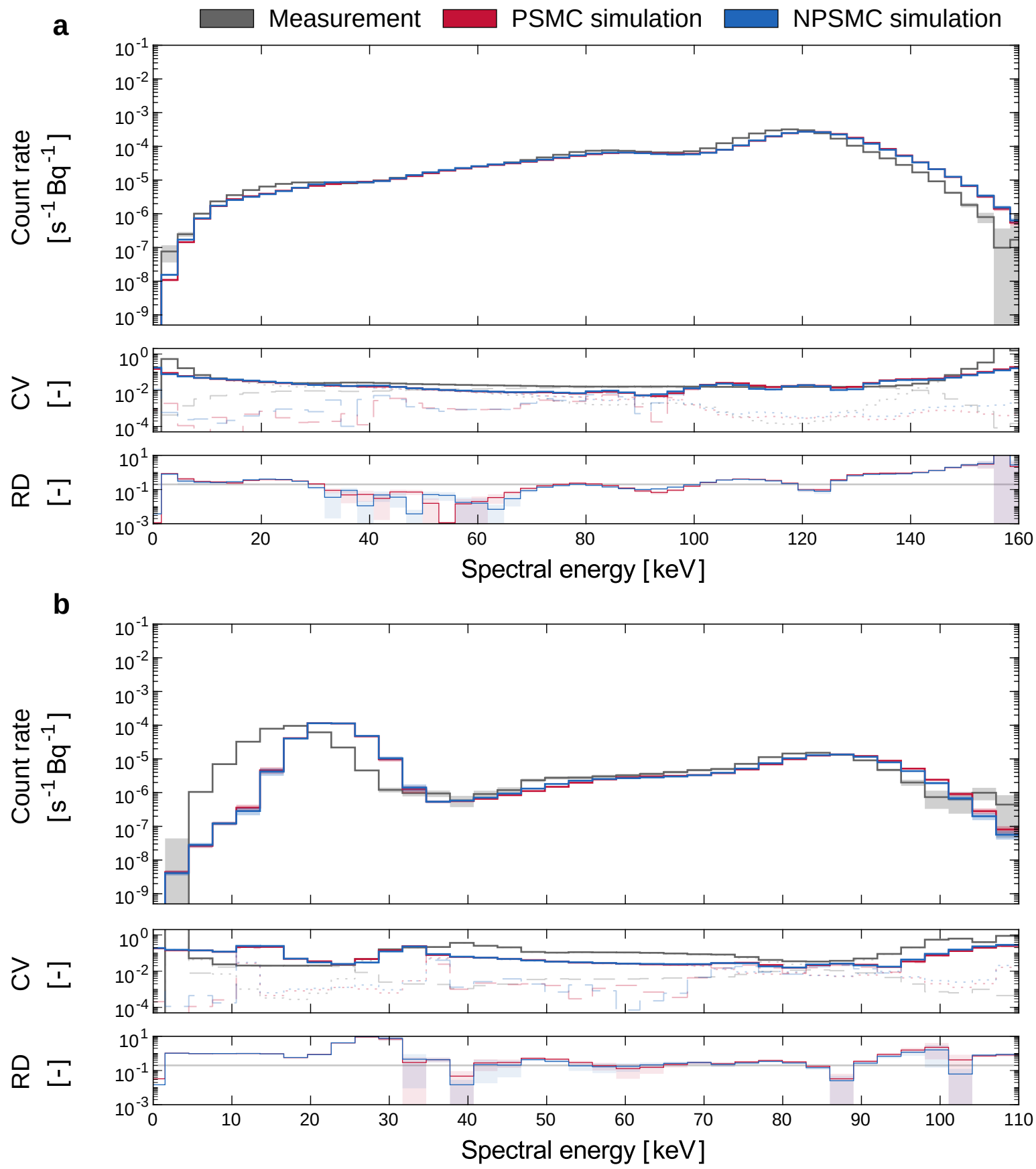

**Figure B.48** The measured ($\hat{c}_{exp}$) and simulated ($\hat{c}_{sim}$) mean spectral signatures for two radionuclides sources and the detector channel `#1` as a function of the spectral energy $E'$ with a spectral energy bin width of $\Delta E' \sim 3$ keV are displayed: **a** $^{57}_{27}$Co ($\mathcal{A} = 1.113(18) \times 10^5$ Bq). **b** $^{109}_{48}$Cd ($\mathcal{A} = 7.38(15) \times 10^4$ Bq). The simulated spectral signatures were obtained by PSMC (in red, cf. Chapter 6) and NPSMC (in blue, using a NPSM derived by Compton edge probing applied to the detector channel `#SUM` as discussed in Chapter 7). Uncertainties ($\hat{\sigma}_{exp}$, $\hat{\sigma}_{sim}$) are provided as 1 standard deviation (SD) shaded areas. In addition, the coefficient of variation (CV) for the measured and simulated signatures (statistical and systematic contributions are indicated by shaded dotted and dashed lines, respectively, cf. Appendix A.8) as well as the relative deviation (RD) computed as $|\hat{c}_{sim} - \hat{c}_{exp}|/\hat{c}_{exp}$ (20 % mark highlighted by a horizontal grey line) are provided.

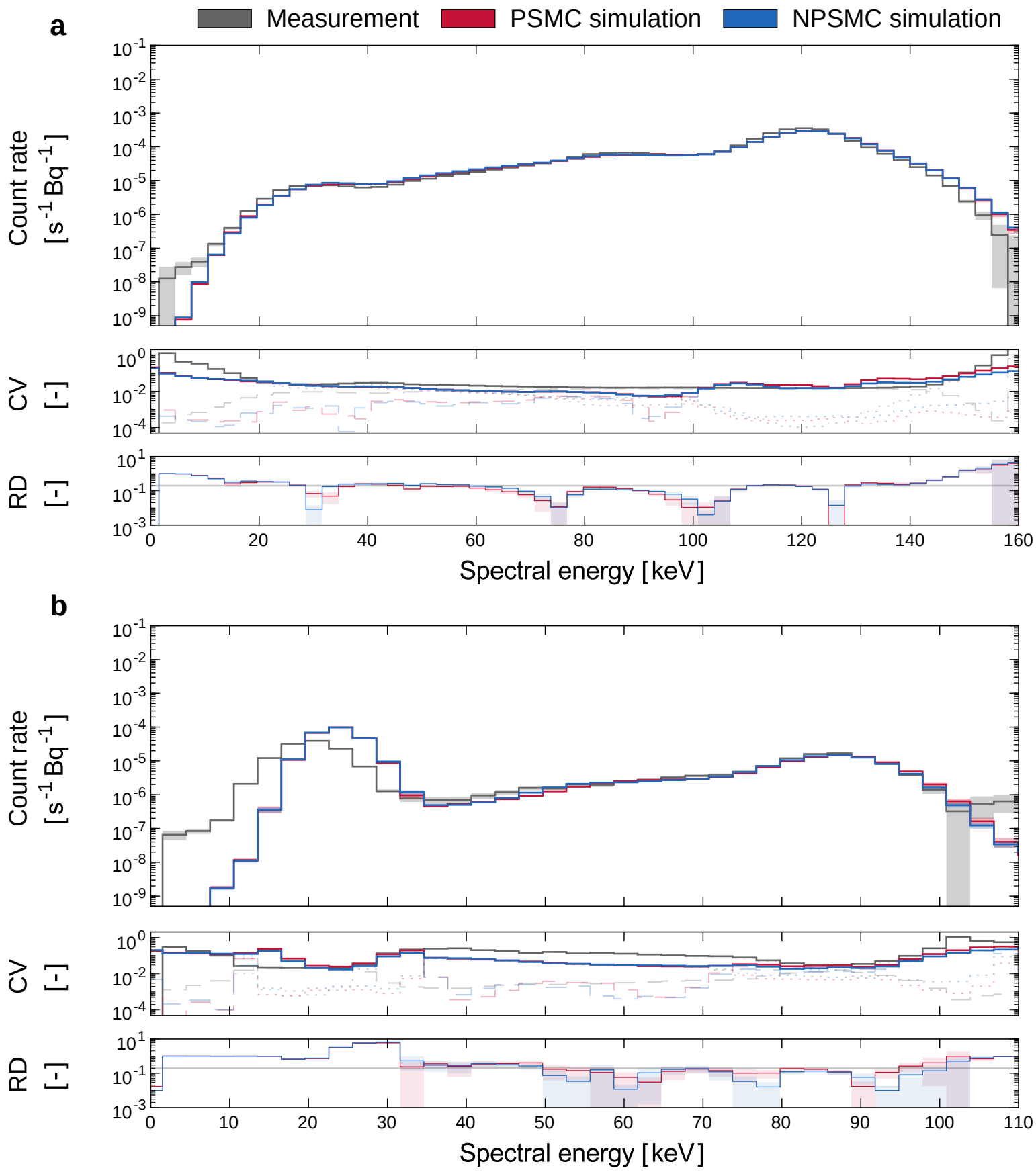

**Figure B.49** The measured ($\hat{c}_{\text{exp}}$) and simulated ($\hat{c}_{\text{sim}}$) mean spectral signatures for two radionuclides sources and the detector channel #2 as a function of the spectral energy $E'$ with a spectral energy bin width of $\Delta E' \sim 3\,\text{keV}$ are displayed: **a** $^{57}_{27}\text{Co}$ ($\mathcal{A} = 1.113(18) \times 10^5\,\text{Bq}$). **b** $^{109}_{48}\text{Cd}$ ($\mathcal{A} = 7.38(15) \times 10^4\,\text{Bq}$). The simulated spectral signatures were obtained by PSMC (in red, cf. Chapter 6) and NPSMC (in blue, using a NPSM derived by Compton edge probing applied to the detector channel #SUM as discussed in Chapter 7). Uncertainties ($\hat{\sigma}_{\text{exp}}$, $\hat{\sigma}_{\text{sim}}$) are provided as 1 standard deviation (SD) shaded areas. In addition, the coefficient of variation (CV) for the measured and simulated signatures (statistical and systematic contributions are indicated by shaded dotted and dashed lines, respectively, cf. Appendix A.8) as well as the relative deviation (RD) computed as $|\hat{c}_{\text{sim}} - \hat{c}_{\text{exp}}|/\hat{c}_{\text{exp}}$ (20 % mark highlighted by a horizontal grey line) are provided.

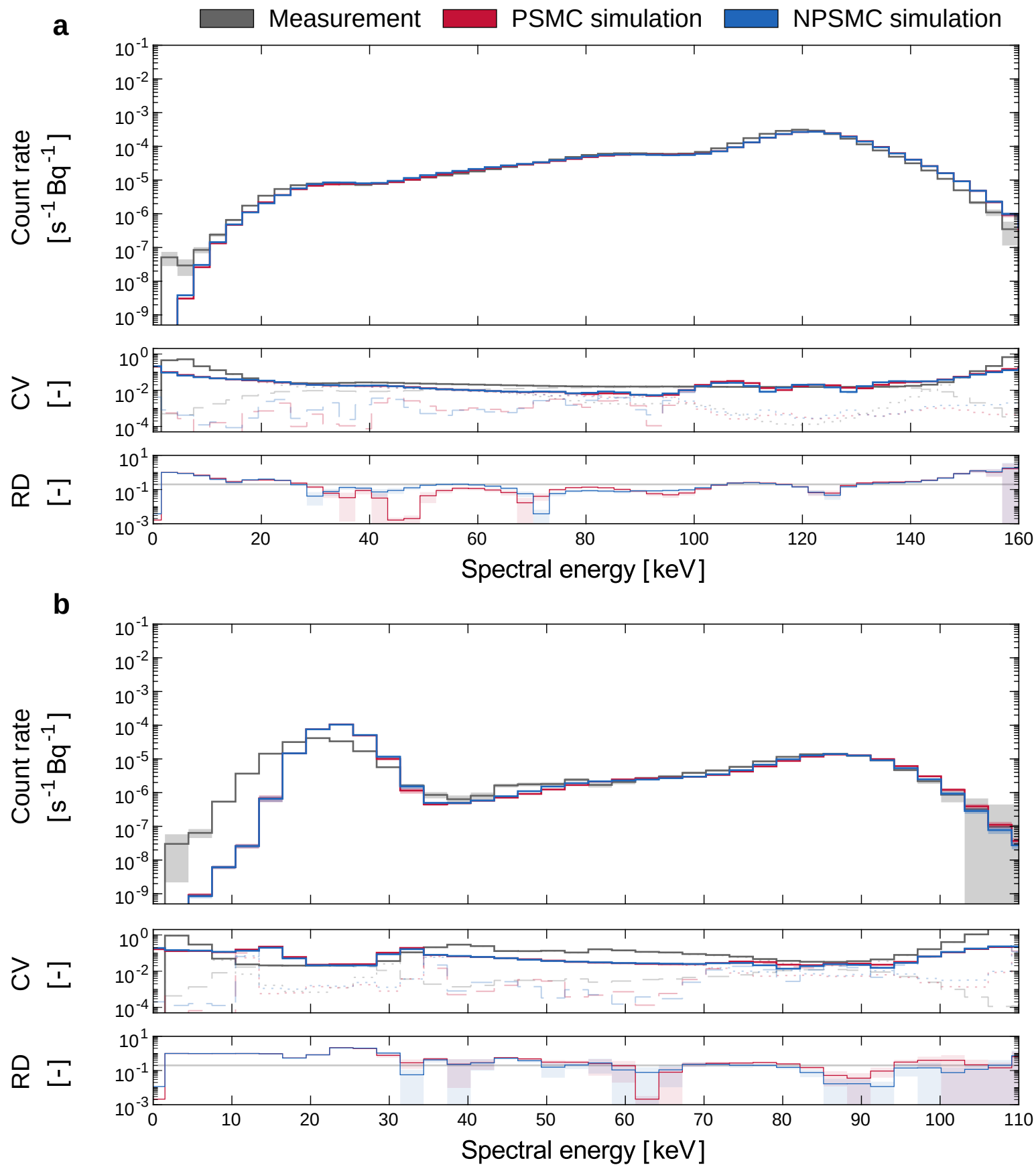

**Figure B.50** The measured ($\hat{c}_{exp}$) and simulated ($\hat{c}_{sim}$) mean spectral signatures for two radionuclides sources and the detector channel #3 as a function of the spectral energy $E'$ with a spectral energy bin width of $\Delta E' \sim 3\,\text{keV}$ are displayed: **a** $^{57}_{27}$Co ($\mathcal{A} = 1.113(18) \times 10^5$ Bq). **b** $^{109}_{48}$Cd ($\mathcal{A} = 7.38(15) \times 10^4$ Bq). The simulated spectral signatures were obtained by PSMC (in red, cf. Chapter 6) and NPSMC (in blue, using a NPSM derived by Compton edge probing applied to the detector channel #SUM as discussed in Chapter 7). Uncertainties ($\hat{\sigma}_{exp}$, $\hat{\sigma}_{sim}$) are provided as 1 standard deviation (SD) shaded areas. In addition, the coefficient of variation (CV) for the measured and simulated signatures (statistical and systematic contributions are indicated by shaded dotted and dashed lines, respectively, cf. Appendix A.8) as well as the relative deviation (RD) computed as $|\hat{c}_{sim} - \hat{c}_{exp}|/\hat{c}_{exp}$ (20 % mark highlighted by a horizontal grey line) are provided.

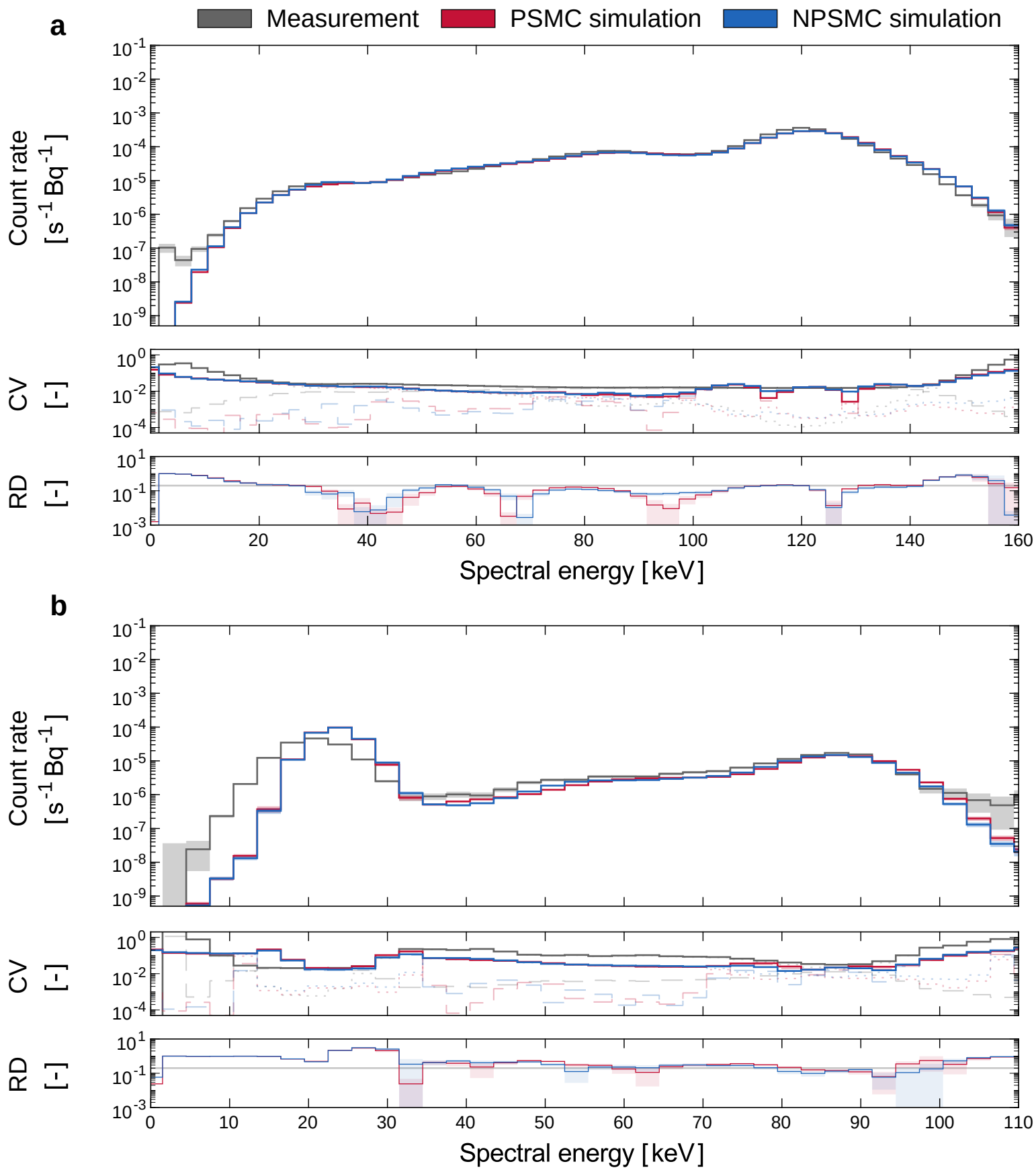

**Figure B.51** The measured ($\hat{c}_{\mathrm{exp}}$) and simulated ($\hat{c}_{\mathrm{sim}}$) mean spectral signatures for two radionuclides sources and the detector channel #4 as a function of the spectral energy $E'$ with a spectral energy bin width of $\Delta E' \sim 3\,\mathrm{keV}$ are displayed: **a** $^{57}_{27}\mathrm{Co}$ ($\mathcal{A} = 1.113(18) \times 10^5\,\mathrm{Bq}$). **b** $^{109}_{48}\mathrm{Cd}$ ($\mathcal{A} = 7.38(15) \times 10^4\,\mathrm{Bq}$). The simulated spectral signatures were obtained by PSMC (in red, cf. Chapter 6) and NPSMC (in blue, using a NPSM derived by Compton edge probing applied to the detector channel #SUM as discussed in Chapter 7). Uncertainties ($\hat{\sigma}_{\mathrm{exp}}$, $\hat{\sigma}_{\mathrm{sim}}$) are provided as 1 standard deviation (SD) shaded areas. In addition, the coefficient of variation (CV) for the measured and simulated signatures (statistical and systematic contributions are indicated by shaded dotted and dashed lines, respectively, cf. Appendix A.8) as well as the relative deviation (RD) computed as $|\hat{c}_{\mathrm{sim}} - \hat{c}_{\mathrm{exp}}|/\hat{c}_{\mathrm{exp}}$ (20 % mark highlighted by a horizontal grey line) are provided.

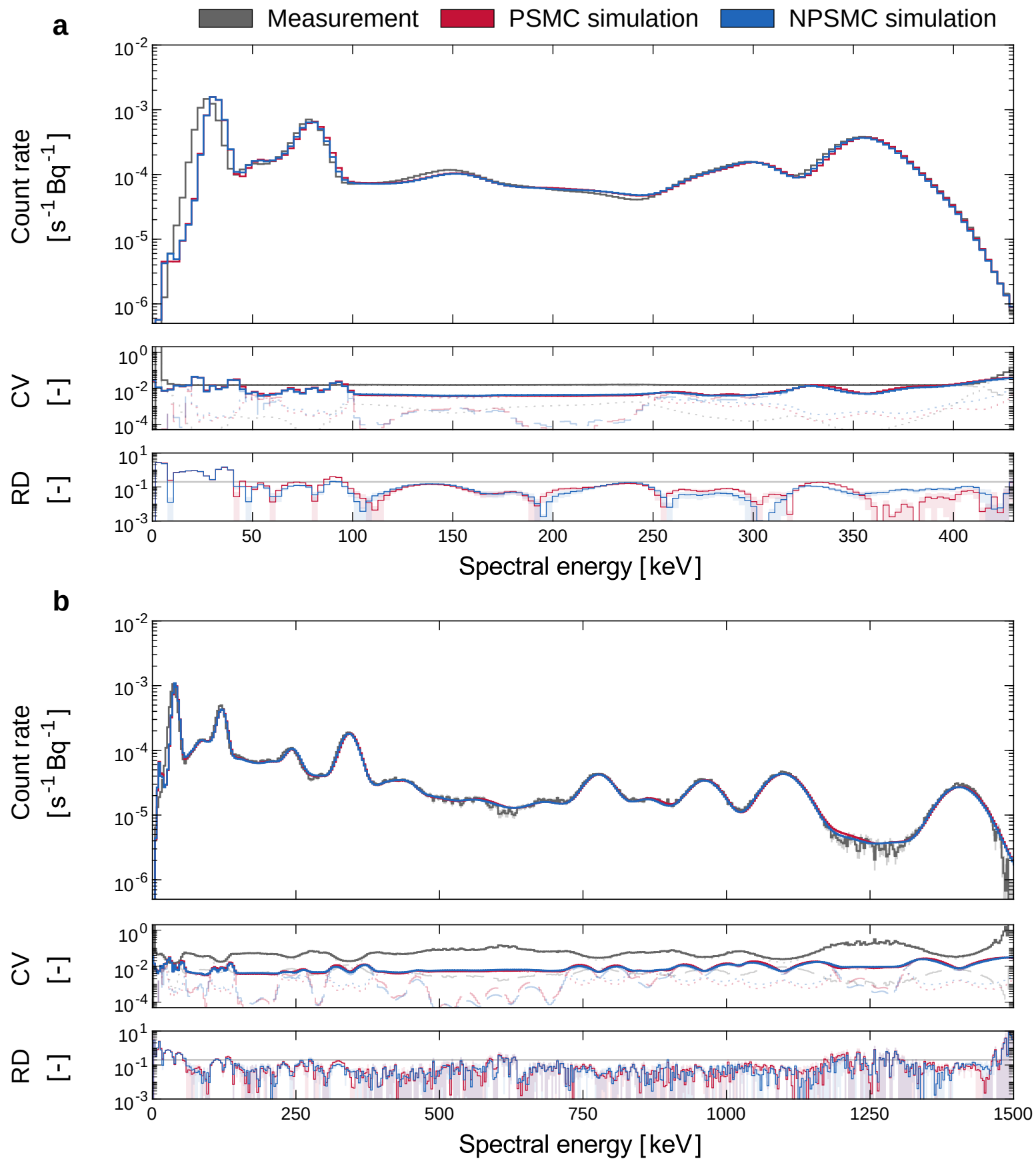

**Figure B.52** The measured ($\hat{c}_{\text{exp}}$) and simulated ($\hat{c}_{\text{sim}}$) mean spectral signatures for two radionuclides sources and the detector channel `#SUM` as a function of the spectral energy $E'$ with a spectral energy bin width of $\Delta E' \sim 3$ keV are displayed: **a** $^{133}_{56}$Ba ($\mathcal{A} = 2.152(32) \times 10^5$ Bq). **b** $^{152}_{63}$Eu ($\mathcal{A} = 1.973(30) \times 10^4$ Bq). The simulated spectral signatures were obtained by PSMC (in red, cf. Chapter 6) and NPSMC (in blue, using a NPSM derived by Compton edge probing applied to the detector channel `#SUM` as discussed in Chapter 7). Uncertainties ($\hat{\sigma}_{\text{exp}}$, $\hat{\sigma}_{\text{sim}}$) are provided as 1 standard deviation (SD) shaded areas. In addition, the coefficient of variation (CV) for the measured and simulated signatures (statistical and systematic contributions are indicated by shaded dotted and dashed lines, respectively, cf. Appendix A.8) as well as the relative deviation (RD) computed as $|\hat{c}_{\text{sim}} - \hat{c}_{\text{exp}}|/\hat{c}_{\text{exp}}$ (20 % mark highlighted by a horizontal grey line) are provided.

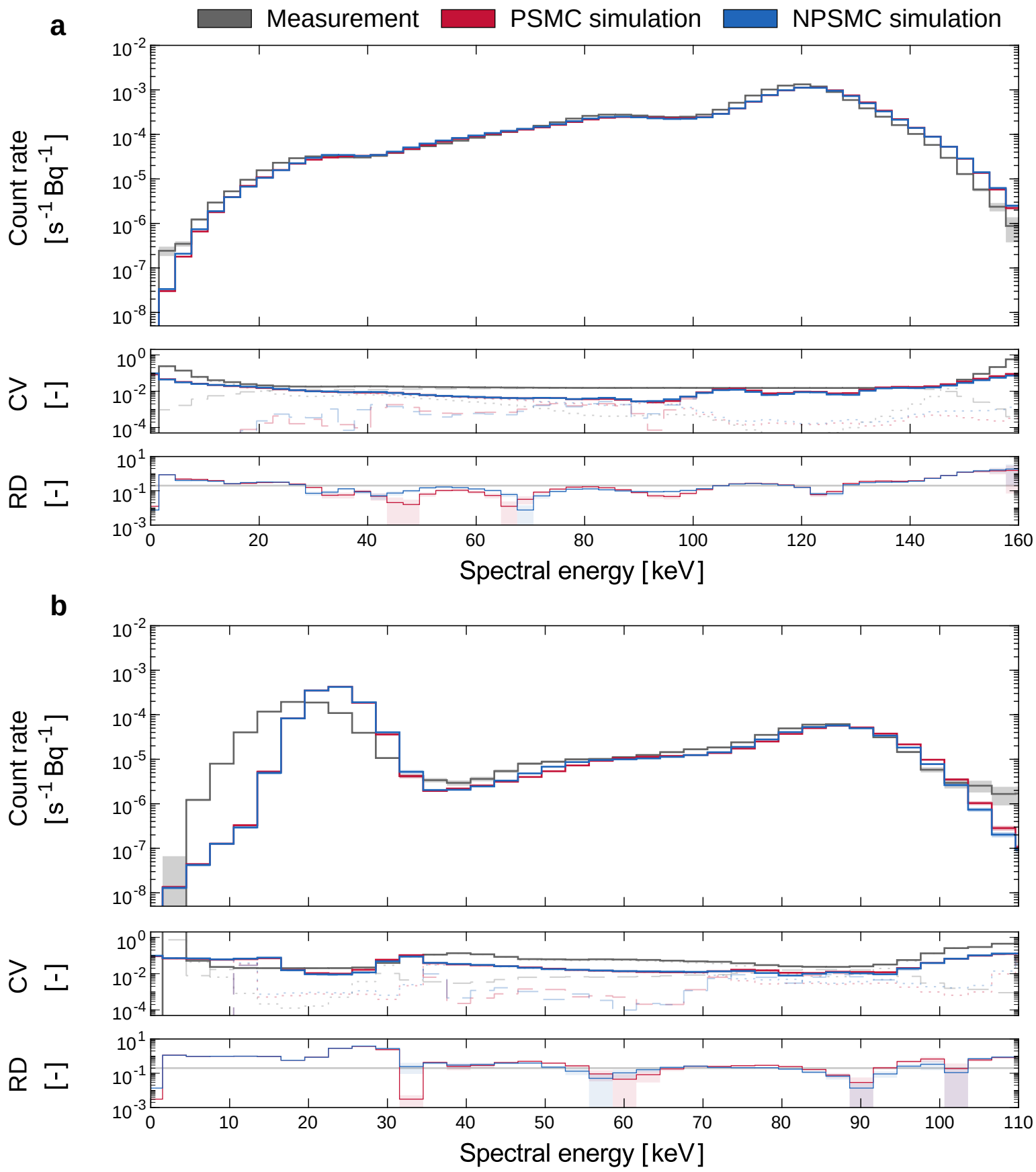

**Figure B.53** The measured ($\hat{c}_{\text{exp}}$) and simulated ($\hat{c}_{\text{sim}}$) mean spectral signatures for two radionuclides sources and the detector channel `#SUM` as a function of the spectral energy $E'$ with a spectral energy bin width of $\Delta E' \sim 3\,\text{keV}$ are displayed: **a** $^{57}_{27}\text{Co}$ ($\mathcal{A} = 1.113(18) \times 10^5\,\text{Bq}$). **b** $^{109}_{48}\text{Cd}$ ($\mathcal{A} = 7.38(15) \times 10^4\,\text{Bq}$). The simulated spectral signatures were obtained by PSMC (in red, cf. Chapter 6) and NPSMC (in blue, using a NPSM derived by Compton edge probing applied to the detector channel `#SUM` as discussed in Chapter 7). Uncertainties ($\hat{\sigma}_{\text{exp}}$, $\hat{\sigma}_{\text{sim}}$) are provided as 1 standard deviation (SD) shaded areas. In addition, the coefficient of variation (CV) for the measured and simulated signatures (statistical and systematic contributions are indicated by shaded dotted and dashed lines, respectively, cf. Appendix A.8) as well as the relative deviation (RD) computed as $|\hat{c}_{\text{sim}} - \hat{c}_{\text{exp}}|/\hat{c}_{\text{exp}}$ (20 % mark highlighted by a horizontal grey line) are provided.

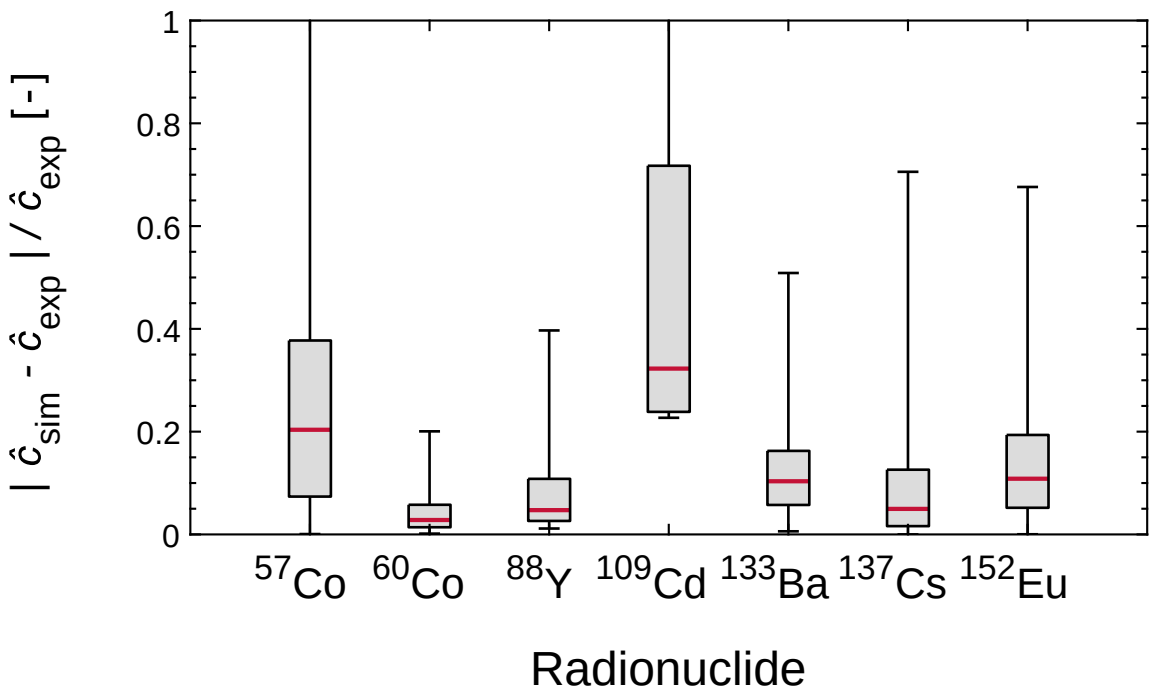

**Figure B.54** Adjusted box plot [721, 722] are displayed for several radionuclides for the detector channel #1 characterizing the statistical distribution of the related relative deviations $|\hat{c}_{\text{sim}} - \hat{c}_{\text{exp}}|/\hat{c}_{\text{exp}}$ between the experimental ($\hat{c}_{\text{exp}}$) and the simulated spectral signatures ($\hat{c}_{\text{sim}}$). The spectral evaluation domain was limited to $n \in \mathcal{D}_{\text{SDOI}}$ (cf. Def. 6.1). The statistical analysis was performed using the Library for Robust Analysis (`LIBRA`) code [723, 724].

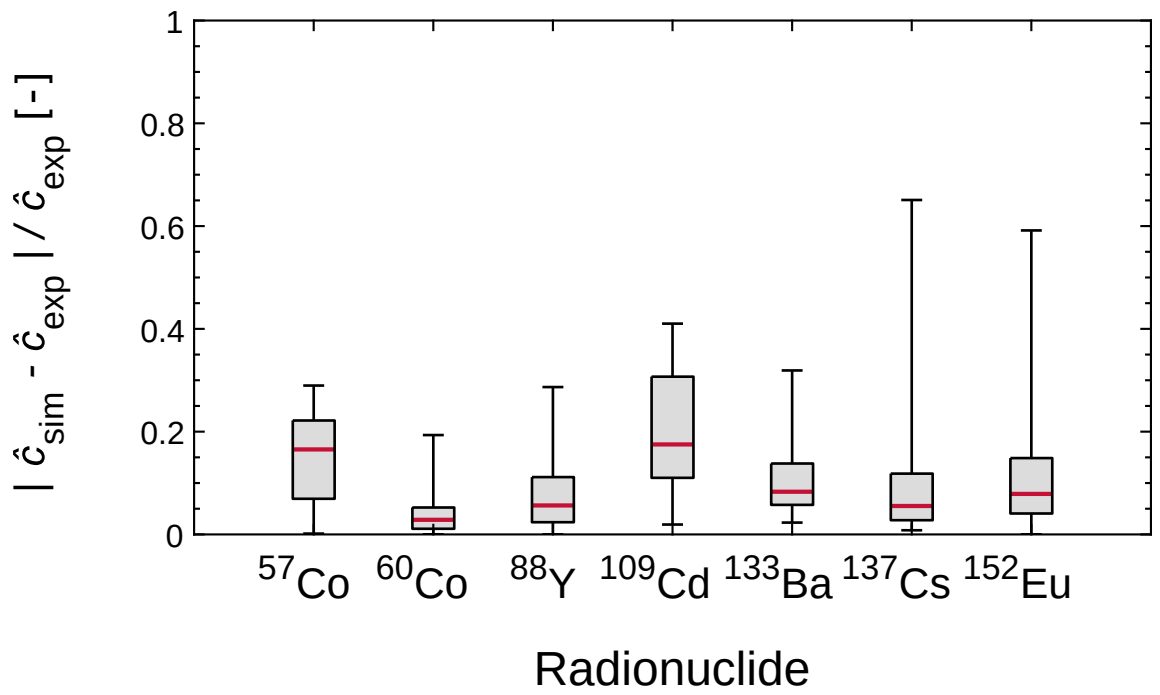

**Figure B.55** Adjusted box plot [721, 722] are displayed for several radionuclides for the detector channel #2 characterizing the statistical distribution of the related relative deviations $|\hat{c}_{\text{sim}} - \hat{c}_{\text{exp}}|/\hat{c}_{\text{exp}}$ between the experimental ($\hat{c}_{\text{exp}}$) and the simulated spectral signatures ($\hat{c}_{\text{sim}}$). The spectral evaluation domain was limited to $n \in \mathcal{D}_{\text{SDOI}}$ (cf. Def. 6.1). The statistical analysis was performed using the Library for Robust Analysis (LIBRA) code [723, 724].

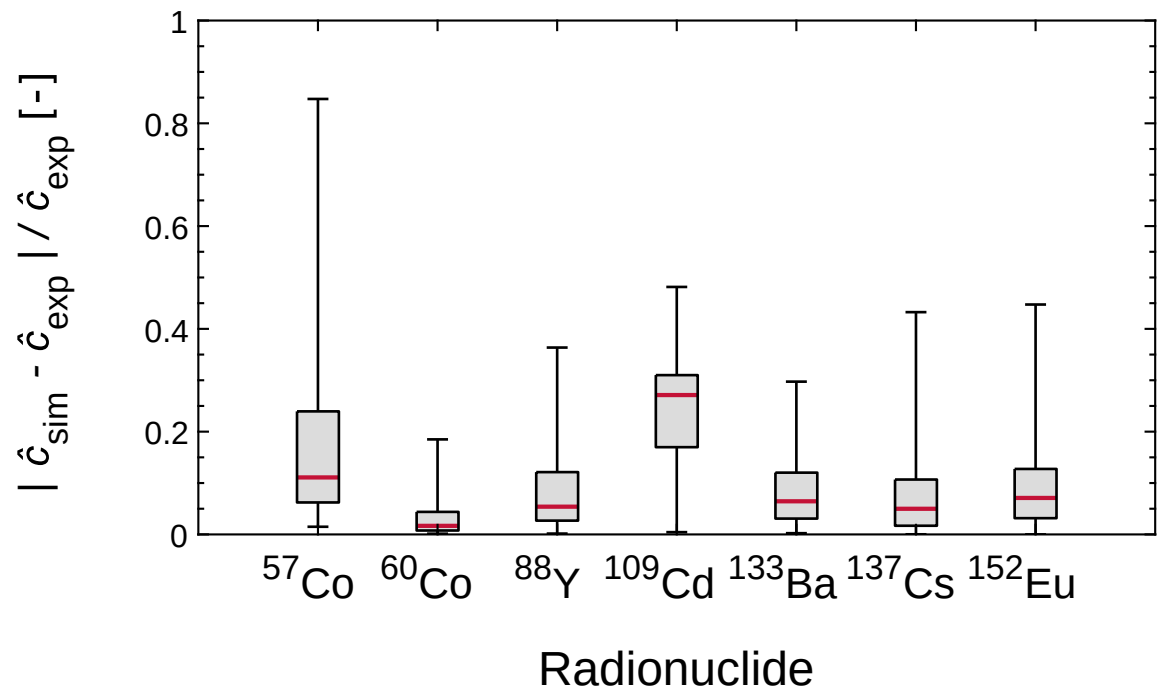

**Figure B.56** Adjusted box plot [721, 722] are displayed for several radionuclides for the detector channel #3 characterizing the statistical distribution of the related relative deviations $|\hat{c}_{\text{sim}} - \hat{c}_{\text{exp}}|/\hat{c}_{\text{exp}}$ between the experimental ($\hat{c}_{\text{exp}}$) and the simulated spectral signatures ($\hat{c}_{\text{sim}}$). The spectral evaluation domain was limited to $n \in \mathcal{D}_{\text{SDOI}}$ (cf. Def. 6.1). The statistical analysis was performed using the Library for Robust Analysis (`LIBRA`) code [723, 724].

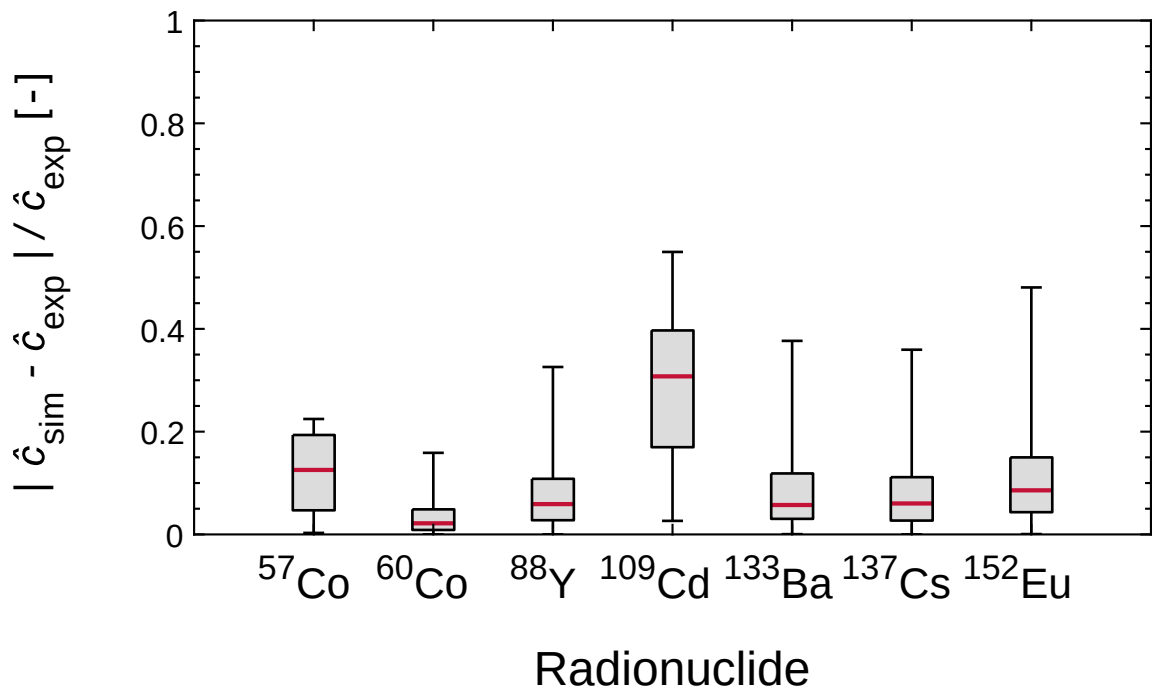

**Figure B.57** Adjusted box plot [721, 722] are displayed for several radionuclides for the detector channel #4 characterizing the statistical distribution of the related relative deviations $|\hat{c}_{\text{sim}} - \hat{c}_{\text{exp}}|/\hat{c}_{\text{exp}}$ between the experimental ($\hat{c}_{\text{exp}}$) and the simulated spectral signatures ($\hat{c}_{\text{sim}}$). The spectral evaluation domain was limited to $n \in \mathcal{D}_{\text{SDOI}}$ (cf. Def. 6.1). The statistical analysis was performed using the Library for Robust Analysis (`LIBRA`) code [723, 724].

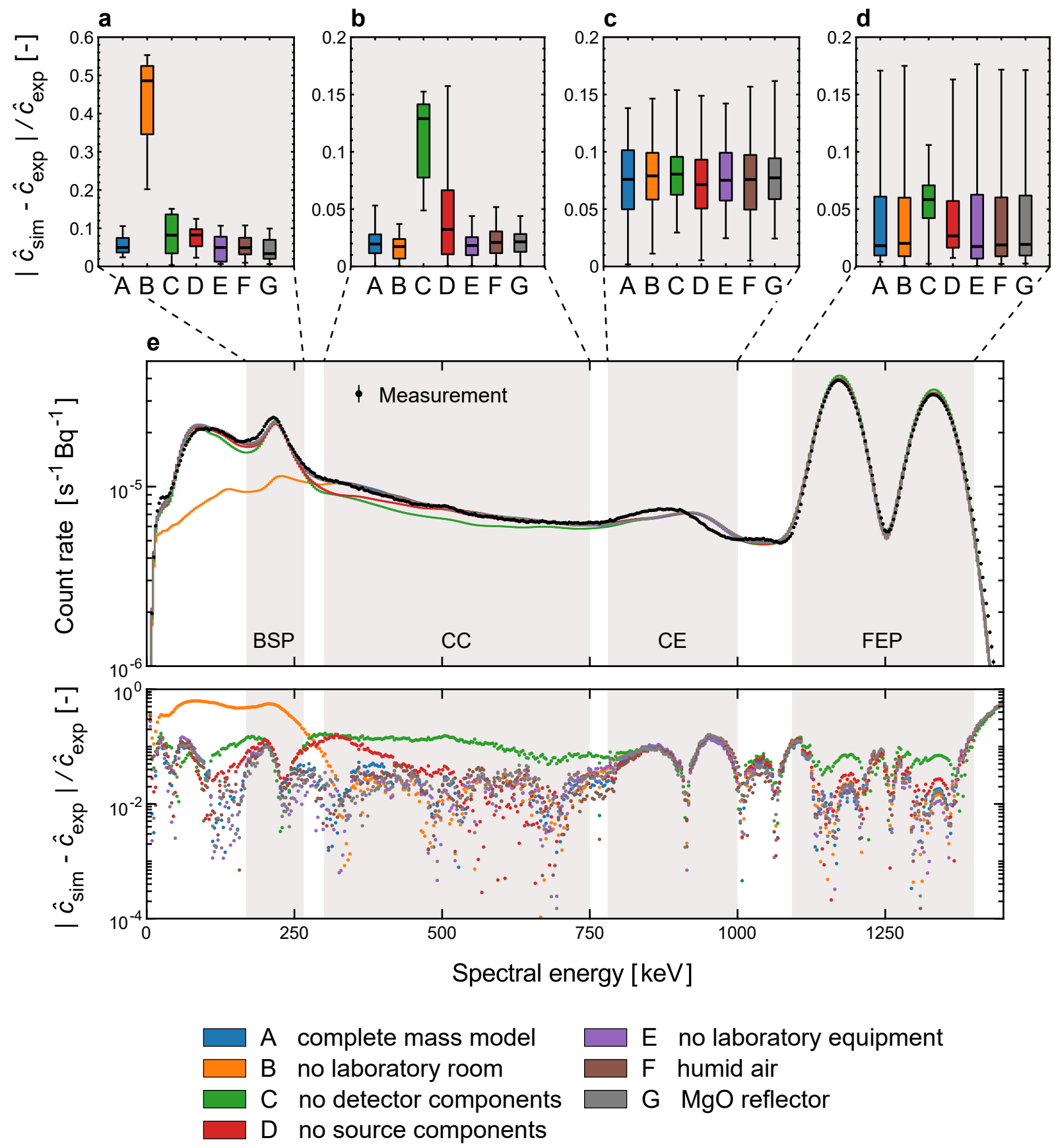

**Figure B.58** Mass model sensitivity analysis for the $^{60}_{27}$Co spectral signature and the detector channel #1. **a–d** Adjusted box plots [721, 722] characterizing the statistical distribution of the relative deviations $|\hat{c}_{sim} - \hat{c}_{exp}|/\hat{c}_{exp}$ between the experimental ($\hat{c}_{exp}$) and the simulated spectral signatures ($\hat{c}_{sim}$) are shown for four different spectral domains (BSP: backscatter peak, CC: medium part of the Compton continuum, CE: Compton edge, FEP: full energy peaks, cf. also Section 4.3.1) and six different mass models A–G (cf. Section 6.3.2). The statistical analysis was performed using the Library for Robust Analysis (LIBRA) [723, 724]. **e** The measured ($\hat{c}_{exp}$) and simulated ($\hat{c}_{sim}$) mean spectral signatures for $^{60}_{27}$Co are displayed as a function of the spectral energy $E'$ with a spectral energy bin width of $\Delta E' \sim 3$ keV. Spectral domains evaluated in **a–d** are highlighted. Uncertainties ($\hat{\sigma}_{exp}$, $\hat{\sigma}_{sim}$) are provided as 1 standard deviation (SD) shaded areas (cf. Appendix A.8). In addition, the relative deviation $|\hat{c}_{sim} - \hat{c}_{exp}|/\hat{c}_{exp}$ as a function of the spectral energy $E'$ is displayed for all six mass models A–G.

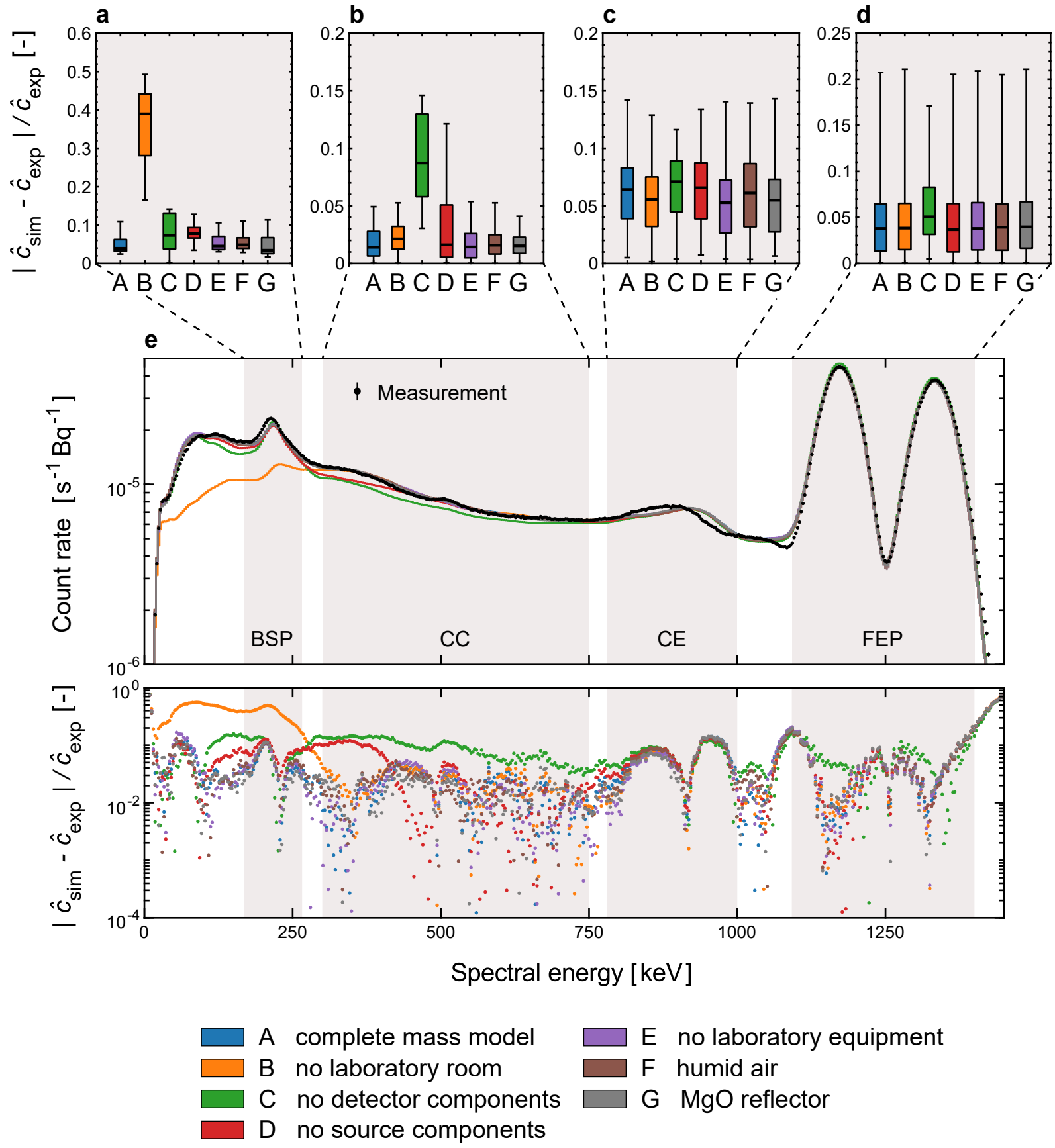

**Figure B.59** Mass model sensitivity analysis for the $^{60}_{27}$Co spectral signature and the detector channel #2. **a–d** Adjusted box plots [721, 722] characterizing the statistical distribution of the relative deviations $|\hat{c}_{sim} - \hat{c}_{exp}|/\hat{c}_{exp}$ between the experimental ($\hat{c}_{exp}$) and the simulated spectral signatures ($\hat{c}_{sim}$) are shown for four different spectral domains (BSP: backscatter peak, CC: medium part of the Compton continuum, CE: Compton edge, FEP: full energy peaks, cf. also Section 4.3.1) and six different mass models A–G (cf. Section 6.3.2). The statistical analysis was performed using the Library for Robust Analysis (LIBRA) [723, 724]. **e** The measured ($\hat{c}_{exp}$) and simulated ($\hat{c}_{sim}$) mean spectral signatures for $^{60}_{27}$Co are displayed as a function of the spectral energy $E'$ with a spectral energy bin width of $\Delta E' \sim 3$ keV. Spectral domains evaluated in **a–d** are highlighted. Uncertainties ($\hat{\sigma}_{exp}$, $\hat{\sigma}_{sim}$) are provided as 1 standard deviation (SD) shaded areas (cf. Appendix A.8). In addition, the relative deviation $|\hat{c}_{sim} - \hat{c}_{exp}|/\hat{c}_{exp}$ as a function of the spectral energy $E'$ is displayed for all six mass models A–G.

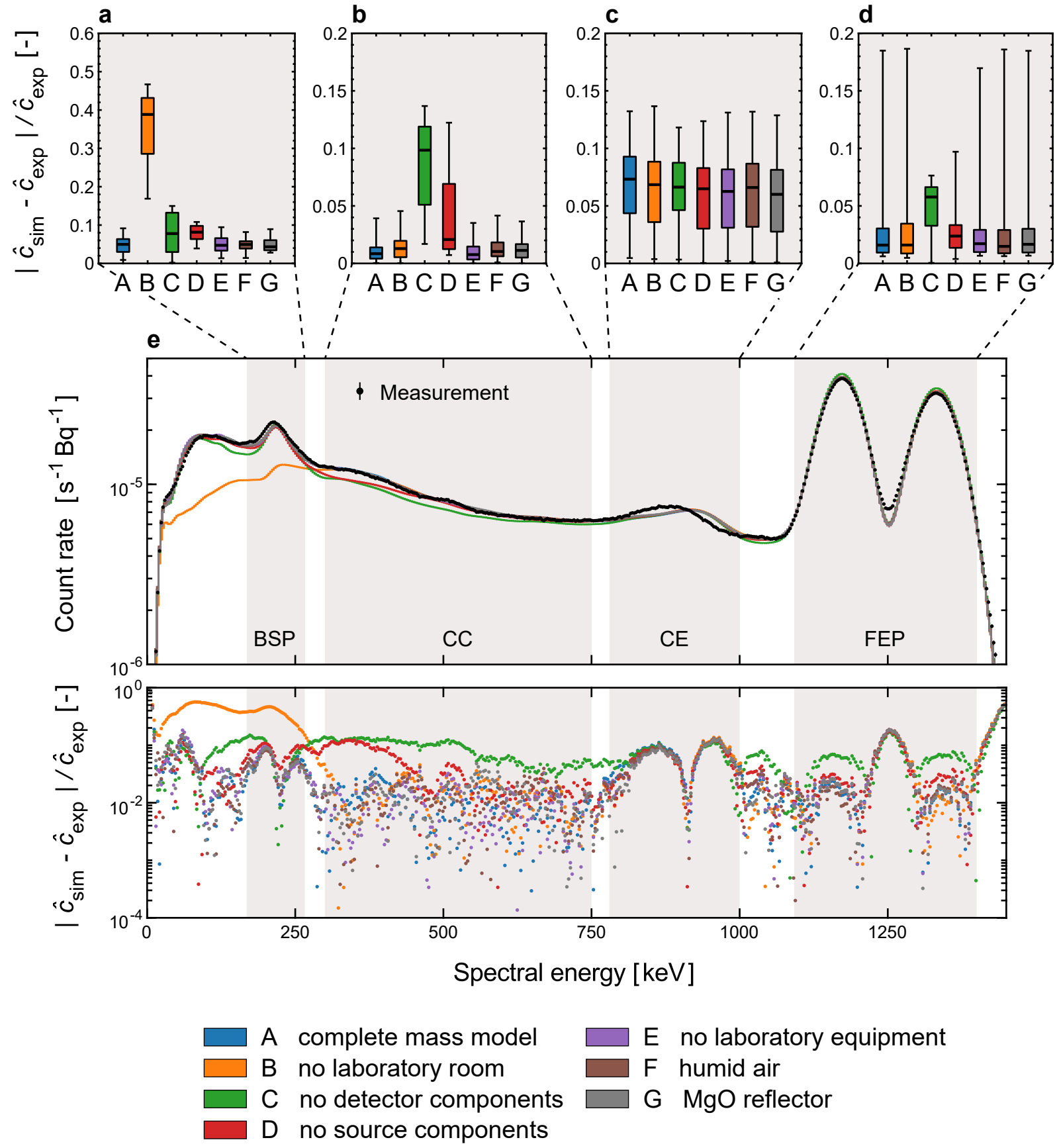

**Figure B.60** Mass model sensitivity analysis for the $^{60}_{27}$Co spectral signature and the detector channel #3. **a–d** Adjusted box plots [721, 722] characterizing the statistical distribution of the relative deviations $|\hat{c}_{\text{sim}} - \hat{c}_{\text{exp}}|/\hat{c}_{\text{exp}}$ between the experimental ($\hat{c}_{\text{exp}}$) and the simulated spectral signatures ($\hat{c}_{\text{sim}}$) are shown for four different spectral domains (BSP: backscatter peak, CC: medium part of the Compton continuum, CE: Compton edge, FEP: full energy peaks, cf. also Section 4.3.1) and six different mass models A–G (cf. Section 6.3.2). The statistical analysis was performed using the Library for Robust Analysis (LIBRA) [723, 724]. **e** The measured ($\hat{c}_{\text{exp}}$) and simulated ($\hat{c}_{\text{sim}}$) mean spectral signatures for $^{60}_{27}$Co are displayed as a function of the spectral energy $E'$ with a spectral energy bin width of $\Delta E' \sim 3$ keV. Spectral domains evaluated in **a–d** are highlighted. Uncertainties ($\hat{\sigma}_{\text{exp}}$, $\hat{\sigma}_{\text{sim}}$) are provided as 1 standard deviation (SD) shaded areas (cf. Appendix A.8). In addition, the relative deviation $|\hat{c}_{\text{sim}} - \hat{c}_{\text{exp}}|/\hat{c}_{\text{exp}}$ as a function of the spectral energy $E'$ is displayed for all six mass models A–G.

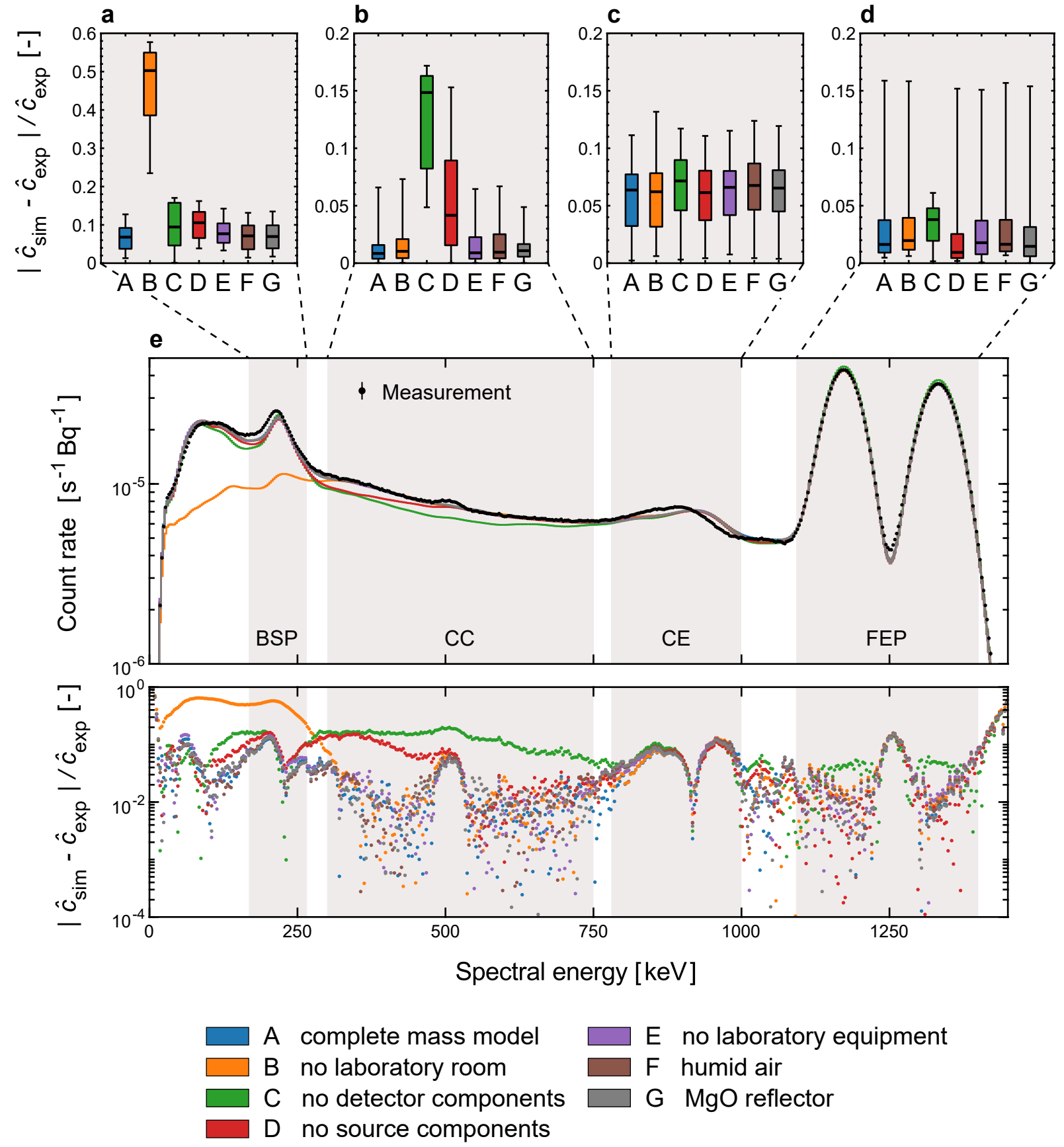

**Figure B.61** Mass model sensitivity analysis for the $^{60}_{27}$Co spectral signature and the detector channel #4. **a–d** Adjusted box plots [721, 722] characterizing the statistical distribution of the relative deviations $|\hat{c}_{sim} - \hat{c}_{exp}|/\hat{c}_{exp}$ between the experimental ($\hat{c}_{exp}$) and the simulated spectral signatures ($\hat{c}_{sim}$) are shown for four different spectral domains (BSP: backscatter peak, CC: medium part of the Compton continuum, CE: Compton edge, FEP: full energy peaks, cf. also Section 4.3.1) and six different mass models A–G (cf. Section 6.3.2). The statistical analysis was performed using the Library for Robust Analysis (LIBRA) [723, 724]. **e** The measured ($\hat{c}_{exp}$) and simulated ($\hat{c}_{sim}$) mean spectral signatures for $^{60}_{27}$Co are displayed as a function of the spectral energy $E'$ with a spectral energy bin width of $\Delta E' \sim 3$ keV. Spectral domains evaluated in **a–d** are highlighted. Uncertainties ($\hat{\sigma}_{exp}$, $\hat{\sigma}_{sim}$) are provided as 1 standard deviation (SD) shaded areas (cf. Appendix A.8). In addition, the relative deviation $|\hat{c}_{sim} - \hat{c}_{exp}|/\hat{c}_{exp}$ as a function of the spectral energy $E'$ is displayed for all six mass models A–G.

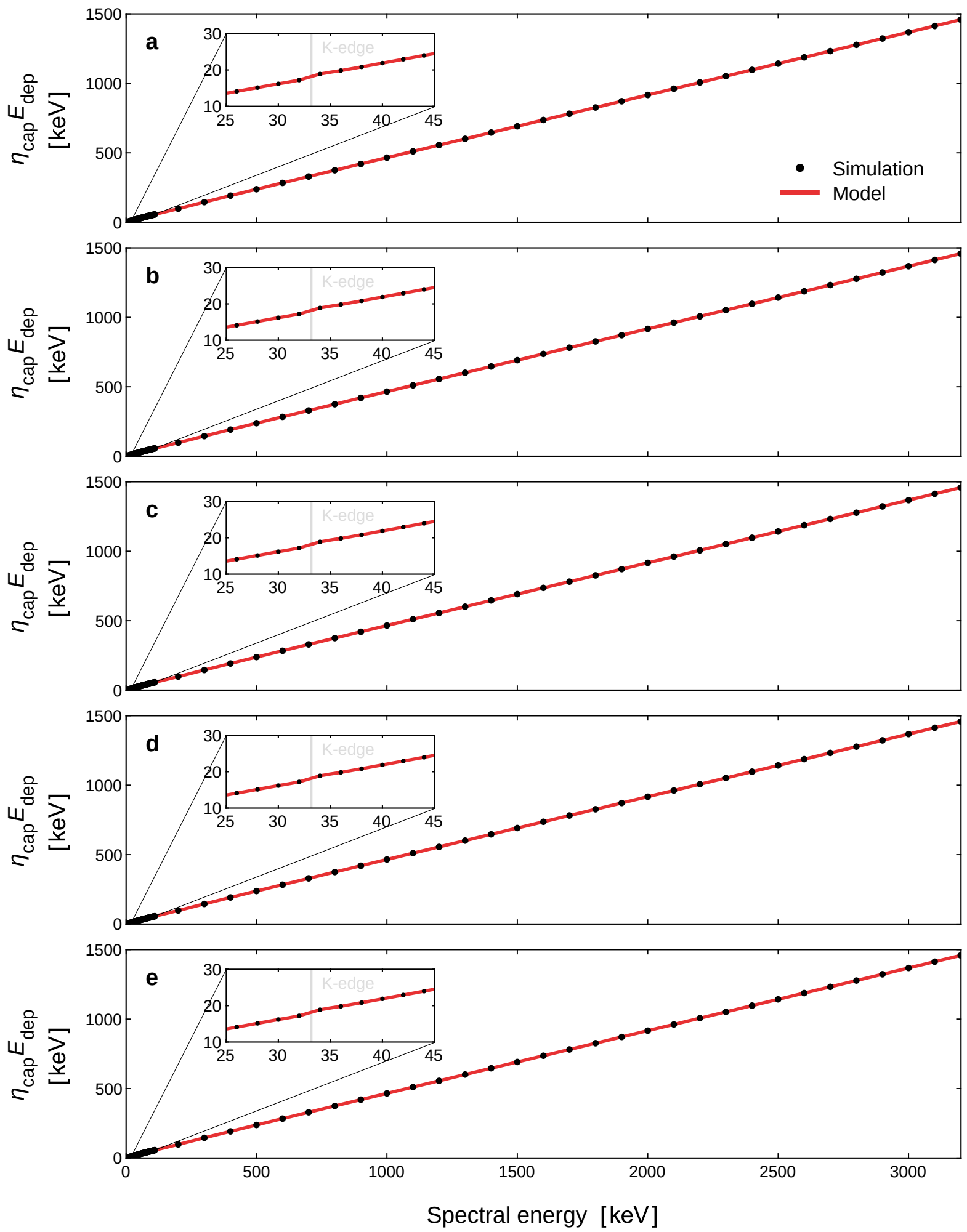

**Figure B.62** This graph presents the scaling of the NPSM response $\eta_{\text{cap}}E_{\text{dep}}$ discussed in Chapter 7 for the four single detector channels `#1` through `#4` (**a–d**) and the detector channel `#SUM` (**e**) as a function of the spectral energy $E'$. The scaling models and simulation data were derived by the `NPScinCal` pipeline discussed in Section 7.2.2. For the zoomed insets, the K-absorption edge for iodine is highlighted [815]. Note that for the reader's convenience, the continuous pulse-height channel numbers $\tilde{n}$ were converted to spectral energies $E'$ using the energy calibration models discussed in Section 6.2.1.3.

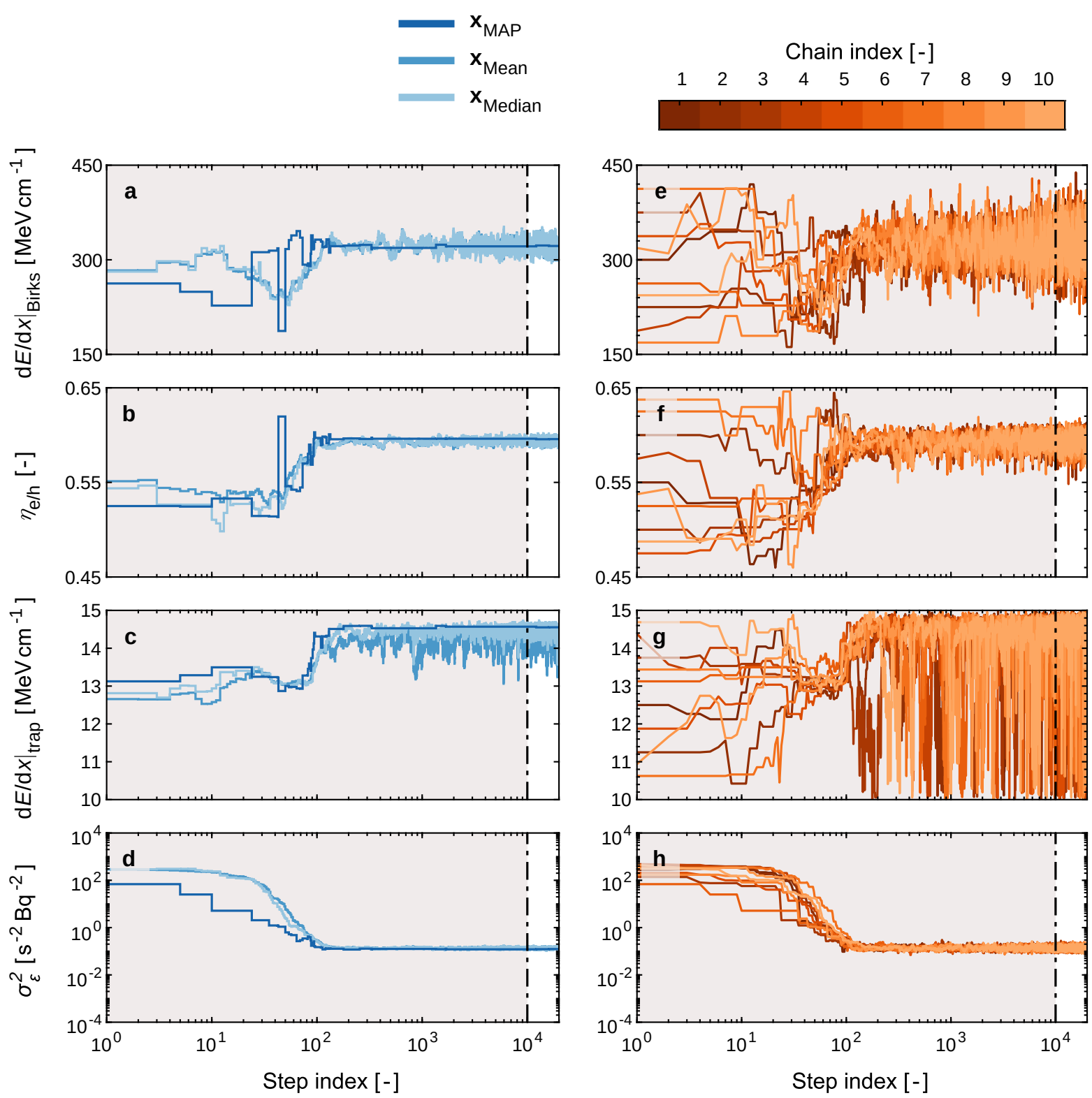

**Figure B.63** These graphs present the convergence (**a–d**) and trace (**e–h**) plots as a function of the MCMC steps for the detector channel #SUM and each individual model parameter, i.e. the Birks stopping parameter $\mathrm{d}E/\mathrm{d}x\,|_{\text{Birks}}$, the electron-hole pair fraction $\eta_{\text{e/h}}$, the trapping stopping parameter $\mathrm{d}E/\mathrm{d}x\,|_{\text{trap}}$ and the discrepancy model variance $\sigma_{\varepsilon}^2$. The convergence is shown for the maximum a posteriori (MAP) probability estimate $\mathbf{x}_{\text{MAP}}$, the posterior mean $\mathbf{x}_{\text{Mean}}$ and the posterior median $\mathbf{x}_{\text{Median}}$. The trace plots are shown for all performed MCMC chains. The burn-in period is highlighted for all subgraphs as gray-shaded areas with the related threshold marked by the dashed-dotted black line.

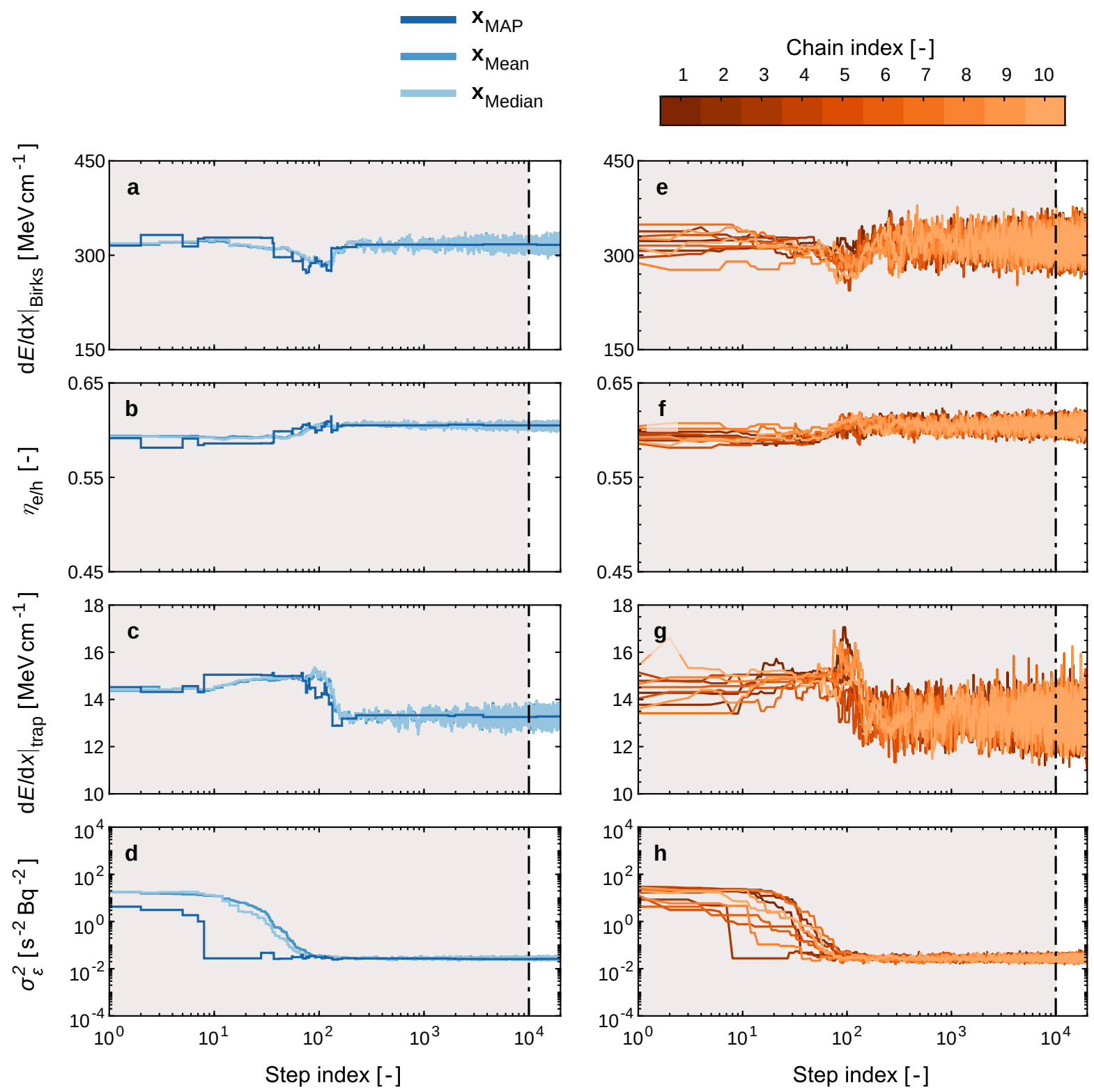

**Figure B.64** These graphs present the convergence (**a–d**) and trace (**e–h**) plots as a function of the MCMC steps for the detector channel #1 and each individual model parameter, i.e. the Birks stopping parameter $dE/dx\,|_{\mathrm{Birks}}$, the electron-hole pair fraction $\eta_{\mathrm{e/h}}$, the trapping stopping parameter $dE/dx\,|_{\mathrm{trap}}$ and the discrepancy model variance $\sigma^2_{\varepsilon}$. The convergence is shown for the maximum a posteriori (MAP) probability estimate $\mathbf{x}_{\mathrm{MAP}}$, the posterior mean $\mathbf{x}_{\mathrm{Mean}}$ and the posterior median $\mathbf{x}_{\mathrm{Median}}$. The trace plots are shown for all performed MCMC chains. The burn-in period is highlighted for all subgraphs as gray-shaded areas with the related threshold marked by the dashed-dotted black line.

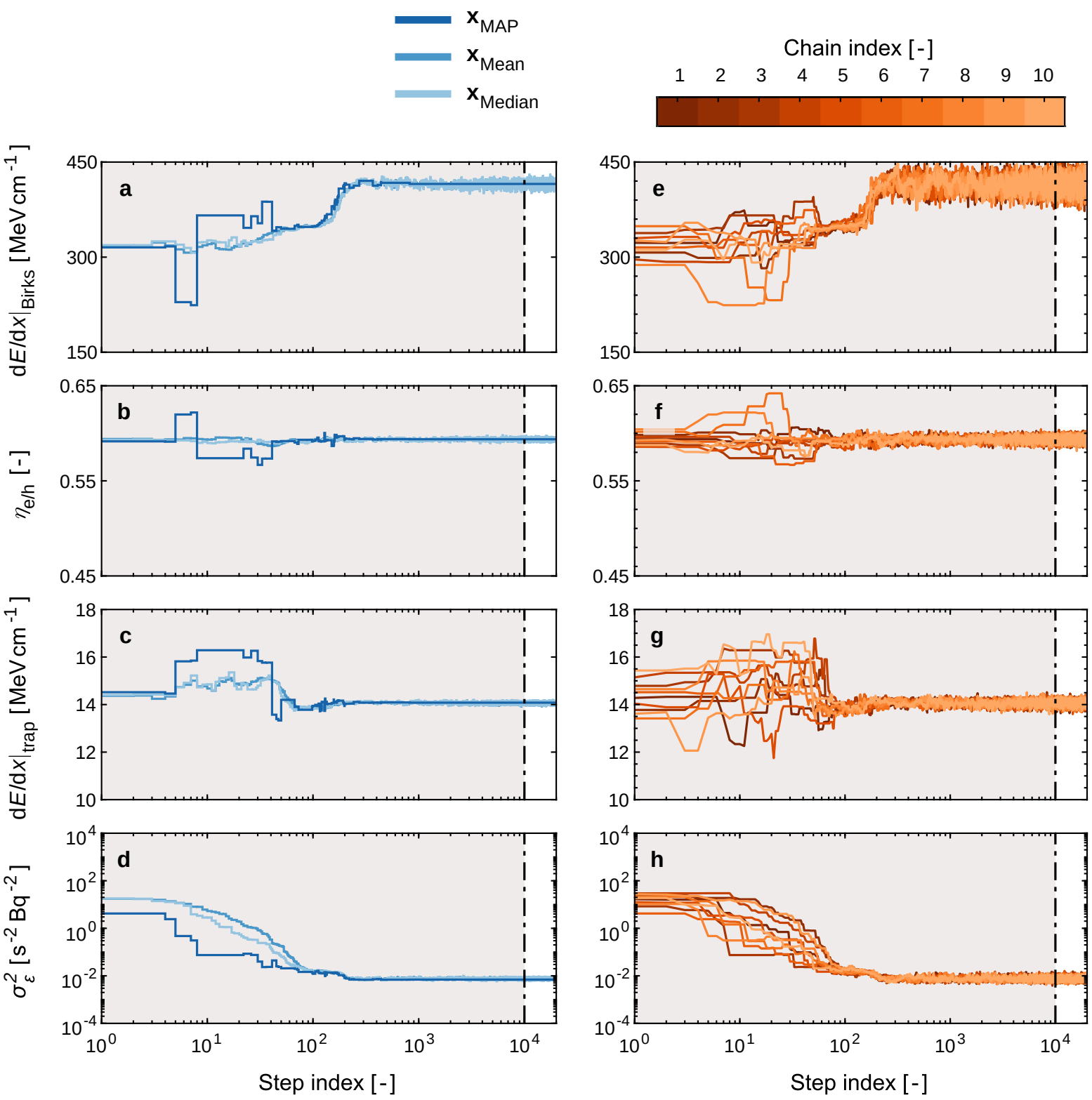

**Figure B.65** These graphs present the convergence (**a–d**) and trace (**e–h**) plots as a function of the MCMC steps for the detector channel #2 and each individual model parameter, i.e. the Birks stopping parameter $\mathrm{d}E/\mathrm{d}x\,|_{\mathrm{Birks}}$, the electron-hole pair fraction $\eta_{\mathrm{e/h}}$, the trapping stopping parameter $\mathrm{d}E/\mathrm{d}x\,|_{\mathrm{trap}}$ and the discrepancy model variance $\sigma_{\varepsilon}^{2}$. The convergence is shown for the maximum a posteriori (MAP) probability estimate $\mathbf{x}_{\mathrm{MAP}}$, the posterior mean $\mathbf{x}_{\mathrm{Mean}}$ and the posterior median $\mathbf{x}_{\mathrm{Median}}$. The trace plots are shown for all performed MCMC chains. The burn-in period is highlighted for all subgraphs as gray-shaded areas with the related threshold marked by the dashed-dotted black line.

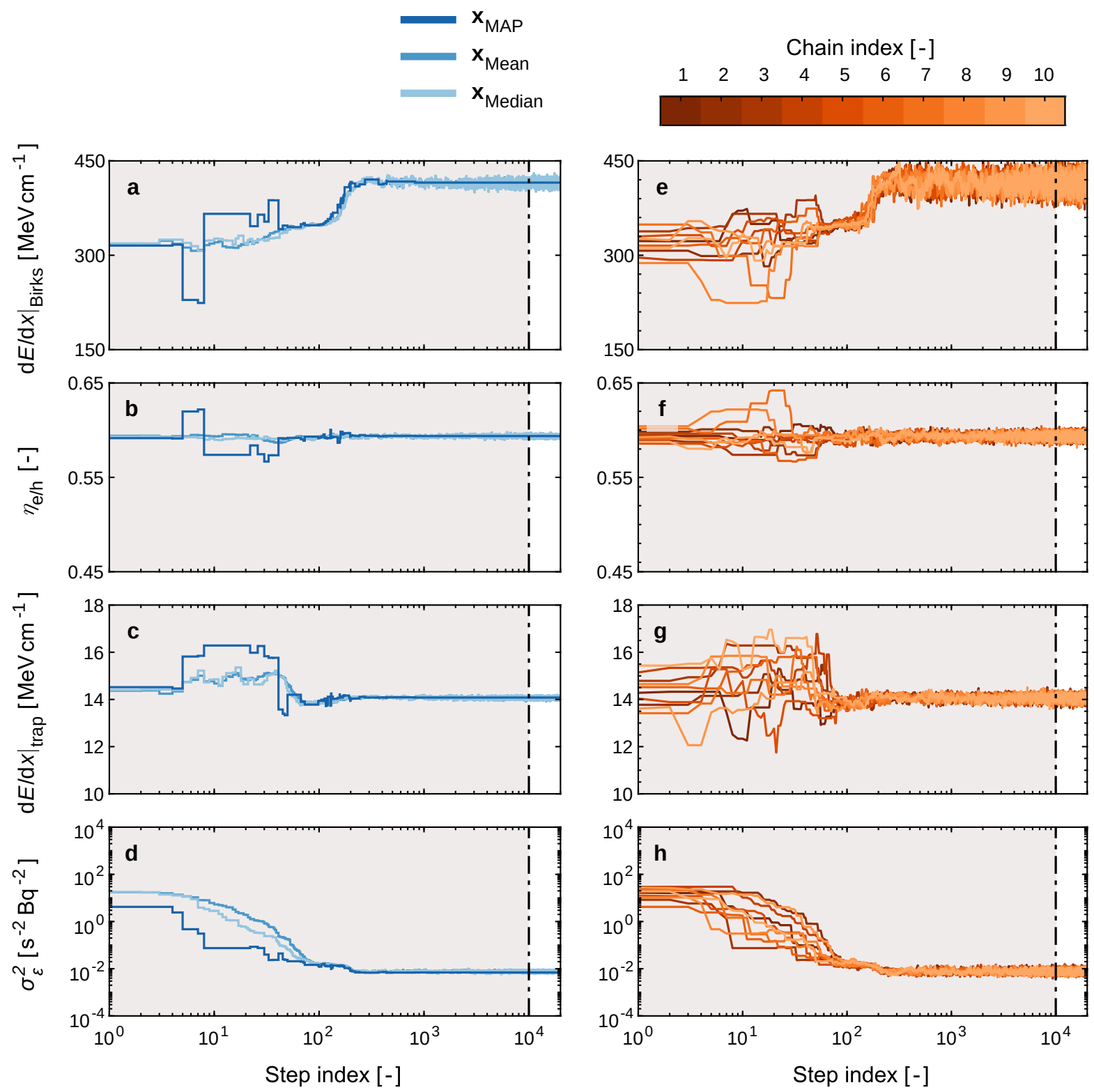

**Figure B.66** These graphs present the convergence (**a–d**) and trace (**e–h**) plots as a function of the MCMC steps for the detector channel #3 and each individual model parameter, i.e. the Birks stopping parameter $dE/dx\,|_{Birks}$, the electron-hole pair fraction $\eta_{e/h}$, the trapping stopping parameter $dE/dx\,|_{trap}$ and the discrepancy model variance $\sigma^2_\varepsilon$. The convergence is shown for the maximum a posteriori (MAP) probability estimate $\mathbf{x}_{MAP}$, the posterior mean $\mathbf{x}_{Mean}$ and the posterior median $\mathbf{x}_{Median}$. The trace plots are shown for all performed MCMC chains. The burn-in period is highlighted for all subgraphs as gray-shaded areas with the related threshold marked by the dashed-dotted black line.

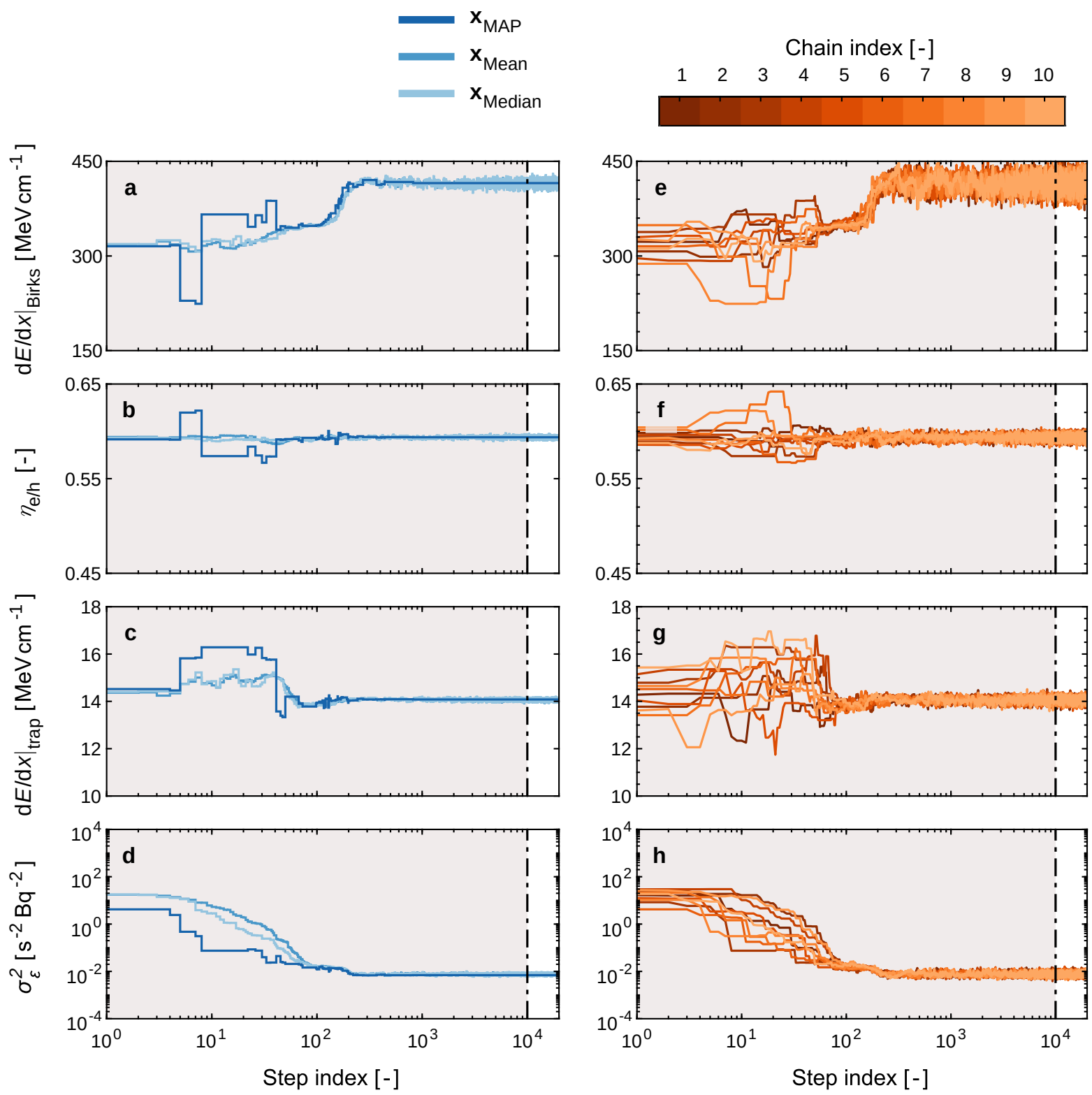

**Figure B.67** These graphs present the convergence (**a–d**) and trace (**e–h**) plots as a function of the MCMC steps for detector channel #4 and each individual model parameter, i.e. the Birks stopping parameter $\mathrm{d}E/\mathrm{d}x\,|_{\mathrm{Birks}}$, the electron-hole pair fraction $\eta_{\mathrm{e/h}}$, the trapping stopping parameter $\mathrm{d}E/\mathrm{d}x\,|_{\mathrm{trap}}$ and the discrepancy model variance $\sigma_{\varepsilon}^{2}$. The convergence is shown for the maximum a posteriori (MAP) probability estimate $\mathbf{x}_{\mathrm{MAP}}$, the posterior mean $\mathbf{x}_{\mathrm{Mean}}$ and the posterior median $\mathbf{x}_{\mathrm{Median}}$. The trace plots are shown for all performed MCMC chains. The burn-in period is highlighted for all subgraphs as gray-shaded areas with the related threshold marked by the dashed-dotted black line.

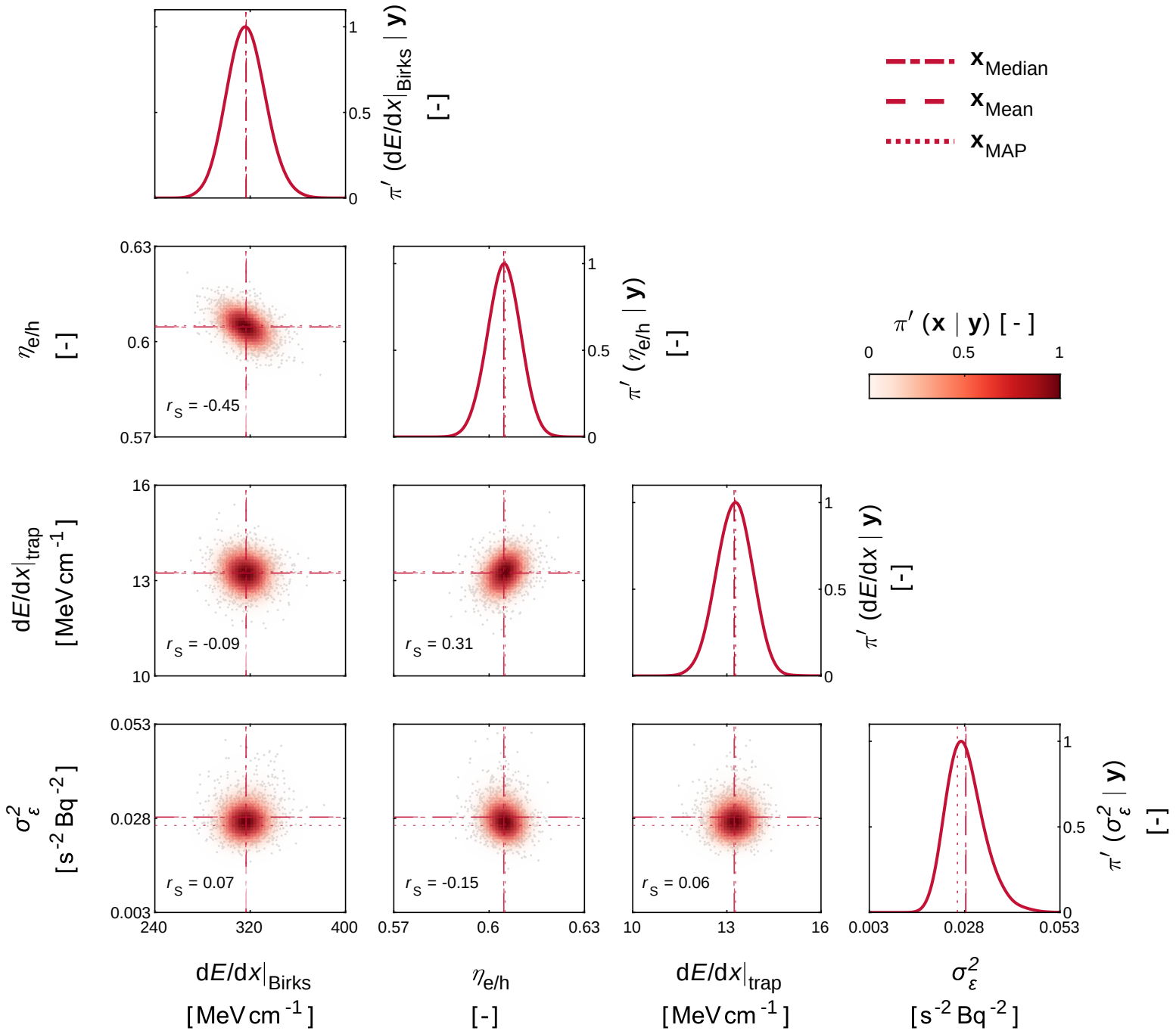

**Figure B.68** Here, I present the Bayesian inversion results for the detector channel #1 in the form of a corner plot. The off-diagonal subfigures display the normalized bivariate posterior marginal estimates (color-encoded) along with a subset of the first $10^3$ MCMC samples (in gray) for the model parameters $\mathbf{x} := ( \mathrm{d}E/\mathrm{d}x|_{\text{Birks}}, \eta_{\text{e/h}}, \mathrm{d}E/\mathrm{d}x|_{\text{trap}}, \sigma_\varepsilon^2)^\top$ and experimental data $\mathbf{y}$. In addition, the Spearman's rank correlation coefficient $r_S$ is provided for the model parameters in the corresponding off-diagonal subfigures. The subfigures on the diagonal axis highlight the normalized univariate posterior marginals for the corresponding model parameter. Both the univariate and multivariate marginals were normalized by their corresponding global maxima. Derived posterior point estimates, i.e. the maximum a posteriori (MAP) probability estimate $\mathbf{x}_{\text{MAP}}$, the posterior mean $\mathbf{x}_{\text{Mean}}$ and the posterior median $\mathbf{x}_{\text{Median}}$, are indicated as well in each subfigure.

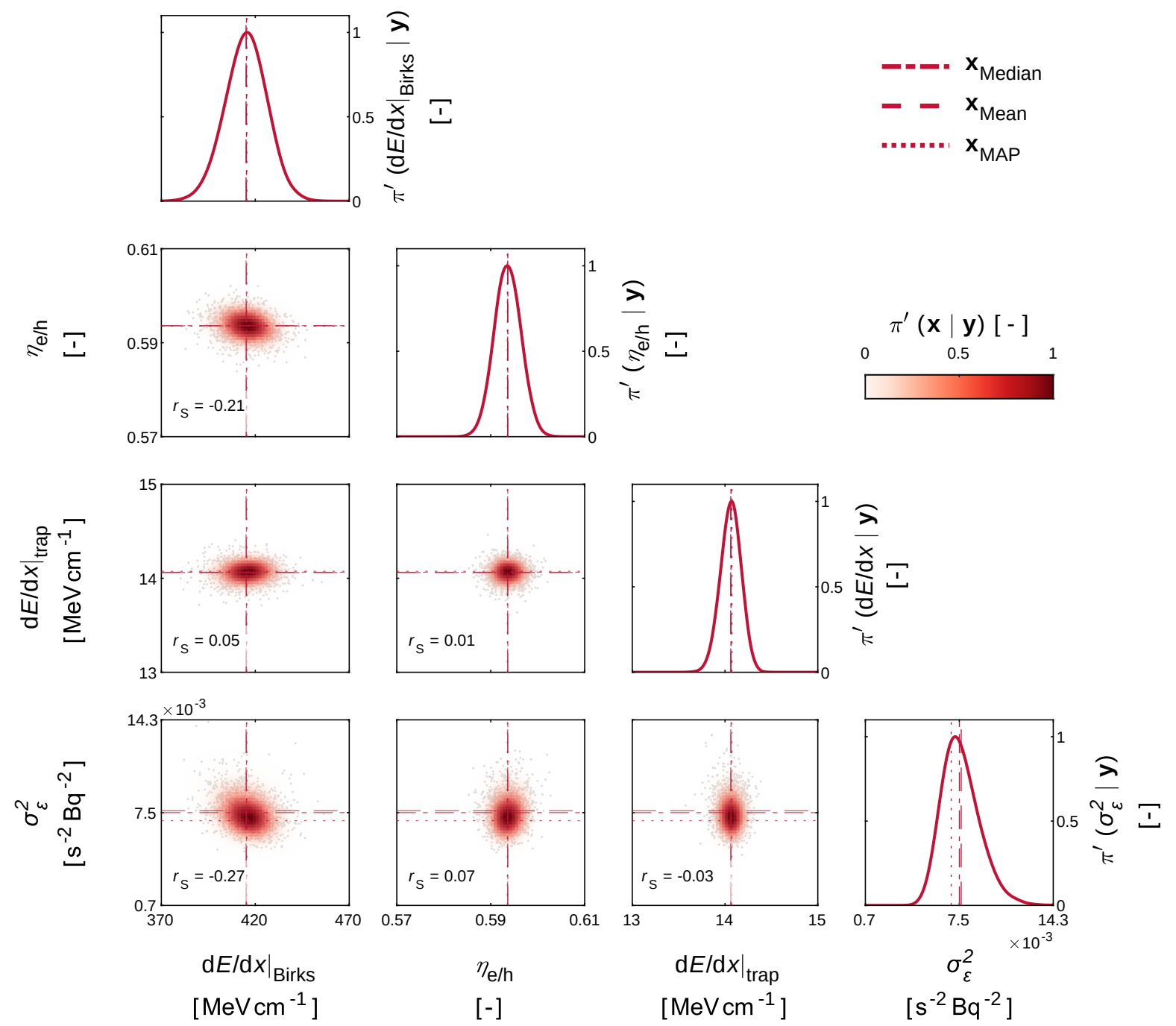

**Figure B.69** Here, I present the Bayesian inversion results for the detector channel #2 in the form of a corner plot. The off-diagonal subfigures display the normalized bivariate posterior marginal estimates (color-encoded) along with a subset of the first $10^3$ MCMC samples (in gray) for the model parameters $\mathbf{x} := (\mathrm{d}E/\mathrm{d}x\,|_{\mathrm{Birks}}, \eta_{\mathrm{e/h}}, \mathrm{d}E/\mathrm{d}x\,|_{\mathrm{trap}}, \sigma_\varepsilon^2)^\top$ and experimental data $\mathbf{y}$. In addition, the Spearman's rank correlation coefficient $r_S$ is provided for the model parameters in the corresponding off-diagonal subfigures. The subfigures on the diagonal axis highlight the normalized univariate posterior marginals for the corresponding model parameter. Both the univariate and multivariate marginals were normalized by their corresponding global maxima. Derived posterior point estimates, i.e. the maximum a posteriori (MAP) probability estimate $\mathbf{x}_{\mathrm{MAP}}$, the posterior mean $\mathbf{x}_{\mathrm{Mean}}$ and the posterior median $\mathbf{x}_{\mathrm{Median}}$, are indicated as well in each subfigure.

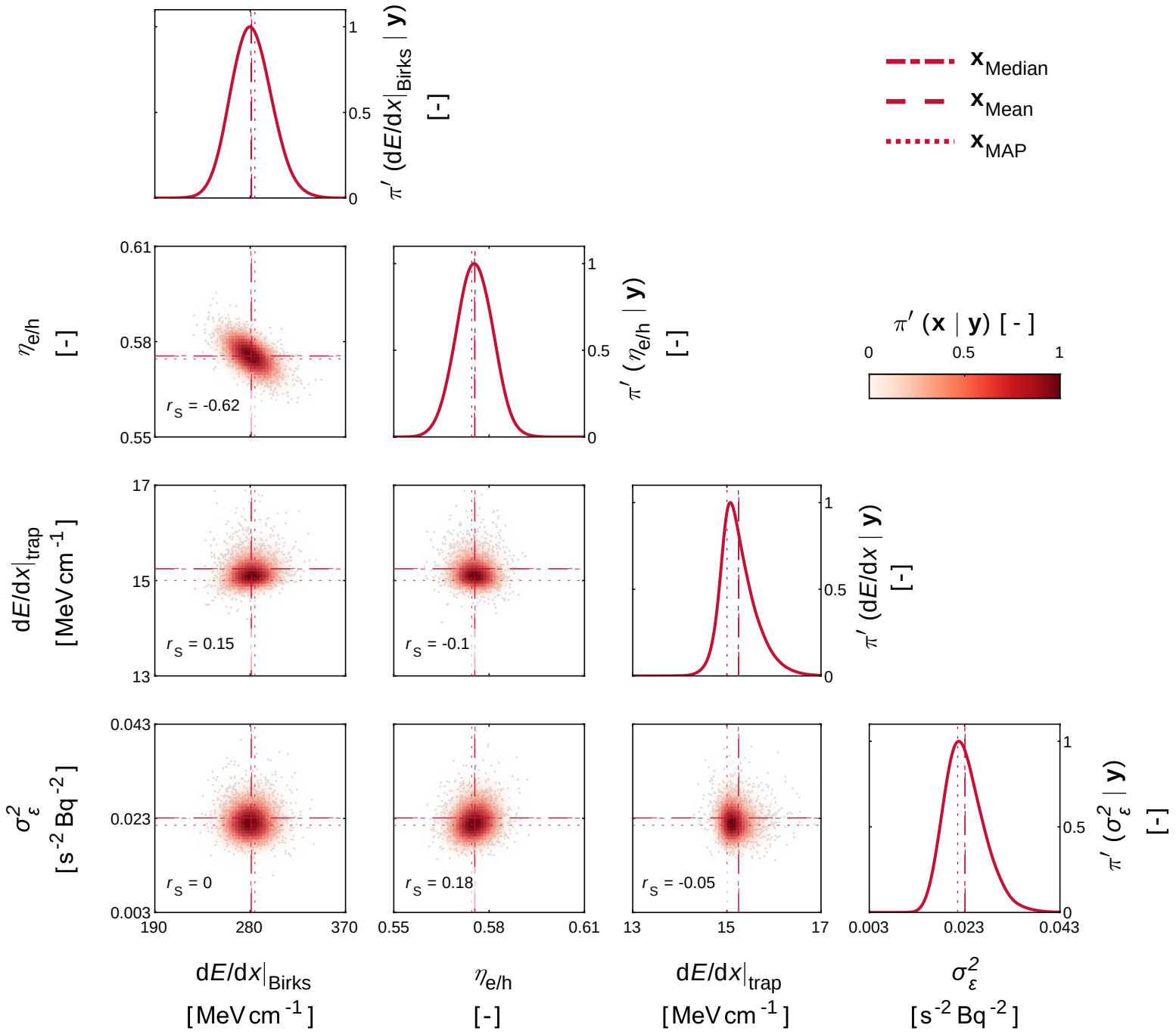

**Figure B.70** Here, I present the Bayesian inversion results for the detector channel #3 in the form of a corner plot. The off-diagonal subfigures display the normalized bivariate posterior marginal estimates (color-encoded) along with a subset of the first $10^3$ MCMC samples (in gray) for the model parameters $\mathbf{x} := (\,\mathrm{d}E/\mathrm{d}x\,|_{\mathrm{Birks}}, \eta_{\mathrm{e/h}},\, \mathrm{d}E/\mathrm{d}x\,|_{\mathrm{trap}}, \sigma_\varepsilon^2)^\top$ and experimental data $\mathbf{y}$. In addition, the Spearman's rank correlation coefficient $r_S$ is provided for the model parameters in the corresponding off-diagonal subfigures. The subfigures on the diagonal axis highlight the normalized univariate posterior marginals for the corresponding model parameter. Both the univariate and multivariate marginals were normalized by their corresponding global maxima. Derived posterior point estimates, i.e. the maximum a posteriori (MAP) probability estimate $\mathbf{x}_{\mathrm{MAP}}$, the posterior mean $\mathbf{x}_{\mathrm{Mean}}$ and the posterior median $\mathbf{x}_{\mathrm{Median}}$, are indicated as well in each subfigure.

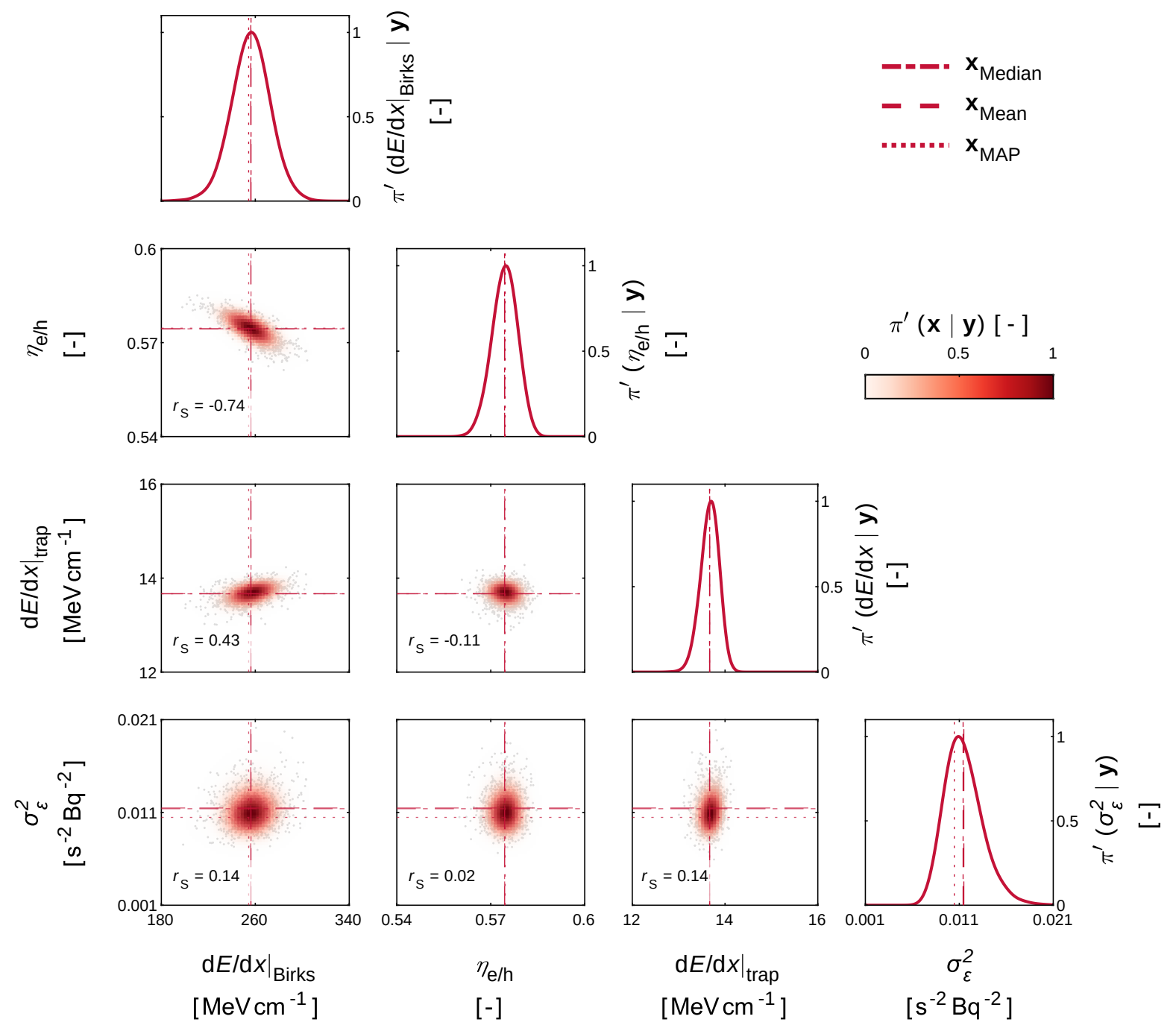

**Figure B.71** Here, I present the Bayesian inversion results for the detector channel #4 in the form of a corner plot. The off-diagonal subfigures display the normalized bivariate posterior marginal estimates (color-encoded) along with a subset of the first $10^3$ MCMC samples (in gray) for the model parameters $\mathbf{x} := ( \mathrm{d}E/\mathrm{d}x\,|_{\mathrm{Birks}}, \eta_{\mathrm{e/h}}, \mathrm{d}E/\mathrm{d}x\,|_{\mathrm{trap}}, \sigma_\varepsilon^2)^\top$ and experimental data $\mathbf{y}$. In addition, the Spearman's rank correlation coefficient $r_S$ is provided for the model parameters in the corresponding off-diagonal subfigures. The subfigures on the diagonal axis highlight the normalized univariate posterior marginals for the corresponding model parameter. Both the univariate and multivariate marginals were normalized by their corresponding global maxima. Derived posterior point estimates, i.e. the maximum a posteriori (MAP) probability estimate $\mathbf{x}_{\mathrm{MAP}}$, the posterior mean $\mathbf{x}_{\mathrm{Mean}}$ and the posterior median $\mathbf{x}_{\mathrm{Median}}$, are indicated as well in each subfigure.

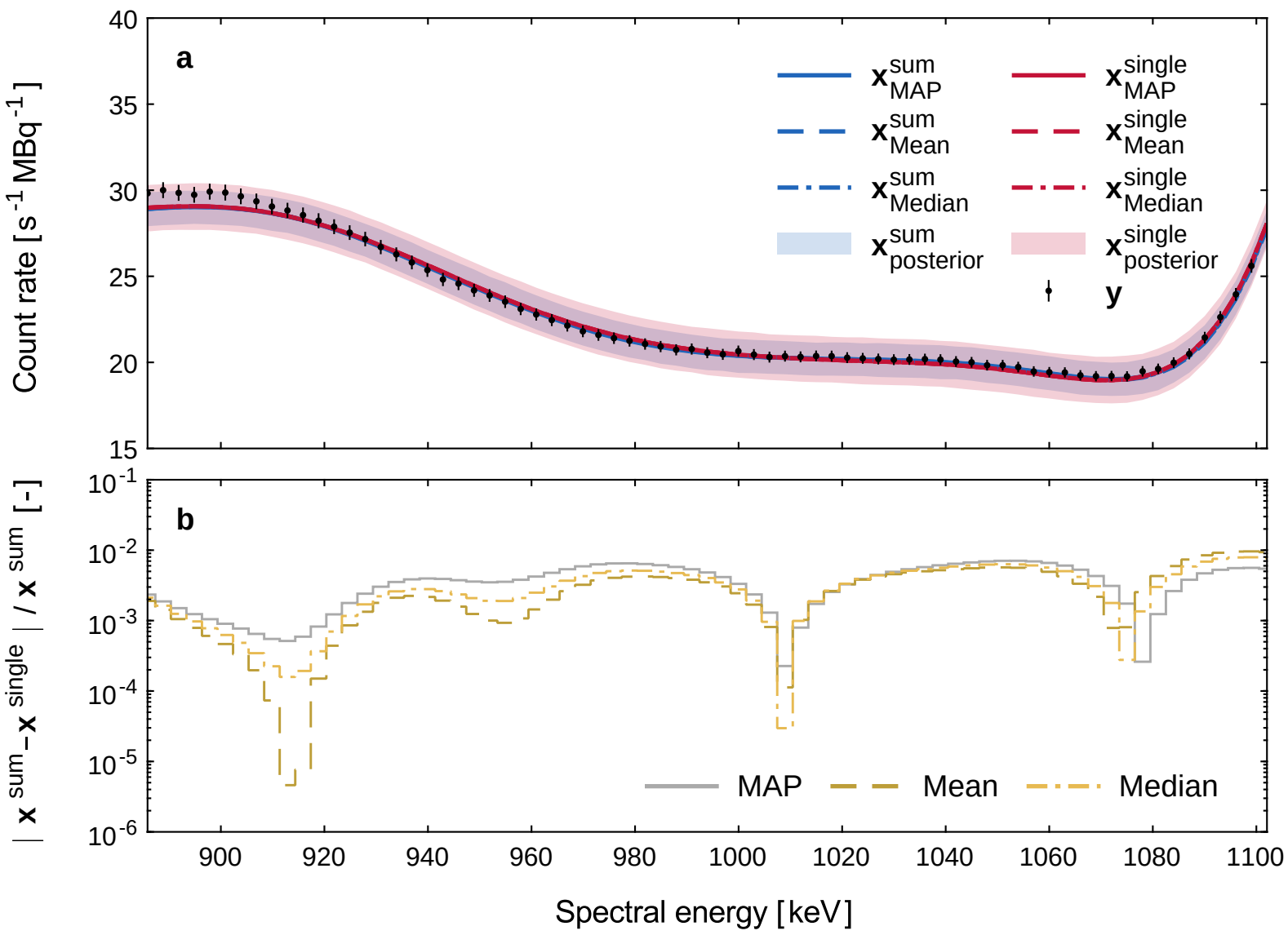

**Figure B.72** Here, I quantitatively compare the Bayesian predictive distributions of calibrated NPSMs derived by the four single detector channels `#1` through `#4` ("single") with that of the detector channel `#SUM` ("sum"). For that purpose, the posterior predictions for the individual scintillation crystals are summed together and compared to the predictions for the detector channel `#SUM`. **a** In this graph, I show the posterior predictive distributions as 99 % central prediction intervals. In addition, the experimental data **y** together with the derived posterior predictions using point estimates, i.e. the maximum a posteriori (MAP) probability estimate $\mathbf{x}_{\mathrm{MAP}}$, the posterior mean $\mathbf{x}_{\mathrm{Mean}}$ and the posterior median $\mathbf{x}_{\mathrm{Median}}$, are indicated. Experimental uncertainties are provided as 1 standard deviation (SD) values. **b** In this subfigure, I present the relative difference between the posterior predictions for the four single detector channels `#1` through `#4` ("single") and the detector channel `#SUM` ("sum") using the three point estimates $\mathbf{x}_{\mathrm{MAP}}$, $\mathbf{x}_{\mathrm{Mean}}$ and $\mathbf{x}_{\mathrm{Median}}$.

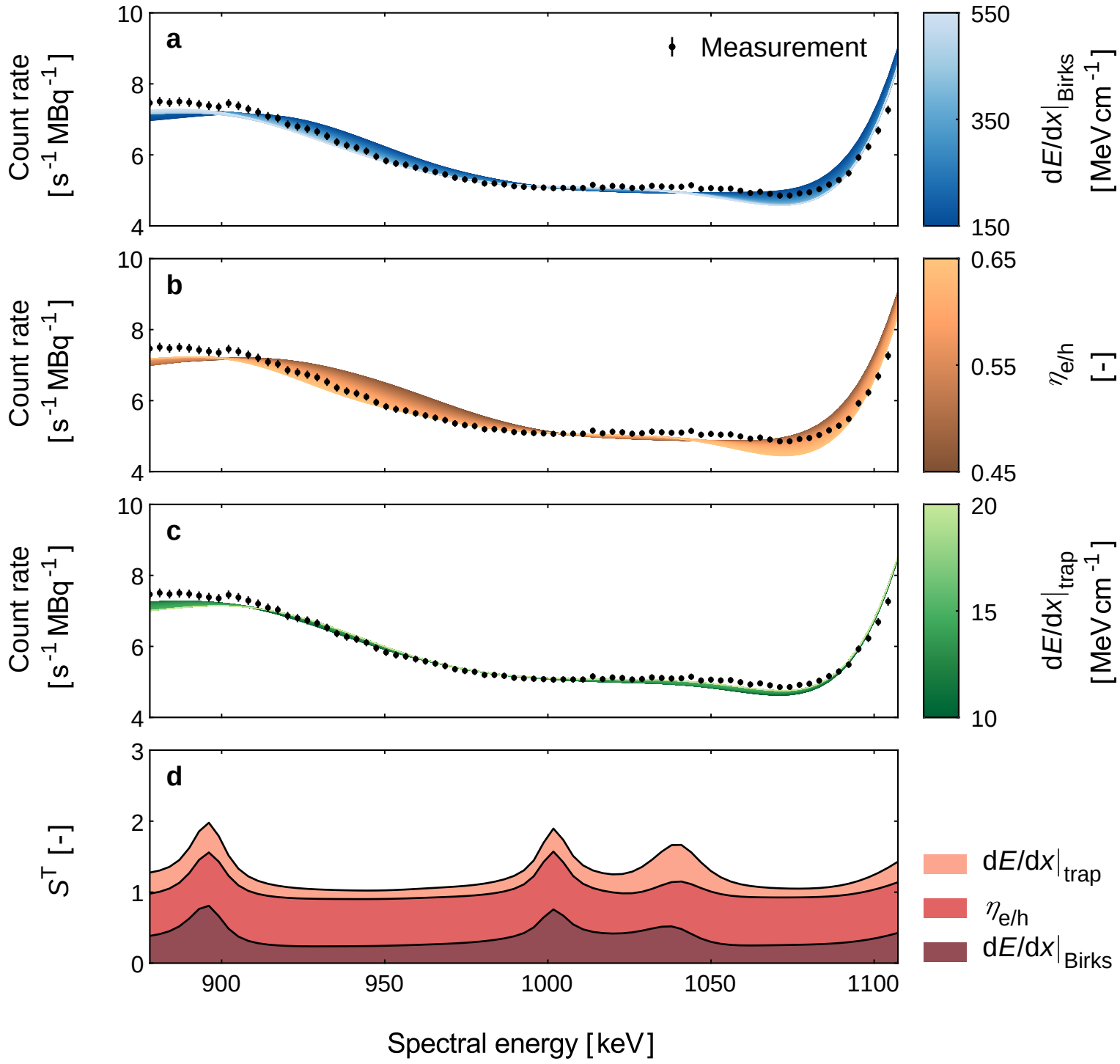

**Figure B.73** Compton edge dynamics characterized by the trained PCE surrogate model for the detector channel #1. **a–c** In these subgraphs, the PCE surrogate model predictions are displayed as a function of the spectral energy $E'$ and the individual NPSM parameters, i.e. the Birks stopping parameter $\mathrm{d}E/\mathrm{d}x|_{\mathrm{Birks}}$, the electron-hole pair fraction $\eta_{\mathrm{e/h}}$ as well as the trapping stopping parameter $\mathrm{d}E/\mathrm{d}x|_{\mathrm{trap}}$. The remaining parameters are fixed at the corresponding maximum a posteriori (MAP) probability estimates. The measured $^{60}_{27}$Co spectral signature for the detector channel #1 (cf. Chapter 6) is indicated as well as a reference. **d** In this subgraph, I show stacked total Sobol' indices $S^{\mathrm{T}}$ computed by the trained PCE surrogate model [742] as a function of the spectral energy $E'$ and the individual NPSM parameters.

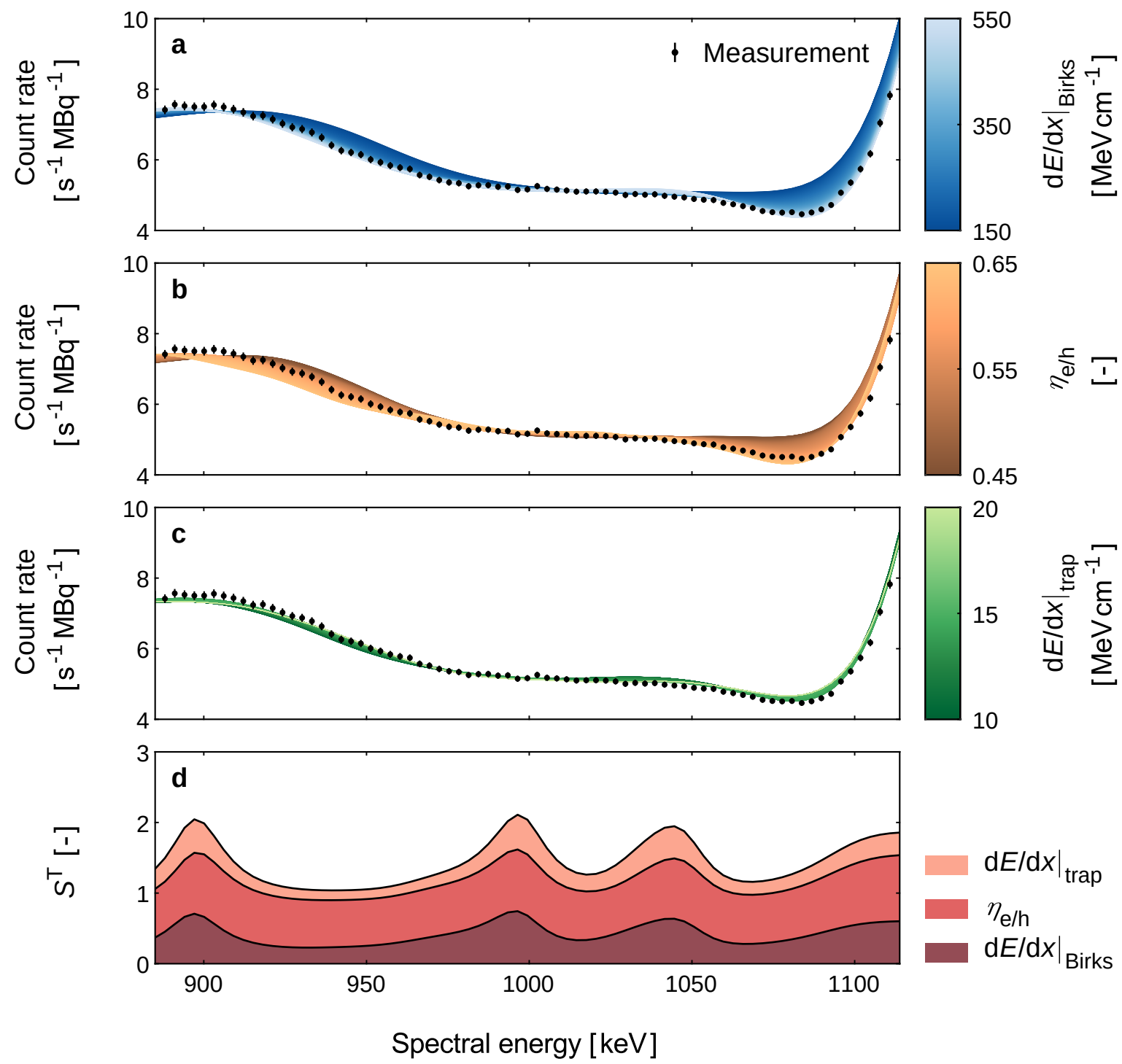

**Figure B.74** Compton edge dynamics characterized by the trained PCE surrogate model for the detector channel #2. **a–c** In these subgraphs, the PCE surrogate model predictions are displayed as a function of the spectral energy $E'$ and the individual NPSM parameters, i.e. the Birks stopping parameter $\mathrm{d}E/\mathrm{d}x\,|_{\mathrm{Birks}}$, the electron-hole pair fraction $\eta_{\mathrm{e/h}}$ as well as the trapping stopping parameter $\mathrm{d}E/\mathrm{d}x\,|_{\mathrm{trap}}$. The remaining parameters are fixed at the corresponding maximum a posteriori (MAP) probability estimates. The measured $^{60}_{27}\mathrm{Co}$ spectral signature for the detector channel #2 (cf. Chapter 6) is indicated as well as a reference. **d** In this subgraph, I show stacked total Sobol′ indices $S^{\mathrm{T}}$ computed by the trained PCE surrogate model [742] as a function of the spectral energy $E'$ and the individual NPSM parameters.

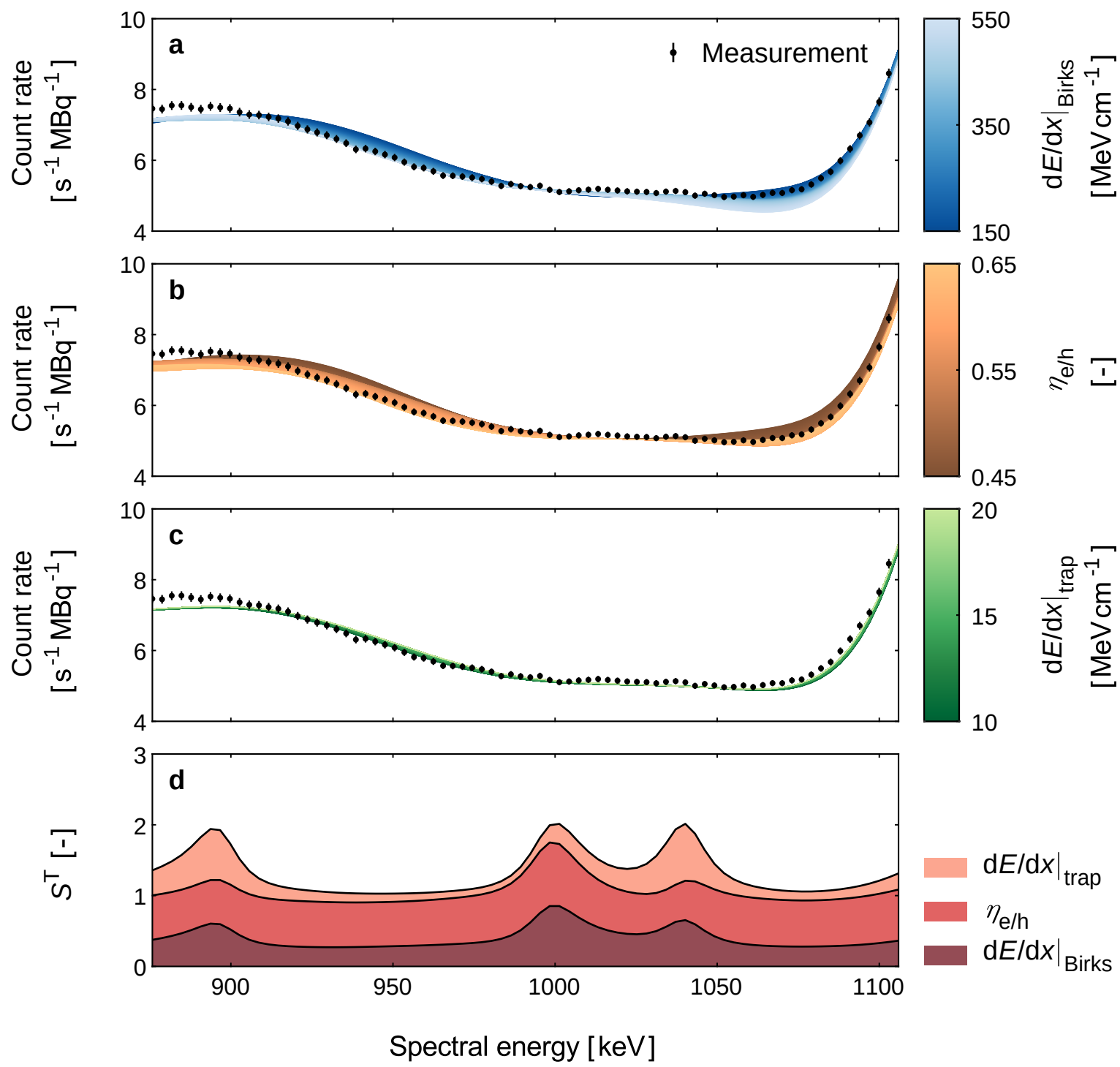

**Figure B.75** Compton edge dynamics characterized by the trained PCE surrogate model for the detector channel #3. **a–c** In these subgraphs, the PCE surrogate model predictions are displayed as a function of the spectral energy $E'$ and the individual NPSM parameters, i.e. the Birks stopping parameter $\mathrm{d}E/\mathrm{d}x|_{\mathrm{Birks}}$, the electron-hole pair fraction $\eta_{\mathrm{e/h}}$ as well as the trapping stopping parameter $\mathrm{d}E/\mathrm{d}x|_{\mathrm{trap}}$. The remaining parameters are fixed at the corresponding maximum a posteriori (MAP) probability estimates. The measured $^{60}_{27}\mathrm{Co}$ spectral signature for the detector channel #3 (cf. Chapter 6) is indicated as well as a reference. **d** In this subgraph, I show stacked total Sobol' indices $S^{\mathrm{T}}$ computed by the trained PCE surrogate model [742] as a function of the spectral energy $E'$ and the individual NPSM parameters.

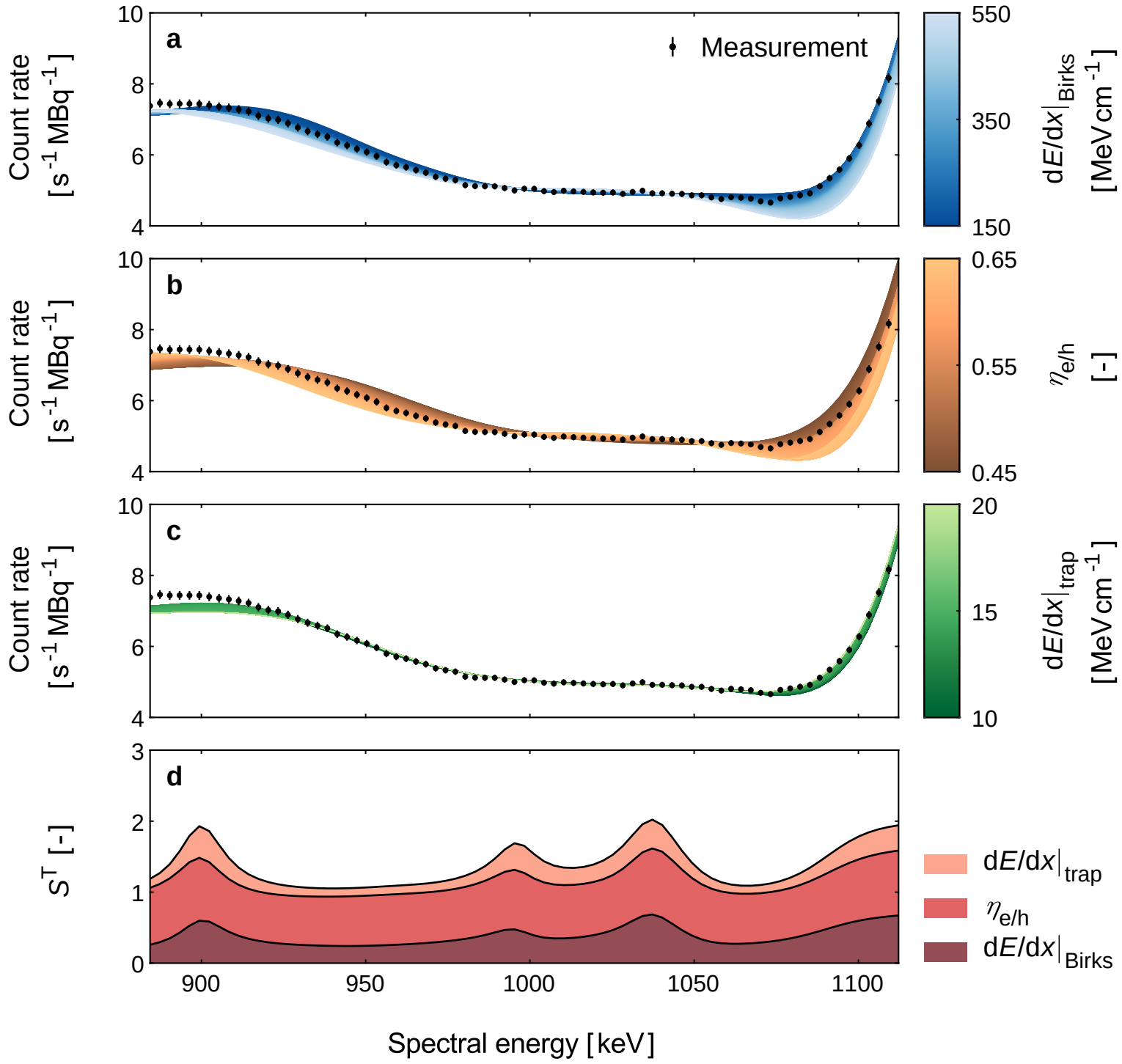

**Figure B.76** Compton edge dynamics characterized by the trained PCE surrogate model for the detector channel #4. **a–c** In these subgraphs, the PCE surrogate model predictions are displayed as a function of the spectral energy $E'$ and the individual NPSM parameters, i.e. the Birks stopping parameter $\mathrm{d}E/\mathrm{d}x|_{\mathrm{Birks}}$, the electron-hole pair fraction $\eta_{\mathrm{e/h}}$ as well as the trapping stopping parameter $\mathrm{d}E/\mathrm{d}x|_{\mathrm{trap}}$. The remaining parameters are fixed at the corresponding maximum a posteriori (MAP) probability estimates. The measured $^{60}_{27}$Co spectral signature for the detector channel #4 (cf. Chapter 6) is indicated as well as a reference. **d** In this subgraph, I show stacked total Sobol′ indices $S^{\mathrm{T}}$ computed by the trained PCE surrogate model [742] as a function of the spectral energy $E'$ and the individual NPSM parameters.

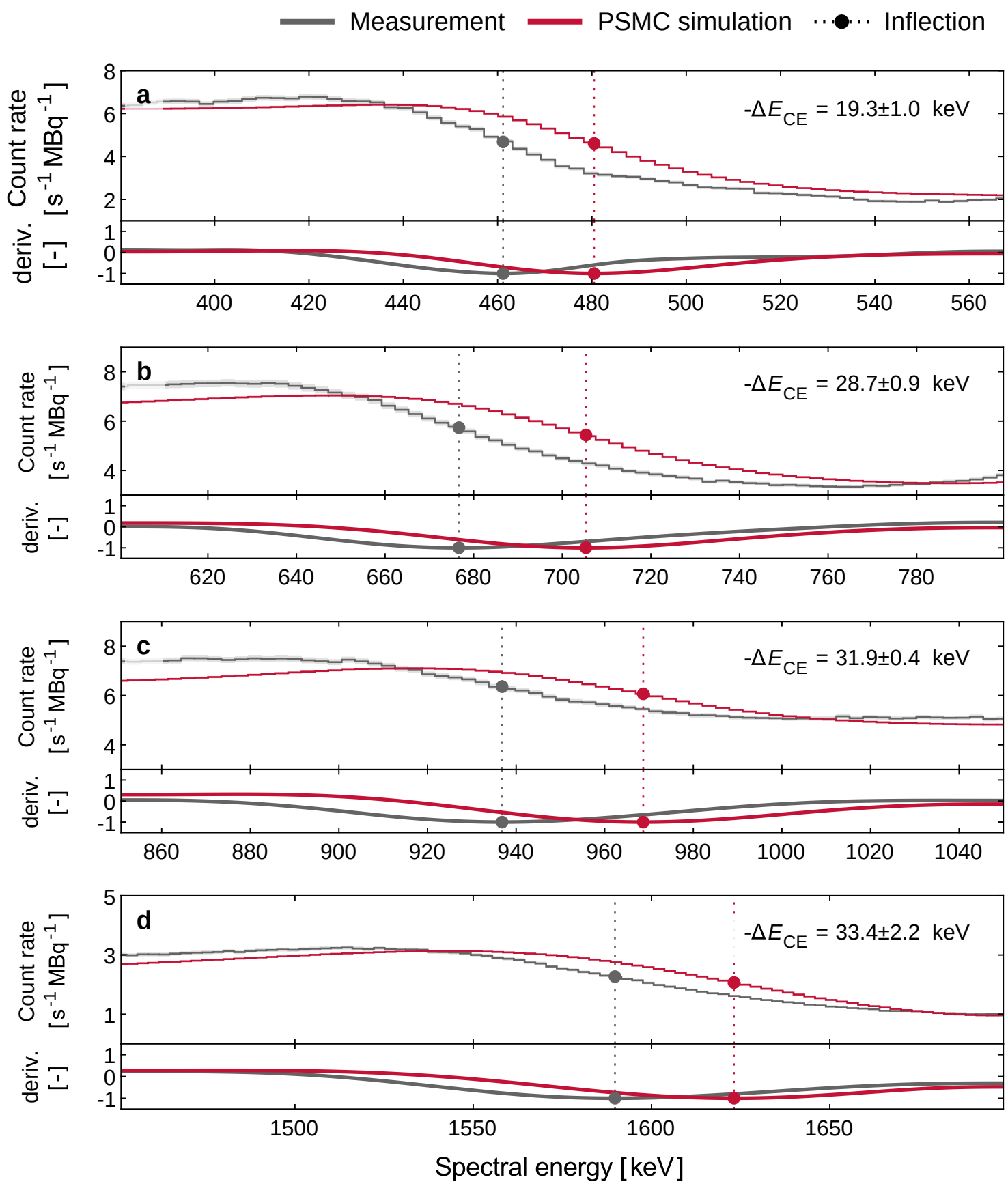

**Figure B.77** Here, the results from the Compton edge shift analysis of four different Compton edges are presented for the detector channel `#1`. These Compton edges are: **a** 477.334(3) keV associated with the $^{137}_{55}$Cs emission line at 661.657(3) keV [68]. **b** 699.133(3) keV associated with the $^{88}_{39}$Y emission line at 898.042(11) keV [51]. **c** 963.419(3) keV associated with the $^{60}_{27}$Co emission line at 1173.228(3) keV [68]. **d** 1611.77(1) keV associated with the $^{88}_{39}$Y emission line at 1836.070(8) keV [51]. The top panels present the measured ($\hat{c}_{exp}$) and simulated ($\hat{c}_{sim}$, derived by PSMC discussed in Chapter 6) mean spectral signatures for the corresponding sources around the Compton edges of interest. The bottom panels show the first derivative (deriv.) of the measured and simulated mean spectral signatures estimated by spline regression [891]. For visualization purposes, the first derivatives were normalized by their corresponding global minima. The Compton edge shift $\Delta E_{CE}$ was estimated as the spectral difference between the indicated inflection points of the measured and simulated spectral signatures (cf. Appendix A.10). Uncertainties are provided as 1 standard deviation (SD) shaded areas (spectral signatures) or numerical values (Compton edge shift)..

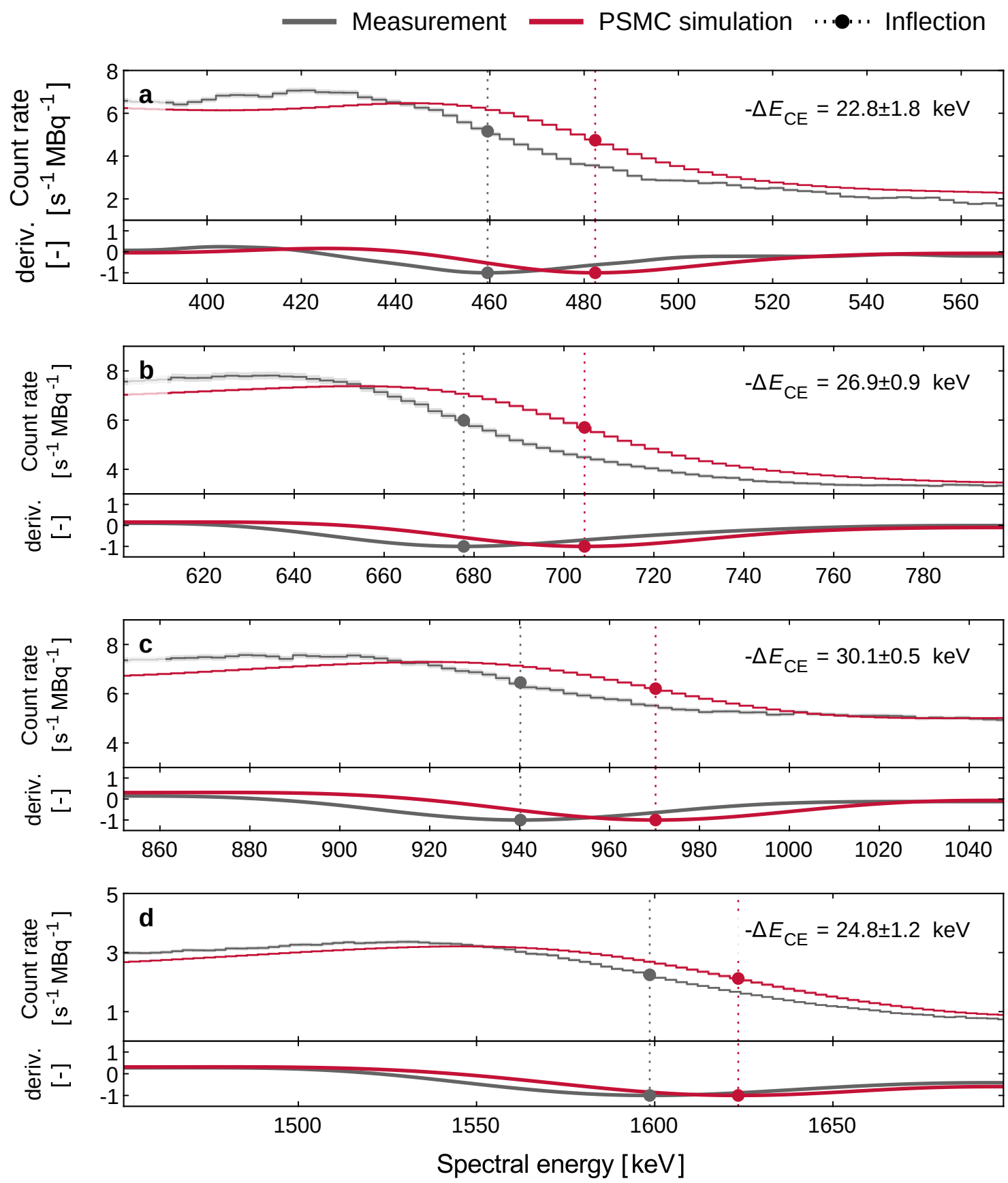

**Figure B.78** Here, the results from the Compton edge shift analysis of four different Compton edges are presented for the detector channel #2. These Compton edges are: **a** 477.334(3) keV associated with the $^{137}_{55}$Cs emission line at 661.657(3) keV [68]. **b** 699.133(3) keV associated with the $^{88}_{39}$Y emission line at 898.042(11) keV [51]. **c** 963.419(3) keV associated with the $^{60}_{27}$Co emission line at 1173.228(3) keV [68]. **d** 1611.77(1) keV associated with the $^{88}_{39}$Y emission line at 1836.070(8) keV [51]. The top panels present the measured ($\hat{c}_{exp}$) and simulated ($\hat{c}_{sim}$, derived by PSMC discussed in Chapter 6) mean spectral signatures for the corresponding sources around the Compton edges of interest. The bottom panels show the first derivative (deriv.) of the measured and simulated mean spectral signatures estimated by spline regression [891]. For visualization purposes, the first derivatives were normalized by their corresponding global minima. The Compton edge shift $\Delta E_{CE}$ was estimated as the spectral difference between the indicated inflection points of the measured and simulated spectral signatures (cf. Appendix A.10). Uncertainties are provided as 1 standard deviation (SD) shaded areas (spectral signatures) or numerical values (Compton edge shift).

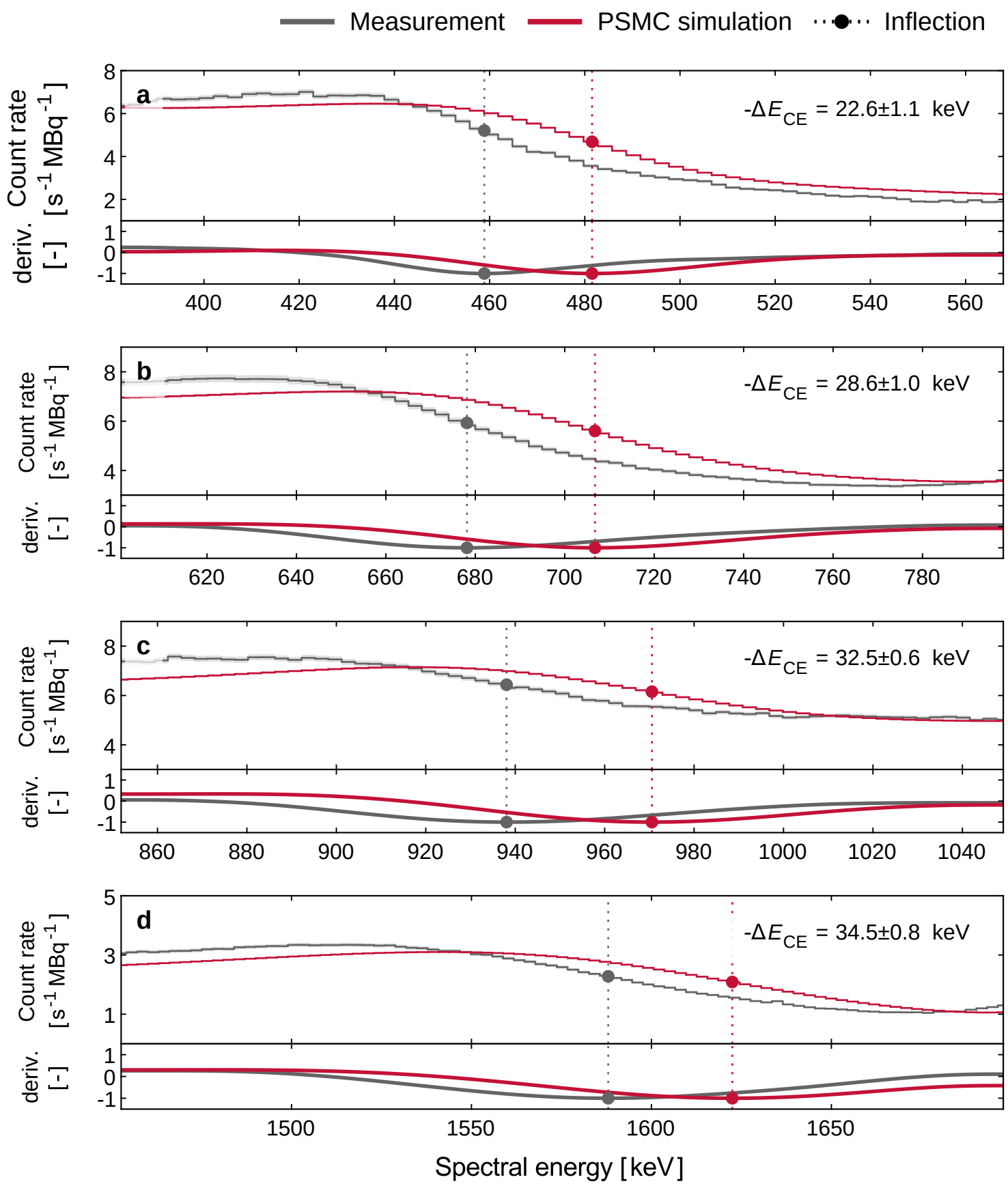

**Figure B.79** Here, the results from the Compton edge shift analysis of four different Compton edges are presented for the detector channel #3. These Compton edges are: **a** 477.334(3) keV associated with the $^{137}_{55}Cs$ emission line at 661.657(3) keV [68]. **b** 699.133(3) keV associated with the $^{88}_{39}Y$ emission line at 898.042(11) keV [51]. **c** 963.419(3) keV associated with the $^{60}_{27}Co$ emission line at 1173.228(3) keV [68]. **d** 1611.77(1) keV associated with the $^{88}_{39}Y$ emission line at 1836.070(8) keV [51]. The top panels present the measured ($\hat{c}_{exp}$) and simulated ($\hat{c}_{sim}$, derived by PSMC discussed in Chapter 6) mean spectral signatures for the corresponding sources around the Compton edges of interest. The bottom panels show the first derivative (deriv.) of the measured and simulated mean spectral signatures estimated by spline regression [891]. For visualization purposes, the first derivatives were normalized by their corresponding global minima. The Compton edge shift $\Delta E_{CE}$ was estimated as the spectral difference between the indicated inflection points of the measured and simulated spectral signatures (cf. Appendix A.10). Uncertainties are provided as 1 standard deviation (SD) shaded areas (spectral signatures) or numerical values (Compton edge shift).

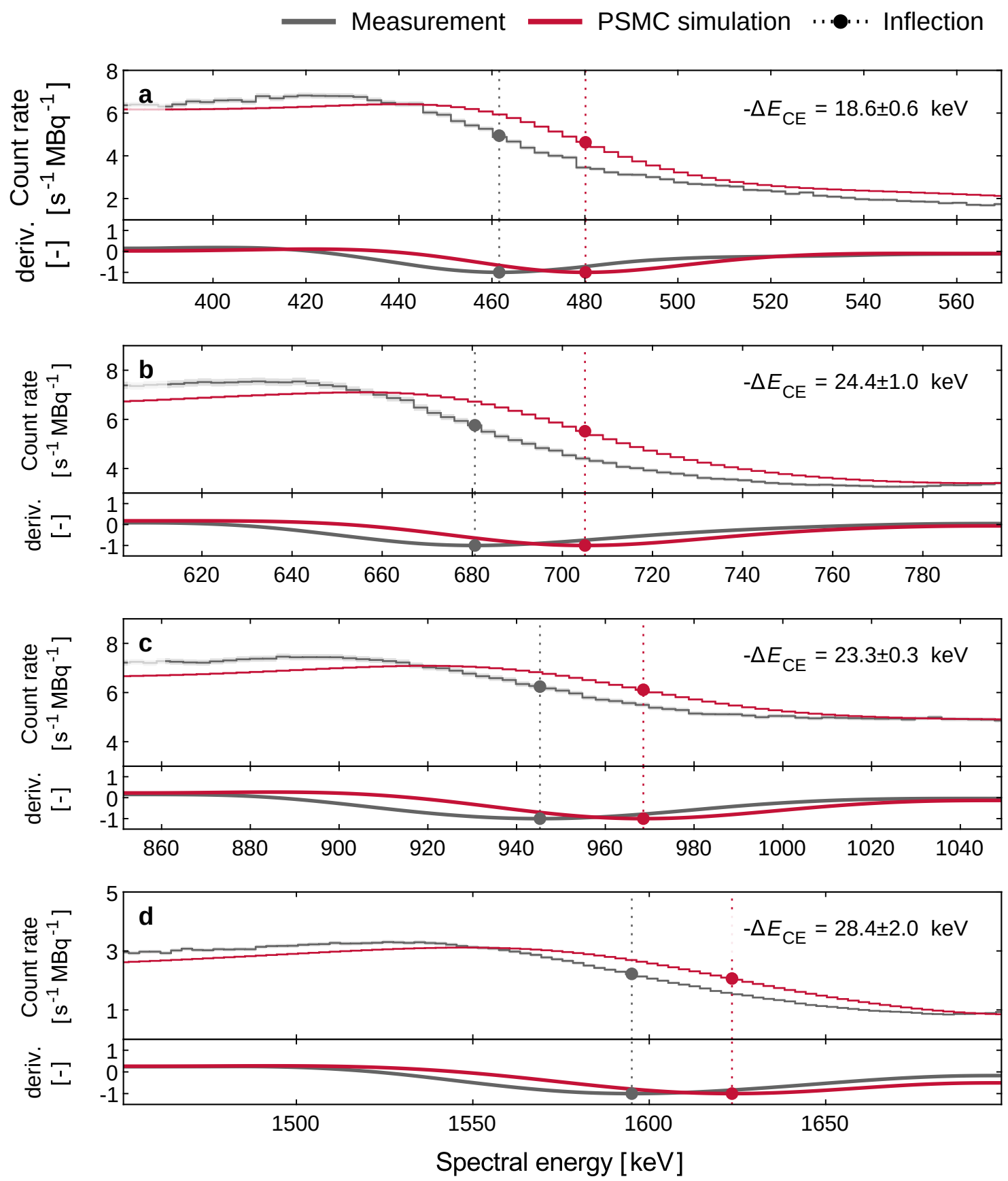

**Figure B.80** Here, the results from the Compton edge shift analysis of four different Compton edges are presented for the detector channel #4. These Compton edges are: **a** 477.334(3) keV associated with the $^{137}_{55}$Cs emission line at 661.657(3) keV [68]. **b** 699.133(3) keV associated with the $^{88}_{39}$Y emission line at 898.042(11) keV [51]. **c** 963.419(3) keV associated with the $^{60}_{27}$Co emission line at 1173.228(3) keV [68]. **d** 1611.77(1) keV associated with the $^{88}_{39}$Y emission line at 1836.070(8) keV [51]. The top panels present the measured ($\hat{c}_{\text{exp}}$) and simulated ($\hat{c}_{\text{sim}}$, derived by PSMC discussed in Chapter 6) mean spectral signatures for the corresponding sources around the Compton edges of interest. The bottom panels show the first derivative (deriv.) of the measured and simulated mean spectral signatures estimated by spline regression [891]. For visualization purposes, the first derivatives were normalized by their corresponding global minima. The Compton edge shift $\Delta E_{\text{CE}}$ was estimated as the spectral difference between the indicated inflection points of the measured and simulated spectral signatures (cf. Appendix A.10). Uncertainties are provided as 1 standard deviation (SD) shaded areas (spectral signatures) or numerical values (Compton edge shift).

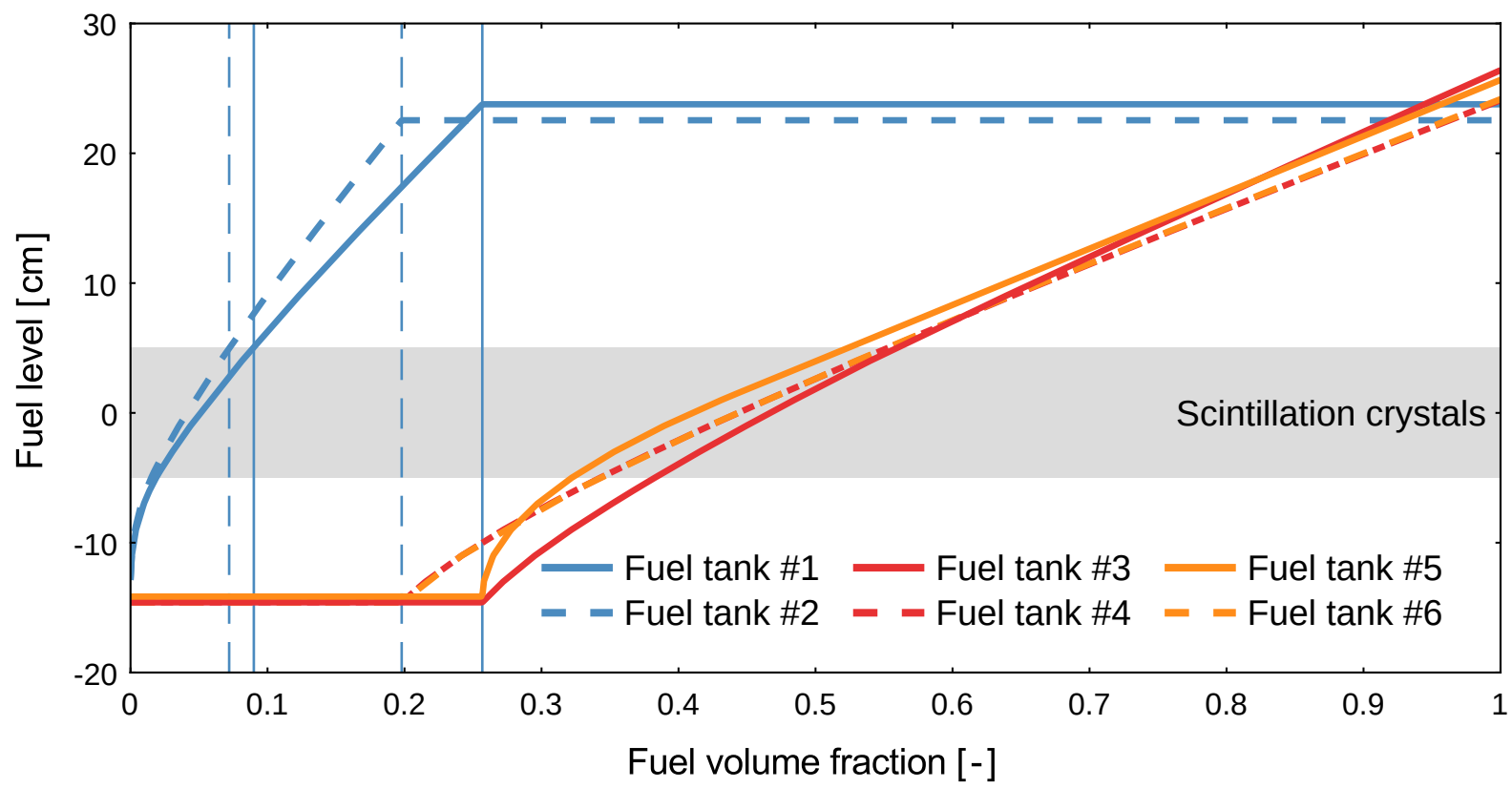

**Figure B.81** In this graph, I present the fuel level dynamics of the individual tanks 1 to 6 of the Aérospatiale AS332M1 Super Puma (TH06) helicopter as a function of the fuel volume fraction $\varrho_{\mathrm{JF}}$ (total capacity $1.988\,\mathrm{m}^3$, cf. Section 8.2). The same numbering of the fuel tanks was used as in Fig. 8.1. In addition, the diagram indicates the relative position of the scintillation crystals with respect to the fuel level in a horizontal flight configuration.

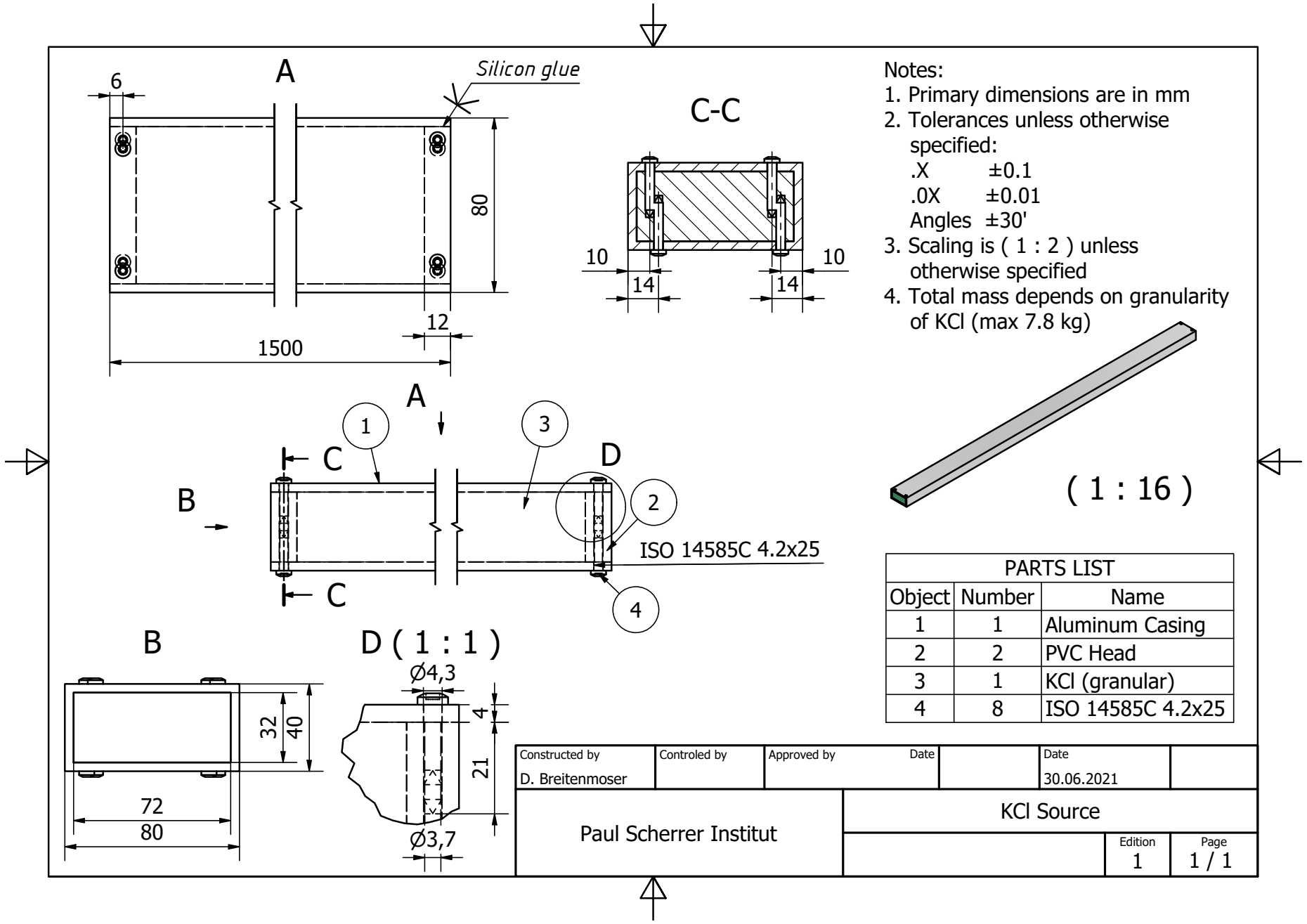

**Figure B.82** Technical drawing of the custom-made $K_{nat}$ sources deployed during the Dübendorf validation campaign (cf. Section 8.3.1.1).

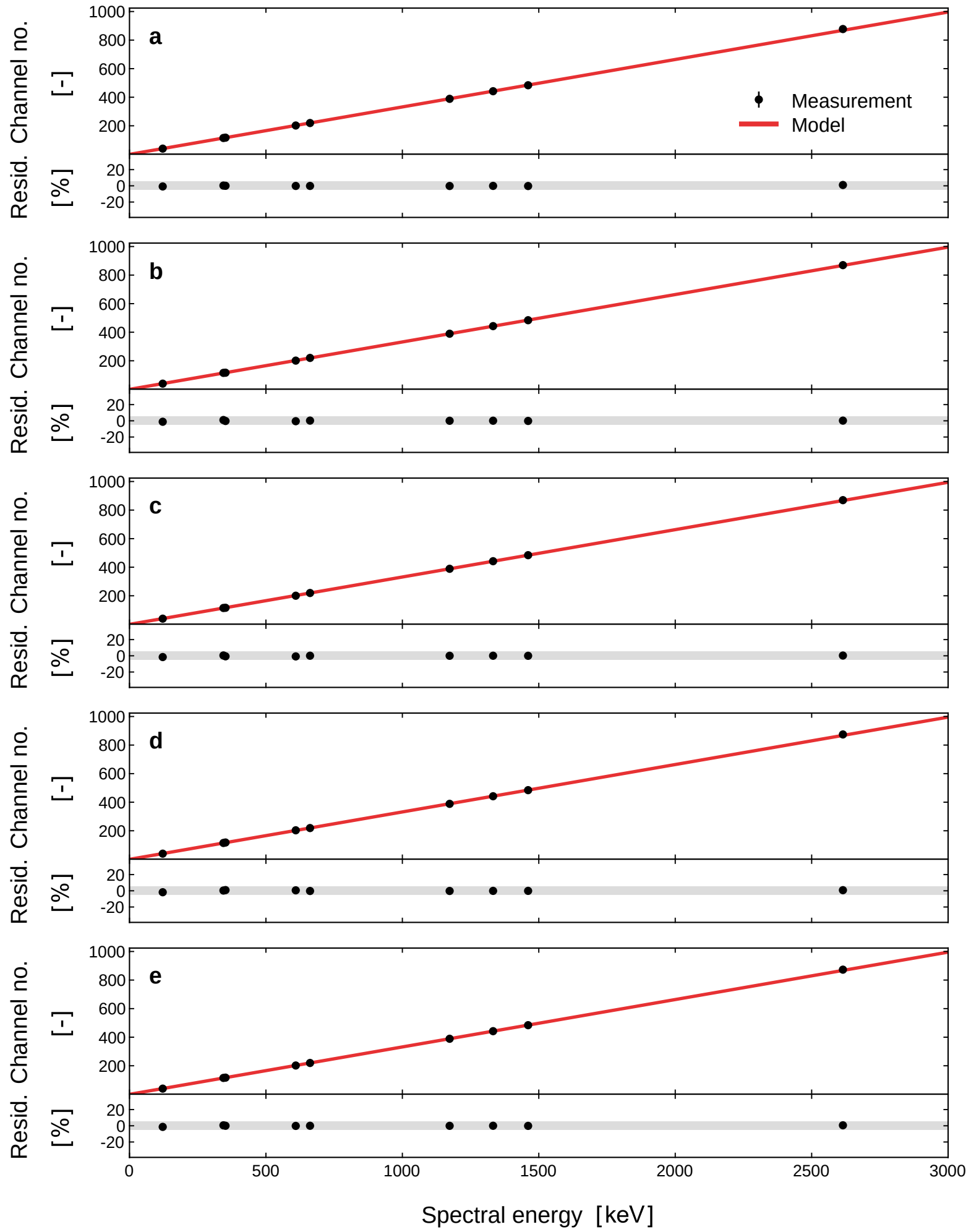

**Figure B.83** Energy calibration models adopted for the field measurements described in Chapter 8 relating the spectral energy $E'$ to the continuous pulse-height channel number $\tilde{n}$ for the four single detector channels `#1` through `#4` (**a–d**) and the detector channel `#SUM` (**e**). Measurement uncertainties are provided as 1 standard deviation (SD) error bars (hidden by the marker size). The model uncertainty is characterized by 99 % prediction intervals displayed as red-shaded areas (hidden by the line width). In addition, the relative residuals (Resid., normalized by the mean measured values) are provided for each detector channel with the 5 % band highlighted.

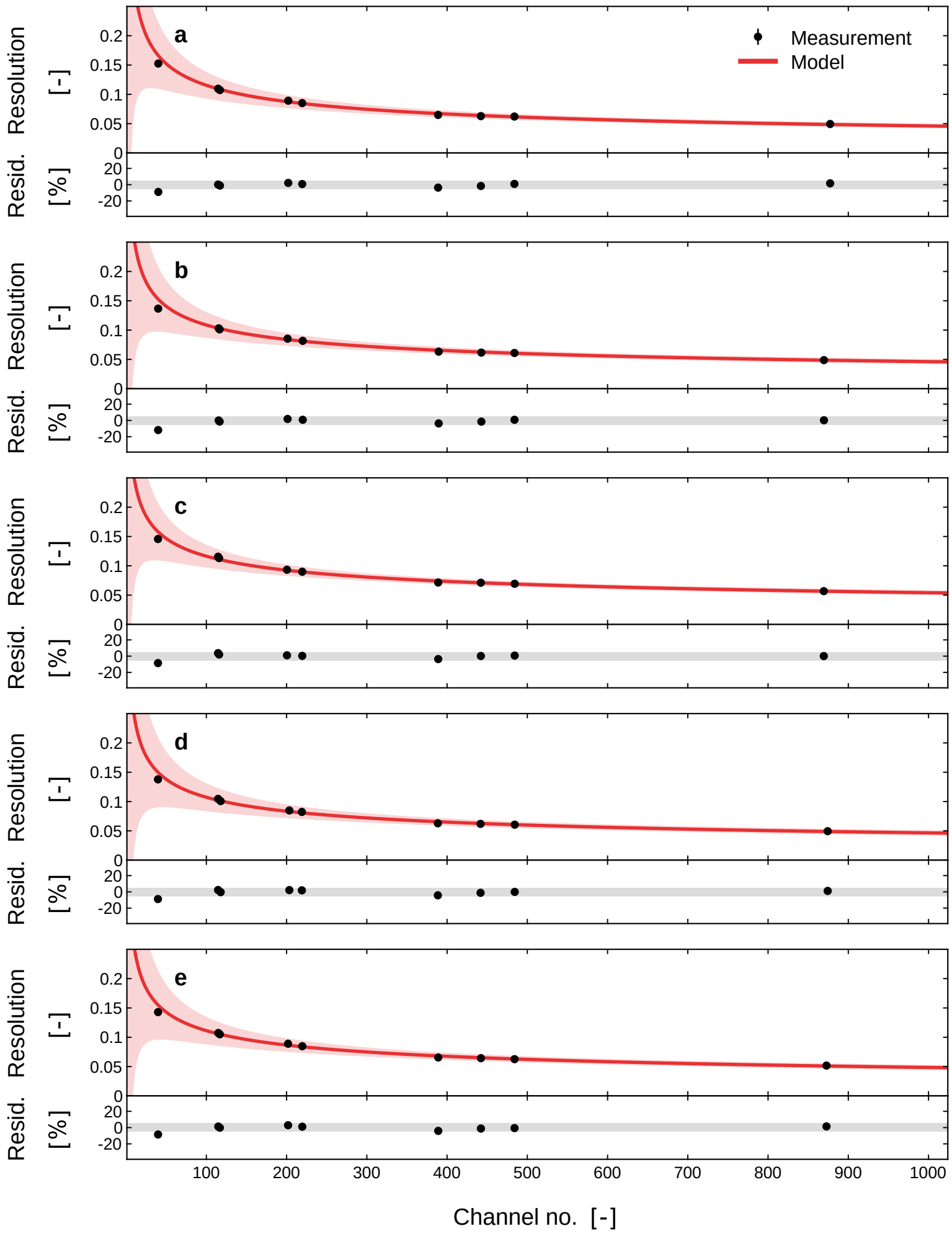

**Figure B.84** Spectral resolution models relating the continuous pulse-height channel number $\tilde{n}$ to the spectral resolution $R_E$ for the four single detector channels `#1` through `#4` (**a**–**d**) and the detector channel `#SUM` (**e**). Measurement uncertainties are provided as 1 standard deviation (SD) error bars (hidden by the marker size). The model uncertainty is characterized by 99 % prediction intervals displayed as red-shaded areas. In addition, the relative residuals (Resid., normalized by the mean measured values) are provided for each detector channel with the 5 % band highlighted.

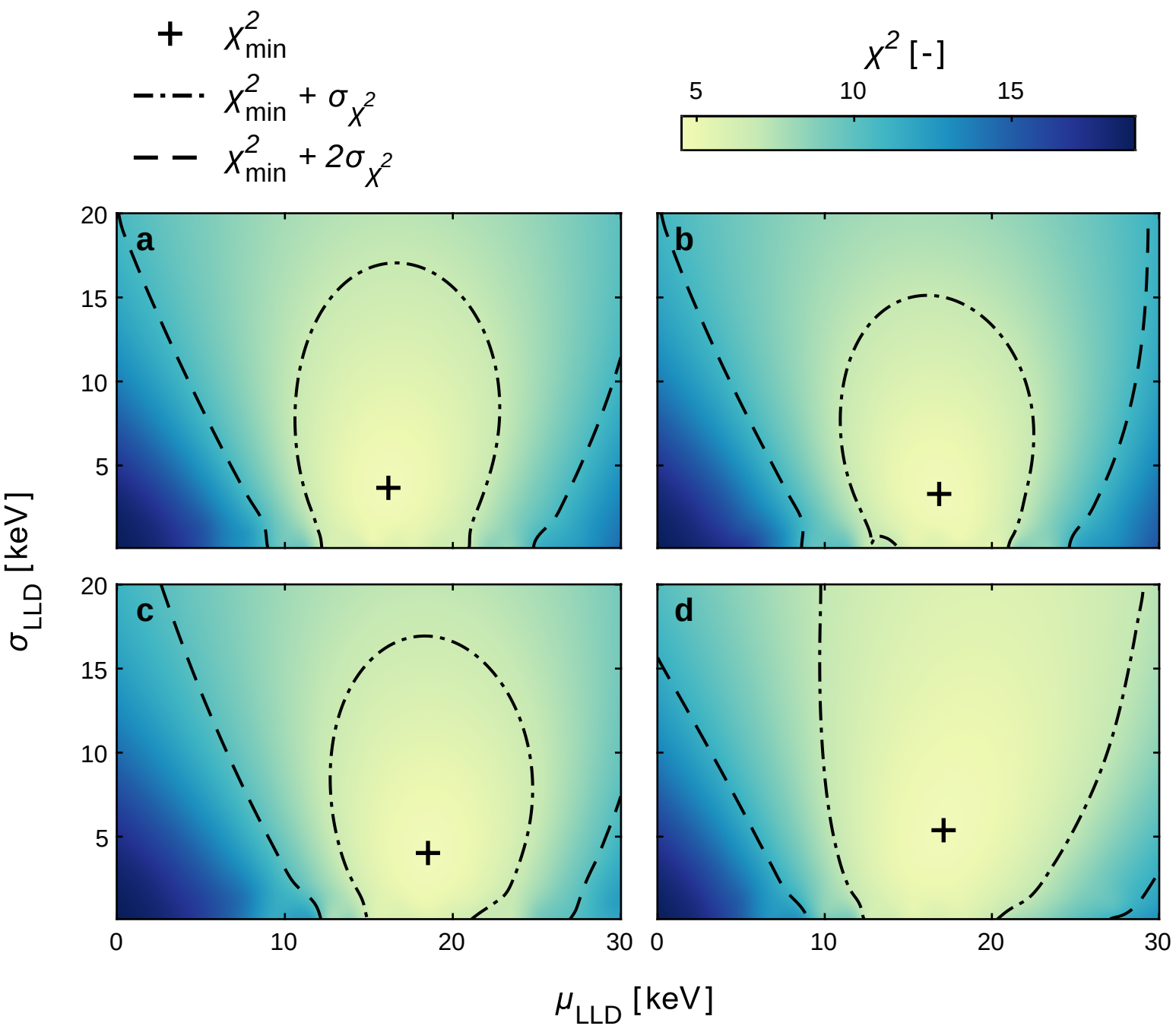

**Figure B.85** Here, the results of the LLD calibration described in Appendix A.9 are presented as chi-squared $\chi^2$ contour maps of the lower level discriminator model parameters $\mu_{LLD}$ and $\sigma_{LLD}$ for the four detector channels: **a** Detector channel #1. **b** Detector channel #2. **c** Detector channel #3. **d** Detector channel #4. The global minima $\chi^2_{min}$ are marked with a plus sign together with the corresponding 1-sigma and 2-sigma confidence regions indicated by the dashed-dotted and dashed contour lines, respectively. For the reader's convenience, the unit of the lower level discriminator model parameters $\mu_{LLD}$ and $\sigma_{LLD}$ was converted from continuous pulse-height channel numbers to spectral energies using the energy calibration models derived by the `RLLCal` pipeline. For the calibration, the laboratory-based radiation measurement with a $^{60}_{27}$Co calibration point source was adopted. The measured and simulated spectral signatures were derived by the `RLLSpec` and `NPScinMC` pipelines, respectively.

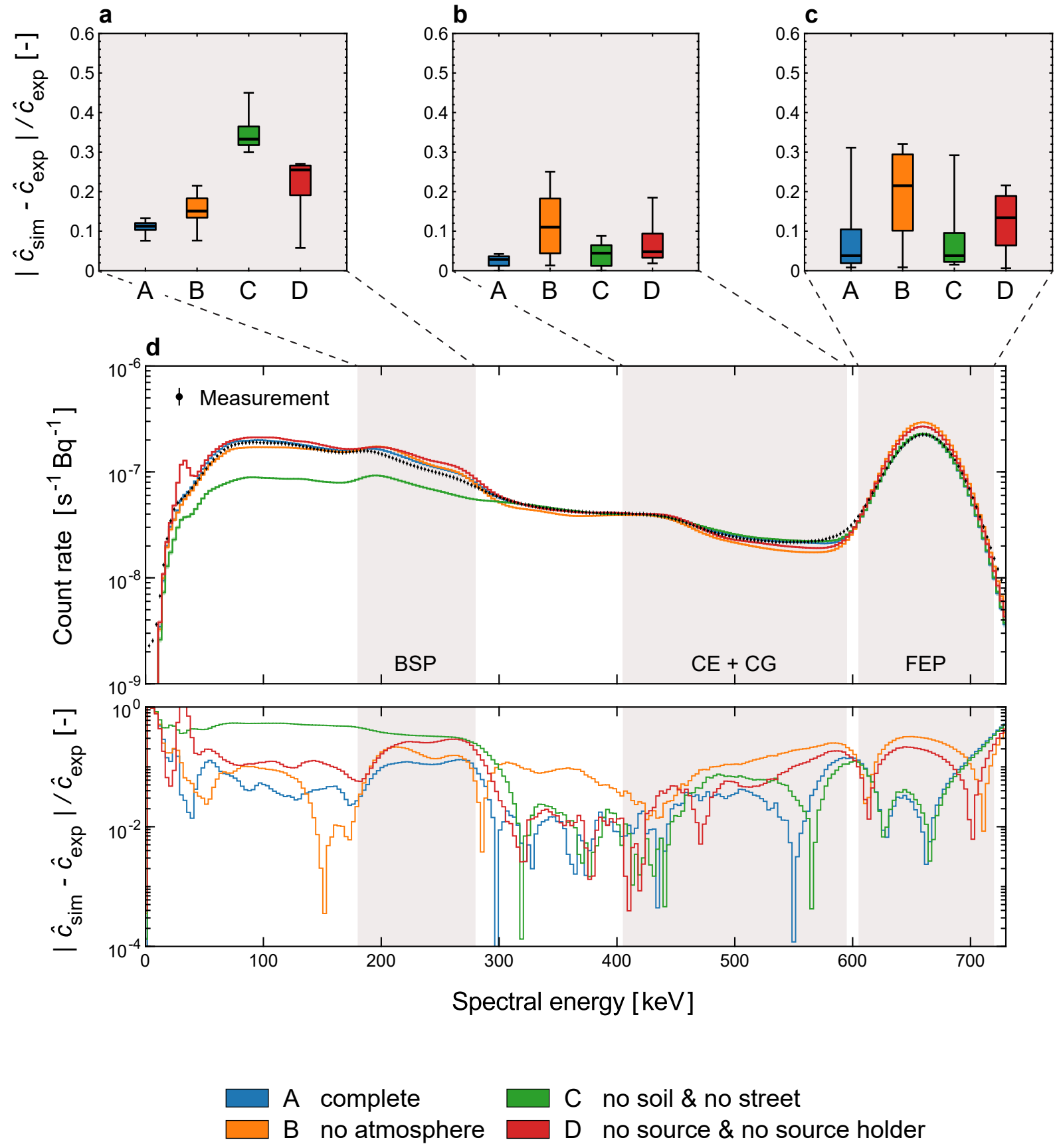

**Figure B.86** Mass model sensitivity analysis for the spectral signature of a $^{137}_{55}$Cs point source ($\mathcal{A}$ = 9.0(5) × $10^9$ Bq) measured by the RLL system (detector channel `#SUM`) in hover flight mode during the ARM22 validation campaign with $h_{air}$ = 31 m. **a–c** Adjusted box plots [721, 722] characterizing the statistical distribution of the relative deviations $|\hat{c}_{sim} - \hat{c}_{exp}|/\hat{c}_{exp}$ between the experimental ($\hat{c}_{exp}$) and the simulated spectral signatures ($\hat{c}_{sim}$) are shown for three different spectral domains (BSP: backscatter peak, CE+CG: Compton edge together with the Compton gap, FEP: full energy peak, cf. also to Section 4.3.1) and four different mass models A–D (cf. Section 8.3.2.3). The statistical analysis was performed using the Library for Robust Analysis (`LIBRA`) [723, 724]. **d** The measured ($\hat{c}_{exp}$) and simulated ($\hat{c}_{sim}$) mean spectral signatures for $^{137}_{55}$Cs are displayed as a function of the spectral energy $E'$ with a spectral energy bin width of $\Delta E' \sim 3$ keV. Spectral domains evaluated in **a–c** are highlighted. Uncertainties ($\hat{\sigma}_{exp}$, $\hat{\sigma}_{sim}$) are provided as 1 standard deviation (SD) shaded areas (cf. Appendix A.8). In addition, the relative deviation $|\hat{c}_{sim} - \hat{c}_{exp}|/\hat{c}_{exp}$ as a function of the spectral energy $E'$ is displayed for all four mass models A–D.

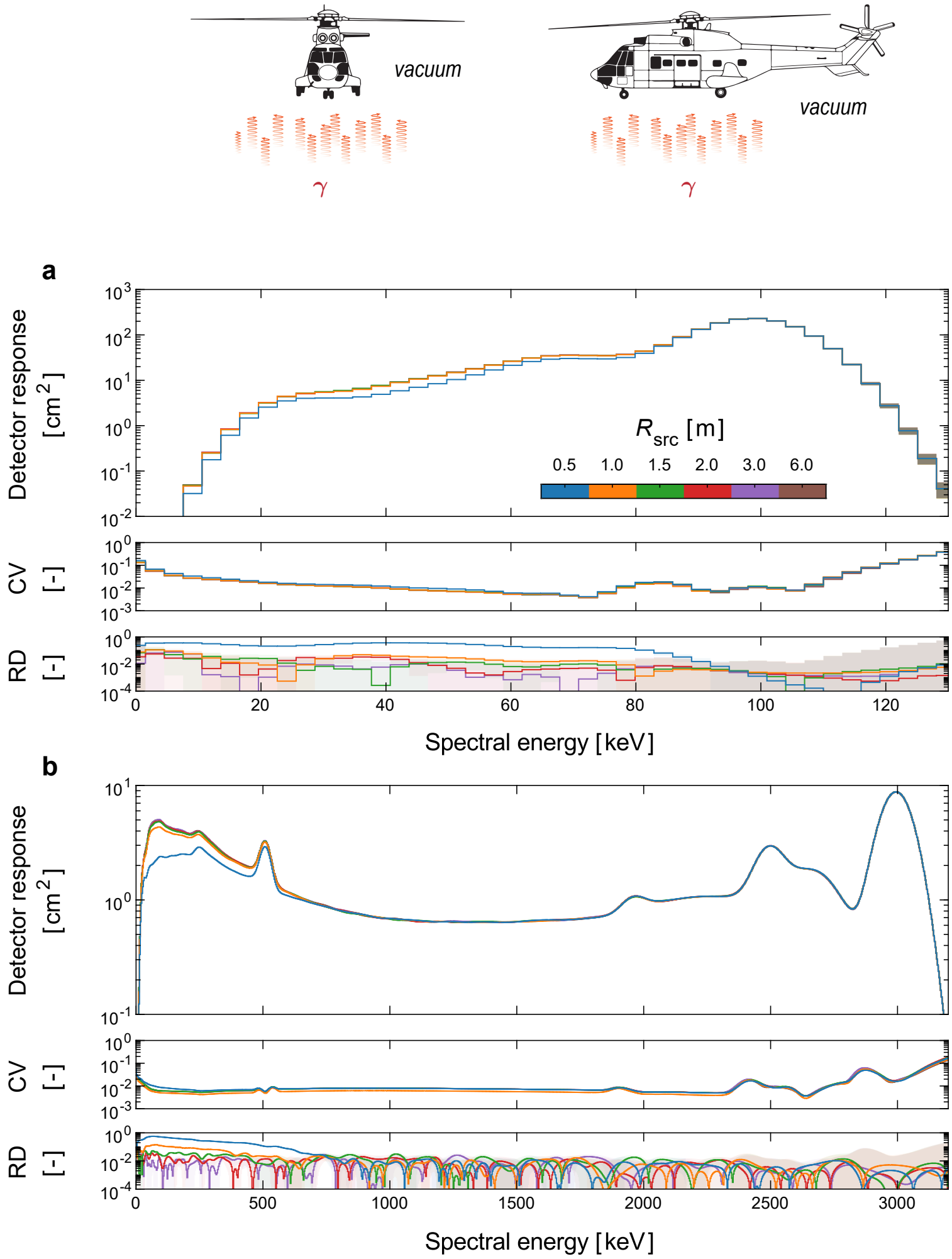

**Figure B.87** Here, I present the results of a sensitivity study conducted to assess the effect of the source radius $R_{src}$ on the simulated detector response $R$ (cf. Section 9.2.2). For that purpose, the detector response was simulated for a monoenergetic circular plane-wave photon source with varying source radius $R_{src}$ oriented along the normal axis $z'$. These simulations were repeated for two photon energies: **a** $E_\gamma = 100\,\text{keV}$. **b** $E_\gamma = 3000\,\text{keV}$. For each photon energy, the simulated mean detector response $R$ in the detector channel `#SUM` is displayed as a function of the spectral energy $E'$ with a spectral energy bin width of $\Delta E' \sim 3\,\text{keV}$. Uncertainties are provided as 1 standard deviation (SD) shaded areas. In addition, the coefficient of variation (CV) (cf. Appendix A.8) as well as the relative deviation (RD) with respect to the detector response at $R_{src} = 6\,\text{m}$ are displayed.

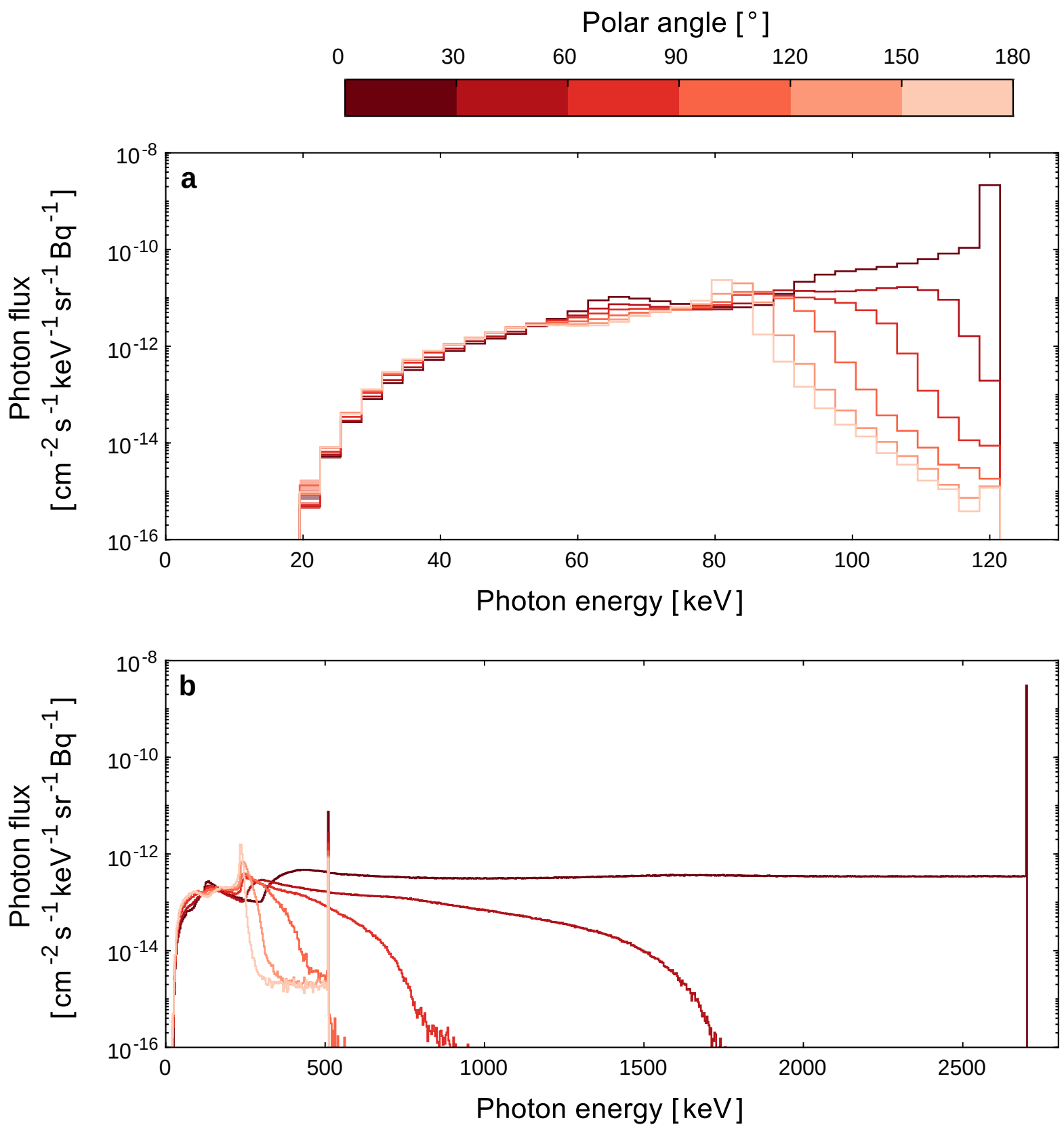

**Figure B.88** Here, I display double differential photon flux signatures at a distance of 30 m from an isotropic monoenergetic point source embedded in homogeneous humid air ($T$ = 20 °C, $p$ = 1013.25 hPa, RH = 50 %). The double differential photon flux was estimated using the `FLUKA` code [20, 216, 281] with the `USRYIELD` card as detailed in Section 9.2.1 for a reference scoring area with a radius of 2 m and two different photon source energies: **a** 120 keV. **b** 2700 keV. For each photon source energy, the mean double differential photon flux signature is displayed as a function of the photon energy $E_\gamma$ and the polar angle $\theta$ with respect to the line-of-sight axis between the source and the reference scoring area. Uncertainties are provided as 1 standard deviation (SD) shaded areas. For visualization purposes, the double differential photon flux signatures were rebinned from a 5° to a 30° polar angle spacing [30].

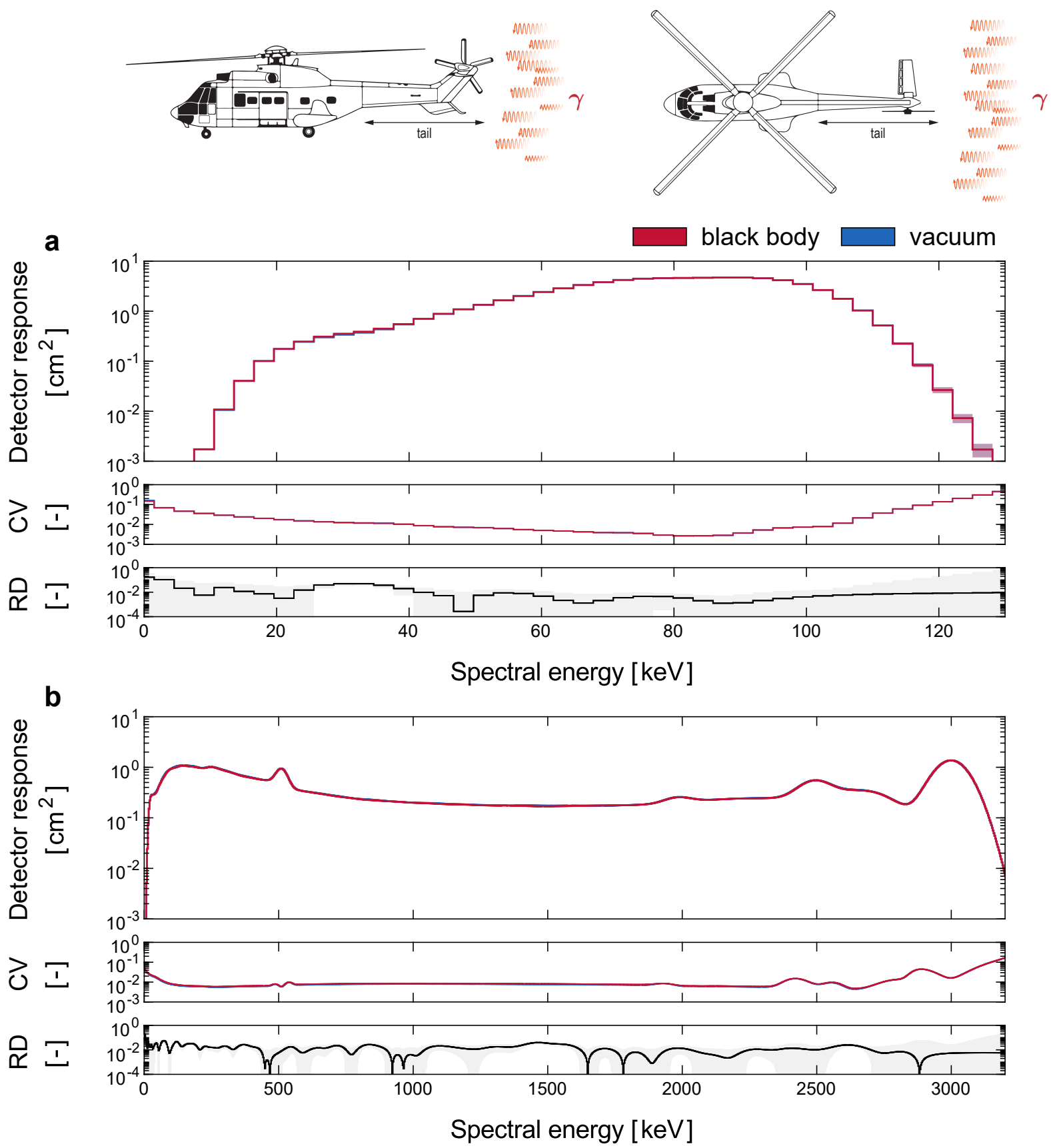

**Figure B.89** Here, I present the results of a sensitivity study conducted to assess the effect of the aircraft tail mass model on the simulated detector response $R$ (cf. Section 9.2.2). For that purpose, the detector response was simulated for a monoenergetic circular plane-wave photon source with a source radius $R_{src}$ of 4.5 m oriented along the aircraft principal axis $y'$. The AGRS Monte Carlo mass model presented in Section 8.2 was adapted by varying the mass model component of the tail between a perfect black body absorber (in red) and vacuum (in blue). In addition, to minimize the attenuation effects, all personnel and the jet fuel ($\varrho_{JF} = 0\,\%$) were excluded from the mass model. These simulations were repeated for two photon energies: **a** $E_\gamma = 100\,\text{keV}$. **b** $E_\gamma = 3000\,\text{keV}$. For each photon energy, the simulated mean detector response $R$ in the detector channel `#SUM` is displayed as a function of the spectral energy $E'$ with a spectral energy bin width of $\Delta E' \sim 3\,\text{keV}$. Uncertainties are provided as 1 standard deviation (SD) shaded areas. In addition, the coefficient of variation (CV) (cf. Appendix A.8) as well as the relative deviation (RD) between the detector response for the two mass models are displayed.

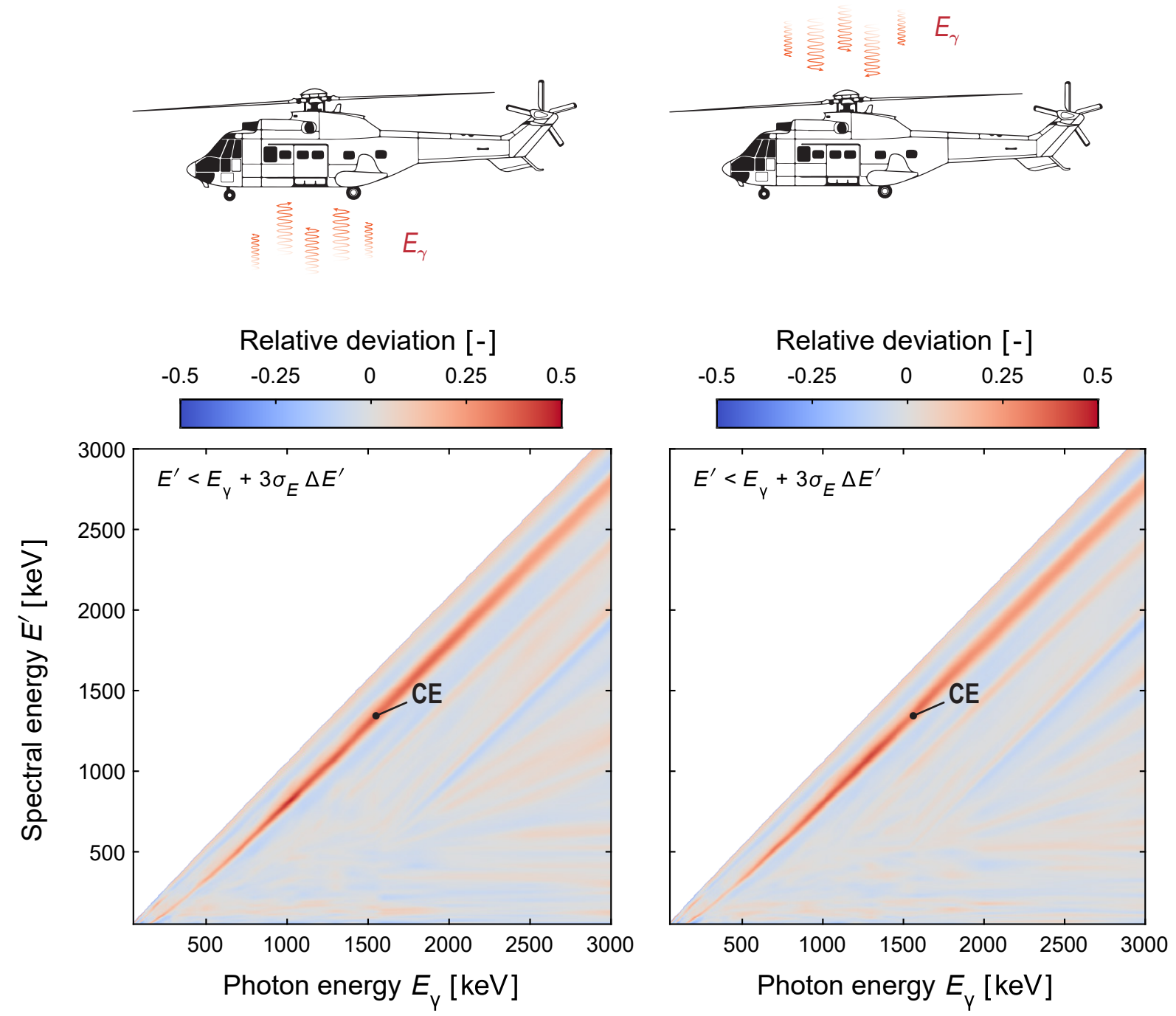

**Figure B.90** Relative deviation in the spectral dispersion between DRFs derived by NPSMC ($R_{\text{nprop}}$) and PSMC ($R_{\text{prop}}$). The relative deviation was computed as $(R_{\text{prop}} - R_{\text{nprop}})/\max(R_{\text{nprop}})$ in the detector channel `#SUM` for the reference mass model (cf. Section 9.4) as a function of the spectral energy $E'$ and the photon energy $E_\gamma$ with a spectral energy bin width of $\Delta E' \sim 3$ keV. The DRF was computed for two different directions $\mathbf{\Omega}'$: **a** Normal ($\theta' = 180°$). **b** Anti-normal ($\theta' = 0°$). For visualization purposes, the DRF is limited to the spectral energy range $E' < E_\gamma + 3\sigma_E \Delta E'$ with $\sigma_E$ being the spectral resolution standard deviation (measured in the pulse-height channel number space, cf. Section 4.3.2) at $E_\gamma/\Delta E'$ and $\Delta E'$ the spectral energy bin width (cf. Section 6.2.1.3). The characteristic Compton edge (CE) is highlighted (cf. Section 4.3.1).

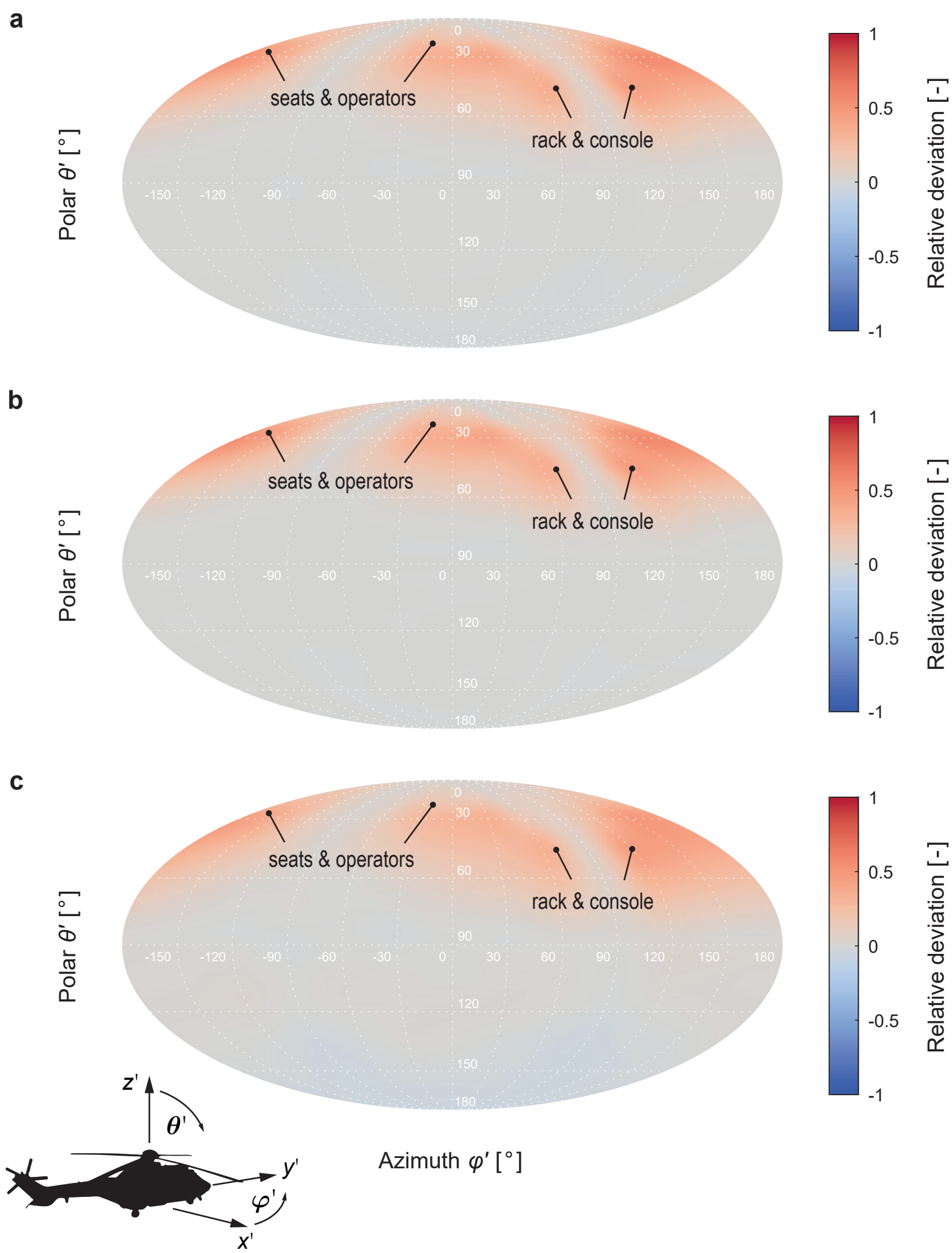

**Figure B.91** Relative deviation in the angular dispersion between the reference DRF ($R_{\text{ref}}$) and a DRF excluding the RLL supporting systems ($R_{\text{RLL}}$) at a photon energy of $E_\gamma = 88\,\text{keV}$. The relative deviation was computed as $(R_{\text{RLL}} - R_{\text{ref}})/\max(R_{\text{ref}})$ in the detector channel `#SUM` as a function of the polar angle $\theta'$ and the azimuthal angle $\varphi'$ with respect to the detector reference frame $x'$-$y'$-$z'$. Three different spectral domains were evaluated: **a** Full spectrum domain $\mathcal{D}_{\text{tot}}$. **b** Full energy peak domain $\mathcal{D}_{\text{FEP}}$. **c** Compton continuum domain $\mathcal{D}_{\text{CC}}$. The graphs are interpolated on a $1° \times 1°$ angular grid (modified Akima spline interpolation [733]) and displayed using the Mollweide projection.

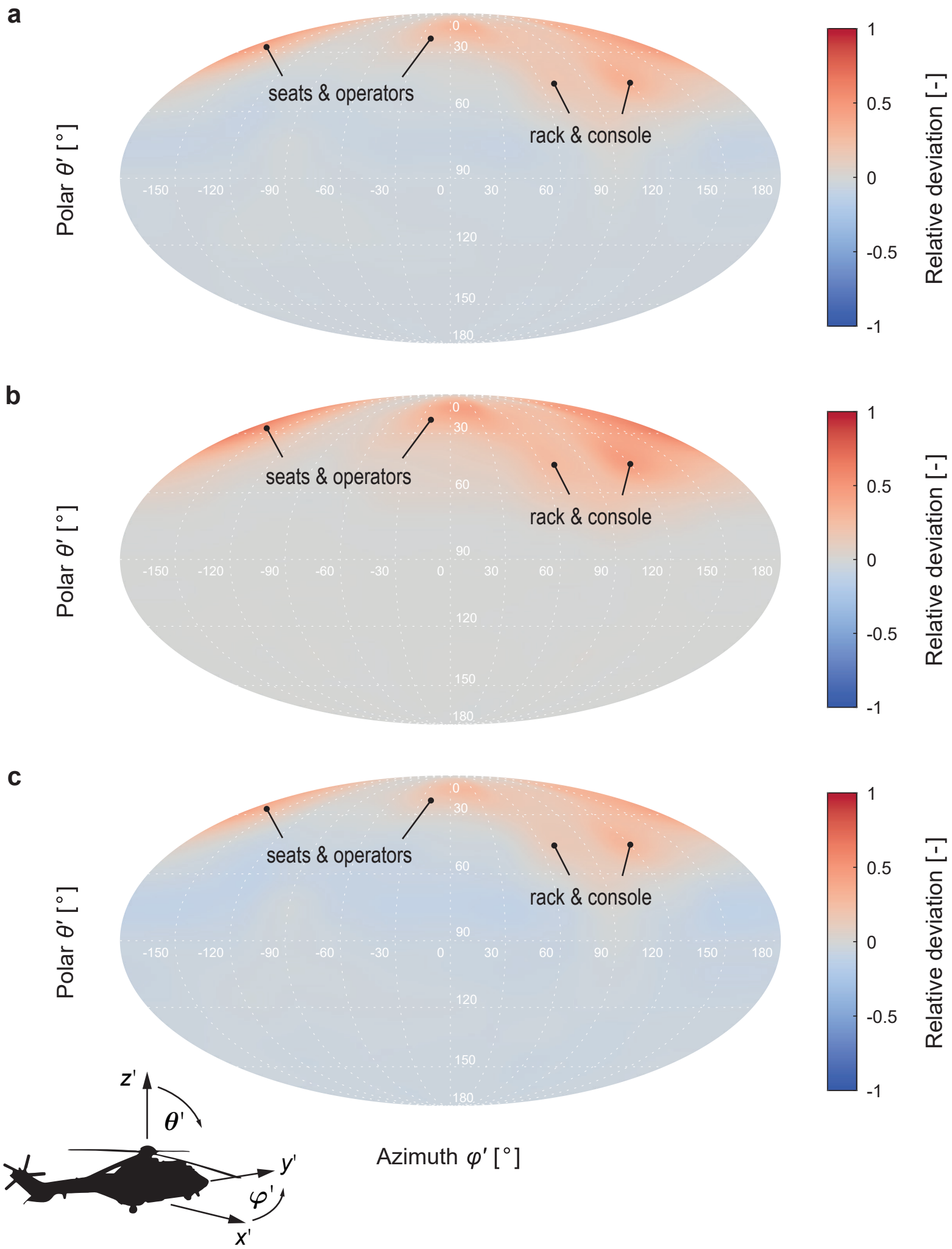

**Figure B.92** Relative deviation in the angular dispersion between the reference DRF ($R_{ref}$) and a DRF excluding the RLL supporting systems ($R_{RLL}$) at a photon energy of $E_\gamma = 2615$ keV. The relative deviation was computed as $(R_{RLL} - R_{ref})/\max(R_{ref})$ in the detector channel `#SUM` as a function of the polar angle $\theta'$ and the azimuthal angle $\varphi'$ with respect to the detector reference frame $x'$-$y'$-$z'$. Three different spectral domains were evaluated: **a** Full spectrum domain $\mathcal{D}_{tot}$. **b** Full energy peak domain $\mathcal{D}_{FEP}$. **c** Compton continuum domain $\mathcal{D}_{CC}$. The graphs are interpolated on a 1° × 1° angular grid (modified Akima spline interpolation [733]) and displayed using the Mollweide projection.

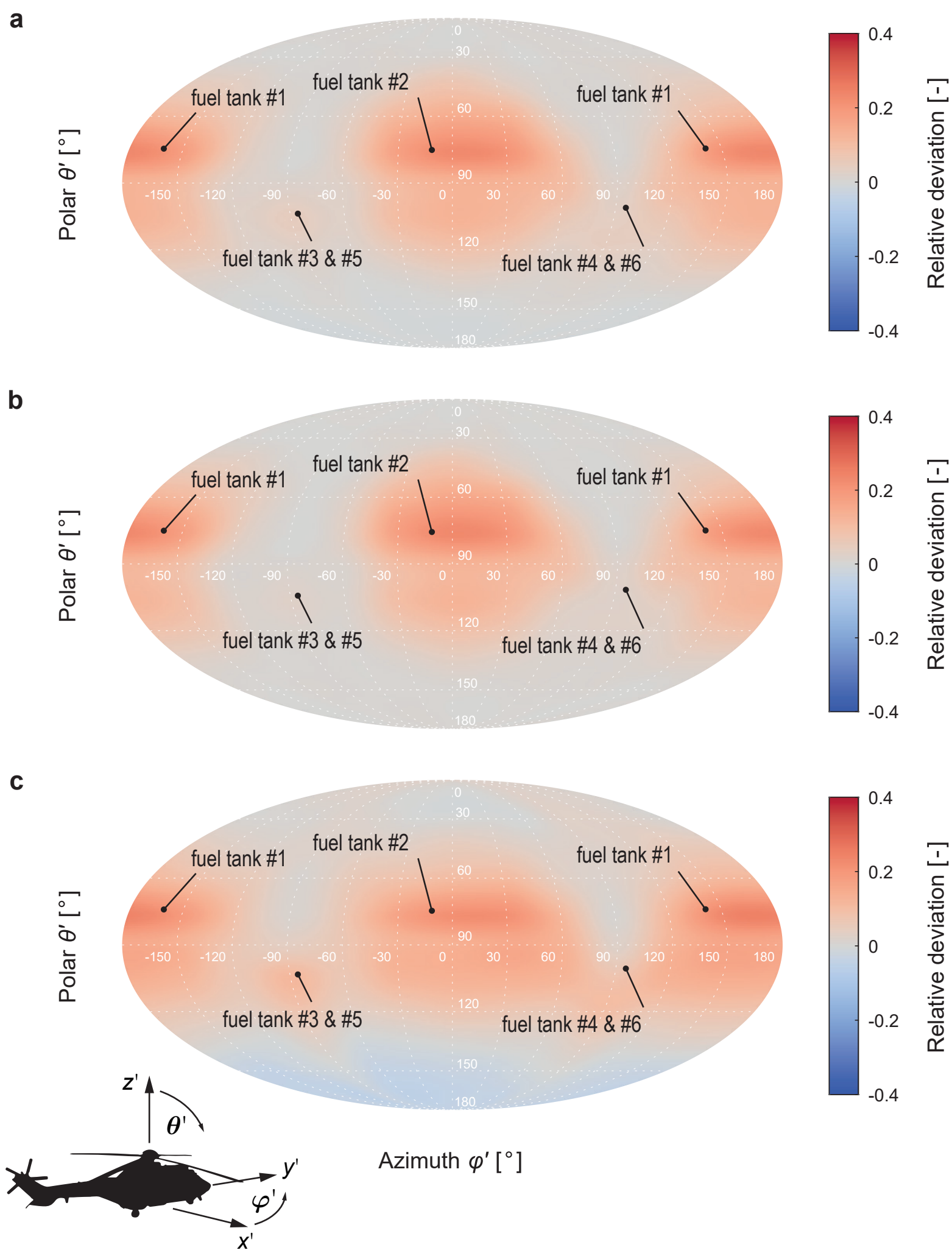

**Figure B.93** Relative deviation in the angular dispersion between the DRF with the empty fuel tanks ($R_{\mathrm{JF0}}$) and the full fuel tanks ($R_{\mathrm{JF100}}$) at a photon energy of $E_\gamma = 88\,\mathrm{keV}$. The relative deviation was computed as $(R_{\mathrm{JF0}} - R_{\mathrm{JF100}})/\max(R_{\mathrm{JF100}})$ in the detector channel `#SUM` as a function of the polar angle $\theta'$ and the azimuthal angle $\varphi'$ with respect to the detector reference frame $x'$-$y'$-$z'$. Three different spectral domains were evaluated: **a** Full spectrum domain $\mathcal{D}_{\mathrm{tot}}$. **b** Full energy peak domain $\mathcal{D}_{\mathrm{FEP}}$. **c** Compton continuum domain $\mathcal{D}_{\mathrm{CC}}$. The graphs are interpolated on a 1° × 1° angular grid (modified Akima spline interpolation [733]) and displayed using the Mollweide projection.

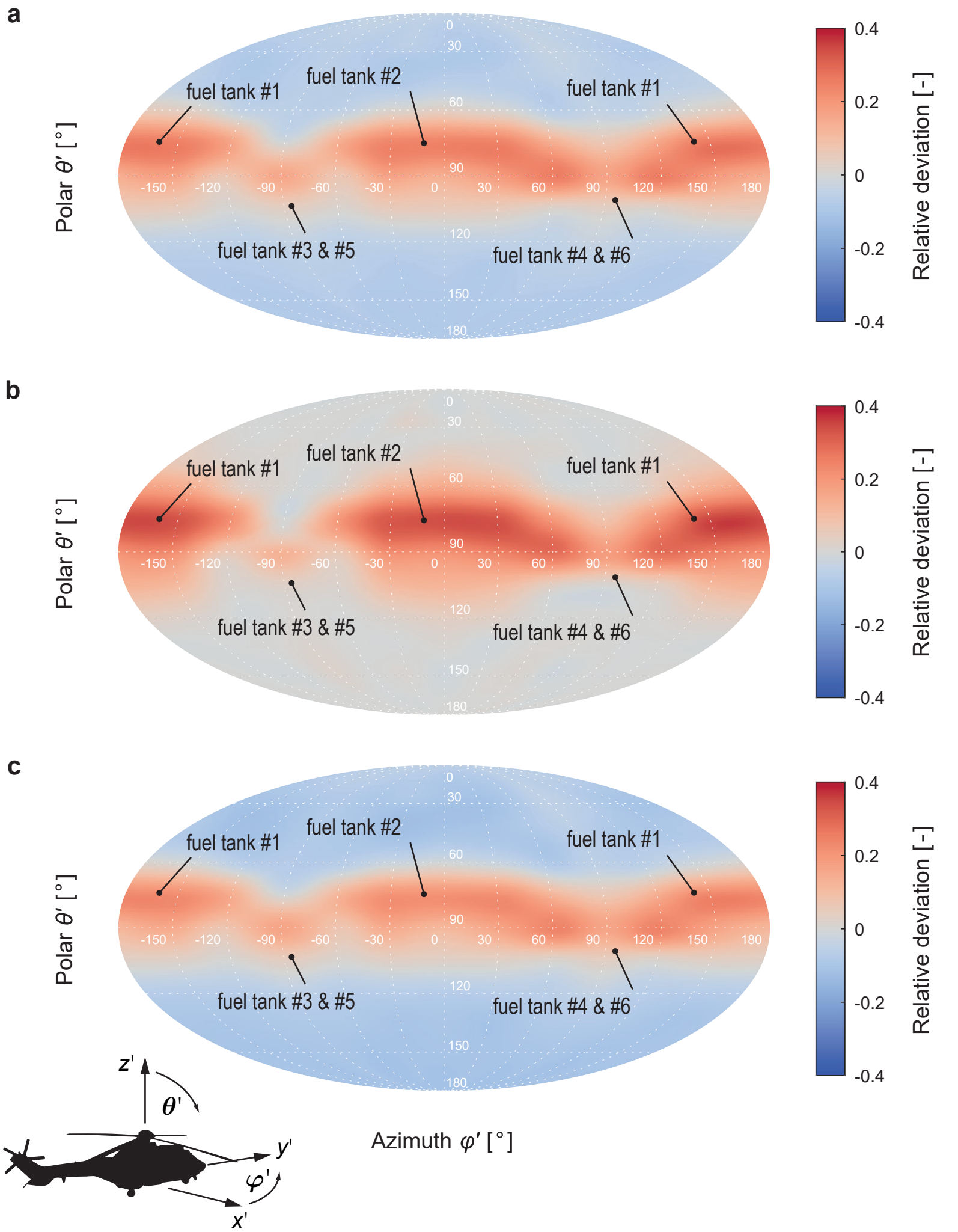

**Figure B.94** Relative deviation in the angular dispersion between the DRF with the empty fuel tanks ($R_{\mathrm{JF0}}$) and the full fuel tanks ($R_{\mathrm{JF100}}$) at a photon energy of $E_\gamma = 2615\,\mathrm{keV}$. The relative deviation was computed as $(R_{\mathrm{JF0}} - R_{\mathrm{JF100}})/\max(R_{\mathrm{JF100}})$ in the detector channel #SUM as a function of the polar angle $\theta'$ and the azimuthal angle $\varphi'$ with respect to the detector reference frame $x'$-$y'$-$z'$. Three different spectral domains were evaluated: **a** Full spectrum domain $\mathcal{D}_{\mathrm{tot}}$. **b** Full energy peak domain $\mathcal{D}_{\mathrm{FEP}}$. **c** Compton continuum domain $\mathcal{D}_{\mathrm{CC}}$. The graphs are interpolated on a 1° × 1° angular grid (modified Akima spline interpolation [733]) and displayed using the Mollweide projection.

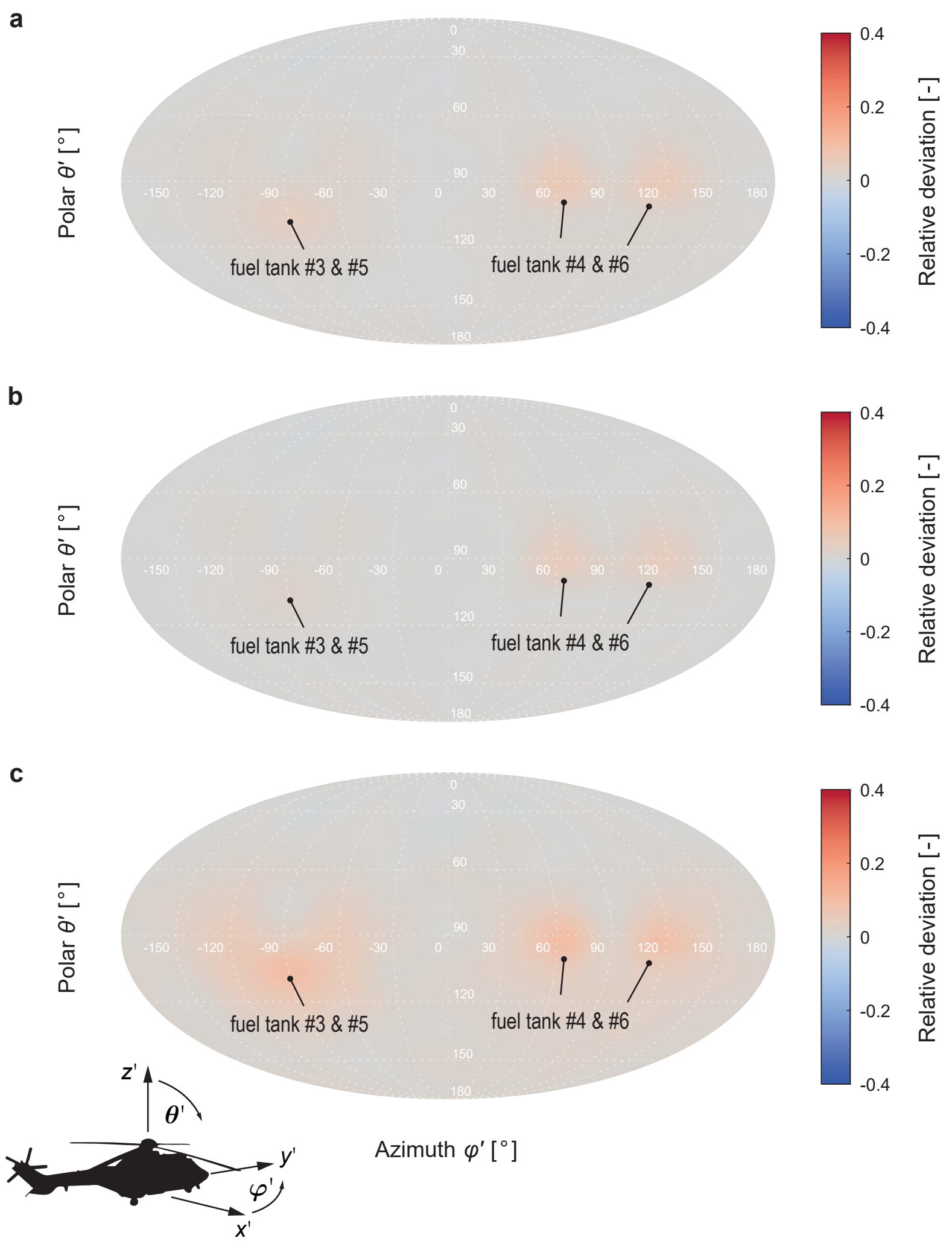

**Figure B.95** Relative deviation in the angular dispersion between the DRF with the quarter-filled fuel tanks ($R_{\mathrm{JF25}}$) and the full fuel tanks ($R_{\mathrm{JF100}}$) at a photon energy of $E_\gamma = 88\,\mathrm{keV}$. The relative deviation was computed as $(R_{\mathrm{JF25}} - R_{\mathrm{JF100}})/\max(R_{\mathrm{JF100}})$ in the detector channel `#SUM` as a function of the polar angle $\theta'$ and the azimuthal angle $\varphi'$ with respect to the detector reference frame $x'$-$y'$-$z'$. Three different spectral domains were evaluated: **a** Full spectrum domain $\mathcal{D}_{\mathrm{tot}}$. **b** Full energy peak domain $\mathcal{D}_{\mathrm{FEP}}$. **c** Compton continuum domain $\mathcal{D}_{\mathrm{CC}}$. The graphs are interpolated on a 1° × 1° angular grid (modified Akima spline interpolation [733]) and displayed using the Mollweide projection.

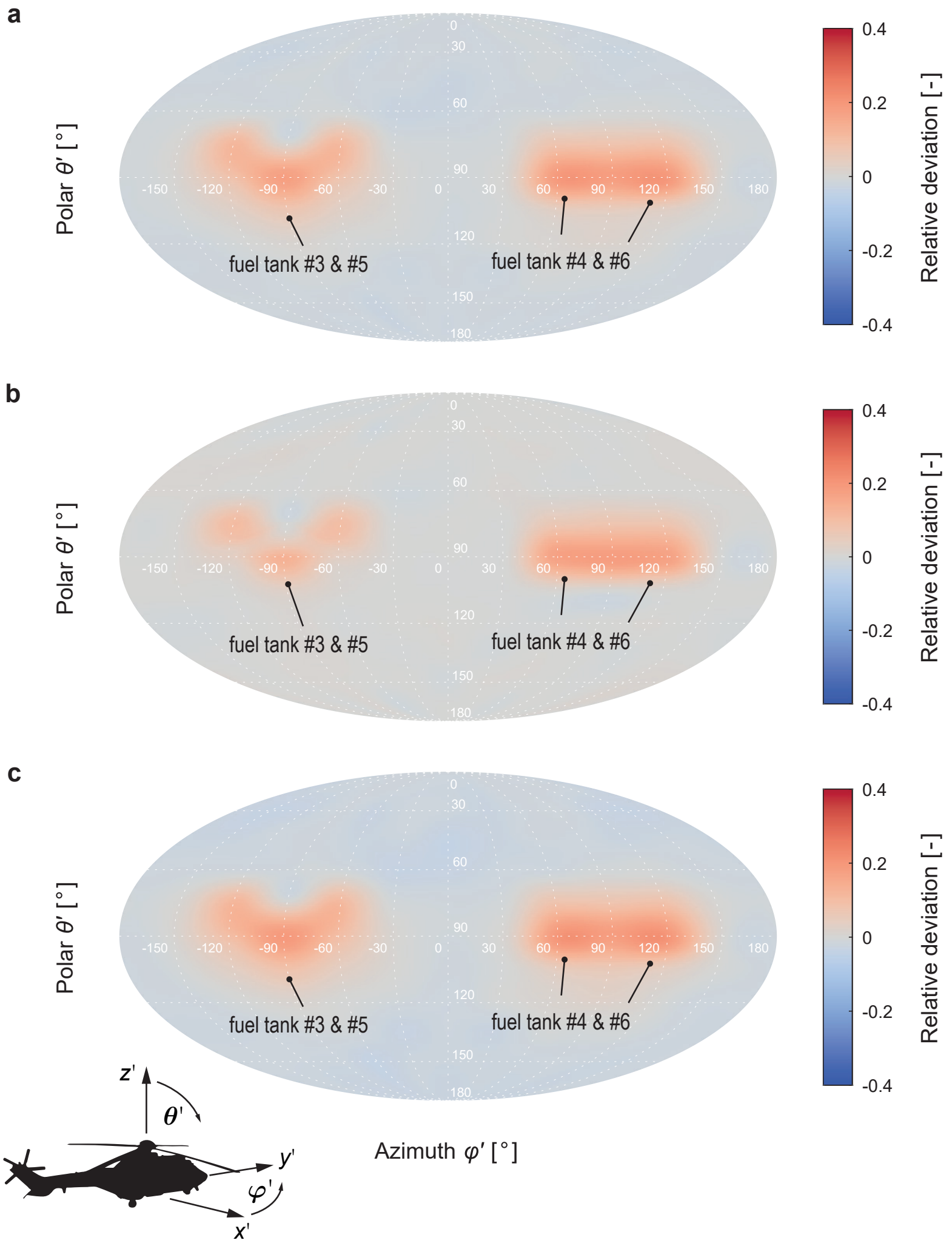

**Figure B.96** Relative deviation in the angular dispersion between the DRF with the quarter-filled fuel tanks ($R_{\mathrm{JF25}}$) and the full fuel tanks ($R_{\mathrm{JF100}}$) at a photon energy of $E_\gamma = 2615\,\mathrm{keV}$. The relative deviation was computed as $(R_{\mathrm{JF25}} - R_{\mathrm{JF100}})/\max(R_{\mathrm{JF100}})$ in the detector channel `#SUM` as a function of the polar angle $\theta'$ and the azimuthal angle $\varphi'$ with respect to the detector reference frame $x'$-$y'$-$z'$. Three different spectral domains were evaluated: **a** Full spectrum domain $\mathcal{D}_{\mathrm{tot}}$. **b** Full energy peak domain $\mathcal{D}_{\mathrm{FEP}}$. **c** Compton continuum domain $\mathcal{D}_{\mathrm{CC}}$. The graphs are interpolated on a $1° \times 1°$ angular grid (modified Akima spline interpolation [733]) and displayed using the Mollweide projection.

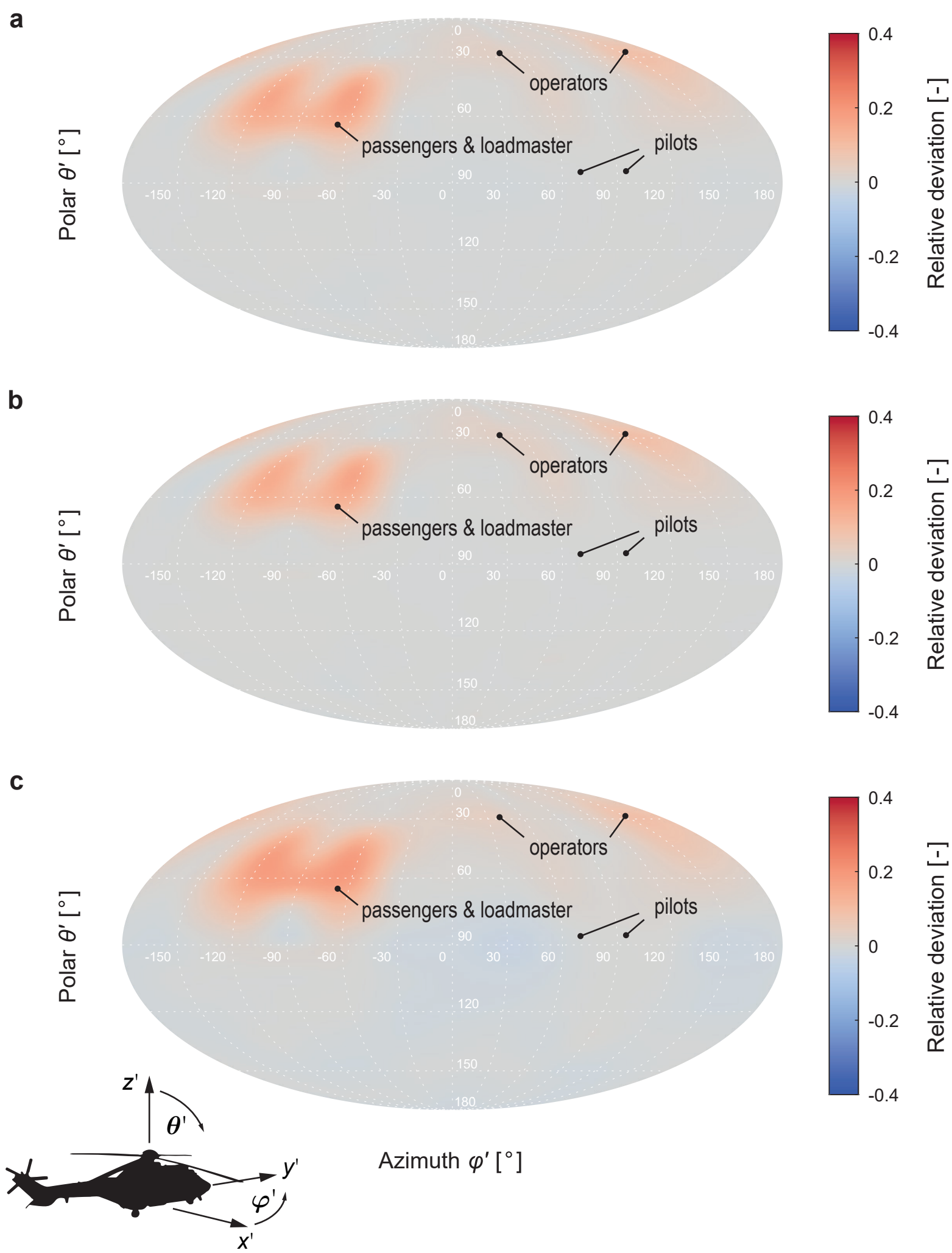

**Figure B.97** Relative deviation in the angular dispersion between the DRF with no personnel on board ($R_{NP}$) and the full crew capacity ($R_{FC}$) at a photon energy of $E_\gamma = 88\,\text{keV}$. The relative deviation was computed as $(R_{NP} - R_{FC})/\max(R_{FC})$ in the detector channel `#SUM` as a function of the polar angle $\theta'$ and the azimuthal angle $\varphi'$ with respect to the detector reference frame $x'$-$y'$-$z'$. Three different spectral domains were evaluated: **a** Full spectrum domain $\mathcal{D}_{tot}$. **b** Full energy peak domain $\mathcal{D}_{FEP}$. **c** Compton continuum domain $\mathcal{D}_{CC}$. The graphs are interpolated on a $1° \times 1°$ angular grid (modified Akima spline interpolation [733]) and displayed using the Mollweide projection.

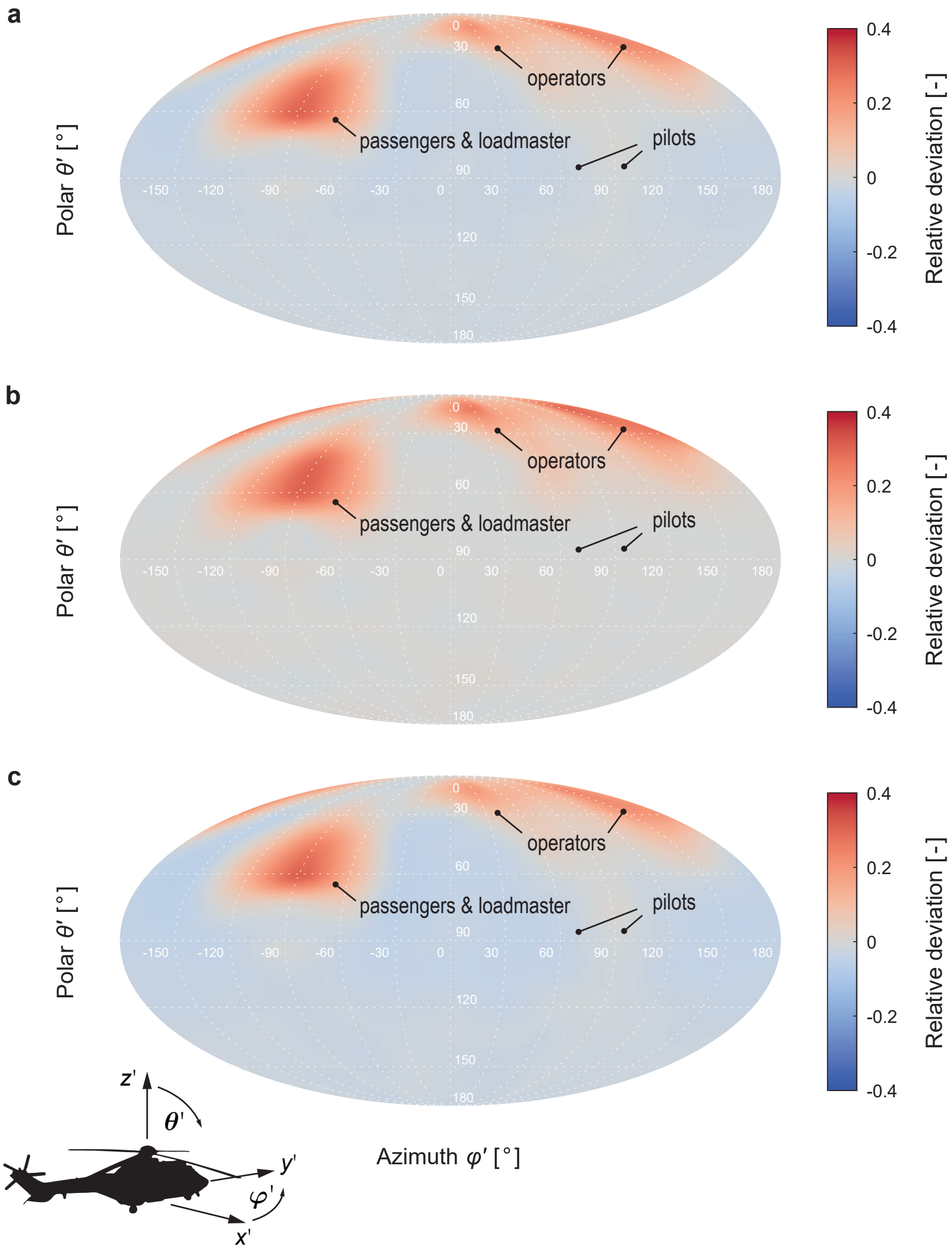

**Figure B.98** Relative deviation in the angular dispersion between the DRF with no personnel on board ($R_{NP}$) and the full crew capacity ($R_{FC}$) at a photon energy of $E_\gamma = 2615\,\text{keV}$. The relative deviation was computed as $(R_{NP} - R_{FC})/\max(R_{FC})$ in the detector channel `#SUM` as a function of the polar angle $\theta'$ and the azimuthal angle $\varphi'$ with respect to the detector reference frame $x'$-$y'$-$z'$. Three different spectral domains were evaluated: **a** Full spectrum domain $\mathcal{D}_{tot}$. **b** Full energy peak domain $\mathcal{D}_{FEP}$. **c** Compton continuum domain $\mathcal{D}_{CC}$. The graphs are interpolated on a $1° \times 1°$ angular grid (modified Akima spline interpolation [733]) and displayed using the Mollweide projection.

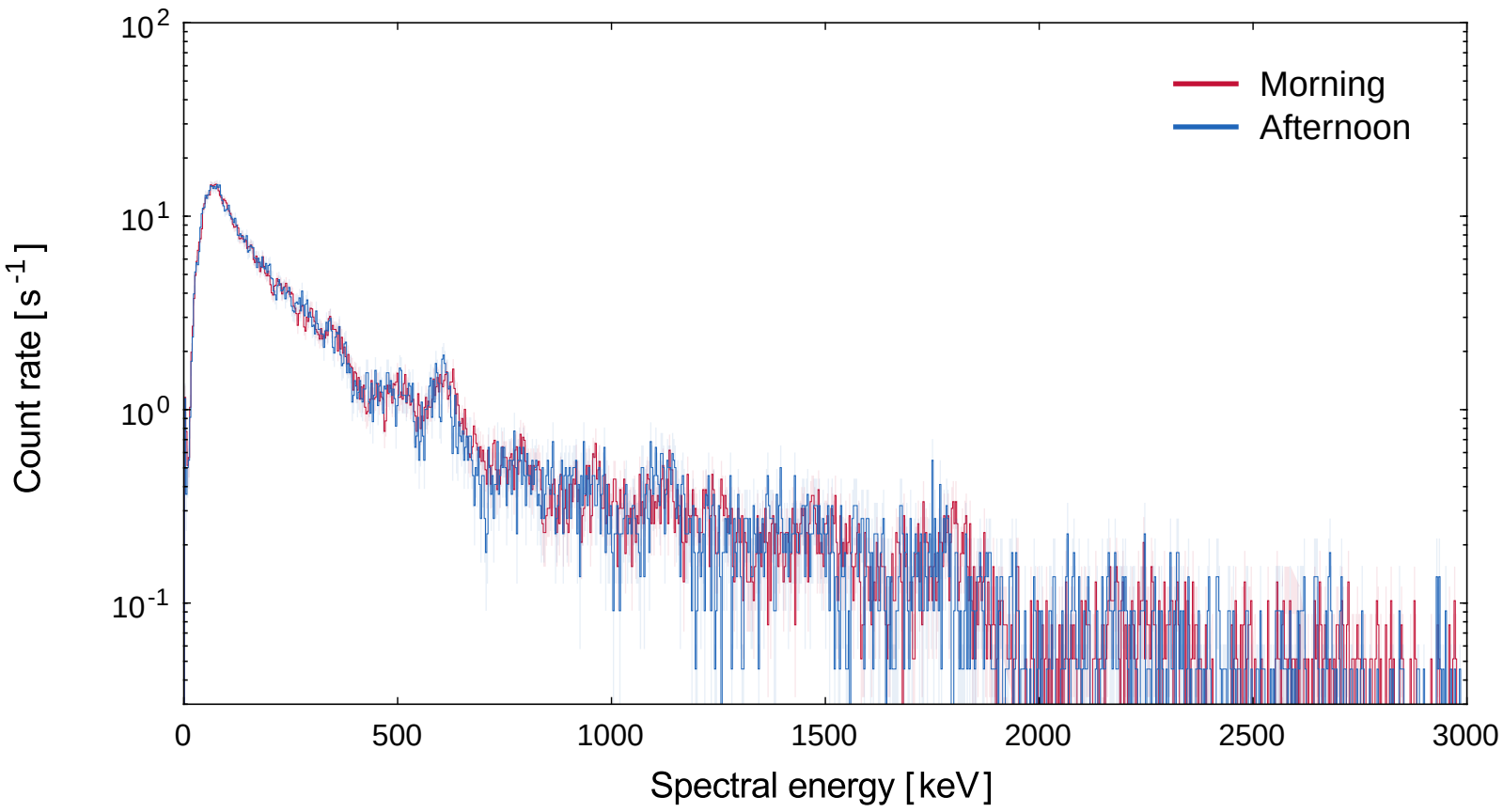

**Figure B.99** Here, I display the measured background pulse-height spectra ($\mathbf{c}_\mathrm{b}$) for the detector channel `#SUM` as a function of the spectral energy $E'$ with a spectral energy bin width of $\Delta E' \sim 3\,\mathrm{keV}$ (cf. Section 10.3.2.2). The data were acquired by the Swiss AGRS system in the morning and afternoon on the 2022-06-16 over Lake Thun during the ARM22 validation campaign at orthometric heights equivalent to the ones of the hover flight measurements used for the validation of the full spectrum Bayesian inversion (FSBI) method presented in Section 10.3, i.e. 659(5) m for the hover flight with the $^{137}_{55}\mathrm{Cs}$ source (morning) and 659(6) m for the hover flight with the $^{133}_{56}\mathrm{Ba}$ source (afternoon), respectively. Uncertainties are provided as 1 standard deviation (SD) shaded areas (cf. Appendix A.8).

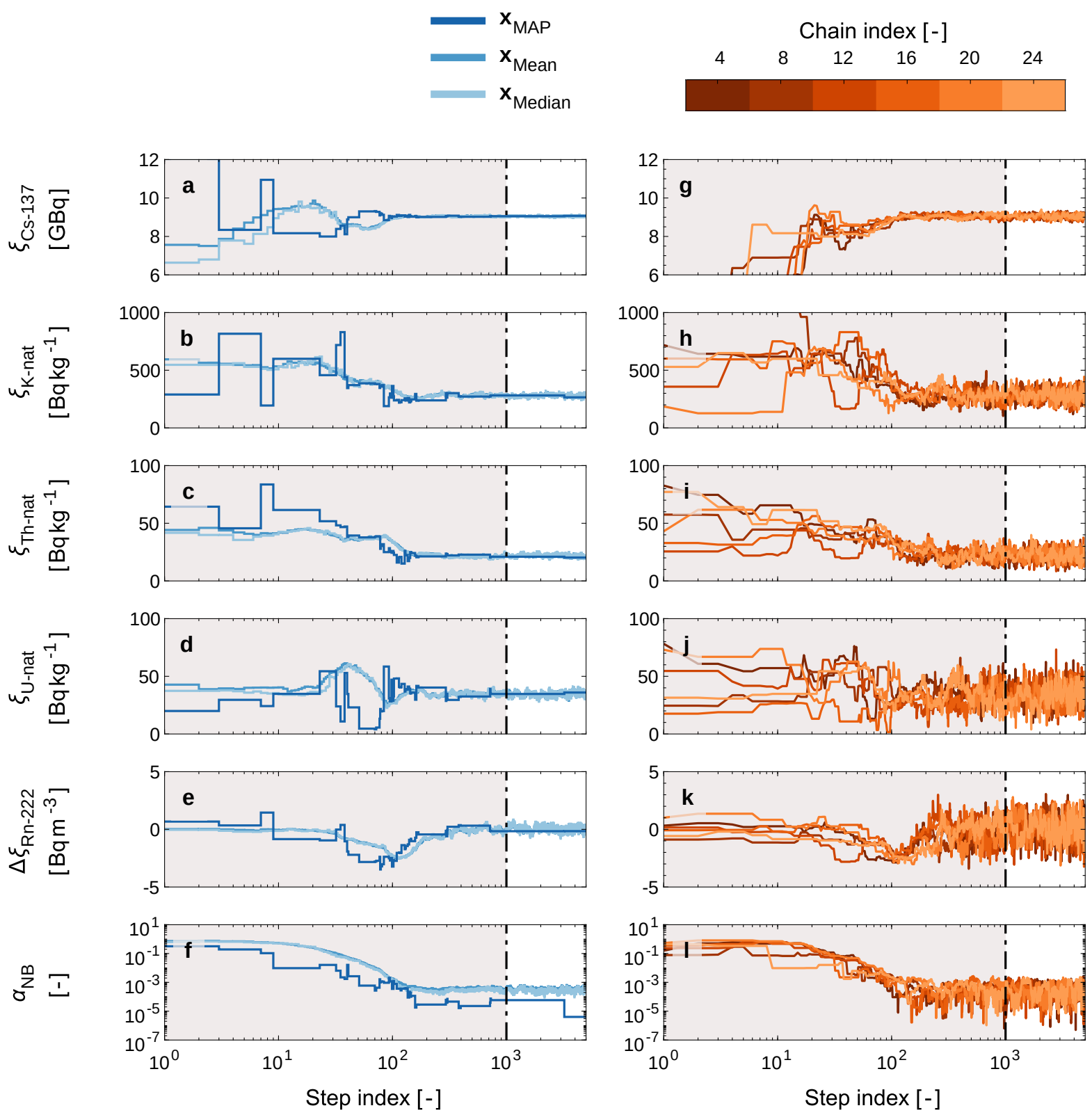

**Figure B.100** These graphs present the convergence (**a–f**) and trace (**g–l**) plots for the MCMC-based FSBI of the dataset `Cs_I` (cf. Table 10.1). The graphs are displayed as a function of the MCMC steps for each model parameter, i.e. the source strength of the sealed $^{137}_{55}$Cs point source ($\xi_{\text{Cs-137}}$), the source strengths of the three natural terrestrial radionuclides $K_{nat}$, $Th_{nat}$ and $U_{nat}$ ($\xi_{\text{K-nat}}$, $\xi_{\text{Th-nat}}$, $\xi_{\text{U-nat}}$), the source strength of the radon source term $\Delta^{222}_{86}$Rn ($\Delta\xi_{\text{Rn-222}}$) and the dispersion parameter of the gamma-Poisson mixture distribution ($\alpha_{NB}$). The convergence is shown for the maximum a posteriori (MAP) probability estimate $\mathbf{x}_{\text{MAP}}$, the posterior mean $\mathbf{x}_{\text{Mean}}$ and the posterior median $\mathbf{x}_{\text{Median}}$. The trace plots display a subset of 6 out of the 24 simulated MCMC chains. The burn-in phase is highlighted for all subgraphs as gray-shaded areas with the related threshold marked by the dashed-dotted black line.

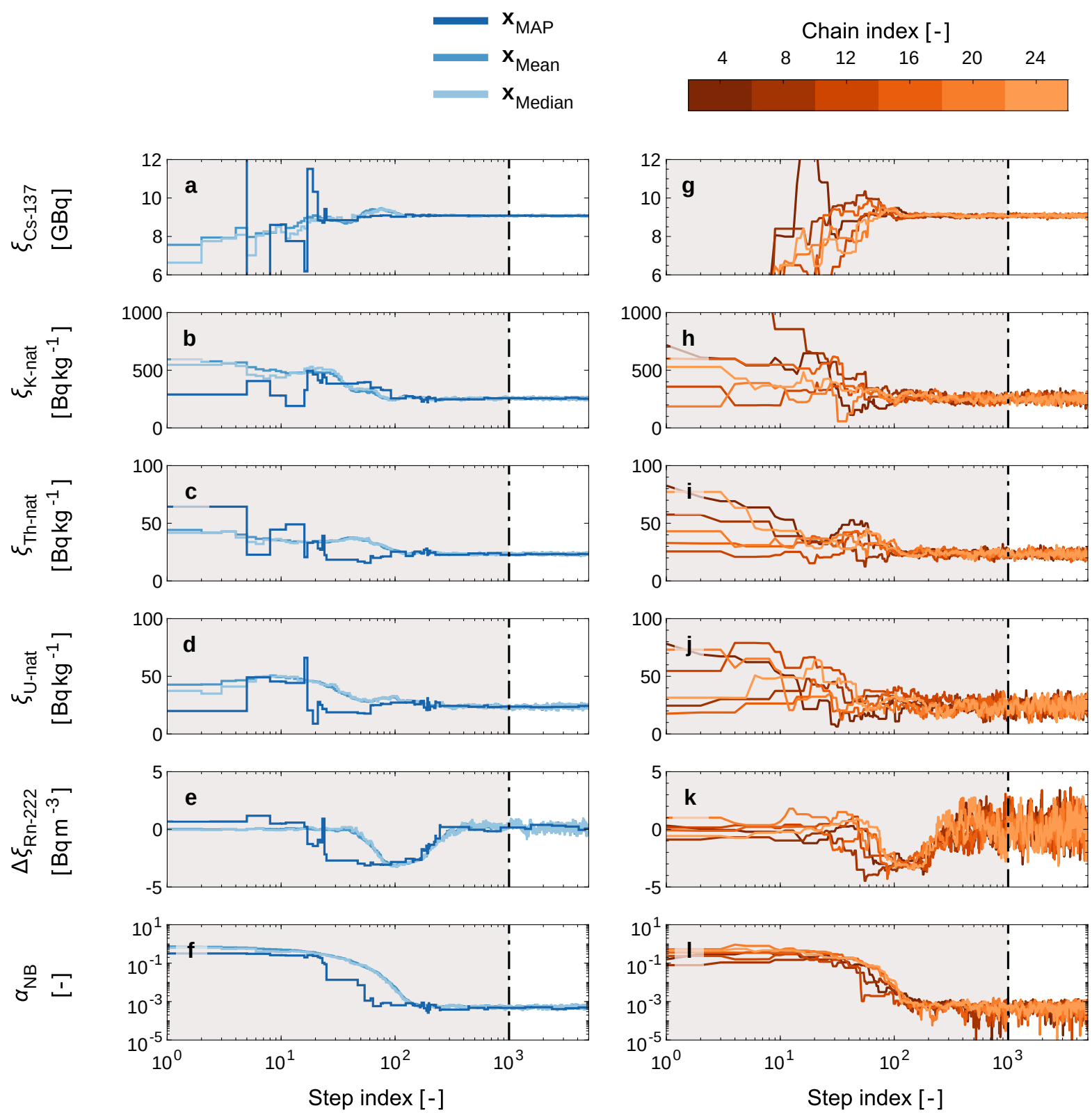

**Figure B.101** These graphs present the convergence (**a–f**) and trace (**g–l**) plots for the MCMC-based FSBI of the dataset `Cs_II` (cf. Table 10.1). The graphs are displayed as a function of the MCMC steps for each model parameter, i.e. the source strength of the sealed $^{137}_{55}$Cs point source ($\xi_{\text{Cs-137}}$), the source strengths of the three natural terrestrial radionuclides $K_{nat}$, $Th_{nat}$ and $U_{nat}$ ($\xi_{\text{K-nat}}$, $\xi_{\text{Th-nat}}$, $\xi_{\text{U-nat}}$), the source strength of the radon source term $\Delta^{222}_{86}$Rn ($\Delta\xi_{\text{Rn-222}}$) and the dispersion parameter of the gamma-Poisson mixture distribution ($\alpha_{NB}$). The convergence is shown for the maximum a posteriori (MAP) probability estimate $\mathbf{x}_{MAP}$, the posterior mean $\mathbf{x}_{Mean}$ and the posterior median $\mathbf{x}_{Median}$. The trace plots display a subset of 6 out of the 24 simulated MCMC chains. The burn-in phase is highlighted for all subgraphs as gray-shaded areas with the related threshold marked by the dashed-dotted black line.

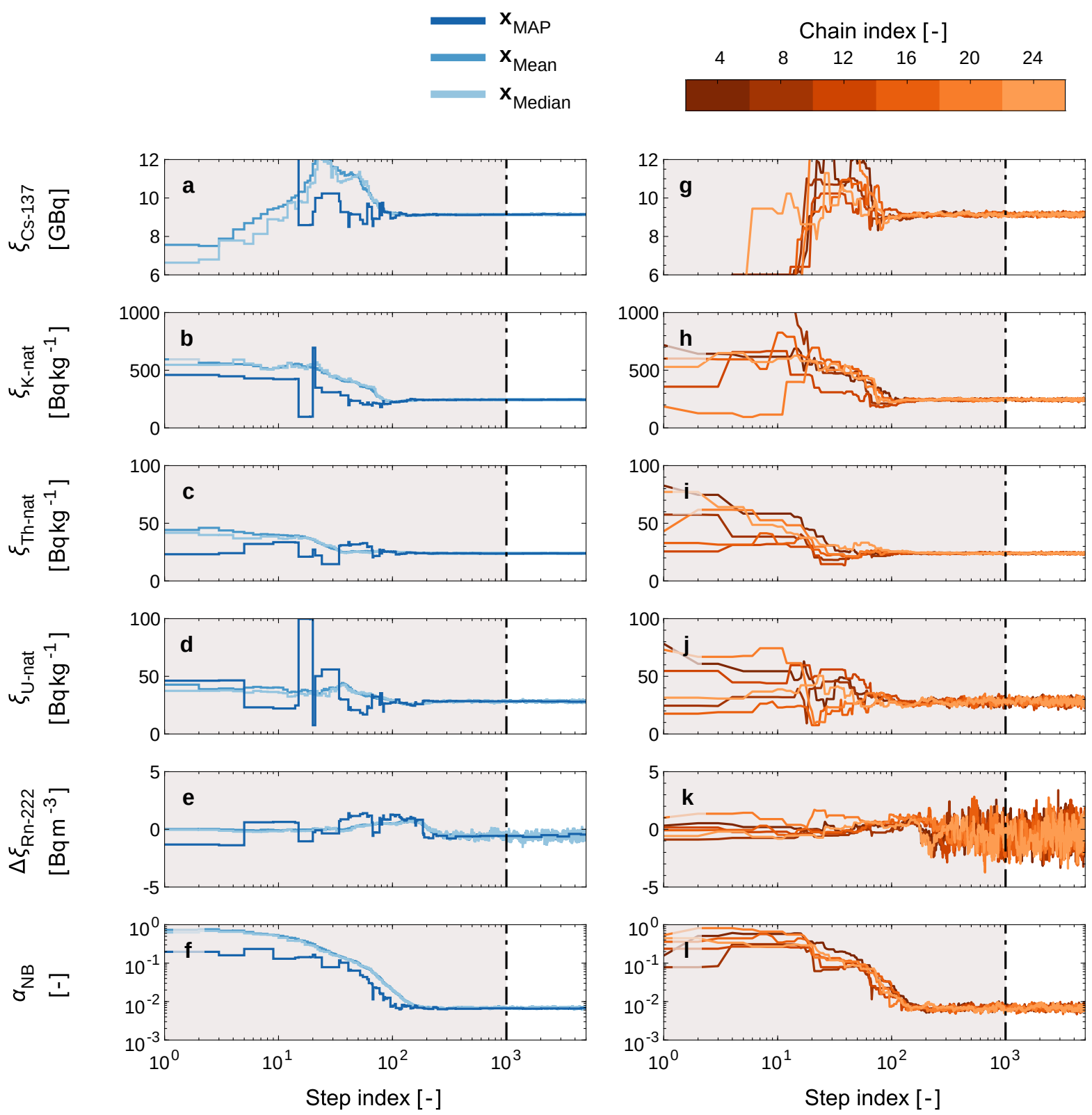

**Figure B.102** These graphs present the convergence (**a–f**) and trace (**g–l**) plots for the MCMC-based FSBI of the dataset `Cs_III` (cf. Table 10.1). The graphs are displayed as a function of the MCMC steps for each model parameter, i.e. the source strength of the sealed $^{137}_{55}$Cs point source ($\xi_{\text{Cs-137}}$), the source strengths of the three natural terrestrial radionuclides $K_{nat}$, $Th_{nat}$ and $U_{nat}$ ($\xi_{\text{K-nat}}$, $\xi_{\text{Th-nat}}$, $\xi_{\text{U-nat}}$), the source strength of the radon source term $\Delta^{222}_{86}$Rn ($\Delta\xi_{\text{Rn-222}}$) and the dispersion parameter of the gamma-Poisson mixture distribution ($\alpha_{NB}$). The convergence is shown for the maximum a posteriori (MAP) probability estimate $\mathbf{x}_{MAP}$, the posterior mean $\mathbf{x}_{Mean}$ and the posterior median $\mathbf{x}_{Median}$. The trace plots display a subset of 6 out of the 24 simulated MCMC chains. The burn-in phase is highlighted for all subgraphs as gray-shaded areas with the related threshold marked by the dashed-dotted black line.

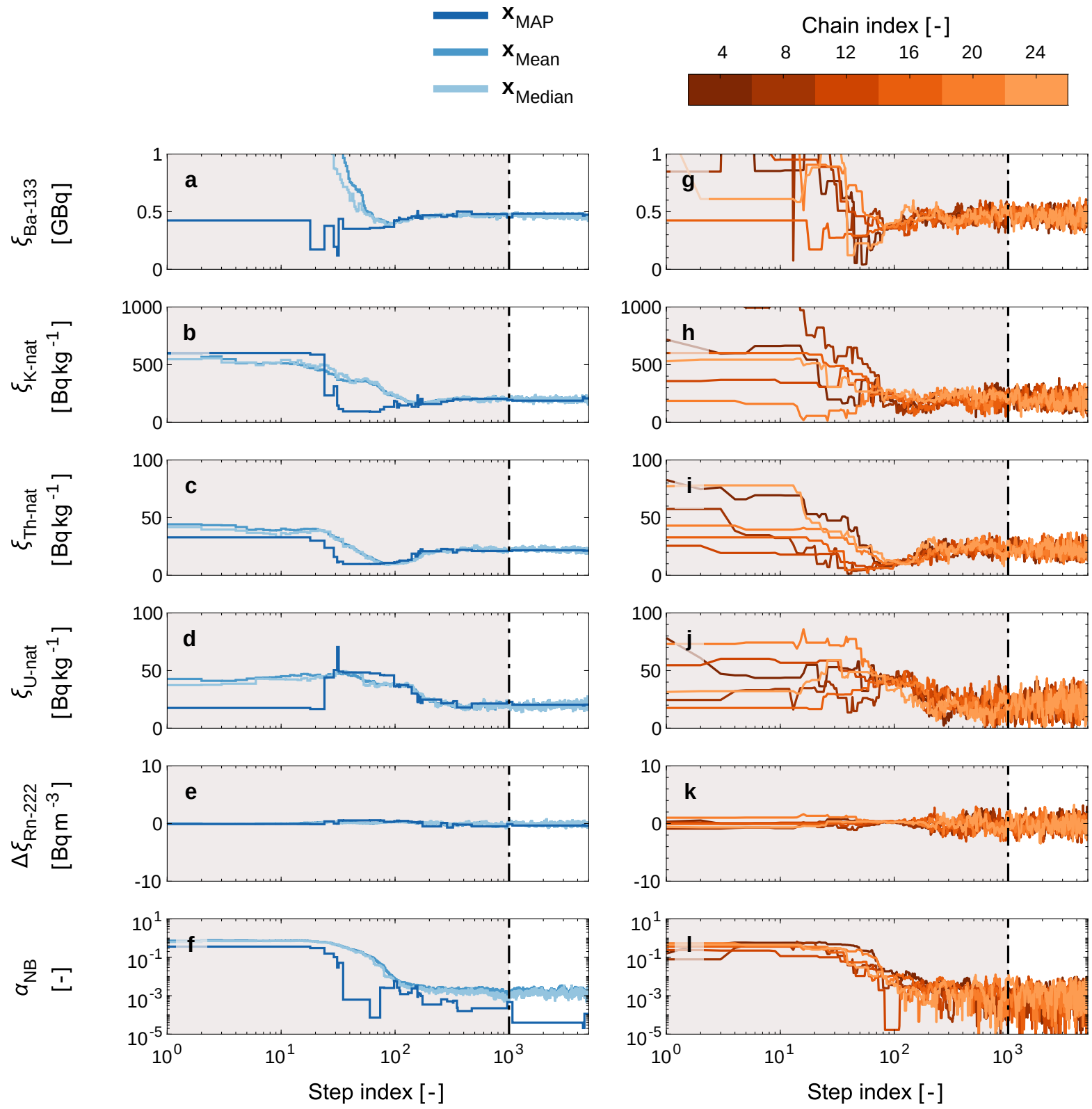

**Figure B.103** These graphs present the convergence (**a–f**) and trace (**g–l**) plots for the MCMC-based FSBI of the dataset `Ba_I` (cf. Table 10.1). The graphs are displayed as a function of the MCMC steps for each model parameter, i.e. the source strength of the sealed $^{133}_{56}$Ba point source ($\xi_{\text{Ba-133}}$), the source strengths of the three natural terrestrial radionuclides $K_{nat}$, $Th_{nat}$ and $U_{nat}$ ($\xi_{\text{K-nat}}$, $\xi_{\text{Th-nat}}$, $\xi_{\text{U-nat}}$), the source strength of the radon source term $\Delta^{222}_{86}$Rn ($\Delta\xi_{\text{Rn-222}}$) and the dispersion parameter of the gamma-Poisson mixture distribution ($\alpha_{NB}$). The convergence is shown for the maximum a posteriori (MAP) probability estimate $\mathbf{x}_{\text{MAP}}$, the posterior mean $\mathbf{x}_{\text{Mean}}$ and the posterior median $\mathbf{x}_{\text{Median}}$. The trace plots display a subset of 6 out of the 24 simulated MCMC chains. The burn-in phase is highlighted for all subgraphs as gray-shaded areas with the related threshold marked by the dashed-dotted black line.

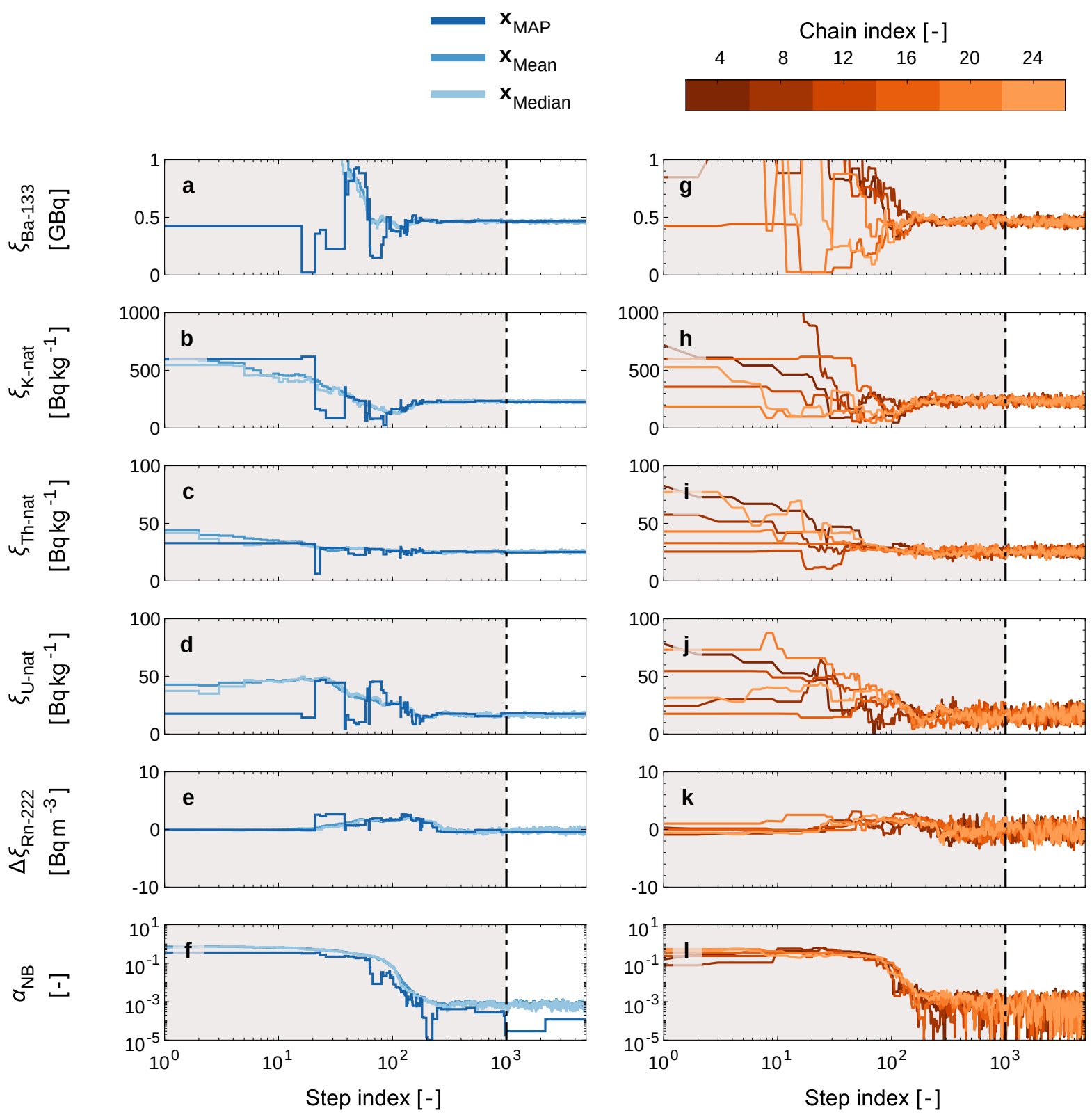

**Figure B.104** These graphs present the convergence (**a–f**) and trace (**g–l**) plots for the MCMC-based FSBI of the dataset `Ba_II` (cf. Table 10.1). The graphs are displayed as a function of the MCMC steps for each model parameter, i.e. the source strength of the sealed $^{133}_{56}$Ba point source ($\xi_{\text{Ba-133}}$), the source strengths of the three natural terrestrial radionuclides $K_{nat}$, $Th_{nat}$ and $U_{nat}$ ($\xi_{\text{K-nat}}$, $\xi_{\text{Th-nat}}$, $\xi_{\text{U-nat}}$), the source strength of the radon source term $\Delta^{222}_{86}$Rn ($\Delta\xi_{\text{Rn-222}}$) and the dispersion parameter of the gamma-Poisson mixture distribution ($\alpha_{NB}$). The convergence is shown for the maximum a posteriori (MAP) probability estimate $\mathbf{x}_{MAP}$, the posterior mean $\mathbf{x}_{Mean}$ and the posterior median $\mathbf{x}_{Median}$. The trace plots display a subset of 6 out of the 24 simulated MCMC chains. The burn-in phase is highlighted for all subgraphs as gray-shaded areas with the related threshold marked by the dashed-dotted black line.

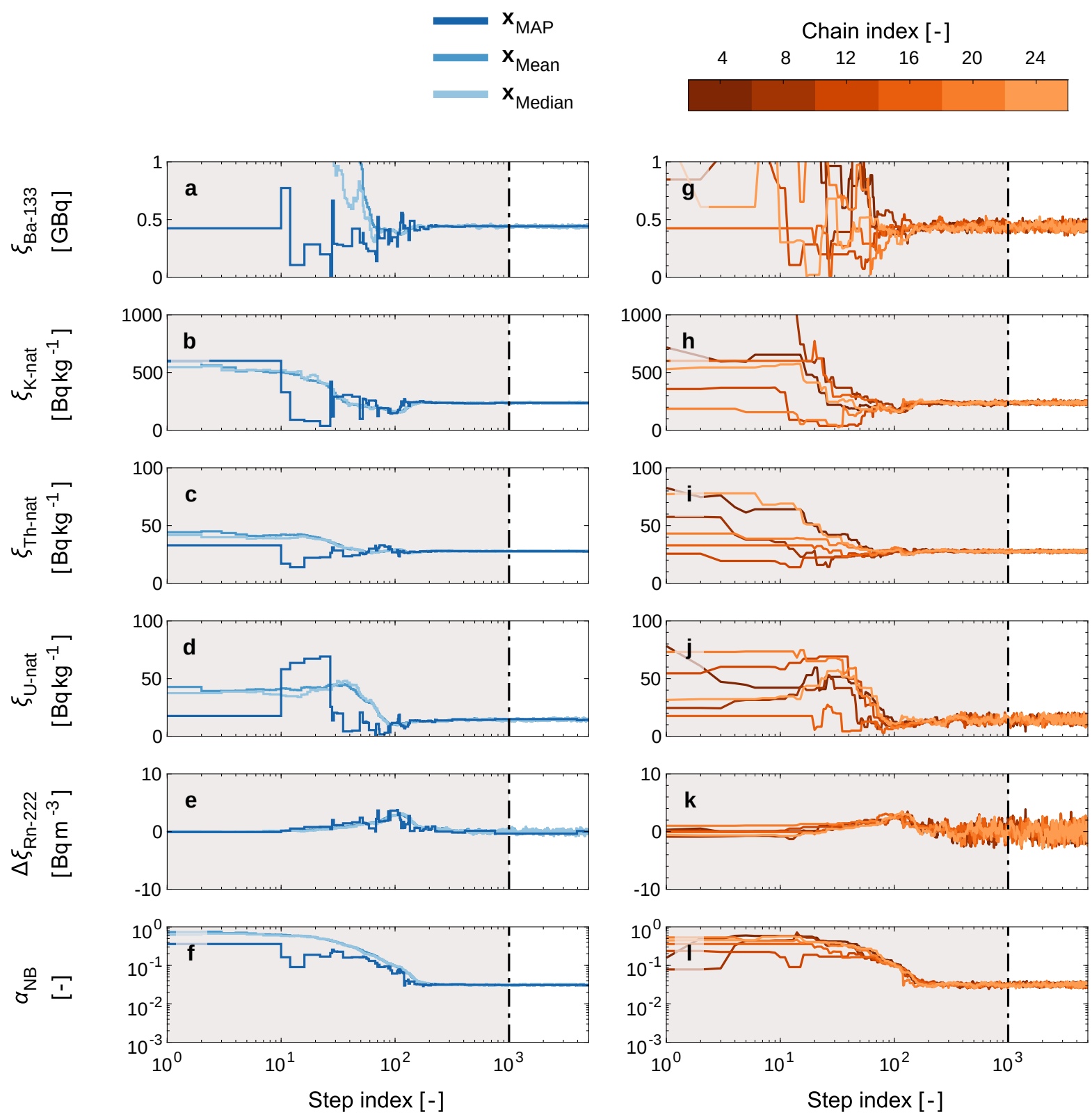

**Figure B.105** These graphs present the convergence (**a–f**) and trace (**g–l**) plots for the MCMC-based FSBI of the dataset `Ba_III` (cf. Table 10.1). The graphs are displayed as a function of the MCMC steps for each model parameter, i.e. the source strength of the sealed $^{133}_{56}$Ba point source ($\xi_{\text{Ba-133}}$), the source strengths of the three natural terrestrial radionuclides $K_{\text{nat}}$, $Th_{\text{nat}}$ and $U_{\text{nat}}$ ($\xi_{\text{K-nat}}$, $\xi_{\text{Th-nat}}$, $\xi_{\text{U-nat}}$), the source strength of the radon source term $\Delta^{222}_{86}$Rn ($\Delta\xi_{\text{Rn-222}}$) and the dispersion parameter of the gamma-Poisson mixture distribution ($\alpha_{\text{NB}}$). The convergence is shown for the maximum a posteriori (MAP) probability estimate $\mathbf{x}_{\text{MAP}}$, the posterior mean $\mathbf{x}_{\text{Mean}}$ and the posterior median $\mathbf{x}_{\text{Median}}$. The trace plots display a subset of 6 out of the 24 simulated MCMC chains. The burn-in phase is highlighted for all subgraphs as gray-shaded areas with the related threshold marked by the dashed-dotted black line.

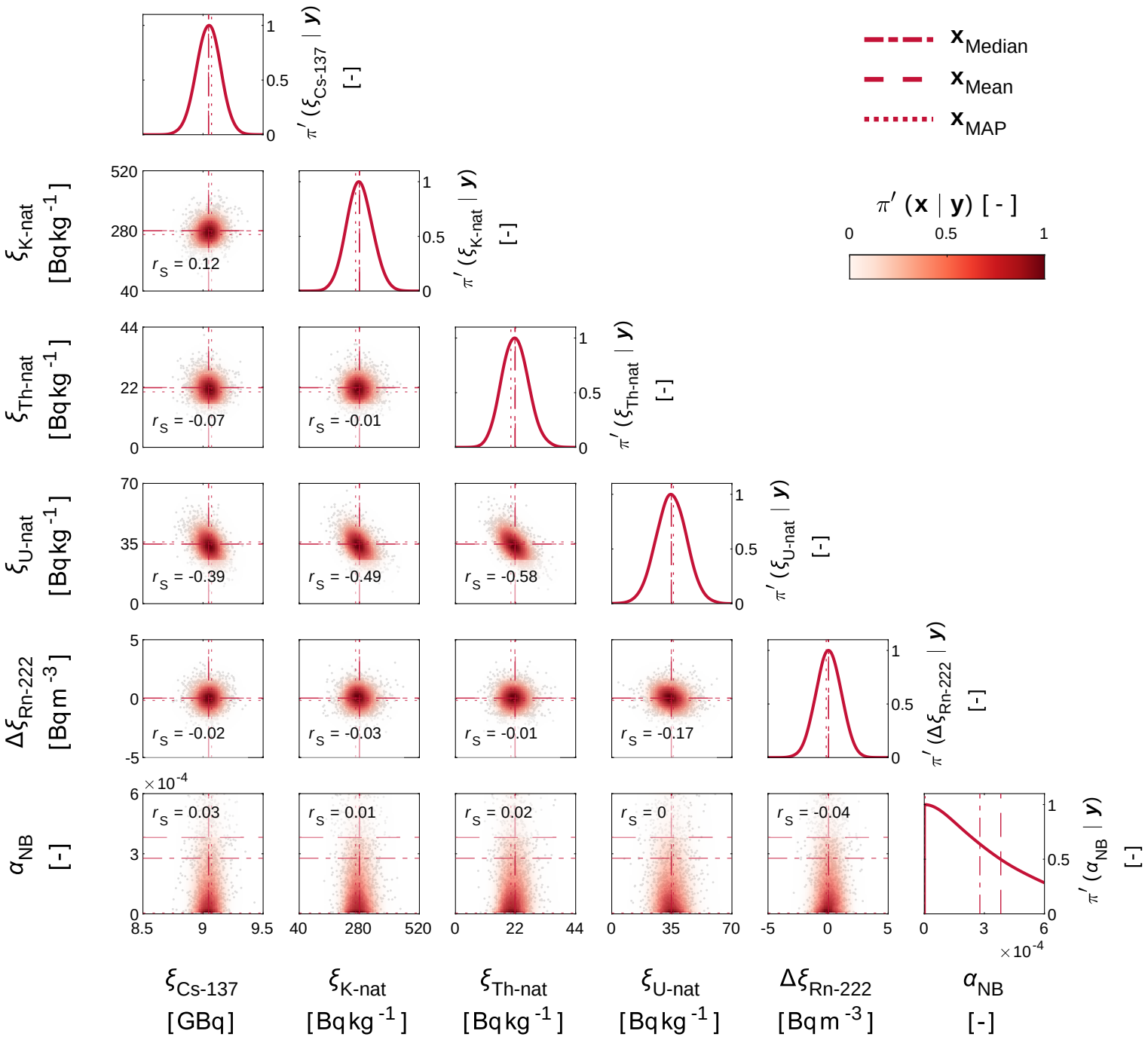

**Figure B.106** Here, I present the Bayesian inversion results for the dataset `Cs_I` (cf. Table 10.1) in the form of a corner plot. The off-diagonal subfigures display the normalized bivariate posterior marginal estimates (color-encoded) along with a subset of the first $10^3$ MCMC samples (in gray) for the experimental data $\mathbf{y}$ and model parameters $\mathbf{x}$, i.e. the source strength of the sealed $^{137}_{55}$Cs point source ($\xi_{\text{Cs-137}}$), the source strengths of the three natural terrestrial radionuclides $K_{nat}$, $Th_{nat}$ and $U_{nat}$ ($\xi_{\text{K-nat}}$, $\xi_{\text{Th-nat}}$, $\xi_{\text{U-nat}}$), the source strength of the radon source term $\Delta^{222}_{86}$Rn ($\Delta\xi_{\text{Rn-222}}$) and the dispersion parameter of the gamma-Poisson mixture distribution ($\alpha_{NB}$) (cf. Section 10.3.2). In addition, the Spearman's rank correlation coefficient $r_S$ is provided for the model parameters in the corresponding off-diagonal subfigures. The subfigures on the diagonal axis highlight the normalized univariate posterior marginals for the corresponding model parameter. Both the univariate and multivariate marginals were normalized by their corresponding global maxima. Derived posterior point estimates, i.e. the maximum a posteriori (MAP) probability estimate $\mathbf{x}_{MAP}$, the posterior mean $\mathbf{x}_{Mean}$ and the posterior median $\mathbf{x}_{Median}$ are indicated as well in each subfigure.

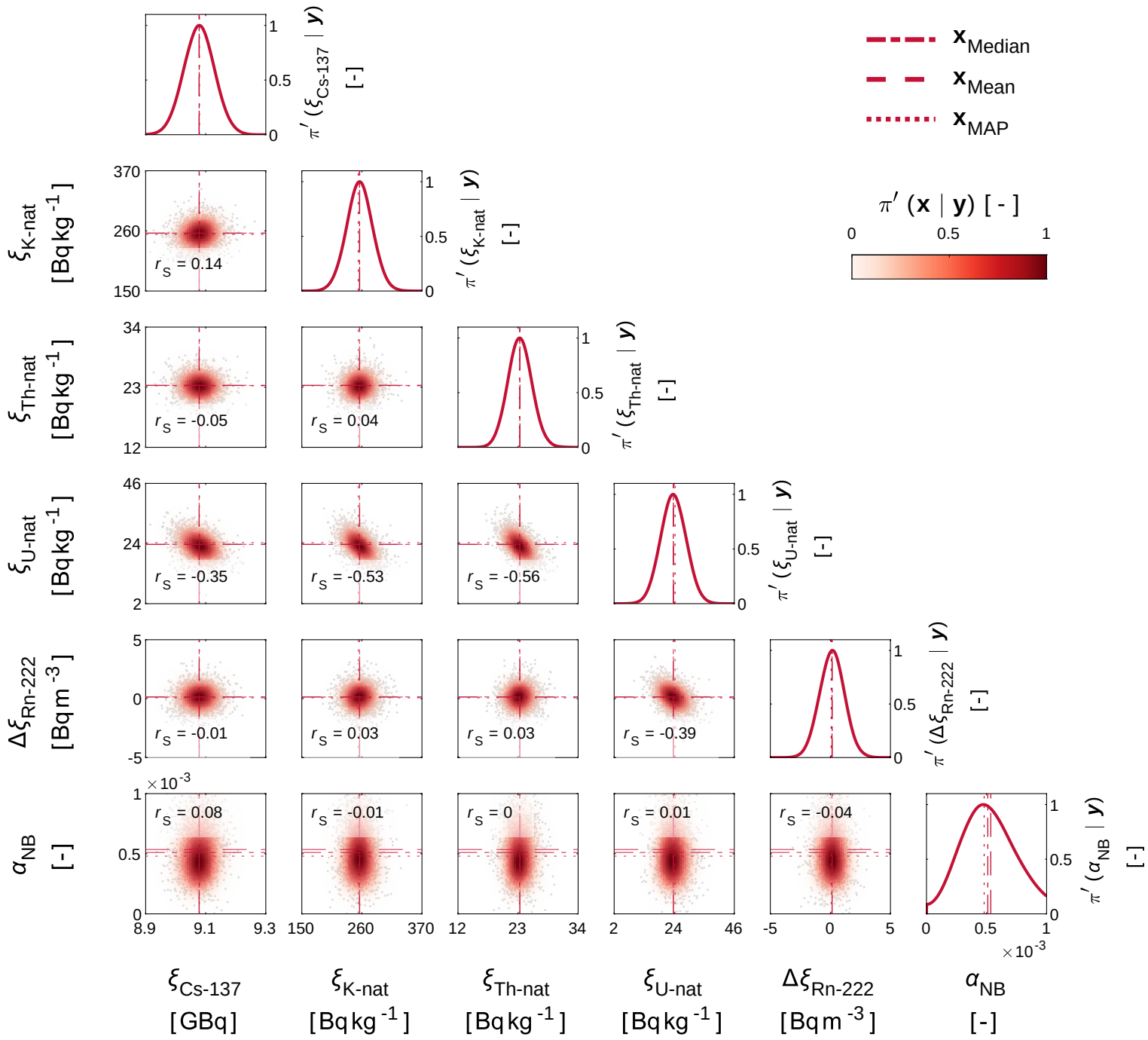

**Figure B.107** Here, I present the Bayesian inversion results for the dataset `Cs_II` (cf. Table 10.1) in the form of a corner plot. The off-diagonal subfigures display the normalized bivariate posterior marginal estimates (color-encoded) along with a subset of the first $10^3$ MCMC samples (in gray) for the experimental data $\mathbf{y}$ and model parameters $\mathbf{x}$, i.e. the source strength of the sealed $^{137}_{55}$Cs point source ($\xi_{\text{Cs-137}}$), the source strengths of the three natural terrestrial radionuclides $K_{\text{nat}}$, $Th_{\text{nat}}$ and $U_{\text{nat}}$ ($\xi_{\text{K-nat}}$, $\xi_{\text{Th-nat}}$, $\xi_{\text{U-nat}}$), the source strength of the radon source term $\Delta^{222}_{86}$Rn ($\Delta\xi_{\text{Rn-222}}$) and the dispersion parameter of the gamma-Poisson mixture distribution ($\alpha_{\text{NB}}$) (cf. Section 10.3.2). In addition, the Spearman's rank correlation coefficient $r_S$ is provided for the model parameters in the corresponding off-diagonal subfigures. The subfigures on the diagonal axis highlight the normalized univariate posterior marginals for the corresponding model parameter. Both the univariate and multivariate marginals were normalized by their corresponding global maxima. Derived posterior point estimates, i.e. the maximum a posteriori (MAP) probability estimate $\mathbf{x}_{\text{MAP}}$, the posterior mean $\mathbf{x}_{\text{Mean}}$ and the posterior median $\mathbf{x}_{\text{Median}}$ are indicated as well in each subfigure.

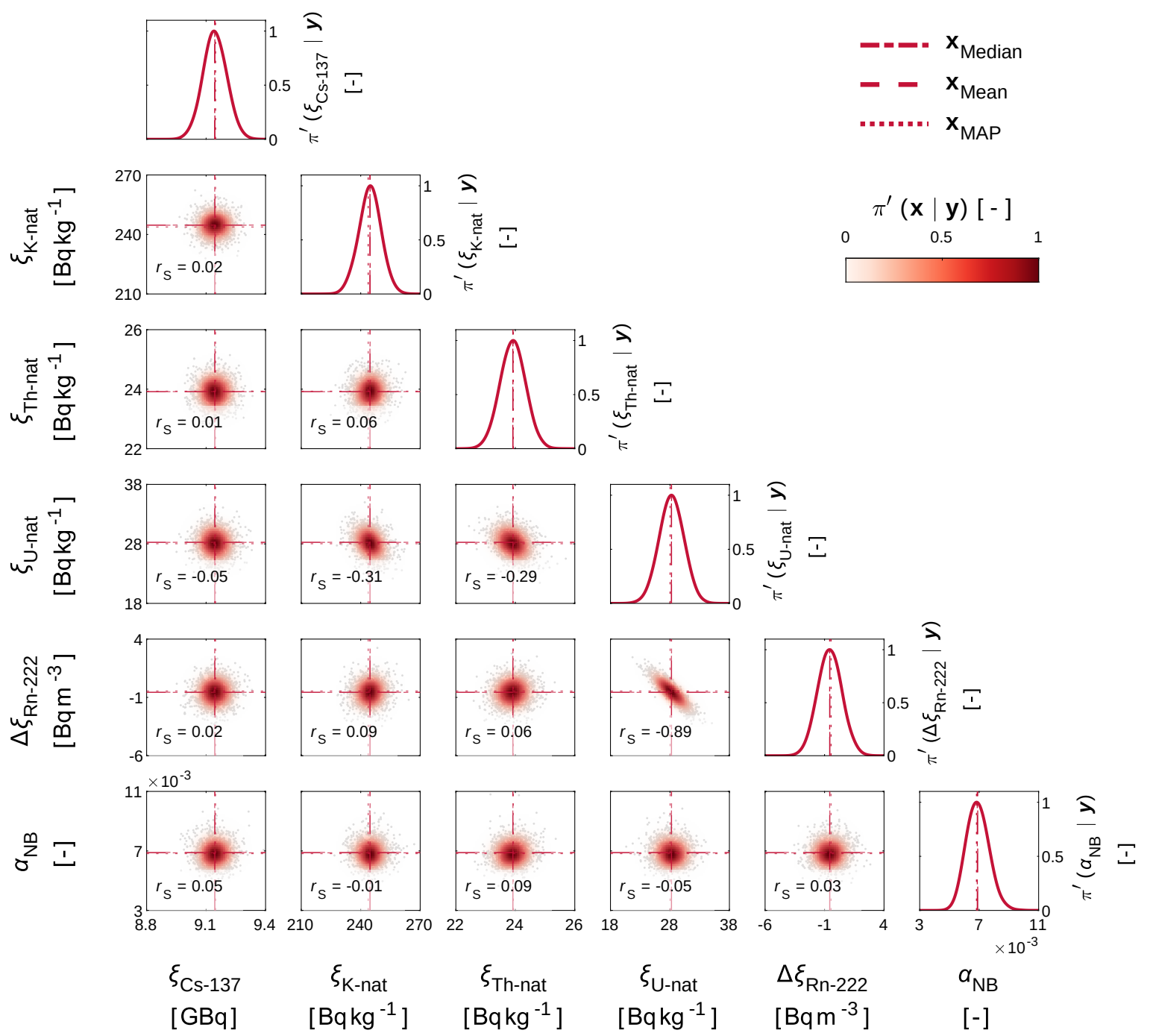

**Figure B.108** Here, I present the Bayesian inversion results for the dataset `Cs_III` (cf. Table 10.1) in the form of a corner plot. The off-diagonal subfigures display the normalized bivariate posterior marginal estimates (color-encoded) along with a subset of the first $10^3$ MCMC samples (in gray) for the experimental data $\mathbf{y}$ and model parameters $\mathbf{x}$, i.e. the source strength of the sealed $^{137}_{55}$Cs point source ($\xi_{\text{Cs-137}}$), the source strengths of the three natural terrestrial radionuclides $K_{nat}$, $Th_{nat}$ and $U_{nat}$ ($\xi_{\text{K-nat}}$, $\xi_{\text{Th-nat}}$, $\xi_{\text{U-nat}}$), the source strength of the radon source term $\Delta^{222}_{86}$Rn ($\Delta\xi_{\text{Rn-222}}$) and the dispersion parameter of the gamma-Poisson mixture distribution ($\alpha_{\text{NB}}$) (cf. Section 10.3.2). In addition, the Spearman's rank correlation coefficient $r_S$ is provided for the model parameters in the corresponding off-diagonal subfigures. The subfigures on the diagonal axis highlight the normalized univariate posterior marginals for the corresponding model parameter. Both the univariate and multivariate marginals were normalized by their corresponding global maxima. Derived posterior point estimates, i.e. the maximum a posteriori (MAP) probability estimate $\mathbf{x}_{\text{MAP}}$, the posterior mean $\mathbf{x}_{\text{Mean}}$ and the posterior median $\mathbf{x}_{\text{Median}}$ are indicated as well in each subfigure.

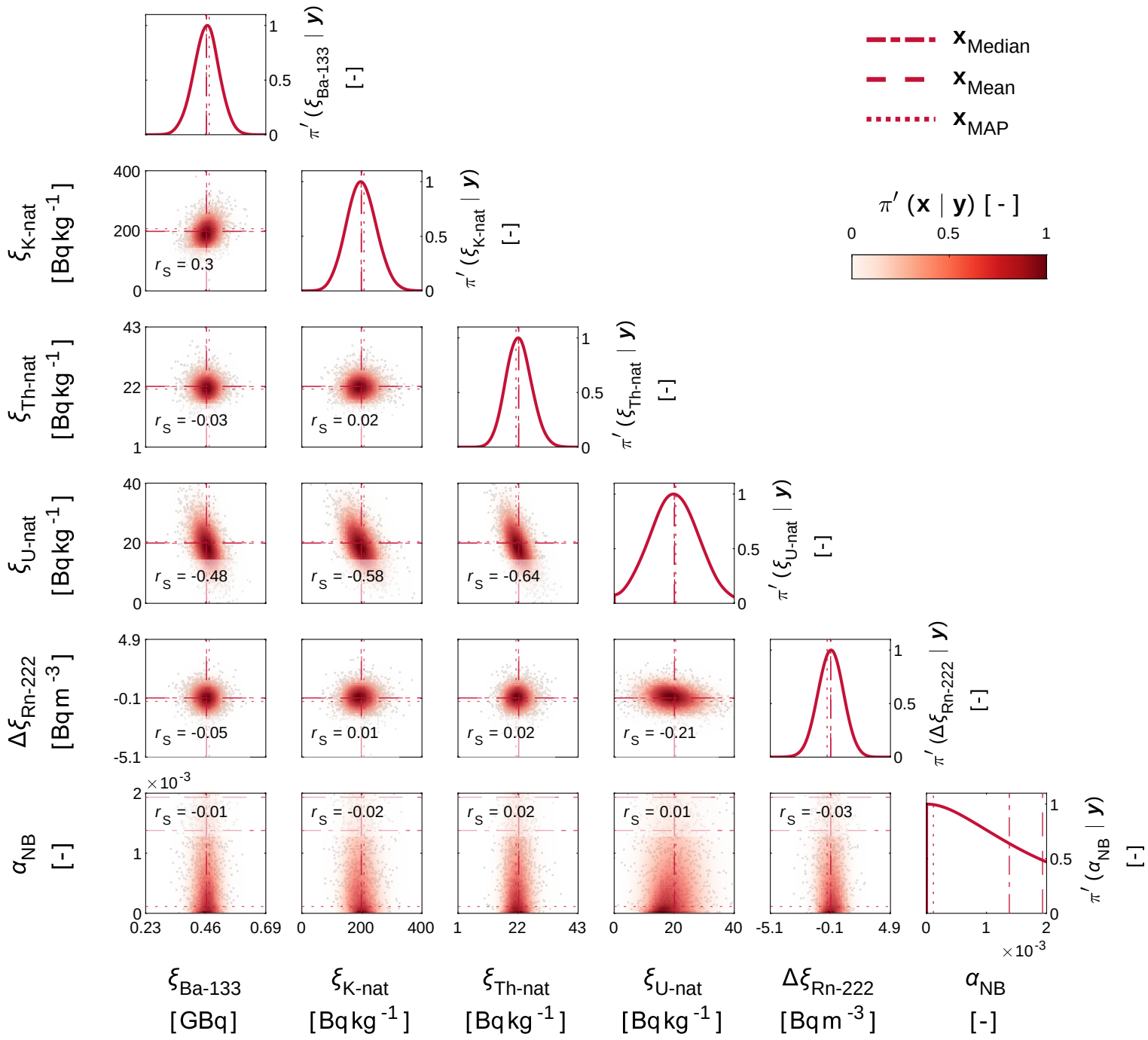

**Figure B.109** Here, I present the Bayesian inversion results for the dataset `Ba_I` (cf. Table 10.1) in the form of a corner plot. The off-diagonal subfigures display the normalized bivariate posterior marginal estimates (color-encoded) along with a subset of the first $10^3$ MCMC samples (in gray) for the experimental data $\mathbf{y}$ and model parameters $\mathbf{x}$, i.e. the source strength of the sealed $^{133}_{56}$Ba point source ($\xi_{\text{Ba-133}}$), the source strengths of the three natural terrestrial radionuclides $K_{\text{nat}}$, $Th_{\text{nat}}$ and $U_{\text{nat}}$ ($\xi_{\text{K-nat}}$, $\xi_{\text{Th-nat}}$, $\xi_{\text{U-nat}}$), the source strength of the radon source term $\Delta^{222}_{86}$Rn ($\Delta\xi_{\text{Rn-222}}$) and the dispersion parameter of the gamma-Poisson mixture distribution ($\alpha_{\text{NB}}$) (cf. Section 10.3.2). In addition, the Spearman's rank correlation coefficient $r_S$ is provided for the model parameters in the corresponding off-diagonal subfigures. The subfigures on the diagonal axis highlight the normalized univariate posterior marginals for the corresponding model parameter. Both the univariate and multivariate marginals were normalized by their corresponding global maxima. Derived posterior point estimates, i.e. the maximum a posteriori (MAP) probability estimate $\mathbf{x}_{\text{MAP}}$, the posterior mean $\mathbf{x}_{\text{Mean}}$ and the posterior median $\mathbf{x}_{\text{Median}}$ are indicated as well in each subfigure.

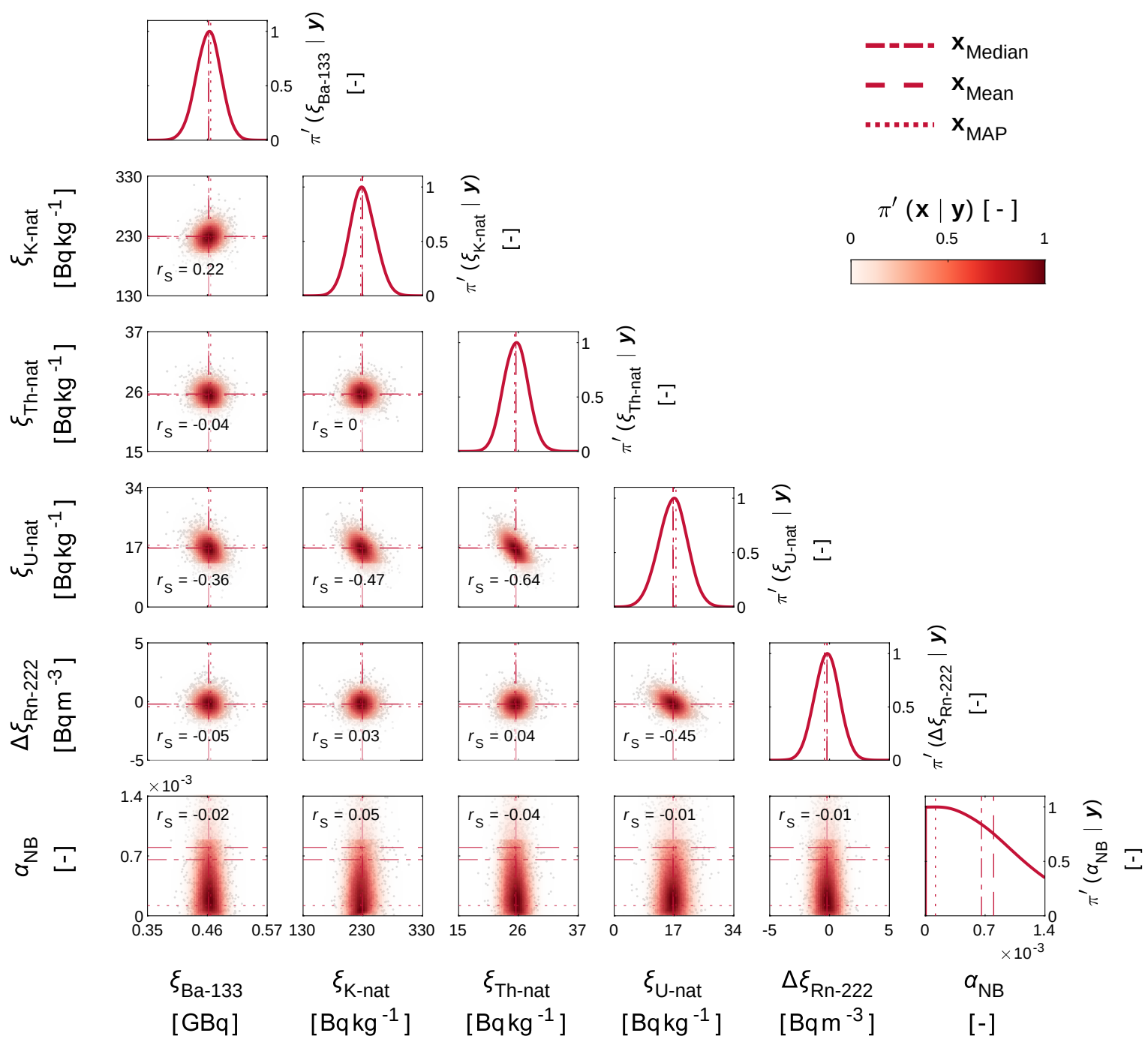

**Figure B.110** Here, I present the Bayesian inversion results for the dataset `Ba_II` (cf. Table 10.1) in the form of a corner plot. The off-diagonal subfigures display the normalized bivariate posterior marginal estimates (color-encoded) along with a subset of the first $10^3$ MCMC samples (in gray) for the experimental data $\mathbf{y}$ and model parameters $\mathbf{x}$, i.e. the source strength of the sealed $^{133}_{56}$Ba point source ($\xi_{\text{Ba-133}}$), the source strengths of the three natural terrestrial radionuclides $K_{\text{nat}}$, $Th_{\text{nat}}$ and $U_{\text{nat}}$ ($\xi_{\text{K-nat}}$, $\xi_{\text{Th-nat}}$, $\xi_{\text{U-nat}}$), the source strength of the radon source term $\Delta^{222}_{86}$Rn ($\Delta\xi_{\text{Rn-222}}$) and the dispersion parameter of the gamma-Poisson mixture distribution ($\alpha_{\text{NB}}$) (cf. Section 10.3.2). In addition, the Spearman's rank correlation coefficient $r_S$ is provided for the model parameters in the corresponding off-diagonal subfigures. The subfigures on the diagonal axis highlight the normalized univariate posterior marginals for the corresponding model parameter. Both the univariate and multivariate marginals were normalized by their corresponding global maxima. Derived posterior point estimates, i.e. the maximum a posteriori (MAP) probability estimate $\mathbf{x}_{\text{MAP}}$, the posterior mean $\mathbf{x}_{\text{Mean}}$ and the posterior median $\mathbf{x}_{\text{Median}}$ are indicated as well in each subfigure.

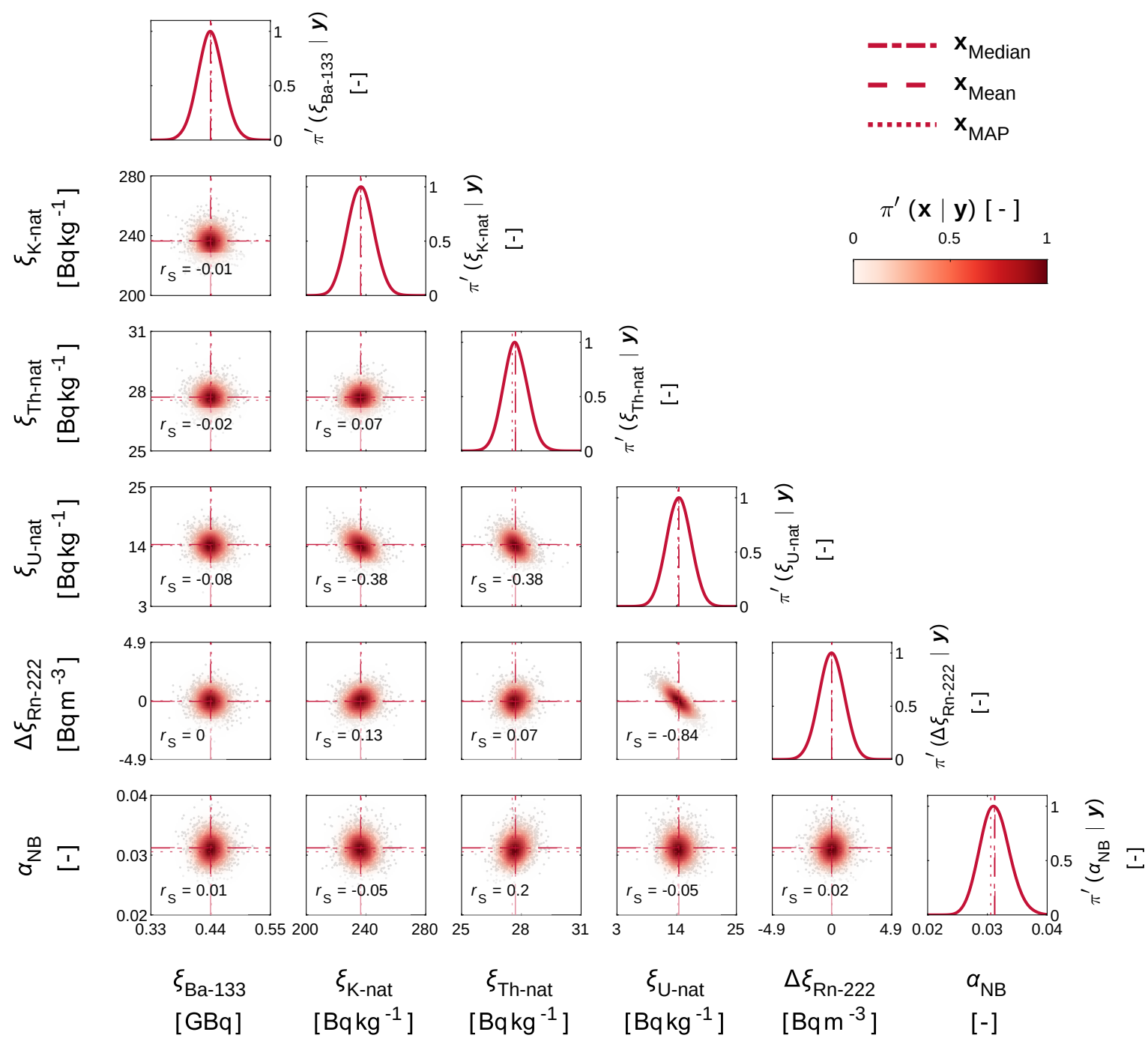

**Figure B.111** Here, I present the Bayesian inversion results for the dataset `Ba_III` (cf. Table 10.1) in the form of a corner plot. The off-diagonal subfigures display the normalized bivariate posterior marginal estimates (color-encoded) along with a subset of the first $10^3$ MCMC samples (in gray) for the experimental data $\mathbf{y}$ and model parameters $\mathbf{x}$, i.e. the source strength of the sealed $^{133}_{56}$Ba point source ($\xi_{\text{Ba-133}}$), the source strengths of the three natural terrestrial radionuclides $K_{nat}$, $Th_{nat}$ and $U_{nat}$ ($\xi_{\text{K-nat}}$, $\xi_{\text{Th-nat}}$, $\xi_{\text{U-nat}}$), the source strength of the radon source term $\Delta^{222}_{86}$Rn ($\Delta\xi_{\text{Rn-222}}$) and the dispersion parameter of the gamma-Poisson mixture distribution ($\alpha_{NB}$) (cf. Section 10.3.2). In addition, the Spearman's rank correlation coefficient $r_S$ is provided for the model parameters in the corresponding off-diagonal subfigures. The subfigures on the diagonal axis highlight the normalized univariate posterior marginals for the corresponding model parameter. Both the univariate and multivariate marginals were normalized by their corresponding global maxima. Derived posterior point estimates, i.e. the maximum a posteriori (MAP) probability estimate $\mathbf{x}_{MAP}$, the posterior mean $\mathbf{x}_{Mean}$ and the posterior median $\mathbf{x}_{Median}$ are indicated as well in each subfigure.

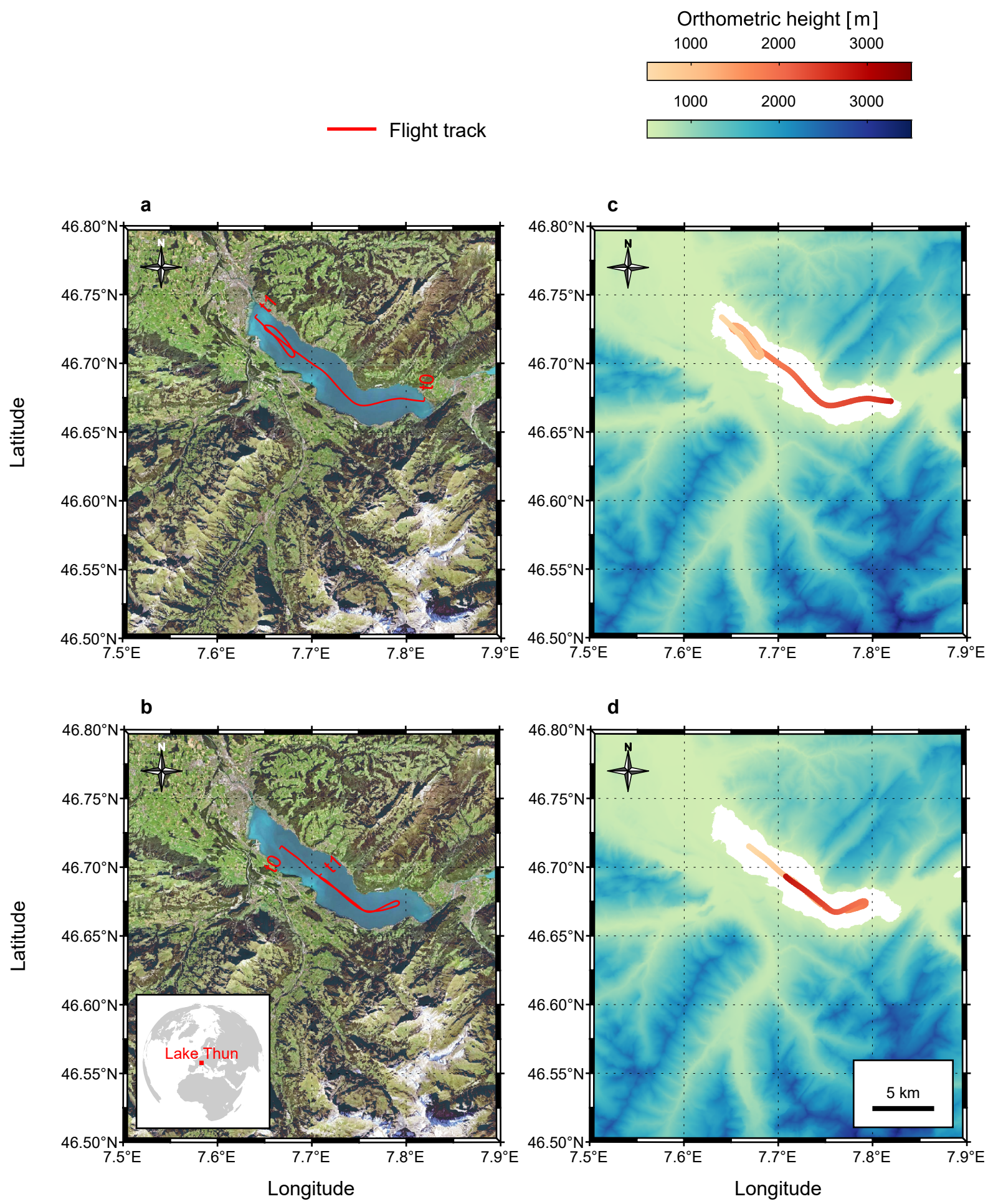

**Figure B.112** Swiss AGRS system's flight path of the ascent and descent flights over Lake Thun performed during the ARM22 validation campaign on 2022-06-16 [120]. **a-b** Orthoimagery of Lake Thun and the surrounding area with the flight path of the `Lake_I` dataset (**a**) and the `Lake_II` dataset (**b**) marked in red (t0: start, t1: end). **c-d** Elevation map of Lake Thun's surrounding area with the color-coded flight path's orthometric height marked for the `Lake_I` dataset (**c**) and the `Lake_II` dataset (**d**). Vertical datum reference: `EGM2008` [675]. Horizontal datum reference: `WGS84`. All displayed maps were created with the `m_map` software package [919]. Attribution: Federal Office of Topography swisstopo (orthoimagery & elevation data).

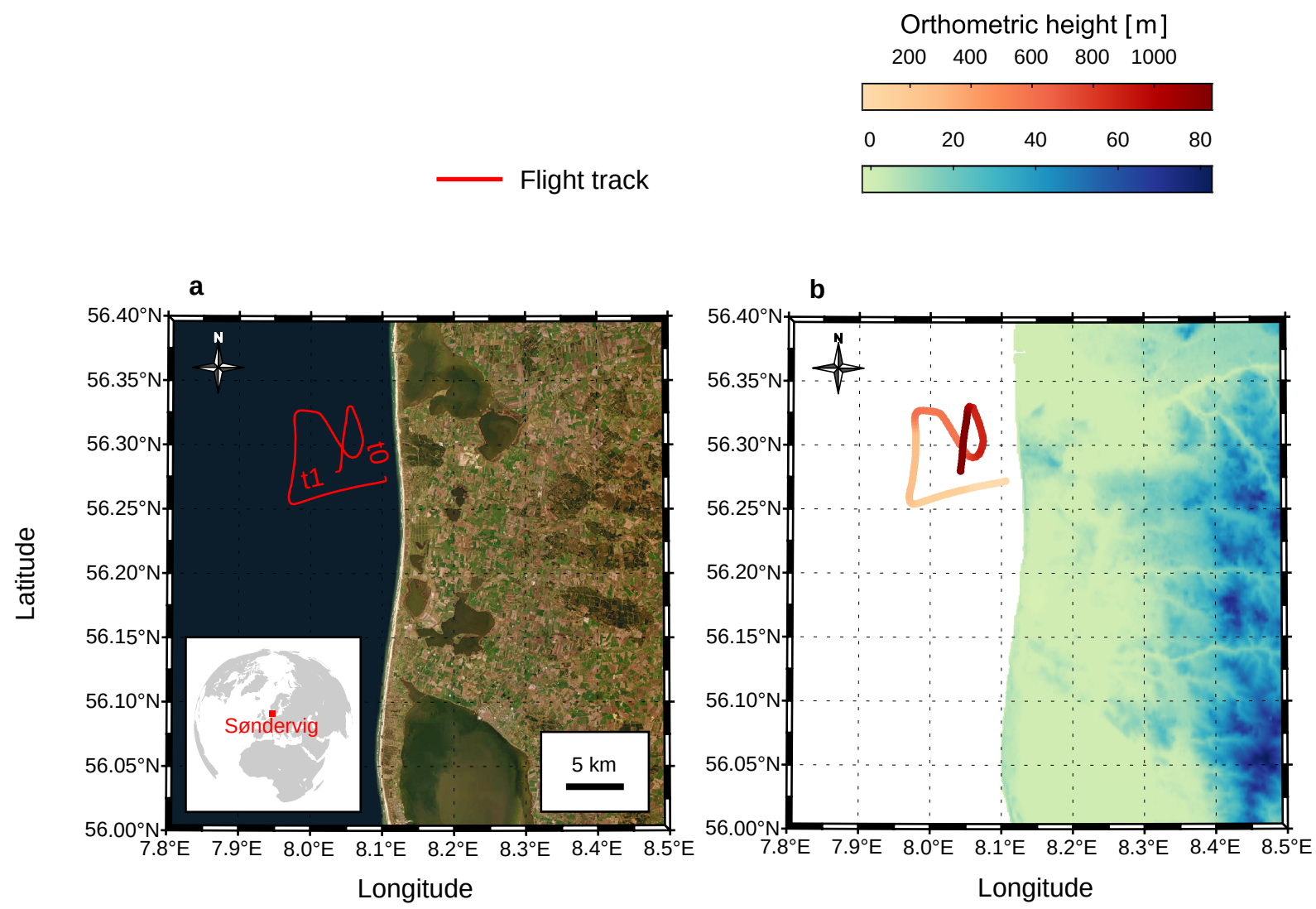

**Figure B.113** Swiss AGRS system's flight path of the ascent flight (`Sea_I` dataset) over the North Sea on 2018-06-19 performed during the International Exercise CONTEX 2018 on the west coast of Denmark close to the city Søndervig [534]. **a** Orthoimagery with the flight path marked in red (t0: start, t1: end). **b** Elevation map with the color-coded flight path's orthometric height marked. Vertical datum reference: `EGM2008` [675]. Horizontal datum reference: `WGS84`. All displayed maps were created with the `m_map` software package [919]. Attributions: Earthstar Geographics (orthoimagery), U.S. Geological Survey (elevation data, `GMTED2010`).

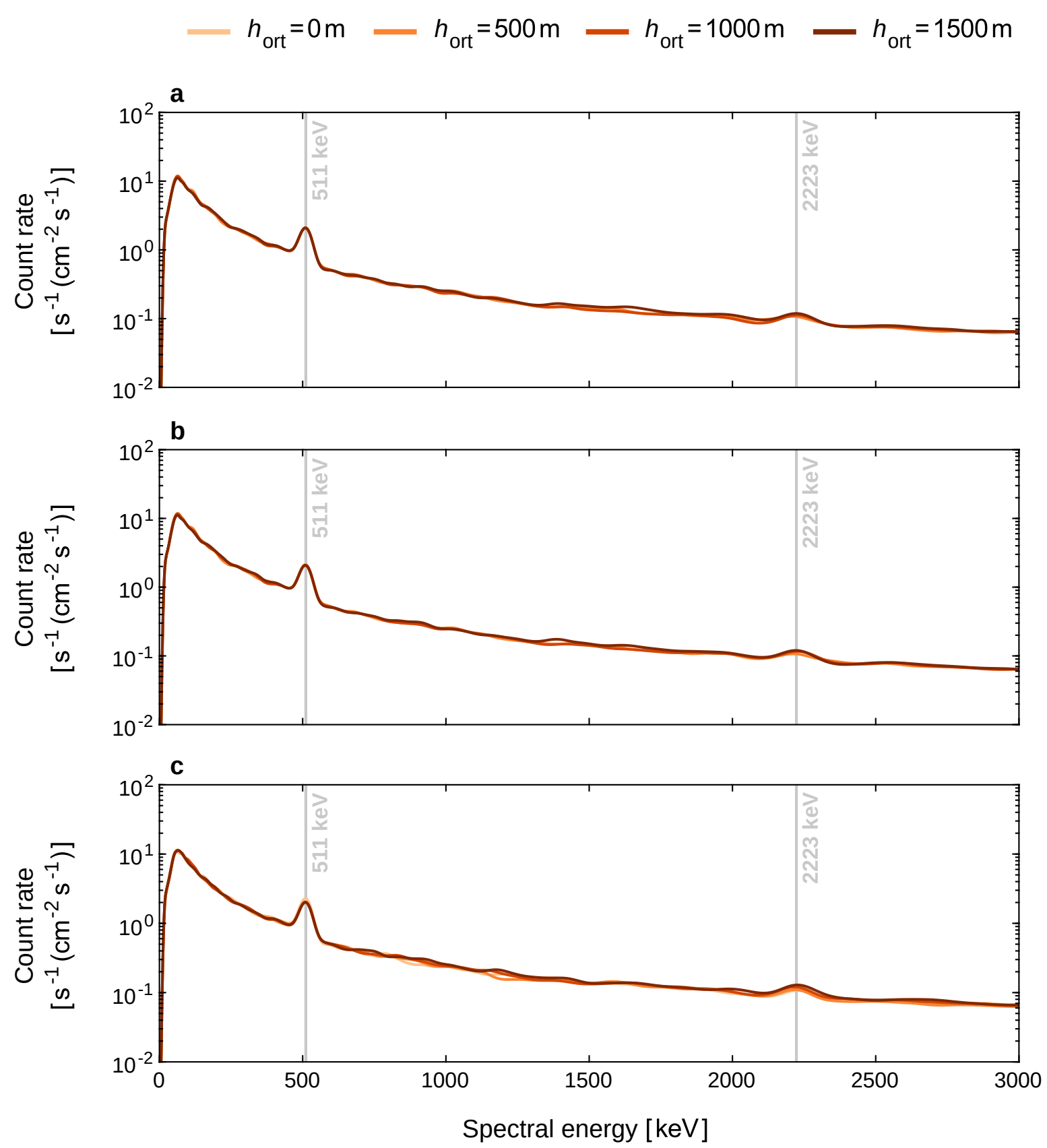

**Figure B.114** Here, I present the cosmic background spectral signature of the Swiss AGRS system in the detector channel `#SUM` (cf. Section 10.4.2.2) as a function of the spectral energy $E'$ with a spectral energy bin width of $\Delta E' \sim 3\,\mathrm{keV}$ and the orthometric height $h_{\mathrm{ort}}$ for the three datasets: **a** `Lake_I`. **b** `Lake_II`. **c** `Sea_I` (cf. Table 10.3). Uncertainties are provided as 1 standard deviation (SD) shaded areas (not visible, cf. Appendix A.8).

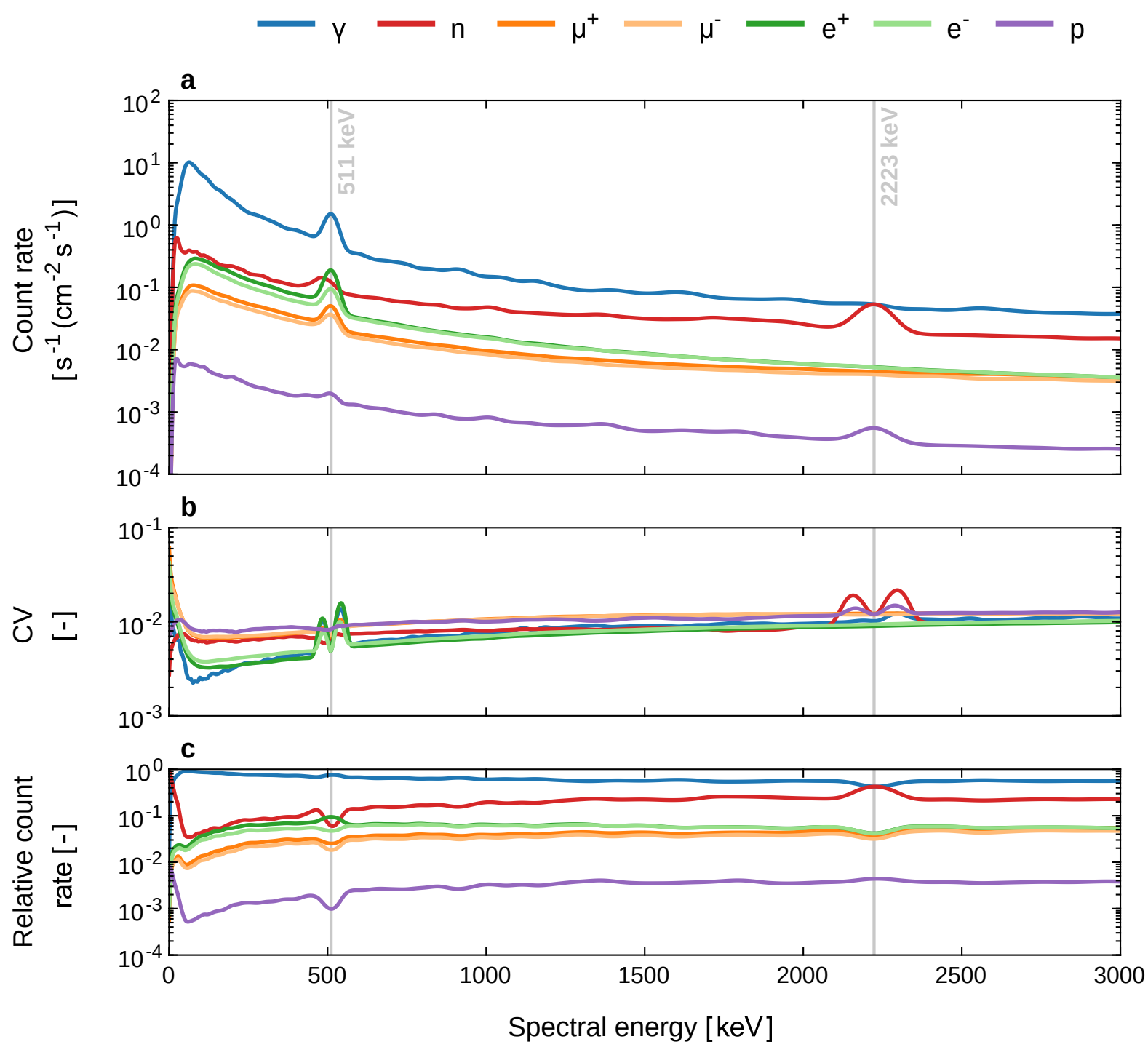

**Figure B.115** Here, I present a spectral analysis of the contribution of individual high-energy ionizing particles to the total cosmic background for the Swiss AGRS system in the detector channel `#SUM` as a function of the spectral energy $E'$ with a spectral energy bin width of $\Delta E' \sim 3$ keV (cf. Section 10.4.2.2). **a** Spectral signature of the individual cosmic background ionizing particles ($\gamma$: photon, n: neutron, $\mu^+$: antimuon, $\mu^-$: muon, $e^+$: positron, $e^-$: electron, p: proton). Uncertainties are provided as 1 standard deviation (SD) shaded areas (not visible, cf. Appendix A.8). **b** Coefficient of variation (CV) (cf. Appendix A.8). **c** Spectral signatures of the individual cosmic background ionizing particles normalized by the total cosmic spectral signature. The cosmic background was computed for a reference orthometric height of 659 m, location: 46.753°N, 7.596°E and date: 2022-06-16 (cf. Section 10.4.2.2).

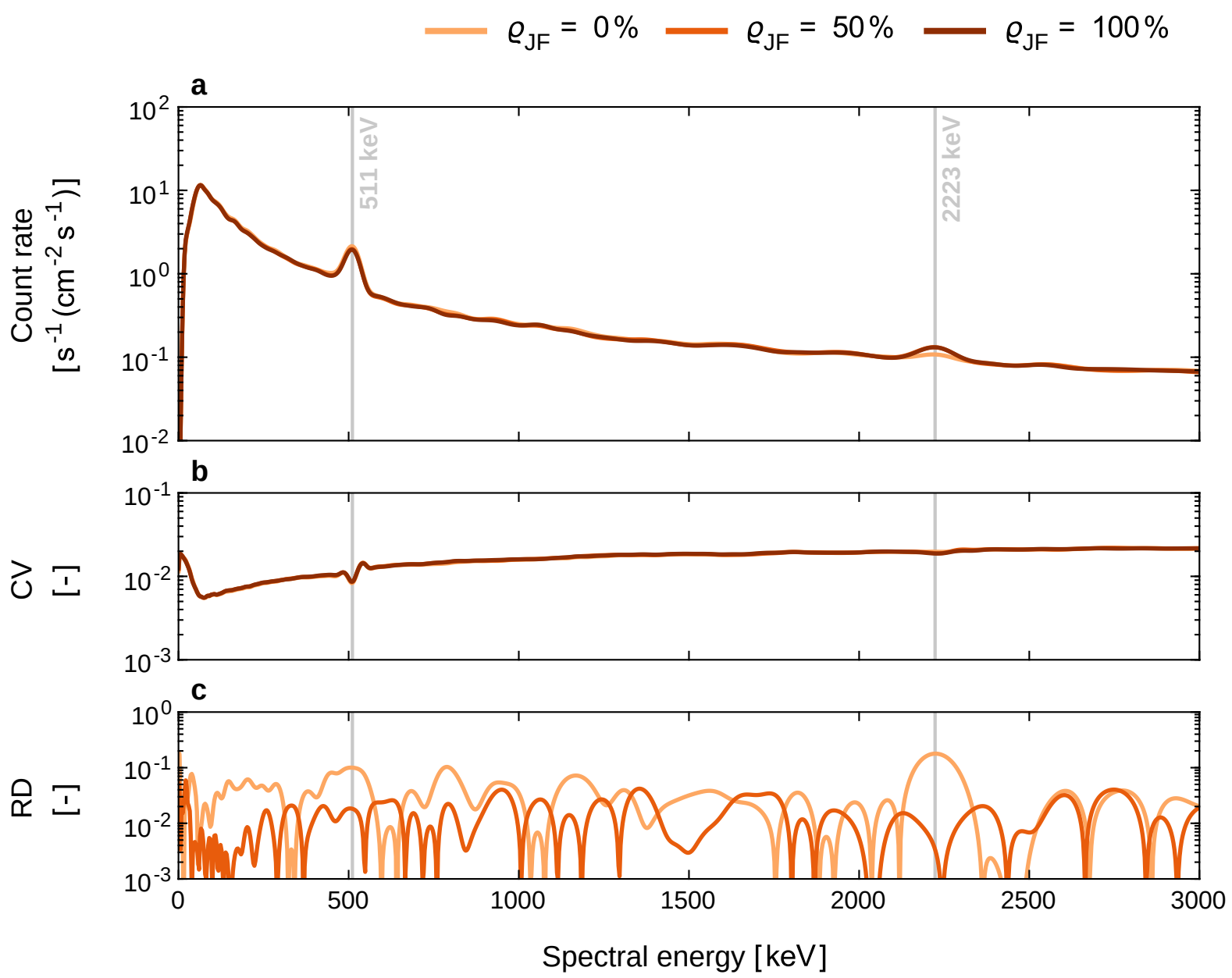

**Figure B.116** Here, I present a sensitivity analysis of the cosmic background for the Swiss AGRS system in the detector channel `#SUM` as a function of the relative fuel volume fraction $\varrho_{\mathrm{JF}}$ and the spectral energy $E'$ with a spectral energy bin width of $\Delta E' \sim 3\,\mathrm{keV}$ (cf. Section 10.4.2.2). **a** Spectral signature with uncertainties provided as 1 standard deviation (SD) shaded areas (cf. Appendix A.8). **b** Coefficient of variation (CV) (cf. Appendix A.8). **c** Relative deviation (RD) with respect to the spectral signature at $\varrho_{\mathrm{JF}} = 100\,\%$. The cosmic background was computed for a reference orthometric height of 659 m, location: 46.753°N, 7.596°E and date: 2022-06-16 (cf. Section 10.4.2.2).

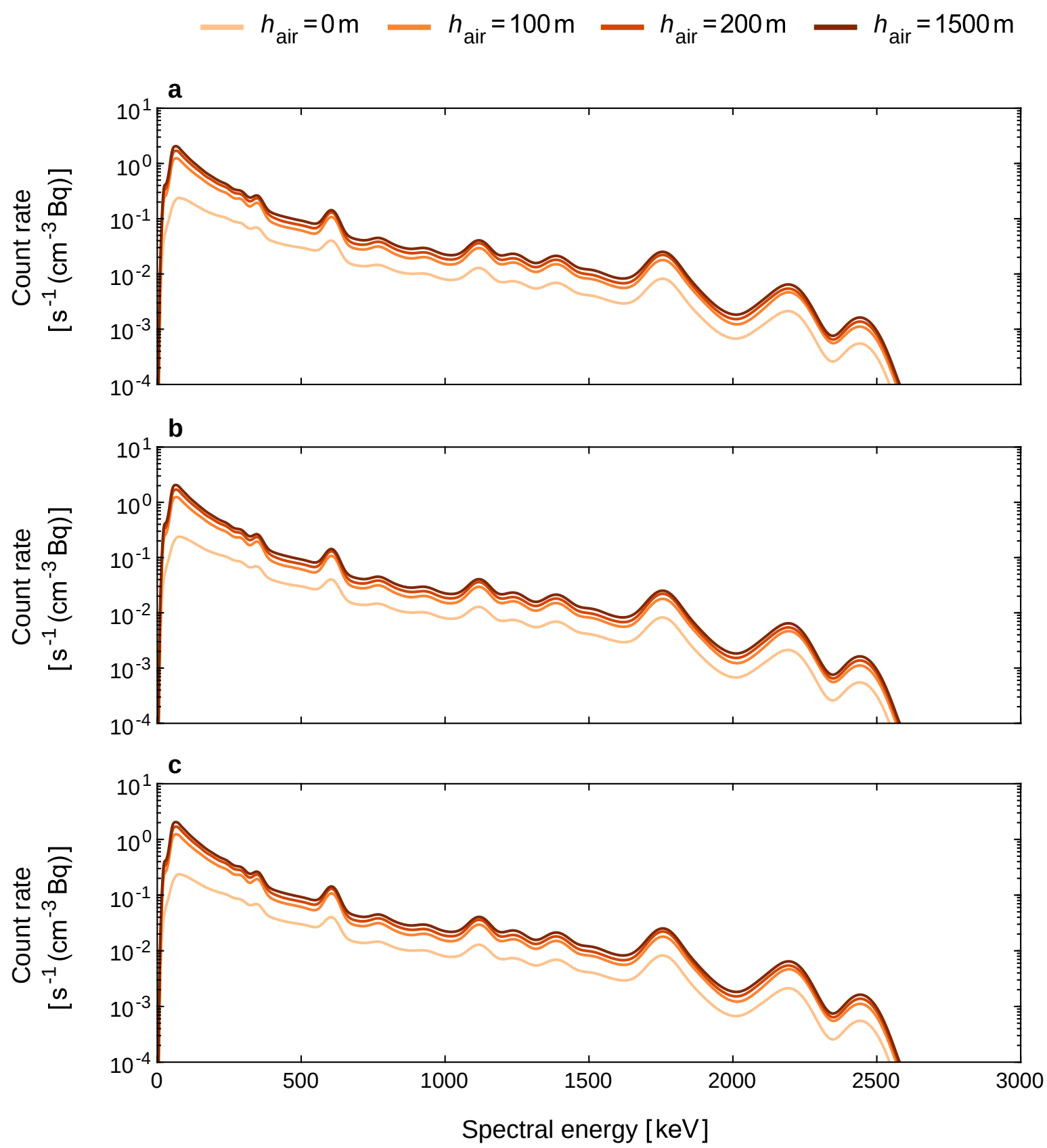

**Figure B.117** Here, I present the radon background spectral signature of the Swiss AGRS system in the detector channel `#SUM` (cf. Section 10.4.2.2) as a function of the spectral energy $E'$ with a spectral energy bin width of $\Delta E' \sim 3\,\mathrm{keV}$ and the ground clearance $h_{\mathrm{air}}$ for the three datasets: **a** `Lake_I`. **b** `Lake_II`. **c** `Sea_I` (cf. Table 10.3). Uncertainties are provided as 1 standard deviation (SD) shaded areas (not visible, cf. Appendix A.8).

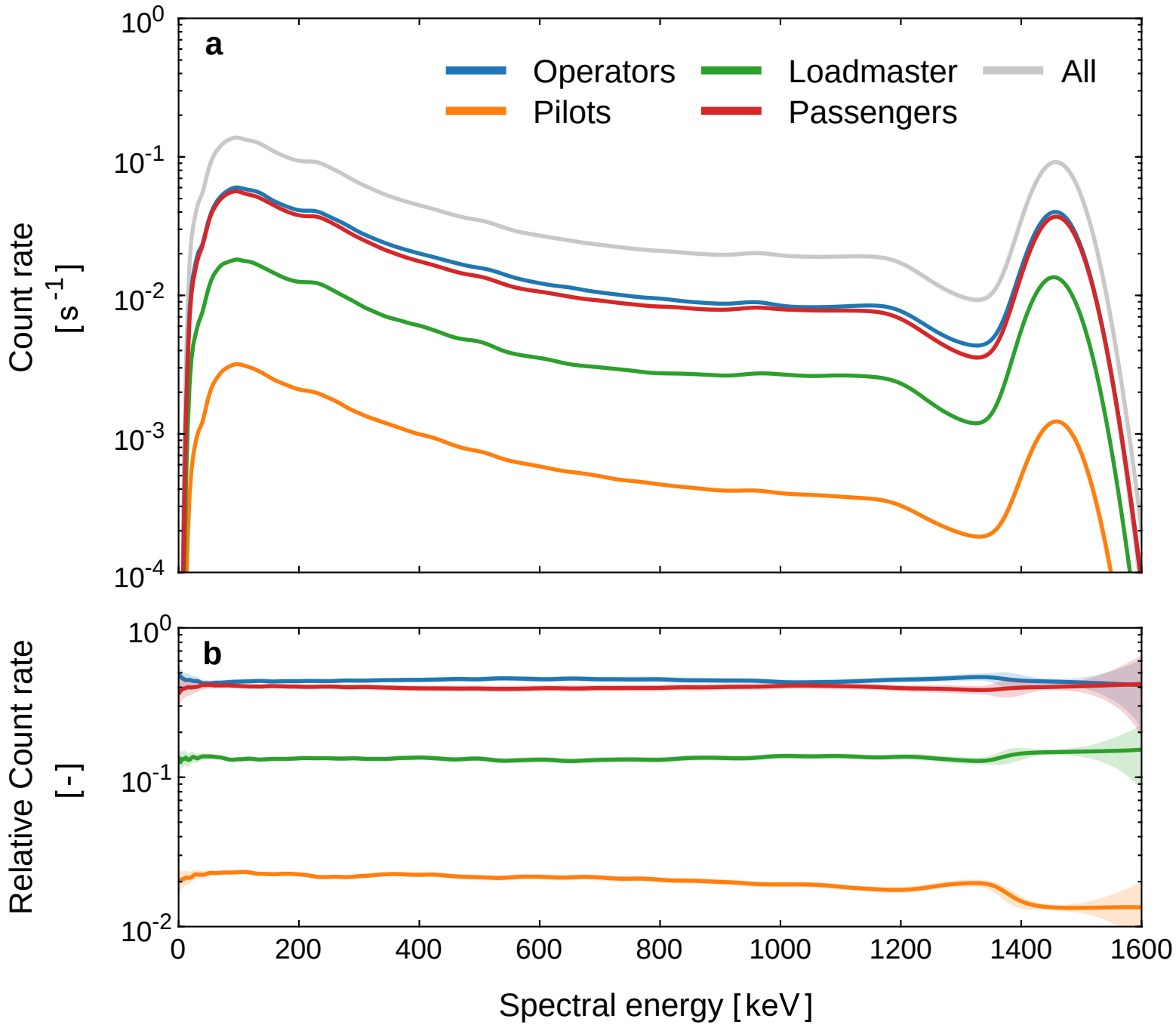

**Figure B.118** Here, I predict the intrinsic $K_{nat}$ background for the Swiss AGRS system considering only the aircraft crew (two pilots, two RLL operators, one loadmaster and seven passengers, cf. Section 8.2). I computed the spectral signature for each group of individuals separately using the methodology outlined in Section 10.4.2.2 and assuming a homogeneous distribution of $K_{nat}$ in the individuals. I then scaled the spectral signatures with the mean $^{40}_{19}K$ activities for a human body (mass of 70 kg) provided in Table 2.2. **a** Mean absolute intrinsic background prediction for the aircraft crew in the detector channel `#SUM` (cf. Section 10.4.2.2) as a function of the spectral energy $E'$ with a spectral energy bin width of $\Delta E' \sim 3$ keV. Uncertainties are provided as 1 standard deviation (SD) shaded areas (not visible, cf. Appendix A.8). **b** Intrinsic background prediction for the individual groups of aircraft crew members normalized by the combined background of all crew members (two pilots, two RLL operators, one loadmaster and seven passengers).

# Appendix C
# Supplementary Tables

**Table C.1** Uranium decay series.

| Nuclide | Half-life | Decay modes★ | Chemical group | References |
|---|---|---|---|---|
| | | (a) Main decay channel | | |
| $^{238}_{92}U$ | $4.468(5) \times 10^9$ a | $^{238}_{92}U \xrightarrow[100\,\%]{\alpha,\gamma} {}^{234}_{90}Th$ | f-block | [68] |
| $^{234}_{90}Th$● | $2.410(3) \times 10^1$ d | $^{234}_{90}Th \xrightarrow[100\,\%]{\beta^-,\gamma,IC} {}^{234}_{91}Pa$ | f-block | [65] |
| $^{234}_{91}Pa$ | 6.70(5) h | $^{234}_{91}Pa \xrightarrow[100\,\%]{\beta^-,\gamma} {}^{234}_{92}U$ | f-block | [85] |
| $^{234}_{92}U$ | $2.455(6) \times 10^5$ a | $^{234}_{92}U \xrightarrow[100\,\%]{\alpha,\gamma} {}^{230}_{90}Th$ | f-block | [68] |
| $^{230}_{90}Th$ | $7.538(30) \times 10^4$ a | $^{230}_{90}Th \xrightarrow[100\,\%]{\alpha,\gamma} {}^{226}_{88}Ra$ | f-block | [921] |
| $^{226}_{88}Ra$ | $1.600(7) \times 10^3$ a | $^{226}_{88}Ra \xrightarrow[100\,\%]{\alpha,\gamma} {}^{222}_{86}Rn$ | 2 | [51] |
| $^{222}_{86}Rn$ | 3.8232(8) d | $^{222}_{86}Rn \xrightarrow[100\,\%]{\alpha,\gamma} {}^{218}_{84}Po$ | 18 | [714] |
| $^{218}_{84}Po$ | 3.071(22) min | $^{218}_{84}Po \xrightarrow[99.978(3)\,\%]{\alpha,\gamma} {}^{214}_{82}Pb$ | 16 | [124] |
| | | $^{218}_{84}Po \xrightarrow[0.022(3)\,\%]{\beta^-} {}^{218}_{85}At$ | | |
| $^{214}_{82}Pb$ | $2.6916(44) \times 10^1$ min | $^{214}_{82}Pb \xrightarrow[100\,\%]{\beta^-,\gamma} {}^{214}_{83}Bi$ | 14 | [124] |
| $^{214}_{83}Bi$ | $1.98(1) \times 10^1$ min | $^{214}_{83}Bi \xrightarrow[99.979(13)\,\%]{\beta^-,\gamma} {}^{214}_{84}Po$ | 15 | [124] |
| | | $^{214}_{83}Bi \xrightarrow[0.0210(13)\,\%]{\alpha,\gamma} {}^{210}_{81}Tl$ | | |
| $^{214}_{84}Po$ | $1.623(12) \times 10^{-4}$ s | $^{214}_{84}Po \xrightarrow[100\,\%]{\alpha,\gamma} {}^{210}_{82}Pb$ | 16 | [124] |
| $^{210}_{82}Pb$ | $2.223(12) \times 10^1$ a | $^{210}_{82}Pb \xrightarrow[99.999\,998\,1\,\%]{\beta^-,\gamma} {}^{210}_{83}Bi$ | 14 | [124] |
| | | $^{210}_{82}Pb \xrightarrow[1.9 \times 10^{-6}\,\%]{\alpha} {}^{206}_{80}Hg$ | | |
| $^{210}_{83}Bi$ | 5.011(5) d | $^{210}_{83}Bi \xrightarrow[99.999\,86\,\%]{\beta^-} {}^{210}_{84}Po$ | 15 | [124] |
| | | $^{210}_{83}Bi \xrightarrow[1.4 \times 10^{-4}\,\%]{\beta^-} {}^{206}_{81}Tl$ | | |
| $^{210}_{84}Po$ | $1.383\,763(17) \times 10^2$ d | $^{210}_{84}Po \xrightarrow[100\,\%]{\alpha,\gamma} {}^{206}_{82}Pb$ | 16 | [124] |
| $^{206}_{82}Pb$ | stable | | 14 | |

*Continued on next page*

**Table C.1** *Continued from previous page*

| Nuclide | Half-life | Decay modes★ | Chemical group | References |
|---|---|---|---|---|
| | | (b) Minor decay channels | | |
| $^{218}_{85}At$ | 1.4(2) s | $^{218}_{85}At \xrightarrow[99.9\,\%]{\alpha,\gamma} {}^{214}_{83}Bi$ | 17 | [124] |
| | | $^{218}_{85}At \xrightarrow[0.1\,\%]{\beta^-} {}^{218}_{86}Rn$ | | |
| $^{218}_{86}Rn$ | $3.60(19) \times 10^{-2}$ s | $^{218}_{86}Rn \xrightarrow[100\,\%]{\alpha,\gamma} {}^{214}_{84}Po$ | 18 | [124] |
| $^{210}_{81}Tl$ | 1.30(3) min | $^{210}_{81}Tl \xrightarrow[100\,\%]{\beta^-,\gamma} {}^{210}_{82}Pb$ | 13 | [124] |
| $^{206}_{80}Hg$ | 8.32(7) min | $^{206}_{80}Hg \xrightarrow[100\,\%]{\beta^-,\gamma} {}^{206}_{81}Tl$ | 12 | [108] |
| $^{206}_{81}Tl$ | 4.202(11) min | $^{206}_{81}Tl \xrightarrow[100\,\%]{\beta^-,\gamma} {}^{206}_{82}Pb$ | 13 | [108] |

★ All decay modes ($\alpha$: radioactive decay with alpha particle emission, $\beta^-$: radioactive beta-decay with electron emission, $\gamma$: gamma-ray emission, IC: internal conversion) are indicated together with the corresponding branching ratios in %. Spontaneous fission decay (SF), heavy cluster Decay (HCD) and other rare decay modes, such as radioactive beta-decay with electron and delayed alpha particle emissions ($\beta^-\alpha$) or radioactive beta-decay with electron and delayed neutron particle emissions ($\beta^-$n), are not included.

• $^{234}_{90}Th$ decays with a probability of $7.78(13) \times 10^1$ % to the metastable state of $^{234}_{91}Pa$, i.e. $^{234m}_{91}Pa$. The rest of the time, $^{234}_{90}Th$ decays directly to $^{234}_{91}Pa$. $^{234m}_{91}Pa$ decays to $^{234}_{91}Pa$ by internal conversion (IC) with a half-life of 1.159(16) min [65, 920].

**Table C.2** Thorium decay series.

| Nuclide | Half-life | Decay modes★ | Chemical group | References |
|---|---|---|---|---|
| | | (a) Main decay channel | | |
| $^{244}_{94}\mathrm{Pu}$ | $8.11(3) \times 10^{7}$ a | $^{244}_{94}\mathrm{Th} \xrightarrow[99.877(6)\,\%]{\alpha,\gamma} {}^{240}_{92}\mathrm{U}$ | f-block | [922] |
| $^{240}_{92}\mathrm{U}$• | $1.41(1) \times 10^{1}$ h | $^{240}_{92}\mathrm{U} \xrightarrow[100\,\%]{\beta^-,\gamma,\mathrm{IC}} {}^{240}_{93}\mathrm{Np}$ | f-block | [922] |
| $^{240}_{93}\mathrm{Np}$ | $6.29(2) \times 10^{1}$ min | $^{240}_{93}\mathrm{Np} \xrightarrow[100\,\%]{\beta^-,\gamma} {}^{240}_{94}\mathrm{Pu}$ | f-block | [922] |
| $^{240}_{94}\mathrm{Pu}$ | $6.561(7) \times 10^{3}$ a | $^{240}_{94}\mathrm{Pu} \xrightarrow[100\,\%]{\alpha,\gamma} {}^{236}_{92}\mathrm{U}$ | f-block | [65] |
| $^{236}_{92}\mathrm{U}$ | $2.243(6) \times 10^{7}$ a | $^{236}_{92}\mathrm{U} \xrightarrow[100\,\%]{\alpha,\gamma} {}^{232}_{90}\mathrm{Th}$ | f-block | [124] |
| $^{232}_{90}\mathrm{Th}$ | $1.402(6) \times 10^{10}$ a | $^{232}_{90}\mathrm{Th} \xrightarrow[100\,\%]{\alpha,\gamma} {}^{228}_{88}\mathrm{Ra}$ | f-block | [65] |
| $^{228}_{88}\mathrm{Ra}$ | 5.75(4) a | $^{228}_{88}\mathrm{Ra} \xrightarrow[100\,\%]{\beta^-,\gamma} {}^{228}_{89}\mathrm{Ac}$ | 2 | [65] |
| $^{228}_{89}\mathrm{Ac}$ | 6.15(3) h | $^{228}_{89}\mathrm{Ac} \xrightarrow[100\,\%]{\beta^-,\gamma} {}^{228}_{90}\mathrm{Th}$ | f-block | [85] |
| $^{228}_{90}\mathrm{Th}$ | 1.9126(9) a | $^{228}_{90}\mathrm{Th} \xrightarrow[100\,\%]{\alpha,\gamma} {}^{224}_{88}\mathrm{Ra}$ | f-block | [108] |
| $^{224}_{88}\mathrm{Ra}$ | 3.631(2) d | $^{224}_{88}\mathrm{Ra} \xrightarrow[100\,\%]{\alpha,\gamma} {}^{220}_{89}\mathrm{Rn}$ | 2 | [714] |
| $^{220}_{86}\mathrm{Rn}$ | $5.58(3) \times 10^{1}$ s | $^{220}_{86}\mathrm{Rn} \xrightarrow[100\,\%]{\alpha,\gamma} {}^{216}_{84}\mathrm{Po}$ | 18 | [714] |
| $^{216}_{84}\mathrm{Po}$ | $1.48(4) \times 10^{-1}$ s | $^{216}_{84}\mathrm{Po} \xrightarrow[100\,\%]{\alpha,\gamma} {}^{212}_{82}\mathrm{Pb}$ | 16 | [714] |
| $^{212}_{82}\mathrm{Pb}$ | $1.064(1) \times 10^{1}$ h | $^{212}_{82}\mathrm{Pb} \xrightarrow[100\,\%]{\beta^-,\gamma} {}^{212}_{83}\mathrm{Bi}$ | 14 | [714] |
| $^{212}_{83}\mathrm{Bi}$ | $6.054(6) \times 10^{1}$ min | $^{212}_{83}\mathrm{Bi} \xrightarrow[64.07(7)\,\%]{\beta^-,\gamma} {}^{212}_{84}\mathrm{Po}$ | 15 | [714] |
| | | $^{212}_{83}\mathrm{Bi} \xrightarrow[35.93(7)\,\%]{\alpha,\gamma} {}^{208}_{81}\mathrm{Tl}$ | | |
| $^{212}_{84}\mathrm{Po}$ | $3.00(2) \times 10^{-7}$ s | $^{212}_{84}\mathrm{Po} \xrightarrow[100\,\%]{\alpha} {}^{208}_{82}\mathrm{Pb}$ | 16 | [714] |
| $^{208}_{82}\mathrm{Pb}$ | stable | | 14 | |
| | | (b) Minor decay channels | | |
| $^{208}_{81}\mathrm{Tl}$ | 3.058(6) min | $^{208}_{81}\mathrm{Tl} \xrightarrow[100\,\%]{\beta^-,\gamma} {}^{208}_{82}\mathrm{Pb}$ | 13 | [108] |

★ All decay modes ($\alpha$: radioactive decay with alpha particle emission, $\beta^-$: radioactive beta-decay with electron emission, $\gamma$: gamma-ray emission) are indicated together with the corresponding branching ratios in %. Spontaneous fission decay (SF), heavy cluster Decay (HCD) and other rare decay modes, such as radioactive beta-decay with electron and delayed alpha particle emissions ($\beta^-\alpha$) or radioactive beta-decay with electron and delayed neutron particle emissions ($\beta^-$n), are not included.

• $^{240}_{92}\mathrm{U}$ decays with a probability of $1.2(1) \times 10^{-1}$ % to the metastable state of $^{240}_{93}\mathrm{Np}$, i.e. $^{240\mathrm{m}}_{93}\mathrm{Np}$. The rest of the time, $^{240}_{92}\mathrm{U}$ decays directly to $^{240}_{93}\mathrm{Np}$. $^{240\mathrm{m}}_{93}\mathrm{Np}$ decays to $^{240}_{93}\mathrm{Np}$ by internal conversion (IC) with a half-life of 7.22(2) min [922].

**Table C.3** Actinium decay series.

| Nuclide | Half-life | Decay modes★ | Chemical group | References |
|---|---|---|---|---|
| | | (a) Main decay channel | | |
| $^{235}_{92}\mathrm{U}$ | $7.04(1) \times 10^{8}$ a | $^{235}_{92}\mathrm{U} \xrightarrow[100\,\%]{\alpha,\gamma} {}^{231}_{90}\mathrm{Th}$ | f-block | [65] |
| $^{231}_{90}\mathrm{Th}$ | $2.5522(10) \times 10^{1}$ h | $^{231}_{90}\mathrm{Th} \xrightarrow[100\,\%]{\beta^-,\gamma} {}^{231}_{91}\mathrm{Pa}$ | f-block | [65] |
| $^{231}_{91}\mathrm{Pa}$ | $3.267(26) \times 10^{4}$ a | $^{231}_{91}\mathrm{Pa} \xrightarrow[100\,\%]{\alpha,\gamma} {}^{227}_{89}\mathrm{Ac}$ | f-block | [85] |
| $^{227}_{89}\mathrm{Ac}$ | $2.1772(3) \times 10^{1}$ a | $^{227}_{89}\mathrm{Ac} \xrightarrow[98.6200(36)\,\%]{\beta^-,\gamma} {}^{227}_{90}\mathrm{Th}$ | f-block | [124, 923] |
| | | $^{227}_{89}\mathrm{Ac} \xrightarrow[1.3800(36)\,\%]{\alpha,\gamma} {}^{223}_{87}\mathrm{Fr}$ | | |
| $^{227}_{90}\mathrm{Th}$ | $1.8697(7) \times 10^{1}$ d | $^{227}_{90}\mathrm{Th} \xrightarrow[100\,\%]{\alpha,\gamma} {}^{223}_{88}\mathrm{Ra}$ | f-block | [923] |
| $^{223}_{88}\mathrm{Ra}$ | $1.143(3) \times 10^{1}$ d | $^{223}_{88}\mathrm{Ra} \xrightarrow[100\,\%]{\alpha,\gamma} {}^{219}_{86}\mathrm{Rn}$ | 2 | [85] |
| $^{219}_{86}\mathrm{Rn}$ | 3.98(3) s | $^{219}_{86}\mathrm{Rn} \xrightarrow[100\,\%]{\alpha,\gamma} {}^{215}_{84}\mathrm{Po}$ | 18 | [85] |
| $^{215}_{84}\mathrm{Po}$ | $1.781(5) \times 10^{-3}$ s | $^{215}_{84}\mathrm{Po} \xrightarrow[99.99977(2)\,\%]{\alpha,\gamma} {}^{211}_{82}\mathrm{Pb}$ | 16 | [85, 924] |
| | | $^{215}_{84}\mathrm{Po} \xrightarrow[2.3(2) \times 10^{-4}\,\%]{\beta^-,\gamma} {}^{215}_{82}\mathrm{At}$ | | |
| $^{211}_{82}\mathrm{Pb}$ | $3.61(2) \times 10^{1}$ min | $^{211}_{82}\mathrm{Pb} \xrightarrow[100\,\%]{\beta^-,\gamma} {}^{211}_{83}\mathrm{Bi}$ | 14 | [108] |
| $^{211}_{83}\mathrm{Bi}$ | 2.15(2) min | $^{211}_{83}\mathrm{Bi} \xrightarrow[99.724(4)\,\%]{\alpha,\gamma} {}^{207}_{81}\mathrm{Tl}$ | 15 | [65] |
| | | $^{211}_{83}\mathrm{Bi} \xrightarrow[0.276(4)\,\%]{\beta^-} {}^{211}_{84}\mathrm{Po}$ | | |
| $^{207}_{81}\mathrm{Tl}$ | 4.774(12) min | $^{207}_{81}\mathrm{Tl} \xrightarrow[100\,\%]{\beta^-,\gamma} {}^{207}_{82}\mathrm{Pb}$ | 13 | [108] |
| $^{207}_{82}\mathrm{Pb}$ | stable | | 14 | |
| | | (b) Minor decay channels | | |
| $^{223}_{87}\mathrm{Fr}$ | $2.200(7) \times 10^{1}$ min | $^{223}_{87}\mathrm{Fr} \xrightarrow[99.980(4)\,\%]{\beta^-,\gamma} {}^{223}_{88}\mathrm{Ra}$ | 1 | [85] |
| | | $^{223}_{87}\mathrm{Fr} \xrightarrow[0.020(4)\,\%]{\alpha,\gamma} {}^{219}_{85}\mathrm{At}$ | | |
| $^{219}_{85}\mathrm{At}$ | $5.6(4) \times 10^{1}$ s | $^{219}_{85}\mathrm{At} \xrightarrow[93.6(10)\,\%]{\alpha,\gamma} {}^{215}_{83}\mathrm{Bi}$ | 17 | [85, 925] |
| | | $^{219}_{85}\mathrm{At} \xrightarrow[6.4(10)\,\%]{\beta^-} {}^{219}_{86}\mathrm{Rn}$ | | |
| $^{215}_{83}\mathrm{Bi}$ | 7.6(2) min | $^{215}_{83}\mathrm{Bi} \xrightarrow[100\,\%]{\beta^-,\gamma} {}^{215}_{84}\mathrm{Po}$ | 15 | [108] |
| $^{215}_{85}\mathrm{At}$ | $1.0(2) \times 10^{-4}$ s | $^{215}_{85}\mathrm{At} \xrightarrow[100\,\%]{\alpha,\gamma} {}^{211}_{83}\mathrm{Bi}$ | 17 | [108] |
| $^{211}_{84}\mathrm{Po}$ | $5.16(3) \times 10^{-1}$ s | $^{211}_{84}\mathrm{Po} \xrightarrow[100\,\%]{\alpha,\gamma} {}^{207}_{82}\mathrm{Pb}$ | 16 | [85] |

★ All decay modes ($\alpha$: radioactive decay with alpha particle emission, $\beta^-$: radioactive beta-decay with electron emission, $\gamma$: gamma-ray emission) are indicated together with the corresponding branching ratios in %. Spontaneous fission decay (SF), heavy cluster Decay (HCD) and other rare decay modes, such as radioactive beta-decay with electron and delayed alpha particle emissions ($\beta^-\alpha$) or radioactive beta-decay with electron and delayed neutron particle emissions ($\beta^-$n), are not included.

**Table C.4** Scintillation non-proportionality parameters of common inorganic scintillators used in gamma-ray spectrometry.

| Material★ | $\mathrm{d}E/\mathrm{d}x\,\vert_{\mathrm{Birks}}$ [MeV cm$^{-1}$]● | $\eta_{\mathrm{e/h}}$ [%]○ | References |
|---|---|---|---|
| **I. Iodides** | | | |
| NaI(Tl) | 166–339 | 44–53 | [313, 324] |
| CsI(Tl) | 556–1338 | 33–36 | [313, 926] |
| CsI(Na) | 667 | 40 | [313] |
| $SrI_2$(Eu) | 454–556 | 21–24 | [313, 324, 423] |
| **II. Bromides** | | | |
| $CeBr_3$† | 133 | 35 | [325] |
| $LaBr_3$(Ce) | 392–500 | 16–19 | [313, 324, 423] |
| **III. Oxides** | | | |
| BGO | 364–455 | 10–16 | [324, 926] |
| GAGG(Ce) | 400 ± 140 | 1 ± 6 | ‡ |
| GSO(Ce) | 556 | 7 | [313] |
| LSO(Ce) | 133–208 | 13–14 | [313, 324] |
| LYSO(Ce) | 45 | 0.76 | [926] |
| YAG(Ce) | 526–541 | 8–10 | [313, 324] |
| YAP(Ce) | 606–741 | 8–12 | [313, 324] |
| YSO(Ce) | 167 | 13 | [313] |
| **IV. Chlorides & Sulfides** | | | |
| $BaF_2$ | 546 | 32 | [926] |
| $CaF_2$(Eu) | 333 | 12 | [313] |
| $LaCl_3$(Ce) | 345–465 | 16–22 | [313, 324] |

★ For the sake of brevity, if available, common scintillator abbreviations are listed. The corresponding chemical formulas are as follows: BGO: $Bi_4Ge_3O_{12}$, GAGG(Ce): $Gd_3Al_2Ga_3O_{12}$(Ce), GSO(Ce): $Gd_2SiO_5$(Ce), LSO(Ce): $Lu_2SiO_5$(Ce), LYSO(Ce): $Lu_{1.8}Y_{0.2}SiO_2$(Ce), YAG(Ce): $Y_3Al_5O_{12}$(Ce), YAP(Ce): $YAlO_3$(Ce), YSO(Ce): $Y_2SiO_5$(Ce).

● Birks stopping parameter.

○ Electron-hole pair fraction.

† Temperature interpolated at 20 °C based on a data set provided by Payne et al. [325].

‡ Because GAGG(Ce) scintillators are still in a comparably early stage in their development, non-proportionality parameters are not readily available. Therefore, I estimate $\mathrm{d}E/\mathrm{d}x\,\vert_{\mathrm{Birks}}$ and $\eta_{\mathrm{e/h}}$ using experimental data provided by Kaewkhao et al. [342]. Uncertainty estimates are provided as 1 standard deviation (SD) values (coverage factor $k = 1$).

**Table C.5** Adopted photon emission lines for laboratory-based spectral calibration using the calibration pipeline `RLLCal`.

| Nuclide⋆ | Photon energy [keV] | References |
|---|---|---|
| $^{40}_{19}$K | 1460.822(6) | [65] |
| $^{60}_{27}$Co | 1173.228(3) | [68] |
| | 1332.492(4) | [68] |
| $^{88}_{39}$Y | 898.042(11) | [51] |
| | 1836.070(8) | [51] |
| $^{109}_{48}$Cd | 88.0336(10) | [102] |
| $^{137}_{55}$Cs | 661.657(3) | [68] |
| $^{152}_{63}$Eu | 121.7817(3) | [714] |
| | 344.2785(12) | [714] |
| $^{208}_{81}$Tl | 2614.511(10) | [108] |
| $^{214}_{82}$Pb• | 351.932(2) | [124] |
| $^{214}_{83}$Bi | 609.312(7) | [124] |

⋆ Additional source properties are included in Table 6.1. For $^{40}_{19}$K, $^{232}_{90}$Th, $^{214}_{82}$Pb and $^{214}_{83}$Bi, natural (uncalibrated) sources ($K_{nat}$, $Th_{nat}$, $U_{nat}$) were used.

• An additional weak photon emission line at 295.224(2) keV was used to improve the peak fit of the 351.932(2) keV FEP (cf. Fig. B.29). However, because of the reduced emission intensity, this additional line was excluded from the spectral calibration.

**Table C.6** Laboratory-based spectral calibration results derived by the RLLCal pipeline.

| Detector channel | $\Delta E'$★ [keV] | $a_1$● [-] | $a_2$○ [-] | corr($\log a_1, a_2$)† [-] |
|---|---|---|---|---|
| #1 | 3.0172(33) | $5.7(4) \times 10^{-1}$ | $6.39(11) \times 10^{-1}$ | $-9.93 \times 10^{-1}$ |
| #2 | 3.010(4) | $5.4(4) \times 10^{-1}$ | $6.31(12) \times 10^{-1}$ | $-9.87 \times 10^{-1}$ |
| #3 | 2.9890(19) | $5.3(4) \times 10^{-1}$ | $6.53(14) \times 10^{-1}$ | $-9.89 \times 10^{-1}$ |
| #4 | 2.9979(17) | $5.0(4) \times 10^{-1}$ | $6.44(12) \times 10^{-1}$ | $-9.92 \times 10^{-1}$ |
| #SUM | 3.0029(26) | $5.7(4) \times 10^{-1}$ | $6.33(11) \times 10^{-1}$ | $-9.92 \times 10^{-1}$ |

★ Spectral energy bin width (cf. Eq. 6.6).
● Scale coefficient of the spectral resolution model (cf. Eq. 4.39).
○ Power coefficient of the spectral resolution model (cf. Eq. 4.39).
† Correlation between the log-transformed scale coefficient $\log a_1$ and the power coefficient $a_2$ (rounded to three significant digits).

**Table C.7** Spectral resolution model selection using the PRESS statistic for the four different detector channels #1 through #4.

| Model⋆ | PRESS [-]• | | | | |
|---|---|---|---|---|---|
| | Channel #1 | Channel #2 | Channel #3 | Channel #4 | Average |
| $a_1 x^{a_2}$ | $1.5 \times 10^{-3}$ | $2.3 \times 10^{-3}$ | $2.3 \times 10^{-3}$ | $1.9 \times 10^{-3}$ | $2.0 \times 10^{-3}$ |
| $a_1 + a_2\sqrt{x + a_3 x^2}$ | $1.9 \times 10^{-3}$ | $2.5 \times 10^{-3}$ | $3.6 \times 10^{-3}$ | $2.7 \times 10^{-3}$ | $2.7 \times 10^{-3}$ |
| $a_1 + a_2 x + a_3\sqrt{x}$ | $2.0 \times 10^{-3}$ | $3.2 \times 10^{-3}$ | $3.7 \times 10^{-3}$ | $2.7 \times 10^{-3}$ | $2.9 \times 10^{-3}$ |
| $a_1 x + a_2\sqrt{x}$ | $3.8 \times 10^{-3}$ | $3.0 \times 10^{-3}$ | $2.8 \times 10^{-3}$ | $2.8 \times 10^{-3}$ | $3.1 \times 10^{-3}$ |
| $a_1 + a_2 x^{a_3}$ | $2.3 \times 10^{-3}$ | $3.8 \times 10^{-3}$ | $4.2 \times 10^{-3}$ | $2.9 \times 10^{-3}$ | $3.3 \times 10^{-3}$ |
| $\sqrt{a_1 + a_2 x + a_3 x^2}$ | $3.3 \times 10^{-3}$ | $2.3 \times 10^{-3}$ | $4.6 \times 10^{-3}$ | $3.2 \times 10^{-3}$ | $3.3 \times 10^{-3}$ |
| $a_1 + a_2 x + a_3 x^2$ | $5.2 \times 10^{-2}$ | $4.5 \times 10^{-2}$ | $1.9 \times 10^{-2}$ | $1.3 \times 10^{-2}$ | $3.2 \times 10^{-2}$ |
| $\sqrt{a_1 + a_2 x}$ | $3.6 \times 10^{-2}$ | $2.4 \times 10^{-2}$ | $3.2 \times 10^{-2}$ | $3.7 \times 10^{-2}$ | $3.2 \times 10^{-2}$ |
| $a_1\sqrt{x}$ | $5.8 \times 10^{-2}$ | $4.3 \times 10^{-2}$ | $4.6 \times 10^{-2}$ | $4.0 \times 10^{-2}$ | $4.7 \times 10^{-2}$ |
| $a_1 + a_2 x$ | $7.7 \times 10^{-2}$ | $5.2 \times 10^{-2}$ | $4.1 \times 10^{-2}$ | $3.4 \times 10^{-2}$ | $5.1 \times 10^{-2}$ |
| $a_1 x + a_2 x^2$ | $2.1 \times 10^{-1}$ | $1.4 \times 10^{-1}$ | $9.2 \times 10^{-2}$ | $6.1 \times 10^{-2}$ | $1.3 \times 10^{-1}$ |

⋆ Empirical spectral resolution model functions with the input variable $x$ (continuous pulse-height channel number $\tilde{n}$ or alternatively the spectral energy $E'$), response variable $R_E$ and model parameter set $\{a_i \subseteq \mathbb{R} \mid i \in \mathbb{N}_+\}$.

• Predicted residual error sum of squares (PRESS) statistic [716, 717] normalized by the empirical response variable variance var $(R_E)$.

**Table C.8** This table summarizes the marginal distributions and associated univariate orthogonal polynomials for the PCE surrogate model adopted in Chapter 7. All parameters are assumed to be statistically independent. In addition, I list the consulted studies, which motivated the range of the adopted marginals.

| Variable⋆ | Marginal• | Marginal parameters | | $\psi_\alpha$∘ | Unit | References |
|---|---|---|---|---|---|---|
| $\mathrm{d}E/\mathrm{d}x\,\vert_{\mathrm{Birks}}$ | $\mathcal{U}$ | $x_l = 1.5 \times 10^2$ | $x_u = 5.5 \times 10^2$ | Legendre | $\mathrm{MeV\,cm^{-1}}$ | [313, 324, 325] |
| $\mathrm{d}E/\mathrm{d}x\,\vert_{\mathrm{trap}}$ | $\mathcal{U}$ | $x_l = 1.0 \times 10^1$ | $x_u = 1.5 \times 10^1$ | Legendre | $\mathrm{MeV\,cm^{-1}}$ | [325] |
| $\eta_{\mathrm{e/h}}$ | $\mathcal{U}$ | $x_l = 4.5 \times 10^{-1}$ | $x_u = 6.5 \times 10^{-1}$ | Legendre | | [313, 324, 325] |

⋆ NPSM parameters, i.e. the Birks stopping parameter $\mathrm{d}E/\mathrm{d}x\,|_{\mathrm{Birks}}$, the electron-hole pair fraction $\eta_{\mathrm{e/h}}$ and the trapping stopping parameter $\mathrm{d}E/\mathrm{d}x\,|_{\mathrm{trap}}$ (cf. Section 7.2.4).

• With $\mathcal{U}$, I denote the continuous uniform distribution $\mathcal{U}(x_l, x_u)$ with the lower and upper boundary parameters $x_l$ and $x_u$.

∘ Here, I list the univariate orthogonal polynomial families associated with the selected marginals. This are the Legendre polynomials $\mathcal{P}_\alpha(x)$ of degree $\alpha$ [748].

**Table C.9** This table summarizes the adopted prior distributions for non-proportional scintillation model (NPSM) inference. In addition, I list the consulted studies, which motivated some of the priors.

| Channel★ | Variable● | Prior○ | Prior parameters† | | Truncation | Unit | References |
|---|---|---|---|---|---|---|---|
| #SUM | $\mathrm{d}E/\mathrm{d}x\,\|_{\mathrm{Birks}}$ | $\mathcal{U}$ | $x_l = 1.5 \times 10^2$ | $x_u = 4.5 \times 10^2$ | | $\mathrm{MeV\,cm^{-1}}$ | [313, 324, 325] |
| | $\mathrm{d}E/\mathrm{d}x\,\|_{\mathrm{trap}}$ | $\mathcal{U}$ | $x_l = 1.0 \times 10^1$ | $x_u = 1.5 \times 10^1$ | | $\mathrm{MeV\,cm^{-1}}$ | [325] |
| | $\eta_{\mathrm{e/h}}$ | $\mathcal{U}$ | $x_l = 4.5 \times 10^{-1}$ | $x_u = 6.5 \times 10^{-1}$ | | | [313, 324, 325] |
| | $\sigma_\varepsilon^2$ | $\mathcal{U}$ | $x_l = 0$ | $x_u = \sigma^2_{\varepsilon,\mathrm{max}}$ | | $\mathrm{s^{-2}\,Bq^{-2}}$ | |
| #1–#4 | $\mathrm{d}E/\mathrm{d}x\,\|_{\mathrm{Birks}}$ | $\mathcal{N}$ | $\mu = 3.23 \times 10^2$ | $\sigma = 2.28 \times 10^1$ | $[1.5, 4.5] \times 10^2$ | $\mathrm{MeV\,cm^{-1}}$ | [313, 324, 325] |
| | $\mathrm{d}E/\mathrm{d}x\,\|_{\mathrm{trap}}$ | $\mathcal{N}$ | $\mu = 1.43 \times 10^1$ | $\sigma = 7.51 \times 10^{-1}$ | $[1.0, 1.8] \times 10^1$ | $\mathrm{MeV\,cm^{-1}}$ | [325] |
| | $\eta_{\mathrm{e/h}}$ | $\mathcal{N}$ | $\mu = 5.94 \times 10^{-1}$ | $\sigma = 6.83 \times 10^{-3}$ | $[4.5, 6.5] \times 10^{-1}$ | | [313, 324, 325] |
| | $\sigma_\varepsilon^2$ | $\mathcal{U}$ | $x_l = 0$ | $x_u = \sigma^2_{\varepsilon,\mathrm{max}}$ | | $\mathrm{s^{-2}\,Bq^{-2}}$ | |

★ Detector channel (cf. Section 6.2.1.2).

● NPSM and model discrepancy parameters, i.e. the Birks stopping parameter $\mathrm{d}E/\mathrm{d}x\,|_{\mathrm{Birks}}$, the electron-hole pair fraction $\eta_{\mathrm{e/h}}$, the trapping stopping parameter $\mathrm{d}E/\mathrm{d}x\,|_{\mathrm{trap}}$ and the discrepancy model variance $\sigma_\varepsilon^2$ (cf. Section 7.2.4).

○ With $\mathcal{U}$, I denote the continuous uniform distribution $\mathcal{U}(x_l, x_u)$ with the lower and upper boundary parameters $x_l$ and $x_u$. With $\mathcal{N}$ on the other hand, I refer to the truncated univariate normal distribution $\mathcal{N}(\mu, \sigma, x_l, x_u)$ with mean $\mu$, standard deviation $\sigma$ and truncation $x \in [x_l, x_u]$.

† I define the upper limit for the discrepancy model variance $\sigma_\varepsilon^2$ as $\langle \hat{c}^2_{\mathrm{exp}} \rangle$ with $\hat{c}_{\mathrm{exp}}$ being the measured spectral signature within the Compton edge domain $\mathcal{D}_{\mathrm{CE}}$ for the corresponding detector channel (cf. Section 7.2.4).

**Table C.10** This table includes posterior point and dispersion estimators for the Bayesian inverted non-proportional scintillation models (NPSMs). The listed estimators are the maximum a posteriori (MAP) probability estimate $\mathbf{x}_{\text{MAP}}$, the posterior mean $\mathbf{x}_{\text{Mean}}$ and the posterior median $\mathbf{x}_{\text{Median}}$ together with the 95 % credible interval and the posterior standard deviation $\boldsymbol{\sigma}_{\mathbf{x}}$ for the parameters $\mathbf{x} := (\,\mathrm{d}E/\mathrm{d}x\,|_{\text{Birks}},\ \mathrm{d}E/\mathrm{d}x\,|_{\text{trap}}, \eta_{\text{e/h}}, \sigma_{\varepsilon}^2)^\top$, i.e. the Birks stopping parameter $\mathrm{d}E/\mathrm{d}x\,|_{\text{Birks}}$, the trapping stopping parameter $\mathrm{d}E/\mathrm{d}x\,|_{\text{trap}}$, the electron-hole pair fraction $\eta_{\text{e/h}}$ and the discrepancy model variance $\sigma_{\varepsilon}^2$. All numerical values displayed are rounded to three significant digits.

| Channel★ | Variable | $\mathbf{x}_{\text{MAP}}$ | $\mathbf{x}_{\text{Mean}}$ | $\mathbf{x}_{\text{Median}}$ | 95 % credible interval● | $\boldsymbol{\sigma}_{\mathbf{x}}$ | Unit |
|---|---|---|---|---|---|---|---|
| #SUM | $\mathrm{d}E/\mathrm{d}x\,\vert_{\text{Birks}}$ | $3.22 \times 10^{2}$ | $3.23 \times 10^{2}$ | $3.22 \times 10^{2}$ | $[2.86, 3.60] \times 10^{2}$ | $2.28 \times 10^{1}$ | $\text{MeV cm}^{-1}$ |
| | $\mathrm{d}E/\mathrm{d}x\,\vert_{\text{trap}}$ | $1.46 \times 10^{1}$ | $1.43 \times 10^{1}$ | $1.44 \times 10^{1}$ | $[1.28, 1.48] \times 10^{1}$ | $7.51 \times 10^{-1}$ | $\text{MeV cm}^{-1}$ |
| | $\eta_{\text{e/h}}$ | $5.96 \times 10^{-1}$ | $5.94 \times 10^{-1}$ | $5.95 \times 10^{-1}$ | $[5.82, 6.04] \times 10^{-1}$ | $6.83 \times 10^{-3}$ | |
| | $\sigma_{\varepsilon}^2$ | $1.24 \times 10^{-1}$ | $1.37 \times 10^{-1}$ | $1.34 \times 10^{-1}$ | $[1.03, 1.80] \times 10^{-1}$ | $2.40 \times 10^{-2}$ | $\text{s}^{-2}\,\text{Bq}^{-2}$ |
| #1 | $\mathrm{d}E/\mathrm{d}x\,\vert_{\text{Birks}}$ | $3.17 \times 10^{2}$ | $3.17 \times 10^{2}$ | $3.16 \times 10^{2}$ | $[2.87, 3.49] \times 10^{2}$ | $1.57 \times 10^{1}$ | $\text{MeV cm}^{-1}$ |
| | $\mathrm{d}E/\mathrm{d}x\,\vert_{\text{trap}}$ | $1.33 \times 10^{1}$ | $1.32 \times 10^{1}$ | $1.32 \times 10^{1}$ | $[1.21, 1.43] \times 10^{1}$ | $5.48 \times 10^{-1}$ | $\text{MeV cm}^{-1}$ |
| | $\eta_{\text{e/h}}$ | $6.05 \times 10^{-1}$ | $6.05 \times 10^{-1}$ | $6.05 \times 10^{-1}$ | $[5.95, 6.14] \times 10^{-1}$ | $4.96 \times 10^{-3}$ | |
| | $\sigma_{\varepsilon}^2$ | $2.61 \times 10^{-2}$ | $2.83 \times 10^{-2}$ | $2.83 \times 10^{-2}$ | $[2.03, 3.94] \times 10^{-2}$ | $4.92 \times 10^{-3}$ | $\text{s}^{-2}\,\text{Bq}^{-2}$ |
| #2 | $\mathrm{d}E/\mathrm{d}x\,\vert_{\text{Birks}}$ | $4.15 \times 10^{2}$ | $4.15 \times 10^{2}$ | $4.15 \times 10^{2}$ | $[3.93, 4.36] \times 10^{2}$ | $1.08 \times 10^{1}$ | $\text{MeV cm}^{-1}$ |
| | $\mathrm{d}E/\mathrm{d}x\,\vert_{\text{trap}}$ | $1.41 \times 10^{1}$ | $1.41 \times 10^{1}$ | $1.41 \times 10^{1}$ | $[1.38, 1.43] \times 10^{1}$ | $1.08 \times 10^{-1}$ | $\text{MeV cm}^{-1}$ |
| | $\eta_{\text{e/h}}$ | $5.94 \times 10^{-1}$ | $5.94 \times 10^{-1}$ | $5.94 \times 10^{-1}$ | $[5.88, 5.99] \times 10^{-1}$ | $2.73 \times 10^{-3}$ | |
| | $\sigma_{\varepsilon}^2$ | $6.92 \times 10^{-3}$ | $7.64 \times 10^{-3}$ | $7.50 \times 10^{-3}$ | $[5.44, 10.7] \times 10^{-3}$ | $1.36 \times 10^{-3}$ | $\text{s}^{-2}\,\text{Bq}^{-2}$ |

*Continued on next page*

**Table C.10** *Continued from previous page*

| Channel★ | Variable | $x_{MAP}$ | $x_{Mean}$ | $x_{Median}$ | 95 % credible interval● | $\sigma_x$ | Unit |
|---|---|---|---|---|---|---|---|
| #3 | $\mathrm{d}E/\mathrm{d}x\,\vert_{\mathrm{Birks}}$ | $2.84 \times 10^{2}$ | $2.81 \times 10^{2}$ | $2.81 \times 10^{2}$ | $[2.47, 3.19] \times 10^{2}$ | $1.85 \times 10^{1}$ | $\mathrm{MeV\,cm^{-1}}$ |
| | $\mathrm{d}E/\mathrm{d}x\,\vert_{\mathrm{trap}}$ | $1.50 \times 10^{1}$ | $1.53 \times 10^{1}$ | $1.52 \times 10^{1}$ | $[1.47, 1.61] \times 10^{1}$ | $3.71 \times 10^{-1}$ | $\mathrm{MeV\,cm^{-1}}$ |
| | $\eta_{\mathrm{e/h}}$ | $5.75 \times 10^{-1}$ | $5.75 \times 10^{-1}$ | $5.76 \times 10^{-1}$ | $[5.65, 5.86] \times 10^{-1}$ | $5.47 \times 10^{-3}$ | |
| | $\sigma_{\varepsilon}^{2}$ | $2.15 \times 10^{-2}$ | $2.31 \times 10^{-2}$ | $2.30 \times 10^{-2}$ | $[1.65, 3.23] \times 10^{-2}$ | $4.08 \times 10^{-3}$ | $\mathrm{s^{-2}\,Bq^{-2}}$ |
| #4 | $\mathrm{d}E/\mathrm{d}x\,\vert_{\mathrm{Birks}}$ | $2.54 \times 10^{2}$ | $2.56 \times 10^{2}$ | $2.56 \times 10^{2}$ | $[2.23, 2.88] \times 10^{2}$ | $1.63 \times 10^{1}$ | $\mathrm{MeV\,cm^{-1}}$ |
| | $\mathrm{d}E/\mathrm{d}x\,\vert_{\mathrm{trap}}$ | $1.37 \times 10^{1}$ | $1.37 \times 10^{1}$ | $1.37 \times 10^{1}$ | $[1.33, 1.40] \times 10^{1}$ | $1.96 \times 10^{-1}$ | $\mathrm{MeV\,cm^{-1}}$ |
| | $\eta_{\mathrm{e/h}}$ | $5.75 \times 10^{-1}$ | $5.74 \times 10^{-1}$ | $5.74 \times 10^{-1}$ | $[5.66, 5.82] \times 10^{-1}$ | $3.95 \times 10^{-3}$ | |
| | $\sigma_{\varepsilon}^{2}$ | $1.05 \times 10^{-2}$ | $1.15 \times 10^{-2}$ | $1.14 \times 10^{-2}$ | $[8.20, 16.0] \times 10^{-3}$ | $2.00 \times 10^{-3}$ | $\mathrm{s^{-2}\,Bq^{-2}}$ |

★ Detector channel (cf. Section 6.2.1.2).
● Central credible interval with a probability mass of 95 %.

**Table C.11** To investigate the sensitivity of the defined Compton edge domain $\mathcal{D}_{\mathrm{CE}}$ (cf. Section 7.2.4) on the Bayesian inversion results, I have altered the domain size by 2.5 % symmetrically with respect to the domain boundaries and performed the emulator training and Bayesian inversion computation on this new domain for the detector channel `#SUM`. This alteration corresponds to ~20 % of the observed Compton edge shift for the $^{60}_{27}\mathrm{Co}$ spectral signature. This table summarizes the posterior point and dispersion estimator results for these additional computations, i.e. the maximum a posteriori (MAP) probability estimate $\mathbf{x}_{\mathrm{MAP}}$, the posterior mean $\mathbf{x}_{\mathrm{Mean}}$ and the posterior median $\mathbf{x}_{\mathrm{Median}}$ together with the 95 % credible interval and the posterior standard deviation $\boldsymbol{\sigma}_{\mathbf{x}}$ for the parameters $\mathbf{x} := (\,\mathrm{d}E/\mathrm{d}x\,|_{\mathrm{Birks}}, \,\mathrm{d}E/\mathrm{d}x\,|_{\mathrm{trap}}, \eta_{\mathrm{e/h}}, \sigma_{\varepsilon}^2)^\top$, i.e. the Birks stopping parameter $\mathrm{d}E/\mathrm{d}x\,|_{\mathrm{Birks}}$, the trapping stopping parameter $\mathrm{d}E/\mathrm{d}x\,|_{\mathrm{trap}}$, the electron-hole pair fraction $\eta_{\mathrm{e/h}}$ and the discrepancy model variance $\sigma_{\varepsilon}^2$. All numerical values displayed are rounded to three significant digits.

| Variable | $\mathbf{x}_{\mathrm{MAP}}$ | $\mathbf{x}_{\mathrm{Mean}}$ | $\mathbf{x}_{\mathrm{Median}}$ | 95 % credible interval★ | $\boldsymbol{\sigma}_{\mathbf{x}}$ | Unit |
|---|---|---|---|---|---|---|
| **(a) 2.5 % decrease in $\mathcal{D}_{\mathrm{CE}}$** | | | | | | |
| $\mathrm{d}E/\mathrm{d}x\,\vert_{\mathrm{Birks}}$ | $3.08 \times 10^{2}$ | $3.10 \times 10^{2}$ | $3.08 \times 10^{2}$ | $[2.74, 3.58] \times 10^{2}$ | $2.14 \times 10^{1}$ | $\mathrm{MeV\,cm^{-1}}$ |
| $\mathrm{d}E/\mathrm{d}x\,\vert_{\mathrm{trap}}$ | $1.50 \times 10^{1}$ | $1.46 \times 10^{1}$ | $1.47 \times 10^{1}$ | $[1.27, 1.50] \times 10^{1}$ | $6.24 \times 10^{-1}$ | $\mathrm{MeV\,cm^{-1}}$ |
| $\eta_{\mathrm{e/h}}$ | $5.93 \times 10^{-1}$ | $5.92 \times 10^{-1}$ | $5.92 \times 10^{-1}$ | $[5.80, 6.03] \times 10^{-1}$ | $5.85 \times 10^{-3}$ | |
| $\sigma_{\varepsilon}^2$ | $1.05 \times 10^{-1}$ | $1.12 \times 10^{-1}$ | $1.18 \times 10^{-1}$ | $[8.55, 16.6] \times 10^{-2}$ | $2.08 \times 10^{-2}$ | $\mathrm{s^{-2}\,Bq^{-2}}$ |
| **(b) 2.5 % increase in $\mathcal{D}_{\mathrm{CE}}$** | | | | | | |
| $\mathrm{d}E/\mathrm{d}x\,\vert_{\mathrm{Birks}}$ | $3.34 \times 10^{2}$ | $3.30 \times 10^{2}$ | $3.31 \times 10^{2}$ | $[2.90, 3.70] \times 10^{2}$ | $2.46 \times 10^{1}$ | $\mathrm{MeV\,cm^{-1}}$ |
| $\mathrm{d}E/\mathrm{d}x\,\vert_{\mathrm{trap}}$ | $1.46 \times 10^{1}$ | $1.42 \times 10^{1}$ | $1.43 \times 10^{1}$ | $[1.22, 1.48] \times 10^{1}$ | $8.70 \times 10^{-1}$ | $\mathrm{MeV\,cm^{-1}}$ |
| $\eta_{\mathrm{e/h}}$ | $5.95 \times 10^{-1}$ | $5.94 \times 10^{-1}$ | $5.95 \times 10^{-1}$ | $[5.82, 6.04] \times 10^{-1}$ | $7.10 \times 10^{-3}$ | |
| $\sigma_{\varepsilon}^2$ | $1.42 \times 10^{-1}$ | $1.58 \times 10^{-1}$ | $1.54 \times 10^{-1}$ | $[1.19, 2.08] \times 10^{-1}$ | $2.75 \times 10^{-2}$ | $\mathrm{s^{-2}\,Bq^{-2}}$ |

★ Central credible interval with a probability mass of 95 %.

**Table C.12** In this table, I summarize all material properties associated with the inorganic scintillator NaI(Tl), which were used for the Compton edge shift analysis in Appendix A.10. Moreover, I list the references, which were consulted to retrieve the individual numerical values.

| Quantity | Symbol | Unit | Reference | |
|---|---|---|---|---|
| atomic number | $Z$ | 64 | | [318] |
| mass density | $\rho$ | 3.667 | $\mathrm{g\,cm^{-3}}$ | [249, 318] |
| mean excitation energy | $I_0$ | 452 | eV | [318] |
| molar mass | $M$ | 149.894 24 | $\mathrm{g\,mol^{-1}}$ | [249] |

**Table C.13** Details of the custom $K_{nat}$ radionuclide sources deployed during the Dübendorf validation campaign.

| Source No.★ | KCl mass [kg]● | Activity [kBq]○ |
|---|---|---|
| 1 | 4.06(2) | 66.6(10) |
| 2 | 4.10(2) | 67.3(10) |
| 3 | 3.96(2) | 65.0(10) |
| 4 | 3.82(2) | 62.7(9) |
| 5 | 3.86(2) | 63.3(10) |
| 6 | 4.02(2) | 66.0(10) |
| 7 | 4.00(2) | 65.6(10) |
| sum | 27.82(6) | 456.6(26) |

★ Custom source identifier.
● Mass of the granular KCl sealed in the corresponding sources, measured with a digital scale (model ICS435s-BC120/f/M).
○ Activity $\mathcal{A}$ of ${}^{40}_{19}\mathrm{K}$. I determined the activity $\mathcal{A}$ following the methodology outlined in Appendix A.1, using nuclear data of ${}^{40}_{19}\mathrm{K}$ provided in Table 2.1 combined with molar mass data from Prohaska et al. [77] and isotopic abundance data from Meija et al. [56]. Uncertainties were computed by the standard error propagation methodology for independent variables [287, 878, 881] (cf. also to Appendix A.8).

**Table C.14** Gross and background measurement live time of the field measurements performed during the Dübendorf validation campaign.

| Measurement⋆ | Gross live time [min]• | Background live time [min]◦ |
|---|---|---|
| 0C | 110 | 28 |
| 1E | 180 | 28 |
| 1W | 180 | 48 |
| 1N | 180 | 28 |
| 1S | 190 | 48 |
| 1NE | 240 | 65 |
| 1SE | 240 | 65 |
| 2E | 780 | 28 |
| 2N | 760 | 48 |
| 2S | 780 | 65 |

⋆ Measurement identifier (cf. Fig. 8.2).
• Gross measurement live time $t_{gr}$ for the corresponding measurement rounded to two significant digits (cf. Section 4.3.1).
◦ Background measurement live time $t_b$ for the corresponding measurement rounded to two significant digits (cf. Section 4.3.1).

**Table C.15** Spectral calibration results adopted in the field measurements presented in Chapter 8 and derived by the `RLLCal` pipeline.

| Detector channel | $\Delta E'$⋆ [keV] | $a_1$• [-] | $a_2$◦ [-] | corr($\log a_1, a_2$)† [-] |
|---|---|---|---|---|
| `#1` | 3.011(4) | $7.2(6) \times 10^{-1}$ | $6.01(16) \times 10^{-1}$ | $-9.89 \times 10^{-1}$ |
| `#2` | 3.0133(32) | $6.0(6) \times 10^{-1}$ | $6.29(17) \times 10^{-1}$ | $-9.91 \times 10^{-1}$ |
| `#3` | 3.0168(29) | $5.4(4) \times 10^{-1}$ | $6.68(14) \times 10^{-1}$ | $-9.90 \times 10^{-1}$ |
| `#4` | 3.013(4) | $5.7(6) \times 10^{-1}$ | $6.37(17) \times 10^{-1}$ | $-9.85 \times 10^{-1}$ |
| `#SUM` | 3.0141(23) | $5.9(5) \times 10^{-1}$ | $6.40(16) \times 10^{-1}$ | $-9.87 \times 10^{-1}$ |

⋆ Spectral energy bin width (cf. Eq. 6.6).
• Scale coefficient of the spectral resolution model (cf. Eq. 4.39).
◦ Power coefficient of the spectral resolution model (cf. Eq. 4.39).
† Correlation between the log-transformed scale coefficient $\log a_1$ and the power coefficient $a_2$ (rounded to three significant digits).

**Table C.16** Details of the hover flight measurements conducted during the ARM22 validation campaign. Provided uncertainties represent the statistical standard deviation for the measurement (coverage factor $k = 1$).

| Source★ | $h_{air}$• [m] | $-\alpha'$○ [°] | $\beta'$† [°] | $\gamma'$‡ [°] | $t_{gr}$§ [s] | $t_b$$ [s] |
|---|---|---|---|---|---|---|
| ${}^{133}_{56}Ba$ | 30.7(1) | 321(3) | 0.3(1) | −2.3(1) | 290 | 314 |
| ${}^{133}_{56}Ba$ | 61.7(1) | 314(3) | 0.6(1) | −2.1(1) | 296 | 285 |
| ${}^{133}_{56}Ba$ | 90.8(2) | 314(3) | 0.3(1) | −1.5(1) | 296 | 260 |
| ${}^{137}_{55}Cs$ | 30.9(1) | 144(3) | −0.6(1) | −1.7(1) | 242 | 263 |
| ${}^{137}_{55}Cs$ | 60.2(1) | 143(3) | −0.8(1) | −2.0(1) | 292 | 307 |
| ${}^{137}_{55}Cs$ | 91.6(2) | 142(3) | −0.6(1) | −2.9(1) | 232 | 298 |

★ Radionuclide source adopted in the corresponding gross measurement.
• Mean ground clearance of the gross measurement determined by the radar altimeter of the TH06 helicopter (cf. Section 5.3.2). Note that the data was corrected for the vertical displacement between the radar altimeter, located at the bottom of the tail, and the RLL spectrometer, mounted in the cargo bay (cf. Fig. 5.1).
○ Mean negative yaw angle relative to geographic north (cf. Fig. 8.1).
† Mean roll angle (cf. Fig. 8.1).
‡ Mean pitch angle (cf. Fig. 8.1).
§ Gross measurement live time (cf. Section 4.3.1).
$ Background measurement live time (cf. Section 4.3.1).

**Table C.17** Details of the ground measurements conducted during the ARM22 validation campaign.

| Fuel volume fraction★ [%] | Gross live time• [s] | Background live time○ [s] |
|---|---|---|
| 9.5 | 679 | 897 |
| 43.7 | 995 | 790 |
| 92.2 | 631 | 814 |

★ Fuel volume fraction of the TH06 helicopter (total capacity: 1.988 $m^3$, cf. Section 8.2).
• Gross measurement live time (cf. Section 4.3.1).
○ Background measurement live time (cf. Section 4.3.1).

**Table C.18** This table summarizes the laboratory-based quantitative gamma-ray spectrometry analysis results of a jet fuel A-1 sample (7.97(1) × $10^{-1}$ kg). The sample was extracted directly from the TH06 aircraft's fuel tanks, sealed into a standard laboratory bottle (borosilicate glass) and then stored for ~5 months to establish secular equilibrium between the short lived progeny radionuclides in the sample. The analysis was performed by the Radioanalytics Group at the Paul Scherrer Institute (PSI) using a HPGe detector (GEM P-type Coaxial by ORTEC/AMETEK Inc.) with a measurement live time of ~69 h. Data postprocessing was performed with the `InterWinner` code, version `8.0`. Uncertainties are reported as 1 standard deviation (SD) values.

| Radionuclide | Activity mass concentration [Bq kg$^{-1}$]⋆ |
|---|---|
| $^{40}_{19}$K• | <1.6 |
| $^{232}_{90}$Th∘ | 1.3(8) × $10^{-1}$ |
| $^{238}_{92}$U† | 3.2(7) × $10^{-1}$ |

⋆ Activity mass concentration $a_m$ (cf. Eq. 2.11a).

• No statistically significant activity of $^{40}_{19}$K was detected. Therefore, I report here the upper limit ($L_U$) equivalent activity mass concentration [296].

∘ Only the progeny $^{212}_{82}$Pb was detected. Therefore, I report its value here as an estimate of the activity mass concentration of "equivalent thorium" $^{232}_{90}$Th [87].

† Only the progeny $^{214}_{82}$Pb and $^{214}_{83}$Bi were detected. Assuming a secular equilibrium, I report here the arithmetic mean of the obtained results as an estimate of the activity mass concentration of "equivalent uranium" $^{238}_{92}$U [87]. The uncertainty was computed by applying the standard error propagation methodology for independent variables [287, 878, 881].

**Table C.19** This table summarizes the adopted marginal prior distributions for the FSBI of the datasets derived from hover flight radiation measurements performed during the ARM22 validation campaign using the Swiss AGRS system (cf. Table 10.1). The marginal prior distributions for the source strengths of the natural terrestrial radionuclides $K_{nat}$, $Th_{nat}$ and $U_{nat}$ were motivated by a study by Bennett [927].

| Variable★ | Prior● | Prior parameters | | Truncation | Unit |
|---|---|---|---|---|---|
| $\xi_{Cs\text{-}137}$ | $\mathcal{N}$ | $\mu = 10^9$ | $\sigma = 10^{10}$ | $[0, \infty]$ | Bq |
| $\xi_{Ba\text{-}133}$ | $\mathcal{N}$ | $\mu = 10^9$ | $\sigma = 10^{10}$ | $[0, \infty]$ | Bq |
| $\xi_{K\text{-}nat}$ | $\mathcal{N}$ | $\mu = 350$ | $\sigma = 600$ | $[0, \infty]$ | $Bq\,kg^{-1}$ |
| $\xi_{Th\text{-}nat}$ | $\mathcal{N}$ | $\mu = 25$ | $\sigma = 43$ | $[0, \infty]$ | $Bq\,kg^{-1}$ |
| $\xi_{U\text{-}nat}$ | $\mathcal{N}$ | $\mu = 25$ | $\sigma = 40$ | $[0, \infty]$ | $Bq\,kg^{-1}$ |
| $\Delta\xi_{Rn\text{-}222}$ | $\mathcal{N}$ | $\mu = 0$ | $\sigma = 1$ | $[-\infty, \infty]$ | $Bq\,m^{-3}$ |
| $\alpha_{NB}$ | $\mathcal{N}$ | $\mu = 0$ | $\sigma = 1$ | $[0, \infty]$ | |

★ FSBI model parameters, i.e. the source strengths of the sealed $^{137}_{55}Cs$ and $^{133}_{56}Ba$ point sources ($\xi_{Cs\text{-}137}$, $\xi_{Ba\text{-}133}$), the source strengths of the three natural terrestrial radionuclides $K_{nat}$, $Th_{nat}$ and $U_{nat}$ ($\xi_{K\text{-}nat}$, $\xi_{Th\text{-}nat}$, $\xi_{U\text{-}nat}$), the source strength of the radon source term $\Delta^{222}_{86}Rn$ ($\Delta\xi_{Rn\text{-}222}$) and the dispersion parameter of the gamma-Poisson mixture distribution ($\alpha_{NB}$) (cf. Section 10.3.2).

● With $\mathcal{N}$, I refer to the truncated univariate normal distribution $\mathcal{N}(\mu, \sigma, x_l, x_u)$ with mean $\mu$, standard deviation $\sigma$ and truncation $x \in [x_l, x_u]$, parametrized by the lower and upper boundary parameters $x_l$ and $x_u$.

**Table C.20** This table summarizes the adopted prior distributions for the FSBI of the AGRS flights conducted over Lake Thun and the North Sea.

| Variable★ | Prior● | Prior parameters | | Truncation | Unit |
|---|---|---|---|---|---|
| $\xi_{cos}$ | $\mathcal{N}$ | $\mu = 10^{-1}$ | $\sigma = 10^{-1}$ | $[0, \infty]$ | $cm^{-2}\,s^{-1}$ |
| $\xi_{Rn\text{-}222}$ | $\mathcal{N}$ | $\mu = 1$ | $\sigma = 1$ | $[0, \infty]$ | $Bq\,m^{-3}$ |
| $\xi_{K\text{-}nat}$ | $\mathcal{N}$ | $\mu = 10^4$ | $\sigma = 10^4$ | $[0, \infty]$ | $Bq\,kg^{-1}$ |
| $\xi_{Th\text{-}nat}$ | $\mathcal{N}$ | $\mu = 10^4$ | $\sigma = 10^4$ | $[0, \infty]$ | $Bq\,kg^{-1}$ |
| $\xi_{U\text{-}nat}$ | $\mathcal{N}$ | $\mu = 10^4$ | $\sigma = 10^4$ | $[0, \infty]$ | $Bq\,kg^{-1}$ |
| $\alpha_{NB}$ | $\mathcal{N}$ | $\mu = 0$ | $\sigma = 1$ | $[0, \infty]$ | |

★ FSBI model parameters, i.e. the dispersion parameter of the gamma-Poisson mixture distribution $\alpha_{NB}$ together with the source strengths of the intrinsic background ($\xi_{K\text{-}nat}$, $\xi_{Th\text{-}nat}$, $\xi_{U\text{-}nat}$), the cosmic background ($\xi_{cos}$) as well as the radon background ($\xi_{Rn\text{-}222}$).

● With $\mathcal{N}$, I refer to the truncated univariate normal distribution $\mathcal{N}(\mu, \sigma, x_l, x_u)$ with mean $\mu$, standard deviation $\sigma$ and truncation $x \in [x_l, x_u]$, parametrized by the lower and upper boundary parameters $x_l$ and $x_u$.

**Table C.21** This table includes posterior point and dispersion estimators for the FSBI of the datasets `Cs_I`, `Cs_II` and `Cs_III` derived from hover flight radiation measurements performed during the ARM22 validation campaign using the Swiss AGRS system (cf. Table 10.1). The listed estimators are the maximum a posteriori (MAP) probability estimate $\mathbf{x}_{\mathrm{MAP}}$, the posterior mean $\mathbf{x}_{\mathrm{Mean}}$ and the posterior median $\mathbf{x}_{\mathrm{Median}}$ together with the 95 % credible interval and the posterior standard deviation $\boldsymbol{\sigma}_{\mathbf{x}}$. All numerical values displayed are rounded to three significant digits.

| ID★ | Variable● | $\mathrm{x}_{\mathrm{MAP}}$ | $\mathrm{x}_{\mathrm{Mean}}$ | $\mathrm{x}_{\mathrm{Median}}$ | 95 % credible interval○ | $\sigma_{\mathrm{x}}$ | Unit |
|---|---|---|---|---|---|---|---|
| `Cs_I` | $\xi_{\mathrm{Cs\text{-}137}}$ | $9.07 \times 10^{9}$ | $9.05 \times 10^{9}$ | $9.05 \times 10^{9}$ | $[8.85, 9.23] \times 10^{9}$ | $9.65 \times 10^{7}$ | Bq |
| | $\xi_{\mathrm{K\text{-}nat}}$ | $2.65 \times 10^{2}$ | $2.81 \times 10^{2}$ | $2.80 \times 10^{2}$ | $[1.87, 3.79] \times 10^{2}$ | $4.90 \times 10^{1}$ | $\mathrm{Bq\,kg^{-1}}$ |
| | $\xi_{\mathrm{Th\text{-}nat}}$ | $2.03 \times 10^{1}$ | $2.19 \times 10^{1}$ | $2.18 \times 10^{1}$ | $[1.32, 3.13] \times 10^{1}$ | 4.63 | $\mathrm{Bq\,kg^{-1}}$ |
| | $\xi_{\mathrm{U\text{-}nat}}$ | $3.59 \times 10^{1}$ | $3.47 \times 10^{1}$ | $3.47 \times 10^{1}$ | $[1.79, 5.14] \times 10^{1}$ | 8.53 | $\mathrm{Bq\,kg^{-1}}$ |
| | $\Delta\xi_{\mathrm{Rn\text{-}222}}$ | $-1.56 \times 10^{-1}$ | $4.28 \times 10^{-2}$ | $4.16 \times 10^{-2}$ | $[-1.90, 1.97]$ | $9.94 \times 10^{-1}$ | $\mathrm{Bq\,m^{-3}}$ |
| | $\alpha_{\mathrm{NB}}$ | $4.02 \times 10^{-6}$ | $3.82 \times 10^{-4}$ | $2.78 \times 10^{-4}$ | $[1.02, 134] \times 10^{-5}$ | $3.60 \times 10^{-4}$ | |
| `Cs_II` | $\xi_{\mathrm{Cs\text{-}137}}$ | $9.08 \times 10^{9}$ | $9.08 \times 10^{9}$ | $9.08 \times 10^{9}$ | $[8.98, 9.17] \times 10^{9}$ | $4.87 \times 10^{7}$ | Bq |
| | $\xi_{\mathrm{K\text{-}nat}}$ | $2.53 \times 10^{2}$ | $2.56 \times 10^{2}$ | $2.56 \times 10^{2}$ | $[2.11, 3.01] \times 10^{2}$ | $2.28 \times 10^{1}$ | $\mathrm{Bq\,kg^{-1}}$ |
| | $\xi_{\mathrm{Th\text{-}nat}}$ | $2.33 \times 10^{1}$ | $2.33 \times 10^{1}$ | $2.33 \times 10^{1}$ | $[1.93, 2.78] \times 10^{1}$ | 2.11 | $\mathrm{Bq\,kg^{-1}}$ |
| | $\xi_{\mathrm{U\text{-}nat}}$ | $2.43 \times 10^{1}$ | $2.37 \times 10^{1}$ | $2.37 \times 10^{1}$ | $[1.51, 3.24] \times 10^{1}$ | 4.41 | $\mathrm{Bq\,kg^{-1}}$ |
| | $\Delta\xi_{\mathrm{Rn\text{-}222}}$ | $6.44 \times 10^{-2}$ | $1.36 \times 10^{-1}$ | $1.47 \times 10^{-1}$ | $[-1.79, 2.08]$ | $9.85 \times 10^{-1}$ | $\mathrm{Bq\,m^{-3}}$ |
| | $\alpha_{\mathrm{NB}}$ | $4.80 \times 10^{-4}$ | $5.35 \times 10^{-4}$ | $5.10 \times 10^{-4}$ | $[1.44, 10.6] \times 10^{-4}$ | $2.33 \times 10^{-4}$ | |

*Continued on next page*

**Table C.21** *Continued from previous page*

| ID★ | Variable● | $x_{MAP}$ | $x_{Mean}$ | $x_{Median}$ | 95 % credible interval○ | $\sigma_x$ | Unit |
|---|---|---|---|---|---|---|---|
| Cs_III | $\xi_{Cs\text{-}137}$ | $9.15 \times 10^{9}$ | $9.14 \times 10^{9}$ | $9.14 \times 10^{9}$ | $[9.03, 9.26] \times 10^{9}$ | $5.82 \times 10^{7}$ | Bq |
| | $\xi_{K\text{-}nat}$ | $2.44 \times 10^{2}$ | $2.45 \times 10^{2}$ | $2.45 \times 10^{2}$ | $[2.34, 2.55] \times 10^{2}$ | 5.50 | $Bq\,kg^{-1}$ |
| | $\xi_{Th\text{-}nat}$ | $2.39 \times 10^{1}$ | $2.39 \times 10^{1}$ | $2.39 \times 10^{1}$ | $[2.31, 2.47] \times 10^{1}$ | $4.13 \times 10^{-1}$ | $Bq\,kg^{-1}$ |
| | $\xi_{U\text{-}nat}$ | $2.80 \times 10^{1}$ | $2.83 \times 10^{1}$ | $2.83 \times 10^{1}$ | $[2.43, 3.21] \times 10^{1}$ | 2.00 | $Bq\,kg^{-1}$ |
| | $\Delta\xi_{Rn\text{-}222}$ | $-4.54 \times 10^{-1}$ | $-5.59 \times 10^{-1}$ | $-5.70 \times 10^{-1}$ | $[-2.45, 1.40]$ | $9.81 \times 10^{-1}$ | $Bq\,m^{-3}$ |
| | $\alpha_{NB}$ | $6.82 \times 10^{-3}$ | $6.90 \times 10^{-3}$ | $6.87 \times 10^{-3}$ | $[5.56, 8.45] \times 10^{-3}$ | $7.43 \times 10^{-4}$ | |

★ Measurement identifier (cf. Table 10.1).

● Model parameters **x** considered in the FSBI, i.e. the source strength of the sealed $^{137}_{55}Cs$ point source ($\xi_{Cs\text{-}137}$), the source strengths of the three natural terrestrial radionuclides $K_{nat}$, $Th_{nat}$ and $U_{nat}$ ($\xi_{K\text{-}nat}$, $\xi_{Th\text{-}nat}$, $\xi_{U\text{-}nat}$), the source strength of the radon source term $\Delta^{222}_{86}Rn$ ($\Delta\xi_{Rn\text{-}222}$) and the dispersion parameter of the gamma-Poisson mixture distribution ($\alpha_{NB}$) (cf. Section 10.3.2).

○ Central credible interval with a probability mass of 95 %.

**Table C.22** This table includes posterior point and dispersion estimators for the FSBI of the datasets `Ba_I`, `Ba_II` and `Ba_III` derived from hover flight radiation measurements performed during the ARM22 validation campaign using the Swiss AGRS system (cf. Table 10.1). The listed estimators are the maximum a posteriori (MAP) probability estimate $\mathbf{x}_{\mathrm{MAP}}$, the posterior mean $\mathbf{x}_{\mathrm{Mean}}$ and the posterior median $\mathbf{x}_{\mathrm{Median}}$ together with the 95 % credible interval and the posterior standard deviation $\boldsymbol{\sigma}_{\mathbf{x}}$. All numerical values displayed are rounded to three significant digits.

| ID★ | Variable● | $\mathbf{x}_{\mathrm{MAP}}$ | $\mathbf{x}_{\mathrm{Mean}}$ | $\mathbf{x}_{\mathrm{Median}}$ | 95 % credible interval○ | $\boldsymbol{\sigma}_{\mathbf{x}}$ | Unit |
|---|---|---|---|---|---|---|---|
| `Ba_I` | $\xi_{\text{Ba-133}}$ | $4.73 \times 10^{8}$ | $4.62 \times 10^{8}$ | $4.63 \times 10^{8}$ | $[3.68, 5.54] \times 10^{8}$ | $4.68 \times 10^{7}$ | Bq |
| | $\xi_{\text{K-nat}}$ | $2.07 \times 10^{2}$ | $1.99 \times 10^{2}$ | $1.98 \times 10^{2}$ | $[1.08, 2.94] \times 10^{2}$ | $4.75 \times 10^{1}$ | $\mathrm{Bq\,kg^{-1}}$ |
| | $\xi_{\text{Th-nat}}$ | $2.13 \times 10^{1}$ | $2.23 \times 10^{1}$ | $2.22 \times 10^{1}$ | $[1.43, 3.10] \times 10^{1}$ | 4.25 | $\mathrm{Bq\,kg^{-1}}$ |
| | $\xi_{\text{U-nat}}$ | $2.05 \times 10^{1}$ | $2.01 \times 10^{1}$ | $2.00 \times 10^{1}$ | [5.42, 35.6] | 7.70 | $\mathrm{Bq\,kg^{-1}}$ |
| | $\Delta\xi_{\text{Rn-222}}$ | $-3.72 \times 10^{-1}$ | $-7.61 \times 10^{-2}$ | $-6.90 \times 10^{-2}$ | [−2.04, 1.87] | $9.99 \times 10^{-1}$ | $\mathrm{Bq\,m^{-3}}$ |
| | $\alpha_{\mathrm{NB}}$ | $1.15 \times 10^{-4}$ | $1.93 \times 10^{-3}$ | $1.38 \times 10^{-3}$ | $[4.83, 681] \times 10^{-5}$ | $1.84 \times 10^{-3}$ | |
| `Ba_II` | $\xi_{\text{Ba-133}}$ | $4.66 \times 10^{8}$ | $4.62 \times 10^{8}$ | $4.63 \times 10^{8}$ | $[4.19, 5.04] \times 10^{8}$ | $2.17 \times 10^{7}$ | Bq |
| | $\xi_{\text{K-nat}}$ | $2.26 \times 10^{2}$ | $2.29 \times 10^{2}$ | $2.29 \times 10^{2}$ | $[1.90, 2.70] \times 10^{2}$ | $2.05 \times 10^{1}$ | $\mathrm{Bq\,kg^{-1}}$ |
| | $\xi_{\text{Th-nat}}$ | $2.53 \times 10^{1}$ | $2.55 \times 10^{1}$ | $2.55 \times 10^{1}$ | $[2.15, 2.97] \times 10^{1}$ | 2.13 | $\mathrm{Bq\,kg^{-1}}$ |
| | $\xi_{\text{U-nat}}$ | $1.76 \times 10^{1}$ | $1.67 \times 10^{1}$ | $1.68 \times 10^{1}$ | [8.89, 24.3] | 3.92 | $\mathrm{Bq\,kg^{-1}}$ |
| | $\Delta\xi_{\text{Rn-222}}$ | $-4.32 \times 10^{-1}$ | $-2.16 \times 10^{-1}$ | $-2.11 \times 10^{-1}$ | [−2.15, 1.71] | $9.89 \times 10^{-1}$ | $\mathrm{Bq\,m^{-3}}$ |
| | $\alpha_{\mathrm{NB}}$ | $1.19 \times 10^{-4}$ | $8.00 \times 10^{-4}$ | $6.57 \times 10^{-4}$ | $[3.53, 243] \times 10^{-5}$ | $6.35 \times 10^{-4}$ | |

*Continued on next page*

**Table C.22** *Continued from previous page*

| ID★ | Variable● | $x_{MAP}$ | $x_{Mean}$ | $x_{Median}$ | 95 % credible interval○ | $\sigma_x$ | Unit |
|---|---|---|---|---|---|---|---|
| `Ba_III` | $\xi_{Ba\text{-}133}$ | $4.41 \times 10^{8}$ | $4.40 \times 10^{8}$ | $4.40 \times 10^{8}$ | $[3.96, 4.85] \times 10^{8}$ | $2.28 \times 10^{7}$ | Bq |
| | $\xi_{K\text{-}nat}$ | $2.37 \times 10^{2}$ | $2.36 \times 10^{2}$ | $2.36 \times 10^{2}$ | $[2.21, 2.53] \times 10^{2}$ | 8.29 | $Bq\,kg^{-1}$ |
| | $\xi_{Th\text{-}nat}$ | $2.75 \times 10^{1}$ | $2.77 \times 10^{1}$ | $2.77 \times 10^{1}$ | $[2.66, 2.88] \times 10^{1}$ | $5.74 \times 10^{-1}$ | $Bq\,kg^{-1}$ |
| | $\xi_{U\text{-}nat}$ | $1.42 \times 10^{1}$ | $1.44 \times 10^{1}$ | $1.44 \times 10^{1}$ | $[1.03, 1.85] \times 10^{1}$ | 2.13 | $Bq\,kg^{-1}$ |
| | $\Delta\xi_{Rn\text{-}222}$ | $3.58 \times 10^{-4}$ | $-3.37 \times 10^{-2}$ | $-3.76 \times 10^{-2}$ | $[-2.95, 1.85]$ | $9.73 \times 10^{-1}$ | $Bq\,m^{-3}$ |
| | $\alpha_{NB}$ | $3.06 \times 10^{-2}$ | $3.13 \times 10^{-2}$ | $3.12 \times 10^{-2}$ | $[2.69, 3.61] \times 10^{-2}$ | $2.34 \times 10^{-3}$ | |

★ Measurement identifier (cf. Table 10.1).

● Model parameters **x** considered in the FSBI, i.e. the source strength of the sealed $^{133}_{56}Ba$ point source ($\xi_{Ba\text{-}133}$), the source strengths of the three natural terrestrial radionuclides $K_{nat}$, $Th_{nat}$ and $U_{nat}$ ($\xi_{K\text{-}nat}$, $\xi_{Th\text{-}nat}$, $\xi_{U\text{-}nat}$), the source strength of the radon source term $\Delta^{222}_{86}Rn$ ($\Delta\xi_{Rn\text{-}222}$) and the dispersion parameter of the gamma-Poisson mixture distribution ($\alpha_{NB}$) (cf. Section 10.3.2).

○ Central credible interval with a probability mass of 95 %.

# Back Matter

# Acknowledgements

The past four years of my PhD journey have been a time of significant growth and learning, both professionally and personally. As I reflect on this period, I am deeply grateful to the many individuals who have supported me throughout this experience. Their contributions have been invaluable, and this work would not have been possible without their support.

First and foremost, I would like to express my deepest gratitude to my supervisor Dr. Gernot Butterweck. His guidance, support and encouragement have been instrumental in shaping my research and academic growth. I am grateful for his unwavering belief in my abilities and his willingness to support me no matter the circumstances. His meticulous attention to detail, coupled with the freedom he provided to explore new ideas, was instrumental in the realization of this work. Gernot's mentorship has served as a constant source of inspiration, and his credo, "Probiers aus!" will continue to guide me in all my future endeavors. I'm honoured to have been his PhD student and I can't imagine having been supervised by anybody else.

I would also like to extend my heartfelt thanks to my academic advisor, Prof. Dr. Klaus Stefan Kirch, for his guidance and constructive feedback in our discussions. In his dedication to teaching and supervision, he has been a true role model for me.

I need to express my gratitude to Prof. Dr. Arturo Vargas Drechsler and Dr. Christopher Strobl for their valuable feedback and support as co-examiners of the dissertation which forms the basis for this book. Their insights have been crucial in refining and enhancing the final outcome of this work. My sincere thanks also go to Prof. Dr. Leonardo De Giorgi for chairing the doctoral examination.

Further, thanks are due to Dr. Alberto Stabilini and Dr. Federico Alejandro Geser, who added invaluable insights in finalizing this book. Their support was especially appreciated during Gernot's well-deserved vacation, when they stepped in to help review the document. Alberto was also a big supporter during the ARM22 validation campaign, where he always had my back.

I would like to acknowledge the Swiss Federal Nuclear Safety Inspectorate (ENSI) and the PSI Department of Radiation Safety and Security for funding this PhD project. I am especially grateful to the department head Dr. Sabine Mayer and the section head Dr. Eduardo Gardenali Yukihara at the PSI for their unwavering advocacy and commitment to my academic growth. They have been strong supporters throughout all the publication processes, consistently available for discussions and providing valuable advice.

A big thank you goes to the Expert Group for Aeroradiometrics (FAR). To begin with, without the FAR initializing the project in which this work is embedded, I would not have had the opportunity to work on this topic. But more importantly, without their support, all the field campaigns presented in this book would not have been possible. I am especially grateful to the president of the FAR, Dr. Benno Bucher, for his unwavering support and encouragement throughout the project. My thanks also go to Prof. em. Dr. Ing. Dr. h.c. Ladislaus Rybach for his invaluable insights and fruitful discussions during the FAR meetings and technical exchanges. He is truly a living legend in the field of AGRS, and I am grateful for the opportunity to have gotten to know him.

I would also like to extend my gratitude to my group members at the PSI for the many moments of laughter and friendship that made my time at PSI truly memorable. I appreciate Dr. Malgorzata Magdalena Kasprzak for her invaluable support with administrative matters, and Giorgio Ambrosini and Dr. Malgorzata Urszula Sliz for always recognizing when it was time for a break, especially when I was too absorbed in work to notice. Special thanks to Jari Rafael Piller for the cultural exchanges across the border between Eastern Switzerland and the rest of our beautiful country. I am also grateful to Sebastian Weiermann and Martin Steffen as well as the former group members Udo Strauch and Rouven Philipp for their technical support in the lab.

Naturally, I could also count on many other individuals from other groups at the PSI during the course of my PhD project. Special thanks go to the Linux guru Dr. Dominik Werthmüller for teaching me the basics of high-performance computing and for his invaluable support with the computer cluster. Everything I know about parallel computing and batch scripting I owe to Dominik. I am also grateful to Dr. Eike Hohmann for the fruitful discussions on data analysis methods, although he couldn't convince me to abandon the Bayesian way. Further thanks go to Dr. Sandra Tanja Baur, Dr. Fabian Matthias Köhler and Dr. Martin Heule from the Radioanalytics Group for their support in analyzing the aircraft fuel sample and enduring the olfactory experience that came with it. Further, thanks go to Jack-of-all-trades Kilian Meier, who seems to know every corner of the PSI and was able to organize everything from a screwdriver to a new height-adjustable office table. I would like to thank also Dr. Jeppe Brage Christensen, Dr. Lilly Bossin, Dr. Mads Lykke Jensen, Dr. Veronika Sabine Heber, Evelyn Dätwyler and Isabelle Schatzmann-Morath for the many good stories shared over lunch.

Throughout my PhD project, I had the privilege of collaborating with numerous third parties. I am particularly grateful for the productive collaboration with RUAG AG, specifically Roland Lörtscher, René Brechbühl and Renato von Ah, who organized technical tours and connected me with all the different experts at RUAG AG. Above all, I would like to express my deepest thanks to Fredy Hess. His invaluable assistance in navigating technical schematics and uncovering crucial information was instrumental in building the aircraft mass model. Fredy's expertise and dedication were pivotal in ensuring the model's accuracy. Additionally, I appreciate the support from Mirion Technologies Ltd., which provided the necessary information to construct the mass model of the gamma-ray spectrometer.

The success of the ARM22 validation campaign was greatly due to the exceptional contributions of many individuals. I am particularly grateful to the National Emergency Operations Centre (NEOC) members Cristina Poretti and Dr. Adrian Hess for their tireless efforts in co-organizing the campaign, as well as to the head of NEOC, Gerald Scharding, for his steadfast support. Cristina's infectious positivity — "Chunt scho guet!" — and her ability to keep the team motivated were vital in ensuring the smooth execution of the fieldwork. My thanks also go to Stéphane Maillard for his technical expertise and support during the campaign. I am especially appreciative of the Swiss Air Force pilots, Oberstlt. Jan "Schwiiz" Schweizer and Hptm. Tamer "Tungo" Tunç. Their exceptional flying skills and dedication in executing the challenging flight missions required for data collection were crucial to the success of the ARM22 validation campaign. Beyond their flying expertise, they also provided valuable support by organizing flight data records and obtaining fuel samples from the helicopter. I am also thankful to the rest of the crew: loadmasters Sasha "Mac" Mackintosh and Manuel "Manu" Wöhrle, and operators Oblt. Pascal Diefenbacher, Oblt. Robin Kämpf and Oblt. Alex Lauber. A special mention goes to Dr. Jean-Marc Vaucher for handling the placement of the radioactive sources in the field. The combined efforts of all these individuals were essential to making the ARM22 validation campaign a success.

Additional thanks go to the wonderful people from the FLUKA.CERN Collaboration who provided invaluable support in case of technical questions and guidance in using the `FLUKA` code. I am particularly grateful to Dr. Francesco Cerutti for his help in developing the `COMSCW` user routine presented in Chapter 7.

Four years of PhD life would have been truly impossible without a group of friends who were always there to make me laugh, offer support and share a drink. To my fellow PhD colleagues, Dr. Silvia Motta, Paul Louis Dutheil and Antonella Mele, to all my colleagues from the academic orchestra Zurich and of course to my chamber music friends Marina Zwimpfer, Dr. Iris Eggenschwiler and Dr. Laura Piveteau — whether it was with a slice of cake, a heartfelt chat or the many unforgettable moments in music we shared — your friendship made all the difference and provided much-needed respite during my PhD journey.

Last but not least, I want to extend my deepest thanks to my family — my parents, Jasmin and Gerold and my brother, Mario. Your unwavering support, encouragement and love have been my rock throughout this entire journey. Through every challenge and triumph, your belief in me kept me going. I am profoundly grateful for your presence in my life and for everything you've done to help me reach this point. Without your support, this achievement would not have been possible. Thank you from the bottom of my heart!

# List of Figures

# List of Tables

# Terminology

## General Notation

Please note the following general conventions used in this book:

- Scalar variables are denoted by italic lowercase or uppercase letters, e.g. $\mathit{\Omega}$.
- Multidimensional quantities (vectors, matrices, ...) are denoted by roman type bold lowercase or uppercase letters, e.g. $\mathbf{\Omega}$. There is one exception: random vectors are denoted by italic type instead of roman type bold uppercase letters, e.g. $\boldsymbol{\mathit{\Omega}}$.
- Quantities denoted as multi-letter symbols, words, abbreviations or acronyms are written in roman type, e.g. $\mathrm{CR_b}$.
- Physical constants are always denoted in roman type letters, e.g. $\mathrm{m_e}$.
- Angles without explicitly specified units are assumed to be measured in radians.
- If not otherwise noted, uncertainties are provided as 1 standard deviation (SD) values (coverage factor $k = 1$) in least significant figure notation, e.g. 12.3(1) corresponds to 12.3 with a standard deviation of 0.1.
- If not otherwise noted, rounding of uncertainty values is performed according to the rounding rules established by the Particle Data Group et al. [134].

- As there is no standard notation for equivalence and similarity relations, I use the following terminology:

| | |
|---|---|
| $\sim$ | approximate equality between physical quantities or assignement of a random variable to a probability distribution |
| $\approx$ | approximate equality between functions |
| $\propto$ | proportional to |
| $\underset{\sim}{\propto}$ | approximatly proportional to |
| $\mathcal{O}$ | order of magnitude |
| $\gtrsim$ | greater than or approximately equal to |
| $\lesssim$ | smaller than or approximately equal to |

- Similarly, there is also no standard notation for several fundamental algebraic operations. I use the following terminology:

| | |
|---|---|
| $\cdot$ | scalar/dot product |
| $\circ$ | Hadamard/element-wise power |
| $\odot$ | Hadamard/element-wise product |
| $\oslash$ | Hadamard/element-wise division |

- Regarding the set of numbers, I use the following notation:

| | |
|---|---|
| $\mathbb{N}$ | Set of natural numbers |
| $\mathbb{N}_+$ | Set of positive natural numbers $\mathbb{N} \setminus \{0\}$ |
| $\mathbb{R}$ | Set of real numbers |
| $\mathbb{R}_+$ | Set of non-negative real numbers $\mathbb{R}_+ = \{x \in \mathbb{R} \mid x \geq 0\}$ |

## Variables

| Symbol | Quantity | Unit |
|---|---|---|
| $\boldsymbol{\alpha}$ | multi-index vector | |
| $\alpha'$ | yaw | rad |
| $\alpha_{\mathrm{NB}}$ | dispersion parameter of the gamma-Poisson mixture distribution | |
| $\alpha_{\gamma}$ | ratio of the photon energy to the energy-equivalent electron rest mass | |
| $\beta$ | ratio of the particle speed to the speed of light in vacuum | |
| $\beta'$ | pitch | rad |
| $\beta_{\mathrm{e}^+}$ | ratio of the positron speed to the speed of light in vacuum | |
| $\beta_{\mathrm{e}^-}$ | ratio of the electron speed to the speed of light in vacuum | |
| $\boldsymbol{\beta}_{\mathcal{GP}}$ | Gaussian process trend function parameters | |
| $\beta_{\mathrm{sci}}$ | ratio between the energy needed to create an electron-hole pair and the bandgap energy | |
| $\gamma$ | Lorentz factor | |
| $\gamma'$ | roll | rad |
| $\delta$ | density effect parameter | |
| $\varepsilon_{\mathrm{PCA}}$ | relative PCA truncation error | |
| $\varepsilon$ | permittivity | F/m |
| $\epsilon_{\mathrm{det}}$ | detector sensitivity coefficient | $1/(\mathrm{s}\,[\xi])$ |
| $\varepsilon_{\mathrm{e}^+}$ | reduced positron energy | |
| $\varepsilon_{\mathrm{e}^-}$ | reduced electron energy | |
| $\eta_{\mathrm{Birks}}$ | Birks survival rate | |
| $\eta_{\mathrm{cap}}$ | energy transfer efficiency | |
| $\eta'_{\mathrm{cap}}$ | local energy transfer efficiency | |
| $\eta^{0}_{\mathrm{cap}}$ | first order non-radiative loss survival efficiency | |
| $\eta_{\mathrm{e/h}}$ | electron-hole pair fraction | |
| $\eta_{\mathrm{gen}}$ | conversion efficiency | |
| $\eta_{\mathrm{lum}}$ | luminescence quantum yield | |
| $\eta_{\mathrm{Ons}}$ | Onsager recombination rate | |

| Symbol | Quantity | Unit |
|---|---|---|
| $\eta_{\text{pe,col}}$ | photoelectron collection efficiency at the first dynode | |
| $\eta_{\text{ph,col}}$ | light collection efficiency | |
| $\eta_{\text{PMT}}$ | photomultiplier transfer efficiency | |
| $\eta_{\text{pe}}$ | photoelectric quantum efficiency | |
| $\eta_{\text{spec}}$ | degree of spectral match | |
| $\theta$ | polar angle | rad |
| $\theta'$ | polar angle in the detector frame | rad |
| $\theta_{\text{e}^+}$ | positron emission angle | rad |
| $\theta_{\text{e}^-}$ | electron emission angle | rad |
| $\theta_{\mathcal{GP}}$ | Gaussian process kernel scale | |
| $\theta_{\text{src}}$ | view angle of the source | rad |
| $\theta_{\gamma}$ | photon scattering angle | rad |
| $\lambda$ | decay constant | 1/s |
| $\Lambda$ | eigenvalue | |
| $\lambda_{\text{D}}$ | Debye length | m |
| $\lambda_{\text{NB}}$ | mean parameter of the gamma-Poisson mixture distribution | |
| $\lambda_{\text{Pois}}$ | mean parameter of the Poisson distribution | |
| $\lambda_{\text{sci}}$ | scintillator emission wavelength | m |
| $\lambda_{\text{sci,max}}$ | scintillator peak emission wavelength | m |
| $\lambda_{\gamma}$ | photon wavelength | m |
| $\mu$ | attenuation coefficient | 1/m |
| $\boldsymbol{\mu}_{\mathbf{Y}}$ | random response vector mean | $[y]$ |
| $\mu_{\text{air}}$ | total attenuation coefficient in air | 1/m |
| $\mu_{\mathcal{H}}$ | centroid of the full energy peak | |
| $\mu_{\text{LLD}}$ | mean lower-level discriminator | |
| $\mu_{\text{src}}$ | total attenuation coefficient in the source | 1/m |
| $\mu_{\text{tot}}$ | total attenuation coefficient | 1/m |
| $\nu_{\text{s}}$ | singular value | |
| $\xi$ | source strength | $[\xi]$ |
| $\boldsymbol{\xi}$ | source strength vector | $[\xi]$ |
| $\Xi$ | integrated source strength | $[\xi]$ s |
| $\rho$ | mass density | $\text{kg/m}^3$ |
| $\rho_{\text{atm}}$ | atmospheric mass density | $\text{kg/m}^3$ |
| $\varrho_{\text{JF}}$ | fuel volume fraction | |
| $\rho_{\text{src}}$ | source mass density | $\text{kg/m}^3$ |
| $\sigma$ | microscopic cross-section | $\text{m}^2$ |

| Symbol | Quantity | Unit |
|---|---|---|
| $\boldsymbol{\sigma}_{Y}$ | random response vector standard deviation | $[y]$ |
| $\sigma_{\mathcal{A}}$ | activity standard deviation | Bq |
| $\sigma_{\mathcal{A}_0}$ | reference activity standard deviation | Bq |
| $\boldsymbol{\sigma}_{c_\mathrm{b}}$ | background standard deviation count rate vector | 1/s |
| $\hat{\boldsymbol{\sigma}}_\mathrm{exp}$ | measured spectral signature standard deviation | $1/(\mathrm{s}\,[\xi])$ |
| $\hat{\sigma}_\mathrm{exp}$ | measured spectral signature element standard deviation | $1/(\mathrm{s}\,[\xi])$ |
| $\boldsymbol{\sigma}_{c_\mathrm{gr}}$ | gross standard deviation count rate vector | 1/s |
| $\sigma_\mathrm{com}$ | Compton scattering cross-section | $\mathrm{m}^2$ |
| $\boldsymbol{\sigma}_{C_\mathrm{b}}$ | background standard deviation count vector | |
| $\boldsymbol{\sigma}_{C_\mathrm{gr}}$ | gross standard deviation count vector | |
| $\hat{\boldsymbol{\sigma}}_\mathrm{exp,stat}$ | measured statistical spectral signature standard deviation | $1/(\mathrm{s}\,[\xi])$ |
| $\hat{\boldsymbol{\sigma}}_\mathrm{exp,sys}$ | measured systematic spectral signature standard deviation | $1/(\mathrm{s}\,[\xi])$ |
| $\sigma_E$ | spectral resolution standard deviation | |
| $\sigma_\mathrm{GDR}$ | giant dipole resonance cross-section | $\mathrm{m}^2$ |
| $\sigma^2_{\mathcal{GP}}$ | Gaussian process noise variance | |
| $\sigma_{\mathcal{H}}$ | full energy peak standard deviation | |
| $\sigma_\mathrm{intr}$ | spectral intrinsic resolution standard deviation | |
| $\sigma_\mathrm{LLD}$ | lower-level discriminator standard deviation | |
| $\boldsymbol{\sigma}_{c_\mathrm{net}}$ | net standard deviation count rate vector | 1/s |
| $\sigma_\mathrm{nPR}$ | degree of non-proportionality to photons | |
| $\sigma_\mathrm{pe}$ | photoelectric cross-section | $\mathrm{m}^2$ |
| $\sigma_\mathrm{pn}$ | photonuclear cross-section | $\mathrm{m}^2$ |
| $\sigma_\mathrm{pp}$ | electron-positron pair production cross-section | $\mathrm{m}^2$ |
| $\sigma_\mathrm{pp,e}$ | electron-positron pair production cross-section in the electric field of the electron | $\mathrm{m}^2$ |
| $\sigma_\mathrm{pp,n}$ | electron-positron pair production cross-section in the electric field of the nucleus | $\mathrm{m}^2$ |

| Symbol | Quantity | Unit |
|---|---|---|
| $\sigma_{\mathrm{QD}}$ | quasi-deuteron cross-section | $\mathrm{m}^2$ |
| $\sigma_{\mathrm{ray}}$ | Rayleigh scattering cross-section | $\mathrm{m}^2$ |
| $\hat{\boldsymbol{\sigma}}_{\mathrm{sim}}$ | simulated spectral signature standard deviation | $1/(\mathrm{s}\,[\xi])$ |
| $\hat{\sigma}_{\mathrm{sim}}$ | simulated spectral signature element standard deviation | $1/(\mathrm{s}\,[\xi])$ |
| $\hat{\boldsymbol{\sigma}}_{\mathrm{sim,stat}}$ | simulated statistical spectral signature standard deviation | $1/(\mathrm{s}\,[\xi])$ |
| $\hat{\boldsymbol{\sigma}}_{\mathrm{sim,sys}}$ | simulated systematic spectral signature standard deviation | $1/(\mathrm{s}\,[\xi])$ |
| $\sigma_S$ | surface density | $\mathrm{kg}/\mathrm{m}^2$ |
| $\sigma_{t_{1/2}}$ | half-life standard deviation | s |
| $\sigma_{\mathrm{tot}}$ | total interaction cross-section | $\mathrm{m}^2$ |
| $\Sigma$ | macroscopic cross-section | $1/\mathrm{m}$ |
| $\sigma_\varepsilon^2$ | discrepancy model variance | $[y]^2$ |
| $\tau$ | mean life | s |
| $\tau_{\mathrm{dead}}$ | dead time constant | s |
| $\tau_{\mathrm{sci,d}}$ | scintillator decay time constant | s |
| $\varphi$ | azimuthal angle | rad |
| $\boldsymbol{\phi}$ | eigenvector | |
| $\varphi'$ | azimuthal angle in the detector frame | rad |
| $\boldsymbol{\Phi}$ | matrix of eigenvectors | |
| $\boldsymbol{\Phi}'$ | matrix of retained eigenvectors | |
| $\phi_{\mathrm{e}^+}$ | positron flux | $1/(\mathrm{s}\,\mathrm{m}^2)$ |
| $\phi_{\mathrm{e}^-}$ | electron flux | $1/(\mathrm{s}\,\mathrm{m}^2)$ |
| $\phi_{\mathrm{n}}$ | neutron flux | $1/(\mathrm{s}\,\mathrm{m}^2)$ |
| $\phi_p$ | particle flux | $1/(\mathrm{s}\,\mathrm{m}^2)$ |
| $\phi_{\mathrm{p}}$ | proton flux | $1/(\mathrm{s}\,\mathrm{m}^2)$ |
| $\phi_{\mathrm{tot}}$ | cosmic-ray flux | $1/(\mathrm{s}\,\mathrm{m}^2)$ |
| $\phi_\pm$ | charged particle flux | $1/(\mathrm{s}\,\mathrm{m}^2)$ |
| $\phi_\alpha$ | alpha particle flux | $1/(\mathrm{s}\,\mathrm{m}^2)$ |
| $\phi_\gamma$ | photon flux | $1/(\mathrm{s}\,\mathrm{m}^2)$ |
| $\hat{\phi}_\gamma$ | double differential photon flux signature | $1/(\mathrm{s}\,\mathrm{m}^2\,\mathrm{eV}\,\mathrm{sr}\,[\xi])$ |
| $\boldsymbol{\phi}_\gamma$ | photon flux vector | $1/(\mathrm{s}\,\mathrm{m}^2)$ |
| $\phi_{\gamma,\mathrm{ref}}$ | reference photon flux | $1/(\mathrm{s}\,\mathrm{m}^2)$ |
| $\phi_{\mu^+}$ | antimuon flux | $1/(\mathrm{s}\,\mathrm{m}^2)$ |
| $\phi_{\mu^-}$ | muon flux | $1/(\mathrm{s}\,\mathrm{m}^2)$ |
| $\chi^2$ | chi-squared objective function | |

| **Symbol** | **Quantity** | **Unit** |
|---|---|---|
| $\boldsymbol{\Psi}$ | multivariate polynomial basis function vector | |
| $\psi_p$ | particle fluence | $1/\mathrm{m}^2$ |
| $\Psi_{\boldsymbol{\alpha}}$ | multivariate polynomial basis function | |
| $\psi_\alpha$ | univariate polynomial basis function | |
| $\psi_\gamma$ | photon fluence | $1/\mathrm{m}^2$ |
| $\boldsymbol{\Omega}$ | direction unit vector | |
| $\boldsymbol{\Omega}'$ | direction unit vector in the detector frame | |
| $\Omega$ | solid angle in the lab frame | sr |
| $\Omega'$ | solid angle in the detector frame | sr |
| $a$ | specific activity | Bq/kg |
| $a_1$ | scale coefficient | |
| $a_2$ | power coefficient | |
| $a_{\mathrm{crust}}$ | activity mass concentration in the Earth's crust | Bq/kg |
| $a_{\mathrm{m}}$ | activity mass concentration | Bq/kg |
| $a_{\mathrm{sea}}$ | activity volume concentration in the sea | Bq/L |
| $a_S$ | surface activity concentration | $\mathrm{Bq/m}^2$ |
| $a_{\mathrm{v}}$ | activity volume concentration | $\mathrm{Bq/m}^3$ |
| $a_{\mathrm{v}}^{\mathrm{eq}}$ | equilibrium activity concentration | $\mathrm{Bq/m}^3$ |
| $\mathbf{a}_{\boldsymbol{\alpha}}$ | expansion coefficient vector | $[y]$ |
| $A$ | mass number | |
| $\mathbf{A}$ | expansion coefficient matrix | |
| $\mathscr{A}$ | set of multi-indices | |
| $\mathcal{A}$ | activity | Bq |
| $\mathcal{A}_0$ | reference activity | Bq |
| $A_{\mathrm{src}}$ | source area | $\mathrm{m}^2$ |
| $b_{i,j}$ | branching ratio from $i$ to $j$ | |
| $\mathcal{B}$ | probability mass | |
| $c$ | count rate | 1/s |
| $\mathbf{c}$ | count rate vector | 1/s |
| $\hat{\mathbf{c}}$ | spectral signature | $1/(\mathrm{s}\,[\xi])$ |
| $\tilde{\mathbf{c}}$ | spectral window count rate vector | 1/s |
| $\check{\mathbf{c}}$ | net peak area count rate vector | 1/s |
| $\mathbf{c}_{\mathrm{b}}$ | background count rate vector | 1/s |
| $\tilde{\mathbf{c}}_{\mathrm{b}}$ | spectral window background count rate vector | 1/s |
| $\check{\mathbf{c}}_{\mathrm{b}}$ | net peak area background count rate vector | 1/s |

| Symbol | Quantity | Unit |
|---|---|---|
| $\hat{c}_{\mathrm{DRM}}$ | mean spectral signature element obtained by the DRM | $1/(\mathrm{s}\,[\xi])$ |
| $\hat{\mathbf{c}}_{\mathrm{exp}}$ | measured mean spectral signature | $1/(\mathrm{s}\,[\xi])$ |
| $\hat{c}_{\mathrm{exp}}$ | measured mean spectral signature element | $1/(\mathrm{s}\,[\xi])$ |
| $\mathbf{c}_{\mathrm{gr}}$ | gross count rate vector | $1/\mathrm{s}$ |
| $\mathbf{c}_{\mathrm{net}}$ | net count rate vector | $1/\mathrm{s}$ |
| $c_{\mathrm{sea}}$ | natural elemental abundance in the sea | $\mathrm{kg}/\mathrm{L}$ |
| $\hat{\mathbf{c}}_{\mathrm{sim}}$ | simulated mean spectral signature | $1/(\mathrm{s}\,[\xi])$ |
| $\hat{c}_{\mathrm{sim}}$ | simulated mean spectral signature element | $1/(\mathrm{s}\,[\xi])$ |
| $C$ | counts | |
| $\mathbf{C}$ | count vector | |
| $C'$ | constant of integration | |
| $\mathbf{C}^{\dagger}$ | prebinned count vector | |
| $C_{\mathrm{b}}$ | background counts | |
| $\mathbf{C}_{\mathrm{b}}$ | background count vector | |
| $\mathbf{C}_{\mathrm{gr}}$ | gross count vector | |
| $\mathrm{CR}_{\mathrm{b}}$ | concentration ratio for biota b in a terrestrial ecosystem | |
| $\mathrm{CV}$ | coefficient of variation | |
| $\mathbf{CV}$ | coefficient of variation vector | |
| $\mathcal{C}_{X}$ | credible region | $[x]$ |
| $\mathcal{C}_{Y}$ | predictive credible region | $[y]$ |
| $\boldsymbol{\mathfrak{d}}$ | experimental condition | $[\mathfrak{d}]$ |
| $d_{\mathrm{atm}}$ | atmospheric depth | $\mathrm{kg}/\mathrm{m}^2$ |
| $\mathrm{d}C/\mathrm{d}\mathcal{H}$ | differential pulse-height spectrum | $1/\mathrm{V}$ |
| $\mathrm{d}E/\mathrm{d}x\,\vert_{\mathrm{Birks}}$ | Birks stopping parameter | $\mathrm{eV}/\mathrm{m}$ |
| $\mathrm{d}E/\mathrm{d}x\,\vert_{\mathrm{Ons}}$ | Onsager stopping parameter | $\mathrm{eV}/\mathrm{m}$ |
| $\mathrm{d}E/\mathrm{d}x\,\vert_{\mathrm{trap}}$ | trapping stopping parameter | $\mathrm{eV}/\mathrm{m}$ |
| $d_{\mathrm{src}}$ | source depth | m |
| $\boldsymbol{\mathfrak{D}}$ | set of experimental conditions | $[\mathfrak{d}]$ |
| $\mathcal{D}_{\mathrm{CC}}$ | Compton continuum domain | |
| $\mathcal{D}_{\mathrm{CE}}$ | Compton edge domain | |
| $\mathcal{D}_{\mathrm{FEP}}$ | full energy peak domain | |
| $\mathcal{D}_{\mathrm{SDOI}}$ | spectral domain of interest | |
| $\mathcal{D}_{\mathrm{tot}}$ | full spectrum domain | |
| $\boldsymbol{E}$ | random discrepancy vector | $[y]$ |
| $E$ | energy | eV |
| $E'$ | spectral energy | eV |

| **Symbol** | **Quantity** | **Unit** |
|---|---|---|
| $E'_0$ | spectral energy offset | eV |
| $\Delta E'$ | spectral energy bin width | eV |
| $E'_{\mathrm{AP}}$ | spectral energy of the annihilation peak | eV |
| $E_{\mathrm{b}}$ | electron binding energy | eV |
| $E'_{\mathrm{BSP}}$ | spectral energy of the backscatter peak | eV |
| $E_{\mathrm{c}}$ | upper core band energy | eV |
| $E'_{\mathrm{CE}}$ | spectral energy of the Compton edge | eV |
| $E_{\mathrm{dep}}$ | deposited energy | eV |
| $\mathbf{E}_{\mathrm{dep}}$ | deposited energy vector | eV |
| $E'_{\mathrm{DEP}}$ | spectral energy of the double escape peak | eV |
| $E_{\mathrm{e^+}}$ | positron energy | eV |
| $E_{\mathrm{e^-}}$ | electron energy | eV |
| $E'_{\mathrm{FEP}}$ | spectral energy of the full energy peak | eV |
| $E_{\mathrm{gap}}$ | band gap energy | eV |
| $E_{\mathrm{k}}$ | kinetic energy | eV |
| $\mathcal{E}_{\mathrm{k}}^{\mathrm{CE}}$ | set of initial kinetic energies of the electrons released in a Compton edge event | |
| $E_{\mathrm{k,e^-,ref}}$ | reference electron kinetic energy | eV |
| $E_{\mathrm{k,e^+}}$ | positron kinetic energy | eV |
| $E_{\mathrm{k,e^-}}$ | electron kinetic energy | eV |
| $E_{\mathrm{k,e^-}}^{\mathrm{CE}}$ | Compton electron kinetic energy in a Compton edge event | eV |
| $E_{\mathrm{k,e^-}}^{\mathrm{COM}}$ | initial kinetic energy of Compton electron | eV |
| $E_{\mathrm{k,e^-}}^{\mathrm{PE}}$ | initial kinetic energy of photoelectron | eV |
| $E_{\mathrm{k,r}}$ | recoil electron kinetic energy | eV |
| $\mathcal{E}_{\mathrm{k}}^{\mathrm{FEP}}$ | set of initial kinetic energies of the electrons released in a full energy peak event | |
| $E'_{\mathrm{SEP}}$ | spectral energy of the single escape peak | eV |
| $E_{\mathrm{v}}$ | valence band energy | eV |
| $\Delta E_{\mathrm{CE}}$ | Compton edge shift | eV |
| $E_{\gamma}$ | photon energy | eV |
| $E_{\gamma}^{0}$ | initial photon energy | eV |
| $E_{\gamma,\mathrm{max}}$ | maximum gamma-ray energy | eV |
| $E_{\gamma,\mathrm{min}}$ | minimum gamma-ray energy | eV |
| $f$ | electromagnetic wave frequency | Hz |
| $F_{\mathrm{src}}$ | source geometry-matter factor | $[F_{\mathrm{src}}]$ |

| Symbol | Quantity | Unit |
|---|---|---|
| $G$ | Gaussian weight | |
| $\mathbf{G}$ | Gaussian weight matrix | |
| $G_{\mathrm{PMT}}$ | photomultiplier total gain | |
| $h$ | altitude | m |
| $h_{\mathrm{air}}$ | ground clearance | m |
| $h_{\mathrm{geo}}$ | geodetic height | m |
| $h_{\mathrm{ort}}$ | orthometric height | m |
| $\mathcal{H}$ | pulse-height | V |
| $\Delta\mathcal{H}$ | pulse-height channel width | V |
| $\mathcal{H}_{\mathrm{max}}$ | upper-level discriminator | V |
| $\mathcal{H}_{\mathrm{min}}$ | lower-level discriminator | V |
| $\mathbb{I}_n$ | $n \times n$ identity matrix | |
| $I_0$ | mean excitation energy | eV |
| $I_{\mathrm{sci}}$ | relative scintillation photon spectrum | |
| $I_\gamma$ | relative photon intensity | |
| $\mathbf{J}_{n,m}$ | all-ones matrix with dimension $n \times m$ | |
| $\mathbf{K}$ | covariance matrix | $[y]^2$ |
| $l$ | length | m |
| $\ell$ | mean free path | m |
| $\mathcal{M}$ | forward model | $[y]$ |
| $\boldsymbol{\mathcal{M}}$ | spectral signature matrix | $1/(\mathrm{s}\,[\xi])$ |
| $M$ | molar mass | kg/mol |
| $\hat{\mathcal{M}}$ | surrogate model | $[y]$ |
| $\tilde{\boldsymbol{\mathcal{M}}}$ | sensitivity matrix (spectral window) | $1/(\mathrm{s}\,[\xi])$ |
| $\check{\boldsymbol{\mathcal{M}}}$ | sensitivity matrix (FEP) | $1/(\mathrm{s}\,[\xi])$ |
| $n$ | pulse-height channel number | |
| $\tilde{n}$ | continuous pulse-height channel number | |
| $\tilde{\mathfrak{N}}_{\mathrm{dep}}$ | random continuous pulse-height channel number of energy deposition events | |
| $\tilde{n}^{\dagger}$ | refined pulse-height channel number | |
| $\tilde{n}_{\mathrm{dep}}$ | continuous pulse-height channel number of an energy deposition event | |
| $\tilde{n}^{*}_{\mathrm{dep}}$ | continuous raw pulse-height channel number of an energy deposition event | |
| $n_{\mathrm{ch}}$ | charge carrier density | $1/\mathrm{m}^3$ |
| $n_{\mathrm{A}}$ | charge center density | $1/\mathrm{m}^3$ |
| $n_{\mathrm{fresh}}$ | natural isotopic abundance in fresh water | kg/L |
| $n_{\mathrm{sea}}$ | natural isotopic abundance in the sea | kg/L |

| Symbol | Quantity | Unit |
|---|---|---|
| $\mathbf{n}_S$ | surface normal vector | m |
| $N_0$ | number of zero-events | |
| $N_{\mathrm{bg}}$ | number of background sources | |
| $N_{\mathrm{b}}$ | number of background pulse-height spectra | |
| $N_{\mathrm{ch}}$ | number of pulse-height channels | |
| $N_{\mathrm{ch}}^{\dagger}$ | number of refined pulse-height channels | |
| $N_{\mathrm{cl}}$ | number of cycles | |
| $N_{\mathrm{COM}}$ | number of Compton scatter events before absorption | eV |
| $N_{\mathfrak{d}}$ | number of experimental condition parameters | |
| $N_{\mathrm{dep}}$ | number of energy deposition events | |
| $N_{E_\gamma}$ | number of photon energy bins | |
| $N_{\mathrm{FEP}}$ | number of full energy peaks | |
| $N_{\mathrm{gr}}$ | number of gross pulse-height spectra | |
| $N_{\mathrm{int}}$ | number of interactions | |
| $N_{\mathrm{n}}$ | neutron number | |
| $N_p$ | number of particles | |
| $N_{\mathrm{pr}}$ | number of primaries | |
| $N_{\mathrm{sci}}$ | number of emitted scintillation photons | |
| $N_{\mathrm{src}}$ | number of sources | |
| $N_t$ | number of time bins | |
| $N_{\mathrm{w}}$ | number of spectral windows | |
| $N_{\hat{\mathcal{X}}}$ | number of MCMC posterior samples | |
| $N_y$ | number of observables | |
| $N_{\mathcal{Y}}$ | number of independent measurements | |
| $\dot{N}_{\mathrm{sci}}$ | scintillation photon intensity | 1/s |
| $N_{\mathbf{\Omega}'}$ | number of directions | |
| $p$ | pressure | Pa |
| $p_{\mathrm{atm}}$ | atmospheric pressure | Pa |
| $q_{\mathrm{cosm}}$ | cosmogenic radionuclide generation rate density | $1/(\mathrm{m}^2\,\mathrm{s})$ |
| $Q_{\mathrm{cosm}}$ | cosmogenic radionuclide generation rate | $1/(\mathrm{m}^3\,\mathrm{s})$ |
| $Q_{\mathrm{ext}}$ | external source strength | $1/(\mathrm{s\,m}^3\,\mathrm{sr\,eV})$ |
| $Q_{\mathrm{int}}$ | internal source strength | $1/(\mathrm{s\,m}^3\,\mathrm{sr\,eV})$ |
| $Q_{\mathrm{iso}}$ | external isotropic source strength | $1/(\mathrm{s\,m}^3\,\mathrm{eV})$ |
| $Q_{\mathrm{p}}$ | external isotropic point source strength | $1/(\mathrm{s\,eV})$ |

| Symbol | Quantity | Unit |
|---|---|---|
| $Q_{\mathrm{PMT}}$ | total charge released in a photomultiplier tube | C |
| $Q_{\mathrm{s}}$ | external isotropic surface source strength | $1/(\mathrm{s\,m^2\,eV})$ |
| $Q_{\mathrm{tot}}$ | total source strength | $1/(\mathrm{s\,m^3\,sr\,eV})$ |
| $Q_{\mathrm{v}}$ | external isotropic volume source strength | $1/(\mathrm{s\,m^3\,eV})$ |
| $\mathbf{r}$ | position vector | m |
| $r$ | radius | m |
| $r_{\mathrm{e/h}}$ | electron-hole separation length | m |
| $r_{\mathrm{Ons}}$ | Onsager radius | m |
| $r_{\mathrm{S}}$ | Spearman's rank correlation coefficient | |
| $\mathcal{R}$ | correlation matrix | |
| $R$ | detector response function | $\mathrm{m^2}$ |
| $\mathbf{R}$ | detector response matrix | $\mathrm{m^2}$ |
| $\mathfrak{R}$ | rigidity | V |
| $\mathscr{R}$ | rotation matrix | |
| $\hat{R}$ | potential scale reduction factor | |
| $\mathfrak{R}_{\mathrm{c}}$ | vertical cut-off rigidity | V |
| $R_E$ | spectral resolution | |
| $R_{E,\mathrm{elec}}$ | spectral resolution induced by electronics | |
| $R_{E,\mathrm{intr}}$ | intrinsic spectral resolution | |
| $R_{E,\mathrm{pmt}}$ | spectral resolution induced by photomultiplier gain variance | |
| $R_{E,\mathrm{stat}}$ | statistical contribution to the spectral resolution | |
| $R_{E,\mathrm{trans}}$ | spectral resolution induced by photon transfer efficiency variance | |
| RH | relative humidity | |
| $R_{\mathrm{src}}$ | source radius | m |
| $s$ | path length | m |
| $S$ | surface area | $\mathrm{m^2}$ |
| $S$ | Sobol' indices | |
| $\mathcal{S}_{\mathrm{e,col}}$ | collisional stopping power of electrons | eV/m |
| $S_{\mathrm{N}}$ | sunspot number | |
| $S^{\mathrm{T}}$ | total Sobol' indices | |
| $t$ | time | s |
| $t_0$ | reference time | s |
| $t_{1/2}$ | half-life | s |
| $t_{\mathrm{b}}$ | background measurement live time | s |

| Symbol | Quantity | Unit |
|---|---|---|
| $t_{\mathrm{gr}}$ | gross measurement live time | s |
| $t_{\mathrm{live}}$ | live time | s |
| $t_{\mathrm{meas}}$ | measurement time | s |
| $t_{\mathrm{s}}$ | sampling time | s |
| $T$ | temperature | K |
| $v$ | speed | m/s |
| $v_{\mathrm{g}}$ | ground speed | m/s |
| $V$ | volume | $\mathrm{m}^3$ |
| VOV | relative variance of the variance | |
| **VOV** | relative variance of the variance vector | |
| $V_{\mathrm{sci}}$ | scintillator total volume | $\mathrm{m}^3$ |
| $V_{\mathrm{src}}$ | source volume | $\mathrm{m}^3$ |
| $w$ | mass fraction | kg/kg |
| $w_{\mathrm{crust}}$ | natural elemental abundance in the Earth's crust | kg/kg |
| $W$ | solar modulation index | |
| $\mathbf{x}$ | model parameter vector | $[x]$ |
| $x_{\mathrm{iso}}$ | natural isotopic abundance | mol/mol |
| $\mathbf{x}_{\mathcal{M}}$ | forward model parameter vector | $[x]$ |
| $\mathbf{x}_{\mathrm{MAP}}$ | maximum a posteriori estimate | $[x]$ |
| $\mathbf{x}_{\mathrm{Mean}}$ | posterior mean | $[x]$ |
| $\mathbf{x}_{\mathrm{Median}}$ | posterior median | $[x]$ |
| $\boldsymbol{\sigma}_{\mathbf{x}}$ | posterior standard deviation | $[x]$ |
| $\mathbf{x}_{\mathrm{MLE}}$ | maximum likelihood estimate | $[x]$ |
| $\mathbf{x}_{\nu}$ | nuisance parameter vector | $[x]$ |
| $\mathbf{x}_{\varepsilon}$ | discrepancy parameter vector | $[x]$ |
| $\mathcal{X}$ | experimental design | $[x]$ |
| $\boldsymbol{\mathcal{X}}$ | experimental design matrix | $[x]$ |
| $\boldsymbol{X}$ | random model parameter vector | $[x]$ |
| $\hat{\mathcal{X}}$ | MCMC posterior sample set | $[x]$ |
| $\mathbf{y}$ | data vector | $[y]$ |
| $\boldsymbol{Y}$ | random response vector | $[y]$ |
| $\mathcal{Y}$ | dataset | $[y]$ |
| $\boldsymbol{\mathcal{Y}}$ | dataset matrix | $[y]$ |
| $\boldsymbol{\mathcal{Z}}$ | transformed dataset matrix | $[y]$ |
| $Y_{\mathrm{sci,a}}$ | absolute light yield | 1/eV |
| $Y_{\mathrm{sci,r}}$ | relative light yield | |
| $Z$ | atomic number | |
| $\boldsymbol{Z}$ | transformed vector | |

## Operators

| Symbol | Quantity |
|---|---|
| $\lvert\cdot\rvert$ | modulus |
| $\Gamma(\cdot)$ | gamma function |
| $\Gamma(\cdot,\cdot)$ | upper incomplete gamma function |
| $\Phi(x)$ | standard normal cumulative distribution function of random variable $X$ |
| $\delta_{i,j}$ | Kronecker delta |
| $\pi(\cdot)$ | probability density function |
| $\varphi(x)$ | standard normal probability density function of random variable $X$ |
| $\mathrm{card}(\cdot)$ | cardinality |
| $\mathrm{cond}(\cdot)$ | condition number of a matrix |
| $\mathrm{corr}(\cdot,\cdot)$ | correlation |
| $\mathcal{C}_{\mathcal{N}}(\cdot)$ | Gaussian copula |
| $\mathrm{Cov}(\cdot,\cdot)$ | covariance |
| $\det(\cdot)$ | determinant |
| $\mathrm{diag}(\cdot)$ | diagonal operator |
| $\mathbb{E}(\cdot)$ | expectation |
| $\mathscr{E}_n(\cdot)$ | exponential integral function of the $n$th order |
| $\mathrm{FWHM}(\cdot)$ | full width at half maximum |
| $\mathcal{GP}(\mu,\kappa)$ | Gaussian process of a function $f(x)$ with mean function $\mu(x)$ and covariance function $\kappa(x,x')$ |
| $\mathcal{L}(\cdot)$ | likelihood function |
| $\mathrm{med}(\cdot)$ | median |
| $\mathcal{N}(\mathbf{x} \mid \boldsymbol{\mu},\mathbf{K})$ | multivariate normal distribution of random vector $\boldsymbol{X}$ with mean vector $\boldsymbol{\mu}$ and covariance matrix $\mathbf{K}$ |
| $\mathcal{P}^+(S)$ | non-empty power set of a given set S |
| $\Pr(\cdot)$ | probability |
| $\mathscr{P}_\alpha(\cdot)$ | Legendre polynomial of degree $\alpha$ |
| $\mathrm{rank}(\cdot)$ | rank of a matrix |
| $\mathrm{span}(\cdot)$ | linear span |
| $\mathscr{U}(a,b)$ | uniform distribution with boundry parameters $a$ and $b$ |
| $\mathrm{var}(\cdot)$ | variance |

## Abbreviations

| Symbol | Full phrase |
| --- | --- |
| $\alpha$ | Radioactive Decay with Alpha Particle Emission |
| $\upalpha$ | Alpha Particle |
| $\beta^+$ | Radioactive Beta-Decay with Positron Emission |
| $2\beta^+$ | Radioactive Double Beta-Decay with Positron Emission |
| $\beta^-$ | Radioactive Beta-Decay with Electron Emission |
| $\beta^-\mathrm{n}$ | Radioactive Beta-Decay with Electron and Delayed Neutron Particle Emissions |
| $\beta^-\alpha$ | Radioactive Beta-Decay with Electron and Delayed Alpha Particle Emissions |
| $\gamma$ | Gamma-Ray Emission |
| $\upgamma$ | Gamma Ray |
| $\upmu^+$ | Antimuon |
| $\upmu^-$ | Muon |
| $\upmu^\pm$ | Muon and Antimuon |
| $\overline{\upnu}_\mathrm{e}$ | Electron Antineutrino |
| $\upnu_\mathrm{e}$ | Electron Neutrino |
| $\uppi^0$ | Neutral Pion |
| $\uppi^\pm$ | Charged Pion |
| $\mathrm{e}^+$ | Positron |
| $\mathrm{e}^-$ | Electron |
| $\mathrm{e}^-/\mathrm{h}$ pair | Electron-Hole Pair |
| $\mathrm{e}^\pm$ | Electron and Positron |
| $\mathrm{K}^\pm$ | Charged Kaon |
| $\mathrm{K}^0$ | Neutral Kaon |
| n | Neutron |
| p | Proton |
| ABC | Approximate Bayesian Computations |
| AČR | Armáda České Republiky (Army of the Czech Republic) |
| AGRS | Airborne Gamma-Ray Spectrometry |
| Ai | Adobe Illustrator |
| AIES | Affine Invariant Ensemble Sampler |
| AKU | Environmental Monitoring Working Group |
| AP | Annihilation Peak |
| APD | Avalanch Photodiode |
| APS | Air Pressure Sensor |
| AT | Austria |

| Symbol | Full phrase |
|---|---|
| AT1 | Austrian AGRS team affiliated with the GSA |
| ATS | Air Temperature Sensor |
| BATSE | Burst and Transient Source Experiment |
| BfS | Bundesamt für Strahlenschutz (Federal Office for Radiation Protection) |
| BGR | Bundesanstalt für Geowissenschaften und Rohstoffe (Federal Institute for Geosciences and Natural Resources) |
| BIMP | Bureau International des Poids et Mesures |
| BM | Bin-Mode |
| BREP | Boundary Representation |
| BSP | Backscatter Peak |
| CA | Canada |
| CA1 | Canadian AGRS team affiliated with the GSC |
| CAD | Computer-Aided Design |
| CAGRC | China Aero Geophysical Survey & Remote Sensing Center for Land and Resources |
| CAGS | Chinese Academy of Geological Sciences |
| CAR | Calibration Range |
| CC | Compton Continuum |
| CCT | Compton Coincidence Technique |
| CDF | Cumulative Distribution Function |
| CE | Compton Edge |
| CEA | Commissariat à L'énergie Atomique et Aux énergies Alternatives (French Alternative Energies and Atomic Energy Commission) |
| CFRP | Carbon Fiber-Reinforced Polymers |
| CG | Compton Gap |
| CGRO | Compton Gamma-Ray Observatory |
| CH | Switzerland |
| CH1 | Swiss AGRS Team |
| CLT | Central Limit Theorem |
| CMGB | China Metallurgical Geology Bureau |
| CN | People's Republic of China |
| CN1 | Chinese AGRS team affiliated with the CAGS |
| CN2 | Chinese AGRS team affiliated with the CAGRC and CMGB |
| COESA | U. S. Committee on Extension To the Standard Atmosphere |
| COM | Compton Scattering |
| CPf | Calibration Pad (fixed) |

| Symbol | Full phrase |
|---|---|
| CPm | Calibration Pad (mobile) |
| CR | Cosmic Rays |
| CSG | Combinatorial Solid Geometry |
| CT | Computed Tomography |
| CXP | Characteristic X-Ray Peak |
| CZ | Czech Republic |
| CZ1 | Czech AGRS team affiliated with the SÚRO and the AČR |
| DAM | Data Acquisition Mode |
| DCAM | Downlooking Camera |
| DE | Germany |
| DE1 | German AGRS team affiliated with the BfS |
| DE2 | German AGRS team affiliated with the BGR |
| DEA | Data Evaluation Approach |
| DEP | Double Escape Peak |
| DGPS | Differential Global Positioning System |
| DICOM | Digital Imaging and Communications in Medicine |
| DRF | Detector Response Function |
| DRM | Detector Response Model |
| DSA | Direktoratet for Strålevern og Atomsikkerhet (Norwegian Radiation and Nuclear Safety Authority) |
| EAS | Extensive Air Shower |
| EC | Electron Capture Decay |
| EM | Expectation-Maximization Algorithm |
| ENSI | Swiss Federal Nuclear Safety Inspectorate |
| EOR | Emergency Organization Radioactivity |
| EPDL97 | Evaluated Photon Data Library, 1997 Version |
| ESS | Effective Sample Size |
| ETHZ | Swiss Federal Institute of Technology Zurich |
| FAR | Expert Group for Aeroradiometrics |
| FEP | Full Energy Peak |
| FLP | Fuel Level Probe |
| FOCP | Federal Office for Civil Protection |
| FR | France |
| FR1 | French AGRS team affiliated with the IRSN |
| FR2 | French AGRS team affiliated with the CEA |
| FS | German-Swiss Association for Radiation Protection |
| FSA | Full Spectrum Analysis |
| FSBI | Full Spectrum Bayesian Inversion |
| FWHM | Full Width At Half Maximum |
| GDR | Giant Dipole Resonance |

| Symbol | Full phrase |
|---|---|
| GLS | Generalized Least-Squares |
| GMF | Geomagnetic Field |
| GMT | Geiger-Müller Tube |
| GNSSr | Global Navigation Satellite System Receiver |
| GP | Gaussian Process |
| GSA | GeoSphere Austria |
| GSC | Geological Survey of Canada |
| HCD | Heavy Cluster Decay |
| HMC | Hamiltonian Monte Carlo |
| HPGe | High-Purity Germanium Detector |
| HYG | Hygrometer |
| i.i.d. | Independent and Identically Distributed |
| IAEA | International Atomic Energy Agency |
| IC | Internal Conversion |
| ICAO | International Civil Aviation Organization |
| ICRP | International Commission on Radiological Protection |
| IGRF | International Geomagnetic Reference Field |
| IMU | Inertial Measurement Unit |
| INFN | Istituto Nazionale di Fisica Nucleare (National Institute for Nuclear Physics) |
| IRSN | Institut de Radioprotection et de Sûreté Nucléaire (Radioprotection and Nuclear Safety Institute) |
| ISA | International Standard Atmosphere |
| ISO | International Organization for Standardization |
| IT | Italy |
| IT1 | Italian AGRS team affiliated with the INFN |
| LA | Laser Altimeter |
| LBTE | Linear Boltzmann Transport Equation |
| LLD | Lower-Level Discriminator |
| LM | List-Mode |
| MAP | Maximum a Posteriori |
| MC | Monte Carlo Simulation |
| MCA | Multichannel Analyzer |
| MCMC | Markov Chain Monte Carlo |
| MEXT | Ministry of Education, Culture, Sports, Science and Technology |
| MH | Metropolis-Hastings |
| MLE | Maximum Likelihood Estimation |
| NASA | National Aeronautics and Space Administration |
| NBC-EOD | Nuclear, Biological, Chemical, Explosive Ordnance Disposal and Mine Action Centre of Competence |

| Symbol | Full phrase |
|---|---|
| NEA | Nuclear Energy Agency |
| NEOC | National Emergency Operations Centre |
| NGU | Norges Geologiske Undersøkelse (Geological Survey of Norway) |
| NIRCAM | Near Infrared Camera |
| NIST | National Institute of Standards and Technology |
| NMDB | Neutron Monitor Database |
| NNSA | National Nuclear Security Administration |
| NO | Norway |
| NO1 | Norwegian AGRS team affiliated with the NGU |
| NO2 | Norwegian AGRS team affiliated with the DSA |
| NPSM | Non-Proportional Scintillation Model |
| NPSMC | Non-Proportional Scintillation Monte Carlo |
| NSC | Federal Nuclear Safety Commission |
| ODE | Ordinary Differential Equation |
| OLS | Ordinary Least-Squares |
| OSL | Optically Stimulated Luminescence |
| PARMA | PHITS-Based Analytical Radiation Model in the Atmosphere |
| PCA | Principal Component Analysis |
| PCE | Polynomial Chaos Expansion |
| PDF | Probability Density Function |
| PE | Polyethylene |
| PE | Photoelectric Absorption |
| PF | Peak Fitting |
| PLA | Polylactide Polymer |
| PMT | Photomultiplier Tube |
| PRESS | Predicted Residual Error Sum of Squares |
| PRNG | Pseudorandom Number Generator |
| PS | Point Source |
| PSI | Paul Scherrer Institute |
| PSMC | Proportional Scintillation Monte Carlo |
| PTFE | Polytetrafluoroethylene |
| PVC | Polyvinyl Chloride |
| QD | Quasi-Deuteron |
| QED | Quantum Electrodynamics |
| RA | Radar Altimeter |
| RD | Relative Deviation |
| RLL | Radiometrie Land-Luft |
| RMC | Radiation Monitoring Center |

| Symbol | Full phrase |
|---|---|
| RREA | Relativistic Runaway Electron Avalanche |
| SD | Standard Deviation |
| SDOI | Spectral Domain of Interest |
| SE | Sweden |
| SE1 | Swedish AGRS team affiliated with the SGU |
| SEP | Single Escape Peak |
| SF | Spontaneous Fission Decay |
| SGPK | Swiss Geophysical Commission |
| SGU | Sveriges Geologiska Undersökning (Geological Survey of Sweden) |
| SiPM | Silicon Photomultiplier |
| SLE | Spectral Likelihood Expansions |
| SNR | Signal-To-Noise Ratio |
| STE | Self-Trapped Exciton |
| SÚRO | Státní Ústav Radiační Ochrany (National Radiation Protection Institute) |
| SVD | Singular Value Decomposition |
| TGF | Terrestrial Gamma-Ray Flash |
| TH06 | Aérospatiale AS332M1 Super Puma |
| TL | Thermoluminescence |
| TMC | Total Monte Carlo |
| TW | Taiwan |
| TW1 | Taiwan AGRS Team |
| UAV | Unmanned Aerial Vehicle |
| ULD | Upper-Level Discriminator |
| UMTS | Universal Mobile Telecommunications System |
| UNSCEAR | United Nations Scientific Committee on the Effects of Atomic Radiation |
| US | United States of America |
| US1 | US AGRS team affiliated with the NNSA |
| UV | Ultraviolet |
| VB | Variational Bayesian Methods |
| WDC-SILSO | World Data Center Sunspot Index and Long-Term Solar Observations |
| WLLS | Weighted Linear Least-Squares |
| WLS | Weighted Least-Squares |
| WM | Window Method |
| WNLLS | Weighted Non-Linear Least-Squares |
| WTPD | Wildlife Transfer Parameter Database |

## Constants

| Symbol | Quantity | Value | Unit | Reference |
|---|---|---|---|---|
| $\alpha$ | fine structure constant (= $\mathfrak{e}^2/(2\varepsilon_0 hc)$) | $1/1.370\,359\,990\,84(21) \times 10^2$ | | [134] |
| $\varepsilon_0$ | electric constant | $8.854\,187\,812\,8(13) \times 10^{-12}$ | $\mathrm{F\,m^{-1}}$ | [134] |
| $\pi$ | Archimedes' constant | $3.141\,592\,653\,589\,793\ldots$ | | [134] |
| $c$ | speed of light in vacuum | $2.997\,924\,58 \times 10^8$ | $\mathrm{m\,s^{-1}}$ | [38] |
| $\mathfrak{e}$ | elementary charge | $1.602\,176\,634 \times 10^{-19}$ | C | [38] |
| $e$ | Euler's number | $2.718\,281\,828\,459\,045\ldots$ | | [134] |
| $g$ | standard gravity | $9.806\,65$ | $\mathrm{m\,s^{-2}}$ | [134] |
| $h$ | Planck constant | $6.626\,070\,15 \times 10^{-34}$ | $\mathrm{J\,s}$ | [38] |
| $k_B$ | Boltzmann constant | $1.380\,649 \times 10^{-23}$ | $\mathrm{J\,K^{-1}}$ | [38] |
| $K$ | stopping power coefficient (= $4\pi N_A r_e^2 m_e c^2$) | $3.070\,75 \times 10^{-1}$ | $\mathrm{MeV\,mol^{-1}\,cm^2}$ | [134] |
| $m_e$ | electron mass | $5.109\,989\,500\,0(15) \times 10^{-1}$ | $\mathrm{MeV\,c^{-2}}$ | [134] |
| $N_A$ | Avogadro constant | $6.022\,140\,76 \times 10^{23}$ | $\mathrm{mol^{-1}}$ | [38] |
| $r_e$ | classical electron radius (= $\mathfrak{e}^2/(4\pi\varepsilon_0 m_e c^2)$) | $2.817\,940\,322\,7(19)$ | fm | [134] |

# List of Publications

Here, I list only the publications that are directly relevant to the work presented in this book and have already been published. A complete publication list is available on the `ORCID` profile: https://orcid.org/0000-0003-0339-6592.

## Scientific Studies

## Scientific Reports

## Conference & Technical Meeting Talks

## Research Data & Software

# Index

## A

## B

## C

## D

## E

## F

## G

## H

## I

## J

## K

## L

## M

## N

## O

## P

## Z